# Abundances in the Local Region I:
# G and K Giants

R. Earle Luck
Department of Astronomy, Case Western Reserve University
10900 Euclid Avenue, Cleveland, OH 44106-7215
rel2@case.edu

## Abstract

Parameters and abundances for 1133 stars of spectral types F, G, and K of luminosity class III have been derived. In terms of stellar parameters, the primary point of interest is the disagreement between gravities derived with masses determined from isochrones, and gravities determined from an ionization balance. This is not a new result *per se*; but the size of this sample emphasizes the severity of the problem. A variety of arguments lead to the selection of the ionization balance gravity as the working value. The derived abundances indicate that the giants in the solar region have Sun-like total abundances and abundance ratios. Stellar evolution indicators have also been investigated with the Li abundances and the [C/Fe] and C/O ratios indicating that standard processing has been operating in these stars. The more salient result for stellar evolution is that the [C/Fe] data across the red-giant clump indicates the presence of mass dependent mixing in accord with standard stellar evolution predictions.

Keywords:   stars: fundamental parameters — stars: abundances — stars: evolution — Galaxy: abundances

## 1. Introduction

This paper and its predecessors (Heiter & Luck 2003, Luck & Heiter 2005, 2006, 2007) are parts of a program to examine the abundance properties of the local region in an effort to determine the standard of normalcy in terms of abundances. The intent on a very local scale — 15 parsecs — is to examine in detail the abundance distribution in dwarfs. On a larger scale — out to 100 pc — the objective is to sample the volume using as probes G/K giants. The primary goal is an increased understanding of the local region about the Sun. The metallicity data should reveal if there are any believable temporal, spatial, or stellar characteristic related variations in the metallicity. If one considers the local region as a typical volume, then these results can be applied to other locations and thus will increase our understanding of galactic evolution. The hope that these studies will be useful in understanding chemical evolution has been borne out by subsequent studies of open clusters (Reddy, Giridhar, & Lambert 2012, 2013) and an examination of carbon and oxygen abundances across the Hertzprung Gap (Adamczak & Lambert 2014) that used the results of Luck & Heiter (2007) – hereafter LH07 – as a comparison benchmark. In LH07, the examination of trends in carbon and oxygen data in the red-giant clump revealed a trend in the carbon data. The effect noted was that higher temperature stars in the clump showed lower carbon abundances. The precise cause of this effect could not be isolated at that time due to sample size. In this paper, a sample of G and K giants is analyzed to re-examine the local abundance patterns and the evolutionary status of G and K giants. A modest sample of F giants has been added to the study in an attempt to widen the temperature and evolutionary status range considered.

At the start of this program, the observing list for giants was set by the desire to sample the G/K giants of the local region out to about 100 pc from the Sun in all directions. The region was subdivided into cubes 25 pc on a side and from each sub-volume, appropriate stars were selected north of declination -30 degrees. This sample yielded the 286 G/K giants found in LH07. This dataset has been augmented for this study by the addition of numerous G/K giants so that the number in the 100-parsec volume is now 594 stars. Since the volume selection criteria used in LH07 formally extended out to 115 parsecs, a more precise comparison is that the current sample has 740 stars out to the older limit. Additional stars from the Bright Star Catalog (Hoffleit & Jaschek 1991) were added driving the sample out to about 200 parsecs. The spectral database has been supplemented using the ELODIE and ESO Archives. The ESO addition adds the southern sky, but with little control over the sampling in distance. Some of the southern stars are ostensibly at large distances, but the parallaxes are uncertain and imply absolute magnitudes that are at odds with their supposed spectral types. Since the number of such objects is small — about 25 stars, the incremental effort is minimal and they are retained in the analysis.

The bulk of the northern stars were observed using the McDonald Observatory Struve Telescope and Sandiford Cassegrain Echelle Spectrograph. For the ELODIE and ESO data archives the process used was to obtain a list of all stars available and then retrieve the spectral type for each from SIMBAD. Stars having a spectral type of F, G, or K III were then processed. The ESO data derives from the HARPS and UVES spectrographs. Basic observational data for the program stars can be found in Table 1 along with some derived quantities such as distance.

## 2. Observational material

The primary source of observational data for this study is a set of high signal-to-noise ratio spectra obtained during numerous observing runs between 1997 and 2010 at McDonald Observatory using the 2.1m Struve Telescope and the Sandiford Cassegrain Echelle Spectrograph (McCarthy et al. 1993). The spectra continuously cover a wavelength range from about 484 to 700 nm, with a resolving power of about 60000. Typical S/N values for the spectra are in excess of 150. To enable cancellation of telluric lines, broad-lined B stars were regularly observed with S/N exceeding that of the program stars. The 726 stars observed with the Sandiford spectrograph have an "S" in column 5 of Table 1.

A further 120 spectra were obtained from the ELODIE Archive (Moultaka et al. 2004). These echelle spectra are fully processed through order co-addition with a continuous wavelength span from about 400 to 680 nm and a resolution of 42000. Only spectra with S/N > 50 were utilized in this analysis. An "E" in Table 1, column 5, marks these stars.

The ESO Archive was used to obtain spectra from the ESO 3.6m telescope and HARPS spectrograph. The HARPS spectra cover a continuous wavelength range from about 400 to 680 nm with a native resolving power of 120000. To match the resolution of the Sandiford data and to increase the signal-to-noise of the data, these spectra were co-added to a resolution of 60000. Typical maximum S/N values (per pixel) for the spectra are in excess of 150. In Table 1, column 5, these stars are marked with an "H". Spectra from the UVES spectrograph and VLT/UT2 were also utilized. These spectra are rather heterogeneous having resolutions of 40000 to 80000 and non-continuous spectral coverages in the range 400 to 700 nm. A number of the spectra from UVES stop at about 625 nm meaning that [O I] 630 nm and Li I 670 nm were not observed. In Table 1, "U" denotes the stars observed with UVES spectrograph.

For both the ESO and ELODIE stars, broad-lined B stars from the archive were used for telluric line cancellation. The total number of spectra from all sources utilized comes to 1154 with 19 stars having two analyzed spectra.

IRAF[1] was used to perform standard CCD processing for the Sandiford dataset including scattered light subtraction and echelle order extraction. All spectra were extracted using a zero-order (i.e., the mean) normalization of the flat field that removes the blaze from the extracted spectra. A Windows based graphical package developed by R. Earle Luck was used to further process the spectra. This includes Beer's law removal of telluric lines, smoothing with a fast Fourier transform procedure, continuum normalization, and wavelength setting.

Continuum setting is done using a natural intelligence system that visually inspects the spectrum plotted against the FFT smoothed version. During this process composite spectra can be detected — either visually, or more effectively, from the power spectrum computed during the spectrum smoothing. A composite spectrum (specifically, a double-line binary) exhibits an anomalous power spectrum in the sense that there is a secondary power maximum separated from the primary maximum by a spatial frequency difference equal to the pixel separation of the two spectra. While there are binaries in this sample — the stars with CCDM (Catalogue of the components of double and multiple stars) designations in Table 1, there are no stars in this analysis with detectable composite spectra.

Equivalent widths from the spectra are measured using the Gaussian approximation. This process demands 1) that a distinct minimum exist in the proper resolution element of the spectrum and 2) that at least one side of the line profile be defined such that the half-width half-maximum depth point be attainable before a new minimum is found. The first criteria centers the resolution element at the wavelength of the line. These criteria eliminate the most egregious blends but the efficacy depends on the spectral resolution and the stellar rotation/macroturbulent velocity. For all species, equivalent width limits in the program stars of 0.5 pm (lower) to 20.0 pm (upper) were applied in the analysis.

## 3. Analysis

### 3.1 Line list and Analysis Resources

The line list used here is that used by Luck (2014) in a study of class I and II stars of types F, G, and K. This line list was created by merging the Cepheid line list of Kovtyukh & Andrievsky (1999) with the G/K giant line list of Luck & Heiter (2006). It was supplemented by lines from the unblended solar line lists of Rutten & van der Zalm (1984a,b) along with lines selected from numerous solar abundance analyses. The final line list has 2943 entries. Solar gf values were derived from equivalent widths measured by direct integration from the Delbouille & Neven (1973) solar intensity atlas.

---

[1] IRAF is distributed by the National Optical Astronomy Observatories, which are operated by the Association of Universities for Research in Astronomy, Inc., under cooperative agreement with the National Science Foundation.

To determine gf values from the solar line measures the solar abundances of Scott et al. (2015a,b) and Grevesse et al. (2015) were adopted. van der Waals damping coefficients were taken from Barklem, Piskunov, & O'Mara (2000) and Barklem & Aspelund-Johanssen (2005) or computed using the van der Waals approximation (Unsöld 1938). Hyperfine data for Mn, Co, and Cu were taken from Kurucz (1992). The solar atmosphere used was from the MARCS model code (Gustafsson et al. 2008). This model uses plane-parallel geometry with effective temperature 5777 K and log g = 4.44, and was used with a microturbulent velocity of 0.8 km s$^{-1}$ for gf determinations.

While the line list will certainly work well for solar-type dwarfs — it is tailored more to work with stars with temperatures in the FGK range. The inclusion of unblended solar lines is predicated upon the idea that an unblended solar line has a better than average chance of being unblended in a G/K giant. The line measurement process and the filtering process applied during abundance/parameter determination pares the initial list by eliminating obvious blends.

Abundances for all program stars were calculated using spherical MARCS model atmospheres (Gustafsson et al. 2008) for gravities below log g = 3.5 and plane-parallel models above that limit. Models were interpolated to the desired parameters using an interpolation code developed by R. Earle Luck. The code was tested by interpolating grid models and in all cases tested the interpolations matched the grid models in temperature to within 5 K and the remaining variables to within 1 to 2%. The line calculations were made using the LINES and MOOG codes (Sneden 1973) as maintained by R. Earle Luck since 1975.

### 3.2 Stellar Parameters and Abundances

Effective temperatures for the program stars were determined using the photometric calibration of Ramìrez & Melèndez (2005). This choice was predicated on the desire to use all photometry extant for these mostly bright stars. The photometry was obtained from the General Catalog of Photometric Data (Mermilliod, Hauck, & Mermilliod 1997), 2MASS photometry (Cutri et al. 2003) from SIMBAD, and Tycho photometry from the Tycho-2 catalog (Hog et al. 2000). Line of sight extinctions have been determined using the code of Hakkila et al. (1997) adopting distances computed from the Hipparcos parallaxes (van Leeuven 2007). The extinctions are principally determined from *(l,b,d)* versus *Av* relations. However, the extinction within 75 pc of Sun (the "Local Bubble") is essentially nil (e.g., Lallement et al. 2003; Breitschwerdt et al. 2000; Sfeir et al. 1999; Leroy 1999; Vergely et al. 1998). To correct for this all stars within 75 pc have had their extinctions (and reddenings) set to 0. For the remaining stars, the extinction out to 75 pc has been subtracted from the total extinction. The parallaxes; and hence the distances, to these giants are well determined. Of the 1087 stars with Hipparcos data, 1080 have parallaxes greater than the expected uncertainty. The median uncertainty in the parallax of these 1080 stars is 4%. It follows that the absolute V magnitudes are good to the ±0.1 mag level.

There are 57 stars from fourteen open clusters in this study that are identified in Table 1 in the column labelled Cluster. For the two northern clusters Mellote 25 (the Hyades) and Mellote 111 (the Coma Berenices cluster), all included stars have measured parallaxes. For the stars in the southern clusters, parallax data is often unavailable. If no parallax data is available for an individual star, the cluster distance given by Kharchenko et al. (2005) was adopted. These stars are easily identified in Table 1 by having a cluster identifier but no parallax data — see for example NGC 6705 MMU 1423.

All available colors were utilized for the temperature determination. The [Fe/H] values used in the Ramìrez & Melèndez (2005) calibration were taken from LH07, the PASTEL database (Soubrian et al. 2010), or assumed to be solar. The individual effective temperatures values were examined if the standard deviation of the mean exceeded 100K. Colors involving 2MASS magnitudes were examined more closely for the brightest stars due to possible saturation effects in the photometry. If there were an obvious outlier, it was eliminated from the final average. Examination of the derived temperatures revealed that if there were an outlier, it most often was $V - J$, where J is the 2MASS J magnitude and hence; it was decided that the better approach would be to eliminate that color in the effective temperature determination. In Table 2, one can find E(B-V) for each star, the adopted photometric temperature, its standard deviation, and the number of colors utilized. These temperatures are on the whole well determined: the mean standard deviation of the temperature for the stars below 6000 K is 54 K and the median is 41 K. The mean and median number of temperatures utilized is 10. However, there are cases that are not so well defined. The worst case is the K5 III star HR 3803 (N Vel). The standard deviation of the 10 colors is nearly 600 K. While the star is not especially variable according to the General Catalog of Variable Stars, it is possible that photometry does exhibit some anomalous behavior making temperature determinations difficult. The mean temperature for this star and the other nine stars with standard deviation in excess of 200 K are consistent with their spectral types, and so they remain in the dataset but caution that their parameters and abundances are not on the same level of reliability as the remainder.

Having photometric temperatures, the surface acceleration due to gravity, commonly called the gravity, is needed next. The gravity can be found using the absolute magnitude ($M_V$) or luminosity, effective temperature, and the mass. To derive the mass, isochrones were utilized in the same fashion as used in Allende-Prieto & Lambert (1999) and LH07. Isochrones were taken from Bertelli et al. (1994), Demarque et al. (2004), and Dotter et al. (2008) and used in conjunction with the photometric effective temperature, the absolute magnitude, and their expected uncertainties to obtain the mass, age, and luminosity for each star. Isochrone fitting is also dependent on the metallicity of the star in question. If the star is found in LH07 or in the PASTEL database (Soubrian et al. 2010), the metallicity given there was used, otherwise a solar metallicity was assumed. This is warranted as more than 90% of the sample has a metallicity within a factor of two of the Sun. The mass and age value from each isochrone set is given in Table 2. The luminosity given in Table 2 is derived from the distance, apparent V magnitude, and the bolometric corrections of Bessell, Castelli, & Plez (1998). These luminosities agree to within 0.02 dex of those derived from the isochrone fit.

The sensitivity of the mass and age on the adopted effective temperature and metallicity is well illustrated by π For (see Table 2) which is present in both the Sandiford and HARPS datasets. The temperatures used in the two cases is different by 4 K and the isochrone input metallicity is -0.5 in one case and -0.6 in the other. The Bertelli et al. isochrone results are 1.40 and 1.49 solar masses respectively. For the ages, the results are 2.5 and 3.2 Gyrs. The difference is due to the assumed metallicity. The fit process has to select an isochrone set based on the metallicity and in this case different metallicity isochrones were used — the first is a -0.4 dex set while the latter is -0.7. What is apparent is that the masses are better determined than the ages.

Inter-comparison of the mass and age data between the isochrone sets in Table 2 reveals no obvious systematic trends, but in some cases, there are significant differences in mass and age estimates between isochrone sets. For example, the mass estimates for 13 Lyn vary from 1.05 to 2.20 solar masses. In other cases, such as δ Eri, the total range is only 0.07 solar masses. The average range in the mass estimate is about 30% of the mean value. The mean mass and age from the Bertelli et al. isochrones are 1.88 solar masses and 2.25 Gyrs; from Dotter et al. 1.65 solar masses and 3.25 Gyr; and from Demarque et al., 1.83 solar masses, 2.28 Gyrs. It appears that the Dotter et al. masses are marginally lower, but the uncertainty in the estimates makes it impossible to favor one estimate over another. Therefore, the masses and ages have been averaged to form the adopted mass and age. The mean adopted mass is 1.83 solar masses and the mean adopted age is 2.6 Gyrs.

The analysis preferentially assumes the photometric temperature, and starts with the gravity determined from the mass determination. This parameter group is the mass-derived gravity set. In Figure 1 (top panel), the mass-derived gravity is plotted against the effective temperature. The distribution is much as expected. Though all of these stars are nominally of luminosity class III, it appears that a few are dwarfs. The stars without parallax data (but having photometry) — denoted in the panel as assumed mass or gravity — sometimes fall as hoped in the bulk of the distribution, other times they do not. A further culling of the sample will be discussed in §4.1.

The next analysis parameter needed is the microturbulent velocity. This is done by demanding that the iron abundance as determined from Fe I lines show no dependence on equivalent width. Initial examination of Fe I data is often accompanied by an editing of the abundance data through an interactive line-editing process. However, in this case individual examination of each star is too time consuming. For Fe I, we assume there is no dependence of abundance on excitation potential and determine the best-fitting microturbulent velocity. The individual Fe I line abundances are then binned and a Gaussian is fit to the Fe I histogram. Using a fit to the histogram deemphasizes the outliers and thus gives a better representation of the actual mean and dispersion. A symmetric 2σ clip is performed and the process repeated except that in the second iteration a 2.5σ clip is done so that only real outliers will be eliminated. All stars are treated in this way and the final clip line list is generated by tallying which lines are eliminated most often. The criterion for line elimination is that if a line is cut in 20% or more of the stars, then it is eliminated in all stars by the final clip process. The number of Fe I lines prior to clipping is of order 700 and after clipping, about 500. For Fe II there are insufficient numbers of lines to pursue such a statistical process. For Fe II, if there are more than 50 lines then the highest ten and the lowest 3 are eliminated, at more than 20 lines the cuts are seven high and two low. After the Fe II cuts are accumulated the overall cut list is assembled in the same fashion as for Fe I. The number of Fe II lines is typically 40 to 60 before trimming and 30 to 50 afterwards. After arriving at the final Fe I and Fe II data for each star, the microturbulent velocity was tweaked to a final value.

The Fe I data provide excitation information that can be interpreted in terms of the effective temperature. After final line trimming and setting of the microturbulent velocity the Fe I line data shows a mean $\Delta\Theta$ ($\Theta = 5040/T$ and $\Delta\Theta$ is the slope of the abundance versus excitation potential relation) of about -0.01 with an uncertainty of about the same value. At 4800 K, this translates into an increased effective temperature of about 50K to achieve excitation balance. If one limits the effective temperature range to 4000 to 5500K – the regime where the excitation analysis has highest sensitivity and accuracy – the mean temperature increase needed falls to about 25 K. These changes fall well within the uncertainties of the changes as well as the accuracy of the photometric effective temperatures; and so, to within the uncertainties, the excitation data is consistent with the photometric effective temperatures.. A caveat here is that the editing of the Fe I lines assumes that there are no intrinsic temperature trends within the data. Inspection of the deleted Fe I lines finds no preference for them to be of any particular excitation energy or equivalent width.

In examining the Fe I and Fe II data, an imbalance in total iron abundance is found in the general sense that the mass-derived gravity is too high; i.e., the Fe II derived total iron abundance exceeds the Fe I value (see Figure 1 – middle panel). To rectify this a new gravity was determined using an ionization balance. This involves forcing the neutral and ionized species of iron to give the same total abundance using the gravity as the free parameter. This was done by interpolating a small grid of models and then determining the best-fit gravity and microturbulence together. Parameter confirmation is done by interpolating a new model at the proper parameters. The iron data relations were recomputed to confirm the ionization balance along with the lack of dependence of iron abundance on line strength.

These stars exhibit a range of metallicities and this is taken into account as the analysis proceeds. Below [Fe/H] of -0.3, models with [M/H] = -0.5 are used, from [Fe/H] of -0.3 to +0.15 solar metallicity models are employed, and above [Fe/H] = +0.15 models with [M/H] = +0.25 are utilized. The preferred models are 2 $M_\odot$, α enhancement proportional to [Fe/H], moderate CN processing (that is, carbon diluted to [C/Fe] = -0.13, nitrogen enhanced to [N/Fe] = +0.31 with $^{12}C/^{13}C = 20$), and 2 km/s Doppler velocity. There is little effect on the abundances due to a change from 2 to 1 $M_\odot$; or from a change of 2 to 5 km/s, so if a preferred model is not available, a change in grid is made.

In Table 3, one can find parameters, iron abundance details, an average macroturbulent/rotational velocity, and lithium, carbon, and oxygen data. Information for both mass and ionization balance derived gravities is presented. There are a few stars which lack parallax and/or photometric data. For these stars, identifiable in Table 3 by the lack of a mass-derived analysis, a traditional excitation and ionization balance analysis was performed. Average abundances for 25 elements with Z > 10 are in Table 4. The averages have been computed in the same manner as used for Fe II except no master kill list was generated for species other than iron. If the neutral and first ionized species are both available for an element, the final average is merely the average of all retained lines. Note that the Mn, Co, and Cu abundances have been computed allowing for hyperfine structure. Details of the abundances (per species average, σ, and number of lines) are available upon request.

### 3.3 Li, C, N, and O Analysis

For lithium, carbon, nitrogen, and oxygen, spectrum syntheses for the features of interest have been performed utilizing laboratory oscillator strengths where available. For the lithium feature, all components of $^{7}$Li (using the data presented by Andersen, Gustafsson, & Lambert (1984)) in the 670.7 nm hyperfine doublet were used to match the observed profiles. There is no evidence in the observed spectra for the presence of $^{6}$Li and therefore, it was not considered in the syntheses. Lithium LTE abundance data are presented in Table 3 including the synthesized equivalent width of the Li blend. Also included are non-LTE lithium abundance corrections from the data of Lind et al. (2009).

Carbon abundances have been derived from C I lines at 505.2 nm and 538.0 nm, and the $C_2$ Swan system lines at 513.5 nm. For the atomic lines, the oscillator strengths of Biémont et al. (1993) or Hibbert et al. (1993) were adopted. These oscillator strengths have been used in determinations of the solar carbon abundance (Asplund et. al. 2005). For the Swan $C_2$ syntheses, f (0,0) = 0.0303 (Grevesse et al. 1991) was adopted with the relative band f values of Danylewych & Nicholls (1974), along with $D_0$ = 6.210 eV (Grevesse et al. 1991) and theoretical line wavelengths (as needed) from C. Amiot (1982, private communication). To form the carbon abundance on a per spectrum basis the individual features are combined as follows: for $T_{eff}$ < 4800K: $C_2$ 513.5 only. At T > 4800 K and less than 5500 K, C I 505.2 and 538.0 have weight 1 and $C_2$ 513.5 has weight 3. For T > 5500 K the two C I lines have equal weight and $C_2$ is not used. The weights are based on relative strength and blending. Typical spreads in abundance for the features are 0.15 dex. For the purpose of abundances with respect to solar values, we adopt log $\varepsilon_C$ = 8.43, the Asplund et al. (2009) recommended solar carbon abundance. Table 3 has the average C and O data on a per star basis.

Oxygen abundance indicators in the available spectral range are rather limited: the O I triplet at 615.6 nm and the [O I] lines at 630.0 and 636.3 nm. The O I 615.6 nm lines are problematic in abundance analyses with only the 615.8 nm line being retained in solar oxygen analyses (Asplund et al. 2004). The O I lines were synthesized using the NIST atomic parameters (Kramida et al. 2012) that were also used by Asplund et al. (2009). For the forbidden oxygen lines, only 630.0 nm is usable as 636.3 nm is weak, heavily blended, and complicated by the presence of the Ca I autoionization feature. In the syntheses of 630.0 nm, the line data presented by Allende Prieto, Lambert, & Asplund (2001) was used except that the experimental oscillator strength for the blending Ni I line (Johansson et al. 2003) was adopted. The syntheses assumed [Ni/Fe] = 0. In G and K giants, the [O I] 630.0 nm equivalent width typically runs from 3 to 8 nm and the Ni I intrusion about 0.5 nm. An uncertainty of 20% in the Ni I strength is therefore not a major concern. To form a final oxygen abundance the data was average in the following manner: for $T_{eff}$ < 5500 K [O I] only is used, while for $T_{eff}$ > 5500 K, O I has weight 1 and [O I] has weight 3. At an effective temperature of 6225 K [O I] is essentially undetectable and the abundance depends only on O I 615.8 nm. For $T_{eff}$ < 5500 K, the C-O dependence has been explicitly taken into account.

## 3.4 Abundance and Parameter Comparisons

### 3.4.1 Internal Comparison

This study affords two type of internal comparisons: there are 19 stars with two analyses, and for each star/spectra there is the mass-derived gravity abundance determination and the ionization balance gravity analysis. The 19 stars with two analyses shall be dealt with first.

Comparing the effective temperature data for the 19 stars, a mean difference of 4 K with a total range of 39 to -20 K is found. The reason why there can be a difference is that the determinations were made for each data source separately without reference any previous determination and thus there can be different colors eliminated in each determination. Note that a difference of 40 K in this temperature range is less than 1% in the temperature. The mass-derived gravities show a mean difference of 0 with a total range of -0.08 to + 0.05 dex. The abundance data associated with the mass-derived gravity is very consistent: the mean differences for iron (Fe I), carbon, and oxygen are -0.01, -0.01, and -0.02 respectively. The standard deviation about the mean is of order 0.1 for three species. For the ionization balance gravities, the mean difference in log $g$ is 0.08 dex ($\sigma$ = 0.12). The mean difference for the iron, carbon, and oxygen abundance between the two analyses is about 0.02 dex for each species, with the standard deviation once again being of order 0.1 dex.

As noted above, in looking at the total dataset there is a systematic difference in gravities derived from the mass and from the ionization balance. The iron abundance difference (Fe I – Fe II) in the mass-derived gravity analysis is dependent on the temperature: at 5000 K the difference is of order 0.2 dex increasing to about 0.5 dex at 4000 K – see Figure 1 (middle panel). In comparing the mass-derived and ionization balance gravities, this difference manifests itself as progressively larger differences between two as a function of effective temperature (bottom panel of Figure 1). Bruntt, Frandsen, & Tygensen (2011) noted this type of variation in Kepler field red giants using astroseismology-determined parameters. The effect is also present in the data of LH07. The dependence of the gravity difference on the effective temperature could indicate a failure in the model atmospheres. However, a brief sanity check using ATLAS9 models (Kurucz 1992) yields the same results to within 0.1 dex in log $g$ and 0.05 dex in total iron abundance. The mass difference implied by the change in gravity is certainly non-physical. A 0.2 dex lower gravity at constant temperature and luminosity yields a 40% lower mass while lowering the gravity by 0.6 dex implies a 75% lower mass. The average mass then would decrease to 0.5 to 1.0 solar masses depending on the temperature – a not especially appealing result.

In Figure 2 (top panel), smoothed (x I – x II) data is presented for Ti, V, Cr, and Fe. The immediate conclusion drawn from this figure is that all of these species suffer from the same problem. Whatever is affecting iron, it also affects the ionized data of other species and yields the same problem relative to the neutral species: the total abundance inferred from the first ionized species is too large. The ionization balance performed using Fe I and Fe II rectifies most of the problems with the difference between the Ti, V, and Cr neutral and ionized total abundances declining but more importantly, the dependence on effective temperature being ameliorated (see bottom panel of Figure 2).

A possible reason behind this problem is that the ionized lines, and specifically the Fe II lines, are affected by progressively stronger blending with decreasing temperature. While possible, this would demand that that all Fe II lines be similarity affected. A weak check is available on this possibility: metal-poor giants should not show the effects of line blending to the extent that solar or above-solar abundance giant do. This of course is not true if the features in the Fe II blend are nearly coincident in wavelength. The situation envisioned here is that lines within 0.015 nm of an Fe II line progressively make the centroid and FWHM of the Fe II line less secure as the overall line strength increases with decreasing temperature. Since metal-poor stars have intrinsically weaker lines: their lines should remain less "blended". In the middle panel of Figure 1, the solid red line is the smoothed differences in Fe I – Fe II computed from stars with [Fe/H] < -0.3. In the smoothing outliers have been excluded though they tend to exhibit large overabundances of Fe II. As can be seen, the metal poor-stars show differences that indicate Fe II is larger than Fe I in the bulk of sample and in the 4500K and below region the differences tail off much as does the entire sample. While not conclusive, this indicates that line-blending in Fe II is not the sole culprit in the problem.

Two other possibilities relative to the gravity are significant non-LTE effects in iron, or that lower temperature, lower gravity stellar atmospheric models do not reflect the actual structure of stars. The later possibility is difficult to assess but the non-LTE alternative has been examined in other studies.

The first study of non-LTE effects in iron in cooler stars is likely that of Tanaka (1971) and these studies have proceeded through the work of Mashonkina et al. (2011) to the review by Mashonkina (2013), and the studies of Ezzedine, Merle, & Plez (2013) and Lapenna et al. (2014). The primary results from the latter two papers relevant to this work are that non-LTE calculations may yield better agreement between Fe I and Fe II using parameters much like the mass-derived parameters used here, and that the non-LTE effects become stronger with decreasing temperature and metallicity. The non-LTE abundances found are closer to the LTE Fe II values, but depend strongly on the poorly known strength of inelastic hydrogen collisions.

A counterpoint to the discussion above is found in arguments presented by Morel et al. (2014) and Adamczak & Lambert (2014). Both works argue that a traditional excitation and ionization analysis to determine stellar parameters is a valid approach. They cite the work of Bergemann et al. (2012) and Lind, Bergemann, & Asplund (2012) as showing that iron is not significantly affected by departures from LTE, especially at higher abundances such as those found in their study and by extension, those determined here. A quandary thus exists at this point as to the importance of non-LTE effects in iron. The resolution of the problem lies beyond the scope of this work and so the priority in this analysis will be the maximization of the reliability of these abundances under the current analysis constraints.

At temperatures above 5500 K, there is a large spread in in the mass-derived and ionization balance gravities (see Figure 1 – bottom panel). This is due to the ionization balance being susceptible to large uncertainties in Fe I and Fe II equivalent widths. The equivalent width uncertainty arises due to the large (>20 km/s) line broadening seen in these stars.

### 3.4.2 External Comparison

To locate previous analyses of the program stars the PASTEL database (Soubrian et al. 2010) was consulted. Over 720 of the program stars have data in PASTEL. The total number of references generated is in excess of 215. This number does not include a number of analyses published post 2012. To investigate the parameter and abundance comparison a subset of the available data has been selected.

For the first comparison, the obvious choice is LH07. There is corresponding data on 288 stars. LH07 also used photometric temperatures but with a different calibration, and mass-derived gravities as part of their analysis. The mean differences relative to this study are small: -42 K in temperature ($\sigma$=78 K), +0.02 in mass-derived log $g$ ($\sigma$=0.14) and +0.10 in [Fe I/H] ($\sigma$=0.07). The sense is this study minus LH07. This study uses a different line list with newly determined solar oscillator strengths. Part of the difference in the iron abundance arises from the selected solar reference point but quantification is difficult as LH07 zero pointed each line abundance individually whereas here a single solar abundance is used and the gf value adjusted to match the solar line strength.

Hekker & Meléndez (2007) performed a traditional spectroscopic parameter determination; i.e., an excitation and ionization balance, on 189 of the stars considered here. The mean offsets in effective temperature, gravity, and [Fe/H] are: -96 K ($\sigma$=78), -0.35 dex ($\sigma$=0.25), and +0.03 dex ($\sigma$=0.11) respectively for the mass-determined gravity and Fe I abundance. The differences for the ionization balance analyses are -0.58 dex ($\sigma$=0.76) in log $g$, and -0.01 dex ($\sigma$=0.09) for Fe I. The differences are in the sense this work minus Hekker & Meléndez. The number of lines used by Hekker & Meléndez was 20 Fe I lines and 6 Fe II lines versus 400 Fe I lines and 33 Fe II lines used on average here. A comparison of the *gf* values between the two studies shows good agreement: a mean difference in log *gf* of -0.05 for Fe I and a difference of -0.02 for Fe II in the sense Hekker & Meléndez minus the work. There is considerable scatter in the gravities but this is expected given the small number of Fe II lines used in Hekker & Meléndez. The systematic offset in gravity cannot be explained by the difference in temperature. In an ionization balance, an effective temperature increase would demand an increase in the gravity, but a 100 K difference does not explain the 0.6 dex shift seen here in the ionization balance gravities. Additionally, the systematic gravity difference is unlikely to be explained by postulating that the masses found here are underestimated. To rectify the mass gravities found here with the Hekker & Meléndez values, an increase of a factor of two in mass would be required yielding a mean mass of about 3.5 solar masses. Note that the mass derivation is relatively insensitive to the temperature, numerical experiment shows that the mass increases by about 0.1 solar mass for a temperature change of +100 K. Given the differences in effective temperature and gravity between these two works, what is remarkable is the small difference in mean iron abundance.

Takeda, Sato, & Murata (2008) analyzed a sample of 322 K giants of which 191 are in common with this work. Their analysis uses an excitation and ionization balance technique to derive parameters. The mean difference in effective temperature (this work – Takeda et al.) is -51 K with a standard deviation of 68 K. The individual temperature differences can be significant: the total spread is -430 to +245 K. Comparing the log $g$ values, the mean difference is -0.12 ($\sigma$=0.10) using the mass gravities given here. For the ionization balance gravities of this work, the difference is -0.31 ($\sigma$=0.19). Once again, the remarkable thing is the agreement of the iron abundances: the mean difference is +0.05 dex ($\sigma$=0.06) for the mass gravity iron abundance derived from Fe I, while for the ionization balance gravities the mean difference in iron is +0.02 dex.

Liu et al. (2014) analyzed lithium in a sample of 378 giants taking the parameters from Takeda et al. (2008) and Liu et al. (2010). The parameters of Takeda et al. relative to this study are discussed above and the parameters of Liu et al. (2010) give similar results. Comparison of the Liu et al. (2014) lithium abundances to the ones found here show that the abundance differences correlate well with the temperature differences. This is expected based upon the sensitivity of the lithium abundance upon the temperature. Where the temperatures agree to within 25 K, the lithium abundances agree to better than 0.1 dex. There is considerable scatter about the mean difference line, but this is also expected, as the lithium feature is very weak in these stars: the median equivalent width in the 200+ stars in common is about 0.75 pm with a maximum of 6.5 pm. Most of the abundances are considered to be limits only in this work.

Another traditional spectroscopic analysis is that of Da Silva, Milone, & Reddy (2011). They analyzed 172 stars of which 98 are included in this work. The comparison of their excitation temperatures with the photometric temperatures of this work shows a mean difference of -93 K with a standard deviation of 125 K. The range is quite large: the differences span -283 K to 900 K in the sense of this work – Da Silva et al.. The gravities differ by -0.22 ($\sigma$ = 0.22) for the mass gravities determined here and by -0.46 (($\sigma$ = 0.31) for the ionization balance values. As seems to be the usual case, the iron abundances are in rather good agreement in the mean: +0.06 ($\sigma$ = 0.11) for the mass determination and +0.01 (($\sigma$ = 0.11) for the ionization balance gravities.

Recent advances in observational techniques; specifically, the photometric capabilities of the Corot and Kepler satellites, have allowed seismic techniques to be applied to stars. This development permits the determination of gravities with high precision. The studies using seismic techniques generally consider samples of 20-40 stars and there is not a large overlap with this work. However, the power of astroseismology warrants some discussion relative to the results determined here. Three studies of particular interest are Morel & Miglio (2012), Creevey et al. (2013), and Morel et al. (2014). The total number of stars in common with this work is twelve of which six appear in two of the cited papers. The first observation is that the temperatures used in the astroseismology work are usually within 100 K of the temperature derived here. Comparing the seismic gravities to the isochrone values the difference is of order -0.05 (this work being the lower) while as expected the spectroscopic gravity values determined here are of order -0.4 dex lower. This means that the differences in this work's parameters noted relative to other techniques/studies are comparable to those found relative determinations using seismic data.

In the past several years there have been a number of other studies (for example: Wang et al. (2011), Mortier et al. (2013), Ramìrez, Alonso-Prieto, & Lambert (2013)) that have considered 5 to 20 of the giants analyzed here. Comparison of the results show no surprises, the differences in temperature and gravity are consistent with those noted above; and, per usual, the [Fe/H] values are in good agreement.

Another compendium of parameters that warrants mention perhaps more as a caveat emptor is that of McDonald, Zijlstra, & Bowyer (2012). Effective temperatures and luminosities are derived therein for the bulk of the Hipparcos stars by fitting spectral energy distributions to a variety of photometry. Over 900 of the stars considered in this work are found in McDonald et al.. The mean temperature difference is -113 K (this work – theirs) with a standard deviation of 133 K. The luminosities agree rather well — the mean difference is +0.02 in log L/L$_\odot$. While the mean offsets are not especially troubling, the range of temperature differences relative to this study, ±600 K, is rather distressing. This range is found after eliminating their temperature of 17296 K for ν Hya (K0/K1 III) and 10700 K for μ Vir (F2 V). There are two immediately apparent problems in McDonald et al.. First, they do not include any reddening, and second, their classification method makes most of the giants analyzed here dwarfs. While the results are sufficiently accurate for their purpose, the temperatures are unsuited for use in an analysis such as this one.

### 4. Discussion

### 4.1 Sample Culling

In Figure 3, an HR diagram of the program stars is shown along with mass tracks from Bressan et al. (1993). The tracks are only to serve as an indicator of mass and evolutionary status. The stars shown in this plot form the first cut on the sample: only stars with a Hipparcos parallax or that are in a cluster are retained. The stars are also identified by broadening velocity (Vm) which can be either a macroturbulent or rotational velocity (see Table 3). The stars with Vm > 40 km s$^{-1}$ are mostly rotating F giants coming off the Main Sequence. In fact, most of the stars blueward of log $T_{eff}$ > 3.78 are fast rotators leading to difficulties in determining accurate line strengths due to velocity smearing of the profiles. As a result, stars with effective temperatures in excess of 6025 K are eliminated from further discussion.

A number of potential dwarfs are seen in Figure 1 (top panel) and these are the stars below the diagonal line starting at log $T_{eff}$ = 3.78 and log L/L$_{Sun}$ = +1.4 in Figure 3 and going towards lower temperatures and luminosities. These stars are eliminated from further discussion. In addition, 12 further stars are removed due to the inability to reach spectroscopic ionization equilibrium inside the model grids. After the culling, there remain 1006 G and K giants. In Table 3 last column, the retained stars are denoted by a "1" and the excluded stars a "2".

### 4.2 Mass versus Ionization Balance Gravity

The next question to be addressed is the question of which analysis — mass or ionization balance — is the more reliable and hence, the basis for further discussion. In Figure 4, a histogram of the ionization balance [Fe/H] data is presented along with a Gaussian fit to the data. Also shown are the Gaussian centroid values for the mass-derived Fe I and Fe II data as well as the simple mean value for the ionization balance [Fe/H] values. The Gaussian fits to the distributions all have similar standard deviations of about 0.17 dex. The slight asymmetry in the [Fe/H] distribution is real as there are more metal-poor than metal-rich stars in the local region. Note that three very metal-poor stars are cut off from the plot but are included in the fit. The metal-poor asymmetry leads to the difference of 0.1 dex between the simple average and the centroid of the ionization balance data.

The metallicity of the local region as evidenced by F, G, and K dwarfs is close to solar (Luck & Heiter 2005, 2006). While the G and K giants of this study are not necessarily coeval with the dwarfs of the local region, one could expect the abundances for larger non-biased samples for each to be commensurate. While the centroid values for the mass-derived gravity Fe I data and the ionization balance [Fe/H] values are not vastly different — 0.09 and 0.05 dex respectively — and fall close to our expectation of solar abundance; the mass-derived Fe II centroid value of +0.18 dex is uncomfortably high.

Another reason for favoring the ionization balance gravity is that central to the discussion will be [x/Fe] ratios where x is a specified element. Such ratios are used to help minimize parameter sensitivity. To form these ratios, one uses total abundances from species with like sensitivity to parameters; in general, ion to ion and neural to neutral. The advantage of the ionization gravities is that neutral and ionized total abundances are equivalent — at least in theory, in practice they closer than yielded by the mass gravity (see Figure 2), but are not perfectly matched. Use of the ionization balance gravity thus allows one to from the ratio without explicitly worrying about parameter sensitivity. Accordingly, the abundances for primary discussion will be those of the ionization balance analysis with the mass values at the periphery. The mass-gravity parameters and abundances are included in the tables so that an interested user can explore that data if so inclined.

### 4.3 Z>10 Abundances

Hinkel et al. (2014) have given an extensive discussion of metallicity trends in the local region. While the stars considered in that work are dwarfs, the trends versus [Fe/H], velocity, and position discussed therein are not significantly different from those found in local giants. The history of nucleosynthesis for both local dwarfs and giants is very similar, and thus here only a brief discussion will be given pointing out salient features in the giant abundances.

In Figure 5, the average [x/Fe] data is presented for the G and K giants of this study. In this notation "x" represents a target element. This distribution is much like that of LH07 (see their Table 10) for most elements. The most striking items in Figure 5 are the behaviors of sulfur and zinc. The zinc lines are badly blended especially at temperatures below 4800K resulting in abundances of poor quality. The difficulty with sulfur is a ferocious temperature dependence in the data — see Figure 6. The span in the abundances is over two orders of magnitude rising from [S/Fe] about +0.2 at 5240 K to greater than +2 at 4000 K. The problem here is once again blends in the relevant lines. In the analysis of LH07, hyperfine structure was not considered for Mn, Co, and Cu. The abundances here account for hyperfine structure but the mean abundances in this analysis are *not* vastly different from those of LH07.

Another way of looking at Z > 10 abundances is to examine either [x/H] or [x/Fe] as a function of [Fe/H]. The overall expectation is that in a sample of disk G and K giants that both [x/H] and [x/Fe] should track [Fe/H]; that is, the slope of [x/H] versus [Fe/H] should be about 1 and the intercept [x/H] should be close to the mean value in the local region. For [x/Fe], the slope should be about 0 with the intercept having the local region [x/Fe] value. There will obviously be exceptions and the strength of the deviations will depend on how metal-poor the stars are. In this sample, we look at the trends down to [Fe/H] = -0.8 throwing out only the four most metal poor stars for this examination. In Table 5, the values and uncertainties in the trends versus [Fe/H] are given along with mean values. The expectations are born out in large part in the data. The exceptions; unsurprisingly, are carbon, oxygen, and the α-elements. Other elements, Mn for example, also show gradients with respect to iron. Carbon and oxygen will be discussed in the following section. Note that the gradients in Table 5 do not allow for different slopes as a function of [Fe/H]. This dataset is too heavily dominated by thin disk stars to discern differences due to the thin versus the thick disk.

The behavior of the α-elements is well documented in the literature (see for example, McWilliam 1997, Allende-Prieto et al. 2004, LH07, Hinkel et al. 2014). Shown in Figure 7 (top panel) is the behavior of [Si/Fe] versus [Fe/H]. As expected, [Si/Fe] increases with decreasing [Fe/H]. While the effect is modest, it is real and persists through Ca. Note that a loess smoothing of the data — the gradients in the figure — yields a result akin to the linear fit. The root cause for the decrease in [α/Fe] as [Fe/H] increases is the growing dominance of Type Ia SN production of the Fe-peak elements over the production of lighter α-elements in Type II SN as the galaxy ages.

In the middle panel of Figure 7, [Mn/Fe] versus [Fe/H] is shown. There is a definite slope in the data with lower [Fe/H] ratios showing lower [Mn/Fe]. This effect was noted by LH07 and was first demonstrated by Wallerstein (1962) and Wallerstein et al. (1963). Gratton (1989) also noted the effect in a sample of metal-poor to solar-metallicity stars. The effect is the reverse of the trends seen in the α-elements and presumably indicates a SN Ia origin for Mn. However, the adjoining odd elements V and Co show no believable trend with [Fe/H] — see the bottom panel of Figure 7 for Co.

The dwarf data from the Hypatia Catalog (Hinkel et al. 2014) shows a curious phenomenon in the Ni abundances. There are two [Ni/Fe] ratios that appear in the data at essentially all [Fe/H] ratios. There is a group at a slightly subsolar ratio: [Ni/Fe] ~ -0.03 and another at [Ni/Fe] ~ -0.2. Hinkel et al. also delineate the abundances based on distance from the Sun. In Figure 8, [Ni/Fe] is presented as a function of [Fe/H] with the stars delineated by heliocentric distance. There is no discernible dependence in the data as a function of distance. There appears to be a downward trend in [Ni/Fe] with decreasing [Fe/H]. The data smoothing shows a flat dependence at [Fe/H] < 0, with increasing [Ni/Fe] for [Fe/H] > 0. There is also a dip in [Ni/Fe] at [Fe/H] about [Fe/H] = 0. However, the actual spread in the data is small. In the bin -0.2 < [Fe/H] < +0.2, the mean [Ni/Fe] ratio is +0.058 with a standard deviation of 0.053 (N = 736). The total spread at [Fe/H] = 0 in [Ni/Fe] is about 0.15 dex. There is no indication of a bimodal distribution of the [Ni/Fe] ratios in the local region giants. All trends in the [Ni/Fe] data are controlled by the stars in the -0.2 < [Fe/H] < +0.2 range. The same is true of all other elements.

In the Figure 9 panels, three s/r process elements are shown. These elements shown much more scatter than the α- or Fe-peak elements. The small number of lines used to determine these abundances means that the abundance on a per star basis is more uncertain. Overabundances of these elements appears to be more common than overabundances in the lighter elements leading to question of whether or not there are Ba or related stars in the sample. There are at least five

known Ba stars in the sample: HD 46407, HD 104979, HD 139195, HD 202109, and HD 205011. While the highest point in the [Zr/Fe] and [Nd/Fe] data is HD 46407 (the Ba II lines are too strong for adequate modelling), the bulk of the high points are not associated with known Ba stars. The suspicion is that small number statistics has corrupted the abundance data in these cases. Once again; however, the discrepant points comprise less than 3% percent of the analyzed stars. They are obvious, but do not compromise the overall data.

A primary point of interest in the s/r data is the behavior of Ba. It appears that [Ba/Fe] decreases once the [Fe/H] ratio goes above solar. LH07 found the same thing in their data and ascribed the effect to either non-LTE or line-formation problems. Both LH07 and this study used a maximum equivalent width cutoff of 20 nm to help control problems in inadequate line formation modelling. Calculations show that non-LTE effects in Ba for more luminous stars (Andrievsky et al. 2013, 2014) are modest contributing an uncertainty/non-LTE correction of about 0.1 dex in the Ba abundances. The primary uncertainty detected was inadequate line modelling; that is, the Ba II line cores go optically thick at very shallow continuum optical depths. The line strength cutoff applied here is well below the values found in the Cepheids considered in the non-LTE studies. The conclusion is that non-LTE is not the primary culprit in the Ba data for the giants. Since these are stronger lines, a more probable cause is that the equivalent widths are somewhat underestimated due to the Gaussian approximation missing the growing line wings. This possibility is given credence as a consideration of computed Ba II profiles at 4950 K, log g = 2.5, and $V_t$ = 1.5 km s$^{-1}$ indicates that the Gaussian approximation underestimates the Voigt profile equivalent width of a true 20.0 pm line by 1.3 pm. While this difference is small, the line is saturated and the abundance needed to match the lesser equivalent width is 0.11 dex lower than that needed to match the Voigt profile value. Any underestimate in equivalent width will lead to potentially significant underabundances.

As a last point, the abundances here are overall in good agreement with those of LH07 and other work. Since the new data in no way contradicts previous analyses, there is no tension between the results of this study and the contention of Reddy et al. (2012, 2013) that field giants are the product of the breakup of open clusters. In fact, the cluster giants that are contained in this work are indistinguishable from the field giants in terms of their abundances.

### 4.4 Li, C, and O in Giants

The discussion of LH07 concerning lithium and C and oxygen abundances in giants is still relevant today and their major points will be revisited briefly here.

#### 4.4.1 Lithium

Lithium abundances have been determined in a range of local objects including dwarfs by Lambert & Reddy (2004) and Luck & Heiter (2006) as well as giants by Brown et al. (1989) and LH07. For the giants, the base findings are that lithium is highly depleted with respect to its assumed initial value, and that the depletion is a function of the effective temperature and mass. Recent studies of lithium in giants have focused on comparisons of thick and thin disk stars and the effects of rotation (Ramírez et al. 2012, Liu et al 2014) with the result that thin disk giants are more lithium depleted than thick disk giants. There appears to be no dependence in giants of lithium on the macroturbulent or rotational velocity (LH07, Liu et al. 2014).

Theoretical predictions by Iben (1967a,b) indicate that lithium should be diluted/depleted during the first giant branch by a factor of 60 at 3 $M_\odot$ decreasing to a factor of 28 at 1 $M_\odot$. However, recent work has shown that lithium abundance predictions in giants are subject to large numerical treatment uncertainties in thermohaline mixing (Lattanzio et al. 2014). While acknowledging that the theoretical predictions are uncertain, the currently available projected lithium dilutions/depletions correspond to decreases in abundance of 1.8 to 1.4 dex relative to the (assumed constant) main sequence initial value. However, the surface lithium content of dwarfs is extremely variable (Lambert & Reddy (2004), Luck & Heiter (2006)) spanning up to two orders of magnitude at any temperature/mass. While this makes the lithium content of any one star (giant or dwarf) impossible to predict under any evolutionary scheme, one can hopefully use overall trends in the data to test generic theoretical predictions.

The work of Lambert & Reddy indicates that the general astration level of lithium in dwarfs is independent of metallicity but dependent upon mass. In Figure 10, we show their LTE abundance data (lithium versus mass) along with the G/K giant data of this study. The data shown in Figure 10 for the giants is from the LTE analysis. The non-LTE corrections from Lind et al. (2009) will increase the lithium abundances by about 0.1 to 0.2 dex for both G/K giants and dwarfs. Figure 10 also shows the level of dilution predicted by standard evolution (a representative factor of 40). What is apparent from the figure and was pointed out by LH07, is that given the scatter in the dwarf lithium abundances and the mass-dependent astration, it is entirely possible that the bulk of giant lithium abundances are consistent with standard evolution. Since the dilution occurs/starts when the star becomes convective at the base of the first giant branch, the appropriate abundance to use for the dilution can be significantly below the local interstellar value of about 3.0 to 3.2 (determined from the dwarfs themselves – see Lambert & Reddy). Detailed consideration of lithium abundances in giants may need additional dilution processes, but certainly not at the level one would need if the initial abundance were 3.0 dex or above.

Further confounding the interpretation of the giant abundances is the "Li-dip." The "Li-dip" is the observed decline in lithium abundance in open cluster dwarfs at a mass of about 1.3 to 1.4 $M_\odot$ (Balachandran 1995). Later work has shown that the dip occurs at different temperatures/masses depending on the metallicity (François et al. 2013). However, the "Li-dip" is not an especially distinct feature in field dwarfs (Lambert & Reddy 2004, Luck & Heiter 2006) and it is thus difficult to ascribe all of the variation in dwarf and giant lithium abundances to its presence.

In Figure 10, several stars show very high abundances of lithium. These stars are HR 7820, HR 8642, HD 188993, HD 9746, HD 95799, and HD 112127. All of these stars have lithium abundances in excess of 3.0 dex. The last three stars are in the super-Li (or Li-rich) list of Charbonnel & Balachandran (2000). The first three are stars in LH07 and all have lithium abundances consistent with the present values in that work. LH07 identified HR 7820 and HR 8642 as K giant super-Li stars. The evolutionary status assumed for a super-Li star is either that it is on the ascent to the first giant branch, or that it resides near the red-giant clump post first dredge-up. In either case, it will be well separated from the main sequence. This means that stars blue-ward of the clump such as HD 188993 while having high Li abundances are not super-Li stars. Another star not plotted in Figure 10 because it fails the culling process being too close to the main sequence is HD 148317. It shows a lithium abundance of 3.04 dex. The idea is that stars such as HD 188993 and HD 148317 have not yet undergone a strong mixing event and thus show their original lithium content. The status of these two stars will be considered further in the discussion of carbon and oxygen.

The statistic for super-Li stars is that about 1% of normal K giants are super-Li stars (Charbonnel & Balachandran 2000). The question is what abundance level constitutes a super-Li star? Some of the stars listed by Charbonnel & Balachandran have lithium abundances as low as 1.5 dex. The number of stars having an abundance at that level or above in this study is 23 or slightly more than 2% of the sample. Note that these stars do not include any of the F giants near the main sequence or stars like HD 148317 and HD 188993. Thus it appears that this sample has too many super-Li stars. However, lithium abundances are very variable and it is certainly possible that a lithium abundance of 1.5 to 1.7 dex — dilution of a factor of 20 to 50 — in a giant merely reflects standard evolution operating on the plateau local interstellar lithium abundance. If the lower abundance to consider as super-Li were taken to be 2.0 dex, this sample then has 15 total candidates.

Super-Li stars are "classified" according to their position in the HR diagram. In Figure 11, we see a group of these stars straddling the red giant clump – the position of one of the two families of such objects (Charbonnel & Balachandran). The four stars including HD 188993 blue-ward of the clump may be pre first dredge-up and thus have not diluted their surface lithium. Charbonnel & Balachandran have given a working hypothesis for the clump super-Li stars. The mechanism is an "extra" mixing that occurs at this position in the HR diagram between the convective envelope and the top of the H-burning shell. The mixing results in the production of $^7$Li that is then convected to the surface. What is still missing is a trigger mechanism for the event. This scenario is complicated by the inconsistency in theoretical lithium predictions found by Lattanzio et al. (2014). The takeaway from their study is that all theoretical lithium predictions should be viewed with caution.

### 4.4.2 C and O

The abundances of carbon, nitrogen, and oxygen in K giants will reflect the effects of both galactic chemical evolution and stellar evolution. The properties of both galactic and stellar CNO evolution are well known having been studied numerous times (see; for example Lambert & Ries 1981, Gustafsson et al. (1999), LH07, Adamczak & Lambert 2014). The primary galactic evolution expectation is that the [O/Fe] (and [C/Fe]) ratio should rise with decreasing [Fe/H]. Stellar evolution will affect the carbon and nitrogen abundances — carbon will be converted to nitrogen with the sum C+N expected to remain constant. The anticipated carbon decrease is modest, perhaps up to 0.2 dex. Oxygen abundances are unchanged during the first dredge-up.

The relationship between [Fe/H] and [C/Fe], [O/Fe], and C/O is shown in Figure 12. As expected, both [C/Fe] and [O/Fe] increase with decreasing [Fe/H]. The C/O ratios are mostly subsolar as projected based on the predictions of standard stellar evolution. Carbon is diluted/depleted because of the first dredge-up while oxygen remains unaffected. Note that C/O is not independent of [Fe/H], but decreases implying an enhancement of O relative to C in the more metal-poor stars.

Our giant data is in accord with the C/O ratios found using the dwarf data of Luck & Heiter (2006). Their data gives solar C/O ratios at solar metallicity with the C/O ratio declining towards lower [Fe/H] values, albeit with significant scatter. Gustafsson et al. (1999) also found this result in dwarfs. This implies that stellar evolution in this mass and metallicity range yields rather uniform results in the carbon dilution/depletion for the first dredge-up.

There are of order 50 C/O ratios at or slightly above the solar ratio in the current data. If these stars were actually dwarfs, then one might expect their [C/Fe] ratio to be near zero, and they should have higher gravities relative to the mean of the sample. However, neither of these expectations are borne out. The real question perhaps is why should a G/K giant be limited to a subsolar C/O ratio? The C/O ratio in local dwarfs spans a considerable range: $0.1 < C/O < 0.9$ (Luck & Heiter 2006). The range seen in the G/K giants is consistent with those initial ratios coupled to standard stellar evolution yielding a dilution/depletion of the surface carbon abundance by about 0.2 dex relative to the initial composition (Iben 1965, 1967, Schaller et al. 1992, Girardi et al. 2000).

LH07 pointed out an interesting result in their clump [C/Fe] abundances. They found that blue side; that is, hotter stars in the clump showed lower [C/Fe] ratios that did the red-side of the clump. The difficulty was that an admixture of [Fe/H] ratios could give rise to a spurious result since [C/Fe] correlates with [Fe/H]. The current sample affords the possibility of exploring this effect further as there are more than 350 stars with [Fe/H] ratios between -0.1 and +0.1 dex that fall in the luminosity range $1.0 < \log (L/L\odot) < 2.5$. Figure 13 (top panel) shows the temperature binned [C/Fe] ratios. Looking at the outermost points, we see that they differ in [C/Fe] by about 0.15 dex with the blue edge having the lower abundance. Note that neither [O/Fe] or [Fe/H] show any dependence on temperature. There is an effect in the C/O ratio but this is expected given the decrease in [C/Fe] and the lack of a variation in the [O/Fe] values against effective temperature.

Accepting the trend in [C/Fe] at face value what could be the cause? In the bottom panel of Figure 13, the abundance ratios are plotted against mass. Here we see that [C/Fe] ratios of -0.3 to -0.4 are associated with higher mass stars (2.5 to 3.5 $M_\odot$), while [C/Fe] increases to about -0.2 with decreasing mass. The obvious explanation is depth dependent mixing as a function of mass: higher mass stars mix deeper bringing more carbon deficient material to the surface than is the case in the lower mass stars. As mass correlates with age, a dependence exists between age and [C/Fe] — the strongest dilutions/depletions exist in the youngest stars. This is in accord both qualitatively and quantitatively with theoretical expectations (Iben 1965, 1967, Schaller et al. 1992, Girardi et al. 2000). More advanced stellar mixing/evolution schemes are available than the ones cited above. However, before elaborating the discussion the nitrogen abundance is needed and it is unavailable for most of the stars considered; and so, the overall discussion will stop here. However, a few specific points do need further elaboration.

In Figure 13, there are three stars that have very low [C/Fe] ratios. These stars must be akin to the weak G band and/or the CN- stars discussed by Adamczak & Lambert (2013) and Luck (1991) respectively. Stars in these classes typically show carbon depletions up to 20 times larger than found in normal giants; but as of now, there is no mechanism to explain them. The stars in this study that show such depletions are HR 6791, HR 40, and HR 1219. HR 6791 is a well-known example included in both studies noted above. The other two have not been mentioned in prior works on carbon-poor giants. HR 1219 and HR 40 are warmer objects, and have lower luminosities placing them nearer the Main Sequence. The rotational/macroturbulent line broadening seen in both is of order 4 to 6 km s$^{-1}$ making the lack of C I lines very obvious. Perhaps these stars are precursors to the more evolved post first giant branch weak G band stars.

In the lithium discussion, two stars were identified with high Li contents: HD 188993 and HD 148317. Both stars are warmer and have higher gravities — about 5700 K and 3.5 — and thus are likely G subgiants. Their carbon abundances are consistent with the idea that they have yet to ascend the first giant branch and undergo strong mixing: the [C/Fe] ratio in both is about +0.06. These two stars are likely on their redward pass towards the first giant branch.

As a last point, it is generally assumed that older stars are more metal-poor than younger stars, and that there significant scatter within the age-metallicity relation. The idea of an age-metallicity relation is further complicated by the strong possibility that stars migrate during their lifetimes and end up far from their point of origin in radial distance from the galactic center (Sellwood & Binney 2002, Schönrich & Binney 2009). The difficulty with this scenario relative to the age-metallicity relation is that stars from different parts of the Galaxy should have different initial compositions due to the galactic metallicity gradient (Luck & Lambert 2011). While this study was not designed to look at this problem, it is worth noting that the giant subset used to explore [C/Fe] in the red-giant clump; some 350 stars, has a total range in [Fe/H] of 0.2 dex centered on the solar value; but has ages spanning more 5 Gyrs – from 0.5 to about 6 Gys. There is little evidence for strong age-metallicity relation in the giants of the local region about the Sun.

## 5. Concluding Remarks

In this study, parameter and abundance data on more than 1130 giants in the local neighborhood is presented. This is likely the largest high-resolution spectroscopic abundance study ever undertaken. The most striking result on parameters is the severe disagreement between isochrone mass derived gravities and ionization balance gravities. There is no accord currently available to rectify the difference. Astroseismology comes down on the side of the isochrone-mass derived gravities, and then proceeds to attribute the difference to NLTE effects. However, the ability to attribute this difference to NLTE is disputed. One perhaps should worry about using 1D or 2D static model atmospheres to model an inherently dynamic spheroid.

In terms of the abundances, the local GK giants look much like the Sun — same total content with similar abundance ratios. Lithium, carbon, and oxygen abundances in the GK giants are substantially as predicted by standard stellar evolution. The lithium abundances are severely diluted but once the "initial" dwarf abundances are taken into account the dilution is most likely not substantially in excess of that predicted by standard evolutionary models. The oxygen abundances are unmodified from the initial value: that is, [O/Fe] = 0 when [Fe/H] is near 0. Carbon abundances are affected by CN processing and subsequent mixing during the first dredge-up. The mean [C/Fe] ratio in the sample is about -0.2 dex. A variation in the [C/Fe] ratio across the red-giant clump is identified as being due to mass dependent mixing during the first giant branch. This result is discernible due to the large number of solar-metallicity (-0.1 < [Fe/H] < +0.1) giants in the sample. Effects of galactic chemical abundance are present: [O/Fe] and [C/Fe] shows their well-known dependence on [Fe/H].

The next step in this overall project will be the analysis of a similarly sized sample of local FGK dwarfs designed to serve as an unevolved comparison sample.

## Acknowledgements

I am grateful to a most conscientious referee whose comments substantially improved this paper. Financial support by Case Western Reserve University made possible the McDonald Observatory observations used in this work. The Sandiford echelle spectra used here are available through the [FGK Spectral Library](). This study is based in part on data obtained from the ESO Science Archive Facility under many request numbers. The ELODIE archive at Observatoire de Haute-Provence (OHP) also provided part of the spectroscopic data. The SIMBAD database, operated at CDS, Strasbourg, France and satellite sites, and NASA's Astrophysics Data System Bibliographic Services were instrumental in this work.

Table 1
Program Stars

| Primary | HD | HIP | HR | CCDM | Cluster | S | KeyName | Spectral Type | Type | V | B-V | Parallax | d (pc) | RV (km/s) |
|---|---|---|---|---|---|---|---|---|---|---|---|---|---|---|
| HR 9101 | 225197 | 343 | 9101 | | | S | hr9101 | K0III | HB* | 5.781 | 1.101 | 11.03 | 90.7 | 25.98 |
| HR 9104 | 225216 | 379 | 9104 | | | S | hr9104 | K1III | Star | 5.691 | 1.047 | 10.79 | 92.7 | -28.77 |
| HD 225292 | 225292 | 410 | | | | S | hd225292 | G8II | Star | 6.475 | 0.931 | 5.95 | 168.1 | 12.10 |
| HR 2 | 6 | 417 | 2 | | | S | hr0002 | G9III: | Star | 6.310 | 1.093 | 7.20 | 138.9 | 15.30 |
| * 86 Peg | 87 | 476 | 4 | | | S | hr0004 | G5III | Star | 5.553 | 0.879 | 8.75 | 114.3 | 1.70 |
| HR 13 | 344 | 655 | 13 | | | H | hd000344 | K1III | HB* | 5.668 | 1.117 | 10.53 | 95.0 | 6.80 |
| HR 16 | 360 | 671 | 16 | | | S | hr0016 | G8III: | HB* | 5.992 | 1.026 | 10.16 | 98.4 | 20.40 |
| HR 19 | 417 | 716 | 19 | | | S | hr0019 | K0III | Star | 6.253 | 0.963 | 7.22 | 138.5 | 16.11 |
| * 87 Peg | 448 | 729 | 22 | | | S | hr0022 | G9III | Star | 5.565 | 1.043 | 11.44 | 87.4 | -20.23 |
| HD 483 | 483 | 759 | | | | H,H | hd000483a,hd000483b | G2III | SB* | 7.050 | 0.640 | 19.39 | 51.6 | -30.34 |
| HD 770 | 770 | 966 | | | | H | hd000770 | K0III | Star | 6.530 | 1.060 | 8.41 | 118.9 | |
| HR 40 | 895 | 1076 | 40 | J00134+2659AB | | S | hr0040 | F8V | SB* | 6.277 | 0.632 | 7.61 | 131.4 | -8.58 |
| HD 1142 | 1142 | 1269 | | | | S | hd001142 | G8III | Star | 6.443 | 0.810 | 7.52 | 133.0 | -17.00 |
| * 36 Psc | 1227 | 1319 | 59 | J00166+0814AB | | S | hr0059 | G8II-III | ** | 6.127 | 0.909 | 8.00 | 125.0 | 0.07 |
| HR 69 | 1419 | 1465 | 69 | | | S | hr0069 | K0III | Star | 6.065 | 1.029 | 6.50 | 153.6 | 9.18 |
| * iot Cet | 1522 | 1562 | 74 | J00194-0850A | | S | hr0074 | K1.5III | V* | 3.550 | 1.220 | 11.88 | 84.2 | 19.35 |
| HD 1638 | 1638 | 1631 | | | | U | hd001638 | K1III | Star | 8.730 | 1.320 | 0.48 | 2083.3 | 26.00 |
| * rho And | 1671 | 1686 | 82 | | | E | hd001671 | F5III | Star | 5.180 | 0.420 | 20.60 | 48.5 | 10.40 |
| HD 1690 | 1690 | 1692 | | | | H | hd001690 | K2III | Star | 9.178 | 1.337 | 3.22 | 310.6 | 18.39 |
| * iot Scl | 1737 | 1708 | 84 | | | S | hr0084 | G5III | RGB* | 5.174 | 0.993 | 9.87 | 101.3 | 20.60 |
| * 44 Psc | 2114 | 2006 | 97 | J00254+0156AB | | H | hd002114 | G5III | ** | 5.778 | 0.828 | 6.94 | 144.1 | -5.40 |
| * 10 Cet | 2273 | 2100 | 101 | J00266-0003AB | | S | hr0101 | G8III | ** | 6.208 | 0.884 | 7.23 | 138.3 | -22.70 |
| * 46 Psc | 2410 | 2213 | | | | S | hd002410 | gG7 | Star | 6.380 | 1.000 | 5.81 | 172.1 | 5.35 |
| HR 111 | 2529 | 2256 | 111 | | | H | hd002529 | K0III | Star | 6.270 | 1.092 | 7.98 | 125.3 | 5.40 |
| * 52 Psc | 2910 | 2568 | 131 | J00326+2018A | | S,E | hr0131,hd002910 | K0III | *in** | 5.374 | 1.074 | 12.50 | 80.0 | -12.07 |
| HD 2954 | 2954 | 2595 | | J00329+1610A | | S | hd002954 | F6III | *in** | 6.900 | 0.460 | 8.96 | 111.6 | -12.50 |
| V* CF Scl | 3001 | | | | | H | cfscl | G9III(e) | RSCVn | 9.927 | 0.703 | | | 20.00 |
| HR 135 | 2952 | 2611 | 135 | | | S | hr0135 | K0III | Star | 5.930 | 1.031 | 8.76 | 114.2 | -34.97 |
| HR 141 | 3166 | 2734 | 141 | | | S | hr0141 | K0 | Star | 6.390 | 1.140 | 9.12 | 109.6 | -19.59 |
| HR 156 | 3411 | 2926 | 156 | | | S | hr0156 | K2III | Star | 6.160 | 1.190 | 8.87 | 112.7 | 0.46 |
| HR 162 | 3488 | 2937 | 162 | | | H | hd003488 | K0II-IIICNv... | Star | 6.410 | 0.975 | 6.81 | 146.8 | -6.40 |
| * eps And | 3546 | 3031 | 163 | | | S | hr0163 | G8III | PM* | 4.380 | 0.870 | 19.91 | 50.2 | -84.43 |
| * del And | 3627 | 3092 | 165 | J00393+3052A | | S | hr0165 | K3III | SB* | 3.280 | 1.280 | 30.91 | 32.4 | -9.88 |
| * alf Cas | 3712 | 3179 | 168 | J00405+5632A | | S | hr0168 | K0IIIa | Star | 2.230 | 1.170 | 14.29 | 70.0 | -4.31 |
| * 32 And | 3817 | 3231 | 175 | | | S | hr0175 | G8III | Star | 5.330 | 0.890 | 9.19 | 108.8 | -5.10 |
| * mu. Phe | 3919 | 3245 | 180 | | | H | hd003919 | G8III | Star | 4.590 | 0.970 | 13.27 | 75.4 | 18.80 |
| * bet Cet | 4128 | 3419 | 188 | | | S | hr0188 | K0III | V* | 2.010 | 1.010 | 33.86 | 29.5 | 13.32 |
| * phi01 Cet | 4188 | 3455 | 194 | | | E | hd004188 | K0III | V* | 4.775 | 1.007 | 13.96 | 71.6 | -0.30 |
| * lam02 Scl | 4211 | 3456 | 195 | | | H | hd004211 | K1III | PM* | 5.960 | 1.100 | 9.63 | 103.8 | 26.50 |
| HD 4388 | 4388 | 3620 | | J00464+3057B | | U | hd004388 | K3III | *in** | 7.300 | 1.150 | 5.68 | 176.1 | -27.54 |
| * 58 Psc | 4482 | 3675 | 213 | | | S | hr0213 | G8II | Star | 5.515 | 0.976 | 11.52 | 86.8 | -2.61 |
| * zet And | 4502 | 3693 | 215 | J00473+2416A | | S,E | hr0215,hd004502 | K1III+KV | EllipVar | 4.060 | 1.120 | 17.24 | 58.0 | -24.43 |
| * 60 Psc | 4526 | 3697 | 216 | | | S | hr0216 | G8III | Star | 5.985 | 0.929 | 7.08 | 141.2 | 14.30 |
| * del Psc | 4656 | 3786 | 224 | J00487+0736A | | S | hr0224 | K4IIIb | *in** | 4.440 | 1.510 | 10.48 | 95.4 | 32.45 |
| HR 229 | 4737 | 3807 | 229 | | | H | hd004737 | G8III | Star | 6.285 | 0.879 | 7.72 | 129.5 | 11.80 |
| HR 228 | 4732 | 3834 | 228 | J00492-2408A | | S | hr0228 | K0IV | *in** | 5.900 | 0.940 | 17.27 | 57.9 | 22.90 |

| Name | HD | HIP | HR | WDS | Src | ID | SpType | ObjType | Vmag | B-V | Plx | RV | ? |
|---|---|---|---|---|---|---|---|---|---|---|---|---|---|
| * 65 Psc B | 4757 | | 230 | J00499+2743B | S | hr0230 | F4III | *in** | 6.276 | 0.416 | 7.00 | 142.9 | 5.00 |
| * 65 Psc A | 4758 | | 231 | | S | hr0231 | F5III | *in** | 6.293 | 0.345 | | | 6.80 |
| HR 249 | 5118 | 4185 | 249 | | S | hr0249 | K3III: | Star | 6.061 | 1.150 | 9.27 | 107.9 | -10.47 |
| * 21 Cet | 5268 | 4257 | 255 | | S | hr0255 | G5IV | Star | 6.162 | 0.905 | 7.61 | 131.4 | 46.30 |
| * ups01 Cas | 5234 | 4292 | 253 | J00551+5858A | E | hd005234 | K2III | Star | 4.829 | 1.231 | 9.93 | 100.7 | -23.00 |
| HD 5418 | 5418 | 4382 | | | S | hd005418 | G8II | Star | 6.440 | 0.980 | 7.99 | 125.2 | 5.50 |
| * ups02 Cas | 5395 | 4422 | 265 | | S | hr0265 | G8IIIb | Pec* | 4.632 | 0.944 | 16.32 | 61.3 | -47.73 |
| * phi04 Cet | 5722 | 4587 | 279 | | S | hr0279 | G7III | HB* | 5.621 | 0.941 | 10.20 | 98.0 | -19.46 |
| HD 6037 | 6037 | 4801 | | | S | hd006037 | K2/K3III | Star | 6.470 | 1.210 | 10.77 | 92.9 | -25.11 |
| HR 295 | 6192 | 4829 | 295 | | H | hd006192 | G8III | Star | 6.109 | 0.908 | 8.29 | 120.6 | 13.00 |
| HR 299 | 6245 | 4890 | 299 | | H | hd006245 | G8III | Star | 5.397 | 0.879 | 13.68 | 73.1 | -1.40 |
| * eps Psc | 6186 | 4906 | 294 | | S | hr0294 | G9III | Star | 4.280 | 0.960 | 17.94 | 55.7 | 7.74 |
| * 25 Cet | 6203 | 4914 | 296 | J01030-0451A | U | hr0296 | K0III | *in** | 5.409 | 1.124 | 9.78 | 102.2 | 15.30 |
| * 27 Cet | 6482 | 5121 | 315 | | S | hr0315 | K0III | Star | 6.101 | 1.008 | 10.76 | 92.9 | 11.69 |
| HR 320 | 6559 | 5170 | 320 | | S,H | hr0320,hd006559 | K1III | Star | 6.125 | 1.054 | 8.52 | 117.4 | 30.30 |
| HR 325 | 6668 | 5259 | 325 | | S | hr0325 | A3III | Star | 6.367 | 0.218 | 14.54 | 68.8 | 9.00 |
| * iot Tuc | 6793 | 5268 | 332 | | H | hd006793 | G5III | semi-regV* | 5.352 | 0.856 | 10.72 | 93.3 | -7.80 |
| * eta Cet | 6805 | 5364 | 334 | J01085-1010A | S | hr0334 | K1III | HB* | 3.450 | 1.160 | 26.32 | 38.0 | 11.74 |
| HR 350 | 7082 | 5477 | 350 | | H | hd007082 | G6II-III | Star | 6.421 | 0.866 | 7.96 | 125.6 | 88.30 |
| * tau Psc | 7106 | 5586 | 352 | | S | hr0352 | K0.5IIIb | Star | 4.523 | 1.096 | 19.32 | 51.8 | 35.20 |
| V* AN Scl | 7280 | | | J01127-2911A | H | anscl | K0III | EB*WUMa | 8.890 | 0.790 | | | -7.50 |
| V* AI Scl | 7312 | 5661 | 359 | | H | aiscl | F0III | PulsV*delSct | 5.946 | 0.266 | 14.41 | 69.4 | 10.90 |
| HR 356 | 7229 | 5679 | 356 | | S | hr0356 | G9III | *in** | 6.234 | 0.976 | 8.62 | 116.0 | 36.40 |
| * 88 Psc | 7446 | 5824 | 367 | | S | hr0367 | G6III: | Star | 6.042 | 1.081 | 6.28 | 159.2 | -9.00 |
| HR 306 | 6319 | 5928 | 306 | | E | hd006319 | K2III: | Star | 6.193 | 1.114 | 9.35 | 107.0 | -9.97 |
| HR 371 | 7578 | 5936 | 371 | | S | hr0371 | K1III | Star | 6.050 | 1.168 | 9.37 | 106.7 | 6.25 |
| * 39 Cet | 7672 | 5951 | 373 | J01167-0230A | H | hd007672 | G5III+DA2.8 | RSCVn | 5.432 | 0.866 | 12.41 | 80.6 | -30.11 |
| * ksi And | 8207 | 6411 | 390 | | S | hr0390 | K0III | Star | 4.875 | 1.078 | 15.21 | 65.7 | -12.87 |
| * tet Cet | 8512 | 6537 | 402 | J01240-0811A | S | hr0402 | K0III | HB* | 3.590 | 1.060 | 28.66 | 34.9 | 17.20 |
| HR 406 | 8599 | 6605 | 406 | | S | hr0406 | G8III | Star | 6.167 | 0.953 | 13.10 | 76.3 | -26.82 |
| HR 407 | 8634 | 6669 | 407 | | S | hr0407 | F6V | SB* | 6.192 | 0.404 | 14.69 | 68.1 | -14.26 |
| * 46 Cet | 8705 | 6670 | 412 | | S | hr0412 | K2III | Star | 4.918 | 1.239 | 11.34 | 88.2 | -22.60 |
| * 94 Psc | 8763 | 6732 | 414 | | S | hr0414 | K1III | Star | 5.505 | 1.107 | 11.69 | 85.5 | -42.53 |
| HR 426 | 8949 | 6868 | 426 | J01284+0758A | S | hr0426 | K1III | V* | 6.214 | 1.113 | 8.73 | 106.7 | 2.47 |
| * 49 And | 9057 | 6999 | 430 | | S | hr0430 | G9III | Star | 5.269 | 0.993 | 11.09 | 90.2 | -11.48 |
| * mu. Psc | 9138 | 7007 | 434 | J01301+0609A | S | hr0434 | K4III | PM* | 4.840 | 1.380 | 10.73 | 93.2 | 34.19 |
| * del Phe | 9362 | 7083 | 440 | | H | hd009362 | G9III | PM* | 3.949 | 0.991 | 22.95 | 43.6 | -7.30 |
| * eta Psc | 9270 | 7097 | 437 | J01315+1521AB | S | hr0437 | G7IIa | ** | 3.620 | 0.980 | 9.33 | 107.2 | 13.78 |
| HD 9611 | 9611 | | | | H | hd009611 | K1/K2III | Star | 9.440 | 1.320 | | | -7.36 |
| HR 467 | 10042 | 7271 | 467 | J01336-7830A | H | hd010042 | K0III | *in** | 6.115 | 0.954 | 8.02 | 124.7 | -1.00 |
| * chi Cas | 9408 | 7294 | 442 | | S | hr0442 | G9IIIb | Star | 4.692 | 0.986 | 15.67 | 63.8 | 6.66 |
| HR 453 | 9742 | 7361 | 453 | | H | hd009742 | K0.5III | Star | 6.120 | 1.099 | 8.74 | 114.4 | -10.90 |
| V* OP And | 9746 | 7493 | 454 | | S | hr0454 | K1III: | RSCVn | 6.220 | 1.250 | 6.36 | 157.2 | -42.67 |
| HR 525 | 11025 | 7568 | 525 | | H | hd011025 | G8III | Star | 5.679 | 0.917 | 8.57 | 116.7 | 17.80 |
| * ups Per | 9927 | 7607 | 464 | | S | hr0464 | K3-III | Star | 3.570 | 1.280 | 18.41 | 54.3 | 16.15 |
| * nu. Psc | 10380 | 7884 | 489 | | E | hd010380 | K3IIIb | Star | 4.440 | 1.370 | 8.98 | 111.4 | 0.76 |
| HR 505 | 10615 | 7921 | 505 | | H | hd010615 | K2.5III | Star | 5.707 | 1.266 | 8.97 | 111.5 | 2.00 |
| HD 232509 | 232509 | 7995 | | | U | hic007995 | F8 | Star | 9.300 | 0.590 | 5.37 | 186.2 | |
| * omi Psc | 10761 | 8198 | 510 | | S | hr0510 | G8III | V* | 4.260 | 0.960 | 11.67 | 85.7 | 15.52 |
| HR 527 | 11037 | 8404 | 527 | | S | hr0527 | G9III | Star | 5.911 | 0.969 | 10.16 | 98.4 | 5.00 |

| Name | HD | HR | FK5 | WDS | Other | Src | Ident | SpType | ObjType | Vmag | B-V | Plx | Dist | RV |
|---|---|---|---|---|---|---|---|---|---|---|---|---|---|---|
| HR 521 | 10975 | 8423 | 521 | | | S | hr0521 | K0III | Star | 5.947 | 0.961 | 9.56 | 104.6 | 34.84 |
| * chi Cet | 11171 | 8497 | 531 | J01495-1042A | | U | hd011171 | F3III | *in** | 4.680 | 0.320 | 43.13 | 23.2 | -1.80 |
| * zet Cet | 11353 | 8645 | 539 | J01515-1019A | | S | hr0539 | K0III | SB* | 3.720 | 1.140 | 13.88 | 72.0 | 7.80 |
| * ksi Psc | 11559 | 8833 | 549 | | | E | hd011559 | K0III | SB* | 4.621 | 0.926 | 18.21 | 54.9 | 26.13 |
| * eta02 Hyi | 11977 | 8928 | 570 | | | H | hd011977 | G8.5III | Star | 4.695 | 0.919 | 14.91 | 67.1 | -16.20 |
| * 56 And | 11749 | 9021 | 557 | J01560+3716A | NGC 752 PLA 446 | S | hr0557 | K0III | *in** | 5.698 | 1.060 | 10.31 | 97.0 | 61.77 |
| HR 574 | 12055 | 9095 | 574 | | | H | hd012055 | G8III | Star | 4.841 | 0.841 | 13.14 | 76.1 | 11.90 |
| HR 584 | 12270 | 9149 | 584 | | | H | hd012270 | G8II-III | Star | 6.365 | 0.889 | 5.89 | 169.8 | 11.60 |
| HD 12116 | 12116 | 9237 | | | | S | hd012116 | G5 | Star | 6.460 | 1.140 | 9.11 | 109.8 | |
| HD 12345 | 12345 | 9404 | | | | H | hd012345 | G8III | PM* | 8.770 | 0.750 | 23.28 | 43.0 | 15.80 |
| * pi. For | 12438 | 9440 | 594 | | | S,H | hr0594,hd012438 | G5III | Star | 5.350 | 0.880 | 11.08 | 90.3 | 24.40 |
| * chi Phe | 12524 | 9459 | 602 | | | H | hd012524 | K5III | Star | 5.140 | 1.490 | 8.69 | 115.1 | -30.60 |
| HR 616 | 12923 | 9827 | 616 | | | S | hr0616 | K0 | SB* | 6.291 | 0.888 | 5.42 | 184.5 | 15.85 |
| HD 13004 | 13004 | 9862 | | | | S | hd013004 | K1III | Star | 6.390 | 1.150 | 11.14 | 89.8 | -5.01 |
| HD 13263 | 13263 | 9870 | | | | H | hd013263 | G8III | Star | 6.710 | 0.880 | 7.13 | 140.3 | 29.70 |
| * alf Ari | 12929 | 9884 | 617 | | | S | hr0617 | K1IIIb | V* | 2.010 | 1.160 | 49.56 | 20.2 | -14.64 |
| HR 619 | 13013 | 9983 | 619 | | | S | hr0619 | G8III-IV | Star | 6.380 | 0.960 | 7.94 | 125.9 | 25.00 |
| * 14 Ari | 13174 | 10053 | 623 | J02094+2557A | | E | hd013174 | F2III | *in** | 4.983 | 0.319 | 11.30 | 88.5 | -3.90 |
| HR 636 | 13423 | 10111 | 636 | | | H | hd013423 | G8III | Star | 6.324 | 0.874 | 8.93 | 112.0 | 40.50 |
| HR 621 | 13137 | 10115 | 621 | | | S | hr0621 | G8III | Star | 6.307 | 0.947 | 5.68 | 176.1 | 12.50 |
| V* ER Eri | | | | | | H | ereri | K1IIIe | RSCVn | 9.850 | 1.100 | | | 9.00 |
| * iot Tri | 13480 | 10280 | 642 | | | E | hd013480 | G0III+G5III | EllipVar | 4.952 | 0.789 | 11.22 | 89.1 | -19.98 |
| HR 651 | 13692 | 10326 | 651 | | | H | hd013692 | K0III | HB* | 5.873 | 1.001 | 9.40 | 106.4 | 38.40 |
| * b And | 13520 | 10340 | 643 | | | E | hd013520 | K4III | SB* | 4.844 | 1.518 | 6.15 | 162.6 | -48.00 |
| HR 659 | 13940 | 10440 | 659 | | | H | hd013940 | G8III | Star | 5.916 | 0.947 | 9.46 | 105.7 | 14.00 |
| * 67 Cet | 14129 | 10642 | 666 | | | S | hr0666 | G8.5III | Star | 5.509 | 0.948 | 9.80 | 102.0 | 6.60 |
| * 8 Per | 13982 | 10718 | 661 | | | S | hr0661 | K3III | Star | 5.757 | 1.203 | 7.94 | 125.9 | 0.72 |
| HD 14703 | 14703 | 10907 | | | | H | hd014703 | K0III | Star | 6.690 | 0.960 | 7.16 | 139.7 | |
| HR 698 | 14832 | 11043 | 698 | | | H | hd014832 | K0III | Star | 6.300 | 0.979 | 7.31 | 136.8 | -6.60 |
| HD 14830 | 14830 | 11108 | 697 | | | S | hr0697 | G9III | PM* | 6.236 | 0.939 | 7.93 | 126.1 | 19.90 |
| * 65 And | 14872 | 11313 | 699 | J02256+5017A | | S | hr0699 | K4III | *in** | 4.734 | 1.572 | 8.83 | 113.3 | -4.93 |
| * 11 Tri | 15176 | 11432 | 712 | | | S | hr0712 | K1III: | Star | 5.564 | 1.115 | 11.73 | 85.3 | -41.80 |
| * 12 Tri | 15257 | 11486 | 717 | | | E | hd015257 | F0III | Star | 5.290 | 0.290 | 20.07 | 49.8 | -24.80 |
| HR 725 | 15453 | 11603 | 725 | J02296+0934AB | | S | hr0725 | K2III | Star | 6.098 | 1.041 | 9.74 | 102.7 | -10.37 |
| HR 726 | 15464 | 11651 | 726 | | | S | hr0726 | K1III | Star | 6.262 | 1.077 | 8.36 | 119.6 | 5.39 |
| * 27 Ari | 15596 | 11698 | 731 | | | E | hd015596 | G5III-IV | Star | 6.238 | 0.889 | 11.30 | 88.5 | -122.71 |
| * mu. Hyi | 16522 | 11757 | 776 | | | H | hd016522 | G8III | Star | 5.278 | 0.975 | 11.49 | 87.0 | -14.50 |
| * 75 Cet | 15779 | 11791 | 739 | | | S | hr0739 | G3III: | HB* | 5.364 | 1.009 | 12.28 | 81.4 | -6.59 |
| HR 738 | 15755 | 11840 | 738 | | | S | hr0738 | K0III | SB* | 5.846 | 1.081 | 12.01 | 83.3 | 2.84 |
| HD 15866 | 15866 | 11924 | | | | S | hd015866 | G0III | Star | 8.010 | 0.620 | 11.02 | 90.7 | -43.94 |
| HD 15533 | 15533 | 11930 | | | | S | hd015533 | K0 | Star | 6.547 | 1.256 | 9.66 | 103.5 | 15.06 |
| HD 16150 | 16150 | 12111 | | | | S | hd016150 | F6III | Star | 7.260 | 0.530 | 5.02 | 199.2 | 14.00 |
| HR 766 | 16247 | 12148 | 766 | | | S | hr0766 | K0III: | V* | 5.819 | 1.026 | 11.02 | 90.7 | -23.81 |
| HD 16314 | 16314 | 12205 | | | | S | hd016314 | F5III | Star | 7.210 | 0.420 | 8.56 | 116.8 | -21.10 |
| * 81 Cet | 16400 | 12247 | 771 | | | S | hr0771 | G5III: | Star | 5.656 | 1.011 | 10.81 | 92.5 | 8.98 |
| HR 743 | 15920 | 12273 | 743 | | | S | hr0743 | G8III | Star | 5.181 | 0.875 | 12.75 | 78.4 | -3.60 |
| HR 768 | 16327 | 12287 | 768 | J02383+3744A | | S | hr0768 | F6III | *in** | 6.190 | 0.437 | 12.15 | 82.3 | 11.60 |
| * iot Eri | 16815 | 12486 | 794 | | | H | hd016815 | K0III | Star | 4.124 | 0.995 | 21.65 | 46.2 | -9.30 |
| HR 805 | 16975 | 12608 | 805 | | | H | hd016975 | G8III | Star | 5.999 | 0.903 | 7.91 | 126.4 | 17.00 |

| Name | HD | HIP | HR | WDS | Src | Ident | SpType | ObjType | Vmag | B-V | Plx | Dist | RV |
|---|---|---|---|---|---|---|---|---|---|---|---|---|---|
| HD 17374 | 17374 | 12715 | | | H | hd017374 | G8III | Star | 6.660 | 1.000 | 6.23 | 160.5 | |
| HD 17001 | 17001 | 12720 | | | S | hd017001 | K0 | Star | 6.480 | 0.980 | 8.32 | 120.2 | |
| * 36 Ari | 17017 | 12784 | 808 | | S | hr0808 | K2III | Star | 6.400 | 1.145 | 9.66 | 103.5 | -34.29 |
| * 39 Ari | 17361 | 13061 | 824 | | S | hr0824 | K1.5III | PM* | 4.510 | 1.120 | 19.01 | 52.6 | -15.53 |
| HR 831 | 17484 | 13176 | 831 | | S | hr0831 | F6III-IV | Star | 6.452 | 0.397 | 10.35 | 96.6 | 6.90 |
| * gam01 For | 17713 | 13197 | 844 | J02499-2434A | S | hr0844 | K0III | *in** | 6.154 | 1.081 | 7.48 | 133.7 | -6.10 |
| * eta02 For | 17793 | 13225 | 848 | | H | hd017793 | K0III | ** | 5.910 | 0.900 | 7.25 | 137.9 | 21.60 |
| * nu. Hyi | 18293 | 13244 | 872 | | H | hd018293 | K3III | Star | 4.754 | 1.357 | 9.61 | 104.1 | 4.70 |
| * 16 Per | 17584 | 13254 | 840 | J02506+3818A | S | hr0840 | F2III | PulsV*delSct | 4.200 | 0.370 | 27.01 | 37.0 | 14.00 |
| * tau02 Eri | 17824 | 13288 | 850 | J02510-2100A | S | hr0850 | K0III | Star | 4.770 | 0.900 | 17.45 | 57.3 | -8.60 |
| HR 856 | 17918 | 13448 | 856 | | S | hr0856 | F5III | Star | 6.300 | 0.440 | 9.33 | 107.2 | 9.00 |
| * eta Eri | 18322 | 13701 | 874 | | H | hd018322 | K1III | V* | 3.870 | 1.120 | 23.89 | 41.9 | -20.32 |
| HR 900 | 18650 | 13907 | 900 | | S | hr0900 | K1III | Star | 6.144 | 1.029 | 7.03 | 142.2 | 14.50 |
| * rho01 Eri | 18784 | 14060 | 907 | | S | hr0907 | K0II | Star | 5.754 | 1.039 | 10.07 | 99.3 | 13.68 |
| HR 908 | 18832 | 14104 | 908 | | S | hr0908 | K0 | Star | 6.258 | 1.039 | 8.36 | 119.6 | -59.40 |
| HR 912 | 18885 | 14110 | 912 | | E | hd018885 | G6III: | Star | 5.834 | 1.105 | 9.89 | 101.1 | 14.33 |
| HR 926 | 19121 | 14315 | 926 | | S | hr0926 | K0III | Star | 6.051 | 1.042 | 7.58 | 131.9 | 1.57 |
| HD 19210 | 19210 | 14355 | | | S | hd019210 | K0 | Star | 6.310 | 1.030 | 7.12 | 140.4 | |
| HR 931 | 19270 | 14439 | 931 | | S | hr0931 | K3III | Star | 5.648 | 1.088 | 10.41 | 96.1 | -15.87 |
| * kap Per | 19476 | 14668 | 941 | J03095+4451A | S | hr0941 | K0III | V* | 3.810 | 0.980 | 28.93 | 34.6 | 29.40 |
| HR 946 | 19637 | 14748 | 946 | | S | hr0946 | K3III | Star | 6.038 | 1.279 | 7.05 | 141.8 | -16.21 |
| HD 20037 | 20037 | 14765 | | | H | hd020037 | G8III | Star | 6.620 | 0.890 | 5.98 | 167.2 | |
| * ome Per | 19656 | 14817 | 947 | J03114+3936A | S | hr0947 | K0III | *in** | 4.614 | 1.122 | 11.32 | 88.3 | 6.61 |
| * del Ari | 19787 | 14838 | 951 | | S | hr0951 | K2III | V* | 4.370 | 1.030 | 19.22 | 52.0 | 22.81 |
| HR 956 | 19845 | 15004 | 956 | | S | hr0956 | G9III | *inCl | 5.921 | 0.986 | 8.58 | 116.6 | -11.00 |
| V* TZ For | 20301 | 15092 | | | H | tzfor | F7III+G8III | EB*Algol | 6.886 | 0.715 | 5.75 | 173.9 | 18.00 |
| * 15 Eri | 20610 | 15382 | 994 | J03184-2231AB | S | hr0994 | K0III | ** | 4.875 | 0.887 | 12.77 | 78.3 | 23.90 |
| * 95 Cet | 20559 | 15383 | 992 | J03184-0056AB | S | hr0992 | G9IV | ** | 5.370 | 1.040 | 14.89 | 67.2 | 31.20 |
| * 60 Ari | 20663 | 15557 | 1000 | | S | hr1000 | K3III: | Star | 6.142 | 1.253 | 10.94 | 91.4 | 23.63 |
| * kap02 Cet | 20791 | 15619 | 1007 | | S | hr1007 | G8.5III | Star | 5.690 | 0.970 | 10.11 | 98.9 | 7.44 |
| HR 1016 | 20894 | 15643 | 1016 | | H | hd020894 | G8III | Star | 5.504 | 0.862 | 8.89 | 112.5 | 7.80 |
| HR 1021 | 21011 | 15655 | 1021 | | H | hd021011 | K0III | Star | 6.405 | 0.980 | 6.85 | 146.0 | -2.20 |
| * 63 Ari | 20893 | 15737 | 1015 | | S | hr1015 | K3III | ** | 5.089 | 1.250 | 10.27 | 97.4 | 2.99 |
| * 64 Ari | 21017 | 15861 | 1022 | | S | hr1022 | K4III | Star | 5.504 | 1.229 | 15.68 | 63.8 | 8.49 |
| * omi Tau | 21120 | 15900 | 1030 | | S | hr1030 | G6III | SB* | 3.600 | 0.890 | 11.21 | 89.2 | -24.40 |
| HR 1045 | 21430 | 16029 | 1045 | | H | hd021430 | G5III | Star | 5.939 | 0.913 | 7.22 | 138.5 | 15.70 |
| HR 1050 | 21530 | 16142 | 1050 | | S | hr1050 | K2III | RGB* | 5.743 | 1.102 | 10.24 | 97.7 | -1.80 |
| HD 21585 | 21585 | 16226 | | | S | hd021585 | G5 | Star | 6.420 | 0.780 | 6.73 | 148.6 | |
| HR 1060 | 21665 | 16266 | 1060 | | S | hr1060 | G5 | Star | 5.996 | 1.009 | 8.42 | 118.8 | 13.70 |
| HD 21760 | 21760 | 16274 | | | S | hd021760 | K1III | Star | 6.780 | 1.090 | 10.05 | 99.5 | |
| HR 1109 | 22676 | 16290 | 1109 | | H | hd022676 | G8III | Star | 5.683 | 0.925 | 10.17 | 98.3 | 10.00 |
| * sig Per | 21552 | 16335 | 1052 | | S | hr1052 | K3III | V* | 4.360 | 1.350 | 9.07 | 110.3 | 14.36 |
| HR 1090 | 22231 | 16509 | 1090 | | H | hd022231 | K2III | Star | 5.666 | 1.103 | 10.44 | 95.8 | 37.00 |
| HR 1098 | 22409 | 16780 | 1098 | | S | hr1098 | G7III: | HB* | 5.589 | 0.892 | 8.63 | 115.9 | 36.50 |
| * y Eri | 22663 | 16870 | 1106 | | H | hd022663 | K1III | Star | 4.580 | 1.040 | 14.13 | 70.8 | 11.50 |
| HR 1108 | 22675 | 16989 | 1108 | | S | hr1108 | G5III: | RGB* | 5.855 | 0.970 | 8.90 | 112.4 | -29.60 |
| HR 1117 | 22799 | 17057 | 1117 | | S | hr1117 | G5 | Star | 6.188 | 1.026 | 9.14 | 109.4 | 22.60 |
| HR 1110 | 22695 | 17058 | 1110 | J03394+1632A | S | hr1110 | K0III | *in** | 6.182 | 0.964 | 7.68 | 130.2 | 14.70 |
| HR 1119 | 22819 | 17120 | 1119 | | S | hr1119 | G5 | Star | 6.128 | 0.980 | 8.38 | 119.3 | 25.20 |
| HR 1112 | 22764 | 17342 | 1112 | J03427+5958A | E | hd022764 | K5III | *in** | 5.780 | 1.760 | 1.58 | 632.9 | -12.53 |

| Name | HD | HIP | HR | WDS | Cluster | Src | ID | SpType | Type | Vmag | B-V | Plx | pmRA | pmDE |
|---|---|---|---|---|---|---|---|---|---|---|---|---|---|---|
| * h Eri | 23319 | 17351 | 1143 | J03428-3719A | | H | hd023319 | K2.5III | HB* | 4.595 | 1.198 | 17.70 | 56.5 | 9.90 |
| * del Eri | 23249 | 17378 | 1136 | | | E | hd023249 | K1III-IV | RSCVn | 3.540 | 0.920 | 110.61 | 9.0 | -6.28 |
| * 14 Tau | 23183 | 17408 | 1132 | | | S | hr1132 | G8III | Star | 6.140 | 1.000 | 7.96 | 125.6 | 78.35 |
| HR 1169 | 23719 | 17534 | 1169 | | | H | hd023719 | K1III | RGB* | 5.724 | 0.946 | 9.66 | 103.5 | -3.00 |
| HR 1159 | 23526 | 17595 | 1159 | | | S | hr1159 | G9III | Star | 5.926 | 0.964 | 9.73 | 102.8 | -23.40 |
| * rho For | 23940 | 17738 | 1184 | | | H | hd023940 | G6III | RGB* | 5.540 | 0.980 | 12.13 | 82.4 | 52.60 |
| * g Eri | 24160 | 17874 | 1195 | | | H | hd024160 | G9III | Star | 4.170 | 0.950 | 15.54 | 64.4 | 2.00 |
| HR 1216 | 24706 | 18199 | 1216 | | | H | hd024706 | K2III | Star | 5.927 | 1.248 | 9.29 | 107.6 | -0.80 |
| HR 1219 | 24744 | 18262 | 1219 | J03544-4021AB | | H | hd024744 | K0III+... | Star | 5.716 | 0.558 | 6.53 | 153.1 | 2.10 |
| HR 1265 | 25723 | 19011 | 1265 | | | S | hr1265 | K1III | Star | 5.622 | 1.067 | 8.48 | 117.9 | 26.75 |
| HD 25627 | 25627 | 19036 | | | | S | hd025627 | K2III | Star | 6.650 | 1.100 | 9.91 | 100.9 | 12.19 |
| HR 1267 | 25803 | 19037 | 1267 | | | S | hr1267 | K1II | Star | 6.119 | 1.167 | 7.99 | 125.2 | -4.40 |
| * 37 Tau | 25604 | 19038 | 1256 | J04047+2204A | | S | hr1256 | K0III | *in** | 4.370 | 1.070 | 17.43 | 57.4 | 9.52 |
| HD 26004 | 26004 | 19149 | | | | S | hd026004 | K0III | Star | 7.630 | 1.200 | 4.93 | 202.8 | -12.60 |
| HR 1255 | 25602 | 19172 | 1255 | | | E | hd025602 | K0III-IV | Star | 6.320 | 0.972 | 12.37 | 80.8 | -8.84 |
| * ome01 Tau | 26162 | 19388 | 1283 | | | S | hr1283 | K2III | Star | 5.510 | 1.086 | 11.49 | 87.0 | 24.83 |
| HD 26625 | 26625 | 19601 | | | | S | hd026625 | K0III | Star | 6.400 | 0.980 | 9.08 | 110.1 | |
| HR 1295 | 26546 | 19641 | 1295 | | | S | hr1295 | K0III | Star | 6.091 | 1.074 | 9.52 | 105.0 | 25.58 |
| HR 1310 | 26703 | 19736 | 1310 | | | S | hr1310 | K0 | Star | 6.250 | 1.150 | 10.06 | 99.4 | 44.81 |
| HR 1301 | 26605 | 19746 | 1301 | | | S | hr1301 | G9III | Star | 6.450 | 1.050 | 6.61 | 151.3 | 29.00 |
| * alf Hor | 26967 | 19747 | 1326 | | | H | hd026967 | K2III | PM* | 3.860 | 1.100 | 28.36 | 35.3 | 21.60 |
| * 39 Eri | 26846 | 19777 | 1318 | J04144-1016AB | | S | hr1318 | K3III | ** | 4.860 | 1.180 | 13.46 | 74.3 | 5.85 |
| * alf Ret | 27256 | 19780 | 1336 | J04144-6228A | | H | hd027256 | G8II-III | *in** | 3.360 | 0.910 | 20.18 | 49.6 | 35.50 |
| * 48 Tau | 26911 | 19877 | 1319 | J04158+1525A | Melotte 25 S 20 | S | hr1319 | F5V | EllipVar | 6.300 | 0.405 | 22.02 | 45.4 | 36.40 |
| * eps Ret | 27442A | 19921 | 1355 | J04165-5918A | | H | epsret | K2III | *in** | 4.440 | 1.080 | 54.83 | 18.2 | 29.30 |
| HR 1313 | 26755 | 19983 | 1313 | | | S | hr1313 | K1III | Star | 5.727 | 1.108 | 12.53 | 79.8 | -38.20 |
| * gam Tau | 27371 | 20205 | 1346 | J04198+1538AB | Melotte 25 MMU 28 | S | hr1346 | K0III | ** | 3.650 | 0.990 | 20.19 | 49.5 | 38.56 |
| * phi Tau | 27382 | 20250 | 1348 | J04203+2721A | | S | hr1348 | K1III | *in** | 4.957 | 1.154 | 10.16 | 98.4 | 1.27 |
| * 54 Per | 27348 | 20252 | 1343 | J04204+3435A | | S | hr1343 | G8III | *in** | 4.933 | 0.936 | 14.94 | 66.9 | -26.82 |
| HR 1327 | 27022 | 20266 | 1327 | | | S | hr1327 | G0II | Star | 5.270 | 0.810 | 10.21 | 97.9 | -19.47 |
| * eta Ret | 28093 | 20384 | 1395 | | | H | hd028093 | G8III | PM* | 5.240 | 0.960 | 8.48 | 117.9 | 45.00 |
| * del Tau | 27697 | 20455 | 1373 | J04229+1733A | Melotte 25 MMU 41 | S | hr1373 | K0III | SB* | 3.760 | 0.980 | 20.96 | 47.7 | 39.42 |
| * d Eri | 28028 | 20535 | 1393 | | | H | hd028028 | K4III | Star | 3.960 | 1.490 | 11.01 | 90.8 | 24.10 |
| HR 1413 | 28322 | 20848 | 1413 | | | S | hr1413 | G9III | Star | 6.154 | 1.007 | 10.61 | 94.3 | 31.70 |
| * 75 Tau | 28292 | 20877 | 1407 | | Melotte 25 HAN 431 | S | hr1407 | K2III | V* | 4.971 | 1.144 | 17.47 | 57.2 | 16.24 |
| * tet01 Tau | 28307 | 20885 | 1411 | J04286+1554B | Melotte 25 S 34 | S | hr1411 | K0IIIb | SB* | 3.840 | 0.940 | 21.13 | 47.3 | 38.79 |
| * eps Tau | 28305 | 20889 | 1409 | J04286+1911A | Melotte 25 S 3 | S | hr1409 | K0III | Star | 3.530 | 1.010 | 22.24 | 45.0 | 38.43 |
| HR 1421 | 28479 | 20892 | 1421 | | | S | hr1421 | K2III | Star | 5.951 | 1.228 | 10.58 | 94.5 | 26.66 |
| HR 1304 | 26659 | 20982 | 1304 | | | E | hd026659 | G8III | SB* | 5.477 | 0.828 | 10.61 | 94.3 | -26.81 |
| HR 1425 | 28505 | 20985 | 1425 | | | S | hr1425 | G8III | Star | 6.460 | 1.040 | 8.37 | 119.5 | -62.40 |
| HR 1431 | 28625 | 20997 | 1431 | | | S | hr1431 | K0III | Star | 6.219 | 1.007 | 6.96 | 143.7 | 13.60 |
| * ups01 Eri | 29085 | 21248 | 1453 | | | S | hr1453 | K0III-IV | V* | 4.510 | 0.980 | 25.67 | 39.0 | 20.60 |
| HR 1475 | 29399 | 21253 | 1475 | J04336-6249A | | H | hd029399 | K1III | V* | 5.787 | 1.030 | 22.15 | 45.1 | 30.80 |
| HD 29751 | 29751 | 21300 | | | | H | hd029751 | G8III | Star | 6.819 | 0.934 | 5.46 | 183.2 | |
| * ups02 Eri | 29291 | 21393 | 1464 | | | H | hd029291 | G8IIIa | Star | 3.820 | 0.980 | 15.25 | 65.6 | -4.00 |
| HR 1455 | 29104 | 21408 | 1455 | J04357+1953AB | | S | hr1455 | G5II-III+... | Star | 6.350 | 0.720 | 7.47 | 133.9 | -2.00 |
| * alf Tau | 29139 | 21421 | 1457 | J04359+1631A | | E | hd029139 | K5III | PulsV* | 0.860 | 1.540 | 48.94 | 20.4 | 54.26 |
| * lam Pic | 30185 | 21914 | 1516 | | | H | hd030185 | K0.5III | Star | 5.307 | 0.961 | 8.71 | 114.8 | 4.60 |
| HR 1517 | 30197 | 22176 | 1517 | | NGC 1647 H 125 | S | hr1517 | K4III | SB* | 5.988 | 1.224 | 12.47 | 80.2 | 44.00 |
| HR 1514 | 30138 | 22220 | 1514 | | | S | hr1514 | G9III | Star | 6.000 | 0.930 | 7.53 | 132.8 | 40.20 |

| Name | HD | HIP | HR | CCDM | Source | ID | SpType | ObjType | Vmag | B-V | Plx | pmRA | pmDE |
|---|---|---|---|---|---|---|---|---|---|---|---|---|---|
| HR 1529 | 30454 | 22393 | 1529 | | S | hr1529 | K1III | Star | 5.577 | 1.134 | 15.19 | 65.8 | 22.22 |
| * 1 Aur | 30504 | 22453 | 1533 | | S | hr1533 | K4III | Star | 4.884 | 1.479 | 6.48 | 154.3 | -24.65 |
| * 60 Eri | 30814 | 22479 | 1549 | | S | hr1549 | K0III | RGB* | 5.037 | 0.988 | 13.83 | 72.3 | 40.56 |
| HR 1535 | 30557 | 22545 | 1535 | | S | hr1535 | G9III | Star | 5.642 | 1.004 | 10.29 | 97.2 | 29.31 |
| * 2 Aur | 30834 | 22678 | 1551 | | E | hd030834 | K3III | Star | 4.787 | 1.424 | 5.42 | 184.5 | -17.24 |
| V* R Eri | 31444 | 22881 | 1581 | | E | hd031444 | G7III | V* | 5.726 | 0.840 | 11.72 | 85.3 | 39.81 |
| * omi02 Ori | 31421 | 22957 | 1580 | J04564+1331A | S | hr1580 | K2IIIb | *in** | 4.060 | 1.150 | 17.54 | 57.0 | 2.54 |
| * ksi Men | 34172 | 23148 | 1716 | | H | hd034172 | G9III | Star | 5.841 | 0.914 | 8.84 | 113.1 | -5.00 |
| HR 1631 | 32453 | 23377 | 1631 | | H | hd032453 | G5III | Star | 6.025 | 0.872 | 8.94 | 111.9 | 6.00 |
| HR 1628 | 32436 | 23430 | 1628 | | S | hr1628 | K0III | Star | 5.016 | 1.065 | 13.99 | 71.5 | 27.40 |
| HR 1635 | 32515 | 23446 | 1635 | | H | hd032515 | G8III | Star | 5.921 | 1.189 | 10.47 | 95.5 | 29.00 |
| * bet Men | 33285 | 23467 | 1677 | | H | hd033285 | G8III | Star | 5.302 | 0.992 | 4.11 | 243.3 | -11.40 |
| HR 1625 | 32393 | 23475 | 1625 | | S | hr1625 | K3 | Star | 5.866 | 1.218 | 10.88 | 91.9 | 38.09 |
| * gam02 Cae | 32846 | 23596 | 1653 | | H | xcae | F1III | PulsV*delSct | 6.335 | 0.270 | 10.17 | 98.3 | 6.40 |
| * eps Lep | 32887 | 23685 | 1654 | | S | hr1654 | K4III | V* | 3.180 | 1.460 | 15.29 | 65.4 | 1.00 |
| HR 1681 | 33419 | 24041 | 1681 | | S,E | hr1681,hd033419 | K0III | Star | 6.118 | 1.089 | 10.03 | 99.7 | 25.17 |
| HD 33844 | 33844 | 24275 | | | S | hd033844 | K0III | RGB* | 7.290 | 1.040 | 9.91 | 100.9 | 36.18 |
| * rho Ori | 33856 | 24331 | 1698 | | S | hr1698 | K0.5III | SB* | 4.440 | 1.190 | 9.32 | 107.3 | 40.50 |
| * tet Dor | 34649 | 24372 | 1744 | | H | hd034649 | K2/3III | Star | 4.812 | 1.309 | 6.64 | 150.6 | 10.50 |
| HD 34253 | 34253 | 24424 | | | H | hd034253 | G8III+... | Star | 9.050 | 0.760 | 19.20 | 52.1 | |
| HR 1721 | 34266 | 24426 | 1721 | | H | hd034266 | G8III | RGB* | 5.743 | 0.998 | 6.56 | 152.4 | 13.20 |
| HR 1688 | 33618 | 24479 | 1688 | | E | hd033618 | K2III-IV | Star | 6.147 | 1.196 | 9.17 | 109.1 | -0.53 |
| * alf Aur | 34029 | 24608 | 1708 | J05168+4559AP | E | hd034029 | G1III+K0III | SB* | 0.080 | 0.800 | 76.20 | 13.1 | 29.19 |
| * 16 Aur | 34334 | 24727 | 1726 | J05182+3322AB | S | hr1726 | K2.5IIIb | SB* | 4.552 | 1.276 | 14.04 | 71.2 | -28.44 |
| * n Tau | 34559 | 24822 | 1739 | | S | hr1739 | G8III | Star | 4.956 | 0.932 | 13.19 | 75.8 | 19.03 |
| * e Ori | 35369 | 25247 | 1784 | | S | hr1784 | G8III | Star | 4.120 | 0.960 | 20.73 | 48.2 | -17.68 |
| * lam Dor | 36189 | 25429 | 1836 | | H | hd036189 | G6III | Star | 5.132 | 0.973 | 7.12 | 140.4 | 10.00 |
| HR 1796 | 35521 | 25475 | 1796 | | S | hr1796 | K0 | Star | 6.175 | 1.156 | 9.92 | 100.8 | -7.52 |
| HD 35929 | 35929 | 25546 | | | H | hd035929 | F2III | Em* | 8.110 | 0.440 | 2.78 | 359.7 | |
| HD 35984 | 35984 | 25730 | 1822 | | S | hr1822 | F6III | TTau* | 6.200 | 0.431 | 11.27 | 88.7 | 13.30 |
| HR 1831 | 36160 | 25806 | 1831 | | S | hr1831 | K0 | Star | 6.299 | 1.178 | 9.13 | 109.5 | 1.89 |
| * eps Col | 36597 | 25859 | 1862 | | H | hd036597 | K1IIIa | Star | 3.870 | 1.140 | 12.39 | 80.7 | -4.90 |
| HR 1870 | 36734 | 25887 | 1870 | | H | hd036734 | K3III | Star | 5.861 | 1.368 | 5.95 | 168.1 | 6.80 |
| * gam Men | 37763 | 25918 | 1953 | J05319-7620A | U | hd037763 | K2III | *in** | 5.200 | 1.130 | 31.89 | 31.4 | 56.70 |
| HR 1877 | 36848 | 25993 | 1877 | | H | hd036848 | K2III | Star | 5.460 | 1.240 | 18.93 | 52.8 | -0.60 |
| HBC 463 | | | | | U | p1955 | G2III | Em* | 11.225 | 0.818 | | | |
| V* TX Pic | 37434 | 26300 | 1927 | | U | txpic | K2III | RSCVn | 6.106 | 1.162 | 6.45 | 155.0 | 15.80 |
| HR 1889 | 36994 | 26332 | 1889 | | S | hr1889 | F5III | Star | 6.516 | 0.408 | 11.66 | 85.8 | 1.50 |
| * phi02 Ori | 37160 | 26366 | 1907 | | E | hd037160 | G9IV | PM* | 4.090 | 0.950 | 27.76 | 36.0 | 98.96 |
| HR 1958 | 37811 | 26649 | 1958 | | H,U | hd037811,hd037811 | G7III | RGB* | 5.441 | 0.904 | 7.73 | 129.4 | -8.30 |
| HR 1954 | 37784 | 26853 | 1954 | | S | hr1954 | K2 | Star | 6.350 | 1.210 | 10.03 | 99.7 | -22.26 |
| * b Ori | 37984 | 26885 | 1963 | | S | hr1963 | K1III | Star | 4.900 | 1.170 | 11.11 | 90.0 | 87.55 |
| * 23 Cam | 37638 | 27046 | 1943 | | E | hd037638 | G5III: | Star | 6.168 | 0.881 | 9.01 | 111.0 | -2.63 |
| HR 1978 | 38309 | 27118 | 1978 | J05450+0400A | E | hd038309 | F0III:n | *in** | 6.105 | 0.297 | 16.99 | 58.9 | 12.50 |
| HR 1986 | 38495 | 27212 | 1986 | | S | hr1986 | K1III+... | *in** | 6.310 | 1.076 | 8.97 | 111.5 | -45.60 |
| HR 1987 | 38527 | 27280 | 1987 | J05469+0931A | S | hr1987 | G8III | *in** | 5.787 | 0.880 | 10.95 | 91.3 | -25.46 |
| * tau Aur | 38656 | 27483 | 1995 | J05492+3911A | S | hr1995 | G8III | *in** | 4.500 | 0.950 | 15.77 | 63.4 | -19.32 |
| CCDM J05496-1429AB | 39070 | 27517 | 2021 | J05496-1429AB | E | hd039070 | G8III | ** | 5.490 | 0.880 | 12.19 | 82.0 | -1.44 |
| HR 2049 | 39640 | 27621 | 2049 | | H | hd039640 | G8III | Star | 5.167 | 0.965 | 12.72 | 78.6 | 1.30 |

| Name | HD | HIP | HR | 2MASS/other | Src | ID | SpType | ObjType | V | B-V | plx | pmRA | pmDE |
|---|---|---|---|---|---|---|---|---|---|---|---|---|---|
| * bet Col | 39425 | 27628 | 2040 | | H | hd039425 | K1IIICN+1 | PM* | 3.120 | 1.160 | 37.41 | 26.7 | 89.40 |
| * nu. Aur | 39003 | 27673 | 2012 | J05515+3909A | S | hr2012 | G9.5III | *in** | 3.950 | 1.140 | 14.16 | 70.6 | 9.92 |
| HD 39833 | 39833 | 27980 | | | E | hd039833 | G0III | Star | 7.660 | 0.620 | 24.26 | 41.2 | 24.63 |
| * ksi Col | 40176 | 28010 | 2087 | | H | hd040176 | K1III | Star | 4.970 | 1.096 | 9.80 | 102.0 | 59.50 |
| HR 2070 | 39910 | 28011 | 2070 | | E | hd039910 | K2III: | Star | 5.874 | 1.167 | 10.31 | 97.0 | 27.49 |
| HR 2076 | 40020 | 28139 | 2076 | | S | hr2076 | K2III | Star | 5.891 | 1.124 | 10.57 | 94.6 | 18.69 |
| * eta Col | 40808 | 28328 | 2120 | | H | hd040808 | K0III | Star | 3.960 | 1.140 | 6.91 | 144.7 | 17.00 |
| * del Aur | 40035 | 28358 | 2077 | J05595+5418A | S | hr2077 | K0III | *in** | 3.720 | 1.010 | 25.88 | 38.6 | 9.75 |
| HR 2080 | 40083 | 28390 | 2080 | | S | hr2080 | K2III | Star | 6.136 | 1.217 | 10.62 | 94.2 | -5.97 |
| HR 2113 | 40657 | 28413 | 2113 | | H | hd040657 | K1.5III | V* | 4.520 | 1.220 | 7.75 | 129.0 | 25.90 |
| HR 2131 | 41047 | 28524 | 2131 | | H | hd041047 | K5III | Star | 5.550 | 1.598 | 5.31 | 188.3 | 19.00 |
| HR 2136 | 41125 | 28622 | 2136 | | S | hr2136 | K0III | Star | 6.196 | 0.927 | 7.98 | 125.3 | 44.70 |
| * 38 Aur | 40801 | 28677 | 2119 | | S,E | hr2119,hd040801 | K0III | PM* | 6.092 | 0.982 | 15.33 | 65.2 | 33.69 |
| HD 42719 | 42719 | 29193 | | | H | hd042719 | G6:III/IV:+... | Star | 7.500 | 0.660 | 14.17 | 70.6 | 34.70 |
| 2MASS J06093315+0441123 | | | | | H | lra03_e2_0678 | G8III | Star~ | | | | | |
| HR 2183 | 42341 | 29205 | 2183 | | S,E | hr2183,hd042341 | K2III | Star | 5.563 | 1.150 | 15.67 | 63.8 | 32.68 |
| * 37 Cam | 41597 | 29246 | 2152 | | S | hr2152 | G8III | Star | 5.359 | 1.099 | 8.56 | 116.8 | 30.91 |
| CD-33 2771 | | | | | H | cd-33_2771 | K5III+K5V | EB* | 9.860 | 1.480 | | | |
| HR 2200 | 42621 | 29294 | 2200 | | S | hr2200 | K1III | V* | 5.708 | 1.074 | 9.89 | 101.1 | 1.00 |
| 2MASS J06135076+0518086 | | | | | H | lra03_e2_1326 | G0III | Star~ | | | | | |
| HR 2218 | 43023 | 29575 | 2218 | | S | hr2218 | G8III | Star | 5.835 | 0.905 | 10.61 | 94.3 | 52.10 |
| HR 2243 | 43429 | 29692 | 2243 | | S | hr2243 | K1III | Star | 5.990 | 1.054 | 15.99 | 62.5 | 64.00 |
| * kap Aur | 43039 | 29696 | 2219 | | S | hr2219 | G8.5IIIb | V* | 4.350 | 1.010 | 18.43 | 54.3 | 20.69 |
| * 8 Gem | 43261 | 29789 | 2230 | | S | hr2230 | G8III | Star | 6.083 | 0.899 | 7.30 | 137.0 | -20.90 |
| * kap Col | 43785 | 29807 | 2256 | | H | hd043785 | K0.5IIIa | V* | 4.370 | 1.000 | 17.87 | 56.0 | 24.20 |
| * 43 Aur | 43380 | 29949 | 2239 | | S | hr2239 | K2III | Star | 6.334 | 1.141 | 8.25 | 121.2 | -3.35 |
| HR 2259 | 43821 | 29982 | 2259 | | S | hr2259 | K0 | SB* | 6.241 | 0.841 | 10.12 | 98.8 | -24.90 |
| HR 2287 | 44497 | 30318 | 2287 | | S | hr2287 | F0III | Star | 6.000 | 0.300 | 14.32 | 69.8 | 20.90 |
| * pi.01 Dor | 45669 | 30321 | 2352 | | H | hd045669 | K5III | Star | 5.566 | 1.536 | 5.27 | 189.8 | 15.70 |
| HR 2302 | 44867 | 30517 | 2302 | | S | hr2302 | G9III | Star | 6.350 | 1.040 | 7.09 | 141.0 | 71.20 |
| * pi.02 Dor | 46116 | 30565 | 2377 | | H | hd046116 | G8III | PM* | 5.380 | 0.970 | 12.08 | 82.8 | 9.10 |
| HR 2333 | 45415 | 30728 | 2333 | | S | hr2333 | G9III | Star | 5.553 | 1.030 | 10.59 | 94.4 | 52.94 |
| HD 46415 | 46415 | 31061 | | | H | hd046415 | G8III | Star | 6.689 | 0.974 | 7.44 | 134.4 | -12.80 |
| HR 2379 | 46184 | 31084 | 2379 | J06314-1224A | S | hr2379 | K1III | *in** | 5.166 | 1.281 | 8.90 | 112.4 | 17.20 |
| HR 2399 | 46568 | 31165 | 2399 | | H | hd046568 | G8III | Star | 5.257 | 0.972 | 11.94 | 83.8 | 39.00 |
| V* HR CMa | 46407 | 31205 | 2392 | | S | hr2392 | G9.5III: | EB*Algol | 6.240 | 1.110 | 8.11 | 123.3 | -3.45 |
| HR 2416 | 47001 | 31265 | 2416 | | H | hd047001 | G8III | Star | 6.190 | 1.075 | 5.44 | 183.8 | 5.40 |
| HR 2391 | 46374 | 31277 | 2391 | | E | hd046374 | K2III: | Star | 5.562 | 1.117 | 12.98 | 77.0 | -11.24 |
| * nu.01 CMa | 47138 | 31564 | 2423 | J06364-1840A | E | hd047138 | G9III+... | *in** | 5.704 | 0.815 | 9.30 | 107.5 | 26.13 |
| * nu.02 CMa | 47205 | 31592 | 2429 | | H | hip031592 | K1III | V* | 3.910 | 1.101 | 50.63 | 19.8 | 2.45 |
| HD 47910 | 47910 | 31661 | | J06376-5521A | H | hd047910 | G8III | *in** | 6.890 | 0.980 | 6.18 | 161.8 | |
| HR 2437 | 47366 | 31674 | 2437 | | E | hd047366 | K1III: | Star | 6.117 | 0.976 | 12.50 | 80.0 | 8.84 |
| HR 2447 | 47536 | 31688 | 2447 | | H | hd047536 | K0III | Star | 5.258 | 1.172 | 8.11 | 123.3 | 78.80 |
| HR 2531 | 49947 | 32222 | 2531 | | H | hd049947 | G8III | Star | 6.360 | 0.960 | 7.31 | 136.8 | 16.00 |
| * 30 Gem | 48433 | 32249 | 2478 | J06440+1314A | S | hr2478 | K1III | *in** | 4.490 | 1.160 | 11.29 | 88.6 | 12.55 |
| 2MASS J06443588+0000283 | | | | | H | lra01_e1_0286 | K6III | Star~ | | | | | |
| CoRoT 102755837 | | | | | H | lra01_e2_2249 | K4III | Star~ | | | | | |

| Name | HD | HIP | HR | WDS | Src | ID | SpType | Type | Vmag | B-V | Plx | Dist | RV |
|---|---|---|---|---|---|---|---|---|---|---|---|---|---|
| HD 49068 | 49068 | 32393 | | | NGC 2287 MMU 75 | H | ngc2287no75 | K0/K1III: | *inCl | 7.430 | 1.280 | 1.56 | 641.0 | 23.94 |
| HD 49050 | 49050 | | | | NGC 2287 MMU 87 | H | ngc2287no87 | K2/K3III: | *inCl | 7.850 | 1.410 | | | 46.00 |
| HD 49091 | 49091 | 32406 | | | NGC 2287 HFMR 21 | H | ngc2287no21 | K3III | SB* | 6.868 | 1.507 | 1.53 | 653.6 | 23.22 |
| HD 49105 | 49105 | 32422 | | | NGC 2287 MMU 97 | H | ngc2287no97 | K0/K1III | SB* | 7.800 | 1.160 | 1.62 | 617.3 | 22.83 |
| HD 49212 | 49212 | 32467 | | | NGC 2287 MMU 107 | H | ngc2287no107 | G9III | SB* | 7.790 | 1.150 | 1.21 | 826.4 | 23.34 |
| * 13 Lyn | 48432 | 32489 | 2477 | | | S | hr2477 | K0III | Star | 5.349 | 0.951 | 15.46 | 64.7 | 18.29 |
| HD 49334 | 49334 | | | | NGC 2287 MMU 204 | H | ngc2287no204 | K1III | SB* | 7.790 | 1.260 | | | 6.57 |
| V* V448 Car | 49877 | 32531 | 2526 | | | H | hd049877 | K5III | semi-regV* | 5.610 | 1.580 | 5.42 | 184.5 | 35.00 |
| * tau Pup | 50310 | 32768 | 2553 | | | H | hd050310 | K1III | SB* | 2.930 | 1.200 | 17.92 | 55.8 | 34.40 |
| HR 2552 | 50282 | 32990 | 2552 | | | S | hr2552 | K0 | Star | 6.298 | 0.952 | 7.09 | 141.0 | 31.10 |
| * tet CMa | 50778 | 33160 | 2574 | | | S | hr2574 | K4III | Star | 4.080 | 1.430 | 12.51 | 79.9 | 96.20 |
| HD 50890 | 50890 | 33243 | 2582 | | | H | hd050890 | K2III | Star | 6.029 | 1.090 | 2.99 | 334.4 | 19.50 |
| HR 2556 | 50384 | 33271 | 2556 | | | S | hr2556 | K0III-IV | Star | 6.325 | 0.912 | 7.26 | 137.7 | 27.66 |
| HR 2573 | 50763 | 33415 | 2573 | | | S | hr2573 | K0III: | Star | 5.884 | 1.092 | 9.33 | 107.2 | 38.56 |
| HR 2586 | 51000 | 33421 | 2586 | | | E | hd051000 | G5III | V* | 5.926 | 0.860 | 7.72 | 129.5 | -9.30 |
| * 15 Lyn | 50522 | 33449 | 2560 | J06573+5825AB | | E | hd050522 | G5III-IV | Star | 4.350 | 0.850 | 18.29 | 54.7 | 1.86 |
| HD 54038 | 54038 | 34201 | | | | H | hd054038 | G8III | Star | 6.720 | 0.920 | 8.57 | 116.7 | |
| HR 2642 | 52708 | 34250 | 2642 | | | S | hr2642 | G8III: | Star | 6.390 | 1.180 | 8.41 | 118.9 | 18.90 |
| HR 2698 | 54732 | 34349 | 2698 | | | H | hd054732 | K0III | Star | 5.968 | 0.977 | 5.99 | 166.9 | 29.00 |
| HR 2682 | 54079 | 34387 | 2682 | | | E | hd054079 | K0III: | Star | 5.748 | 1.172 | 5.48 | 182.5 | 24.30 |
| * 45 Gem | 54131 | 34440 | 2684 | J07084+1556A | | S | hr2684 | G8III | *in** | 5.490 | 1.018 | 10.11 | 98.9 | -18.48 |
| * 20 Mon | 54810 | 34622 | 2701 | J07102-0414A | | S | hr2701 | K0III | PM* | 4.920 | 1.030 | 16.08 | 62.2 | 77.52 |
| HR 2728 | 55730 | 35005 | 2728 | | | S | hr2728 | G6III | Star | 5.710 | 1.003 | 11.08 | 90.3 | 29.40 |
| * 18 Lyn | 55280 | 35146 | 2715 | | | E | hd055280 | K2III | PM* | 5.200 | 1.070 | 16.39 | 61.0 | 23.92 |
| * 145 CMa | 56577 | 35210 | 2764 | J07166-2319A | | S | hr2764 | K4III | *in** | 4.790 | 1.710 | 2.30 | 434.8 | 29.00 |
| * 65 Aur | 57264 | 35710 | 2793 | J07220+3646A | | S | hr2793 | G8III | *in** | 5.140 | 1.079 | 13.82 | 72.4 | 21.81 |
| * 57 Gem | 57727 | 35846 | 2808 | | | E | hd057727 | G8III | Star | 5.035 | 0.889 | 20.52 | 48.7 | 5.55 |
| * iot Gem | 58207 | 36046 | 2821 | | | E | hd058207 | G9IIIb | V* | 3.790 | 1.040 | 27.10 | 36.9 | 7.26 |
| HR 2862 | 59219 | 36114 | 2862 | | | H | hd059219 | K0III | Star | 5.087 | 1.035 | 5.39 | 185.5 | 7.80 |
| * eta CMi | 58923 | 36265 | 2851 | J07280+0657AB | | E | hd058923 | F0III | Star | 5.239 | 0.209 | 10.25 | 97.6 | 17.20 |
| HD 58898 | 58898 | 36325 | | | | S | hd058898 | K2III | Star | 6.360 | 1.220 | 8.73 | 114.5 | 20.89 |
| * sig Pup | 59717 | 36377 | 2878 | J07292-4318A | | H | sigpup | K5III | EllipVar | 3.250 | 1.520 | 16.84 | 59.4 | 87.30 |
| HD 59894 | 59894 | 36474 | | | | H | hd059894 | G8III | Star | 6.550 | 0.970 | 6.81 | 146.8 | |
| HR 2877 | 59686 | 36616 | 2877 | | | S | hr2877 | K2III | Star | 5.450 | 1.135 | 10.32 | 96.9 | -32.29 |
| HR 2908 | 60574 | 36729 | 2908 | | | H | hd060574 | G8III | Star | 6.544 | 0.882 | 7.14 | 140.1 | -47.60 |
| HR 2899 | 60341 | 36732 | 2899 | | | S | hr2899 | K0III | Star | 5.637 | 1.135 | 11.96 | 83.6 | 18.43 |
| HR 2896 | 60318 | 36896 | 2896 | J07351+3057AB | | S | hr2896 | K0III | Star | 5.348 | 1.005 | 8.76 | 114.2 | -4.04 |
| * ups Gem | 60522 | 36962 | 2905 | J07359+2654A | | E | hd060522 | M0III | V* | 4.060 | 1.540 | 12.04 | 83.1 | -21.61 |
| HR 2894 | 60294 | 37046 | 2894 | | | E | hd060294 | K2III | Star | 5.941 | 1.124 | 12.24 | 81.7 | 0.30 |
| HD 61191 | 61191 | 37069 | | | | S | hd061191 | K1III | Star | 6.880 | 1.060 | 9.09 | 110.0 | |
| * 70 Gem | 60986 | 37204 | 2924 | J07386+3503A | | S | hr2924 | K0III | *in** | 5.588 | 0.904 | 8.77 | 114.0 | -36.33 |
| * 140 Pup | 61772 | 37379 | 2959 | | | H | hr2959 | K3III | Star | 4.981 | 1.584 | 4.69 | 213.2 | 0.10 |
| HR 2951 | 61603 | 37428 | 2951 | | | U | hd061603 | K5III | Star | 5.932 | 1.578 | 2.14 | 467.3 | 39.80 |
| HR 2939 | 61363 | 37441 | 2939 | | | E | hd061363 | K0III | Star | 5.596 | 0.986 | 10.38 | 96.3 | 39.32 |
| * alf Mon | 61935 | 37447 | 2970 | | | S | hr2970 | G9III | Star | 3.930 | 1.020 | 22.07 | 45.3 | 11.66 |
| * zet Vol | 63295 | 37504 | 3024 | J07418-7236A | | H | hd063295 | K0III | *in** | 3.960 | 1.022 | 23.13 | 43.2 | 48.10 |
| HR 2988 | 62412 | 37590 | 2988 | | | U | hd062412 | K1III | Star | 5.635 | 0.980 | 10.33 | 96.8 | -18.00 |
| HR 3012 | 62897 | 37599 | 3012 | | | H | hd062897 | K0III | Star | 6.214 | 1.036 | 5.05 | 198.0 | -22.50 |
| * 192 Gem | 62141 | 37636 | 2978 | | | E | hd062141 | K0III | Star | 6.244 | 0.913 | 10.18 | 98.2 | 1.13 |
| HD 62849 | 62849 | | | J07434-5231A | | H | hd062849 | G8III: | *in** | 9.770 | 1.000 | | | |

| Name | HD | SAO | HR | 2MASS | Other | Src | Ident | SpType | VarType | V | B-V | Plx | Dist | RV |
|---|---|---|---|---|---|---|---|---|---|---|---|---|---|---|
| HD 62713 | 62713 | 37664 | 3002 | | | H | hd062713 | K1III | PM* | 5.170 | 1.100 | 15.03 | 66.5 | 53.20 |
| V* AZ CMi | 62437 | 37705 | 2989 | | | E | hd062437 | F0III: | PulsV*delSct | 6.470 | 0.200 | 8.11 | 123.3 | 14.90 |
| CPD-23 2745 | | | | | NGC 2447 MMU 41 | U | ngc2447No41 | G8III | *inCl | 10.160 | 0.890 | | | 21.51 |
| * kap Gem | 62345 | 37740 | 2985 | | | S | hr2985 | G8IIIa | Star | 3.570 | 0.920 | 23.07 | 43.3 | 20.15 |
| * bet Gem | 62509 | 37826 | 2990 | J07454+2802A | | S | hr2990 | K0III | V* | 1.140 | 1.000 | 96.54 | 10.4 | 3.23 |
| * g Gem | 62721 | 37908 | 3003 | | | E | hd062721 | K4III | SB* | 4.870 | 1.460 | 9.61 | 104.1 | 83.13 |
| * 51 Cam | 62066 | 37949 | 2975 | | | S | hr2975 | K2III: | Star | 5.944 | 1.189 | 9.93 | 100.7 | -30.87 |
| * 6 Pup | 63697 | 38211 | 3044 | | | S | hr3044 | K3III | Star | 5.179 | 1.310 | 12.81 | 78.1 | 48.84 |
| HR 3069 | 64181 | 38292 | 3069 | | | H | hd064181 | G6III | Star | 6.462 | 0.880 | 5.98 | 167.2 | 33.00 |
| HR 3054 | 63894 | 38307 | 3054 | | | S | hr3054 | K0 | Star | 6.163 | 1.135 | 8.11 | 123.3 | 43.40 |
| HR 3068 | 64152 | 38375 | 3068 | | | E | hd064152 | G9III | Star | 5.619 | 0.947 | 10.67 | 93.7 | 32.50 |
| HD 65354 | 65354 | 38851 | | | | U | hd065354 | K3III | Star | 6.820 | 1.590 | 0.61 | 1639.3 | 39.00 |
| HR 3097 | 65066 | 38868 | 3097 | | | S | hr3097 | K0III | Star | 6.033 | 0.983 | 9.69 | 103.2 | -36.78 |
| HR 3094 | 64958 | 38959 | 3094 | | | S | hr3094 | K0 | Star | 6.360 | 1.050 | 7.74 | 129.2 | -51.05 |
| * 14 CMi | 65345 | 38962 | 3110 | J07584+0213A | | S | hr3110 | K0III | *in** | 5.297 | 0.916 | 13.50 | 74.1 | 42.61 |
| HR 3140 | 65925 | 39061 | 3140 | | | U | hd065925 | F5III | Star | 5.237 | 0.348 | 18.00 | 55.6 | -8.20 |
| * 27 Mon | 65695 | 39079 | 3122 | | | S | hr3122 | K2III | Star | 4.942 | 1.207 | 11.87 | 84.2 | -28.02 |
| HR 3125 | 65735 | 39180 | 3125 | | | S | hr3125 | K1III | Star | 6.306 | 1.104 | 8.30 | 120.5 | 29.12 |
| * ome Cnc | 65714 | 39191 | 3124 | | | U | hd065714 | G8III | Star | 5.868 | 1.013 | 4.92 | 203.3 | 1.90 |
| HR 3145 | 66141 | 39311 | 3145 | J08022+0221A | | S | hr3145 | K2III | *in** | 4.380 | 1.250 | 12.84 | 77.9 | 71.57 |
| HR 3150 | 66242 | 39326 | 3150 | | | S | hr3150 | G0III | Star | 6.339 | 0.572 | 8.71 | 114.8 | -15.90 |
| * chi Gem | 66216 | 39424 | 3149 | J08036+2747A | | S | hr3149 | K2III | SB* | 4.944 | 1.143 | 12.73 | 78.6 | -3.83 |
| * 19 Pup | 68290 | 40084 | 3211 | J08112-1256AB | | S | hr3211 | K0III | ** | 4.720 | 0.960 | 18.46 | 54.2 | 36.07 |
| HR 3212 | 68312 | 40107 | 3212 | | | S | hr3212 | G6III | Star | 5.351 | 0.898 | 11.22 | 89.1 | -11.18 |
| HR 3222 | 68461 | 40231 | 3222 | | | S | hr3222 | G8III | SB* | 6.047 | 0.856 | 6.78 | 147.5 | -18.30 |
| HD 68667 | 68667 | 40263 | | | | S | hd068667 | K0 | Star | 6.470 | 0.930 | 6.54 | 152.9 | |
| HR 3253 | 69511 | 40485 | 3253 | | | U | hd069511 | K2III | Star | 6.158 | 1.570 | 1.97 | 507.6 | 30.20 |
| HD 69836 | 69836 | | | | | U | hd069836 | K1III | Star | 8.580 | 1.100 | | | |
| HR 3216 | 68375 | 40793 | 3216 | | | S | hr3216 | G8III | Star | 5.560 | 0.887 | 11.63 | 86.0 | 4.33 |
| HR 3281 | 70523 | 40990 | 3281 | | | S | hr3281 | K0III | Star | 5.720 | 1.050 | 10.50 | 95.2 | 70.26 |
| HD 70522 | 70522 | 40997 | | J08220-1346AB | | S | hd070522 | F7III | Star | 7.630 | 0.500 | 9.50 | 105.3 | 20.90 |
| HR 3263 | 69976 | 41060 | 3263 | | | S | hr3263 | K0III | Star | 6.390 | 0.980 | 8.34 | 119.9 | -5.74 |
| * 22 Pup | 70673 | 41067 | 3289 | | | S | hr3289 | K0III | Star | 6.116 | 0.999 | 7.93 | 126.1 | -18.80 |
| HD 71160 | 71160 | 41222 | | J08246-3256A | | U | hd071160 | K3/K4III | *in** | 7.960 | 1.480 | 2.01 | 497.5 | 3.00 |
| HR 3324 | 71377 | 41395 | 3324 | | | S | hr3324 | K1/K2III | V* | 5.520 | 1.178 | 11.86 | 84.3 | 60.63 |
| HR 3303 | 71088 | 41676 | 3303 | | | S | hr3303 | G8III | Star | 5.889 | 0.961 | 10.20 | 98.0 | -3.15 |
| HD 72320 | 72320 | 41724 | | | | U | hd072320 | K0III | Star | 8.980 | 0.970 | 2.14 | 467.3 | |
| * eta Cnc | 72292 | 41909 | 3366 | | | S | hr3366 | K3III | *in** | 5.343 | 1.265 | 10.93 | 91.5 | 22.46 |
| * ups02 Cnc | 72324 | 41940 | 3369 | | | S,U | hr3369,hd072324 | G9III | Star | 6.341 | 1.022 | 5.87 | 170.4 | 73.70 |
| HR 3376 | 72505 | 42010 | 3376 | | | S | hr3376 | K0III | Star | 6.256 | 1.206 | 9.52 | 105.0 | 30.31 |
| HR 3409 | 73192 | 42365 | 3409 | | | S | hr3409 | K2III: | Star | 5.961 | 1.111 | 8.02 | 124.7 | 4.19 |
| HD 74088 | 74088 | 42381 | | | | U | hd074088 | K4III | Star | 6.710 | 1.590 | 2.72 | 367.6 | |
| HD 73829 | 73829 | | | | | U | hd073829 | K0III | Star | 9.410 | 1.120 | | | |
| * sig Hya | 73471 | 42402 | 3418 | | | S | hr3418 | K1III | Star | 4.430 | 1.200 | 8.75 | 114.3 | 27.28 |
| HR 3424 | 73599 | 42462 | 3424 | | | S | hr3424 | K1III | Star | 6.450 | 1.100 | 9.81 | 101.9 | 18.64 |
| * zet Pyx | 73898 | 42483 | 3433 | J08397-2934A | | S | hr3433 | G4III | V* | 4.886 | 0.875 | 13.35 | 74.9 | -30.10 |
| HD 73598 | 73598 | 42497 | | J08399+1933D | NGC 2632 MMU 212 | H | hd073598 | K0III | ** | 6.600 | 0.930 | 3.62 | 276.2 | 34.97 |
| * 39 Cnc | 73665 | 42516 | 3427 | J08401+2000A | NGC 2632 MMU 253 | H | hd073665 | G8III | *inCl | 6.390 | 0.980 | 5.36 | 186.6 | 33.88 |
| * pi.02 UMa | 73108 | 42527 | 3403 | | | S | hr3403 | K1III | Star | 4.610 | 1.170 | 12.74 | 78.5 | 14.62 |
| HD 74212 | 74212 | | | | IC 2391 L 26 | U | ic2391-0022 | K0III | *inCl | 8.690 | 1.060 | | | 3.00 |

| Name | HD | HR col2 | HR col3 | J-ID | Cluster/Other | Flag | ID | SpType | Class | V | B-V | col12 | col13 | col14 |
|---|---|---|---|---|---|---|---|---|---|---|---|---|---|---|
| HR 3423 | 73596 | 42538 | 3423 | | | S | hr3423 | F5III | Star | 6.220 | 0.390 | 8.07 | 123.9 | 23.30 |
| HR 3428 | 73710 | 42549 | 3428 | J08404+1941A | NGC 2632 MMU 283 | H | hd073710 | G9III | *in** | 6.390 | 1.020 | 4.58 | 218.3 | 34.32 |
| HD 74166 | 74166 | | | | | U | hd074166 | K2III | Star | 7.670 | 1.300 | | | |
| HD 74165 | 74165 | | | | | U | hd074165 | K1III | Star | 9.090 | 1.140 | | | |
| HD 74387 | 74387 | | | | IC 2391 H 31 | U | ic2391-0026 | K0/K1III | *inCl | 9.280 | 1.050 | | | 48.99 |
| HD 74529 | 74529 | 42747 | | | | U | hd074529 | K0III | Star | 8.410 | 1.190 | 2.74 | 365.0 | |
| * del Cnc | 74442 | 42911 | 3461 | J08447+1809A | | S | hr3461 | K0III | PM* | 3.940 | 1.080 | 24.98 | 40.0 | 17.14 |
| HD 74900 | 74900 | 42944 | | | | U | hd074900 | K1III | Star | 7.950 | 1.160 | 4.25 | 235.3 | |
| * 46 Cnc | 74485 | 42954 | 3464 | | | S | hr3464 | G5III | Star | 6.122 | 0.912 | 5.52 | 181.2 | -13.10 |
| HD 75066 | 75066 | | | | | U | ic2391-0044 | F4III | *inCl | 9.280 | 0.370 | | | 19.00 |
| HR 3478 | 74794 | 43026 | 3478 | | | E | hd074794 | K0III: | Star | 5.698 | 1.101 | 10.80 | 92.6 | 7.84 |
| HD 75058 | 75058 | 43066 | | | | U | hd075058 | K0III | Star | 8.930 | 1.060 | 3.67 | 272.5 | |
| * D Hya | 74918 | 43067 | 3484 | J08464-1333A | | S | hr3484 | G8III | *in** | 4.320 | 0.900 | 13.31 | 75.1 | -8.50 |
| * iot Cnc A | 74739 | 43103 | 3475 | J08467+2846A | | S | hr3475 | G8Iab: | *in** | 4.028 | 0.981 | 9.85 | 101.5 | 15.74 |
| * eps Hya | 74874 | 43109 | 3482 | J08468+0625ABC | | H | hd074874 | G1III+A8V | BYDra | 3.380 | 0.680 | 25.23 | 39.6 | 36.40 |
| BD+12 1917 | | | | | NGC 2682 MMU 6470 | H | sand364 | K3III | *inCl | 9.770 | 1.450 | | | 33.16 |
| * gam Pyx | 75691 | 43409 | 3518 | | | S | hr3518 | K3III | Star | 4.010 | 1.260 | 15.73 | 63.6 | 24.50 |
| 2MASS J08511269+1152423 | | | | | NGC 2682 YBP 1479 | H | sand1074 | G8III | RGB* | 10.400 | 1.000 | | | 35.20 |
| Cl* NGC 2682 SAB 5 | | | | | NGC 2682 SAB 5 | H | sand1054 | K0III/IV | *inCl | 11.210 | 1.070 | | | 33.40 |
| 2MASS J08511710+1148160 | | | | | NGC 2682 SAB 4 | H | sand1016 | K2III | RGB* | 10.310 | 1.260 | | | 34.00 |
| NSV 4275 | | | | | NGC 2682 ZTP 894 | H | sand978 | K4III | RGB* | 9.711 | 1.373 | | | 34.53 |
| 2MASS J08512156+1146061 | | | | | NGC 2682 YBP 891 | H | sand989 | K2III | RGB* | 11.440 | 1.060 | | | 34.72 |
| 2MASS J08512618+1153520 | | | | | NGC 2682 SAB 9 | H | sand1084 | G8III | RGB* | 10.500 | 1.000 | | | 35.00 |
| 2MASS J08512898+1150330 | | | | | NGC 2682 YBP 1362 | H,U | sand1279,ngc2682No164 | K1III | RGB* | 10.441 | 0.859 | | | 34.30 |
| 2MASS J08513938+1151456 | | | | | NGC 2682 SAB 14 | U | NGC2682ESIII-35 | K0III | RGB* | 12.117 | 1.063 | | | 33.59 |
| 2MASS J08515020+1146069 | | | | | NGC 2682 MMU 244 | H | sand1237 | G8III-IV | RGB* | 10.700 | 0.800 | | | 33.55 |
| * 35 Lyn | 75506 | 43531 | 3508 | | | S | hr3508 | K0III | Star | 5.160 | 0.958 | 12.31 | 81.2 | 13.49 |
| 2MASS J08521856+1144263 | | | | | NGC 2682 SAB 21 | U | ngc2682No286 | K0III | RGB* | 10.500 | 1.000 | | | 33.64 |
| HR 3529 | 75916 | 43580 | 3529 | J08525-1314A | | S | hr3529 | K1III | *in** | 6.119 | 1.152 | 9.62 | 104.0 | 13.00 |
| HD 76128 | 76128 | 43601 | | J08528-4126A | | U | hd076128 | K2III | *in** | 8.120 | 1.200 | 0.32 | 3125.0 | |
| * zet Hya | 76294 | 43813 | 3547 | | | S | hr3547 | G9II-III | Star | 3.100 | 1.000 | 19.51 | 51.3 | 22.30 |
| * 6 UMa | 75958 | 43903 | 3531 | | | S | hr3531 | G6III | SB* | 5.571 | 0.863 | 10.63 | 94.1 | 5.31 |
| * sig03 Cnc | 76813 | 44154 | 3575 | J08595+3225A | | S | hr3575 | G9III | *in** | 5.233 | 0.895 | 11.03 | 90.7 | 20.82 |
| HD 77232 | 77232 | 44246 | | | | S | hd077232 | F2III | Star | 7.100 | 0.360 | 9.31 | 107.4 | |
| HD 78002 | 78002 | | | | | U | hd078002 | K0III | Star | 9.450 | 1.030 | | | |
| HD 78528 | 78528 | | | | | U | hd078528 | K3/K4III | Star | 8.830 | 1.600 | | | |
| HD 78964 | 78964 | 44814 | | J09080-6402B | | H | hd078964b | G9III: | *in** | 9.750 | 0.930 | 12.17 | 82.2 | 6.00 |
| * tau Cnc | 78235 | 44818 | 3621 | | | S | hr3621 | G8III | Star | 5.427 | 0.882 | 11.92 | 83.9 | -13.90 |
| * 78 Cnc | 78479 | 44918 | | | | U | hd078479 | K3III | Star | 7.189 | 1.230 | 4.62 | 216.5 | 77.45 |
| HD 78959 | 78959 | 44952 | | | | U | hd078959 | K3III | Star | 7.770 | 1.710 | 0.73 | 1369.9 | 44.00 |
| * 79 Cnc | 78715 | 45033 | 3640 | J09103+2200AB | | S | hr3640 | G5III | Star | 6.030 | 0.871 | 8.86 | 112.9 | -13.00 |

| Name | HD | HIP | HR | WDS | Src | ID | SpType | Type | Vmag | B-V | plx | pm | RV |
|---|---|---|---|---|---|---|---|---|---|---|---|---|---|
| HR 3653 | 79181 | 45158 | 3653 | | S | hr3653 | G8III | Star | 5.727 | 0.954 | 10.73 | 93.2 | 1.10 |
| HR 3664 | 79452 | 45412 | 3664 | J09152+3438A | S | hr3664 | G6III | *in** | 5.992 | 0.810 | 7.69 | 130.0 | 55.30 |
| * 23 Hya | 79910 | 45527 | 3681 | J09167-0621AB | S | hr3681 | K2III | SB* | 5.239 | 1.190 | 13.29 | 75.2 | -17.58 |
| HR 3687 | 80050 | 45559 | 3687 | | S | hr3687 | K0III | Star | 5.842 | 1.049 | 9.04 | 110.6 | -36.60 |
| HD 80571 | 80571 | 45711 | | | U | hd080571 | K0III | Star | 7.880 | 1.090 | 3.08 | 324.7 | |
| * 26 Hya | 80499 | 45751 | 3706 | J09198-1159AB | S | hr3706 | G8III | Star | 4.783 | 0.916 | 9.68 | 103.3 | -1.18 |
| * 27 Hya | 80586 | 45811 | 3709 | J09204-0934A | S | hr3709 | G8III-IV+... | *in** | 4.818 | 0.917 | 14.66 | 68.2 | 25.60 |
| * k Car | 81101 | 45856 | 3728 | | H | hr3728 | G6III | Star | 4.804 | 0.923 | 13.96 | 71.6 | 50.80 |
| HR 3707 | 80546 | 45896 | 3707 | | S | hr3707 | K3III | Star | 6.180 | 1.090 | 11.12 | 89.9 | 33.53 |
| HD 81278 | 81278 | | | | U | hd081278 | K1III | Star | 8.480 | 1.070 | | | |
| * lam Pyx | 81169 | 46026 | 3733 | | S | hr3733 | G7III | Star | 4.680 | 0.910 | 16.98 | 58.9 | 10.20 |
| * kap Leo | 81146 | 46146 | 3731 | J09247+2611AB | S | hr3731 | K2III | Star | 4.460 | 1.230 | 16.20 | 61.7 | 27.94 |
| * alf Hya | 81797 | 46390 | 3748 | J09277-0841A | S | hr3748 | K3II-III | V* | 1.970 | 1.450 | 18.09 | 55.3 | -4.38 |
| HR 3743 | 81688 | 46471 | 3743 | J09287+4536A | E | hd081688 | K0III-IV | *in** | 5.413 | 0.983 | 11.65 | 85.8 | 38.58 |
| HD 82403 | 82403 | | | | U | hd082403 | K2III | Star | 8.350 | 1.210 | | | 20.00 |
| HR 3772 | 82232 | 46618 | 3772 | | S | hr3772 | K2III | Star | 5.857 | 1.211 | 13.68 | 73.1 | 19.30 |
| * N Vel | 82668 | 46701 | 3803 | | U | hd082668 | K5III | V* | 3.179 | 1.573 | 13.65 | 73.3 | -13.90 |
| HR 3788 | 82477 | 46768 | 3788 | | S | hr3788 | K0 | Star | 6.128 | 1.188 | 9.13 | 109.5 | 63.90 |
| * ksi Leo | 82395 | 46771 | 3782 | | U | hd082395 | K0III | V* | 4.973 | 1.051 | 15.13 | 66.1 | 35.83 |
| HR 3801 | 82638 | 46869 | 3801 | | S | hr3801 | K0 | Star | 6.128 | 0.947 | 7.99 | 125.2 | 36.10 |
| HR 3808 | 82734 | 46880 | 3808 | | S | hr3808 | K0III | Star | 5.008 | 1.016 | 9.00 | 111.1 | 15.50 |
| HR 3805 | 82674 | 46893 | 3805 | | S | hr3805 | K0 | SB* | 6.256 | 1.177 | 8.01 | 124.8 | 28.70 |
| * 9 LMi | 82522 | 46904 | 3791 | | S | hr3791 | K4III: | Star | 6.192 | 1.277 | 6.71 | 149.0 | -14.38 |
| * 10 LMi | 82635 | 46952 | 3800 | | S | hr3800 | G8III | RSCVn | 4.600 | 0.870 | 17.63 | 56.7 | -11.94 |
| HD 83155 | 83155 | | | | U | hd083155 | K0III | Star | 8.220 | 0.980 | | | 28.00 |
| HR 3809 | 82741 | 47029 | 3809 | | S | hr3809 | G9.5III | Star | 4.815 | 0.980 | 15.14 | 66.1 | -13.57 |
| HD 83234 | 83234 | | | | U | hd083234 | K2III | Star | 8.550 | 1.430 | | | |
| HD 83087 | 83087 | 47056 | | | S | hd083087 | K0/K1III | Star | 7.000 | 1.020 | 10.94 | 91.4 | |
| * 11 LMi | 82885 | 47080 | 3815 | J09357+3549AB | E | hd082885 | G8+V | RSCVn | 5.340 | 0.770 | 87.96 | 11.4 | 14.40 |
| * 10 Leo | 83240 | 47205 | 3827 | | S | hr3827 | K1III | SB* | 5.012 | 1.039 | 13.28 | 75.3 | 12.16 |
| * 2 Sex | 83425 | 47310 | 3834 | | S | hr3834 | K3III | PM* | 4.680 | 1.320 | 11.04 | 90.6 | 44.85 |
| * iot Hya | 83618 | 47431 | 3845 | | S | hr3845 | K2.5III | V* | 3.910 | 1.320 | 12.39 | 80.7 | 24.19 |
| * 43 Lyn | 83805 | 47570 | 3851 | | S | hr3851 | G8III | Star | 5.620 | 0.937 | 9.39 | 106.5 | 28.60 |
| HD 84598 | 84598 | 47824 | | J09449-5001A | U | hd084598 | K0III | *in** | 7.450 | 0.970 | 4.27 | 234.2 | |
| * ups01 Hya | 85444 | 48356 | 3903 | | S | hr3903 | G7III | Star | 4.110 | 0.920 | 12.36 | 80.9 | -14.00 |
| HD 85552 | 85552 | 48369 | | | U | hd085552 | G1II/III | Star | 7.770 | 0.730 | 5.11 | 195.7 | |
| HR 3908 | 85519 | 48396 | 3908 | | S | hr3908 | K0III | Star | 6.081 | 1.017 | 8.35 | 119.8 | 7.50 |
| HR 3907 | 85505 | 48413 | 3907 | | S | hr3907 | K2III | Star | 6.335 | 0.919 | 7.19 | 139.1 | 19.10 |
| HD 85440 | 85440 | 48433 | | | S | hd085440 | G8III | Star | 7.810 | 0.880 | 8.86 | 112.9 | -4.16 |
| * mu. Leo | 85503 | 48455 | 3905 | | S | hr3905 | K2III | PM* | 3.880 | 1.220 | 26.28 | 38.1 | 13.63 |
| HD 85859 | 85859 | 48559 | 3919 | | H | hr3919 | K2III | PM* | 4.880 | 1.230 | 9.54 | 104.8 | 50.50 |
| HR 3911 | 85583 | 48638 | 3911 | J09551+6107A | S | hr3911 | K0 | *in** | 6.303 | 1.030 | 8.35 | 119.8 | -9.86 |
| HR 3929 | 86166 | 48861 | 3929 | | S | hr3929 | K0III | Star | 6.335 | 1.095 | 8.25 | 121.2 | 2.17 |
| HD 86757 | 86757 | | | | U | hd086757 | K2III | Star | 8.360 | 1.590 | | | |
| HR 3942 | 86513 | 48982 | 3942 | | S | hr3942 | G9III: | Star | 5.753 | 1.050 | 9.51 | 105.2 | -6.62 |
| HD 87566 | 87566 | | | NGC 3114 MMU 181 | U | ngc3114no181 | K1III | *inCl | 8.310 | 1.240 | | | -2.18 |
| HR 3972 | 87606 | 49418 | 3972 | | U | hip049418 | K1III | Star | 6.271 | 1.107 | 7.13 | 140.3 | 28.00 |
| * 14 Sex | 87682 | 49530 | 3973 | | S | hr3973 | K1III | Star | 6.193 | 0.916 | 9.16 | 109.2 | 21.37 |
| * lam Hya | 88284 | 49841 | 3994 | J10106-1222A | S | hr3994 | K0IIICN+1 | SB* | 3.610 | 1.000 | 28.98 | 34.5 | 19.40 |
| HR 4006 | 88639 | 50109 | 4006 | | S | hr4006 | G3III | SB* | 6.056 | 0.810 | 6.36 | 157.2 | 7.60 |

| Name | HD | HIP | HR | WDS | Src | Ident | SpType | ObjType | Vmag | B-V | plx | pm | RV |
|---|---|---|---|---|---|---|---|---|---|---|---|---|---|
| * zet Leo | 89025 | 50335 | 4031 | J10166+2327A | E | hd089025 | F0III | V* | 3.410 | 0.310 | 11.90 | 84.0 | -21.40 |
| HR 4032 | 89024 | 50336 | 4032 | | S | hr4032 | K2III | *in** | 5.840 | 1.220 | 9.18 | 108.9 | 34.19 |
| HD 89280 | 89280 | 50417 | | | S | hd089280 | A2III | Star | 7.310 | 0.230 | 11.07 | 90.3 | |
| * gam01 Leo | 89484 | HR4057 | | J10199+1951A | S | hr4057 | K1-III | PM* | 1.980 | 1.150 | 25.96 | 38.5 | -36.70 |
| HR 4066 | 89736 | 50609 | 4066 | | U | hd089736 | K7III | Star | 5.652 | 1.710 | 1.94 | 515.5 | 16.00 |
| HR 4052 | 89414 | 50635 | 4052 | | S | hr4052 | K3III: | Star | 6.023 | 1.130 | 6.15 | 162.6 | 8.94 |
| * 43 Leo | 89962 | 50851 | 4077 | | S | hr4077 | K3III | Star | 6.080 | 1.120 | 14.06 | 71.1 | -26.64 |
| HR 4078 | 89993 | 50904 | 4078 | | S | hr4078 | G8III | Star | 6.378 | 1.080 | 8.70 | 114.9 | -13.80 |
| HR 4085 | 90125 | 50939 | 4085 | J10243+0223A | S | hr4085 | G9V | *in** | 6.323 | 0.979 | 10.63 | 94.1 | -14.14 |
| * 29 LMi | 90250 | 51047 | | | S | hd090250 | K1III | Star | 6.480 | 1.097 | 7.93 | 126.1 | 11.70 |
| HR 4099 | 90518 | 51077 | 4099 | | U | hip051077 | K1III | Star | 6.137 | 1.129 | 10.20 | 98.0 | 23.20 |
| HR 4097 | 90472 | 51161 | 4097 | | S | hr4097 | K0 | Star | 6.142 | 1.151 | 9.54 | 104.8 | 32.24 |
| * bet LMi | 90537 | 51233 | 4100 | J10279+3642AB | S | hr4100 | G9IIIb | SB* | 4.210 | 0.900 | 21.19 | 47.2 | 6.34 |
| * 35 UMa | 90633 | 51401 | 4106 | | S | hr4106 | K2III: | Star | 6.320 | 1.144 | 9.88 | 101.2 | -26.08 |
| HD 91267 | 91267 | 51510 | | | H | hd091267 | K1/K2III: | PM* | 9.810 | 0.920 | 20.20 | 49.5 | 10.00 |
| * 48 Leo | 91612 | 51775 | 4146 | | S | hr4146 | G8.5III | Star | 5.081 | 0.918 | 12.26 | 81.6 | 5.60 |
| HR 4126 | 91190 | 51808 | 4126 | | S | hr4126 | K0III | Star | 4.857 | 0.944 | 13.19 | 75.8 | 16.60 |
| * phi Hya | 92214 | 52085 | 4171 | | S | hr4171 | G8III | SB* | 4.913 | 0.899 | 15.49 | 64.6 | 16.58 |
| * 38 UMa | 92424 | 52353 | 4178 | | S | hr4178 | K2III | V* | 5.114 | 1.221 | 14.38 | 69.5 | -16.86 |
| * m Leo | 93257 | 52686 | 4208 | | S | hr4208 | K3III | Star | 5.501 | 1.154 | 17.48 | 57.2 | -7.68 |
| * k Leo | 93291 | 52689 | 4209 | | S | hr4209 | G4III: | Star | 5.488 | 0.894 | 11.35 | 88.1 | 29.59 |
| * nu. Hya | 93813 | 52943 | 4232 | | S | hr4232 | K0/K1III | PM* | 3.110 | 1.240 | 22.69 | 44.1 | -1.37 |
| HR 4233 | 93833 | 52948 | 4233 | | S | hr4233 | G8III: | Star | 5.850 | 1.066 | 9.57 | 104.5 | 40.30 |
| * 43 UMa | 93859 | 53043 | 4235 | | S | hr4235 | K2III | Star | 5.669 | 1.135 | 9.13 | 109.5 | 11.20 |
| * 42 UMa | 93875 | 53064 | 4236 | | E | hd093875 | K2III | Star | 5.564 | 1.164 | 11.97 | 83.5 | -20.45 |
| HR 4242 | 94084 | 53157 | 4242 | | S | hr4242 | K2III | Star | 6.430 | 1.130 | 10.31 | 97.0 | -7.63 |
| * p01 Leo | 94402 | 53273 | 4253 | J10537-0208AB | S | hr4253 | G8III | ** | 5.454 | 0.958 | 10.43 | 95.9 | 20.28 |
| HR 4255 | 94481 | 53316 | 4255 | | S | hr4255 | K0III+... | Star | 5.660 | 0.802 | 6.62 | 151.1 | -1.70 |
| HR 4256 | 94497 | 53377 | 4256 | | S | hr4256 | G7III: | Star | 5.734 | 1.026 | 11.25 | 88.9 | -27.90 |
| * 55 Leo | 94672 | 53423 | 4265 | J10557+0044AB | E | hd094672 | F2III | Star | 5.927 | 0.385 | 22.29 | 44.9 | 9.30 |
| * 46 UMa | 94600 | 53426 | 4258 | | S | hr4258 | K1III | PM* | 5.035 | 1.098 | 13.77 | 72.6 | -23.47 |
| HR 4264 | 94669 | 53465 | 4264 | | S | hr4264 | K2III | Star | 6.037 | 1.132 | 10.47 | 95.5 | -56.04 |
| * iot Ant | 94890 | 53502 | 4273 | | U | hip053502 | K1III | Star | 4.600 | 1.030 | 17.16 | 58.3 | -0.20 |
| * alf Crt | 95272 | 53740 | 4287 | | S | hr4287 | K1III | PM* | 4.070 | 1.090 | 20.49 | 48.8 | 47.54 |
| HR 4283 | 95233 | 53798 | 4283 | | S | hr4283 | G9III | Star | 6.390 | 1.020 | 5.69 | 175.7 | 1.60 |
| * 58 Leo | 95345 | 53807 | 4291 | | E | hd095345 | K1III | Pec* | 4.852 | 1.163 | 9.05 | 110.5 | 5.98 |
| HD 95799 | 95799 | | | NGC 3532 MMU 649 | H | ngc3532no649 | G8III | *inCl | 7.940 | 1.020 | | | -6.76 |
| HD 95879 | 95879 | | | NGC 3532 MMU 596 | H | ngc3532no596 | G8III | *inCl | 7.869 | 0.985 | | | 2.50 |
| HR 4305 | 95808 | 54029 | 4305 | J11032-1118AB | S | hr4305 | G7III... | Star | 5.513 | 0.925 | 9.63 | 103.8 | -8.90 |
| * p03 Leo | 95849 | 54049 | 4306 | | U | hd095849 | K3III | Star | 5.943 | 1.241 | 5.38 | 185.9 | 0.10 |
| * alf UMa | 95689 | 54061 | 4301 | J11037+6145AB | S | hr4301 | G9III+A7.5 | SB* | 1.790 | 1.070 | 26.54 | 37.7 | -9.40 |
| CPD-58 3077 | | | | NGC 3532 MMU 122 | H | ngc3532no122 | G6III | *inCl | 8.189 | 0.941 | | | 3.34 |
| HD 96445 | 96445 | | | NGC 3532 MMU 19 | H | ngc3532no19 | G6II-III | *inCl | 7.702 | 0.975 | | | 2.94 |
| CPD-58 3092 | | | | NGC 3532 MMU 100 | H | ngc3532no100 | G9III | Star | 7.457 | 1.120 | | | 4.49 |
| HR 4321 | 96484 | 54291 | 4321 | | U | hr4321 | K2III | Star | 6.311 | 1.156 | 8.54 | 117.1 | 9.30 |
| * psi UMa | 96833 | 54539 | 4335 | | S | hr4335 | K1III | Star | 3.010 | 1.140 | 22.57 | 44.3 | -3.39 |
| HR 4351 | 97501 | 54842 | 4351 | J11137+4105AB | S | hr4351 | K2III | Star | 6.347 | 1.163 | 8.77 | 114.0 | 12.69 |
| * nu. UMa | 98262 | 55219 | 4377 | J11185+3306A | E | hd098262 | K3III | *in** | 3.490 | 1.400 | 8.17 | 122.4 | -9.63 |
| * 76 Leo | 98366 | 55249 | 4381 | | E | hd098366 | K0III: | Star | 5.908 | 1.041 | 11.08 | 90.3 | 4.73 |
| * del Crt | 98430 | 55282 | 4382 | | S | hr4382 | K0III | PM* | 3.560 | 1.110 | 17.56 | 56.9 | -4.94 |

| Name | HD | HIP | HR | 2MASS/other | Cluster | Src | ID | SpType | Type | V | B-V | π | d | RV |
|---|---|---|---|---|---|---|---|---|---|---|---|---|---|---|
| HD 98579 | 98579 | 55374 | | J11203-2820A | | S | hd098579 | K1III | PM* | 6.700 | 1.130 | 10.90 | 91.7 | |
| HR 4383 | 98499 | 55412 | 4383 | | | S | hr4383 | G8 | Star | 6.209 | 1.010 | 7.86 | 127.2 | -56.20 |
| * 79 Leo | 99055 | 55650 | 4400 | | | S | hr4400 | G8IIICN... | Star | 5.392 | 0.924 | 9.22 | 108.5 | -9.90 |
| HR 4409 | 99322 | 55756 | 4409 | | | U | hd099322 | K0III | Star | 5.225 | 0.976 | 11.90 | 84.0 | 4.30 |
| CD-42 6977 | | | | | NGC 3680 MMU 26 | U | ngc3680no26 | K1III | *inCl | 10.800 | 1.300 | | | 0.32 |
| HR 4407 | 99283 | 55797 | 4407 | | | S | hr4407 | K0III | Star | 5.700 | 1.000 | 10.63 | 94.1 | -6.35 |
| HR 4419 | 99651 | 55941 | 4419 | J11279-0142AB | | S | hr4419 | K2III: | Star | 6.267 | 0.998 | 6.21 | 161.0 | -9.60 |
| * 86 Leo | 100006 | 56146 | 4433 | | | S | hr4433 | K0III | Star | 5.559 | 1.058 | 10.36 | 96.5 | 24.37 |
| * ksi Hya | 100407 | 56343 | 4450 | J11330-3152A | | H | ksihya | G7III | PM* | 3.540 | 0.930 | 25.16 | 39.7 | -4.90 |
| HR 4452 | 100470 | 56410 | 4452 | | | S | hr4452 | K0III | Star | 6.386 | 1.048 | 7.18 | 139.3 | 20.50 |
| HR 4459 | 100655 | 56508 | 4459 | | | S | hr4459 | G9III | Star | 6.438 | 1.030 | 8.18 | 122.2 | -5.20 |
| * 2 Dra | 100696 | 56583 | 4461 | | | S | hr4461 | K0III | PM* | 5.188 | 0.972 | 13.73 | 72.8 | -3.47 |
| * ups Leo | 100920 | 56647 | 4471 | | | S | hr4471 | G9III | Star | 4.300 | 1.010 | 17.97 | 55.6 | 1.79 |
| HD 101063 | 101063 | 56713 | | | | H | hip056713 | K5/M0III | PM* | 9.450 | | 0.90 | 1111.1 | 183.40 |
| HR 4474 | 101013 | 56731 | 4474 | | | S | hr4474 | G9III: | SB* | 6.124 | 1.065 | 7.75 | 129.0 | -13.70 |
| HR 4478 | 101112 | 56756 | 4478 | | | S | hr4478 | K1III | Star | 6.190 | 1.100 | 8.67 | 115.3 | 11.81 |
| * 60 UMa | 101133 | 56789 | 4480 | | | S | hr4480 | F5IIIs | Star | 6.087 | 0.361 | 10.08 | 99.2 | -24.70 |
| HD 101321 | 101321 | 56864 | | | | S | hd101321 | K0III | Star | 6.800 | 1.010 | 10.27 | 97.4 | -4.07 |
| * 92 Leo | 101484 | 56975 | 4495 | | | S | hr4495 | K0III | Star | 5.266 | 0.962 | 12.08 | 82.8 | 9.38 |
| HR 4510 | 101933 | 57214 | 4510 | | | S | hr4510 | G8III: | Star | 6.059 | 0.941 | 8.09 | 123.6 | -3.00 |
| * chi UMa | 102224 | 57399 | 4518 | | | S | hr4518 | K0.5IIIb | V* | 3.720 | 1.180 | 17.76 | 56.3 | -9.02 |
| HR 4521 | 102328 | 57477 | 4521 | | | S | hr4521 | K3III | Star | 5.260 | 1.293 | 15.13 | 66.1 | 0.55 |
| HR 4544 | 102928 | 57791 | 4544 | | | S | hr4544 | K0III | SB* | 5.635 | 1.038 | 10.94 | 91.4 | 14.12 |
| HD 103295 | 103295 | 57982 | | | | U | hd103295 | G5/G6III | Star | 9.560 | 0.780 | 1.10 | 909.1 | 3.40 |
| HR 4558 | 103462 | 58082 | 4558 | | | S | hr4558 | G8III | Star | 5.283 | 0.861 | 11.40 | 87.7 | -10.30 |
| * 66 UMa | 103605 | 58181 | 4566 | | | S | hr4566 | K1III | Star | 5.838 | 1.101 | 10.54 | 94.9 | 16.91 |
| HR 4593 | 104438 | 58654 | 4593 | | | S | hr4593 | K0III | SB* | 5.570 | 1.020 | 9.73 | 102.8 | 25.00 |
| HD 104760 | 104760 | 58813 | | J12038-4407A | | H | hd104760a | G2/G3III | *in** | 8.070 | 0.610 | 18.34 | 54.5 | 6.20 |
| HD 104819 | 104819 | 58849 | | | | S | hd104819 | K2III | PM* | 7.910 | 1.180 | 8.55 | 117.0 | 65.00 |
| HD 104883 | 104883 | 58893 | | | | S | hd104883 | F5III | Star | 7.700 | 0.470 | 6.88 | 145.3 | |
| * omi Vir | 104979 | 58948 | 4608 | | | S | hr4608 | G8IIIa | PM* | 4.120 | 0.990 | 19.98 | 50.1 | -29.62 |
| HD 104985 | 104985 | 58952 | 4609 | | | S | hr4609 | G9III | PM* | 5.797 | 1.022 | 10.30 | 97.1 | -20.19 |
| HR 4610 | 105043 | 58989 | 4610 | J12057+6256AB | | S | hr4610 | K2III | Star | 6.143 | 1.208 | 8.48 | 117.9 | -27.19 |
| * 10 Vir | 105639 | 59285 | 4626 | J12097+0154A | | S | hr4626 | K3III | PM* | 5.961 | 1.134 | 13.69 | 73.0 | 1.80 |
| * eps Crv | 105707 | 59316 | 4630 | | | S | hr4630 | K2III | V* | 2.980 | 1.340 | 10.26 | 97.5 | 5.00 |
| HD 105740 | 105740 | 59334 | | | | U | hd105740 | G9III | Star | 8.400 | 0.990 | 4.91 | 203.7 | -2.30 |
| HR 4654 | 106478 | 59708 | 4654 | | | S | hr4654 | K0III: | Star | 6.184 | 1.057 | 6.79 | 147.3 | -1.70 |
| HR 4655 | 106485 | 59728 | 4655 | | | S | hr4655 | K0IIICN | Star | 5.815 | 1.044 | 12.07 | 82.9 | 16.00 |
| HD 106572 | 106572 | 59785 | 4658 | | | U | hip059785 | K0III | PM* | 6.230 | 1.040 | 8.12 | 123.2 | 25.20 |
| * 7 Com | 106714 | 59847 | 4667 | | Melotte 111 AV 782 | S | hr4667 | G8III | *inCl | 4.860 | 0.950 | 13.08 | 76.5 | -27.89 |
| HR 4668 | 106760 | 59856 | 4668 | | | S | hr4668 | K1III | SB* | 4.940 | 1.140 | 10.31 | 97.0 | -40.40 |
| HD 106972 | 106972 | 59980 | | J12181+1826A | | S | hd106972 | F6III | *in** | 7.580 | 0.490 | 8.68 | 115.2 | -24.30 |
| * c Vir | 107328 | 60172 | 4695 | J12204+0319A | | S | hr4695 | K0IIIb | V* | 4.960 | 1.160 | 10.59 | 94.4 | 36.40 |
| * 11 Com | 107383 | 60202 | 4697 | J12207+1748A | | S | hr4697 | G8II-III | *in** | 4.740 | 1.000 | 11.25 | 88.9 | 43.60 |
| HD 107415 | 107415 | 60210 | | | | S | hd107415 | G7II-III | Star | 6.450 | 1.000 | 8.29 | 120.6 | -25.20 |
| HR 4699 | 107418 | 60221 | 4699 | J12209-1334A | | S | hr4699 | K0III | V* | 5.154 | 1.043 | 16.21 | 61.7 | 14.60 |
| * eps Cru | 107446 | 60260 | 4700 | | | U | hd107446 | K3.5III | V* | 3.590 | 1.420 | 14.19 | 70.5 | -4.60 |
| HD 107569 | 107569 | 60290 | | | | S | hd107569 | F8III | Star | 7.420 | 0.510 | 7.48 | 133.7 | -6.20 |
| HD 107610 | 107610 | 60305 | | | | S | hd107610 | K2III | Star | 6.328 | 1.116 | 10.25 | 97.6 | -8.39 |
| * 12 Com | 107700 | 60351 | 4707 | J12225+2551A | Melotte 111 MMU 91 | E | hd107700 | F6III+A3V | SB* | 4.810 | 0.490 | 11.07 | 90.3 | 0.50 |

| Name | HD | HIP | HR | 2MASS/Other | Melotte | Src | Ident | SpType | ObjType | V | B-V | plx | pm | RV |
|---|---|---|---|---|---|---|---|---|---|---|---|---|---|---|
| * 5 CVn | 107950 | 60485 | 4716 | | | E | hd107950 | G6III | Star | 4.767 | 0.868 | 8.44 | 118.5 | -13.90 |
| Cl* NGC 4349 MMU 127 | | | | | NGC 4349 MMU 127 | H | ngc4349no127 | KIII | *inCl | 10.820 | 2.030 | | | -12.12 |
| HR 4721 | 108063 | 60591 | 4721 | | | H | hd108063 | G5III+... | Star | 6.114 | 0.613 | 18.53 | 54.0 | 34.60 |
| * 6 CVn | 108225 | 60646 | 4728 | | | S | hr4728 | G8III | Star | 5.010 | 0.940 | 12.47 | 80.2 | -4.30 |
| * gam Com | 108381 | 60742 | 4737 | | Melotte 111 AV 1665 | S | hr4737 | K1III | *inCl | 4.340 | 1.130 | 19.50 | 51.3 | 3.38 |
| HD 108570 | 108570 | 60870 | 4749 | J12285-5624A | | H | hd108570 | K0.5III | PM* | 6.130 | 0.940 | 23.60 | 42.4 | 7.50 |
| * 18 Com | 108722 | 60941 | 4753 | | Melotte 111 AV 1866 | S | hr4753 | F5IV | Star | 5.481 | 0.401 | 16.23 | 61.6 | 25.10 |
| HR 4772 | 109014 | 61134 | 4772 | | | S | hr4772 | G9III: | Star | 6.187 | 1.048 | 8.06 | 124.1 | 2.20 |
| * 20 Vir | 109217 | 61246 | 4777 | | | S | hr4777 | G8III | Star | 6.290 | 0.932 | 6.90 | 144.9 | -0.70 |
| HR 4783 | 109317 | 61309 | 4783 | | | S | hr4783 | K0III | Star | 5.380 | 1.010 | 11.19 | 89.4 | -21.41 |
| HR 4784 | 109345 | 61320 | 4784 | | | S | hr4784 | K0III | Star | 6.230 | 1.050 | 7.86 | 127.2 | -42.34 |
| * bet Crv | 109379 | 61359 | 4786 | | | S | hr4786 | G5II | V* | 2.640 | 0.880 | 22.39 | 44.7 | -7.60 |
| HR 4812 | 109996 | 61719 | 4812 | J12390+2240A | Melotte 111 AV 2383 | S | hr4812 | K1III | *in** | 6.370 | 1.100 | 7.51 | 133.2 | -26.84 |
| * 26 Com | 110024 | 61724 | 4815 | | | S | hr4815 | G9III | SB* | 5.490 | 0.970 | 11.93 | 83.8 | -20.30 |
| * chi Vir | 110014 | 61740 | 4813 | J12393-0800A | | S | hr4813 | K2III | *in** | 4.657 | 1.247 | 11.11 | 90.0 | -18.11 |
| HD 110291 | 110291 | 61940 | | | | H | hd110291 | K1/K2III+... | Star | 9.160 | 0.730 | 18.04 | 55.4 | |
| * w Cen | 110458 | 62012 | 4831 | | | U | hd110458 | K0III | Star | 4.660 | 1.100 | 17.11 | 58.4 | -11.70 |
| HR 4840 | 110678 | 62046 | 4840 | J12431+6109A | | S | hr4840 | K0 | *in** | 6.387 | 1.282 | 6.09 | 164.2 | -5.56 |
| * 27 Com | 111067 | 62356 | 4851 | | | S | hr4851 | K4III | Star | 5.120 | 1.340 | 9.60 | 104.2 | 50.58 |
| HR 4860 | 111295 | 62500 | 4860 | | | S | hr4860 | G5III-IV | Star | 5.664 | 0.923 | 10.18 | 98.2 | -10.80 |
| HD 111464 | 111464 | 62651 | | | | U | hd111464 | K3III | Star | 6.630 | 1.470 | 4.10 | 243.9 | 28.00 |
| HR 4873 | 111591 | 62653 | 4873 | | | S | hr4873 | K0III | Star | 6.430 | 1.000 | 9.22 | 108.5 | 5.67 |
| HR 4877 | 111720 | 62743 | 4877 | J12514-1021A | | S | hr4877 | G8III | V* | 6.473 | 1.029 | 6.41 | 156.0 | -11.60 |
| HD 111721 | 111721 | 62747 | | | | U | hd111721 | G6III | PM* | 7.970 | 0.810 | 4.33 | 230.9 | 21.40 |
| * 31 Com | 111812 | 62763 | 4883 | | Melotte 111 MGM 399 | S | hr4883 | G0IIIp | RotV* | 4.940 | 0.670 | 11.22 | 89.1 | -0.08 |
| * 35 Com | 112033 | 62886 | 4894 | J12533+2115AB | | S | hr4894 | G5III | ** | 4.910 | 0.890 | 11.52 | 86.8 | -6.09 |
| HR 4896 | 112048 | 62915 | 4896 | | | S | hr4896 | K0 | SB* | 6.446 | 1.090 | 10.11 | 98.9 | 37.50 |
| HD 112127 | 112127 | 62944 | | | | E | hd112127 | K3III | C* | 6.880 | 1.260 | 9.45 | 105.8 | 5.83 |
| HD 112357 | 112357 | 63142 | | | | S | hd112357 | K0III | Star | 7.200 | 0.910 | 12.15 | 82.3 | -53.54 |
| HD 113002 | 113002 | 63484 | | | | U | hd113002 | G2II-III | PM* | 8.740 | 0.750 | 3.95 | 253.2 | -93.70 |
| * 46 Vir | 112992 | 63494 | 4925 | J13006-0322AB | | S | hr4925 | K2III | Star | 5.994 | 1.139 | 9.97 | 100.3 | 24.70 |
| * 38 Com | 113095 | 63533 | 4929 | | | S | hr4929 | G9III | Star | 5.930 | 0.970 | 8.02 | 124.7 | -5.87 |
| * eps Vir | 113226 | 63608 | 4932 | J13022+1057A | | S | hr4932 | G8III | PM* | 2.790 | 0.920 | 29.76 | 33.6 | -14.22 |
| HR 4953 | 113994 | 63952 | 4953 | | | S | hr4953 | G7III | Star | 6.158 | 0.988 | 8.60 | 116.3 | 13.90 |
| HR 4956 | 114092 | 64077 | 4956 | | | S | hr4956 | K4III | Star | 6.140 | 1.360 | 7.54 | 132.6 | -12.31 |
| * 49 Vir | 114038 | 64078 | 4955 | | | S | hr4955 | K1III | Star | 5.164 | 1.126 | 10.61 | 94.3 | -9.53 |
| HR 4960 | 114256 | 64179 | 4960 | | | S | hr4960 | K0III | Star | 5.798 | 1.018 | 10.58 | 94.5 | -0.63 |
| HR 4959 | 114203 | 64181 | 4959 | | | S | hr4959 | K0 | Star | 6.320 | 1.010 | 7.82 | 127.9 | 0.00 |
| HR 4964 | 114357 | 64212 | 4964 | | | S | hr4964 | K2III | SB* | 5.990 | 1.170 | 9.39 | 106.5 | -18.00 |
| HR 4984 | 114724 | 64417 | 4984 | | | S | hr4984 | K0III | Star | 6.290 | 1.000 | 7.59 | 131.8 | -18.60 |
| HD 114747 | 114747 | 64475 | | | | H | hd114747 | K2III-IV | PM* | 8.050 | 0.950 | 37.28 | 26.8 | -16.10 |
| HD 114889 | 114889 | 64496 | 4992 | | | S | hr4992 | K3III | PM* | 6.080 | 1.210 | 9.41 | 106.3 | -22.40 |
| * 57 Vir | 115202 | 64725 | 5001 | | | S,H | hr5001,hd115202 | K1III | PM* | 5.220 | 1.030 | 25.99 | 38.5 | 35.41 |
| HR 5007 | 115319 | 64751 | 5007 | | | S | hr5007 | K0III-IV | Star | 6.450 | 0.970 | 9.86 | 101.4 | -47.70 |
| * 20 CVn | 115604 | 64844 | 5017 | | | E | hd115604 | F3III | PulsV*delSct | 4.715 | 0.300 | 12.34 | 81.0 | 8.62 |
| * gam Hya | 115659 | 64962 | 5020 | J13190-2310A | | S | hr5020 | G8III | V* | 3.000 | 0.920 | 24.37 | 41.0 | -3.60 |
| V* BM CVn | 116204 | 65187 | | | | S | hd116204 | K2III | RSCVn | 7.360 | 1.200 | 8.86 | 112.9 | 7.80 |
| * 63 Vir | 116292 | 65301 | 5044 | | | S,E | hr5044,hd116292 | K0III | V* | 5.367 | 0.972 | 10.66 | 93.8 | -25.85 |

| Name | HD | HIP | HR | WDS | Src | ID | SpType | ObjType | Vmag | B-V | plx | dist | RV |
|---|---|---|---|---|---|---|---|---|---|---|---|---|---|
| HD 116515 | 116515 | 65366 | | | E | hd116515 | K0III | Star | 7.400 | 1.030 | 3.25 | 307.7 | -5.26 |
| HR 5053 | 116594 | 65417 | 5053 | | S | hr5053 | K0III | SB* | 6.440 | 1.060 | 5.24 | 190.8 | -8.90 |
| HR 5067 | 116957 | 65550 | 5067 | | S | hr5067 | K0III: | Star | 5.885 | 0.974 | 7.92 | 126.3 | 5.34 |
| * 69 Vir | 116976 | 65639 | 5068 | | S | hr5068 | K1IIICN | V* | 4.758 | 1.095 | 11.64 | 85.9 | -13.12 |
| * 71 Vir | 117304 | 65790 | 5081 | | S | hr5081 | K0III | Star | 5.656 | 1.065 | 10.18 | 98.2 | 0.69 |
| HR 5096 | 117710 | 65951 | 5096 | | E | hd117710 | K2III | Star | 6.071 | 1.087 | 12.44 | 80.4 | -22.09 |
| * h Vir | 117818 | 66098 | 5100 | | S | hr5100 | K0III | Star | 5.210 | 0.960 | 13.87 | 72.1 | -0.26 |
| * 80 Vir | 118219 | 66320 | 5111 | | S | hr5111 | G6III | Star | 5.715 | 0.937 | 10.29 | 97.2 | -8.20 |
| HR 5126 | 118536 | 66385 | 5126 | | S | hr5126 | K1III | Star | 6.460 | 1.200 | 6.40 | 156.3 | -9.90 |
| HR 5143 | 119035 | 66690 | 5143 | | S | hr5143 | G8II-III | Star | 6.200 | 0.970 | 6.04 | 165.6 | -18.70 |
| * 2 Boo | 119126 | 66763 | 5149 | | S | hr5149 | G9III | Star | 5.580 | 1.010 | 9.71 | 103.0 | 4.17 |
| HR 5184 | 120084 | 66903 | 5184 | | E | hd120084 | G7III: | Star | 5.910 | 0.991 | 9.93 | 100.7 | -8.84 |
| HR 5161 | 119458 | 66907 | 5161 | | S | hr5161 | G5III | SB* | 5.960 | 0.860 | 6.81 | 146.8 | -7.70 |
| * 84 Vir | 119425 | 66936 | 5159 | J13431+0332AB | U | hip066936 | K2III | Star | 5.360 | 1.110 | 13.68 | 73.1 | -41.54 |
| HR 5180 | 120048 | 67210 | 5180 | | S | hr5180 | G9III | Star | 5.924 | 0.937 | 7.65 | 130.7 | -14.70 |
| HR 5186 | 120164A | 67250 | 5186 | J13470+3832A | S | hr5186 | K0III | Star | 5.507 | 1.040 | 10.29 | 97.2 | -11.05 |
| HR 5176 | 119971 | 67304 | 5176 | | U | hd119971 | K2III | Star | 5.467 | 1.354 | 6.55 | 152.7 | 28.10 |
| HR 5195 | 120420 | 67384 | 5195 | | S | hr5195 | K0III | Star | 5.570 | 1.020 | 11.47 | 87.2 | 10.93 |
| HR 5213 | 120787 | 67485 | 5213 | | S | hr5213 | G3V | Star | 5.970 | 0.943 | 7.95 | 125.8 | -12.70 |
| * 89 Vir | 120452 | 67494 | 5196 | | S | hr5196 | K0III | Star | 4.970 | 1.049 | 13.35 | 74.9 | -39.52 |
| HR 5205 | 120602 | 67545 | 5205 | | S | hr5205 | K0 | Star | 6.008 | 0.863 | 6.46 | 154.8 | -25.31 |
| HD 121146 | 121146 | 67589 | 5227 | J13511+6819A | S | hr5227 | K2IV | PM* | 6.402 | 1.186 | 9.77 | 102.4 | -46.60 |
| * eta Boo | 121370 | 67927 | 5235 | J13547+1824A | S | hr5235 | G0IV | SB* | 2.680 | 0.570 | 87.75 | 11.4 | 0.70 |
| * p Vir | 121299 | 67929 | 5232 | | S | hr5232 | K2III | Star | 5.164 | 1.092 | 13.23 | 75.6 | -6.95 |
| HD 121416 | 121416 | 68079 | 5239 | | H | hd121416 | K1III | PM* | 5.822 | 1.142 | 10.25 | 97.6 | -1.00 |
| HR 5276 | 122815 | 68739 | 5276 | | S | hr5276 | K0 | Star | 6.401 | 1.354 | 9.91 | 100.9 | -10.15 |
| HD 122721 | 122721 | 68740 | | | U | hd122721 | K0III | Star | 9.130 | 1.090 | 4.52 | 221.2 | |
| HR 5277 | 122837 | 68763 | 5277 | | S | hr5277 | K1III+... | Star | 6.353 | 1.089 | 7.94 | 125.9 | -15.00 |
| * pi. Hya | 123123 | 68895 | 5287 | | H | pihya | K1III-IV | Star | 3.280 | 1.130 | 32.30 | 31.0 | 27.20 |
| HR 5302 | 123977 | 69107 | 5302 | | S | hr5302 | K0III | Star | 6.475 | 1.011 | 8.35 | 119.8 | 7.63 |
| HR 5310 | 124186 | 69316 | 5310 | | U | hd124186 | K4III | Star | 6.139 | 1.274 | 9.03 | 110.7 | -20.58 |
| * kap Vir | 124294 | 69427 | 5315 | | S | hr5315 | K2.5IIIb | V* | 4.210 | 1.320 | 12.80 | 78.1 | -4.38 |
| HD 123517 | 123517 | | | | H | hd123517 | F9III/IV | Star | 9.550 | 0.690 | | | |
| * alf Boo | 124897 | 69673 | 5340 | | S | hr5340 | K0III | RGB* | -0.050 | 1.230 | 88.83 | 11.3 | -5.19 |
| HR 5344 | 124990 | 69792 | 5344 | | S | hr5344 | K0III | Star | 6.226 | 0.972 | 7.96 | 125.6 | -12.60 |
| * ups Vir | 125454 | 70012 | 5366 | | S | hr5366 | G8III | V* | 5.140 | 1.020 | 12.47 | 80.2 | -26.68 |
| * 20 Boo | 125560 | 70027 | 5370 | | S | hr5370 | K3III | V* | 4.850 | 1.250 | 17.44 | 57.3 | -7.64 |
| * k Hya | 125932 | 70306 | 5381 | | U | hip070306 | K5III | V* | 4.750 | 1.310 | 20.72 | 48.3 | 20.40 |
| * 2 Lib | 126035 | 70336 | 5383 | | S | hr5383 | G7III | Star | 6.225 | 0.980 | 9.74 | 102.7 | -1.20 |
| HD 126265 | 126265 | 70344 | | | S | hd126265 | G2III | PM* | 7.240 | 0.560 | 14.20 | 70.4 | -18.40 |
| HR 5394 | 126271 | 70414 | 5394 | | S | hr5394 | K4III | V* | 6.180 | 1.217 | 9.09 | 110.0 | -30.80 |
| * 5 UMi | 127700 | 70692 | 5430 | J14276+7542A | S | hr5430 | K4III | V* | 4.253 | 1.457 | 9.09 | 110.0 | 9.34 |
| * rho Boo | 127665 | 71053 | 5429 | J14318+3022A | S | hr5429 | K3III | V* | 3.590 | 1.300 | 20.37 | 49.1 | -13.57 |
| HD 127740 | 127740 | 71132 | | | S | hd127740 | F5III | Star | 7.030 | 0.450 | 11.28 | 88.7 | -14.40 |
| HR 5454 | 128402 | 71406 | 5454 | | S | hr5454 | K0 | Star | 6.370 | 1.050 | 8.57 | 116.7 | 5.80 |
| HD 128279 | 128279 | 71458 | | | U | hd128279 | G0III | PM* | 8.000 | 0.630 | 6.09 | 164.2 | -74.90 |
| HD 128853 | 128853 | 71697 | | | S | hd128853 | G8III | Star | 7.130 | 0.970 | 9.76 | 102.5 | |
| * 31 Boo | 129312 | 71832 | 5480 | | H | hr5480 | G7III | V* | 4.865 | 0.994 | 6.07 | 164.7 | -22.30 |
| * 32 Boo | 129336 | 71837 | 5481 | | S | hr5481 | G8III | PM* | 5.560 | 0.940 | 8.09 | 123.6 | -23.30 |
| * mu. Vir | 129502 | 71957 | 5487 | | S | hr5487 | F2V | PM* | 3.880 | 0.380 | 54.73 | 18.3 | 5.10 |

| Name | HD | HIP | HR | 2MASS | Src | ID | SpType | ObjType | Vmag | B-V | Plx | Dist | RV |
|---|---|---|---|---|---|---|---|---|---|---|---|---|---|
| * omi Boo | 129972 | 72125 | 5502 | | S | hr5502 | G8.5III | Star | 4.600 | 0.980 | 13.42 | 74.5 | -9.18 |
| HR 5518 | 130325 | 72373 | 5518 | | S | hr5518 | K0III | Star | 6.320 | 1.110 | 8.87 | 112.7 | 32.40 |
| * bet UMi | 131873 | 72607 | 5563 | J14506+7410A | E | hd131873 | K4III | V* | 2.080 | 1.470 | 24.91 | 40.1 | 16.96 |
| * 11 Lib | 130952 | 72631 | 5535 | | E | hd130952 | G8III | Star | 4.940 | 0.980 | 14.92 | 67.0 | 83.93 |
| * 1 Ser | 132132 | 73193 | 5573 | J14575-0011A | S | hr5573 | K1III | *in** | 5.516 | 1.132 | 11.19 | 89.4 | 20.12 |
| * bet Boo | 133208 | 73555 | 5602 | | S | hr5602 | G8IIIa | Flare* | 3.520 | 0.950 | 14.48 | 69.1 | -18.40 |
| * ome Boo | 133124 | 73568 | 5600 | | S | hr5600 | K4III | Star | 4.810 | 1.500 | 8.78 | 113.9 | 12.47 |
| * 110 Vir | 133165 | 73620 | 5601 | | S | hr5601 | K0.5IIIb | V* | 4.400 | 1.040 | 16.69 | 59.9 | -15.92 |
| HR 5609 | 133392 | 73634 | 5609 | | S | hr5609 | G8III: | V* | 5.521 | 1.016 | 11.65 | 85.8 | -30.60 |
| V* HY Lib | 133194 | 73710 | | | S | hd133194 | F5III+... | PulsV*delSct | 7.720 | 0.490 | 5.48 | 182.5 | |
| * psi Boo | 133582 | 73745 | 5616 | | S | hr5616 | K2III | PM* | 4.550 | 1.230 | 13.26 | 75.4 | -25.72 |
| HR 5635 | 134190 | 73909 | 5635 | | S | hr5635 | G7.5III | Star | 5.250 | 0.960 | 12.67 | 78.9 | 15.56 |
| HR 5620 | 133670 | 73927 | 5620 | | S | hr5620 | K0III | V* | 6.140 | 1.020 | 17.22 | 58.1 | 4.70 |
| HR 5648 | 134493 | 74083 | 5648 | | S | hr5648 | K0III | Star | 6.331 | 1.046 | 7.19 | 139.1 | -26.28 |
| HR 5673 | 135402 | 74514 | 5673 | | S | hr5673 | K2III: | Star | 6.199 | 1.227 | 7.93 | 126.1 | -68.54 |
| * del Boo | 135722 | 74666 | 5681 | J15156+3319A | S | hr5681a | G8IV | *in** | 3.490 | 0.950 | 26.78 | 37.3 | -12.29 |
| BD+33 2562 | | 74674 | | J15156+3319B | S | hr5681b | G0Vv | *in** | 7.810 | 0.580 | 27.05 | 37.0 | -12.30 |
| * 5 Ser | 136202 | 74975 | 5694 | J15193+0146A | E | hd136202 | F8III-IV | BYDra | 5.100 | 0.500 | 39.40 | 25.4 | 54.30 |
| * omi CrB | 136512 | 75049 | 5709 | J15201+2937A | S | hr5709 | K0III | V* | 5.510 | 1.020 | 12.08 | 82.8 | -54.15 |
| * 6 Ser | 136514 | 75119 | 5710 | J15210+0043AB | E | hd136514 | K3III | Star | 5.382 | 1.224 | 13.63 | 73.4 | 9.55 |
| HR 5707 | 136479 | 75127 | 5707 | J15211-0550A | S | hr5707 | K1III | *in** | 5.543 | 1.047 | 10.79 | 92.7 | -32.29 |
| * iot Dra | 137759 | 75458 | 5744 | J15250+5859A | S | hr5744 | K2III | V* | 3.290 | 1.160 | 32.23 | 31.0 | -10.71 |
| HD 138085 | 138085 | 75810 | | | S | hd138085 | G8III-IV | Star | 6.383 | 0.942 | 7.70 | 129.9 | |
| HR 5785 | 138852 | 75974 | 5785 | | S | hr5785 | K0III-IV | Star | 5.750 | 0.964 | 10.57 | 94.6 | 4.73 |
| HR 5769 | 138525 | 76006 | 5769 | | S | hr5769 | F6III | SB* | 6.390 | 0.500 | 13.37 | 74.8 | -46.39 |
| HD 138686 | 138686 | 76209 | | | S | hd138686 | F5III | Star | 7.250 | 0.500 | 5.08 | 196.9 | -19.10 |
| * 36 Lib | 138688 | 76259 | 5775 | | U | hd138688 | K4III | Star | 5.142 | 1.320 | 10.25 | 97.6 | 12.30 |
| HR 5811 | 139357 | 76311 | 5811 | | S | hr5811 | K4III: | Star | 5.977 | 1.191 | 8.47 | 118.1 | -8.75 |
| * 16 Ser | 139195 | 76425 | 5802 | | S | hr5802 | K0III:CNs... | SB* | 5.261 | 0.937 | 14.11 | 70.9 | 6.32 |
| * ups Lib | 139063 | 76470 | 5794 | J15370-2808AB | S | hr5794 | K5III | Star | 3.609 | 1.408 | 14.58 | 68.6 | -24.90 |
| HR 5828 | 139778 | 76509 | 5828 | | S | hr5828 | K1III: | Star | 5.860 | 1.097 | 9.16 | 109.2 | -25.09 |
| HR 5806 | 139254 | 76532 | 5806 | | E | hd139254 | K0III | HB* | 5.796 | 1.067 | 11.87 | 84.2 | 2.49 |
| HR 5810 | 139329 | 76569 | 5810 | | S | hr5810 | K0III | HB* | 5.829 | 1.063 | 11.45 | 87.3 | 34.26 |
| HR 5835 | 139906 | 76594 | 5835 | | S | hr5835 | G8III | Star | 5.840 | 0.830 | 8.66 | 115.5 | -14.00 |
| HR 5841 | 140117 | 76651 | 5841 | | S | hr5841 | K1III | Rapid_Irreg_V* | 6.447 | 1.102 | 7.10 | 140.8 | -7.30 |
| * tau06 Ser | 140027 | 76810 | 5840 | | S | hr5840 | G8III | Star | 6.000 | 0.890 | 5.38 | 185.9 | 3.60 |
| * pi. CrB | 140716 | 77048 | 5855 | | S | hr5855 | G9III: | Star | 5.578 | 1.074 | 13.40 | 74.6 | -4.94 |
| * ome Ser | 141680 | 77578 | 5888 | | S | hr5888 | G8III | Star | 5.230 | 1.009 | 11.93 | 83.8 | -3.79 |
| HR 5893 | 141832 | 77729 | 5893 | J15522-2953A | S | hr5893 | K1III | V* | 6.400 | 0.981 | 12.91 | 77.5 | -16.50 |
| HR 5922 | 142531 | 77738 | 5922 | | S | hr5922 | G8III: | Star | 5.809 | 0.960 | 9.72 | 102.9 | -29.70 |
| HD 142527 | 142527 | 78092 | | | H | hd142527 | F6III | ** | 8.340 | 0.700 | 4.29 | 233.1 | -3.10 |
| * phi Ser | 142980 | 78132 | 5940 | | S | hr5940 | K1IV | Star | 5.547 | 1.144 | 13.52 | 74.0 | -70.98 |
| * eps CrB | 143107 | 78159 | 5947 | J15576+2652AB | S | hr5947 | K2III | ** | 4.130 | 1.230 | 14.73 | 67.9 | -32.42 |
| HR 5963 | 143553 | 78442 | 5963 | | E | hd143553 | K0III: | Star | 5.823 | 0.989 | 13.32 | 75.1 | -7.85 |
| * r Her | 143666 | 78481 | 5966 | | S | hr5966 | G8IIIb | SB* | 5.112 | 0.990 | 13.00 | 76.9 | -17.85 |
| HR 5969 | 143787 | 78650 | 5969 | | U | hip078650 | K3III | Star | 4.973 | 1.245 | 15.71 | 63.7 | -39.00 |
| * 43 Ser | 144046 | 78685 | 5976 | J16038+0459A | S | hr5976 | G9III | *in** | 6.076 | 0.947 | 9.04 | 110.6 | -42.60 |
| V* V1045 Sco | 144377 | 78880 | | | H | v1045sco | K5III | PulsV* | 8.040 | 1.690 | 2.39 | 418.4 | |
| HD 144589 | 144589 | | | | H | hd144589 | K0III:+... | Star | 9.800 | 0.540 | | | |
| HR 6038 | 145694 | 79164 | 6038 | | S | hr6038 | K0 | Star | 6.400 | 1.090 | 7.63 | 131.1 | -15.40 |

| Name | HD | HIP | HR | CCDM | Src | ID | SpType | Type | Vmag | B-V | Plx | RA | Dec |
|---|---|---|---|---|---|---|---|---|---|---|---|---|---|
| HR 6016 | 145206 | 79195 | 6016 | | U | hd145206 | K4III | SB* | 5.401 | 1.445 | 6.81 | 146.8 | -36.94 |
| HR 6057 | 146084 | 79581 | 6057 | | S | hr6057 | K2III | Star | 6.293 | 1.171 | 9.15 | 109.3 | -22.01 |
| * 16 Her | 146388 | 79666 | 6065 | | S | hr6065 | K3III | Star | 5.690 | 1.120 | 9.38 | 106.6 | -19.07 |
| * eps Oph | 146791 | 79882 | 6075 | J16183-0442A | S | hr6075 | G9.5IIIb | *in** | 3.230 | 0.980 | 30.64 | 32.6 | -9.18 |
| HR 6077 | 146836 | 79980 | 6077 | J16195-3054A | U | hd146836 | F6III | *in** | 5.506 | 0.421 | 20.72 | 48.3 | -0.90 |
| V* V1054 Sco | 147135 | 80060 | | | H | hd147135 | F0III/IV | EB* | 6.570 | 0.350 | 10.56 | 94.7 | -29.40 |
| HR 6126 | 148293 | 80161 | 6126 | | S | hr6126 | K2III | Star | 5.272 | 1.116 | 11.12 | 89.9 | -9.41 |
| CD-30 13092B | | 80242 | | J16229-3104B | H | hip080242b | K0III | *in** | 9.660 | 1.330 | 9.40 | 106.4 | |
| CCDM J16238+6142AB | 148374 | 80309 | 6130 | J16238+6142AB | S | hr6130 | G8III | ** | 5.670 | 0.960 | 6.44 | 155.3 | -24.80 |
| * eta Dra | 148387 | 80331 | 6132 | J16239+6130A | S | hr6132 | G8III-IV | V* | 2.740 | 0.910 | 35.42 | 28.2 | -15.20 |
| * psi Oph | 147700 | 80343 | 6104 | | S,U | hr6104,hip080343 | K0III | Star | 4.500 | 1.030 | 16.35 | 61.2 | 0.60 |
| HR 6121 | 148228 | 80514 | 6121 | | S | hr6121 | G8III | Star | 6.112 | 1.061 | 9.14 | 109.4 | -15.90 |
| HD 148317 | 148317 | 80543 | | | S | hd148317 | G0III | Star | 6.690 | 0.630 | 12.56 | 79.6 | -37.60 |
| HR 6124 | 148287 | 80558 | 6124 | | S | hr6124 | G8III | Star | 6.065 | 0.895 | 7.00 | 142.9 | 5.60 |
| HR 6136 | 148513 | 80693 | 6136 | | U | hd148513 | K4III | Star | 5.390 | 1.460 | 8.01 | 124.8 | 7.72 |
| HR 6150 | 148880 | 80710 | 6150 | J16287+5124A | S | hr6150 | G9III | *in** | 6.307 | 1.047 | 7.70 | 129.6 | -16.00 |
| HR 6140 | 148604 | 80793 | 6140 | | E | hd148604 | G5III/IV | Star | 5.675 | 0.795 | 10.95 | 91.3 | -19.87 |
| * bet Her | 148856 | 80816 | 6148 | J16302+2129A | S,E | hr6148,hd148856 | G7IIIa | SB* | 2.770 | 0.930 | 23.44 | 42.7 | -25.91 |
| * s Her | 148897 | 80843 | 6152 | | E | hd148897 | G8III | V* | 5.250 | 1.258 | 5.19 | 192.7 | 18.40 |
| * phi Oph | 148786 | 80894 | 6147 | J16311-1636A | S | hr6147 | G9III | *in** | 4.270 | 0.920 | 13.39 | 74.7 | -33.46 |
| HR 6145 | 148760 | 80910 | 6145 | | S | hr6145 | K1III | Star | 6.090 | 1.088 | 14.58 | 68.6 | 39.40 |
| * h Her | 149161 | 81008 | 6159 | | E | hd149161 | K4III | V* | 4.850 | 1.490 | 9.29 | 107.6 | 2.92 |
| * gam Aps | 147675 | 81065 | 6102 | | H | gamaps | G8III | Star | 3.872 | 0.905 | 20.87 | 47.9 | 5.40 |
| HD 149216 | 149216 | 81098 | | | S | hd149216 | K0III | Star | 7.770 | 1.230 | 8.06 | 124.1 | |
| * H Sco | 149447 | 81304 | 6166 | | U | hd149447 | K6III | V* | 4.184 | 1.592 | 9.52 | 105.0 | -2.10 |
| HR 6199 | 150449 | 81437 | 6199 | | S | hr6199 | K1III | Star | 5.287 | 1.050 | 12.04 | 83.1 | -24.30 |
| HR 6190 | 150259 | 81632 | 6190 | | S | hr6190 | K0III | Star | 6.258 | 1.074 | 7.26 | 137.7 | 32.80 |
| * 14 Oph | 150557 | 81734 | 6205 | | E | hd150557 | F2III | Star | 5.734 | 0.301 | 17.99 | 55.6 | -48.80 |
| * eta Her | 150997 | 81833 | 6220 | J16428+3855A | S | hr6220 | G7.5IIIb | *in** | 3.500 | 0.920 | 30.02 | 33.3 | 8.27 |
| * alf TrA | 150798 | 82273 | 6217 | | U | hd150798 | K2II-III | Star | 1.920 | 1.440 | 8.35 | 119.8 | -3.00 |
| * eps Sco | 151680 | 82396 | 6241 | | U | hip082396 | K1III | V* | 2.290 | 1.160 | 51.19 | 19.5 | -2.50 |
| HR 6259 | 152224 | 82426 | 6259 | | S | hr6259 | K0III | Star | 6.168 | 1.005 | 10.63 | 94.1 | -23.38 |
| * 23 Oph | 152601 | 82730 | 6280 | | S | hr6280 | K2III | Star | 5.235 | 1.099 | 13.55 | 73.8 | -14.55 |
| HR 6287 | 152815 | 82764 | 6287 | | S | hr6287 | G8III | Star | 5.401 | 0.951 | 12.08 | 82.8 | -2.24 |
| * 56 Her | 152863 | 82780 | 6292 | J16550+2544A | S | hr6292 | G5III | *in** | 6.073 | 0.908 | 7.31 | 136.8 | -0.50 |
| * 57 Her | 153287 | 82987 | 6305 | | S | hr6305 | G5III: | Star | 6.290 | 0.910 | 5.90 | 169.5 | 8.60 |
| * kap Oph | 153210 | 83000 | 6299 | | E | hd153210 | K2III | PulsV* | 3.200 | 1.160 | 35.66 | 28.0 | -55.85 |
| HR 6307 | 153312 | 83007 | 6307 | | S | hr6307 | K0III | Star | 6.336 | 1.093 | 6.61 | 151.3 | -22.40 |
| * zet Ara | 152786 | 83081 | 6285 | | U | hd152786 | K3III | Star | 3.127 | 1.643 | 6.71 | 149.0 | -6.00 |
| HR 6330 | 153956 | 83138 | 6330 | | E | hd153956 | K1III: | SB* | 6.050 | 1.171 | 10.74 | 93.1 | -13.65 |
| HR 6360 | 154633 | 83359 | 6360 | | S | hr6360 | G5V | Star | 6.099 | 0.955 | 8.64 | 115.7 | -23.50 |
| HR 6333 | 154084 | 83367 | 6333 | | S | hr6333 | G7III: | Star | 5.774 | 1.003 | 10.42 | 96.0 | -50.40 |
| HR 6342 | 154278 | 83504 | 6342 | J17038+1335B | S | hr6342 | K1III | *in** | 6.063 | 1.046 | 10.21 | 97.9 | 45.71 |
| HR 6363 | 154732 | 83575 | 6363 | | S | hr6363 | K1III | SB* | 6.114 | 1.092 | 8.91 | 112.2 | 12.51 |
| HR 6359 | 154619 | 83688 | 6359 | | S | hr6359 | G8III-IV | Star | 6.380 | 0.890 | 8.40 | 119.0 | -22.20 |
| HR 6364 | 154733 | 83692 | 6364 | | S | hr6364 | K3III | Star | 5.578 | 1.315 | 9.54 | 104.8 | -96.88 |
| HR 6365 | 154779 | 83854 | 6365 | | S | hr6365 | K0III | Star | 5.985 | 1.000 | 8.70 | 114.9 | -13.70 |
| HR 6388 | 155410 | 83947 | 6388 | | S | hr6388 | K3III | SB* | 5.075 | 1.290 | 11.56 | 86.5 | -57.60 |
| HR 6390 | 155500 | 84113 | 6390 | | S | hr6390 | K0III | Star | 6.344 | 1.030 | 7.35 | 136.1 | -3.76 |

| Name | HD | HIP | HR | WDS | Cluster | Src | ID | SpType | Type | V | B-V | Plx | pmRA | RV |
|---|---|---|---|---|---|---|---|---|---|---|---|---|---|---|
| HR 6394 | 155646 | 84217 | 6394 | | | S | hr6394 | F6III | Star | 6.655 | 0.457 | 14.39 | 69.5 | 60.80 |
| HR 6404 | 155970 | 84402 | 6404 | J17153-1435AB | | S | hr6404 | K1III | SB* | 6.006 | 1.096 | 11.19 | 89.4 | -7.41 |
| HD 148451 | 148451 | 84461 | 6133 | | | U | hd148451 | G5III | PM* | 6.601 | 0.873 | 5.14 | 194.6 | -4.00 |
| V* VW Dra | 156947 | 84496 | 6448 | | | S | hr6448 | K1.5IIIb | semi-regV* | 6.328 | 1.089 | 6.63 | 150.8 | 14.18 |
| * 41 Oph | 156266 | 84514 | 6415 | J17166-0027AB | | S | hr6415 | K2III | ** | 4.743 | 1.169 | 15.85 | 63.1 | -0.16 |
| HR 6444 | 156891 | 84656 | 6444 | | | S | hr6444 | G7III: | Star | 5.974 | 1.003 | 10.30 | 97.1 | -40.61 |
| HR 6443 | 156874 | 84691 | 6443 | | | S | hr6443 | K0III | Star | 5.688 | 0.970 | 10.30 | 97.1 | -12.36 |
| HR 6488 | 157853 | 85149 | 6488 | J17240+3835AB | | S | hr6488 | F8IV | Star | 6.545 | 0.633 | 5.70 | 175.4 | -23.70 |
| CD-49 11401 | | | | | IC 4651 MMU 7646 | U,U | ic4651-E12,ic4651no7646 | G9III | *inCl | 10.400 | 1.100 | | | -31.01 |
| HR 6472 | 157527 | 85207 | 6472 | J17247-2126AB | | S | hr6472 | K0III | Star | 5.837 | 0.923 | 10.81 | 92.5 | -55.90 |
| CD-49 11402 | | | | | IC 4651 MMU 8540 | H | ic4651no8540 | K0III | *inCl | 10.940 | 1.170 | | | -30.03 |
| Cl* IC 4651 MMU 9025 | | | | | IC 4651 MMU 9025 | H,U | ic4651no9025,ic4651-E60 | K0III | *inCl | 10.900 | 1.130 | | | -30.46 |
| CD-49 11404 | | | | | IC 4651 MMU 9122 | U | ic4651no9122 | K1/K2III | *inCl | 10.700 | 0.800 | | | -30.58 |
| CD-49 11415 | | | | | IC 4651 MMU 9791 | H | ic4651no9791 | K2III | *inCl | 10.440 | 1.320 | | | -31.44 |
| Cl* IC 4651 MMU 14527 | | | | | IC 4651 MMU 14527 | H | ic4651no14527 | K0III | *inCl | 10.940 | 1.140 | | | -31.85 |
| HD 157935 | 157935 | 85289 | | | | S | hd157935 | F5III | Star | 6.620 | 0.380 | 7.41 | 135.0 | -54.50 |
| * kap Ara | 157457 | 85312 | 6468 | J17260-5038A | | U | hd157457 | G8III | *in** | 5.204 | 1.039 | 7.16 | 139.7 | 17.80 |
| HR 6524 | 158837 | 85749 | 6524 | J17314+0243AB | | S | hr6524 | G8III | SB* | 5.586 | 0.815 | 8.96 | 111.6 | -42.22 |
| * f Dra | 159966 | 85805 | 6566 | J17319+6809A | | S | hr6566 | G9IIIb | Star | 5.083 | 1.050 | 15.46 | 64.7 | -73.96 |
| HR 6542 | 159353 | 85930 | 6542 | | | E | hd159353 | K0III: | Star | 5.691 | 0.993 | 10.76 | 92.9 | -24.61 |
| HR 6564 | 159926 | 86130 | 6564 | | | S | hr6564 | K5 | Star | 6.390 | 1.370 | 7.49 | 133.5 | -35.26 |
| HR 6606 | 161178 | 86219 | 6606 | | | S | hr6606 | G9III | Star | 5.881 | 1.008 | 8.99 | 111.2 | 6.32 |
| * ksi Ser | 159876 | 86263 | 6561 | J17376-1524A | | E | hd159876 | A9IIIpSr: | SB* | 3.539 | 0.256 | 30.98 | 32.3 | -42.80 |
| HR 6579 | 160507 | 86362 | 6579 | | | E | hd160507 | G5III: | Star | 6.560 | 0.990 | 7.48 | 133.7 | -14.53 |
| HD 160314 | 160314 | 86379 | | J17391+0202B | | S | hr6575b | F0 | gammaDor | 7.740 | 0.360 | 8.70 | 114.9 | 0.50 |
| HR 6575 | 160315 | 86391 | 6575 | J17391+0202A | | S | hr6575a | K0III+... | *in** | 6.262 | 1.008 | 8.85 | 113.0 | -1.50 |
| HR 6591 | 160822 | 86506 | 6591 | J17407+3117A | | S | hr6591 | K0III | *in** | 6.314 | 1.071 | 7.67 | 130.4 | -5.41 |
| HR 6607 | 161193 | 86561 | 6607 | | | S | hr6607 | K0III: | Star | 5.999 | 1.065 | 10.76 | 92.9 | -6.51 |
| * 83 Her | 161074 | 86667 | 6602 | J17424+2434A | | E | hd161074 | K4III | *in** | 5.570 | 1.463 | 8.04 | 124.4 | -28.60 |
| * bet Oph | 161096 | 86742 | 6603 | | | S | hr6603 | K2III | V* | 2.750 | 1.180 | 39.85 | 25.1 | -12.53 |
| HD 161502 | 161502 | 86906 | | | | S | hd161502 | G5III | Star | 6.960 | 0.890 | 8.39 | 119.2 | -21.99 |
| HR 6638 | 162076 | 87158 | 6638 | | | S | hr6638 | G5IV | Star | 5.700 | 0.933 | 12.58 | 79.5 | -26.66 |
| * 87 Her | 162211 | 87194 | 6644 | | | S | hr6644 | K2III | Star | 5.104 | 1.150 | 15.39 | 65.0 | -25.24 |
| HR 6639 | 162113 | 87224 | 6639 | | | S | hr6639 | K0III | Star | 6.458 | 1.230 | 8.65 | 115.6 | -59.85 |
| HR 6654 | 162555 | 87308 | 6654 | | | S | hr6654 | K1III | Star | 5.526 | 1.068 | 12.86 | 77.8 | -14.84 |
| HR 6659 | 162596 | 87428 | 6659 | | | S | hr6659 | K0 | SB* | 6.322 | 1.128 | 9.02 | 110.9 | -42.80 |
| HR 6648 | 162391 | 87472 | 6648 | | NGC 6475 MMU 134 | U | hd162391 | G8III | SB* | 5.850 | 1.100 | 4.12 | 242.7 | -15.57 |
| HR 6666 | 162757 | 87540 | 6666 | | | E | hd162757 | K1III: | Star | 6.192 | 1.103 | 10.87 | 92.0 | -35.60 |
| HR 6658 | 162587 | 87569 | 6658 | J17534-3454AB | NGC 6475 MMU 58 | U | hd162587 | K3III | SB* | 5.600 | 1.090 | 3.96 | 252.5 | -11.94 |
| * ksi Dra | 163588 | 87585 | 6688 | J17535+5653A | | E | hd163588 | K2III | *in** | 3.750 | 1.180 | 28.98 | 34.5 | -26.46 |
| * gam Dra | 164058 | 87833 | 6705 | J17566+5129A | | E | hd164058 | K5III | *in** | 2.230 | 1.530 | 21.14 | 47.3 | -27.91 |
| * ksi Her | 163993 | 87933 | 6703 | | | S | hr6703 | G8III | semi-regV* | 3.700 | 0.940 | 23.84 | 41.9 | -1.72 |
| HR 6711 | 164280 | 88020 | 6711 | | | S | hr6711 | G5III: | Star | 6.035 | 0.935 | 8.83 | 113.3 | 9.80 |
| HR 6691 | 163652 | 88038 | 6691 | J17589-3652A | | U | hd163652 | G8III | Star | 5.744 | 0.897 | 8.21 | 121.8 | -87.30 |
| * nu. Oph | 163917 | 88048 | 6698 | | | S | hr6698 | G9III | Star | 3.340 | 0.990 | 21.64 | 46.2 | 12.95 |
| HR 6757 | 165462 | 88671 | 6757 | | | S | hr6757 | G8IIp | Star | 6.340 | 1.060 | 4.65 | 215.1 | -9.50 |
| * 71 Oph | 165760 | 88765 | 6770 | | | S | hr6770 | G8III | Star | 4.647 | 0.951 | 11.96 | 83.6 | -3.00 |

| Name | HD | HIP | HR | 2MASS | Src | Simbad ID | SpType | ObjType | Vmag | B-V | Plx | Dist | RV |
|---|---|---|---|---|---|---|---|---|---|---|---|---|---|
| HR 6791 | 166208 | 88788 | 6791 | | E | hd166208 | G8IIICN... | SB* | 5.011 | 0.893 | 7.96 | 125.6 | -14.72 |
| HD 166229 | 166229 | 88836 | 6793 | | S | hr6793 | K2.5III | PM* | 5.480 | 1.170 | 15.72 | 63.6 | -7.93 |
| HD 166411 | 166411 | 88929 | 6799 | | S | hr6799 | K1III: | PM* | 6.365 | 1.215 | 8.10 | 123.5 | -83.83 |
| * 11 Sgr | 166464 | 89153 | 6801 | J18117-2342A | S | hr6801 | K0III | *in** | 4.980 | 1.050 | 12.74 | 78.5 | 4.40 |
| * 37 Dra | 168653 | 89448 | 6865 | | S | hr6865 | K1III: | Star | 5.969 | 1.050 | 11.81 | 84.7 | -13.10 |
| HD 167768 | 167768 | 89587 | 6840 | | S | hr6840 | G3III | PM* | 6.000 | 0.890 | 9.67 | 103.4 | 1.60 |
| HD 167576 | 167576 | 89592 | | | S | hd167576 | K3III | Star | 6.660 | 1.280 | 8.09 | 123.6 | -10.40 |
| HD 168322 | 168322 | 89604 | 6853 | | S | hr6853 | G8.5IIIb | PM* | 6.138 | 0.956 | 8.15 | 122.7 | -73.20 |
| HR 6842 | 167818 | 89678 | 6842 | | U | hd167818 | K5III | Star | 4.662 | 1.670 | 4.28 | 233.6 | -16.90 |
| * kap Lyr | 168775 | 89826 | 6872 | | S | hr6872 | K2III | V* | 4.340 | 1.170 | 12.96 | 77.2 | -24.36 |
| * 74 Oph | 168656 | 89918 | 6866 | J18209+0323A | S | hr6866 | G8III | *in** | 4.857 | 0.897 | 12.52 | 79.9 | 4.35 |
| * del Sgr | 168454 | 89931 | 6859 | J18210-2950A | S | hr6859 | K3IIIa | *in** | 2.710 | 1.414 | 9.38 | 106.6 | -20.40 |
| * eta Ser | 168723 | 89962 | 6869 | J18214-0253A | E | hd168723 | K0III-IV | V* | 3.250 | 0.940 | 53.93 | 18.5 | 9.83 |
| HR 6885 | 169191 | 90067 | 6885 | | U | hd169191 | K3III | Star | 5.261 | 1.263 | 8.99 | 111.2 | -19.24 |
| V* V4393 Sgr | 168988 | 90113 | | | H | v4393sgr | K7III: | PulsV*~ | 7.680 | 2.060 | 1.28 | 781.3 | -12.43 |
| * zet Sct | 169156 | 90135 | 6884 | | S | hr6884 | G9IIIb | SB* | 4.673 | 0.945 | 15.78 | 63.4 | -5.02 |
| * 109 Her | 169414 | 90139 | 6895 | J18236+2147A | S | hr6895 | K2III | V* | 3.840 | 1.180 | 27.42 | 36.5 | -58.16 |
| HR 6890 | 169268 | 90174 | 6890 | | E | hd169268a | F6III-IV | SB* | 6.380 | 0.340 | 16.15 | 61.9 | -17.55 |
| HD 169689 | 169690 | 90313 | 6902 | | H | hd169689 | G9II+B9V | EB*Algol | 5.668 | 0.861 | 4.14 | 241.5 | -19.30 |
| * 42 Dra | 170693 | 90344 | 6945 | | S | hr6945 | K1.5III_Fe-1 | Star | 4.833 | 1.187 | 10.36 | 96.5 | 32.17 |
| * c Ser | 170474 | 90642 | 6935 | | S | hr6935 | K0III | Star | 5.390 | 0.945 | 14.20 | 70.4 | 28.36 |
| V* X Sct | | 90791 | | | U | XSct | G5III | deltaCep | 10.070 | 1.040 | 2.46 | 406.5 | 10.90 |
| HR 6970 | 171391 | 91105 | 6970 | | S | hr6970 | G8III | Star | 5.130 | 0.896 | 9.91 | 100.9 | 7.59 |
| * alf Sct | 171443 | 91117 | 6973 | | S | hr6973 | K3III | V* | 3.830 | 1.340 | 16.38 | 61.1 | 36.50 |
| HD 172052 | 172052 | 91446 | | | S | hd172052 | F5II | Star | 6.700 | 0.660 | -0.33 | | 15.40 |
| HR 7010 | 172424 | 91523 | 7010 | | S | hr7010 | G8III | *in** | 6.272 | 0.948 | 6.78 | 147.5 | -40.80 |
| HR 7042 | 173398 | 91606 | 7042 | | S | hr7042 | K0III | Star | 6.096 | 0.966 | 9.00 | 111.1 | -25.02 |
| * del Sct | 172748 | 91726 | 7020 | J18423-0903A | E | hd172748 | F2IIIp | PulsV*delSct | 4.710 | 0.330 | 16.11 | 62.1 | -45.10 |
| HD 173378 | 173378 | 92086 | | | S | hd173378 | K0III | Star | 6.620 | 0.940 | 9.92 | 100.8 | |
| HR 7064 | 173780 | 92088 | 7064 | | S | hr7064 | K3III | Star | 4.840 | 1.200 | 12.80 | 78.1 | -16.80 |
| HD 175305 | 175305 | 92167 | | | E | hd175305 | G5III | PM* | 7.180 | 0.750 | 6.39 | 156.5 | -184.33 |
| Cl* NGC 6705 MMU 1423 | | | | NGC 6705 MMU 1423 | H | ngc6705no1423 | KIII: | *inCl | 11.440 | 1.610 | | | 36.13 |
| Cl* NGC 6705 MMU 1286 | | | | NGC 6705 MMU 1286 | H | ngc6705no1286 | G8II-III | *inCl | 11.790 | 0.980 | | | 34.29 |
| Cl* NGC 6705 MMU 779 | | | | NGC 6705 PPM 306 | H | ngc6705no779 | K3II-III | *inCl | | 0.000 | | | 33.69 |
| * omi Dra | 175306 | 92512 | 7125 | J18512+5924A | S | hr7125 | G9III_Fe-0.5 | RSCVn | 4.642 | 1.171 | 9.54 | 104.8 | -19.52 |
| Cl* NGC 6705 MMU 660 | | | | NGC 6705 STR 674 | H | ngc6705no660 | K1II-III | *inCl | | 0.000 | | | 35.51 |
| Cl* NGC 6705 MMU 411 | | | | NGC 6705 MMU 411 | H | ngc6705no411 | K2II-III | *inCl | 11.640 | 1.670 | | | 34.35 |
| HR 7137 | 175535 | 92689 | 7137 | | S | hr7137 | G7IIIa | Star | 4.920 | 0.900 | 7.15 | 139.9 | 8.50 |
| * ups Dra | 176524 | 92782 | 7180 | | S | hr7180 | K0III | SB* | 4.827 | 1.151 | 9.48 | 105.5 | -11.10 |
| HR 7146 | 175740 | 92831 | 7146 | J18549+4136A | S | hr7146 | G8III | *in** | 5.440 | 1.050 | 12.20 | 82.0 | -9.18 |
| * nu.02 Sgr | 175190 | 92845 | 7120 | | S | hr7120 | K1Ib/II | Star | 4.980 | 1.320 | 11.91 | 84.0 | -109.60 |
| HR 7135 | 175515 | 92872 | 7135 | | S | hr7135 | G9III | SB* | 5.580 | 1.035 | 11.33 | 88.3 | 27.74 |
| HD 175545 | 175545 | 92914 | | | U | hd175545 | K2III | Star | 7.400 | 1.200 | 8.95 | 111.7 | -19.00 |

| Name | HD | HIP | HR | WDS | Src | ID2 | SpType | ObjType | Vmag | B-V | Plx | pmRA | pmDE |
|---|---|---|---|---|---|---|---|---|---|---|---|---|---|
| HD 175743 | 175743 | 92937 | 7148 | | S | hr7148 | K1III | PM* | 5.706 | 1.090 | 10.84 | 92.3 | 45.69 |
| HD 175679 | 175679 | 92968 | 7144 | J18564+0229A | S | hr7144 | G8III | Star | 6.143 | 0.967 | 6.39 | 156.5 | 16.20 |
| HR 7187 | 176598 | 92969 | 7187 | | S | hr7187 | G8III | Star | 5.632 | 0.923 | 10.40 | 96.2 | -8.11 |
| HD 175940 | 175940 | 92986 | | | S | hd175940 | K2III | Star | 6.950 | 1.130 | 8.55 | 117.0 | -35.15 |
| * 48 Dra | 176408 | 92997 | 7175 | | E | hd176408 | K1III | Star | 5.676 | 1.160 | 11.81 | 84.7 | -35.51 |
| * ksi02 Sgr | 175775 | 93085 | 7150 | | H | hr7150 | G9II/III | Star | 3.510 | 1.180 | 8.93 | 112.0 | -20.10 |
| HR 7196 | 176707 | 93197 | 7196 | | S | hr7196 | G8III | Star | 6.342 | 0.969 | 7.59 | 131.8 | -21.20 |
| * eps Aql | 176411 | 93244 | 7176 | J18596+1503A | S | hr7176 | K1III | SB* | 4.020 | 1.080 | 21.05 | 47.5 | -45.90 |
| HR 7204 | 176896 | 93354 | 7204 | | S | hr7204 | K0III: | Star | 6.050 | 0.972 | 8.46 | 118.2 | -28.52 |
| HR 7186 | 176593 | 93418 | 7186 | | S | hr7186 | K0III | Star | 6.320 | 1.000 | 6.07 | 164.7 | 20.30 |
| * i Aql | 176678 | 93429 | 7193 | | S | hr7193 | K1III | V* | 4.020 | 1.090 | 22.66 | 44.1 | -43.92 |
| HD 176704 | 176704 | 93498 | 7195 | | U | hip093498 | K2III | PM* | 5.640 | 1.227 | 12.66 | 79.0 | 2.80 |
| HD 176354 | 176354 | 93547 | | | H | hd176354 | K0III | Star | 7.070 | 0.870 | 21.68 | 46.1 | -29.50 |
| * omi Sgr | 177241 | 93683 | 7217 | J19047-2144A | S | hr7217 | K0III | *in** | 3.770 | 1.000 | 22.96 | 43.6 | 26.10 |
| HD 177442 | 177442 | 93716 | | J19050-0402B | S | hr7225b | K0 | *in** | 6.810 | 1.440 | 5.90 | 169.5 | -59.00 |
| * h Aql | 177463 | 93717 | 7225 | J19050-0402A | S | hr7225a | K1III | *in** | 5.409 | 1.121 | 11.27 | 88.7 | -22.17 |
| * tau Sgr | 177716 | 93864 | 7234 | | S | hr7234 | K1III | PM* | 3.310 | 1.200 | 26.82 | 37.3 | 45.40 |
| * 19 Aql | 178596 | 94068 | 7266 | | E | hd178596 | F0III-IV | V* | 5.230 | 0.340 | 21.84 | 4.87 | -51.60 |
| * 53 Dra | 180006 | 94302 | 7295 | | S | hr7295 | G8III | V* | 5.140 | 1.000 | 9.48 | 105.5 | -16.10 |
| * del Dra | 180711 | 94376 | 7310 | J19126+6740A | S | hr7310 | G9III | PM* | 3.070 | 1.000 | 33.48 | 29.9 | 24.71 |
| * tau Dra | 181984 | 94648 | 7352 | | E | hd181984 | K2III: | V* | 4.450 | 1.250 | 22.28 | 44.9 | -33.70 |
| HR 7325 | 181122 | 94916 | 7325 | | S | hr7325 | G9III | Star | 6.316 | 1.061 | 6.99 | 143.1 | -10.70 |
| HD 181214 | 181214 | 94954 | | | E | hd181214 | F8III | Star | 7.740 | 0.510 | 7.11 | 140.6 | -36.10 |
| * 28 Aql | 181333 | 94982 | 7331 | J19197+1222AC | S | hr7331 | F0III | ** | 5.531 | 0.237 | 9.67 | 103.4 | 3.10 |
| HD 181907 | 181907 | 95222 | 7349 | | S | hr7349 | G8III | Star | 5.824 | 1.090 | 9.64 | 103.7 | -10.90 |
| HR 7359 | 182272 | 95235 | 7359 | | S | hr7359 | K0III | Star | 6.094 | 1.047 | 9.10 | 109.9 | -13.86 |
| HD 181517 | 181517 | 95322 | | | H | hd181517 | K0III | Star | 6.610 | 0.990 | 6.97 | 143.5 | |
| HR 7376 | 182635 | 95370 | 7376 | | S | hr7376 | K1III | Star | 6.430 | 1.080 | 8.22 | 121.7 | -33.51 |
| HD 181433 | 181433 | 95467 | | | H | hd181433 | K3III-IV | PM* | 8.380 | 1.040 | 37.37 | 26.8 | 40.21 |
| * 4 Vul | 182762 | 95498 | 7385 | J19255+1948A | S | hr7385 | K0III | *in** | 5.160 | 0.980 | 12.01 | 83.3 | 0.65 |
| HR 7389 | 182900 | 95572 | 7389 | | S | hr7389 | F6III | Star | 5.770 | 0.429 | 17.86 | 56.0 | -35.50 |
| HD 182901 | 182901 | 95586 | | | S | hd182901 | F5III | Star | 6.914 | 0.396 | 15.71 | 63.7 | -44.80 |
| HR 7407 | 183492 | 95822 | 7407 | | S | hr7407 | K0III | Star | 5.576 | 1.045 | 10.90 | 91.7 | -41.58 |
| HR 7398 | 183275 | 95865 | 7398 | J19299-2659AB | U | hr183275 | K3III | *in** | 5.499 | 1.135 | 13.94 | 71.7 | -31.60 |
| HR 7388 | 182893 | 95866 | 7388 | | H | hd182893 | K0.5III | Star | 6.141 | 0.965 | 10.42 | 96.0 | -27.10 |
| * mu. Aql | 184406 | 96229 | 7429 | J19341+0723A | E | hd184406 | K3IIIb | V* | 4.450 | 1.180 | 30.31 | 33.0 | -24.73 |
| HR 7433 | 184574 | 96365 | 7433 | | S | hr7433 | K0III | Star | 6.287 | 1.089 | 9.73 | 102.8 | 5.60 |
| HR 7465 | 185264 | 96396 | 7465 | | S | hr7465 | G9III | Star | 6.460 | 1.070 | 6.01 | 166.4 | 8.70 |
| HR 7449 | 184944 | 96428 | 7449 | | S | hr7449 | K0II-III | Star | 6.364 | 1.048 | 8.40 | 119.0 | -43.90 |
| HR 7468 | 185351 | 96459 | 7468 | | E | hd185351 | G9IIIbCN... | Star | 5.180 | 0.925 | 24.49 | 40.8 | -5.90 |
| * phi Cyg | 185734 | 96683 | 7478 | | S | hr7478b | G8III | SB* | 4.685 | 0.963 | 12.25 | 81.6 | 4.50 |
| HR 7487 | 185955 | 96706 | 7487 | | S | hr7487 | G9II-III | Star | 6.252 | 0.914 | 7.91 | 126.4 | -11.86 |
| * e01 Sgr | 185644 | 96808 | 7476 | J19407-1618A | E | hd185644 | K1III | *in** | 5.302 | 1.126 | 13.56 | 73.7 | -59.16 |
| HR 7526 | 186815 | 97070 | 7526 | | S | hr7526 | K2III | Star | 6.293 | 0.865 | 12.86 | 77.8 | -27.29 |
| * 10 Vul | 186486 | 97077 | 7506 | | S | hr7506 | G8III | Star | 5.497 | 0.923 | 9.83 | 101.7 | -9.90 |
| * 15 Cyg | 186675 | 97118 | 7517 | | S | hr7517 | G7III | Star | 4.899 | 0.931 | 11.28 | 88.7 | -23.62 |
| HD 186535 | 186535 | 97144 | | | S | hd186535 | K0 | Star | 6.438 | 0.925 | 6.75 | 148.1 | 4.44 |
| V* CN Dra | 187764 | 97326 | 7563 | | E | hd187764 | F0III | PulsV*delSct | 6.339 | 0.276 | 5.71 | 175.1 | -10.90 |
| HR 7540 | 187193 | 97402 | 7540 | | S | hr7540 | K0II-III | Star | 6.013 | 0.983 | 7.88 | 126.9 | -18.03 |
| * eps Dra | 188119 | 97433 | 7582 | J19482+7016AB | E | hd188119 | G8III | V* | 3.840 | 0.890 | 22.04 | 45.4 | 2.69 |

| Name | HD | HIP | HR | WDS | Src | ID | SpType | ObjType | Vmag | B-V | Plx | pmRA | pmDE |
|---|---|---|---|---|---|---|---|---|---|---|---|---|---|
| HR 7541 | 187195 | 97499 | 7541 | | S | hr7541 | K5III | Star | 6.040 | 1.240 | 10.37 | 96.4 | -33.30 |
| * d Cyg | 188056 | 97635 | 7576 | | S | hr7576 | K3III | V* | 5.030 | 1.280 | 16.11 | 62.1 | -21.75 |
| * 57 Sgr | 187739 | 97783 | 7561 | | S | hr7561 | K0III | Star | 5.905 | 0.957 | 8.63 | 115.9 | -21.59 |
| HD 187669 | 187669 | | | | H | hd187669a | K0III | Star | 8.880 | 1.240 | | | |
| * iot Sgr | 188114 | 98032 | 7581 | | H | hd188114 | K0II-III | Star | 4.130 | 1.080 | 17.94 | 55.7 | 35.80 |
| * ome Sgr | 188376 | 98066 | 7597 | | S | hr7597 | G5IV | PM* | 4.700 | 0.750 | 38.48 | 26.0 | -21.00 |
| * eta Cyg | 188947 | 98110 | 7615 | J19563+3505A | E | hd188947 | K0III | *in** | 3.880 | 1.030 | 24.17 | 41.4 | -25.87 |
| HD 188993 | 188993 | 98138 | | | S | hd188993 | G2III | Star | 6.800 | 0.640 | 10.74 | 93.1 | -21.23 |
| HD 189186 | 189186 | 98314 | | | S | hd189186 | K0 | Star | 6.770 | 0.900 | 11.40 | 87.7 | 56.76 |
| * gam Sge | 189319 | 98337 | 7635 | | E | hd189319 | M0III | V* | 3.470 | 1.570 | 12.62 | 79.2 | -34.00 |
| HR 7659 | 190056 | 98842 | 7659 | J20043-3203A | H | hd190056 | K1III | V* | 4.990 | 1.210 | 11.22 | 89.1 | -11.80 |
| HR 7681 | 190664 | 99024 | 7681 | | S | hr7681 | K0 | Star | 6.476 | 1.146 | 8.95 | 111.7 | -2.10 |
| HR 7713 | 191814 | 99445 | 7713 | | S | hr7713 | K0 | Star | 6.245 | 0.873 | 6.69 | 149.5 | -4.25 |
| * ksi01 Cap | 191753 | 99529 | 7712 | | S | hr7712 | K0III | Star | 6.340 | 1.210 | 6.95 | 143.9 | 0.90 |
| HR 7743 | 192787 | 99841 | 7743 | J20154+3344AB | E | hd192787 | K0III | Star | 5.711 | 0.908 | 10.26 | 97.5 | -9.13 |
| * 18 Sge | 192836 | 99913 | 7746 | | E | hd192836 | K1III | Star | 6.127 | 1.040 | 9.84 | 101.6 | -4.81 |
| * 4 Cap | 192879 | 100062 | 7748 | | S | hr7748 | G8III | HB* | 5.862 | 0.988 | 10.31 | 97.0 | -18.30 |
| HR 7778 | 193556 | 100274 | 7778 | | S | hr7778 | G8III | Star | 6.173 | 0.898 | 5.88 | 170.1 | 11.70 |
| HR 7802 | 194220 | 100515 | 7802 | J20230+4259AB | S | hr7802 | K0IIIv | Star | 6.252 | 0.975 | 9.71 | 103.0 | -19.98 |
| HR 7788 | 193896 | 100524 | 7788 | | S | hr7788 | G5IIIa | Star | 6.300 | 0.877 | 6.25 | 160.1 | -16.40 |
| HR 7794 | 194013 | 100541 | 7794 | | S | hr7794 | G8III-IV | Star | 5.305 | 0.972 | 13.23 | 75.6 | -11.81 |
| * 39 Cyg | 194317 | 100587 | 7806 | | S | hr7806 | K3III | SB* | 4.440 | 1.330 | 13.05 | 76.6 | -17.18 |
| HD 194708 | 194708 | 100736 | | | S | hd194708 | F6III | Star | 6.889 | 0.431 | 10.38 | 96.3 | -27.00 |
| HR 7820 | 194937 | 100953 | 7820 | | S | hr7820 | G9III | Star | 6.234 | 1.060 | 10.58 | 94.5 | -10.80 |
| HR 7824 | 194953 | 100969 | 7824 | | S | hr7824 | G8III | Star | 6.210 | 0.877 | 8.05 | 124.2 | -28.00 |
| * 69 Aql | 195135 | 101101 | 7831 | | S | hr7831 | K2III | Star | 4.912 | 1.169 | 16.32 | 61.3 | -22.58 |
| HR 7854 | 195820 | 101245 | 7854 | | S | hr7854 | K0III | Star | 6.217 | 1.010 | 8.68 | 115.2 | -10.46 |
| HR 7867 | 196134 | 101467 | 7867 | | E | hd196134 | K0III-IV | Star | 6.513 | 0.984 | 10.86 | 92.1 | 1.19 |
| * alf Ind | 196171 | 101772 | 7869 | J20376-4717A | H | hd196171 | K0III-IV | *in** | 3.110 | 1.000 | 33.17 | 30.1 | -1.30 |
| HR 7904 | 196852 | 101899 | 7904 | | S | hr7904 | K2III | Star | 5.686 | 1.086 | 9.21 | 108.6 | 11.52 |
| * 1 Aqr | 196758 | 101936 | 7897 | J20394+0029A | S | hr7897 | K1III | *in** | 5.154 | 1.057 | 13.99 | 71.5 | -42.81 |
| HR 7905 | 196857 | 102026 | 7905 | | S | hr7905 | K0III | Star | 5.800 | 0.986 | 9.96 | 100.4 | -3.70 |
| HD 196983 | 196983 | | | | U | hd196983 | K2III | Star | 9.080 | 1.170 | | | -8.00 |
| V* LU Del | 197249 | 102158 | 7923 | | S | hr7923 | G8III | semi-regV* | 6.247 | 0.932 | 6.72 | 148.8 | 1.60 |
| * 30 Vul | 197752 | 102388 | 7939 | | S | hr7939 | K2III | SB* | 4.910 | 1.190 | 10.04 | 99.6 | 30.00 |
| * 52 Cyg | 197912 | 102453 | 7942 | J20457+3043A | S | hr7942 | G9.5III | *in** | 4.230 | 1.060 | 16.22 | 61.7 | -0.72 |
| CD-38 14203B | 197557B | | | J20457-3736B | H | cd-3814203b | G5:III:+... | *in** | 10.500 | 0.440 | | | 46.80 |
| * eps Cyg | 197989 | 102488 | 7949 | J20462+3358A | E | hd197989 | K0III-IV | PM* | 2.480 | 1.040 | 44.86 | 22.3 | -12.41 |
| HR 7962 | 198181 | 102499 | 7962 | | S | hr7962 | K0 | Star | 6.355 | 1.133 | 7.32 | 136.6 | -27.27 |
| * alf Mic | 198232 | 102831 | 7965 | J20500-3347A | H | hd198232 | G7III | *in** | 4.900 | 0.980 | 8.62 | 116.0 | -14.50 |
| HR 7976 | 198431 | 102891 | 7976 | | E | hd198431 | K1III | Star | 5.880 | 1.060 | 12.35 | 81.0 | -46.11 |
| * 31 Vul | 198809 | 103004 | 7995 | | S | hr7995 | G7III | V* | 4.576 | 0.813 | 17.30 | 57.8 | 2.25 |
| V* V1794 Cyg | 199178 | 103144 | | | E | hd199178 | G5III-IV | RotV* | 7.240 | 0.740 | 9.87 | 101.3 | -26.65 |
| * 19 Cap | 199012 | 103226 | 8000 | | S | hr8000 | K0III | Star | 5.784 | 1.115 | 8.94 | 111.9 | -38.90 |
| HR 8017 | 199442 | 103414 | 8017 | J20572+0028A | S | hr8017 | K2III | *in** | 6.065 | 1.227 | 10.64 | 94.0 | -25.98 |
| HR 8035 | 199870 | 103519 | 8035 | | S | hr8035 | K0IIIbCN... | SB* | 5.562 | 0.965 | 12.61 | 79.3 | -28.70 |
| * 18 Del | 199665 | 103527 | 8030 | | S | hr8030 | G6III: | *in** | 5.522 | 0.918 | 13.28 | 75.3 | 3.81 |
| * eps Equ B | 199766B | | | J20591+0418B | S | hr8034b | F6(V) | Star | 6.310 | 0.460 | | | |
| * eps Equ A | 199766A | | | J20591+0418A | S | hr8034a | F5(V) | Star | 5.960 | 0.520 | | | |
| HD 199642 | 199642 | 103693 | | | U | hd199642 | K7III | Star | 6.260 | 1.580 | 4.31 | 232.0 | |

| Name | HD | HIP | HR | WDS | Src | Ident | SpType | ObjType | V | B-V | plx | pmRA | pmDE |
|---|---|---|---|---|---|---|---|---|---|---|---|---|---|
| * gam Mic | 199951 | 103738 | 8039 | J21013-3215A | H,U | hd199951,hip103738 | G6III | *in** | 4.677 | 0.866 | 14.24 | 70.2 | 17.60 |
| HR 8072 | 200740 | 103929 | 8072 | | S | hr8072 | K0 | Star | 6.379 | 0.975 | 8.94 | 111.9 | -20.08 |
| V* V1719 Cyg | 200925 | 104029 | | | S | hd200925 | F5III | PulsV*delSct | 7.990 | 0.420 | 3.79 | 263.9 | 15.00 |
| HR 8082 | 201051 | 104172 | 8082 | | S | hr8082 | K0II-III | Star | 6.135 | 1.019 | 7.93 | 126.1 | -3.44 |
| HR 8076 | 200763 | 104174 | 8076 | | H | hd200763 | K3III | Star | 5.203 | 1.101 | 9.94 | 100.6 | 3.10 |
| * nu. Aqr | 201381 | 104459 | 8093 | | S | hr8093 | G8III | Star | 4.520 | 0.940 | 20.47 | 48.9 | -11.23 |
| HR 8096 | 201567 | 104557 | 8096 | | S | hr8096 | K0III | Star | 6.274 | 1.170 | 9.65 | 103.6 | -39.10 |
| * zet Cyg | 202109 | 104732 | 8115 | J21129+3014A | S,E | hr8115,hd202109 | G8IIIp+DA4.2 | SB* | 3.210 | 0.990 | 22.79 | 43.9 | 16.72 |
| HR 8108 | 201852 | 104752 | 8108 | | H | hd201852 | G8III: | Star | 5.967 | 0.962 | 8.92 | 112.1 | 0.40 |
| * phi Cap | 202320 | 104963 | 8127 | | U | hd202320 | K0II/III | Star | 5.163 | 1.167 | 5.07 | 197.2 | -4.50 |
| HR 8179 | 203574 | 105370 | 8179 | | S | hr8179 | G5III | Star | 6.115 | 0.996 | 8.04 | 124.4 | -25.50 |
| * 34 Vul | 203344 | 105411 | 8165 | | S | hr8165 | K1III | PM* | 5.570 | 1.050 | 12.06 | 82.9 | -88.40 |
| HR 8185 | 203644 | 105497 | 8185 | | S | hr8185 | K0III | Star | 5.684 | 1.097 | 9.91 | 100.9 | -4.33 |
| * 1 Peg | 203504 | 105502 | 8173 | J21221+1949A | S | hr8173 | K1III | SB* | 4.090 | 1.110 | 20.93 | 47.8 | -77.16 |
| * iot Cap | 203387 | 105515 | 8167 | | S | hr8167 | G8III | BYDra | 4.270 | 0.910 | 16.58 | 60.3 | 12.58 |
| * 33 Cap | 203638 | 105665 | 8183 | | U | hd203638 | K0III | V* | 5.372 | 1.182 | 13.63 | 73.4 | 22.10 |
| HR 8191 | 203842 | 105695 | 8191 | | S | hr8191 | F5III | Star | 6.328 | 0.448 | 8.80 | 113.6 | -33.20 |
| * b Cap | 204381 | 106039 | 8213 | | U | hip106039 | K0III | Star | 4.500 | 0.900 | 19.06 | 52.5 | -22.20 |
| HD 204642 | 204642 | 106081 | | | S | hd204642 | K2III | Star | 6.757 | 1.090 | 10.22 | 97.8 | 20.42 |
| * g Cyg | 204771 | 106093 | 8228 | | S | hr8228 | K0III | Star | 5.231 | 0.952 | 15.05 | 66.4 | -21.54 |
| HD 205011 | 205011 | 106306 | | | S | hd205011 | G9IIIa | SB* | 6.420 | 1.080 | 7.25 | 137.9 | 11.98 |
| * rho Cyg | 205435 | 106481 | 8252 | | S | hr8252 | G5III | V* | 4.020 | 0.890 | 26.39 | 37.9 | 6.88 |
| * 72 Cyg | 205512 | 106551 | 8255 | | S | hr8255 | K1III | PM* | 4.910 | 1.080 | 14.09 | 71.0 | -68.12 |
| HR 8274 | 206027 | 106872 | 8274 | | S | hr8274 | G9III | Star | 6.205 | 1.010 | 7.73 | 129.4 | -9.10 |
| HD 205972 | 205972 | 106922 | | | S | hd205972 | K0III | RGB* | 7.240 | 1.060 | 8.22 | 121.7 | |
| * d Aqr | 206067 | 106944 | 8277 | J21396+0215A | S | hr8277 | K0III | *in** | 5.106 | 1.029 | 14.40 | 69.4 | -34.63 |
| * 11 Cep | 206952 | 107119 | 8317 | | S | hr8317 | K1III | PM* | 4.544 | 1.120 | 17.88 | 55.9 | -38.36 |
| * kap Cap | 206453 | 107188 | 8288 | | S | hr8288 | G8III | Star | 4.738 | 0.866 | 11.09 | 90.2 | -2.87 |
| * 78 Dra | 207130 | 107230 | 8324 | | S | hr8324 | K1III | Star | 5.182 | 1.059 | 13.34 | 75.0 | -40.62 |
| HD 206642 | 206642 | 107344 | 8299 | | H | hd206642 | G5III | PM* | 6.286 | 1.132 | 3.76 | 266.0 | -58.00 |
| HR 8320 | 207088 | 107445 | 8320 | | S | hr8320 | G8III | Star | 6.430 | 1.010 | 6.60 | 151.5 | -4.30 |
| HD 207134 | 207134 | 107502 | 8325 | | S | hr8325 | K3III: | PM* | 6.292 | 1.225 | 8.99 | 111.2 | -47.52 |
| HR 8360 | 208111 | 108102 | 8360 | | S | hr8360 | K2III | Star | 5.716 | 1.194 | 11.15 | 89.7 | -43.50 |
| HR 8352 | 207964 | 108195 | 8352 | J21552-6153AB | U | hd207964 | F1III | ** | 5.924 | 0.353 | 21.52 | 46.5 | 1.00 |
| HR 8391 | 209149 | 108632 | 8391 | | S | hr8391 | F5III | Star | 6.470 | 0.420 | 14.54 | 68.8 | -1.30 |
| HR 8394 | 209240 | 108784 | 8394 | | S | hr8394 | K0III | Star | 6.293 | 0.995 | 11.12 | 89.9 | -13.98 |
| * 30 Aqr | 209396 | 108868 | 8401 | | S | hr8401 | K0III | Star | 5.558 | 0.951 | 10.74 | 93.1 | 39.74 |
| HD 209449 | 209449 | 108950 | | | H | hd209449 | G9III+... | Star | 7.240 | 0.700 | 19.83 | 50.4 | -17.10 |
| * 20 Cep | 209960 | 109005 | 8426 | | E | hd209960 | K4III | V* | 5.277 | 1.462 | 10.76 | 92.9 | -23.27 |
| * nu. Peg | 209747 | 109068 | 8413 | | E | hd209747 | K4III | V* | 4.840 | 1.440 | 12.01 | 83.3 | -18.90 |
| HR 8442 | 210220 | 109190 | 8442 | | S | hr8442 | G6III | Star | 6.355 | 0.853 | 4.80 | 208.3 | -9.10 |
| * pi. Peg | 210459 | 109410 | 8454 | | S | hr8454 | F5III | Star | 4.290 | 0.460 | 12.40 | 80.6 | 5.10 |
| HR 8456 | 210461 | 109445 | 8456 | J22104+1438A | S | hr8456 | K0III | *in** | 6.362 | 1.059 | 7.37 | 135.7 | -44.50 |
| HR 8453 | 210434 | 109466 | 8453 | J22106-0416A | S | hr8453 | K0III-IV | *in** | 5.979 | 0.973 | 9.81 | 101.9 | -20.33 |
| HR 8461 | 210702 | 109577 | 8461 | | S | hr8461 | K1III | Star | 5.939 | 0.946 | 18.20 | 54.9 | 15.98 |
| HD 210905 | 210905 | 109585 | 8476 | | S | hr8476 | K0III | PM* | 6.296 | 1.126 | 8.48 | 117.9 | -30.33 |
| HR 8482 | 211006 | 109730 | 8482 | | S | hr8482 | K2III | Star | 5.878 | 1.174 | 13.36 | 74.9 | -19.24 |
| HD 211173 | 211173 | 109933 | | | U | hd211173 | G8II/III | Pec* | 8.490 | 0.940 | 5.16 | 193.8 | -24.40 |
| * tet Aqr | 211391 | 110003 | 8499 | | S | hr8499 | G8III | Star | 4.160 | 0.990 | 17.40 | 57.5 | -13.77 |
| HR 8500 | 211392 | 110009 | 8500 | | S | hr8500 | K3III: | V* | 5.799 | 1.169 | 8.83 | 113.3 | 12.10 |

| Name | HD | HIP | HR | WDS | Src | Simbad ID | SpType | ObjType | Vmag | B-V | plx | pm | RV |
|---|---|---|---|---|---|---|---|---|---|---|---|---|---|
| HD 211607 | 211607 | 110089 | | | S | hd211607 | K0 | Star | 6.850 | 0.970 | 8.42 | 118.8 | -29.25 |
| HD 211317 | 211317 | 110161 | | J22189-6819A | H | hd211317 | G5III/IV | *in** | 7.250 | 0.650 | 20.35 | 49.1 | 27.80 |
| * 47 Aqr | 212010 | 110391 | 8516 | | S | hr8516 | K0III | RGB* | 5.135 | 1.054 | 19.07 | 52.4 | 48.20 |
| HD 212334 | 212334 | 110487 | | | S | hd212334 | K0 | Star | 6.463 | 1.047 | 7.30 | 137.0 | -9.42 |
| HR 8530 | 212320 | 110532 | 8530 | | U | hd212320 | G6III | Star | 5.922 | 0.990 | 6.99 | 143.1 | -14.20 |
| * bet Lac | 212496 | 110538 | 8538 | | S | hr8538 | G8.5IIIb | ** | 4.440 | 1.020 | 19.19 | 52.1 | -11.42 |
| * 35 Peg | 212943 | 110882 | 8551 | J22278+0441A | E | hd212943 | K0III | PM* | 4.800 | 1.060 | 21.99 | 45.5 | 54.16 |
| HR 8568 | 213242 | 110919 | 8568 | | S | hr8568 | K0 | Star | 6.294 | 1.114 | 8.88 | 112.6 | -27.65 |
| HD 213619 | 213619 | 111282 | | | S | hd213619 | F2III | Star | 6.559 | 0.309 | 11.72 | 85.3 | 15.30 |
| HR 8594 | 213930 | 111362 | 8594 | | S | hr8594 | K0III | PulsV* | 5.732 | 0.954 | 9.47 | 105.6 | -8.10 |
| HR 8596 | 213986 | 111515 | 8596 | | S | hr8596 | K1III | RGB* | 5.978 | 0.980 | 10.76 | 92.9 | -3.00 |
| * 31 Cep | 214470 | 111532 | 8615 | | E | hd214470 | F3III-IV | PM* | 5.084 | 0.374 | 18.22 | 54.9 | 0.10 |
| HR 8617 | 214558 | 111753 | 8617 | | S | hr8617 | G2III+... | Star | 6.370 | 0.760 | 5.89 | 169.8 | 7.80 |
| HD 214448 | 214448 | 111761 | 8612 | J22384-0754AB | E | hd214448 | G0III+... | Star | 6.256 | 0.754 | 8.38 | 119.3 | -18.34 |
| * 11 Lac | 214868 | 111944 | 8632 | | S | hr8632 | K2.5III | Star | 4.460 | 1.330 | 9.80 | 102.0 | -10.91 |
| HD 215030 | 215030 | 112041 | 8643 | | S | hr8643 | G9III | PM* | 5.920 | 1.020 | 10.90 | 91.7 | -14.29 |
| HR 8642 | 214995 | 112067 | 8642 | | S | hr8642 | K0III: | Star | 5.927 | 1.101 | 11.77 | 85.0 | -28.87 |
| * 13 Lac | 215373 | 112242 | 8656 | J22441+4149A | S | hr8656 | K0III | *in** | 5.080 | 0.960 | 12.71 | 78.7 | 12.73 |
| * 45 Peg | 215510 | 112358 | 8660 | | S | hr8660 | G6III: | Star | 6.271 | 1.054 | 8.82 | 113.4 | -21.80 |
| * 68 Aqr | 215721 | 112529 | 8670 | | S | hr8670 | G8III | PM* | 5.260 | 0.940 | 12.84 | 77.9 | 25.42 |
| HR 8678 | 215943 | 112590 | 8678 | | S | hr8678 | G8III: | V* | 5.817 | 1.017 | 9.68 | 103.3 | -25.10 |
| * iot Cep | 216228 | 112724 | 8694 | | E | hd216228 | K0III | V* | 3.540 | 1.060 | 28.29 | 35.3 | -12.59 |
| * 48 Peg | 216131 | 112748 | 8684 | | E | hd216131 | G8+III | PM* | 3.480 | 0.940 | 30.74 | 32.5 | 13.54 |
| V* IM Peg | 216489 | 112997 | 8703 | | S | hr8703 | K2III | RSCVn | 5.892 | 1.134 | 11.17 | 89.5 | -14.43 |
| HR 8712 | 216646 | 113084 | 8712 | | S | hr8712 | K0III | Star | 5.821 | 1.145 | 10.73 | 93.2 | -7.81 |
| * 1 Psc | 216701 | 113167 | 8715 | | S | hr8715 | A7III | Star | 6.110 | 0.200 | 9.24 | 108.2 | 12.80 |
| * del PsA | 216763 | 113246 | 8720 | J22559-3232A | U | hip113246 | G8III | *in** | 4.226 | 0.953 | 21.16 | 47.3 | -11.60 |
| HR 8730 | 217019 | 113360 | 8730 | | S | hr8730 | K1III | Star | 6.285 | 1.150 | 8.75 | 114.3 | 12.29 |
| * 2 Psc | 217264 | 113521 | 8742 | J22595+0058AB | S | hr8742 | K1III: | ** | 5.431 | 0.983 | 10.91 | 91.7 | -14.23 |
| HD 217590 | 217590 | 113705 | | | S | hd217590 | G5 | Star | 6.450 | 0.980 | 7.92 | 126.3 | |
| * 3 And | 218031 | 113919 | 8780 | | S | hr8780 | K0IIIb | PM* | 4.650 | 1.060 | 18.46 | 54.2 | -34.90 |
| * c01 Aqr | 218240 | 114119 | 8789 | J23067-2345AB | U | hip114119 | G8III | ** | 4.470 | 0.900 | 15.08 | 66.3 | 15.20 |
| * 5 Psc | 218527 | 114273 | 8807 | | S | hr8807 | G8III-IV | PM* | 5.431 | 0.889 | 12.27 | 81.5 | -12.10 |
| * c02 Aqr | 218594 | 114341 | 8812 | | S | hr8812 | K1III | Star | 3.640 | 1.230 | 12.05 | 83.0 | 21.30 |
| HR 8839 | 219310 | 114742 | 8839 | | S | hr8839 | K2III | Star | 6.340 | 1.170 | 5.68 | 176.1 | -27.79 |
| HD 219418 | 219418 | 114809 | | | E | hd219418 | G5III | Star | 6.810 | 0.830 | 6.22 | 160.8 | 39.41 |
| HD 219409 | 219409 | 114842 | | | S | hd219409 | K1III | Star | 6.531 | 1.058 | 9.91 | 100.9 | -47.70 |
| * psi01 Aqr | 219449 | 114855 | 8841 | J23159-0905A | S | hr8841 | K0III | *in** | 4.250 | 1.110 | 21.77 | 45.9 | -25.88 |
| * gam Psc | 219615 | 114971 | 8852 | | S | hr8852 | G9III: | PM* | 3.700 | 0.920 | 23.64 | 42.3 | -14.42 |
| * omi Cep | 219916 | 115088 | 8872 | J23186+6807AB | E | hd219916 | K0III | ** | 4.868 | 0.882 | 16.06 | 62.3 | -18.36 |
| * gam Scl | 219784 | 115102 | 8863 | | U | hip115102 | G8III | Star | 4.415 | 1.130 | 17.90 | 55.9 | 15.60 |
| HD 219962 | 219962 | 115171 | 8875 | J23197+4823A | S | hr8875 | K1III | PM* | 6.320 | 1.120 | 8.70 | 114.9 | 19.78 |
| * 7 Psc | 220009 | 115227 | 8878 | | S | hr8878 | K2III | Star | 5.077 | 1.207 | 7.54 | 132.6 | 40.34 |
| V* BP Psc | | | | | H | bppsc | G9IIIe | TTau* | 11.850 | 1.470 | | | -5.80 |
| * b01 Aqr | 220321 | 115438 | 8892 | | S | hr8892 | K0III | Star | 3.980 | 1.100 | 19.96 | 50.1 | -6.10 |
| * 9 Psc | 220858 | 115768 | 8912 | J23272+0107AB | S | hr8912 | G7III | V* | 6.250 | 1.020 | 8.85 | 113.0 | -8.30 |
| * tet Psc | 220954 | 115830 | 8916 | | E | hd220954 | K1III | Star | 4.300 | 1.080 | 21.96 | 45.5 | 6.05 |
| HR 8922 | 221113 | 115915 | 8922 | | S | hr8922 | G9III | Star | 6.440 | 1.100 | 7.35 | 136.1 | 21.70 |
| * 70 Peg | 221115 | 115919 | 8923 | | S | hr8923 | G7III | Star | 4.560 | 0.940 | 18.65 | 53.6 | -14.80 |
| HR 8924 | 221148 | 115953 | 8924 | | E | hd221148 | K3IIIv | V* | 6.250 | 1.090 | 21.14 | 47.3 | -26.53 |

| Name | HD | HIP | HR | CCDM | S | Keyname | SpType | Otype | Vmag | B-V | plx | dist | RV |
|---|---|---|---|---|---|---|---|---|---|---|---|---|---|
| * 14 And | 221345 | 116076 | 8930 | | S | hr8930 | G8III | V* | 5.220 | 1.020 | 12.63 | 79.2 | -59.99 |
| HR 8941 | 221661 | 116292 | 8941 | | S | hr8941 | G8II | Star | 6.255 | 0.986 | 5.44 | 183.8 | 8.40 |
| * 73 Peg | 221758 | 116355 | 8948 | | S | hr8948 | K0III: | Star | 5.635 | 1.036 | 11.24 | 89.0 | -4.12 |
| HR 8946 | 221745 | 116368 | 8946 | | S | hr8946 | K4III | Star | 5.957 | 1.359 | 8.78 | 113.9 | -36.80 |
| HR 8958 | 222093 | 116591 | 8958 | J23377-1304A | S | hr8958 | K0III | Star | 5.668 | 1.011 | 12.21 | 81.9 | -12.53 |
| HR 8979 | 222493 | 116853 | 8979 | | U | hip116853 | K0III | Star | 5.898 | 0.971 | 9.34 | 107.1 | -9.40 |
| * 76 Peg | 222683 | 116972 | | | S | hd222683 | K0 | Star | 6.297 | 0.958 | 8.87 | 112.7 | -2.39 |
| HD 223094 | 223094 | 117246 | | | U | hd223094 | K5III | Star | 6.970 | 1.630 | 0.74 | 1351.4 | 20.13 |
| * tau Cas | 223165 | 117301 | 9008 | | S | hr9008 | K1III | V* | 4.881 | 1.124 | 18.75 | 53.3 | -20.87 |
| HR 9009 | 223170 | 117314 | 9009 | | S | hr9009 | K0III | HB* | 5.739 | 1.075 | 10.69 | 93.5 | 11.20 |
| * 20 Psc | 223252 | 117375 | 9012 | | S | hr9012 | G8III | *in** | 5.509 | 0.924 | 10.30 | 97.1 | -15.19 |
| HD 223869 | 223869 | 117778 | | | S | hd223869 | K1III | Star | 7.510 | 0.990 | 10.87 | 92.0 | 15.60 |
| HD 224349 | 224349 | 118081 | | | S | hd224349 | G9III | Star | 6.440 | 0.990 | 6.86 | 145.8 | |
| * 27 Psc | 224533 | 118209 | 9067 | J23587-0333AB | S | hr9067 | G9III | ** | 4.890 | 0.918 | 13.91 | 71.9 | -0.20 |

All information except columns 5 (S) and 6 (Keyname) from SIMBAD. CCDM is the Catalog of Double and Multiple Stars. Cluster is the first cluster designation from SIMBAD.

S: Source of spectroscopic material:
S is McDonald Observatory Struve Reflector and Sandiford echelle spectrograph.
E is Observatoire d'Haute Provence ELODIE spectrograph.
H is the European Southern Observatory HARPS spectrograph.
U is the European Southern Observatory UVES spectrograph.

Some stars have multiple sources.

Keyname: Identifier used in subsequent tables to identify stars. The tag in some cases provides an alternate identification for the object. If a star has multiple sources, there will be a matching keyname for each.

Table 2
Reddening, Temperature, Mass, and Age for the Program Stars

| Primary ID | S | E(B-V) | $T_{eff}$ | σ | N | log(L/L☉) | Mass in Solar Masses | | | | Age in Gyr | | | |
|---|---|---|---|---|---|---|---|---|---|---|---|---|---|---|
| | | | | | | | B | D | Y | <Mass> | B | D | Y | <Age> |
| * 1 Aqr  | S | 0.000 | 4715 | 15  | 10 | 1.72  | 1.71 | 1.29 | 1.91 | 1.64 | 2.52 | 6.35 | 1.73 | 3.53 |
| * 1 Aur  | S | 0.029 | 4043 | 30  | 8  | 2.75  | 1.31 | 1.23 | 1.94 | 1.49 | 3.91 | 6.25 | 1.62 | 3.93 |
| * 1 Peg  | S | 0.000 | 4620 | 67  | 13 | 1.82  | 1.50 | 1.71 | 2.11 | 1.77 | 3.81 | 1.92 | 1.20 | 2.31 |
| * 1 Psc  | S | 0.002 | 7728 | 89  | 7  | 1.50  | 1.94 | 1.93 | 2.00 | 1.96 |      | 1.05 | 1.07 | 1.06 |
| * 1 Ser  | S | 0.002 | 4627 | 69  | 12 | 1.79  | 1.65 | 1.78 | 1.95 | 1.79 | 2.69 | 1.75 | 1.83 | 2.09 |
| * 2 Aur  | E | 0.040 | 4153 | 50  | 11 | 2.92  | 3.74 | 1.97 |      | 2.86 | 0.26 | 3.34 |      | 1.80 |
| * 2 Boo  | S | 0.002 | 4836 | 84  | 10 | 1.83  | 1.55 | 1.86 | 2.25 | 1.89 | 2.46 | 1.70 | 0.90 | 1.69 |
| * 2 Dra  | S | 0.000 | 4802 | 45  | 10 | 1.70  | 1.06 | 1.08 | 1.89 | 1.34 | 6.18 | 6.25 | 1.50 | 4.64 |
| * 2 Lib  | S | 0.003 | 4874 | 40  | 8  | 1.58  |      | 1.31 | 1.89 | 1.60 |      | 4.81 | 1.50 | 3.16 |
| * 2 Psc  | S | 0.001 | 4831 | 119 | 12 | 1.80  | 1.57 | 1.31 |      | 1.44 | 2.68 | 3.25 |      | 2.96 |
| * 2 Sex  | S | 0.000 | 4188 | 33  | 14 | 2.28  | 1.30 | 1.45 | 1.21 | 1.32 | 3.69 | 3.38 | 6.67 | 4.58 |
| * 3 And  | S | 0.000 | 4668 | 45  | 13 | 1.69  | 1.76 | 1.62 | 1.76 | 1.71 |      |      | 2.27 | 2.27 |
| * 4 Cap  | S | 0.016 | 4782 | 25  | 10 | 1.71  | 1.56 | 1.24 | 1.68 | 1.49 | 2.97 | 5.65 | 2.33 | 3.65 |
| * 4 Vul  | S | 0.003 | 4763 | 26  | 8  | 1.83  | 1.60 | 1.52 | 2.04 | 1.72 | 3.29 | 3.39 | 1.20 | 2.63 |
| * 5 CVn  | E | 0.004 | 5098 | 75  | 12 | 2.24  | 3.49 | 2.42 |      | 2.96 |      | 0.53 |      | 0.53 |
| * 5 Psc  | S | 0.000 | 4984 | 51  | 13 | 1.67  | 1.97 | 2.16 | 2.25 | 2.13 |      | 1.08 | 0.83 | 0.96 |
| * 5 Ser  | E | 0.000 | 6025 | 177 | 10 | 0.70  | 1.22 | 1.09 | 1.18 | 1.16 | 4.50 | 6.05 |      | 5.27 |
| * 5 UMi  | S | 0.007 | 4095 | 39  | 7  | 2.65  | 1.66 | 1.80 | 2.12 | 1.86 | 2.59 | 2.19 | 1.23 | 2.00 |
| * 6 CVn  | S | 0.001 | 4938 | 21  | 10 | 1.83  | 2.18 | 1.32 | 2.61 | 2.04 | 1.04 | 4.50 | 0.60 | 2.05 |
| * 6 Pup  | S | 0.000 | 4330 | 25  | 10 | 1.90  | 1.61 | 1.31 | 1.57 | 1.50 | 2.41 | 5.78 | 3.70 | 3.96 |
| * 6 Ser  | E | 0.000 | 4417 | 17  | 10 | 1.74  | 1.37 | 1.09 | 1.35 | 1.27 | 4.16 | 8.75 | 5.17 | 6.03 |
| * 6 UMa  | S | 0.004 | 4989 | 31  | 8  | 1.73  | 1.58 | 1.70 | 2.20 | 1.83 | 3.40 | 3.68 | 0.88 | 2.65 |
| * 7 Com  | S | 0.000 | 4861 | 33  | 13 | 1.83  | 2.04 | 2.18 | 2.14 | 2.12 | 1.14 | 1.08 | 1.10 | 1.11 |
| * 7 Psc  | S | 0.004 | 4246 | 48  | 12 | 2.45  | 1.17 | 1.54 | 1.39 | 1.37 | 5.22 | 4.44 | 4.07 | 4.58 |
| * 8 Gem  | S | 0.025 | 5072 | 31  | 11 | 1.88  | 2.53 | 1.81 | 3.00 | 2.45 | 0.60 | 1.65 | 0.40 | 0.88 |
| * 8 Per  | S | 0.034 | 4502 | 53  | 10 | 2.06  | 1.98 | 1.57 | 1.95 | 1.83 | 1.46 | 3.97 | 1.66 | 2.36 |
| * 9 LMi  | S | 0.011 | 4402 | 27  | 8  | 2.04  | 1.55 | 1.54 | 1.72 | 1.60 | 3.49 | 2.75 | 3.00 | 3.08 |
| * 9 Psc  | S | 0.003 | 4828 | 50  | 5  | 1.66  | 2.09 | 1.25 | 2.02 | 1.79 | 1.07 | 5.25 | 1.35 | 2.56 |
| * 10 Cet | S | 0.006 | 5001 | 40  | 11 | 1.83  | 2.40 | 1.66 | 2.74 | 2.27 | 0.69 | 3.06 | 0.50 | 1.42 |
| * 10 Leo | S | 0.000 | 4708 | 27  | 10 | 1.82  | 1.65 | 1.18 | 2.13 | 1.65 | 2.56 | 6.46 | 1.20 | 3.41 |
| * 10 LMi | S | 0.000 | 5003 | 77  | 11 | 1.70  | 1.98 | 2.08 | 2.35 | 2.14 | 1.02 | 2.49 | 0.75 | 1.42 |
| * 10 Vir | S | 0.000 | 4553 | 59  | 12 | 1.46  | 1.49 |      | 1.36 | 1.43 | 3.30 |      | 5.00 | 4.15 |
| * 10 Vul | S | 0.029 | 5008 | 33  | 13 | 1.86  | 2.43 | 1.87 | 2.74 | 2.35 | 0.70 | 2.26 | 0.50 | 1.15 |
| * 11 Cep | S | 0.000 | 4588 | 104 | 13 | 1.78  | 1.54 | 1.68 | 1.71 | 1.64 | 2.93 | 2.13 | 2.67 | 2.57 |
| * 11 Com | S | 0.001 | 4711 | 40  | 10 | 2.08  | 1.54 | 2.16 | 1.88 | 1.86 | 3.01 | 1.17 | 1.30 | 1.83 |
| * 11 Lac | S | 0.008 | 4256 | 44  | 13 | 2.45  | 1.49 | 1.27 | 1.78 | 1.51 | 3.83 | 6.16 | 2.35 | 4.11 |
| * 11 Lib | E | 0.000 | 4745 | 31  | 13 | 1.74  | 1.40 | 1.11 | 1.39 | 1.30 | 4.56 | 7.18 | 3.25 | 5.00 |
| * 11 LMi | E | 0.000 | 5412 | 123 | 11 | -0.09 | 0.94 | 0.97 | 0.99 | 0.97 |      | 7.25 | 6.50 | 6.88 |
| * 11 Sgr | S | 0.001 | 4687 | 17  | 13 | 1.88  | 1.63 | 1.05 | 2.01 | 1.56 | 2.05 | 6.75 | 1.30 | 3.37 |
| * 11 Tri | S | 0.004 | 4572 | 32  | 9  | 1.75  | 1.37 | 1.50 | 1.49 | 1.45 | 3.81 | 3.00 | 4.18 | 3.66 |
| * 12 Com | E | 0.000 | 6115 | 129 | 8  | 1.90  | 2.52 | 2.31 | 2.52 | 2.45 | 0.50 | 0.62 | 0.60 | 0.57 |
| * 12 Tri | E | 0.000 | 7079 | 20  | 5  | 1.15  | 1.52 | 1.37 |      | 1.45 | 2.00 | 2.38 |      | 2.19 |
| * 13 Lac | S | 0.001 | 4904 | 17  | 11 | 1.78  | 1.87 | 2.18 | 2.32 | 2.12 | 1.39 | 1.08 | 0.80 | 1.09 |
| * 13 Lyn | S | 0.000 | 4883 | 71  | 10 | 1.52  | 1.80 | 1.05 | 2.20 | 1.68 | 1.50 | 6.75 | 1.00 | 3.08 |
| * 14 And | S | 0.001 | 4613 | 146 | 13 | 1.81  | 1.33 | 1.68 | 1.30 | 1.44 | 4.17 | 2.00 | 5.65 | 3.94 |
| * 14 Ari | E | 0.006 | 6710 | 177 | 7  | 1.79  | 2.36 | 2.09 | 2.30 | 2.25 | 0.63 | 0.80 | 0.80 | 0.74 |
| * 14 CMi | S | 0.000 | 5008 | 60  | 5  | 1.63  | 2.52 | 1.42 | 1.91 | 1.95 |      | 5.31 | 1.20 | 3.26 |
| * 14 Oph | E | 0.000 | 6744 | 27  | 7  | 1.08  | 1.52 | 1.31 | 1.50 | 1.44 |      | 2.75 | 2.26 | 2.51 |
| * 14 Sex | S | 0.000 | 4959 | 28  | 11 | 1.62  | 2.47 | 1.90 | 2.29 | 2.22 | 0.57 | 1.51 | 0.87 | 0.98 |
| * 14 Tau | S | 0.022 | 4801 | 21  | 13 | 1.82  | 1.63 | 1.63 | 1.81 | 1.69 | 2.87 | 2.62 | 1.58 | 2.36 |
| * 15 Cyg | S | 0.004 | 4920 | 61  | 12 | 1.97  | 2.42 | 1.76 | 2.73 | 2.30 | 0.76 | 3.24 | 0.50 | 1.50 |
| * 15 Eri | S | 0.000 | 4960 | 31  | 8  | 1.86  | 2.51 | 1.79 | 2.67 | 2.32 | 0.63 | 3.13 | 0.55 | 1.44 |
| * 15 Lyn | E | 0.000 | 5095 | 47  | 9  | 1.74  |      | 1.71 |      | 1.71 |      | 2.33 |      | 2.33 |
| * 16 Aur | S | 0.000 | 4264 | 58  | 10 | 2.10  | 1.23 | 1.28 | 1.39 | 1.30 | 5.14 | 5.00 |      | 5.07 |
| * 16 Her | S | 0.000 | 4620 | 27  | 9  | 1.87  | 1.87 | 1.80 | 1.80 | 1.82 | 2.02 | 1.63 | 1.87 | 1.84 |
| * 16 Per | S | 0.000 | 6753 | 54  | 7  | 1.33  | 1.86 | 1.69 | 1.86 | 1.80 | 1.26 | 1.63 | 1.45 | 1.44 |

| Name | Type | | | | | | | | | | | | | |
|---|---|---|---|---|---|---|---|---|---|---|---|---|---|---|
| * 16 Ser | S | 0.000 | 4889 | 58 | 16 | 1.63 | 2.09 | 1.18 | 1.82 | 1.70 | 1.04 | 4.50 | 1.67 | 2.40 |
| * 18 Com | S | 0.000 | 6388 | 64 | 5 | 1.29 | 1.65 | 1.74 | 1.86 | 1.75 | | | 1.52 | 1.52 |
| * 18 Del | S | 0.000 | 4966 | 26 | 12 | 1.58 | 1.81 | 1.93 | 2.26 | 2.00 | | 1.49 | 0.93 | 1.21 |
| * 18 Lyn | E | 0.000 | 4639 | 24 | 10 | 1.58 | | 1.42 | 1.52 | 1.47 | | | 4.13 | 4.13 |
| * 18 Sge | E | 0.009 | 4805 | 61 | 8 | 1.63 | 2.29 | 1.14 | 1.77 | 1.73 | 0.75 | 6.92 | 1.85 | 3.17 |
| * 19 Aql | E | 0.000 | 6784 | 53 | 5 | 1.11 | 1.61 | 1.46 | | 1.54 | | 2.25 | | 2.25 |
| * 19 Cap | S | 0.004 | 4525 | 100 | 10 | 1.90 | 1.67 | 1.65 | 1.81 | 1.71 | 2.66 | 2.15 | 2.06 | 2.29 |
| * 19 Pup | S | 0.000 | 4917 | 30 | 13 | 1.61 | 2.38 | 1.21 | 2.61 | 2.07 | 0.65 | 4.13 | | 2.39 |
| * 20 Boo | S | 0.000 | 4433 | 20 | 10 | 1.72 | 1.42 | 1.19 | 1.43 | 1.35 | 4.60 | 7.25 | 4.25 | 5.37 |
| * 20 Cep | E | 0.007 | 4112 | 42 | 10 | 2.10 | 1.42 | 1.11 | | 1.27 | 2.84 | 9.00 | | 5.92 |
| * 20 CVn | E | 0.000 | 7314 | 42 | 7 | 1.80 | 2.33 | 2.38 | 2.59 | 2.43 | | 0.75 | | 0.75 |
| * 20 Mon | S | 0.000 | 4692 | 20 | 5 | 1.70 | | 1.68 | 1.46 | 1.57 | | | 4.30 | 4.30 |
| * 20 Psc | S | 0.002 | 4926 | 12 | 7 | 1.80 | 1.99 | 1.69 | 2.13 | 1.94 | 1.03 | 4.10 | | 2.56 |
| * 20 Vir | S | 0.010 | 5003 | 69 | 10 | 1.83 | 2.34 | 1.73 | 2.74 | 2.27 | 0.73 | 2.87 | 0.50 | 1.37 |
| * 21 Cet | S | 0.005 | 4842 | 33 | 12 | 1.82 | 1.55 | 1.89 | 1.64 | 1.69 | 2.99 | 1.67 | 2.09 | 2.25 |
| * 22 Pup | S | 0.007 | 4863 | 71 | 10 | 1.81 | 1.71 | 1.97 | 2.39 | 2.02 | 1.92 | 1.48 | 0.80 | 1.40 |
| * 23 Cam | E | 0.012 | 5088 | 122 | 11 | 1.64 | 2.23 | 1.75 | 2.40 | 2.13 | 0.77 | 3.03 | 0.70 | 1.50 |
| * 23 Hya | S | 0.000 | 4501 | 20 | 8 | 1.78 | 1.48 | 1.39 | 1.50 | 1.46 | 3.91 | 3.88 | 4.60 | 4.13 |
| * 23 Oph | S | 0.000 | 4641 | 14 | 10 | 1.72 | 1.88 | 1.62 | 1.75 | 1.75 | 1.57 | | 2.45 | 2.01 |
| * 25 Cet | U | 0.003 | 4500 | 61 | 13 | 1.97 | 1.26 | 1.69 | 1.39 | 1.45 | 3.83 | | 4.88 | 4.36 |
| * 26 Com | S | 0.001 | 4943 | 155 | 10 | 1.68 | 2.27 | 1.72 | 1.93 | 1.97 | 0.75 | 1.85 | | 1.30 |
| * 26 Hya | S | 0.005 | 5003 | 82 | 13 | 2.13 | 2.97 | 2.46 | | 2.72 | 0.40 | 0.63 | | 0.51 |
| * 27 Ari | E | 0.004 | 4788 | 49 | 13 | 1.46 | 1.26 | 1.59 | 1.02 | 1.29 | | 2.38 | 8.50 | 5.44 |
| * 27 Cet | S | 0.001 | 4753 | 25 | 10 | 1.56 | 1.65 | 1.65 | 1.76 | 1.69 | 2.59 | 2.13 | 2.27 | 2.33 |
| * 27 Com | S | 0.005 | 4191 | 32 | 9 | 2.23 | 1.64 | 1.35 | | 1.50 | 2.73 | 4.25 | | 3.49 |
| * 27 Hya | S | 0.000 | 4965 | 26 | 12 | 1.76 | 2.34 | 1.99 | | 2.17 | 0.74 | 3.08 | | 1.91 |
| * 27 Mon | S | 0.000 | 4422 | 18 | 10 | 2.03 | 1.61 | 1.58 | 1.58 | 1.59 | 2.94 | | 3.00 | 2.97 |
| * 27 Psc | S | 0.000 | 5026 | 88 | 10 | 1.78 | 2.74 | 1.63 | | 2.19 | | 3.08 | | 3.08 |
| * 28 Aql | S | 0.019 | 7262 | 20 | 2 | 1.72 | 2.13 | 2.00 | 2.09 | 2.07 | | 0.92 | 0.90 | 0.91 |
| * 29 LMi | S | 0.006 | 4691 | 79 | 10 | 1.70 | 1.76 | 1.72 | 1.95 | 1.81 | 2.23 | 1.88 | 1.83 | 1.98 |
| * 30 Aqr | S | 0.002 | 4944 | 79 | 10 | 1.75 | 2.13 | 1.66 | 2.25 | 2.01 | 1.01 | 4.09 | 0.83 | 1.98 |
| * 30 Gem | S | 0.003 | 4502 | 31 | 11 | 2.22 | 1.94 | 1.44 | 1.81 | 1.73 | 1.75 | 3.78 | 1.47 | 2.33 |
| * 30 Vul | S | 0.008 | 4473 | 19 | 8 | 2.17 | 1.97 | 1.02 | 1.67 | 1.55 | 2.28 | 7.75 | 2.57 | 4.20 |
| * 31 Boo | H | 0.000 | 4807 | 40 | 14 | 2.54 | 3.30 | 2.73 | 3.77 | 3.27 | 0.33 | 0.41 | | 0.37 |
| * 31 Cep | E | 0.000 | 6668 | 47 | 7 | 1.33 | 1.86 | 1.78 | 1.86 | 1.83 | 1.26 | 1.50 | 1.45 | 1.40 |
| * 31 Com | S | 0.002 | 5564 | 63 | 13 | 1.87 | 2.54 | 2.48 | | 2.51 | | 0.52 | | 0.52 |
| * 31 Vul | S | 0.000 | 5155 | 106 | 11 | 1.69 | 2.43 | 1.60 | 2.16 | 2.06 | 0.57 | 1.98 | | 1.27 |
| * 32 And | S | 0.008 | 5020 | 51 | 10 | 1.96 | 2.08 | 2.29 | 2.40 | 2.26 | 0.96 | 1.14 | | 1.05 |
| * 32 Boo | S | 0.000 | 4903 | 30 | 12 | 1.99 | 1.99 | 2.03 | 2.42 | 2.15 | 1.40 | 2.31 | 0.68 | 1.46 |
| * 33 Cap | U | 0.000 | 4532 | 45 | 10 | 1.70 | 1.35 | 1.53 | 1.89 | 1.59 | 4.79 | 3.13 | | 3.96 |
| * 34 Vul | S | 0.001 | 4689 | 61 | 10 | 1.69 | 1.27 | 1.68 | 1.31 | 1.42 | | | 5.15 | 5.15 |
| * 35 Com | S | 0.001 | 4982 | 112 | 12 | 1.93 | 2.49 | 1.89 | 2.73 | 2.37 | 0.66 | 2.16 | 0.50 | 1.11 |
| * 35 Lyn | S | 0.001 | 4761 | 67 | 10 | 1.82 | 1.78 | 1.16 | 1.82 | 1.59 | 2.00 | 4.94 | 1.80 | 2.91 |
| * 35 Peg | E | 0.000 | 4631 | 45 | 16 | 1.49 | 1.43 | 1.28 | 1.38 | 1.36 | | 4.94 | | 4.94 |
| * 35 UMa | S | 0.005 | 4635 | 104 | 10 | 1.58 | 1.30 | 1.50 | 1.61 | 1.47 | 5.23 | 3.13 | 3.50 | 3.95 |
| * 36 Ari | S | 0.009 | 4624 | 54 | 7 | 1.57 | 1.53 | 1.40 | 1.38 | 1.44 | 3.30 | 3.75 | 5.40 | 4.15 |
| * 36 Lib | U | 0.017 | 4191 | 74 | 8 | 2.18 | 1.55 | 1.25 | 1.30 | 1.37 | 3.29 | 5.85 | 5.40 | 4.85 |
| * 36 Psc | S | 0.005 | 5061 | 22 | 13 | 1.76 | 2.69 | 2.73 | | 2.71 | 0.45 | 0.51 | | 0.48 |
| * 37 Cam | S | 0.014 | 4609 | 70 | 13 | 2.10 | 1.53 | 1.36 | 1.39 | 1.43 | 3.16 | 3.60 | 3.73 | 3.50 |
| * 37 Dra | S | 0.001 | 4646 | 83 | 9 | 1.56 | 1.30 | 1.41 | 1.61 | 1.44 | 5.23 | 3.63 | 3.50 | 4.12 |
| * 37 Tau | S | 0.000 | 4732 | 26 | 10 | 1.84 | 1.68 | 2.33 | 1.96 | 1.99 | 1.98 | 0.83 | 1.35 | 1.39 |
| * 38 Aur | S | 0.000 | 4786 | 47 | 10 | 1.25 | 1.29 | 1.34 | 1.05 | 1.23 | 5.23 | 4.25 | 8.50 | 5.99 |
| * 38 Aur | E | 0.000 | 4787 | 47 | 10 | 1.25 | 1.24 | 1.36 | 1.05 | 1.22 | 5.59 | 4.00 | 8.50 | 6.03 |
| * 38 Com | S | 0.003 | 4916 | 73 | 10 | 1.85 | 2.15 | 1.52 | 2.45 | 2.04 | 0.95 | 4.68 | 0.70 | 2.11 |
| * 38 UMa | S | 0.000 | 4425 | 60 | 10 | 1.78 | 1.49 | 1.21 | 1.53 | 1.41 | 3.06 | 6.63 | 3.50 | 4.40 |
| * 39 Ari | S | 0.000 | 4628 | 67 | 10 | 1.73 | 1.61 | 1.60 | 1.73 | 1.65 | 2.05 | 2.38 | 2.45 | 2.29 |
| * 39 Cet | H | 0.000 | 4937 | 48 | 13 | 1.67 | 2.42 | 1.42 | 1.78 | 1.87 | | | 1.50 | 1.50 |
| * 39 Cnc | H | 0.015 | 4954 | 25 | 13 | 2.02 | 2.75 | 3.00 | | 2.88 | 0.51 | 0.41 | | 0.46 |
| * 39 Cyg | S | 0.000 | 4211 | 41 | 10 | 2.22 | 1.78 | 1.39 | 1.55 | 1.57 | 2.10 | 3.88 | 4.40 | 3.46 |
| * 39 Eri | S | 0.000 | 4549 | 35 | 10 | 1.91 | 1.62 | 1.86 | 1.95 | 1.81 | 2.72 | 1.62 | 1.74 | 2.03 |
| * 41 Oph | S | 0.000 | 4509 | 65 | 9 | 1.83 | 1.40 | 1.48 | 1.51 | 1.46 | 4.11 | 3.13 | 3.95 | 3.73 |
| * 42 Dra | S | 0.003 | 4413 | 17 | 10 | 2.19 | 1.52 | 1.59 | 1.28 | 1.46 | 2.42 | 4.08 | 4.50 | 3.67 |
| * 42 UMa | | E | 0.002 | 4556 | 41 | 7 | 1.73 | 1.31 | 1.55 | 1.54 | 1.47 | 4.88 | 2.81 | 3.73 | 3.81 |
| * 43 Aur | | S | 0.015 | 4552 | 62 | 10 | 1.74 | 1.35 | 1.43 | 1.51 | 1.43 | 4.48 | 3.50 | 4.12 | 4.03 |
| * 43 Leo | | S | 0.000 | 4652 | 37 | 10 | 1.37 | 1.60 | 1.30 | 1.47 | 1.46 | 2.62 | 5.63 | 3.75 | 4.00 |
| * 43 Lyn | | S | 0.006 | 4945 | 23 | 10 | 1.84 | 2.26 | 1.56 | 2.61 | 2.14 | 0.88 | 4.14 | 0.60 | 1.87 |
| * 43 Ser | | S | 0.007 | 4933 | 15 | 10 | 1.69 | 2.27 | 2.02 | 1.93 | 2.07 | 0.75 | 1.38 | | 1.07 |
| * 43 UMa | | S | 0.006 | 4527 | 68 | 10 | 1.93 | 1.67 | 1.70 | 1.88 | 1.75 | 2.56 | 1.96 | 1.76 | 2.09 |
| * 44 Psc | | H | 0.006 | 5148 | 61 | 10 | 2.01 | 2.94 | 2.35 | 2.87 | 2.72 | 0.36 | 0.64 | | 0.50 |
| * 45 Gem | | S | 0.005 | 4740 | 31 | 12 | 1.88 | 1.44 | 1.45 | 1.85 | 1.58 | 3.67 | 3.57 | 1.63 | 2.96 |
| * 45 Peg | | S | 0.007 | 4738 | 58 | 9 | 1.68 | 1.78 | 1.17 | 1.91 | 1.62 | 2.02 | 7.28 | 1.73 | 3.68 |
| * 46 Cet | | S | 0.002 | 4316 | 89 | 11 | 2.12 | 1.34 | 1.47 | 1.32 | 1.38 | 3.91 | 3.25 | 5.36 | 4.17 |
| * 46 Cnc | | S | 0.010 | 4966 | 74 | 10 | 2.10 | 2.88 | 2.19 | 2.87 | 2.65 | 0.43 | 1.06 | | 0.74 |
| * 46 Psc | | S | 0.167 | 5281 | 161 | 5 | 2.10 | 3.12 | 2.62 | | 2.87 | 0.32 | 0.45 | | 0.38 |
| * 46 UMa | | S | 0.000 | 4569 | 65 | 10 | 1.82 | 1.25 | 1.61 | 1.71 | 1.52 | 4.73 | 2.33 | 2.67 | 3.24 |
| * 46 Vir | | S | 0.001 | 4580 | 21 | 10 | 1.71 | 1.49 | 1.43 | 1.44 | 1.45 | 3.49 | 3.52 | 4.76 | 3.92 |
| * 47 Aqr | | S | 0.000 | 4552 | 140 | 10 | 1.49 | 1.26 | 1.07 | 1.22 | 1.18 | | 9.38 | | 9.38 |
| * 48 Dra | | E | 0.001 | 4517 | 35 | 9 | 1.71 | 1.33 | 1.33 | 1.89 | 1.52 | 5.08 | 4.54 | | 4.81 |
| * 48 Leo | | S | 0.000 | 4900 | 48 | 16 | 1.82 | 1.46 | 1.67 | 1.95 | 1.69 | 3.40 | 4.35 | 1.15 | 2.97 |
| * 48 Peg | | E | 0.000 | 4934 | 55 | 13 | 1.66 | 2.42 | 1.40 | 1.93 | 1.92 | | 2.63 | | 2.63 |
| * 48 Tau | | S | 0.000 | 6606 | 65 | 7 | 0.68 | 1.29 | | 1.51 | 1.40 | 2.92 | | | 2.92 |
| * 49 And | | S | 0.004 | 4879 | 106 | 10 | 1.85 | 1.77 | 1.86 | 2.57 | 2.07 | 1.42 | 3.19 | 0.65 | 1.75 |
| * 49 Vir | | S | 0.001 | 4585 | 41 | 10 | 1.98 | 2.07 | 1.23 | 1.52 | 1.61 | 1.22 | 6.69 | | 3.95 |
| * 51 Cam | | S | 0.007 | 4357 | 183 | 7 | 1.81 | 1.38 | 1.08 | 1.35 | 1.27 | 3.97 | 9.25 | 5.30 | 6.17 |
| * 52 Cyg | | S | 0.000 | 4701 | 18 | 7 | 1.97 | 1.54 | 1.21 | 1.65 | 1.47 | 2.15 | 4.25 | 1.95 | 2.78 |
| * 52 Psc | | S | 0.002 | 4701 | 32 | 10 | 1.74 | 1.74 | 2.00 | 1.68 | 1.81 | 1.58 | | 2.20 | 1.89 |
| * 52 Psc | | E | 0.002 | 4696 | 32 | 10 | 1.74 | 1.82 | 1.88 | 1.91 | 1.87 | 1.42 | 1.50 | 1.73 | 1.55 |
| * 53 Dra | | S | 0.004 | 4875 | 78 | 7 | 2.04 | 2.51 | 2.90 | | 2.71 | 0.63 | 0.46 | | 0.54 |
| * 54 Per | | S | 0.000 | 4926 | 31 | 13 | 1.71 | 2.32 | 1.72 | | 2.02 | | 1.85 | | 1.85 |
| * 55 Leo | | E | 0.000 | 6446 | 75 | 8 | 0.84 | 1.28 | 1.11 | 1.36 | 1.25 | | 4.75 | | 4.75 |
| * 56 And | | S | 0.005 | 4509 | 105 | 5 | 1.82 | 1.17 | 1.46 | 1.40 | 1.34 | 5.59 | 3.25 | 4.50 | 4.45 |
| * 56 Her | | S | 0.005 | 4979 | 30 | 9 | 1.86 | 2.38 | 1.73 | 2.74 | 2.28 | 0.73 | 2.87 | 0.50 | 1.37 |
| * 57 Gem | | E | 0.000 | 5007 | 22 | 7 | 1.37 | | 1.55 | 1.98 | 1.77 | | 3.13 | 1.35 | 2.24 |
| * 57 Her | | S | 0.005 | 4991 | 84 | 7 | 1.96 | 2.57 | 1.99 | 2.99 | 2.52 | 0.61 | 1.69 | | 1.15 |
| * 57 Sgr | | S | 0.030 | 4779 | 67 | 8 | 1.86 | 1.59 | 1.65 | 1.86 | 1.70 | 2.82 | 2.55 | 1.55 | 2.31 |
| * 57 Vir | | S | 0.000 | 4741 | 37 | 14 | 1.15 | 1.53 | 1.09 | 1.38 | 1.33 | 2.78 | 8.75 | 4.50 | 5.34 |
| * 57 Vir | | H | 0.000 | 4742 | 38 | 14 | 1.15 | 1.53 | 1.09 | 1.22 | 1.28 | 2.78 | 8.75 | | 5.77 |
| * 58 Leo | | E | 0.000 | 4519 | 52 | 16 | 2.26 | 1.89 | 1.88 | 1.85 | 1.87 | 1.66 | 2.02 | 1.40 | 1.69 |
| * 58 Psc | | S | 0.002 | 4871 | 47 | 9 | 1.72 | 2.12 | 1.82 | 2.08 | 2.01 | 0.96 | 2.05 | 1.10 | 1.37 |
| * 60 Ari | | S | 0.007 | 4449 | 34 | 10 | 1.62 | 1.46 | | 1.26 | 1.36 | 3.95 | | 6.67 | 5.31 |
| * 60 Eri | | S | 0.000 | 4818 | 12 | 10 | 1.75 | 1.44 | 1.58 | 2.04 | 1.69 | 4.01 | 3.07 | 1.20 | 2.76 |
| * 60 Psc | | S | 0.008 | 5014 | 135 | 10 | 1.93 | 2.58 | 1.99 | 2.82 | 2.46 | 0.57 | 1.69 | 0.47 | 0.91 |
| * 60 UMa | | S | 0.005 | 6672 | 37 | 7 | 1.45 | 1.80 | 1.95 | 2.03 | 1.93 | | | 1.18 | 1.18 |
| * 63 Ari | | S | 0.009 | 4385 | 36 | 13 | 2.11 | 1.74 | 1.63 | 1.71 | 1.69 | 2.43 | 2.25 | 2.76 | 2.48 |
| * 63 Vir | | S | 0.003 | 4820 | 11 | 10 | 1.85 | 1.59 | 1.69 | 2.26 | 1.85 | 2.44 | 2.02 | 0.90 | 1.79 |
| * 63 Vir | | E | 0.003 | 4840 | 19 | 10 | 1.85 | 1.91 | 2.03 | 2.58 | 2.17 | 1.48 | 1.36 | | 1.42 |
| * 64 Ari | | S | 0.000 | 4486 | 29 | 9 | 1.53 | 1.18 | 1.16 | 1.09 | 1.14 | | 7.50 | 9.00 | 8.25 |
| * 65 And | | S | 0.010 | 3927 | 24 | 8 | 2.57 | 1.49 | 1.64 | 1.77 | 1.63 | 3.10 | 2.49 | 3.45 | 3.01 |
| * 65 Aur | | S | 0.000 | 4575 | 17 | 8 | 1.77 | 1.35 | 1.51 | 1.16 | 1.34 | 3.75 | 2.88 | | 3.31 |
| * 65 Psc A | | S | 0.000 | 6732 | 70 | 4 | 1.97 | | | | 2.00 | | | | |
| * 65 Psc B | | S | 0.000 | 6671 | 103 | 5 | 1.97 | | | | 2.00 | | | | |
| * 66 UMa | | S | 0.002 | 4565 | 88 | 9 | 1.71 | 1.25 | 1.43 | 1.44 | 1.37 | 5.27 | 3.50 | 4.78 | 4.52 |
| * 67 Cet | | S | 0.000 | 4884 | 14 | 10 | 1.85 | 2.02 | 1.55 | 2.57 | 2.05 | 1.11 | 4.57 | 0.65 | 2.11 |
| * 68 Aqr | | S | 0.000 | 4829 | 30 | 10 | 1.73 | 1.23 | 1.51 | 1.43 | 1.39 | 4.60 | 3.39 | 3.37 | 3.79 |
| * 69 Aql | | S | 0.000 | 4592 | 66 | 10 | 1.71 | 1.65 | 1.56 | 1.40 | 1.54 | 2.62 | 2.75 | 4.94 | 3.44 |
| * 69 Vir | | S | 0.001 | 4730 | 71 | 13 | 2.04 | 2.00 | 2.22 | 2.06 | 2.09 | 1.25 | 1.18 | | 1.21 |
| * 70 Gem | | S | 0.014 | 4997 | 63 | 9 | 1.92 | 2.49 | 1.91 | 2.73 | 2.38 | 0.66 | 1.99 | 0.50 | 1.05 |
| * 70 Peg | | S | 0.000 | 4938 | 20 | 13 | 1.66 | 2.42 | 1.83 | 1.93 | 2.06 | | 1.73 | | 1.73 |
| * 71 Oph | | S | 0.005 | 4921 | 33 | 14 | 2.02 | 2.52 | 2.01 | | 2.27 | 0.67 | 1.90 | | 1.28 |
| * 71 Vir | | S | 0.000 | 4658 | 13 | 10 | 1.81 | 1.54 | 1.84 | 1.68 | 1.69 | 2.89 | | 2.20 | 2.54 |
| * 72 Cyg | | S | 0.000 | 4640 | 25 | 13 | 1.84 | 1.59 | 1.84 | 1.68 | 1.70 | 3.04 | | 2.20 | 2.62 |
| * 73 Peg | | S | 0.004 | 4761 | 54 | 12 | 1.71 | 1.48 | 1.26 | 1.77 | 1.50 | 3.82 | 5.94 | 1.85 | 3.87 |
| * 74 Oph | | S | 0.003 | 4982 | 27 | 13 | 1.88 | 1.90 | 1.73 | 2.31 | 1.98 | 1.43 | 3.07 | 0.70 | 1.73 |
| * 75 Cet | | S | 0.000 | 4803 | 15 | 11 | 1.73 | 1.23 | 1.12 | 2.18 | 1.51 | 5.35 | 5.50 | 1.10 | 3.98 |

| Name | Type | | | | | | | | | | | | |
|---|---|---|---|---|---|---|---|---|---|---|---|---|---|
| * 75 Tau | S | 0.000 | 4516 | 20 | 8 | 1.64 | | 1.16 | 1.30 | 1.23 | | 7.00 | 7.00 |
| * 76 Leo | E | 0.000 | 4697 | 34 | 12 | 1.63 | 2.05 | 1.62 | 1.73 | 1.80 | 1.11 | 2.45 | 1.78 |
| * 76 Peg | S | 0.005 | 4936 | 67 | 7 | 1.62 | 2.25 | 1.69 | 2.26 | 2.07 | 0.80 | 2.27 0.93 | 1.33 |
| * 78 Cnc | U | 0.012 | 4479 | 58 | 6 | 1.94 | 1.61 | 1.59 | 1.66 | 1.62 | 3.02 | 2.57 3.14 | 2.91 |
| * 78 Dra | S | 0.000 | 4710 | 40 | 13 | 1.75 | 1.54 | 1.40 | 2.03 | 1.66 | 3.17 | 5.50 | 4.34 |
| * 79 Cnc | S | 0.001 | 5076 | 47 | 10 | 1.71 | 2.38 | 2.21 | | 2.30 | 0.64 | 0.90 | 0.77 |
| * 79 Leo | S | 0.000 | 4984 | 116 | 12 | 1.93 | 2.56 | 1.99 | 2.99 | 2.51 | 0.61 | 1.63 | 1.12 |
| * 80 Vir | S | 0.001 | 4881 | 66 | 10 | 1.73 | 1.94 | 1.46 | 1.75 | 1.72 | 1.35 | 2.85 1.60 | 1.93 |
| * 81 Cet | S | 0.003 | 4758 | 31 | 10 | 1.74 | 1.44 | 1.22 | 2.13 | 1.60 | 3.64 | 6.05 1.20 | 3.63 |
| * 83 Her | E | 0.016 | 4033 | 64 | 8 | 2.29 | 1.29 | 1.09 | 1.18 | 1.19 | 4.87 | 9.00 7.00 | 6.96 |
| * 84 Vir | U | 0.000 | 4589 | 37 | 14 | 1.69 | 1.55 | 1.43 | 1.43 | 1.47 | 3.23 | 3.45 4.89 | 3.86 |
| * 86 Leo | S | 0.001 | 4678 | 52 | 13 | 1.84 | 1.51 | 1.15 | 1.58 | 1.41 | 3.09 | 7.29 2.90 | 4.43 |
| * 86 Peg | S | 0.003 | 5105 | 189 | 10 | 1.91 | 2.32 | 2.12 | 2.99 | 2.48 | | 0.91 | 0.91 |
| * 87 Her | S | 0.000 | 4555 | 40 | 8 | 1.69 | 1.41 | 1.38 | 1.48 | 1.42 | 4.03 | 3.88 4.53 | 4.15 |
| * 87 Peg | S | 0.003 | 4774 | 75 | 10 | 1.73 | 1.42 | 1.37 | 2.13 | 1.64 | 3.58 | 4.58 1.20 | 3.12 |
| * 88 Psc | S | 0.010 | 4732 | 38 | 14 | 2.07 | 2.21 | 1.95 | 2.35 | 2.17 | 1.03 | 2.44 0.78 | 1.42 |
| * 89 Vir | S | 0.000 | 4706 | 13 | 10 | 1.84 | 1.95 | 1.03 | 2.13 | 1.70 | 1.74 | 7.25 1.20 | 3.40 |
| * 92 Leo | S | 0.000 | 4863 | 30 | 10 | 1.77 | 1.67 | 1.67 | 1.93 | 1.76 | 2.22 | 2.18 | 2.20 |
| * 94 Psc | S | 0.002 | 4617 | 26 | 10 | 1.76 | 1.65 | 1.62 | 1.73 | 1.67 | 2.15 | 2.25 2.60 | 2.33 |
| * 95 Cet | S | 0.000 | 4621 | 153 | 9 | 1.59 | 1.64 | 1.40 | 1.21 | 1.42 | 2.88 | 3.75 6.67 | 4.43 |
| * 109 Her | S | 0.000 | 4484 | 33 | 13 | 1.72 | 1.40 | 1.18 | 1.79 | 1.46 | 4.27 | 6.50 1.90 | 4.22 |
| * 110 Vir | S | 0.000 | 4664 | 19 | 13 | 1.88 | 1.61 | 1.24 | 2.16 | 1.67 | 2.73 | 6.31 | 4.52 |
| * 140 Pup | H | 0.020 | 3916 | 48 | 8 | 3.04 | 1.40 | 1.44 | 1.68 | 1.51 | 3.94 | 3.18 1.78 | 2.96 |
| * 145 CMa | S | 0.024 | 3937 | 141 | 2 | 3.72 | 1.90 | | | 1.90 | 1.05 | | 1.05 |
| * 192 Gem | E | 0.003 | 4922 | 11 | 7 | 1.53 | 1.94 | 1.30 | 1.82 | 1.69 | 1.19 | 4.50 1.60 | 2.43 |
| * alf Ari | S | 0.000 | 4512 | 18 | 13 | 1.92 | 1.69 | 1.63 | 1.64 | 1.65 | 2.49 | 2.60 | 2.55 |
| * alf Aur | E | 0.000 | 5155 | 41 | 9 | 2.20 | 2.79 | 2.67 | | 2.73 | 0.47 | 0.45 | 0.46 |
| * alf Boo | S | 0.000 | 4281 | 36 | 15 | 2.33 | 1.30 | 1.92 | 1.06 | 1.43 | 3.76 | 1.38 7.50 | 4.21 |
| * alf Cas | S | 0.000 | 4555 | 55 | 5 | 2.91 | 3.98 | 2.93 | | 3.46 | 0.21 | 0.38 | 0.29 |
| * alf Crt | S | 0.000 | 4645 | 14 | 14 | 1.84 | 1.57 | 1.84 | 1.43 | 1.61 | 3.45 | | 3.45 |
| * alf Hor | H | 0.000 | 4614 | 30 | 14 | 1.65 | 2.17 | 1.46 | 1.56 | 1.73 | | 3.63 | 3.63 |
| * alf Hya | S | 0.000 | 4097 | 47 | 5 | 2.97 | 3.87 | 1.97 | | 2.92 | 0.22 | 1.26 | 0.74 |
| * alf Ind | H | 0.000 | 4810 | 18 | 13 | 1.77 | 1.53 | 1.73 | 2.04 | 1.77 | 3.28 | 2.48 1.20 | 2.32 |
| * alf Mic | H | 0.005 | 4750 | 145 | 13 | 2.24 | 2.76 | 2.00 | 3.16 | 2.64 | 0.56 | 2.33 0.40 | 1.09 |
| * alf Mon | S | 0.000 | 4749 | 24 | 16 | 1.81 | 1.95 | 1.10 | 1.96 | 1.67 | 1.63 | 5.92 1.35 | 2.97 |
| * alf Ret | H | 0.000 | 5048 | 39 | 13 | 2.06 | 2.69 | 2.67 | | 2.68 | 0.51 | 0.48 | 0.50 |
| * alf Sct | S | 0.000 | 4241 | 33 | 15 | 2.25 | 1.74 | 1.52 | 1.69 | 1.65 | 2.76 | 2.88 3.73 | 3.12 |
| * alf Tau | E | 0.000 | 3903 | 27 | 8 | 2.62 | 1.34 | 1.89 | 1.46 | 1.56 | 3.44 | 1.97 4.74 | 3.38 |
| * alf TrA | U | 0.009 | 4135 | 91 | 13 | 3.66 | 1.96 | 2.70 | | 2.33 | 1.37 | | 1.37 |
| * alf UMa | S | 0.000 | 4636 | 39 | 13 | 2.53 | 3.10 | 2.38 | | 2.74 | 0.39 | 1.58 | 0.98 |
| * b And | E | 0.020 | 4010 | 22 | 8 | 2.81 | 2.16 | 1.08 | 1.90 | 1.71 | 2.02 | 6.62 1.92 | 3.52 |
| * b Cap | U | 0.000 | 5017 | 18 | 13 | 1.65 | 2.06 | 1.64 | 2.13 | 1.94 | 0.90 | 3.94 | 2.42 |
| * b Ori | S | 0.004 | 4389 | 34 | 13 | 2.11 | 1.23 | 1.66 | 1.18 | 1.36 | 4.05 | 2.13 6.00 | 4.06 |
| * b01 Aqr | S | 0.000 | 4545 | 26 | 13 | 1.93 | 1.29 | 1.73 | 1.40 | 1.47 | 3.83 | 1.83 4.53 | 3.40 |
| * bet Boo | S | 0.000 | 4920 | 33 | 12 | 2.30 | 2.81 | 2.46 | 3.85 | 3.04 | | 0.58 | 0.58 |
| * bet Cet | S | 0.000 | 4792 | 35 | 13 | 2.18 | 2.76 | 1.88 | | 2.32 | 0.61 | 2.37 | 1.49 |
| * bet Col | H | 0.000 | 4566 | 15 | 14 | 1.71 | 1.22 | 1.65 | 2.09 | 1.65 | 5.62 | 2.42 1.15 | 3.06 |
| * bet Crv | S | 0.000 | 5090 | 42 | 12 | 2.24 | | 2.73 | | 2.73 | | 0.42 | 0.42 |
| * bet Gem | S | 0.000 | 4821 | 34 | 13 | 1.62 | 2.29 | 1.03 | | 1.66 | 0.75 | 7.25 | 4.00 |
| * bet Her | S | 0.000 | 4912 | 21 | 10 | 2.18 | 2.83 | 2.04 | | 2.44 | 0.48 | 1.18 | 0.83 |
| * bet Her | E | 0.000 | 4903 | 19 | 10 | 2.18 | 2.52 | 1.95 | | 2.24 | 0.67 | 1.36 | 1.02 |
| * bet Lac | S | 0.000 | 4711 | 24 | 13 | 1.73 | 1.36 | 1.84 | 1.28 | 1.49 | 4.76 | 4.50 | 4.63 |
| * bet LMi | S | 0.000 | 4986 | 48 | 11 | 1.68 | | 1.61 | 2.05 | 1.83 | | 4.39 | 4.39 |
| * bet Men | H | 0.011 | 4825 | 64 | 10 | 2.71 | 4.17 | 2.99 | | 3.58 | 0.17 | 0.37 | 0.27 |
| * bet Oph | S | 0.000 | 4553 | 40 | 16 | 1.80 | 1.65 | 1.63 | 1.44 | 1.57 | 2.54 | 2.38 4.00 | 2.97 |
| * bet UMi | E | 0.000 | 4005 | 22 | 11 | 2.68 | 1.31 | 1.94 | 1.69 | 1.65 | 4.85 | 0.90 3.64 | 3.13 |
| * c Ser | S | 0.000 | 4833 | 108 | 13 | 1.58 | 2.05 | 1.23 | 1.77 | 1.68 | | 6.00 1.85 | 3.93 |
| * c Vir | S | 0.003 | 4423 | 32 | 16 | 2.12 | 1.53 | 1.76 | 1.58 | 1.62 | 2.60 | 3.84 | 3.22 |
| * c01 Aqr | U | 0.000 | 4963 | 37 | 13 | 1.88 | 2.47 | 1.76 | 2.47 | 2.23 | 0.70 | 3.12 | 1.91 |
| * c02 Aqr | S | 0.001 | 4414 | 30 | 13 | 2.52 | 1.74 | 2.09 | 2.01 | 1.95 | 2.06 | 0.86 1.10 | 1.34 |
| * chi Cas | S | 0.000 | 4736 | 27 | 10 | 1.80 | 1.43 | 1.06 | 1.89 | 1.46 | 3.17 | 6.58 1.50 | 3.75 |
| * chi Cet | U | 0.000 | 6746 | 258 | 6 | 0.75 | 1.52 | 1.03 | 1.52 | 1.36 | | 1.20 | 1.20 |

| Name | Type | | | | | | | | | | | | |
|---|---|---|---|---|---|---|---|---|---|---|---|---|---|
| * chi Gem | S | 0.000 | 4567 | 16 | 4 | 1.92 | 1.76 | 1.80 | 1.93 | 1.83 | 2.29 | 1.68 1.80 | 1.92 |
| * chi Phe | H | 0.002 | 3960 | 44 | 11 | 2.39 | 1.63 | 1.30 | 1.39 | 1.44 | 2.47 | 7.06 5.00 | 4.84 |
| * chi UMa | S | 0.000 | 4392 | 22 | 13 | 2.18 | 1.63 | 1.76 | 1.08 | 1.49 | | | |
| * chi Vir | S | 0.001 | 4404 | 18 | 2 | 2.20 | 2.10 | 1.69 | 1.89 | 1.89 | 1.45 | 3.38 1.65 | 2.16 |
| * d Aqr | S | 0.000 | 4701 | 21 | 10 | 1.72 | 1.93 | 1.75 | 1.49 | 1.72 | 1.64 | 3.47 | 2.55 |
| * d Cyg | S | 0.000 | 4337 | 67 | 13 | 1.76 | 1.12 | | 1.43 | 1.28 | | | |
| * d Eri | H | 0.002 | 3983 | 17 | 9 | 2.65 | 1.46 | 2.03 | 1.74 | 1.74 | 3.20 | 1.20 3.18 | 2.53 |
| * D Hya | S | 0.000 | 4968 | 13 | 11 | 2.04 | 2.31 | 2.02 | 2.64 | 2.32 | 0.79 | 1.45 0.50 | 0.91 |
| * del And | S | 0.000 | 4330 | 34 | 12 | 1.89 | 1.64 | 1.22 | 1.23 | 1.36 | 2.39 | 6.75 | 4.57 |
| * del Ari | S | 0.000 | 4769 | 19 | 7 | 1.76 | 1.45 | 1.08 | 2.13 | 1.55 | 3.26 | 6.25 1.20 | 3.57 |
| * del Aur | S | 0.000 | 4768 | 11 | 5 | 1.75 | 1.65 | 1.19 | 2.06 | 1.63 | 2.81 | 5.72 1.26 | 3.26 |
| * del Boo | S | 0.000 | 4821 | 40 | 7 | 0.06 | | | | 1.50 | | | |
| * del Cnc | S | 0.000 | 4637 | 27 | 13 | 1.72 | 1.77 | 1.62 | 1.73 | 1.71 | | 2.45 | 2.45 |
| * del Crt | S | 0.000 | 4510 | 15 | 13 | 2.21 | 1.35 | 1.76 | 1.58 | 1.56 | 3.47 | 2.52 2.68 | 2.89 |
| * del Dra | S | 0.000 | 4776 | 14 | 12 | 1.78 | 1.84 | 1.14 | 1.77 | 1.58 | 2.29 | 5.25 1.90 | 3.15 |
| * del Eri | E | 0.000 | 4966 | 59 | 16 | 0.52 | 1.22 | 1.16 | 1.15 | 1.18 | 5.15 | 7.00 | 6.07 |
| * del Phe | H | 0.000 | 4762 | 26 | 14 | 1.76 | 1.42 | 1.08 | 1.89 | 1.46 | 3.33 | 6.25 1.50 | 3.69 |
| * del PsA | U | 0.000 | 4828 | 29 | 14 | 1.72 | 1.10 | 1.16 | 1.99 | 1.42 | 5.10 | 4.88 1.25 | 3.74 |
| * del Psc | S | 0.003 | 3955 | 23 | 13 | 2.51 | 1.50 | 1.31 | 2.13 | 1.65 | 3.16 | 4.63 1.15 | 2.98 |
| * del Sct | E | 0.000 | 6832 | 60 | 7 | 1.58 | 1.95 | | 2.23 | 2.09 | | 0.95 | 0.95 |
| * del Sgr | S | 0.011 | 4203 | 69 | 5 | 3.22 | 3.86 | 2.55 | | 3.21 | 0.26 | | 0.26 |
| * del Tau | S | 0.000 | 4883 | 32 | 13 | 1.89 | 2.55 | 2.33 | 2.22 | 2.37 | 0.60 | 0.93 | 0.76 |
| * e Ori | S | 0.000 | 4839 | 24 | 16 | 1.76 | 1.33 | 1.61 | 1.83 | 1.59 | 4.20 | 2.54 | 3.37 |
| * e01 Sgr | E | 0.000 | 4579 | 34 | 11 | 1.72 | 1.24 | 1.56 | 1.59 | 1.46 | 4.52 | 2.75 3.57 | 3.61 |
| * eps And | S | 0.000 | 4870 | 77 | 13 | 1.69 | | 1.83 | 1.48 | 1.66 | | 2.02 2.97 | 2.49 |
| * eps Aql | S | 0.000 | 4692 | 19 | 7 | 1.82 | 1.75 | 1.19 | 2.07 | 1.67 | 2.25 | 6.85 1.55 | 3.55 |
| * eps Col | H | 0.001 | 4573 | 33 | 13 | 2.37 | 2.30 | 1.95 | 3.16 | 2.47 | 0.81 | 3.39 0.40 | 1.53 |
| * eps CrB | S | 0.000 | 4365 | 28 | 10 | 2.18 | 1.57 | 1.69 | 1.30 | 1.52 | 2.60 | 5.65 | 4.13 |
| * eps Cru | U | 0.000 | 4121 | 27 | 12 | 2.52 | 2.34 | 1.94 | 1.96 | 2.08 | 0.97 | 1.22 2.00 | 1.40 |
| * eps Crv | S | 0.004 | 4285 | 59 | 13 | 2.98 | 2.60 | 1.25 | | 1.93 | | 3.13 | 3.13 |
| * eps Cyg | E | 0.000 | 4713 | 25 | 13 | 1.78 | 1.61 | 0.99 | 1.77 | 1.46 | 2.77 | 8.50 1.85 | 4.37 |
| * eps Dra | E | 0.000 | 4945 | 53 | 13 | 1.80 | 1.70 | 1.53 | 2.05 | 1.76 | 2.40 | 4.40 1.00 | 2.60 |
| * eps Equ A | S | 0.000 | 6223 | 73 | 4 | 0.99 | 1.53 | 1.28 | 1.58 | 1.46 | 2.24 | 3.40 2.48 | 2.70 |
| * eps Equ B | S | 0.000 | 6399 | 76 | 2 | 0.84 | 1.38 | 1.15 | 1.41 | 1.31 | 2.87 | 4.47 2.75 | 3.36 |
| * eps Hya | H | 0.000 | 5384 | 199 | 11 | 1.81 | 2.59 | 2.28 | | 2.44 | 0.50 | 0.63 | 0.56 |
| * eps Lep | S | 0.000 | 4005 | 14 | 11 | 2.67 | 1.68 | 1.96 | 1.61 | 1.75 | 2.36 | 0.91 4.03 | 2.43 |
| * eps Oph | S | 0.000 | 4837 | 20 | 13 | 1.78 | 1.50 | 1.29 | 1.83 | 1.54 | 3.15 | 3.38 | 3.26 |
| * eps Psc | S | 0.000 | 4814 | 40 | 15 | 1.83 | 1.57 | 1.58 | 1.58 | 1.58 | 2.68 | 2.45 | 2.56 |
| * eps Ret | H | 0.000 | 4702 | 97 | 14 | 0.82 | | 1.21 | 1.20 | 1.21 | | 7.06 | 7.06 |
| * eps Sco | U | 0.000 | 4522 | 28 | 13 | 1.78 | 1.42 | 1.44 | 1.07 | 1.31 | 4.47 | 3.38 | 3.92 |
| * eps Tau | S | 0.000 | 4836 | 30 | 13 | 1.94 | 2.67 | 2.50 | 2.13 | 2.43 | | 0.79 | 0.79 |
| * eps Vir | S | 0.000 | 4989 | 41 | 12 | 1.94 | 2.57 | 2.01 | 2.99 | 2.52 | 0.66 | 1.66 | 1.16 |
| * eta Boo | S | 0.000 | 6028 | 101 | 10 | 0.96 | | 1.47 | 1.61 | 1.54 | | 3.00 2.33 | 2.66 |
| * eta Cet | S | 0.000 | 4543 | 24 | 13 | 1.89 | 1.77 | 1.78 | 1.97 | 1.84 | 2.16 | 1.79 1.43 | 1.80 |
| * eta CMi | E | 0.001 | 7505 | 66 | 6 | 1.76 | | 2.00 | 2.32 | 2.16 | | 0.88 | 0.88 |
| * eta Cnc | S | 0.001 | 4415 | 57 | 12 | 1.94 | 1.45 | 1.55 | 1.54 | 1.51 | 4.28 | 2.96 4.52 | 3.92 |
| * eta Col | H | 0.012 | 4620 | 54 | 13 | 2.85 | 3.66 | 3.00 | | 3.33 | | | |
| * eta Cyg | E | 0.000 | 4783 | 20 | 13 | 1.74 | 1.53 | 1.10 | 2.13 | 1.59 | 2.93 | 5.75 1.20 | 3.29 |
| * eta Dra | S | 0.000 | 4962 | 46 | 12 | 1.83 | 2.51 | 1.54 | 3.00 | 2.35 | 0.60 | 4.68 | 2.64 |
| * eta Eri | H | 0.000 | 4614 | 27 | 16 | 1.79 | 1.50 | 1.65 | 1.79 | 1.65 | 3.13 | 2.13 2.23 | 2.50 |
| * eta Her | S | 0.000 | 4932 | 39 | 12 | 1.67 | 2.42 | 1.42 | 1.78 | 1.87 | | 1.50 | 1.50 |
| * eta Psc | S | 0.005 | 4875 | 48 | 13 | 2.66 | 4.04 | 2.86 | | 3.45 | 0.18 | 0.40 | 0.29 |
| * eta Ret | H | 0.005 | 4951 | 48 | 10 | 2.08 | 2.90 | 1.85 | 2.88 | 2.54 | 0.44 | 1.90 | 1.17 |
| * eta Ser | E | 0.000 | 4858 | 27 | 15 | 1.28 | | 1.52 | 1.22 | 1.37 | | | |
| * eta02 For | H | 0.001 | 5042 | 22 | 7 | 1.92 | 2.58 | 2.02 | 2.99 | 2.53 | 0.59 | 1.42 | 1.00 |
| * eta02 Hyi | H | 0.000 | 4878 | 32 | 13 | 1.82 | 1.91 | 2.18 | 2.32 | 2.14 | 1.35 | 1.08 0.80 | 1.07 |
| * f Dra | S | 0.000 | 4616 | 73 | 8 | 1.69 | 1.91 | 1.52 | 1.62 | 1.68 | 1.39 | 2.85 3.60 | 2.61 |
| * g Cyg | S | 0.000 | 4909 | 58 | 10 | 1.58 | 1.94 | 1.76 | 1.99 | 1.90 | 1.19 | 2.54 1.25 | 1.66 |
| * g Eri | H | 0.000 | 4948 | 45 | 13 | 1.91 | 2.65 | 1.87 | 2.47 | 2.33 | 0.62 | 2.90 | 1.76 |
| * g Gem | E | 0.002 | 4002 | 27 | 8 | 2.40 | 1.06 | 1.30 | 1.30 | 1.22 | 6.60 | 6.04 | 6.32 |
| * gam Aps | H | 0.000 | 4957 | 32 | 13 | 1.83 | 2.42 | 1.54 | 3.00 | 2.32 | 0.65 | 4.68 | 2.67 |
| * gam Com | S | 0.000 | 4634 | 32 | 13 | 1.77 | 1.78 | 1.78 | 1.81 | 1.79 | 2.02 | 1.75 2.17 | 1.98 |

| Name | Type | | | | | | | | | | | | |
|---|---|---|---|---|---|---|---|---|---|---|---|---|---|
| * gam Dra | E | 0.000 | 3925 | 34 | 11 | 2.79 | 1.63 | 1.86 | 2.20 | 1.90 | 4.08 | 1.30 | 2.69 |
| * gam Hya | S | 0.000 | 5019 | 28 | 13 | 2.04 | 3.00 | 2.64 | 2.87 | 2.84 | 0.36 | 0.51 | 0.43 |
| * gam Men | U | 0.000 | 4620 | 25 | 11 | 1.02 | | 1.34 | 1.16 | 1.25 | | | |
| * gam Mic | U | 0.000 | 5009 | 26 | 13 | 1.83 | 1.94 | 1.88 | 2.74 | 2.19 | | 2.65 | 0.50 | 1.58 |
| * gam Mic | H | 0.000 | 5009 | 26 | 13 | 1.83 | 1.94 | 1.88 | 2.74 | 2.19 | | 2.65 | 0.50 | 1.58 |
| * gam Psc | S | 0.000 | 4810 | 39 | 16 | 1.83 | 1.39 | 1.29 | 1.40 | 1.36 | 2.99 | 3.38 | 3.00 | 3.12 |
| * gam Pyx | S | 0.000 | 4322 | 23 | 13 | 2.19 | 1.53 | 1.60 | 1.78 | 1.64 | 3.25 | 2.38 | 2.40 | 2.67 |
| * gam Scl | U | 0.000 | 4578 | 24 | 13 | 1.83 | 1.72 | 1.64 | 1.76 | 1.71 | 2.65 | 2.19 | 2.57 | 2.47 |
| * gam Sge | E | 0.000 | 3862 | 40 | 9 | 2.77 | 1.40 | 1.99 | 1.92 | 1.77 | 4.15 | 1.08 | 1.82 | 2.35 |
| * gam Tau | S | 0.000 | 4901 | 36 | 13 | 1.96 | 2.32 | 2.65 | 2.73 | 2.57 | 0.84 | 0.60 | 0.50 | 0.65 |
| * gam01 For | S | 0.000 | 4657 | 56 | 11 | 1.89 | 1.72 | 1.21 | 2.03 | 1.65 | 2.19 | 6.53 | 1.37 | 3.36 |
| * gam01 Leo | S | 0.000 | 4305 | 100 | 4 | 2.45 | 1.18 | 1.82 | 1.89 | 1.63 | 4.45 | 2.38 | 1.53 | 2.79 |
| * gam02 Cae | H | 0.003 | 6973 | 65 | 5 | 1.33 | 1.88 | 1.66 | 1.77 | 1.77 | 1.17 | 1.63 | 1.50 | 1.43 |
| * h Aql | S | 0.006 | 4496 | 58 | 10 | 1.86 | 1.25 | 1.50 | 1.36 | 1.37 | 4.93 | 3.00 | 4.75 | 4.23 |
| * h Eri | H | 0.000 | 4539 | 31 | 13 | 1.78 | 1.48 | 1.94 | 1.83 | 1.75 | 3.70 | 1.61 | 1.90 | 2.40 |
| * h Her | E | 0.000 | 3958 | 25 | 10 | 2.45 | 1.11 | 1.17 | 1.30 | 1.19 | 6.07 | 7.00 | | 6.53 |
| * H Sco | U | 0.015 | 3875 | 21 | 9 | 2.75 | 1.65 | 1.92 | 1.86 | 1.81 | 3.31 | 0.99 | 2.25 | 2.18 |
| * h Vir | S | 0.000 | 4837 | 66 | 10 | 1.68 | 1.84 | 1.13 | 1.77 | 1.58 | 1.93 | 5.25 | 1.90 | 3.03 |
| * i Aql | S | 0.000 | 4601 | 23 | 13 | 1.78 | 1.53 | 1.61 | 1.73 | 1.62 | 2.86 | 2.31 | 2.60 | 2.59 |
| * iot Ant | U | 0.000 | 4786 | 22 | 13 | 1.75 | 1.58 | 1.10 | 1.96 | 1.55 | 2.85 | 5.75 | 1.35 | 3.32 |
| * iot Cap | S | 0.000 | 5011 | 22 | 13 | 1.86 | 2.35 | 1.66 | 2.47 | 2.16 | | 2.53 | | 2.53 |
| * iot Cep | E | 0.000 | 4743 | 56 | 13 | 1.76 | 1.47 | 1.13 | 2.61 | 1.74 | 3.78 | 6.75 | | 5.27 |
| * iot Cet | S | 0.001 | 4479 | 54 | 15 | 2.56 | 3.19 | 2.37 | | 2.78 | | 2.23 | | 2.23 |
| * iot Cnc A | S | 0.000 | 4849 | 48 | 11 | 2.45 | 3.55 | 2.40 | | 2.98 | 0.25 | 0.60 | | 0.43 |
| * iot Dra | S | 0.000 | 4531 | 18 | 12 | 1.78 | 1.39 | 1.58 | 1.46 | 1.48 | 4.55 | 2.70 | 4.68 | 3.98 |
| * iot Eri | H | 0.000 | 4683 | 35 | 13 | 1.76 | 1.34 | | 1.49 | 1.42 | 4.76 | | 3.47 | 4.12 |
| * iot Gem | E | 0.000 | 4750 | 43 | 11 | 1.68 | 2.52 | 1.26 | | 1.89 | | 6.67 | | 6.67 |
| * iot Hya | S | 0.000 | 4244 | 32 | 16 | 2.47 | 2.21 | 1.56 | 1.98 | 1.92 | 1.53 | 4.42 | 1.47 | 2.47 |
| * iot Scl | S | 0.000 | 4888 | 57 | 11 | 1.98 | 2.47 | 2.71 | 2.73 | 2.64 | 0.73 | 0.56 | 0.50 | 0.60 |
| * iot Sgr | H | 0.000 | 4594 | 41 | 16 | 1.94 | 1.67 | 0.96 | 1.58 | 1.40 | 2.08 | 9.25 | 2.90 | 4.74 |
| * iot Tri | E | 0.005 | 5082 | 185 | 6 | 1.93 | 2.62 | 1.96 | 2.99 | 2.52 | 0.54 | 1.18 | | 0.86 |
| * iot Tuc | H | 0.003 | 5039 | 63 | 11 | 1.81 | 2.55 | 1.59 | 2.47 | 2.20 | 0.51 | 2.86 | | 1.69 |
| * k Car | H | 0.000 | 4888 | 54 | 13 | 1.83 | 1.85 | 1.96 | 2.05 | 1.95 | 1.39 | 2.69 | | 2.04 |
| * k Hya | U | 0.000 | 4255 | 31 | 13 | 1.67 | 1.31 | | | 1.31 | | | | |
| * k Leo | S | 0.000 | 4985 | 19 | 12 | 1.71 | 1.98 | 1.68 | 2.35 | 2.00 | 1.02 | 4.17 | 0.75 | 1.98 |
| * kap Ara | U | 0.054 | 4911 | 311 | 8 | 2.30 | 3.20 | 2.22 | 3.85 | 3.09 | 0.33 | 0.75 | | 0.54 |
| * kap Aur | S | 0.000 | 4685 | 29 | 13 | 1.81 | 1.43 | 0.97 | 1.35 | 1.25 | 4.01 | 9.00 | 3.75 | 5.59 |
| * kap Cap | S | 0.002 | 4960 | 52 | 14 | 2.04 | 2.28 | 2.11 | 2.40 | 2.26 | 0.90 | 1.48 | | 1.19 |
| * kap Col | H | 0.000 | 4876 | 33 | 13 | 1.79 | 1.78 | 1.74 | | 1.76 | 1.52 | 1.83 | | 1.68 |
| * kap Gem | S | 0.000 | 4954 | 47 | 12 | 1.87 | 2.53 | 1.61 | | 2.07 | 0.57 | 3.58 | | 2.07 |
| * kap Leo | S | 0.000 | 4403 | 24 | 10 | 1.95 | 1.52 | 1.58 | 1.21 | 1.44 | 3.52 | 2.81 | 6.67 | 4.33 |
| * kap Lyr | S | 0.000 | 4549 | 40 | 10 | 2.15 | 2.00 | 1.83 | 2.07 | 1.97 | 2.08 | 2.26 | 1.30 | 1.88 |
| * kap Oph | E | 0.000 | 4559 | 40 | 15 | 1.72 | 1.30 | 1.43 | 1.52 | 1.42 | 5.13 | | 4.70 | 4.92 |
| * kap Per | S | 0.000 | 4857 | 69 | 13 | 1.60 | | 1.37 | 1.62 | 1.50 | | 4.58 | | 4.58 |
| * kap Vir | S | 0.000 | 4155 | 27 | 13 | 2.37 | 1.39 | 1.52 | 1.11 | 1.34 | 3.03 | 2.88 | | 2.95 |
| * kap02 Cet | S | 0.005 | 4948 | 18 | 9 | 1.75 | 2.08 | 2.27 | 2.25 | 2.20 | 1.05 | 0.94 | 0.83 | 0.94 |
| * ksi And | S | 0.000 | 4704 | 60 | 11 | 1.76 | 1.47 | 1.26 | 2.03 | 1.59 | 3.70 | 6.67 | | 5.19 |
| * ksi Col | H | 0.004 | 4659 | 74 | 10 | 2.12 | 2.11 | 2.07 | 2.06 | 2.08 | 1.25 | 1.35 | | 1.30 |
| * ksi Dra | E | 0.000 | 4459 | 36 | 13 | 1.71 | 1.47 | 1.12 | 1.30 | 1.30 | 3.69 | 8.00 | | 5.85 |
| * ksi Her | S | 0.000 | 4931 | 25 | 9 | 1.79 | 2.21 | 1.69 | 2.13 | 2.01 | 0.85 | 4.10 | | 2.48 |
| * ksi Hya | H | 0.000 | 4951 | 18 | 14 | 1.80 | 2.09 | 1.53 | 2.22 | 1.95 | 1.11 | 4.40 | | 2.75 |
| * ksi Leo | U | 0.000 | 4686 | 19 | 10 | 1.73 | 1.93 | 1.75 | 1.50 | 1.73 | 1.64 | | 4.20 | 2.92 |
| * ksi Men | H | 0.006 | 4990 | 21 | 10 | 1.79 | 2.41 | 1.41 | | 1.91 | 0.61 | 5.01 | | 2.81 |
| * ksi Psc | E | 0.000 | 4947 | 25 | 13 | 1.66 | 2.19 | 1.89 | 1.93 | 2.00 | 0.82 | 1.63 | | 1.22 |
| * ksi Ser | E | 0.000 | 7217 | 41 | 5 | 1.48 | 2.00 | 2.09 | 2.08 | 2.06 | 1.00 | | 1.05 | 1.03 |
| * ksi01 Cap | S | 0.024 | 4439 | 42 | 8 | 1.95 | 1.55 | 1.50 | 1.59 | 1.55 | 3.59 | 3.00 | 3.47 | 3.35 |
| * ksi02 Sgr | H | 0.017 | 4541 | 64 | 11 | 2.83 | 3.66 | 2.64 | 3.77 | 3.36 | 0.26 | 0.50 | | 0.38 |
| * lam Dor | H | 0.011 | 4911 | 31 | 10 | 2.29 | 3.08 | 2.75 | 3.85 | 3.23 | 0.38 | 0.45 | | 0.41 |
| * lam Hya | S | 0.000 | 4851 | 26 | 13 | 1.68 | 2.16 | 1.69 | 2.07 | 1.97 | 0.90 | 2.90 | 1.23 | 1.67 |
| * lam Pic | H | 0.005 | 4851 | 45 | 10 | 2.05 | 2.34 | 1.78 | 2.47 | 2.20 | 0.79 | 3.70 | | 2.24 |
| * lam Pyx | S | 0.000 | 4959 | 22 | 13 | 1.69 | 1.96 | 2.34 | 1.87 | 2.06 | | | 1.30 | 1.30 |
| * lam02 Scl | H | 0.000 | 4531 | 25 | 11 | 1.80 | 1.49 | 1.46 | 1.51 | 1.49 | 3.53 | 3.25 | 3.95 | 3.58 |

| Name | Type | col3 | col4 | col5 | col6 | col7 | col8 | col9 | col10 | col11 | col12 | col13 | col14 | col15 |
|---|---|---|---|---|---|---|---|---|---|---|---|---|---|---|
| * m Leo    | S | 0.000 | 4612 | 32  | 12 | 1.41 |      | 1.29 | 1.54 | 1.42 |      | 5.85 | 3.43 | 4.64 |
| * mu. Aql  | E | 0.000 | 4487 | 43  | 16 | 1.38 | 1.50 |      | 1.26 | 1.38 |      |      |      |      |
| * mu. Hyi  | H | 0.001 | 4833 | 29  | 10 | 1.82 | 1.66 | 1.67 | 1.93 | 1.75 | 2.46 | 2.18 |      | 2.32 |
| * mu. Leo  | S | 0.000 | 4471 | 40  | 13 | 1.74 | 1.36 | 1.66 | 1.82 | 1.61 | 4.19 | 2.54 |      | 3.37 |
| * mu. Phe  | H | 0.000 | 4952 | 177 | 13 | 1.94 | 2.69 | 1.82 | 2.99 | 2.50 | 0.46 | 2.34 |      | 1.40 |
| * mu. Psc  | S | 0.004 | 4126 | 27  | 16 | 2.27 | 1.13 | 1.24 | 1.39 | 1.25 | 5.58 | 5.55 |      | 5.57 |
| * mu. Vir  | S | 0.000 | 6487 | 150 | 7  | 0.87 | 1.51 | 1.26 | 1.42 | 1.40 |      | 3.67 |      | 3.67 |
| * n Tau    | S | 0.001 | 4954 | 31  | 10 | 1.81 | 2.42 | 1.21 | 2.61 | 2.08 | 0.65 | 5.33 | 0.60 | 2.19 |
| * N Vel    | U | 0.000 | 4105 | 592 | 10 | 2.74 | 2.53 | 1.15 | 2.05 | 1.91 | 1.20 | 5.08 | 1.13 | 2.47 |
| * nu. Aqr  | S | 0.000 | 4938 | 19  | 16 | 1.60 | 2.17 | 1.66 | 2.26 | 2.03 | 0.88 | 2.37 | 0.93 | 1.40 |
| * nu. Aur  | S | 0.000 | 4590 | 44  | 13 | 2.22 | 2.18 | 2.07 | 2.10 | 2.12 | 1.04 | 1.35 | 0.95 | 1.11 |
| * nu. Hya  | S | 0.000 | 4340 | 21  | 13 | 2.22 | 1.59 | 1.73 | 1.43 | 1.58 | 2.76 | 1.88 | 4.80 | 3.14 |
| * nu. Hyi  | H | 0.004 | 4262 | 33  | 13 | 2.35 | 1.31 | 1.77 | 2.28 | 1.79 | 3.76 |      | 1.00 | 2.38 |
| * nu. Oph  | S | 0.000 | 4874 | 25  | 13 | 2.03 | 2.51 | 2.89 |      | 2.70 | 0.63 | 0.46 |      | 0.54 |
| * nu. Peg  | E | 0.001 | 4068 | 25  | 13 | 2.19 | 1.33 | 1.18 | 1.26 | 1.26 | 4.63 | 7.25 | 6.67 | 6.18 |
| * nu. Psc  | E | 0.008 | 4154 | 23  | 16 | 2.58 | 1.66 | 1.68 | 1.65 | 1.66 | 2.59 | 4.85 | 2.80 | 3.41 |
| * nu. UMa  | E | 0.000 | 4113 | 39  | 12 | 3.05 | 3.19 | 1.56 |      | 2.38 |      | 5.34 |      | 5.34 |
| * nu.01 CMa | E | 0.004 | 6091 | 822 | 7  | 0.92 | 1.45 | 1.22 | 1.56 | 1.41 | 2.60 | 3.95 | 2.64 | 3.07 |
| * nu.02 CMa | H | 0.000 | 4735 | 34  | 16 | 1.08 | 1.25 |      | 1.17 | 1.21 |      |      |      |      |
| * nu.02 Sgr | S | 0.004 | 4244 | 57  | 13 | 2.08 | 1.51 | 1.16 | 1.64 | 1.44 | 3.69 | 7.00 | 2.88 | 4.52 |
| * ome Boo  | S | 0.001 | 3962 | 35  | 8  | 2.51 | 1.50 | 1.31 | 2.13 | 1.65 | 3.16 | 4.67 | 1.15 | 2.99 |
| * ome Cnc  | U | 0.057 | 5070 | 177 | 10 | 2.35 | 3.45 | 2.96 | 3.83 | 3.41 | 0.26 | 0.34 | 0.20 | 0.27 |
| * ome Per  | S | 0.006 | 4586 | 18  | 7  | 2.16 | 2.38 | 1.47 | 2.26 | 2.04 | 0.99 | 3.06 | 0.90 | 1.65 |
| * ome Ser  | S | 0.000 | 4715 | 21  | 10 | 1.83 | 1.65 | 1.05 | 1.68 | 1.46 | 2.14 | 6.75 | 2.33 | 3.74 |
| * ome Sgr  | S | 0.000 | 5307 | 41  | 13 | 0.92 | 1.58 | 1.22 | 1.57 | 1.46 | 2.17 | 5.03 | 2.50 | 3.23 |
| * ome01 Tau | S | 0.007 | 4737 | 77  | 10 | 1.76 | 1.47 | 1.08 | 2.05 | 1.53 | 3.80 | 7.40 | 1.43 | 4.21 |
| * omi Boo  | S | 0.000 | 4864 | 25  | 10 | 1.93 | 2.39 | 1.54 | 2.22 | 2.05 | 0.80 | 4.64 |      | 2.72 |
| * omi Cep  | E | 0.000 | 5124 | 61  | 8  | 1.69 | 2.33 | 2.65 |      | 2.49 |      |      |      |      |
| * omi CrB  | S | 0.001 | 4705 | 21  | 11 | 1.71 | 1.63 | 1.75 | 1.39 | 1.59 | 2.21 |      | 4.35 | 3.28 |
| * omi Dra  | S | 0.004 | 4354 | 86  | 6  | 2.36 | 1.61 | 1.05 | 1.29 | 1.32 | 3.08 | 7.00 | 5.00 | 5.03 |
| * omi Psc  | S | 0.002 | 4903 | 26  | 13 | 2.20 | 3.01 | 2.14 |      | 2.58 | 0.40 | 1.22 |      | 0.81 |
| * omi Sgr  | S | 0.000 | 4766 | 30  | 13 | 1.83 | 1.48 | 1.88 | 2.04 | 1.80 | 3.97 | 1.99 | 1.20 | 2.39 |
| * omi Tau  | S | 0.003 | 4982 | 86  | 16 | 2.48 | 3.24 | 2.69 | 3.74 | 3.22 | 0.32 | 0.41 | 0.20 | 0.31 |
| * omi Vir  | S | 0.000 | 4762 | 114 | 16 | 1.81 | 1.92 | 1.14 | 1.41 | 1.49 | 1.50 | 5.25 | 3.53 | 3.43 |
| * omi02 Ori | S | 0.000 | 4466 | 29  | 10 | 2.02 | 1.84 | 1.69 | 1.39 | 1.64 | 2.00 |      | 4.87 | 3.43 |
| * p Vir    | S | 0.000 | 4703 | 52  | 11 | 1.77 | 1.38 | 1.26 | 1.77 | 1.47 | 3.99 | 6.67 | 1.85 | 4.17 |
| * p01 Leo  | S | 0.000 | 4922 | 25  | 11 | 1.81 | 1.99 | 2.31 | 2.13 | 2.14 | 1.03 | 0.89 |      | 0.96 |
| * p03 Leo  | U | 0.000 | 4472 | 40  | 12 | 2.29 | 1.93 | 1.70 | 2.00 | 1.88 | 1.83 | 2.69 | 1.20 | 1.90 |
| * phi Cap  | U | 0.009 | 4490 | 25  | 10 | 2.65 | 2.76 | 2.08 | 3.05 | 2.63 | 0.73 | 2.60 | 0.40 | 1.24 |
| * phi Cyg  | S | 0.003 | 4872 | 22  | 13 | 1.99 | 2.41 | 1.48 | 2.99 | 2.29 | 0.76 | 4.63 |      | 2.70 |
| * phi Hya  | S | 0.000 | 4952 | 17  | 10 | 1.68 | 2.20 | 1.87 |      | 2.04 | 0.82 | 1.53 |      | 1.17 |
| * phi Oph  | S | 0.000 | 5038 | 20  | 13 | 2.05 | 3.00 | 2.32 |      | 2.66 | 0.36 |      |      | 0.36 |
| * phi Ser  | S | 0.000 | 4593 | 77  | 13 | 1.62 | 1.77 | 1.37 | 1.48 | 1.54 | 1.76 | 3.97 | 4.53 | 3.42 |
| * phi Tau  | S | 0.000 | 4457 | 23  | 10 | 2.14 | 1.66 | 1.30 | 1.36 | 1.44 | 2.06 | 6.58 | 3.75 | 4.13 |
| * phi01 Cet | E | 0.000 | 4793 | 22  | 13 | 1.86 | 1.60 | 1.90 |      | 1.75 | 2.58 | 1.84 |      | 2.21 |
| * phi02 Ori | E | 0.000 | 4742 | 61  | 16 | 1.55 | 1.51 | 1.62 | 1.00 | 1.38 | 2.78 |      |      | 2.78 |
| * phi04 Cet | S | 0.002 | 4841 | 19  | 10 | 1.78 | 1.29 | 1.65 | 1.74 | 1.56 | 4.06 | 2.39 | 1.60 | 2.68 |
| * pi. CrB  | S | 0.000 | 4626 | 27  | 9  | 1.61 | 2.17 | 1.44 | 1.23 | 1.61 |      | 3.38 | 6.00 | 4.69 |
| * pi. For  | S | 0.000 | 4880 | 19  | 10 | 1.81 | 1.49 | 2.18 | 1.70 | 1.79 | 2.50 | 1.08 | 1.87 | 1.81 |
| * pi. For  | H | 0.000 | 4884 | 19  | 10 | 1.81 | 1.40 | 2.18 | 1.70 | 1.76 | 3.19 | 1.08 | 1.87 | 2.04 |
| * pi. Hya  | H | 0.000 | 4563 | 20  | 13 | 1.78 | 1.54 | 1.51 | 1.16 | 1.40 | 3.10 | 2.88 |      | 2.99 |
| * pi. Peg  | S | 0.001 | 6253 | 58  | 8  | 2.00 |      | 2.48 |      | 2.48 |      | 0.53 |      | 0.53 |
| * pi.01 Dor | H | 0.021 | 3984 | 26  | 8  | 2.68 | 1.67 | 1.98 | 1.95 | 1.87 | 2.76 | 0.94 | 1.62 | 1.77 |
| * pi.02 Dor | H | 0.002 | 4853 | 38  | 10 | 1.73 | 1.81 | 1.71 | 2.08 | 1.87 |      | 2.35 | 1.10 | 1.72 |
| * pi.02 UMa | S | 0.001 | 4438 | 28  | 10 | 2.09 | 1.53 | 1.76 | 1.71 | 1.67 | 2.60 |      | 3.05 | 2.82 |
| * psi Boo  | S | 0.000 | 4302 | 22  | 10 | 2.13 | 1.42 | 1.45 | 1.26 | 1.38 | 3.27 | 3.38 | 5.83 | 4.16 |
| * psi Oph  | U | 0.000 | 4759 | 26  | 14 | 1.84 | 1.60 | 1.39 | 2.04 | 1.68 | 3.29 | 3.96 | 1.20 | 2.82 |
| * psi Oph  | S | 0.000 | 4757 | 27  | 14 | 1.84 | 1.60 | 1.39 | 2.04 | 1.68 | 3.29 | 3.96 | 1.20 | 2.82 |
| * psi UMa  | S | 0.000 | 4523 | 40  | 13 | 2.20 | 2.28 | 1.19 | 1.84 | 1.77 | 1.35 | 4.50 | 1.40 | 2.42 |
| * psi01 Aqr | S | 0.000 | 4624 | 31  | 15 | 1.72 | 1.77 | 1.57 | 1.71 | 1.68 | 2.41 | 2.50 | 2.67 | 2.53 |
| * r Her    | S | 0.000 | 4798 | 73  | 10 | 1.78 | 1.33 | 1.55 | 1.81 | 1.56 | 4.24 | 3.09 | 1.71 | 3.01 |
| * rho And  | E | 0.000 | 6323 | 93  | 8  | 1.20 | 1.76 | 1.50 | 1.74 | 1.67 | 1.42 | 2.13 | 1.80 | 1.78 |

| Name | Type | | | | | | | | | | | | |
|---|---|---|---|---|---|---|---|---|---|---|---|---|---|
| * rho Boo | S | 0.000 | 4281 | 26 | 12 | 2.15 | 1.64 | 1.45 | 1.57 | 1.55 | 2.52 | 3.38 | 4.60 | 3.50 |
| * rho Cyg | S | 0.000 | 5010 | 23 | 13 | 1.56 | 1.96 | 1.85 | 1.78 | 1.86 | | 1.55 | 1.50 | 1.53 |
| * rho For | H | 0.001 | 4776 | 26 | 11 | 1.68 | 1.57 | 1.00 | 1.77 | 1.45 | 3.65 | 8.00 | 1.85 | 4.50 |
| * rho Ori | S | 0.009 | 4533 | 45 | 13 | 2.40 | 2.60 | 2.68 | 2.74 | 2.67 | 0.67 | 0.69 | 0.58 | 0.65 |
| * rho01 Eri | S | 0.004 | 4751 | 52 | 11 | 1.76 | 1.60 | 1.21 | 2.13 | 1.65 | 2.99 | 6.18 | 1.20 | 3.46 |
| * s Her | E | 0.000 | 4167 | 108 | 11 | 2.72 | 1.32 | 1.38 | 1.76 | 1.49 | 3.88 | 3.50 | 2.06 | 3.15 |
| * sig Hya | S | 0.005 | 4491 | 51 | 16 | 2.47 | 3.38 | 2.75 | | 3.07 | 0.28 | 0.57 | | 0.43 |
| * sig Per | S | 0.012 | 4163 | 20 | 9 | 2.60 | 2.34 | 1.77 | 1.91 | 2.01 | 1.48 | 2.77 | 1.74 | 2.00 |
| * sig Pup | H | 0.000 | 4081 | 452 | 11 | 0.30 | | | | 1.00 | | | | |
| * sig03 Cnc | S | 0.000 | 4974 | 31 | 12 | 1.84 | 2.34 | 1.62 | 1.99 | 1.98 | | 3.33 | | 3.33 |
| * tau Aur | S | 0.000 | 4876 | 13 | 13 | 1.84 | 2.15 | 2.23 | 1.91 | 2.10 | 0.92 | 0.99 | 1.20 | 1.04 |
| * tau Cas | S | 0.000 | 4617 | 77 | 10 | 1.60 | 1.42 | 1.41 | 1.49 | 1.44 | 3.87 | 3.63 | 4.20 | 3.90 |
| * tau Cnc | S | 0.000 | 4965 | 62 | 10 | 1.68 | 1.96 | 1.76 | 1.87 | 1.86 | | 3.26 | 1.30 | 2.28 |
| * tau Dra | E | 0.000 | 4413 | 77 | 10 | 1.68 | 1.46 | 1.19 | 1.11 | 1.25 | 3.95 | 7.00 | 8.50 | 6.48 |
| * tau Psc | S | 0.000 | 4658 | 41 | 10 | 1.71 | | 1.62 | 1.76 | 1.69 | | | 2.27 | 2.27 |
| * tau Pup | H | 0.000 | 4489 | 33 | 13 | 2.45 | 3.19 | 2.53 | 3.85 | 3.19 | 0.36 | 0.73 | | 0.54 |
| * tau Sgr | S | 0.000 | 4444 | 25 | 11 | 1.96 | 1.33 | 1.52 | 1.46 | 1.44 | 4.20 | 2.88 | 3.90 | 3.66 |
| * tau02 Eri | S | 0.000 | 4976 | 21 | 13 | 1.63 | 2.30 | 1.87 | 1.93 | 2.03 | 0.71 | 1.53 | | 1.12 |
| * tau06 Ser | S | 0.000 | 4982 | 67 | 10 | 2.15 | 3.05 | 2.44 | | 2.75 | 0.36 | 0.70 | | 0.53 |
| * tet Aqr | S | 0.000 | 4894 | 29 | 16 | 1.89 | 2.55 | 2.43 | 2.22 | 2.40 | 0.65 | 0.81 | | 0.73 |
| * tet Cet | S | 0.000 | 4660 | 17 | 16 | 1.73 | 1.90 | 1.68 | 1.81 | 1.80 | 2.24 | | 2.17 | 2.20 |
| * tet CMa | S | 0.001 | 4044 | 22 | 14 | 2.47 | 1.36 | 1.44 | 1.37 | 1.39 | 3.33 | 3.50 | 5.12 | 3.98 |
| * tet Dor | H | 0.010 | 4320 | 59 | 13 | 2.63 | 1.77 | 1.91 | 3.02 | 2.23 | 1.42 | 1.62 | 0.47 | 1.17 |
| * tet Psc | E | 0.000 | 4684 | 23 | 16 | 1.69 | 1.27 | 1.68 | 1.79 | 1.58 | | | 2.45 | 2.45 |
| * tet01 Tau | S | 0.000 | 4949 | 25 | 13 | 1.84 | 2.42 | 2.66 | 3.00 | 2.69 | 0.65 | 0.57 | | 0.61 |
| * ups Dra | S | 0.004 | 4561 | 61 | 12 | 2.23 | 2.12 | 1.84 | 2.19 | 2.05 | 1.17 | 2.00 | 0.93 | 1.37 |
| * ups Gem | E | 0.000 | 3884 | 32 | 10 | 2.57 | 1.30 | 1.47 | 1.79 | 1.52 | 4.31 | 4.22 | 2.07 | 3.53 |
| * ups Leo | S | 0.000 | 4786 | 19 | 14 | 1.83 | 1.25 | 1.75 | 1.65 | 1.55 | 4.59 | 2.36 | 1.95 | 2.97 |
| * ups Lib | S | 0.000 | 4135 | 20 | 12 | 2.49 | 1.74 | 1.69 | 1.58 | 1.67 | 2.38 | | 3.90 | 3.14 |
| * ups Per | S | 0.000 | 4310 | 53 | 13 | 2.23 | 2.00 | 1.78 | 2.01 | 1.93 | 1.51 | 1.83 | 1.40 | 1.58 |
| * ups Vir | S | 0.001 | 4730 | 31 | 11 | 1.81 | 1.73 | 1.07 | 1.89 | 1.56 | 2.48 | 6.42 | 1.50 | 3.46 |
| * ups01 Cas | E | 0.014 | 4422 | 14 | 10 | 2.24 | 1.39 | 1.02 | 1.77 | 1.39 | 4.17 | 7.75 | 2.31 | 4.75 |
| * ups01 Eri | S | 0.000 | 4773 | 29 | 13 | 1.44 | | 1.52 | | 1.52 | | | | |
| * ups01 Hya | S | 0.001 | 4955 | 26 | 10 | 2.19 | 3.07 | 2.52 | 2.80 | 2.80 | | 0.64 | | 0.64 |
| * ups02 Cas | S | 0.000 | 4829 | 78 | 13 | 1.77 | 1.15 | 1.65 | 1.83 | 1.54 | 4.62 | 2.51 | | 3.56 |
| * ups02 Cnc | U | 0.013 | 4781 | 52 | 13 | 2.00 | 2.16 | 1.82 | 2.40 | 2.13 | 1.14 | 2.56 | 0.74 | 1.48 |
| * ups02 Cnc | S | 0.013 | 4781 | 52 | 13 | 2.00 | 2.16 | 1.82 | 2.40 | 2.13 | 1.14 | 2.56 | 0.74 | 1.48 |
| * ups02 Eri | H | 0.000 | 4897 | 30 | 13 | 2.14 | 2.80 | 1.77 | 2.88 | 2.48 | 0.50 | 2.09 | | 1.30 |
| * w Cen | U | 0.000 | 4682 | 26 | 13 | 1.75 | 1.48 | 1.75 | 1.91 | 1.71 | 3.49 | | 1.73 | 2.61 |
| * y Eri | H | 0.000 | 4708 | 113 | 13 | 1.94 | 1.60 | 2.46 | 3.00 | 2.35 | 2.80 | 0.72 | | 1.76 |
| * zet And | S | 0.000 | 4584 | 33 | 5 | 2.00 | 2.11 | 1.02 | 2.16 | 1.76 | 1.25 | 7.50 | | 4.37 |
| * zet And | E | 0.000 | 4571 | 39 | 8 | 2.01 | 2.10 | 1.19 | 1.84 | 1.71 | 1.16 | 6.85 | 1.75 | 3.25 |
| * zet Ara | U | 0.020 | 3851 | 66 | 11 | 3.48 | 1.28 | 2.49 | | 1.89 | 2.48 | 0.53 | | 1.50 |
| * zet Cet | S | 0.000 | 4579 | 35 | 13 | 2.33 | 2.32 | 2.10 | 2.61 | 2.34 | 0.88 | 2.23 | 0.60 | 1.24 |
| * zet Cyg | S | 0.000 | 4891 | 54 | 5 | 2.04 | 2.54 | 1.87 | | 2.21 | | 2.43 | | 2.43 |
| * zet Cyg | E | 0.000 | 4893 | 63 | 13 | 2.04 | 2.54 | 1.87 | | 2.21 | | 2.43 | | 2.43 |
| * zet Hya | S | 0.000 | 4795 | 50 | 13 | 2.23 | 2.89 | 2.23 | 2.40 | 2.51 | 0.40 | 1.80 | | 1.10 |
| * zet Leo | E | 0.001 | 6977 | 58 | 7 | 2.35 | | 3.13 | | 3.13 | | 0.30 | | 0.30 |
| * zet Pyx | S | 0.000 | 4876 | 8 | 7 | 1.84 | 1.61 | 2.23 | 2.05 | 1.96 | 2.78 | 0.99 | | 1.88 |
| * zet Sct | S | 0.000 | 4894 | 26 | 11 | 1.77 | 1.98 | 1.92 | 1.98 | 1.96 | 1.28 | 1.54 | 1.24 | 1.36 |
| * zet Vol | H | 0.000 | 4736 | 24 | 13 | 1.77 | 1.47 | 1.13 | 2.61 | 1.74 | 3.78 | 6.75 | | 5.27 |
| 2MASS J08511269+1152423 | H | 0.060 | 4718 | 39 | 16 | 1.83 | 1.73 | 1.48 | 2.02 | 1.74 | 2.29 | 4.41 | 1.43 | 2.71 |
| 2MASS J08511710+1148160 | H | 0.060 | 4431 | 33 | 16 | 2.01 | 1.66 | 1.35 | 1.76 | 1.59 | 2.93 | 4.89 | 2.52 | 3.45 |
| 2MASS J08512156+1146061 | H | 0.060 | 4793 | 35 | 13 | 1.47 | 1.67 | 1.37 | 1.69 | 1.58 | 2.64 | 4.61 | 2.90 | 3.38 |
| 2MASS J08512618+1153520 | H | 0.060 | 4745 | 43 | 14 | 1.86 | 1.81 | 1.40 | 2.08 | 1.76 | 2.13 | 4.38 | 1.26 | 2.59 |
| 2MASS J08512898+1150330 | U | 0.060 | 4735 | 127 | 14 | 1.84 | 1.77 | 1.37 | 2.03 | 1.72 | 2.24 | 4.71 | 1.41 | 2.79 |
| 2MASS J08512898+1150330 | H | 0.060 | 4732 | 132 | 13 | 1.84 | 1.77 | 1.37 | 2.03 | 1.72 | 2.24 | 4.74 | 1.41 | 2.80 |
| 2MASS J08513938+1151456 | U | 0.060 | 5047 | 388 | 10 | 1.13 | 1.75 | 1.30 | 1.79 | 1.61 | 1.64 | 4.90 | 1.87 | 2.81 |
| 2MASS J08515020+1146069 | H | 0.060 | 5067 | 59 | 15 | 1.69 | 2.37 | 1.67 | 2.43 | 2.16 | 0.67 | 2.99 | 0.73 | 1.46 |
| 2MASS J08521856+1144263 | U | 0.060 | 4779 | 70 | 13 | 1.88 | 1.87 | 2.06 | 2.16 | 2.03 | 2.05 | 1.94 | 1.12 | 1.70 |
| BD+12 1917 | H | 0.060 | 4245 | 182 | 8 | 2.30 | 1.85 | 1.53 | 1.79 | 1.72 | 2.38 | 3.96 | 2.68 | 3.01 |
| BD+33 2562 | S | 0.000 | 5812 | 67 | 5 | -0.06 | | 1.03 | 0.92 | 0.98 | | 0.68 | 5.50 | 3.09 |

| Name | Src | E(B-V) | Teff | N | n | [Fe/H] | A1 | A2 | A3 | A | B1 | B2 | B3 | B |
|---|---|---|---|---|---|---|---|---|---|---|---|---|---|---|
| CCDM J05496-1429AB | E | 0.001 | 5086 | 36 | 9 | 1.63 | 2.23 | 2.27 | 2.39 | 2.30 | 0.77 | 0.85 | 0.73 | 0.78 |
| CCDM J16238+6142AB | S | 0.006 | 4946 | 123 | 8 | 2.14 | 2.70 | 2.20 | | 2.45 | 0.51 | 0.99 | | 0.75 |
| CD-30 13092B | H | 0.000 | 4315 | 50 | 5 | 0.38 | 0.82 | 0.63 | 0.71 | 0.72 | 2.49 | 3.08 | 3.39 | 2.99 |
| CD-33 2771 | H | | 4002 | 1 | 2 | | | | | | | | | |
| CD-38 14203B | H | 0.000 | 6074 | 236 | 6 | | | | | | | | | |
| CD-42 6977 | U | 0.070 | 4662 | 67 | 10 | 1.74 | 1.63 | 1.41 | 1.77 | 1.60 | 3.01 | 4.41 | 2.63 | 3.35 |
| CD-49 11401 | U | 0.120 | 4829 | 193 | 11 | 1.93 | 2.08 | 2.34 | 2.24 | 2.22 | 1.93 | 1.53 | 1.11 | 1.53 |
| CD-49 11401 | U | 0.120 | 4829 | 193 | 11 | 1.93 | 2.08 | 2.34 | 2.24 | 2.22 | 1.93 | 1.53 | 1.11 | 1.53 |
| CD-49 11402 | H | 0.120 | 4766 | 79 | 11 | 1.71 | 1.85 | 1.72 | 2.03 | 1.87 | 2.29 | 3.34 | 1.49 | 2.37 |
| CD-49 11404 | U | 0.120 | 4694 | 168 | 11 | 1.77 | 1.77 | 1.84 | 1.87 | 1.83 | 2.38 | 2.57 | 2.26 | 2.40 |
| CD-49 11415 | H | 0.120 | 4414 | 74 | 11 | 2.00 | 1.82 | 1.51 | 1.77 | 1.70 | 2.41 | 4.00 | 2.66 | 3.02 |
| Cl* IC 4651 MMU 9025 | U | 0.120 | 4761 | 24 | 8 | 1.74 | 1.85 | 1.54 | 1.92 | 1.77 | 2.10 | 3.48 | 2.04 | 2.54 |
| Cl* IC 4651 MMU 9025 | H | 0.120 | 4761 | 24 | 8 | 1.73 | 1.85 | 1.68 | 2.00 | 1.84 | 2.37 | 3.27 | 1.63 | 2.43 |
| Cl* IC 4651 MMU 14527 | H | 0.120 | 4823 | 164 | 11 | 1.72 | 1.81 | 1.68 | 2.00 | 1.83 | 2.47 | 3.30 | 1.63 | 2.47 |
| Cl* NGC 2682 SAB 5 | H | 0.060 | 4721 | 62 | 10 | 1.59 | 1.62 | 1.23 | 1.73 | 1.53 | 2.89 | 5.96 | 2.81 | 3.89 |
| Cl* NGC 4349 MMU 127 | H | 0.380 | 4479 | 27 | 6 | 2.93 | 3.91 | 2.24 | 3.77 | 3.31 | 0.22 | 1.15 | | 0.69 |
| Cl* NGC 6705 MMU 411 | H | 0.430 | 4445 | 96 | 3 | 2.55 | 2.93 | 2.09 | 2.81 | 2.61 | 0.58 | 2.07 | 0.56 | 1.07 |
| Cl* NGC 6705 MMU 660 | H | 0.430 | 4788 | 130 | 5 | 2.49 | 3.49 | 2.50 | 3.82 | 3.27 | 0.28 | 0.69 | 0.20 | 0.39 |
| Cl* NGC 6705 MMU 779 | H | 0.430 | 4307 | 118 | 5 | 2.65 | 2.77 | 1.84 | 2.50 | 2.37 | 0.80 | 2.39 | 0.75 | 1.31 |
| Cl* NGC 6705 MMU 1286 | H | 0.430 | 4929 | 63 | 5 | 2.35 | 3.39 | 2.53 | 3.83 | 3.25 | 0.28 | 0.58 | 0.20 | 0.35 |
| Cl* NGC 6705 MMU 1423 | H | 0.430 | 4521 | 26 | 3 | 2.61 | 3.05 | 1.98 | 3.08 | 2.70 | 0.45 | 2.22 | 0.40 | 1.02 |
| CPD-23 2745 | U | 0.050 | 5058 | 58 | 10 | 2.09 | 2.92 | 2.27 | 3.32 | 2.84 | 0.43 | 0.91 | 0.32 | 0.55 |
| CPD-58 3077 | H | 0.040 | 4963 | 48 | 10 | 2.19 | 2.96 | 2.28 | 3.26 | 2.83 | 0.44 | 1.25 | 0.34 | 0.68 |
| CPD-58 3092 | H | 0.040 | 4740 | 29 | 7 | 2.51 | 3.30 | 2.28 | 3.37 | 2.98 | 0.37 | 1.45 | 0.32 | 0.71 |
| HBC 463 | U | 0.000 | 5211 | 0 | 1 | | | | | | | | | |
| HD 483 | H | 0.000 | 5757 | 87 | 5 | 0.54 | 1.09 | 1.05 | 1.18 | 1.11 | 7.13 | 7.75 | 5.67 | 6.85 |
| HD 483 | H | 0.000 | 5718 | 78 | 9 | 0.54 | 1.09 | 1.05 | 1.21 | 1.12 | 7.13 | 7.75 | 6.00 | 6.96 |
| HD 770 | H | 0.000 | 4710 | 17 | 11 | 1.61 | 1.79 | 1.62 | 1.64 | 1.68 | 2.07 | | 2.76 | 2.41 |
| HD 1142 | S | 0.041 | 5187 | 159 | 6 | 1.71 | 2.50 | 1.87 | 2.52 | 2.30 | 0.54 | 1.38 | 0.60 | 0.84 |
| HD 1638 | U | 0.000 | 4225 | 34 | 6 | 3.37 | | | | 1.30 | | | | |
| HD 1690 | H | 0.010 | 4219 | 51 | 5 | 1.55 | 1.23 | | 1.26 | 1.25 | 4.91 | | 6.50 | 5.71 |
| HD 2954 | S | 0.008 | 6308 | 43 | 5 | 1.25 | 1.70 | 1.49 | 1.77 | 1.65 | 1.58 | 2.00 | 1.72 | 1.77 |
| HD 4388 | U | 0.044 | 4640 | 13 | 9 | 1.71 | 1.73 | 1.58 | 1.79 | 1.70 | 2.43 | 2.50 | 2.43 | 2.45 |
| HD 5418 | S | 0.008 | 4977 | 163 | 9 | 1.65 | 2.30 | 1.72 | 2.20 | 2.07 | 0.71 | 3.15 | 0.93 | 1.60 |
| HD 6037 | S | 0.001 | 4566 | 84 | 9 | 1.45 | 1.37 | 1.26 | 1.51 | 1.38 | 4.16 | 6.30 | 3.72 | 4.73 |
| HD 9611 | H | 0.000 | 4196 | 58 | 5 | | | | | | | | | |
| HD 12116 | S | 0.000 | 4462 | 144 | 9 | 1.63 | 1.49 | 1.07 | 1.37 | 1.31 | 3.30 | | 5.13 | 4.21 |
| HD 12345 | H | 0.000 | 5327 | 38 | 8 | -0.27 | 0.92 | 0.88 | 0.90 | 0.90 | 2.65 | 5.29 | 4.28 | 4.07 |
| HD 13004 | S | 0.000 | 4584 | 27 | 8 | 1.45 | 1.73 | 1.32 | 1.58 | 1.54 | | 5.50 | 3.08 | 4.29 |
| HD 13263 | H | 0.005 | 5038 | 43 | 8 | 1.62 | 2.26 | 1.54 | 2.26 | 2.02 | 0.73 | 4.63 | 0.83 | 2.06 |
| HD 14703 | H | 0.005 | 4930 | 13 | 9 | 1.65 | 2.26 | 1.70 | 2.17 | 2.04 | 0.80 | 2.13 | 1.00 | 1.31 |
| HD 14830 | S | 0.000 | 4865 | 54 | 11 | 1.75 | 1.96 | 1.67 | 1.88 | 1.84 | 1.43 | 2.28 | 1.30 | 1.67 |
| HD 15533 | S | 0.017 | 4370 | 38 | 5 | 1.60 | 1.51 | | 1.26 | 1.39 | 11.62 | | | |
| HD 15866 | S | 0.006 | 5718 | 52 | 6 | 0.66 | 1.20 | 1.18 | 1.24 | 1.21 | 4.86 | 5.52 | 5.00 | 5.13 |
| HD 16150 | S | 0.045 | 6145 | 65 | 6 | 1.66 | 2.17 | 2.03 | 2.21 | 2.14 | 0.79 | 0.92 | 0.90 | 0.87 |
| HD 16314 | S | 0.003 | 6533 | 34 | 3 | 1.15 | 1.65 | 1.50 | 1.70 | 1.62 | 1.79 | 2.25 | 1.80 | 1.95 |
| HD 17001 | S | 0.007 | 4863 | 58 | 8 | 1.62 | 2.04 | 1.22 | 2.01 | 1.76 | 1.11 | 5.06 | 1.32 | 2.50 |
| HD 17374 | H | 0.007 | 4824 | 42 | 8 | 1.81 | 1.51 | 1.66 | 1.88 | 1.68 | 3.34 | 2.43 | 1.30 | 2.36 |
| HD 19210 | S | 0.081 | 4952 | 48 | 8 | 1.89 | 2.55 | 1.78 | 2.82 | 2.38 | 0.59 | 2.93 | 0.47 | 1.33 |
| HD 20037 | H | 0.005 | 5077 | 39 | 9 | 1.81 | 2.44 | 1.66 | 2.74 | 2.28 | 0.63 | 2.17 | 0.50 | 1.10 |
| HD 21585 | S | 0.015 | 5204 | 104 | 9 | 1.78 | 2.59 | 2.05 | | 2.32 | 0.50 | 0.87 | | 0.68 |
| HD 21760 | S | 0.002 | 4627 | 35 | 5 | 1.38 | 1.53 | 1.10 | 1.39 | 1.34 | 2.78 | 8.50 | 4.63 | 5.30 |
| HD 25627 | S | 0.016 | 4613 | 49 | 6 | 1.46 | 1.60 | 1.23 | 1.47 | 1.43 | 2.62 | 6.11 | 3.75 | 4.16 |
| HD 26004 | S | 0.007 | 4490 | 47 | 10 | 1.74 | 1.44 | 1.28 | 1.51 | 1.41 | 3.96 | 5.59 | 4.13 | 4.56 |
| HD 26625 | S | 0.005 | 4898 | 11 | 5 | 1.56 | 2.22 | 1.54 | 1.82 | 1.86 | | 3.50 | 1.60 | 2.55 |
| HD 29751 | H | 0.008 | 4855 | 57 | 10 | 1.85 | 1.89 | 2.09 | 2.44 | 2.14 | 1.46 | 1.26 | 0.73 | 1.15 |
| HD 33844 | S | 0.004 | 4792 | 36 | 6 | 1.15 | 1.45 | 1.35 | 1.62 | 1.47 | 4.12 | 4.52 | 2.58 | 3.74 |
| HD 34253 | H | 0.000 | 5490 | 119 | 5 | -0.23 | 0.97 | 0.94 | 0.94 | 0.95 | 1.40 | 1.75 | 2.59 | 1.91 |
| HD 35929 | H | 0.030 | 6345 | 230 | 7 | 1.83 | 2.40 | 2.17 | 2.31 | 2.29 | 0.62 | 0.77 | 0.81 | 0.73 |
| HD 35984 | S | 0.006 | 6273 | 107 | 7 | 1.31 | 1.74 | 1.72 | 1.81 | 1.76 | 1.42 | 1.67 | 1.60 | 1.56 |
| HD 39833 | E | 0.000 | 5751 | 32 | 6 | 0.10 | 1.08 | 1.01 | 1.08 | 1.06 | 2.84 | 7.20 | 4.67 | 4.90 |
| HD 42719 | H | 0.000 | 5699 | 90 | 6 | 0.63 | 1.19 | 1.05 | 1.19 | 1.14 | 5.10 | 7.08 | 5.50 | 5.89 |

| Star | Src | E(B-V) | Teff | v | n | A | B | C | D | Avg | P | Q | R | S |
|---|---|---|---|---|---|---|---|---|---|---|---|---|---|---|
| HD 46415 | H | 0.007 | 4818 | 33 | 16 | 1.64 | 2.04 | 1.22 | 1.94 | 1.73 | 1.17 | 6.05 | 1.40 | 2.87 |
| HD 47910 | H | 0.019 | 4869 | 72 | 5 | 1.73 | 1.97 | 1.66 | 2.03 | 1.89 | 1.34 | 2.46 | 1.13 | 1.64 |
| HD 49050 | H | 0.030 | 4199 | 45 | 10 | 2.81 | 2.58 | 1.69 | 2.26 | 2.18 | 1.27 | 2.98 | 1.32 | 1.86 |
| HD 49068 | H | 0.034 | 4423 | 51 | 11 | 2.83 | 2.61 | 2.04 | 2.35 | 2.33 | 1.37 | 2.04 | 0.96 | 1.46 |
| HD 49091 | H | 0.034 | 4020 | 60 | 10 | 3.21 | 2.38 | 1.73 | 2.83 | 2.31 | 1.82 | 3.15 | 0.78 | 1.92 |
| HD 49105 | H | 0.033 | 4596 | 39 | 11 | 2.60 | 2.92 | 2.04 | 2.59 | 2.52 | 0.78 | 2.04 | 0.78 | 1.20 |
| HD 49212 | H | 0.030 | 4602 | 38 | 11 | 2.86 | 2.73 | 2.06 | 2.59 | 2.46 | 1.30 | 2.01 | 0.78 | 1.36 |
| HD 49334 | H | 0.030 | 4382 | 64 | 7 | 2.77 | 2.91 | 1.99 | 2.53 | 2.48 | 0.95 | 2.21 | 0.85 | 1.34 |
| HD 50890 | H | 0.039 | 4733 | 64 | 10 | 2.75 | 4.02 | 2.96 |  | 3.49 | 0.19 | 0.35 |  | 0.27 |
| HD 54038 | H | 0.006 | 5028 | 63 | 6 | 1.46 | 2.12 | 1.53 | 2.14 | 1.93 | 0.89 | 2.61 | 1.03 | 1.51 |
| HD 58898 | S | 0.009 | 4387 | 158 | 5 | 1.75 | 1.39 | 1.21 | 1.30 | 1.30 | 4.11 | 6.81 | 5.92 | 5.61 |
| HD 59894 | H | 0.010 | 4904 | 55 | 6 | 1.76 | 2.06 | 2.07 | 2.32 | 2.15 | 1.16 | 1.27 | 0.80 | 1.07 |
| HD 61191 | S | 0.005 | 4689 | 25 | 5 | 1.42 | 1.53 | 1.38 | 1.50 | 1.47 | 3.82 | 4.02 | 4.20 | 4.01 |
| HD 62713 | H | 0.000 | 4624 | 14 | 11 | 1.69 |  | 1.53 | 1.62 | 1.58 |  | 2.75 | 3.60 | 3.18 |
| HD 62849 | H | 0.000 | 4871 | 255 | 6 |  |  |  |  |  |  |  |  |  |
| HD 65354 | U | 0.000 | 3832 | 93 | 4 | 4.07 | 2.05 | 1.91 | 3.79 | 2.58 | 1.84 | 2.67 | 0.20 | 1.57 |
| HD 68667 | S | 0.007 | 5010 | 61 | 9 | 1.78 | 2.36 | 1.81 | 2.61 | 2.26 | 0.72 | 3.04 | 0.60 | 1.45 |
| HD 69836 | U | 0.000 | 4604 | 35 | 5 |  |  |  |  |  |  |  |  |  |
| HD 70522 | S | 0.004 | 6122 | 42 | 8 | 0.91 | 1.44 | 1.21 | 1.53 | 1.39 | 2.61 | 4.00 | 2.72 | 3.11 |
| HD 71160 | U | 0.057 | 4096 | 16 | 5 | 2.55 | 1.99 | 1.60 | 1.89 | 1.83 | 2.35 | 3.19 | 2.29 | 2.61 |
| HD 72320 | U | 0.057 | 5028 | 45 | 5 | 1.82 | 2.54 | 2.47 | 2.47 | 2.49 | 0.59 | 0.71 | 0.70 | 0.67 |
| HD 73598 | H | 0.033 | 5013 | 17 | 10 | 2.29 | 3.29 | 2.90 | 3.59 | 3.26 | 0.31 | 0.40 | 0.25 | 0.32 |
| HD 73829 | U | 0.000 | 4489 | 85 | 5 |  |  |  |  |  |  |  |  |  |
| HD 74088 | U | 0.036 | 3840 | 12 | 3 | 2.86 | 0.99 | 1.69 | 2.04 | 1.57 |  | 2.74 | 0.76 | 1.75 |
| HD 74165 | U | 0.000 | 4539 | 41 | 5 |  |  |  |  |  |  |  |  |  |
| HD 74166 | U | 0.000 | 4230 | 60 | 5 |  |  |  |  |  |  |  |  |  |
| HD 74212 | U | 0.010 | 4610 | 75 | 5 | 1.13 | 1.36 |  | 1.27 | 1.32 | 4.38 |  | 6.00 | 5.19 |
| HD 74387 | U | 0.010 | 4585 | 75 | 3 | 0.90 | 1.22 |  |  | 1.22 | 5.10 |  |  | 5.10 |
| HD 74529 | U | 0.031 | 4598 | 46 | 5 | 1.90 | 1.77 | 1.67 | 1.84 | 1.76 | 2.40 | 2.98 | 2.28 | 2.56 |
| HD 74900 | U | 0.021 | 4572 | 16 | 5 | 1.69 | 1.54 | 1.34 | 1.58 | 1.49 | 3.31 | 4.59 | 3.58 | 3.82 |
| HD 75058 | U | 0.023 | 4650 | 76 | 5 | 1.41 | 1.56 | 1.26 | 1.42 | 1.41 | 3.23 | 5.75 | 4.75 | 4.58 |
| HD 75066 | U | 0.010 | 6699 | 56 | 3 | 0.66 | 1.41 | 1.23 | 1.44 | 1.36 | 1.03 | 2.75 | 0.88 | 1.55 |
| HD 76128 | U | 0.000 | 4487 | 27 | 5 |  |  |  |  |  |  |  |  |  |
| HD 77232 | S | 0.006 | 6773 | 14 | 3 | 1.11 | 1.61 | 1.40 | 1.59 | 1.53 | 1.64 | 2.44 | 2.05 | 2.04 |
| HD 78002 | U | 0.000 | 4771 | 40 | 5 |  |  |  |  |  |  |  |  |  |
| HD 78528 | U | 0.000 | 3829 | 62 | 3 |  |  |  |  |  |  |  |  |  |
| HD 78959 | U | 0.000 |  |  | 0 |  |  |  |  |  |  |  |  |  |
| HD 78964 | H | 0.002 | 5118 | 90 | 6 | -0.07 |  |  |  | 2.00 |  |  |  |  |
| HD 80571 | U | 0.029 | 4659 | 39 | 5 | 1.99 | 1.80 | 1.86 | 2.10 | 1.92 | 2.11 | 2.71 | 1.21 | 2.01 |
| HD 81278 | U | 0.000 | 4654 | 60 | 5 |  |  |  |  |  |  |  |  |  |
| HD 82403 | U | 0.000 | 4387 | 54 | 5 |  |  |  |  |  |  |  |  |  |
| HD 83087 | S | 0.003 | 4777 | 48 | 5 | 1.19 | 1.64 | 1.24 | 1.38 | 1.42 | 2.21 | 5.45 |  | 3.83 |
| HD 83155 | U | 0.000 | 4760 | 77 | 5 |  |  |  |  |  |  |  |  |  |
| HD 83234 | U | 0.000 | 4185 | 77 | 5 |  |  |  |  |  |  |  |  |  |
| HD 84598 | U | 0.025 | 4896 | 10 | 3 | 1.82 | 2.13 | 1.72 | 2.32 | 2.06 | 1.08 | 3.37 | 0.82 | 1.76 |
| HD 85440 | S | 0.005 | 5041 | 10 | 5 | 0.99 | 1.57 | 1.41 | 1.63 | 1.54 | 2.17 | 4.25 | 2.41 | 2.94 |
| HD 85552 | U | 0.020 | 5426 | 52 | 6 | 1.46 | 2.09 | 2.04 | 2.22 | 2.12 | 0.90 | 0.98 | 0.90 | 0.92 |
| HD 85859 | H | 0.007 | 4415 | 25 | 13 | 2.25 | 1.62 | 1.02 | 2.01 | 1.55 | 3.98 | 7.75 | 1.30 | 4.34 |
| HD 86757 | U | 0.000 | 3867 | 64 | 3 |  |  |  |  |  |  |  |  |  |
| HD 87566 | U | 0.080 | 4456 | 45 | 10 | 2.83 | 3.61 | 2.01 | 3.09 | 2.90 | 0.31 | 2.39 | 0.40 | 1.03 |
| HD 89280 | S | 0.004 | 7384 | 54 | 4 | 0.87 | 1.60 | 1.48 | 1.58 | 1.55 | 0.71 | 1.56 | 0.93 | 1.07 |
| HD 91267 | H | 0.000 | 4873 | 101 | 9 | -0.48 | 0.85 |  | 0.85 | 0.85 | 2.01 |  | 3.28 | 2.64 |
| HD 95799 | H | 0.040 | 4849 | 81 | 8 | 2.28 | 3.05 | 2.15 | 3.15 | 2.78 | 0.45 | 2.15 | 0.40 | 1.00 |
| HD 95879 | H | 0.040 | 4955 | 36 | 8 | 2.29 | 3.16 | 2.80 | 3.47 | 3.14 | 0.37 | 0.46 | 0.28 | 0.37 |
| HD 96445 | H | 0.040 | 4845 | 270 | 13 | 2.40 | 3.27 | 2.87 | 3.37 | 3.17 | 0.37 | 0.46 | 0.32 | 0.38 |
| HD 98579 | S | 0.003 | 4665 | 66 | 8 | 1.32 | 1.64 | 1.22 | 1.37 | 1.41 | 2.28 | 6.55 | 4.70 | 4.51 |
| HD 101321 | S | 0.000 | 4757 | 40 | 8 | 1.32 | 1.24 | 1.36 | 1.20 | 1.27 | 5.59 | 4.00 | 6.67 | 5.42 |
| HD 103295 | U | 0.000 | 4912 | 51 | 12 |  |  |  |  |  |  |  |  |  |
| HD 104760 | H | 0.000 | 5824 | 43 | 6 | 0.17 | 1.10 | 1.04 | 1.09 | 1.08 | 3.90 | 6.50 | 4.63 | 5.01 |
| HD 104819 | S | 0.006 | 4618 | 28 | 8 | 1.07 | 1.45 | 1.27 | 1.29 | 1.34 | 2.84 | 6.08 | 5.05 | 4.66 |
| HD 104883 | S | 0.003 | 6203 | 66 | 4 | 1.16 | 1.64 | 1.42 | 1.74 | 1.60 | 1.79 | 2.39 | 1.87 | 2.02 |
| HD 104985 | S | 0.003 | 4677 | 56 | 8 | 1.74 | 1.61 | 1.75 | 1.08 | 1.48 | 2.10 |  |  | 2.10 |

| Star | Type | col3 | col4 | col5 | col6 | col7 | col8 | col9 | col10 | col11 | col12 | col13 | col14 | col15 |
|---|---|---|---|---|---|---|---|---|---|---|---|---|---|---|
| HD 105740 | U | 0.004 | 4668 | 89 | 10 | 1.35 | 1.32 | 1.22 | 1.04 | 1.19 | 4.52 | 6.25 | | 5.39 |
| HD 106572 | U | 0.009 | 4793 | 57 | 11 | 1.75 | 1.51 | 1.32 | 2.06 | 1.63 | 4.12 | 4.44 | 1.24 | 3.27 |
| HD 106972 | S | 0.002 | 6166 | 66 | 4 | 1.00 | 1.57 | 1.27 | 1.58 | 1.47 | 2.03 | 3.38 | 2.48 | 2.63 |
| HD 107415 | S | 0.002 | 4797 | 108 | 7 | 1.64 | 2.01 | 1.26 | 1.77 | 1.68 | 1.34 | 6.11 | 1.85 | 3.10 |
| HD 107569 | S | 0.002 | 6101 | 49 | 5 | 1.20 | 1.66 | 1.44 | 1.72 | 1.61 | 1.72 | 2.17 | 1.92 | 1.94 |
| HD 107610 | S | 0.002 | 4693 | 59 | 5 | 1.52 | 1.65 | 1.53 | 1.47 | 1.55 | 3.03 | 2.90 | 3.98 | 3.30 |
| HD 108570 | H | 0.000 | 4984 | 51 | 11 | 0.81 | | 1.34 | | 1.34 | | | | |
| HD 110291 | H | 0.000 | 5392 | 52 | 8 | -0.21 | 0.96 | 0.90 | 0.97 | 0.94 | 2.42 | 5.59 | 2.90 | 3.64 |
| HD 111464 | U | 0.075 | 4170 | 55 | 5 | 2.46 | 1.99 | 1.81 | 1.82 | 1.87 | 2.07 | 2.40 | 2.27 | 2.25 |
| HD 111721 | U | 0.001 | 4893 | 39 | 16 | 1.58 | 1.22 | | 1.08 | 1.15 | 4.20 | | 6.50 | 5.35 |
| HD 112127 | E | 0.002 | 4383 | 46 | 13 | 1.46 | 1.31 | | 1.23 | 1.27 | | | 6.33 | 6.33 |
| HD 112357 | S | 0.001 | 4969 | 29 | 9 | 0.97 | 1.68 | 1.44 | 1.41 | 1.51 | 1.79 | 3.25 | 3.83 | 2.96 |
| HD 113002 | U | 0.005 | 5274 | 163 | 7 | 1.29 | 1.97 | 1.46 | 2.03 | 1.82 | 1.07 | 3.27 | 1.21 | 1.85 |
| HD 114747 | H | 0.000 | 5073 | 51 | 9 | -0.35 | | 0.88 | 0.89 | 0.89 | | 7.75 | 6.00 | 6.88 |
| HD 114889 | S | 0.000 | 4398 | 53 | 10 | 1.78 | 1.37 | 1.09 | 1.35 | 1.27 | 4.16 | 8.75 | 5.17 | 6.03 |
| HD 116515 | E | 0.010 | 4728 | 38 | 7 | 2.10 | 2.25 | 1.87 | 2.42 | 2.18 | 1.02 | 2.67 | 0.74 | 1.48 |
| HD 121146 | S | 0.005 | 4464 | 56 | 10 | 1.62 | 1.49 | | 1.29 | 1.39 | 3.30 | | 6.00 | 4.65 |
| HD 121416 | H | 0.006 | 4576 | 16 | 11 | 1.76 | 1.43 | 1.51 | 1.65 | 1.53 | 3.60 | 2.88 | 3.07 | 3.18 |
| HD 122721 | U | 0.028 | 4684 | 7 | 5 | 1.15 | 1.46 | 1.22 | 1.35 | 1.34 | 3.72 | 6.25 | 5.29 | 5.09 |
| HD 123517 | H | 0.000 | 5554 | 97 | 6 | | | | | | | | | |
| HD 126265 | S | 0.000 | 5868 | 42 | 6 | 0.72 | 1.34 | 1.12 | 1.36 | 1.27 | 3.50 | 5.72 | 4.00 | 4.41 |
| HD 127740 | S | 0.001 | 6241 | 65 | 4 | 0.99 | 1.52 | 1.30 | 1.49 | 1.44 | 2.29 | 3.28 | 2.63 | 2.73 |
| HD 128279 | U | 0.021 | 5162 | 74 | 12 | 1.25 | 1.06 | | | 1.06 | 5.81 | | | 5.81 |
| HD 128853 | S | 0.008 | 4819 | 58 | 6 | 1.23 | 1.55 | 1.40 | 1.52 | 1.49 | 3.27 | 3.63 | 3.90 | 3.60 |
| HD 138085 | S | 0.000 | 4799 | 77 | 7 | 1.73 | 1.45 | 1.41 | 2.06 | 1.64 | 4.10 | 4.24 | 1.24 | 3.19 |
| HD 138686 | S | 0.050 | 6257 | 64 | 6 | 1.65 | 2.14 | 2.02 | 2.21 | 2.12 | 0.84 | 0.91 | 0.90 | 0.88 |
| HD 142527 | H | 0.500 | 6632 | 1011 | 8 | 1.89 | 2.46 | 2.21 | 2.38 | 2.35 | 0.57 | 0.71 | 0.73 | 0.67 |
| HD 144589 | H | 0.000 | 5904 | 95 | 5 | | | | | | | | | |
| HD 148317 | S | 0.000 | 5759 | 47 | 6 | 1.06 | 1.58 | 1.52 | | 1.55 | 2.00 | 2.31 | | 2.15 |
| HD 148451 | U | 0.024 | 5015 | 52 | 8 | 1.99 | 2.77 | 2.19 | 2.87 | 2.61 | 0.47 | 1.16 | | 0.81 |
| HD 149216 | S | 0.034 | 4295 | 123 | 5 | 1.32 | 1.12 | | | 1.12 | 7.13 | | | 7.13 |
| HD 157935 | S | 0.029 | 6769 | 57 | 3 | 1.53 | 1.99 | 1.79 | 2.15 | 1.98 | 1.00 | 1.25 | 1.01 | 1.09 |
| HD 160314 | S | 0.021 | 6616 | 65 | 5 | 0.94 | 1.49 | 1.21 | 1.44 | 1.38 | 2.03 | 3.67 | 2.60 | 2.77 |
| HD 161502 | S | 0.023 | 4963 | 102 | 6 | 1.42 | 1.98 | 1.32 | 1.99 | 1.76 | 1.21 | 4.68 | 1.34 | 2.41 |
| HD 166229 | S | 0.000 | 4516 | 80 | 7 | 1.53 | 1.49 | 1.34 | 1.36 | 1.40 | 3.30 | 4.75 | 5.00 | 4.35 |
| HD 166411 | S | 0.008 | 4487 | 35 | 9 | 1.77 | 1.56 | 1.31 | 1.64 | 1.50 | 2.70 | 4.67 | 2.80 | 3.39 |
| HD 167576 | S | 0.031 | 4506 | 47 | 11 | 1.67 | 1.46 | 1.76 | 1.92 | 1.71 | 3.74 | 2.04 | 1.39 | 2.39 |
| HD 167768 | S | 0.015 | 4899 | 24 | 12 | 1.68 | 1.60 | 1.69 | 1.40 | 1.56 | 1.89 | 2.27 | 3.00 | 2.38 |
| HD 168322 | S | 0.010 | 4798 | 44 | 13 | 1.79 | 1.36 | 1.67 | 1.46 | 1.50 | 3.97 | 2.72 | 3.30 | 3.33 |
| HD 169689 | H | 0.115 | 4989 | 249 | 11 | 2.66 | 3.96 | 2.93 | | 3.45 | 0.18 | 0.33 | | 0.26 |
| HD 172052 | S | 0.000 | 5544 | 109 | 9 | | | | | 2.70 | | | | |
| HD 173378 | S | 0.009 | 4855 | 80 | 6 | 1.41 | 1.42 | 1.64 | 1.37 | 1.48 | 4.12 | 2.13 | 4.49 | 3.58 |
| HD 175305 | E | 0.010 | 4981 | 31 | 10 | 1.55 | 1.18 | | 1.22 | 1.20 | 5.23 | | | 5.23 |
| HD 175545 | U | 0.012 | 4471 | 26 | 10 | 1.28 | 1.36 | | 1.14 | 1.25 | 3.57 | | 8.50 | 6.04 |
| HD 175679 | S | 0.041 | 4951 | 61 | 10 | 2.00 | 2.73 | 1.94 | | 2.34 | 0.52 | 1.90 | | 1.21 |
| HD 175743 | S | 0.012 | 4710 | 64 | 7 | 1.74 | 1.54 | 1.40 | 1.68 | 1.54 | 3.17 | 5.50 | 2.20 | 3.62 |
| HD 175940 | S | 0.010 | 4629 | 57 | 7 | 1.46 | 1.64 | 1.30 | 1.48 | 1.47 | 2.21 | 5.10 | 3.67 | 3.66 |
| HD 176354 | H | 0.000 | 5165 | 107 | 8 | 0.49 | 1.18 | 1.17 | 1.19 | 1.18 | 5.66 | 6.60 | 6.50 | 6.25 |
| HD 176704 | U | 0.002 | 4482 | 39 | 11 | 1.67 | 1.60 | 1.55 | 1.84 | 1.66 | 2.62 | 3.00 | | 2.81 |
| HD 177442 | S | 0.054 | 4156 | 33 | 5 | 2.05 | 1.40 | 1.08 | 1.33 | 1.27 | 3.78 | 9.25 | 5.56 | 6.20 |
| HD 181214 | E | 0.030 | 6238 | 59 | 4 | 1.14 | 1.64 | 1.40 | 1.74 | 1.59 | 1.79 | 2.53 | 1.87 | 2.06 |
| HD 181433 | H | 0.000 | 4800 | 113 | 14 | | | | | | | | | |
| HD 181517 | H | 0.012 | 4895 | 52 | 5 | 1.72 | 2.10 | 1.74 | 2.17 | 2.00 | 1.07 | 2.08 | 1.00 | 1.38 |
| HD 181907 | S | 0.008 | 4662 | 49 | 11 | 1.80 | 1.45 | 1.80 | 1.68 | 1.64 | 3.46 | 1.63 | 2.20 | 2.43 |
| HD 182901 | S | 0.000 | 6491 | 127 | 3 | 0.74 | 1.34 | 1.30 | | 1.32 | 2.62 | 3.50 | | 3.06 |
| HD 186535 | S | 0.031 | 5012 | 33 | 10 | 1.82 | 2.45 | 1.85 | 2.82 | 2.37 | 0.64 | 2.47 | 0.47 | 1.19 |
| HD 187669 | H | 0.000 | 4240 | 131 | 5 | | | | | | | | | |
| HD 188993 | S | 0.007 | 5673 | 35 | 6 | 1.17 | 1.71 | 1.50 | 1.76 | 1.66 | 1.58 | 1.92 | 1.77 | 1.76 |
| HD 189186 | S | 0.008 | 4899 | 45 | 5 | 1.22 | 1.51 | 1.54 | 1.31 | 1.45 | 2.78 | 2.63 | 5.33 | 3.58 |
| HD 194708 | S | 0.004 | 6313 | 75 | 6 | 1.12 | 1.71 | 1.37 | 1.72 | 1.60 | 1.60 | 2.63 | 2.00 | 2.08 |
| HD 196983 | U | 0.000 | 4518 | 24 | 5 | | | | | | | | | |
| HD 199642 | U | 0.003 | 3828 | 21 | 5 | 2.60 | 1.46 | 1.59 | 1.79 | 1.61 | 3.54 | 4.07 | 2.95 | 3.52 |

| Name | | | | | | | | | | | | | |
|---|---|---|---|---|---|---|---|---|---|---|---|---|---|
| HD 204642 | S | 0.004 | 4660 | 88 | 10 | 1.37 | 1.54 | 1.23 | 1.47 | 1.41 | 2.80 | 6.17 | 3.75 | 4.24 |
| HD 205011 | S | 0.014 | 4770 | 118 | 8 | 1.79 | 1.65 | 1.44 | 2.06 | 1.72 | 2.89 | 3.94 | 1.24 | 2.69 |
| HD 205972 | S | 0.004 | 4742 | 40 | 8 | 1.35 | 1.43 | 1.40 | 1.44 | 1.42 | 4.44 | 3.82 | 4.55 | 4.27 |
| HD 206642 | H | 0.011 | 4326 | 86 | 10 | 2.53 | 1.35 | 1.70 | 1.56 | 1.54 | 3.60 | 1.92 | 2.63 | 2.71 |
| HD 207134 | S | 0.007 | 4484 | 77 | 10 | 1.71 | 1.49 | 1.22 | 1.56 | 1.42 | 3.50 | 6.08 | 3.20 | 4.26 |
| HD 209449 | H | 0.000 | 5685 | 43 | 9 | 0.45 | 1.08 | 1.21 | 1.21 | 1.17 | 7.94 | 5.75 | 5.00 | 6.23 |
| HD 210905 | S | 0.022 | 4734 | 118 | 13 | 1.72 | 1.48 | 1.05 | 1.77 | 1.43 | 3.82 | 8.23 | 1.85 | 4.63 |
| HD 211173 | U | 0.005 | 4883 | 42 | 8 | 1.23 | 1.64 | 1.51 | 1.66 | 1.60 | 2.37 | 2.88 | 2.79 | 2.68 |
| HD 211317 | H | 0.000 | 5780 | 46 | 9 | 0.41 | 1.14 | 1.09 | | 1.12 | 5.96 | 7.75 | | 6.86 |
| HD 211607 | S | 0.009 | 4944 | 157 | 7 | 1.44 | 1.98 | 1.44 | 2.07 | 1.83 | 1.21 | 4.14 | 1.23 | 2.19 |
| HD 212334 | S | 0.011 | 4722 | 16 | 5 | 1.79 | 1.64 | 1.06 | 1.99 | 1.56 | 2.96 | 7.63 | 1.46 | 4.02 |
| HD 213619 | S | 0.001 | 6884 | 95 | 3 | 1.14 | 1.60 | 1.33 | 1.67 | 1.53 | 1.75 | 2.50 | | 2.12 |
| HD 214448 | E | 0.004 | 5272 | 50 | 9 | 1.65 | 2.38 | 1.99 | 2.55 | 2.31 | 0.63 | 1.02 | 0.60 | 0.75 |
| HD 215030 | S | 0.003 | 4754 | 96 | 10 | 1.63 | 1.90 | 1.75 | 1.21 | 1.62 | | | 5.33 | 5.33 |
| HD 217590 | S | 0.003 | 4950 | 146 | 9 | 1.65 | 2.30 | 1.83 | 2.29 | 2.14 | 0.71 | 1.68 | 0.87 | 1.09 |
| HD 219409 | S | 0.003 | 4634 | 77 | 11 | 1.49 | 1.72 | 1.26 | 1.50 | 1.49 | 2.08 | 5.28 | 4.20 | 3.85 |
| HD 219418 | E | 0.013 | 5077 | 25 | 5 | 1.71 | 2.25 | 1.55 | 2.22 | 2.01 | 0.73 | 3.37 | 0.83 | 1.64 |
| HD 219962 | S | 0.018 | 4639 | 70 | 11 | 1.71 | 1.65 | 1.59 | 1.73 | 1.66 | 2.45 | 2.38 | 2.45 | 2.43 |
| HD 223094 | U | 0.000 | 3764 | 30 | 5 | 3.87 | 2.03 | 1.90 | 3.78 | 2.57 | 1.87 | 2.69 | 0.20 | 1.59 |
| HD 223869 | S | 0.003 | 4830 | 36 | 10 | 0.98 | | 1.13 | 1.41 | 1.27 | | 7.50 | 4.00 | 5.75 |
| HD 224349 | S | 0.000 | 4830 | 41 | 8 | 1.80 | 1.47 | 1.69 | 2.17 | 1.78 | 3.53 | 2.36 | 1.00 | 2.30 |
| HD 225292 | S | 0.019 | 4940 | 27 | 6 | 1.91 | 2.60 | 1.81 | 2.73 | 2.38 | 0.60 | 2.98 | 0.50 | 1.36 |
| HD 232509 | U | 0.041 | 6036 | 87 | 3 | 0.78 | 1.35 | 1.14 | 1.44 | 1.31 | 3.34 | 5.19 | 3.27 | 3.94 |
| HR 2 | S | 0.004 | 4648 | 38 | 9 | 1.86 | 1.77 | 1.34 | 1.95 | 1.69 | 2.24 | 5.81 | 1.62 | 3.22 |
| HR 13 | H | 0.000 | 4584 | 38 | 5 | 1.79 | 1.64 | 1.67 | 1.73 | 1.68 | 2.51 | 2.16 | 2.45 | 2.37 |
| HR 16 | S | 0.002 | 4738 | 15 | 10 | 1.66 | 1.78 | 1.35 | 1.91 | 1.68 | 2.02 | 5.63 | 1.73 | 3.13 |
| HR 19 | S | 0.011 | 4898 | 84 | 11 | 1.83 | 2.05 | 1.76 | 2.57 | 2.13 | 1.03 | 3.70 | 0.65 | 1.79 |
| HR 40 | S | 0.010 | 5642 | 44 | 6 | 1.67 | 2.31 | 2.12 | 2.30 | 2.24 | 0.67 | 0.77 | 0.80 | 0.75 |
| HR 69 | S | 0.011 | 4888 | 149 | 8 | 2.00 | 2.45 | 1.83 | 2.73 | 2.34 | 0.71 | 2.76 | 0.50 | 1.32 |
| HR 111 | H | 0.000 | 4621 | 40 | 10 | 1.78 | 1.60 | 1.67 | 1.86 | 1.71 | 2.66 | 2.07 | 2.16 | 2.30 |
| HR 135 | S | 0.016 | 4792 | 35 | 10 | 1.82 | 1.64 | 1.60 | 1.97 | 1.74 | 3.15 | 2.75 | 1.27 | 2.39 |
| HR 141 | S | 0.008 | 4646 | 109 | 8 | 1.62 | 1.65 | 1.49 | 1.58 | 1.57 | 3.03 | 3.00 | 3.62 | 3.22 |
| HR 156 | S | 0.008 | 4609 | 56 | 5 | 1.74 | 1.80 | 1.62 | 2.06 | 1.83 | 1.84 | 2.39 | 1.28 | 1.84 |
| HR 162 | H | 0.000 | 4827 | 43 | 10 | 1.82 | 1.51 | 1.53 | 2.11 | 1.72 | 2.85 | 2.70 | 1.07 | 2.21 |
| HR 228 | S | 0.000 | 4898 | 67 | 11 | 1.20 | 1.26 | 1.52 | 1.22 | 1.33 | | | | |
| HR 229 | H | 0.000 | 5031 | 20 | 10 | 1.72 | 2.38 | 1.56 | 2.40 | 2.11 | 0.64 | 3.63 | 0.70 | 1.66 |
| HR 249 | S | 0.009 | 4613 | 51 | 7 | 1.75 | 1.63 | 1.59 | 1.73 | 1.65 | 2.50 | 2.43 | 2.45 | 2.46 |
| HR 295 | H | 0.000 | 4992 | 50 | 10 | 1.73 | 2.21 | 1.73 | 2.18 | 2.04 | 0.85 | 3.69 | 0.85 | 1.80 |
| HR 299 | H | 0.000 | 5001 | 35 | 10 | 1.59 | 2.19 | 1.46 | 2.61 | 2.09 | 0.82 | | | 0.82 |
| HR 306 | E | 0.006 | 4755 | 116 | 5 | 1.63 | 1.90 | 1.87 | 1.91 | 1.89 | | 1.50 | 1.73 | 1.62 |
| HR 320 | S | 0.000 | 4682 | 30 | 10 | 1.77 | 1.39 | 1.64 | 1.94 | 1.66 | 4.05 | 3.50 | 1.60 | 3.05 |
| HR 320 | H | 0.000 | 4681 | 29 | 10 | 1.77 | 1.39 | 1.64 | 1.94 | 1.66 | 4.05 | 3.50 | 1.60 | 3.05 |
| HR 325 | S | 0.000 | 7461 | 46 | 5 | 1.01 | 1.77 | 1.19 | 1.76 | 1.57 | 0.36 | | 0.65 | 0.50 |
| HR 350 | H | 0.005 | 4969 | 86 | 10 | 1.66 | 2.19 | 1.90 | 1.99 | 2.03 | 0.82 | 1.45 | 1.10 | 1.12 |
| HR 356 | S | 0.014 | 4885 | 4 | 2 | 1.69 | 2.06 | 1.61 | 2.13 | 1.93 | 1.05 | 2.59 | 1.10 | 1.58 |
| HR 371 | S | 0.008 | 4573 | 41 | 10 | 1.76 | 1.42 | 1.66 | 1.98 | 1.69 | 3.62 | 2.36 | 1.50 | 2.49 |
| HR 406 | S | 0.000 | 4845 | 40 | 10 | 1.35 | 1.51 | 1.56 | 1.35 | 1.47 | 2.78 | | 5.50 | 4.14 |
| HR 407 | S | 0.000 | 6455 | 142 | 4 | 1.09 | 1.65 | 1.64 | 1.62 | 1.64 | 1.79 | 2.14 | 1.98 | 1.97 |
| HR 426 | S | 0.005 | 4689 | 29 | 14 | 1.72 | 2.06 | 2.25 | 2.19 | 2.17 | 1.12 | 0.96 | 1.06 | 1.04 |
| HR 453 | H | 0.001 | 4629 | 31 | 10 | 1.76 | 1.65 | 1.67 | 1.88 | 1.73 | 2.69 | 2.06 | 1.86 | 2.20 |
| HR 467 | H | 0.004 | 4799 | 16 | 10 | 1.81 | 1.41 | 1.74 | 1.88 | 1.68 | 3.72 | 2.36 | 1.57 | 2.55 |
| HR 505 | H | 0.005 | 4420 | 59 | 10 | 1.97 | 1.36 | 1.48 | 1.47 | 1.44 | 5.01 | 3.13 | 4.68 | 4.27 |
| HR 521 | S | 0.007 | 4855 | 78 | 13 | 1.71 | 1.95 | 1.65 | 1.84 | 1.81 | 1.16 | 2.68 | 1.58 | 1.80 |
| HR 525 | H | 0.006 | 4961 | 46 | 10 | 1.89 | 2.54 | 1.88 | 2.73 | 2.38 | 0.62 | 2.69 | 0.50 | 1.27 |
| HR 527 | S | 0.005 | 4837 | 32 | 12 | 1.68 | 2.15 | 1.13 | 1.82 | 1.70 | 0.90 | 5.31 | 1.74 | 2.65 |
| HR 574 | H | 0.000 | 5035 | 79 | 13 | 1.84 | 2.63 | 1.79 | 2.74 | 2.39 | 0.48 | 2.25 | 0.50 | 1.08 |
| HR 584 | H | 0.007 | 4992 | 123 | 10 | 1.94 | 2.63 | 1.95 | 2.99 | 2.52 | 0.54 | 1.79 | | 1.17 |
| HR 616 | S | 0.019 | 5061 | 39 | 11 | 2.04 | 2.83 | 2.18 | 2.87 | 2.63 | 0.46 | 0.82 | | 0.64 |
| HR 619 | S | 0.012 | 4916 | 20 | 8 | 1.69 | 2.43 | 1.69 | 1.93 | 2.02 | 0.63 | 2.03 | | 1.33 |
| HR 621 | S | 0.076 | 5110 | 66 | 10 | 2.06 | 2.84 | 2.44 | | 2.64 | 0.45 | 0.59 | | 0.52 |
| HR 636 | H | 0.001 | 5037 | 50 | 10 | 1.58 | 2.19 | 1.50 | 2.33 | 2.01 | 0.82 | 4.61 | 0.80 | 2.08 |
| HR 651 | H | 0.000 | 4726 | 107 | 11 | 1.78 | 1.64 | 1.08 | 2.10 | 1.61 | 3.07 | 7.38 | 1.37 | 3.94 |

| ID | Type | col3 | col4 | col5 | col6 | col7 | col8 | col9 | col10 | col11 | col12 | col13 | col14 | col15 |
|---|---|---|---|---|---|---|---|---|---|---|---|---|---|---|
| HR 659 | H | 0.001 | 4883 | 24 | 10 | 1.72 | 1.98 | 1.46 | 1.88 | 1.77 | 1.39 | 2.85 | 1.30 | 1.85 |
| HR 698 | H | 0.001 | 4804 | 50 | 10 | 1.81 | 1.57 | 1.60 | 1.97 | 1.71 | 3.13 | 2.75 | 1.27 | 2.38 |
| HR 725 | S | 0.006 | 4780 | 33 | 10 | 1.66 | 2.01 | 1.22 | 1.77 | 1.67 | 1.34 | 6.53 | 1.85 | 3.24 |
| HR 726 | S | 0.017 | 4723 | 70 | 7 | 1.75 | 1.60 | 1.09 | 1.90 | 1.53 | 2.82 | 7.59 | 1.83 | 4.08 |
| HR 738 | S | 0.003 | 4622 | 48 | 10 | 1.60 | 1.57 | 1.42 | 1.43 | 1.47 | 3.25 | 3.50 | 5.00 | 3.92 |
| HR 743 | S | 0.002 | 5084 | 133 | 10 | 1.73 | 2.65 | 1.62 | 2.07 | 2.11 |  | 3.10 |  | 3.10 |
| HR 766 | S | 0.002 | 4679 | 33 | 12 | 1.67 | 1.52 | 1.65 | 1.28 | 1.48 | 2.78 | 2.13 | 5.70 | 3.54 |
| HR 768 | S | 0.003 | 6268 | 52 | 6 | 1.28 | 1.70 | 1.54 | 1.81 | 1.68 | 1.58 | 1.81 | 1.65 | 1.68 |
| HR 805 | H | 0.001 | 5014 | 49 | 11 | 1.82 | 2.33 | 2.57 | 2.74 | 2.55 | 0.83 | 0.63 | 0.50 | 0.65 |
| HR 831 | S | 0.009 | 6504 | 34 | 6 | 1.29 | 1.84 | 1.58 | 1.81 | 1.74 | 1.28 | 1.81 | 1.60 | 1.56 |
| HR 856 | S | 0.005 | 6413 | 19 | 3 | 1.45 | 1.84 | 1.95 | 1.93 | 1.91 | 1.26 |  | 1.33 | 1.30 |
| HR 900 | S | 0.000 | 4735 | 34 | 10 | 1.92 | 1.75 | 1.61 | 2.24 | 1.87 | 2.14 | 2.98 | 1.00 | 2.04 |
| HR 908 | S | 0.004 | 4702 | 40 | 10 | 1.74 | 1.67 | 1.30 | 1.85 | 1.61 | 2.61 | 6.05 | 1.75 | 3.47 |
| HR 912 | E | 0.003 | 4671 | 57 | 11 | 1.77 | 1.59 | 2.04 | 2.17 | 1.93 | 2.87 | 1.23 | 1.10 | 1.73 |
| HR 926 | S | 0.009 | 4705 | 29 | 10 | 1.91 | 1.62 | 1.57 | 1.79 | 1.66 | 2.72 | 3.43 | 1.75 | 2.64 |
| HR 931 | S | 0.002 | 4697 | 55 | 10 | 1.80 | 1.71 | 1.29 | 2.05 | 1.68 | 2.75 | 6.29 | 1.43 | 3.49 |
| HR 946 | S | 0.026 | 4348 | 32 | 10 | 2.10 | 1.62 | 1.52 | 1.63 | 1.59 | 3.11 | 2.88 | 3.54 | 3.18 |
| HR 956 | S | 0.022 | 4959 | 74 | 10 | 1.82 | 2.51 | 2.48 | 2.61 | 2.53 | 0.60 | 0.73 | 0.60 | 0.64 |
| HR 1016 | H | 0.002 | 5060 | 63 | 11 | 1.91 | 2.58 | 2.06 | 3.00 | 2.55 | 0.59 | 1.30 | 0.40 | 0.76 |
| HR 1021 | H | 0.004 | 4799 | 91 | 10 | 1.83 | 1.42 | 1.64 | 1.83 | 1.63 | 3.44 | 2.57 |  | 3.00 |
| HR 1045 | H | 0.002 | 4979 | 78 | 10 | 1.93 | 2.59 | 1.92 | 2.73 | 2.41 | 0.57 | 2.04 | 0.50 | 1.04 |
| HR 1050 | S | 0.003 | 4652 | 27 | 10 | 1.78 | 1.61 | 1.83 | 1.97 | 1.80 | 2.71 | 1.63 | 1.54 | 1.96 |
| HR 1060 | S | 0.006 | 4851 | 89 | 11 | 1.81 | 1.43 | 1.85 | 2.25 | 1.84 | 3.04 | 1.70 | 0.90 | 1.88 |
| HR 1090 | H | 0.002 | 4591 | 70 | 10 | 1.80 | 1.52 | 1.61 | 1.87 | 1.67 | 3.52 | 2.31 | 2.18 | 2.67 |
| HR 1098 | S | 0.005 | 4985 | 100 | 11 | 1.92 | 1.96 | 1.79 | 2.54 | 2.10 | 1.30 | 2.29 | 0.60 | 1.40 |
| HR 1108 | S | 0.005 | 4899 | 38 | 11 | 1.80 | 1.97 | 2.33 | 2.26 | 2.19 | 1.22 | 0.90 | 0.83 | 0.98 |
| HR 1109 | H | 0.003 | 4957 | 19 | 10 | 1.74 | 1.97 | 1.67 | 2.36 | 2.00 |  | 4.39 | 0.75 | 2.57 |
| HR 1110 | S | 0.012 | 4948 | 108 | 11 | 1.80 | 2.11 | 1.60 | 2.45 | 2.05 | 1.02 | 4.45 | 0.70 | 2.06 |
| HR 1112 | E | 0.415 | 4205 | 63 | 4 | 2.84 | 2.86 | 1.68 | 2.55 | 2.36 | 0.82 | 3.03 | 0.78 | 1.54 |
| HR 1117 | S | 0.004 | 4783 | 64 | 11 | 1.67 | 1.70 | 1.26 | 1.77 | 1.58 | 2.79 | 6.11 | 1.85 | 3.58 |
| HR 1119 | S | 0.008 | 4901 | 96 | 11 | 1.75 | 2.03 | 1.88 | 2.25 | 2.05 | 1.20 | 1.58 | 0.90 | 1.23 |
| HR 1159 | S | 0.006 | 4800 | 18 | 10 | 1.72 | 1.24 | 1.09 | 1.82 | 1.38 | 5.12 | 6.00 | 1.74 | 4.29 |
| HR 1169 | H | 0.003 | 4939 | 18 | 10 | 1.77 | 2.02 | 2.28 | 2.36 | 2.22 | 1.13 | 0.94 | 0.75 | 0.94 |
| HR 1216 | H | 0.003 | 4427 | 18 | 14 | 1.84 | 1.55 | 1.35 | 1.77 | 1.56 | 3.00 | 4.55 | 2.20 | 3.25 |
| HR 1219 | H | 0.004 | 5846 | 74 | 8 | 2.02 | 2.81 | 2.58 |  | 2.70 | 0.40 | 0.48 |  | 0.44 |
| HR 1255 | E | 0.004 | 4743 | 104 | 12 | 1.36 | 1.33 | 1.38 | 1.20 | 1.30 | 4.41 | 3.88 | 6.67 | 4.98 |
| HR 1265 | S | 0.006 | 4655 | 70 | 11 | 1.99 | 1.64 | 1.65 | 2.19 | 1.83 | 2.99 | 3.05 | 1.15 | 2.40 |
| HR 1267 | S | 0.002 | 4605 | 78 | 10 | 1.85 | 1.68 | 1.74 | 1.72 | 1.71 | 2.87 | 1.82 | 2.46 | 2.39 |
| HR 1295 | S | 0.018 | 4732 | 50 | 11 | 1.70 | 1.59 | 1.22 | 2.00 | 1.60 | 3.03 | 6.55 | 1.50 | 3.69 |
| HR 1301 | S | 0.072 | 4934 | 128 | 10 | 1.89 | 2.37 | 1.80 | 2.56 | 2.24 | 0.78 | 3.12 | 0.62 | 1.51 |
| HR 1304 | E | 0.004 | 5170 | 143 | 7 | 1.76 |  | 1.81 |  | 1.81 |  | 1.46 |  | 1.46 |
| HR 1310 | S | 0.012 | 4627 | 75 | 10 | 1.60 | 1.59 | 1.43 | 1.49 | 1.50 | 3.43 | 3.50 | 4.41 | 3.78 |
| HR 1313 | S | 0.003 | 4693 | 69 | 7 | 1.60 | 1.58 | 1.68 | 1.67 | 1.64 | 3.55 | 2.13 | 2.73 | 2.80 |
| HR 1327 | S | 0.013 | 5211 | 57 | 13 | 1.88 | 2.81 | 2.46 |  | 2.64 | 0.40 | 0.60 |  | 0.50 |
| HR 1413 | S | 0.002 | 4753 | 18 | 10 | 1.56 | 1.68 | 1.65 | 1.37 | 1.57 | 2.34 | 2.13 | 4.80 | 3.09 |
| HR 1421 | S | 0.003 | 4462 | 21 | 10 | 1.71 | 1.39 | 1.38 | 1.61 | 1.46 | 4.22 | 4.89 | 3.20 | 4.10 |
| HR 1425 | S | 0.020 | 4749 | 26 | 10 | 1.65 | 2.21 | 1.38 | 1.97 | 1.85 | 0.88 | 5.44 | 1.54 | 2.62 |
| HR 1431 | S | 0.008 | 4814 | 51 | 11 | 1.89 | 1.92 | 2.03 | 2.39 | 2.11 | 1.63 | 1.36 | 0.80 | 1.26 |
| HR 1455 | S | 0.039 | 5455 | 102 | 9 | 1.72 | 2.45 | 2.23 | 2.55 | 2.41 | 0.58 | 0.67 | 0.60 | 0.62 |
| HR 1475 | H | 0.000 | 4822 | 113 | 9 | -0.40 |  |  |  | 2.00 |  |  |  |  |
| HR 1514 | S | 0.021 | 4969 | 92 | 7 | 1.90 | 2.54 | 1.81 | 2.73 | 2.36 | 0.62 | 2.40 | 0.50 | 1.17 |
| HR 1517 | S | 0.003 | 4382 | 68 | 5 | 1.58 | 1.51 | 1.19 | 1.37 | 1.36 |  | 7.25 | 4.67 | 5.96 |
| HR 1529 | S | 0.000 | 4416 | 159 | 10 | 1.56 |  |  |  | 1.40 |  |  |  |  |
| HR 1535 | S | 0.015 | 4868 | 99 | 10 | 1.77 | 1.98 | 1.69 | 1.93 | 1.87 | 1.51 | 2.02 |  | 1.76 |
| HR 1625 | S | 0.004 | 4476 | 31 | 10 | 1.72 | 1.35 | 1.19 | 1.57 | 1.37 | 4.35 | 6.63 | 3.20 | 4.72 |
| HR 1628 | S | 0.000 | 4708 | 18 | 13 | 1.78 | 1.61 | 0.97 | 1.77 | 1.45 | 2.77 | 9.00 | 1.85 | 4.54 |
| HR 1631 | H | 0.004 | 5032 | 61 | 10 | 1.70 | 2.16 | 1.73 | 2.40 | 2.10 |  | 3.25 | 0.70 | 1.98 |
| HR 1635 | H | 0.003 | 4529 | 38 | 10 | 1.71 | 1.32 | 1.35 | 1.89 | 1.52 | 4.93 | 4.25 |  | 4.59 |
| HR 1681 | S | 0.006 | 4671 | 15 | 11 | 1.65 | 1.64 | 1.66 | 1.72 | 1.67 | 2.64 | 2.22 | 2.74 | 2.53 |
| HR 1681 | E | 0.006 | 4674 | 15 | 11 | 1.65 | 1.65 | 1.66 | 1.73 | 1.68 | 2.80 | 2.22 | 2.61 | 2.54 |
| HR 1688 | E | 0.020 | 4559 | 38 | 10 | 1.75 | 1.38 | 1.61 | 1.56 | 1.52 | 4.34 | 2.50 | 4.02 | 3.62 |
| HR 1721 | H | 0.013 | 4881 | 37 | 10 | 2.12 | 2.92 | 2.84 |  | 2.88 | 0.41 | 0.49 |  | 0.45 |

| | | | | | | | | | | | | | |
|---|---|---|---|---|---|---|---|---|---|---|---|---|---|
| HR 1796 | S | 0.015 | 4624 | 29 | 10 | 1.65 | 1.87 | 1.57 | 1.48 | 1.64 | 2.14 | 2.65 | 4.18 | 2.99 |
| HR 1831 | S | 0.018 | 4554 | 41 | 7 | 1.70 | 1.31 | 1.39 | 1.50 | 1.40 | 4.81 | 3.88 | 4.60 | 4.43 |
| HR 1870 | H | 0.012 | 4244 | 79 | 10 | 2.34 | 1.73 | 1.70 | 1.87 | 1.77 | 2.29 | 2.00 | 2.17 | 2.15 |
| HR 1877 | H | 0.000 | 4476 | 17 | 10 | 1.39 | 1.50 | 1.18 | 1.32 | 1.33 | | 7.33 | 5.00 | 6.17 |
| HR 1889 | S | 0.005 | 6379 | 83 | 5 | 1.17 | 1.67 | 1.48 | 1.76 | 1.64 | 1.69 | 2.25 | 1.82 | 1.92 |
| HR 1954 | S | 0.009 | 4549 | 106 | 6 | 1.58 | 1.60 | 1.15 | 1.53 | 1.43 | 2.62 | 7.25 | 3.50 | 4.46 |
| HR 1958 | U | 0.008 | 5023 | 26 | 10 | 2.07 | 2.82 | 2.83 | | 2.83 | | 0.44 | | 0.44 |
| HR 1958 | H | 0.008 | 5023 | 26 | 10 | 2.07 | 2.82 | 2.83 | | 2.83 | | 0.44 | | 0.44 |
| HR 1978 | E | 0.000 | 6927 | 20 | 5 | 0.98 | | 1.26 | 1.69 | 1.48 | | 3.13 | 1.23 | 2.18 |
| HR 1986 | S | 0.007 | 4701 | 61 | 10 | 1.64 | 1.85 | 1.62 | 1.76 | 1.74 | 1.97 | 2.25 | 2.33 | 2.18 |
| HR 1987 | S | 0.002 | 4988 | 24 | 10 | 1.62 | 1.95 | 1.75 | 1.82 | 1.84 | 1.21 | 1.68 | 1.40 | 1.43 |
| HR 2049 | H | 0.001 | 4850 | 29 | 13 | 1.77 | 1.51 | 1.31 | | 1.41 | 2.93 | 3.25 | | 3.09 |
| HR 2070 | E | 0.002 | 4577 | 28 | 10 | 1.73 | 1.44 | 1.63 | 2.04 | 1.70 | 3.49 | 2.54 | 1.35 | 2.46 |
| HR 2076 | S | 0.002 | 4640 | 14 | 10 | 1.69 | 1.77 | 1.63 | 1.67 | 1.69 | 2.41 | 2.31 | 3.04 | 2.59 |
| HR 2080 | S | 0.011 | 4540 | 20 | 10 | 1.62 | 1.34 | 1.38 | 1.63 | 1.45 | | 4.46 | 2.84 | 3.65 |
| HR 2113 | H | 0.011 | 4288 | 27 | 13 | 2.62 | 1.36 | 1.80 | 1.87 | 1.68 | 3.53 | 1.69 | 1.58 | 2.27 |
| HR 2131 | H | 0.020 | 3891 | 37 | 8 | 2.70 | 1.63 | 1.93 | 1.88 | 1.81 | 2.89 | 1.21 | 2.48 | 2.19 |
| HR 2136 | S | 0.006 | 4950 | 59 | 11 | 1.75 | 2.12 | 1.73 | 2.25 | 2.03 | 1.00 | 3.82 | 0.83 | 1.88 |
| HR 2183 | S | 0.000 | 4635 | 33 | 10 | 1.48 | 1.43 | 1.39 | 1.57 | 1.46 | | 4.11 | 3.00 | 3.55 |
| HR 2183 | E | 0.000 | 4635 | 33 | 10 | 1.48 | 1.43 | 1.39 | 1.57 | 1.46 | | 4.11 | 3.00 | 3.55 |
| HR 2200 | S | 0.004 | 4690 | 18 | 10 | 1.81 | 1.71 | 1.29 | 2.05 | 1.68 | 2.75 | 6.29 | 1.43 | 3.49 |
| HR 2218 | S | 0.002 | 4926 | 21 | 13 | 1.65 | 2.16 | 1.73 | 1.91 | 1.93 | 0.89 | 2.10 | 1.40 | 1.46 |
| HR 2243 | S | 0.000 | 4701 | 17 | 10 | 1.27 | 1.59 | 1.21 | 1.46 | 1.42 | 2.62 | 6.54 | 3.75 | 4.30 |
| HR 2259 | S | 0.005 | 5076 | 75 | 10 | 1.50 | 2.17 | 1.64 | 2.10 | 1.97 | 0.81 | 3.58 | 1.00 | 1.80 |
| HR 2287 | S | 0.000 | 6986 | 23 | 5 | 1.16 | | 1.37 | 1.60 | 1.49 | | 2.38 | 2.00 | 2.19 |
| HR 2302 | S | 0.019 | 4729 | 60 | 10 | 1.86 | 1.64 | 1.39 | 1.96 | 1.66 | 2.83 | 4.37 | 1.35 | 2.85 |
| HR 2333 | S | 0.005 | 4736 | 11 | 10 | 1.81 | 1.68 | 1.07 | 2.16 | 1.64 | 2.34 | 6.38 | 1.13 | 3.28 |
| HR 2379 | S | 0.005 | 4345 | 28 | 10 | 2.22 | 1.90 | 1.86 | 2.02 | 1.93 | 1.79 | 1.56 | 1.60 | 1.65 |
| HR 2391 | E | 0.000 | 4656 | 77 | 12 | 1.65 | 1.77 | 1.55 | 1.54 | 1.62 | 2.82 | 2.63 | 3.55 | 3.00 |
| HR 2399 | H | 0.001 | 4808 | 60 | 10 | 1.80 | 1.07 | 1.76 | 1.83 | 1.55 | 5.81 | 2.26 | | 4.03 |
| HR 2416 | H | 0.022 | 4675 | 41 | 10 | 2.16 | 2.17 | 1.99 | 2.35 | 2.17 | 1.01 | 2.72 | 0.78 | 1.50 |
| HR 2437 | E | 0.001 | 4857 | 74 | 10 | 1.40 | 2.17 | 1.68 | 1.94 | 1.93 | | | 1.47 | 1.47 |
| HR 2447 | H | 0.009 | 4353 | 20 | 8 | 2.27 | 1.34 | 1.91 | 1.28 | 1.51 | 3.25 | 1.38 | 4.50 | 3.04 |
| HR 2531 | H | 0.012 | 4885 | 30 | 10 | 1.78 | 1.90 | 2.07 | 2.19 | 2.05 | 1.47 | 1.27 | 0.93 | 1.22 |
| HR 2552 | S | 0.004 | 4960 | 84 | 10 | 1.80 | 2.37 | 1.53 | 2.61 | 2.17 | 0.71 | 4.52 | 0.60 | 1.94 |
| HR 2556 | S | 0.015 | 4994 | 49 | 7 | 1.78 | 2.34 | 1.61 | 2.61 | 2.19 | 0.72 | 3.71 | 0.60 | 1.67 |
| HR 2573 | S | 0.003 | 4738 | 110 | 3 | 1.79 | 1.63 | 1.10 | 2.06 | 1.60 | 2.66 | 6.60 | 1.26 | 3.51 |
| HR 2586 | E | 0.025 | 5102 | 31 | 6 | 1.89 | 2.59 | 2.04 | 3.00 | 2.54 | 0.56 | 1.11 | 0.40 | 0.69 |
| HR 2642 | S | 0.014 | 4615 | 92 | 8 | 1.71 | 1.79 | 1.53 | 1.66 | 1.66 | 2.59 | 2.75 | 2.95 | 2.76 |
| HR 2682 | E | 0.017 | 4454 | 17 | 10 | 2.38 | 1.79 | 1.81 | 1.71 | 1.77 | 2.29 | 1.94 | 2.01 | 2.08 |
| HR 2698 | H | 0.019 | 4902 | 36 | 10 | 2.12 | 2.96 | 1.87 | 2.88 | 2.57 | 0.42 | 2.03 | | 1.22 |
| HR 2728 | S | 0.003 | 4787 | 64 | 12 | 1.72 | 1.22 | 1.08 | 1.89 | 1.40 | 5.45 | 6.25 | 1.50 | 4.40 |
| HR 2862 | H | 0.022 | 4803 | 54 | 13 | 2.58 | 3.88 | 2.82 | | 3.35 | 0.20 | 0.39 | | 0.29 |
| HR 2877 | S | 0.002 | 4620 | 26 | 12 | 1.90 | 1.91 | 2.43 | 2.06 | 2.13 | 1.73 | 0.73 | 1.43 | 1.30 |
| HR 2894 | E | 0.001 | 4580 | 14 | 2 | 1.56 | | 1.24 | 1.57 | 1.41 | | 5.56 | 3.00 | 4.28 |
| HR 2896 | S | 0.014 | 4829 | 38 | 9 | 2.04 | 2.44 | 2.71 | 2.56 | 2.57 | 0.74 | 0.57 | 0.60 | 0.64 |
| HR 2899 | S | 0.001 | 4600 | 13 | 10 | 1.69 | 1.78 | 1.53 | 1.40 | 1.57 | 2.13 | 2.88 | 4.94 | 3.31 |
| HR 2908 | H | 0.009 | 4929 | 72 | 10 | 1.72 | 2.23 | 1.90 | 2.32 | 2.15 | 0.81 | 1.50 | 0.80 | 1.04 |
| HR 2939 | E | 0.002 | 4713 | 66 | 10 | 1.82 | 1.47 | 1.19 | 1.68 | 1.45 | 3.59 | 6.39 | 2.38 | 4.12 |
| HR 2951 | U | 0.093 | 4012 | 32 | 8 | 3.40 | 1.74 | 1.87 | | 1.81 | 3.12 | 2.11 | | 2.61 |
| HR 2988 | U | 0.005 | 4861 | 14 | 10 | 1.77 | 1.74 | 1.67 | 1.93 | 1.78 | 2.06 | 2.18 | | 2.12 |
| HR 3012 | H | 0.024 | 4762 | 35 | 10 | 2.20 | 2.71 | 2.00 | 2.47 | 2.39 | 0.61 | 2.34 | | 1.47 |
| HR 3054 | S | 0.002 | 4660 | 84 | 10 | 1.80 | 1.59 | 1.38 | 1.94 | 1.64 | 2.99 | 5.44 | 1.60 | 3.34 |
| HR 3068 | E | 0.003 | 4928 | 21 | 10 | 1.73 | 2.20 | 2.03 | 2.05 | 2.09 | 0.82 | 1.27 | | 1.04 |
| HR 3069 | H | 0.013 | 5018 | 53 | 10 | 1.89 | 2.58 | 1.91 | 2.82 | 2.44 | 0.56 | 1.98 | 0.47 | 1.00 |
| HR 3094 | S | 0.010 | 4772 | 64 | 5 | 1.75 | 1.65 | 1.24 | 2.06 | 1.65 | 2.83 | 5.21 | 1.26 | 3.10 |
| HR 3097 | S | 0.000 | 4865 | 45 | 10 | 1.65 | 2.15 | 1.76 | 2.13 | 2.01 | 0.93 | 2.67 | 1.12 | 1.57 |
| HR 3125 | S | 0.003 | 4711 | 63 | 7 | 1.74 | 1.50 | 1.50 | 1.86 | 1.62 | 3.25 | 4.60 | 1.78 | 3.21 |
| HR 3140 | U | 0.000 | 6585 | 49 | 5 | 1.29 | 1.78 | 1.82 | 1.81 | 1.80 | 1.37 | | 1.65 | 1.51 |
| HR 3145 | S | 0.000 | 4280 | 32 | 16 | 2.23 | 1.48 | 1.58 | 1.45 | 1.50 | 2.56 | | 4.00 | 3.28 |
| HR 3150 | S | 0.002 | 5702 | 71 | 13 | 1.53 | 1.87 | 2.00 | 2.11 | 1.99 | | 0.98 | 0.96 | 0.97 |
| HR 3212 | S | 0.001 | 4988 | 18 | 10 | 1.77 | 2.45 | 1.52 | 2.22 | 2.06 | 0.60 | 4.70 | | 2.65 |

| | | | | | | | | | | | | | |
|---|---|---|---|---|---|---|---|---|---|---|---|---|---|
| HR 3216 | S | 0.001 | 4997 | 38 | 10 | 1.66 | | 1.32 | 1.91 | 1.62 | | 5.95 | 1.20 | 3.58 |
| HR 3222 | S | 0.003 | 4971 | 160 | 7 | 1.96 | 2.58 | 1.84 | 2.99 | 2.47 | 0.56 | 2.18 | | 1.37 |
| HR 3253 | U | 0.070 | 4021 | 100 | 8 | 3.33 | 2.05 | 1.77 | | 1.91 | 3.52 | 2.36 | | 2.94 |
| HR 3263 | S | 0.011 | 4930 | 36 | 7 | 1.64 | 2.26 | 1.66 | 2.17 | 2.03 | 0.80 | 2.29 | 1.00 | 1.36 |
| HR 3281 | S | 0.003 | 4660 | 17 | 11 | 1.76 | 1.50 | 1.72 | 1.44 | 1.55 | 2.99 | 1.88 | 4.17 | 3.01 |
| HR 3303 | S | 0.006 | 4934 | 71 | 5 | 1.66 | 2.42 | 1.68 | 1.71 | 1.94 | | 2.03 | 1.67 | 1.85 |
| HR 3324 | S | 0.001 | 4490 | 26 | 11 | 1.77 | 1.53 | 1.32 | 1.57 | 1.47 | 2.89 | 4.54 | 4.00 | 3.81 |
| HR 3376 | S | 0.000 | 4553 | 45 | 10 | 1.64 | 1.83 | 1.37 | 1.80 | 1.67 | 1.61 | 4.32 | 2.00 | 2.64 |
| HR 3409 | S | 0.001 | 4673 | 60 | 7 | 1.89 | 1.63 | 1.16 | 2.06 | 1.62 | 2.52 | 6.28 | 1.26 | 3.35 |
| HR 3423 | S | 0.001 | 6574 | 49 | 5 | 1.59 | 2.05 | 1.94 | 2.13 | 2.04 | 0.90 | 1.06 | 1.00 | 0.99 |
| HR 3424 | S | 0.000 | 4577 | 140 | 10 | 1.54 | 1.60 | 1.19 | 1.53 | 1.44 | 2.62 | 6.40 | 3.50 | 4.17 |
| HR 3428 | H | 0.023 | 4900 | 33 | 12 | 2.17 | 3.23 | 3.23 | 3.67 | 3.38 | 0.32 | 0.34 | | 0.33 |
| HR 3478 | E | 0.003 | 4648 | 14 | 11 | 1.75 | 1.71 | 1.68 | 1.80 | 1.73 | 2.09 | | 2.08 | 2.08 |
| HR 3529 | S | 0.005 | 4552 | 10 | 10 | 1.70 | 1.31 | 1.39 | 1.50 | 1.40 | 4.83 | 3.88 | 4.60 | 4.43 |
| HR 3653 | S | 0.003 | 4825 | 33 | 10 | 1.70 | 1.38 | 1.15 | 1.77 | 1.43 | 3.92 | 5.08 | 1.90 | 3.63 |
| HR 3664 | S | 0.005 | 5019 | 49 | 13 | 1.85 | 1.81 | 1.79 | 2.04 | 1.88 | 1.51 | 2.44 | 0.98 | 1.64 |
| HR 3687 | S | 0.007 | 4665 | 66 | 10 | 1.85 | 1.72 | 1.29 | 2.05 | 1.69 | 2.31 | 6.29 | 1.43 | 3.34 |
| HR 3707 | S | 0.000 | 4666 | 52 | 10 | 1.52 | 1.53 | 1.42 | 1.49 | 1.48 | | 3.50 | 4.60 | 4.05 |
| HR 3743 | E | 0.002 | 4760 | 32 | 10 | 1.77 | 1.45 | 1.09 | 1.89 | 1.48 | 3.17 | 6.00 | 1.50 | 3.56 |
| HR 3772 | S | 0.000 | 4426 | 42 | 10 | 1.53 | | | 1.33 | 1.33 | | | | |
| HR 3788 | S | 0.006 | 4419 | 80 | 13 | 1.79 | 1.43 | 1.13 | 1.55 | 1.37 | 3.50 | 7.75 | 3.46 | 4.90 |
| HR 3801 | S | 0.001 | 4861 | 28 | 10 | 1.79 | 1.68 | 1.70 | 2.25 | 1.88 | 2.03 | 2.01 | 0.90 | 1.65 |
| HR 3805 | S | 0.001 | 4490 | 38 | 10 | 1.82 | 1.51 | 1.38 | 1.58 | 1.49 | 3.60 | 4.02 | 3.73 | 3.78 |
| HR 3808 | S | 0.006 | 4883 | 49 | 14 | 2.14 | 2.80 | 2.93 | | 2.87 | 0.50 | 0.44 | | 0.47 |
| HR 3809 | S | 0.000 | 4809 | 51 | 10 | 1.77 | 1.35 | 1.67 | 1.85 | 1.62 | 3.81 | 2.78 | 1.63 | 2.74 |
| HR 3907 | S | 0.001 | 4988 | 70 | 10 | 1.76 | 2.26 | 1.67 | 2.31 | 2.08 | 0.77 | 3.65 | 0.77 | 1.73 |
| HR 3908 | S | 0.008 | 4779 | 23 | 10 | 1.79 | 1.72 | 1.61 | 2.00 | 1.78 | 2.54 | 3.15 | 1.28 | 2.32 |
| HR 3911 | S | 0.008 | 4868 | 105 | 5 | 1.70 | 2.05 | 1.60 | 1.97 | 1.87 | 1.05 | 2.75 | 1.27 | 1.69 |
| HR 3929 | S | 0.004 | 4630 | 32 | 7 | 1.74 | 1.75 | 1.62 | 1.76 | 1.71 | 2.19 | 2.25 | 2.33 | 2.26 |
| HR 3942 | S | 0.003 | 4702 | 24 | 7 | 1.83 | 1.80 | 1.20 | 2.13 | 1.71 | 2.21 | 6.45 | 1.20 | 3.29 |
| HR 3972 | U | 0.017 | 4661 | 16 | 10 | 1.89 | 1.73 | 1.15 | 2.03 | 1.64 | 2.21 | 6.82 | 1.40 | 3.48 |
| HR 4006 | S | 0.012 | 5169 | 65 | 10 | 1.98 | 2.88 | 2.35 | 2.87 | 2.70 | 0.36 | 0.62 | | 0.49 |
| HR 4032 | S | 0.004 | 4388 | 22 | 7 | 1.91 | 1.30 | 1.28 | 1.39 | 1.32 | 4.54 | 5.00 | | 4.77 |
| HR 4052 | S | 0.013 | 4520 | 139 | 12 | 2.15 | 1.92 | 1.49 | 2.02 | 1.81 | 2.37 | 4.18 | 1.34 | 2.63 |
| HR 4066 | U | 0.037 | 3830 | 161 | 3 | 3.58 | 1.95 | 2.60 | | 2.28 | 1.26 | 0.48 | | 0.87 |
| HR 4078 | S | 0.004 | 4583 | 83 | 7 | 1.67 | 1.70 | 1.42 | 1.46 | 1.53 | 3.06 | 3.59 | 4.64 | 3.77 |
| HR 4085 | S | 0.000 | 4803 | 34 | 5 | 1.47 | 1.56 | 1.65 | 1.76 | 1.66 | 2.37 | 2.13 | 2.27 | 2.26 |
| HR 4097 | S | 0.003 | 4509 | 40 | 7 | 1.70 | 1.38 | 1.27 | 1.70 | 1.45 | 4.62 | 5.21 | 2.47 | 4.10 |
| HR 4099 | U | 0.003 | 4589 | 15 | 10 | 1.63 | 1.65 | 1.38 | 1.37 | 1.47 | 3.11 | 3.88 | 5.57 | 4.19 |
| HR 4126 | S | 0.000 | 4965 | 106 | 12 | 1.84 | 2.56 | 1.60 | 3.00 | 2.39 | 0.54 | 3.58 | | 2.06 |
| HR 4233 | S | 0.003 | 4675 | 22 | 10 | 1.78 | 1.47 | 1.40 | 1.68 | 1.52 | 3.51 | 5.50 | 2.20 | 3.74 |
| HR 4242 | S | 0.004 | 4736 | 153 | 8 | 1.47 | 1.67 | 1.66 | 1.57 | 1.63 | 2.60 | 2.30 | 3.30 | 2.73 |
| HR 4255 | S | 0.006 | 5193 | 41 | 11 | 2.09 | 3.12 | 2.87 | | 3.00 | 0.32 | 0.40 | | 0.36 |
| HR 4256 | S | 0.000 | 4673 | 49 | 8 | 1.69 | 1.52 | 1.65 | 1.28 | 1.48 | 2.78 | 2.13 | 5.70 | 3.54 |
| HR 4264 | S | 0.001 | 4524 | 71 | 10 | 1.66 | 1.57 | 1.24 | 1.65 | 1.49 | 3.34 | 5.61 | 2.60 | 3.85 |
| HR 4283 | S | 0.014 | 4893 | 80 | 10 | 1.99 | 2.53 | 1.91 | 2.73 | 2.39 | 0.62 | 2.67 | 0.50 | 1.26 |
| HR 4305 | S | 0.003 | 4967 | 30 | 10 | 1.85 | 2.34 | 1.69 | 2.74 | 2.26 | 0.77 | 3.47 | 0.50 | 1.58 |
| HR 4321 | U | 0.005 | 4569 | 22 | 10 | 1.72 | 1.33 | 1.46 | 1.56 | 1.45 | 4.50 | 3.25 | 3.85 | 3.87 |
| HR 4351 | S | 0.001 | 4556 | 19 | 8 | 1.69 | 1.37 | 1.34 | 1.50 | 1.40 | 4.47 | 4.36 | 4.60 | 4.48 |
| HR 4383 | S | 0.004 | 4825 | 44 | 8 | 1.78 | 1.44 | 1.63 | 2.08 | 1.72 | 3.80 | 2.58 | 1.10 | 2.49 |
| HR 4407 | S | 0.003 | 4837 | 71 | 12 | 1.72 | 1.33 | 1.38 | 1.95 | 1.55 | 4.04 | 3.88 | 1.45 | 3.12 |
| HR 4409 | U | 0.002 | 4857 | 18 | 10 | 1.80 | 1.54 | 2.01 | 1.93 | 1.83 | 2.33 | 1.35 | | 1.84 |
| HR 4419 | S | 0.000 | 4638 | 97 | 9 | 2.01 | 1.80 | 1.46 | 2.11 | 1.79 | 2.25 | 3.90 | 1.17 | 2.44 |
| HR 4452 | S | 0.007 | 4648 | 40 | 13 | 1.83 | 1.47 | 1.41 | 1.27 | 1.38 | 3.33 | 5.25 | 4.70 | 4.43 |
| HR 4459 | S | 0.000 | 4814 | 87 | 7 | 1.65 | 1.76 | 1.29 | 1.99 | 1.68 | 2.86 | 5.39 | 1.46 | 3.24 |
| HR 4474 | S | 0.008 | 4718 | 127 | 13 | 1.85 | 1.57 | 1.24 | 1.80 | 1.54 | 2.92 | 5.49 | 1.89 | 3.43 |
| HR 4478 | S | 0.002 | 4691 | 18 | 10 | 1.73 | 1.69 | 1.83 | 1.97 | 1.83 | 2.89 | 1.60 | 1.54 | 2.01 |
| HR 4510 | S | 0.000 | 4914 | 42 | 10 | 1.79 | 2.20 | 1.78 | 2.44 | 2.14 | 0.88 | 3.30 | 0.73 | 1.64 |
| HR 4521 | S | 0.000 | 4411 | 40 | 7 | 1.68 | 1.46 | 1.31 | 1.31 | 1.36 | 3.95 | 5.63 | 6.33 | 5.30 |
| HR 4544 | S | 0.000 | 4668 | 29 | 9 | 1.75 | 1.50 | 1.72 | 1.46 | 1.56 | 2.99 | 1.88 | 4.20 | 3.02 |
| HR 4558 | S | 0.002 | 4970 | 100 | 10 | 1.79 | 2.45 | 1.54 | 2.22 | 2.07 | 0.60 | 4.15 | | 2.37 |
| HR 4593 | S | 0.001 | 4670 | 86 | 10 | 1.88 | 1.73 | 1.15 | 2.03 | 1.64 | 2.21 | 6.82 | 1.40 | 3.48 |

| | | | | | | | | | | | | | |
|---|---|---|---|---|---|---|---|---|---|---|---|---|---|
| HR 4610 | S | 0.005 | 4433 | 34 | 8 | 1.84 | 1.57 | 1.27 | 1.62 | 1.49 | 2.85 | 5.28 | 3.56 | 3.89 |
| HR 4654 | S | 0.005 | 4771 | 62 | 8 | 1.94 | 1.90 | 1.67 | 2.39 | 1.99 | 1.55 | 2.33 | 0.80 | 1.56 |
| HR 4655 | S | 0.002 | 4765 | 27 | 9 | 1.57 | 1.73 | 1.68 | 1.89 | 1.77 | 2.12 | 2.00 | 1.81 | 1.98 |
| HR 4668 | S | 0.001 | 4463 | 74 | 13 | 2.11 | 1.62 | 1.69 | 1.53 | 1.61 | 2.13 | 3.05 | 3.40 | 2.86 |
| HR 4699 | S | 0.000 | 4724 | 20 | 10 | 1.59 | 1.66 | 1.62 | 1.99 | 1.76 | 2.54 | | 1.40 | 1.97 |
| HR 4721 | H | 0.000 | 5859 | 121 | 11 | 0.95 | 1.48 | 1.51 | 1.48 | 1.49 | | 3.00 | | 3.00 |
| HR 4772 | S | 0.001 | 4719 | 27 | 10 | 1.79 | 1.73 | 1.10 | 2.08 | 1.64 | 2.46 | 7.43 | 1.33 | 3.74 |
| HR 4783 | S | 0.003 | 4785 | 11 | 7 | 1.80 | 1.79 | 1.57 | 2.04 | 1.80 | 3.07 | 3.22 | 1.20 | 2.50 |
| HR 4784 | S | 0.009 | 4705 | 32 | 10 | 1.80 | 1.66 | 1.14 | 2.08 | 1.63 | 2.76 | 7.25 | 1.33 | 3.78 |
| HR 4812 | S | 0.010 | 4719 | 39 | 7 | 1.78 | 2.06 | 2.44 | 2.27 | 2.26 | 1.14 | 0.74 | 0.93 | 0.94 |
| HR 4840 | S | 0.008 | 4372 | 20 | 5 | 2.05 | 1.57 | 1.50 | 1.66 | 1.58 | 3.37 | 3.00 | 3.58 | 3.31 |
| HR 4860 | S | 0.004 | 4814 | 29 | 11 | 1.78 | 1.27 | 1.73 | 1.55 | 1.52 | 4.33 | 2.47 | 2.63 | 3.14 |
| HR 4873 | S | 0.006 | 4808 | 23 | 9 | 1.56 | 1.90 | 1.23 | 2.00 | 1.71 | 1.47 | 6.75 | 1.53 | 3.25 |
| HR 4877 | S | 0.001 | 4744 | 35 | 10 | 1.85 | 1.65 | 1.33 | 1.97 | 1.65 | 2.66 | 4.23 | 1.30 | 2.73 |
| HR 4896 | S | 0.001 | 4702 | 86 | 8 | 1.51 | 1.44 | 1.47 | 1.55 | 1.49 | 3.81 | 3.13 | 3.84 | 3.59 |
| HR 4953 | S | 0.005 | 4816 | 48 | 12 | 1.73 | 1.38 | 1.42 | 2.00 | 1.60 | 4.17 | 3.79 | 1.28 | 3.08 |
| HR 4956 | S | 0.003 | 4198 | 36 | 13 | 2.01 | 1.33 | 1.95 | 1.26 | 1.51 | 4.36 | | 6.67 | 5.51 |
| HR 4959 | S | 0.001 | 4802 | 30 | 8 | 1.74 | 1.45 | 1.44 | 2.06 | 1.65 | 4.25 | 3.94 | 1.24 | 3.14 |
| HR 4960 | S | 0.001 | 4765 | 31 | 10 | 1.70 | 1.47 | 1.20 | 1.77 | 1.48 | 3.49 | 6.45 | 1.85 | 3.93 |
| HR 4964 | S | 0.003 | 4464 | 68 | 9 | 1.79 | 1.53 | 1.29 | 1.70 | 1.51 | 2.89 | 4.86 | 2.47 | 3.40 |
| HR 4984 | S | 0.004 | 4825 | 20 | 10 | 1.77 | 1.35 | 1.60 | 2.01 | 1.65 | 4.16 | 2.77 | 1.18 | 2.70 |
| HR 5007 | S | 0.000 | 4807 | 17 | 7 | 1.49 | 1.85 | 1.65 | 1.40 | 1.63 | 1.90 | 2.13 | 4.48 | 2.83 |
| HR 5053 | S | 0.000 | 4643 | 49 | 8 | 2.07 | 1.86 | 1.76 | 2.21 | 1.94 | 1.89 | 2.42 | 1.00 | 1.77 |
| HR 5067 | S | 0.006 | 4851 | 56 | 9 | 1.90 | 2.09 | 1.32 | 2.05 | 1.82 | 1.17 | 6.27 | 1.03 | 2.82 |
| HR 5096 | E | 0.000 | 4661 | 23 | 12 | 1.47 | 1.53 | 1.32 | 1.60 | 1.48 | 4.47 | 4.42 | 2.90 | 3.93 |
| HR 5126 | S | 0.010 | 4519 | 26 | 10 | 1.93 | 1.69 | 1.70 | 1.79 | 1.73 | 2.60 | 1.96 | 2.10 | 2.22 |
| HR 5143 | S | 0.001 | 4819 | 29 | 13 | 2.01 | 2.42 | 1.85 | 2.16 | 2.14 | 0.75 | 3.77 | 0.95 | 1.83 |
| HR 5161 | S | 0.006 | 5155 | 106 | 9 | 1.94 | 2.88 | 2.13 | 2.93 | 2.65 | 0.36 | 0.78 | 0.40 | 0.51 |
| HR 5176 | U | 0.019 | 4071 | 40 | 10 | 2.49 | 1.19 | 1.99 | 1.85 | 1.68 | 5.55 | | | 5.55 |
| HR 5180 | S | 0.006 | 4935 | 20 | 9 | 1.89 | 2.49 | 2.60 | | 2.55 | 0.67 | 0.62 | | 0.64 |
| HR 5184 | E | 0.004 | 4806 | 66 | 9 | 1.70 | 0.92 | 1.63 | 2.13 | 1.56 | 7.94 | 3.38 | 1.20 | 4.17 |
| HR 5186 | S | 0.003 | 4698 | 13 | 9 | 1.86 | 1.57 | 1.05 | 1.82 | 1.48 | 2.34 | 6.75 | 1.98 | 3.69 |
| HR 5195 | S | 0.001 | 4683 | 32 | 12 | 1.72 | 1.63 | 1.75 | 1.46 | 1.61 | 2.21 | | 4.30 | 3.26 |
| HR 5205 | S | 0.000 | 5049 | 94 | 12 | 1.98 | 2.71 | 2.06 | 2.87 | 2.55 | 0.50 | 1.11 | | 0.80 |
| HR 5213 | S | 0.005 | 4921 | 73 | 11 | 1.85 | 2.10 | 1.53 | 2.61 | 2.08 | 0.99 | 4.43 | 0.60 | 2.00 |
| HR 5276 | S | 0.003 | 4265 | 19 | 10 | 1.67 | 1.36 | | 1.15 | 1.26 | 3.57 | | 8.50 | 6.04 |
| HR 5277 | S | 0.010 | 4684 | 30 | 8 | 1.77 | 1.58 | 1.38 | 1.94 | 1.63 | 3.46 | 5.44 | 1.60 | 3.50 |
| HR 5302 | S | 0.003 | 4770 | 18 | 10 | 1.63 | 1.96 | 1.38 | 1.91 | 1.75 | 1.42 | 5.44 | 1.73 | 2.86 |
| HR 5310 | U | 0.001 | 4417 | 91 | 12 | 1.80 | 1.46 | 1.15 | 1.63 | 1.41 | 3.34 | 7.25 | 2.93 | 4.51 |
| HR 5344 | S | 0.010 | 4833 | 35 | 9 | 1.77 | 1.38 | 1.63 | 1.96 | 1.66 | 3.86 | 2.62 | 1.25 | 2.58 |
| HR 5394 | S | 0.000 | 4440 | 31 | 16 | 1.75 | 1.40 | 1.14 | 1.53 | 1.36 | 3.92 | 7.50 | 3.70 | 5.04 |
| HR 5454 | S | 0.003 | 4737 | 47 | 7 | 1.66 | 1.70 | 1.35 | 1.97 | 1.67 | 2.25 | 5.63 | 1.54 | 3.14 |
| HR 5518 | S | 0.012 | 4641 | 23 | 8 | 1.68 | 1.74 | 1.55 | 1.67 | 1.65 | 2.31 | 2.63 | 3.04 | 2.66 |
| HR 5609 | S | 0.001 | 4800 | 17 | 9 | 1.72 | 1.35 | 1.63 | 2.02 | 1.67 | 5.32 | 3.38 | 1.35 | 3.35 |
| HR 5620 | S | 0.000 | 4721 | 41 | 8 | 1.13 | | | 1.17 | 1.17 | | | | |
| HR 5635 | S | 0.000 | 4812 | 24 | 10 | 1.75 | 1.20 | 1.50 | 1.75 | 1.48 | 4.76 | 3.55 | | 4.15 |
| HR 5648 | S | 0.005 | 4722 | 37 | 10 | 1.81 | 1.67 | 1.14 | 2.13 | 1.65 | 2.69 | 6.58 | 1.20 | 3.49 |
| HR 5673 | S | 0.002 | 4399 | 48 | 7 | 1.88 | 1.65 | 1.27 | 1.70 | 1.54 | 2.14 | 5.21 | 2.47 | 3.27 |
| HR 5707 | S | 0.005 | 4783 | 28 | 13 | 1.78 | 1.79 | 1.83 | 2.00 | 1.87 | 2.36 | 2.46 | 1.28 | 2.03 |
| HR 5769 | S | 0.000 | 6040 | 63 | 6 | 1.11 | 1.57 | 1.36 | 1.61 | 1.51 | 2.00 | 2.67 | 2.25 | 2.30 |
| HR 5785 | S | 0.002 | 4798 | 28 | 10 | 1.69 | 1.19 | 1.51 | 1.89 | 1.53 | 4.99 | 3.88 | 1.50 | 3.46 |
| HR 5806 | E | 0.006 | 4676 | 24 | 8 | 1.62 | 1.77 | 1.55 | 1.54 | 1.62 | 2.82 | 2.63 | 3.55 | 3.00 |
| HR 5810 | S | 0.008 | 4633 | 25 | 8 | 1.65 | 1.54 | 1.51 | 1.29 | 1.45 | 3.53 | 2.88 | 5.17 | 3.86 |
| HR 5811 | S | 0.004 | 4517 | 53 | 9 | 1.88 | 1.56 | 1.55 | 1.62 | 1.58 | 3.44 | 2.63 | 3.00 | 3.02 |
| HR 5828 | S | 0.004 | 4625 | 46 | 9 | 1.83 | 1.77 | 1.71 | 1.91 | 1.80 | 2.71 | 1.92 | 1.73 | 2.12 |
| HR 5835 | S | 0.004 | 5172 | 67 | 6 | 1.78 | 2.64 | 1.86 | 2.47 | 2.32 | 0.45 | 1.28 | | 0.87 |
| HR 5841 | S | 0.005 | 4659 | 35 | 7 | 1.81 | 1.41 | 1.40 | 1.80 | 1.54 | 3.66 | 5.50 | 1.87 | 3.67 |
| HR 5893 | S | 0.002 | 4873 | 50 | 8 | 1.26 | 1.63 | 1.54 | 1.63 | 1.60 | 2.49 | 2.63 | 2.80 | 2.64 |
| HR 5922 | S | 0.003 | 4823 | 117 | 9 | 1.76 | 1.35 | 1.91 | 2.04 | 1.77 | 4.24 | 1.96 | 1.20 | 2.47 |
| HR 5963 | E | 0.000 | 4697 | 66 | 12 | 1.50 | 1.13 | 1.44 | 1.11 | 1.23 | | 3.38 | | 3.38 |
| HR 5969 | U | 0.000 | 4370 | 22 | 10 | 1.79 | 1.33 | | 1.17 | 1.25 | 4.46 | | | 4.46 |
| HR 6016 | U | 0.018 | 4029 | 50 | 6 | 2.49 | 1.73 | 1.45 | 1.48 | 1.55 | 2.23 | 3.38 | 4.53 | 3.38 |

| | | | | | | | | | | | | | |
|---|---|---|---|---|---|---|---|---|---|---|---|---|---|
| HR 6038 | S | 0.004 | 4744 | 142 | 5 | 1.74 | 1.47 | 1.09 | 1.77 | 1.44 | 3.80 | 7.27 | 1.85 | 4.31 |
| HR 6057 | S | 0.013 | 4560 | 18 | 10 | 1.68 | 1.46 | 1.51 | 1.50 | 1.49 | 3.99 | 3.11 | 4.60 | 3.90 |
| HR 6077 | U | 0.000 | 6285 | 68 | 6 | 1.15 | 1.69 | 1.44 | 1.77 | 1.63 | 1.62 | 2.38 | 1.83 | 1.94 |
| HR 6121 | S | 0.000 | 4682 | 39 | 10 | 1.71 | 1.76 | 1.72 | 1.95 | 1.81 | 2.23 | 1.88 | 1.83 | 1.98 |
| HR 6124 | S | 0.021 | 5030 | 35 | 12 | 1.92 | 2.62 | 2.09 | 2.99 | 2.57 | 0.54 | 1.45 | | 1.00 |
| HR 6126 | S | 0.001 | 4636 | 51 | 7 | 1.90 | 1.82 | 1.23 | 2.07 | 1.71 | 1.92 | 6.69 | 1.55 | 3.38 |
| HR 6136 | U | 0.014 | 4065 | 34 | 10 | 2.34 | 1.28 | 1.29 | 1.68 | 1.42 | 4.88 | 5.60 | 2.76 | 4.41 |
| HR 6140 | E | 0.012 | 5167 | 16 | 8 | 1.66 | 2.48 | 1.69 | 2.25 | 2.14 | 0.57 | 2.14 | 0.78 | 1.16 |
| HR 6145 | S | 0.000 | 4696 | 41 | 10 | 1.31 | 1.85 | 1.26 | 1.46 | 1.52 | | 5.75 | 3.75 | 4.75 |
| HR 6150 | S | 0.005 | 4735 | 2 | 2 | 1.78 | 1.60 | 1.09 | 1.98 | 1.56 | 2.99 | 6.95 | 1.53 | 3.82 |
| HR 6190 | S | 0.000 | 4718 | 99 | 8 | 1.85 | 1.70 | 1.33 | 2.06 | 1.70 | 2.54 | 4.96 | 1.24 | 2.91 |
| HR 6199 | S | 0.001 | 4641 | 81 | 10 | 1.81 | 1.56 | 1.75 | 1.91 | 1.74 | 3.22 | | 1.73 | 2.48 |
| HR 6259 | S | 0.001 | 4756 | 49 | 7 | 1.56 | 1.85 | 1.65 | 1.37 | 1.62 | 1.90 | 2.13 | 4.60 | 2.87 |
| HR 6287 | S | 0.002 | 4838 | 22 | 10 | 1.72 | 1.28 | 1.73 | 1.85 | 1.62 | 4.53 | 2.48 | 1.63 | 2.88 |
| HR 6307 | S | 0.005 | 4634 | 38 | 10 | 1.92 | 1.87 | 1.21 | 2.03 | 1.70 | 2.10 | 6.53 | 1.37 | 3.33 |
| HR 6330 | E | 0.001 | 4541 | 74 | 9 | 1.63 | 1.77 | 1.40 | 1.78 | 1.65 | 1.81 | 4.23 | 2.08 | 2.70 |
| HR 6333 | S | 0.004 | 4779 | 25 | 9 | 1.72 | 1.42 | 1.29 | 2.13 | 1.61 | 3.58 | 5.19 | 1.20 | 3.32 |
| HR 6342 | S | 0.012 | 4702 | 15 | 5 | 1.65 | 1.52 | 1.65 | 1.39 | 1.52 | 2.78 | 2.13 | 4.88 | 3.26 |
| HR 6359 | S | 0.022 | 5078 | 35 | 12 | 1.62 | 2.26 | 1.71 | 2.39 | 2.12 | 0.73 | 3.47 | 0.73 | 1.64 |
| HR 6360 | S | 0.003 | 4886 | 53 | 10 | 1.73 | 2.29 | 1.36 | 1.93 | 1.86 | 0.82 | 2.88 | | 1.85 |
| HR 6363 | S | 0.005 | 4603 | 58 | 5 | 1.76 | 1.57 | 1.59 | 1.62 | 1.59 | 2.71 | 2.43 | 3.16 | 2.77 |
| HR 6364 | S | 0.005 | 4281 | 41 | 9 | 2.02 | 1.50 | 1.16 | 1.63 | 1.43 | 3.24 | 7.00 | 2.88 | 4.37 |
| HR 6365 | S | 0.020 | 4889 | 80 | 10 | 1.79 | 1.95 | 2.05 | 2.19 | 2.06 | 1.28 | 1.28 | 0.93 | 1.16 |
| HR 6388 | S | 0.002 | 4329 | 25 | 6 | 2.03 | 1.33 | 1.36 | | 1.35 | 4.34 | 4.13 | | 4.23 |
| HR 6390 | S | 0.030 | 4697 | 103 | 10 | 1.84 | 1.70 | 1.14 | 1.98 | 1.61 | 2.42 | 6.72 | 1.43 | 3.52 |
| HR 6394 | S | 0.000 | 6069 | 52 | 9 | 0.94 | 1.48 | 1.24 | 1.40 | 1.37 | 2.34 | 3.79 | 3.00 | 3.04 |
| HR 6404 | S | 0.006 | 4658 | 14 | 8 | 1.60 | 1.54 | 1.56 | 1.61 | 1.57 | 3.97 | 2.76 | 3.51 | 3.42 |
| HR 6443 | S | 0.002 | 4870 | 65 | 7 | 1.76 | 1.98 | 1.67 | 1.71 | 1.79 | 1.51 | 2.18 | 1.67 | 1.79 |
| HR 6444 | S | 0.003 | 4773 | 38 | 4 | 1.66 | 1.71 | 1.20 | 1.58 | 1.50 | 2.01 | 6.75 | 2.90 | 3.89 |
| HR 6472 | S | 0.008 | 4980 | 45 | 10 | 1.62 | 2.30 | 1.72 | 2.27 | 2.10 | 0.71 | 1.85 | 0.90 | 1.15 |
| HR 6488 | S | 0.006 | 5461 | 83 | 6 | 1.85 | 2.61 | 2.40 | | 2.51 | 0.48 | 0.56 | | 0.52 |
| HR 6524 | S | 0.019 | 5162 | 47 | 9 | 1.88 | 2.73 | 2.20 | 2.99 | 2.64 | 0.45 | 0.76 | 0.40 | 0.54 |
| HR 6542 | E | 0.010 | 4833 | 64 | 7 | 1.72 | 1.43 | 1.51 | 2.11 | 1.68 | 3.66 | 3.39 | 1.10 | 2.72 |
| HR 6564 | S | 0.010 | 4191 | 36 | 5 | 1.94 | 1.29 | 1.94 | 1.39 | 1.54 | 4.26 | | 4.50 | 4.38 |
| HR 6575 | S | 0.020 | 4813 | 51 | 14 | 1.68 | 1.59 | 1.53 | 2.06 | 1.73 | 3.76 | 4.08 | 1.26 | 3.03 |
| HR 6579 | E | 0.010 | 4921 | 108 | 5 | 1.67 | 2.16 | 1.71 | 2.17 | 2.01 | 0.89 | 2.17 | 1.00 | 1.36 |
| HR 6591 | S | 0.009 | 4705 | 16 | 4 | 1.79 | 1.61 | 1.22 | 2.08 | 1.64 | 3.02 | 6.53 | 1.33 | 3.62 |
| HR 6606 | S | 0.004 | 4819 | 56 | 5 | 1.80 | 1.40 | 1.44 | 1.83 | 1.56 | 3.75 | 3.06 | | 3.40 |
| HR 6607 | S | 0.003 | 4745 | 39 | 7 | 1.61 | 1.63 | 1.72 | 1.88 | 1.74 | 2.21 | 1.88 | 1.86 | 1.98 |
| HR 6638 | S | 0.002 | 4944 | 31 | 9 | 1.55 | 1.94 | 1.19 | 1.99 | 1.71 | 1.19 | 4.38 | 1.25 | 2.27 |
| HR 6639 | S | 0.031 | 4468 | 71 | 9 | 1.71 | 1.42 | 1.42 | 1.70 | 1.51 | 3.91 | 4.41 | 2.71 | 3.68 |
| HR 6648 | U | 0.067 | 4724 | 72 | 9 | 2.57 | 3.49 | 2.78 | 3.77 | 3.35 | 0.29 | 0.43 | | 0.36 |
| HR 6654 | S | 0.000 | 4661 | 40 | 10 | 1.67 | 2.06 | 1.57 | 1.71 | 1.78 | 1.11 | | 2.67 | 1.89 |
| HR 6658 | U | 0.070 | 4764 | 54 | 11 | 2.72 | 4.03 | 3.01 | | 3.52 | 0.19 | 0.34 | | 0.26 |
| HR 6659 | S | 0.027 | 4577 | 53 | 9 | 1.69 | 1.47 | 1.41 | 1.53 | 1.47 | 3.61 | 3.64 | 4.21 | 3.82 |
| HR 6666 | E | 0.012 | 4621 | 39 | 11 | 1.56 | 1.53 | 1.37 | 1.46 | 1.45 | 3.30 | 4.00 | 4.68 | 3.99 |
| HR 6691 | U | 0.015 | 4968 | 52 | 8 | 1.91 | 2.54 | 1.86 | 2.73 | 2.38 | 0.62 | 2.28 | 0.50 | 1.13 |
| HR 6711 | S | 0.003 | 4999 | 62 | 7 | 1.72 | 2.21 | 1.71 | 2.40 | 2.11 | 0.85 | 3.65 | 0.70 | 1.73 |
| HR 6757 | S | 0.128 | 4943 | 121 | 13 | 2.31 | 3.29 | 2.78 | 3.86 | 3.31 | 0.31 | 0.43 | 0.20 | 0.31 |
| HR 6791 | E | 0.009 | 5040 | 95 | 10 | 2.21 | 3.12 | 2.83 | | 2.98 | 0.35 | 0.42 | | 0.38 |
| HR 6842 | U | 0.096 | 3909 | 117 | 11 | 3.33 | 1.75 | 1.58 | | 1.67 | 3.10 | 3.02 | | 3.06 |
| HR 6885 | U | 0.019 | 4357 | 35 | 12 | 2.19 | 1.86 | 1.69 | 1.70 | 1.75 | 1.92 | | 2.33 | 2.13 |
| HR 6890 | E | 0.000 | 6812 | 63 | 5 | 0.92 | 1.36 | 1.19 | | 1.28 | | 3.44 | | 3.44 |
| HR 6970 | S | 0.025 | 5008 | 27 | 10 | 2.00 | 2.58 | 2.12 | | 2.35 | 0.60 | 1.28 | | 0.94 |
| HR 7010 | S | 0.035 | 5085 | 99 | 5 | 1.87 | 2.66 | 1.84 | 3.00 | 2.50 | 0.47 | 1.58 | 0.40 | 0.81 |
| HR 7042 | S | 0.005 | 4943 | 94 | 7 | 1.68 | 1.96 | 1.87 | | 1.92 | | 1.53 | | 1.53 |
| HR 7064 | S | 0.001 | 4438 | 33 | 13 | 2.00 | 1.74 | 1.56 | 1.80 | 1.70 | 2.32 | 2.63 | 2.32 | 2.42 |
| HR 7135 | S | 0.006 | 4666 | 51 | 12 | 1.75 | 1.46 | 1.76 | 1.39 | 1.54 | 3.22 | 1.75 | 4.64 | 3.21 |
| HR 7137 | S | 0.009 | 5024 | 74 | 13 | 2.34 | 3.46 | 2.51 | 3.83 | 3.27 | 0.26 | 0.50 | 0.20 | 0.32 |
| HR 7146 | S | 0.001 | 4803 | 128 | 9 | 1.71 | 0.95 | 1.10 | 2.13 | 1.39 | 7.40 | 5.75 | 1.20 | 4.78 |
| HR 7186 | S | 0.079 | 4991 | 30 | 12 | 2.02 | 2.72 | 2.06 | 2.87 | 2.55 | 0.50 | 1.30 | | 0.90 |
| HR 7187 | S | 0.003 | 4953 | 41 | 6 | 1.74 | 1.75 | 1.66 | 2.36 | 1.92 | 1.39 | 4.09 | 0.75 | 2.08 |

| HR | | | | | | | | | | | | | |
|---|---|---|---|---|---|---|---|---|---|---|---|---|---|
| HR 7196 | S | 0.008 | 4824 | 60 | 7 | 1.78 | 1.44 | 1.63 | 2.08 | 1.72 | 3.70 | 2.58 | 1.10 | 2.46 |
| HR 7204 | S | 0.011 | 4869 | 38 | 7 | 1.79 | 1.67 | 1.88 | 2.25 | 1.93 | 1.86 | 1.58 | 0.90 | 1.45 |
| HR 7325 | S | 0.031 | 4753 | 32 | 13 | 1.89 | 1.69 | 1.40 | 2.24 | 1.78 | 2.39 | 3.44 | 1.00 | 2.28 |
| HR 7359 | S | 0.010 | 4838 | 104 | 7 | 1.72 | 1.28 | 1.66 | 2.11 | 1.68 | 4.53 | 2.79 | 1.10 | 2.81 |
| HR 7376 | S | 0.012 | 4679 | 23 | 7 | 1.69 | 1.90 | 1.68 | 1.88 | 1.82 | 2.24 | | 1.86 | 2.05 |
| HR 7388 | H | 0.006 | 4864 | 38 | 8 | 1.56 | 1.80 | 1.04 | 2.03 | 1.62 | 1.50 | 7.00 | | 4.25 |
| HR 7389 | S | 0.000 | 6281 | 56 | 6 | 1.09 | | 1.23 | 1.63 | 1.43 | | | 2.10 | 2.10 |
| HR 7398 | U | 0.000 | 4743 | 228 | 4 | | | | | | | | | |
| HR 7407 | S | 0.006 | 4740 | 53 | 12 | 1.77 | 1.71 | 1.11 | 2.13 | 1.65 | 2.57 | 6.85 | 1.20 | 3.54 |
| HR 7433 | S | 0.011 | 4631 | 27 | 8 | 1.62 | 1.65 | 1.46 | 1.45 | 1.52 | 3.03 | 3.25 | 4.38 | 3.55 |
| HR 7449 | S | 0.007 | 4810 | 95 | 10 | 1.66 | 1.42 | 1.52 | 2.03 | 1.66 | 4.33 | 4.40 | 1.35 | 3.36 |
| HR 7465 | S | 0.023 | 4730 | 60 | 8 | 1.94 | 1.78 | 1.53 | 2.26 | 1.86 | 1.94 | 2.95 | 1.00 | 1.97 |
| HR 7468 | E | 0.000 | 4941 | 33 | 10 | 1.18 | 1.55 | 1.56 | 1.66 | 1.59 | 2.37 | | 2.47 | 2.42 |
| HR 7487 | S | 0.009 | 5052 | 36 | 7 | 1.74 | 2.49 | 2.68 | | 2.59 | 0.57 | 0.53 | | 0.55 |
| HR 7526 | S | 0.000 | 5035 | 57 | 5 | 1.28 | | 1.35 | 1.62 | 1.49 | | 4.33 | | 4.33 |
| HR 7540 | S | 0.056 | 4959 | 123 | 9 | 1.91 | 2.61 | 1.85 | 2.73 | 2.40 | 0.58 | 2.50 | 0.50 | 1.20 |
| HR 7541 | S | 0.009 | 4421 | 26 | 8 | 1.72 | 1.27 | 1.33 | 1.53 | 1.38 | 4.98 | 5.33 | 3.78 | 4.70 |
| HR 7659 | H | 0.005 | 4275 | 29 | 10 | 2.12 | | 1.36 | | 1.36 | | 4.13 | | 4.13 |
| HR 7681 | S | 0.014 | 4519 | 50 | 8 | 1.64 | 1.50 | 1.20 | 1.56 | 1.42 | 3.52 | 6.33 | 3.20 | 4.35 |
| HR 7713 | S | 0.023 | 5060 | 33 | 8 | 1.89 | 2.62 | 2.11 | 3.00 | 2.58 | 0.54 | 1.36 | 0.40 | 0.77 |
| HR 7743 | E | 0.005 | 4888 | 123 | 9 | 1.75 | 2.45 | 1.69 | 1.93 | 2.02 | 0.68 | 2.02 | | 1.35 |
| HR 7778 | S | 0.011 | 5080 | 95 | 8 | 2.01 | 2.77 | 2.65 | 2.87 | 2.76 | 0.48 | 0.49 | | 0.48 |
| HR 7788 | S | 0.028 | 4969 | 107 | 8 | 1.94 | 2.49 | 1.92 | 2.73 | 2.38 | 0.66 | 2.11 | 0.50 | 1.09 |
| HR 7794 | S | 0.000 | 4866 | 43 | 12 | 1.68 | 1.95 | 1.44 | 2.23 | 1.87 | 1.19 | 3.60 | 0.98 | 1.92 |
| HR 7802 | S | 0.005 | 4827 | 105 | 9 | 1.60 | 2.04 | 1.31 | 1.77 | 1.71 | 1.17 | 5.63 | 1.85 | 2.88 |
| HR 7820 | S | 0.002 | 4703 | 40 | 10 | 1.53 | 1.48 | 1.51 | 1.70 | 1.56 | 3.57 | 2.88 | 2.78 | 3.07 |
| HR 7824 | S | 0.013 | 5029 | 45 | 10 | 1.73 | 2.38 | 1.61 | 2.40 | 2.13 | 0.64 | 3.71 | 0.70 | 1.68 |
| HR 7854 | S | 0.005 | 4800 | 33 | 9 | 1.71 | 1.37 | 1.09 | 2.02 | 1.49 | 4.85 | 6.00 | 1.35 | 4.07 |
| HR 7867 | E | 0.003 | 4770 | 68 | 7 | 1.40 | 1.67 | 1.47 | 1.38 | 1.51 | 2.60 | 3.13 | 4.25 | 3.33 |
| HR 7904 | S | 0.007 | 4728 | 87 | 7 | 1.88 | 1.58 | 1.61 | 2.04 | 1.74 | 3.25 | 3.22 | 1.20 | 2.56 |
| HR 7905 | S | 0.005 | 4760 | 100 | 10 | 1.75 | 1.56 | 1.24 | 2.13 | 1.64 | 3.19 | 5.65 | 1.20 | 3.35 |
| HR 7962 | S | 0.039 | 4623 | 22 | 6 | 1.88 | 1.87 | 1.26 | 2.03 | 1.72 | 2.02 | 6.67 | 1.60 | 3.43 |
| HR 7976 | E | 0.001 | 4652 | 57 | 10 | 1.55 | | 1.42 | 1.29 | 1.36 | | 3.50 | 5.50 | 4.50 |
| HR 8017 | S | 0.006 | 4441 | 57 | 10 | 1.67 | 1.49 | 1.36 | 1.55 | 1.47 | 3.30 | 4.92 | 3.70 | 3.97 |
| HR 8035 | S | 0.001 | 4926 | 67 | 9 | 1.60 | 2.19 | 1.51 | 2.26 | 1.99 | 0.85 | 2.94 | 0.93 | 1.58 |
| HR 8072 | S | 0.012 | 4887 | 50 | 12 | 1.59 | 2.37 | 1.14 | 2.02 | 1.84 | 0.65 | 5.08 | 1.23 | 2.32 |
| HR 8076 | H | 0.003 | 4632 | 24 | 10 | 2.03 | 1.50 | 1.88 | 2.36 | 1.91 | 2.95 | 2.06 | 0.93 | 1.98 |
| HR 8082 | S | 0.012 | 4819 | 94 | 9 | 1.82 | 1.54 | 1.54 | 1.88 | 1.65 | 2.87 | 2.74 | 1.30 | 2.30 |
| HR 8096 | S | 0.000 | 4563 | 54 | 10 | 1.63 | 1.72 | 1.34 | 1.70 | 1.59 | 2.08 | 4.44 | 2.47 | 3.00 |
| HR 8108 | H | 0.004 | 4883 | 30 | 10 | 1.75 | 1.87 | 1.69 | 1.93 | 1.83 | 1.58 | 2.02 | | 1.80 |
| HR 8179 | S | 0.031 | 4805 | 102 | 11 | 1.84 | 1.48 | 1.44 | 1.88 | 1.60 | 3.10 | 3.01 | 1.30 | 2.47 |
| HR 8185 | S | 0.009 | 4691 | 54 | 9 | 1.82 | 1.85 | 1.19 | 2.07 | 1.70 | 2.04 | 6.85 | 1.55 | 3.48 |
| HR 8191 | S | 0.004 | 6271 | 42 | 6 | 1.48 | 1.94 | 1.85 | 2.00 | 1.93 | 1.09 | 1.25 | 1.20 | 1.18 |
| HR 8274 | S | 0.010 | 4779 | 42 | 10 | 1.82 | 1.50 | 1.58 | 1.97 | 1.68 | 3.52 | 2.90 | 1.27 | 2.56 |
| HR 8320 | S | 0.013 | 4892 | 69 | 7 | 1.84 | 1.98 | 1.73 | 2.40 | 2.04 | 1.22 | 3.63 | 0.74 | 1.86 |
| HR 8352 | U | 0.000 | 6538 | 56 | 4 | 0.87 | | 1.21 | 1.46 | 1.34 | | 3.94 | 2.50 | 3.22 |
| HR 8360 | S | 0.001 | 4527 | 21 | 10 | 1.73 | 1.26 | 1.57 | 1.88 | 1.57 | 4.92 | 2.83 | 1.75 | 3.17 |
| HR 8391 | S | 0.000 | 6397 | 98 | 6 | 0.99 | 1.46 | 1.51 | 1.60 | 1.52 | 2.51 | 2.50 | 2.03 | 2.35 |
| HR 8394 | S | 0.002 | 4811 | 40 | 11 | 1.45 | 1.46 | 1.62 | 1.59 | 1.56 | 2.91 | | 2.60 | 2.75 |
| HR 8442 | S | 0.085 | 5261 | 40 | 12 | 2.19 | 3.24 | 2.70 | | 2.97 | 0.28 | 0.41 | | 0.35 |
| HR 8453 | S | 0.003 | 4946 | 112 | 13 | 1.64 | 2.47 | 2.03 | 2.27 | 2.26 | 0.57 | 1.39 | 0.90 | 0.95 |
| HR 8456 | S | 0.012 | 4811 | 124 | 9 | 1.80 | 1.47 | 1.63 | 1.97 | 1.69 | 3.59 | 2.62 | 1.27 | 2.49 |
| HR 8461 | S | 0.000 | 4950 | 98 | 10 | 1.12 | | | 1.88 | 1.88 | | | | |
| HR 8482 | S | 0.000 | 4503 | 50 | 9 | 1.52 | | 1.20 | 1.11 | 1.16 | | 6.68 | 8.67 | 7.67 |
| HR 8500 | S | 0.004 | 4544 | 19 | 10 | 1.91 | 1.70 | 1.70 | 1.92 | 1.77 | 2.45 | 1.96 | 1.82 | 2.08 |
| HR 8530 | U | 0.006 | 4851 | 113 | 10 | 1.99 | 1.82 | 1.40 | 2.27 | 1.83 | 2.15 | 5.59 | 0.82 | 2.85 |
| HR 8568 | S | 0.023 | 4695 | 64 | 13 | 1.69 | 1.72 | 1.72 | 1.95 | 1.80 | 2.55 | 1.88 | 1.83 | 2.08 |
| HR 8594 | S | 0.014 | 4942 | 43 | 10 | 1.81 | 1.91 | 2.39 | 2.22 | 2.17 | 1.19 | 0.80 | | 0.99 |
| HR 8596 | S | 0.002 | 4875 | 67 | 10 | 1.59 | 2.37 | 1.31 | 1.82 | 1.83 | 0.65 | 4.81 | 1.60 | 2.35 |
| HR 8617 | S | 0.029 | 5338 | 114 | 11 | 1.92 | 2.81 | 2.42 | 2.95 | 2.73 | 0.40 | 0.55 | 0.40 | 0.45 |
| HR 8642 | S | 0.001 | 4585 | 69 | 12 | 1.60 | 1.72 | 1.38 | 1.82 | 1.64 | 2.08 | 4.00 | | 3.04 |
| HR 8678 | S | 0.005 | 4803 | 101 | 9 | 1.71 | 1.35 | 1.09 | 1.82 | 1.42 | 4.56 | 6.00 | 1.60 | 4.05 |

| Name | Type | E(B-V) | Teff | σ | N | Log(L/L☉) | Mass B | Mass D | Mass Y | <Mass> | Age B | Age D | Age Y | <Age> |
|---|---|---|---|---|---|---|---|---|---|---|---|---|---|---|
| HR 8712 | S | 0.004 | 4641 | 99 | 8 | 1.71 | 1.77 | 1.59 | 1.73 | 1.70 | 2.41 | 2.38 | 2.45 | 2.41 |
| HR 8730 | S | 0.003 | 4586 | 11 | 8 | 1.71 | 1.53 | 1.45 | 1.50 | 1.49 | 3.22 | 3.32 | 4.35 | 3.63 |
| HR 8839 | S | 0.015 | 4594 | 62 | 5 | 2.07 | 1.93 | 1.51 | 2.07 | 1.84 | 2.30 | 3.73 | 1.27 | 2.43 |
| HR 8922 | S | 0.008 | 4727 | 83 | 7 | 1.77 | 1.65 | 1.09 | 1.99 | 1.58 | 2.75 | 7.33 | 1.46 | 3.85 |
| HR 8924 | E | 0.000 | 4660 | 52 | 16 | 0.93 | | | 1.20 | 1.20 | | | | |
| HR 8941 | S | 0.035 | 4962 | 92 | 13 | 2.09 | 2.83 | 2.17 | 2.87 | 2.62 | 0.46 | 1.17 | | 0.82 |
| HR 8946 | S | 0.000 | 4208 | 9 | 8 | 1.96 | 1.40 | | 1.36 | 1.38 | 3.66 | | 5.33 | 4.49 |
| HR 8958 | S | 0.001 | 4708 | 20 | 9 | 1.64 | 1.52 | 1.65 | 1.37 | 1.51 | 2.78 | 2.13 | 4.60 | 3.17 |
| HR 8979 | U | 0.002 | 4887 | 36 | 12 | 1.74 | 1.93 | 1.67 | 1.93 | 1.84 | 1.39 | 2.18 | | 1.79 |
| HR 9009 | S | 0.001 | 4661 | 48 | 10 | 1.73 | 1.74 | 1.68 | 1.81 | 1.74 | 2.12 | | 2.17 | 2.14 |
| HR 9101 | S | 0.001 | 4663 | 45 | 10 | 1.69 | 2.06 | 1.70 | 1.99 | 1.92 | 1.11 | 2.00 | 1.43 | 1.51 |
| HR 9104 | S | 0.008 | 4701 | 67 | 7 | 1.75 | 1.54 | 1.40 | 1.43 | 1.46 | 3.17 | 5.50 | | 4.34 |
| NSV 4275 | H | 0.060 | 4257 | 36 | 16 | 2.32 | 1.82 | 1.57 | 1.81 | 1.73 | 2.46 | 3.68 | 2.53 | 2.89 |
| V* AI Scl | H | 0.000 | 7128 | 29 | 5 | 1.20 | | 1.58 | | 1.58 | | 1.75 | | 1.75 |
| V* AN Scl | H | 0.000 | 4883 | 89 | 11 | | | | | | | | | |
| V* AZ CMi | E | 0.002 | 7600 | 31 | 5 | 1.47 | 1.94 | 1.91 | 1.88 | 1.91 | 0.63 | 1.08 | 1.25 | 0.99 |
| V* BM CVn | S | 0.004 | 4472 | 105 | 7 | 1.36 | 1.41 | | 1.21 | 1.31 | | | 7.00 | 7.00 |
| V* BP Psc | H | 0.000 | 4019 | 1 | 2 | | | | | | | | | |
| V* CF Scl | H | 0.000 | 5002 | 357 | 8 | | | | | | | | | |
| V* CN Dra | E | 0.012 | 6919 | 21 | 5 | 1.84 | 2.33 | 2.18 | | 2.26 | 0.63 | 0.73 | | 0.68 |
| V* ER Eri | H | 0.000 | 4558 | 71 | 5 | | | | | | | | | |
| V* HR CMa | S | 0.006 | 4725 | 112 | 14 | 1.76 | 1.66 | 1.09 | 2.08 | 1.61 | 2.80 | 7.51 | 1.33 | 3.88 |
| V* HY Lib | S | 0.027 | 6312 | 45 | 6 | 1.37 | 1.78 | 1.67 | 1.89 | 1.78 | 1.39 | 1.57 | 1.44 | 1.47 |
| V* IM Peg | S | 0.002 | 4603 | 93 | 5 | 1.75 | 1.34 | 1.55 | 1.37 | 1.42 | 3.46 | 2.63 | 4.33 | 3.47 |
| V* LU Del | S | 0.014 | 4917 | 74 | 12 | 1.91 | 2.41 | 1.89 | 2.73 | 2.34 | 0.76 | 2.95 | 0.50 | 1.41 |
| V* OP And | S | 0.030 | 4357 | 140 | 13 | 2.11 | 1.68 | 1.56 | 1.72 | 1.65 | 2.89 | 2.63 | 3.00 | 2.84 |
| V* R Eri | E | 0.003 | 5051 | 11 | 10 | 1.58 | 2.18 | 1.50 | 1.91 | 1.86 | 0.81 | 4.40 | 1.20 | 2.14 |
| V* TX Pic | U | 0.011 | 4435 | 183 | 9 | 2.10 | 1.88 | 1.73 | 1.81 | 1.81 | 1.77 | 1.88 | 1.86 | 1.84 |
| V* TZ For | H | 0.003 | 5332 | 71 | 9 | 1.71 | 2.50 | 2.12 | 2.55 | 2.39 | 0.54 | 0.76 | 0.60 | 0.63 |
| V* V1045 Sco | H | 0.106 | 3789 | 40 | 2 | 2.54 | 1.38 | 1.72 | 1.75 | 1.62 | 4.15 | 2.73 | 3.31 | 3.40 |
| V* V1054 Sco | H | 0.015 | 6832 | 55 | 3 | 1.23 | 1.69 | 1.54 | 1.70 | 1.64 | 1.58 | 1.97 | 1.68 | 1.74 |
| V* V1719 Cyg | S | 0.066 | 6643 | 123 | 5 | 1.59 | 2.08 | 1.96 | 2.14 | 2.06 | 0.90 | 1.04 | 1.01 | 0.98 |
| V* V1794 Cyg | E | 0.004 | 5180 | 92 | 8 | 1.11 | 1.75 | 1.41 | 1.80 | 1.65 | 1.42 | 2.96 | 1.67 | 2.02 |
| V* V448 Car | H | 0.022 | 3911 | 25 | 6 | 2.64 | 1.57 | 1.88 | 1.74 | 1.73 | 3.21 | 1.84 | 3.48 | 2.85 |
| V* VW Dra | S | 0.006 | 4650 | 45 | 7 | 1.92 | 1.65 | 1.22 | 2.01 | 1.63 | 2.06 | 6.05 | 1.30 | 3.14 |
| V* X Sct | U | 0.205 | 4763 | 291 | 5 | 1.50 | 1.71 | 1.42 | 1.76 | 1.63 | 2.46 | 4.07 | 2.75 | 3.09 |

| | |
|---|---|
| E(B-V): | Computed using the reddening maps of Hakkila et al. (1997), the parallax distance, and a correction for the lack of reddening within 75 pc. |
| Teff: | Adopted effective temperature derived using the calibration of Ramìrez & Melèndez (2005). |
| σ: | Standard deviation of the mean effective temperature. |
| N: | Number of colors used to determine the effective temperature. |
| Log(L/L☉): | Logarithm of the luminosity in solar units. Derived from the distance, apparent V magnitude, and the bolometric corrections of Bessell, Castelli, & Plez (1998) |
| Mass: | B = Mass determined from the Bertelli et al. (1994) isochrones. <br> D = Mass determined from the Dotter et al. (2008) isochrones. <br> Y = Mass determined from the Demarque et al. (2003) isochrones. |
| <Mass>: | Mean value of the mass determinations. |
| Age: | B = Age determined from the Bertelli et al. (1994) isochrones. <br> D = Age determined from the Dotter et al. (2008) isochrones. <br> Y = Age determined from the Demarque et al. (2003) isochrones. |
| <Age>: | Mean value of the age determinations. |

Table 3
Stellar Parameters, Fe, C, O, Li Abundance Data

| KeyName | S | T | g | Vt | FeI | S | N | FeII | S | N | C | O | Li | EW | N1 | T | G | Vt | FeI | S | N | FeII | S | N | C | O | Li | N2 | Vm | B | Cl |
|---|---|---|---|---|---|---|---|---|---|---|---|---|---|---|---|---|---|---|---|---|---|---|---|---|---|---|---|---|---|---|---|
| hd001671 | E | 6323 | 3.61 | 2.61 | 0.02 | 0.22 | 85 | 0.26 | 0.14 | 7 | 7.95 | 8.93 | 2.77 | 5.24 | 0.02 | 6323 | 2.95 | 2.71 | 0.00 | 0.22 | 85 | 0.01 | 0.15 | 7 | 7.75 | 8.74 | 2.77 | 0.05 | 44.0 | R | 2 |
| hd002910 | E | 4696 | 2.60 | 1.36 | 0.15 | 0.14 | 534 | 0.35 | 0.21 | 50 | 8.39 | 8.82 | 0.45 | 1.28 | 0.23 | 4696 | 2.10 | 1.50 | 0.05 | 0.14 | 534 | 0.05 | 0.23 | 50 | 8.23 | 8.59 | 0.46 | 0.28 | 5.1 | G | 1 |
| hd004188 | E | 4793 | 2.49 | 1.39 | 0.07 | 0.12 | 549 | 0.19 | 0.19 | 54 | 8.15 | 8.74 | 0.31 | 0.68 | 0.21 | 4793 | 2.20 | 1.46 | 0.02 | 0.12 | 549 | 0.03 | 0.20 | 54 | 8.07 | 8.60 | 0.31 | 0.23 | 5.0 | G | 1 |
| hd004502 | E | 4570 | 2.25 | 2.88 | -0.31 | 0.41 | 52 | 1.14 | 1.24 | 6 | | | 1.05 | 6.75 | 0.29 | 4570 | 0.10 | 2.61 | -0.53 | 0.46 | 52 | 0.09 | 1.36 | 6 | | | 0.96 | 0.47 | 40.0 | R | 2 |
| hd005234 | E | 4422 | 1.87 | 1.58 | 0.02 | 0.17 | 506 | 0.14 | 0.21 | 41 | 8.10 | 8.67 | 0.38 | 2.75 | 0.34 | 4422 | 1.58 | 1.62 | -0.04 | 0.18 | 506 | -0.04 | 0.22 | 41 | 8.02 | 8.54 | 0.37 | 0.37 | 5.7 | G | 1 |
| hd006319 | E | 4755 | 2.74 | 1.38 | 0.24 | 0.14 | 531 | 0.36 | 0.18 | 45 | 8.48 | 9.00 | 0.24 | 0.67 | 0.20 | 4755 | 2.44 | 1.48 | 0.17 | 0.14 | 531 | 0.17 | 0.19 | 45 | 8.39 | 8.86 | 0.25 | 0.22 | 5.0 | G | 1 |
| hd010380 | E | 4154 | 1.50 | 1.83 | -0.11 | 0.19 | 470 | 0.14 | 0.23 | 36 | 8.18 | 8.71 | -0.49 | 1.08 | 0.38 | 4154 | 0.90 | 1.86 | -0.30 | 0.20 | 470 | -0.30 | 0.26 | 36 | 7.86 | 8.33 | -0.53 | 0.47 | 5.4 | G | 1 |
| hd011559 | E | 4947 | 2.80 | 1.33 | 0.17 | 0.11 | 550 | 0.33 | 0.20 | 58 | 8.24 | 8.83 | 0.42 | 0.55 | 0.16 | 4947 | 2.54 | 1.37 | 0.12 | 0.12 | 550 | 0.12 | 0.21 | 58 | 8.06 | 8.61 | 0.42 | 0.17 | 5.5 | G | 1 |
| hd013174 | E | 6710 | 3.25 | 6.00 | 0.52 | 0.43 | 24 | -0.02 | 0.02 | 2 | 7.62 | 8.70 | 2.66 | 2.40 | 0.07 | 6710 | 3.49 | 7.75 | 0.47 | 0.39 | 24 | -0.01 | 0.12 | 2 | 7.66 | 8.74 | 2.66 | 0.05 | 80.0 | R | 2 |
| hd013480 | E | 5082 | 2.68 | 2.89 | -0.02 | 0.37 | 104 | 0.39 | 0.67 | 10 | 8.24 | 9.25 | 1.82 | 7.61 | 0.14 | 5082 | 1.72 | 2.92 | -0.06 | 0.37 | 104 | -0.06 | 0.67 | 10 | 8.01 | 8.80 | 1.83 | 0.19 | 36.0 | R | 1 |
| hd013520 | E | 4010 | 1.22 | 1.96 | -0.15 | 0.24 | 448 | 0.20 | 0.26 | 32 | 8.21 | 8.76 | -0.96 | 0.65 | 0.41 | 4010 | 0.19 | 1.96 | -0.43 | 0.27 | 448 | -0.43 | 0.30 | 32 | 7.76 | 8.19 | -1.11 | 0.67 | 6.2 | G | 1 |
| hd015257 | E | 7079 | 3.79 | 4.31 | 0.38 | 0.27 | 32 | -0.01 | 0.28 | 3 | 8.08 | | | | 0.02 | 7079 | 4.80 | 4.16 | 0.36 | 0.27 | 32 | 0.36 | 0.29 | 3 | 8.34 | | | -0.46 | 60.0 | R | 2 |
| hd015596 | E | 4788 | 2.76 | 1.08 | -0.59 | 0.09 | 563 | -0.44 | 0.17 | 57 | 7.99 | 8.67 | 0.31 | 0.71 | 0.20 | 4788 | 2.38 | 1.21 | -0.65 | 0.09 | 563 | -0.65 | 0.18 | 57 | 7.87 | 8.49 | 0.31 | 0.22 | 4.1 | G | 1 |
| hd018885 | E | 4670 | 2.58 | 1.40 | 0.26 | 0.15 | 522 | 0.44 | 0.20 | 41 | 8.46 | 8.91 | 0.33 | 1.06 | 0.23 | 4670 | 2.14 | 1.51 | 0.16 | 0.15 | 522 | 0.16 | 0.21 | 41 | 8.33 | 8.71 | 0.34 | 0.28 | 5.2 | G | 1 |
| hd022764 | E | 4205 | 1.41 | 2.54 | 0.11 | 0.28 | 346 | 0.24 | 0.27 | 22 | 8.21 | 8.92 | -0.33 | 1.30 | 0.39 | 4205 | 1.08 | 2.52 | 0.04 | 0.28 | 346 | 0.04 | 0.27 | 22 | 8.10 | 8.78 | -0.35 | 0.44 | 8.3 | G | 1 |
| hd023249 | E | 4966 | 3.72 | 0.50 | 0.19 | 0.13 | 529 | 0.40 | 0.25 | 60 | 8.50 | 8.74 | 1.15 | 2.70 | 0.15 | 4966 | 3.27 | 0.84 | 0.13 | 0.10 | 529 | 0.13 | 0.25 | 60 | 8.36 | 8.51 | 1.15 | 0.15 | 4.1 | G | 1 |
| hd025602 | E | 4743 | 2.84 | 1.07 | -0.11 | 0.10 | 558 | 0.16 | 0.21 | 59 | 8.24 | 8.70 | 0.07 | 0.47 | 0.20 | 4743 | 2.16 | 1.29 | -0.24 | 0.11 | 558 | -0.24 | 0.22 | 59 | 8.05 | 8.38 | 0.08 | 0.26 | 4.6 | G | 1 |
| hd026659 | E | 5170 | 2.74 | 1.42 | -0.05 | 0.11 | 538 | -0.04 | 0.15 | 59 | 7.97 | 8.58 | 0.09 | 0.14 | 0.13 | 5170 | 2.71 | 1.43 | -0.06 | 0.11 | 538 | -0.06 | 0.16 | 59 | 7.96 | 8.56 | 0.09 | 0.13 | 6.4 | G | 1 |
| hd029139 | E | 3903 | 1.32 | 2.06 | -0.01 | 0.38 | 430 | 0.43 | 0.47 | 40 | 8.55 | 8.99 | -1.17 | 0.58 | 0.38 | 3903 | -0.15 | 2.03 | -0.37 | 0.39 | 430 | -0.36 | 0.47 | 40 | 8.00 | 8.34 | -1.35 | 0.86 | 7.0 | G | 1 |
| hd030834 | E | 4153 | 1.39 | 1.86 | -0.17 | 0.18 | 471 | 0.12 | 0.24 | 41 | 7.97 | 8.60 | 2.17 | 37.38 | 0.39 | 4153 | 0.66 | 1.88 | -0.37 | 0.19 | 471 | -0.37 | 0.27 | 41 | 7.61 | 8.16 | 2.09 | 0.51 | 5.6 | G | 1 |
| hd031444 | E | 5051 | 2.89 | 1.21 | -0.03 | 0.11 | 561 | 0.11 | 0.15 | 59 | 8.04 | 8.58 | 0.94 | 1.35 | 0.14 | 5051 | 2.58 | 1.33 | -0.07 | 0.12 | 561 | -0.07 | 0.16 | 59 | 7.96 | 8.44 | 0.95 | 0.15 | 5.5 | G | 1 |
| hd033419 | E | 4674 | 2.64 | 1.40 | 0.28 | 0.15 | 523 | 0.51 | 0.21 | 45 | 8.45 | 8.99 | 0.33 | 1.06 | 0.23 | 4674 | 2.07 | 1.55 | 0.15 | 0.16 | 523 | 0.15 | 0.23 | 45 | 8.28 | 8.73 | 0.34 | 0.28 | 5.2 | G | 1 |
| hd033618 | E | 4558 | 2.45 | 1.25 | 0.38 | 0.15 | 441 | 0.49 | 0.24 | 38 | 8.58 | 8.97 | 0.18 | 1.12 | 0.27 | 4558 | 2.17 | 1.36 | 0.29 | 0.15 | 441 | 0.29 | 0.25 | 38 | 8.48 | 8.84 | 0.19 | 0.30 | 5.3 | G | 1 |
| hd034029 | E | 5155 | 2.47 | 2.45 | -0.48 | 0.69 | 186 | -0.26 | 0.46 | 19 | | | 1.41 | 2.92 | 0.14 | 5155 | 1.93 | 2.47 | -0.49 | 0.69 | 186 | -0.49 | 0.46 | 19 | | | 1.41 | 0.16 | 4.5 | G | 1 |
| hd037160 | E | 4742 | 2.68 | 1.10 | -0.49 | 0.09 | 562 | -0.29 | 0.18 | 56 | 8.06 | 8.61 | 0.52 | 1.30 | 0.21 | 4742 | 2.20 | 1.25 | -0.56 | 0.10 | 562 | -0.56 | 0.20 | 56 | 7.91 | 8.39 | 0.52 | 0.26 | 4.5 | G | 1 |
| hd037638 | E | 5088 | 2.90 | 1.32 | 0.03 | 0.10 | 562 | 0.16 | 0.17 | 61 | 8.08 | 8.67 | 0.73 | 0.78 | 0.14 | 5088 | 2.62 | 1.41 | 0.00 | 0.10 | 562 | 0.00 | 0.18 | 61 | 8.01 | 8.53 | 0.74 | 0.15 | 5.3 | G | 1 |
| hd038309 | E | 6927 | 3.94 | 3.91 | -0.24 | 0.29 | 131 | -0.31 | 0.20 | 7 | 7.92 | 8.95 | 2.64 | 1.81 | -0.03 | 6927 | 4.16 | 3.90 | -0.24 | 0.29 | 131 | -0.24 | 0.20 | 7 | 7.98 | 9.01 | 2.63 | -0.10 | 20.0 | R | 2 |
| hd039070 | E | 5086 | 2.94 | 1.19 | 0.11 | 0.12 | 552 | 0.17 | 0.16 | 54 | 8.17 | 8.68 | 0.36 | 0.34 | 0.15 | 5086 | 2.80 | 1.26 | 0.09 | 0.12 | 552 | 0.09 | 0.16 | 54 | 8.13 | 8.61 | 0.36 | 0.14 | 5.7 | G | 1 |
| hd039833 | E | 5751 | 4.35 | 0.68 | 0.25 | 0.08 | 511 | 0.40 | 0.15 | 80 | 8.42 | 8.99 | 2.32 | 5.45 | 0.04 | 5751 | 4.03 | 1.04 | 0.21 | 0.08 | 511 | 0.21 | 0.16 | 80 | 8.30 | 8.85 | 2.33 | 0.04 | 4.7 | G | 2 |
| hd039910 | E | 4577 | 2.53 | 1.56 | 0.31 | 0.19 | 497 | 0.52 | 0.28 | 41 | 8.61 | 9.03 | 0.34 | 1.49 | 0.26 | 4577 | 2.02 | 1.67 | 0.19 | 0.19 | 497 | 0.19 | 0.29 | 41 | 8.44 | 8.79 | 0.34 | 0.31 | 5.3 | G | 1 |
| hd040801 | E | 4787 | 2.94 | 1.07 | -0.08 | 0.10 | 560 | 0.15 | 0.23 | 54 | 8.30 | 8.73 | 0.43 | 0.93 | 0.19 | 4787 | 2.38 | 1.29 | -0.19 | 0.11 | 560 | -0.19 | 0.25 | 54 | 8.14 | 8.46 | 0.44 | 0.22 | 4.5 | G | 1 |
| hd042341 | E | 4635 | 2.73 | 1.42 | 0.40 | 0.19 | 506 | 0.56 | 0.25 | 43 | 8.55 | 8.99 | 1.02 | 5.14 | 0.23 | 4635 | 2.32 | 1.57 | 0.28 | 0.19 | 506 | 0.28 | 0.27 | 43 | 8.42 | 8.80 | 1.02 | 0.27 | 5.2 | G | 1 |
| hd046374 | E | 4656 | 2.62 | 1.36 | 0.16 | 0.14 | 522 | 0.35 | 0.18 | 45 | 8.40 | 8.91 | 1.04 | 5.07 | 0.23 | 4656 | 2.15 | 1.51 | 0.05 | 0.14 | 522 | 0.05 | 0.19 | 45 | 8.24 | 8.69 | 1.05 | 0.28 | 4.9 | G | 1 |
| hd047138 | E | 6091 | 3.75 | 1.38 | 0.56 | 0.14 | 523 | 0.15 | 0.17 | 64 | 7.70 | 9.13 | 2.03 | 1.59 | 0.01 | 6091 | 4.61 | 0.81 | 0.63 | 0.17 | 523 | 0.63 | 0.19 | 64 | 7.99 | 9.00 | 2.02 | -0.04 | 5.7 | G | 2 |
| hd047366 | E | 4857 | 3.02 | 0.94 | 0.14 | 0.12 | 558 | 0.30 | 0.18 | 53 | 8.23 | 8.76 | 0.47 | 0.81 | 0.17 | 4857 | 2.63 | 1.23 | 0.02 | 0.11 | 558 | 0.02 | 0.19 | 53 | 8.12 | 8.58 | 0.47 | 0.19 | 4.6 | G | 1 |
| hd050522 | E | 5095 | 2.71 | 1.03 | 0.26 | 0.16 | 553 | 0.22 | 0.30 | 55 | 8.30 | 8.77 | 0.95 | 1.19 | 0.14 | 5095 | 2.81 | 0.97 | 0.28 | 0.17 | 553 | 0.28 | 0.29 | 55 | 8.33 | 8.81 | 0.95 | 0.14 | 5.9 | G | 1 |
| hd051000 | E | 5102 | 2.73 | 1.40 | -0.01 | 0.12 | 549 | 0.07 | 0.15 | 56 | 8.07 | 8.54 | 1.20 | 2.05 | 0.14 | 5102 | 2.56 | 1.44 | -0.02 | 0.13 | 549 | -0.02 | 0.16 | 56 | 8.03 | 8.46 | 1.20 | 0.14 | 6.5 | G | 1 |
| hd054079 | E | 4454 | 1.85 | 1.60 | -0.34 | 0.13 | 532 | -0.20 | 0.18 | 45 | 7.92 | 8.52 | -1.12 | 0.08 | 0.34 | 4454 | 1.49 | 1.63 | -0.39 | 0.13 | 532 | -0.39 | 0.19 | 45 | 7.81 | 8.35 | -1.13 | 0.37 | 5.1 | G | 1 |
| hd055280 | E | 4639 | 2.64 | 1.22 | 0.24 | 0.15 | 532 | 0.41 | 0.21 | 46 | 8.38 | 8.85 | 0.31 | 1.13 | 0.24 | 4639 | 2.31 | 1.30 | 0.13 | 0.14 | 532 | 0.13 | 0.22 | 46 | 8.18 | 8.61 | 0.31 | 0.27 | 4.9 | G | 1 |
| hd057727 | E | 5007 | 3.06 | 1.12 | 0.04 | 0.10 | 567 | 0.21 | 0.16 | 61 | 8.23 | 8.68 | 1.35 | 3.62 | 0.15 | 5007 | 2.64 | 1.32 | -0.04 | 0.10 | 567 | -0.03 | 0.18 | 61 | 8.10 | 8.47 | 1.35 | 0.16 | 4.8 | G | 1 |
| hd058207 | E | 4750 | 2.69 | 1.30 | 0.03 | 0.12 | 550 | 0.29 | 0.17 | 51 | 8.29 | 8.78 | 0.27 | 0.73 | 0.21 | 4750 | 2.08 | 1.47 | -0.07 | 0.12 | 550 | -0.07 | 0.18 | 51 | 8.12 | 8.49 | 0.28 | 0.27 | 5.0 | G | 1 |
| hd058923 | E | 7505 | 3.46 | 3.18 | 0.50 | 0.27 | 66 | 0.32 | 0.17 | 5 | 8.33 | | 3.21 | 2.33 | 0.18 | 7505 | 3.94 | 3.09 | 0.51 | 0.27 | 66 | 0.51 | 0.18 | 5 | 8.41 | 8.85 | 3.21 | -0.02 | 50.0 | R | 2 |
| hd060294 | E | 4580 | 2.62 | 1.23 | 0.26 | 0.16 | 527 | 0.48 | 0.22 | 42 | 8.46 | 8.90 | 0.09 | 0.84 | 0.25 | 4580 | 2.16 | 1.37 | 0.10 | 0.16 | 527 | 0.10 | 0.24 | 42 | 8.20 | 8.60 | 0.09 | 0.30 | 5.0 | G | 1 |
| hd060522 | E | 3884 | 1.35 | 2.13 | 0.17 | 0.37 | 389 | 0.59 | 0.34 | 22 | 8.63 | 8.95 | -0.38 | 3.48 | 0.37 | 3884 | 0.10 | 2.15 | -0.19 | 0.39 | 389 | -0.19 | 0.37 | 22 | 8.11 | 8.39 | -0.53 | 0.75 | 7.9 | G | 1 |
| hd061363 | E | 4713 | 2.42 | 1.38 | -0.18 | 0.10 | 553 | 0.08 | 0.16 | 53 | 8.09 | 8.67 | 0.00 | 0.43 | 0.25 | 4713 | 1.92 | 1.45 | -0.29 | 0.10 | 553 | -0.29 | 0.18 | 53 | 7.80 | 8.30 | -0.01 | 0.29 | 5.0 | G | 1 |
| hd062141 | E | 4922 | 2.85 | 1.23 | -0.01 | 0.11 | 554 | 0.15 | 0.16 | 52 | 8.11 | 8.66 | 1.11 | 2.72 | 0.17 | 4922 | 2.50 | 1.37 | -0.07 | 0.11 | 554 | -0.06 | 0.16 | 52 | 8.03 | 8.49 | 1.11 | 0.18 | 5.2 | G | 1 |
| hd062437 | E | 7600 | 3.72 | 2.60 | 0.47 | 0.18 | 54 | 0.44 | 0.19 | 20 | 8.54 | 8.85 | 3.45 | 3.28 | 0.11 | 7600 | 3.80 | 2.59 | 0.47 | 0.18 | 54 | 0.47 | 0.19 | 20 | 8.55 | 8.84 | 3.45 | 0.07 | 44.0 | R | 2 |

| | | | | | | | | | | | | | | | | | | | | | | | | | | | |
|---|---|---|---|---|---|---|---|---|---|---|---|---|---|---|---|---|---|---|---|---|---|---|---|---|---|---|---|
| hd062721 | E | 4002 | 1.48 | 1.76 | 0.00 | 0.30 | 441 | 0.39 | 0.29 | 32 | 8.55 | 8.93 | -1.43 | 0.22 | 0.37 | 4002 | 0.35 | 1.81 | -0.30 | 0.32 | 441 | -0.30 | 0.32 | 32 | 8.11 | 8.43 | -1.55 | 0.62 | 6.2 | G | 1 |
| hd064152 | E | 4928 | 2.75 | 1.32 | 0.11 | 0.12 | 549 | 0.29 | 0.23 | 56 | 8.23 | 8.75 | 0.75 | 1.23 | 0.17 | 4928 | 2.32 | 1.46 | 0.05 | 0.12 | 549 | 0.05 | 0.25 | 56 | 8.13 | 8.55 | 0.76 | 0.19 | 5.0 | G | 1 |
| hd074794 | E | 4648 | 2.54 | 1.40 | 0.22 | 0.15 | 525 | 0.44 | 0.22 | 45 | 8.45 | 8.91 | 0.04 | 0.60 | 0.24 | 4648 | 2.09 | 1.48 | 0.10 | 0.15 | 525 | 0.10 | 0.23 | 45 | 8.20 | 8.61 | 0.05 | 0.29 | 5.1 | G | 1 |
| hd081688 | E | 4760 | 2.49 | 1.43 | -0.21 | 0.10 | 554 | 0.03 | 0.14 | 55 | 8.22 | 8.71 | -0.26 | 0.21 | 0.22 | 4760 | 2.07 | 1.49 | -0.29 | 0.10 | 554 | -0.29 | 0.16 | 55 | 7.95 | 8.37 | -0.26 | 0.24 | 5.1 | G | 1 |
| hd082885 | E | 5412 | 4.39 | 0.50 | 0.36 | 0.09 | 494 | 0.56 | 0.15 | 55 | 8.56 | 8.52 | 0.78 | 0.40 | 0.10 | 5412 | 3.99 | 1.06 | 0.29 | 0.08 | 494 | 0.29 | 0.15 | 55 | 8.48 | 8.31 | 0.79 | 0.08 | 4.2 | G | 2 |
| hd089025 | E | 6977 | 2.91 | 6.50 | -0.29 | 0.23 | 46 | -0.35 | 0.14 | 6 | 8.11 | 8.82 | 2.87 | 2.77 | 0.13 | 6977 | 3.11 | 7.21 | -0.31 | 0.23 | 46 | -0.31 | 0.15 | 6 | 8.14 | 8.86 | 2.86 | 0.12 | 60.0 | R | 2 |
| hd093875 | E | 4556 | 2.46 | 1.25 | 0.35 | 0.15 | 468 | 0.44 | 0.20 | 36 | 8.61 | 9.01 | 0.15 | 1.04 | 0.27 | 4556 | 2.22 | 1.35 | 0.27 | 0.15 | 468 | 0.27 | 0.21 | 36 | 8.52 | 8.90 | 0.15 | 0.29 | 5.2 | G | 1 |
| hd094672 | E | 6446 | 3.88 | 1.87 | -0.11 | 0.09 | 265 | 0.00 | 0.18 | 52 | 8.17 | 8.84 | 1.97 | 0.84 | -0.01 | 6446 | 3.59 | 1.86 | -0.10 | 0.09 | 265 | -0.10 | 0.18 | 52 | 8.07 | 8.73 | 1.99 | 0.02 | 15.0 | R | 2 |
| hd095345 | E | 4519 | 2.02 | 1.56 | -0.04 | 0.15 | 520 | 0.08 | 0.20 | 44 | 8.11 | 8.69 | -0.05 | 0.77 | 0.32 | 4519 | 1.73 | 1.59 | -0.09 | 0.15 | 520 | -0.09 | 0.21 | 44 | 8.02 | 8.56 | -0.05 | 0.34 | 5.5 | G | 1 |
| hd098262 | E | 4113 | 1.17 | 1.86 | -0.09 | 0.20 | 453 | 0.14 | 0.25 | 39 | 8.05 | 8.64 | -0.09 | 3.13 | 0.42 | 4113 | 0.54 | 1.85 | -0.22 | 0.21 | 453 | -0.22 | 0.27 | 39 | 7.86 | 8.37 | -0.14 | 0.55 | 6.1 | G | 1 |
| hd098366 | E | 4697 | 2.70 | 1.12 | 0.06 | 0.13 | 545 | 0.27 | 0.23 | 55 | 8.31 | 8.81 | 0.53 | 1.51 | 0.22 | 4697 | 2.18 | 1.33 | -0.06 | 0.13 | 545 | -0.06 | 0.25 | 55 | 8.15 | 8.57 | 0.53 | 0.27 | 5.1 | G | 1 |
| hd107700 | E | 6115 | 3.02 | 0.50 | 0.38 | 0.18 | 475 | -0.38 | 0.19 | 65 | 7.34 | 8.63 | 2.18 | 2.15 | 0.04 | 6115 | 3.50 | 0.50 | 0.36 | 0.20 | 475 | -0.20 | 0.19 | 65 | 7.49 | 8.58 | 2.18 | 0.03 | 6.9 | G | 2 |
| hd107950 | E | 5098 | 2.45 | 1.78 | 0.07 | 0.14 | 514 | 0.15 | 0.15 | 57 | 7.95 | 8.59 | 0.20 | 0.22 | 0.15 | 5098 | 2.27 | 1.80 | 0.07 | 0.15 | 514 | 0.07 | 0.15 | 57 | 7.91 | 8.50 | 0.20 | 0.16 | 7.4 | G | 1 |
| hd112127 | E | 4383 | 2.60 | 1.73 | 0.49 | 0.24 | 439 | 0.82 | 0.25 | 31 | 9.01 | 9.22 | 3.24 | 44.83 | 0.29 | 4383 | 1.79 | 1.94 | 0.24 | 0.25 | 439 | 0.24 | 0.25 | 31 | 8.62 | 8.80 | 3.11 | 0.35 | 5.9 | G | 1 |
| hd115604 | E | 7314 | 3.43 | 2.09 | 0.54 | 0.10 | 398 | 0.57 | 0.10 | 37 | 8.44 | 9.07 | 1.99 | 0.21 | 0.15 | 7314 | 3.34 | 2.09 | 0.54 | 0.10 | 398 | 0.54 | 0.10 | 37 | 8.42 | 9.04 | 1.99 | 0.17 | 5.7 | G | 2 |
| hd116292 | E | 4840 | 2.61 | 1.36 | -0.03 | 0.12 | 551 | 0.14 | 0.23 | 65 | 8.32 | 9.10 | 1.36 | 5.69 | 0.19 | 4840 | 2.21 | 1.45 | -0.08 | 0.12 | 551 | -0.08 | 0.24 | 65 | 8.20 | 8.91 | 1.37 | 0.22 | 5.2 | G | 1 |
| hd116515 | E | 4728 | 2.32 | 1.44 | -0.08 | 0.12 | 541 | 0.05 | 0.20 | 51 | 8.16 | 8.65 | 0.43 | 1.08 | 0.25 | 4728 | 2.01 | 1.49 | -0.13 | 0.12 | 541 | -0.13 | 0.21 | 51 | 8.08 | 8.50 | 0.43 | 0.28 | 5.0 | G | 1 |
| hd117710 | E | 4661 | 2.76 | 1.25 | 0.36 | 0.18 | 520 | 0.53 | 0.24 | 46 | 8.46 | 8.81 | 0.08 | 0.62 | 0.22 | 4661 | 2.29 | 1.46 | 0.21 | 0.18 | 520 | 0.21 | 0.25 | 46 | 8.31 | 8.59 | 0.08 | 0.27 | 5.1 | G | 1 |
| hd120084 | E | 4806 | 2.61 | 1.34 | 0.15 | 0.13 | 541 | 0.28 | 0.20 | 54 | 8.21 | 8.69 | 0.41 | 0.83 | 0.20 | 4806 | 2.29 | 1.44 | 0.09 | 0.13 | 541 | 0.09 | 0.21 | 54 | 8.12 | 8.54 | 0.41 | 0.22 | 5.2 | G | 1 |
| hd130952 | E | 4744 | 2.46 | 1.44 | -0.29 | 0.12 | 532 | -0.14 | 0.15 | 50 | 8.12 | 8.71 | -0.25 | 0.23 | 0.24 | 4744 | 2.10 | 1.52 | -0.33 | 0.12 | 532 | -0.33 | 0.16 | 50 | 8.00 | 8.55 | -0.25 | 0.27 | 6.3 | G | 1 |
| hd131873 | E | 4005 | 1.33 | 1.83 | -0.02 | 0.22 | 450 | 0.26 | 0.24 | 31 | 8.28 | 8.72 | -0.82 | 0.90 | 0.39 | 4005 | 0.58 | 1.86 | -0.21 | 0.24 | 450 | -0.21 | 0.26 | 31 | 8.02 | 8.40 | -0.87 | 0.55 | 6.0 | G | 1 |
| hd136202 | E | 6025 | 3.87 | 1.34 | -0.02 | 0.06 | 367 | 0.06 | 0.11 | 68 | 8.39 | 8.74 | 0.37 | 0.04 | 0.01 | 6025 | 3.69 | 1.41 | -0.02 | 0.06 | 367 | -0.02 | 0.11 | 68 | 8.33 | 8.81 | 0.37 | 0.02 | 5.9 | G | 2 |
| hd136514 | E | 4417 | 2.33 | 1.38 | 0.16 | 0.19 | 507 | 0.33 | 0.28 | 44 | 8.37 | 8.78 | -0.03 | 1.12 | 0.30 | 4417 | 1.86 | 1.54 | 0.01 | 0.19 | 507 | 0.01 | 0.29 | 44 | 8.20 | 8.57 | -0.04 | 0.34 | 5.3 | G | 1 |
| hd139254 | E | 4676 | 2.65 | 1.20 | 0.11 | 0.14 | 532 | 0.29 | 0.17 | 48 | 8.28 | 8.77 | 1.14 | 5.76 | 0.23 | 4676 | 2.19 | 1.39 | -0.01 | 0.14 | 532 | -0.01 | 0.18 | 48 | 8.14 | 8.56 | 1.14 | 0.27 | 5.0 | G | 1 |
| hd143553 | E | 4697 | 2.66 | 1.14 | -0.14 | 0.11 | 550 | 0.08 | 0.20 | 56 | 8.22 | 8.63 | 0.35 | 1.01 | 0.22 | 4697 | 2.26 | 1.21 | -0.25 | 0.11 | 550 | -0.25 | 0.22 | 56 | 7.95 | 8.31 | 0.35 | 0.26 | 4.6 | G | 1 |
| hd148604 | E | 5167 | 2.91 | 1.04 | -0.04 | 0.10 | 566 | -0.04 | 0.16 | 62 | 8.02 | 8.68 | 1.14 | 1.57 | 0.12 | 5167 | 2.91 | 1.04 | -0.04 | 0.10 | 566 | -0.04 | 0.16 | 62 | 8.02 | 8.68 | 1.14 | 0.12 | 5.1 | G | 1 |
| hd148856 | E | 4903 | 2.32 | 1.57 | -0.04 | 0.12 | 533 | 0.12 | 0.17 | 53 | 8.11 | 8.59 | 0.72 | 1.21 | 0.20 | 4903 | 1.95 | 1.62 | -0.07 | 0.13 | 533 | -0.07 | 0.18 | 53 | 8.04 | 8.41 | 0.72 | 0.21 | 6.2 | G | 1 |
| hd148897 | E | 4167 | 1.32 | 1.83 | -1.07 | 0.11 | 505 | -0.56 | 0.18 | 50 | 7.36 | 8.27 | -1.24 | 0.19 | 0.40 | 4167 | -0.60 | 1.68 | -1.27 | 0.20 | 505 | -1.14 | 0.36 | 50 | 6.88 | 7.43 | -1.61 | 0.83 | 5.6 | G | 2 |
| hd149161 | E | 3958 | 1.40 | 1.87 | -0.04 | 0.30 | 433 | 0.48 | 0.34 | 33 | 8.48 | 8.86 | -0.77 | 1.17 | 0.37 | 3958 | -0.18 | 1.87 | -0.42 | 0.34 | 433 | -0.42 | 0.38 | 33 | 7.94 | 8.18 | -0.97 | 0.84 | 6.6 | G | 1 |
| hd150557 | E | 6744 | 3.78 | 3.83 | 0.10 | 0.25 | 40 | -0.02 | 0.06 | 4 | | | 2.92 | 4.04 | 0.01 | 6744 | 4.10 | 3.83 | 0.10 | 0.25 | 40 | 0.10 | 0.06 | 4 | 8.33 | 8.49 | 2.92 | -0.07 | 70.0 | R | 2 |
| hd153210 | E | 4559 | 2.45 | 1.48 | 0.29 | 0.18 | 502 | 0.52 | 0.24 | 44 | 8.58 | 9.03 | 0.04 | 0.80 | 0.27 | 4559 | 1.87 | 1.60 | 0.16 | 0.18 | 502 | 0.16 | 0.26 | 44 | 8.39 | 8.77 | 0.04 | 0.32 | 5.5 | G | 1 |
| hd153956 | E | 4541 | 2.60 | 1.61 | 0.34 | 0.20 | 473 | 0.57 | 0.24 | 39 | 8.65 | 9.03 | 0.29 | 1.49 | 0.26 | 4541 | 2.03 | 1.73 | 0.20 | 0.20 | 473 | 0.20 | 0.26 | 39 | 8.45 | 8.77 | 0.29 | 0.31 | 5.9 | G | 1 |
| hd159353 | E | 4832 | 2.63 | 1.31 | 0.07 | 0.11 | 550 | 0.20 | 0.18 | 54 | 8.22 | 8.73 | 0.81 | 1.85 | 0.19 | 4832 | 2.32 | 1.40 | 0.02 | 0.11 | 550 | 0.02 | 0.19 | 54 | 8.14 | 8.59 | 0.81 | 0.21 | 5.2 | G | 1 |
| hd159876 | E | 7217 | 3.65 | 3.61 | 0.00 | 0.13 | 190 | 0.06 | 0.14 | 47 | 8.18 | 8.80 | 2.87 | 1.85 | 0.08 | 7217 | 3.47 | 3.59 | 0.01 | 0.13 | 190 | 0.01 | 0.14 | 47 | 8.15 | 8.75 | 2.87 | 0.12 | 24.0 | R | 2 |
| hd160507 | E | 4921 | 2.79 | 1.33 | 0.07 | 0.13 | 542 | 0.26 | 0.22 | 52 | 8.41 | 8.72 | 0.38 | 0.54 | 0.17 | 4921 | 2.34 | 1.46 | 0.01 | 0.14 | 542 | 0.00 | 0.23 | 52 | 8.30 | 8.50 | 0.38 | 0.19 | 5.3 | G | 1 |
| hd161074 | E | 4033 | 1.59 | 1.72 | 0.15 | 0.28 | 434 | 0.56 | 0.35 | 30 | 8.62 | 8.99 | -0.67 | 1.10 | 0.36 | 4033 | 0.44 | 1.81 | -0.17 | 0.31 | 434 | -0.17 | 0.37 | 30 | 8.15 | 8.46 | -0.78 | 0.59 | 5.9 | G | 1 |
| hd162757 | E | 4621 | 2.65 | 1.21 | 0.05 | 0.14 | 532 | 0.30 | 0.18 | 46 | 8.27 | 8.77 | 0.16 | 0.85 | 0.24 | 4621 | 2.01 | 1.42 | -0.10 | 0.14 | 532 | -0.11 | 0.21 | 46 | 8.07 | 8.48 | 0.16 | 0.30 | 4.7 | G | 1 |
| hd163588 | E | 4459 | 2.38 | 1.30 | 0.20 | 0.18 | 509 | 0.39 | 0.24 | 39 | 8.36 | 8.75 | -0.15 | 0.74 | 0.29 | 4459 | 1.87 | 1.50 | 0.03 | 0.18 | 509 | 0.04 | 0.26 | 39 | 8.18 | 8.52 | -0.16 | 0.34 | 5.0 | G | 1 |
| hd164058 | E | 3925 | 1.25 | 2.02 | 0.15 | 0.29 | 404 | 0.66 | 0.34 | 29 | 8.46 | 8.83 | -0.61 | 1.90 | 0.39 | 3925 | -0.26 | 2.01 | -0.22 | 0.32 | 404 | -0.23 | 0.36 | 29 | 7.94 | 8.18 | -0.81 | 0.89 | 6.7 | G | 1 |
| hd166208 | E | 5040 | 2.46 | 1.58 | 0.17 | 0.14 | 532 | 0.23 | 0.23 | 56 | 7.62 | 8.73 | 1.81 | 7.95 | 0.16 | 5040 | 2.33 | 1.60 | 0.16 | 0.14 | 532 | 0.16 | 0.23 | 56 | 7.59 | 8.67 | 1.82 | 0.17 | 6.2 | G | 1 |
| hd168723 | E | 4858 | 2.99 | 1.02 | -0.11 | 0.10 | 569 | 0.11 | 0.18 | 61 | 8.19 | 8.67 | 0.09 | 0.35 | 0.18 | 4858 | 2.47 | 1.23 | -0.20 | 0.10 | 569 | -0.20 | 0.20 | 61 | 8.05 | 8.42 | 0.10 | 0.20 | 4.5 | G | 1 |
| hd169268a | E | 6812 | 3.90 | 1.48 | -0.53 | 0.20 | 218 | -0.51 | 0.28 | 53 | 8.17 | 8.66 | 2.66 | 2.26 | -0.02 | 6812 | 3.86 | 1.48 | -0.53 | 0.20 | 218 | -0.53 | 0.28 | 53 | 8.16 | 8.65 | 2.66 | -0.01 | 20.0 | R | 2 |
| hd172748 | E | 6832 | 3.46 | 4.09 | 0.04 | 0.10 | 187 | 0.11 | 0.11 | 33 | 8.30 | 8.75 | 2.41 | 1.18 | 0.07 | 6832 | 3.25 | 4.08 | 0.04 | 0.10 | 187 | 0.04 | 0.11 | 33 | 8.24 | 8.69 | 2.41 | 0.09 | 24.0 | R | 2 |
| hd175305 | E | 4981 | 2.70 | 1.30 | -1.40 | 0.10 | 404 | -1.24 | 0.13 | 51 | 7.12 | 7.87 | 0.99 | 1.82 | 0.16 | 4981 | 2.33 | 1.40 | -1.41 | 0.11 | 404 | -1.41 | 0.13 | 51 | 7.07 | 7.71 | 1.00 | 0.18 | 5.0 | G | 1 |
| hd176408 | E | 4517 | 2.48 | 1.37 | 0.24 | 0.17 | 515 | 0.45 | 0.25 | 40 | 8.51 | 8.94 | 0.21 | 1.37 | 0.28 | 4517 | 2.05 | 1.46 | 0.10 | 0.17 | 515 | 0.10 | 0.27 | 40 | 8.27 | 8.65 | 0.21 | 0.31 | 5.2 | G | 1 |
| hd178596 | E | 6784 | 3.79 | 3.38 | -0.16 | 0.22 | 48 | -0.24 | 0.04 | 3 | 8.33 | 9.04 | 2.53 | 1.74 | 0.01 | 6784 | 4.01 | 3.35 | -0.16 | 0.21 | 48 | -0.16 | 0.04 | 3 | 8.39 | 8.86 | 2.53 | -0.05 | 40.0 | R | 2 |
| hd181214 | E | 6238 | 3.63 | 2.61 | 0.03 | 0.18 | 289 | 0.13 | 0.16 | 36 | 8.41 | 8.82 | 3.03 | 9.24 | 0.02 | 6238 | 3.38 | 2.66 | 0.02 | 0.18 | 289 | 0.02 | 0.15 | 36 | 8.32 | 8.74 | 3.02 | 0.03 | 26.0 | R | 2 |
| hd181984 | E | 4413 | 2.38 | 1.75 | 0.43 | 0.25 | 456 | 0.70 | 0.35 | 40 | 8.79 | 9.15 | 0.07 | 1.43 | 0.30 | 4413 | 1.69 | 1.88 | 0.24 | 0.25 | 456 | 0.24 | 0.36 | 40 | 8.52 | 8.83 | 0.06 | 0.36 | 5.9 | G | 1 |
| hd184406 | E | 4487 | 2.75 | 1.16 | 0.39 | 0.20 | 503 | 0.71 | 0.33 | 46 | 8.69 | 9.04 | -0.05 | 0.84 | 0.26 | 4487 | 1.96 | 1.51 | 0.07 | 0.21 | 503 | 0.07 | 0.35 | 46 | 8.29 | 8.58 | -0.06 | 0.33 | 5.2 | G | 1 |
| hd185351 | E | 4941 | 3.18 | 0.99 | 0.12 | 0.11 | 557 | 0.31 | 0.18 | 57 | 8.24 | 8.63 | 0.79 | 1.32 | 0.16 | 4941 | 2.72 | 1.31 | 0.00 | 0.11 | 557 | 0.00 | 0.21 | 57 | 8.13 | 8.40 | 0.80 | 0.17 | 4.7 | G | 1 |

| ID | Type | | | | | | | | | | | | | | | | | | | | | | | | | | | | | | |
|---|---|---|---|---|---|---|---|---|---|---|---|---|---|---|---|---|---|---|---|---|---|---|---|---|---|---|---|---|---|---|---|
| hd185644 | E | 4579 | 2.47 | 1.23 | 0.19 | 0.15 | 524 | 0.30 | 0.20 | 41 | 8.26 | 8.71 | 0.13 | 0.92 | 0.27 | 4579 | 2.18 | 1.38 | 0.10 | 0.15 | 524 | 0.09 | 0.22 | 41 | 8.17 | 8.58 | 0.13 | 0.29 | 5.0 | G | 1 |
| hd187764 | E | 6919 | 3.26 | 6.50 | -0.88 | 0.27 | 31 | 0.06 | 0.29 | 3 | 7.73 | 9.03 | 2.33 | 1.01 | 0.10 | 6919 | 1.20 | 5.75 | -0.77 | 0.27 | 31 | -0.52 | 0.32 | 3 | 7.41 | 9.00 | 2.44 | 0.00 | 50.0 | R | 2 |
| hd188119 | E | 4945 | 2.61 | 1.39 | -0.34 | 0.08 | 559 | -0.22 | 0.14 | 59 | 7.91 | 8.49 | 0.14 | 0.30 | 0.17 | 4945 | 2.33 | 1.45 | -0.36 | 0.08 | 559 | -0.36 | 0.15 | 59 | 7.85 | 8.36 | 0.14 | 0.19 | 4.9 | G | 1 |
| hd188947 | E | 4783 | 2.56 | 1.41 | 0.13 | 0.14 | 535 | 0.22 | 0.19 | 48 | 8.22 | 8.67 | 0.45 | 0.97 | 0.21 | 4783 | 2.35 | 1.47 | 0.09 | 0.14 | 535 | 0.09 | 0.19 | 48 | 8.17 | 8.57 | 0.45 | 0.22 | 5.3 | G | 1 |
| hd189319 | E | 3862 | 1.21 | 2.25 | 0.19 | 0.33 | 370 | 0.69 | 0.33 | 22 | 8.46 | 8.82 | -0.10 | 6.56 | 0.39 | 3862 | -0.30 | 2.25 | -0.21 | 0.36 | 370 | -0.19 | 0.36 | 22 | 7.93 | 8.17 | -0.32 | 0.96 | 7.7 | G | 1 |
| hd192787 | E | 4888 | 2.70 | 1.27 | -0.06 | 0.10 | 558 | 0.16 | 0.17 | 62 | 8.13 | 8.64 | 0.50 | 0.79 | 0.18 | 4888 | 2.17 | 1.40 | -0.13 | 0.11 | 558 | -0.13 | 0.18 | 62 | 8.02 | 8.39 | 0.51 | 0.21 | 5.1 | G | 1 |
| hd192836 | E | 4805 | 2.72 | 1.25 | 0.24 | 0.14 | 544 | 0.33 | 0.22 | 44 | 8.38 | 8.84 | 1.46 | 7.41 | 0.19 | 4805 | 2.51 | 1.33 | 0.20 | 0.14 | 544 | 0.19 | 0.22 | 44 | 8.32 | 8.74 | 1.46 | 0.20 | 4.9 | G | 1 |
| hd196134 | E | 4770 | 2.88 | 1.04 | -0.01 | 0.11 | 562 | 0.24 | 0.22 | 56 | 8.18 | 8.74 | 0.15 | 0.52 | 0.19 | 4770 | 2.27 | 1.30 | -0.14 | 0.12 | 562 | -0.14 | 0.25 | 56 | 8.01 | 8.45 | 0.16 | 0.23 | 4.6 | G | 1 |
| hd197989 | E | 4713 | 2.46 | 1.38 | -0.06 | 0.12 | 550 | 0.12 | 0.19 | 54 | 8.19 | 8.71 | -0.01 | 0.43 | 0.24 | 4713 | 2.02 | 1.47 | -0.13 | 0.12 | 550 | -0.13 | 0.21 | 54 | 8.07 | 8.51 | -0.01 | 0.28 | 5.1 | G | 1 |
| hd198431 | E | 4652 | 2.64 | 1.20 | -0.04 | 0.13 | 546 | 0.16 | 0.22 | 53 | 8.30 | 8.74 | 0.15 | 0.75 | 0.23 | 4652 | 2.15 | 1.38 | -0.15 | 0.13 | 546 | -0.15 | 0.23 | 53 | 8.14 | 8.51 | 0.15 | 0.28 | 4.7 | G | 1 |
| hd199178 | E | 5180 | 3.35 | 3.61 | 0.34 | 0.38 | 67 | 0.27 | 0.30 | 7 | 8.16 | 8.83 | 1.76 | 5.64 | 0.12 | 5180 | 3.53 | 3.61 | 0.35 | 0.38 | 67 | 0.35 | 0.30 | 7 | | | 1.76 | 0.12 | 90.0 | R | 1 |
| hd202109 | E | 4892 | 2.45 | 1.61 | 0.05 | 0.12 | 533 | 0.21 | 0.19 | 52 | 8.29 | 8.68 | 0.59 | 0.95 | 0.19 | 4892 | 2.08 | 1.67 | 0.02 | 0.12 | 533 | 0.02 | 0.20 | 52 | 8.21 | 8.50 | 0.60 | 0.21 | 5.5 | G | 1 |
| hd209747 | E | 4068 | 1.73 | 2.16 | 0.21 | 0.30 | 411 | 0.70 | 0.46 | 36 | 8.72 | 8.99 | -0.62 | 1.07 | 0.35 | 4068 | 0.37 | 2.19 | -0.13 | 0.33 | 411 | -0.13 | 0.45 | 36 | 8.17 | 8.38 | -0.75 | 0.60 | 6.5 | G | 1 |
| hd209960 | E | 4112 | 1.84 | 1.84 | 0.32 | 0.27 | 423 | 0.78 | 0.43 | 34 | 8.70 | 9.06 | -0.64 | 0.83 | 0.34 | 4112 | 0.69 | 1.92 | -0.04 | 0.29 | 423 | -0.04 | 0.43 | 34 | 8.17 | 8.47 | -0.73 | 0.51 | 6.1 | G | 1 |
| hd212943 | E | 4630 | 2.69 | 1.09 | -0.06 | 0.13 | 550 | 0.17 | 0.21 | 48 | 8.35 | 8.78 | 0.10 | 0.72 | 0.23 | 4630 | 2.12 | 1.31 | -0.20 | 0.13 | 550 | -0.20 | 0.23 | 48 | 8.17 | 8.51 | 0.10 | 0.29 | 4.8 | G | 1 |
| hd214448 | E | 5272 | 2.99 | 0.88 | 0.29 | 0.15 | 559 | 0.15 | 0.22 | 62 | 8.21 | 8.80 | 0.95 | 0.77 | 0.10 | 5272 | 3.33 | 0.47 | 0.40 | 0.17 | 559 | 0.40 | 0.22 | 62 | 8.30 | 8.96 | 0.94 | 0.10 | 5.6 | G | 1 |
| hd214470 | E | 6668 | 3.61 | 2.79 | 0.51 | 0.13 | 15 | 0.22 | 0.12 | 2 | | | 2.74 | 2.88 | 0.03 | 6668 | 4.40 | 2.69 | 0.50 | 0.13 | 15 | 0.50 | 0.13 | 2 | 8.64 | 9.24 | 2.72 | -0.14 | 62.0 | R | 2 |
| hd216131 | E | 4934 | 2.78 | 1.25 | 0.02 | 0.10 | 563 | 0.21 | 0.17 | 61 | 8.14 | 8.68 | 0.37 | 0.52 | 0.17 | 4934 | 2.33 | 1.39 | -0.04 | 0.10 | 563 | -0.05 | 0.19 | 61 | 8.05 | 8.47 | 0.38 | 0.19 | 4.8 | G | 1 |
| hd216228 | E | 4742 | 2.57 | 1.42 | 0.09 | 0.13 | 538 | 0.24 | 0.21 | 52 | 8.24 | 8.76 | 0.29 | 0.77 | 0.22 | 4742 | 2.15 | 1.51 | 0.02 | 0.13 | 538 | 0.03 | 0.22 | 52 | 8.13 | 8.59 | 0.29 | 0.26 | 5.0 | G | 1 |
| hd219418 | E | 5077 | 2.80 | 1.40 | -0.33 | 0.10 | 557 | -0.14 | 0.13 | 64 | 7.94 | 8.52 | -0.29 | 0.08 | 0.14 | 5077 | 2.36 | 1.51 | -0.35 | 0.10 | 557 | -0.36 | 0.14 | 64 | 7.83 | 8.31 | -0.29 | 0.16 | 5.4 | G | 1 |
| hd219916 | E | 5124 | 2.93 | 1.27 | 0.10 | 0.10 | 561 | 0.10 | 0.15 | 58 | 8.06 | 8.73 | 0.38 | 0.32 | 0.13 | 5124 | 2.93 | 1.27 | 0.10 | 0.10 | 561 | 0.10 | 0.15 | 58 | 8.06 | 8.73 | 0.38 | 0.13 | 5.0 | G | 1 |
| hd220954 | E | 4684 | 2.58 | 1.35 | 0.25 | 0.15 | 526 | 0.46 | 0.22 | 49 | 8.38 | 8.87 | 0.22 | 0.80 | 0.23 | 4684 | 2.06 | 1.49 | 0.14 | 0.15 | 526 | 0.14 | 0.24 | 49 | 8.23 | 8.63 | 0.23 | 0.28 | 5.4 | G | 1 |
| hd221148 | E | 4660 | 3.21 | 1.00 | 0.59 | 0.23 | 489 | 0.77 | 0.27 | 39 | 8.79 | 9.09 | 0.40 | 1.29 | 0.20 | 4660 | 2.72 | 1.51 | 0.34 | 0.21 | 489 | 0.34 | 0.30 | 39 | 8.57 | 8.84 | 0.40 | 0.23 | 4.8 | G | 1 |
| aiscl | H | 7128 | 3.80 | 5.97 | -0.63 | 0.30 | 15 | -0.93 | 0.35 | 4 | 7.72 | 9.03 | 2.78 | 1.94 | 0.02 | 7128 | 4.62 | 6.14 | -0.67 | 0.30 | 15 | -0.67 | 0.35 | 4 | 7.95 | 9.28 | 2.74 | -0.37 | 50.0 | R | 2 |
| anscl | H | 4968 | 2.50 | 2.53 | -0.10 | 0.20 | 224 | -0.46 | 0.18 | 16 | 8.07 | 8.29 | 1.94 | 11.68 | 0.18 | 4968 | 3.32 | 2.44 | -0.02 | 0.19 | 224 | -0.02 | 0.17 | 16 | 8.24 | 8.69 | 1.93 | 0.15 | 14.5 | G | 2 |
| bppsc | H | 4050 | 2.50 | 3.54 | -0.43 | 0.39 | 54 | 1.45 | 0.91 | 6 | | 0.28 | 7.56 | 0.32 | 4050 | 0.20 | 3.70 | -1.05 | 0.43 | 54 | 0.05 | 0.83 | 6 | | | 0.09 | 0.65 | 32.0 | R | 2 |
| cd-33_2771 | H | 4002 | 2.50 | 1.68 | 0.25 | 0.40 | 428 | 1.08 | 0.31 | 24 | 8.98 | 9.26 | -1.17 | 0.32 | 0.32 | 4002 | 0.20 | 1.91 | -0.53 | 0.46 | 428 | -0.52 | 0.37 | 24 | 7.82 | 8.02 | -1.43 | 0.67 | 2.8 | G | 2 |
| cd-3814203b | H | 6297 | 2.50 | 1.58 | 0.39 | 0.07 | 550 | -0.40 | 0.13 | 95 | 7.95 | 8.29 | 1.70 | 0.52 | 0.03 | 6297 | 4.31 | 1.12 | 0.42 | 0.08 | 550 | 0.42 | 0.14 | 95 | 8.49 | 8.81 | 1.68 | -0.05 | 4.0 | G | 2 |
| cfscl | H | 5127 | 2.50 | 1.82 | -0.13 | 0.25 | 311 | -0.81 | 0.36 | 19 | 7.85 | 8.18 | 1.96 | 8.72 | 0.15 | 5127 | 4.00 | 0.92 | 0.13 | 0.28 | 311 | 0.13 | 0.32 | 19 | 8.21 | 8.90 | 1.95 | 0.13 | 12.5 | G | 2 |
| epsret | H | 4702 | 3.34 | 0.50 | 0.46 | 0.09 | 463 | 0.68 | 0.23 | 66 | 8.68 | 8.94 | 0.23 | 0.78 | 0.19 | 4702 | 2.86 | 0.94 | 0.24 | 0.07 | 463 | 0.25 | 0.24 | 66 | 8.39 | 8.61 | 0.23 | 0.21 | 2.6 | G | 1 |
| ereri | H | 4935 | 2.50 | 3.54 | -0.75 | 0.11 | 13 | 0.35 | 0.27 | 7 | | | 1.95 | 13.14 | 0.18 | 4935 | 0.10 | 3.96 | -0.75 | 0.11 | 13 | -0.64 | 0.23 | 7 | | | 1.94 | 0.32 | 33.0 | R | 2 |
| gamaps | H | 4957 | 2.70 | 1.29 | -0.01 | 0.08 | 612 | 0.07 | 0.12 | 60 | 8.04 | 8.60 | 0.38 | 0.49 | 0.17 | 4957 | 2.52 | 1.35 | -0.03 | 0.08 | 612 | -0.04 | 0.12 | 60 | 8.01 | 8.52 | 0.38 | 0.17 | 4.0 | G | 1 |
| hd000344 | H | 4584 | 2.47 | 1.27 | 0.12 | 0.10 | 592 | 0.30 | 0.14 | 59 | 8.31 | 8.79 | 0.10 | 0.85 | 0.27 | 4584 | 2.00 | 1.43 | 0.00 | 0.10 | 592 | 0.00 | 0.16 | 59 | 8.16 | 8.58 | 0.10 | 0.31 | 3.5 | G | 1 |
| hd000483a | H | 5718 | 3.92 | 0.50 | -0.40 | 0.25 | 420 | -0.41 | 0.33 | 47 | 8.19 | 8.79 | 1.32 | 0.71 | 0.04 | 5718 | 3.90 | 0.50 | -0.39 | 0.25 | 420 | -0.41 | 0.33 | 47 | 8.18 | 8.78 | 1.32 | 0.04 | 7.5 | G | 2 |
| hd000483b | H | 5757 | 3.93 | 0.50 | -0.24 | 0.26 | 466 | -0.27 | 0.27 | 59 | 8.15 | 8.75 | 1.35 | 0.70 | 0.20 | 5757 | 3.95 | 0.50 | -0.24 | 0.26 | 466 | -0.26 | 0.27 | 59 | 8.16 | 8.75 | 1.35 | 0.04 | 7.7 | G | 2 |
| hd000770 | H | 4710 | 2.69 | 1.30 | -0.03 | 0.08 | 600 | 0.23 | 0.15 | 61 | 8.34 | 8.84 | 0.27 | 0.82 | 0.22 | 4710 | 2.05 | 1.46 | -0.14 | 0.09 | 600 | -0.14 | 0.16 | 61 | 8.15 | 8.54 | 0.28 | 0.28 | 3.6 | G | 1 |
| hd001690 | H | 4219 | 2.43 | 1.29 | 0.13 | 0.13 | 556 | 0.61 | 0.18 | 48 | 8.56 | 9.05 | -0.97 | 0.27 | 0.31 | 4219 | 1.10 | 1.61 | -0.27 | 0.14 | 556 | -0.27 | 0.22 | 48 | 8.06 | 8.46 | -1.01 | 0.43 | 3.9 | G | 1 |
| hd002114 | H | 5148 | 2.66 | 1.59 | -0.02 | 0.07 | 603 | 0.14 | 0.11 | 63 | 8.18 | 8.72 | 0.83 | 0.81 | 0.13 | 5148 | 2.28 | 1.65 | -0.04 | 0.08 | 603 | -0.04 | 0.11 | 63 | 8.10 | 8.54 | 0.83 | 0.15 | 4.4 | G | 1 |
| hd002529 | H | 4621 | 2.50 | 1.15 | 0.05 | 0.09 | 601 | 0.17 | 0.17 | 60 | 8.26 | 8.74 | 0.02 | 0.63 | 0.26 | 4621 | 2.19 | 1.27 | -0.03 | 0.09 | 601 | -0.03 | 0.18 | 60 | 8.16 | 8.59 | 0.02 | 0.29 | 3.3 | G | 1 |
| hd003488 | H | 4827 | 2.53 | 1.31 | -0.05 | 0.08 | 607 | 0.03 | 0.13 | 62 | 8.10 | 8.66 | 0.36 | 0.69 | 0.20 | 4827 | 2.36 | 1.36 | -0.07 | 0.08 | 607 | -0.07 | 0.14 | 62 | 8.06 | 8.58 | 0.36 | 0.21 | 3.5 | G | 1 |
| hd003919 | H | 4952 | 2.62 | 1.22 | 0.12 | 0.08 | 609 | 0.03 | 0.13 | 61 | 8.19 | 8.73 | 0.54 | 0.72 | 0.17 | 4952 | 2.71 | 1.24 | 0.14 | 0.08 | 609 | 0.14 | 0.13 | 61 | 8.32 | 8.86 | 0.55 | 0.17 | 3.8 | G | 1 |
| hd004211 | H | 4531 | 2.38 | 1.32 | 0.14 | 0.13 | 568 | 0.29 | 0.18 | 51 | 8.51 | 8.96 | -0.15 | 0.58 | 0.28 | 4531 | 1.99 | 1.45 | 0.04 | 0.12 | 568 | 0.04 | 0.19 | 51 | 8.36 | 8.78 | -0.15 | 0.32 | 4.0 | G | 1 |
| hd004737 | H | 5031 | 2.80 | 1.30 | -0.03 | 0.08 | 609 | 0.06 | 0.13 | 60 | 8.12 | 8.67 | 1.22 | 2.62 | 0.15 | 5031 | 2.60 | 1.37 | -0.06 | 0.08 | 609 | -0.06 | 0.13 | 60 | 8.07 | 8.57 | 1.23 | 0.16 | 4.1 | G | 1 |
| hd006192 | H | 4992 | 2.76 | 1.28 | 0.04 | 0.07 | 610 | 0.12 | 0.13 | 66 | 8.12 | 8.66 | 0.57 | 0.68 | 0.16 | 4992 | 2.57 | 1.34 | 0.02 | 0.07 | 610 | 0.02 | 0.13 | 66 | 8.08 | 8.57 | 0.57 | 0.16 | 3.6 | G | 1 |
| hd006245 | H | 5001 | 2.91 | 1.18 | 0.01 | 0.08 | 603 | 0.16 | 0.13 | 63 | 8.09 | 8.61 | 1.01 | 1.78 | 0.15 | 5001 | 2.55 | 1.32 | -0.04 | 0.09 | 603 | -0.04 | 0.14 | 63 | 7.99 | 8.44 | 1.01 | 0.16 | 4.4 | G | 1 |
| hd006559 | H | 4681 | 2.52 | 1.31 | 0.11 | 0.10 | 578 | 0.22 | 0.17 | 55 | 8.32 | 8.82 | 0.18 | 0.73 | 0.24 | 4681 | 2.25 | 1.41 | 0.05 | 0.11 | 578 | 0.05 | 0.18 | 55 | 8.23 | 8.70 | 0.18 | 0.27 | 4.2 | G | 1 |
| hd006793 | H | 5039 | 2.73 | 1.58 | -0.03 | 0.12 | 540 | 0.07 | 0.13 | 49 | 8.18 | 8.55 | 0.87 | 1.19 | 0.15 | 5039 | 2.50 | 1.64 | -0.05 | 0.13 | 540 | -0.05 | 0.14 | 49 | 8.13 | 8.44 | 0.88 | 0.16 | 6.9 | G | 1 |
| hd007082 | H | 4969 | 2.82 | 1.38 | -0.72 | 0.09 | 548 | -0.43 | 0.12 | 62 | 7.98 | 8.68 | 0.33 | 0.44 | 0.16 | 4969 | 2.30 | 1.49 | -0.77 | 0.09 | 548 | -0.77 | 0.15 | 62 | 7.71 | 8.31 | 0.33 | 0.18 | 3.6 | G | 1 |
| hd007672 | H | 4937 | 2.76 | 1.53 | -0.40 | 0.09 | 576 | -0.27 | 0.16 | 55 | 7.87 | 8.33 | 0.04 | 0.25 | 0.17 | 4937 | 2.46 | 1.61 | -0.43 | 0.09 | 576 | -0.43 | 0.17 | 55 | 7.81 | 8.18 | 0.04 | 0.18 | 5.0 | G | 1 |
| hd009362 | H | 4762 | 2.50 | 1.31 | -0.30 | 0.07 | 608 | -0.11 | 0.14 | 66 | 8.05 | 8.56 | 0.05 | 0.43 | 0.22 | 4762 | 2.04 | 1.43 | -0.36 | 0.08 | 608 | -0.36 | 0.15 | 66 | 7.92 | 8.34 | 0.06 | 0.24 | 3.4 | G | 1 |

| ID | | T | | | | | | | | | | | | | | | | | | | | | | | | | | | | |
|---|---|---|---|---|---|---|---|---|---|---|---|---|---|---|---|---|---|---|---|---|---|---|---|---|---|---|---|---|---|---|---|
| hd009611 | H | 4627 | 2.50 | 3.00 | 0.08 | 0.25 | 148 | 0.42 | 0.43 | 10 | 8.16 | 8.68 | 0.65 | 2.48 | 0.26 | 4627 | 1.67 | 2.91 | -0.02 | 0.26 | 148 | -0.02 | 0.44 | 10 | 7.94 | 8.30 | 0.65 | 0.33 | 15.0 | G | 2 |
| hd009742 | H | 4629 | 2.53 | 1.19 | 0.05 | 0.10 | 593 | 0.17 | 0.17 | 59 | 8.29 | 8.77 | 0.16 | 0.82 | 0.25 | 4629 | 2.23 | 1.32 | -0.03 | 0.10 | 593 | -0.03 | 0.18 | 59 | 8.19 | 8.63 | 0.16 | 0.28 | 3.4 | G | 1 |
| hd010042 | H | 4799 | 2.52 | 1.49 | -0.30 | 0.09 | 576 | -0.13 | 0.13 | 49 | 8.12 | 8.76 | 0.11 | 0.44 | 0.21 | 4799 | 2.12 | 1.58 | -0.35 | 0.09 | 576 | -0.35 | 0.14 | 49 | 8.00 | 8.57 | 0.12 | 0.23 | 5.0 | G | 1 |
| hd010615 | H | 4420 | 2.15 | 1.35 | 0.32 | 0.14 | 547 | 0.43 | 0.23 | 47 | 8.43 | 8.91 | 0.20 | 1.87 | 0.32 | 4420 | 1.85 | 1.43 | 0.23 | 0.14 | 547 | 0.23 | 0.24 | 47 | 8.33 | 8.78 | 0.20 | 0.34 | 3.8 | G | 1 |
| hd011025 | H | 4961 | 2.65 | 1.62 | 0.03 | 0.14 | 512 | 0.12 | 0.14 | 43 | 8.16 | 8.68 | 0.34 | 0.45 | 0.17 | 4961 | 2.45 | 1.66 | 0.01 | 0.14 | 512 | 0.01 | 0.15 | 43 | 8.11 | 8.59 | 0.35 | 0.18 | 7.6 | G | 1 |
| hd011977 | H | 4878 | 2.65 | 1.30 | -0.20 | 0.06 | 621 | 0.00 | 0.13 | 66 | 8.08 | 8.70 | 0.24 | 0.45 | 0.18 | 4878 | 2.32 | 1.35 | -0.26 | 0.06 | 621 | -0.26 | 0.15 | 66 | 7.84 | 8.41 | 0.23 | 0.20 | 3.4 | G | 1 |
| hd012055 | H | 5035 | 2.73 | 1.40 | -0.07 | 0.10 | 567 | 0.06 | 0.11 | 55 | 8.03 | 8.64 | 0.33 | 0.35 | 0.15 | 5035 | 2.45 | 1.47 | -0.09 | 0.10 | 567 | -0.09 | 0.11 | 55 | 7.96 | 8.50 | 0.33 | 0.16 | 5.7 | G | 1 |
| hd012270 | H | 4992 | 2.64 | 1.33 | -0.02 | 0.08 | 612 | 0.09 | 0.12 | 67 | 8.06 | 8.57 | 0.49 | 0.57 | 0.16 | 4992 | 2.38 | 1.39 | -0.05 | 0.08 | 612 | -0.05 | 0.12 | 67 | 8.02 | 8.45 | 0.50 | 0.18 | 3.8 | G | 1 |
| hd012345 | H | 5326 | 4.51 | 0.00 | -0.17 | 0.09 | 676 | -0.03 | 0.10 | 59 | 8.26 | 8.53 | 0.05 | 0.10 | 0.12 | 5326 | 4.19 | 0.30 | -0.16 | 0.07 | 676 | -0.16 | 0.10 | 59 | 8.17 | 8.37 | 0.05 | 0.10 | 2.0 | G | 2 |
| hd012438 | H | 4884 | 2.58 | 1.38 | -0.63 | 0.07 | 587 | -0.34 | 0.10 | 68 | 7.83 | 8.51 | 0.13 | 0.35 | 0.18 | 4884 | 1.89 | 1.46 | -0.67 | 0.08 | 587 | -0.67 | 0.12 | 68 | 7.69 | 8.19 | 0.13 | 0.22 | 3.5 | G | 1 |
| hd012524 | H | 3960 | 1.54 | 1.50 | 0.10 | 0.25 | 494 | 0.50 | 0.40 | 41 | 8.74 | 9.11 | -0.73 | 1.23 | 0.35 | 3960 | 0.29 | 1.65 | -0.29 | 0.28 | 494 | -0.29 | 0.40 | 41 | 8.20 | 8.52 | -0.86 | 0.65 | 4.9 | G | 1 |
| hd013263 | H | 5038 | 2.88 | 1.24 | -0.03 | 0.09 | 602 | 0.11 | 0.15 | 72 | 8.06 | 8.65 | 0.62 | 0.68 | 0.15 | 5038 | 2.54 | 1.34 | -0.07 | 0.09 | 602 | -0.07 | 0.15 | 72 | 7.97 | 8.49 | 0.62 | 0.16 | 3.8 | G | 2 |
| hd013423 | H | 5037 | 2.92 | 1.23 | -0.03 | 0.07 | 611 | 0.10 | 0.13 | 65 | 8.09 | 8.72 | 0.54 | 0.57 | 0.15 | 5037 | 2.62 | 1.33 | -0.07 | 0.07 | 611 | -0.07 | 0.14 | 65 | 8.00 | 8.58 | 0.55 | 0.15 | 3.7 | G | 1 |
| hd013692 | H | 4726 | 2.51 | 1.32 | -0.13 | 0.07 | 614 | 0.10 | 0.14 | 67 | 8.19 | 8.71 | 0.14 | 0.58 | 0.23 | 4726 | 1.95 | 1.44 | -0.21 | 0.09 | 614 | -0.21 | 0.16 | 67 | 8.05 | 8.45 | 0.15 | 0.28 | 3.5 | G | 1 |
| hd013940 | H | 4883 | 2.67 | 1.26 | -0.03 | 0.07 | 616 | 0.08 | 0.15 | 65 | 8.17 | 8.72 | 0.76 | 1.43 | 0.18 | 4883 | 2.42 | 1.33 | -0.06 | 0.07 | 616 | -0.06 | 0.15 | 65 | 8.11 | 8.60 | 0.76 | 0.20 | 3.4 | G | 1 |
| hd014703 | H | 4930 | 2.82 | 1.16 | 0.12 | 0.08 | 609 | 0.19 | 0.13 | 59 | 8.18 | 8.65 | 0.09 | 0.28 | 0.17 | 4930 | 2.66 | 1.25 | 0.09 | 0.08 | 609 | 0.09 | 0.13 | 59 | 8.14 | 8.57 | 0.10 | 0.17 | 3.6 | G | 1 |
| hd014832 | H | 4804 | 2.54 | 1.37 | -0.19 | 0.07 | 607 | -0.02 | 0.14 | 65 | 8.21 | 8.73 | 0.21 | 0.53 | 0.20 | 4804 | 2.27 | 1.40 | -0.25 | 0.07 | 607 | -0.25 | 0.16 | 65 | 7.99 | 8.47 | 0.21 | 0.22 | 3.6 | G | 1 |
| hd016522 | H | 4833 | 2.55 | 1.35 | -0.14 | 0.07 | 615 | -0.02 | 0.13 | 65 | 8.28 | 8.75 | -0.04 | 0.27 | 0.20 | 4833 | 2.25 | 1.42 | -0.18 | 0.08 | 615 | -0.18 | 0.14 | 65 | 8.21 | 8.60 | -0.04 | 0.21 | 3.5 | G | 1 |
| hd016815 | H | 4683 | 2.46 | 1.16 | -0.35 | 0.07 | 611 | -0.20 | 0.17 | 62 | 7.96 | 8.47 | -0.12 | 0.37 | 0.25 | 4683 | 2.11 | 1.28 | -0.41 | 0.07 | 611 | -0.41 | 0.19 | 62 | 7.86 | 8.31 | -0.12 | 0.28 | 3.1 | G | 1 |
| hd016975 | H | 5014 | 2.77 | 1.29 | 0.05 | 0.07 | 612 | 0.14 | 0.12 | 67 | 8.16 | 8.70 | 0.78 | 1.05 | 0.15 | 5014 | 2.55 | 1.36 | 0.02 | 0.07 | 612 | 0.02 | 0.13 | 67 | 8.09 | 8.59 | 0.79 | 0.16 | 3.5 | G | 1 |
| hd017374 | H | 4824 | 2.53 | 1.34 | -0.06 | 0.08 | 605 | 0.03 | 0.14 | 64 | 8.19 | 8.69 | 0.28 | 0.58 | 0.20 | 4824 | 2.32 | 1.40 | -0.09 | 0.08 | 605 | -0.09 | 0.14 | 64 | 8.14 | 8.59 | 0.28 | 0.21 | 3.5 | G | 1 |
| hd017793 | H | 5042 | 2.68 | 1.34 | 0.05 | 0.08 | 597 | 0.08 | 0.13 | 58 | 8.04 | 8.64 | 0.36 | 0.37 | 0.15 | 5042 | 2.60 | 1.37 | 0.03 | 0.08 | 597 | 0.03 | 0.13 | 58 | 8.02 | 8.60 | 0.36 | 0.15 | 4.6 | G | 1 |
| hd018293 | H | 4262 | 1.81 | 1.21 | 0.38 | 0.14 | 516 | 0.59 | 0.29 | 54 | 8.46 | 8.91 | -0.15 | 1.49 | 0.35 | 4262 | 1.19 | 1.46 | 0.13 | 0.14 | 516 | 0.13 | 0.30 | 54 | 8.24 | 8.63 | -0.18 | 0.42 | 4.2 | G | 1 |
| hd018322 | H | 4614 | 2.47 | 1.19 | 0.00 | 0.08 | 603 | 0.11 | 0.16 | 57 | 8.24 | 8.74 | 0.62 | 2.39 | 0.26 | 4614 | 2.19 | 1.29 | -0.07 | 0.09 | 603 | -0.07 | 0.17 | 57 | 8.15 | 8.61 | 0.62 | 0.29 | 3.3 | G | 1 |
| hd020037 | H | 5077 | 2.76 | 1.32 | -0.06 | 0.09 | 607 | 0.04 | 0.14 | 67 | 8.07 | 8.65 | 1.05 | 1.59 | 0.14 | 5077 | 2.55 | 1.38 | -0.08 | 0.09 | 607 | -0.08 | 0.14 | 67 | 8.02 | 8.55 | 1.05 | 0.15 | 4.4 | G | 1 |
| hd020894 | H | 5060 | 2.70 | 1.44 | -0.04 | 0.08 | 601 | 0.06 | 0.12 | 62 | 8.06 | 8.69 | 0.63 | 0.66 | 0.15 | 5060 | 2.46 | 1.49 | -0.06 | 0.08 | 601 | -0.06 | 0.12 | 62 | 8.00 | 8.58 | 0.64 | 0.16 | 4.6 | G | 1 |
| hd021011 | H | 4799 | 2.49 | 1.32 | -0.09 | 0.08 | 611 | 0.02 | 0.14 | 66 | 8.16 | 8.66 | 0.16 | 0.47 | 0.21 | 4799 | 2.22 | 1.39 | -0.13 | 0.08 | 611 | -0.13 | 0.14 | 66 | 8.10 | 8.54 | 0.16 | 0.22 | 3.4 | G | 1 |
| hd021430 | H | 4979 | 2.63 | 1.40 | -0.31 | 0.06 | 608 | -0.23 | 0.13 | 66 | 7.86 | 8.55 | 0.12 | 0.27 | 0.16 | 4979 | 2.45 | 1.45 | -0.33 | 0.06 | 608 | -0.33 | 0.14 | 66 | 7.82 | 8.46 | 0.13 | 0.18 | 3.5 | G | 1 |
| hd022231 | H | 4591 | 2.45 | 1.31 | 0.25 | 0.11 | 570 | 0.49 | 0.20 | 55 | 8.49 | 8.92 | 0.62 | 2.63 | 0.27 | 4591 | 1.95 | 1.42 | 0.11 | 0.12 | 570 | 0.11 | 0.21 | 55 | 8.23 | 8.60 | 0.63 | 0.31 | 4.1 | G | 1 |
| hd022663 | H | 4708 | 2.51 | 0.91 | 0.19 | 0.12 | 597 | 0.12 | 0.16 | 57 | 8.29 | 8.83 | 1.30 | 7.09 | 0.23 | 4708 | 2.60 | 0.92 | 0.24 | 0.12 | 597 | 0.24 | 0.16 | 57 | 8.41 | 8.96 | 1.30 | 0.23 | 3.9 | G | 1 |
| hd022676 | H | 4957 | 2.73 | 1.32 | 0.08 | 0.10 | 593 | 0.15 | 0.14 | 59 | 8.19 | 8.64 | 1.34 | 4.03 | 0.17 | 4957 | 2.56 | 1.38 | 0.06 | 0.10 | 593 | 0.06 | 0.14 | 59 | 8.15 | 8.55 | 1.35 | 0.17 | 4.6 | G | 1 |
| hd023319 | H | 4539 | 2.48 | 1.18 | 0.42 | 0.14 | 564 | 0.54 | 0.19 | 48 | 8.61 | 9.02 | 0.71 | 3.67 | 0.27 | 4539 | 2.16 | 1.37 | 0.29 | 0.13 | 564 | 0.29 | 0.19 | 48 | 8.49 | 8.87 | 0.70 | 0.30 | 3.6 | G | 1 |
| hd023719 | H | 4939 | 2.74 | 1.29 | 0.14 | 0.08 | 611 | 0.24 | 0.11 | 63 | 8.26 | 8.77 | 0.66 | 0.97 | 0.17 | 4939 | 2.51 | 1.38 | 0.11 | 0.08 | 611 | 0.11 | 0.12 | 63 | 8.20 | 8.66 | 0.66 | 0.18 | 3.5 | G | 1 |
| hd023940 | H | 4776 | 2.58 | 1.38 | -0.31 | 0.09 | 559 | -0.10 | 0.13 | 52 | 8.13 | 8.69 | 0.26 | 0.66 | 0.21 | 4776 | 2.08 | 1.50 | -0.37 | 0.10 | 559 | -0.37 | 0.14 | 52 | 7.98 | 8.45 | 0.27 | 0.24 | 5.6 | G | 1 |
| hd024160 | H | 4948 | 2.62 | 1.37 | 0.03 | 0.07 | 608 | 0.16 | 0.12 | 67 | 8.15 | 8.68 | 0.69 | 1.02 | 0.17 | 4948 | 2.33 | 1.45 | 0.00 | 0.07 | 608 | 0.00 | 0.12 | 67 | 8.09 | 8.55 | 0.70 | 0.19 | 3.8 | G | 1 |
| hd024706 | H | 4427 | 2.32 | 1.17 | 0.32 | 0.14 | 567 | 0.47 | 0.20 | 49 | 8.54 | 8.95 | 0.24 | 1.99 | 0.30 | 4427 | 1.90 | 1.36 | 0.16 | 0.13 | 567 | 0.16 | 0.21 | 49 | 8.39 | 8.75 | 0.24 | 0.34 | 3.7 | G | 1 |
| hd024744 | H | 5846 | 2.86 | 0.68 | 0.48 | 0.14 | 600 | -0.26 | 0.17 | 65 | | | 1.73 | 1.30 | 0.05 | 5846 | 4.33 | 0.25 | 0.45 | 0.19 | 600 | 0.45 | 0.17 | 65 | 7.97 | | 1.73 | 0.02 | 5.0 | G | 1 |
| hd026967 | H | 4614 | 2.63 | 1.04 | 0.09 | 0.09 | 602 | 0.21 | 0.13 | 57 | 8.27 | 8.75 | -0.03 | 0.58 | 0.24 | 4614 | 2.32 | 1.22 | -0.01 | 0.09 | 602 | -0.01 | 0.14 | 57 | 8.17 | 8.61 | -0.03 | 0.28 | 3.2 | G | 1 |
| hd027256 | H | 5048 | 2.57 | 1.57 | 0.12 | 0.11 | 555 | 0.19 | 0.13 | 54 | 8.20 | 8.66 | 1.40 | 3.62 | 0.15 | 5048 | 2.43 | 1.60 | 0.11 | 0.11 | 555 | 0.11 | 0.13 | 54 | 8.16 | 8.60 | 1.41 | 0.16 | 6.0 | G | 1 |
| hd028028 | H | 3982 | 1.38 | 1.55 | 0.01 | 0.19 | 518 | 0.40 | 0.28 | 43 | 8.44 | 8.86 | -0.04 | 5.09 | 0.38 | 3982 | 0.23 | 1.62 | -0.30 | 0.21 | 518 | -0.30 | 0.29 | 43 | 8.02 | 8.35 | -0.18 | 0.67 | 4.5 | G | 1 |
| hd028093 | H | 4951 | 2.49 | 1.45 | -0.10 | 0.07 | 609 | -0.05 | 0.13 | 66 | 8.17 | 8.73 | 0.43 | 0.55 | 0.18 | 4951 | 2.38 | 1.47 | -0.11 | 0.07 | 609 | -0.11 | 0.13 | 66 | 8.15 | 8.68 | 0.43 | 0.18 | 3.7 | G | 1 |
| hd029291 | H | 4897 | 2.40 | 1.49 | -0.04 | 0.08 | 602 | 0.07 | 0.11 | 60 | 8.12 | 8.64 | 0.54 | 0.83 | 0.19 | 4897 | 2.23 | 1.52 | -0.02 | 0.08 | 602 | -0.02 | 0.12 | 60 | 8.09 | 8.56 | 0.54 | 0.20 | 4.0 | G | 1 |
| hd029399 | H | 4822 | 4.82 | 0.50 | 0.43 | 0.19 | 482 | 1.23 | 0.27 | 66 | 9.24 | 9.58 | 0.38 | 0.78 | 0.19 | 4822 | 2.84 | 1.04 | 0.20 | 0.10 | 482 | 0.20 | 0.22 | 66 | 8.45 | 8.65 | 0.40 | 0.27 | 2.0 | G | 2 |
| hd029751 | H | 4855 | 2.61 | 1.36 | -0.20 | 0.08 | 612 | 0.09 | 0.12 | 67 | 8.12 | 8.72 | 0.33 | 0.60 | 0.19 | 4855 | 2.06 | 1.46 | -0.28 | 0.08 | 612 | -0.29 | 0.14 | 67 | 7.83 | 8.33 | 0.33 | 0.22 | 3.7 | G | 1 |
| hd030185 | H | 4851 | 2.42 | 1.32 | -0.09 | 0.07 | 620 | -0.05 | 0.15 | 68 | 8.09 | 8.64 | 0.58 | 1.04 | 0.20 | 4851 | 2.32 | 1.35 | -0.11 | 0.07 | 620 | -0.11 | 0.15 | 68 | 8.07 | 8.60 | 0.58 | 0.21 | 3.6 | G | 1 |
| hd032453 | H | 5032 | 2.81 | 1.30 | -0.08 | 0.07 | 615 | 0.07 | 0.13 | 63 | 8.11 | 8.71 | 0.53 | 0.56 | 0.15 | 5032 | 2.47 | 1.38 | -0.11 | 0.08 | 615 | -0.11 | 0.14 | 63 | 8.02 | 8.54 | 0.53 | 0.16 | 3.9 | G | 1 |
| hd032515 | H | 4529 | 2.48 | 1.19 | 0.33 | 0.13 | 574 | 0.54 | 0.20 | 51 | 8.59 | 8.98 | 0.22 | 1.34 | 0.28 | 4529 | 1.96 | 1.39 | 0.17 | 0.12 | 574 | 0.17 | 0.20 | 51 | 8.41 | 8.74 | 0.22 | 0.32 | 3.7 | G | 1 |
| hd033285 | H | 4825 | 1.96 | 1.88 | -0.03 | 0.13 | 515 | 0.08 | 0.15 | 46 | 8.07 | 8.58 | 0.54 | 1.03 | 0.23 | 4825 | 1.70 | 1.88 | -0.05 | 0.13 | 515 | -0.04 | 0.15 | 46 | 8.03 | 8.46 | 0.55 | 0.25 | 7.1 | G | 1 |
| hd034172 | H | 4990 | 2.67 | 1.32 | 0.05 | 0.08 | 607 | 0.06 | 0.13 | 60 | 8.22 | 8.67 | 1.04 | 1.94 | 0.16 | 4990 | 2.64 | 1.33 | 0.04 | 0.08 | 607 | 0.04 | 0.13 | 60 | 8.21 | 8.66 | 1.04 | 0.16 | 4.0 | G | 1 |
| hd034253 | H | 5490 | 4.55 | 0.50 | -0.13 | 0.10 | 623 | 0.00 | 0.11 | 57 | 8.46 | 8.54 | 0.75 | 0.32 | 0.09 | 5490 | 4.21 | 0.54 | -0.06 | 0.07 | 623 | -0.06 | 0.11 | 57 | 8.43 | 8.43 | 0.76 | 0.08 | 2.5 | G | 2 |

| ID | | T1 | | | | | | | | | | | T2 | | | | | | | | | | | | |
|---|---|---|---|---|---|---|---|---|---|---|---|---|---|---|---|---|---|---|---|---|---|---|---|---|---|
| hd034266 | H | 4881 | 2.48 | 1.53 | 0.06 | 0.09 | 591 | 0.12 | 0.14 | 60 | 8.22 | 8.73 | 0.89 | 1.88 | 0.19 | 4881 | 2.33 | 1.56 | 0.04 | 0.10 | 591 | 0.04 | 0.15 | 60 | 8.18 | 8.66 | 0.89 | 0.20 | 4.4 | G | 1 |
| hd034649 | H | 4320 | 1.65 | 1.59 | -0.01 | 0.12 | 555 | 0.14 | 0.17 | 49 | 8.20 | 8.71 | -0.11 | 1.39 | 0.37 | 4320 | 1.24 | 1.62 | -0.10 | 0.12 | 555 | -0.10 | 0.18 | 49 | 8.07 | 8.53 | -0.12 | 0.41 | 4.2 | G | 1 |
| hd035929 | H | 6345 | 3.13 | 4.44 | 0.15 | 0.41 | 32 | 0.53 | 0.07 | 2 | 8.20 | 8.37 | 2.80 | 5.55 | 0.05 | 6345 | 2.06 | 4.25 | 0.19 | 0.41 | 32 | 0.19 | 0.07 | 2 | 7.90 | 8.04 | 2.84 | 0.02 | 65.0 | R | 2 |
| hd036189 | H | 4911 | 2.37 | 1.78 | -0.01 | 0.11 | 540 | 0.12 | 0.12 | 52 | 8.11 | 8.66 | 0.80 | 1.42 | 0.19 | 4911 | 2.06 | 1.81 | -0.04 | 0.11 | 540 | -0.04 | 0.13 | 52 | 8.05 | 8.52 | 0.80 | 0.21 | 6.2 | G | 1 |
| hd036597 | H | 4573 | 2.05 | 1.52 | -0.03 | 0.10 | 576 | 0.09 | 0.16 | 55 | 8.16 | 8.74 | 0.17 | 1.03 | 0.31 | 4573 | 1.76 | 1.55 | -0.07 | 0.10 | 576 | -0.07 | 0.17 | 55 | 8.08 | 8.61 | 0.17 | 0.33 | 4.0 | G | 1 |
| hd036734 | H | 4244 | 1.80 | 1.47 | -0.02 | 0.13 | 554 | 0.16 | 0.21 | 49 | 8.34 | 8.78 | 0.35 | 4.72 | 0.35 | 4244 | 1.31 | 1.54 | -0.14 | 0.14 | 554 | -0.15 | 0.21 | 49 | 8.17 | 8.56 | 0.33 | 0.41 | 3.9 | G | 1 |
| hd036848 | H | 4476 | 2.72 | 0.75 | 0.67 | 0.21 | 614 | 0.79 | 0.29 | 52 | 8.74 | 9.02 | 0.28 | 1.80 | 0.26 | 4476 | 2.33 | 1.15 | 0.44 | 0.17 | 614 | 0.44 | 0.29 | 52 | 8.56 | 8.82 | 0.28 | 0.30 | 3.3 | G | 1 |
| hd037811 | H | 5023 | 2.57 | 1.45 | -0.01 | 0.08 | 607 | 0.03 | 0.12 | 59 | 8.05 | 8.62 | 0.83 | 1.11 | 0.16 | 5023 | 2.48 | 1.47 | -0.02 | 0.08 | 607 | -0.01 | 0.12 | 59 | 8.05 | 8.60 | 0.83 | 0.17 | 4.4 | G | 1 |
| hd039425 | H | 4566 | 2.53 | 1.23 | 0.30 | 0.12 | 578 | 0.48 | 0.17 | 51 | 8.60 | 9.00 | 0.80 | 4.07 | 0.26 | 4566 | 2.09 | 1.39 | 0.18 | 0.11 | 578 | 0.18 | 0.18 | 51 | 8.45 | 8.79 | 0.79 | 0.30 | 3.6 | G | 1 |
| hd039640 | H | 4850 | 2.51 | 1.34 | -0.10 | 0.07 | 620 | -0.01 | 0.13 | 68 | 8.13 | 8.67 | 0.37 | 0.66 | 0.19 | 4850 | 2.31 | 1.38 | -0.12 | 0.07 | 620 | -0.12 | 0.14 | 68 | 8.08 | 8.58 | 0.38 | 0.21 | 3.4 | G | 1 |
| hd040176 | H | 4659 | 2.26 | 1.53 | 0.29 | 0.12 | 556 | 0.38 | 0.17 | 52 | 8.33 | 8.91 | 0.42 | 1.32 | 0.27 | 4659 | 2.03 | 1.57 | 0.24 | 0.12 | 556 | 0.24 | 0.17 | 52 | 8.26 | 8.81 | 0.42 | 0.29 | 4.2 | G | 1 |
| hd040657 | H | 4288 | 1.52 | 1.51 | -0.58 | 0.10 | 573 | -0.44 | 0.20 | 53 | 7.94 | 8.59 | -0.75 | 0.37 | 0.38 | 4288 | 1.15 | 1.52 | -0.64 | 0.11 | 573 | -0.64 | 0.21 | 53 | 7.82 | 8.43 | -0.76 | 0.42 | 3.7 | G | 1 |
| hd040808 | H | 4620 | 1.72 | 1.80 | -0.01 | 0.11 | 550 | 0.03 | 0.15 | 51 | 8.12 | 8.62 | 0.43 | 1.55 | 0.32 | 4620 | 1.64 | 1.80 | -0.02 | 0.11 | 550 | -0.01 | 0.15 | 51 | 8.10 | 8.59 | 0.43 | 0.33 | 4.6 | G | 1 |
| hd041047 | H | 3891 | 1.30 | 1.87 | 0.09 | 0.28 | 470 | 0.56 | 0.38 | 36 | 8.81 | 9.23 | -0.71 | 1.69 | 0.38 | 3891 | -0.17 | 1.90 | -0.29 | 0.31 | 470 | -0.30 | 0.39 | 36 | 8.15 | 8.51 | -0.89 | 0.87 | 5.7 | G | 1 |
| hd042719 | H | 5699 | 3.84 | 0.94 | 0.29 | 0.07 | 591 | 0.43 | 0.11 | 82 | 8.83 | 8.98 | 0.83 | 0.23 | 0.04 | 5699 | 3.56 | 1.16 | 0.26 | 0.07 | 591 | 0.26 | 0.10 | 82 | 8.72 | 8.86 | 0.83 | 0.05 | 4.2 | G | 2 |
| hd043785 | H | 4876 | 2.59 | 1.33 | 0.11 | 0.08 | 602 | 0.17 | 0.14 | 68 | 8.24 | 8.71 | 0.44 | 0.71 | 0.19 | 4876 | 2.45 | 1.38 | 0.09 | 0.08 | 602 | 0.09 | 0.15 | 68 | 8.20 | 8.64 | 0.44 | 0.20 | 3.5 | G | 1 |
| hd045669 | H | 3984 | 1.38 | 1.64 | 0.10 | 0.22 | 503 | 0.44 | 0.32 | 38 | 8.49 | 8.93 | -0.15 | 4.06 | 0.38 | 3984 | 0.39 | 1.71 | -0.18 | 0.24 | 503 | -0.18 | 0.34 | 38 | 8.10 | 8.48 | -0.26 | 0.61 | 4.8 | G | 1 |
| hd046116 | H | 4853 | 2.67 | 1.35 | -0.28 | 0.07 | 609 | -0.13 | 0.12 | 62 | 8.02 | 8.63 | 0.34 | 0.63 | 0.19 | 4853 | 2.32 | 1.45 | -0.32 | 0.07 | 609 | -0.32 | 0.13 | 62 | 7.92 | 8.46 | 0.35 | 0.21 | 3.4 | G | 1 |
| hd046415 | H | 4818 | 2.72 | 1.29 | -0.17 | 0.07 | 618 | 0.06 | 0.16 | 62 | 8.20 | 8.75 | 0.44 | 0.86 | 0.19 | 4818 | 2.30 | 1.37 | -0.25 | 0.07 | 618 | -0.25 | 0.19 | 62 | 7.93 | 8.42 | 0.43 | 0.22 | 3.4 | G | 1 |
| hd046568 | H | 4808 | 2.50 | 1.35 | -0.20 | 0.07 | 617 | -0.03 | 0.14 | 65 | 8.14 | 8.71 | 0.18 | 0.49 | 0.21 | 4808 | 2.22 | 1.39 | -0.27 | 0.07 | 617 | -0.27 | 0.15 | 65 | 7.91 | 8.44 | 0.17 | 0.22 | 3.5 | G | 1 |
| hd047001 | H | 4675 | 2.24 | 1.47 | -0.28 | 0.08 | 602 | -0.18 | 0.17 | 61 | 7.91 | 8.52 | 0.00 | 0.50 | 0.27 | 4675 | 1.98 | 1.52 | -0.32 | 0.08 | 602 | -0.32 | 0.17 | 61 | 7.83 | 8.40 | 0.00 | 0.29 | 3.6 | G | 1 |
| hd047536 | H | 4353 | 1.85 | 1.43 | -0.62 | 0.09 | 588 | -0.42 | 0.18 | 54 | 8.02 | 8.59 | -0.48 | 0.52 | 0.35 | 4353 | 1.32 | 1.48 | -0.72 | 0.10 | 588 | -0.71 | 0.20 | 54 | 7.85 | 8.36 | -0.49 | 0.40 | 3.3 | G | 1 |
| hd047910 | H | 4869 | 2.68 | 1.25 | 0.05 | 0.08 | 608 | 0.11 | 0.14 | 62 | 8.25 | 8.78 | 0.42 | 0.70 | 0.18 | 4869 | 2.52 | 1.31 | 0.02 | 0.08 | 608 | 0.02 | 0.14 | 62 | 8.21 | 8.70 | 0.42 | 0.19 | 3.6 | G | 1 |
| hd049877 | H | 3911 | 1.35 | 4.06 | -0.22 | 0.31 | 37 | 0.61 | 0.72 | 4 | | | 0.45 | 15.48 | 0.37 | 3911 | -0.30 | 3.82 | -0.64 | 0.36 | 37 | -0.28 | 1.01 | 4 | | | 0.18 | 0.92 | 22.9 | G | 2 |
| hd049947 | H | 4885 | 2.67 | 1.34 | -0.17 | 0.07 | 606 | 0.01 | 0.14 | 66 | 8.17 | 8.76 | 0.11 | 0.33 | 0.18 | 4885 | 2.24 | 1.43 | -0.22 | 0.08 | 606 | -0.22 | 0.15 | 66 | 8.07 | 8.55 | 0.12 | 0.20 | 3.6 | G | 1 |
| hd050310 | H | 4489 | 2.05 | 1.53 | 0.04 | 0.11 | 565 | 0.16 | 0.15 | 51 | 8.35 | 8.81 | -0.01 | 0.92 | 0.32 | 4489 | 1.74 | 1.57 | -0.02 | 0.11 | 565 | -0.02 | 0.16 | 51 | 8.25 | 8.67 | -0.01 | 0.34 | 3.9 | G | 1 |
| hd050890 | H | 4733 | 1.88 | 1.98 | 0.10 | 0.17 | 314 | 0.10 | 0.20 | 29 | 8.10 | 8.53 | 1.29 | 6.48 | 0.29 | 4733 | 1.88 | 1.98 | 0.10 | 0.17 | 314 | 0.10 | 0.20 | 29 | 8.10 | 8.53 | 1.29 | 0.29 | 11.1 | G | 1 |
| hd054038 | H | 5028 | 3.02 | 1.25 | 0.06 | 0.09 | 601 | 0.20 | 0.15 | 61 | 8.30 | 8.64 | 1.48 | 4.50 | 0.15 | 5028 | 2.69 | 1.40 | 0.01 | 0.10 | 601 | 0.01 | 0.16 | 61 | 8.21 | 8.47 | 1.49 | 0.15 | 4.6 | G | 1 |
| hd054732 | H | 4902 | 2.44 | 1.27 | -0.04 | 0.16 | 576 | -0.10 | 0.18 | 49 | 8.06 | 8.71 | 0.65 | 1.05 | 0.19 | 4902 | 2.58 | 1.23 | -0.02 | 0.16 | 576 | -0.02 | 0.17 | 49 | 8.10 | 8.78 | 0.65 | 0.18 | 6.4 | G | 1 |
| hd059219 | H | 4803 | 2.06 | 2.19 | 0.10 | 0.17 | 409 | 0.18 | 0.16 | 33 | 8.13 | 8.62 | 1.24 | 4.92 | 0.23 | 4803 | 1.85 | 2.19 | 0.08 | 0.17 | 409 | 0.08 | 0.17 | 33 | 8.08 | 8.52 | 1.24 | 0.24 | 9.4 | G | 1 |
| hd059894 | H | 4904 | 2.72 | 1.29 | -0.10 | 0.07 | 617 | 0.04 | 0.14 | 65 | 8.16 | 8.77 | 0.84 | 1.62 | 0.18 | 4904 | 2.38 | 1.38 | -0.15 | 0.07 | 617 | -0.15 | 0.15 | 65 | 8.08 | 8.61 | 0.85 | 0.19 | 3.4 | G | 1 |
| hd060574 | H | 4929 | 2.77 | 1.33 | -0.40 | 0.07 | 612 | -0.17 | 0.12 | 64 | 8.01 | 8.62 | 0.12 | 0.31 | 0.17 | 4929 | 2.25 | 1.47 | -0.44 | 0.07 | 612 | -0.44 | 0.13 | 64 | 7.88 | 8.37 | 0.13 | 0.19 | 3.4 | G | 1 |
| hd062713 | H | 4624 | 2.55 | 1.24 | 0.12 | 0.09 | 594 | 0.29 | 0.13 | 59 | 8.43 | 8.90 | 0.20 | 0.93 | 0.25 | 4624 | 2.12 | 1.40 | 0.02 | 0.09 | 594 | 0.02 | 0.13 | 59 | 8.28 | 8.70 | 0.20 | 0.29 | 3.5 | G | 1 |
| hd062849 | H | 4871 | 2.50 | 0.96 | -0.72 | 0.17 | 567 | -0.32 | 0.14 | 80 | 7.69 | 8.02 | -0.10 | 0.22 | 0.19 | 4871 | 1.71 | 1.11 | -0.78 | 0.18 | 567 | -0.78 | 0.16 | 80 | 7.46 | 7.53 | -0.10 | 0.23 | 3.7 | G | 2 |
| hd062897 | H | 4762 | 2.28 | 1.54 | -0.01 | 0.12 | 554 | 0.05 | 0.14 | 48 | 8.18 | 8.70 | 0.52 | 1.20 | 0.23 | 4762 | 2.15 | 1.56 | -0.03 | 0.12 | 554 | -0.03 | 0.16 | 48 | 8.14 | 8.64 | 0.52 | 0.24 | 5.5 | G | 1 |
| hd063295 | H | 4736 | 2.56 | 1.34 | -0.10 | 0.07 | 609 | 0.10 | 0.15 | 62 | 8.27 | 8.77 | 0.06 | 0.47 | 0.22 | 4736 | 2.07 | 1.45 | -0.18 | 0.08 | 609 | -0.18 | 0.16 | 62 | 8.13 | 8.54 | 0.07 | 0.27 | 3.5 | G | 1 |
| hd064181 | H | 5018 | 2.68 | 1.31 | -0.05 | 0.07 | 607 | 0.05 | 0.12 | 64 | 8.06 | 8.66 | 0.62 | 0.72 | 0.16 | 5018 | 2.47 | 1.36 | -0.07 | 0.07 | 607 | -0.07 | 0.12 | 64 | 8.00 | 8.56 | 0.62 | 0.17 | 4.3 | G | 1 |
| hd073598 | H | 5012 | 2.41 | 1.49 | 0.21 | 0.09 | 597 | 0.08 | 0.14 | 60 | 8.25 | 8.74 | 1.31 | 3.19 | 0.17 | 5012 | 2.71 | 1.42 | 0.25 | 0.09 | 597 | 0.25 | 0.14 | 60 | 8.34 | 8.89 | 1.30 | 0.16 | 3.9 | G | 1 |
| hd073665 | H | 4954 | 2.60 | 1.48 | 0.22 | 0.08 | 596 | 0.28 | 0.12 | 60 | 8.32 | 8.86 | 0.71 | 1.03 | 0.17 | 4954 | 2.48 | 1.51 | 0.21 | 0.08 | 596 | 0.21 | 0.12 | 60 | 8.30 | 8.80 | 0.71 | 0.18 | 3.9 | G | 1 |
| hd073710 | H | 4900 | 2.50 | 1.55 | 0.24 | 0.11 | 574 | 0.29 | 0.16 | 56 | 8.40 | 8.86 | 1.14 | 3.06 | 0.19 | 4900 | 2.39 | 1.57 | 0.23 | 0.11 | 574 | 0.23 | 0.16 | 56 | 8.37 | 8.80 | 1.15 | 0.19 | 4.8 | G | 1 |
| hd074874 | H | 5384 | 2.89 | 0.72 | 0.08 | 0.10 | 612 | -0.21 | 0.14 | 66 | 7.98 | 8.59 | 1.49 | 2.01 | 0.09 | 5384 | 3.45 | 0.50 | 0.12 | 0.11 | 612 | 0.13 | 0.12 | 66 | 8.15 | 8.94 | 1.49 | 0.08 | 4.8 | G | 1 |
| hd078964b | H | 5118 | 4.59 | 0.25 | 0.18 | 0.10 | 596 | 0.35 | 0.19 | 62 | 8.39 | 8.47 | 1.33 | 2.81 | 0.16 | 5118 | 4.15 | 1.11 | 0.14 | 0.07 | 596 | 0.14 | 0.19 | 62 | 8.30 | 8.31 | 1.34 | 0.14 | 2.0 | G | 2 |
| hd091267 | H | 4873 | 4.55 | 0.25 | 0.02 | 0.16 | 614 | 0.23 | 0.25 | 56 | 8.85 | 9.17 | -0.15 | 0.20 | 0.18 | 4873 | 4.07 | 0.50 | 0.01 | 0.12 | 614 | 0.01 | 0.26 | 56 | 8.64 | 8.95 | -0.14 | 0.17 | 2.0 | G | 2 |
| hd104760a | H | 5824 | 4.31 | 0.50 | 0.16 | 0.08 | 629 | 0.26 | 0.10 | 77 | 8.54 | 8.79 | 2.12 | 3.25 | 0.03 | 5824 | 4.04 | 0.86 | 0.15 | 0.06 | 629 | 0.16 | 0.09 | 77 | 8.42 | 8.72 | 2.13 | 0.03 | 3.4 | G | 2 |
| hd108063 | H | 5859 | 3.68 | 1.43 | 0.52 | 0.09 | 616 | 0.57 | 0.13 | 83 | 8.84 | 9.12 | 0.71 | 0.13 | 0.03 | 5859 | 3.56 | 1.48 | 0.51 | 0.09 | 616 | 0.51 | 0.13 | 83 | 8.79 | 9.07 | 0.71 | 0.03 | 5.3 | G | 2 |
| hd108570 | H | 4984 | 3.49 | 0.50 | 0.08 | 0.10 | 670 | 0.21 | 0.18 | 76 | 8.35 | 8.70 | -0.28 | 0.10 | 0.15 | 4984 | 3.20 | 0.80 | 0.02 | 0.07 | 670 | 0.02 | 0.18 | 76 | 8.26 | 8.56 | -0.27 | 0.15 | 2.6 | G | 1 |
| hd110291 | H | 5392 | 4.50 | 0.50 | 0.02 | 0.11 | 658 | 0.31 | 0.17 | 77 | 8.39 | 8.57 | 0.75 | 0.40 | 0.10 | 5392 | 3.87 | 0.76 | 0.03 | 0.07 | 658 | 0.03 | 0.18 | 77 | 8.25 | 8.25 | 0.76 | 0.08 | 2.0 | G | 2 |
| hd114747 | H | 5073 | 4.50 | 0.50 | 0.30 | 0.12 | 637 | 0.55 | 0.23 | 61 | 8.83 | 9.04 | 0.41 | 0.41 | 0.16 | 5073 | 3.96 | 0.50 | 0.34 | 0.08 | 637 | 0.34 | 0.20 | 61 | 8.63 | 8.76 | 0.43 | 0.14 | 2.0 | G | 2 |
| hd115202 | H | 4742 | 3.05 | 0.50 | 0.16 | 0.12 | 607 | 0.30 | 0.17 | 69 | 8.42 | 8.75 | 0.14 | 0.56 | 0.19 | 4742 | 2.70 | 0.91 | 0.02 | 0.08 | 607 | 0.02 | 0.17 | 69 | 8.29 | 8.58 | 0.15 | 0.21 | 2.9 | G | 1 |
| hd121416 | H | 4576 | 2.45 | 1.25 | 0.18 | 0.11 | 578 | 0.34 | 0.17 | 53 | 8.45 | 8.85 | 0.13 | 0.93 | 0.27 | 4576 | 2.07 | 1.39 | 0.08 | 0.11 | 578 | 0.09 | 0.18 | 53 | 8.32 | 8.67 | 0.13 | 0.30 | 3.8 | G | 1 |

| ID | | | | | | | | | | | | | | | | | | | | | | | | | | |
|---|---|---|---|---|---|---|---|---|---|---|---|---|---|---|---|---|---|---|---|---|---|---|---|---|---|---|
| hd123517 | H | 6340 | 2.50 | 1.56 | 0.37 | 0.10 | 531 | -0.43 | 0.10 | 67 | 7.82 | 8.24 | 2.28 | 1.79 | 0.03 | 6340 | 4.45 | 1.09 | 0.40 | 0.11 | 531 | 0.40 | 0.11 | 67 | 8.41 | 8.84 | 2.26 | -0.08 | 5.9 | G | 2 |
| hd142527 | H | 6632 | 3.15 | 3.64 | 0.33 | 0.20 | 22 | | | | 6.87 | 8.53 | 3.00 | 5.26 | 0.07 | | | | | | | | | | | | | | 49.0 | R | 2 |
| hd144589 | H | 6486 | 2.50 | 1.58 | 0.13 | 0.09 | 496 | -0.57 | 0.14 | 75 | 7.82 | 8.24 | 2.01 | 0.81 | 0.15 | 6486 | 4.29 | 1.31 | 0.14 | 0.09 | 496 | 0.14 | 0.16 | 75 | 8.32 | 8.75 | 2.00 | -0.08 | 6.1 | G | 2 |
| hd147135 | H | 6832 | 3.71 | 3.45 | -0.25 | 0.10 | 285 | -0.12 | 0.16 | 69 | 8.28 | 8.72 | 3.03 | 4.60 | 0.02 | 6832 | 3.35 | 3.37 | -0.23 | 0.11 | 285 | -0.23 | 0.16 | 69 | 8.18 | 8.60 | 3.05 | 0.08 | 14.0 | G | 2 |
| hd169689 | H | 4989 | 2.05 | 1.94 | 0.15 | 0.18 | 455 | 0.01 | 0.20 | 40 | 8.17 | 8.76 | 1.26 | 3.07 | 0.19 | 4989 | 2.25 | 1.91 | 0.18 | 0.19 | 455 | 0.18 | 0.20 | 40 | 8.32 | 8.95 | 1.26 | 0.18 | 9.3 | G | 1 |
| hd176354 | H | 5165 | 3.82 | 0.50 | 0.39 | 0.10 | 529 | 0.60 | 0.18 | 68 | 8.72 | 8.93 | 1.84 | 6.62 | 0.12 | 5165 | 3.40 | 0.86 | 0.33 | 0.06 | 529 | 0.33 | 0.17 | 68 | 8.59 | 8.70 | 1.85 | 0.12 | 2.7 | G | 2 |
| hd181433 | H | 4800 | 2.50 | 1.48 | -0.03 | 0.17 | 561 | -0.44 | 0.36 | 57 | 8.29 | 8.31 | 0.86 | 2.28 | 0.21 | 4800 | 3.57 | 0.82 | 0.25 | 0.14 | 561 | 0.25 | 0.34 | 57 | 8.67 | 8.87 | 0.86 | 0.17 | 2.7 | G | 2 |
| hd181517 | H | 4895 | 2.73 | 1.27 | 0.10 | 0.09 | 598 | 0.18 | 0.16 | 68 | 8.17 | 8.63 | 0.57 | 0.91 | 0.18 | 4895 | 2.54 | 1.35 | 0.07 | 0.09 | 598 | 0.07 | 0.16 | 68 | 8.13 | 8.54 | 0.58 | 0.18 | 3.6 | G | 1 |
| hd182893 | H | 4864 | 2.78 | 1.15 | 0.15 | 0.09 | 609 | 0.23 | 0.14 | 62 | 8.21 | 8.70 | 0.49 | 0.84 | 0.18 | 4864 | 2.59 | 1.26 | 0.11 | 0.09 | 609 | 0.11 | 0.14 | 62 | 8.15 | 8.61 | 0.50 | 0.19 | 3.3 | G | 1 |
| hd187669a | H | 4754 | 2.50 | 1.97 | -0.22 | 0.32 | 277 | -0.15 | 0.31 | 18 | 8.35 | 8.81 | 0.82 | 2.40 | 0.22 | 4754 | 2.50 | 1.96 | -0.26 | 0.31 | 277 | -0.26 | 0.31 | 18 | 8.21 | 8.68 | 0.82 | 0.22 | 12.0 | G | 1 |
| hd188114 | H | 4594 | 2.24 | 1.32 | -0.30 | 0.08 | 610 | -0.17 | 0.17 | 63 | 7.93 | 8.52 | 0.83 | 3.96 | 0.29 | 4594 | 1.93 | 1.40 | -0.35 | 0.08 | 610 | -0.35 | 0.18 | 63 | 7.83 | 8.38 | 0.83 | 0.31 | 3.3 | G | 1 |
| hd190056 | H | 4275 | 1.92 | 1.37 | -0.52 | 0.10 | 586 | -0.28 | 0.19 | 55 | 8.17 | 8.74 | 0.02 | 2.12 | 0.34 | 4275 | 1.26 | 1.47 | -0.66 | 0.11 | 586 | -0.66 | 0.21 | 55 | 7.94 | 8.44 | 0.00 | 0.41 | 3.5 | G | 1 |
| hd196171 | H | 4810 | 2.59 | 1.28 | -0.04 | 0.07 | 618 | 0.07 | 0.13 | 65 | 8.18 | 8.72 | 0.40 | 0.79 | 0.20 | 4810 | 2.31 | 1.36 | -0.09 | 0.08 | 618 | -0.09 | 0.14 | 65 | 8.11 | 8.58 | 0.40 | 0.22 | 3.4 | G | 1 |
| hd198232 | H | 4750 | 2.27 | 1.44 | -0.05 | 0.09 | 591 | 0.11 | 0.14 | 57 | 8.13 | 8.61 | 0.99 | 3.40 | 0.25 | 4750 | 1.89 | 1.49 | -0.09 | 0.10 | 591 | -0.09 | 0.15 | 57 | 8.04 | 8.44 | 0.99 | 0.28 | 4.5 | G | 1 |
| hd199951 | H | 5009 | 2.69 | 1.49 | -0.04 | 0.12 | 545 | 0.10 | 0.13 | 50 | 8.02 | 8.57 | 1.16 | 2.43 | 0.16 | 5009 | 2.36 | 1.57 | -0.07 | 0.12 | 545 | -0.07 | 0.14 | 50 | 7.94 | 8.41 | 1.17 | 0.17 | 6.8 | G | 1 |
| hd200763 | H | 4632 | 2.30 | 1.36 | 0.07 | 0.09 | 587 | 0.15 | 0.15 | 58 | 8.22 | 8.73 | 1.20 | 7.32 | 0.27 | 4632 | 2.11 | 1.42 | 0.03 | 0.09 | 587 | 0.03 | 0.16 | 58 | 8.16 | 8.64 | 1.20 | 0.29 | 3.7 | G | 1 |
| hd201852 | H | 4883 | 2.65 | 1.25 | -0.01 | 0.07 | 617 | 0.08 | 0.14 | 66 | 8.15 | 8.69 | 0.49 | 0.78 | 0.18 | 4883 | 2.44 | 1.32 | -0.04 | 0.07 | 617 | -0.04 | 0.14 | 66 | 8.10 | 8.59 | 0.49 | 0.19 | 3.4 | G | 1 |
| hd206642 | H | 4326 | 1.59 | 1.58 | -1.04 | 0.08 | 578 | -0.73 | 0.14 | 53 | 7.49 | 8.22 | -0.59 | 0.45 | 0.37 | 4326 | 0.67 | 1.59 | -1.15 | 0.10 | 578 | -1.15 | 0.17 | 53 | 7.22 | 7.81 | -0.63 | 0.48 | 3.6 | G | 1 |
| hd209449 | H | 5685 | 4.02 | 0.78 | 0.49 | 0.08 | 631 | 0.59 | 0.12 | 89 | 8.81 | 9.13 | 2.18 | 4.51 | 0.05 | 5685 | 3.79 | 1.01 | 0.44 | 0.07 | 631 | 0.45 | 0.11 | 89 | 8.71 | 9.03 | 2.18 | 0.05 | 3.2 | G | 2 |
| hd211317 | H | 5780 | 4.07 | 0.86 | 0.28 | 0.06 | 587 | 0.35 | 0.09 | 77 | 8.68 | 9.00 | 2.49 | 6.96 | 0.03 | 5780 | 3.92 | 1.00 | 0.27 | 0.06 | 587 | 0.27 | 0.09 | 77 | 8.62 | 8.93 | 2.49 | 0.03 | 3.8 | G | 2 |
| hip031592 | H | 4735 | 3.09 | 0.50 | 0.43 | 0.13 | 573 | 0.53 | 0.17 | 54 | 8.60 | 8.88 | 0.41 | 1.05 | 0.19 | 4735 | 2.84 | 0.88 | 0.30 | 0.09 | 573 | 0.30 | 0.16 | 54 | 8.50 | 8.75 | 0.41 | 0.20 | 2.9 | G | 1 |
| hip056713 | H | | | | | | | | | | 1.26 | 3.30 | | | | 4993 | 2.73 | 1.00 | -1.31 | 0.07 | 407 | -1.31 | 0.07 | 47 | 7.14 | 7.69 | 0.85 | 0.16 | 2.7 | G | 2 |
| hip080242b | H | 4315 | 3.40 | 0.75 | 0.47 | 0.18 | 519 | 1.62 | 0.20 | 68 | 8.59 | 9.09 | -0.16 | 1.16 | 0.25 | 4315 | 0.70 | 1.41 | -0.20 | 0.19 | 519 | -0.12 | 0.19 | 68 | 7.63 | 7.83 | -0.24 | 0.48 | 4.0 | G | 2 |
| hr2959 | H | 3916 | 2.50 | 1.25 | 0.67 | 0.20 | 477 | 1.45 | 0.21 | 32 | 8.99 | 9.36 | 0.00 | 5.21 | 0.31 | 3916 | 0.30 | 1.73 | -0.16 | 0.19 | 477 | -0.15 | 0.24 | 32 | 7.91 | 8.23 | -0.22 | 0.66 | 4.4 | G | 1 |
| hr3728 | H | 4888 | 2.60 | 1.34 | -0.38 | 0.06 | 613 | -0.23 | 0.12 | 63 | 7.94 | 8.55 | 0.06 | 0.30 | 0.18 | 4888 | 2.25 | 1.43 | -0.41 | 0.06 | 613 | -0.42 | 0.13 | 63 | 7.85 | 8.38 | 0.06 | 0.20 | 3.2 | G | 1 |
| hr3919 | H | 4415 | 1.90 | 1.41 | 0.01 | 0.12 | 574 | 0.05 | 0.19 | 51 | 8.32 | 8.70 | -0.09 | 1.01 | 0.34 | 4415 | 1.79 | 1.43 | -0.02 | 0.12 | 574 | -0.02 | 0.19 | 51 | 8.29 | 8.65 | -0.09 | 0.35 | 3.7 | G | 1 |
| hr5480 | H | 4807 | 2.09 | 1.76 | 0.13 | 0.13 | 521 | 0.15 | 0.16 | 47 | 8.12 | 8.63 | 0.79 | 1.90 | 0.13 | 4807 | 1.76 | 1.78 | -0.02 | 0.14 | 521 | -0.02 | 0.16 | 47 | 8.05 | 8.48 | 0.80 | 0.25 | 6.9 | G | 1 |
| hr7150 | H | 4541 | 1.71 | 1.82 | 0.01 | 0.13 | 516 | 0.11 | 0.17 | 44 | 8.17 | 8.66 | 0.17 | 1.16 | 0.34 | 4541 | 1.47 | 1.82 | -0.02 | 0.13 | 516 | -0.02 | 0.17 | 44 | 8.11 | 8.55 | 0.17 | 0.36 | 5.6 | G | 1 |
| ic4651no14527 | H | 4803 | 2.65 | 1.34 | 0.20 | 0.10 | 585 | 0.33 | 0.14 | 58 | 8.46 | 8.75 | -0.09 | 0.27 | 0.20 | 4803 | 2.35 | 1.42 | 0.15 | 0.11 | 585 | 0.15 | 0.15 | 58 | 8.38 | 8.61 | -0.08 | 0.22 | 4.0 | G | 1 |
| ic4651no8540 | H | 4826 | 2.68 | 1.29 | 0.24 | 0.10 | 596 | 0.27 | 0.15 | 59 | 8.42 | 8.92 | 1.02 | 2.92 | 0.19 | 4826 | 2.61 | 1.32 | 0.22 | 0.10 | 596 | 0.22 | 0.16 | 59 | 8.40 | 8.89 | 1.02 | 0.20 | 3.6 | G | 1 |
| ic4651no9025 | H | 4799 | 2.65 | 1.26 | 0.15 | 0.11 | 594 | 0.21 | 0.16 | 59 | 8.36 | 8.85 | 0.66 | 1.48 | 0.20 | 4799 | 2.51 | 1.32 | 0.12 | 0.10 | 594 | 0.12 | 0.17 | 59 | 8.32 | 8.78 | 0.66 | 0.21 | 3.4 | G | 1 |
| ic4651no9791 | H | 4463 | 2.21 | 1.29 | 0.14 | 0.12 | 576 | 0.30 | 0.21 | 51 | 8.36 | 8.81 | 1.09 | 9.44 | 0.31 | 4463 | 1.79 | 1.41 | 0.02 | 0.12 | 576 | 0.03 | 0.21 | 51 | 8.21 | 8.62 | 1.08 | 0.34 | 3.7 | G | 1 |
| ksihya | H | 4950 | 2.65 | 1.32 | 0.09 | 0.08 | 596 | 0.17 | 0.14 | 59 | 8.18 | 8.62 | 1.18 | 2.93 | 0.17 | 4950 | 2.46 | 1.38 | 0.07 | 0.09 | 596 | 0.07 | 0.14 | 59 | 8.15 | 8.53 | 1.18 | 0.18 | 4.5 | G | 1 |
| lra01_e1_0286 | H | | | | | | | | | | 0.24 | 3.30 | | | | 4950 | 2.74 | 1.18 | -0.13 | 0.10 | 620 | -0.13 | 0.21 | 56 | 8.02 | 8.56 | 0.05 | 0.17 | 3.1 | G | 2 |
| lra01_e2_2249 | H | | | | | | | | | | 2.47 | 3.30 | | | | 5250 | 3.86 | 0.27 | 0.39 | 0.24 | 568 | 0.39 | 0.27 | 57 | 8.56 | 9.24 | 1.46 | 0.11 | 3.4 | G | 2 |
| lra03_e2_0678 | H | | | | | | | | | | 1.03 | 3.30 | | | | 5235 | 4.89 | 0.74 | 0.20 | 0.23 | 488 | -0.09 | 0.29 | 35 | 8.80 | 9.10 | 1.00 | 0.15 | 1.7 | G | 2 |
| lra03_e2_1326 | H | | | | | | | | | | 0.39 | 3.30 | | | | 4970 | 3.36 | 1.04 | 0.14 | 0.22 | 623 | 0.14 | 0.16 | 46 | 8.48 | 8.99 | 0.29 | 0.15 | 3.4 | G | 2 |
| ngc2287no107 | H | 4602 | 1.57 | 1.84 | -0.13 | 0.12 | 549 | -0.12 | 0.14 | 47 | 8.07 | 8.52 | 0.82 | 3.86 | 0.34 | 4602 | 1.54 | 1.84 | -0.14 | 0.12 | 549 | -0.14 | 0.14 | 47 | 8.06 | 8.51 | 0.82 | 0.34 | 4.9 | G | 1 |
| ngc2287no204 | H | 4382 | 1.58 | 1.52 | -0.32 | 0.11 | 569 | -0.27 | 0.18 | 52 | 7.90 | 8.50 | -0.41 | 0.56 | 0.37 | 4382 | 1.44 | 1.53 | -0.34 | 0.11 | 569 | -0.34 | 0.18 | 52 | 7.85 | 8.44 | -0.41 | 0.38 | 3.9 | G | 1 |
| ngc2287no21 | H | 4020 | 2.50 | 1.58 | 0.59 | 0.19 | 476 | 1.45 | 0.24 | 39 | 8.89 | 9.37 | 0.60 | 11.76 | 0.32 | 4020 | 0.20 | 2.00 | -0.21 | 0.22 | 476 | -0.20 | 0.29 | 39 | 7.88 | 8.23 | 0.33 | 0.66 | 5.0 | G | 1 |
| ngc2287no75 | H | 4423 | 1.51 | 1.91 | -0.03 | 0.14 | 505 | 0.07 | 0.17 | 42 | 8.12 | 8.63 | 0.71 | 5.53 | 0.37 | 4423 | 1.28 | 1.90 | -0.06 | 0.14 | 505 | -0.06 | 0.17 | 42 | 8.05 | 8.53 | 0.71 | 0.40 | 5.6 | G | 1 |
| ngc2287no87 | H | 4199 | 1.41 | 1.56 | -0.27 | 0.15 | 557 | -0.20 | 0.25 | 46 | 8.08 | 8.53 | -0.52 | 0.86 | 0.39 | 4199 | 1.20 | 1.59 | -0.32 | 0.16 | 557 | -0.32 | 0.25 | 46 | 8.00 | 8.44 | -0.53 | 0.42 | 3.9 | G | 1 |
| ngc2287no97 | H | 4596 | 1.84 | 1.81 | -0.04 | 0.13 | 548 | 0.09 | 0.16 | 49 | 8.14 | 8.63 | 0.40 | 1.59 | 0.30 | 4596 | 1.56 | 1.82 | -0.06 | 0.13 | 548 | -0.06 | 0.16 | 49 | 8.07 | 8.50 | 0.40 | 0.34 | 5.1 | G | 1 |
| ngc3532no100 | H | 4740 | 2.05 | 1.75 | -0.02 | 0.12 | 556 | 0.09 | 0.19 | 57 | 8.12 | 8.61 | 0.93 | 3.09 | 0.27 | 4740 | 1.79 | 1.76 | -0.05 | 0.12 | 556 | -0.04 | 0.19 | 57 | 8.06 | 8.49 | 0.93 | 0.29 | 5.3 | G | 1 |
| ngc3532no122 | H | 4963 | 2.43 | 1.80 | 0.02 | 0.17 | 461 | 0.10 | 0.15 | 40 | 8.08 | 8.50 | 1.21 | 3.00 | 0.18 | 4963 | 2.26 | 1.83 | 0.01 | 0.17 | 461 | 0.01 | 0.15 | 40 | 8.06 | 8.42 | 1.21 | 0.19 | 9.1 | G | 1 |
| ngc3532no19 | H | 4844 | 2.23 | 1.55 | 0.00 | 0.11 | 572 | 0.13 | 0.13 | 57 | 8.09 | 8.60 | 1.23 | 4.30 | 0.21 | 4844 | 1.92 | 1.58 | -0.02 | 0.12 | 572 | -0.02 | 0.14 | 57 | 8.04 | 8.45 | 1.24 | 0.23 | 5.4 | G | 1 |
| ngc3532no596 | H | 4955 | 2.37 | 1.89 | -0.03 | 0.14 | 520 | 0.10 | 0.17 | 51 | 8.19 | 8.70 | 1.01 | 2.00 | 0.18 | 4955 | 2.07 | 1.92 | -0.05 | 0.14 | 520 | -0.05 | 0.17 | 51 | 8.14 | 8.55 | 1.02 | 0.20 | 6.9 | G | 1 |
| ngc3532no649 | H | 4849 | 2.29 | 1.41 | -0.15 | 0.09 | 612 | -0.15 | 0.14 | 66 | 8.16 | 8.67 | 3.34 | 35.53 | 0.21 | 4849 | 2.27 | 1.41 | -0.15 | 0.09 | 612 | -0.16 | 0.14 | 66 | 8.16 | 8.66 | 3.34 | 0.21 | 3.7 | G | 1 |
| ngc4349no127 | H | 4479 | 1.58 | 1.76 | -0.05 | 0.14 | 537 | -0.08 | 0.18 | 45 | 8.11 | 8.58 | 1.17 | 10.73 | 0.36 | 4479 | 1.67 | 1.76 | -0.04 | 0.14 | 537 | -0.03 | 0.18 | 45 | 8.14 | 8.62 | 1.17 | 0.35 | 4.9 | G | 1 |
| ngc6705no1286 | H | 4929 | 2.32 | 2.18 | 0.28 | 0.20 | 396 | 0.26 | 0.21 | 35 | 8.43 | 8.90 | 1.62 | 7.27 | 0.19 | 4929 | 2.35 | 2.18 | 0.28 | 0.20 | 396 | 0.28 | 0.21 | 35 | 8.43 | 8.91 | 1.62 | 0.19 | 10.0 | G | 1 |

| ID | Type | | | | | | | | | | | | | | | | | | | | | | | | | |
|---|---|---|---|---|---|---|---|---|---|---|---|---|---|---|---|---|---|---|---|---|---|---|---|---|---|---|
| ngc6705no1423 | H | 4521 | 1.83 | 1.92 | 0.25 | 0.17 | 493 | 0.23 | 0.19 | 38 | 8.52 | 8.91 | 1.28 | 11.49 | 0.33 | 4521 | 1.89 | 1.92 | 0.26 | 0.17 | 493 | 0.26 | 0.19 | 38 | 8.54 | 8.94 | 1.28 | 0.33 | 5.4 | G | 1 |
| ngc6705no411 | H | 4445 | 1.84 | 2.00 | 0.22 | 0.19 | 463 | 0.30 | 0.24 | 38 | 8.44 | 8.89 | 1.37 | 15.67 | 0.34 | 4445 | 1.65 | 2.01 | 0.18 | 0.19 | 463 | 0.18 | 0.24 | 38 | 8.38 | 8.80 | 1.36 | 0.36 | 6.4 | G | 1 |
| ngc6705no660 | H | 4788 | 2.13 | 1.95 | 0.29 | 0.15 | 496 | 0.27 | 0.23 | 48 | 8.40 | 8.87 | 1.54 | 8.86 | 0.23 | 4788 | 2.17 | 1.95 | 0.29 | 0.15 | 496 | 0.29 | 0.23 | 48 | 8.41 | 8.89 | 1.54 | 0.23 | 5.9 | G | 1 |
| ngc6705no779 | H | 4307 | 1.65 | 1.98 | 0.15 | 0.19 | 479 | 0.30 | 0.24 | 39 | 8.30 | 8.73 | 1.26 | 18.31 | 0.37 | 4307 | 1.26 | 2.00 | 0.07 | 0.20 | 479 | 0.07 | 0.25 | 39 | 8.17 | 8.56 | 1.24 | 0.41 | 5.6 | G | 1 |
| pihya | H | 4563 | 2.39 | 1.20 | 0.00 | 0.09 | 599 | 0.12 | 0.14 | 57 | 8.25 | 8.74 | 0.23 | 1.21 | 0.28 | 4563 | 2.11 | 1.30 | -0.07 | 0.09 | 599 | -0.07 | 0.15 | 57 | 8.16 | 8.61 | 0.23 | 0.30 | 3.3 | G | 1 |
| sand1016 | H | 4431 | 2.16 | 1.32 | 0.09 | 0.13 | 562 | 0.22 | 0.21 | 53 | 8.31 | 8.76 | -0.04 | 1.07 | 0.31 | 4431 | 1.82 | 1.42 | 0.00 | 0.13 | 562 | 0.00 | 0.22 | 53 | 8.19 | 8.61 | -0.04 | 0.34 | 3.9 | G | 1 |
| sand1054 | H | 4721 | 2.67 | 1.11 | 0.08 | 0.11 | 592 | 0.17 | 0.17 | 59 | 8.29 | 8.71 | 0.42 | 1.11 | 0.22 | 4721 | 2.45 | 1.23 | 0.02 | 0.10 | 592 | 0.02 | 0.17 | 59 | 8.23 | 8.60 | 0.43 | 0.24 | 3.6 | G | 1 |
| sand1074 | H | 4718 | 2.49 | 1.37 | 0.07 | 0.10 | 584 | 0.18 | 0.14 | 56 | 8.30 | 8.79 | -0.09 | 0.35 | 0.24 | 4718 | 2.24 | 1.44 | 0.02 | 0.10 | 584 | 0.03 | 0.15 | 56 | 8.22 | 8.68 | -0.08 | 0.26 | 3.8 | G | 1 |
| sand1084 | H | 4745 | 2.48 | 1.37 | 0.07 | 0.10 | 590 | 0.14 | 0.14 | 55 | 8.28 | 8.73 | 0.21 | 0.63 | 0.24 | 4745 | 2.30 | 1.43 | 0.03 | 0.09 | 590 | 0.03 | 0.15 | 55 | 8.22 | 8.64 | 0.21 | 0.25 | 3.9 | G | 1 |
| sand1237 | H | 5067 | 2.85 | 1.35 | 0.06 | 0.09 | 600 | 0.05 | 0.13 | 59 | 8.17 | 8.59 | 0.42 | 0.40 | 0.14 | 5067 | 2.87 | 1.37 | 0.07 | 0.09 | 600 | 0.07 | 0.12 | 59 | 8.17 | 8.61 | 0.42 | 0.14 | 4.2 | G | 1 |
| sand1279 | H | 4732 | 2.48 | 1.34 | 0.11 | 0.10 | 585 | 0.15 | 0.13 | 57 | 8.24 | 8.77 | -0.02 | 0.39 | 0.24 | 4732 | 2.38 | 1.37 | 0.09 | 0.10 | 585 | 0.09 | 0.14 | 57 | 8.21 | 8.73 | -0.02 | 0.27 | 3.8 | G | 1 |
| sand364 | H | 4245 | 1.83 | 1.36 | 0.05 | 0.13 | 558 | 0.24 | 0.21 | 46 | 8.35 | 8.78 | -0.33 | 1.10 | 0.35 | 4245 | 1.31 | 1.46 | -0.09 | 0.14 | 558 | -0.09 | 0.22 | 46 | 8.16 | 8.55 | -0.35 | 0.41 | 4.0 | G | 1 |
| sand978 | H | 4257 | 1.82 | 1.48 | 0.04 | 0.16 | 547 | 0.13 | 0.21 | 42 | 8.36 | 8.75 | -0.65 | 0.51 | 0.35 | 4257 | 1.56 | 1.53 | -0.04 | 0.16 | 547 | -0.04 | 0.22 | 42 | 8.26 | 8.64 | -0.66 | 0.38 | 4.1 | G | 1 |
| sand989 | H | 4793 | 2.84 | 1.14 | 0.19 | 0.12 | 594 | 0.31 | 0.20 | 59 | 8.41 | 8.96 | 0.00 | 0.34 | 0.19 | 4793 | 2.67 | 1.13 | 0.14 | 0.12 | 594 | 0.14 | 0.20 | 59 | 8.25 | 8.79 | 0.00 | 0.20 | 3.8 | G | 1 |
| sigpup | H | 4081 | 3.53 | 1.25 | 0.64 | 0.31 | 483 | 1.45 | 0.29 | 25 | 9.91 | 10.21 | -0.51 | 0.93 | 0.28 | 4081 | 1.55 | 1.11 | 0.31 | 0.25 | 483 | 0.30 | 0.31 | 25 | 8.64 | 9.12 | -0.65 | 0.37 | 4.5 | G | 1 |
| tzfor | H | 5332 | 2.96 | 0.52 | 0.14 | 0.20 | 568 | -0.16 | 0.21 | 51 | 8.20 | 8.64 | 0.90 | 0.61 | 0.09 | 5332 | 3.53 | 0.50 | 0.14 | 0.21 | 568 | 0.14 | 0.20 | 51 | 8.39 | 8.99 | 0.90 | 0.09 | 6.5 | G | 1 |
| v1045sco | H | 3789 | 1.37 | 2.80 | -0.21 | 0.37 | 272 | 0.62 | 0.40 | 16 | 8.46 | 8.87 | 2.10 | 44.28 | 0.34 | 3789 | -0.30 | 2.69 | -0.66 | 0.41 | 272 | -0.34 | 0.41 | 16 | 7.77 | 8.06 | 1.71 | 1.02 | 10.0 | G | 2 |
| v4393sgr | H | | | | | | | | | | | | | | | 4000 | 0.31 | 3.04 | -0.13 | 0.24 | 152 | -0.13 | 0.12 | 4 | 7.95 | 8.48 | -0.40 | 0.63 | 15.0 | G | 2 |
| xcae | H | 6973 | 3.68 | 5.37 | -0.35 | 0.24 | 65 | -0.18 | 0.27 | 6 | 7.89 | 8.83 | 2.81 | 2.45 | 0.04 | 6973 | 3.15 | 5.30 | -0.36 | 0.24 | 65 | -0.36 | 0.27 | 6 | 7.77 | 8.69 | 2.81 | 0.12 | 56.0 | R | 2 |
| hd001142 | S | 5186 | 2.90 | 1.07 | 0.15 | 0.12 | 396 | 0.00 | 0.18 | 29 | 8.13 | 8.69 | 1.22 | 1.78 | 0.12 | 5186 | 3.12 | 0.99 | 0.19 | 0.12 | 396 | 0.19 | 0.17 | 29 | 8.25 | 8.89 | 1.23 | 0.12 | 6.4 | G | 1 |
| hd002410 | S | 5281 | 2.63 | 1.58 | 0.26 | 0.09 | 401 | -0.20 | 0.10 | 26 | 8.25 | 8.78 | 1.00 | 0.84 | 0.11 | 5281 | 3.59 | 1.15 | 0.37 | 0.12 | 401 | 0.37 | 0.11 | 26 | 8.35 | 9.22 | 0.99 | 0.10 | 4.2 | G | 1 |
| hd002954 | S | 6308 | 3.55 | 3.00 | 0.01 | 0.22 | 98 | 0.03 | 0.13 | 12 | | 8.75 | 2.08 | 1.29 | 0.02 | 6308 | 3.51 | 3.01 | 0.01 | 0.22 | 98 | 0.01 | 0.13 | 12 | | 8.73 | 2.08 | 0.03 | 33.5 | R | 2 |
| hd005418 | S | 4977 | 2.84 | 1.33 | 0.02 | 0.07 | 409 | 0.05 | 0.09 | 29 | 8.17 | 8.77 | 0.30 | 0.39 | 0.16 | 4977 | 2.78 | 1.35 | 0.01 | 0.07 | 409 | 0.01 | 0.10 | 29 | 8.15 | 8.74 | 0.30 | 0.16 | 4.1 | G | 1 |
| hd006037 | S | 4566 | 2.71 | 1.26 | 0.36 | 0.16 | 363 | 0.53 | 0.22 | 27 | 8.66 | 9.00 | 0.30 | 1.42 | 0.25 | 4566 | 2.27 | 1.46 | 0.21 | 0.16 | 363 | 0.21 | 0.23 | 27 | 8.49 | 8.79 | 0.30 | 0.29 | 4.1 | G | 1 |
| hd012116 | S | 4462 | 2.47 | 1.23 | -0.10 | 0.10 | 394 | 0.16 | 0.12 | 24 | 8.43 | 8.80 | -0.18 | 0.68 | 0.29 | 4462 | 2.00 | 1.34 | -0.27 | 0.10 | 394 | -0.27 | 0.13 | 24 | 8.13 | 8.46 | -0.18 | 0.33 | 4.3 | G | 1 |
| hd013004 | S | 4584 | 2.77 | 1.08 | 0.33 | 0.13 | 388 | 0.55 | 0.26 | 27 | 8.61 | 8.99 | 0.28 | 1.28 | 0.24 | 4584 | 2.23 | 1.37 | 0.15 | 0.14 | 388 | 0.14 | 0.28 | 27 | 8.42 | 8.74 | 0.28 | 0.29 | 4.1 | G | 1 |
| hd015533 | S | 4370 | 2.49 | 1.24 | 0.38 | 0.14 | 224 | 0.65 | 0.17 | 14 | | -0.05 | 1.26 | 0.30 | | 4370 | 1.81 | 1.35 | 0.19 | 0.14 | 224 | 0.19 | 0.18 | 14 | | -0.06 | 0.35 | | 4.4 | G | 1 |
| hd015866 | S | 5718 | 3.84 | 1.29 | 0.25 | 0.09 | 430 | 0.38 | 0.11 | 36 | 8.77 | 8.95 | 1.81 | 2.03 | 0.04 | 5718 | 3.57 | 1.44 | 0.23 | 0.10 | 430 | 0.23 | 0.11 | 36 | 8.66 | 8.84 | 1.81 | 0.04 | 4.7 | G | 2 |
| hd016150 | S | 6145 | 3.21 | 3.19 | -0.06 | 0.26 | 118 | 0.06 | 0.34 | 5 | 7.73 | 8.00 | 2.90 | 8.67 | 0.04 | 6145 | 2.87 | 3.00 | -0.02 | 0.26 | 118 | -0.02 | 0.36 | 5 | 7.63 | 8.20 | 2.91 | 0.04 | 42.0 | R | 2 |
| hd016314 | S | 6533 | 3.70 | 3.58 | 0.07 | 0.15 | 201 | 0.22 | 0.11 | 21 | 8.59 | 9.17 | 1.67 | 0.36 | 0.01 | 6533 | 3.29 | 3.59 | 0.07 | 0.15 | 201 | 0.07 | 0.11 | 21 | 8.47 | 9.05 | 1.67 | 0.05 | 26.6 | R | 2 |
| hd017001 | S | 4863 | 2.76 | 1.32 | -0.07 | 0.08 | 403 | 0.08 | 0.10 | 27 | 8.24 | 8.77 | 0.48 | 0.81 | 0.18 | 4863 | 2.44 | 1.43 | -0.12 | 0.09 | 403 | -0.12 | 0.10 | 27 | 8.14 | 8.62 | 0.48 | 0.20 | 4.7 | G | 1 |
| hd019210 | S | 4952 | 2.65 | 1.39 | 0.13 | 0.08 | 400 | 0.01 | 0.12 | 30 | 8.23 | 8.75 | 0.97 | 1.84 | 0.17 | 4952 | 2.81 | 1.37 | 0.16 | 0.08 | 400 | 0.17 | 0.11 | 30 | 8.35 | 8.92 | 0.97 | 0.16 | 4.0 | G | 1 |
| hd021585 | S | 5204 | 2.84 | 1.62 | -0.69 | 0.09 | 369 | -0.56 | 0.06 | 24 | 7.80 | 8.61 | 0.30 | 0.22 | 0.12 | 5204 | 2.54 | 1.68 | -0.70 | 0.09 | 369 | -0.70 | 0.06 | 24 | 7.73 | 8.48 | 0.31 | 0.13 | 6.0 | G | 1 |
| hd021760 | S | 4627 | 2.79 | 1.03 | 0.12 | 0.11 | 408 | 0.22 | 0.17 | 26 | 8.40 | 8.75 | 0.22 | 0.97 | 0.23 | 4627 | 2.52 | 1.26 | 0.01 | 0.11 | 408 | 0.01 | 0.18 | 26 | 8.31 | 8.62 | 0.23 | 0.25 | 3.4 | G | 1 |
| hd025627 | S | 4613 | 2.74 | 1.14 | 0.23 | 0.12 | 388 | 0.33 | 0.22 | 22 | 8.56 | 9.00 | 0.07 | 0.73 | 0.23 | 4613 | 2.61 | 1.11 | 0.18 | 0.12 | 388 | 0.18 | 0.22 | 22 | 8.42 | 8.85 | 0.08 | 0.24 | 4.5 | G | 1 |
| hd026004 | S | 4490 | 2.40 | 1.27 | -0.04 | 0.10 | 394 | 0.19 | 0.16 | 29 | 8.19 | 8.78 | 0.21 | 1.51 | 0.29 | 4490 | 1.83 | 1.44 | -0.19 | 0.10 | 394 | -0.19 | 0.18 | 29 | 8.01 | 8.53 | 0.21 | 0.34 | 4.1 | G | 1 |
| hd026625 | S | 4898 | 2.85 | 1.22 | 0.11 | 0.09 | 429 | 0.12 | 0.11 | 26 | 8.26 | 8.75 | 0.50 | 0.78 | 0.17 | 4898 | 2.83 | 1.23 | 0.10 | 0.09 | 429 | 0.10 | 0.11 | 26 | 8.25 | 8.74 | 0.50 | 0.17 | 4.1 | G | 1 |
| hd033844 | S | 4792 | 3.13 | 0.79 | 0.37 | 0.12 | 399 | 0.51 | 0.16 | 27 | 8.52 | 8.94 | 0.46 | 0.98 | 0.18 | 4792 | 2.89 | 1.02 | 0.23 | 0.10 | 399 | 0.23 | 0.18 | 27 | 8.33 | 8.73 | 0.46 | 0.19 | 3.9 | G | 1 |
| hd058898 | S | 4387 | 2.32 | 1.27 | 0.16 | 0.14 | 371 | 0.34 | 0.19 | 25 | 8.48 | 8.83 | -0.16 | 0.95 | 0.31 | 4387 | 1.85 | 1.45 | 0.00 | 0.14 | 371 | 0.00 | 0.21 | 25 | 8.30 | 8.61 | -0.16 | 0.34 | 4.4 | G | 1 |
| hd061191 | S | 4688 | 2.82 | 1.01 | 0.16 | 0.11 | 395 | 0.27 | 0.14 | 27 | 8.28 | 8.78 | 0.13 | 0.64 | 0.21 | 4688 | 2.57 | 1.24 | 0.07 | 0.10 | 395 | 0.07 | 0.14 | 27 | 8.20 | 8.67 | 0.13 | 0.23 | 4.0 | G | 1 |
| hd068667 | S | 5010 | 2.76 | 1.33 | 0.03 | 0.08 | 409 | 0.04 | 0.09 | 31 | 8.12 | 8.62 | 0.77 | 1.03 | 0.15 | 5010 | 2.74 | 1.34 | 0.02 | 0.08 | 409 | 0.02 | 0.09 | 31 | 8.12 | 8.61 | 0.77 | 0.16 | 4.9 | G | 1 |
| hd070522 | S | 6122 | 3.77 | 1.94 | 0.01 | 0.14 | 298 | 0.04 | 0.10 | 25 | 8.57 | 8.28 | 3.13 | 12.16 | 0.01 | 6122 | 3.71 | 1.97 | 0.01 | 0.14 | 298 | 0.01 | 0.10 | 25 | 8.54 | 8.64 | 3.13 | 0.02 | 10.0 | G | 1 |
| hd077232 | S | 6773 | 3.78 | 6.50 | 0.00 | 0.22 | 20 | 0.05 | 0.08 | 2 | 7.29 | 9.91 | 2.74 | 2.72 | 0.05 | 6773 | 3.64 | 7.75 | 0.00 | 0.22 | 20 | 0.00 | 0.08 | 2 | 8.14 | | 2.74 | 0.03 | 50.0 | R | 2 |
| hd083087 | S | 4777 | 3.06 | 0.82 | 0.10 | 0.10 | 406 | 0.27 | 0.14 | 29 | 8.26 | 8.68 | 0.05 | 0.40 | 0.19 | 4777 | 2.69 | 1.19 | -0.02 | 0.09 | 406 | -0.02 | 0.15 | 29 | 8.14 | 8.51 | 0.05 | 0.20 | 3.7 | G | 1 |
| hd085440 | S | 5041 | 3.39 | 1.01 | -0.09 | 0.08 | 410 | 0.08 | 0.10 | 22 | 8.29 | 8.65 | 1.06 | 1.84 | 0.14 | 5041 | 3.07 | 1.25 | -0.15 | 0.09 | 410 | -0.15 | 0.11 | 22 | 8.19 | 8.49 | 1.07 | 0.14 | 4.0 | G | 1 |
| hd089280 | S | 7384 | 4.18 | 5.28 | -0.87 | 0.33 | 24 | -0.72 | 0.22 | 4 | 7.96 | 8.67 | 2.63 | 1.04 | -0.15 | 7384 | 3.78 | 5.18 | -0.85 | 0.33 | 24 | -0.85 | 0.22 | 4 | 7.87 | 8.55 | 2.65 | 0.05 | 40.0 | R | 2 |
| hd090250 | S | 4691 | 2.63 | 1.33 | 0.10 | 0.09 | 382 | 0.23 | 0.14 | 28 | 8.37 | 8.86 | 0.25 | 0.84 | 0.23 | 4691 | 2.34 | 1.45 | 0.03 | 0.09 | 382 | 0.04 | 0.14 | 28 | 8.27 | 8.72 | 0.26 | 0.26 | 4.4 | G | 1 |
| hd098579 | S | 4665 | 2.89 | 1.05 | 0.26 | 0.12 | 394 | 0.46 | 0.19 | 27 | 8.58 | 8.98 | 0.36 | 1.16 | 0.21 | 4665 | 2.53 | 1.26 | 0.11 | 0.12 | 394 | 0.11 | 0.21 | 27 | 8.35 | 8.71 | 0.36 | 0.24 | 4.1 | G | 1 |
| hd101321 | S | 4757 | 2.88 | 1.08 | -0.12 | 0.07 | 400 | 0.00 | 0.14 | 28 | 8.24 | 8.68 | 0.33 | 0.81 | 0.20 | 4757 | 2.59 | 1.24 | -0.19 | 0.08 | 400 | -0.19 | 0.15 | 28 | 8.15 | 8.54 | 0.33 | 0.21 | 3.6 | G | 1 |
| hd104819 | S | 4618 | 3.10 | 1.20 | 0.46 | 0.16 | 359 | 0.72 | 0.27 | 24 | 8.73 | 9.04 | 0.42 | 1.55 | 0.21 | 4618 | 2.50 | 1.55 | 0.25 | 0.17 | 359 | 0.25 | 0.28 | 24 | 8.50 | 8.75 | 0.42 | 0.26 | 4.1 | G | 1 |

| ID | | | | | | | | | | | | | | | | | | | | | | | | | | | | | |
|---|---|---|---|---|---|---|---|---|---|---|---|---|---|---|---|---|---|---|---|---|---|---|---|---|---|---|---|---|---|
| hd104883 | S | 6203 | 3.60 | 4.19 | -0.12 | 0.26 | 53 | -0.08 | 0.02 | 3 | | | 1.64 | 0.59 | 0.02 | 6203 | 3.46 | 4.11 | -0.11 | 0.26 | 53 | -0.11 | 0.02 | 3 | 7.93 | 9.17 | 1.65 | 0.03 | 55.0 | R | 2 |
| hd106972 | S | 6166 | 3.71 | 2.01 | -0.02 | 0.09 | 264 | 0.04 | 0.09 | 22 | 8.34 | 8.62 | 0.87 | 0.11 | 0.01 | 6166 | 3.58 | 2.07 | -0.02 | 0.09 | 264 | -0.02 | 0.09 | 22 | 8.29 | 8.70 | 0.87 | 0.01 | 15.3 | R | 2 |
| hd107415 | S | 4797 | 2.69 | 1.27 | -0.14 | 0.08 | 403 | -0.04 | 0.09 | 29 | 7.94 | 8.67 | 0.16 | 0.48 | 0.20 | 4797 | 2.47 | 1.34 | -0.17 | 0.08 | 403 | -0.18 | 0.10 | 29 | 7.88 | 8.57 | 0.16 | 0.21 | 4.2 | G | 1 |
| hd107569 | S | 6101 | 3.53 | 2.91 | 0.07 | 0.16 | 167 | 0.06 | 0.08 | 12 | 8.41 | 8.85 | 1.07 | 0.19 | 0.02 | 6101 | 3.55 | 2.90 | 0.07 | 0.16 | 167 | 0.07 | 0.08 | 12 | 8.41 | 8.85 | 1.08 | 0.02 | 30.3 | R | 2 |
| hd107610 | S | 4693 | 2.74 | 1.18 | 0.31 | 0.12 | 398 | 0.37 | 0.22 | 25 | 8.51 | 8.90 | 0.39 | 1.13 | 0.22 | 4693 | 2.59 | 1.27 | 0.26 | 0.12 | 398 | 0.26 | 0.22 | 25 | 8.46 | 8.83 | 0.39 | 0.23 | 4.2 | G | 1 |
| hd112357 | S | 4969 | 3.38 | 0.92 | -0.15 | 0.07 | 411 | -0.01 | 0.11 | 26 | 8.27 | 8.66 | 0.00 | 0.21 | 0.15 | 4969 | 3.08 | 1.16 | -0.22 | 0.08 | 411 | -0.22 | 0.12 | 26 | 8.18 | 8.51 | 0.01 | 0.15 | 3.1 | G | 1 |
| hd116204 | S | 4472 | 2.74 | 2.93 | -0.01 | 0.24 | 194 | 0.65 | 0.44 | 12 | 8.21 | 8.58 | 0.31 | 2.00 | 0.26 | 4472 | 1.09 | 2.72 | -0.25 | 0.26 | 194 | -0.26 | 0.47 | 12 | 7.77 | 7.82 | 0.29 | 0.41 | 16.0 | R | 1 |
| hd126265 | S | 5868 | 3.84 | 1.35 | -0.01 | 0.10 | 389 | 0.09 | 0.09 | 33 | 8.48 | 8.70 | 0.78 | 0.15 | 0.03 | 5868 | 3.64 | 1.49 | -0.02 | 0.10 | 389 | -0.02 | 0.08 | 33 | 8.40 | 8.62 | 0.78 | 0.03 | 6.4 | G | 2 |
| hd127740 | S | 6241 | 3.73 | 3.40 | -0.15 | 0.24 | 104 | -0.14 | 0.13 | 9 | 8.17 | 8.78 | 2.16 | 1.77 | 0.01 | 6241 | 3.70 | 3.41 | -0.15 | 0.24 | 104 | -0.15 | 0.13 | 9 | 8.16 | 8.76 | 2.16 | 0.01 | 40.0 | R | 2 |
| hd128853 | S | 4819 | 3.06 | 0.97 | -0.13 | 0.08 | 412 | 0.09 | 0.09 | 24 | 8.19 | 8.62 | 0.99 | 2.89 | 0.18 | 4819 | 2.73 | 1.05 | -0.22 | 0.08 | 412 | -0.22 | 0.09 | 24 | 7.98 | 8.34 | 0.99 | 0.19 | 3.6 | G | 1 |
| hd133194 | S | 6312 | 3.47 | 3.64 | -0.07 | 0.16 | 281 | 0.23 | 0.16 | 30 | 8.69 | 9.16 | 2.09 | 1.33 | 0.03 | 6312 | 2.68 | 3.65 | -0.07 | 0.17 | 281 | -0.06 | 0.15 | 30 | 8.44 | 8.92 | 2.10 | 0.02 | 12.0 | G | 2 |
| hd138085 | S | 4799 | 2.59 | 1.35 | -0.41 | 0.07 | 416 | -0.22 | 0.07 | 25 | 7.89 | 8.46 | 0.13 | 0.45 | 0.20 | 4799 | 2.17 | 1.45 | -0.46 | 0.07 | 416 | -0.46 | 0.07 | 25 | 7.78 | 8.27 | 0.13 | 0.23 | 4.2 | G | 1 |
| hd138686 | S | 6257 | 3.25 | 6.50 | -0.21 | 0.41 | 34 | 0.07 | 0.07 | 2 | | 8.72 | 2.72 | 5.62 | 0.04 | 6257 | 2.43 | 7.75 | -0.23 | 0.39 | 34 | -0.23 | 0.06 | 2 | | 8.46 | 2.73 | 0.03 | 46.0 | R | 2 |
| hd148317 | S | 5758 | 3.56 | 1.43 | 0.10 | 0.07 | 403 | 0.15 | 0.07 | 36 | 8.55 | 8.83 | 3.07 | 16.19 | 0.04 | 5758 | 3.43 | 1.49 | 0.09 | 0.07 | 403 | 0.08 | 0.07 | 36 | 8.50 | 8.78 | 3.07 | 0.04 | 4.6 | G | 2 |
| hd149216 | S | 4295 | 2.65 | 0.68 | -0.05 | 0.17 | 418 | 0.92 | 0.11 | 27 | 8.13 | 8.72 | -0.49 | 0.61 | 0.30 | 4295 | 0.36 | 1.39 | -0.69 | 0.22 | 418 | -0.69 | 0.17 | 27 | 7.32 | 7.57 | -0.59 | 0.54 | 3.9 | G | 1 |
| hd157935 | S | 6768 | 3.47 | 2.53 | -0.21 | 0.18 | 136 | -0.24 | 0.10 | 14 | | 8.81 | 1.71 | 0.27 | 0.06 | 6768 | 3.55 | 2.60 | -0.22 | 0.18 | 136 | -0.22 | 0.10 | 14 | | 8.84 | 1.69 | 0.05 | 15.3 | R | 2 |
| hd161502 | S | 4963 | 2.99 | 1.30 | -0.31 | 0.08 | 426 | -0.17 | 0.10 | 26 | 8.00 | 8.35 | 0.70 | 1.05 | 0.16 | 4963 | 2.69 | 1.45 | -0.35 | 0.09 | 426 | -0.35 | 0.10 | 26 | 7.92 | 8.20 | 0.71 | 0.17 | 3.9 | G | 1 |
| hd167576 | S | 4506 | 2.56 | 1.58 | 0.41 | 0.17 | 355 | 0.60 | 0.24 | 24 | 8.74 | 9.07 | 0.26 | 1.59 | 0.27 | 4506 | 2.10 | 1.70 | 0.28 | 0.17 | 355 | 0.28 | 0.25 | 24 | 8.56 | 8.86 | 0.26 | 0.31 | 4.9 | G | 1 |
| hd172052 | S | 5544 | 4.79 | 1.47 | -0.46 | 0.05 | 78 | 1.53 | 0.18 | 16 | | 9.48 | 0.51 | 0.17 | 0.09 | 5544 | 0.10 | 2.74 | -0.49 | 0.08 | 78 | -0.46 | 0.15 | 16 | | 7.56 | 0.60 | 0.15 | 5.0 | G | 2 |
| hd173378 | S | 4855 | 2.89 | 1.11 | -0.29 | 0.07 | 409 | -0.23 | 0.12 | 29 | 7.98 | 8.42 | -0.01 | 0.28 | 0.18 | 4855 | 2.76 | 1.19 | -0.31 | 0.07 | 409 | -0.31 | 0.12 | 29 | 7.94 | 8.35 | -0.01 | 0.18 | 3.7 | G | 1 |
| hd175940 | S | 4629 | 2.76 | 1.06 | 0.16 | 0.11 | 395 | 0.27 | 0.16 | 26 | 8.37 | 8.81 | 0.00 | 0.58 | 0.23 | 4629 | 2.49 | 1.27 | 0.06 | 0.10 | 395 | 0.06 | 0.18 | 26 | 8.27 | 8.68 | 0.00 | 0.26 | 3.8 | G | 1 |
| hd182901 | S | 6491 | 4.01 | 2.33 | 0.17 | 0.13 | 210 | 0.09 | 0.11 | 19 | 8.44 | 8.71 | 1.74 | 0.44 | -0.03 | 6491 | 4.19 | 2.24 | 0.18 | 0.13 | 210 | 0.18 | 0.12 | 19 | 8.49 | 8.77 | 1.74 | -0.06 | 21.0 | R | 2 |
| hd186535 | S | 5012 | 2.74 | 1.34 | 0.04 | 0.07 | 405 | 0.01 | 0.09 | 26 | 8.16 | 8.68 | 0.35 | 0.40 | 0.15 | 5012 | 2.80 | 1.32 | 0.04 | 0.07 | 405 | 0.05 | 0.09 | 26 | 8.18 | 8.71 | 0.35 | 0.15 | 4.2 | G | 1 |
| hd188993 | S | 5673 | 3.45 | 1.59 | -0.01 | 0.09 | 394 | 0.06 | 0.08 | 31 | 8.49 | 8.68 | 3.20 | 19.67 | 0.05 | 5673 | 3.30 | 1.65 | -0.02 | 0.09 | 394 | -0.02 | 0.08 | 31 | 8.44 | 8.61 | 3.20 | 0.05 | 6.6 | G | 1 |
| hd189186 | S | 4899 | 3.09 | 0.91 | -0.37 | 0.07 | 422 | -0.22 | 0.07 | 25 | 7.93 | 8.47 | 0.10 | 0.32 | 0.17 | 4899 | 2.78 | 1.12 | -0.43 | 0.07 | 422 | -0.42 | 0.06 | 25 | 7.83 | 8.32 | 0.10 | 0.17 | 3.6 | G | 1 |
| hd194708 | S | 6313 | 3.67 | 3.00 | 0.24 | 0.39 | 55 | 0.21 | 0.26 | 4 | 8.32 | 8.98 | 3.13 | 10.00 | 0.01 | 6313 | 3.73 | 2.99 | 0.24 | 0.39 | 55 | 0.24 | 0.27 | 4 | 8.34 | 8.99 | 3.13 | 0.01 | 50.0 | R | 2 |
| hd200925 | S | 6642 | 3.40 | 3.10 | -0.00 | 0.19 | 248 | 0.04 | 0.17 | 32 | 8.42 | 8.77 | 2.89 | 4.30 | 0.05 | 6642 | 3.03 | 3.11 | -0.09 | 0.20 | 248 | -0.09 | 0.17 | 32 | 8.32 | 8.66 | 2.89 | 0.07 | 17.3 | R | 2 |
| hd204642 | S | 4660 | 2.84 | 1.09 | 0.20 | 0.10 | 401 | 0.32 | 0.12 | 23 | 8.50 | 8.91 | 0.26 | 0.94 | 0.22 | 4660 | 2.68 | 1.12 | 0.13 | 0.10 | 401 | 0.12 | 0.12 | 23 | 8.35 | 8.74 | 0.26 | 0.23 | 3.7 | G | 1 |
| hd205011 | S | 4770 | 2.54 | 1.43 | -0.03 | 0.12 | 426 | 0.14 | 0.13 | 29 | 8.41 | 8.72 | 0.53 | 1.21 | 0.21 | 4770 | 2.16 | 1.52 | -0.09 | 0.13 | 426 | -0.09 | 0.14 | 29 | 8.30 | 8.54 | 0.54 | 0.23 | 5.4 | G | 1 |
| hd205972 | S | 4742 | 2.89 | 1.02 | 0.13 | 0.10 | 394 | 0.11 | 0.11 | 22 | 8.33 | 8.73 | 0.17 | 0.59 | 0.20 | 4742 | 2.82 | 1.21 | 0.11 | 0.10 | 394 | 0.11 | 0.11 | 22 | 8.41 | 8.79 | 0.17 | 0.20 | 3.4 | G | 1 |
| hd211607 | S | 4944 | 2.98 | 1.19 | 0.20 | 0.08 | 414 | 0.27 | 0.12 | 29 | 8.37 | 8.91 | 0.64 | 0.91 | 0.16 | 4944 | 2.84 | 1.27 | 0.17 | 0.08 | 414 | 0.17 | 0.12 | 29 | 8.33 | 8.84 | 0.64 | 0.16 | 4.5 | G | 1 |
| hd212334 | S | 4722 | 2.49 | 1.38 | -0.06 | 0.08 | 395 | 0.07 | 0.12 | 35 | 8.27 | 8.75 | 0.46 | 1.20 | 0.24 | 4722 | 2.19 | 1.45 | -0.11 | 0.09 | 395 | -0.11 | 0.13 | 35 | 8.18 | 8.61 | 0.47 | 0.26 | 4.0 | G | 1 |
| hd213619 | S | 6884 | 3.78 | 5.80 | 0.09 | 0.29 | 20 | -0.16 | 0.10 | 2 | 7.62 | 9.88 | 1.36 | 0.10 | 0.01 | 6884 | 4.50 | 5.79 | 0.08 | 0.29 | 20 | 0.08 | 0.09 | 2 | 7.82 | | 1.35 | -0.23 | 45.0 | R | 2 |
| hd217590 | S | 4950 | 2.84 | 1.27 | 0.13 | 0.08 | 409 | 0.15 | 0.12 | 31 | 8.21 | 8.69 | 0.00 | 0.21 | 0.16 | 4950 | 2.80 | 1.29 | 0.13 | 0.08 | 409 | 0.13 | 0.12 | 31 | 8.20 | 8.67 | 0.00 | 0.16 | 4.1 | G | 1 |
| hd219409 | S | 4634 | 2.73 | 1.08 | 0.03 | 0.10 | 417 | 0.16 | 0.15 | 27 | 8.31 | 8.74 | 0.25 | 1.00 | 0.23 | 4634 | 2.42 | 1.27 | -0.06 | 0.10 | 417 | -0.06 | 0.15 | 27 | 8.21 | 8.59 | 0.25 | 0.26 | 4.0 | G | 1 |
| hd222683 | S | 4936 | 2.85 | 1.23 | 0.13 | 0.08 | 398 | 0.18 | 0.13 | 34 | 8.25 | 8.74 | 0.70 | 1.08 | 0.17 | 4936 | 2.76 | 1.28 | 0.11 | 0.08 | 398 | 0.12 | 0.13 | 34 | 8.22 | 8.70 | 0.70 | 0.17 | 4.3 | G | 1 |
| hd223869 | S | 4830 | 3.24 | 0.68 | 0.09 | 0.10 | 414 | 0.22 | 0.16 | 25 | 8.35 | 8.67 | 0.07 | 0.37 | 0.17 | 4830 | 2.98 | 1.08 | -0.02 | 0.09 | 414 | -0.02 | 0.17 | 25 | 8.25 | 8.54 | 0.08 | 0.18 | 3.1 | G | 1 |
| hd224349 | S | 4830 | 2.57 | 1.37 | -0.15 | 0.07 | 404 | -0.03 | 0.07 | 29 | 8.17 | 8.70 | 0.33 | 0.64 | 0.20 | 4830 | 2.28 | 1.44 | -0.19 | 0.07 | 404 | -0.19 | 0.08 | 29 | 8.09 | 8.56 | 0.34 | 0.21 | 4.2 | G | 1 |
| hd225292 | S | 4940 | 2.63 | 1.36 | -0.10 | 0.07 | 418 | -0.07 | 0.09 | 29 | 8.10 | 8.69 | 0.29 | 0.42 | 0.17 | 4940 | 2.56 | 1.37 | -0.10 | 0.07 | 418 | -0.11 | 0.09 | 29 | 8.08 | 8.66 | 0.29 | 0.17 | 4.8 | G | 1 |
| hr0002 | S | 4648 | 2.42 | 1.39 | 0.22 | 0.11 | 382 | 0.37 | 0.19 | 26 | 8.51 | 8.90 | 0.37 | 1.25 | 0.26 | 4648 | 2.18 | 1.42 | 0.14 | 0.11 | 382 | 0.15 | 0.19 | 26 | 8.33 | 8.69 | 0.38 | 0.28 | 4.6 | G | 1 |
| hr0004 | S | 5104 | 2.70 | 1.42 | 0.00 | 0.08 | 403 | -0.02 | 0.08 | 32 | 8.07 | 8.61 | 1.02 | 1.39 | 0.14 | 5104 | 2.73 | 1.41 | 0.00 | 0.08 | 403 | 0.00 | 0.08 | 32 | 8.08 | 8.63 | 1.02 | 0.14 | 4.9 | G | 1 |
| hr0016 | S | 4738 | 2.65 | 1.29 | -0.04 | 0.08 | 401 | 0.13 | 0.10 | 26 | 8.24 | 8.75 | 0.17 | 0.60 | 0.22 | 4738 | 2.28 | 1.41 | -0.11 | 0.08 | 401 | -0.11 | 0.11 | 26 | 8.14 | 8.58 | 0.17 | 0.25 | 4.2 | G | 1 |
| hr0019 | S | 4898 | 2.64 | 1.38 | -0.11 | 0.08 | 414 | -0.04 | 0.11 | 29 | 8.19 | 8.74 | -0.10 | 0.20 | 0.18 | 4898 | 2.48 | 1.42 | -0.13 | 0.08 | 414 | -0.13 | 0.11 | 29 | 8.15 | 8.67 | -0.10 | 0.19 | 4.1 | G | 1 |
| hr0022 | S | 4774 | 2.59 | 1.32 | 0.12 | 0.09 | 395 | 0.13 | 0.13 | 27 | 8.26 | 8.80 | 0.25 | 0.63 | 0.21 | 4774 | 2.57 | 1.33 | 0.11 | 0.09 | 395 | 0.12 | 0.13 | 27 | 8.25 | 8.79 | 0.25 | 0.21 | 4.2 | G | 1 |
| hr0040 | S | 5642 | 3.07 | 0.64 | 0.25 | 0.14 | 523 | -0.22 | 0.15 | 43 | 7.56 | 8.69 | 1.01 | 0.39 | 0.06 | 5642 | 3.94 | 0.20 | 0.25 | 0.17 | 523 | 0.25 | 0.15 | 43 | 7.84 | 9.10 | 1.01 | 0.05 | 4.0 | G | 1 |
| hr0059 | S | 5061 | 2.87 | 1.35 | 0.11 | 0.08 | 409 | 0.13 | 0.12 | 30 | 8.18 | 8.69 | 0.78 | 0.91 | 0.14 | 5061 | 2.83 | 1.37 | 0.11 | 0.08 | 409 | 0.11 | 0.12 | 30 | 8.16 | 8.67 | 0.78 | 0.14 | 4.3 | G | 1 |
| hr0069 | S | 4888 | 2.51 | 1.42 | 0.18 | 0.09 | 390 | 0.10 | 0.14 | 31 | 8.35 | 8.87 | 0.80 | 1.52 | 0.19 | 4888 | 2.69 | 1.37 | 0.21 | 0.09 | 390 | 0.21 | 0.14 | 31 | 8.41 | 8.96 | 0.80 | 0.18 | 4.2 | G | 1 |
| hr0074 | S | 4479 | 1.87 | 1.71 | 0.12 | 0.13 | 350 | 0.11 | 0.17 | 24 | 8.25 | 8.76 | 0.29 | 1.87 | 0.33 | 4479 | 1.89 | 1.71 | 0.12 | 0.13 | 350 | 0.12 | 0.17 | 24 | 8.25 | 8.77 | 0.29 | 0.33 | 5.1 | G | 1 |
| hr0084 | S | 4888 | 2.58 | 1.51 | 0.19 | 0.10 | 382 | 0.27 | 0.14 | 28 | 8.40 | 8.85 | 1.46 | 6.01 | 0.18 | 4888 | 2.40 | 1.55 | 0.16 | 0.11 | 382 | 0.16 | 0.15 | 28 | 8.34 | 8.76 | 1.46 | 0.20 | 5.0 | G | 1 |
| hr0101 | S | 5001 | 2.71 | 1.36 | -0.32 | 0.06 | 411 | -0.31 | 0.08 | 26 | 7.87 | 8.52 | 0.31 | 0.39 | 0.16 | 5001 | 2.69 | 1.36 | -0.32 | 0.06 | 411 | -0.32 | 0.08 | 26 | 7.86 | 8.51 | 0.31 | 0.16 | 4.1 | G | 1 |

| ID | S | | | | | | | | | | | | | | | | | | | | | | | | | | | |
|---|---|---|---|---|---|---|---|---|---|---|---|---|---|---|---|---|---|---|---|---|---|---|---|---|---|---|---|---|
| hr0131 | S | 4701 | 2.59 | 1.29 | 0.13 | 0.09 | 241 | 0.30 | 0.14 | 17 | | | 0.39 | 1.09 | 0.23 | 4701 | 2.18 | 1.35 | 0.06 | 0.09 | 241 | 0.06 | 0.14 | 17 | | | 0.39 | 0.27 | 4.2 | G | 1 |
| hr0135 | S | 4792 | 2.53 | 1.41 | 0.01 | 0.08 | 398 | 0.03 | 0.13 | 28 | 8.26 | 8.77 | 0.20 | 0.54 | 0.21 | 4792 | 2.48 | 1.42 | 0.00 | 0.08 | 398 | 0.00 | 0.13 | 28 | 8.24 | 8.75 | 0.20 | 0.21 | 4.2 | G | 1 |
| hr0141 | S | 4646 | 2.63 | 1.23 | 0.08 | 0.10 | 392 | 0.15 | 0.17 | 25 | 8.34 | 8.86 | 0.29 | 1.05 | 0.24 | 4646 | 2.47 | 1.31 | 0.04 | 0.10 | 392 | 0.04 | 0.17 | 25 | 8.29 | 8.79 | 0.29 | 0.26 | 3.9 | G | 1 |
| hr0156 | S | 4608 | 2.56 | 1.47 | 0.40 | 0.13 | 368 | 0.48 | 0.22 | 24 | 8.56 | 9.02 | 0.36 | 1.40 | 0.25 | 4608 | 2.36 | 1.53 | 0.34 | 0.13 | 368 | 0.34 | 0.22 | 24 | 8.49 | 8.92 | 0.36 | 0.27 | 4.5 | G | 1 |
| hr0163 | S | 4870 | 2.66 | 1.43 | -0.61 | 0.08 | 418 | -0.34 | 0.05 | 24 | 7.96 | 8.67 | 0.00 | 0.28 | 0.18 | 4870 | 2.05 | 1.53 | -0.66 | 0.09 | 418 | -0.66 | 0.05 | 24 | 7.77 | 8.38 | 0.00 | 0.22 | 5.0 | G | 1 |
| hr0165 | S | 4330 | 2.18 | 1.37 | 0.35 | 0.14 | 216 | 0.66 | 0.25 | 17 | | | 0.20 | 2.49 | 0.32 | 4330 | 1.39 | 1.41 | 0.15 | 0.14 | 216 | 0.15 | 0.26 | 17 | | | 0.17 | 0.39 | 4.4 | G | 1 |
| hr0168 | S | 4555 | 1.65 | 1.99 | -0.01 | 0.13 | 338 | 0.05 | 0.18 | 26 | 8.11 | 8.79 | 0.16 | 1.07 | 0.34 | 4555 | 1.49 | 1.99 | -0.03 | 0.13 | 338 | -0.03 | 0.18 | 26 | 8.07 | 8.72 | 0.16 | 0.36 | 6.9 | G | 1 |
| hr0175 | S | 5020 | 2.58 | 1.44 | -0.07 | 0.07 | 411 | -0.02 | 0.08 | 28 | 8.04 | 8.60 | 0.72 | 0.88 | 0.16 | 5020 | 2.45 | 1.46 | -0.08 | 0.08 | 411 | -0.08 | 0.08 | 28 | 8.01 | 8.54 | 0.72 | 0.17 | 4.7 | G | 1 |
| hr0188 | S | 4792 | 2.29 | 1.59 | 0.03 | 0.11 | 377 | 0.08 | 0.15 | 28 | 8.01 | 8.60 | 0.38 | 0.79 | 0.22 | 4792 | 2.16 | 1.61 | 0.01 | 0.11 | 377 | 0.01 | 0.15 | 28 | 7.98 | 8.54 | 0.38 | 0.23 | 5.6 | G | 1 |
| hr0213 | S | 4871 | 2.72 | 1.33 | 0.07 | 0.09 | 391 | 0.17 | 0.10 | 26 | 8.26 | 8.67 | 0.95 | 2.23 | 0.18 | 4871 | 2.50 | 1.42 | 0.03 | 0.09 | 391 | 0.03 | 0.11 | 26 | 8.19 | 8.56 | 0.95 | 0.19 | 4.6 | G | 1 |
| hr0215 | S | 4584 | 2.28 | 2.81 | -0.02 | 0.29 | 37 | 1.09 | 0.53 | 4 | | | 1.15 | 7.79 | 0.28 | 4584 | 0.01 | 2.65 | -0.20 | 0.29 | 37 | 0.06 | 0.55 | 4 | | | 1.06 | 0.47 | 40.0 | R | 2 |
| hr0216 | S | 5014 | 2.65 | 1.28 | 0.12 | 0.07 | 243 | 0.04 | 0.08 | 15 | | | 0.86 | 1.24 | 0.16 | 5014 | 2.73 | 1.27 | 0.15 | 0.07 | 243 | 0.15 | 0.08 | 15 | | | 0.87 | 0.15 | 4.7 | G | 1 |
| hr0224 | S | 3955 | 1.48 | 1.58 | 0.12 | 0.22 | 355 | 0.68 | 0.34 | 24 | 8.51 | 8.85 | -0.76 | 1.20 | 0.36 | 3955 | -0.05 | 1.60 | -0.30 | 0.25 | 355 | -0.30 | 0.34 | 24 | 7.96 | 8.19 | -0.94 | 0.78 | 5.6 | G | 1 |
| hr0228 | S | 4898 | 3.07 | 0.98 | 0.06 | 0.08 | 441 | 0.09 | 0.11 | 26 | 8.28 | 8.71 | 0.45 | 0.69 | 0.17 | 4898 | 3.01 | 1.04 | 0.04 | 0.08 | 441 | 0.04 | 0.11 | 26 | 8.26 | 8.68 | 0.45 | 0.17 | 3.7 | G | 1 |
| hr0230 | S | 6671 | 3.01 | 2.20 | 0.45 | 0.29 | 17 | -0.48 | 0.15 | 2 | | 7.98 | 2.94 | 4.46 | 0.08 | 6671 | 4.89 | 1.79 | 0.46 | 0.28 | 17 | 0.23 | 0.13 | 2 | | 8.71 | 2.93 | -0.31 | 95.0 | R | 2 |
| hr0231 | S | 6732 | 3.03 | 6.50 | 0.78 | 0.26 | 13 | -0.70 | 0.00 | 1 | | | 2.76 | 2.92 | 0.09 | 6732 | 4.89 | 7.63 | 0.76 | 0.26 | 13 | -0.08 | | 1 | | | 2.74 | -0.34 | 95.0 | R | 2 |
| hr0249 | S | 4612 | 2.51 | 1.42 | 0.22 | 0.11 | 378 | 0.42 | 0.19 | 27 | 8.49 | 8.93 | 0.20 | 0.97 | 0.25 | 4612 | 2.14 | 1.49 | 0.11 | 0.11 | 378 | 0.11 | 0.20 | 27 | 8.27 | 8.67 | 0.20 | 0.29 | 4.4 | G | 1 |
| hr0255 | S | 4842 | 2.53 | 1.43 | -0.52 | 0.07 | 406 | -0.32 | 0.06 | 27 | 7.93 | 8.56 | -0.61 | 0.07 | 0.20 | 4842 | 2.08 | 1.51 | -0.56 | 0.07 | 406 | -0.55 | 0.07 | 27 | 7.79 | 8.35 | -0.61 | 0.22 | 4.0 | G | 1 |
| hr0265 | S | 4829 | 2.54 | 1.29 | -0.31 | 0.06 | 234 | -0.15 | 0.14 | 16 | | | 0.45 | 0.86 | 0.20 | 4829 | 2.18 | 1.34 | -0.34 | 0.06 | 234 | -0.34 | 0.15 | 16 | | | 0.45 | 0.24 | 4.2 | G | 1 |
| hr0279 | S | 4841 | 2.54 | 1.36 | -0.21 | 0.07 | 429 | -0.11 | 0.07 | 26 | 8.02 | 8.64 | 0.06 | 0.33 | 0.20 | 4841 | 2.46 | 1.35 | -0.25 | 0.08 | 429 | -0.25 | 0.08 | 26 | 7.89 | 8.47 | 0.05 | 0.20 | 4.3 | G | 1 |
| hr0294 | S | 4814 | 2.48 | 1.34 | -0.27 | 0.07 | 428 | -0.19 | 0.09 | 29 | 7.94 | 8.58 | 0.00 | 0.32 | 0.21 | 4814 | 2.30 | 1.40 | -0.30 | 0.07 | 428 | -0.30 | 0.10 | 29 | 7.88 | 8.49 | 0.00 | 0.22 | 4.3 | G | 1 |
| hr0315 | S | 4753 | 2.76 | 1.16 | 0.03 | 0.08 | 403 | 0.13 | 0.12 | 27 | 8.21 | 8.69 | 0.52 | 1.24 | 0.20 | 4753 | 2.55 | 1.27 | -0.02 | 0.08 | 403 | -0.02 | 0.12 | 27 | 8.15 | 8.60 | 0.52 | 0.21 | 3.8 | G | 1 |
| hr0320 | S | 4682 | 2.52 | 1.39 | 0.10 | 0.11 | 384 | 0.25 | 0.21 | 27 | 8.25 | 8.71 | 0.35 | 1.06 | 0.24 | 4682 | 2.17 | 1.51 | 0.02 | 0.11 | 384 | 0.02 | 0.22 | 27 | 8.15 | 8.54 | 0.35 | 0.27 | 5.0 | G | 1 |
| hr0325 | S | 7461 | 4.06 | 2.86 | 0.53 | 0.16 | 25 | 0.18 | 0.01 | 2 | 6.82 | 8.19 | 3.01 | 1.67 | -0.09 | 7461 | 4.97 | 2.68 | 0.50 | 0.16 | 25 | 0.47 | 0.02 | 2 | 7.02 | 8.43 | 2.97 | -0.85 | 15.0 | R | 2 |
| hr0334 | S | 4543 | 2.39 | 1.38 | 0.21 | 0.11 | 381 | 0.37 | 0.19 | 26 | 8.43 | 8.85 | 1.17 | 8.83 | 0.28 | 4543 | 2.10 | 1.43 | 0.10 | 0.11 | 381 | 0.10 | 0.20 | 26 | 8.24 | 8.63 | 1.17 | 0.31 | 4.2 | G | 1 |
| hr0352 | S | 4658 | 2.58 | 1.30 | 0.11 | 0.10 | 383 | 0.28 | 0.17 | 30 | 8.40 | 8.83 | 0.22 | 0.85 | 0.24 | 4658 | 2.19 | 1.44 | 0.02 | 0.10 | 383 | 0.02 | 0.18 | 30 | 8.27 | 8.65 | 0.22 | 0.28 | 4.2 | G | 1 |
| hr0356 | S | 4884 | 2.74 | 1.39 | -0.13 | 0.07 | 405 | -0.04 | 0.07 | 28 | 8.26 | 8.89 | 0.11 | 0.33 | 0.18 | 4884 | 2.55 | 1.45 | -0.15 | 0.07 | 405 | -0.15 | 0.07 | 28 | 8.20 | 8.80 | 0.11 | 0.19 | 4.3 | G | 1 |
| hr0367 | S | 4732 | 2.35 | 1.44 | 0.14 | 0.10 | 398 | 0.13 | 0.19 | 27 | 8.31 | 8.76 | 1.50 | 9.53 | 0.25 | 4732 | 2.26 | 1.50 | 0.14 | 0.10 | 398 | 0.14 | 0.19 | 27 | 8.39 | 8.81 | 1.50 | 0.26 | 4.6 | G | 1 |
| hr0371 | S | 4573 | 2.49 | 1.41 | 0.32 | 0.12 | 374 | 0.50 | 0.21 | 28 | 8.54 | 8.99 | 0.30 | 1.38 | 0.27 | 4573 | 2.06 | 1.52 | 0.21 | 0.12 | 374 | 0.21 | 0.21 | 28 | 8.40 | 8.79 | 0.30 | 0.30 | 4.3 | G | 1 |
| hr0390 | S | 4704 | 2.52 | 1.38 | 0.21 | 0.09 | 381 | 0.34 | 0.16 | 28 | 8.39 | 8.85 | 0.36 | 1.02 | 0.23 | 4704 | 2.34 | 1.39 | 0.15 | 0.09 | 381 | 0.16 | 0.16 | 28 | 8.24 | 8.68 | 0.36 | 0.26 | 4.1 | G | 1 |
| hr0402 | S | 4660 | 2.58 | 1.38 | -0.03 | 0.09 | 415 | 0.15 | 0.12 | 25 | 8.36 | 8.88 | -0.06 | 0.46 | 0.24 | 4660 | 2.17 | 1.50 | -0.11 | 0.10 | 415 | -0.11 | 0.12 | 25 | 8.22 | 8.70 | -0.05 | 0.28 | 4.5 | G | 1 |
| hr0406 | S | 4845 | 2.94 | 1.10 | -0.15 | 0.07 | 415 | -0.05 | 0.09 | 27 | 8.21 | 8.72 | -0.06 | 0.26 | 0.18 | 4845 | 2.73 | 1.20 | -0.19 | 0.07 | 415 | -0.19 | 0.10 | 27 | 8.14 | 8.62 | -0.05 | 0.19 | 3.3 | G | 1 |
| hr0407 | S | 6455 | 3.75 | 3.09 | -0.04 | 0.23 | 145 | 0.08 | 0.24 | 18 | 8.32 | 8.70 | 3.24 | 10.08 | 0.01 | 6455 | 3.43 | 3.14 | -0.05 | 0.23 | 145 | -0.05 | 0.24 | 18 | 8.22 | 8.60 | 3.23 | 0.04 | 32.4 | R | 2 |
| hr0412 | S | 4316 | 1.94 | 1.44 | -0.20 | 0.12 | 374 | -0.03 | 0.19 | 27 | 8.13 | 8.54 | -0.50 | 0.57 | 0.34 | 4316 | 1.50 | 1.55 | -0.32 | 0.13 | 374 | -0.32 | 0.21 | 27 | 7.97 | 8.34 | -0.51 | 0.38 | 4.2 | G | 1 |
| hr0414 | S | 4617 | 2.50 | 1.34 | 0.12 | 0.10 | 384 | 0.27 | 0.15 | 29 | 8.39 | 8.93 | 0.00 | 0.61 | 0.22 | 4617 | 2.13 | 1.46 | 0.03 | 0.10 | 384 | 0.03 | 0.16 | 29 | 8.27 | 8.76 | 0.00 | 0.29 | 4.4 | G | 1 |
| hr0426 | S | 4688 | 2.68 | 1.35 | 0.29 | 0.10 | 381 | 0.48 | 0.18 | 28 | 8.49 | 8.97 | 0.34 | 1.03 | 0.22 | 4688 | 2.25 | 1.50 | 0.19 | 0.11 | 381 | 0.19 | 0.19 | 28 | 8.36 | 8.77 | 0.35 | 0.27 | 4.2 | G | 1 |
| hr0430 | S | 4879 | 2.60 | 1.41 | 0.11 | 0.08 | 401 | 0.10 | 0.11 | 26 | 8.27 | 8.71 | 0.46 | 0.74 | 0.18 | 4879 | 2.63 | 1.39 | 0.12 | 0.08 | 401 | 0.12 | 0.11 | 26 | 8.27 | 8.72 | 0.46 | 0.18 | 4.3 | G | 1 |
| hr0434 | S | 4126 | 1.68 | 1.48 | -0.28 | 0.16 | 386 | 0.00 | 0.27 | 26 | 8.32 | 8.78 | -0.07 | 2.97 | 0.36 | 4126 | 0.93 | 1.59 | -0.48 | 0.18 | 386 | -0.48 | 0.27 | 26 | 8.02 | 8.44 | -0.11 | 0.47 | 4.5 | G | 1 |
| hr0437 | S | 4875 | 2.01 | 1.97 | -0.04 | 0.13 | 335 | -0.02 | 0.16 | 29 | 8.08 | 8.46 | 1.21 | 3.78 | 0.22 | 4875 | 1.96 | 1.98 | -0.05 | 0.13 | 335 | -0.05 | 0.16 | 29 | 8.06 | 8.44 | 1.21 | 0.22 | 8.5 | G | 1 |
| hr0442 | S | 4736 | 2.45 | 1.36 | -0.28 | 0.08 | 410 | -0.18 | 0.09 | 28 | 8.05 | 8.57 | -0.20 | 0.26 | 0.24 | 4736 | 2.23 | 1.43 | -0.31 | 0.08 | 410 | -0.31 | 0.09 | 28 | 7.98 | 8.46 | -0.20 | 0.26 | 4.1 | G | 1 |
| hr0454 | S | 4357 | 2.05 | 1.96 | -0.08 | 0.18 | 307 | 0.23 | 0.11 | 17 | 8.32 | 8.65 | 3.56 | 50.80 | 0.33 | 4357 | 1.29 | 1.97 | -0.23 | 0.19 | 307 | -0.23 | 0.11 | 17 | 8.08 | 8.32 | 3.53 | 0.40 | 8.7 | G | 1 |
| hr0464 | S | 4310 | 1.98 | 1.65 | 0.12 | 0.15 | 348 | 0.29 | 0.18 | 24 | 8.47 | 8.86 | -0.45 | 0.66 | 0.34 | 4310 | 1.57 | 1.69 | 0.03 | 0.15 | 348 | 0.02 | 0.19 | 24 | 8.31 | 8.68 | -0.46 | 0.38 | 4.7 | G | 1 |
| hr0510 | S | 4903 | 2.36 | 1.55 | -0.04 | 0.08 | 420 | -0.03 | 0.09 | 26 | 8.15 | 8.63 | 0.10 | 0.30 | 0.19 | 4903 | 2.34 | 1.55 | -0.04 | 0.08 | 420 | -0.04 | 0.09 | 26 | 8.14 | 8.62 | 0.10 | 0.19 | 4.7 | G | 1 |
| hr0521 | S | 4855 | 2.68 | 1.37 | -0.11 | 0.08 | 403 | -0.01 | 0.09 | 28 | 8.16 | 8.76 | -0.10 | 0.22 | 0.19 | 4855 | 2.46 | 1.43 | -0.14 | 0.08 | 403 | -0.14 | 0.09 | 28 | 8.10 | 8.66 | -0.10 | 0.20 | 4.2 | G | 1 |
| hr0527 | S | 4837 | 2.67 | 1.33 | -0.09 | 0.07 | 411 | 0.08 | 0.11 | 28 | 8.15 | 8.70 | 0.45 | 0.82 | 0.19 | 4837 | 2.30 | 1.43 | -0.14 | 0.08 | 411 | -0.14 | 0.12 | 28 | 8.04 | 8.52 | 0.45 | 0.21 | 4.2 | G | 1 |
| hr0539 | S | 4579 | 2.07 | 1.60 | -0.02 | 0.10 | 376 | 0.05 | 0.14 | 27 | 8.10 | 8.65 | 0.15 | 0.96 | 0.30 | 4579 | 1.90 | 1.62 | -0.05 | 0.11 | 376 | -0.05 | 0.14 | 27 | 8.05 | 8.57 | 0.15 | 0.32 | 4.8 | G | 1 |
| hr0557 | S | 4509 | 2.31 | 1.38 | -0.16 | 0.09 | 452 | 0.15 | 0.10 | 27 | 8.23 | 8.73 | -0.20 | 0.56 | 0.29 | 4509 | 1.69 | 1.48 | -0.33 | 0.10 | 452 | -0.33 | 0.12 | 27 | 7.89 | 8.32 | -0.21 | 0.35 | 4.2 | G | 1 |
| hr0594 | S | 4880 | 2.58 | 1.39 | -0.63 | 0.07 | 430 | -0.42 | 0.05 | 22 | 7.84 | 8.53 | -0.65 | 0.06 | 0.19 | 4880 | 2.09 | 1.47 | -0.67 | 0.08 | 430 | -0.67 | 0.04 | 22 | 7.71 | 8.30 | -0.65 | 0.21 | 4.1 | G | 1 |
| hr0616 | S | 5061 | 2.58 | 1.30 | 0.19 | 0.10 | 406 | 0.04 | 0.13 | 27 | 8.18 | 8.77 | 0.76 | 0.85 | 0.15 | 5061 | 2.90 | 1.17 | 0.24 | 0.11 | 406 | 0.24 | 0.14 | 27 | 8.27 | 8.93 | 0.75 | 0.14 | 5.2 | G | 1 |
| hr0617 | S | 4512 | 2.30 | 1.33 | -0.07 | 0.10 | 413 | 0.03 | 0.14 | 26 | 8.22 | 8.71 | -0.12 | 0.67 | 0.29 | 4512 | 2.06 | 1.39 | -0.12 | 0.11 | 413 | -0.12 | 0.14 | 26 | 8.15 | 8.61 | -0.12 | 0.31 | 4.2 | G | 1 |

| ID | S | | | | | | | | | | | | | | | | | | | | | | | | | |
|---|---|---|---|---|---|---|---|---|---|---|---|---|---|---|---|---|---|---|---|---|---|---|---|---|---|---|
| hr0619 | S | 4916 | 2.77 | 1.34 | -0.28 | 0.06 | 409 | -0.19 | 0.08 | 26 | 8.00 | 8.55 | 0.40 | 0.60 | 0.17 | 4916 | 2.56 | 1.42 | -0.31 | 0.06 | 409 | -0.31 | 0.09 | 26 | 7.94 | 8.44 | 0.40 | 0.18 | 3.8 | G | 1 |
| hr0621 | S | 5110 | 2.58 | 1.49 | 0.08 | 0.08 | 406 | -0.10 | 0.09 | 26 | 8.15 | 8.61 | 1.09 | 1.60 | 0.14 | 5110 | 2.96 | 1.34 | 0.12 | 0.08 | 406 | 0.12 | 0.09 | 26 | 8.25 | 8.80 | 1.09 | 0.13 | 4.5 | G | 1 |
| hr0661 | S | 4502 | 2.20 | 1.41 | 0.10 | 0.11 | 380 | 0.17 | 0.15 | 26 | 8.22 | 8.75 | 0.09 | 1.09 | 0.30 | 4502 | 2.04 | 1.46 | 0.06 | 0.11 | 380 | 0.06 | 0.16 | 26 | 8.17 | 8.68 | 0.09 | 0.32 | 4.1 | G | 1 |
| hr0666 | S | 4884 | 2.60 | 1.35 | 0.04 | 0.08 | 427 | 0.08 | 0.11 | 30 | 8.18 | 8.68 | 0.34 | 0.55 | 0.18 | 4884 | 2.50 | 1.38 | 0.02 | 0.08 | 427 | 0.03 | 0.11 | 30 | 8.15 | 8.63 | 0.34 | 0.19 | 4.6 | G | 1 |
| hr0697 | S | 4865 | 2.65 | 1.44 | -0.39 | 0.07 | 402 | -0.32 | 0.09 | 25 | 8.11 | 8.80 | -0.10 | 0.22 | 0.19 | 4865 | 2.47 | 1.50 | -0.42 | 0.07 | 402 | -0.42 | 0.09 | 25 | 8.04 | 8.71 | -0.10 | 0.20 | 4.6 | G | 1 |
| hr0699 | S | 3927 | 1.40 | 1.65 | 0.18 | 0.21 | 330 | 0.74 | 0.28 | 23 | 8.45 | 8.84 | 0.30 | 10.88 | 0.37 | 3927 | -0.18 | 1.68 | -0.26 | 0.24 | 330 | -0.26 | 0.29 | 23 | 7.90 | 8.16 | 0.08 | 0.86 | 5.6 | G | 1 |
| hr0712 | S | 4572 | 2.44 | 1.29 | 0.00 | 0.10 | 391 | 0.11 | 0.14 | 24 | 8.27 | 8.74 | 0.11 | 0.90 | 0.27 | 4572 | 2.20 | 1.38 | -0.06 | 0.10 | 391 | -0.06 | 0.15 | 24 | 8.19 | 8.63 | 0.11 | 0.29 | 3.9 | G | 1 |
| hr0725 | S | 4780 | 2.66 | 1.25 | 0.03 | 0.08 | 398 | 0.01 | 0.14 | 27 | 8.22 | 8.72 | 1.19 | 4.74 | 0.20 | 4780 | 2.70 | 1.22 | 0.04 | 0.08 | 398 | 0.04 | 0.13 | 27 | 8.23 | 8.74 | 1.19 | 0.20 | 3.8 | G | 1 |
| hr0726 | S | 4723 | 2.52 | 1.35 | 0.14 | 0.09 | 390 | 0.18 | 0.13 | 28 | 8.24 | 8.72 | 0.39 | 1.01 | 0.23 | 4723 | 2.45 | 1.38 | 0.13 | 0.09 | 390 | 0.13 | 0.13 | 28 | 8.21 | 8.69 | 0.39 | 0.24 | 4.0 | G | 1 |
| hr0738 | S | 4622 | 2.61 | 1.15 | 0.05 | 0.10 | 418 | 0.15 | 0.12 | 25 | 8.26 | 8.76 | 0.46 | 1.67 | 0.24 | 4622 | 2.39 | 1.26 | -0.01 | 0.10 | 418 | -0.01 | 0.13 | 25 | 8.19 | 8.66 | 0.46 | 0.27 | 4.3 | G | 1 |
| hr0739 | S | 4803 | 2.56 | 1.33 | 0.08 | 0.08 | 402 | 0.09 | 0.12 | 29 | 8.24 | 8.72 | 0.48 | 0.97 | 0.20 | 4803 | 2.45 | 1.37 | 0.02 | 0.09 | 402 | 0.02 | 0.12 | 29 | 8.21 | 8.66 | 0.48 | 0.21 | 4.1 | G | 1 |
| hr0743 | S | 5084 | 2.80 | 1.38 | 0.00 | 0.08 | 427 | -0.04 | 0.07 | 26 | 8.08 | 8.59 | 1.13 | 1.87 | 0.14 | 5084 | 2.88 | 1.35 | 0.01 | 0.08 | 427 | 0.01 | 0.07 | 26 | 8.10 | 8.63 | 1.13 | 0.14 | 5.0 | G | 1 |
| hr0766 | S | 4678 | 2.57 | 1.18 | -0.13 | 0.08 | 406 | -0.05 | 0.09 | 26 | 8.23 | 8.68 | 0.06 | 0.56 | 0.23 | 4678 | 2.38 | 1.26 | -0.17 | 0.08 | 406 | -0.17 | 0.09 | 26 | 8.17 | 8.59 | 0.06 | 0.26 | 3.8 | G | 1 |
| hr0768 | S | 6268 | 3.52 | 3.00 | 0.11 | 0.26 | 110 | -0.01 | 0.12 | 7 | 8.35 | 8.79 | 1.50 | 0.37 | 0.02 | 6268 | 3.80 | 2.90 | 0.11 | 0.26 | 110 | 0.11 | 0.13 | 7 | 8.44 | 8.87 | 1.50 | 0.00 | 40.0 | R | 2 |
| hr0771 | S | 4758 | 2.56 | 1.34 | 0.04 | 0.08 | 399 | 0.19 | 0.12 | 29 | 8.22 | 8.72 | 0.22 | 0.63 | 0.21 | 4758 | 2.21 | 1.45 | -0.03 | 0.09 | 399 | -0.03 | 0.13 | 29 | 8.12 | 8.56 | 0.23 | 0.23 | 4.2 | G | 1 |
| hr0808 | S | 4624 | 2.63 | 1.20 | 0.25 | 0.10 | 383 | 0.43 | 0.19 | 29 | 8.46 | 8.91 | 0.31 | 1.20 | 0.24 | 4624 | 2.29 | 1.29 | 0.13 | 0.10 | 383 | 0.13 | 0.21 | 29 | 8.25 | 8.66 | 0.32 | 0.28 | 4.1 | G | 1 |
| hr0824 | S | 4628 | 2.53 | 1.37 | 0.14 | 0.11 | 391 | 0.25 | 0.17 | 27 | 8.34 | 8.81 | 0.40 | 1.42 | 0.25 | 4628 | 2.26 | 1.46 | 0.07 | 0.11 | 391 | 0.07 | 0.17 | 27 | 8.25 | 8.68 | 0.40 | 0.28 | 4.2 | G | 1 |
| hr0831 | S | 6504 | 3.59 | 2.85 | 0.03 | 0.07 | 237 | 0.12 | 0.07 | 24 | 8.60 | 8.95 | 3.29 | 10.30 | 0.02 | 6504 | 3.34 | 2.87 | 0.03 | 0.08 | 237 | 0.03 | 0.07 | 24 | 8.52 | 8.87 | 3.29 | 0.05 | 8.7 | G | 2 |
| hr0840 | S | 6752 | 3.63 | 6.50 | 0.40 | 0.21 | 15 | 0.06 | 0.04 | 2 | | 9.33 | 1.97 | 0.49 | 0.03 | 6752 | 4.51 | 7.75 | 0.40 | 0.21 | 15 | 0.40 | 0.04 | 2 | | 9.69 | 1.97 | -0.20 | 70.0 | R | 2 |
| hr0844 | S | 4657 | 2.39 | 1.40 | 0.02 | 0.10 | 414 | 0.08 | 0.15 | 27 | 8.37 | 8.82 | 0.08 | 0.62 | 0.26 | 4657 | 2.25 | 1.43 | -0.01 | 0.10 | 414 | -0.01 | 0.15 | 27 | 8.32 | 8.75 | 0.08 | 0.27 | 4.4 | G | 1 |
| hr0850 | S | 4976 | 2.85 | 1.24 | 0.01 | 0.08 | 404 | 0.10 | 0.10 | 24 | 8.13 | 8.65 | 1.11 | 2.36 | 0.16 | 4976 | 2.64 | 1.33 | -0.03 | 0.09 | 404 | -0.03 | 0.10 | 24 | 8.07 | 8.54 | 1.11 | 0.16 | 4.7 | G | 1 |
| hr0856 | S | 6413 | 3.44 | 3.61 | 0.33 | 0.35 | 25 | 0.60 | 0.25 | 2 | 8.10 | 8.59 | 2.70 | 4.11 | 0.03 | 6413 | 2.67 | 3.44 | 0.35 | 0.35 | 25 | 0.35 | 0.25 | 2 | 7.88 | 8.34 | 2.73 | 0.05 | 70.0 | R | 2 |
| hr0900 | S | 4735 | 2.44 | 1.36 | 0.11 | 0.10 | 392 | 0.17 | 0.14 | 29 | 8.22 | 8.71 | 0.38 | 0.96 | 0.24 | 4735 | 2.31 | 1.41 | 0.09 | 0.10 | 392 | 0.09 | 0.15 | 29 | 8.18 | 8.65 | 0.38 | 0.25 | 4.3 | G | 1 |
| hr0907 | S | 4751 | 2.55 | 1.42 | -0.03 | 0.08 | 396 | 0.08 | 0.12 | 27 | 8.32 | 8.80 | 0.40 | 0.95 | 0.21 | 4751 | 2.30 | 1.49 | -0.07 | 0.09 | 396 | -0.07 | 0.13 | 27 | 8.25 | 8.69 | 0.40 | 0.23 | 4.4 | G | 1 |
| hr0908 | S | 4702 | 2.54 | 1.28 | -0.30 | 0.07 | 401 | -0.29 | 0.10 | 26 | 8.06 | 8.54 | 0.22 | 0.76 | 0.23 | 4702 | 2.52 | 1.28 | -0.30 | 0.07 | 401 | -0.30 | 0.10 | 26 | 8.06 | 8.53 | 0.22 | 0.23 | 3.5 | G | 1 |
| hr0926 | S | 4705 | 2.39 | 1.45 | -0.16 | 0.08 | 403 | -0.04 | 0.10 | 27 | 8.15 | 8.72 | -0.30 | 0.23 | 0.25 | 4705 | 2.13 | 1.50 | -0.20 | 0.09 | 403 | -0.20 | 0.10 | 27 | 8.08 | 8.59 | -0.30 | 0.27 | 4.2 | G | 1 |
| hr0931 | S | 4697 | 2.50 | 1.35 | 0.21 | 0.11 | 390 | 0.31 | 0.16 | 27 | 8.49 | 8.89 | 0.47 | 1.32 | 0.24 | 4697 | 2.29 | 1.42 | 0.16 | 0.11 | 390 | 0.17 | 0.16 | 27 | 8.43 | 8.79 | 0.47 | 0.26 | 4.1 | G | 1 |
| hr0941 | S | 4857 | 2.71 | 1.28 | 0.18 | 0.09 | 400 | 0.24 | 0.13 | 29 | 8.39 | 8.84 | 0.54 | 0.95 | 0.18 | 4857 | 2.67 | 1.23 | 0.16 | 0.09 | 400 | 0.16 | 0.13 | 29 | 8.31 | 8.74 | 0.54 | 0.19 | 4.1 | G | 1 |
| hr0946 | S | 4348 | 2.04 | 1.41 | -0.06 | 0.12 | 379 | 0.12 | 0.18 | 27 | 8.33 | 8.77 | -0.39 | 0.65 | 0.33 | 4348 | 1.59 | 1.50 | -0.18 | 0.13 | 379 | -0.18 | 0.18 | 27 | 8.17 | 8.57 | -0.40 | 0.37 | 4.3 | G | 1 |
| hr0947 | S | 4586 | 2.18 | 1.54 | -0.02 | 0.10 | 375 | 0.11 | 0.15 | 29 | 8.16 | 8.65 | 0.07 | 0.78 | 0.29 | 4586 | 1.88 | 1.59 | -0.07 | 0.10 | 375 | -0.07 | 0.16 | 29 | 8.07 | 8.52 | 0.07 | 0.32 | 4.4 | G | 1 |
| hr0951 | S | 4769 | 2.53 | 1.37 | 0.13 | 0.09 | 392 | 0.18 | 0.14 | 27 | 8.22 | 8.68 | 0.69 | 1.71 | 0.21 | 4769 | 2.42 | 1.41 | 0.11 | 0.09 | 392 | 0.11 | 0.14 | 27 | 8.19 | 8.62 | 0.69 | 0.22 | 4.3 | G | 1 |
| hr0956 | S | 4959 | 2.75 | 1.38 | 0.20 | 0.09 | 417 | 0.15 | 0.11 | 27 | 8.31 | 8.77 | 0.52 | 0.67 | 0.16 | 4959 | 2.85 | 1.34 | 0.22 | 0.09 | 417 | 0.21 | 0.11 | 27 | 8.34 | 8.82 | 0.52 | 0.16 | 4.3 | G | 1 |
| hr0992 | S | 4621 | 2.61 | 1.12 | -0.03 | 0.09 | 400 | 0.10 | 0.13 | 25 | 8.33 | 8.73 | 0.32 | 1.23 | 0.24 | 4621 | 2.32 | 1.27 | -0.11 | 0.10 | 400 | -0.11 | 0.14 | 25 | 8.24 | 8.59 | 0.32 | 0.27 | 3.7 | G | 1 |
| hr0994 | S | 4960 | 2.67 | 1.19 | 0.10 | 0.10 | 410 | -0.01 | 0.14 | 26 | 8.24 | 8.74 | 0.55 | 0.72 | 0.17 | 4960 | 2.80 | 1.19 | 0.13 | 0.10 | 410 | 0.13 | 0.13 | 26 | 8.35 | 8.90 | 0.56 | 0.16 | 4.9 | G | 1 |
| hr1000 | S | 4449 | 2.49 | 1.69 | 0.24 | 0.16 | 350 | 0.52 | 0.29 | 26 | 8.65 | 9.05 | 0.09 | 1.32 | 0.29 | 4449 | 1.75 | 1.76 | 0.13 | 0.16 | 350 | 0.13 | 0.29 | 26 | 8.46 | 8.81 | 0.10 | 0.35 | 4.7 | G | 1 |
| hr1007 | S | 4948 | 2.76 | 1.37 | 0.13 | 0.08 | 398 | 0.20 | 0.09 | 29 | 8.26 | 8.74 | 0.40 | 0.53 | 0.17 | 4948 | 2.60 | 1.43 | 0.11 | 0.08 | 398 | 0.11 | 0.10 | 29 | 8.21 | 8.66 | 0.40 | 0.17 | 4.3 | G | 1 |
| hr1015 | S | 4385 | 2.07 | 1.50 | 0.16 | 0.15 | 350 | 0.28 | 0.23 | 24 | 8.42 | 8.91 | -0.03 | 1.26 | 0.33 | 4385 | 1.77 | 1.56 | 0.08 | 0.15 | 350 | 0.08 | 0.23 | 24 | 8.31 | 8.77 | -0.04 | 0.35 | 4.6 | G | 1 |
| hr1022 | S | 4486 | 2.52 | 1.20 | 0.34 | 0.14 | 372 | 0.51 | 0.19 | 24 | 8.62 | 8.97 | 0.14 | 1.28 | 0.28 | 4486 | 2.12 | 1.37 | 0.20 | 0.14 | 372 | 0.20 | 0.20 | 24 | 8.47 | 8.78 | 0.13 | 0.31 | 4.2 | G | 1 |
| hr1030 | S | 4982 | 2.20 | 1.66 | -0.16 | 0.10 | 387 | -0.12 | 0.10 | 29 | 7.91 | 8.47 | 0.39 | 0.46 | 0.19 | 4982 | 2.11 | 1.67 | -0.17 | 0.10 | 387 | -0.16 | 0.10 | 29 | 7.90 | 8.43 | 0.39 | 0.19 | 7.0 | G | 1 |
| hr1050 | S | 4652 | 2.53 | 1.39 | 0.27 | 0.12 | 404 | 0.38 | 0.18 | 27 | 8.44 | 8.91 | 0.45 | 1.47 | 0.24 | 4652 | 2.28 | 1.46 | 0.21 | 0.12 | 404 | 0.21 | 0.18 | 27 | 8.37 | 8.80 | 0.45 | 0.27 | 4.4 | G | 1 |
| hr1052 | S | 4163 | 1.57 | 1.59 | -0.14 | 0.14 | 365 | 0.17 | 0.27 | 29 | 8.26 | 8.75 | 0.07 | 3.55 | 0.37 | 4163 | 0.84 | 1.64 | -0.36 | 0.15 | 365 | -0.36 | 0.29 | 29 | 7.89 | 8.32 | 0.03 | 0.48 | 4.7 | G | 1 |
| hr1060 | S | 4851 | 2.58 | 1.36 | 0.08 | 0.08 | 398 | 0.06 | 0.10 | 26 | 8.29 | 8.76 | 0.46 | 0.81 | 0.19 | 4851 | 2.63 | 1.33 | 0.10 | 0.08 | 398 | 0.09 | 0.10 | 26 | 8.30 | 8.78 | 0.46 | 0.19 | 4.0 | G | 1 |
| hr1098 | S | 4984 | 2.58 | 1.36 | -0.23 | 0.06 | 412 | -0.26 | 0.11 | 32 | 7.89 | 8.45 | 0.67 | 0.91 | 0.18 | 4984 | 2.51 | 1.40 | -0.21 | 0.07 | 412 | -0.21 | 0.10 | 32 | 7.98 | 8.54 | 0.69 | 0.18 | 3.8 | G | 1 |
| hr1108 | S | 4899 | 2.69 | 1.35 | 0.04 | 0.08 | 407 | 0.11 | 0.11 | 31 | 8.26 | 8.77 | 0.68 | 1.14 | 0.18 | 4899 | 2.53 | 1.41 | 0.01 | 0.08 | 407 | 0.01 | 0.12 | 31 | 8.21 | 8.69 | 0.68 | 0.18 | 4.0 | G | 1 |
| hr1110 | S | 4948 | 2.68 | 1.33 | 0.07 | 0.08 | 404 | 0.05 | 0.10 | 27 | 8.19 | 8.62 | 0.60 | 0.83 | 0.17 | 4948 | 2.73 | 1.31 | 0.08 | 0.08 | 404 | 0.08 | 0.10 | 27 | 8.21 | 8.64 | 0.60 | 0.17 | 4.2 | G | 1 |
| hr1117 | S | 4783 | 2.63 | 1.36 | -0.08 | 0.08 | 397 | -0.04 | 0.11 | 26 | 8.39 | 9.00 | 0.26 | 0.63 | 0.20 | 4783 | 2.54 | 1.39 | -0.10 | 0.08 | 397 | -0.10 | 0.12 | 26 | 8.36 | 8.96 | 0.26 | 0.21 | 3.9 | G | 1 |
| hr1119 | S | 4901 | 2.71 | 1.35 | 0.01 | 0.07 | 404 | 0.04 | 0.10 | 28 | 8.20 | 8.75 | 0.56 | 0.87 | 0.18 | 4901 | 2.63 | 1.38 | 0.00 | 0.07 | 404 | -0.01 | 0.10 | 28 | 8.18 | 8.71 | 0.56 | 0.18 | 3.8 | G | 1 |
| hr1132 | S | 4801 | 2.52 | 1.42 | -0.19 | 0.08 | 405 | -0.06 | 0.08 | 27 | 8.18 | 8.67 | 0.00 | 0.33 | 0.20 | 4801 | 2.35 | 1.43 | -0.25 | 0.08 | 405 | -0.25 | 0.09 | 27 | 8.03 | 8.47 | -0.01 | 0.22 | 4.2 | G | 1 |
| hr1159 | S | 4800 | 2.53 | 1.36 | -0.12 | 0.08 | 407 | -0.02 | 0.10 | 28 | 8.14 | 8.62 | 0.66 | 1.47 | 0.20 | 4800 | 2.30 | 1.41 | -0.15 | 0.08 | 407 | -0.15 | 0.10 | 28 | 8.08 | 8.51 | 0.67 | 0.22 | 4.2 | G | 1 |
| hr1256 | S | 4732 | 2.54 | 1.22 | -0.04 | 0.08 | 227 | -0.13 | 0.20 | 15 | | | 0.33 | 0.86 | 0.22 | 4732 | 2.76 | 1.16 | 0.00 | 0.09 | 227 | 0.00 | 0.19 | 15 | | | 0.32 | 0.21 | 3.9 | G | 1 |

| ID | | | | | | | | | | | | | | | | | | | | | | | | | | | | | |
|---|---|---|---|---|---|---|---|---|---|---|---|---|---|---|---|---|---|---|---|---|---|---|---|---|---|---|---|---|---|
| hr1265 | S | 4654 | 2.33 | 1.41 | 0.01 | 0.09 | 387 | 0.11 | 0.13 | 30 | 8.18 | 8.69 | 0.13 | 0.71 | 0.27 | 4654 | 2.10 | 1.47 | -0.03 | 0.09 | 387 | -0.03 | 0.13 | 30 | 8.11 | 8.58 | 0.13 | 0.29 | 4.1 | G | 1 |
| hr1267 | S | 4604 | 2.42 | 1.45 | 0.21 | 0.11 | 377 | 0.36 | 0.19 | 26 | 8.46 | 8.96 | 0.33 | 1.32 | 0.27 | 4604 | 2.19 | 1.47 | 0.13 | 0.11 | 377 | 0.14 | 0.20 | 26 | 8.28 | 8.76 | 0.33 | 0.29 | 4.3 | G | 1 |
| hr1283 | S | 4736 | 2.51 | 1.38 | 0.13 | 0.09 | 391 | 0.16 | 0.15 | 28 | 8.30 | 8.79 | 0.36 | 0.91 | 0.23 | 4736 | 2.45 | 1.40 | 0.12 | 0.09 | 391 | 0.12 | 0.15 | 28 | 8.28 | 8.76 | 0.36 | 0.24 | 4.2 | G | 1 |
| hr1295 | S | 4732 | 2.59 | 1.34 | 0.12 | 0.09 | 392 | 0.21 | 0.16 | 27 | 8.28 | 8.80 | 0.42 | 1.07 | 0.22 | 4732 | 2.39 | 1.42 | 0.07 | 0.09 | 392 | 0.07 | 0.17 | 27 | 8.22 | 8.70 | 0.42 | 0.25 | 4.3 | G | 1 |
| hr1301 | S | 4934 | 2.62 | 1.44 | 0.00 | 0.08 | 403 | -0.11 | 0.12 | 27 | 8.22 | 8.77 | 0.38 | 0.52 | 0.17 | 4934 | 2.85 | 1.35 | 0.04 | 0.08 | 403 | 0.03 | 0.12 | 27 | 8.30 | 8.88 | 0.37 | 0.17 | 4.1 | G | 1 |
| hr1310 | S | 4627 | 2.62 | 1.34 | 0.28 | 0.12 | 375 | 0.47 | 0.15 | 26 | 8.64 | 9.05 | 0.44 | 1.56 | 0.24 | 4627 | 2.18 | 1.47 | 0.17 | 0.12 | 375 | 0.17 | 0.15 | 26 | 8.48 | 8.85 | 0.44 | 0.28 | 4.4 | G | 1 |
| hr1313 | S | 4693 | 2.69 | 1.18 | 0.22 | 0.11 | 385 | 0.29 | 0.17 | 26 | 8.43 | 8.91 | 0.18 | 0.71 | 0.22 | 4693 | 2.52 | 1.27 | 0.17 | 0.10 | 385 | 0.17 | 0.17 | 26 | 8.38 | 8.83 | 0.18 | 0.23 | 3.9 | G | 1 |
| hr1318 | S | 4549 | 2.36 | 1.35 | 0.25 | 0.12 | 375 | 0.32 | 0.18 | 26 | 8.40 | 8.92 | 0.17 | 1.11 | 0.28 | 4549 | 2.20 | 1.39 | 0.21 | 0.12 | 375 | 0.21 | 0.18 | 26 | 8.35 | 8.85 | 0.17 | 0.30 | 4.1 | G | 1 |
| hr1319 | S | 6606 | 4.13 | 3.89 | 0.33 | 0.25 | 17 | 0.57 | 0.15 | 2 | 8.50 | | | | -0.06 | 6606 | 3.55 | 4.01 | 0.33 | 0.24 | 17 | 0.33 | 0.11 | 2 | 8.32 | | | 0.03 | 60.0 | R | 2 |
| hr1327 | S | 5211 | 2.79 | 1.43 | -0.07 | 0.08 | 409 | -0.02 | 0.08 | 30 | 8.07 | 8.49 | 1.21 | 1.64 | 0.12 | 5211 | 2.69 | 1.46 | -0.07 | 0.08 | 409 | -0.07 | 0.08 | 30 | 8.05 | 8.44 | 1.22 | 0.12 | 4.7 | G | 1 |
| hr1343 | S | 4926 | 2.75 | 1.31 | 0.09 | 0.09 | 398 | 0.19 | 0.12 | 29 | 8.25 | 8.76 | 0.53 | 0.76 | 0.17 | 4926 | 2.51 | 1.41 | 0.05 | 0.09 | 398 | 0.05 | 0.13 | 29 | 8.18 | 8.65 | 0.53 | 0.18 | 4.7 | G | 1 |
| hr1346 | S | 4901 | 2.60 | 1.49 | 0.17 | 0.10 | 399 | 0.20 | 0.11 | 25 | 8.38 | 8.81 | 1.14 | 3.06 | 0.18 | 4901 | 2.54 | 1.50 | 0.16 | 0.10 | 399 | 0.16 | 0.11 | 25 | 8.36 | 8.79 | 1.14 | 0.18 | 4.7 | G | 1 |
| hr1348 | S | 4457 | 2.00 | 1.39 | -0.33 | 0.09 | 393 | -0.27 | 0.15 | 24 | 8.01 | 8.53 | 0.04 | 1.14 | 0.33 | 4457 | 1.86 | 1.43 | -0.36 | 0.10 | 393 | -0.36 | 0.15 | 24 | 7.97 | 8.47 | 0.04 | 0.34 | 4.2 | G | 1 |
| hr1373 | S | 4883 | 2.62 | 1.47 | 0.17 | 0.09 | 391 | 0.26 | 0.15 | 27 | 8.38 | 8.83 | 1.01 | 2.44 | 0.18 | 4883 | 2.51 | 1.46 | 0.14 | 0.10 | 391 | 0.13 | 0.15 | 27 | 8.27 | 8.69 | 1.01 | 0.19 | 5.2 | G | 1 |
| hr1407 | S | 4516 | 2.45 | 1.18 | 0.06 | 0.11 | 390 | 0.21 | 0.16 | 26 | 8.37 | 8.76 | 0.06 | 0.99 | 0.28 | 4516 | 2.08 | 1.36 | -0.06 | 0.11 | 390 | -0.06 | 0.17 | 26 | 8.24 | 8.59 | 0.06 | 0.31 | 3.9 | G | 1 |
| hr1409 | S | 4836 | 2.57 | 1.49 | 0.22 | 0.10 | 382 | 0.29 | 0.11 | 27 | 8.40 | 8.86 | 0.82 | 1.88 | 0.19 | 4836 | 2.39 | 1.54 | 0.18 | 0.10 | 382 | 0.18 | 0.11 | 27 | 8.34 | 8.78 | 0.83 | 0.21 | 5.2 | G | 1 |
| hr1411 | S | 4948 | 2.75 | 1.40 | 0.18 | 0.09 | 402 | 0.23 | 0.10 | 27 | 8.36 | 8.86 | 1.25 | 3.37 | 0.17 | 4948 | 2.63 | 1.44 | 0.16 | 0.09 | 402 | 0.16 | 0.11 | 27 | 8.32 | 8.80 | 1.25 | 0.17 | 4.6 | G | 1 |
| hr1413 | S | 4753 | 2.73 | 1.26 | -0.06 | 0.08 | 404 | 0.18 | 0.12 | 29 | 8.26 | 8.80 | 0.40 | 0.95 | 0.20 | 4753 | 2.18 | 1.44 | -0.16 | 0.09 | 404 | -0.16 | 0.14 | 29 | 8.10 | 8.54 | 0.40 | 0.24 | 4.0 | G | 1 |
| hr1421 | S | 4462 | 2.44 | 1.48 | 0.35 | 0.14 | 357 | 0.57 | 0.25 | 25 | 8.66 | 9.04 | 0.00 | 1.03 | 0.29 | 4462 | 1.92 | 1.60 | 0.20 | 0.15 | 357 | 0.20 | 0.26 | 25 | 8.47 | 8.81 | -0.01 | 0.33 | 4.5 | G | 1 |
| hr1425 | S | 4748 | 2.71 | 1.32 | -0.06 | 0.08 | 399 | 0.12 | 0.09 | 26 | 8.37 | 8.86 | -0.15 | 0.28 | 0.21 | 4748 | 2.32 | 1.45 | -0.13 | 0.08 | 399 | -0.13 | 0.09 | 26 | 8.25 | 8.68 | -0.15 | 0.25 | 4.1 | G | 1 |
| hr1431 | S | 4814 | 2.55 | 1.38 | 0.05 | 0.08 | 400 | 0.12 | 0.11 | 30 | 8.22 | 8.73 | 0.56 | 1.13 | 0.20 | 4814 | 2.39 | 1.43 | 0.03 | 0.08 | 400 | 0.02 | 0.12 | 30 | 8.17 | 8.66 | 0.57 | 0.21 | 4.0 | G | 1 |
| hr1453 | S | 4773 | 2.84 | 1.15 | -0.18 | 0.08 | 406 | -0.04 | 0.10 | 24 | 8.20 | 8.63 | -0.06 | 0.32 | 0.20 | 4773 | 2.66 | 1.11 | -0.23 | 0.08 | 406 | -0.23 | 0.11 | 24 | 8.00 | 8.41 | -0.06 | 0.20 | 3.7 | G | 1 |
| hr1455 | S | 5455 | 2.99 | 0.74 | 0.15 | 0.14 | 421 | -0.23 | 0.13 | 29 | 8.19 | 8.83 | 1.10 | 0.72 | 0.08 | 5455 | 3.78 | 0.50 | 0.16 | 0.17 | 421 | 0.16 | 0.14 | 29 | 8.36 | 9.19 | 1.09 | 0.07 | 5.3 | G | 1 |
| hr1514 | S | 4969 | 2.64 | 1.30 | 0.23 | 0.10 | 393 | 0.29 | 0.12 | 27 | 8.38 | 8.83 | 1.12 | 2.45 | 0.17 | 4969 | 2.50 | 1.35 | 0.21 | 0.10 | 393 | 0.21 | 0.13 | 27 | 8.34 | 8.76 | 1.13 | 0.18 | 5.4 | G | 1 |
| hr1517 | S | 4382 | 2.50 | 1.31 | 0.36 | 0.15 | 365 | 0.65 | 0.20 | 23 | 8.60 | 8.96 | 1.12 | 12.02 | 0.29 | 4382 | 1.79 | 1.53 | 0.13 | 0.16 | 365 | 0.13 | 0.21 | 23 | 8.34 | 8.64 | 1.09 | 0.35 | 4.5 | G | 1 |
| hr1529 | S | 4416 | 2.55 | 1.00 | 0.13 | 0.22 | 207 | 0.26 | 0.12 | 10 | | | | | 0.29 | 4416 | 2.27 | 1.42 | -0.07 | 0.19 | 207 | -0.07 | 0.11 | 10 | | | | 0.31 | 4.0 | G | 1 |
| hr1533 | S | 4043 | 1.24 | 1.64 | -0.34 | 0.16 | 371 | -0.11 | 0.24 | 25 | 8.13 | 8.57 | 0.37 | 9.58 | 0.41 | 4043 | 0.62 | 1.68 | -0.49 | 0.17 | 371 | -0.49 | 0.23 | 25 | 7.91 | 8.31 | 0.33 | 0.54 | 4.7 | G | 1 |
| hr1535 | S | 4868 | 2.64 | 1.37 | 0.03 | 0.08 | 397 | 0.05 | 0.12 | 29 | 8.18 | 8.83 | 0.35 | 0.59 | 0.19 | 4868 | 2.60 | 1.38 | 0.03 | 0.08 | 397 | 0.03 | 0.11 | 29 | 8.17 | 8.81 | 0.35 | 0.19 | 4.1 | G | 1 |
| hr1549 | S | 4818 | 2.59 | 1.38 | 0.06 | 0.09 | 400 | 0.11 | 0.11 | 25 | 8.26 | 8.77 | 0.57 | 1.15 | 0.20 | 4818 | 2.47 | 1.42 | 0.04 | 0.09 | 400 | 0.04 | 0.11 | 25 | 8.22 | 8.71 | 0.57 | 0.21 | 4.5 | G | 1 |
| hr1580 | S | 4466 | 2.18 | 1.37 | -0.15 | 0.10 | 384 | 0.01 | 0.15 | 28 | 8.21 | 8.72 | -0.29 | 0.54 | 0.31 | 4466 | 1.93 | 1.39 | -0.25 | 0.10 | 384 | -0.25 | 0.15 | 28 | 8.00 | 8.48 | -0.28 | 0.33 | 4.1 | G | 1 |
| hr1625 | S | 4476 | 2.41 | 1.45 | 0.39 | 0.14 | 358 | 0.61 | 0.20 | 26 | 8.68 | 9.03 | 0.12 | 1.28 | 0.29 | 4476 | 1.90 | 1.58 | 0.25 | 0.14 | 358 | 0.25 | 0.21 | 26 | 8.49 | 8.80 | 0.12 | 0.33 | 4.7 | G | 1 |
| hr1628 | S | 4708 | 2.46 | 1.37 | 0.18 | 0.10 | 414 | 0.23 | 0.13 | 25 | 8.29 | 8.81 | 0.58 | 1.61 | 0.25 | 4708 | 2.45 | 1.31 | 0.16 | 0.10 | 414 | 0.16 | 0.13 | 25 | 8.18 | 8.71 | 0.58 | 0.25 | 4.4 | G | 1 |
| hr1654 | S | 4005 | 1.37 | 1.73 | 0.05 | 0.19 | 325 | 0.45 | 0.24 | 23 | 8.30 | 8.71 | 0.08 | 6.09 | 0.38 | 4005 | 0.30 | 1.73 | -0.23 | 0.21 | 325 | -0.23 | 0.24 | 23 | 7.94 | 8.25 | -0.04 | 0.64 | 5.9 | G | 1 |
| hr1681 | S | 4671 | 2.64 | 1.30 | 0.26 | 0.10 | 379 | 0.45 | 0.15 | 26 | 8.48 | 8.97 | 0.38 | 1.19 | 0.23 | 4671 | 2.22 | 1.44 | 0.16 | 0.10 | 379 | 0.16 | 0.16 | 26 | 8.35 | 8.78 | 0.39 | 0.27 | 4.2 | G | 1 |
| hr1698 | S | 4533 | 2.04 | 1.73 | 0.25 | 0.14 | 361 | 0.24 | 0.20 | 24 | 8.45 | 8.94 | 0.92 | 5.84 | 0.31 | 4533 | 2.07 | 1.72 | 0.26 | 0.14 | 361 | 0.26 | 0.20 | 24 | 8.46 | 8.96 | 0.92 | 0.31 | 5.3 | G | 1 |
| hr1726 | S | 4264 | 1.92 | 1.38 | -0.28 | 0.13 | 373 | -0.11 | 0.18 | 27 | 8.25 | 8.76 | -0.57 | 0.59 | 0.34 | 4264 | 1.45 | 1.49 | -0.41 | 0.13 | 373 | -0.41 | 0.18 | 27 | 8.07 | 8.55 | -0.58 | 0.39 | 4.5 | G | 1 |
| hr1739 | S | 4954 | 2.67 | 1.35 | 0.07 | 0.09 | 406 | 0.08 | 0.09 | 27 | 8.19 | 8.58 | 0.65 | 0.90 | 0.17 | 4954 | 2.66 | 1.35 | 0.07 | 0.09 | 406 | 0.07 | 0.09 | 27 | 8.19 | 8.58 | 0.65 | 0.17 | 4.9 | G | 1 |
| hr1784 | S | 4839 | 2.57 | 1.37 | -0.18 | 0.08 | 399 | -0.04 | 0.10 | 29 | 8.07 | 8.68 | 0.23 | 0.50 | 0.19 | 4839 | 2.25 | 1.44 | -0.22 | 0.08 | 399 | -0.22 | 0.11 | 29 | 7.98 | 8.53 | 0.24 | 0.21 | 4.4 | G | 1 |
| hr1796 | S | 4624 | 2.61 | 1.38 | 0.29 | 0.12 | 383 | 0.44 | 0.19 | 27 | 8.58 | 9.07 | 0.45 | 1.62 | 0.24 | 4624 | 2.23 | 1.50 | 0.19 | 0.12 | 383 | 0.19 | 0.20 | 27 | 8.46 | 8.89 | 0.45 | 0.28 | 4.3 | G | 1 |
| hr1822 | S | 6273 | 3.51 | 3.16 | -0.04 | 0.21 | 118 | 0.06 | 0.07 | 3 | 8.17 | 8.75 | 2.64 | 4.40 | 0.02 | 6273 | 3.22 | 3.20 | -0.05 | 0.21 | 118 | -0.05 | 0.08 | 3 | 8.08 | 8.66 | 2.63 | 0.04 | 44.0 | R | 2 |
| hr1831 | S | 4554 | 2.46 | 1.18 | 0.11 | 0.11 | 385 | 0.12 | 0.19 | 24 | 8.37 | 8.79 | -0.01 | 0.73 | 0.27 | 4554 | 2.43 | 1.20 | 0.10 | 0.11 | 385 | 0.10 | 0.19 | 24 | 8.36 | 8.77 | -0.01 | 0.28 | 3.8 | G | 1 |
| hr1889 | S | 6379 | 3.65 | 3.63 | 0.00 | 0.22 | 22 | -0.30 | 0.14 | 2 | 7.34 | 8.34 | 3.00 | 7.57 | 0.02 | 6379 | 4.45 | 3.40 | 0.01 | 0.22 | 22 | 0.01 | 0.12 | 2 | 7.60 | 8.65 | 3.00 | -0.09 | 70.0 | R | 2 |
| hr1954 | S | 4549 | 2.59 | 1.35 | 0.39 | 0.13 | 362 | 0.52 | 0.22 | 25 | 8.58 | 9.05 | 0.16 | 1.10 | 0.26 | 4549 | 2.27 | 1.46 | 0.29 | 0.13 | 362 | 0.29 | 0.23 | 25 | 8.46 | 8.90 | 0.16 | 0.29 | 4.1 | G | 1 |
| hr1963 | S | 4389 | 1.98 | 1.42 | -0.44 | 0.09 | 393 | -0.25 | 0.12 | 26 | 8.02 | 8.52 | -0.16 | 0.93 | 0.33 | 4389 | 1.52 | 1.50 | -0.53 | 0.10 | 393 | -0.53 | 0.13 | 26 | 7.86 | 8.31 | -0.17 | 0.38 | 4.2 | G | 1 |
| hr1986 | S | 4701 | 2.68 | 1.16 | 0.13 | 0.09 | 396 | 0.18 | 0.15 | 27 | 8.27 | 8.76 | 0.20 | 0.72 | 0.22 | 4701 | 2.57 | 1.24 | 0.09 | 0.09 | 396 | 0.09 | 0.15 | 27 | 8.24 | 8.71 | 0.20 | 0.23 | 3.7 | G | 1 |
| hr1987 | S | 4988 | 2.82 | 1.27 | -0.10 | 0.08 | 407 | 0.00 | 0.08 | 27 | 8.10 | 8.57 | 0.24 | 0.33 | 0.16 | 4988 | 2.59 | 1.34 | -0.13 | 0.08 | 407 | -0.13 | 0.08 | 27 | 8.05 | 8.46 | 0.25 | 0.16 | 4.4 | G | 1 |
| hr1995 | S | 4876 | 2.62 | 1.36 | -0.10 | 0.08 | 409 | 0.03 | 0.10 | 29 | 8.13 | 8.68 | 0.01 | 0.27 | 0.18 | 4876 | 2.34 | 1.43 | -0.13 | 0.08 | 409 | -0.13 | 0.11 | 29 | 8.05 | 8.55 | 0.01 | 0.20 | 4.6 | G | 1 |
| hr2012 | S | 4590 | 2.14 | 1.50 | 0.02 | 0.10 | 376 | 0.08 | 0.17 | 28 | 8.23 | 8.77 | 0.23 | 1.11 | 0.30 | 4590 | 1.99 | 1.53 | -0.01 | 0.10 | 376 | -0.01 | 0.17 | 28 | 8.18 | 8.70 | 0.23 | 0.31 | 4.8 | G | 1 |
| hr2076 | S | 4640 | 2.59 | 1.43 | 0.24 | 0.12 | 383 | 0.43 | 0.21 | 29 | 8.51 | 8.98 | 0.36 | 1.26 | 0.24 | 4640 | 2.23 | 1.50 | 0.13 | 0.12 | 383 | 0.13 | 0.21 | 29 | 8.29 | 8.72 | 0.37 | 0.28 | 4.5 | G | 1 |
| hr2077 | S | 4768 | 2.56 | 1.34 | 0.00 | 0.08 | 396 | 0.10 | 0.10 | 29 | 8.24 | 8.80 | 0.31 | 0.74 | 0.21 | 4768 | 2.32 | 1.42 | -0.05 | 0.09 | 396 | -0.05 | 0.10 | 29 | 8.17 | 8.69 | 0.31 | 0.23 | 4.2 | G | 1 |

| | | | | | | | | | | | | | | | | | | | | | | | | | | |
|---|---|---|---|---|---|---|---|---|---|---|---|---|---|---|---|---|---|---|---|---|---|---|---|---|---|---|
| hr2080 | S | 4540 | 2.55 | 1.47 | 0.38 | 0.15 | 357 | 0.58 | 0.27 | 26 | 8.69 | 9.10 | 0.47 | 2.25 | 0.26 | 4540 | 2.10 | 1.60 | 0.26 | 0.15 | 357 | 0.26 | 0.28 | 26 | 8.53 | 8.89 | 0.47 | 0.31 | 4.6 | G | 1 |
| hr2119 | S | 4786 | 2.94 | 1.07 | -0.12 | 0.08 | 403 | -0.03 | 0.11 | 25 | 8.34 | 8.86 | 0.34 | 0.76 | 0.19 | 4786 | 2.75 | 1.18 | -0.16 | 0.08 | 403 | -0.16 | 0.11 | 25 | 8.28 | 8.77 | 0.34 | 0.20 | 3.6 | G | 1 |
| hr2136 | S | 4950 | 2.72 | 1.34 | -0.04 | 0.07 | 405 | 0.04 | 0.09 | 28 | 8.17 | 8.68 | 0.50 | 0.66 | 0.17 | 4950 | 2.55 | 1.39 | -0.06 | 0.08 | 405 | -0.06 | 0.09 | 28 | 8.13 | 8.60 | 0.50 | 0.17 | 4.5 | G | 1 |
| hr2152 | S | 4609 | 2.09 | 1.58 | -0.45 | 0.08 | 419 | -0.42 | 0.07 | 20 | 7.94 | 8.54 | -0.68 | 0.13 | 0.30 | 4609 | 2.01 | 1.59 | -0.46 | 0.08 | 419 | -0.46 | 0.07 | 20 | 7.91 | 8.50 | -0.68 | 0.30 | 4.5 | G | 1 |
| hr2183 | S | 4635 | 2.73 | 1.40 | 0.36 | 0.13 | 371 | 0.49 | 0.23 | 25 | 8.56 | 9.00 | 1.04 | 5.37 | 0.23 | 4635 | 2.41 | 1.52 | 0.27 | 0.14 | 371 | 0.27 | 0.24 | 25 | 8.45 | 8.85 | 1.04 | 0.26 | 4.1 | G | 1 |
| hr2200 | S | 4690 | 2.49 | 1.36 | 0.06 | 0.10 | 388 | 0.13 | 0.13 | 25 | 8.18 | 8.71 | 0.10 | 0.59 | 0.25 | 4690 | 2.34 | 1.42 | 0.03 | 0.10 | 388 | 0.03 | 0.13 | 25 | 8.14 | 8.64 | 0.10 | 0.26 | 4.5 | G | 1 |
| hr2218 | S | 4926 | 2.79 | 1.27 | -0.08 | 0.08 | 401 | 0.07 | 0.11 | 28 | 8.12 | 8.77 | 0.49 | 0.69 | 0.17 | 4926 | 2.47 | 1.37 | -0.12 | 0.08 | 401 | -0.12 | 0.11 | 28 | 8.03 | 8.61 | 0.49 | 0.18 | 4.2 | G | 1 |
| hr2219 | S | 4685 | 2.35 | 1.41 | -0.26 | 0.07 | 402 | -0.17 | 0.10 | 27 | 8.05 | 8.57 | 0.05 | 0.54 | 0.26 | 4685 | 2.13 | 1.47 | -0.30 | 0.08 | 402 | -0.30 | 0.10 | 27 | 7.98 | 8.47 | 0.05 | 0.28 | 4.1 | G | 1 |
| hr2230 | S | 5072 | 2.71 | 1.49 | 0.04 | 0.08 | 406 | 0.06 | 0.10 | 29 | 8.18 | 8.69 | 0.55 | 0.53 | 0.15 | 5072 | 2.66 | 1.50 | 0.03 | 0.08 | 406 | 0.03 | 0.10 | 29 | 8.16 | 8.67 | 0.55 | 0.15 | 5.4 | G | 1 |
| hr2239 | S | 4552 | 2.43 | 1.37 | 0.13 | 0.12 | 376 | 0.24 | 0.16 | 25 | 8.53 | 8.96 | 0.36 | 1.70 | 0.28 | 4552 | 2.16 | 1.47 | 0.06 | 0.12 | 376 | 0.06 | 0.17 | 25 | 8.43 | 8.84 | 0.36 | 0.30 | 4.9 | G | 1 |
| hr2243 | S | 4701 | 2.96 | 0.91 | 0.23 | 0.11 | 418 | 0.39 | 0.17 | 30 | 8.48 | 8.87 | 0.07 | 0.54 | 0.20 | 4701 | 2.68 | 1.04 | 0.12 | 0.10 | 418 | 0.12 | 0.18 | 30 | 8.28 | 8.65 | 0.07 | 0.22 | 4.2 | G | 1 |
| hr2259 | S | 5076 | 3.00 | 0.96 | 0.00 | 0.07 | 410 | -0.06 | 0.11 | 27 | 8.10 | 8.69 | 1.38 | 3.27 | 0.14 | 5076 | 3.13 | 0.83 | 0.03 | 0.08 | 410 | 0.03 | 0.11 | 27 | 8.14 | 8.75 | 1.38 | 0.14 | 4.1 | G | 1 |
| hr2287 | S | 6986 | 3.77 | 3.43 | 0.14 | 0.33 | 34 | -0.26 | 0.08 | 3 | 7.85 | 8.61 | 2.21 | 0.62 | 0.02 | 6986 | 4.91 | 3.03 | 0.14 | 0.33 | 34 | 0.14 | 0.11 | 3 | 8.16 | 8.91 | 2.21 | -0.48 | 44.0 | R | 2 |
| hr2302 | S | 4729 | 2.44 | 1.38 | -0.03 | 0.08 | 399 | 0.02 | 0.12 | 30 | 8.24 | 8.75 | 0.53 | 1.37 | 0.24 | 4729 | 2.31 | 1.42 | -0.06 | 0.08 | 399 | -0.06 | 0.12 | 30 | 8.20 | 8.69 | 0.53 | 0.25 | 4.2 | G | 1 |
| hr2333 | S | 4736 | 2.49 | 1.38 | -0.04 | 0.09 | 395 | 0.06 | 0.14 | 30 | 8.19 | 8.73 | 0.15 | 0.57 | 0.24 | 4736 | 2.27 | 1.44 | -0.08 | 0.09 | 395 | -0.08 | 0.15 | 30 | 8.12 | 8.63 | 0.15 | 0.25 | 4.4 | G | 1 |
| hr2379 | S | 4345 | 2.00 | 1.55 | 0.14 | 0.14 | 378 | 0.27 | 0.20 | 24 | 8.33 | 8.81 | -0.18 | 1.07 | 0.33 | 4345 | 1.68 | 1.60 | 0.06 | 0.15 | 378 | 0.06 | 0.20 | 24 | 8.22 | 8.67 | -0.18 | 0.36 | 4.8 | G | 1 |
| hr2392 | S | 4725 | 2.53 | 1.42 | -0.08 | 0.17 | 370 | 0.16 | 0.14 | 22 | 8.54 | 8.81 | 0.69 | 1.97 | 0.23 | 4725 | 1.98 | 1.54 | -0.17 | 0.18 | 370 | -0.17 | 0.15 | 22 | 8.37 | 8.54 | 0.70 | 0.28 | 6.4 | G | 1 |
| hr2477 | S | 4883 | 2.85 | 1.15 | -0.02 | 0.07 | 407 | 0.02 | 0.09 | 26 | 8.09 | 8.65 | 0.22 | 0.43 | 0.17 | 4883 | 2.77 | 1.20 | -0.04 | 0.07 | 407 | -0.04 | 0.09 | 26 | 8.07 | 8.61 | 0.22 | 0.18 | 4.3 | G | 1 |
| hr2478 | S | 4502 | 2.02 | 1.55 | -0.10 | 0.10 | 390 | -0.03 | 0.11 | 25 | 8.09 | 8.72 | 0.00 | 0.90 | 0.32 | 4502 | 1.86 | 1.58 | -0.13 | 0.10 | 390 | -0.13 | 0.11 | 25 | 8.04 | 8.64 | -0.01 | 0.33 | 4.6 | G | 1 |
| hr2552 | S | 4960 | 2.70 | 1.40 | 0.14 | 0.09 | 397 | 0.19 | 0.12 | 28 | 8.23 | 8.69 | 0.95 | 1.75 | 0.17 | 4960 | 2.59 | 1.44 | 0.12 | 0.09 | 397 | 0.12 | 0.12 | 28 | 8.19 | 8.63 | 0.95 | 0.17 | 5.0 | G | 1 |
| hr2556 | S | 4994 | 2.74 | 1.33 | 0.06 | 0.07 | 405 | 0.03 | 0.11 | 25 | 8.18 | 8.63 | 1.00 | 1.77 | 0.16 | 4994 | 2.81 | 1.29 | 0.08 | 0.07 | 405 | 0.08 | 0.11 | 25 | 8.20 | 8.66 | 1.00 | 0.16 | 4.3 | G | 1 |
| hr2573 | S | 4738 | 2.50 | 1.38 | 0.10 | 0.09 | 391 | 0.11 | 0.14 | 27 | 8.26 | 8.73 | 0.30 | 0.80 | 0.24 | 4738 | 2.47 | 1.39 | 0.09 | 0.09 | 391 | 0.09 | 0.14 | 27 | 8.25 | 8.72 | 0.30 | 0.24 | 4.2 | G | 1 |
| hr2574 | S | 4044 | 1.49 | 1.57 | -0.15 | 0.17 | 354 | 0.26 | 0.30 | 25 | 8.39 | 8.85 | -0.78 | 0.85 | 0.37 | 4044 | 0.44 | 1.63 | -0.47 | 0.20 | 354 | -0.47 | 0.32 | 25 | 7.90 | 8.28 | -0.86 | 0.58 | 5.4 | G | 1 |
| hr2642 | S | 4614 | 2.55 | 1.43 | 0.33 | 0.12 | 364 | 0.41 | 0.21 | 25 | 8.58 | 9.04 | 0.34 | 1.31 | 0.25 | 4614 | 2.36 | 1.49 | 0.28 | 0.12 | 364 | 0.28 | 0.22 | 25 | 8.51 | 8.96 | 0.34 | 0.27 | 4.7 | G | 1 |
| hr2684 | S | 4740 | 2.41 | 1.42 | -0.10 | 0.08 | 392 | -0.04 | 0.13 | 28 | 8.16 | 8.72 | 0.70 | 1.93 | 0.24 | 4740 | 2.28 | 1.45 | -0.12 | 0.08 | 392 | -0.12 | 0.13 | 28 | 8.12 | 8.66 | 0.70 | 0.25 | 4.6 | G | 1 |
| hr2701 | S | 4692 | 2.57 | 1.12 | -0.24 | 0.08 | 403 | -0.17 | 0.11 | 26 | 7.99 | 8.53 | 0.07 | 0.55 | 0.23 | 4692 | 2.41 | 1.21 | -0.28 | 0.08 | 403 | -0.28 | 0.11 | 26 | 7.94 | 8.46 | 0.07 | 0.25 | 3.9 | G | 1 |
| hr2728 | S | 4787 | 2.53 | 1.42 | -0.11 | 0.09 | 400 | -0.02 | 0.10 | 26 | 8.18 | 8.72 | -0.20 | 0.22 | 0.21 | 4787 | 2.35 | 1.47 | -0.14 | 0.09 | 400 | -0.13 | 0.11 | 26 | 8.13 | 8.63 | -0.20 | 0.22 | 4.2 | G | 1 |
| hr2764 | S | 3937 | 0.32 | 2.98 | -0.10 | 0.26 | 250 | 0.21 | 0.27 | 15 | 8.16 | 8.76 | 0.61 | 19.17 | 0.65 | 3937 | -0.49 | 2.82 | -0.26 | 0.27 | 250 | -0.26 | 0.29 | 15 | 7.87 | 8.35 | 0.46 | 1.00 | 12.0 | G | 1 |
| hr2793 | S | 4575 | 2.38 | 1.30 | -0.33 | 0.08 | 395 | -0.24 | 0.11 | 25 | 8.04 | 8.54 | 0.15 | 0.97 | 0.28 | 4575 | 2.16 | 1.38 | -0.37 | 0.09 | 395 | -0.38 | 0.12 | 25 | 7.97 | 8.44 | 0.15 | 0.30 | 3.9 | G | 1 |
| hr2877 | S | 4620 | 2.47 | 1.39 | 0.19 | 0.11 | 378 | 0.31 | 0.19 | 29 | 8.32 | 8.82 | -0.03 | 0.56 | 0.26 | 4620 | 2.20 | 1.48 | 0.12 | 0.11 | 378 | 0.13 | 0.19 | 29 | 8.24 | 8.70 | -0.03 | 0.28 | 4.6 | G | 1 |
| hr2896 | S | 4829 | 2.49 | 1.53 | 0.13 | 0.11 | 366 | 0.17 | 0.20 | 30 | 8.26 | 8.71 | 0.44 | 0.82 | 0.20 | 4829 | 2.41 | 1.55 | 0.12 | 0.11 | 366 | 0.12 | 0.20 | 30 | 8.24 | 8.67 | 0.44 | 0.21 | 7.0 | G | 1 |
| hr2899 | S | 4600 | 2.54 | 1.38 | 0.20 | 0.11 | 376 | 0.34 | 0.13 | 23 | 8.51 | 8.98 | 0.37 | 1.48 | 0.25 | 4600 | 2.31 | 1.41 | 0.11 | 0.11 | 376 | 0.11 | 0.15 | 23 | 8.34 | 8.79 | 0.38 | 0.28 | 4.3 | G | 1 |
| hr2924 | S | 4997 | 2.64 | 1.40 | 0.05 | 0.08 | 429 | 0.03 | 0.10 | 28 | 8.10 | 8.71 | 0.42 | 0.48 | 0.16 | 4997 | 2.68 | 1.39 | 0.05 | 0.08 | 429 | 0.05 | 0.10 | 28 | 8.11 | 8.72 | 0.42 | 0.16 | 4.7 | G | 1 |
| hr2970 | S | 4749 | 2.50 | 1.36 | -0.01 | 0.08 | 403 | 0.06 | 0.11 | 26 | 8.21 | 8.70 | 0.40 | 0.96 | 0.23 | 4749 | 2.35 | 1.41 | -0.03 | 0.08 | 403 | -0.03 | 0.11 | 26 | 8.17 | 8.63 | 0.40 | 0.25 | 4.3 | G | 1 |
| hr2975 | S | 4357 | 2.24 | 1.42 | 0.25 | 0.12 | 368 | 0.55 | 0.18 | 26 | 8.51 | 8.88 | -0.20 | 0.95 | 0.32 | 4357 | 1.59 | 1.53 | 0.03 | 0.13 | 368 | 0.03 | 0.18 | 26 | 8.19 | 8.50 | -0.22 | 0.37 | 4.5 | G | 1 |
| hr2985 | S | 4954 | 2.61 | 1.45 | 0.00 | 0.08 | 410 | 0.04 | 0.10 | 27 | 8.16 | 8.66 | 0.60 | 0.82 | 0.17 | 4954 | 2.52 | 1.47 | -0.01 | 0.08 | 410 | -0.01 | 0.10 | 27 | 8.13 | 8.61 | 0.61 | 0.17 | 4.7 | G | 1 |
| hr2990 | S | 4821 | 2.72 | 1.21 | 0.13 | 0.09 | 405 | 0.16 | 0.14 | 29 | 8.23 | 8.63 | 0.83 | 2.02 | 0.19 | 4821 | 2.65 | 1.25 | 0.11 | 0.09 | 405 | 0.11 | 0.14 | 29 | 8.20 | 8.59 | 0.83 | 0.19 | 4.4 | G | 1 |
| hr3044 | S | 4330 | 2.21 | 1.49 | 0.25 | 0.15 | 355 | 0.45 | 0.23 | 22 | 8.59 | 9.00 | -0.19 | 1.06 | 0.32 | 4330 | 1.83 | 1.54 | 0.10 | 0.15 | 355 | 0.10 | 0.23 | 22 | 8.35 | 8.74 | -0.21 | 0.35 | 4.5 | G | 1 |
| hr3054 | S | 4660 | 2.47 | 1.46 | 0.14 | 0.10 | 382 | 0.18 | 0.16 | 27 | 8.31 | 8.86 | 0.20 | 0.82 | 0.25 | 4660 | 2.38 | 1.49 | 0.12 | 0.10 | 382 | 0.12 | 0.16 | 27 | 8.28 | 8.82 | 0.20 | 0.26 | 4.3 | G | 1 |
| hr3094 | S | 4772 | 2.57 | 1.38 | 0.03 | 0.08 | 397 | 0.07 | 0.13 | 30 | 8.24 | 8.75 | 0.20 | 0.57 | 0.21 | 4772 | 2.48 | 1.41 | 0.01 | 0.08 | 397 | 0.01 | 0.13 | 30 | 8.21 | 8.70 | 0.20 | 0.22 | 4.1 | G | 1 |
| hr3097 | S | 4865 | 2.79 | 1.29 | 0.13 | 0.09 | 417 | 0.18 | 0.13 | 28 | 8.28 | 8.79 | 0.46 | 0.78 | 0.18 | 4865 | 2.67 | 1.35 | 0.10 | 0.09 | 417 | 0.10 | 0.13 | 28 | 8.24 | 8.73 | 0.46 | 0.18 | 4.4 | G | 1 |
| hr3110 | S | 5008 | 2.84 | 1.27 | 0.03 | 0.07 | 409 | 0.02 | 0.10 | 28 | 7.98 | 8.69 | 0.34 | 0.39 | 0.15 | 5008 | 2.88 | 1.25 | 0.04 | 0.07 | 409 | 0.04 | 0.09 | 28 | 7.99 | 8.71 | 0.33 | 0.15 | 4.2 | G | 1 |
| hr3122 | S | 4422 | 2.14 | 1.41 | -0.10 | 0.10 | 387 | 0.07 | 0.15 | 27 | 8.23 | 8.76 | 0.71 | 5.36 | 0.23 | 4422 | 1.72 | 1.49 | -0.20 | 0.11 | 387 | -0.20 | 0.15 | 27 | 8.09 | 8.57 | 0.70 | 0.35 | 4.2 | G | 1 |
| hr3125 | S | 4710 | 2.55 | 1.38 | 0.09 | 0.09 | 393 | 0.16 | 0.17 | 26 | 8.30 | 8.81 | 0.20 | 0.70 | 0.23 | 4710 | 2.40 | 1.43 | 0.06 | 0.09 | 393 | 0.06 | 0.17 | 26 | 8.25 | 8.74 | 0.20 | 0.25 | 4.4 | G | 1 |
| hr3145 | S | 4280 | 1.86 | 1.46 | -0.32 | 0.11 | 373 | -0.17 | 0.15 | 25 | 8.12 | 8.62 | -0.31 | 1.00 | 0.35 | 4280 | 1.48 | 1.54 | -0.42 | 0.12 | 373 | -0.41 | 0.15 | 25 | 7.99 | 8.45 | -0.32 | 0.39 | 4.7 | G | 1 |
| hr3149 | S | 4567 | 2.37 | 1.30 | 0.07 | 0.10 | 385 | 0.18 | 0.19 | 28 | 8.28 | 8.79 | 0.83 | 4.31 | 0.28 | 4567 | 2.11 | 1.39 | 0.01 | 0.10 | 385 | 0.01 | 0.20 | 28 | 8.20 | 8.67 | 0.83 | 0.30 | 4.4 | G | 1 |
| hr3150 | S | 5702 | 3.18 | 1.70 | -0.21 | 0.09 | 377 | -0.12 | 0.09 | 33 | 8.31 | 8.63 | 0.50 | 0.11 | 0.05 | 5702 | 2.99 | 1.75 | -0.21 | 0.10 | 377 | -0.21 | 0.09 | 33 | 8.24 | 8.55 | 0.50 | 0.05 | 5.8 | G | 1 |
| hr3211 | S | 4917 | 2.86 | 1.26 | 0.05 | 0.08 | 409 | 0.10 | 0.09 | 25 | 8.22 | 8.66 | 0.91 | 1.80 | 0.17 | 4917 | 2.75 | 1.32 | 0.03 | 0.08 | 409 | 0.03 | 0.10 | 25 | 8.19 | 8.60 | 0.91 | 0.17 | 4.1 | G | 1 |
| hr3212 | S | 4988 | 2.72 | 1.29 | -0.08 | 0.07 | 406 | -0.01 | 0.09 | 31 | 8.10 | 8.62 | 0.31 | 0.39 | 0.16 | 4988 | 2.57 | 1.34 | -0.10 | 0.07 | 406 | -0.10 | 0.10 | 31 | 8.06 | 8.55 | 0.31 | 0.17 | 4.2 | G | 1 |
| hr3216 | S | 4997 | 2.73 | 1.30 | -0.06 | 0.08 | 429 | -0.04 | 0.09 | 29 | 8.07 | 8.56 | 0.30 | 0.37 | 0.16 | 4997 | 2.67 | 1.32 | -0.07 | 0.08 | 429 | -0.07 | 0.09 | 29 | 8.05 | 8.53 | 0.30 | 0.16 | 4.1 | G | 1 |

| ID | | Teff | logg | [Fe/H] | | | N | A1 | A2 | A3 | A4 | | Teff | logg | [Fe/H] | | | N | B1 | B2 | B3 | B4 | | | |
|---|---|---|---|---|---|---|---|---|---|---|---|---|---|---|---|---|---|---|---|---|---|---|---|---|---|
| hr3222 | S | 4971 | 2.60 | 1.23 | 0.03 | 0.10 | 406 | -0.11 | 0.11 | 26 | 8.05 | 8.78 | 0.45 | 0.55 | 0.17 | 4971 | 2.89 | 1.06 | 0.09 | 0.10 | 406 | 0.09 | 0.11 | 26 | 8.14 | 8.92 | 0.44 | 0.16 | 4.6 | G | 1 |
| hr3263 | S | 4930 | 2.82 | 1.29 | 0.07 | 0.08 | 403 | 0.11 | 0.09 | 28 | 8.23 | 8.72 | 0.70 | 1.10 | 0.17 | 4930 | 2.74 | 1.33 | 0.06 | 0.08 | 403 | 0.06 | 0.09 | 28 | 8.20 | 8.68 | 0.70 | 0.17 | 4.4 | G | 1 |
| hr3281 | S | 4660 | 2.49 | 1.33 | -0.10 | 0.09 | 393 | -0.01 | 0.12 | 26 | 8.46 | 8.97 | -0.15 | 0.37 | 0.25 | 4660 | 2.27 | 1.39 | -0.14 | 0.09 | 393 | -0.14 | 0.12 | 26 | 8.39 | 8.87 | -0.15 | 0.27 | 4.3 | G | 1 |
| hr3289 | S | 4863 | 2.63 | 1.37 | -0.05 | 0.07 | 406 | -0.02 | 0.10 | 31 | 8.21 | 8.72 | 0.57 | 1.01 | 0.19 | 4863 | 2.55 | 1.40 | -0.07 | 0.07 | 406 | -0.07 | 0.11 | 31 | 8.18 | 8.68 | 0.58 | 0.19 | 4.2 | G | 1 |
| hr3303 | S | 4934 | 2.78 | 1.31 | 0.00 | 0.08 | 408 | 0.00 | 0.10 | 27 | 8.16 | 8.70 | 0.72 | 1.14 | 0.17 | 4934 | 2.79 | 1.30 | 0.01 | 0.08 | 408 | 0.01 | 0.10 | 27 | 8.17 | 8.70 | 0.72 | 0.17 | 4.7 | G | 1 |
| hr3324 | S | 4490 | 2.39 | 1.26 | 0.03 | 0.11 | 385 | 0.20 | 0.17 | 28 | 8.37 | 8.83 | -0.03 | 0.87 | 0.29 | 4490 | 1.99 | 1.40 | -0.08 | 0.11 | 385 | -0.08 | 0.19 | 28 | 8.23 | 8.65 | -0.03 | 0.32 | 4.2 | G | 1 |
| hr3366 | S | 4415 | 2.20 | 1.52 | 0.23 | 0.14 | 354 | 0.38 | 0.20 | 24 | 8.45 | 8.92 | 0.15 | 1.69 | 0.31 | 4415 | 1.96 | 1.54 | 0.14 | 0.14 | 354 | 0.14 | 0.20 | 24 | 8.27 | 8.72 | 0.15 | 0.33 | 4.8 | G | 1 |
| hr3369 | S | 4781 | 2.43 | 1.53 | 0.03 | 0.09 | 391 | 0.14 | 0.09 | 29 | 8.13 | 8.77 | 0.98 | 3.09 | 0.22 | 4781 | 2.18 | 1.58 | -0.01 | 0.10 | 391 | 0.00 | 0.10 | 29 | 8.06 | 8.66 | 0.99 | 0.23 | 5.2 | G | 1 |
| hr3376 | S | 4553 | 2.60 | 1.49 | 0.32 | 0.15 | 362 | 0.48 | 0.19 | 23 | 8.64 | 9.00 | 0.98 | 6.09 | 0.26 | 4553 | 2.24 | 1.58 | 0.23 | 0.14 | 362 | 0.23 | 0.20 | 23 | 8.51 | 8.83 | 0.98 | 0.29 | 5.0 | G | 1 |
| hr3403 | S | 4438 | 2.11 | 1.45 | -0.12 | 0.11 | 390 | -0.03 | 0.13 | 22 | 8.17 | 8.65 | 0.49 | 3.28 | 0.32 | 4438 | 1.90 | 1.49 | -0.17 | 0.11 | 390 | -0.17 | 0.13 | 22 | 8.10 | 8.55 | 0.49 | 0.34 | 4.7 | G | 1 |
| hr3409 | S | 4673 | 2.38 | 1.37 | 0.06 | 0.09 | 390 | 0.05 | 0.17 | 28 | 8.23 | 8.69 | 1.33 | 8.16 | 0.26 | 4673 | 2.42 | 1.35 | 0.07 | 0.09 | 390 | 0.07 | 0.17 | 28 | 8.24 | 8.71 | 1.33 | 0.26 | 4.2 | G | 1 |
| hr3418 | S | 4491 | 2.01 | 1.74 | 0.14 | 0.13 | 360 | 0.22 | 0.17 | 25 | 8.26 | 8.74 | 0.47 | 2.65 | 0.32 | 4491 | 1.83 | 1.75 | 0.11 | 0.13 | 360 | 0.11 | 0.17 | 25 | 8.20 | 8.66 | 0.47 | 0.34 | 5.5 | G | 1 |
| hr3423 | S | 6574 | 3.38 | 2.19 | 0.03 | 0.34 | 237 | -0.02 | 0.20 | 21 | 8.02 | 8.74 | 1.76 | 0.41 | 0.05 | 6574 | 3.52 | 2.18 | 0.04 | 0.34 | 237 | 0.03 | 0.20 | 21 | 8.06 | 8.77 | 1.76 | 0.04 | 29.1 | R | 2 |
| hr3424 | S | 4577 | 2.65 | 1.19 | 0.14 | 0.10 | 391 | 0.43 | 0.13 | 27 | 8.36 | 8.86 | 0.14 | 0.95 | 0.25 | 4577 | 2.00 | 1.49 | -0.04 | 0.11 | 391 | -0.04 | 0.13 | 27 | 8.14 | 8.55 | 0.14 | 0.31 | 4.2 | G | 1 |
| hr3433 | S | 4876 | 2.59 | 1.40 | -0.52 | 0.06 | 409 | -0.33 | 0.07 | 23 | 7.89 | 8.52 | -0.30 | 0.14 | 0.19 | 4876 | 2.16 | 1.48 | -0.55 | 0.07 | 409 | -0.56 | 0.06 | 23 | 7.77 | 8.31 | -0.30 | 0.21 | 4.2 | G | 1 |
| hr3461 | S | 4636 | 2.56 | 1.34 | 0.08 | 0.10 | 384 | 0.23 | 0.11 | 24 | 8.36 | 8.83 | 0.43 | 1.48 | 0.24 | 4636 | 2.20 | 1.47 | -0.01 | 0.10 | 384 | -0.01 | 0.12 | 24 | 8.24 | 8.66 | 0.43 | 0.28 | 4.8 | G | 1 |
| hr3464 | S | 4966 | 2.49 | 1.56 | -0.04 | 0.10 | 381 | 0.09 | 0.15 | 33 | 8.15 | 8.64 | 1.28 | 3.46 | 0.18 | 4966 | 2.20 | 1.61 | -0.06 | 0.11 | 381 | -0.06 | 0.16 | 33 | 8.08 | 8.50 | 1.29 | 0.19 | 7.2 | G | 1 |
| hr3475 | S | 4849 | 2.15 | 1.98 | 0.04 | 0.15 | 349 | 0.14 | 0.15 | 28 | 8.11 | 8.59 | 0.90 | 2.11 | 0.22 | 4849 | 1.89 | 1.99 | 0.01 | 0.15 | 349 | 0.00 | 0.15 | 28 | 8.05 | 8.47 | 0.90 | 0.23 | 9.5 | G | 1 |
| hr3484 | S | 4968 | 2.50 | 1.41 | -0.15 | 0.08 | 404 | -0.14 | 0.09 | 29 | 7.97 | 8.47 | 0.10 | 0.25 | 0.18 | 4968 | 2.46 | 1.42 | -0.16 | 0.08 | 404 | -0.16 | 0.09 | 29 | 7.96 | 8.45 | 0.10 | 0.18 | 4.8 | G | 1 |
| hr3508 | S | 4761 | 2.48 | 1.33 | -0.30 | 0.07 | 407 | -0.15 | 0.09 | 29 | 7.95 | 8.54 | 0.18 | 0.57 | 0.22 | 4761 | 2.14 | 1.43 | -0.35 | 0.08 | 407 | -0.35 | 0.10 | 29 | 7.85 | 8.38 | 0.18 | 0.24 | 4.2 | G | 1 |
| hr3518 | S | 4322 | 1.95 | 1.47 | 0.01 | 0.13 | 372 | 0.22 | 0.22 | 27 | 8.27 | 8.76 | -0.27 | 0.94 | 0.34 | 4322 | 1.43 | 1.56 | -0.12 | 0.14 | 372 | -0.12 | 0.23 | 27 | 8.09 | 8.53 | -0.29 | 0.39 | 4.7 | G | 1 |
| hr3529 | S | 4552 | 2.46 | 1.28 | 0.10 | 0.11 | 387 | 0.23 | 0.18 | 27 | 8.31 | 8.75 | 0.50 | 2.30 | 0.27 | 4552 | 2.16 | 1.40 | 0.02 | 0.11 | 387 | 0.02 | 0.19 | 27 | 8.21 | 8.62 | 0.50 | 0.30 | 4.1 | G | 1 |
| hr3531 | S | 4989 | 2.71 | 1.27 | -0.14 | 0.08 | 415 | -0.10 | 0.10 | 27 | 8.06 | 8.63 | 0.21 | 0.31 | 0.16 | 4989 | 2.62 | 1.30 | -0.15 | 0.08 | 415 | -0.15 | 0.10 | 27 | 8.04 | 8.59 | 0.21 | 0.16 | 4.4 | G | 1 |
| hr3547 | S | 4795 | 2.28 | 1.52 | -0.07 | 0.09 | 396 | -0.04 | 0.09 | 27 | 8.12 | 8.59 | 0.45 | 0.93 | 0.22 | 4795 | 2.20 | 1.53 | -0.08 | 0.09 | 396 | -0.08 | 0.09 | 27 | 8.10 | 8.55 | 0.45 | 0.23 | 4.9 | G | 1 |
| hr3575 | S | 4974 | 2.63 | 1.37 | -0.04 | 0.08 | 426 | 0.01 | 0.07 | 28 | 8.04 | 8.57 | 0.49 | 0.60 | 0.17 | 4974 | 2.52 | 1.40 | -0.05 | 0.08 | 426 | -0.06 | 0.07 | 28 | 8.01 | 8.52 | 0.49 | 0.17 | 4.3 | G | 1 |
| hr3621 | S | 4965 | 2.76 | 1.31 | -0.07 | 0.09 | 426 | 0.04 | 0.09 | 27 | 8.06 | 8.57 | 0.62 | 0.83 | 0.16 | 4965 | 2.53 | 1.39 | -0.10 | 0.10 | 426 | -0.10 | 0.09 | 27 | 8.00 | 8.46 | 0.63 | 0.17 | 5.0 | G | 1 |
| hr3640 | S | 5076 | 2.86 | 1.33 | -0.06 | 0.07 | 416 | 0.03 | 0.08 | 28 | 8.07 | 8.64 | 0.87 | 1.09 | 0.16 | 5076 | 2.67 | 1.39 | -0.08 | 0.08 | 416 | -0.08 | 0.07 | 28 | 8.02 | 8.54 | 0.88 | 0.16 | 4.6 | G | 1 |
| hr3653 | S | 4825 | 2.58 | 1.32 | -0.24 | 0.07 | 426 | -0.18 | 0.08 | 27 | 8.02 | 8.56 | 0.00 | 0.31 | 0.20 | 4825 | 2.45 | 1.37 | -0.26 | 0.07 | 426 | -0.25 | 0.08 | 27 | 7.98 | 8.50 | 0.00 | 0.21 | 4.2 | G | 1 |
| hr3664 | S | 5019 | 2.61 | 1.55 | -0.65 | 0.08 | 387 | -0.49 | 0.05 | 24 | 7.86 | 8.52 | 0.00 | 0.18 | 0.16 | 5019 | 2.25 | 1.61 | -0.67 | 0.08 | 387 | -0.67 | 0.05 | 24 | 7.77 | 8.35 | 0.00 | 0.18 | 5.9 | G | 1 |
| hr3681 | S | 4500 | 2.38 | 1.31 | 0.13 | 0.12 | 380 | 0.27 | 0.16 | 27 | 8.32 | 8.81 | 0.08 | 1.09 | 0.29 | 4500 | 2.03 | 1.45 | 0.03 | 0.12 | 380 | 0.03 | 0.17 | 27 | 8.20 | 8.66 | 0.08 | 0.32 | 4.4 | G | 1 |
| hr3687 | S | 4665 | 2.44 | 1.28 | -0.32 | 0.07 | 397 | -0.26 | 0.11 | 28 | 8.00 | 8.61 | -0.16 | 0.36 | 0.26 | 4665 | 2.31 | 1.33 | -0.34 | 0.07 | 397 | -0.34 | 0.11 | 28 | 7.96 | 8.55 | -0.16 | 0.27 | 4.1 | G | 1 |
| hr3706 | S | 5003 | 2.49 | 1.59 | -0.05 | 0.08 | 402 | 0.00 | 0.09 | 29 | 8.13 | 8.65 | 1.04 | 1.87 | 0.17 | 5003 | 2.38 | 1.61 | -0.06 | 0.08 | 402 | -0.06 | 0.09 | 29 | 8.10 | 8.59 | 1.04 | 0.17 | 5.2 | G | 1 |
| hr3707 | S | 4666 | 2.71 | 1.13 | 0.22 | 0.10 | 389 | 0.31 | 0.17 | 24 | 8.49 | 8.91 | 0.38 | 1.21 | 0.23 | 4666 | 2.59 | 1.09 | 0.17 | 0.10 | 389 | 0.17 | 0.18 | 24 | 8.36 | 8.76 | 0.38 | 0.23 | 3.9 | G | 1 |
| hr3709 | S | 4965 | 2.74 | 1.34 | 0.02 | 0.08 | 406 | 0.06 | 0.12 | 28 | 8.17 | 8.58 | 0.92 | 1.62 | 0.16 | 4965 | 2.61 | 1.37 | 0.01 | 0.08 | 406 | 0.01 | 0.12 | 28 | 8.15 | 8.54 | 0.92 | 0.17 | 4.8 | G | 1 |
| hr3731 | S | 4402 | 2.17 | 1.43 | 0.02 | 0.12 | 375 | 0.14 | 0.20 | 23 | 8.36 | 8.83 | 0.24 | 2.18 | 0.32 | 4402 | 1.87 | 1.51 | -0.06 | 0.12 | 375 | -0.06 | 0.20 | 23 | 8.25 | 8.70 | 0.24 | 0.34 | 4.4 | G | 1 |
| hr3733 | S | 4959 | 2.79 | 1.31 | -0.10 | 0.07 | 410 | 0.02 | 0.12 | 29 | 8.13 | 8.62 | 0.43 | 0.55 | 0.16 | 4959 | 2.52 | 1.40 | -0.14 | 0.08 | 410 | -0.14 | 0.12 | 29 | 8.06 | 8.49 | 0.43 | 0.17 | 4.6 | G | 1 |
| hr3748 | S | 4097 | 1.33 | 1.91 | 0.02 | 0.16 | 332 | 0.34 | 0.26 | 24 | 8.23 | 8.69 | 0.22 | 6.10 | 0.40 | 4097 | 0.52 | 1.88 | -0.16 | 0.17 | 332 | -0.16 | 0.27 | 24 | 7.98 | 8.35 | 0.16 | 0.55 | 5.7 | G | 1 |
| hr3772 | S | 4426 | 2.56 | 1.19 | 0.33 | 0.15 | 365 | 0.58 | 0.22 | 26 | 8.65 | 8.96 | -0.05 | 1.04 | 0.29 | 4426 | 1.93 | 1.43 | 0.12 | 0.15 | 365 | 0.12 | 0.24 | 26 | 8.41 | 8.67 | -0.06 | 0.33 | 4.2 | G | 1 |
| hr3788 | S | 4419 | 2.31 | 1.23 | -0.05 | 0.11 | 388 | 0.15 | 0.16 | 27 | 8.35 | 8.83 | -0.08 | 1.00 | 0.30 | 4419 | 1.81 | 1.39 | -0.19 | 0.12 | 388 | -0.19 | 0.17 | 27 | 8.18 | 8.61 | -0.09 | 0.35 | 4.0 | G | 1 |
| hr3791 | S | 4402 | 2.12 | 1.40 | 0.13 | 0.12 | 367 | 0.22 | 0.22 | 26 | 8.32 | 8.82 | -0.05 | 1.13 | 0.32 | 4402 | 1.87 | 1.47 | 0.06 | 0.12 | 367 | 0.05 | 0.22 | 26 | 8.24 | 8.70 | -0.06 | 0.34 | 4.3 | G | 1 |
| hr3800 | S | 5003 | 2.81 | 1.43 | 0.00 | 0.10 | 391 | 0.11 | 0.13 | 32 | 8.09 | 8.58 | 0.52 | 0.60 | 0.15 | 5003 | 2.56 | 1.51 | -0.03 | 0.10 | 391 | -0.03 | 0.13 | 32 | 8.02 | 8.46 | 0.53 | 0.16 | 6.5 | G | 1 |
| hr3801 | S | 4861 | 2.62 | 1.37 | -0.11 | 0.07 | 407 | 0.01 | 0.10 | 28 | 8.15 | 8.72 | 0.46 | 0.78 | 0.19 | 4861 | 2.35 | 1.43 | -0.14 | 0.07 | 407 | -0.14 | 0.10 | 28 | 8.08 | 8.59 | 0.46 | 0.20 | 4.1 | G | 1 |
| hr3805 | S | 4490 | 2.35 | 1.28 | 0.13 | 0.12 | 383 | 0.27 | 0.20 | 26 | 8.28 | 8.83 | -0.17 | 0.63 | 0.20 | 4490 | 2.01 | 1.42 | 0.02 | 0.12 | 383 | 0.03 | 0.20 | 26 | 8.17 | 8.68 | -0.18 | 0.32 | 4.3 | G | 1 |
| hr3808 | S | 4883 | 2.46 | 1.72 | 0.23 | 0.12 | 369 | 0.29 | 0.20 | 29 | 8.39 | 8.80 | 1.31 | 4.51 | 0.19 | 4883 | 2.34 | 1.73 | 0.22 | 0.12 | 369 | 0.22 | 0.20 | 29 | 8.36 | 8.74 | 1.31 | 0.20 | 6.2 | G | 1 |
| hr3809 | S | 4808 | 2.55 | 1.36 | -0.12 | 0.07 | 405 | -0.03 | 0.08 | 28 | 8.14 | 8.67 | 0.45 | 0.89 | 0.20 | 4808 | 2.36 | 1.41 | -0.15 | 0.08 | 405 | -0.14 | 0.09 | 28 | 8.08 | 8.58 | 0.45 | 0.21 | 4.3 | G | 1 |
| hr3827 | S | 4708 | 2.47 | 1.28 | 0.02 | 0.09 | 396 | 0.05 | 0.14 | 27 | 8.19 | 8.71 | 1.31 | 7.19 | 0.24 | 4708 | 2.40 | 1.31 | 0.01 | 0.09 | 396 | 0.01 | 0.14 | 27 | 8.17 | 8.68 | 1.31 | 0.25 | 4.5 | G | 1 |
| hr3834 | S | 4188 | 1.71 | 1.47 | -0.28 | 0.13 | 371 | -0.08 | 0.24 | 26 | 8.13 | 8.59 | -0.19 | 1.82 | 0.36 | 4188 | 1.16 | 1.57 | -0.42 | 0.14 | 371 | -0.42 | 0.25 | 26 | 7.93 | 8.34 | -0.22 | 0.43 | 4.3 | G | 1 |
| hr3845 | S | 4244 | 1.71 | 1.58 | -0.05 | 0.13 | 357 | 0.12 | 0.16 | 24 | 8.18 | 8.72 | 0.71 | 9.34 | 0.36 | 4244 | 1.28 | 1.63 | -0.16 | 0.13 | 357 | -0.16 | 0.17 | 24 | 8.04 | 8.54 | 0.69 | 0.41 | 4.7 | G | 1 |
| hr3851 | S | 4945 | 2.65 | 1.43 | 0.04 | 0.08 | 402 | 0.08 | 0.09 | 25 | 8.18 | 8.62 | 0.57 | 0.79 | 0.17 | 4945 | 2.57 | 1.45 | 0.03 | 0.08 | 402 | 0.03 | 0.09 | 25 | 8.16 | 8.58 | 0.58 | 0.17 | 4.5 | G | 1 |
| hr3903 | S | 4955 | 2.42 | 1.55 | -0.02 | 0.10 | 397 | 0.01 | 0.11 | 30 | 8.09 | 8.54 | 1.29 | 3.58 | 0.18 | 4955 | 2.35 | 1.57 | -0.03 | 0.10 | 397 | -0.03 | 0.11 | 30 | 8.07 | 8.51 | 1.29 | 0.18 | 5.4 | G | 1 |

| ID | | | | | | | | | | | | | | | | | | | | | | | | | | | | | |
|---|---|---|---|---|---|---|---|---|---|---|---|---|---|---|---|---|---|---|---|---|---|---|---|---|---|---|---|---|---|
| hr3905 | S | 4471 | 2.45 | 1.57 | 0.42 | 0.16 | 381 | 0.69 | 0.36 | 28 | 8.70 | 9.07 | 0.41 | 2.44 | 0.29 | 4471 | 1.79 | 1.68 | 0.25 | 0.17 | 381 | 0.25 | 0.36 | 28 | 8.46 | 8.77 | 0.40 | 0.34 | 4.8 | G | 1 |
| hr3907 | S | 4988 | 2.73 | 1.47 | 0.04 | 0.09 | 390 | 0.11 | 0.13 | 30 | 8.12 | 8.65 | 0.61 | 0.75 | 0.16 | 4988 | 2.58 | 1.52 | 0.02 | 0.10 | 390 | 0.02 | 0.13 | 30 | 8.08 | 8.58 | 0.61 | 0.16 | 6.0 | G | 1 |
| hr3908 | S | 4779 | 2.56 | 1.40 | 0.10 | 0.09 | 388 | 0.14 | 0.13 | 32 | 8.18 | 8.75 | 0.42 | 0.92 | 0.21 | 4779 | 2.46 | 1.43 | 0.08 | 0.09 | 388 | 0.08 | 0.14 | 32 | 8.15 | 8.70 | 0.42 | 0.22 | 4.6 | G | 1 |
| hr3911 | S | 4868 | 2.71 | 1.42 | 0.27 | 0.10 | 380 | 0.27 | 0.16 | 28 | 8.41 | 8.89 | 0.49 | 0.82 | 0.18 | 4868 | 2.71 | 1.42 | 0.27 | 0.10 | 380 | 0.27 | 0.16 | 28 | 8.41 | 8.89 | 0.49 | 0.18 | 4.5 | G | 1 |
| hr3929 | S | 4630 | 2.54 | 1.30 | 0.24 | 0.10 | 386 | 0.44 | 0.17 | 29 | 8.47 | 8.90 | 0.49 | 1.72 | 0.25 | 4630 | 2.16 | 1.39 | 0.12 | 0.10 | 386 | 0.12 | 0.18 | 29 | 8.24 | 8.63 | 0.49 | 0.29 | 4.2 | G | 1 |
| hr3942 | S | 4702 | 2.48 | 1.35 | 0.05 | 0.09 | 394 | 0.13 | 0.11 | 28 | 8.26 | 8.74 | 0.53 | 1.49 | 0.25 | 4702 | 2.29 | 1.41 | 0.01 | 0.09 | 394 | 0.01 | 0.12 | 28 | 8.20 | 8.66 | 0.53 | 0.26 | 4.2 | G | 1 |
| hr3973 | S | 4958 | 2.89 | 1.31 | 0.07 | 0.08 | 410 | 0.21 | 0.12 | 29 | 8.21 | 8.68 | 0.88 | 1.51 | 0.16 | 4958 | 2.58 | 1.44 | 0.02 | 0.09 | 410 | 0.02 | 0.13 | 29 | 8.12 | 8.53 | 0.88 | 0.17 | 4.7 | G | 1 |
| hr3994 | S | 4851 | 2.74 | 1.35 | 0.25 | 0.09 | 389 | 0.30 | 0.13 | 26 | 8.45 | 8.90 | 0.39 | 0.68 | 0.18 | 4851 | 2.63 | 1.39 | 0.23 | 0.09 | 389 | 0.23 | 0.14 | 26 | 8.42 | 8.84 | 0.39 | 0.19 | 4.3 | G | 1 |
| hr4006 | S | 5169 | 2.69 | 1.65 | -0.10 | 0.13 | 361 | -0.10 | 0.13 | 29 | 7.84 | 8.54 | 0.30 | 0.23 | 0.13 | 5169 | 2.68 | 1.65 | -0.10 | 0.13 | 361 | -0.11 | 0.13 | 29 | 7.84 | 8.54 | 0.30 | 0.13 | 9.6 | G | 1 |
| hr4032 | S | 4388 | 2.17 | 1.25 | 0.04 | 0.11 | 375 | 0.22 | 0.18 | 27 | 8.36 | 8.74 | -0.45 | 0.48 | 0.32 | 4388 | 1.74 | 1.39 | -0.08 | 0.12 | 375 | -0.08 | 0.19 | 27 | 8.21 | 8.55 | -0.46 | 0.35 | 4.4 | G | 1 |
| hr4052 | S | 4520 | 2.11 | 1.43 | -0.18 | 0.09 | 392 | -0.12 | 0.10 | 22 | 8.13 | 8.70 | -0.20 | 0.54 | 0.31 | 4520 | 1.97 | 1.46 | -0.21 | 0.09 | 392 | -0.21 | 0.09 | 22 | 8.09 | 8.64 | -0.20 | 0.32 | 4.1 | G | 1 |
| hr4057 | S | 4304 | 1.68 | 1.58 | -0.39 | 0.09 | 413 | -0.16 | 0.13 | 26 | 7.93 | 8.51 | -0.60 | 0.48 | 0.36 | 4304 | 1.08 | 1.61 | -0.50 | 0.10 | 413 | -0.50 | 0.13 | 26 | 7.74 | 8.25 | -0.62 | 0.43 | 4.4 | G | 1 |
| hr4077 | S | 4652 | 2.85 | 1.09 | 0.32 | 0.13 | 385 | 0.45 | 0.21 | 24 | 8.60 | 8.99 | 0.26 | 0.96 | 0.22 | 4652 | 2.54 | 1.32 | 0.19 | 0.13 | 385 | 0.20 | 0.22 | 24 | 8.49 | 8.84 | 0.26 | 0.24 | 3.9 | G | 1 |
| hr4078 | S | 4583 | 2.54 | 1.22 | -0.11 | 0.08 | 389 | 0.08 | 0.15 | 27 | 8.30 | 8.76 | 0.04 | 0.74 | 0.26 | 4583 | 2.11 | 1.36 | -0.21 | 0.09 | 389 | -0.21 | 0.16 | 27 | 8.16 | 8.56 | 0.04 | 0.30 | 4.1 | G | 1 |
| hr4085 | S | 4803 | 2.86 | 1.15 | -0.16 | 0.07 | 468 | -0.08 | 0.11 | 29 | 8.24 | 8.74 | 0.56 | 1.17 | 0.19 | 4803 | 2.68 | 1.23 | -0.20 | 0.07 | 468 | -0.20 | 0.11 | 29 | 8.18 | 8.66 | 0.56 | 0.20 | 3.8 | G | 1 |
| hr4097 | S | 4509 | 2.46 | 1.41 | 0.24 | 0.12 | 376 | 0.50 | 0.21 | 26 | 8.58 | 8.96 | 0.09 | 1.08 | 0.28 | 4509 | 1.95 | 1.51 | 0.08 | 0.12 | 376 | 0.08 | 0.23 | 26 | 8.30 | 8.63 | 0.09 | 0.32 | 4.5 | G | 1 |
| hr4100 | S | 4986 | 2.76 | 1.24 | 0.20 | 0.10 | 395 | 0.15 | 0.15 | 29 | 8.34 | 8.82 | 0.58 | 0.71 | 0.16 | 4986 | 2.86 | 1.19 | 0.22 | 0.10 | 395 | 0.22 | 0.15 | 29 | 8.37 | 8.87 | 0.58 | 0.16 | 4.9 | G | 1 |
| hr4106 | S | 4635 | 2.64 | 1.39 | 0.24 | 0.11 | 380 | 0.40 | 0.15 | 25 | 8.62 | 9.03 | 2.11 | 23.68 | 0.24 | 4635 | 2.38 | 1.44 | 0.15 | 0.11 | 380 | 0.15 | 0.16 | 25 | 8.43 | 8.82 | 2.13 | 0.27 | 4.4 | G | 1 |
| hr4126 | S | 4965 | 2.71 | 1.38 | 0.07 | 0.08 | 396 | 0.12 | 0.10 | 26 | 8.18 | 8.67 | 0.71 | 1.01 | 0.16 | 4965 | 2.60 | 1.42 | 0.05 | 0.08 | 396 | 0.05 | 0.10 | 26 | 8.15 | 8.62 | 0.71 | 0.17 | 4.8 | G | 1 |
| hr4146 | S | 4900 | 2.55 | 1.41 | -0.18 | 0.07 | 402 | -0.11 | 0.08 | 28 | 8.06 | 8.57 | -0.05 | 0.22 | 0.18 | 4900 | 2.39 | 1.44 | -0.20 | 0.08 | 402 | -0.20 | 0.08 | 28 | 8.02 | 8.49 | -0.05 | 0.19 | 4.2 | G | 1 |
| hr4171 | S | 4952 | 2.79 | 1.34 | -0.01 | 0.08 | 402 | 0.11 | 0.09 | 27 | 8.21 | 8.74 | 0.39 | 0.51 | 0.16 | 4952 | 2.54 | 1.43 | -0.04 | 0.08 | 402 | -0.05 | 0.10 | 27 | 8.13 | 8.62 | 0.39 | 0.17 | 4.6 | G | 1 |
| hr4178 | S | 4425 | 2.34 | 1.43 | 0.30 | 0.14 | 362 | 0.54 | 0.20 | 24 | 8.64 | 8.99 | 0.04 | 1.28 | 0.30 | 4425 | 1.77 | 1.54 | 0.15 | 0.14 | 362 | 0.15 | 0.21 | 24 | 8.44 | 8.73 | 0.03 | 0.35 | 4.6 | G | 1 |
| hr4208 | S | 4612 | 2.78 | 1.24 | 0.28 | 0.13 | 384 | 0.40 | 0.19 | 25 | 8.60 | 8.95 | 0.37 | 1.43 | 0.23 | 4612 | 2.51 | 1.39 | 0.19 | 0.13 | 384 | 0.19 | 0.21 | 25 | 8.50 | 8.83 | 0.37 | 0.25 | 4.2 | G | 1 |
| hr4209 | S | 4985 | 2.77 | 1.27 | -0.03 | 0.07 | 410 | 0.07 | 0.09 | 32 | 8.15 | 8.63 | 0.31 | 0.39 | 0.16 | 4985 | 2.56 | 1.35 | -0.06 | 0.07 | 410 | -0.06 | 0.10 | 32 | 8.09 | 8.53 | 0.32 | 0.17 | 4.6 | G | 1 |
| hr4232 | S | 4340 | 1.91 | 1.57 | -0.12 | 0.11 | 371 | 0.09 | 0.18 | 27 | 8.25 | 8.76 | -0.36 | 0.73 | 0.34 | 4340 | 1.50 | 1.60 | -0.26 | 0.11 | 371 | -0.26 | 0.21 | 27 | 7.98 | 8.46 | -0.36 | 0.38 | 4.7 | G | 1 |
| hr4233 | S | 4675 | 2.46 | 1.39 | -0.03 | 0.08 | 398 | 0.11 | 0.10 | 29 | 8.27 | 8.78 | 0.00 | 0.50 | 0.25 | 4675 | 2.15 | 1.46 | -0.08 | 0.09 | 398 | -0.08 | 0.11 | 29 | 8.18 | 8.64 | 0.00 | 0.28 | 4.4 | G | 1 |
| hr4235 | S | 4526 | 2.32 | 1.34 | -0.07 | 0.09 | 382 | 0.11 | 0.15 | 28 | 8.20 | 8.72 | 0.82 | 4.87 | 0.25 | 4526 | 1.90 | 1.44 | -0.16 | 0.10 | 382 | -0.16 | 0.15 | 28 | 8.07 | 8.53 | 0.82 | 0.33 | 4.3 | G | 1 |
| hr4242 | S | 4736 | 2.83 | 1.55 | 0.23 | 0.12 | 367 | 0.37 | 0.10 | 22 | 8.64 | 9.04 | 1.09 | 4.44 | 0.20 | 4736 | 2.51 | 1.63 | 0.16 | 0.12 | 367 | 0.16 | 0.11 | 22 | 8.53 | 8.89 | 1.09 | 0.23 | 5.4 | G | 1 |
| hr4253 | S | 4922 | 2.67 | 1.41 | 0.10 | 0.09 | 389 | 0.18 | 0.11 | 27 | 8.23 | 8.71 | 0.75 | 1.24 | 0.17 | 4922 | 2.51 | 1.47 | 0.08 | 0.09 | 389 | 0.08 | 0.11 | 27 | 8.19 | 8.64 | 0.75 | 0.18 | 4.6 | G | 1 |
| hr4255 | S | 5193 | 2.63 | 1.61 | -0.12 | 0.07 | 407 | -0.06 | 0.07 | 28 | 7.97 | 8.60 | 0.21 | 0.18 | 0.13 | 5193 | 2.51 | 1.63 | -0.13 | 0.08 | 407 | -0.12 | 0.08 | 28 | 7.95 | 8.55 | 0.21 | 0.13 | 4.8 | G | 1 |
| hr4256 | S | 4673 | 2.54 | 1.27 | -0.12 | 0.08 | 394 | 0.03 | 0.11 | 26 | 8.21 | 8.68 | 0.06 | 0.58 | 0.24 | 4673 | 2.21 | 1.38 | -0.19 | 0.09 | 394 | -0.19 | 0.11 | 26 | 8.11 | 8.53 | 0.07 | 0.27 | 3.9 | G | 1 |
| hr4258 | S | 4569 | 2.39 | 1.31 | -0.07 | 0.09 | 408 | 0.05 | 0.12 | 23 | 8.23 | 8.73 | -0.08 | 0.59 | 0.28 | 4569 | 2.13 | 1.39 | -0.13 | 0.10 | 408 | -0.13 | 0.12 | 23 | 8.15 | 8.61 | -0.08 | 0.30 | 4.1 | G | 1 |
| hr4264 | S | 4524 | 2.52 | 1.20 | -0.01 | 0.10 | 400 | 0.15 | 0.15 | 27 | 8.29 | 8.73 | -0.01 | 0.82 | 0.27 | 4524 | 2.13 | 1.36 | -0.12 | 0.10 | 400 | -0.12 | 0.15 | 27 | 8.16 | 8.55 | -0.01 | 0.31 | 4.0 | G | 1 |
| hr4283 | S | 4892 | 2.53 | 1.53 | 0.21 | 0.09 | 381 | 0.25 | 0.13 | 27 | 8.36 | 8.82 | 0.77 | 1.39 | 0.19 | 4892 | 2.42 | 1.55 | 0.19 | 0.09 | 381 | 0.19 | 0.13 | 27 | 8.33 | 8.76 | 0.77 | 0.19 | 4.7 | G | 1 |
| hr4287 | S | 4645 | 2.42 | 1.37 | 0.04 | 0.09 | 389 | 0.12 | 0.13 | 26 | 8.28 | 8.76 | -0.10 | 0.44 | 0.26 | 4645 | 2.23 | 1.43 | 0.00 | 0.09 | 389 | 0.00 | 0.14 | 26 | 8.22 | 8.67 | -0.10 | 0.28 | 4.4 | G | 1 |
| hr4301 | S | 4636 | 1.96 | 1.57 | -0.05 | 0.10 | 383 | -0.05 | 0.14 | 27 | 8.12 | 8.68 | 1.01 | 4.99 | 0.30 | 4636 | 1.94 | 1.57 | -0.06 | 0.10 | 383 | -0.06 | 0.14 | 27 | 8.11 | 8.67 | 1.01 | 0.30 | 4.9 | G | 1 |
| hr4305 | S | 4967 | 2.67 | 1.36 | -0.01 | 0.07 | 397 | -0.03 | 0.08 | 27 | 8.03 | 8.74 | 0.43 | 0.54 | 0.17 | 4967 | 2.70 | 1.35 | -0.01 | 0.07 | 397 | -0.01 | 0.08 | 27 | 8.04 | 8.76 | 0.43 | 0.16 | 4.1 | G | 1 |
| hr4335 | S | 4523 | 2.05 | 1.50 | -0.01 | 0.10 | 381 | 0.06 | 0.12 | 28 | 8.22 | 8.74 | 1.16 | 9.21 | 0.31 | 4523 | 1.89 | 1.53 | -0.04 | 0.10 | 381 | -0.04 | 0.12 | 28 | 8.17 | 8.67 | 1.16 | 0.33 | 4.5 | G | 1 |
| hr4351 | S | 4556 | 2.48 | 1.37 | 0.22 | 0.11 | 377 | 0.41 | 0.20 | 25 | 8.52 | 8.93 | 0.42 | 1.89 | 0.27 | 4556 | 2.16 | 1.43 | 0.11 | 0.11 | 377 | 0.11 | 0.21 | 25 | 8.31 | 8.69 | 0.42 | 0.30 | 4.3 | G | 1 |
| hr4382 | S | 4510 | 1.99 | 1.55 | -0.45 | 0.08 | 398 | -0.34 | 0.09 | 25 | 7.89 | 8.53 | -0.25 | 0.49 | 0.32 | 4510 | 1.72 | 1.58 | -0.49 | 0.08 | 398 | -0.49 | 0.09 | 25 | 7.80 | 8.41 | -0.26 | 0.34 | 4.4 | G | 1 |
| hr4383 | S | 4825 | 2.57 | 1.31 | 0.20 | 0.11 | 382 | 0.14 | 0.16 | 24 | 8.41 | 8.95 | 0.35 | 0.68 | 0.20 | 4825 | 2.70 | 1.25 | 0.23 | 0.11 | 382 | 0.23 | 0.15 | 24 | 8.45 | 9.02 | 0.35 | 0.19 | 5.0 | G | 1 |
| hr4400 | S | 4984 | 2.65 | 1.50 | -0.05 | 0.08 | 400 | 0.04 | 0.09 | 32 | 8.10 | 8.64 | 0.75 | 1.05 | 0.16 | 4984 | 2.46 | 1.54 | -0.06 | 0.08 | 400 | -0.06 | 0.10 | 32 | 8.05 | 8.55 | 0.75 | 0.17 | 4.9 | G | 1 |
| hr4407 | S | 4836 | 2.59 | 1.35 | -0.09 | 0.08 | 418 | 0.01 | 0.10 | 29 | 8.15 | 8.71 | 0.43 | 0.79 | 0.19 | 4836 | 2.38 | 1.41 | -0.12 | 0.08 | 418 | -0.12 | 0.10 | 29 | 8.08 | 8.61 | 0.43 | 0.21 | 4.6 | G | 1 |
| hr4419 | S | 4638 | 2.29 | 1.31 | -0.03 | 0.11 | 388 | -0.03 | 0.14 | 24 | 8.23 | 8.81 | 1.10 | 5.94 | 0.27 | 4638 | 2.30 | 1.31 | -0.03 | 0.11 | 388 | -0.03 | 0.13 | 24 | 8.24 | 8.82 | 1.10 | 0.27 | 4.8 | G | 1 |
| hr4433 | S | 4678 | 2.38 | 1.34 | -0.11 | 0.08 | 410 | -0.07 | 0.11 | 26 | 8.11 | 8.63 | 1.01 | 4.41 | 0.26 | 4678 | 2.29 | 1.36 | -0.12 | 0.08 | 410 | -0.12 | 0.11 | 26 | 8.09 | 8.59 | 1.01 | 0.27 | 4.1 | G | 1 |
| hr4452 | S | 4648 | 2.36 | 1.37 | -0.10 | 0.09 | 399 | 0.01 | 0.12 | 27 | 8.37 | 8.79 | 0.10 | 0.69 | 0.27 | 4648 | 2.13 | 1.42 | -0.14 | 0.09 | 399 | -0.14 | 0.12 | 27 | 8.30 | 8.69 | 0.11 | 0.28 | 4.3 | G | 1 |
| hr4459 | S | 4814 | 2.69 | 1.30 | 0.10 | 0.08 | 388 | 0.20 | 0.12 | 28 | 8.29 | 8.77 | 0.73 | 1.66 | 0.19 | 4814 | 2.48 | 1.39 | 0.06 | 0.08 | 388 | 0.06 | 0.13 | 28 | 8.22 | 8.67 | 0.74 | 0.21 | 4.5 | G | 1 |
| hr4461 | S | 4802 | 2.54 | 1.34 | -0.29 | 0.08 | 417 | -0.20 | 0.07 | 24 | 8.03 | 8.54 | 0.07 | 0.40 | 0.20 | 4802 | 2.33 | 1.41 | -0.32 | 0.08 | 417 | -0.32 | 0.08 | 24 | 7.96 | 8.44 | 0.08 | 0.22 | 4.3 | G | 1 |
| hr4471 | S | 4786 | 2.47 | 1.36 | -0.14 | 0.07 | 411 | -0.03 | 0.10 | 27 | 8.10 | 8.63 | 0.20 | 0.55 | 0.21 | 4786 | 2.22 | 1.41 | -0.17 | 0.08 | 411 | -0.17 | 0.10 | 27 | 8.04 | 8.51 | 0.20 | 0.23 | 4.1 | G | 1 |
| hr4474 | S | 4718 | 2.42 | 1.41 | -0.12 | 0.15 | 376 | 0.10 | 0.13 | 24 | 8.44 | 8.75 | 0.45 | 1.19 | 0.25 | 4718 | 1.93 | 1.50 | -0.19 | 0.16 | 376 | -0.19 | 0.13 | 24 | 8.30 | 8.51 | 0.46 | 0.29 | 5.8 | G | 1 |

| | | | | | | | | | | | | | | | | | | | | | | | | | | |
|---|---|---|---|---|---|---|---|---|---|---|---|---|---|---|---|---|---|---|---|---|---|---|---|---|---|---|
| hr4478 | S | 4691 | 2.60 | 1.39 | 0.25 | 0.11 | 371 | 0.41 | 0.16 | 29 | 8.42 | 8.91 | 0.65 | 2.01 | 0.23 | 4691 | 2.24 | 1.49 | 0.17 | 0.11 | 371 | 0.17 | 0.16 | 29 | 8.31 | 8.75 | 0.65 | 0.27 | 5.0 | G | 1 |
| hr4480 | S | 6672 | 3.52 | 4.72 | -0.08 | 0.14 | 115 | -0.04 | 0.06 | 12 | 8.28 | 8.69 | 2.39 | 1.46 | 0.04 | 6672 | 3.41 | 4.70 | -0.08 | 0.14 | 115 | -0.08 | 0.06 | 12 | 8.25 | 8.66 | 2.39 | 0.05 | 34.0 | R | 2 |
| hr4495 | S | 4863 | 2.61 | 1.36 | 0.04 | 0.09 | 394 | 0.06 | 0.13 | 29 | 8.12 | 8.61 | 0.48 | 0.81 | 0.19 | 4863 | 2.56 | 1.37 | 0.03 | 0.09 | 394 | 0.03 | 0.14 | 29 | 8.11 | 8.59 | 0.48 | 0.19 | 4.0 | G | 1 |
| hr4510 | S | 4914 | 2.69 | 1.31 | -0.02 | 0.07 | 409 | 0.04 | 0.09 | 29 | 8.12 | 8.65 | 0.82 | 1.49 | 0.17 | 4914 | 2.55 | 1.36 | -0.04 | 0.07 | 409 | -0.04 | 0.09 | 29 | 8.08 | 8.59 | 0.82 | 0.18 | 4.2 | G | 1 |
| hr4518 | S | 4392 | 1.95 | 1.56 | -0.36 | 0.09 | 379 | -0.19 | 0.09 | 24 | 8.06 | 8.65 | -0.41 | 0.52 | 0.34 | 4392 | 1.54 | 1.61 | -0.44 | 0.10 | 379 | -0.44 | 0.09 | 24 | 7.92 | 8.47 | -0.42 | 0.37 | 4.4 | G | 1 |
| hr4521 | S | 4411 | 2.42 | 1.69 | 0.42 | 0.19 | 350 | 0.69 | 0.29 | 25 | 8.76 | 9.09 | 0.22 | 1.97 | 0.30 | 4411 | 1.75 | 1.80 | 0.24 | 0.19 | 350 | 0.24 | 0.30 | 25 | 8.50 | 8.78 | 0.21 | 0.35 | 4.9 | G | 1 |
| hr4544 | S | 4668 | 2.50 | 1.22 | -0.07 | 0.08 | 405 | 0.01 | 0.13 | 26 | 8.18 | 8.70 | 0.40 | 1.26 | 0.25 | 4668 | 2.34 | 1.29 | -0.11 | 0.08 | 405 | -0.10 | 0.13 | 26 | 8.13 | 8.62 | 0.40 | 0.26 | 3.9 | G | 1 |
| hr4558 | S | 4970 | 2.70 | 1.45 | -0.53 | 0.06 | 407 | -0.38 | 0.06 | 24 | 7.84 | 8.50 | -0.14 | 0.15 | 0.16 | 4970 | 2.39 | 1.52 | -0.55 | 0.07 | 407 | -0.55 | 0.06 | 24 | 7.76 | 8.35 | -0.14 | 0.18 | 4.2 | G | 1 |
| hr4566 | S | 4565 | 2.45 | 1.42 | 0.04 | 0.12 | 387 | 0.22 | 0.16 | 25 | 8.45 | 8.89 | 0.41 | 1.81 | 0.27 | 4565 | 2.00 | 1.53 | -0.06 | 0.12 | 387 | -0.06 | 0.17 | 25 | 8.30 | 8.68 | 0.41 | 0.31 | 5.3 | G | 1 |
| hr4593 | S | 4670 | 2.40 | 1.36 | -0.06 | 0.09 | 397 | 0.05 | 0.10 | 26 | 8.30 | 8.73 | 0.20 | 0.79 | 0.26 | 4670 | 2.16 | 1.43 | -0.10 | 0.10 | 397 | -0.10 | 0.11 | 26 | 8.22 | 8.62 | 0.20 | 0.28 | 4.5 | G | 1 |
| hr4608 | S | 4762 | 2.46 | 1.33 | -0.41 | 0.09 | 409 | -0.25 | 0.06 | 23 | 8.01 | 8.45 | -0.02 | 0.37 | 0.22 | 4762 | 2.10 | 1.42 | -0.46 | 0.10 | 409 | -0.46 | 0.06 | 23 | 7.90 | 8.28 | -0.01 | 0.24 | 4.5 | G | 1 |
| hr4609 | S | 4677 | 2.50 | 1.43 | -0.27 | 0.09 | 392 | -0.14 | 0.13 | 27 | 8.23 | 8.79 | 0.03 | 0.53 | 0.25 | 4677 | 2.20 | 1.52 | -0.32 | 0.09 | 392 | -0.32 | 0.14 | 27 | 8.12 | 8.65 | 0.03 | 0.27 | 4.8 | G | 1 |
| hr4610 | S | 4433 | 2.30 | 1.34 | -0.04 | 0.11 | 376 | 0.14 | 0.18 | 25 | 8.38 | 8.85 | -0.27 | 0.62 | 0.30 | 4433 | 1.87 | 1.46 | -0.15 | 0.12 | 376 | -0.15 | 0.19 | 25 | 8.23 | 8.66 | -0.28 | 0.34 | 4.2 | G | 2 |
| hr4626 | S | 4553 | 2.71 | 1.12 | 0.20 | 0.12 | 393 | 0.48 | 0.20 | 26 | 8.54 | 8.90 | 0.00 | 0.75 | 0.25 | 4553 | 2.16 | 1.34 | 0.00 | 0.12 | 393 | 0.00 | 0.22 | 26 | 8.25 | 8.55 | 0.00 | 0.30 | 4.0 | G | 1 |
| hr4630 | S | 4285 | 1.22 | 1.98 | -0.01 | 0.14 | 331 | 0.00 | 0.15 | 22 | 8.17 | 8.55 | 0.62 | 7.41 | 0.42 | 4285 | 1.18 | 1.97 | -0.02 | 0.15 | 331 | -0.02 | 0.15 | 22 | 8.16 | 8.53 | 0.62 | 0.42 | 6.0 | G | 1 |
| hr4654 | S | 4771 | 2.46 | 1.42 | -0.01 | 0.08 | 388 | -0.01 | 0.10 | 28 | 8.19 | 8.76 | 0.32 | 0.75 | 0.22 | 4771 | 2.46 | 1.41 | -0.01 | 0.08 | 388 | -0.01 | 0.10 | 28 | 8.19 | 8.76 | 0.32 | 0.22 | 4.4 | G | 1 |
| hr4655 | S | 4765 | 2.77 | 1.32 | 0.30 | 0.11 | 372 | 0.48 | 0.16 | 26 | 8.46 | 8.97 | 0.52 | 1.21 | 0.20 | 4765 | 2.37 | 1.47 | 0.21 | 0.11 | 372 | 0.21 | 0.16 | 26 | 8.34 | 8.78 | 0.53 | 0.22 | 4.5 | G | 1 |
| hr4667 | S | 4861 | 2.63 | 1.35 | -0.13 | 0.07 | 399 | 0.00 | 0.08 | 31 | 8.13 | 8.67 | 0.27 | 0.51 | 0.19 | 4861 | 2.34 | 1.42 | -0.17 | 0.07 | 399 | -0.17 | 0.08 | 31 | 8.05 | 8.53 | 0.28 | 0.20 | 4.2 | G | 1 |
| hr4668 | S | 4463 | 2.08 | 1.39 | -0.22 | 0.09 | 384 | -0.16 | 0.14 | 25 | 8.07 | 8.58 | -0.15 | 0.72 | 0.32 | 4463 | 1.93 | 1.44 | -0.26 | 0.09 | 384 | -0.26 | 0.14 | 25 | 8.01 | 8.51 | -0.16 | 0.33 | 4.3 | G | 1 |
| hr4695 | S | 4422 | 2.06 | 1.71 | -0.39 | 0.09 | 376 | -0.24 | 0.12 | 23 | 8.18 | 8.81 | -0.65 | 0.27 | 0.32 | 4422 | 1.70 | 1.75 | -0.45 | 0.10 | 376 | -0.45 | 0.12 | 23 | 8.05 | 8.65 | -0.65 | 0.36 | 4.7 | G | 1 |
| hr4697 | S | 4711 | 2.27 | 1.43 | -0.35 | 0.08 | 406 | -0.31 | 0.10 | 26 | 7.89 | 8.47 | 0.36 | 1.01 | 0.26 | 4711 | 2.18 | 1.45 | -0.36 | 0.08 | 406 | -0.36 | 0.10 | 26 | 7.86 | 8.43 | 0.36 | 0.27 | 4.2 | G | 1 |
| hr4699 | S | 4724 | 2.74 | 1.20 | 0.06 | 0.09 | 390 | 0.13 | 0.13 | 26 | 8.22 | 8.73 | 0.95 | 3.46 | 0.21 | 4724 | 2.59 | 1.29 | 0.02 | 0.09 | 390 | 0.02 | 0.13 | 26 | 8.17 | 8.66 | 0.95 | 0.22 | 3.9 | G | 1 |
| hr4728 | S | 4938 | 2.64 | 1.40 | 0.09 | 0.08 | 394 | 0.10 | 0.11 | 32 | 8.17 | 8.63 | 0.55 | 0.76 | 0.17 | 4938 | 2.61 | 1.41 | 0.08 | 0.08 | 394 | 0.08 | 0.11 | 32 | 8.16 | 8.61 | 0.55 | 0.17 | 4.6 | G | 1 |
| hr4737 | S | 4634 | 2.53 | 1.49 | 0.31 | 0.12 | 369 | 0.45 | 0.23 | 26 | 8.53 | 8.97 | 0.49 | 1.68 | 0.25 | 4634 | 2.21 | 1.56 | 0.24 | 0.12 | 369 | 0.24 | 0.23 | 26 | 8.42 | 8.83 | 0.49 | 0.28 | 4.8 | G | 1 |
| hr4753 | S | 6388 | 3.56 | 6.50 | 0.24 | 0.65 | 23 | -0.10 | 0.21 | 3 | 6.93 | 9.76 | 2.83 | 5.58 | 0.02 | 6388 | 4.43 | 7.75 | 0.22 | 0.62 | 23 | 0.21 | 0.22 | 3 | 7.19 | 10.09 | 2.81 | -0.09 | 60.0 | R | 2 |
| hr4772 | S | 4719 | 2.50 | 1.46 | -0.16 | 0.08 | 394 | -0.04 | 0.08 | 26 | 8.24 | 8.77 | -0.40 | 0.17 | 0.24 | 4719 | 2.23 | 1.51 | -0.20 | 0.08 | 394 | -0.20 | 0.08 | 26 | 8.16 | 8.65 | -0.40 | 0.26 | 4.4 | G | 1 |
| hr4777 | S | 5003 | 2.71 | 1.38 | 0.12 | 0.08 | 404 | 0.10 | 0.11 | 28 | 8.13 | 8.66 | 0.64 | 0.78 | 0.16 | 5003 | 2.75 | 1.37 | 0.12 | 0.08 | 404 | 0.12 | 0.11 | 28 | 8.14 | 8.68 | 0.64 | 0.16 | 4.4 | G | 1 |
| hr4783 | S | 4784 | 2.56 | 1.40 | -0.02 | 0.08 | 396 | 0.08 | 0.11 | 32 | 8.21 | 8.75 | 0.28 | 0.66 | 0.21 | 4784 | 2.34 | 1.46 | -0.06 | 0.08 | 396 | -0.06 | 0.11 | 32 | 8.15 | 8.65 | 0.28 | 0.22 | 4.3 | G | 1 |
| hr4784 | S | 4705 | 2.49 | 1.40 | -0.02 | 0.08 | 385 | 0.08 | 0.11 | 26 | 8.23 | 8.76 | 0.20 | 0.71 | 0.24 | 4705 | 2.26 | 1.46 | -0.06 | 0.08 | 385 | -0.06 | 0.11 | 26 | 8.16 | 8.65 | 0.20 | 0.26 | 4.2 | G | 1 |
| hr4786 | S | 5090 | 2.41 | 1.82 | -0.01 | 0.10 | 382 | 0.06 | 0.11 | 30 | 8.13 | 8.54 | 0.96 | 1.25 | 0.16 | 5090 | 2.26 | 1.84 | -0.02 | 0.10 | 382 | -0.02 | 0.11 | 30 | 8.10 | 8.46 | 0.96 | 0.16 | 6.7 | G | 1 |
| hr4812 | S | 4719 | 2.65 | 1.35 | 0.21 | 0.09 | 391 | 0.28 | 0.16 | 29 | 8.38 | 8.89 | 0.36 | 0.98 | 0.22 | 4719 | 2.47 | 1.43 | 0.16 | 0.09 | 391 | 0.16 | 0.16 | 29 | 8.33 | 8.81 | 0.37 | 0.24 | 4.2 | G | 1 |
| hr4813 | S | 4404 | 2.04 | 1.71 | 0.33 | 0.15 | 340 | 0.52 | 0.26 | 24 | 8.57 | 8.99 | 0.17 | 1.83 | 0.33 | 4404 | 1.58 | 1.74 | 0.22 | 0.16 | 340 | 0.22 | 0.26 | 24 | 8.42 | 8.79 | 0.16 | 0.37 | 5.3 | G | 1 |
| hr4815 | S | 4943 | 2.78 | 1.35 | 0.12 | 0.09 | 415 | 0.10 | 0.11 | 26 | 8.23 | 8.70 | 0.71 | 1.08 | 0.17 | 4943 | 2.71 | 1.42 | 0.12 | 0.09 | 415 | 0.12 | 0.11 | 26 | 8.29 | 8.76 | 0.72 | 0.17 | 4.7 | G | 1 |
| hr4840 | S | 4372 | 2.10 | 1.44 | 0.14 | 0.13 | 352 | 0.26 | 0.22 | 25 | 8.36 | 8.82 | 0.01 | 1.45 | 0.32 | 4372 | 1.82 | 1.50 | 0.07 | 0.13 | 352 | 0.07 | 0.22 | 25 | 8.26 | 8.70 | 0.00 | 0.35 | 4.4 | G | 1 |
| hr4851 | S | 4190 | 1.82 | 1.45 | 0.03 | 0.15 | 357 | 0.32 | 0.26 | 26 | 8.46 | 8.86 | -0.60 | 0.73 | 0.35 | 4190 | 1.09 | 1.55 | -0.17 | 0.15 | 357 | -0.17 | 0.27 | 26 | 8.19 | 8.53 | -0.63 | 0.44 | 4.6 | G | 1 |
| hr4860 | S | 4814 | 2.52 | 1.39 | -0.42 | 0.07 | 425 | -0.29 | 0.05 | 26 | 7.93 | 8.49 | -0.08 | 0.27 | 0.20 | 4814 | 2.22 | 1.46 | -0.45 | 0.08 | 425 | -0.45 | 0.06 | 26 | 7.84 | 8.34 | -0.08 | 0.22 | 4.1 | G | 1 |
| hr4873 | S | 4808 | 2.79 | 1.16 | -0.05 | 0.07 | 411 | 0.06 | 0.11 | 28 | 8.23 | 8.75 | 1.12 | 3.86 | 0.19 | 4808 | 2.56 | 1.27 | -0.10 | 0.07 | 411 | -0.10 | 0.12 | 28 | 8.15 | 8.64 | 1.13 | 0.20 | 3.9 | G | 1 |
| hr4877 | S | 4744 | 2.46 | 1.40 | -0.14 | 0.08 | 410 | -0.01 | 0.09 | 26 | 8.22 | 8.71 | 0.23 | 0.66 | 0.24 | 4744 | 2.15 | 1.47 | -0.19 | 0.08 | 410 | -0.19 | 0.10 | 26 | 8.14 | 8.57 | 0.23 | 0.26 | 4.1 | G | 1 |
| hr4883 | S | 5564 | 2.90 | 2.86 | 0.16 | 0.40 | 84 | 0.10 | 0.28 | 5 | | | 2.63 | 12.30 | 0.07 | 5564 | 3.04 | 2.85 | 0.17 | 0.40 | 84 | 0.17 | 0.28 | 5 | | | 2.63 | 0.06 | 64.0 | R | 1 |
| hr4894 | S | 4982 | 2.62 | 1.48 | 0.23 | 0.11 | 387 | 0.21 | 0.14 | 28 | 8.37 | 8.83 | 1.17 | 2.60 | 0.16 | 4982 | 2.67 | 1.46 | 0.24 | 0.11 | 387 | 0.24 | 0.14 | 28 | 8.39 | 8.85 | 1.17 | 0.16 | 5.3 | G | 1 |
| hr4896 | S | 4702 | 2.74 | 1.15 | 0.09 | 0.10 | 395 | 0.12 | 0.17 | 27 | 8.36 | 8.78 | 0.16 | 0.66 | 0.22 | 4702 | 2.67 | 1.20 | 0.06 | 0.10 | 395 | 0.06 | 0.17 | 27 | 8.33 | 8.74 | 0.16 | 0.22 | 3.8 | G | 1 |
| hr4925 | S | 4580 | 2.48 | 1.36 | 0.15 | 0.11 | 382 | 0.28 | 0.18 | 25 | 8.45 | 8.86 | 0.06 | 0.79 | 0.27 | 4580 | 2.18 | 1.46 | 0.08 | 0.11 | 382 | 0.08 | 0.18 | 25 | 8.34 | 8.72 | 0.06 | 0.29 | 4.4 | G | 1 |
| hr4929 | S | 4916 | 2.61 | 1.38 | 0.00 | 0.07 | 396 | -0.01 | 0.10 | 28 | 8.21 | 8.73 | 0.59 | 0.89 | 0.18 | 4916 | 2.62 | 1.37 | 0.00 | 0.07 | 396 | 0.00 | 0.10 | 28 | 8.22 | 8.74 | 0.59 | 0.18 | 4.1 | G | 1 |
| hr4932 | S | 4988 | 2.64 | 1.49 | 0.10 | 0.08 | 400 | 0.15 | 0.11 | 28 | 8.21 | 8.82 | 0.57 | 0.69 | 0.18 | 4988 | 2.52 | 1.52 | 0.09 | 0.09 | 400 | 0.08 | 0.11 | 28 | 8.17 | 8.76 | 0.57 | 0.17 | 4.7 | G | 1 |
| hr4953 | S | 4816 | 2.59 | 1.24 | -0.28 | 0.07 | 396 | -0.34 | 0.09 | 26 | 8.01 | 8.55 | 0.42 | 0.83 | 0.20 | 4816 | 2.72 | 1.17 | -0.26 | 0.08 | 396 | -0.26 | 0.09 | 26 | 8.06 | 8.61 | 0.42 | 0.19 | 3.7 | G | 1 |
| hr4955 | S | 4585 | 2.26 | 1.39 | 0.10 | 0.10 | 389 | 0.12 | 0.14 | 27 | 8.19 | 8.74 | 0.16 | 0.97 | 0.29 | 4585 | 2.20 | 1.41 | 0.08 | 0.10 | 389 | 0.08 | 0.14 | 27 | 8.17 | 8.71 | 0.16 | 0.29 | 4.3 | G | 1 |
| hr4956 | S | 4198 | 2.05 | 1.56 | 0.35 | 0.18 | 334 | 0.71 | 0.30 | 23 | 8.61 | 8.98 | -0.14 | 1.84 | 0.33 | 4198 | 1.15 | 1.68 | 0.08 | 0.19 | 334 | 0.07 | 0.31 | 23 | 8.28 | 8.58 | -0.18 | 0.43 | 4.7 | G | 1 |
| hr4959 | S | 4802 | 2.59 | 1.37 | -0.05 | 0.08 | 387 | 0.06 | 0.10 | 28 | 8.23 | 8.73 | 0.27 | 0.61 | 0.20 | 4802 | 2.34 | 1.44 | -0.09 | 0.08 | 387 | -0.09 | 0.10 | 28 | 8.15 | 8.61 | 0.28 | 0.22 | 4.4 | G | 1 |
| hr4960 | S | 4765 | 2.57 | 1.33 | 0.07 | 0.09 | 399 | 0.13 | 0.10 | 27 | 8.13 | 8.65 | 0.89 | 2.67 | 0.21 | 4765 | 2.44 | 1.37 | 0.04 | 0.09 | 399 | 0.05 | 0.10 | 27 | 8.09 | 8.59 | 0.89 | 0.22 | 4.1 | G | 1 |
| hr4964 | S | 4464 | 2.37 | 1.48 | 0.24 | 0.14 | 362 | 0.41 | 0.24 | 23 | 8.49 | 8.93 | 0.15 | 1.42 | 0.29 | 4464 | 2.06 | 1.52 | 0.12 | 0.14 | 362 | 0.13 | 0.24 | 23 | 8.29 | 8.70 | 0.15 | 0.32 | 4.7 | G | 1 |

| ID | | | | | | | | | | | | | | | | | | | | | | | | | | |
|---|---|---|---|---|---|---|---|---|---|---|---|---|---|---|---|---|---|---|---|---|---|---|---|---|---|---|
| hr4984 | S | 4825 | 2.57 | 1.36 | 0.10 | 0.08 | 400 | 0.09 | 0.11 | 28 | 8.30 | 8.78 | 0.53 | 1.03 | 0.20 | 4825 | 2.58 | 1.35 | 0.10 | 0.08 | 400 | 0.10 | 0.11 | 28 | 8.31 | 8.78 0.53 0.20 | 4.5 G 1 |
| hr4992 | S | 4398 | 2.28 | 1.32 | -0.03 | 0.11 | 373 | 0.19 | 0.18 | 25 | 8.32 | 8.79 | -0.30 | 0.66 | 0.31 | 4398 | 1.77 | 1.46 | -0.17 | 0.12 | 373 | -0.16 | 0.19 | 25 | 8.15 | 8.56 -0.31 0.35 | 4.2 G 1 |
| hr5001 | S | 4741 | 3.06 | 0.71 | 0.13 | 0.11 | 408 | 0.22 | 0.14 | 25 | 8.40 | 8.74 | 0.31 | 0.82 | 0.19 | 4741 | 2.84 | 1.04 | 0.02 | 0.09 | 408 | 0.02 | 0.14 | 25 | 8.31 | 8.63 0.31 0.20 | 3.4 G 1 |
| hr5007 | S | 4807 | 2.84 | 1.17 | -0.16 | 0.07 | 417 | -0.03 | 0.11 | 26 | 8.02 | 8.68 | 0.98 | 2.92 | 0.19 | 4807 | 2.55 | 1.28 | -0.21 | 0.08 | 417 | -0.21 | 0.12 | 26 | 7.93 | 8.55 0.99 0.20 | 3.8 G 1 |
| hr5020 | S | 5019 | 2.60 | 1.58 | 0.02 | 0.09 | 397 | 0.07 | 0.12 | 27 | 8.11 | 8.61 | 1.36 | 3.59 | 0.16 | 5019 | 2.49 | 1.60 | 0.02 | 0.09 | 397 | 0.01 | 0.12 | 27 | 8.09 | 8.56 1.37 0.17 | 5.5 G 1 |
| hr5044 | S | 4820 | 2.53 | 1.40 | -0.06 | 0.09 | 401 | 0.06 | 0.12 | 31 | 8.11 | 8.45 | 1.44 | 6.93 | 0.20 | 4820 | 2.26 | 1.47 | -0.10 | 0.09 | 401 | -0.10 | 0.12 | 31 | 8.03 | 8.32 1.45 0.22 | 4.6 G 1 |
| hr5053 | S | 4643 | 2.27 | 1.44 | -0.11 | 0.10 | 388 | 0.01 | 0.13 | 31 | 8.18 | 8.71 | -0.04 | 0.50 | 0.27 | 4643 | 1.99 | 1.48 | -0.15 | 0.10 | 388 | -0.15 | 0.13 | 31 | 8.09 | 8.58 -0.04 0.30 | 4.9 G 1 |
| hr5067 | S | 4851 | 2.49 | 1.42 | -0.08 | 0.08 | 405 | -0.06 | 0.11 | 27 | 8.10 | 8.61 | 0.62 | 1.15 | 0.20 | 4851 | 2.45 | 1.42 | -0.08 | 0.08 | 405 | -0.08 | 0.11 | 27 | 8.09 | 8.60 0.62 0.20 | 4.2 G 1 |
| hr5068 | S | 4730 | 2.37 | 1.59 | 0.23 | 0.11 | 371 | 0.25 | 0.19 | 25 | 8.38 | 8.85 | 0.67 | 1.85 | 0.25 | 4730 | 2.31 | 1.60 | 0.22 | 0.11 | 371 | 0.22 | 0.19 | 25 | 8.36 | 8.83 0.67 0.25 | 5.0 G 1 |
| hr5081 | S | 4658 | 2.47 | 1.30 | -0.09 | 0.08 | 398 | -0.04 | 0.11 | 25 | 8.21 | 8.66 | 1.17 | 6.41 | 0.25 | 4658 | 2.35 | 1.34 | -0.11 | 0.09 | 398 | -0.11 | 0.11 | 25 | 8.17 | 8.61 1.17 0.26 | 4.4 G 1 |
| hr5100 | S | 4837 | 2.64 | 1.29 | -0.23 | 0.07 | 398 | -0.11 | 0.08 | 27 | 8.04 | 8.58 | 0.35 | 0.67 | 0.20 | 4837 | 2.39 | 1.40 | -0.27 | 0.07 | 398 | -0.27 | 0.08 | 27 | 7.96 | 8.46 0.35 0.21 | 4.0 G 1 |
| hr5111 | S | 4881 | 2.64 | 1.37 | -0.25 | 0.06 | 400 | -0.21 | 0.07 | 25 | 8.00 | 8.62 | 0.25 | 0.47 | 0.18 | 4881 | 2.54 | 1.40 | -0.26 | 0.06 | 400 | -0.26 | 0.07 | 25 | 7.96 | 8.57 0.25 0.19 | 4.1 G 1 |
| hr5126 | S | 4519 | 2.31 | 1.44 | 0.20 | 0.12 | 365 | 0.37 | 0.22 | 25 | 8.47 | 8.92 | 1.28 | 11.34 | 0.29 | 4519 | 2.00 | 1.47 | 0.10 | 0.12 | 365 | 0.10 | 0.23 | 25 | 8.27 | 8.69 1.28 0.32 | 4.6 G 1 |
| hr5143 | S | 4819 | 2.44 | 1.42 | -0.31 | 0.07 | 398 | -0.26 | 0.07 | 25 | 7.99 | 8.56 | 0.21 | 0.51 | 0.21 | 4819 | 2.32 | 1.46 | -0.33 | 0.07 | 398 | -0.33 | 0.07 | 25 | 7.95 | 8.50 0.21 0.21 | 4.1 G 1 |
| hr5149 | S | 4836 | 2.57 | 1.41 | 0.00 | 0.08 | 393 | 0.02 | 0.08 | 26 | 8.20 | 8.80 | 0.43 | 0.79 | 0.19 | 4836 | 2.52 | 1.42 | 0.00 | 0.08 | 393 | -0.01 | 0.08 | 26 | 8.18 | 8.78 0.43 0.20 | 4.3 G 1 |
| hr5161 | S | 5155 | 2.72 | 1.43 | 0.03 | 0.08 | 393 | -0.03 | 0.09 | 28 | 8.08 | 8.67 | 1.35 | 2.52 | 0.13 | 5155 | 2.85 | 1.38 | 0.05 | 0.09 | 393 | 0.05 | 0.09 | 28 | 8.11 | 8.73 1.35 0.13 | 6.1 G 1 |
| hr5180 | S | 4935 | 2.67 | 1.42 | 0.11 | 0.09 | 386 | 0.16 | 0.13 | 29 | 8.22 | 8.65 | 1.30 | 3.93 | 0.17 | 4935 | 2.56 | 1.46 | 0.09 | 0.09 | 386 | 0.09 | 0.13 | 29 | 8.19 | 8.59 1.30 0.18 | 5.2 G 1 |
| hr5186 | S | 4698 | 2.38 | 1.43 | -0.08 | 0.09 | 395 | -0.03 | 0.09 | 24 | 8.21 | 8.72 | 0.60 | 1.76 | 0.25 | 4698 | 2.27 | 1.46 | -0.10 | 0.09 | 395 | -0.10 | 0.09 | 24 | 8.18 | 8.67 0.60 0.26 | 4.2 G 1 |
| hr5195 | S | 4683 | 2.55 | 1.21 | -0.13 | 0.08 | 414 | 0.02 | 0.10 | 27 | 8.17 | 8.69 | 0.43 | 1.28 | 0.23 | 4683 | 2.21 | 1.33 | -0.20 | 0.09 | 414 | -0.20 | 0.11 | 27 | 8.07 | 8.53 0.44 0.27 | 4.1 G 1 |
| hr5196 | S | 4706 | 2.47 | 1.37 | 0.05 | 0.09 | 393 | 0.12 | 0.14 | 27 | 8.25 | 8.74 | 0.31 | 0.91 | 0.25 | 4706 | 2.31 | 1.43 | 0.01 | 0.09 | 393 | 0.01 | 0.14 | 27 | 8.20 | 8.67 0.32 0.26 | 4.3 G 1 |
| hr5205 | S | 5049 | 2.62 | 1.45 | -0.07 | 0.09 | 411 | 0.04 | 0.12 | 30 | 7.94 | 8.56 | 2.05 | 11.51 | 0.15 | 5049 | 2.38 | 1.50 | -0.09 | 0.09 | 411 | -0.09 | 0.12 | 30 | 7.88 | 8.45 2.05 0.16 | 5.7 G 1 |
| hr5213 | S | 4921 | 2.62 | 1.39 | -0.27 | 0.07 | 424 | -0.25 | 0.08 | 27 | 7.97 | 8.58 | 0.20 | 0.37 | 0.18 | 4921 | 2.57 | 1.41 | -0.27 | 0.07 | 424 | -0.27 | 0.08 | 27 | 7.96 | 8.55 0.20 0.18 | 4.2 G 1 |
| hr5227 | S | 4464 | 2.51 | 1.06 | 0.16 | 0.13 | 374 | 0.29 | 0.23 | 26 | 8.55 | 8.98 | -0.57 | 0.28 | 0.28 | 4464 | 2.16 | 1.31 | 0.01 | 0.12 | 374 | 0.01 | 0.23 | 26 | 8.41 | 8.81 -0.58 0.31 | 3.9 G 1 |
| hr5232 | S | 4703 | 2.47 | 1.45 | 0.16 | 0.11 | 388 | 0.17 | 0.18 | 26 | 8.31 | 8.77 | 0.28 | 0.86 | 0.25 | 4703 | 2.33 | 1.53 | 0.15 | 0.11 | 388 | 0.15 | 0.18 | 26 | 8.37 | 8.80 0.29 0.26 | 4.3 G 1 |
| hr5235 | S | 6028 | 3.73 | 1.85 | 0.38 | 0.12 | 268 | 0.34 | 0.12 | 30 | 8.65 | 9.05 | 1.49 | 0.55 | 0.02 | 6028 | 3.74 | 1.83 | 0.39 | 0.12 | 268 | 0.39 | 0.12 | 30 | 8.63 | 9.01 1.50 0.02 | 14.0 R 2 |
| hr5276 | S | 4265 | 2.33 | 1.38 | 0.41 | 0.17 | 338 | 0.80 | 0.33 | 23 | 8.78 | 9.09 | -0.16 | 1.40 | 0.31 | 4265 | 1.36 | 1.61 | 0.09 | 0.19 | 338 | 0.09 | 0.34 | 23 | 8.38 | 8.64 -0.19 0.40 | 4.4 G 1 |
| hr5277 | S | 4684 | 2.51 | 1.35 | 0.13 | 0.11 | 376 | 0.26 | 0.15 | 27 | 8.36 | 8.71 | 0.63 | 1.95 | 0.24 | 4684 | 2.21 | 1.46 | 0.06 | 0.11 | 376 | 0.07 | 0.15 | 27 | 8.27 | 8.57 0.63 0.27 | 4.7 G 1 |
| hr5302 | S | 4770 | 2.71 | 1.38 | -0.10 | 0.08 | 391 | 0.08 | 0.07 | 28 | 8.34 | 8.89 | 0.18 | 0.56 | 0.20 | 4770 | 2.29 | 1.49 | -0.17 | 0.09 | 391 | -0.17 | 0.09 | 28 | 8.21 | 8.70 0.19 0.23 | 4.2 G 1 |
| hr5315 | S | 4155 | 1.62 | 1.59 | -0.36 | 0.13 | 374 | -0.12 | 0.20 | 26 | 8.01 | 8.51 | 0.10 | 3.84 | 0.37 | 4155 | 0.99 | 1.65 | -0.51 | 0.14 | 374 | -0.51 | 0.20 | 26 | 7.79 | 8.24 0.07 0.45 | 4.3 G 1 |
| hr5340 | S | 4281 | 1.74 | 1.62 | -0.49 | 0.10 | 417 | -0.28 | 0.15 | 21 | 8.12 | 8.73 | -1.18 | 0.14 | 0.36 | 4281 | 1.20 | 1.65 | -0.60 | 0.11 | 417 | -0.60 | 0.15 | 21 | 7.94 | 8.49 -1.20 0.42 | 4.4 G 1 |
| hr5344 | S | 4833 | 2.57 | 1.35 | -0.10 | 0.08 | 403 | -0.02 | 0.09 | 29 | 8.12 | 8.64 | 0.31 | 0.61 | 0.20 | 4833 | 2.38 | 1.40 | -0.13 | 0.08 | 403 | -0.13 | 0.09 | 29 | 8.07 | 8.55 0.32 0.21 | 4.2 G 1 |
| hr5366 | S | 4730 | 2.47 | 1.40 | -0.07 | 0.09 | 403 | 0.05 | 0.12 | 27 | 8.18 | 8.66 | 0.19 | 0.64 | 0.24 | 4730 | 2.19 | 1.47 | -0.12 | 0.09 | 403 | -0.12 | 0.12 | 27 | 8.10 | 8.53 0.19 0.26 | 4.4 G 1 |
| hr5370 | S | 4433 | 2.38 | 1.47 | 0.35 | 0.15 | 347 | 0.55 | 0.20 | 24 | 8.66 | 9.03 | 0.75 | 5.53 | 0.30 | 4433 | 1.90 | 1.60 | 0.21 | 0.15 | 347 | 0.21 | 0.21 | 24 | 8.48 | 8.81 0.74 0.34 | 4.5 G 1 |
| hr5383 | S | 4874 | 2.76 | 1.27 | 0.12 | 0.09 | 405 | 0.19 | 0.12 | 29 | 8.24 | 8.69 | 0.48 | 0.79 | 0.18 | 4874 | 2.60 | 1.34 | 0.09 | 0.09 | 405 | 0.09 | 0.12 | 29 | 8.19 | 8.61 0.48 0.19 | 4.5 G 1 |
| hr5394 | S | 4440 | 2.36 | 1.39 | 0.07 | 0.12 | 374 | 0.22 | 0.17 | 24 | 8.38 | 8.88 | -0.15 | 0.79 | 0.30 | 4440 | 1.99 | 1.51 | -0.03 | 0.13 | 374 | -0.03 | 0.18 | 24 | 8.25 | 8.71 -0.16 0.33 | 4.2 G 1 |
| hr5429 | S | 4281 | 1.95 | 1.55 | 0.06 | 0.13 | 366 | 0.31 | 0.23 | 26 | 8.37 | 8.83 | -0.26 | 1.11 | 0.34 | 4281 | 1.31 | 1.61 | -0.10 | 0.13 | 366 | -0.10 | 0.23 | 26 | 8.15 | 8.56 -0.29 0.40 | 4.7 G 1 |
| hr5430 | S | 4095 | 1.45 | 1.76 | -0.01 | 0.16 | 350 | 0.30 | 0.25 | 25 | 8.20 | 8.65 | -0.54 | 1.21 | 0.38 | 4095 | 0.65 | 1.75 | -0.20 | 0.17 | 350 | -0.20 | 0.26 | 25 | 7.95 | 8.32 -0.59 0.52 | 5.2 G 1 |
| hr5454 | S | 4736 | 2.65 | 1.33 | 0.01 | 0.08 | 390 | 0.16 | 0.09 | 25 | 8.27 | 8.81 | 0.30 | 0.80 | 0.22 | 4736 | 2.31 | 1.44 | -0.05 | 0.08 | 390 | -0.05 | 0.10 | 25 | 8.17 | 8.65 0.30 0.25 | 4.1 G 1 |
| hr5481 | S | 4903 | 2.49 | 1.39 | -0.19 | 0.07 | 411 | -0.16 | 0.08 | 26 | 8.13 | 8.77 | 0.29 | 0.47 | 0.19 | 4903 | 2.42 | 1.40 | -0.20 | 0.07 | 411 | -0.20 | 0.08 | 26 | 8.11 | 8.73 0.29 0.19 | 4.3 G 1 |
| hr5487 | S | 6487 | 3.91 | 4.23 | -0.31 | 0.26 | 71 | -0.18 | 0.15 | 3 | 7.89 | 8.85 | 2.77 | 4.36 | -0.02 | 6487 | 3.56 | 4.24 | -0.31 | 0.26 | 71 | -0.31 | 0.16 | 3 | 7.78 | 8.73 2.77 0.03 | 48.0 R 2 |
| hr5502 | S | 4864 | 2.51 | 1.46 | -0.01 | 0.08 | 399 | 0.06 | 0.09 | 26 | 8.13 | 8.67 | 0.64 | 1.15 | 0.19 | 4864 | 2.35 | 1.49 | -0.03 | 0.09 | 399 | -0.03 | 0.09 | 26 | 8.09 | 8.60 0.64 0.20 | 4.8 G 1 |
| hr5518 | S | 4641 | 2.59 | 1.39 | -0.03 | 0.10 | 394 | 0.18 | 0.13 | 28 | 8.46 | 8.95 | 0.10 | 0.70 | 0.24 | 4641 | 2.10 | 1.52 | -0.13 | 0.10 | 394 | -0.13 | 0.14 | 28 | 8.29 | 8.72 0.10 0.29 | 4.5 G 1 |
| hr5573 | S | 4627 | 2.51 | 1.41 | 0.20 | 0.11 | 385 | 0.36 | 0.17 | 29 | 8.38 | 8.88 | 0.36 | 1.30 | 0.25 | 4627 | 2.23 | 1.45 | 0.11 | 0.11 | 385 | 0.11 | 0.17 | 29 | 8.19 | 8.66 0.36 0.28 | 4.6 G 1 |
| hr5600 | S | 3962 | 1.48 | 1.78 | 0.12 | 0.21 | 333 | 0.61 | 0.32 | 21 | 8.42 | 8.83 | -0.49 | 2.09 | 0.36 | 3962 | 0.16 | 1.81 | -0.25 | 0.23 | 333 | -0.25 | 0.33 | 21 | 7.96 | 8.25 -0.64 0.70 | 5.5 G 1 |
| hr5601 | S | 4664 | 2.40 | 1.31 | -0.14 | 0.08 | 412 | -0.07 | 0.08 | 26 | 8.12 | 8.67 | 0.30 | 1.00 | 0.26 | 4664 | 2.22 | 1.35 | -0.17 | 0.09 | 412 | -0.18 | 0.08 | 26 | 8.07 | 8.59 0.30 0.27 | 4.0 G 1 |
| hr5602 | S | 4920 | 2.34 | 1.70 | -0.06 | 0.08 | 393 | 0.08 | 0.12 | 30 | 8.14 | 8.65 | 1.11 | 2.73 | 0.19 | 4920 | 2.05 | 1.73 | -0.08 | 0.09 | 393 | -0.08 | 0.12 | 30 | 8.07 | 8.51 1.12 0.21 | 5.0 G 1 |
| hr5609 | S | 4800 | 2.61 | 1.34 | 0.12 | 0.08 | 389 | 0.17 | 0.10 | 26 | 8.26 | 8.75 | 0.48 | 0.99 | 0.20 | 4800 | 2.50 | 1.39 | 0.09 | 0.08 | 389 | 0.09 | 0.10 | 26 | 8.22 | 8.70 0.48 0.21 | 4.3 G 1 |
| hr5616 | S | 4302 | 1.93 | 1.51 | -0.09 | 0.09 | 233 | 0.19 | 0.16 | 15 | 8.30 | 8.80 | -0.43 | 0.71 | 0.34 | 4302 | 1.23 | 1.49 | -0.23 | 0.10 | 233 | -0.23 | 0.17 | 15 | 8.08 | 8.50 -0.45 0.41 | 4.5 G 1 |
| hr5620 | S | 4720 | 3.02 | 0.80 | 0.08 | 0.10 | 413 | 0.28 | 0.17 | 29 | 8.38 | 8.78 | 0.00 | 0.43 | 0.20 | 4720 | 2.57 | 1.24 | -0.08 | 0.10 | 413 | -0.08 | 0.20 | 29 | 8.22 | 8.57 0.00 0.22 | 3.6 G 1 |
| hr5635 | S | 4812 | 2.54 | 1.32 | -0.34 | 0.07 | 422 | -0.19 | 0.08 | 30 | 8.05 | 8.56 | 0.05 | 0.37 | 0.20 | 4812 | 2.21 | 1.42 | -0.38 | 0.07 | 422 | -0.38 | 0.08 | 30 | 7.95 | 8.40 0.06 0.22 | 4.2 G 1 |

| ID | S | | | | | | | | | | | | | | | | | | | | | | | | | | | |
|---|---|---|---|---|---|---|---|---|---|---|---|---|---|---|---|---|---|---|---|---|---|---|---|---|---|---|---|---|
| hr5648 | S | 4722 | 2.49 | 1.37 | 0.06 | 0.09 | 386 | 0.13 | 0.12 | 26 | 8.23 | 8.70 | 0.35 | 0.94 | 0.24 | 4722 | 2.33 | 1.42 | 0.03 | 0.09 | 386 | 0.03 | 0.12 | 26 | 8.19 | 8.62 | 0.35 | 0.25 | 4.4 | G | 1 |
| hr5673 | S | 4399 | 2.27 | 1.37 | -0.03 | 0.11 | 385 | 0.20 | 0.15 | 24 | 8.30 | 8.77 | -0.31 | 0.64 | 0.31 | 4399 | 1.73 | 1.50 | -0.17 | 0.12 | 385 | -0.16 | 0.16 | 24 | 8.12 | 8.53 | -0.32 | 0.35 | 4.1 | G | 1 |
| hr5681a | S | 4821 | 4.23 | 0.51 | 0.03 | 0.10 | 291 | 0.91 | 0.15 | 29 | 8.58 | 9.31 | 0.75 | 1.75 | 0.18 | 4821 | 2.18 | 1.30 | -0.30 | 0.07 | 291 | -0.30 | 0.10 | 29 | 7.86 | 8.38 | 0.77 | 0.22 | 3.1 | G | 2 |
| hr5681b | S | 5812 | 4.49 | 0.78 | -0.42 | 0.05 | 312 | -0.41 | 0.04 | 20 | 7.98 | 8.48 | 2.16 | 3.71 | 0.03 | 5812 | 4.47 | 0.82 | -0.42 | 0.04 | 312 | -0.42 | 0.04 | 20 | 7.97 | 8.47 | 2.16 | 0.03 | 2.8 | G | 2 |
| hr5707 | S | 4783 | 2.60 | 1.44 | 0.29 | 0.11 | 370 | 0.39 | 0.16 | 27 | 8.46 | 8.88 | 0.45 | 0.96 | 0.20 | 4783 | 2.36 | 1.50 | 0.24 | 0.11 | 370 | 0.24 | 0.16 | 27 | 8.39 | 8.76 | 0.45 | 0.22 | 4.7 | G | 1 |
| hr5709 | S | 4705 | 2.57 | 1.30 | -0.16 | 0.08 | 416 | 0.05 | 0.07 | 24 | 8.27 | 8.83 | 0.34 | 0.97 | 0.23 | 4705 | 2.23 | 1.36 | -0.25 | 0.08 | 416 | -0.25 | 0.08 | 24 | 8.02 | 8.54 | 0.34 | 0.26 | 4.2 | G | 1 |
| hr5744 | S | 4531 | 2.40 | 1.37 | 0.17 | 0.13 | 412 | 0.28 | 0.23 | 28 | 8.35 | 8.81 | 0.38 | 1.90 | 0.28 | 4531 | 2.13 | 1.47 | 0.09 | 0.13 | 412 | 0.09 | 0.23 | 28 | 8.26 | 8.69 | 0.38 | 0.31 | 4.2 | G | 1 |
| hr5769 | S | 6040 | 3.58 | 1.79 | -0.08 | 0.09 | 268 | -0.13 | 0.09 | 25 | 8.23 | 8.66 | 1.92 | 1.44 | 0.02 | 6040 | 3.68 | 1.74 | -0.08 | 0.09 | 268 | -0.08 | 0.09 | 25 | 8.26 | 8.81 | 1.92 | 0.02 | 10.0 | R | 2 |
| hr5785 | S | 4798 | 2.60 | 1.34 | -0.17 | 0.07 | 405 | 0.00 | 0.10 | 28 | 8.13 | 8.74 | 0.06 | 0.39 | 0.20 | 4798 | 2.33 | 1.38 | -0.24 | 0.07 | 405 | -0.24 | 0.12 | 28 | 7.91 | 8.48 | 0.05 | 0.22 | 4.2 | G | 1 |
| hr5794 | S | 4135 | 1.58 | 1.61 | -0.02 | 0.15 | 352 | 0.27 | 0.21 | 26 | 8.23 | 8.69 | 0.41 | 7.50 | 0.37 | 4135 | 0.83 | 1.65 | -0.21 | 0.16 | 352 | -0.21 | 0.22 | 26 | 7.98 | 8.37 | 0.36 | 0.48 | 4.7 | G | 1 |
| hr5802 | S | 4889 | 2.74 | 1.23 | -0.02 | 0.08 | 403 | 0.09 | 0.11 | 31 | 8.28 | 8.66 | 0.41 | 0.64 | 0.18 | 4889 | 2.50 | 1.33 | -0.06 | 0.09 | 403 | -0.06 | 0.11 | 31 | 8.21 | 8.54 | 0.41 | 0.19 | 4.4 | G | 1 |
| hr5810 | S | 4633 | 2.56 | 1.24 | -0.12 | 0.09 | 410 | 0.06 | 0.11 | 25 | 8.31 | 8.72 | 0.00 | 0.58 | 0.24 | 4633 | 2.13 | 1.39 | -0.21 | 0.10 | 410 | -0.22 | 0.11 | 25 | 8.17 | 8.52 | 0.01 | 0.29 | 4.1 | G | 1 |
| hr5811 | S | 4517 | 2.32 | 1.51 | 0.31 | 0.13 | 353 | 0.47 | 0.25 | 25 | 8.45 | 8.88 | 0.15 | 1.20 | 0.29 | 4517 | 1.96 | 1.57 | 0.23 | 0.13 | 353 | 0.23 | 0.25 | 25 | 8.33 | 8.72 | 0.15 | 0.32 | 4.5 | G | 1 |
| hr5828 | S | 4625 | 2.47 | 1.26 | 0.01 | 0.09 | 397 | 0.08 | 0.13 | 26 | 8.22 | 8.72 | 1.10 | 6.18 | 0.26 | 4625 | 2.30 | 1.33 | -0.03 | 0.09 | 397 | -0.04 | 0.13 | 26 | 8.17 | 8.64 | 1.10 | 0.28 | 4.0 | G | 1 |
| hr5835 | S | 5172 | 2.83 | 1.30 | 0.08 | 0.08 | 414 | 0.01 | 0.12 | 29 | 8.02 | 8.61 | 0.69 | 0.57 | 0.13 | 5172 | 2.97 | 1.23 | 0.10 | 0.08 | 414 | 0.10 | 0.12 | 29 | 8.06 | 8.68 | 0.69 | 0.12 | 5.4 | G | 1 |
| hr5840 | S | 4982 | 2.46 | 1.41 | -0.10 | 0.08 | 402 | -0.04 | 0.09 | 29 | 8.05 | 8.54 | 0.72 | 0.98 | 0.17 | 4982 | 2.32 | 1.44 | -0.11 | 0.08 | 402 | -0.11 | 0.08 | 29 | 8.02 | 8.47 | 0.72 | 0.18 | 4.4 | G | 1 |
| hr5841 | S | 4659 | 2.43 | 1.35 | 0.15 | 0.10 | 392 | 0.19 | 0.16 | 27 | 8.29 | 8.76 | 0.31 | 1.05 | 0.26 | 4659 | 2.22 | 1.46 | 0.12 | 0.10 | 392 | 0.12 | 0.16 | 27 | 8.32 | 8.75 | 0.31 | 0.28 | 4.2 | G | 1 |
| hr5855 | S | 4626 | 2.64 | 1.13 | 0.10 | 0.09 | 383 | 0.28 | 0.14 | 28 | 8.24 | 8.70 | 0.50 | 1.78 | 0.24 | 4626 | 2.21 | 1.35 | -0.02 | 0.10 | 383 | -0.02 | 0.15 | 28 | 8.10 | 8.51 | 0.50 | 0.28 | 4.0 | G | 1 |
| hr5888 | S | 4715 | 2.41 | 1.41 | -0.16 | 0.09 | 415 | 0.01 | 0.14 | 31 | 8.13 | 8.67 | 0.17 | 0.63 | 0.25 | 4715 | 2.02 | 1.48 | -0.21 | 0.09 | 415 | -0.22 | 0.14 | 31 | 8.03 | 8.49 | 0.17 | 0.28 | 4.3 | G | 1 |
| hr5893 | S | 4873 | 3.08 | 1.06 | 0.18 | 0.09 | 414 | 0.30 | 0.14 | 28 | 8.46 | 8.79 | 0.39 | 0.66 | 0.17 | 4873 | 2.92 | 1.08 | 0.13 | 0.09 | 414 | 0.13 | 0.15 | 28 | 8.33 | 8.62 | 0.39 | 0.17 | 4.0 | G | 1 |
| hr5922 | S | 4823 | 2.61 | 1.36 | 0.08 | 0.09 | 400 | 0.16 | 0.12 | 28 | 8.22 | 8.70 | 0.33 | 0.65 | 0.20 | 4823 | 2.43 | 1.43 | 0.04 | 0.09 | 400 | 0.04 | 0.13 | 28 | 8.17 | 8.61 | 0.33 | 0.21 | 4.4 | G | 1 |
| hr5940 | S | 4593 | 2.60 | 1.21 | 0.24 | 0.12 | 370 | 0.32 | 0.19 | 22 | 8.57 | 9.01 | 0.16 | 0.95 | 0.25 | 4593 | 2.40 | 1.32 | 0.17 | 0.12 | 370 | 0.17 | 0.19 | 22 | 8.50 | 8.92 | 0.16 | 0.27 | 4.0 | G | 1 |
| hr5947 | S | 4364 | 1.95 | 1.55 | -0.10 | 0.10 | 376 | 0.06 | 0.14 | 24 | 8.19 | 8.77 | -0.31 | 0.74 | 0.34 | 4364 | 1.57 | 1.59 | -0.18 | 0.11 | 376 | -0.18 | 0.14 | 24 | 8.07 | 8.61 | -0.32 | 0.37 | 4.6 | G | 1 |
| hr5966 | S | 4798 | 2.52 | 1.39 | -0.13 | 0.08 | 414 | 0.00 | 0.11 | 30 | 8.23 | 8.72 | 0.35 | 0.73 | 0.21 | 4798 | 2.23 | 1.46 | -0.17 | 0.08 | 414 | -0.17 | 0.12 | 30 | 8.16 | 8.58 | 0.35 | 0.22 | 4.5 | G | 1 |
| hr5976 | S | 4933 | 2.78 | 1.32 | 0.10 | 0.08 | 401 | 0.18 | 0.10 | 29 | 8.21 | 8.71 | 0.81 | 1.38 | 0.17 | 4933 | 2.60 | 1.41 | 0.07 | 0.08 | 401 | 0.07 | 0.11 | 29 | 8.15 | 8.63 | 0.81 | 0.17 | 4.2 | G | 1 |
| hr6038 | S | 4744 | 2.51 | 1.38 | 0.07 | 0.08 | 397 | 0.09 | 0.13 | 27 | 8.26 | 8.74 | 0.32 | 0.81 | 0.22 | 4744 | 2.47 | 1.39 | 0.07 | 0.08 | 397 | 0.06 | 0.13 | 27 | 8.24 | 8.72 | 0.32 | 0.24 | 4.2 | G | 1 |
| hr6057 | S | 4560 | 2.51 | 1.26 | 0.13 | 0.11 | 399 | 0.22 | 0.20 | 26 | 8.36 | 8.74 | 0.08 | 0.88 | 0.24 | 4560 | 2.27 | 1.37 | 0.06 | 0.11 | 399 | 0.06 | 0.21 | 26 | 8.27 | 8.63 | 0.08 | 0.29 | 4.0 | G | 1 |
| hr6065 | S | 4620 | 2.43 | 1.42 | 0.16 | 0.11 | 391 | 0.22 | 0.16 | 27 | 8.32 | 8.76 | 0.67 | 2.64 | 0.27 | 4620 | 2.28 | 1.47 | 0.12 | 0.11 | 391 | 0.12 | 0.16 | 27 | 8.26 | 8.69 | 0.67 | 0.28 | 4.3 | G | 1 |
| hr6075 | S | 4837 | 2.53 | 1.37 | -0.07 | 0.08 | 415 | -0.02 | 0.11 | 28 | 8.09 | 8.61 | 0.15 | 0.41 | 0.20 | 4837 | 2.42 | 1.40 | -0.08 | 0.08 | 415 | -0.08 | 0.11 | 28 | 8.05 | 8.56 | 0.15 | 0.21 | 4.4 | G | 1 |
| hr6104 | S | 4757 | 2.48 | 1.37 | -0.07 | 0.08 | 414 | 0.00 | 0.09 | 23 | 8.28 | 8.70 | 0.30 | 0.74 | 0.22 | 4757 | 2.32 | 1.42 | -0.10 | 0.09 | 414 | -0.10 | 0.09 | 23 | 8.23 | 8.62 | 0.30 | 0.23 | 4.4 | G | 1 |
| hr6121 | S | 4682 | 2.61 | 1.36 | -0.02 | 0.09 | 406 | 0.18 | 0.12 | 27 | 8.26 | 8.77 | 0.20 | 0.76 | 0.23 | 4682 | 2.16 | 1.49 | -0.11 | 0.10 | 406 | -0.10 | 0.12 | 27 | 8.13 | 8.57 | 0.20 | 0.28 | 4.2 | G | 1 |
| hr6124 | S | 5030 | 2.68 | 1.42 | -0.05 | 0.07 | 406 | 0.00 | 0.09 | 29 | 8.12 | 8.66 | 0.92 | 1.36 | 0.15 | 5030 | 2.57 | 1.44 | -0.06 | 0.07 | 406 | -0.06 | 0.09 | 29 | 8.09 | 8.60 | 0.93 | 0.16 | 4.3 | G | 1 |
| hr6126 | S | 4636 | 2.38 | 1.42 | 0.27 | 0.12 | 369 | 0.38 | 0.19 | 25 | 8.34 | 8.88 | 2.03 | 22.37 | 0.27 | 4636 | 2.13 | 1.48 | 0.21 | 0.12 | 369 | 0.21 | 0.20 | 25 | 8.26 | 8.77 | 2.02 | 0.29 | 4.9 | G | 1 |
| hr6130 | S | 4946 | 2.41 | 1.49 | 0.03 | 0.08 | 408 | -0.07 | 0.09 | 28 | 8.13 | 8.51 | 0.53 | 0.71 | 0.18 | 4946 | 2.63 | 1.43 | 0.06 | 0.08 | 408 | 0.06 | 0.09 | 28 | 8.19 | 8.62 | 0.53 | 0.17 | 4.7 | G | 1 |
| hr6132 | S | 4962 | 2.71 | 1.34 | -0.04 | 0.07 | 401 | 0.06 | 0.10 | 29 | 8.02 | 8.68 | 0.15 | 0.29 | 0.16 | 4962 | 2.50 | 1.41 | -0.07 | 0.07 | 401 | -0.07 | 0.11 | 29 | 7.97 | 8.58 | 0.15 | 0.18 | 4.4 | G | 1 |
| hr6145 | S | 4696 | 2.94 | 1.06 | 0.31 | 0.11 | 394 | 0.47 | 0.17 | 26 | 8.52 | 8.92 | 0.20 | 0.74 | 0.21 | 4696 | 2.54 | 1.35 | 0.17 | 0.12 | 394 | 0.16 | 0.19 | 26 | 8.38 | 8.74 | 0.20 | 0.23 | 4.0 | G | 1 |
| hr6147 | S | 5038 | 2.57 | 1.64 | 0.21 | 0.10 | 375 | 0.33 | 0.16 | 32 | 8.29 | 8.81 | 0.73 | 0.85 | 0.16 | 5038 | 2.29 | 1.68 | 0.18 | 0.11 | 375 | 0.18 | 0.16 | 32 | 8.22 | 8.67 | 0.74 | 0.17 | 6.8 | R | 1 |
| hr6148 | S | 4912 | 2.36 | 1.56 | -0.07 | 0.09 | 400 | 0.02 | 0.11 | 29 | 8.12 | 8.69 | 0.76 | 1.29 | 0.19 | 4912 | 2.15 | 1.60 | -0.09 | 0.09 | 400 | -0.09 | 0.11 | 29 | 8.07 | 8.59 | 0.76 | 0.20 | 5.4 | G | 1 |
| hr6150 | S | 4735 | 2.50 | 1.38 | 0.00 | 0.09 | 394 | 0.05 | 0.13 | 28 | 8.30 | 8.77 | 0.20 | 0.64 | 0.24 | 4735 | 2.38 | 1.42 | -0.02 | 0.09 | 394 | -0.02 | 0.13 | 28 | 8.26 | 8.71 | 0.20 | 0.25 | 4.1 | G | 1 |
| hr6190 | S | 4718 | 2.46 | 1.40 | -0.03 | 0.09 | 392 | 0.02 | 0.12 | 27 | 8.32 | 8.77 | 0.28 | 0.82 | 0.24 | 4718 | 2.36 | 1.43 | -0.05 | 0.09 | 392 | -0.05 | 0.12 | 27 | 8.29 | 8.72 | 0.29 | 0.25 | 4.3 | G | 1 |
| hr6199 | S | 4641 | 2.48 | 1.54 | 0.02 | 0.13 | 365 | 0.30 | 0.28 | 27 | 8.24 | 8.75 | 0.28 | 1.06 | 0.26 | 4641 | 1.82 | 1.66 | -0.09 | 0.14 | 365 | -0.09 | 0.29 | 27 | 8.05 | 8.44 | 0.29 | 0.31 | 7.1 | G | 1 |
| hr6220 | S | 4932 | 2.76 | 1.30 | -0.17 | 0.07 | 407 | -0.01 | 0.10 | 29 | 8.13 | 8.68 | 0.89 | 1.65 | 0.17 | 4932 | 2.41 | 1.40 | -0.21 | 0.09 | 407 | -0.21 | 0.10 | 29 | 8.04 | 8.52 | 0.89 | 0.19 | 4.4 | G | 1 |
| hr6259 | S | 4756 | 2.74 | 1.14 | -0.08 | 0.08 | 413 | -0.02 | 0.11 | 27 | 8.22 | 8.72 | 0.10 | 0.49 | 0.21 | 4756 | 2.62 | 1.21 | -0.11 | 0.08 | 413 | -0.11 | 0.11 | 27 | 8.18 | 8.67 | 0.11 | 0.21 | 3.8 | G | 1 |
| hr6280 | S | 4640 | 2.57 | 1.29 | 0.22 | 0.10 | 390 | 0.37 | 0.17 | 27 | 8.41 | 8.87 | 0.53 | 1.84 | 0.24 | 4640 | 2.30 | 1.34 | 0.13 | 0.11 | 390 | 0.13 | 0.19 | 27 | 8.22 | 8.66 | 0.54 | 0.27 | 4.2 | G | 1 |
| hr6287 | S | 4838 | 2.61 | 1.30 | -0.11 | 0.08 | 420 | 0.01 | 0.09 | 27 | 8.12 | 8.68 | 0.32 | 0.62 | 0.19 | 4838 | 2.34 | 1.38 | -0.15 | 0.08 | 420 | -0.15 | 0.09 | 27 | 8.05 | 8.55 | 0.33 | 0.21 | 4.2 | G | 1 |
| hr6292 | S | 4979 | 2.67 | 1.35 | -0.14 | 0.07 | 414 | -0.12 | 0.07 | 28 | 8.07 | 8.65 | 0.32 | 0.40 | 0.16 | 4979 | 2.64 | 1.36 | -0.14 | 0.07 | 414 | -0.14 | 0.07 | 28 | 8.06 | 8.63 | 0.32 | 0.16 | 4.5 | G | 1 |
| hr6305 | S | 4991 | 2.62 | 1.47 | -0.16 | 0.08 | 410 | -0.07 | 0.08 | 29 | 7.98 | 8.61 | 1.16 | 2.53 | 0.16 | 4991 | 2.41 | 1.51 | -0.18 | 0.08 | 410 | -0.18 | 0.08 | 29 | 7.93 | 8.51 | 1.16 | 0.17 | 5.5 | G | 1 |
| hr6307 | S | 4634 | 2.36 | 1.40 | 0.03 | 0.09 | 381 | 0.19 | 0.13 | 27 | 8.24 | 8.72 | 0.00 | 0.57 | 0.27 | 4634 | 2.00 | 1.49 | -0.04 | 0.10 | 381 | -0.04 | 0.13 | 27 | 8.14 | 8.55 | 0.00 | 0.30 | 4.3 | G | 1 |
| hr6333 | S | 4779 | 2.59 | 1.39 | -0.11 | 0.07 | 392 | 0.01 | 0.09 | 28 | 8.26 | 8.75 | 0.17 | 0.53 | 0.21 | 4779 | 2.32 | 1.45 | -0.15 | 0.08 | 392 | -0.15 | 0.09 | 28 | 8.19 | 8.63 | 0.18 | 0.22 | 4.1 | G | 1 |
| hr6342 | S | 4702 | 2.61 | 1.22 | -0.14 | 0.08 | 416 | -0.03 | 0.11 | 25 | 8.17 | 8.67 | 0.20 | 0.71 | 0.23 | 4702 | 2.37 | 1.31 | -0.19 | 0.08 | 416 | -0.19 | 0.12 | 25 | 8.10 | 8.56 | 0.20 | 0.25 | 4.0 | G | 1 |

| ID | S | | | | | | | | | | | | | | | | | | | | | | | | | | | | |
|---|---|---|---|---|---|---|---|---|---|---|---|---|---|---|---|---|---|---|---|---|---|---|---|---|---|---|---|---|---|
| hr6359 | S | 5078 | 2.91 | 1.25 | -0.03 | 0.07 | 414 | -0.01 | 0.12 | 28 | 8.02 | 8.60 | 0.46 | 0.43 | 0.14 | 5078 | 2.87 | 1.27 | -0.03 | 0.07 | 414 | -0.03 | 0.12 | 28 | 8.01 | 8.58 | 0.46 | 0.14 | 4.4 G 1 |
| hr6360 | S | 4886 | 2.68 | 1.29 | 0.21 | 0.10 | 413 | 0.30 | 0.13 | 30 | 8.33 | 8.75 | 1.14 | 3.20 | 0.18 | 4886 | 2.49 | 1.36 | 0.18 | 0.10 | 413 | 0.18 | 0.13 | 30 | 8.28 | 8.66 | 1.15 | 0.19 | 4.4 G 1 |
| hr6363 | S | 4602 | 2.48 | 1.16 | 0.08 | 0.11 | 400 | 0.14 | 0.15 | 26 | 8.29 | 8.73 | 0.28 | 1.19 | 0.26 | 4602 | 2.35 | 1.23 | 0.04 | 0.11 | 400 | 0.05 | 0.16 | 26 | 8.24 | 8.67 | 0.28 | 0.28 | 4.1 G 1 |
| hr6364 | S | 4281 | 2.05 | 1.36 | 0.13 | 0.15 | 372 | 0.39 | 0.25 | 27 | 8.41 | 8.91 | -0.46 | 0.71 | 0.33 | 4281 | 1.38 | 1.50 | -0.06 | 0.15 | 372 | -0.06 | 0.26 | 27 | 8.17 | 8.61 | -0.48 | 0.40 | 4.4 G 1 |
| hr6365 | S | 4889 | 2.67 | 1.39 | 0.14 | 0.09 | 391 | 0.33 | 0.11 | 31 | 8.37 | 8.89 | 0.66 | 1.12 | 0.18 | 4889 | 2.24 | 1.53 | 0.07 | 0.10 | 391 | 0.08 | 0.11 | 31 | 8.23 | 8.68 | 0.67 | 0.20 | 4.9 G 1 |
| hr6388 | S | 4329 | 2.03 | 1.40 | 0.02 | 0.14 | 370 | 0.14 | 0.20 | 24 | 8.39 | 8.77 | -0.38 | 0.72 | 0.33 | 4329 | 1.74 | 1.48 | -0.06 | 0.14 | 370 | -0.07 | 0.20 | 24 | 8.28 | 8.64 | -0.39 | 0.36 | 4.5 G 1 |
| hr6390 | S | 4697 | 2.44 | 1.29 | -0.20 | 0.08 | 411 | -0.17 | 0.10 | 27 | 8.18 | 8.65 | -0.15 | 0.32 | 0.25 | 4697 | 2.38 | 1.30 | -0.21 | 0.08 | 411 | -0.21 | 0.10 | 27 | 8.16 | 8.63 | -0.16 | 0.25 | 3.9 G 1 |
| hr6394 | S | 6068 | 3.71 | 1.68 | -0.17 | 0.05 | 254 | -0.04 | 0.07 | 28 | 8.36 | 8.75 | 2.97 | 10.26 | 0.02 | 6068 | 3.37 | 1.76 | -0.18 | 0.06 | 254 | -0.18 | 0.06 | 28 | 8.23 | 8.72 | 2.97 | 0.03 | 6.3 G 2 |
| hr6404 | S | 4658 | 2.65 | 1.37 | 0.25 | 0.12 | 393 | 0.46 | 0.18 | 29 | 8.49 | 8.98 | 0.31 | 1.06 | 0.23 | 4658 | 2.25 | 1.45 | 0.14 | 0.12 | 393 | 0.14 | 0.20 | 29 | 8.26 | 8.70 | 0.32 | 0.27 | 4.9 G 1 |
| hr6415 | S | 4508 | 2.34 | 1.46 | 0.28 | 0.13 | 357 | 0.44 | 0.23 | 25 | 8.58 | 8.99 | 0.22 | 1.44 | 0.29 | 4508 | 1.93 | 1.55 | 0.17 | 0.14 | 357 | 0.17 | 0.23 | 25 | 8.43 | 8.80 | 0.22 | 0.33 | 4.6 G 1 |
| hr6443 | S | 4870 | 2.63 | 1.37 | -0.02 | 0.08 | 414 | 0.02 | 0.08 | 25 | 8.18 | 8.68 | 0.29 | 0.52 | 0.19 | 4870 | 2.55 | 1.40 | -0.03 | 0.08 | 414 | -0.03 | 0.08 | 25 | 8.15 | 8.64 | 0.29 | 0.19 | 4.3 G 1 |
| hr6444 | S | 4773 | 2.62 | 1.33 | 0.11 | 0.09 | 388 | 0.27 | 0.13 | 26 | 8.23 | 8.62 | 0.74 | 1.89 | 0.21 | 4773 | 2.26 | 1.47 | 0.03 | 0.10 | 388 | 0.03 | 0.14 | 26 | 8.13 | 8.45 | 0.74 | 0.23 | 4.7 G 1 |
| hr6448 | S | 4650 | 2.35 | 1.44 | -0.02 | 0.09 | 396 | 0.05 | 0.10 | 26 | 8.29 | 8.79 | 0.10 | 0.68 | 0.27 | 4650 | 2.19 | 1.48 | -0.05 | 0.09 | 396 | -0.05 | 0.11 | 26 | 8.24 | 8.72 | 0.10 | 0.28 | 4.5 G 1 |
| hr6472 | S | 4980 | 2.88 | 1.31 | 0.10 | 0.09 | 414 | 0.19 | 0.13 | 32 | 8.22 | 8.68 | 1.14 | 2.53 | 0.16 | 4980 | 2.66 | 1.40 | 0.06 | 0.09 | 414 | 0.06 | 0.14 | 32 | 8.16 | 8.57 | 1.15 | 0.16 | 4.8 G 1 |
| hr6488 | S | 5461 | 2.88 | 0.00 | -0.19 | 0.23 | 476 | -0.33 | 0.26 | 33 | 8.10 | 8.73 | 1.66 | 2.47 | 0.08 | 5461 | 3.18 | 0.50 | -0.24 | 0.25 | 476 | -0.24 | 0.25 | 33 | 8.15 | 8.86 | 1.66 | 0.07 | 6.6 G 1 |
| hr6524 | S | 5162 | 2.78 | 1.14 | 0.25 | 0.10 | 399 | 0.09 | 0.11 | 30 | 8.23 | 8.86 | 1.32 | 2.26 | 0.13 | 5162 | 3.11 | 0.92 | 0.31 | 0.11 | 399 | 0.31 | 0.11 | 30 | 8.32 | 9.02 | 1.31 | 0.12 | 5.3 G 1 |
| hr6564 | S | 4191 | 2.12 | 1.45 | 0.37 | 0.18 | 333 | 0.76 | 0.33 | 23 | 8.71 | 9.03 | -0.28 | 1.39 | 0.33 | 4191 | 1.16 | 1.61 | 0.07 | 0.20 | 333 | 0.07 | 0.34 | 23 | 8.33 | 8.60 | -0.32 | 0.43 | 4.6 G 1 |
| hr6566 | S | 4616 | 2.58 | 1.30 | -0.02 | 0.09 | 408 | 0.22 | 0.15 | 29 | 8.33 | 8.83 | 0.20 | 0.95 | 0.25 | 4616 | 2.03 | 1.47 | -0.14 | 0.10 | 408 | -0.14 | 0.15 | 29 | 8.15 | 8.57 | 0.20 | 0.30 | 4.4 G 1 |
| hr6575a | S | 4813 | 2.67 | 1.27 | 0.11 | 0.08 | 409 | 0.22 | 0.10 | 31 | 8.25 | 8.68 | 0.25 | 0.56 | 0.19 | 4813 | 2.43 | 1.38 | 0.06 | 0.09 | 409 | 0.06 | 0.11 | 31 | 8.17 | 8.56 | 0.25 | 0.21 | 4.1 G 1 |
| hr6575b | S | 6616 | 3.87 | 3.62 | 0.32 | 0.34 | 22 | 0.08 | 0.00 | 1 | 8.38 | | | | -0.01 | 6616 | 4.36 | 3.41 | 0.33 | 0.34 | 22 | 0.33 | | 1 | 8.54 | | | -0.12 | 50.0 R 2 |
| hr6591 | S | 4705 | 2.50 | 1.27 | -0.11 | 0.09 | 409 | -0.14 | 0.13 | 29 | 8.24 | 8.73 | 0.28 | 0.84 | 0.24 | 4705 | 2.55 | 1.25 | -0.10 | 0.09 | 409 | -0.10 | 0.13 | 29 | 8.26 | 8.75 | 0.28 | 0.23 | 4.0 G 1 |
| hr6603 | S | 4553 | 2.41 | 1.53 | 0.31 | 0.14 | 358 | 0.40 | 0.20 | 24 | 8.63 | 9.01 | 0.37 | 1.73 | 0.28 | 4553 | 2.20 | 1.57 | 0.26 | 0.15 | 358 | 0.26 | 0.20 | 24 | 8.55 | 8.91 | 0.37 | 0.30 | 4.5 G 1 |
| hr6606 | S | 4819 | 2.51 | 1.42 | -0.09 | 0.08 | 401 | -0.08 | 0.09 | 25 | 8.18 | 8.69 | 0.14 | 0.43 | 0.20 | 4819 | 2.48 | 1.42 | -0.09 | 0.08 | 401 | -0.09 | 0.09 | 25 | 8.17 | 8.68 | 0.14 | 0.21 | 4.2 G 1 |
| hr6607 | S | 4745 | 2.72 | 1.24 | 0.34 | 0.11 | 380 | 0.49 | 0.19 | 29 | 8.24 | 8.95 | 0.47 | 1.13 | 0.21 | 4745 | 2.38 | 1.40 | 0.25 | 0.11 | 380 | 0.25 | 0.19 | 29 | 8.14 | 8.80 | 0.47 | 0.24 | 4.3 G 1 |
| hr6638 | S | 4944 | 2.84 | 1.26 | 0.08 | 0.08 | 392 | 0.17 | 0.13 | 28 | 8.23 | 8.74 | 1.20 | 3.14 | 0.16 | 4944 | 2.64 | 1.36 | 0.04 | 0.09 | 392 | 0.04 | 0.13 | 28 | 8.17 | 8.64 | 1.21 | 0.17 | 4.7 G 1 |
| hr6639 | S | 4468 | 2.46 | 1.50 | 0.19 | 0.13 | 390 | 0.50 | 0.21 | 27 | 8.59 | 8.99 | 0.43 | 2.56 | 0.29 | 4468 | 1.79 | 1.61 | -0.01 | 0.14 | 390 | -0.01 | 0.22 | 27 | 8.26 | 8.60 | 0.42 | 0.34 | 4.4 G 1 |
| hr6644 | S | 4555 | 2.48 | 1.43 | 0.21 | 0.13 | 365 | 0.39 | 0.19 | 25 | 8.65 | 9.08 | 0.12 | 0.98 | 0.27 | 4555 | 2.14 | 1.49 | 0.09 | 0.13 | 365 | 0.09 | 0.20 | 25 | 8.43 | 8.84 | 0.12 | 0.30 | 4.9 G 1 |
| hr6654 | S | 4661 | 2.64 | 1.20 | 0.01 | 0.09 | 403 | 0.12 | 0.16 | 26 | 8.28 | 8.79 | -0.03 | 0.48 | 0.23 | 4661 | 2.39 | 1.32 | -0.06 | 0.09 | 403 | -0.06 | 0.16 | 26 | 8.20 | 8.67 | -0.03 | 0.26 | 3.8 G 1 |
| hr6659 | S | 4577 | 2.50 | 1.27 | -0.05 | 0.10 | 403 | 0.34 | 0.11 | 30 | 8.22 | 8.66 | 0.47 | 1.99 | 0.27 | 4577 | 1.59 | 1.48 | -0.22 | 0.13 | 403 | -0.23 | 0.12 | 30 | 7.96 | 8.24 | 0.47 | 0.34 | 4.9 G 1 |
| hr6698 | S | 4874 | 2.54 | 1.57 | 0.12 | 0.09 | 386 | 0.15 | 0.15 | 29 | 8.14 | 8.68 | 0.42 | 0.69 | 0.19 | 4874 | 2.49 | 1.58 | 0.12 | 0.09 | 386 | 0.12 | 0.15 | 29 | 8.12 | 8.66 | 0.43 | 0.19 | 4.8 G 1 |
| hr6703 | S | 4930 | 2.67 | 1.41 | 0.03 | 0.09 | 397 | 0.13 | 0.12 | 30 | 8.19 | 8.60 | 1.18 | 3.09 | 0.17 | 4930 | 2.45 | 1.47 | 0.00 | 0.10 | 397 | 0.01 | 0.13 | 30 | 8.13 | 8.50 | 1.18 | 0.18 | 5.2 G 1 |
| hr6711 | S | 4999 | 2.78 | 1.37 | 0.08 | 0.08 | 412 | 0.11 | 0.11 | 31 | 8.22 | 8.72 | 0.60 | 0.73 | 0.16 | 4999 | 2.73 | 1.39 | 0.08 | 0.08 | 412 | 0.08 | 0.11 | 31 | 8.21 | 8.70 | 0.60 | 0.16 | 4.6 G 1 |
| hr6757 | S | 4943 | 2.37 | 1.71 | 0.07 | 0.14 | 348 | 0.08 | 0.18 | 28 | 7.52 | 8.58 | 2.03 | 13.63 | 0.19 | 4943 | 2.35 | 1.71 | 0.07 | 0.14 | 348 | 0.07 | 0.18 | 28 | | | 2.03 | 0.19 | 8.6 G 1 |
| hr6770 | S | 4921 | 2.49 | 1.49 | -0.01 | 0.08 | 391 | 0.03 | 0.10 | 28 | 8.11 | 8.62 | 0.35 | 0.51 | 0.18 | 4921 | 2.45 | 1.50 | 0.00 | 0.08 | 391 | 0.00 | 0.10 | 28 | 8.10 | 8.60 | 0.35 | 0.19 | 4.6 G 1 |
| hr6793 | S | 4516 | 2.62 | 1.22 | 0.30 | 0.13 | 385 | 0.51 | 0.20 | 24 | 8.59 | 8.93 | 0.17 | 1.26 | 0.26 | 4516 | 2.13 | 1.45 | 0.13 | 0.14 | 385 | 0.13 | 0.22 | 24 | 8.41 | 8.70 | 0.17 | 0.31 | 4.0 G 1 |
| hr6799 | S | 4487 | 2.40 | 1.31 | 0.17 | 0.13 | 378 | 0.32 | 0.23 | 27 | 8.40 | 8.87 | 0.15 | 1.33 | 0.29 | 4487 | 2.04 | 1.46 | 0.06 | 0.13 | 378 | 0.06 | 0.24 | 27 | 8.27 | 8.71 | 0.15 | 0.32 | 4.0 G 1 |
| hr6801 | S | 4687 | 2.38 | 1.44 | -0.07 | 0.08 | 405 | 0.06 | 0.13 | 27 | 8.15 | 8.68 | -0.06 | 0.41 | 0.26 | 4687 | 2.11 | 1.50 | -0.11 | 0.09 | 405 | -0.11 | 0.13 | 27 | 8.07 | 8.56 | -0.06 | 0.28 | 4.3 G 1 |
| hr6840 | S | 4898 | 2.66 | 1.53 | -0.67 | 0.09 | 394 | -0.37 | 0.10 | 28 | 7.83 | 8.62 | 0.00 | 0.25 | 0.18 | 4898 | 1.97 | 1.63 | -0.71 | 0.10 | 394 | -0.71 | 0.11 | 28 | 7.65 | 8.30 | 0.00 | 0.21 | 5.8 G 1 |
| hr6853 | S | 4798 | 2.49 | 1.48 | -0.40 | 0.08 | 401 | -0.35 | 0.10 | 24 | 8.11 | 8.78 | -0.35 | 0.15 | 0.21 | 4798 | 2.37 | 1.51 | -0.42 | 0.08 | 401 | -0.42 | 0.10 | 24 | 8.07 | 8.73 | -0.35 | 0.22 | 4.7 G 1 |
| hr6859 | S | 4202 | 1.16 | 2.11 | -0.07 | 0.15 | 327 | 0.08 | 0.25 | 25 | 8.12 | 8.69 | -0.18 | 1.87 | 0.43 | 4202 | 0.78 | 2.09 | -0.14 | 0.16 | 327 | -0.14 | 0.25 | 25 | 8.01 | 8.53 | -0.21 | 0.49 | 6.1 G 1 |
| hr6865 | S | 4646 | 2.65 | 1.11 | 0.02 | 0.08 | 402 | 0.13 | 0.14 | 27 | 8.22 | 8.69 | -0.01 | 0.53 | 0.23 | 4646 | 2.38 | 1.25 | -0.06 | 0.09 | 402 | -0.06 | 0.15 | 27 | 8.14 | 8.56 | -0.01 | 0.26 | 3.8 G 1 |
| hr6866 | S | 4982 | 2.59 | 1.36 | -0.06 | 0.08 | 400 | -0.01 | 0.08 | 27 | 8.12 | 8.57 | 0.82 | 1.22 | 0.17 | 4982 | 2.48 | 1.38 | -0.07 | 0.08 | 400 | -0.07 | 0.08 | 27 | 8.10 | 8.51 | 0.82 | 0.17 | 4.4 G 1 |
| hr6872 | S | 4549 | 2.16 | 1.54 | 0.14 | 0.12 | 365 | 0.18 | 0.18 | 27 | 8.30 | 8.79 | 0.18 | 1.14 | 0.30 | 4549 | 2.06 | 1.56 | 0.12 | 0.12 | 365 | 0.12 | 0.18 | 27 | 8.27 | 8.74 | 0.18 | 0.31 | 4.9 G 1 |
| hr6884 | S | 4894 | 2.67 | 1.33 | -0.03 | 0.08 | 403 | 0.05 | 0.12 | 29 | 8.17 | 8.65 | 0.62 | 1.01 | 0.18 | 4894 | 2.50 | 1.38 | -0.06 | 0.08 | 403 | -0.06 | 0.13 | 29 | 8.12 | 8.57 | 0.62 | 0.19 | 4.6 G 1 |
| hr6895 | S | 4484 | 2.43 | 1.19 | 0.04 | 0.11 | 374 | 0.21 | 0.18 | 27 | 8.42 | 8.85 | -0.08 | 0.81 | 0.29 | 4484 | 2.02 | 1.37 | -0.08 | 0.11 | 374 | -0.08 | 0.19 | 27 | 8.27 | 8.66 | -0.08 | 0.32 | 3.9 G 1 |
| hr6935 | S | 4833 | 2.77 | 1.18 | 0.06 | 0.08 | 401 | 0.21 | 0.11 | 31 | 8.25 | 8.68 | 0.30 | 0.60 | 0.19 | 4833 | 2.44 | 1.33 | -0.01 | 0.09 | 401 | 0.00 | 0.11 | 31 | 8.15 | 8.52 | 0.31 | 0.21 | 3.9 G 1 |
| hr6945 | S | 4413 | 1.94 | 1.45 | -0.40 | 0.09 | 389 | -0.33 | 0.15 | 26 | 7.94 | 8.50 | -0.04 | 1.13 | 0.33 | 4413 | 1.77 | 1.48 | -0.43 | 0.09 | 389 | -0.43 | 0.15 | 26 | 7.88 | 8.42 | -0.04 | 0.35 | 4.4 G 1 |
| hr6970 | S | 5008 | 2.55 | 1.38 | -0.16 | 0.19 | 444 | -0.07 | 0.12 | 34 | 8.10 | 8.51 | 1.26 | 2.96 | 0.16 | 5008 | 2.35 | 1.42 | -0.18 | 0.19 | 444 | -0.18 | 0.12 | 34 | 8.05 | 8.42 | 1.26 | 0.17 | 5.0 G 1 |
| hr6973 | S | 4241 | 1.86 | 1.62 | 0.04 | 0.14 | 346 | 0.25 | 0.21 | 22 | 8.43 | 8.82 | -0.43 | 0.88 | 0.35 | 4241 | 1.34 | 1.66 | -0.09 | 0.15 | 346 | -0.09 | 0.22 | 22 | 8.24 | 8.59 | -0.45 | 0.40 | 4.7 G 1 |
| hr7010 | S | 5085 | 2.74 | 1.46 | -0.03 | 0.07 | 403 | -0.12 | 0.08 | 26 | 8.14 | 8.80 | 0.57 | 0.54 | 0.14 | 5085 | 2.93 | 1.39 | -0.01 | 0.07 | 403 | -0.01 | 0.08 | 26 | 8.20 | 8.89 | 0.57 | 0.14 | 4.2 G 1 |

| ID | | | | | | | | | | | | | | | | | | | | | | | | | | | | | |
|---|---|---|---|---|---|---|---|---|---|---|---|---|---|---|---|---|---|---|---|---|---|---|---|---|---|---|---|---|---|
| hr7042 | S | 4943 | 2.76 | 1.36 | 0.21 | 0.09 | 400 | 0.24 | 0.13 | 32 | 8.39 | 8.85 | 0.56 | 0.76 | 0.17 | 4943 | 2.70 | 1.39 | 0.20 | 0.09 | 400 | 0.20 | 0.13 | 32 | 8.37 | 8.82 | 0.56 | 0.17 | 4.6 | G | 1 |
| hr7064 | S | 4438 | 2.20 | 1.42 | 0.02 | 0.11 | 387 | 0.15 | 0.15 | 25 | 8.24 | 8.74 | 0.05 | 1.26 | 0.31 | 4438 | 1.88 | 1.51 | -0.07 | 0.12 | 387 | -0.07 | 0.16 | 25 | 8.14 | 8.60 | 0.05 | 0.34 | 4.3 | G | 1 |
| hr7120 | S | 4244 | 1.97 | 1.69 | 0.03 | 0.18 | 347 | 0.37 | 0.24 | 23 | 8.64 | 8.96 | -0.10 | 1.81 | 0.34 | 4244 | 1.11 | 1.74 | -0.18 | 0.18 | 347 | -0.17 | 0.26 | 23 | 8.29 | 8.57 | -0.13 | 0.43 | 5.2 | G | 1 |
| hr7125 | S | 4354 | 1.70 | 2.19 | -0.49 | 0.21 | 241 | -0.36 | 0.20 | 15 | 7.72 | 8.40 | -0.69 | 0.32 | 0.36 | 4354 | 1.39 | 2.18 | -0.54 | 0.22 | 241 | -0.53 | 0.20 | 15 | 7.63 | 8.27 | -0.70 | 0.39 | 16.0 | R | 1 |
| hr7135 | S | 4666 | 2.50 | 1.32 | -0.14 | 0.08 | 405 | -0.03 | 0.09 | 25 | 8.17 | 8.66 | 1.10 | 5.45 | 0.25 | 4666 | 2.24 | 1.40 | -0.19 | 0.09 | 405 | -0.19 | 0.09 | 25 | 8.09 | 8.54 | 1.10 | 0.27 | 4.2 | G | 1 |
| hr7137 | S | 5024 | 2.36 | 1.48 | -0.02 | 0.09 | 405 | -0.10 | 0.10 | 28 | 8.03 | 8.45 | 1.32 | 3.20 | 0.17 | 5024 | 2.54 | 1.44 | 0.00 | 0.09 | 405 | 0.00 | 0.10 | 28 | 8.07 | 8.53 | 1.32 | 0.16 | 5.2 | G | 1 |
| hr7144 | S | 4951 | 2.53 | 1.57 | -0.01 | 0.09 | 174 | 0.05 | 0.08 | 14 | | | | | 0.17 | 4951 | 2.40 | 1.61 | -0.02 | 0.09 | 174 | -0.02 | 0.08 | 14 | | | | 0.18 | 4.0 | G | 1 |
| hr7146 | S | 4803 | 2.55 | 1.38 | 0.09 | 0.09 | 416 | 0.08 | 0.14 | 27 | 8.25 | 8.69 | 0.51 | 1.04 | 0.20 | 4803 | 2.58 | 1.37 | 0.10 | 0.09 | 416 | 0.10 | 0.14 | 27 | 8.26 | 8.71 | 0.51 | 0.20 | 4.3 | G | 1 |
| hr7148 | S | 4710 | 2.52 | 1.41 | 0.14 | 0.10 | 390 | 0.17 | 0.15 | 25 | 8.33 | 8.81 | 0.32 | 0.90 | 0.23 | 4710 | 2.35 | 1.51 | 0.13 | 0.10 | 390 | 0.13 | 0.15 | 25 | 8.38 | 8.82 | 0.32 | 0.25 | 4.5 | G | 1 |
| hr7176 | S | 4692 | 2.47 | 1.47 | 0.20 | 0.11 | 381 | 0.27 | 0.18 | 28 | 8.38 | 8.88 | 0.42 | 1.20 | 0.25 | 4692 | 2.31 | 1.51 | 0.17 | 0.11 | 381 | 0.17 | 0.18 | 28 | 8.34 | 8.81 | 0.42 | 0.26 | 4.8 | G | 1 |
| hr7180 | S | 4561 | 2.10 | 1.55 | -0.01 | 0.10 | 388 | -0.01 | 0.11 | 24 | 8.15 | 8.67 | 0.20 | 1.14 | 0.30 | 4561 | 2.11 | 1.54 | -0.01 | 0.10 | 388 | -0.01 | 0.11 | 24 | 8.15 | 8.68 | 0.20 | 0.30 | 4.8 | G | 1 |
| hr7186 | S | 4991 | 2.56 | 1.59 | 0.09 | 0.08 | 392 | 0.14 | 0.12 | 31 | 8.18 | 8.67 | 0.71 | 0.94 | 0.16 | 4991 | 2.46 | 1.61 | 0.08 | 0.09 | 392 | 0.08 | 0.12 | 31 | 8.16 | 8.62 | 0.71 | 0.17 | 5.1 | G | 1 |
| hr7187 | S | 4952 | 2.71 | 1.42 | 0.05 | 0.10 | 398 | 0.10 | 0.14 | 29 | 8.22 | 8.69 | 1.25 | 3.36 | 0.17 | 4952 | 2.61 | 1.46 | 0.03 | 0.10 | 398 | 0.03 | 0.14 | 29 | 8.19 | 8.64 | 1.25 | 0.17 | 5.9 | G | 2 |
| hr7193 | S | 4601 | 2.47 | 1.27 | 0.02 | 0.09 | 401 | 0.15 | 0.17 | 29 | 8.25 | 8.71 | 0.35 | 1.41 | 0.27 | 4601 | 2.15 | 1.38 | -0.05 | 0.10 | 401 | -0.05 | 0.17 | 29 | 8.15 | 8.56 | 0.35 | 0.29 | 4.0 | G | 1 |
| hr7196 | S | 4824 | 2.57 | 1.39 | -0.15 | 0.07 | 409 | -0.06 | 0.09 | 29 | 8.17 | 8.70 | 0.31 | 0.62 | 0.20 | 4824 | 2.37 | 1.44 | -0.18 | 0.08 | 409 | -0.18 | 0.09 | 29 | 8.11 | 8.61 | 0.31 | 0.21 | 4.1 | G | 1 |
| hr7204 | S | 4869 | 2.63 | 1.34 | 0.08 | 0.08 | 407 | 0.11 | 0.09 | 30 | 8.20 | 8.64 | 0.35 | 0.59 | 0.19 | 4869 | 2.56 | 1.37 | 0.06 | 0.08 | 407 | 0.06 | 0.09 | 30 | 8.18 | 8.60 | 0.35 | 0.19 | 4.0 | G | 1 |
| hr7217 | S | 4766 | 2.52 | 1.41 | -0.03 | 0.09 | 409 | 0.06 | 0.11 | 27 | 8.21 | 8.70 | 0.34 | 0.81 | 0.21 | 4766 | 2.31 | 1.46 | -0.06 | 0.09 | 409 | -0.06 | 0.11 | 27 | 8.15 | 8.60 | 0.35 | 0.23 | 4.2 | G | 1 |
| hr7225a | S | 4496 | 2.27 | 1.35 | -0.17 | 0.08 | 400 | 0.08 | 0.11 | 26 | 8.15 | 8.74 | 0.91 | 6.33 | 0.30 | 4496 | 1.80 | 1.42 | -0.32 | 0.09 | 400 | -0.32 | 0.12 | 26 | 7.86 | 8.40 | 0.92 | 0.34 | 4.1 | G | 1 |
| hr7225b | S | 4156 | 1.91 | 1.63 | 0.30 | 0.17 | 335 | 0.71 | 0.31 | 24 | 8.71 | 9.08 | -0.37 | 1.28 | 0.34 | 4156 | 0.97 | 1.69 | 0.00 | 0.18 | 335 | 0.00 | 0.32 | 24 | 8.25 | 8.57 | -0.44 | 0.46 | 5.2 | G | 1 |
| hr7234 | S | 4444 | 2.17 | 1.37 | -0.09 | 0.10 | 379 | 0.06 | 0.17 | 26 | 8.21 | 8.71 | -0.32 | 0.54 | 0.31 | 4444 | 1.79 | 1.46 | -0.18 | 0.11 | 379 | -0.18 | 0.18 | 26 | 8.09 | 8.53 | -0.32 | 0.35 | 4.1 | G | 1 |
| hr7295 | S | 4875 | 2.53 | 1.64 | 0.16 | 0.12 | 377 | 0.24 | 0.15 | 28 | 8.40 | 8.87 | 1.21 | 3.82 | 0.19 | 4875 | 2.46 | 1.63 | 0.14 | 0.12 | 377 | 0.14 | 0.16 | 28 | 8.30 | 8.75 | 1.21 | 0.20 | 6.4 | G | 1 |
| hr7310 | S | 4776 | 2.52 | 1.37 | -0.12 | 0.08 | 409 | -0.02 | 0.09 | 26 | 8.13 | 8.65 | 0.56 | 1.27 | 0.21 | 4776 | 2.30 | 1.42 | -0.15 | 0.08 | 409 | -0.15 | 0.09 | 26 | 8.07 | 8.54 | 0.57 | 0.23 | 4.4 | G | 1 |
| hr7325 | S | 4753 | 2.45 | 1.44 | -0.11 | 0.08 | 389 | -0.05 | 0.10 | 27 | 8.11 | 8.67 | 0.24 | 0.66 | 0.22 | 4753 | 2.32 | 1.47 | -0.13 | 0.08 | 389 | -0.13 | 0.10 | 27 | 8.07 | 8.61 | 0.24 | 0.23 | 4.1 | G | 1 |
| hr7331 | S | 7262 | 3.43 | 3.45 | 0.37 | 0.38 | 79 | 0.06 | 0.12 | 5 | 8.36 | 8.17 | 2.47 | 0.67 | 0.14 | 7262 | 4.30 | 3.35 | 0.35 | 0.38 | 79 | 0.35 | 0.12 | 5 | 8.55 | 8.89 | 2.44 | -0.21 | 46.0 | R | 2 |
| hr7349 | S | 4662 | 2.47 | 1.39 | -0.01 | 0.09 | 389 | 0.15 | 0.15 | 31 | 8.24 | 8.79 | 0.05 | 0.58 | 0.25 | 4662 | 2.10 | 1.49 | -0.08 | 0.09 | 389 | -0.09 | 0.16 | 31 | 8.13 | 8.62 | 0.06 | 0.28 | 4.7 | G | 1 |
| hr7359 | S | 4838 | 2.63 | 1.31 | 0.15 | 0.09 | 403 | 0.13 | 0.15 | 30 | 8.28 | 8.76 | 0.92 | 2.30 | 0.19 | 4838 | 2.57 | 1.39 | 0.15 | 0.09 | 403 | 0.16 | 0.15 | 30 | 8.33 | 8.82 | 0.92 | 0.19 | 4.0 | G | 1 |
| hr7376 | S | 4679 | 2.64 | 1.17 | -0.08 | 0.09 | 408 | -0.02 | 0.16 | 29 | 8.32 | 8.79 | 0.31 | 0.99 | 0.23 | 4679 | 2.50 | 1.24 | -0.11 | 0.09 | 408 | -0.12 | 0.16 | 29 | 8.28 | 8.73 | 0.31 | 0.25 | 3.8 | G | 1 |
| hr7385 | S | 4763 | 2.50 | 1.36 | -0.07 | 0.08 | 412 | 0.01 | 0.09 | 24 | 8.13 | 8.65 | 0.32 | 0.77 | 0.22 | 4763 | 2.31 | 1.41 | -0.10 | 0.09 | 412 | -0.10 | 0.09 | 24 | 8.08 | 8.57 | 0.32 | 0.23 | 4.3 | G | 1 |
| hr7389 | S | 6281 | 3.64 | 3.27 | -0.01 | 0.20 | 171 | 0.00 | 0.09 | 15 | 8.11 | 8.84 | 2.70 | 4.93 | 0.02 | 6281 | 3.61 | 3.27 | -0.01 | 0.20 | 171 | -0.01 | 0.09 | 15 | 8.10 | 8.83 | 2.70 | 0.02 | 22.0 | R | 2 |
| hr7407 | S | 4740 | 2.54 | 1.38 | 0.09 | 0.09 | 400 | 0.15 | 0.14 | 31 | 8.19 | 8.76 | 2.06 | 19.45 | 0.22 | 4740 | 2.40 | 1.43 | 0.06 | 0.09 | 400 | 0.06 | 0.14 | 31 | 8.15 | 8.70 | 2.06 | 0.24 | 4.1 | G | 1 |
| hr7433 | S | 4631 | 2.61 | 1.27 | -0.01 | 0.09 | 394 | 0.26 | 0.11 | 27 | 8.26 | 8.82 | -0.13 | 0.43 | 0.24 | 4631 | 2.00 | 1.47 | -0.14 | 0.10 | 394 | -0.14 | 0.12 | 27 | 8.07 | 8.54 | -0.13 | 0.30 | 4.2 | G | 1 |
| hr7449 | S | 4810 | 2.67 | 1.38 | 0.10 | 0.11 | 403 | 0.17 | 0.14 | 25 | 8.45 | 8.83 | 0.55 | 1.13 | 0.20 | 4810 | 2.53 | 1.43 | 0.08 | 0.11 | 403 | 0.08 | 0.15 | 25 | 8.40 | 8.76 | 0.56 | 0.20 | 4.6 | G | 1 |
| hr7465 | S | 4730 | 2.41 | 1.34 | -0.13 | 0.08 | 408 | -0.12 | 0.10 | 26 | 8.10 | 8.65 | 0.52 | 1.33 | 0.24 | 4730 | 2.37 | 1.34 | -0.14 | 0.08 | 408 | -0.14 | 0.10 | 26 | 8.09 | 8.63 | 0.52 | 0.25 | 4.1 | G | 1 |
| hr7478b | S | 4872 | 2.51 | 0.80 | -0.27 | 0.30 | 393 | -0.19 | 0.38 | 29 | 8.05 | 8.57 | 0.38 | 0.64 | 0.19 | 4872 | 2.33 | 0.95 | -0.32 | 0.31 | 393 | -0.32 | 0.38 | 29 | 7.99 | 8.48 | 0.38 | 0.20 | 7.3 | G | 1 |
| hr7487 | S | 5052 | 2.87 | 1.33 | 0.06 | 0.07 | 414 | 0.11 | 0.09 | 33 | 8.17 | 8.70 | 0.71 | 0.80 | 0.15 | 5052 | 2.75 | 1.37 | 0.04 | 0.07 | 414 | 0.04 | 0.09 | 33 | 8.13 | 8.64 | 0.71 | 0.16 | 4.4 | G | 1 |
| hr7506 | S | 5008 | 2.69 | 1.48 | 0.05 | 0.09 | 409 | 0.06 | 0.11 | 30 | 8.07 | 8.65 | 0.64 | 0.77 | 0.16 | 5008 | 2.66 | 1.48 | 0.05 | 0.09 | 409 | 0.05 | 0.11 | 30 | 8.06 | 8.64 | 0.64 | 0.16 | 5.6 | G | 1 |
| hr7517 | S | 4920 | 2.55 | 1.40 | 0.01 | 0.08 | 401 | 0.09 | 0.12 | 32 | 8.12 | 8.63 | 0.31 | 0.47 | 0.18 | 4920 | 2.39 | 1.45 | -0.01 | 0.08 | 401 | -0.01 | 0.12 | 32 | 8.08 | 8.55 | 0.32 | 0.19 | 4.9 | G | 1 |
| hr7526 | S | 5035 | 3.08 | 0.99 | -0.23 | 0.07 | 411 | -0.27 | 0.07 | 25 | 7.92 | 8.43 | 1.23 | 2.70 | 0.14 | 5035 | 3.04 | 1.13 | -0.22 | 0.07 | 411 | -0.22 | 0.06 | 25 | 7.98 | 8.54 | 1.24 | 0.14 | 3.5 | G | 1 |
| hr7540 | S | 4959 | 2.64 | 1.40 | -0.06 | 0.07 | 403 | -0.14 | 0.10 | 28 | 8.17 | 8.74 | 0.77 | 1.18 | 0.17 | 4959 | 2.82 | 1.34 | -0.04 | 0.07 | 403 | -0.04 | 0.10 | 28 | 8.23 | 8.82 | 0.77 | 0.16 | 3.9 | G | 1 |
| hr7541 | S | 4421 | 2.39 | 1.42 | 0.26 | 0.15 | 364 | 0.43 | 0.25 | 25 | 8.61 | 8.92 | 0.16 | 1.68 | 0.30 | 4421 | 2.06 | 1.47 | 0.13 | 0.15 | 364 | 0.13 | 0.26 | 25 | 8.39 | 8.68 | 0.15 | 0.32 | 4.6 | G | 1 |
| hr7561 | S | 4779 | 2.47 | 0.98 | -0.09 | 0.10 | 410 | -0.18 | 0.17 | 28 | 8.12 | 8.56 | 0.42 | 0.91 | 0.22 | 4779 | 2.68 | 0.83 | -0.04 | 0.11 | 410 | -0.03 | 0.16 | 28 | 8.19 | 8.66 | 0.42 | 0.20 | 4.5 | G | 1 |
| hr7576 | S | 4337 | 2.28 | 1.69 | 0.42 | 0.19 | 335 | 0.75 | 0.33 | 26 | 8.79 | 9.10 | 0.26 | 2.79 | 0.31 | 4337 | 1.48 | 1.79 | 0.20 | 0.20 | 335 | 0.21 | 0.34 | 26 | 8.47 | 8.73 | 0.24 | 0.38 | 4.9 | G | 1 |
| hr7597 | S | 5306 | 3.53 | 0.94 | -0.05 | 0.06 | 365 | 0.03 | 0.06 | 28 | 8.30 | 8.58 | 2.04 | 7.17 | 0.09 | 5306 | 3.38 | 1.07 | -0.07 | 0.06 | 365 | -0.07 | 0.05 | 28 | 8.26 | 8.51 | 2.04 | 0.09 | 4.2 | G | 1 |
| hr7681 | S | 4519 | 2.52 | 1.25 | 0.11 | 0.11 | 167 | 0.52 | 0.15 | 14 | | | | | 0.27 | 4519 | 1.73 | 1.58 | -0.15 | 0.12 | 167 | -0.15 | 0.17 | 14 | | | | 0.34 | 4.0 | G | 1 |
| hr7712 | S | 4439 | 2.21 | 1.42 | -0.03 | 0.10 | 389 | 0.28 | 0.11 | 27 | 8.22 | 8.74 | -0.13 | 0.84 | 0.31 | 4439 | 1.48 | 1.53 | -0.18 | 0.11 | 389 | -0.18 | 0.12 | 27 | 7.99 | 8.41 | -0.14 | 0.37 | 4.6 | G | 1 |
| hr7713 | S | 5060 | 2.72 | 1.41 | -0.40 | 0.06 | 421 | -0.39 | 0.07 | 30 | 7.86 | 8.54 | 0.12 | 0.21 | 0.15 | 5060 | 2.71 | 1.42 | -0.40 | 0.06 | 421 | -0.40 | 0.07 | 30 | 7.86 | 8.53 | 0.12 | 0.15 | 4.0 | G | 1 |
| hr7748 | S | 4782 | 2.57 | 1.33 | -0.08 | 0.08 | 429 | 0.03 | 0.09 | 27 | 8.16 | 8.72 | 0.36 | 0.79 | 0.21 | 4782 | 2.34 | 1.40 | -0.12 | 0.09 | 429 | -0.12 | 0.09 | 27 | 8.10 | 8.61 | 0.36 | 0.22 | 4.0 | G | 1 |
| hr7778 | S | 5080 | 2.64 | 1.45 | 0.04 | 0.07 | 329 | -0.01 | 0.10 | 24 | 8.11 | 8.71 | 0.30 | 0.29 | 0.15 | 5080 | 2.75 | 1.41 | 0.05 | 0.07 | 329 | 0.05 | 0.10 | 24 | 8.15 | 8.76 | 0.30 | 0.14 | 4.7 | G | 1 |
| hr7788 | S | 4969 | 2.61 | 1.51 | -0.12 | 0.10 | 402 | 0.17 | 0.09 | 29 | 8.09 | 8.59 | 0.44 | 0.55 | 0.17 | 4969 | 1.96 | 1.60 | -0.17 | 0.12 | 402 | -0.17 | 0.10 | 29 | 7.96 | 8.28 | 0.45 | 0.20 | 6.3 | G | 1 |
| hr7794 | S | 4866 | 2.73 | 1.25 | 0.03 | 0.08 | 401 | 0.12 | 0.08 | 28 | 8.23 | 8.74 | 0.20 | 0.43 | 0.18 | 4866 | 2.53 | 1.34 | -0.01 | 0.08 | 401 | -0.01 | 0.09 | 28 | 8.17 | 8.64 | 0.20 | 0.19 | 4.2 | G | 1 |

| ID | | | | | | | | | | | | | | | | | | | | | | | | | | | | |
|---|---|---|---|---|---|---|---|---|---|---|---|---|---|---|---|---|---|---|---|---|---|---|---|---|---|---|---|---|
| hr7802 | S | 4827 | 2.75 | 1.25 | 0.25 | 0.09 | 400 | 0.34 | 0.15 | 29 | 8.40 | 8.82 | 0.44 | 0.84 | 0.19 | 4827 | 2.56 | 1.35 | 0.20 | 0.09 | 400 | 0.20 | 0.15 | 29 | 8.34 | 8.73 | 0.45 | 0.20 | 4.2 | G | 1 |
| hr7806 | S | 4211 | 1.86 | 1.55 | 0.08 | 0.15 | 346 | 0.31 | 0.21 | 25 | 8.30 | 8.75 | 0.64 | 9.00 | 0.35 | 4211 | 1.28 | 1.62 | -0.08 | 0.15 | 346 | -0.08 | 0.22 | 25 | 8.10 | 8.50 | 0.61 | 0.41 | 4.7 | G | 1 |
| hr7820 | S | 4703 | 2.74 | 1.25 | 0.03 | 0.09 | 421 | 0.22 | 0.12 | 26 | 8.23 | 8.79 | 3.15 | 35.81 | 0.22 | 4703 | 2.31 | 1.42 | -0.07 | 0.09 | 421 | -0.07 | 0.12 | 26 | 8.10 | 8.59 | 3.10 | 0.26 | 4.1 | G | 1 |
| hr7824 | S | 5029 | 2.79 | 1.29 | -0.10 | 0.07 | 417 | -0.04 | 0.06 | 28 | 8.07 | 8.68 | 0.47 | 0.49 | 0.15 | 5029 | 2.66 | 1.33 | -0.11 | 0.07 | 417 | -0.11 | 0.06 | 28 | 8.03 | 8.62 | 0.47 | 0.15 | 4.6 | G | 1 |
| hr7831 | S | 4592 | 2.51 | 1.48 | 0.28 | 0.13 | 365 | 0.39 | 0.18 | 26 | 8.62 | 9.04 | 0.35 | 1.45 | 0.26 | 4592 | 2.24 | 1.55 | 0.21 | 0.13 | 365 | 0.21 | 0.18 | 26 | 8.53 | 8.92 | 0.35 | 0.29 | 4.5 | G | 1 |
| hr7854 | S | 4800 | 2.57 | 1.28 | -0.04 | 0.08 | 410 | -0.05 | 0.12 | 29 | 8.18 | 8.63 | 1.32 | 5.80 | 0.20 | 4800 | 2.59 | 1.26 | -0.03 | 0.08 | 410 | -0.03 | 0.11 | 29 | 8.19 | 8.64 | 1.32 | 0.20 | 4.1 | G | 1 |
| hr7897 | S | 4715 | 2.57 | 1.37 | 0.12 | 0.10 | 381 | 0.18 | 0.13 | 26 | 8.25 | 8.73 | 0.44 | 1.17 | 0.23 | 4715 | 2.43 | 1.42 | 0.09 | 0.09 | 381 | 0.09 | 0.14 | 26 | 8.21 | 8.66 | 0.44 | 0.25 | 4.6 | G | 1 |
| hr7904 | S | 4728 | 2.44 | 1.39 | 0.03 | 0.09 | 399 | 0.05 | 0.13 | 29 | 8.21 | 8.76 | 0.29 | 0.81 | 0.24 | 4728 | 2.40 | 1.40 | 0.03 | 0.09 | 399 | 0.03 | 0.13 | 29 | 8.20 | 8.74 | 0.29 | 0.25 | 4.4 | G | 1 |
| hr7905 | S | 4760 | 2.56 | 1.32 | -0.27 | 0.07 | 418 | -0.18 | 0.08 | 26 | 8.06 | 8.58 | 0.88 | 2.69 | 0.21 | 4760 | 2.34 | 1.40 | -0.31 | 0.07 | 418 | -0.31 | 0.08 | 26 | 7.99 | 8.47 | 0.88 | 0.23 | 3.9 | G | 1 |
| hr7923 | S | 4916 | 2.61 | 1.35 | -0.30 | 0.07 | 410 | -0.29 | 0.06 | 29 | 7.89 | 8.47 | 0.40 | 0.60 | 0.18 | 4916 | 2.60 | 1.35 | -0.30 | 0.07 | 410 | -0.30 | 0.06 | 29 | 7.89 | 8.47 | 0.40 | 0.18 | 4.1 | G | 1 |
| hr7939 | S | 4473 | 2.01 | 1.54 | -0.06 | 0.10 | 400 | 0.01 | 0.14 | 26 | 8.22 | 8.76 | -0.11 | 0.78 | 0.32 | 4473 | 1.86 | 1.56 | -0.09 | 0.11 | 400 | -0.09 | 0.15 | 26 | 8.17 | 8.69 | -0.11 | 0.34 | 4.5 | G | 1 |
| hr7942 | S | 4701 | 2.27 | 1.42 | -0.07 | 0.09 | 417 | -0.05 | 0.10 | 31 | 8.09 | 8.63 | 0.60 | 1.73 | 0.26 | 4701 | 2.24 | 1.43 | -0.07 | 0.09 | 417 | -0.07 | 0.10 | 31 | 8.08 | 8.61 | 0.60 | 0.26 | 4.5 | G | 1 |
| hr7962 | S | 4623 | 2.40 | 1.36 | -0.12 | 0.10 | 398 | -0.13 | 0.11 | 27 | 8.30 | 8.79 | 0.13 | 0.79 | 0.27 | 4623 | 2.41 | 1.35 | -0.12 | 0.10 | 398 | -0.12 | 0.11 | 27 | 8.30 | 8.80 | 0.13 | 0.27 | 3.9 | G | 1 |
| hr7995 | S | 5155 | 2.86 | 1.51 | -0.08 | 0.10 | 402 | -0.02 | 0.09 | 28 | 8.14 | 8.65 | 0.00 | 0.12 | 0.13 | 5155 | 2.72 | 1.55 | -0.09 | 0.10 | 402 | -0.10 | 0.09 | 28 | 8.11 | 8.58 | 0.00 | 0.13 | 6.7 | G | 1 |
| hr8000 | S | 4525 | 2.34 | 1.35 | -0.47 | 0.08 | 401 | -0.39 | 0.12 | 24 | 8.10 | 8.78 | -0.50 | 0.26 | 0.29 | 4525 | 2.14 | 1.41 | -0.51 | 0.09 | 401 | -0.51 | 0.12 | 24 | 8.02 | 8.69 | -0.50 | 0.31 | 3.8 | G | 1 |
| hr8017 | S | 4441 | 2.47 | 1.45 | 0.35 | 0.14 | 371 | 0.58 | 0.21 | 23 | 8.63 | 8.98 | 0.31 | 2.16 | 0.29 | 4441 | 1.90 | 1.59 | 0.18 | 0.15 | 371 | 0.18 | 0.21 | 23 | 8.42 | 8.73 | 0.30 | 0.34 | 4.5 | G | 1 |
| hr8030 | S | 4966 | 2.89 | 1.20 | 0.03 | 0.07 | 417 | 0.06 | 0.09 | 29 | 8.18 | 8.62 | 0.61 | 0.81 | 0.16 | 4966 | 2.81 | 1.24 | 0.02 | 0.07 | 417 | 0.01 | 0.09 | 29 | 8.16 | 8.58 | 0.61 | 0.16 | 4.1 | G | 1 |
| hr8034a | S | 6223 | 3.74 | 2.46 | -0.05 | 0.26 | 90 | -0.10 | 0.07 | | | | 1.20 | 0.20 | 0.01 | 6223 | 3.87 | 2.41 | -0.05 | 0.26 | 90 | -0.05 | 0.08 | 4 | 7.81 | 9.18 | 1.19 | 0.00 | 50.0 | R | 2 |
| hr8034b | S | 6399 | 3.89 | 1.59 | 0.01 | 0.07 | 240 | -0.16 | 0.07 | 26 | 8.26 | 8.63 | 2.11 | 1.20 | -0.01 | 6399 | 4.29 | 1.44 | 0.02 | 0.07 | 240 | 0.02 | 0.08 | 26 | 8.39 | 8.75 | 2.11 | -0.07 | 8.7 | G | 2 |
| hr8035 | S | 4926 | 2.85 | 1.29 | 0.18 | 0.09 | 391 | 0.25 | 0.13 | 29 | 8.35 | 8.82 | 0.63 | 0.94 | 0.17 | 4926 | 2.80 | 1.23 | 0.16 | 0.09 | 391 | 0.16 | 0.13 | 29 | 8.26 | 8.70 | 0.62 | 0.17 | 4.5 | G | 1 |
| hr8072 | S | 4887 | 2.82 | 1.33 | 0.22 | 0.10 | 397 | 0.34 | 0.17 | 28 | 8.43 | 8.83 | 1.41 | 5.47 | 0.18 | 4887 | 2.55 | 1.43 | 0.17 | 0.10 | 397 | 0.17 | 0.17 | 28 | 8.35 | 8.70 | 1.41 | 0.19 | 5.3 | G | 1 |
| hr8082 | S | 4819 | 2.52 | 1.39 | 0.00 | 0.08 | 407 | -0.03 | 0.10 | 27 | 8.16 | 8.75 | 0.44 | 0.85 | 0.20 | 4819 | 2.57 | 1.37 | 0.01 | 0.08 | 407 | 0.01 | 0.10 | 27 | 8.17 | 8.77 | 0.44 | 0.20 | 4.0 | G | 1 |
| hr8093 | S | 4938 | 2.87 | 1.24 | 0.04 | 0.08 | 405 | 0.13 | 0.11 | 28 | 8.16 | 8.59 | 0.41 | 0.56 | 0.16 | 4938 | 2.70 | 1.32 | 0.01 | 0.08 | 405 | 0.02 | 0.11 | 28 | 8.11 | 8.51 | 0.41 | 0.17 | 4.4 | G | 1 |
| hr8096 | S | 4563 | 2.59 | 1.41 | 0.28 | 0.13 | 372 | 0.49 | 0.19 | 26 | 8.62 | 9.06 | 0.16 | 1.04 | 0.26 | 4563 | 2.10 | 1.55 | 0.16 | 0.13 | 372 | 0.15 | 0.20 | 26 | 8.45 | 8.84 | 0.16 | 0.30 | 4.3 | G | 1 |
| hr8115 | S | 4891 | 2.45 | 1.50 | 0.01 | 0.10 | 395 | 0.09 | 0.13 | 29 | 8.31 | 8.67 | 0.56 | 0.89 | 0.19 | 4891 | 2.26 | 1.54 | -0.02 | 0.10 | 395 | -0.01 | 0.13 | 29 | 8.26 | 8.58 | 0.56 | 0.20 | 4.8 | G | 1 |
| hr8165 | S | 4689 | 2.53 | 1.44 | -0.10 | 0.10 | 397 | -0.02 | 0.13 | 24 | 8.43 | 8.95 | 0.84 | 3.06 | 0.23 | 4689 | 2.33 | 1.49 | -0.14 | 0.10 | 397 | -0.14 | 0.13 | 24 | 8.36 | 8.86 | 0.85 | 0.26 | 4.6 | G | 1 |
| hr8167 | S | 5011 | 2.66 | 1.49 | -0.04 | 0.10 | 397 | 0.04 | 0.14 | 29 | 8.10 | 8.57 | 0.40 | 0.44 | 0.15 | 5011 | 2.48 | 1.53 | -0.06 | 0.10 | 397 | -0.06 | 0.14 | 29 | 8.05 | 8.49 | 0.40 | 0.17 | 6.6 | G | 1 |
| hr8173 | S | 4620 | 2.47 | 1.32 | 0.13 | 0.10 | 389 | 0.26 | 0.12 | 28 | 8.37 | 8.84 | 0.17 | 0.87 | 0.26 | 4620 | 2.15 | 1.44 | 0.05 | 0.10 | 389 | 0.05 | 0.14 | 28 | 8.26 | 8.69 | 0.17 | 0.29 | 4.0 | G | 1 |
| hr8179 | S | 4805 | 2.48 | 1.43 | -0.08 | 0.08 | 409 | -0.01 | 0.11 | 30 | 8.19 | 8.74 | 0.36 | 0.74 | 0.21 | 4805 | 2.32 | 1.46 | -0.11 | 0.08 | 409 | -0.11 | 0.11 | 30 | 8.14 | 8.67 | 0.37 | 0.22 | 4.1 | G | 1 |
| hr8185 | S | 4691 | 2.48 | 1.37 | 0.14 | 0.10 | 387 | 0.18 | 0.16 | 25 | 8.33 | 8.83 | 0.30 | 0.93 | 0.25 | 4691 | 2.39 | 1.40 | 0.12 | 0.10 | 387 | 0.12 | 0.16 | 25 | 8.30 | 8.79 | 0.30 | 0.25 | 4.3 | G | 1 |
| hr8191 | S | 6271 | 3.38 | 4.59 | 0.00 | 0.29 | 71 | 0.08 | 0.14 | 8 | 8.04 | 8.60 | 3.20 | 12.26 | 0.03 | 6271 | 3.15 | 4.60 | 0.00 | 0.29 | 71 | -0.01 | 0.14 | 8 | 7.96 | 8.51 | 3.20 | 0.04 | 100.0 | R | 2 |
| hr8228 | S | 4909 | 2.85 | 1.20 | 0.14 | 0.08 | 414 | 0.17 | 0.12 | 31 | 8.27 | 8.71 | 0.51 | 0.77 | 0.17 | 4909 | 2.78 | 1.25 | 0.12 | 0.08 | 414 | 0.12 | 0.13 | 31 | 8.25 | 8.67 | 0.52 | 0.17 | 3.9 | G | 1 |
| hr8252 | S | 5010 | 2.89 | 1.23 | -0.09 | 0.07 | 410 | 0.02 | 0.09 | 30 | 8.08 | 8.59 | 1.03 | 1.84 | 0.15 | 5010 | 2.66 | 1.32 | -0.12 | 0.08 | 410 | -0.12 | 0.10 | 30 | 8.02 | 8.48 | 1.04 | 0.16 | 4.4 | G | 1 |
| hr8255 | S | 4640 | 2.44 | 1.37 | 0.09 | 0.10 | 406 | 0.26 | 0.16 | 29 | 8.30 | 8.66 | 0.26 | 1.00 | 0.26 | 4640 | 2.04 | 1.49 | 0.00 | 0.10 | 406 | 0.00 | 0.16 | 29 | 8.18 | 8.47 | 0.26 | 0.29 | 4.5 | G | 1 |
| hr8274 | S | 4779 | 2.51 | 1.36 | 0.01 | 0.08 | 403 | 0.05 | 0.13 | 29 | 8.19 | 8.71 | 0.28 | 0.67 | 0.21 | 4779 | 2.43 | 1.38 | 0.00 | 0.08 | 403 | 0.00 | 0.13 | 29 | 8.16 | 8.67 | 0.28 | 0.22 | 4.0 | G | 1 |
| hr8277 | S | 4701 | 2.59 | 1.36 | 0.00 | 0.09 | 433 | 0.17 | 0.12 | 27 | 8.35 | 8.81 | 0.19 | 0.70 | 0.23 | 4701 | 2.21 | 1.46 | -0.07 | 0.09 | 433 | -0.07 | 0.12 | 27 | 8.23 | 8.63 | 0.19 | 0.27 | 4.2 | G | 1 |
| hr8288 | S | 4960 | 2.48 | 1.49 | -0.41 | 0.07 | 401 | -0.35 | 0.07 | 28 | 7.77 | 8.44 | 0.33 | 0.45 | 0.18 | 4960 | 2.33 | 1.52 | -0.42 | 0.08 | 401 | -0.42 | 0.07 | 28 | 7.73 | 8.37 | 0.33 | 0.18 | 4.4 | G | 1 |
| hr8317 | S | 4588 | 2.47 | 1.38 | 0.26 | 0.12 | 375 | 0.41 | 0.17 | 27 | 8.46 | 8.83 | 0.19 | 1.04 | 0.27 | 4588 | 2.10 | 1.49 | 0.16 | 0.12 | 375 | 0.16 | 0.18 | 27 | 8.34 | 8.66 | 0.19 | 0.30 | 4.3 | G | 1 |
| hr8320 | S | 4892 | 2.61 | 1.41 | -0.21 | 0.07 | 416 | -0.16 | 0.07 | 27 | 8.11 | 8.71 | 0.25 | 0.45 | 0.18 | 4892 | 2.64 | 1.36 | -0.23 | 0.07 | 416 | -0.23 | 0.08 | 27 | 8.02 | 8.60 | 0.24 | 0.18 | 4.2 | G | 1 |
| hr8324 | S | 4710 | 2.55 | 1.30 | 0.21 | 0.10 | 386 | 0.34 | 0.17 | 28 | 8.38 | 8.77 | 0.50 | 1.37 | 0.23 | 4710 | 2.36 | 1.31 | 0.15 | 0.10 | 386 | 0.15 | 0.18 | 28 | 8.22 | 8.59 | 0.51 | 0.25 | 4.2 | G | 1 |
| hr8325 | S | 4484 | 2.43 | 1.29 | 0.17 | 0.13 | 410 | 0.33 | 0.22 | 27 | 8.36 | 8.81 | 0.01 | 0.98 | 0.29 | 4484 | 2.03 | 1.44 | 0.05 | 0.13 | 410 | 0.05 | 0.23 | 27 | 8.22 | 8.63 | 0.01 | 0.32 | 4.2 | G | 1 |
| hr8360 | S | 4527 | 2.47 | 1.48 | 0.31 | 0.14 | 381 | 0.56 | 0.27 | 26 | 8.63 | 9.03 | 0.26 | 1.46 | 0.28 | 4527 | 1.86 | 1.60 | 0.16 | 0.15 | 381 | 0.16 | 0.28 | 26 | 8.42 | 8.75 | 0.25 | 0.33 | 4.7 | G | 1 |
| hr8391 | S | 6397 | 3.80 | 3.50 | 0.07 | 0.34 | 89 | 0.06 | 0.05 | 6 | 7.72 | 8.90 | 3.11 | 8.94 | 0.00 | 6397 | 3.81 | 3.50 | 0.07 | 0.34 | 89 | 0.07 | 0.05 | 6 | 7.72 | 8.90 | 3.11 | 0.00 | 52.0 | R | 2 |
| hr8394 | S | 4811 | 2.86 | 1.16 | -0.03 | 0.08 | 414 | 0.04 | 0.12 | 27 | 8.27 | 8.73 | 1.04 | 3.22 | 0.19 | 4811 | 2.73 | 1.23 | -0.05 | 0.08 | 414 | -0.05 | 0.13 | 27 | 8.22 | 8.67 | 1.04 | 0.19 | 3.5 | G | 1 |
| hr8401 | S | 4944 | 2.72 | 1.38 | 0.10 | 0.09 | 423 | 0.08 | 0.10 | 24 | 8.33 | 8.75 | 0.58 | 0.80 | 0.17 | 4944 | 2.76 | 1.36 | 0.11 | 0.09 | 423 | 0.11 | 0.10 | 24 | 8.35 | 8.77 | 0.58 | 0.17 | 4.8 | G | 1 |
| hr8442 | S | 5260 | 2.55 | 1.47 | 0.23 | 0.10 | 437 | -0.07 | 0.12 | 33 | 8.15 | 8.74 | 1.24 | 1.50 | 0.12 | 5260 | 3.21 | 1.23 | 0.30 | 0.11 | 437 | 0.30 | 0.12 | 33 | 8.31 | 9.05 | 1.23 | 0.10 | 5.0 | G | 1 |
| hr8453 | S | 4946 | 2.88 | 1.26 | 0.25 | 0.08 | 406 | 0.25 | 0.13 | 30 | 8.38 | 8.81 | 0.54 | 0.73 | 0.16 | 4946 | 2.88 | 1.26 | 0.25 | 0.08 | 406 | 0.25 | 0.13 | 30 | 8.38 | 8.81 | 0.54 | 0.16 | 4.3 | G | 1 |
| hr8454 | S | 6253 | 2.96 | 6.50 | -0.11 | 0.56 | 31 | -0.31 | 0.00 | 1 | 6.43 | 9.45 | 2.62 | 4.58 | 0.05 | 6253 | 3.56 | 7.75 | -0.12 | 0.56 | 31 | -0.12 | | 1 | 6.60 | 9.67 | 2.62 | 0.02 | 85.0 | R | 2 |
| hr8456 | S | 4811 | 2.54 | 1.36 | 0.14 | 0.09 | 398 | 0.12 | 0.17 | 29 | 8.18 | 8.58 | 0.49 | 0.98 | 0.20 | 4811 | 2.48 | 1.43 | 0.14 | 0.09 | 398 | 0.14 | 0.17 | 29 | 8.24 | 8.64 | 0.50 | 0.21 | 4.2 | G | 1 |
| hr8461 | S | 4950 | 3.32 | 0.84 | 0.18 | 0.09 | 409 | 0.29 | 0.09 | 28 | 8.39 | 8.77 | 0.46 | 0.62 | 0.15 | 4950 | 3.18 | 0.89 | 0.12 | 0.09 | 409 | 0.12 | 0.10 | 28 | 8.27 | 8.61 | 0.46 | 0.16 | 3.3 | G | 1 |

| ID | Type | | | | | | | | | | | | | | | | | | | | | | | | | | |
|---|---|---|---|---|---|---|---|---|---|---|---|---|---|---|---|---|---|---|---|---|---|---|---|---|---|---|---|
| hr8476 | S | 4734 | 2.52 | 1.44 | 0.21 | 0.11 | 405 | 0.21 | 0.17 | 26 | 8.51 | 8.93 | 0.51 | 1.28 | 0.23 | 4734 | 2.53 | 1.43 | 0.22 | 0.11 | 405 | 0.22 | 0.17 | 26 | 8.51 | 8.94 | 0.51 | 0.23 | 4.5 | G | 1 |
| hr8482 | S | 4503 | 2.54 | 1.21 | 0.30 | 0.13 | 403 | 0.49 | 0.20 | 26 | 8.58 | 8.92 | 0.03 | 0.96 | 0.27 | 4503 | 2.09 | 1.39 | 0.16 | 0.14 | 403 | 0.16 | 0.21 | 26 | 8.42 | 8.71 | 0.03 | 0.31 | 4.4 | G | 1 |
| hr8499 | S | 4894 | 2.63 | 1.46 | 0.13 | 0.09 | 388 | 0.18 | 0.11 | 29 | 8.25 | 8.64 | 0.40 | 0.62 | 0.18 | 4894 | 2.52 | 1.50 | 0.12 | 0.09 | 388 | 0.12 | 0.11 | 29 | 8.22 | 8.59 | 0.40 | 0.19 | 4.4 | G | 1 |
| hr8500 | S | 4544 | 2.35 | 1.30 | 0.08 | 0.11 | 413 | 0.18 | 0.18 | 27 | 8.29 | 8.77 | 0.88 | 5.09 | 0.28 | 4544 | 2.11 | 1.39 | 0.01 | 0.11 | 413 | 0.01 | 0.19 | 27 | 8.21 | 8.66 | 0.87 | 0.31 | 4.1 | G | 1 |
| hr8516 | S | 4552 | 2.60 | 1.12 | -0.09 | 0.09 | 426 | 0.11 | 0.14 | 27 | 8.31 | 8.71 | 0.05 | 0.85 | 0.26 | 4552 | 2.13 | 1.33 | -0.22 | 0.11 | 426 | -0.22 | 0.15 | 27 | 8.15 | 8.49 | 0.05 | 0.30 | 4.0 | G | 1 |
| hr8538 | S | 4711 | 2.52 | 1.18 | -0.26 | 0.08 | 434 | -0.21 | 0.12 | 30 | 7.97 | 8.51 | 0.13 | 0.60 | 0.23 | 4711 | 2.40 | 1.23 | -0.28 | 0.07 | 434 | -0.28 | 0.12 | 30 | 7.93 | 8.45 | 0.13 | 0.25 | 3.6 | G | 1 |
| hr8568 | S | 4695 | 2.64 | 1.33 | 0.28 | 0.10 | 411 | 0.40 | 0.15 | 27 | 8.42 | 8.92 | 0.30 | 0.92 | 0.23 | 4695 | 2.37 | 1.42 | 0.22 | 0.10 | 411 | 0.21 | 0.15 | 27 | 8.34 | 8.80 | 0.31 | 0.26 | 4.1 | G | 1 |
| hr8594 | S | 4942 | 2.69 | 1.48 | 0.20 | 0.10 | 382 | 0.26 | 0.17 | 32 | 8.35 | 8.75 | 1.49 | 5.51 | 0.17 | 4942 | 2.56 | 1.52 | 0.18 | 0.10 | 382 | 0.18 | 0.17 | 32 | 8.31 | 8.69 | 1.49 | 0.17 | 5.5 | G | 1 |
| hr8596 | S | 4875 | 2.81 | 1.22 | 0.14 | 0.09 | 411 | 0.15 | 0.12 | 29 | 8.27 | 8.68 | 0.55 | 0.93 | 0.18 | 4875 | 2.67 | 1.36 | 0.12 | 0.09 | 411 | 0.11 | 0.11 | 29 | 8.30 | 8.70 | 0.55 | 0.18 | 4.0 | G | 1 |
| hr8617 | S | 5338 | 2.81 | 1.18 | 0.05 | 0.09 | 422 | -0.10 | 0.10 | 31 | 7.99 | 8.59 | 0.89 | 0.59 | 0.10 | 5338 | 3.11 | 1.04 | 0.08 | 0.10 | 422 | 0.08 | 0.10 | 31 | 8.05 | 8.73 | 0.88 | 0.09 | 5.5 | G | 1 |
| hr8632 | S | 4256 | 1.63 | 1.62 | -0.14 | 0.13 | 366 | 0.06 | 0.10 | 26 | 8.11 | 8.62 | 0.75 | 9.69 | 0.37 | 4256 | 1.23 | 1.65 | -0.28 | 0.13 | 366 | -0.28 | 0.21 | 26 | 7.85 | 8.32 | 0.74 | 0.42 | 4.6 | G | 1 |
| hr8642 | S | 4584 | 2.64 | 1.51 | 0.11 | 0.13 | 371 | 0.39 | 0.17 | 24 | 8.49 | 8.84 | 3.11 | 38.57 | 0.25 | 4584 | 1.98 | 1.65 | -0.03 | 0.14 | 371 | -0.03 | 0.18 | 24 | 8.27 | 8.53 | 3.04 | 0.31 | 6.5 | G | 1 |
| hr8643 | S | 4754 | 2.67 | 1.14 | -0.36 | 0.07 | 435 | -0.26 | 0.08 | 25 | 8.01 | 8.55 | -0.36 | 0.17 | 0.21 | 4754 | 2.46 | 1.23 | -0.39 | 0.07 | 435 | -0.39 | 0.08 | 25 | 7.95 | 8.45 | -0.36 | 0.22 | 3.7 | G | 1 |
| hr8656 | S | 4904 | 2.69 | 1.40 | 0.11 | 0.10 | 385 | 0.20 | 0.12 | 28 | 8.17 | 8.71 | 0.47 | 0.70 | 0.18 | 4904 | 2.47 | 1.48 | 0.07 | 0.10 | 385 | 0.07 | 0.12 | 28 | 8.10 | 8.60 | 0.47 | 0.19 | 5.0 | G | 1 |
| hr8660 | S | 4738 | 2.62 | 1.33 | -0.02 | 0.08 | 388 | 0.13 | 0.10 | 33 | 8.29 | 8.81 | 0.41 | 1.01 | 0.22 | 4738 | 2.28 | 1.43 | -0.08 | 0.08 | 388 | -0.08 | 0.10 | 33 | 8.19 | 8.65 | 0.41 | 0.25 | 4.0 | G | 1 |
| hr8670 | S | 4829 | 2.53 | 1.40 | -0.40 | 0.07 | 431 | -0.30 | 0.06 | 25 | 7.98 | 8.48 | -0.09 | 0.25 | 0.20 | 4829 | 2.29 | 1.46 | -0.43 | 0.08 | 431 | -0.43 | 0.06 | 25 | 7.90 | 8.37 | -0.09 | 0.21 | 4.2 | G | 1 |
| hr8678 | S | 4803 | 2.55 | 1.36 | -0.02 | 0.08 | 394 | 0.04 | 0.10 | 27 | 8.16 | 8.71 | 0.06 | 0.38 | 0.20 | 4803 | 2.43 | 1.41 | -0.04 | 0.08 | 394 | -0.04 | 0.10 | 27 | 8.13 | 8.65 | 0.06 | 0.21 | 4.1 | G | 1 |
| hr8703 | S | 4603 | 2.44 | 2.48 | -0.03 | 0.34 | 123 | 0.07 | 0.26 | 10 | 8.19 | 8.61 | 1.11 | 6.81 | 0.27 | 4603 | 2.21 | 2.48 | -0.06 | 0.34 | 123 | -0.06 | 0.26 | 10 | 8.12 | 8.50 | 1.11 | 0.29 | 25.0 | R | 1 |
| hr8712 | S | 4641 | 2.57 | 1.40 | 0.25 | 0.11 | 375 | 0.37 | 0.19 | 27 | 8.59 | 8.99 | 0.33 | 1.17 | 0.24 | 4641 | 2.29 | 1.48 | 0.18 | 0.11 | 375 | 0.18 | 0.20 | 27 | 8.50 | 8.86 | 0.33 | 0.27 | 4.2 | G | 1 |
| hr8715 | S | 7728 | 3.73 | 4.14 | 0.41 | 0.15 | 21 | 0.20 | 0.05 | 3 | 7.73 | 8.20 | 2.70 | 0.56 | 0.12 | 7728 | 4.29 | 4.04 | 0.39 | 0.15 | 21 | 0.39 | 0.05 | 3 | 7.82 | 8.32 | 2.68 | -0.28 | 30.0 | R | 2 |
| hr8730 | S | 4586 | 2.49 | 1.31 | 0.12 | 0.10 | 391 | 0.32 | 0.17 | 28 | 8.36 | 8.79 | 0.18 | 1.00 | 0.27 | 4586 | 2.02 | 1.47 | 0.01 | 0.11 | 391 | 0.01 | 0.19 | 28 | 8.21 | 8.57 | 0.18 | 0.31 | 4.1 | G | 1 |
| hr8742 | S | 4831 | 2.48 | 1.34 | 0.10 | 0.09 | 413 | 0.06 | 0.14 | 30 | 8.17 | 8.63 | 0.48 | 0.90 | 0.20 | 4831 | 2.56 | 1.30 | 0.12 | 0.09 | 413 | 0.12 | 0.14 | 30 | 8.20 | 8.67 | 0.48 | 0.20 | 4.3 | G | 1 |
| hr8780 | S | 4668 | 2.61 | 1.32 | -0.05 | 0.08 | 421 | 0.16 | 0.10 | 28 | 8.33 | 8.85 | -0.09 | 0.42 | 0.23 | 4668 | 2.13 | 1.45 | -0.14 | 0.09 | 421 | -0.14 | 0.10 | 28 | 8.17 | 8.63 | -0.08 | 0.28 | 4.2 | G | 1 |
| hr8807 | S | 4984 | 2.83 | 1.31 | -0.19 | 0.09 | 438 | -0.10 | 0.10 | 27 | 8.16 | 8.80 | 0.00 | 0.19 | 0.16 | 4984 | 2.63 | 1.37 | -0.22 | 0.09 | 438 | -0.21 | 0.10 | 27 | 8.10 | 8.71 | 0.00 | 0.16 | 5.2 | G | 1 |
| hr8812 | S | 4414 | 1.73 | 1.62 | -0.11 | 0.11 | 369 | -0.04 | 0.15 | 25 | 8.10 | 8.62 | -0.09 | 1.02 | 0.35 | 4414 | 1.57 | 1.63 | -0.14 | 0.11 | 369 | -0.14 | 0.15 | 25 | 8.05 | 8.55 | -0.09 | 0.37 | 4.6 | G | 1 |
| hr8839 | S | 4594 | 2.23 | 1.52 | 0.34 | 0.14 | 370 | 0.38 | 0.21 | 27 | 8.59 | 8.95 | 0.42 | 1.68 | 0.29 | 4594 | 2.15 | 1.53 | 0.33 | 0.14 | 370 | 0.33 | 0.21 | 27 | 8.56 | 8.92 | 0.42 | 0.29 | 4.7 | G | 1 |
| hr8841 | S | 4624 | 2.55 | 1.38 | 0.10 | 0.11 | 409 | 0.21 | 0.14 | 26 | 8.40 | 8.87 | 0.00 | 0.59 | 0.25 | 4624 | 2.28 | 1.48 | 0.03 | 0.11 | 409 | 0.04 | 0.14 | 26 | 8.30 | 8.74 | 0.00 | 0.28 | 4.6 | G | 1 |
| hr8852 | S | 4810 | 2.42 | 1.40 | -0.59 | 0.07 | 414 | -0.47 | 0.05 | 26 | 7.89 | 8.57 | 0.56 | 1.17 | 0.21 | 4810 | 2.15 | 1.45 | -0.61 | 0.07 | 414 | -0.61 | 0.05 | 26 | 7.81 | 8.45 | 0.57 | 0.23 | 3.9 | G | 1 |
| hr8875 | S | 4639 | 2.56 | 1.37 | 0.09 | 0.10 | 393 | 0.21 | 0.14 | 25 | 8.43 | 8.89 | 0.24 | 0.97 | 0.24 | 4639 | 2.28 | 1.48 | 0.02 | 0.10 | 393 | 0.02 | 0.15 | 25 | 8.33 | 8.76 | 0.25 | 0.27 | 4.0 | G | 1 |
| hr8878 | S | 4246 | 1.58 | 1.48 | -0.67 | 0.11 | 417 | -0.49 | 0.14 | 25 | 7.96 | 8.55 | -0.75 | 0.42 | 0.37 | 4246 | 1.28 | 1.49 | -0.78 | 0.11 | 417 | -0.78 | 0.15 | 25 | 7.74 | 8.30 | -0.76 | 0.41 | 4.4 | G | 1 |
| hr8892 | S | 4545 | 2.25 | 1.33 | -0.36 | 0.08 | 391 | -0.29 | 0.10 | 24 | 7.99 | 8.44 | -0.25 | 0.43 | 0.29 | 4545 | 2.09 | 1.38 | -0.39 | 0.09 | 391 | -0.39 | 0.10 | 24 | 7.93 | 8.37 | -0.25 | 0.31 | 3.7 | G | 1 |
| hr8912 | S | 4828 | 2.71 | 1.30 | -0.03 | 0.07 | 416 | 0.08 | 0.11 | 31 | 8.23 | 8.83 | 0.66 | 1.35 | 0.19 | 4828 | 2.46 | 1.39 | -0.07 | 0.07 | 416 | -0.07 | 0.11 | 31 | 8.15 | 8.71 | 0.66 | 0.21 | 4.1 | G | 1 |
| hr8922 | S | 4727 | 2.51 | 1.38 | -0.03 | 0.09 | 399 | -0.02 | 0.11 | 26 | 8.34 | 8.78 | 1.09 | 4.56 | 0.23 | 4727 | 2.48 | 1.38 | -0.03 | 0.09 | 399 | -0.03 | 0.11 | 26 | 8.33 | 8.77 | 1.09 | 0.24 | 4.1 | G | 1 |
| hr8923 | S | 4938 | 2.81 | 1.32 | 0.11 | 0.09 | 395 | 0.20 | 0.11 | 27 | 8.21 | 8.71 | 0.71 | 1.11 | 0.17 | 4938 | 2.62 | 1.40 | 0.08 | 0.09 | 395 | 0.08 | 0.11 | 27 | 8.16 | 8.62 | 0.72 | 0.17 | 4.9 | G | 1 |
| hr8930 | S | 4613 | 2.39 | 1.39 | -0.29 | 0.09 | 415 | -0.17 | 0.12 | 25 | 8.15 | 8.69 | -0.26 | 0.34 | 0.27 | 4613 | 2.10 | 1.48 | -0.34 | 0.10 | 415 | -0.35 | 0.13 | 25 | 8.05 | 8.56 | -0.27 | 0.29 | 4.4 | G | 1 |
| hr8941 | S | 4962 | 2.50 | 1.48 | 0.10 | 0.09 | 406 | 0.02 | 0.11 | 31 | 8.16 | 8.63 | 0.55 | 0.71 | 0.18 | 4962 | 2.66 | 1.43 | 0.12 | 0.09 | 406 | 0.12 | 0.11 | 31 | 8.20 | 8.71 | 0.55 | 0.17 | 4.9 | G | 1 |
| hr8946 | S | 4208 | 2.06 | 1.44 | 0.21 | 0.17 | 358 | 0.55 | 0.30 | 23 | 8.52 | 8.87 | -0.40 | 1.05 | 0.33 | 4208 | 1.21 | 1.60 | -0.05 | 0.18 | 358 | -0.05 | 0.32 | 23 | 8.19 | 8.49 | -0.43 | 0.42 | 4.5 | G | 2 |
| hr8948 | S | 4761 | 2.56 | 1.25 | 0.11 | 0.08 | 231 | 0.13 | 0.07 | 14 | | | 0.70 | 1.82 | 0.21 | 4761 | 2.51 | 1.26 | 0.10 | 0.08 | 231 | 0.10 | 0.07 | 14 | | | 0.70 | 0.21 | 4.0 | G | 1 |
| hr8958 | S | 4708 | 2.62 | 1.35 | -0.10 | 0.08 | 408 | 0.11 | 0.09 | 29 | 8.29 | 8.81 | 0.30 | 0.87 | 0.22 | 4708 | 2.16 | 1.47 | -0.18 | 0.09 | 408 | -0.17 | 0.10 | 29 | 8.15 | 8.60 | 0.30 | 0.27 | 4.3 | G | 1 |
| hr9008 | S | 4617 | 2.60 | 1.46 | 0.20 | 0.10 | 388 | 0.47 | 0.11 | 24 | 8.56 | 8.93 | 0.07 | 0.71 | 0.24 | 4617 | 2.08 | 1.56 | 0.06 | 0.11 | 388 | 0.06 | 0.11 | 24 | 8.29 | 8.59 | 0.07 | 0.30 | 4.6 | G | 1 |
| hr9009 | S | 4661 | 2.57 | 1.29 | 0.12 | 0.10 | 388 | 0.29 | 0.17 | 27 | 8.30 | 8.76 | 0.10 | 0.65 | 0.24 | 4661 | 2.20 | 1.44 | 0.03 | 0.10 | 388 | 0.03 | 0.18 | 27 | 8.19 | 8.59 | 0.10 | 0.28 | 4.3 | G | 1 |
| hr9012 | S | 4926 | 2.64 | 1.35 | 0.00 | 0.08 | 399 | 0.09 | 0.12 | 31 | 8.18 | 8.58 | 0.66 | 1.01 | 0.17 | 4926 | 2.44 | 1.41 | -0.03 | 0.09 | 399 | -0.03 | 0.12 | 31 | 8.13 | 8.48 | 0.67 | 0.19 | 4.6 | G | 1 |
| hr9067 | S | 5026 | 2.75 | 1.36 | 0.05 | 0.08 | 403 | 0.02 | 0.11 | 34 | 8.09 | 8.68 | 0.60 | 0.66 | 0.15 | 5026 | 2.82 | 1.33 | 0.06 | 0.08 | 403 | 0.06 | 0.11 | 34 | 8.11 | 8.71 | 0.59 | 0.15 | 4.3 | G | 1 |
| hr9101 | S | 4663 | 2.65 | 1.33 | 0.28 | 0.12 | 405 | 0.51 | 0.18 | 28 | 8.51 | 8.99 | 0.25 | 0.91 | 0.23 | 4663 | 2.12 | 1.49 | 0.16 | 0.12 | 405 | 0.15 | 0.19 | 28 | 8.35 | 8.75 | 0.26 | 0.28 | 4.6 | G | 1 |
| hr9104 | S | 4701 | 2.49 | 1.38 | 0.01 | 0.09 | 414 | 0.08 | 0.13 | 27 | 8.23 | 8.72 | 0.29 | 0.87 | 0.24 | 4701 | 2.32 | 1.44 | -0.03 | 0.09 | 414 | -0.03 | 0.13 | 27 | 8.18 | 8.64 | 0.29 | 0.26 | 4.1 | G | 1 |
| hd001638 | U | 4225 | 2.50 | 1.64 | -0.38 | 0.12 | 514 | 0.26 | 0.13 | 48 | 8.30 | 9.00 | -1.24 | 0.14 | 0.31 | 4225 | 0.69 | 1.83 | -0.76 | 0.14 | 514 | -0.76 | 0.20 | 48 | 7.66 | 8.21 | -1.31 | 0.50 | 4.9 | G | 1 |
| hd004388 | U | 4640 | 2.57 | 1.23 | 0.15 | 0.13 | 435 | 0.27 | 0.18 | 41 | | | | | 0.24 | 4640 | 2.26 | 1.39 | 0.06 | 0.12 | 435 | 0.06 | 0.19 | 41 | | | | 0.28 | 2.3 | G | 1 |
| hd011171 | U | 6746 | 4.08 | 3.91 | 0.09 | 0.17 | 24 | -0.09 | 0.14 | 2 | 8.07 | 9.07 | 2.79 | 3.11 | -0.06 | 6746 | 4.51 | 3.76 | 0.09 | 0.17 | 24 | 0.09 | 0.13 | 2 | 8.20 | 9.19 | 2.78 | -0.19 | 58.0 | R | 2 |
| hd037763 | U | 4620 | 3.12 | 0.75 | 0.48 | 0.12 | 410 | 0.67 | 0.23 | 42 | 8.78 | 9.07 | 0.28 | 1.12 | 0.21 | 4620 | 2.66 | 1.07 | 0.34 | 0.10 | 410 | 0.34 | 0.24 | 42 | 8.59 | 8.85 | 0.28 | 0.24 | 4.0 | G | 1 |
| hd037811 | U | 5023 | 2.57 | 1.42 | -0.01 | 0.07 | 489 | 0.01 | 0.08 | 37 | 8.07 | 8.65 | 0.74 | 0.92 | 0.16 | 5023 | 2.54 | 1.43 | -0.01 | 0.07 | 489 | -0.01 | 0.08 | 37 | 8.06 | 8.63 | 0.74 | 0.16 | 4.7 | G | 1 |

| ID | | | | | | | | | | | | | | | | | | | | | | | | | | | | |
|---|---|---|---|---|---|---|---|---|---|---|---|---|---|---|---|---|---|---|---|---|---|---|---|---|---|---|---|---|
| hd061603 | U | 4012 | 0.65 | 2.19 | 0.04 | 0.28 | 373 | 0.13 | 0.31 | 26 | 8.28 | 8.68 | -0.30 | 3.06 | 0.53 | 4012 | 0.39 | 2.19 | -0.02 | 0.28 | 373 | -0.02 | 0.31 | 26 | 8.19 | 8.57 | -0.32 | 0.61 | 7.6 | G | 1 |
| hd062412 | U | 4861 | 2.61 | 1.38 | 0.07 | 0.10 | 556 | 0.20 | 0.17 | 57 | 8.25 | 8.69 | 0.62 | 1.11 | 0.19 | 4861 | 2.31 | 1.47 | 0.02 | 0.10 | 556 | 0.02 | 0.17 | 57 | 8.17 | 8.55 | 0.62 | 0.21 | 5.0 | G | 1 |
| hd065354 | U | 3832 | 0.06 | 2.28 | -0.14 | 0.23 | 391 | 0.11 | 0.22 | 30 | 7.96 | 8.28 | -0.22 | 7.11 | 0.79 | 3832 | -0.99 | 2.36 | -0.30 | 0.29 | 391 | -0.23 | 0.35 | 30 | 7.62 | 7.79 | -0.51 | 1.45 | 7.7 | G | 2 |
| hd065714 | U | 5070 | 2.39 | 1.75 | 0.29 | 0.12 | 535 | 0.18 | 0.19 | 57 | 8.05 | 8.81 | 1.29 | 2.64 | 0.16 | 5070 | 2.65 | 1.70 | 0.32 | 0.12 | 535 | 0.32 | 0.19 | 57 | 8.12 | 8.94 | 1.29 | 0.15 | 5.8 | G | 1 |
| hd065925 | U | 6585 | 3.63 | 6.50 | -0.52 | 0.26 | 103 | -0.39 | 0.21 | 11 | 7.75 | 9.96 | 3.03 | 6.73 | 0.02 | 6585 | 3.30 | 7.01 | -0.53 | 0.26 | 103 | -0.53 | 0.23 | 11 | 7.65 | 8.49 | 3.03 | 0.06 | 80.0 | R | 2 |
| hd069511 | U | 4020 | 0.75 | 2.45 | -0.09 | 0.19 | 330 | 0.17 | 0.14 | 19 | | | | | 0.51 | 4020 | -0.02 | 2.40 | -0.22 | 0.19 | 330 | -0.22 | 0.16 | 19 | | | | 0.74 | 7.1 | G | 1 |
| hd069836 | U | 4788 | 2.50 | 1.43 | 0.14 | 0.11 | 470 | 0.11 | 0.15 | 40 | | | | | 0.21 | 4788 | 2.45 | 1.49 | 0.14 | 0.11 | 470 | 0.14 | 0.16 | 40 | | | | 0.21 | 4.3 | G | 2 |
| hd071160 | U | 4096 | 1.55 | 1.77 | 0.12 | 0.21 | 452 | 0.37 | 0.30 | 37 | 8.43 | 8.85 | -0.59 | 1.08 | 0.37 | 4096 | 0.91 | 1.82 | -0.05 | 0.23 | 452 | -0.05 | 0.30 | 37 | 8.19 | 8.57 | -0.62 | 0.47 | 5.3 | G | 1 |
| hd072320 | U | 5028 | 2.77 | 1.34 | 0.02 | 0.08 | 491 | 0.08 | 0.12 | 46 | | | | | 0.15 | 5028 | 2.64 | 1.38 | 0.00 | 0.08 | 491 | 0.00 | 0.12 | 46 | | | | 0.16 | 4.2 | G | 1 |
| hd072324 | U | 4781 | 2.43 | 1.54 | 0.06 | 0.11 | 533 | 0.23 | 0.17 | 50 | 8.13 | 8.70 | 0.93 | 2.79 | 0.22 | 4781 | 2.03 | 1.61 | 0.00 | 0.12 | 533 | 0.00 | 0.17 | 50 | 8.03 | 8.51 | 0.94 | 0.24 | 5.9 | G | 1 |
| hd073829 | U | 4962 | 2.50 | 1.50 | 0.23 | 0.13 | 465 | -0.01 | 0.20 | 41 | | | | | 0.18 | 4962 | 3.06 | 1.27 | 0.34 | 0.14 | 465 | 0.34 | 0.19 | 41 | | | | 0.16 | 4.4 | G | 2 |
| hd074088 | U | 3840 | 1.06 | 1.74 | -0.26 | 0.22 | 454 | 0.26 | 0.26 | 40 | 8.25 | 8.66 | -1.25 | 0.66 | 0.42 | 3840 | -0.80 | 1.79 | -0.61 | 0.30 | 454 | -0.53 | 0.34 | 40 | 7.65 | 7.90 | -1.55 | 1.30 | 5.8 | G | 2 |
| hd074165 | U | 4675 | 2.50 | 1.43 | 0.28 | 0.14 | 454 | 0.38 | 0.18 | 35 | | | | | 0.25 | 4675 | 2.26 | 1.51 | 0.22 | 0.14 | 454 | 0.23 | 0.19 | 35 | | | | 0.27 | 4.4 | G | 2 |
| hd074166 | U | 4629 | 2.50 | 1.78 | 0.40 | 0.19 | 409 | 0.37 | 0.16 | 25 | | | | | 0.26 | 4629 | 2.58 | 1.76 | 0.42 | 0.19 | 409 | 0.42 | 0.16 | 25 | | | | 0.24 | 5.3 | G | 2 |
| hd074529 | U | 4598 | 2.38 | 1.40 | 0.25 | 0.13 | 459 | 0.39 | 0.19 | 37 | | | | | 0.27 | 4598 | 2.03 | 1.50 | 0.17 | 0.13 | 459 | 0.17 | 0.20 | 37 | | | | 0.30 | 4.6 | G | 1 |
| hd074900 | U | 4572 | 2.51 | 1.35 | 0.23 | 0.13 | 464 | 0.45 | 0.18 | 38 | | | | | 0.26 | 4572 | 2.04 | 1.46 | 0.08 | 0.13 | 464 | 0.08 | 0.18 | 38 | | | | 0.31 | 4.4 | G | 1 |
| hd075058 | U | 4650 | 2.79 | 1.00 | 0.15 | 0.12 | 478 | 0.30 | 0.17 | 39 | | | | | 0.22 | 4650 | 2.43 | 1.26 | 0.02 | 0.11 | 478 | 0.03 | 0.18 | 39 | | | | 0.26 | 4.1 | G | 1 |
| hd076128 | U | 4487 | 2.50 | 1.39 | 0.30 | 0.14 | 451 | 0.58 | 0.17 | 35 | | | | | 0.28 | 4487 | 1.78 | 1.60 | 0.10 | 0.15 | 451 | 0.10 | 0.19 | 35 | | | | 0.34 | 4.6 | G | 2 |
| hd078002 | U | 4859 | 2.50 | 1.41 | 0.18 | 0.10 | 476 | 0.22 | 0.14 | 42 | | | | | 0.20 | 4859 | 2.40 | 1.44 | 0.16 | 0.10 | 476 | 0.16 | 0.15 | 42 | | | | 0.20 | 4.5 | G | 2 |
| hd078479 | U | 4479 | 2.26 | 1.64 | 0.33 | 0.19 | 486 | 0.49 | 0.26 | 43 | 8.65 | 8.95 | 0.10 | 1.22 | 0.30 | 4479 | 1.85 | 1.70 | 0.23 | 0.19 | 486 | 0.23 | 0.27 | 43 | 8.50 | 8.76 | 0.10 | 0.34 | 5.6 | G | 1 |
| hd078528 | U | 4012 | 2.50 | 1.00 | 0.77 | 0.20 | 299 | 1.27 | 0.25 | 17 | | | | | 0.32 | 4012 | 1.01 | 1.66 | 0.05 | 0.23 | 299 | 0.05 | 0.25 | 17 | | | | 0.45 | 5.0 | G | 2 |
| hd078959 | U | 4150 | 1.20 | 2.16 | -0.02 | 0.23 | 363 | 0.06 | 0.19 | 20 | | | | | 0.42 | 4150 | 0.99 | 2.16 | -0.07 | 0.23 | 363 | -0.07 | 0.20 | 20 | | | | 0.45 | 6.1 | G | 2 |
| hd080571 | U | 4659 | 2.35 | 1.41 | 0.06 | 0.11 | 470 | 0.11 | 0.13 | 38 | | | | | 0.26 | 4659 | 2.23 | 1.44 | 0.04 | 0.11 | 470 | 0.04 | 0.14 | 38 | | | | 0.27 | 4.4 | G | 1 |
| hd081278 | U | 4936 | 2.50 | 1.43 | 0.19 | 0.11 | 479 | -0.10 | 0.22 | 40 | | | | | 0.18 | 4936 | 3.20 | 1.00 | 0.38 | 0.14 | 479 | 0.38 | 0.22 | 40 | | | | 0.16 | 4.1 | G | 2 |
| hd082395 | U | 4686 | 2.58 | 1.37 | 0.03 | 0.11 | 553 | 0.28 | 0.18 | 56 | 8.37 | 8.78 | -0.06 | 0.42 | 0.23 | 4686 | 1.98 | 1.51 | -0.08 | 0.11 | 553 | -0.08 | 0.19 | 56 | 8.19 | 8.50 | -0.05 | 0.29 | 5.0 | G | 1 |
| hd082403 | U | 4724 | 2.50 | 1.33 | 0.23 | 0.14 | 453 | 0.10 | 0.15 | 38 | 8.50 | 8.98 | | | 0.24 | 4724 | 2.85 | 1.11 | 0.35 | 0.16 | 453 | 0.35 | 0.14 | 38 | 8.63 | 9.14 | | 0.21 | 4.5 | G | 2 |
| hd082668 | U | 4105 | 1.38 | 1.70 | -0.06 | 0.24 | 391 | 0.14 | 0.33 | 27 | 8.44 | 8.97 | -0.55 | 1.16 | 0.39 | 4105 | 0.88 | 1.72 | -0.18 | 0.25 | 391 | -0.18 | 0.33 | 27 | 8.25 | 8.75 | -0.58 | 0.48 | 5.7 | G | 1 |
| hd083155 | U | 5166 | 2.50 | 1.62 | -0.01 | 0.10 | 478 | -0.33 | 0.09 | 38 | 7.99 | 8.60 | | | 0.14 | 5166 | 3.21 | 1.36 | 0.06 | 0.11 | 478 | 0.06 | 0.10 | 38 | 8.17 | 8.94 | | 0.12 | 4.4 | G | 2 |
| hd083234 | U | 4185 | 2.50 | 1.31 | 0.51 | 0.18 | 424 | 1.00 | 0.15 | 26 | | | | | 0.31 | 4185 | 1.24 | 1.74 | -0.01 | 0.19 | 424 | -0.01 | 0.19 | 26 | | | | 0.42 | 4.9 | G | 2 |
| hd084598 | U | 4896 | 2.64 | 1.40 | 0.03 | 0.10 | 479 | 0.17 | 0.11 | 39 | | | | | 0.18 | 4896 | 2.32 | 1.49 | -0.01 | 0.10 | 479 | -0.01 | 0.11 | 39 | | | | 0.20 | 4.9 | G | 1 |
| hd085552 | U | 5426 | 3.19 | 2.15 | -0.07 | 0.17 | 334 | 0.32 | 0.16 | 39 | 8.23 | 8.75 | 2.30 | 9.15 | 0.08 | 5426 | 2.29 | 2.28 | -0.10 | 0.18 | 334 | -0.10 | 0.18 | 39 | 8.09 | 8.34 | 2.31 | 0.11 | 14.0 | G | 1 |
| hd086757 | U | 4081 | 2.50 | 2.42 | 0.45 | 0.19 | 329 | 1.27 | 0.17 | 19 | | | | | 0.31 | 4081 | 0.32 | 2.46 | -0.14 | 0.20 | 329 | -0.15 | 0.20 | 19 | | | | 0.61 | 7.1 | G | 2 |
| hd089736 | U | 3830 | 0.49 | 2.58 | -0.03 | 0.26 | 342 | 0.33 | 0.19 | 23 | 8.25 | 8.71 | 0.32 | 16.13 | 0.60 | 3830 | -0.77 | 2.52 | -0.27 | 0.29 | 342 | -0.27 | 0.25 | 23 | 7.88 | 8.18 | 0.06 | 1.29 | 9.2 | G | 1 |
| hd095849 | U | 4472 | 1.97 | 1.64 | 0.22 | 0.17 | 505 | 0.30 | 0.23 | 46 | 8.38 | 8.82 | 0.07 | 1.17 | 0.33 | 4472 | 1.76 | 1.66 | 0.18 | 0.17 | 505 | 0.18 | 0.24 | 46 | 8.32 | 8.73 | 0.07 | 0.34 | 5.0 | G | 1 |
| hd099322 | U | 4857 | 2.59 | 1.32 | 0.04 | 0.07 | 594 | 0.09 | 0.12 | 62 | 8.18 | 8.68 | 0.17 | 0.41 | 0.19 | 4857 | 2.47 | 1.36 | 0.02 | 0.07 | 594 | 0.02 | 0.12 | 62 | 8.15 | 8.62 | 0.17 | 0.20 | 3.7 | G | 1 |
| hd103295 | U | | | | | | | | | | | | | 0.16 | 3.30 | 4912 | 2.29 | 1.16 | -1.06 | 0.10 | 385 | -1.06 | 0.08 | 31 | 7.44 | 7.95 | -0.19 | 0.20 | 3.4 | G | 2 |
| hd105740 | U | 4668 | 2.79 | 1.04 | -0.54 | 0.08 | 475 | -0.32 | 0.11 | 32 | 8.10 | 8.73 | -0.06 | 0.45 | 0.22 | 4668 | 2.26 | 1.25 | -0.64 | 0.09 | 475 | -0.64 | 0.12 | 32 | 7.92 | 8.48 | -0.06 | 0.27 | 3.9 | G | 1 |
| hd107446 | U | 4121 | 1.64 | 1.76 | -0.14 | 0.18 | 520 | 0.19 | 0.25 | 47 | 8.21 | 8.62 | -1.01 | 0.38 | 0.36 | 4121 | 0.81 | 1.82 | -0.39 | 0.20 | 520 | -0.39 | 0.27 | 47 | 7.81 | 8.14 | -1.06 | 0.49 | 4.4 | G | 1 |
| hd110458 | U | 4682 | 2.55 | 1.32 | 0.22 | 0.10 | 570 | 0.36 | 0.16 | 58 | 8.44 | 8.90 | 0.19 | 0.75 | 0.23 | 4682 | 2.22 | 1.42 | 0.15 | 0.10 | 570 | 0.14 | 0.16 | 58 | 8.34 | 8.75 | 0.20 | 0.27 | 3.8 | G | 1 |
| hd111464 | U | 4170 | 1.68 | 1.56 | 0.01 | 0.17 | 509 | 0.27 | 0.22 | 45 | 8.29 | 8.75 | 0.21 | 4.60 | 0.36 | 4170 | 0.98 | 1.63 | -0.17 | 0.18 | 509 | -0.17 | 0.24 | 45 | 8.05 | 8.45 | 0.18 | 0.45 | 5.2 | G | 1 |
| hd111721 | U | 4892 | 2.62 | 1.22 | -1.42 | 0.07 | 330 | -1.26 | 0.08 | 30 | 7.14 | 7.83 | 0.87 | 1.78 | 0.18 | 4892 | 2.24 | 1.33 | -1.43 | 0.07 | 330 | -1.43 | 0.07 | 30 | 7.07 | 7.67 | 0.88 | 0.20 | 3.0 | G | 1 |
| hd113002 | U | 5274 | 3.24 | 1.57 | -0.80 | 0.08 | 473 | -0.43 | 0.11 | 67 | 7.66 | 8.59 | 0.15 | 0.14 | 0.10 | 5274 | 2.36 | 1.78 | -0.82 | 0.09 | 473 | -0.82 | 0.13 | 67 | 7.47 | 8.21 | 0.17 | 0.13 | 5.0 | G | 1 |
| hd119971 | U | 4071 | 1.56 | 1.68 | -0.55 | 0.13 | 512 | -0.12 | 0.21 | 48 | 8.11 | 8.72 | -1.54 | 0.14 | 0.37 | 4071 | 0.20 | 1.68 | -0.82 | 0.16 | 512 | -0.82 | 0.25 | 48 | 7.66 | 8.13 | -1.69 | 0.65 | 5.0 | G | 1 |
| hd122721 | U | 4684 | 3.05 | 0.50 | 0.36 | 0.15 | 431 | 0.61 | 0.12 | 34 | | | | | 0.20 | 4684 | 2.48 | 1.19 | 0.04 | 0.09 | 431 | 0.04 | 0.11 | 34 | | | | 0.25 | 4.0 | G | 1 |
| hd124186 | U | 4417 | 2.32 | 1.53 | 0.42 | 0.19 | 492 | 0.58 | 0.25 | 39 | 8.68 | 9.01 | 0.30 | 2.31 | 0.30 | 4417 | 1.91 | 1.63 | 0.30 | 0.19 | 492 | 0.29 | 0.26 | 39 | 8.52 | 8.82 | 0.29 | 0.34 | 5.1 | G | 1 |
| hd128279 | U | 5162 | 3.01 | 1.59 | -2.28 | 0.09 | 152 | -2.13 | 0.05 | 12 | | | 0.97 | 1.12 | 0.12 | 5162 | 2.64 | 1.69 | -2.28 | 0.09 | 152 | -2.28 | 0.06 | 12 | | | 0.98 | 0.13 | 2.4 | G | 1 |
| hd138688 | U | 4191 | 1.83 | 1.38 | 0.11 | 0.17 | 520 | 0.32 | 0.26 | 46 | 8.41 | 8.90 | 0.09 | 3.26 | 0.35 | 4191 | 1.28 | 1.51 | -0.06 | 0.18 | 520 | -0.06 | 0.26 | 46 | 8.20 | 8.65 | 0.06 | 0.41 | 4.8 | G | 1 |
| hd145206 | U | 4029 | 1.51 | 1.47 | 0.18 | 0.17 | 510 | 0.53 | 0.24 | 45 | 8.46 | 8.85 | 0.16 | 6.03 | 0.37 | 4029 | 0.59 | 1.54 | -0.12 | 0.19 | 510 | -0.12 | 0.26 | 45 | 8.05 | 8.36 | 0.06 | 0.55 | 4.9 | G | 1 |
| hd146836 | U | 6285 | 3.64 | 2.38 | -0.18 | 0.11 | 198 | -0.16 | 0.13 | 33 | 8.19 | 8.71 | 1.88 | 0.88 | 0.02 | 6285 | 3.61 | 2.39 | -0.18 | 0.11 | 198 | -0.18 | 0.13 | 33 | 8.18 | 8.69 | 1.88 | 0.02 | 19.0 | R | 2 |
| hd148451 | U | 5015 | 2.61 | 1.41 | -0.54 | 0.07 | 580 | -0.47 | 0.11 | 65 | 7.56 | 8.65 | 1.07 | 2.01 | 0.16 | 5015 | 2.45 | 1.45 | -0.55 | 0.07 | 580 | -0.55 | 0.11 | 65 | 7.51 | 8.57 | 1.07 | 0.17 | 3.7 | G | 1 |

| ID | | | | | | | | | | | | | | | | | | | | | | | | | | | | | |
|---|---|---|---|---|---|---|---|---|---|---|---|---|---|---|---|---|---|---|---|---|---|---|---|---|---|---|---|---|---|
| hd148513 | U | 4065 | 1.63 | 1.83 | 0.18 | 0.23 | 454 | 0.51 | 0.28 | 33 | 8.54 | 8.87 | 0.07 | 4.89 | 0.36 | 4065 | 0.75 | 1.91 | -0.07 | 0.25 | 454 | -0.07 | 0.29 | 33 | 8.19 | 8.48 | 0.02 | 0.50 | 5.5 | G | 1 |
| hd149447 | U | 3875 | 1.25 | 1.92 | 0.08 | 0.31 | 444 | 0.56 | 0.37 | 37 | 8.58 | 8.96 | -0.78 | 1.53 | 0.38 | 3875 | -0.25 | 1.94 | -0.31 | 0.34 | 444 | -0.31 | 0.39 | 37 | 8.01 | 8.29 | -0.97 | 0.93 | 6.2 | G | 1 |
| hd150798 | U | 4135 | 2.50 | 2.46 | 0.42 | 0.19 | 329 | 1.25 | 0.10 | 19 | 8.84 | 9.39 | 0.83 | 13.78 | 0.31 | 4135 | 0.54 | 2.42 | -0.10 | 0.19 | 329 | -0.11 | 0.13 | 19 | 8.02 | 8.46 | 0.68 | 0.54 | 7.8 | G | 1 |
| hd152786 | U | 3851 | 0.52 | 2.17 | -0.08 | 0.21 | 404 | 0.20 | 0.19 | 27 | 8.15 | 8.52 | 0.57 | 20.60 | 0.59 | 3851 | -0.36 | 2.12 | -0.26 | 0.23 | 404 | -0.26 | 0.23 | 27 | 7.89 | 8.16 | 0.39 | 1.01 | 6.9 | G | 1 |
| hd157457 | U | 4911 | 2.34 | 1.72 | 0.07 | 0.13 | 411 | -0.03 | 0.15 | 29 | 8.20 | 8.71 | 1.52 | 6.36 | 0.19 | 4911 | 2.58 | 1.67 | 0.10 | 0.13 | 411 | 0.11 | 0.15 | 29 | 8.26 | 8.83 | 1.52 | 0.18 | 7.6 | G | 1 |
| hd162391 | U | 4724 | 2.04 | 2.17 | 0.01 | 0.15 | 426 | 0.26 | 0.14 | 38 | 8.09 | 8.65 | 0.25 | 0.73 | 0.28 | 4724 | 1.45 | 2.16 | -0.03 | 0.16 | 426 | -0.03 | 0.15 | 38 | 7.96 | 8.38 | 0.25 | 0.32 | 8.6 | G | 1 |
| hd162587 | U | 4764 | 1.92 | 1.76 | -0.01 | 0.11 | 549 | 0.03 | 0.14 | 57 | 8.10 | 8.57 | 0.84 | 2.37 | 0.25 | 4764 | 1.83 | 1.76 | -0.01 | 0.11 | 549 | -0.01 | 0.14 | 57 | 8.08 | 8.52 | 0.84 | 0.26 | 5.1 | G | 1 |
| hd163652 | U | 4968 | 2.64 | 1.41 | -0.36 | 0.06 | 594 | -0.26 | 0.13 | 68 | 7.91 | 8.56 | 0.16 | 0.30 | 0.17 | 4968 | 2.39 | 1.47 | -0.38 | 0.06 | 594 | -0.38 | 0.14 | 68 | 7.86 | 8.44 | 0.16 | 0.18 | 3.7 | G | 1 |
| hd167818 | U | 3909 | 0.64 | 2.25 | -0.12 | 0.18 | 443 | 0.29 | 0.17 | 36 | 8.16 | 8.51 | -0.75 | 1.64 | 0.54 | 3909 | -0.90 | 2.28 | -0.40 | 0.25 | 443 | -0.34 | 0.32 | 36 | 7.66 | 7.81 | -1.09 | 1.27 | 6.5 | G | 2 |
| hd169191 | U | 4357 | 1.99 | 1.49 | -0.02 | 0.14 | 526 | 0.20 | 0.19 | 47 | 8.21 | 8.72 | 0.40 | 3.61 | 0.33 | 4357 | 1.44 | 1.56 | -0.14 | 0.15 | 526 | -0.14 | 0.21 | 47 | 8.04 | 8.48 | 0.39 | 0.39 | 4.1 | G | 1 |
| hd175545 | U | 4471 | 2.80 | 0.92 | 0.43 | 0.17 | 522 | 0.69 | 0.19 | 44 | 8.64 | 9.00 | -0.09 | 0.80 | 0.26 | 4471 | 2.20 | 1.36 | 0.11 | 0.16 | 522 | 0.12 | 0.22 | 44 | 8.30 | 8.63 | -0.10 | 0.31 | 3.5 | G | 1 |
| hd183275 | U | 4743 | 2.50 | 1.56 | 0.33 | 0.15 | 420 | 0.27 | 0.20 | 28 | 8.49 | 8.93 | 0.48 | 1.16 | 0.24 | 4743 | 2.63 | 1.52 | 0.36 | 0.15 | 420 | 0.36 | 0.20 | 28 | 8.53 | 8.99 | 0.47 | 0.22 | 4.5 | G | 2 |
| hd196983 | U | 4593 | 2.50 | 1.39 | 0.26 | 0.15 | 425 | 0.37 | 0.19 | 27 | 8.48 | 8.90 | 0.56 | 2.29 | 0.26 | 4593 | 2.24 | 1.46 | 0.19 | 0.15 | 425 | 0.20 | 0.19 | 27 | 8.39 | 8.78 | 0.56 | 0.29 | 4.6 | G | 2 |
| hd199642 | U | 3828 | 1.32 | 1.94 | 0.04 | 0.31 | 427 | 0.57 | 0.42 | 34 | 8.75 | 9.20 | -1.16 | 0.74 | 0.36 | 3828 | -0.38 | 1.95 | -0.40 | 0.33 | 427 | -0.40 | 0.42 | 34 | 8.08 | 8.42 | -1.39 | 1.04 | 6.9 | G | 1 |
| hd202320 | U | 4490 | 1.76 | 1.79 | -0.07 | 0.13 | 517 | 0.12 | 0.18 | 50 | 8.26 | 8.62 | -0.04 | 0.87 | 0.34 | 4490 | 1.29 | 1.79 | -0.13 | 0.13 | 517 | -0.13 | 0.18 | 50 | 8.13 | 8.41 | -0.04 | 0.39 | 5.5 | G | 1 |
| hd203638 | U | 4532 | 2.51 | 1.47 | 0.30 | 0.16 | 531 | 0.54 | 0.20 | 45 | 8.65 | 9.02 | 0.23 | 1.36 | 0.27 | 4532 | 1.90 | 1.62 | 0.14 | 0.16 | 531 | 0.14 | 0.21 | 45 | 8.44 | 8.73 | 0.23 | 0.32 | 5.0 | G | 1 |
| hd207964 | U | 6538 | 3.90 | 3.23 | -0.03 | 0.21 | 9 | 0.35 | 0.00 | 1 | 7.95 | 8.50 | 3.29 | 9.99 | -0.02 | 6538 | 2.82 | 3.32 | -0.03 | 0.22 | 9 | -0.03 | | 1 | 7.65 | 8.17 | 3.29 | 0.07 | 150.0 | R | 2 |
| hd211173 | U | 4883 | 3.11 | 1.00 | -0.17 | 0.07 | 454 | 0.09 | 0.07 | 39 | 8.35 | 8.79 | | | 0.17 | 4883 | 2.66 | 1.15 | -0.27 | 0.07 | 454 | -0.27 | 0.10 | 39 | 8.07 | 8.44 | | 0.18 | 3.5 | G | 1 |
| hd212320 | U | 4851 | 2.40 | 1.47 | -0.17 | 0.12 | 558 | 0.10 | 0.16 | 57 | 8.27 | 8.59 | 0.31 | 0.58 | 0.20 | 4851 | 1.76 | 1.54 | -0.22 | 0.13 | 558 | -0.22 | 0.18 | 57 | 8.16 | 8.28 | 0.32 | 0.24 | 5.2 | G | 1 |
| hd223094 | U | 3764 | 0.23 | 1.94 | -0.37 | 0.24 | 306 | -0.01 | 0.29 | 17 | | | | | 0.73 | 3764 | -0.95 | 2.00 | -0.54 | 0.28 | 306 | -0.45 | 0.31 | 17 | | | | 1.53 | 5.7 | G | 2 |
| hic007995 | U | 6036 | 3.85 | 0.93 | 0.79 | 0.20 | 540 | 0.11 | 0.18 | 34 | 7.52 | 8.46 | | | 0.01 | 6036 | 4.89 | 0.50 | 0.73 | 0.26 | 540 | 0.54 | 0.20 | 34 | 7.87 | 8.76 | | -0.05 | 3.2 | G | 2 |
| hip049418 | U | 4660 | 2.38 | 1.33 | 0.13 | 0.10 | 475 | 0.20 | 0.17 | 44 | | | | | 0.26 | 4660 | 2.21 | 1.39 | 0.09 | 0.10 | 475 | 0.09 | 0.18 | 44 | | | | 0.28 | 4.0 | G | 1 |
| hip051077 | U | 4589 | 2.57 | 1.29 | 0.23 | 0.11 | 469 | 0.46 | 0.15 | 43 | | | | | 0.25 | 4589 | 2.09 | 1.43 | 0.08 | 0.11 | 469 | 0.08 | 0.16 | 43 | | | | 0.30 | 4.1 | G | 1 |
| hip053502 | U | 4786 | 2.54 | 1.35 | 0.00 | 0.08 | 489 | 0.10 | 0.13 | 46 | | | | | 0.21 | 4786 | 2.29 | 1.43 | -0.05 | 0.08 | 489 | -0.05 | 0.13 | 46 | | | | 0.22 | 4.0 | G | 1 |
| hip059785 | U | 4793 | 2.57 | 1.52 | -0.39 | 0.08 | 479 | -0.25 | 0.10 | 44 | | | | | 0.20 | 4793 | 2.24 | 1.60 | -0.42 | 0.08 | 479 | -0.42 | 0.11 | 44 | | | | 0.22 | 4.3 | G | 1 |
| hip066936 | U | 4589 | 2.51 | 1.32 | 0.29 | 0.13 | 467 | 0.45 | 0.20 | 41 | | | | | 0.26 | 4589 | 2.11 | 1.48 | 0.17 | 0.13 | 467 | 0.17 | 0.22 | 41 | | | | 0.30 | 4.2 | G | 1 |
| hip070306 | U | 4255 | 2.35 | 1.91 | -0.16 | 0.29 | 441 | 0.15 | 0.29 | 31 | | | | | 0.31 | 4255 | 1.63 | 2.00 | -0.40 | 0.30 | 441 | -0.40 | 0.30 | 31 | | | | 0.37 | 3.5 | G | 1 |
| hip078650 | U | 4370 | 2.25 | 1.53 | -0.07 | 0.19 | 466 | 0.12 | 0.22 | 40 | | | | | 0.31 | 4370 | 1.75 | 1.65 | -0.20 | 0.19 | 466 | -0.20 | 0.22 | 40 | | | | 0.35 | 3.7 | G | 1 |
| hip080343 | U | 4758 | 2.48 | 1.64 | -0.33 | 0.17 | 484 | -0.18 | 0.15 | 42 | | | | | 0.22 | 4758 | 2.11 | 1.72 | -0.37 | 0.17 | 484 | -0.37 | 0.16 | 42 | | | | 0.24 | 3.3 | G | 1 |
| hip082396 | U | 4522 | 2.34 | 1.30 | 0.14 | 0.11 | 471 | 0.29 | 0.16 | 40 | | | | | 0.29 | 4522 | 1.95 | 1.44 | 0.03 | 0.11 | 471 | 0.03 | 0.16 | 40 | | | | 0.32 | 4.0 | G | 1 |
| hip093498 | U | 4482 | 2.54 | 1.24 | 0.48 | 0.17 | 444 | 0.64 | 0.18 | 35 | | | | | 0.28 | 4482 | 2.10 | 1.50 | 0.29 | 0.16 | 444 | 0.29 | 0.19 | 35 | | | | 0.31 | 4.3 | G | 1 |
| hip103738 | U | 5009 | 2.69 | 1.60 | -0.09 | 0.12 | 426 | 0.05 | 0.13 | 38 | | | | | 0.16 | 5009 | 2.38 | 1.67 | -0.12 | 0.13 | 426 | -0.12 | 0.13 | 38 | | | | 0.17 | 7.0 | G | 1 |
| hip106039 | U | 5017 | 2.83 | 1.32 | -0.07 | 0.07 | 496 | 0.05 | 0.11 | 51 | | | | | 0.15 | 5017 | 2.55 | 1.40 | -0.09 | 0.07 | 496 | -0.10 | 0.12 | 51 | | | | 0.16 | 4.1 | G | 1 |
| hip113246 | U | 4828 | 2.55 | 1.36 | -0.17 | 0.07 | 496 | -0.02 | 0.13 | 50 | | | | | 0.20 | 4828 | 2.18 | 1.44 | -0.22 | 0.08 | 496 | -0.22 | 0.14 | 50 | | | | 0.22 | 4.0 | G | 1 |
| hip114119 | U | 4962 | 2.64 | 1.09 | 0.04 | 0.08 | 494 | -0.02 | 0.13 | 46 | | | | | 0.17 | 4962 | 2.77 | 1.01 | 0.07 | 0.08 | 494 | 0.07 | 0.13 | 46 | | | | 0.16 | 4.4 | G | 1 |
| hip115102 | U | 4578 | 2.43 | 1.36 | 0.04 | 0.10 | 475 | 0.24 | 0.15 | 45 | | | | | 0.27 | 4578 | 1.93 | 1.49 | -0.07 | 0.10 | 475 | -0.07 | 0.17 | 45 | | | | 0.31 | 4.1 | G | 1 |
| hip116853 | U | 4886 | 2.67 | 1.33 | 0.19 | 0.09 | 478 | 0.27 | 0.14 | 42 | | | | | 0.18 | 4886 | 2.47 | 1.40 | 0.15 | 0.09 | 478 | 0.15 | 0.15 | 42 | | | | 0.19 | 4.5 | G | 1 |
| hr0296 | U | 4500 | 2.19 | 1.38 | -0.23 | 0.11 | 434 | -0.08 | 0.15 | 39 | | | | | 0.30 | 4500 | 1.78 | 1.50 | -0.32 | 0.11 | 434 | -0.32 | 0.16 | 39 | | | | 0.34 | 2.7 | G | 1 |
| hr4321 | U | 4569 | 2.47 | 1.32 | 0.33 | 0.14 | 457 | 0.50 | 0.21 | 31 | 8.57 | 9.02 | 0.34 | 1.52 | 0.27 | 4569 | 2.06 | 1.44 | 0.22 | 0.14 | 457 | 0.23 | 0.22 | 31 | 8.43 | 8.83 | 0.34 | 0.31 | 4.9 | G | 1 |
| ic2391-0022 | U | 4610 | 3.03 | 0.90 | 0.18 | 0.12 | 576 | 0.81 | 0.13 | 58 | 8.49 | 9.01 | 0.09 | 0.76 | 0.22 | 4610 | 1.56 | 1.46 | -0.22 | 0.14 | 576 | -0.22 | 0.18 | 58 | 7.96 | 8.24 | 0.09 | 0.34 | 4.4 | G | 1 |
| ic2391-0026 | U | 4585 | 3.22 | 0.44 | 0.02 | 0.15 | 596 | 0.74 | 0.15 | 65 | 8.38 | 8.90 | -0.21 | 0.42 | 0.21 | 4585 | 1.51 | 1.44 | -0.46 | 0.11 | 596 | -0.46 | 0.18 | 65 | 7.69 | 7.98 | -0.21 | 0.35 | 3.4 | G | 1 |
| ic2391-0044 | U | 6699 | 4.16 | 5.45 | 1.12 | 0.22 | 7 | 1.00 | 0.00 | 1 | 8.16 | 8.57 | 3.12 | 6.06 | -0.08 | 6699 | 4.48 | 5.52 | 1.11 | 0.23 | 7 | 1.11 | | 1 | 8.26 | 8.65 | 3.11 | -0.17 | 100.0 | R | 2 |
| ic4651-E12 | U | 4829 | 2.54 | 1.37 | 0.06 | 0.07 | 486 | 0.15 | 0.12 | 40 | 8.21 | 8.80 | 0.38 | 0.72 | 0.20 | 4829 | 2.32 | 1.43 | 0.02 | 0.08 | 486 | 0.02 | 0.12 | 40 | 8.15 | 8.70 | 0.39 | 0.21 | 3.8 | G | 1 |
| ic4651-E60 | U | 4760 | 2.60 | 1.23 | 0.12 | 0.08 | 482 | 0.19 | 0.15 | 40 | 8.29 | 8.77 | 0.61 | 1.49 | 0.21 | 4760 | 2.44 | 1.30 | 0.09 | 0.08 | 482 | 0.09 | 0.13 | 40 | 8.24 | 8.69 | 0.61 | 0.22 | 3.7 | G | 1 |
| ic4651no7646 | U | 4829 | 2.54 | 1.47 | 0.08 | 0.10 | 449 | 0.17 | 0.13 | 36 | 8.18 | 8.71 | 0.28 | 0.58 | 0.20 | 4829 | 2.31 | 1.53 | 0.04 | 0.10 | 449 | 0.04 | 0.14 | 36 | 8.12 | 8.61 | 0.29 | 0.21 | 4.2 | G | 1 |
| ic4651no9122 | U | 4694 | 2.56 | 1.36 | 0.20 | 0.12 | 438 | 0.27 | 0.16 | 33 | 8.43 | 8.95 | 0.29 | 0.90 | 0.23 | 4694 | 2.40 | 1.42 | 0.16 | 0.12 | 438 | 0.16 | 0.17 | 33 | 8.38 | 8.88 | 0.29 | 0.25 | 4.0 | G | 1 |
| ngc2447No41 | U | 5058 | 2.56 | 1.54 | -0.02 | 0.10 | 454 | 0.03 | 0.14 | 38 | 8.09 | 8.64 | 0.54 | 0.54 | 0.15 | 5058 | 2.45 | 1.56 | -0.03 | 0.10 | 454 | -0.03 | 0.14 | 38 | 8.07 | 8.59 | 0.55 | 0.16 | 5.7 | G | 1 |
| ngc2682esiii-35 | U | 5046 | 3.27 | 0.99 | 0.21 | 0.09 | 473 | 0.22 | 0.15 | 37 | 8.30 | 8.85 | 0.52 | 0.53 | 0.14 | 5046 | 3.23 | 1.02 | 0.20 | 0.09 | 473 | 0.20 | 0.15 | 37 | 8.29 | 8.83 | 0.52 | 0.14 | 3.8 | G | 1 |
| ngc2682No164 | U | 4734 | 2.48 | 1.37 | 0.11 | 0.11 | 437 | 0.14 | 0.17 | 32 | 8.30 | 8.79 | 0.17 | 0.60 | 0.24 | 4734 | 2.41 | 1.39 | 0.10 | 0.11 | 437 | 0.10 | 0.17 | 32 | 8.27 | 8.76 | 0.17 | 0.24 | 3.8 | G | 1 |
| ngc2682No286 | U | 4778 | 2.53 | 1.41 | 0.12 | 0.11 | 435 | 0.11 | 0.13 | 30 | 8.31 | 8.82 | 0.31 | 0.72 | 0.21 | 4778 | 2.45 | 1.47 | 0.13 | 0.11 | 435 | 0.13 | 0.13 | 30 | 8.39 | 8.87 | 0.31 | 0.22 | 3.9 | G | 1 |

| KeyName | S | T | G | Vt | FeI | S | N | FeII | S | N | C | O | Li | EW | N1 | T | G | Vt | FeI | S | N | FeII | S | N | C | O | Li | EW | S |
|---|---|---|---|---|---|---|---|---|---|---|---|---|---|---|---|---|---|---|---|---|---|---|---|---|---|---|---|---|---|
| ngc3114no181 | U | 4456 | 1.61 | 1.83 | -0.01 | 0.14 | 395 | 0.09 | 0.18 | 30 | 8.14 | 8.61 | 0.35 | 2.30 | 0.36 | 4456 | 1.38 | 1.82 | -0.03 | 0.14 | 395 | -0.04 | 0.19 | 30 | 8.08 | 8.51 | 0.34 | 0.38 | 5.0 G 1 |
| ngc3680no26 | U | 4662 | 2.52 | 1.22 | 0.07 | 0.11 | 462 | 0.15 | 0.16 | 33 | 8.23 | 8.74 | 1.11 | 5.67 | 0.24 | 4662 | 2.35 | 1.29 | 0.03 | 0.11 | 462 | 0.03 | 0.16 | 33 | 8.18 | 8.66 | 1.11 | 0.26 | 4.7 G 1 |
| p1955 | U | 5211 | 2.50 | 3.34 | 0.12 | 0.37 | 9 | 0.94 | 0.00 | 1 | | | 2.82 | 23.42 | 0.13 | 5211 | 0.46 | 3.26 | 0.11 | 0.36 | 9 | 0.11 | | 1 | | | 2.78 | 0.21 | 90.0 R 2 |
| txpic | U | 4435 | 2.13 | 4.25 | | | | -0.76 | 0.00 | 1 | | | 0.52 | 3.57 | 0.32 | 4435 | 2.13 | 3.52 | | | | -0.69 | | 1 | | | 0.53 | 0.32 | 46.0 R 2 |
| XSct | U | 4763 | 2.81 | 3.19 | -0.69 | 0.23 | 50 | 1.37 | 0.61 | 6 | | | | | 0.20 | 4763 | 0.20 | 3.18 | -0.76 | 0.28 | 50 | 0.33 | 0.73 | 6 | | | | 0.37 | 5.0 G 2 |

| | |
|---|---|
| KeyName | Identifier |
| S | Source of Spectra |

Group 1: Data for the mass determined gravity:

| | |
|---|---|
| T | Effective temperature (Kelvins) |
| G | Logarithm of the surface gravity (cm/s^2) |
| Vt | Mictoturbulent Velocity (km/s) |
| FeI | [Fe/H] determined from Fe I – logarithmic with respect to the solar value = 7.47 |
| S | Standard Deviation of Fe I derived [Fe/H] ratio |
| N | Number of Fe I lines used |
| FeII | [Fe/H] determined from Fe II – logarithmic with respect to the solar value = 7.47 |
| S | Standard Deviation of Fe II derived [Fe/H] ratio |
| N | Number of Fe II lines used |
| C | Carbon abundance with respect to log $\varepsilon_H$ = 12. |
| O | Oxygen abundance with respect to log $\varepsilon_H$ = 12. |
| Li | Lithium abundance with respect to log $\varepsilon_H$ = 12. |
| EW | Equivalent width in picometers of the lithium 670.7 nm feature. EW < 1.00 pm corresponds to an upper limit. |
| N1 | Lithium non-LTE abundance correction from the data of Lind et al. (2009) |

Group 2: Data for the ionization balance determined gravity:

| | |
|---|---|
| T | Effective temperature (Kelvins) |
| G | Logarithm of the surface gravity (cm/s^2) |
| Vt | Mictoturbulent Velocity (km/s) |
| FeI | [Fe/H] determined from Fe I – logarithmic with respect to the solar value = 7.47 |
| S | Standard Deviation of Fe I derived [Fe/H] ratio |
| N | Number of Fe I lines used |
| FeII | [Fe/H] determined from Fe II – logarithmic with respect to the solar value = 7.47 |
| S | Standard Deviation of Fe II derived [Fe/H] ratio |
| N | Number of Fe II lines used |
| C | Carbon abundance with respect to log $\varepsilon_H$ = 12. |
| O | Oxygen abundance with respect to log $\varepsilon_H$ = 12. |
| Li | Lithium abundance with respect to log $\varepsilon_H$ = 12. |
| EW | Equivalent width of the lithium 670.7 nm feature |

N2          Lithium non-LTE abundance correction from the data of Lind et al. (2009)

Vm          Broadening velocity (km/s)
B           Type of broadening: G = Gaussian Macroturbulence   R = Rotational
Cl          Clump Star:   1 = True           2 = False

Table 4
Average Element Abundances [x/H]

| Keyname | S | T | G | Vt | Na | Mg | Al | Si | S | Ca | Sc | Ti | V | Cr | Mn | Fe | Co | Ni | Cu | Zn | Sr | Y | Zr | Ba | La | Ce | Nd | Sm | Eu |
|---|---|---|---|---|---|---|---|---|---|---|---|---|---|---|---|---|---|---|---|---|---|---|---|---|---|---|---|---|---|
| hd001671 | E | 6323 | 3.61 | 2.61 | 0.11 | | | 0.17 | 0.55 | 0.22 | 0.33 | 0.33 | 0.82 | 0.44 | 0.06 | 0.03 | 0.39 | 0.33 | -0.59 | -0.35 | 1.46 | 0.91 | 1.97 | | 0.25 | 0.29 | 0.41 | 0.78 | |
| hd002910 | E | 4696 | 2.60 | 1.36 | 0.30 | 0.26 | 0.22 | 0.41 | 1.02 | 0.17 | 0.13 | 0.12 | 0.27 | 0.22 | 0.19 | 0.17 | 0.11 | 0.26 | 0.30 | 0.71 | 0.02 | 0.22 | -0.02 | 0.12 | 0.56 | 0.86 | 0.50 | 0.43 | 0.21 |
| hd004188 | E | 4793 | 2.49 | 1.39 | 0.29 | 0.22 | 0.26 | 0.25 | 0.70 | 0.11 | 0.03 | 0.08 | 0.17 | 0.16 | 0.11 | 0.08 | 0.03 | 0.12 | 0.12 | 0.35 | 0.12 | 0.20 | 0.04 | 0.12 | 0.44 | 0.65 | 0.38 | 0.40 | 0.14 |
| hd004502 | E | 4570 | 2.25 | 2.88 | | | 0.08 | 0.69 | | -0.07 | 0.06 | 0.00 | -0.04 | -0.01 | -0.16 | -0.16 | 0.15 | -0.02 | | | -0.03 | 0.82 | | 0.85 | | 0.48 | 0.84 | 0.44 | |
| hd005234 | E | 4422 | 1.87 | 1.58 | 0.27 | 0.09 | 0.21 | 0.41 | 0.81 | -0.01 | -0.08 | 0.02 | 0.06 | 0.18 | 0.04 | 0.03 | -0.06 | 0.08 | 0.12 | -0.29 | 0.06 | 0.04 | -0.11 | 0.13 | 0.50 | 0.81 | 0.52 | 0.79 | 0.06 |
| hd006319 | E | 4755 | 2.74 | 1.38 | 0.38 | 0.29 | 0.42 | 0.37 | 0.95 | 0.18 | 0.30 | 0.32 | 0.48 | 0.31 | 0.30 | 0.25 | 0.25 | 0.35 | 0.48 | 0.46 | 0.17 | 0.26 | 0.17 | 0.28 | 0.68 | 0.90 | 0.75 | 0.67 | 0.31 |
| hd010380 | E | 4154 | 1.50 | 1.83 | 0.19 | 0.26 | 0.13 | 0.33 | 1.16 | -0.20 | -0.13 | -0.09 | -0.02 | 0.01 | -0.14 | -0.09 | 0.01 | 0.03 | 0.18 | 0.23 | -0.06 | -0.11 | -0.25 | -0.23 | 0.31 | 0.30 | 0.37 | 0.33 | -0.03 |
| hd011559 | E | 4947 | 2.80 | 1.33 | 0.46 | 0.27 | 0.31 | 0.37 | 0.86 | 0.21 | 0.12 | 0.18 | 0.27 | 0.27 | 0.18 | 0.19 | 0.14 | 0.22 | 0.23 | 0.22 | 0.19 | 0.29 | 0.09 | 0.23 | 0.45 | 0.75 | 0.53 | 0.47 | 0.35 |
| hd013174 | E | 6710 | 3.25 | 6.00 | -1.12 | | | 0.30 | | 1.30 | 0.75 | 1.34 | 1.36 | 0.37 | -0.09 | 0.48 | 1.58 | 0.42 | | | 0.55 | -0.11 | 2.23 | | 0.11 | 0.72 | 1.32 | 0.72 | |
| hd013480 | E | 5082 | 2.68 | 2.89 | | | 0.18 | 0.47 | 0.08 | 0.02 | 0.28 | 0.37 | 0.59 | -0.26 | 0.01 | 0.39 | 0.09 | -0.55 | | 2.30 | 0.23 | 1.29 | -0.15 | 0.65 | 0.66 | 0.98 | 0.56 | 0.89 | |
| hd013520 | E | 4010 | 1.22 | 1.96 | 0.00 | -0.07 | 0.01 | 0.28 | 1.48 | -0.32 | -0.22 | -0.18 | -0.09 | -0.04 | -0.33 | -0.12 | -0.03 | -0.04 | 0.42 | 0.28 | 0.24 | -0.01 | -0.38 | -0.24 | 0.07 | 0.12 | 0.28 | 0.45 | 0.19 |
| hd015257 | E | 7079 | 3.79 | 4.31 | 0.76 | | | 0.16 | | 0.51 | 1.02 | 1.14 | 1.22 | 0.69 | 0.07 | 0.34 | 1.23 | 1.03 | | 0.23 | | 1.25 | 0.29 | | 1.14 | | 1.74 | | |
| hd015596 | E | 4788 | 2.76 | 1.08 | -0.43 | -0.26 | -0.20 | -0.23 | 0.23 | -0.28 | -0.42 | -0.31 | -0.43 | -0.51 | -0.78 | -0.58 | -0.48 | -0.53 | -0.53 | -0.37 | -0.31 | -0.45 | -0.43 | -0.56 | -0.34 | -0.21 | -0.13 | -0.30 | 0.05 |
| hd018885 | E | 4670 | 2.58 | 1.40 | 0.53 | 0.45 | 0.31 | 0.50 | 1.11 | 0.23 | 0.19 | 0.23 | 0.41 | 0.33 | 0.35 | 0.27 | 0.26 | 0.39 | 0.49 | 0.86 | 0.13 | 0.27 | 0.17 | 0.23 | 0.65 | 0.87 | 0.59 | 0.65 | 0.27 |
| hd022764 | E | 4205 | 1.41 | 2.54 | 0.81 | 0.08 | 0.49 | 0.58 | 1.99 | 0.15 | 0.04 | 0.17 | 0.31 | 0.39 | 0.10 | 0.11 | 0.18 | 0.21 | 0.63 | -0.17 | 0.47 | 0.11 | 0.22 | | 0.61 | 0.64 | 0.49 | 0.68 | 0.58 |
| hd023249 | E | 4966 | 3.72 | 0.50 | 0.33 | 0.28 | 0.36 | 0.32 | 0.84 | 0.21 | 0.29 | 0.27 | 0.41 | 0.29 | 0.26 | 0.21 | 0.21 | 0.36 | 0.51 | 0.61 | 0.16 | 0.35 | 0.05 | 0.03 | 0.56 | 0.75 | 0.74 | 0.72 | 0.45 |
| hd025602 | E | 4743 | 2.84 | 1.07 | 0.00 | -0.01 | 0.04 | 0.10 | 0.52 | -0.06 | -0.09 | -0.10 | -0.06 | -0.08 | -0.13 | -0.09 | -0.11 | -0.02 | 0.06 | 0.15 | -0.12 | -0.03 | -0.27 | -0.02 | 0.25 | 0.34 | 0.34 | 0.16 | 0.35 |
| hd026659 | E | 5170 | 2.74 | 1.42 | 0.17 | 0.09 | 0.09 | 0.03 | 0.13 | 0.03 | -0.12 | -0.03 | -0.02 | 0.03 | -0.13 | -0.05 | -0.07 | -0.07 | -0.18 | -0.21 | 0.06 | 0.03 | -0.02 | 0.15 | 0.08 | 0.31 | 0.15 | -0.05 | 0.46 |
| hd029139 | E | 3903 | 1.32 | 2.06 | 0.21 | -0.01 | 0.31 | 0.75 | 2.55 | -0.37 | -0.06 | -0.16 | 0.03 | 0.10 | -0.52 | 0.03 | 0.12 | 0.12 | 0.88 | 0.49 | 0.22 | -0.02 | -0.35 | -0.14 | -0.04 | 0.12 | 0.34 | 0.14 | -0.98 |
| hd030834 | E | 4153 | 1.39 | 1.86 | 0.08 | 0.06 | -0.03 | 0.16 | 1.07 | -0.27 | -0.22 | -0.20 | -0.20 | -0.09 | -0.33 | -0.15 | -0.12 | -0.09 | 0.22 | -0.29 | -0.13 | -0.19 | -0.37 | 0.02 | 0.31 | 0.38 | 0.31 | 0.38 | 0.20 |
| hd031444 | E | 5051 | 2.89 | 1.21 | 0.18 | 0.13 | 0.12 | 0.11 | 0.50 | 0.09 | -0.02 | -0.03 | 0.02 | 0.08 | -0.08 | -0.01 | -0.08 | -0.02 | -0.22 | 0.05 | 0.24 | 0.13 | -0.03 | 0.24 | 0.27 | 0.55 | 0.38 | 0.19 | 0.45 |
| hd033419 | E | 4674 | 2.64 | 1.40 | 0.51 | 0.47 | 0.36 | 0.50 | 1.21 | 0.23 | 0.23 | 0.28 | 0.40 | 0.36 | 0.37 | 0.30 | 0.27 | 0.41 | 0.62 | 0.61 | 0.16 | 0.33 | 0.13 | 0.27 | 0.68 | 0.82 | 0.71 | 0.68 | 0.30 |
| hd033618 | E | 4558 | 2.45 | 1.25 | 0.66 | 0.46 | 0.51 | 0.57 | 1.42 | 0.41 | 0.35 | 0.51 | 0.68 | 0.53 | 0.57 | 0.39 | 0.43 | 0.60 | 0.80 | 0.86 | 0.32 | 0.39 | 0.19 | 0.35 | 0.88 | 1.18 | 0.96 | 0.97 | 0.54 |
| hd034029 | E | 5155 | 2.47 | 2.45 | | 0.43 | -0.39 | 0.09 | 0.51 | -0.51 | -0.10 | -0.22 | 0.03 | 0.36 | -0.45 | -0.46 | -0.27 | -0.11 | -1.59 | -0.31 | 1.00 | 0.06 | 0.08 | -1.69 | 0.36 | 0.17 | -0.18 | -0.10 | -0.18 |
| hd037160 | E | 4742 | 2.68 | 1.10 | -0.33 | -0.22 | -0.14 | -0.17 | 0.34 | -0.28 | -0.31 | -0.27 | -0.30 | -0.41 | -0.62 | -0.47 | -0.40 | -0.41 | -0.33 | -0.18 | -0.40 | -0.43 | -0.56 | -0.53 | -0.24 | -0.18 | -0.16 | -0.07 | 0.27 |
| hd037638 | E | 5088 | 2.90 | 1.32 | 0.25 | 0.10 | 0.17 | 0.17 | 0.51 | 0.13 | 0.03 | 0.05 | 0.06 | 0.11 | 0.03 | 0.04 | -0.04 | 0.03 | -0.07 | 0.20 | 0.34 | 0.16 | 0.08 | 0.27 | 0.41 | 0.74 | 0.52 | 0.34 | 0.25 |
| hd038309 | E | 6927 | 3.94 | 3.91 | -0.03 | 1.04 | | 0.02 | | -0.20 | 0.16 | 0.62 | 0.72 | 0.36 | 0.05 | -0.24 | 1.42 | 0.17 | -0.34 | | | 0.77 | 2.23 | -0.87 | | 0.31 | 0.93 | 0.88 | 0.43 |
| hd039070 | E | 5086 | 2.94 | 1.19 | 0.35 | 0.19 | 0.19 | 0.24 | 0.42 | 0.22 | 0.11 | 0.15 | 0.18 | 0.21 | 0.06 | 0.12 | 0.07 | 0.12 | 0.07 | 0.22 | 0.39 | 0.28 | 0.46 | 0.28 | 0.40 | 0.79 | 0.57 | 0.40 | 0.62 |
| hd039833 | E | 5751 | 4.35 | 0.68 | 0.21 | 0.36 | 0.25 | 0.26 | 0.30 | 0.27 | 0.30 | 0.29 | 0.30 | 0.35 | 0.21 | 0.27 | 0.21 | 0.27 | 0.15 | 0.34 | 0.59 | 0.47 | 0.46 | 0.38 | 0.72 | 0.75 | 0.61 | 0.55 | 0.94 |
| hd039910 | E | 4577 | 2.53 | 1.56 | 0.49 | 0.52 | 0.45 | 0.60 | 1.32 | 0.25 | 0.33 | 0.33 | 0.48 | 0.38 | 0.35 | 0.33 | 0.39 | 0.47 | 0.73 | 0.60 | 0.11 | 0.20 | 0.06 | 0.10 | 0.69 | 1.02 | 0.67 | 0.81 | 0.40 |
| hd040801 | E | 4787 | 2.94 | 1.07 | 0.00 | 0.10 | 0.10 | 0.14 | 0.58 | 0.00 | -0.03 | -0.03 | 0.08 | -0.05 | -0.08 | -0.06 | -0.07 | 0.01 | 0.12 | 0.29 | -0.16 | -0.08 | -0.27 | -0.08 | 0.26 | 0.54 | 0.45 | 0.38 | 0.39 |
| hd042341 | E | 4635 | 2.73 | 1.42 | 0.81 | 0.60 | 0.53 | 0.60 | 1.41 | 0.44 | 0.39 | 0.42 | 0.65 | 0.52 | 0.47 | 0.41 | 0.45 | 0.54 | 0.58 | 0.95 | 0.26 | 0.28 | 0.30 | 0.10 | 0.62 | 1.00 | 0.82 | 1.09 | 0.47 |
| hd046374 | E | 4656 | 2.62 | 1.36 | 0.34 | 0.38 | 0.38 | 0.42 | 1.00 | 0.14 | 0.18 | 0.18 | 0.35 | 0.24 | 0.19 | 0.17 | 0.15 | 0.27 | 0.43 | 0.37 | 0.13 | 0.26 | 0.01 | 0.14 | 0.56 | 0.88 | 0.60 | 0.46 | 0.18 |
| hd047138 | E | 6091 | 3.75 | 1.38 | 0.62 | 0.56 | 0.44 | 0.23 | -0.04 | 0.51 | 0.54 | 0.75 | 0.93 | 0.61 | 0.57 | 0.52 | 0.70 | 0.51 | 0.50 | 0.28 | 1.04 | 0.73 | 0.84 | 0.80 | 0.96 | 1.16 | 1.07 | 0.85 | 0.86 |
| hd047366 | E | 4857 | 3.02 | 0.94 | 0.24 | 0.17 | 0.22 | 0.25 | 0.68 | 0.21 | 0.10 | 0.21 | 0.31 | 0.20 | 0.15 | 0.15 | 0.09 | 0.20 | 0.30 | 0.38 | 0.21 | 0.22 | 0.19 | 0.25 | 0.60 | 0.86 | 0.88 | 0.58 | 0.22 |
| hd050522 | E | 5095 | 2.71 | 1.03 | 0.55 | 0.35 | 0.37 | 0.33 | 0.65 | 0.33 | 0.21 | 0.33 | 0.55 | 0.41 | 0.29 | 0.25 | 0.21 | 0.31 | 0.21 | 0.13 | 0.30 | 0.30 | 0.35 | 0.21 | 0.41 | 0.71 | 0.39 | 0.58 | 0.74 |
| hd051000 | E | 5102 | 2.73 | 1.40 | 0.29 | 0.05 | 0.06 | 0.12 | 0.36 | 0.06 | -0.08 | -0.01 | 0.00 | 0.10 | -0.08 | 0.00 | -0.05 | 0.00 | -0.21 | -0.04 | 0.27 | 0.15 | 0.07 | 0.27 | 0.39 | 0.59 | 0.25 | 0.15 | 0.21 |
| hd054079 | E | 4454 | 1.85 | 1.60 | -0.17 | -0.15 | -0.07 | 0.04 | 0.66 | -0.33 | -0.28 | -0.28 | -0.24 | -0.26 | -0.39 | -0.33 | -0.34 | -0.27 | -0.15 | -0.14 | -0.21 | -0.22 | -0.45 | -0.21 | 0.03 | 0.29 | 0.18 | 0.02 | -0.06 |
| hd055280 | E | 4639 | 2.64 | 1.22 | 0.42 | 0.34 | 0.40 | 0.44 | 0.99 | 0.23 | 0.15 | 0.22 | 0.37 | 0.31 | 0.32 | 0.26 | 0.19 | 0.34 | 0.50 | 0.72 | 0.18 | 0.22 | 0.07 | 0.22 | 0.71 | 0.92 | 0.81 | 0.73 | 0.42 |
| hd057727 | E | 5007 | 3.06 | 1.12 | 0.12 | 0.06 | 0.10 | 0.16 | 0.27 | 0.09 | -0.02 | 0.05 | 0.06 | 0.10 | 0.03 | 0.05 | -0.04 | 0.05 | -0.03 | 0.16 | 0.15 | 0.13 | 0.14 | 0.24 | 0.43 | 0.62 | 0.50 | 0.37 | 0.43 |
| hd058207 | E | 4750 | 2.69 | 1.30 | 0.17 | 0.21 | 0.19 | 0.29 | 0.91 | 0.03 | 0.05 | 0.02 | 0.11 | 0.09 | 0.06 | -0.01 | 0.13 | 0.19 | 0.41 | 0.00 | 0.11 | -0.06 | 0.18 | 0.47 | 0.71 | 0.48 | 0.37 | 0.16 | |
| hd058923 | E | 7505 | 3.46 | 3.18 | | 0.48 | | 0.86 | | 0.76 | 0.62 | 0.94 | 1.33 | 1.18 | 0.70 | 0.49 | 1.48 | 0.78 | | 0.90 | | 1.84 | 1.64 | | 0.61 | 0.68 | 2.20 | 0.76 | |
| hd060294 | E | 4580 | 2.62 | 1.23 | 0.47 | 0.38 | 0.42 | 0.49 | 1.21 | 0.27 | 0.22 | 0.26 | 0.42 | 0.35 | 0.31 | 0.27 | 0.26 | 0.40 | 0.60 | 0.79 | 0.17 | 0.24 | 0.08 | 0.19 | 0.71 | 0.91 | 0.73 | 0.74 | 0.37 |
| hd060522 | E | 3884 | 1.35 | 2.13 | 0.41 | 0.13 | 0.49 | 0.93 | 2.96 | -0.01 | 0.03 | 0.05 | 0.28 | 0.29 | -0.20 | 0.19 | 0.31 | 0.35 | 1.04 | 0.10 | 0.83 | 0.14 | -0.03 | -0.08 | 0.42 | 0.69 | 0.46 | 0.41 | 0.09 |
| hd061363 | E | 4713 | 2.42 | 1.38 | -0.03 | 0.00 | -0.03 | 0.11 | 0.46 | -0.15 | -0.18 | -0.19 | -0.19 | -0.13 | -0.23 | -0.16 | -0.19 | -0.11 | -0.09 | 0.14 | -0.15 | -0.17 | -0.27 | 0.00 | 0.20 | 0.52 | 0.32 | 0.16 | 0.00 |
| hd062141 | E | 4922 | 2.85 | 1.23 | 0.20 | 0.05 | 0.19 | 0.15 | 0.52 | 0.10 | -0.02 | -0.03 | 0.01 | 0.09 | 0.00 | 0.01 | -0.10 | 0.00 | -0.07 | 0.07 | 0.15 | 0.19 | 0.05 | 0.23 | 0.39 | 0.66 | 0.35 | 0.23 | 0.10 |

| ID | Type | T | c1 | c2 | c3 | c4 | c5 | c6 | c7 | c8 | c9 | c10 | c11 | c12 | c13 | c14 | c15 | c16 | c17 | c18 | c19 | c20 | c21 | c22 | c23 | c24 | c25 | c26 | c27 |
|---|---|---|---|---|---|---|---|---|---|---|---|---|---|---|---|---|---|---|---|---|---|---|---|---|---|---|---|---|---|
| hd062437 | E | 7600 | 3.72 | 2.60 | 0.44 | 0.47 |  | 0.71 | 0.36 | 0.86 | 0.63 | 1.17 | 1.33 | 1.03 | 0.99 | 0.46 | 1.39 | 0.82 |  |  | 1.65 | 1.74 | 0.80 | 0.22 | 0.86 | 1.31 | 0.85 | 2.70 | 2.21 |
| hd062721 | E | 4002 | 1.48 | 1.76 | 0.21 | 0.17 | 0.32 | 0.56 | 2.23 | -0.01 | 0.06 | 0.07 | 0.25 | 0.13 | -0.24 | 0.03 | 0.19 | 0.14 | 1.00 | -0.11 | 0.42 | 0.16 | -0.11 | -0.24 | 0.21 | 0.59 | 0.67 | 0.78 | 0.10 |
| hd064152 | E | 4928 | 2.75 | 1.32 | 0.40 | 0.26 | 0.25 | 0.28 | 0.82 | 0.20 | 0.07 | 0.10 | 0.18 | 0.21 | 0.15 | 0.13 | 0.06 | 0.16 | 0.04 | 0.25 | 0.27 | 0.21 | 0.03 | 0.19 | 0.46 | 0.64 | 0.46 | 0.44 | 0.46 |
| hd074794 | E | 4648 | 2.54 | 1.40 | 0.38 | 0.39 | 0.32 | 0.46 | 1.19 | 0.20 | 0.19 | 0.19 | 0.35 | 0.27 | 0.31 | 0.24 | 0.22 | 0.34 | 0.55 | 0.53 | 0.08 | 0.21 | 0.00 | 0.18 | 0.63 | 0.82 | 0.64 | 0.63 | 0.28 |
| hd081688 | E | 4760 | 2.49 | 1.43 | -0.19 | 0.01 | 0.00 | 0.08 | 0.44 | -0.22 | -0.16 | -0.16 | -0.18 | -0.17 | -0.26 | -0.18 | -0.18 | -0.15 | -0.11 | -0.07 | -0.31 | -0.26 | -0.36 | -0.14 | 0.07 | 0.26 | 0.18 | 0.12 | 0.00 |
| hd082885 | E | 5412 | 4.39 | 0.50 | 0.53 | 0.43 | 0.45 | 0.43 | 0.81 | 0.39 | 0.44 | 0.43 | 0.51 | 0.50 | 0.49 | 0.38 | 0.39 | 0.50 | 0.58 | 0.75 | 0.49 | 0.67 | 0.53 | 0.26 | 0.84 | 1.10 | 0.85 | 1.30 | 0.97 |
| hd089025 | E | 6977 | 2.91 | 6.50 | 0.84 |  |  | 0.29 | 0.08 | -0.60 | 0.27 | 0.69 | 0.62 | 0.27 | -0.15 | -0.30 | 1.54 | 0.32 |  |  | -0.30 | 0.06 |  | 0.00 | 0.06 | 0.46 | 1.04 |  |  |
| hd093875 | E | 4556 | 2.46 | 1.25 | 0.64 | 0.49 | 0.46 | 0.60 | 1.44 | 0.37 | 0.36 | 0.43 | 0.60 | 0.48 | 0.47 | 0.36 | 0.40 | 0.58 | 1.20 | 0.77 | 0.19 | 0.37 | 0.20 | 0.31 | 0.77 | 1.14 | 0.92 | 0.92 | 0.52 |
| hd094672 | E | 6446 | 3.88 | 1.87 | 0.10 | -0.01 | -0.25 | 0.03 | -0.10 | 0.17 | 0.10 | 0.13 | 0.00 | 0.00 | -0.21 | -0.09 | 0.17 | -0.05 | -0.34 | -0.32 |  | 0.03 | 0.46 | 0.24 | 0.00 | 0.14 | 0.16 | 0.42 | 0.89 |
| hd095345 | E | 4519 | 2.02 | 1.56 | 0.13 | 0.15 | 0.13 | 0.25 | 0.90 | -0.05 | -0.06 | -0.03 | 0.03 | 0.06 | -0.12 | -0.03 | -0.05 | 0.01 | 0.04 | 0.08 | 0.00 | 0.06 | -0.24 | 0.13 | 0.43 | 0.72 | 0.46 | 0.46 | 0.15 |
| hd098262 | E | 4113 | 1.17 | 1.86 | 0.39 | -0.04 | 0.07 | 0.35 | 1.28 | -0.21 | -0.24 | -0.14 | -0.12 | 0.06 | -0.19 | -0.07 | -0.08 | -0.01 | 0.06 | -0.26 | 0.09 | -0.08 | -0.27 | 0.18 | 0.34 | 0.44 | 0.32 | 0.43 | 0.23 |
| hd098366 | E | 4697 | 2.70 | 1.12 | 0.21 | 0.20 | 0.20 | 0.28 | 0.73 | 0.06 | 0.10 | 0.09 | 0.21 | 0.12 | 0.06 | 0.08 | 0.02 | 0.16 | 0.25 | 0.37 | 0.09 | 0.19 | 0.02 | 0.12 | 0.47 | 0.85 | 0.60 | 0.56 | 0.16 |
| hd107700 | E | 6115 | 3.02 | 0.50 | 0.42 | 0.16 | 0.38 | 0.07 | -0.31 | 0.36 | 0.23 | 0.50 | 0.77 | 0.48 | 0.24 | 0.29 | 0.63 | 0.38 | 0.01 | -0.34 | 0.68 | 0.42 | 0.64 | 0.50 | 1.04 | 0.53 | 0.52 | 0.39 | 0.43 |
| hd107950 | E | 5098 | 2.45 | 1.78 | 0.43 | 0.19 | 0.20 | 0.25 | 0.51 | 0.13 | 0.02 | 0.08 | 0.04 | 0.17 | -0.02 | 0.08 | 0.01 | 0.07 | -0.21 | -0.12 | 0.48 | 0.24 | 0.10 | 0.28 | 0.25 | 0.46 | 0.40 | 0.15 | 0.08 |
| hd112127 | E | 4383 | 2.60 | 1.73 | 0.64 | 0.69 | 0.47 | 0.90 | 1.96 | 0.26 | 0.44 | 0.36 | 0.60 | 0.53 | 0.57 | 0.51 | 0.65 | 0.71 | 0.71 | 0.90 | 0.17 | 0.40 | 0.14 | 0.06 | 0.89 | 1.19 | 0.86 | 0.92 | 0.89 |
| hd115604 | E | 7314 | 3.43 | 2.09 | 0.71 | 0.64 | 0.67 | 0.62 | 0.49 | 0.69 | 0.89 | 0.72 | 0.86 | 0.69 | 0.46 | 0.54 | 0.82 | 0.67 | 0.57 | 0.65 | 1.11 | 0.94 | 0.98 | 1.69 | 1.12 | 0.98 | 0.88 | 0.71 | 1.27 |
| hd116292 | E | 4840 | 2.61 | 1.36 | 0.14 | 0.04 | 0.11 | 0.21 | 0.43 | 0.01 | -0.06 | -0.08 | -0.05 | 0.03 | -0.14 | -0.01 | -0.13 | 0.01 | -0.12 | 0.01 | -0.06 | -0.01 | -0.12 | 0.14 | 0.22 | 0.29 | 0.29 | 0.05 | 0.34 |
| hd116515 | E | 4728 | 2.32 | 1.44 | 0.07 | 0.05 | 0.15 | 0.12 | 0.45 | -0.05 | -0.07 | -0.06 | -0.01 | 0.00 | -0.08 | -0.07 | -0.08 | -0.01 | 0.00 | -0.07 | -0.07 | 0.06 | -0.16 | -0.06 | 0.19 | 0.52 | 0.28 | 0.22 | -0.03 |
| hd117710 | E | 4661 | 2.76 | 1.25 | 0.64 | 0.51 | 0.59 | 0.62 | 1.21 | 0.37 | 0.28 | 0.35 | 0.54 | 0.46 | 0.41 | 0.37 | 0.35 | 0.53 | 0.56 | 0.95 | 0.21 | 0.35 | 0.18 | 0.10 | 0.82 | 0.95 | 0.75 | 0.84 | 0.46 |
| hd120084 | E | 4806 | 2.61 | 1.34 | 0.41 | 0.26 | 0.25 | 0.34 | 0.90 | 0.14 | 0.09 | 0.10 | 0.24 | 0.25 | 0.20 | 0.16 | 0.05 | 0.22 | 0.29 | 0.19 | 0.13 | 0.30 | 0.05 | 0.13 | 0.46 | 0.66 | 0.50 | 0.43 | 0.18 |
| hd130952 | E | 4744 | 2.46 | 1.44 | -0.17 | 0.05 | 0.08 | 0.09 | 0.39 | -0.17 | -0.24 | -0.11 | -0.13 | -0.19 | -0.38 | -0.27 | -0.20 | -0.20 | -0.21 | -0.09 | -0.29 | -0.33 | -0.34 | -0.37 | -0.17 | 0.11 | 0.09 | -0.04 | 0.06 |
| hd131873 | E | 4005 | 1.33 | 1.83 | 0.03 | 0.09 | 0.10 | 0.40 | 1.95 | -0.19 | -0.15 | -0.07 | 0.02 | 0.10 | -0.14 | 0.00 | 0.06 | 0.11 | 0.46 | -0.28 | 0.21 | 0.03 | -0.23 | 0.08 | 0.28 | 0.53 | 0.54 | 0.64 | 0.32 |
| hd136202 | E | 6025 | 3.87 | 1.34 | 0.14 | 0.06 | 0.05 | 0.10 | 0.17 | 0.10 | 0.12 | 0.04 | 0.04 | 0.06 | -0.02 | -0.01 | 0.06 | 0.00 | -0.15 | -0.06 | 0.20 | 0.19 | 0.41 | 0.11 | 0.17 | 0.12 | 0.16 | -0.04 | 0.51 |
| hd136514 | E | 4417 | 2.33 | 1.38 | 0.31 | 0.39 | 0.45 | 0.48 | 1.28 | 0.12 | 0.17 | 0.22 | 0.43 | 0.27 | 0.16 | 0.17 | 0.20 | 0.32 | 0.51 | 0.54 | 0.08 | 0.04 | -0.01 | -0.04 | 0.43 | 0.74 | 0.74 | 0.79 | 0.36 |
| hd139254 | E | 4676 | 2.65 | 1.20 | 0.31 | 0.26 | 0.14 | 0.30 | 0.88 | 0.08 | 0.12 | 0.12 | 0.21 | 0.24 | 0.15 | 0.13 | 0.04 | 0.19 | 0.25 | 0.67 | 0.17 | 0.30 | 0.04 | 0.21 | 0.35 | 0.89 | 0.42 | 0.69 | 0.16 |
| hd143553 | E | 4697 | 2.66 | 1.14 | -0.06 | -0.02 | 0.05 | 0.09 | 0.42 | -0.10 | -0.11 | -0.10 | -0.05 | -0.10 | -0.15 | -0.12 | -0.14 | -0.05 | 0.02 | 0.20 | -0.16 | -0.06 | -0.26 | -0.13 | 0.22 | 0.29 | 0.35 | 0.26 | -0.08 |
| hd148604 | E | 5167 | 2.91 | 1.04 | 0.13 | 0.05 | 0.16 | 0.01 | 0.19 | 0.04 | -0.08 | 0.00 | 0.03 | 0.08 | -0.12 | -0.04 | -0.09 | -0.07 | -0.28 | 0.02 | 0.24 | 0.13 | 0.09 | 0.22 | 0.20 | 0.35 | 0.38 | 0.12 | 0.40 |
| hd148856 | E | 4903 | 2.32 | 1.57 | 0.25 | 0.09 | 0.08 | 0.19 | 0.74 | 0.06 | -0.10 | -0.07 | -0.07 | 0.06 | -0.13 | -0.02 | -0.12 | -0.01 | -0.14 | -0.02 | 0.06 | 0.12 | -0.11 | 0.23 | 0.32 | 0.68 | 0.34 | 0.21 | 0.10 |
| hd148897 | E | 4167 | 1.32 | 1.83 | -1.36 | -0.75 | -1.00 | -0.48 | 0.12 | -0.99 | -1.06 | -1.00 | -1.19 | -1.12 | -1.53 | -1.03 | -1.00 | -1.02 | -1.38 | -0.80 | -0.91 | -0.92 | -1.01 | -0.86 | -0.82 | -0.70 | -0.70 | -0.81 | -0.58 |
| hd149161 | E | 3958 | 1.40 | 1.87 | 0.18 | 0.04 | 0.17 | 0.49 | 2.09 | -0.21 | -0.06 | 0.00 | 0.11 | 0.12 | -0.32 | 0.00 | 0.11 | 0.12 | 0.51 | -0.05 | 0.37 | 0.08 | -0.15 | -0.20 | 0.20 | 0.48 | 0.56 | 0.70 | 0.12 |
| hd150557 | E | 6744 | 3.78 | 3.83 |  | 0.47 |  | 0.34 | 0.74 | 0.40 | 0.53 | 0.77 | 1.05 | 0.64 | 0.63 | 0.09 | 1.07 | 0.61 |  | -0.36 |  | 1.50 | 0.35 |  | 0.82 |  | 0.51 | 1.32 |  |
| hd153210 | E | 4559 | 2.45 | 1.48 | 0.40 | 0.35 | 0.39 | 0.55 | 1.45 | 0.15 | 0.29 | 0.26 | 0.42 | 0.34 | 0.33 | 0.31 | 0.31 | 0.43 | 0.70 | 0.59 | 0.01 | 0.18 | -0.01 | 0.08 | 0.62 | 0.97 | 0.75 | 0.73 | 0.47 |
| hd153956 | E | 4541 | 2.60 | 1.61 | 0.74 | 0.46 | 0.40 | 0.66 | 1.57 | 0.23 | 0.29 | 0.30 | 0.47 | 0.43 | 0.34 | 0.36 | 0.41 | 0.50 | 0.77 | 0.51 | 0.13 | 0.29 | 0.07 | 0.15 | 0.67 | 0.98 | 0.70 | 0.88 | 0.53 |
| hd159353 | E | 4832 | 2.63 | 1.31 | 0.21 | 0.18 | 0.16 | 0.26 | 0.79 | 0.14 | -0.01 | 0.06 | 0.12 | 0.13 | 0.10 | 0.08 | -0.01 | 0.12 | 0.06 | 0.13 | 0.11 | 0.12 | 0.06 | 0.18 | 0.42 | 0.67 | 0.48 | 0.23 | 0.11 |
| hd159876 | E | 7217 | 3.65 | 3.61 | 0.04 | -0.07 | 0.27 | 0.23 | 0.20 | 0.14 | 0.20 | 0.40 | 0.45 | 0.22 | -0.12 | 0.01 | 0.96 | 0.37 | 0.32 | 0.18 |  | 0.99 | 1.06 | 0.56 | 0.68 | 0.79 | 0.80 | 0.63 | 0.83 |
| hd160507 | E | 4921 | 2.79 | 1.33 | 0.31 | 0.16 | 0.18 | 0.29 | 0.77 | 0.15 | 0.04 | 0.05 | 0.07 | 0.19 | 0.08 | 0.09 | 0.00 | 0.15 | 0.19 | 0.71 | 0.64 | 0.54 | 0.51 | 0.68 | 0.87 | 1.35 | 0.80 | 0.62 | 0.28 |
| hd161074 | E | 4033 | 1.59 | 1.72 | 0.52 | 0.37 | 0.50 | 0.65 | 2.21 | 0.15 | 0.15 | 0.24 | 0.37 | 0.36 | 0.02 | 0.18 | 0.30 | 0.33 | 0.84 | 0.99 | 0.47 | 0.28 | 0.08 | -0.13 | 0.39 | 0.84 | 0.65 | 0.81 | 0.14 |
| hd162757 | E | 4621 | 2.65 | 1.21 | 0.16 | 0.15 | 0.18 | 0.29 | 0.91 | 0.06 | 0.04 | 0.04 | 0.14 | 0.12 | 0.04 | 0.07 | -0.02 | 0.14 | 0.22 | 0.36 | 0.11 | 0.08 | -0.02 | 0.10 | 0.54 | 0.66 | 0.53 | 0.54 | 0.05 |
| hd163588 | E | 4459 | 2.38 | 1.30 | 0.45 | 0.35 | 0.46 | 0.51 | 1.26 | 0.23 | 0.16 | 0.25 | 0.43 | 0.29 | 0.22 | 0.21 | 0.21 | 0.33 | 0.54 | 0.45 | 0.11 | 0.07 | 0.08 | 0.03 | 0.60 | 0.88 | 0.71 | 0.81 | 0.29 |
| hd164058 | E | 3925 | 1.25 | 2.02 | 0.76 | 0.22 | 0.35 | 0.68 | 2.49 | 0.01 | 0.01 | 0.01 | 0.18 | 0.27 | 0.10 | 0.18 | 0.20 | 0.34 | 0.91 | 0.92 | 0.70 | 0.09 | -0.04 | -0.01 | 0.36 | 0.68 | 0.42 | 0.49 | 0.04 |
| hd166208 | E | 5040 | 2.46 | 1.58 | 0.70 | 0.35 | 0.33 | 0.30 | 0.65 | 0.22 | 0.11 | 0.18 | 0.26 | 0.31 | 0.11 | 0.18 | 0.17 | 0.19 | 0.07 | 0.06 | 0.30 | 0.15 | 0.13 | 0.21 | 0.38 | 0.59 | 0.34 | 0.26 | 0.29 |
| hd168723 | E | 4858 | 2.99 | 1.02 | -0.03 | -0.01 | 0.01 | 0.09 | 0.39 | -0.04 | -0.09 | -0.07 | -0.03 | -0.08 | -0.12 | -0.09 | -0.13 | -0.05 | -0.01 | 0.07 | -0.08 | -0.06 | -0.19 | 0.02 | 0.27 | 0.47 | 0.39 | 0.19 | 0.35 |
| hd169268a | E | 6812 | 3.90 | 1.48 | -0.13 | -0.68 | -0.37 | -0.11 | -0.23 | -0.57 | 0.06 | 0.13 | 0.56 | 0.04 | -0.11 | -0.52 | 0.60 | -0.15 | -0.39 | -0.46 | 1.19 | 0.04 | 0.98 | -0.27 | 0.59 | 0.79 | 0.60 | 0.70 | 1.28 |
| hd172748 | E | 6832 | 3.46 | 4.09 | 0.34 | 0.04 | 0.48 | 0.23 | 0.39 | 0.23 | 0.40 | 0.33 | 0.57 | 0.14 | 0.13 | 0.05 | 0.58 | 0.43 | 0.33 | 0.27 |  | 0.78 | 0.79 | 0.40 |  | 0.81 | 0.59 | 0.74 | 0.61 |
| hd175305 | E | 4981 | 2.70 | 1.30 | -1.49 | -1.11 |  | -1.04 | -0.44 | -1.16 | -1.25 | -1.20 | -1.33 | -1.38 | -1.82 | -1.38 | -1.24 | -1.45 | -1.93 | -1.46 | -1.05 | -1.40 | -0.94 | -1.26 | -1.06 | -1.05 | -0.96 | -0.98 | -0.13 |
| hd176408 | E | 4517 | 2.48 | 1.37 | 0.54 | 0.38 | 0.52 | 0.54 | 1.31 | 0.18 | 0.26 | 0.23 | 0.42 | 0.30 | 0.27 | 0.26 | 0.33 | 0.39 | 0.54 | 0.72 | 0.04 | 0.12 | 0.04 | 0.09 | 0.62 | 0.89 | 0.81 | 0.75 | 0.39 |
| hd178596 | E | 6784 | 3.79 | 3.38 | -0.04 |  |  | 0.14 |  | 0.27 | 0.10 | 0.45 | 0.54 | 0.70 | 0.35 | -0.16 | 0.55 | 0.41 |  |  | 1.70 | 0.31 | 1.38 | -0.28 | 0.67 | 1.17 | 0.82 |  |  |
| hd181214 | E | 6238 | 3.63 | 2.61 | 0.29 | 0.23 | 0.00 | 0.23 | 0.21 | 0.15 | 0.17 | 0.28 | 0.15 | 0.19 | -0.03 | 0.04 | 0.42 | 0.08 | -0.24 | -0.33 |  | 0.17 | 0.77 | -0.13 | 0.43 | -0.01 | 0.14 | 0.71 | 0.81 |
| hd181984 | E | 4413 | 2.38 | 1.75 | 0.77 | 0.60 | 0.60 | 0.82 | 1.89 | 0.33 | 0.39 | 0.45 | 0.60 | 0.60 | 0.55 | 0.46 | 0.55 | 0.59 | 0.74 | 0.81 | 0.29 | 0.31 | 0.21 | -0.06 | 0.71 | 0.96 | 1.06 | 0.95 | 0.59 |
| hd184406 | E | 4487 | 2.75 | 1.16 | 0.55 | 0.50 | 0.66 | 0.65 | 1.45 | 0.32 | 0.42 | 0.48 | 0.68 | 0.51 | 0.39 | 0.41 | 0.46 | 0.56 | 0.77 | 1.19 | 0.32 | 0.38 | 0.29 | 0.19 | 0.93 | 1.25 | 1.16 | 1.39 | 0.44 |

| ID | Type | T | logg | ? | ? | ? | ? | ? | ? | ? | ? | ? | ? | ? | ? | ? | ? | ? | ? | ? | ? | ? | ? | ? | ? | ? | ? | ? | ? | ? |
|---|---|---|---|---|---|---|---|---|---|---|---|---|---|---|---|---|---|---|---|---|---|---|---|---|---|---|---|---|---|---|
| hd185351 | E | 4941 | 3.18 | 0.99 | 0.32 | 0.20 | 0.28 | 0.26 | 0.77 | 0.16 | 0.07 | 0.10 | 0.19 | 0.20 | 0.13 | 0.14 | 0.07 | 0.18 | 0.17 | 0.50 | 0.13 | 0.17 | 0.07 | 0.16 | 0.42 | 0.52 | 0.43 | 0.34 | 0.51 |
| hd185644 | E | 4579 | 2.47 | 1.23 | 0.38 | 0.29 | 0.43 | 0.39 | 1.04 | 0.17 | 0.17 | 0.23 | 0.35 | 0.32 | 0.24 | 0.20 | 0.13 | 0.29 | 0.49 | 0.38 | 0.10 | 0.24 | 0.05 | 0.14 | 0.55 | 0.87 | 0.72 | 0.65 | 0.25 |
| hd187764 | E | 6919 | 3.26 | 6.50 | -0.39 | | | 0.04 | -0.45 | -0.37 | 0.62 | 0.15 | 0.53 | 0.18 | 0.39 | -0.79 | 0.56 | 0.21 | -0.24 | -0.83 | 0.83 | | 1.41 | | | -0.38 | 0.49 | 0.24 | -0.26 |
| hd188119 | E | 4945 | 2.61 | 1.39 | -0.15 | -0.16 | -0.12 | -0.10 | 0.10 | -0.21 | -0.26 | -0.26 | -0.28 | -0.29 | -0.42 | -0.32 | -0.32 | -0.31 | -0.34 | -0.18 | -0.16 | -0.28 | -0.31 | -0.19 | -0.06 | 0.16 | 0.06 | -0.01 | -0.11 |
| hd188947 | E | 4783 | 2.56 | 1.41 | 0.38 | 0.26 | 0.25 | 0.33 | 0.78 | 0.16 | 0.09 | 0.13 | 0.23 | 0.23 | 0.18 | 0.14 | 0.08 | 0.18 | 0.18 | 0.22 | 0.20 | 0.19 | 0.08 | 0.13 | 0.51 | 0.77 | 0.51 | 0.39 | 0.25 |
| hd189319 | E | 3862 | 1.21 | 2.25 | 0.78 | 0.19 | 0.21 | 0.92 | 2.76 | 0.27 | -0.02 | 0.00 | 0.21 | 0.31 | 0.11 | 0.21 | 0.23 | 0.34 | 1.19 | -0.02 | 0.57 | 0.08 | -0.22 | -0.10 | 0.27 | 0.71 | 0.23 | 0.36 | 0.11 |
| hd192787 | E | 4888 | 2.70 | 1.27 | 0.12 | 0.03 | 0.09 | 0.16 | 0.44 | 0.03 | -0.07 | -0.09 | -0.08 | 0.01 | -0.08 | -0.04 | -0.11 | -0.02 | -0.14 | 0.20 | 0.01 | 0.02 | -0.05 | 0.14 | 0.34 | 0.62 | 0.42 | 0.19 | 0.08 |
| hd192836 | E | 4805 | 2.72 | 1.25 | 0.43 | 0.37 | 0.41 | 0.38 | 0.79 | 0.25 | 0.18 | 0.24 | 0.41 | 0.31 | 0.30 | 0.25 | 0.16 | 0.31 | 0.43 | 0.49 | 0.29 | 0.33 | 0.21 | 0.29 | 0.67 | 0.86 | 0.72 | 0.61 | 0.26 |
| hd196134 | E | 4770 | 2.88 | 1.04 | 0.12 | 0.06 | 0.13 | 0.18 | 0.54 | 0.02 | 0.01 | 0.01 | 0.12 | 0.04 | -0.03 | 0.02 | -0.05 | 0.06 | 0.13 | 0.19 | -0.02 | 0.06 | -0.19 | 0.09 | 0.40 | 0.72 | 0.67 | 0.36 | 0.12 |
| hd197989 | E | 4713 | 2.46 | 1.38 | 0.07 | 0.09 | 0.16 | 0.20 | 0.64 | -0.08 | -0.09 | -0.06 | 0.01 | 0.01 | -0.06 | -0.04 | -0.06 | 0.02 | 0.08 | 0.18 | -0.01 | -0.03 | -0.23 | -0.03 | 0.25 | 0.56 | 0.32 | 0.25 | 0.05 |
| hd198431 | E | 4652 | 2.64 | 1.20 | 0.05 | 0.09 | 0.13 | 0.16 | 0.64 | -0.02 | 0.06 | 0.03 | 0.13 | 0.03 | -0.03 | -0.02 | -0.05 | 0.05 | 0.19 | 0.25 | -0.14 | 0.00 | -0.19 | -0.07 | 0.33 | 0.63 | 0.55 | 0.51 | -0.01 |
| hd199178 | E | 5180 | 3.35 | 3.61 | 0.60 | | 0.40 | 0.48 | 0.67 | 0.92 | 0.44 | 0.45 | 0.28 | 1.03 | 0.42 | 0.33 | 0.50 | 0.47 | | | 2.15 | 0.67 | 1.13 | | 1.82 | 1.13 | 1.33 | 0.37 | | |
| hd202109 | E | 4892 | 2.45 | 1.61 | 0.36 | 0.20 | 0.21 | 0.31 | 0.85 | 0.14 | -0.02 | 0.04 | 0.04 | 0.17 | 0.09 | 0.07 | -0.02 | 0.12 | 0.13 | 0.53 | 0.50 | 0.40 | 0.34 | 0.45 | 0.60 | 1.06 | 0.64 | 0.48 | 0.17 |
| hd209747 | E | 4068 | 1.73 | 2.16 | 0.51 | 0.43 | 0.41 | 0.76 | 1.88 | 0.23 | 0.14 | 0.11 | 0.30 | 0.39 | 0.16 | 0.25 | 0.40 | 0.43 | 0.75 | 0.70 | 0.51 | 0.16 | -0.05 | -0.37 | 0.30 | 0.77 | 0.41 | 0.73 | 0.25 |
| hd209960 | E | 4112 | 1.84 | 1.84 | 0.79 | 0.51 | 0.51 | 0.78 | 2.11 | 0.26 | 0.25 | 0.32 | 0.50 | 0.53 | 0.36 | 0.35 | 0.49 | 0.56 | 0.98 | 0.81 | 0.46 | 0.47 | 0.13 | -0.17 | 0.52 | 1.06 | 0.85 | 1.03 | 0.29 |
| hd212943 | E | 4630 | 2.69 | 1.09 | -0.05 | 0.10 | 0.10 | 0.14 | 0.57 | -0.04 | 0.02 | -0.01 | 0.10 | 0.01 | -0.11 | -0.04 | -0.07 | 0.02 | 0.25 | 0.04 | -0.12 | 0.02 | -0.17 | -0.05 | 0.29 | 0.68 | 0.51 | 0.40 | 0.06 |
| hd214448 | E | 5272 | 2.99 | 0.88 | 0.54 | 0.25 | 0.38 | 0.21 | 0.39 | 0.33 | 0.23 | 0.35 | 0.50 | 0.39 | 0.25 | 0.28 | 0.25 | 0.32 | 0.15 | 0.22 | 0.35 | 0.44 | 0.41 | 0.37 | 0.48 | 0.88 | 0.67 | 0.57 | 0.80 |
| hd214470 | E | 6668 | 3.61 | 2.79 | | | 1.00 | | 0.74 | | 1.52 | 1.83 | 1.34 | 2.57 | 0.47 | 0.93 | 0.92 | | | 2.78 | 2.29 | | | | | | | 2.18 | | |
| hd216131 | E | 4934 | 2.78 | 1.25 | 0.21 | 0.11 | 0.15 | 0.19 | 0.59 | 0.14 | -0.01 | 0.00 | 0.02 | 0.09 | 0.00 | 0.04 | -0.07 | 0.03 | -0.04 | 0.07 | 0.12 | 0.14 | 0.03 | 0.22 | 0.40 | 0.72 | 0.46 | 0.28 | 0.18 |
| hd216228 | E | 4742 | 2.57 | 1.42 | 0.27 | 0.23 | 0.27 | 0.30 | 0.73 | 0.11 | 0.06 | 0.09 | 0.18 | 0.18 | 0.13 | 0.10 | 0.06 | 0.16 | 0.23 | 0.15 | 0.05 | 0.10 | -0.03 | 0.13 | 0.49 | 0.74 | 0.49 | 0.37 | 0.13 |
| hd219418 | E | 5077 | 2.80 | 1.40 | 0.04 | -0.17 | -0.05 | -0.08 | 0.07 | -0.16 | -0.25 | -0.28 | -0.36 | -0.26 | -0.39 | -0.31 | -0.31 | -0.29 | -0.37 | -0.17 | -0.13 | -0.16 | -0.11 | -0.16 | -0.17 | 0.08 | -0.04 | -0.20 | 0.37 |
| hd219916 | E | 5124 | 2.93 | 1.27 | 0.28 | 0.16 | 0.25 | 0.15 | 0.42 | 0.19 | 0.04 | 0.13 | 0.18 | 0.19 | 0.09 | 0.10 | 0.06 | 0.10 | 0.05 | 0.11 | 0.24 | 0.24 | 0.21 | 0.21 | 0.40 | 0.59 | 0.49 | 0.22 | 0.57 |
| hd220954 | E | 4684 | 2.58 | 1.35 | 0.42 | 0.47 | 0.34 | 0.46 | 1.04 | 0.22 | 0.20 | 0.22 | 0.36 | 0.36 | 0.28 | 0.27 | 0.19 | 0.36 | 0.51 | 0.57 | 0.27 | 0.26 | 0.16 | 0.27 | 0.65 | 0.93 | 0.61 | 0.70 | 0.23 |
| hd221148 | E | 4660 | 3.21 | 1.00 | 1.01 | 0.71 | 0.78 | 0.77 | 1.65 | 0.57 | 0.67 | 0.66 | 0.99 | 0.79 | 0.68 | 0.61 | 0.73 | 0.83 | 0.80 | 1.33 | 0.41 | 0.72 | 0.50 | 0.24 | 1.16 | 1.50 | 1.28 | 1.64 | 0.89 |
| aiscl | H | 7128 | 3.80 | 5.97 | | 0.70 | | 0.19 | | 0.00 | 0.10 | 0.57 | 1.31 | 0.29 | 1.54 | -0.70 | 1.48 | 0.17 | | | 1.48 | 2.74 | | 0.83 | 1.05 | 0.84 | | | | |
| anscl | H | 4968 | 2.50 | 2.53 | 0.45 | -0.11 | | 0.07 | 0.39 | 0.22 | -0.17 | 0.11 | 0.13 | 0.15 | 0.05 | -0.13 | -0.03 | -0.02 | | -0.41 | | -0.28 | -0.23 | -0.48 | -0.33 | -0.22 | 0.13 | 0.82 | -0.46 | | |
| bppsc | H | 4050 | 2.50 | 3.54 | | | 0.99 | | | -1.25 | -1.09 | -1.13 | -0.44 | -0.85 | -0.24 | -0.10 | 0.11 | | 0.53 | | -0.27 | -2.03 | | -0.78 | | | -0.23 | 0.28 | | | |
| cd-33_2771 | H | 4002 | 2.50 | 1.68 | 0.15 | 0.09 | 0.34 | 0.64 | | -0.18 | 0.35 | 0.16 | 0.45 | 0.46 | | 0.30 | | 0.54 | | 0.71 | 0.62 | 0.20 | 0.40 | 0.09 | 0.62 | 1.09 | 1.11 | 0.81 | 0.47 | |
| cd-3814203b | H | 6297 | 2.50 | 1.58 | 0.64 | 0.45 | 0.44 | 0.32 | -0.14 | 0.47 | -0.14 | 0.24 | 0.36 | 0.34 | 0.38 | 0.27 | 0.45 | 0.40 | 0.26 | 0.09 | 0.71 | 0.05 | 0.19 | -0.09 | -0.23 | -0.24 | -0.21 | -0.20 | 0.81 |
| cfscl | H | 5127 | 2.50 | 1.82 | 0.44 | 0.31 | 0.44 | -0.20 | -0.11 | 0.20 | -0.07 | 0.27 | 0.39 | 0.18 | -0.07 | -0.17 | -0.17 | -0.17 | -0.13 | -0.42 | | 0.13 | 0.41 | -0.47 | 0.05 | 0.14 | 0.25 | 0.64 | 1.21 | |
| epsret | H | 4702 | 3.34 | 0.50 | 0.59 | 0.59 | 0.56 | 0.61 | 1.23 | 0.44 | 0.55 | 0.56 | 0.98 | 0.60 | 0.59 | 0.49 | 0.48 | 0.63 | 0.81 | 0.64 | 0.44 | 0.51 | 0.45 | 0.35 | 0.77 | 0.94 | 1.08 | 1.04 | 0.56 |
| ereri | H | 4935 | 2.50 | 3.54 | | | 0.08 | | -0.03 | -0.97 | -0.68 | | 0.19 | | -0.36 | -0.25 | 0.05 | | -0.07 | | -0.52 | -0.18 | | -0.35 | -0.06 | 0.07 | 0.15 | | | | |
| gamaps | H | 4957 | 2.70 | 1.29 | 0.38 | 0.09 | 0.19 | 0.16 | 0.53 | 0.10 | -0.11 | -0.02 | 0.04 | 0.09 | 0.01 | 0.00 | -0.06 | -0.01 | -0.10 | -0.10 | 0.13 | 0.01 | -0.04 | 0.17 | 0.16 | 0.30 | 0.12 | 0.04 | 0.41 |
| hd000344 | H | 4584 | 2.47 | 1.27 | 0.25 | 0.22 | 0.27 | 0.36 | 0.74 | 0.09 | 0.03 | 0.12 | 0.33 | 0.17 | 0.21 | 0.13 | 0.08 | 0.21 | 0.35 | 0.24 | 0.11 | 0.17 | -0.08 | 0.16 | 0.24 | 0.51 | 0.35 | 0.28 | 0.15 |
| hd000483a | H | 5718 | 3.92 | 0.50 | 0.05 | -0.51 | -0.23 | -0.24 | 0.12 | -0.43 | -0.08 | -0.24 | -0.06 | -0.04 | -0.32 | -0.40 | -0.10 | -0.20 | -0.40 | -0.30 | 0.59 | -0.03 | 0.25 | -0.86 | 0.51 | 0.66 | 1.07 | 0.83 | 0.91 |
| hd000483b | H | 5757 | 3.93 | 0.50 | 0.06 | -0.05 | -0.04 | -0.12 | 0.09 | -0.30 | 0.04 | -0.20 | 0.26 | 0.14 | -0.30 | -0.24 | 0.02 | -0.14 | -0.43 | -0.34 | -0.53 | 0.13 | 0.55 | -0.59 | 0.28 | 0.59 | 0.71 | 1.28 | 0.36 |
| hd000770 | H | 4710 | 2.69 | 1.30 | 0.02 | 0.07 | 0.13 | 0.24 | 0.56 | -0.04 | -0.04 | -0.02 | 0.04 | -0.01 | 0.03 | -0.01 | -0.02 | 0.05 | 0.17 | 0.23 | -0.08 | -0.05 | -0.21 | 0.05 | 0.20 | 0.36 | 0.22 | 0.20 | 0.14 |
| hd001690 | H | 4219 | 2.43 | 1.29 | -0.05 | 0.20 | 0.14 | 0.42 | 1.06 | -0.06 | 0.18 | 0.17 | 0.39 | 0.09 | 0.09 | 0.16 | 0.17 | 0.26 | 0.26 | 0.47 | -0.01 | 0.15 | -0.11 | 0.35 | 0.48 | 0.68 | 0.66 | 0.61 | 0.07 |
| hd002114 | H | 5148 | 2.66 | 1.59 | 0.29 | 0.08 | 0.02 | 0.15 | 0.42 | 0.09 | 0.01 | -0.04 | -0.10 | 0.05 | -0.06 | -0.01 | -0.09 | -0.06 | -0.18 | -0.05 | 0.30 | 0.22 | 0.04 | 0.40 | 0.24 | 0.49 | 0.25 | 0.17 | 0.16 |
| hd002529 | H | 4621 | 2.50 | 1.15 | 0.18 | 0.13 | 0.21 | 0.21 | 0.53 | 0.12 | -0.04 | 0.06 | 0.25 | 0.11 | 0.13 | 0.06 | -0.03 | 0.10 | 0.23 | 0.12 | 0.12 | 0.11 | -0.03 | 0.13 | 0.34 | 0.54 | 0.39 | 0.23 | 0.05 |
| hd003488 | H | 4827 | 2.53 | 1.31 | 0.09 | 0.08 | 0.16 | 0.11 | 0.44 | 0.08 | -0.07 | 0.01 | 0.05 | 0.02 | -0.03 | -0.04 | -0.07 | -0.04 | 0.01 | -0.03 | 0.14 | 0.10 | -0.05 | 0.19 | 0.32 | 0.39 | 0.37 | 0.20 | 0.05 |
| hd003919 | H | 4952 | 2.62 | 1.22 | 0.38 | 0.16 | 0.19 | 0.19 | 0.36 | 0.22 | 0.02 | 0.16 | 0.30 | 0.19 | 0.19 | 0.11 | 0.06 | 0.11 | 0.11 | -0.10 | 0.31 | 0.19 | 0.17 | 0.23 | 0.36 | 0.46 | 0.31 | 0.24 | 0.22 |
| hd004211 | H | 4531 | 2.38 | 1.32 | 0.29 | 0.39 | 0.45 | 0.37 | 0.84 | 0.21 | 0.11 | 0.29 | 0.51 | 0.22 | 0.20 | 0.15 | 0.20 | 0.26 | 0.83 | 0.52 | 0.13 | 0.10 | -0.15 | 0.06 | 0.37 | 0.49 | 0.45 | 0.47 | 0.16 |
| hd004737 | H | 5031 | 2.80 | 1.30 | 0.26 | 0.08 | 0.12 | 0.07 | 0.29 | 0.08 | -0.09 | -0.03 | -0.01 | 0.07 | -0.01 | -0.02 | -0.11 | -0.06 | -0.14 | -0.06 | 0.22 | 0.04 | -0.02 | 0.23 | 0.13 | 0.34 | 0.16 | 0.11 | 0.14 |
| hd006192 | H | 4992 | 2.76 | 1.28 | 0.27 | 0.13 | 0.06 | 0.15 | 0.32 | 0.12 | -0.04 | 0.04 | 0.07 | 0.11 | 0.06 | 0.05 | -0.05 | 0.03 | 0.00 | 0.02 | 0.40 | 0.13 | 0.10 | 0.26 | 0.32 | 0.46 | 0.35 | 0.20 | 0.19 |
| hd006245 | H | 5001 | 2.91 | 1.18 | 0.21 | 0.05 | 0.14 | 0.13 | 0.40 | 0.09 | -0.03 | 0.00 | 0.00 | 0.08 | -0.01 | 0.02 | -0.09 | 0.00 | -0.07 | -0.04 | 0.20 | 0.13 | 0.09 | 0.29 | 0.35 | 0.47 | 0.31 | 0.18 | 0.48 |
| hd006559 | H | 4681 | 2.52 | 1.31 | 0.37 | 0.28 | 0.36 | 0.33 | 0.65 | 0.17 | 0.04 | 0.15 | 0.36 | 0.20 | 0.24 | 0.12 | 0.10 | 0.20 | 0.40 | 0.25 | 0.24 | 0.17 | 0.00 | 0.12 | 0.21 | 0.33 | 0.24 | 0.29 | 0.10 |
| hd006793 | H | 5039 | 2.73 | 1.58 | 0.34 | 0.05 | 0.09 | 0.14 | 0.36 | 0.08 | -0.11 | -0.05 | -0.02 | 0.10 | -0.08 | -0.02 | -0.11 | -0.04 | -0.15 | -0.17 | 0.11 | 0.15 | -0.03 | 0.16 | 0.14 | 0.41 | 0.21 | 0.09 | 0.07 |
| hd007082 | H | 4969 | 2.82 | 1.38 | -0.58 | -0.39 | -0.32 | -0.31 | -0.16 | -0.50 | -0.44 | -0.44 | -0.61 | -0.68 | -0.93 | -0.69 | -0.57 | -0.65 | -0.73 | -0.38 | -0.07 | -0.16 | -0.20 | 0.08 | 0.19 | 0.27 | 0.14 | -0.08 | -0.13 |
| hd007672 | H | 4937 | 2.76 | 1.53 | -0.17 | -0.27 | -0.11 | -0.21 | 0.11 | -0.19 | -0.37 | -0.29 | -0.27 | -0.24 | -0.42 | -0.39 | -0.39 | -0.43 | -0.49 | -0.40 | -0.19 | -0.36 | -0.34 | -0.15 | -0.27 | -0.16 | -0.20 | -0.35 | 0.09 |

| ID | Type | T | logg | [Fe/H] | | | | | | | | | | | | | | | | | | | | | | | | |
|---|---|---|---|---|---|---|---|---|---|---|---|---|---|---|---|---|---|---|---|---|---|---|---|---|---|---|---|---|
| hd009362 | H | 4762 | 2.50 | 1.31 | -0.16 | -0.11 | -0.05 | -0.03 | -0.03 | -0.18 | -0.29 | -0.24 | -0.24 | -0.25 | -0.31 | -0.28 | -0.31 | -0.24 | -0.15 | -0.16 | -0.14 | -0.28 | -0.44 | -0.21 | -0.13 | 0.00 | -0.09 | -0.22 | -0.16 |
| hd009611 | H | 4627 | 2.50 | 3.00 | 0.54 | -0.07 | 0.88 | 0.54 | 1.38 | 0.43 | 0.18 | 0.34 | 0.38 | 0.51 | 0.35 | 0.10 | 0.24 | 0.24 | | -0.56 | | 0.54 | 0.41 | -0.37 | 0.39 | 0.60 | 0.63 | 1.16 |
| hd009742 | H | 4629 | 2.53 | 1.19 | 0.20 | 0.17 | 0.21 | 0.18 | 0.70 | 0.09 | 0.05 | 0.12 | 0.31 | 0.14 | 0.16 | 0.07 | 0.02 | 0.11 | 0.29 | 0.03 | 0.16 | 0.13 | 0.01 | 0.19 | 0.47 | 0.61 | 0.50 | 0.29 | 0.09 |
| hd010042 | H | 4799 | 2.52 | 1.49 | -0.15 | 0.04 | 0.04 | 0.04 | 0.16 | -0.16 | -0.18 | -0.09 | -0.16 | -0.24 | -0.41 | -0.29 | -0.21 | -0.22 | -0.15 | 0.15 | -0.23 | -0.33 | -0.33 | -0.36 | -0.20 | 0.06 | -0.01 | 0.02 | -0.04 |
| hd010615 | H | 4420 | 2.15 | 1.35 | 0.65 | 0.41 | 0.53 | 0.57 | 0.78 | 0.40 | 0.28 | 0.34 | 0.73 | 0.45 | 0.43 | 0.32 | 0.33 | 0.41 | 0.32 | 0.36 | 0.38 | 0.28 | 0.17 | 0.19 | 0.51 | 0.63 | 0.51 | 0.48 | 0.15 |
| hd011025 | H | 4961 | 2.65 | 1.62 | 0.34 | 0.17 | 0.26 | 0.22 | 0.54 | 0.06 | -0.03 | 0.08 | 0.10 | 0.16 | -0.01 | 0.04 | -0.02 | 0.05 | -0.11 | 0.03 | 0.12 | 0.14 | 0.02 | 0.12 | 0.28 | 0.39 | 0.31 | 0.26 | 0.25 |
| hd011977 | H | 4878 | 2.65 | 1.30 | -0.06 | -0.09 | -0.04 | -0.01 | 0.20 | -0.11 | -0.20 | -0.18 | -0.21 | -0.15 | -0.24 | -0.18 | -0.23 | -0.19 | -0.23 | -0.22 | -0.01 | -0.06 | -0.14 | 0.18 | 0.25 | 0.36 | 0.31 | 0.15 | 0.10 |
| hd012055 | H | 5035 | 2.73 | 1.40 | 0.20 | 0.06 | 0.14 | 0.12 | 0.46 | 0.04 | -0.09 | -0.06 | -0.07 | 0.03 | -0.11 | -0.06 | -0.11 | -0.06 | -0.29 | 0.06 | 0.08 | 0.01 | -0.13 | 0.20 | 0.19 | 0.40 | 0.26 | 0.03 | 0.45 |
| hd012270 | H | 4992 | 2.64 | 1.33 | 0.29 | 0.08 | 0.14 | 0.12 | 0.27 | 0.07 | -0.07 | -0.06 | -0.07 | 0.04 | -0.02 | -0.01 | -0.10 | -0.04 | -0.13 | -0.07 | 0.16 | 0.17 | 0.07 | 0.22 | 0.22 | 0.48 | 0.24 | 0.15 | 0.17 |
| hd012345 | H | 5326 | 4.51 | 0.00 | -0.18 | -0.17 | -0.02 | -0.10 | 0.03 | -0.20 | -0.10 | -0.11 | -0.12 | -0.12 | -0.22 | -0.16 | -0.14 | -0.14 | -0.07 | -0.14 | -0.05 | -0.04 | -0.07 | -0.21 | 0.11 | 0.19 | 0.23 | 0.14 | 0.45 |
| hd012438 | H | 4884 | 2.58 | 1.38 | -0.57 | -0.45 | -0.37 | -0.32 | -0.22 | -0.49 | -0.47 | -0.55 | -0.67 | -0.59 | -0.78 | -0.60 | -0.60 | -0.60 | -0.66 | -0.50 | -0.55 | -0.53 | -0.62 | -0.43 | -0.46 | -0.35 | -0.33 | -0.40 | -0.11 |
| hd012524 | H | 3960 | 1.54 | 1.50 | 0.26 | 0.19 | 0.41 | 0.60 | 1.34 | 0.26 | 0.16 | 0.27 | 0.48 | 0.18 | -0.04 | 0.13 | 0.25 | 0.24 | 0.42 | 0.80 | 0.61 | 0.19 | -0.03 | 0.20 | 0.28 | 0.34 | 0.67 | 0.56 | -0.32 |
| hd013263 | H | 5038 | 2.88 | 1.24 | 0.14 | 0.02 | 0.06 | 0.09 | 0.10 | 0.05 | -0.04 | 0.00 | -0.05 | 0.04 | -0.08 | -0.02 | -0.11 | -0.04 | -0.08 | -0.07 | 0.33 | 0.12 | 0.08 | 0.23 | 0.29 | 0.55 | 0.46 | 0.29 | 0.56 |
| hd013423 | H | 5037 | 2.92 | 1.23 | 0.26 | 0.05 | 0.11 | 0.11 | 0.29 | 0.06 | -0.05 | 0.00 | 0.00 | 0.04 | -0.02 | -0.02 | -0.08 | -0.02 | -0.05 | -0.02 | 0.17 | 0.08 | -0.02 | 0.16 | 0.18 | 0.44 | 0.28 | 0.10 | 0.17 |
| hd013692 | H | 4726 | 2.51 | 1.32 | -0.03 | 0.00 | 0.03 | 0.12 | 0.46 | -0.07 | -0.16 | -0.14 | -0.12 | -0.11 | -0.14 | -0.11 | -0.16 | -0.09 | -0.06 | -0.08 | -0.02 | -0.04 | -0.20 | 0.12 | 0.17 | 0.45 | 0.20 | 0.11 | 0.03 |
| hd013940 | H | 4883 | 2.67 | 1.26 | 0.16 | 0.06 | 0.11 | 0.13 | 0.29 | 0.05 | -0.09 | -0.04 | 0.00 | 0.03 | 0.02 | -0.02 | -0.11 | -0.01 | -0.02 | -0.03 | 0.17 | 0.11 | 0.02 | 0.19 | 0.22 | 0.51 | 0.31 | 0.14 | 0.15 |
| hd014703 | H | 4930 | 2.82 | 1.16 | 0.32 | 0.12 | 0.25 | 0.24 | 0.41 | 0.20 | 0.06 | 0.12 | 0.18 | 0.21 | 0.20 | 0.13 | 0.00 | 0.14 | 0.19 | 0.15 | 0.29 | 0.25 | 0.11 | 0.23 | 0.37 | 0.59 | 0.37 | 0.24 | 0.19 |
| hd014832 | H | 4804 | 2.54 | 1.37 | -0.07 | -0.02 | 0.01 | 0.04 | 0.20 | -0.11 | -0.15 | -0.14 | -0.12 | -0.15 | -0.16 | -0.17 | -0.16 | -0.14 | -0.07 | -0.01 | 0.04 | -0.12 | -0.25 | -0.03 | 0.13 | 0.23 | 0.17 | 0.08 | -0.04 |
| hd016522 | H | 4833 | 2.55 | 1.35 | -0.04 | 0.03 | 0.07 | 0.07 | 0.22 | -0.07 | -0.15 | -0.11 | -0.09 | -0.11 | -0.12 | -0.13 | -0.12 | -0.09 | 0.01 | -0.05 | -0.05 | -0.12 | -0.21 | -0.06 | 0.01 | 0.22 | 0.09 | 0.05 | 0.04 |
| hd016815 | H | 4683 | 2.46 | 1.16 | -0.17 | -0.24 | -0.08 | -0.11 | 0.21 | -0.21 | -0.35 | -0.27 | -0.24 | -0.29 | -0.34 | -0.33 | -0.35 | -0.30 | -0.21 | -0.22 | -0.19 | -0.32 | -0.41 | -0.23 | -0.14 | 0.03 | -0.02 | -0.20 | -0.27 |
| hd016975 | H | 5014 | 2.77 | 1.29 | 0.32 | 0.13 | 0.19 | 0.17 | 0.26 | 0.16 | 0.00 | 0.04 | 0.06 | 0.12 | 0.08 | 0.05 | -0.04 | 0.04 | 0.03 | -0.04 | 0.21 | 0.19 | 0.07 | 0.26 | 0.36 | 0.48 | 0.31 | 0.18 | 0.19 |
| hd017374 | H | 4824 | 2.53 | 1.34 | 0.07 | 0.07 | 0.17 | 0.11 | 0.26 | 0.04 | -0.07 | 0.00 | 0.04 | 0.02 | -0.02 | -0.05 | -0.09 | -0.04 | 0.04 | 0.00 | 0.07 | 0.03 | -0.08 | 0.15 | 0.30 | 0.36 | 0.29 | 0.14 | 0.36 |
| hd017793 | H | 5042 | 2.68 | 1.34 | 0.41 | 0.16 | 0.12 | 0.16 | 0.50 | 0.14 | -0.05 | 0.04 | 0.07 | 0.14 | 0.08 | 0.05 | -0.01 | 0.05 | -0.06 | 0.01 | 0.25 | 0.15 | 0.01 | 0.16 | 0.07 | 0.35 | 0.16 | 0.13 | 0.21 |
| hd018293 | H | 4262 | 1.81 | 1.21 | 0.63 | 0.46 | 0.54 | 0.66 | 1.04 | 0.51 | 0.35 | 0.47 | 0.85 | 0.51 | 0.55 | 0.40 | 0.34 | 0.53 | 0.55 | 0.77 | 0.43 | 0.49 | 0.34 | 0.57 | 0.72 | 0.94 | 0.73 | 0.77 | 0.09 |
| hd018322 | H | 4614 | 2.47 | 1.19 | 0.14 | 0.13 | 0.19 | 0.19 | 0.52 | 0.07 | -0.05 | 0.03 | 0.20 | 0.05 | 0.07 | 0.01 | -0.04 | 0.05 | 0.15 | 0.00 | 0.10 | 0.04 | -0.09 | 0.14 | 0.25 | 0.54 | 0.39 | 0.19 | 0.04 |
| hd020037 | H | 5077 | 2.76 | 1.32 | 0.21 | 0.03 | 0.00 | 0.06 | 0.18 | 0.05 | -0.07 | -0.05 | -0.06 | 0.02 | -0.10 | -0.05 | -0.14 | -0.08 | -0.19 | -0.15 | 0.13 | 0.12 | -0.01 | 0.22 | 0.14 | 0.43 | 0.22 | 0.19 | 0.49 |
| hd020894 | H | 5060 | 2.70 | 1.44 | 0.19 | 0.02 | 0.02 | 0.11 | 0.33 | 0.05 | -0.06 | -0.05 | -0.08 | 0.02 | -0.10 | -0.03 | -0.10 | -0.07 | -0.15 | -0.09 | 0.21 | 0.08 | 0.05 | 0.23 | 0.24 | 0.48 | 0.26 | 0.15 | 0.11 |
| hd021011 | H | 4799 | 2.49 | 1.32 | 0.08 | -0.03 | 0.09 | 0.11 | 0.27 | -0.01 | -0.16 | -0.09 | -0.06 | -0.02 | -0.07 | -0.08 | -0.16 | -0.08 | -0.04 | -0.11 | 0.04 | -0.07 | -0.11 | 0.12 | 0.12 | 0.38 | 0.24 | 0.11 | -0.05 |
| hd021430 | H | 4979 | 2.63 | 1.40 | -0.12 | -0.15 | -0.10 | -0.15 | 0.02 | -0.18 | -0.28 | -0.21 | -0.23 | -0.25 | -0.33 | -0.30 | -0.29 | -0.30 | -0.28 | -0.34 | 0.05 | -0.26 | -0.22 | -0.13 | -0.08 | 0.11 | 0.05 | -0.11 | -0.09 |
| hd022231 | H | 4591 | 2.45 | 1.31 | 0.44 | 0.37 | 0.43 | 0.52 | 1.22 | 0.28 | 0.17 | 0.23 | 0.44 | 0.34 | 0.41 | 0.27 | 0.25 | 0.38 | 0.59 | 0.53 | 0.21 | 0.31 | 0.03 | 0.27 | 0.45 | 0.54 | 0.44 | 0.43 | 0.18 |
| hd022663 | H | 4708 | 2.51 | 0.91 | 0.42 | 0.18 | 0.41 | 0.25 | 0.58 | 0.32 | 0.17 | 0.34 | 0.72 | 0.35 | 0.29 | 0.19 | 0.12 | 0.25 | 0.39 | 0.13 | 0.43 | 0.26 | 0.28 | 0.30 | 0.55 | 0.61 | 0.62 | 0.52 | 0.25 |
| hd022676 | H | 4957 | 2.73 | 1.32 | 0.34 | 0.14 | 0.23 | 0.19 | 0.49 | 0.14 | 0.00 | 0.07 | 0.14 | 0.19 | 0.08 | 0.09 | -0.02 | 0.08 | 0.00 | 0.06 | 0.26 | 0.16 | 0.03 | 0.22 | 0.36 | 0.57 | 0.27 | 0.15 | 0.18 |
| hd023319 | H | 4539 | 2.48 | 1.18 | 0.69 | 0.54 | 0.69 | 0.69 | 1.21 | 0.45 | 0.37 | 0.48 | 0.91 | 0.52 | 0.59 | 0.43 | 0.48 | 0.59 | 1.23 | 0.75 | 0.39 | 0.42 | 0.18 | 0.34 | 0.56 | 0.61 | 0.58 | 0.53 | 0.34 |
| hd023719 | H | 4939 | 2.74 | 1.29 | 0.43 | 0.21 | 0.26 | 0.30 | 0.70 | 0.24 | 0.06 | 0.13 | 0.19 | 0.22 | 0.23 | 0.15 | 0.00 | 0.17 | 0.18 | 0.15 | 0.29 | 0.26 | 0.12 | 0.27 | 0.29 | 0.57 | 0.33 | 0.25 | 0.23 |
| hd023940 | H | 4776 | 2.58 | 1.38 | -0.23 | -0.07 | -0.01 | 0.00 | 0.18 | -0.20 | -0.23 | -0.16 | -0.24 | -0.25 | -0.41 | -0.29 | -0.28 | -0.26 | -0.20 | 0.08 | -0.28 | -0.35 | -0.47 | -0.27 | -0.11 | 0.11 | -0.04 | -0.12 | -0.16 |
| hd024160 | H | 4948 | 2.62 | 1.37 | 0.30 | 0.13 | 0.18 | 0.20 | 0.52 | 0.13 | -0.04 | 0.00 | 0.01 | 0.10 | 0.05 | 0.04 | -0.05 | 0.03 | -0.03 | 0.00 | 0.17 | 0.15 | 0.04 | 0.24 | 0.25 | 0.44 | 0.26 | 0.19 | 0.20 |
| hd024706 | H | 4427 | 2.32 | 1.17 | 0.54 | 0.40 | 0.50 | 0.55 | 1.06 | 0.37 | 0.31 | 0.37 | 0.78 | 0.40 | 0.42 | 0.33 | 0.34 | 0.47 | 0.38 | 0.58 | 0.35 | 0.31 | 0.18 | 0.28 | 0.57 | 0.57 | 0.57 | 0.50 | 0.15 |
| hd024744 | H | 5846 | 2.86 | 0.68 | 0.45 | 0.23 | 0.49 | 0.14 | -0.22 | 0.40 | 0.23 | 0.51 | 0.77 | 0.46 | 0.29 | 0.41 | 0.47 | 0.37 | 0.24 | -0.25 | 0.75 | 0.40 | 0.51 | 0.47 | 0.56 | 0.63 | 0.48 | 0.46 | 0.61 |
| hd026967 | H | 4614 | 2.63 | 1.04 | 0.20 | 0.18 | 0.17 | 0.25 | 0.55 | 0.13 | 0.06 | 0.15 | 0.37 | 0.15 | 0.15 | 0.10 | 0.03 | 0.16 | 0.34 | 0.12 | 0.16 | 0.18 | -0.02 | 0.18 | 0.28 | 0.62 | 0.48 | 0.34 | 0.10 |
| hd027256 | H | 5048 | 2.57 | 1.57 | 0.47 | 0.26 | 0.28 | 0.28 | 0.68 | 0.24 | 0.06 | 0.11 | 0.13 | 0.21 | 0.17 | 0.13 | 0.05 | 0.15 | 0.09 | 0.25 | 0.22 | 0.20 | 0.13 | 0.20 | 0.33 | 0.51 | 0.32 | 0.23 | 0.23 |
| hd028028 | H | 3982 | 1.38 | 1.55 | 0.14 | 0.11 | 0.15 | 0.46 | 1.38 | 0.15 | -0.01 | 0.07 | 0.28 | 0.09 | -0.04 | 0.04 | 0.08 | 0.12 | 0.29 | 0.26 | 0.32 | 0.02 | -0.05 | 0.19 | 0.22 | 0.36 | 0.44 | 0.31 | -0.26 |
| hd028093 | H | 4951 | 2.49 | 1.45 | 0.21 | 0.07 | 0.15 | 0.09 | 0.14 | 0.00 | -0.05 | -0.03 | 0.02 | -0.06 | -0.08 | -0.09 | -0.05 | -0.07 | -0.02 | -0.14 | 0.02 | -0.12 | -0.16 | -0.02 | -0.01 | 0.18 | 0.08 | 0.03 | 0.06 |
| hd029291 | H | 4897 | 2.40 | 1.49 | 0.31 | 0.12 | 0.15 | 0.16 | 0.54 | 0.10 | -0.11 | -0.04 | -0.03 | 0.06 | 0.02 | 0.00 | -0.10 | 0.00 | -0.04 | 0.01 | 0.11 | 0.16 | -0.04 | 0.15 | 0.30 | 0.32 | 0.16 | 0.11 | 0.06 |
| hd029399 | H | 4822 | 4.82 | 0.50 | 0.03 | 0.12 | 0.30 | 0.62 | 1.33 | -0.13 | 0.68 | 0.40 | 0.59 | 0.40 | 0.28 | 0.53 | 0.65 | 0.57 | 0.47 | 0.55 | 0.18 | 0.72 | 0.46 | 0.28 | 1.11 | 1.24 | 1.20 | 1.31 | 0.80 |
| hd029751 | H | 4855 | 2.61 | 1.36 | 0.11 | -0.08 | 0.01 | 0.07 | 0.31 | -0.14 | -0.12 | -0.19 | -0.24 | -0.15 | -0.21 | -0.17 | -0.21 | -0.15 | -0.14 | 0.02 | -0.08 | -0.08 | -0.22 | 0.08 | 0.11 | 0.28 | 0.18 | 0.08 | 0.15 |
| hd030185 | H | 4851 | 2.42 | 1.32 | 0.19 | 0.04 | 0.12 | 0.08 | 0.30 | 0.02 | -0.15 | -0.08 | -0.04 | -0.05 | -0.08 | -0.09 | -0.14 | -0.09 | -0.09 | -0.13 | 0.05 | -0.06 | -0.13 | 0.07 | 0.11 | 0.23 | 0.13 | -0.02 | -0.01 |
| hd032453 | H | 5032 | 2.81 | 1.30 | 0.17 | 0.03 | 0.06 | 0.06 | 0.23 | 0.02 | -0.08 | -0.06 | -0.09 | -0.01 | -0.10 | -0.07 | -0.15 | -0.09 | -0.16 | -0.06 | 0.17 | 0.06 | 0.03 | 0.21 | 0.29 | 0.44 | 0.28 | 0.21 | 0.56 |
| hd032515 | H | 4529 | 2.48 | 1.19 | 0.43 | 0.43 | 0.48 | 0.57 | 1.17 | 0.30 | 0.23 | 0.35 | 0.68 | 0.40 | 0.48 | 0.35 | 0.35 | 0.48 | 0.75 | 0.81 | 0.24 | 0.34 | 0.07 | 0.34 | 0.53 | 0.64 | 0.51 | 0.56 | 0.26 |
| hd033285 | H | 4825 | 1.96 | 1.88 | 0.28 | 0.10 | 0.14 | 0.20 | 0.60 | -0.01 | -0.16 | -0.09 | -0.11 | 0.08 | -0.20 | -0.02 | -0.15 | -0.02 | -0.24 | -0.11 | 0.20 | 0.08 | -0.06 | 0.19 | 0.33 | 0.70 | 0.37 | 0.09 | 0.11 |
| hd034172 | H | 4990 | 2.67 | 1.32 | 0.33 | 0.15 | 0.20 | 0.17 | 0.36 | 0.15 | -0.10 | 0.04 | 0.07 | 0.14 | 0.12 | 0.05 | -0.03 | 0.04 | 0.02 | -0.02 | 0.19 | 0.10 | 0.04 | 0.16 | 0.10 | 0.37 | 0.21 | 0.13 | 0.16 |

| ID | Type | T | logg | [Fe/H] | v | c1 | c2 | c3 | c4 | c5 | c6 | c7 | c8 | c9 | c10 | c11 | c12 | c13 | c14 | c15 | c16 | c17 | c18 | c19 | c20 | c21 | c22 | c23 | c24 |
|---|---|---|---|---|---|---|---|---|---|---|---|---|---|---|---|---|---|---|---|---|---|---|---|---|---|---|---|---|---|
| hd034253 | H | 5490 | 4.55 | 0.50 | -0.09 | -0.08 | -0.14 | -0.05 | 0.14 | -0.13 | -0.07 | -0.05 | -0.02 | -0.02 | -0.13 | -0.12 | -0.04 | -0.10 | 0.00 | 0.12 | 0.82 | 0.80 | 0.71 | 0.42 | 0.68 | 0.85 | 0.68 | 0.51 | 0.87 |
| hd034266 | H | 4881 | 2.48 | 1.53 | 0.31 | 0.11 | 0.19 | 0.24 | 0.42 | 0.16 | 0.02 | 0.06 | 0.11 | 0.15 | 0.05 | 0.06 | -0.02 | 0.04 | 0.01 | 0.04 | 0.30 | 0.11 | 0.03 | 0.26 | 0.42 | 0.66 | 0.35 | 0.33 | 0.14 |
| hd034649 | H | 4320 | 1.65 | 1.59 | 0.18 | 0.09 | 0.12 | 0.31 | 0.75 | 0.06 | -0.17 | -0.04 | 0.15 | 0.08 | -0.03 | 0.00 | -0.09 | 0.01 | 0.58 | -0.02 | 0.15 | -0.04 | -0.15 | 0.26 | 0.42 | 0.50 | 0.34 | 0.29 | -0.08 |
| hd035929 | H | 6345 | 3.13 | 4.44 | -1.34 |  |  | 0.13 | -0.08 | -0.03 | 0.37 | 0.38 | 0.60 | 0.17 | -0.49 | 0.17 | 0.94 | -0.12 |  |  | 2.02 | 0.45 |  |  |  | 0.12 | -0.49 | 1.09 |
| hd036189 | H | 4911 | 2.37 | 1.78 | 0.32 | 0.12 | 0.14 | 0.19 | 0.52 | 0.04 | -0.02 | -0.03 | -0.04 | 0.07 | -0.08 | 0.00 | -0.12 | -0.01 | -0.07 | 0.10 | 0.16 | 0.10 | 0.03 | 0.27 | 0.43 | 0.65 | 0.43 | 0.24 | 0.18 |
| hd036597 | H | 4573 | 2.05 | 1.52 | 0.28 | 0.13 | 0.10 | 0.25 | 0.58 | 0.02 | -0.13 | -0.02 | 0.10 | 0.06 | 0.03 | -0.02 | -0.06 | 0.00 | 0.10 | 0.01 | 0.14 | 0.07 | -0.13 | 0.14 | 0.35 | 0.40 | 0.28 | 0.22 | -0.02 |
| hd036734 | H | 4244 | 1.80 | 1.47 | 0.16 | 0.06 | 0.14 | 0.24 | 0.46 | 0.03 | -0.03 | 0.05 | 0.37 | 0.10 | 0.05 | 0.00 | -0.01 | 0.07 | 0.08 | 0.13 | 0.20 | -0.06 | -0.08 | 0.09 | 0.36 | 0.56 | 0.43 | 0.31 | -0.12 |
| hd036848 | H | 4476 | 2.72 | 0.75 | 0.99 | 0.74 | 0.94 | 0.84 | 1.59 | 0.63 | 0.82 | 0.79 | 1.23 | 0.82 | 0.78 | 0.68 | 0.70 | 0.87 | 0.92 | 1.07 | 0.61 | 0.73 | 0.69 | 0.45 | 1.01 | 1.12 | 1.10 | 1.10 | 0.47 |
| hd037811 | H | 5023 | 2.57 | 1.45 | 0.26 | 0.09 | 0.17 | 0.12 | 0.45 | 0.10 | -0.06 | -0.01 | 0.00 | 0.07 | -0.02 | 0.00 | -0.07 | -0.03 | -0.12 | -0.14 | 0.21 | 0.06 | 0.06 | 0.21 | 0.21 | 0.41 | 0.24 | 0.21 | 0.16 |
| hd039425 | H | 4566 | 2.53 | 1.23 | 0.42 | 0.41 | 0.39 | 0.54 | 0.93 | 0.29 | 0.25 | 0.34 | 0.65 | 0.37 | 0.46 | 0.32 | 0.33 | 0.45 | 0.70 | 0.80 | 0.25 | 0.35 | 0.09 | 0.31 | 0.52 | 0.51 | 0.54 | 0.42 | 0.24 |
| hd039640 | H | 4850 | 2.51 | 1.34 | 0.07 | 0.04 | 0.07 | 0.10 | 0.33 | 0.00 | -0.15 | -0.09 | -0.05 | -0.05 | -0.08 | -0.09 | -0.15 | -0.09 | -0.11 | -0.14 | 0.04 | -0.05 | -0.08 | 0.15 | 0.17 | 0.38 | 0.25 | 0.11 | 0.07 |
| hd040176 | H | 4659 | 2.26 | 1.53 | 0.91 | 0.47 | 0.56 | 0.57 | 1.03 | 0.34 | 0.16 | 0.20 | 0.41 | 0.36 | 0.42 | 0.29 | 0.26 | 0.39 | 0.52 | 0.36 | 0.21 | 0.20 | -0.01 | 0.10 | 0.34 | 0.39 | 0.30 | 0.26 | 0.28 |
| hd040657 | H | 4288 | 1.52 | 1.51 | -0.35 | -0.19 | -0.17 | -0.17 | 0.12 | -0.31 | -0.44 | -0.32 | -0.29 | -0.51 | -0.72 | -0.57 | -0.47 | -0.52 | -0.18 | -0.51 | -0.31 | -0.67 | -0.73 | -0.58 | -0.46 | -0.35 | -0.27 | -0.31 | -0.43 |
| hd040808 | H | 4620 | 1.72 | 1.80 | 0.47 | 0.11 | 0.21 | 0.27 | 0.61 | 0.11 | -0.10 | -0.04 | 0.01 | 0.11 | 0.05 | -0.01 | -0.08 | 0.00 | 0.02 | 0.00 | 0.20 | 0.11 | -0.13 | 0.18 | 0.26 | 0.39 | 0.18 | 0.13 | -0.04 |
| hd041047 | H | 3891 | 1.30 | 1.87 | 0.38 | 0.21 | 0.31 | 0.62 | 1.72 | 0.29 | 0.07 | 0.06 | 0.31 | 0.15 | -0.10 | 0.13 | 0.11 | 0.18 | 0.52 | 0.27 | 0.81 | -0.10 | -0.09 | 0.29 | 0.25 | 0.17 | 0.53 | 0.29 | 0.14 |
| hd042719 | H | 5699 | 3.84 | 0.94 | 0.47 | 0.46 | 0.45 | 0.37 | 0.60 | 0.36 | 0.39 | 0.34 | 0.30 | 0.36 | 0.33 | 0.31 | 0.30 | 0.37 | 0.54 | 0.54 | 0.36 | 0.41 | 0.28 | 0.31 | 0.46 | 0.47 | 0.40 | 0.36 | 1.29 |
| hd043785 | H | 4876 | 2.59 | 1.33 | 0.50 | 0.22 | 0.30 | 0.31 | 0.53 | 0.24 | 0.01 | 0.08 | 0.21 | 0.19 | 0.24 | 0.12 | 0.03 | 0.15 | 0.18 | 0.13 | 0.19 | 0.13 | -0.02 | 0.10 | 0.16 | 0.24 | 0.15 | 0.08 | 0.14 |
| hd045669 | H | 3984 | 1.38 | 1.64 | 0.32 | 0.18 | 0.28 | 0.52 | 1.42 | 0.29 | 0.03 | 0.15 | 0.36 | 0.20 | 0.05 | 0.13 | 0.11 | 0.22 | 0.46 | 0.29 | 0.50 | 0.20 | 0.08 | 0.26 | 0.32 | 0.35 | 0.51 | 0.41 | 0.24 |
| hd046116 | H | 4853 | 2.67 | 1.35 | -0.19 | -0.09 | -0.01 | -0.04 | 0.10 | -0.16 | -0.21 | -0.16 | -0.17 | -0.24 | -0.29 | -0.27 | -0.26 | -0.24 | -0.16 | -0.19 | -0.16 | -0.28 | -0.37 | -0.25 | -0.12 | -0.03 | -0.05 | -0.15 | -0.14 |
| hd046415 | H | 4818 | 2.72 | 1.29 | -0.09 | -0.05 | -0.01 | 0.05 | 0.24 | -0.12 | -0.11 | -0.10 | -0.08 | -0.14 | -0.16 | -0.14 | -0.13 | -0.10 | -0.04 | -0.06 | -0.08 | -0.09 | -0.18 | 0.06 | 0.12 | 0.33 | 0.21 | 0.09 | 0.07 |
| hd046568 | H | 4808 | 2.50 | 1.35 | -0.06 | -0.06 | 0.00 | 0.02 | 0.28 | -0.12 | -0.17 | -0.18 | -0.17 | -0.16 | -0.18 | -0.18 | -0.20 | -0.18 | -0.14 | -0.14 | -0.03 | -0.07 | -0.26 | 0.06 | 0.18 | 0.24 | 0.13 | 0.03 | -0.06 |
| hd047001 | H | 4675 | 2.24 | 1.47 | -0.08 | -0.11 | -0.05 | -0.06 | 0.36 | -0.15 | -0.25 | -0.19 | -0.13 | -0.21 | -0.30 | -0.27 | -0.29 | -0.28 | -0.21 | -0.29 | 0.00 | -0.19 | -0.28 | 0.00 | 0.12 | 0.23 | 0.15 | -0.06 | -0.16 |
| hd047536 | H | 4353 | 1.85 | 1.43 | -0.46 | -0.26 | -0.24 | -0.21 | 0.09 | -0.42 | -0.44 | -0.37 | -0.36 | -0.58 | -0.75 | -0.60 | -0.50 | -0.54 | -0.37 | -0.49 | -0.38 | -0.60 | -0.71 | -0.61 | -0.42 | -0.34 | -0.23 | -0.26 | -0.45 |
| hd047910 | H | 4869 | 2.68 | 1.25 | 0.24 | 0.14 | 0.25 | 0.17 | 0.57 | 0.14 | 0.01 | 0.13 | 0.26 | 0.13 | 0.14 | 0.05 | 0.02 | 0.07 | 0.14 | 0.06 | 0.17 | 0.11 | 0.04 | 0.17 | 0.22 | 0.32 | 0.28 | 0.15 | 0.21 |
| hd049877 | H | 3911 | 1.35 | 4.06 |  |  | -0.42 | 0.39 | 1.08 | -0.41 | -0.36 | -0.66 | -0.40 | 0.20 | -0.38 | -0.14 | 0.18 | -0.03 |  |  | -0.65 | -1.44 |  | 0.34 | 0.44 | 0.36 |  |  |  |
| hd049947 | H | 4885 | 2.67 | 1.34 | -0.06 | -0.08 | 0.00 | 0.01 | 0.19 | -0.10 | -0.15 | -0.11 | -0.13 | -0.12 | -0.18 | -0.15 | -0.16 | -0.14 | -0.18 | -0.13 | 0.06 | 0.01 | -0.12 | 0.12 | 0.22 | 0.37 | 0.27 | 0.14 | 0.12 |
| hd050310 | H | 4489 | 2.05 | 1.53 | 0.34 | 0.18 | 0.25 | 0.36 | 0.84 | 0.08 | -0.04 | 0.04 | 0.26 | 0.11 | 0.16 | 0.05 | 0.04 | 0.12 | 0.30 | 0.24 | 0.09 | 0.02 | -0.15 | 0.07 | 0.29 | 0.39 | 0.23 | 0.20 | 0.00 |
| hd050890 | H | 4733 | 1.88 | 1.98 | 0.99 | 0.28 | 0.48 | 0.37 | 0.72 | 0.18 | 0.00 | 0.17 | 0.18 | 0.37 | 0.07 | 0.10 | 0.04 | 0.17 |  | 0.14 | 0.75 | 0.37 | 0.24 | 0.27 | 0.40 | 0.10 | 0.25 | 0.49 |  |
| hd054038 | H | 5028 | 3.02 | 1.25 | 0.12 | 0.12 | 0.18 | 0.14 | 0.38 | 0.11 | -0.04 | 0.04 | 0.09 | 0.16 | 0.08 | 0.07 | -0.01 | 0.07 | 0.00 | 0.08 | 0.15 | 0.12 | 0.07 | 0.21 | 0.33 | 0.50 | 0.27 | 0.14 | 0.49 |
| hd054732 | H | 4902 | 2.44 | 1.27 | 0.39 | 0.24 | 0.14 | 0.16 | 0.50 | 0.07 | -0.13 | 0.02 | 0.07 | 0.12 | 0.10 | -0.04 | -0.03 | -0.01 | 0.64 | -0.44 | 0.15 | 0.09 | -0.02 | 0.03 | 0.20 | 0.28 | 0.16 | 0.10 | 0.12 |
| hd059219 | H | 4803 | 2.06 | 2.19 | 0.71 | 0.32 | 0.29 | 0.37 | 0.87 | 0.11 | -0.02 | 0.07 | 0.05 | 0.27 | 0.03 | 0.10 | 0.01 | 0.12 | -0.05 | 0.09 | 0.11 | 0.11 | 0.14 | 0.07 | 0.40 | 0.30 | 0.39 | 0.24 | 0.14 |
| hd059894 | H | 4904 | 2.72 | 1.29 | 0.00 | 0.01 | -0.01 | 0.06 | 0.17 | -0.02 | -0.10 | -0.05 | -0.04 | -0.05 | -0.10 | -0.09 | -0.15 | -0.10 | -0.04 | -0.11 | 0.10 | 0.05 | -0.04 | 0.18 | 0.30 | 0.45 | 0.34 | 0.20 | 0.09 |
| hd060574 | H | 4929 | 2.77 | 1.33 | -0.31 | -0.22 | -0.16 | -0.15 | 0.04 | -0.29 | -0.27 | -0.29 | -0.33 | -0.38 | -0.42 | -0.38 | -0.35 | -0.35 | -0.31 | -0.25 | -0.25 | -0.35 | -0.38 | -0.24 | -0.16 | -0.08 | -0.11 | -0.13 | 0.12 |
| hd062713 | H | 4624 | 2.55 | 1.24 | 0.16 | 0.25 | 0.26 | 0.33 | 0.78 | 0.11 | 0.01 | 0.13 | 0.33 | 0.17 | 0.21 | 0.14 | 0.05 | 0.20 | 0.32 | 0.33 | 0.06 | 0.07 | -0.15 | 0.15 | 0.26 | 0.55 | 0.41 | 0.34 | 0.12 |
| hd062849 | H | 4871 | 2.50 | 0.96 | -0.49 | -0.35 | -0.43 | -0.24 | 0.11 | -0.44 | -0.72 | -0.83 | -1.06 | -0.63 | -0.84 | -0.67 | -0.87 | -0.60 | -0.93 | -0.36 | -0.73 | -0.65 | -0.90 | -0.51 | -0.67 | -0.66 | -0.70 | -0.92 | 0.12 |
| hd062897 | H | 4762 | 2.28 | 1.54 | 0.19 | 0.16 | 0.25 | 0.18 | 0.60 | 0.05 | -0.02 | 0.03 | 0.10 | 0.11 | -0.02 | -0.01 | -0.04 | 0.00 | 0.02 | 0.03 | 0.10 | 0.03 | -0.08 | 0.15 | 0.37 | 0.68 | 0.25 | 0.25 | 0.02 |
| hd063295 | H | 4736 | 2.56 | 1.34 | -0.01 | 0.04 | 0.09 | 0.15 | 0.30 | -0.05 | -0.07 | -0.07 | 0.00 | -0.08 | -0.04 | -0.08 | -0.08 | -0.04 | 0.08 | 0.00 | -0.03 | -0.04 | -0.23 | 0.04 | 0.09 | 0.31 | 0.18 | 0.12 | 0.05 |
| hd064181 | H | 5018 | 2.68 | 1.31 | 0.19 | 0.03 | 0.10 | 0.09 | 0.39 | 0.06 | -0.12 | -0.04 | -0.05 | 0.03 | -0.07 | -0.04 | -0.11 | -0.06 | -0.14 | -0.11 | 0.20 | 0.05 | 0.05 | 0.21 | 0.28 | 0.45 | 0.27 | 0.20 | 0.10 |
| hd073598 | H | 5012 | 2.41 | 1.49 | 0.63 | 0.35 | 0.39 | 0.28 | 0.48 | 0.30 | 0.09 | 0.21 | 0.36 | 0.30 | 0.37 | 0.20 | 0.17 | 0.21 | 0.18 | 0.09 | 0.34 | 0.16 | 0.11 | 0.19 | 0.23 | 0.26 | 0.19 | 0.13 | 0.26 |
| hd073665 | H | 4954 | 2.60 | 1.48 | 0.69 | 0.32 | 0.38 | 0.39 | 0.60 | 0.29 | 0.16 | 0.20 | 0.31 | 0.30 | 0.32 | 0.23 | 0.18 | 0.26 | 0.25 | 0.18 | 0.29 | 0.28 | 0.14 | 0.29 | 0.33 | 0.41 | 0.29 | 0.22 | 0.29 |
| hd073710 | H | 4900 | 2.50 | 1.55 | 0.64 | 0.37 | 0.41 | 0.39 | 0.83 | 0.33 | 0.14 | 0.23 | 0.38 | 0.34 | 0.33 | 0.24 | 0.20 | 0.29 | 0.26 | 0.27 | 0.30 | 0.24 | 0.17 | 0.26 | 0.51 | 0.44 | 0.33 | 0.38 | 0.35 |
| hd074874 | H | 5384 | 2.89 | 0.72 | 0.16 | -0.03 | 0.18 | -0.03 | 0.02 | 0.12 | -0.07 | 0.12 | 0.23 | 0.14 | -0.05 | 0.05 | 0.04 | 0.02 | -0.14 | -0.38 | 0.31 | 0.04 | 0.17 | 0.35 | 0.19 | 0.34 | 0.22 | 0.16 | 0.46 |
| hd078964b | H | 5118 | 4.59 | 0.25 | 0.15 | 0.08 | 0.24 | 0.21 | 0.54 | 0.19 | 0.22 | 0.27 | 0.42 | 0.31 | 0.21 | 0.19 | 0.14 | 0.22 | 0.18 | 0.12 | 0.32 | 0.39 | 0.39 | 0.21 | 0.56 | 0.61 | 0.75 | 0.50 | 0.79 |
| hd091267 | H | 4873 | 4.55 | 0.25 | 0.14 | -0.04 | 0.11 | 0.14 | 0.43 | 0.02 | 0.18 | 0.14 | 0.35 | 0.15 | 0.09 | 0.03 | 0.13 | 0.10 | 0.22 | 0.13 | 0.15 | 0.02 | 0.09 | -0.18 | 0.39 | 0.48 | 0.92 | 0.42 | 0.40 |
| hd104760a | H | 5824 | 4.31 | 0.50 | 0.21 | 0.21 | 0.10 | 0.16 | 0.32 | 0.17 | 0.23 | 0.18 | 0.14 | 0.23 | 0.18 | 0.17 | 0.13 | 0.20 | 0.24 | 0.26 | 0.27 | 0.34 | 0.33 | 0.17 | 0.36 | 0.38 | 0.25 | 0.25 | 0.66 |
| hd108063 | H | 5859 | 3.68 | 1.43 | 0.92 | 0.72 | 0.70 | 0.57 | 0.72 | 0.55 | 0.63 | 0.53 | 0.54 | 0.60 | 0.63 | 0.53 | 0.56 | 0.63 | 0.69 | 0.65 | 0.61 | 0.55 | 0.55 | 0.43 | 0.62 | 0.51 | 0.41 | 0.40 | 1.56 |
| hd108570 | H | 4984 | 3.49 | 0.50 | 0.09 | 0.11 | 0.12 | 0.13 | 0.24 | 0.10 | 0.08 | 0.17 | 0.26 | 0.14 | 0.12 | 0.09 | 0.01 | 0.14 | 0.26 | 0.12 | 0.22 | 0.24 | 0.16 | 0.16 | 0.43 | 0.57 | 0.65 | 0.29 | 0.61 |
| hd110291 | H | 5392 | 4.50 | 0.50 | -0.01 | 0.04 | 0.06 | 0.10 | 0.28 | -0.06 | 0.10 | 0.07 | 0.06 | 0.11 | 0.00 | 0.05 | 0.05 | 0.06 | 0.14 | 0.09 | 0.22 | 0.23 | 0.17 | -0.04 | 0.39 | 0.58 | 0.39 | 0.35 | 0.71 |
| hd114747 | H | 5073 | 4.50 | 0.50 | 0.68 | 0.34 | 0.44 | 0.46 | 0.89 | 0.28 | 0.51 | 0.40 | 0.67 | 0.44 | 0.44 | 0.32 | 0.48 | 0.45 | 0.55 | 0.45 | 0.30 | 0.40 | 0.34 | 0.06 | 0.63 | 0.82 | 0.81 | 0.68 | 0.96 |
| hd115202 | H | 4742 | 3.05 | 0.50 | 0.25 | 0.19 | 0.31 | 0.27 | 0.51 | 0.20 | 0.21 | 0.28 | 0.51 | 0.24 | 0.26 | 0.17 | 0.09 | 0.27 | 0.49 | 0.51 | 0.12 | 0.23 | -0.03 | 0.17 | 0.41 | 0.62 | 0.60 | 0.42 | 0.20 |

| Name | | T | log g | vt | [Fe/H] | [Na/H] | [Mg/H] | [Al/H] | [Si/H] | [Ca/H] | [Sc/H] | [Ti/H] | [V/H] | [Cr/H] | [Mn/H] | [Co/H] | [Ni/H] | [Cu/H] | [Zn/H] | [Y/H] | [Zr/H] | [Ba/H] | [La/H] | [Ce/H] | [Nd/H] | [Eu/H] |
|---|---|---|---|---|---|---|---|---|---|---|---|---|---|---|---|---|---|---|---|---|---|---|---|---|---|---|
| hd121416 | H | 4576 | 2.45 | 1.25 | 0.37 | 0.32 | 0.38 | 0.42 | 0.75 | 0.22 | 0.12 | 0.22 | 0.49 | 0.29 | 0.33 | 0.20 | 0.17 | 0.29 | 0.53 | 0.51 | 0.14 | 0.10 | -0.05 | 0.14 | 0.33 | 0.32 | 0.39 | 0.32 | 0.06 |
| hd123517 | H | 6340 | 2.50 | 1.56 | 0.42 | 0.43 | 0.33 | 0.23 | -0.30 | 0.38 | -0.01 | 0.24 | 0.31 | 0.28 | 0.34 | 0.28 | 0.37 | 0.32 | 0.22 | 0.08 | 0.55 | -0.09 | -0.05 | 0.04 | -0.23 | -0.13 | -0.12 | -0.27 | 1.17 |
| hd142527 | H | 6632 | 3.15 | 3.64 | 0.65 | | | 0.73 | | | 0.40 | 0.07 | 1.25 | 1.10 | 0.75 | 0.81 | 0.33 | 1.16 | 0.54 | | 0.67 | | 1.64 | | | 0.64 | 1.70 | 1.36 | 1.52 | 1.86 |
| hd144589 | H | 6486 | 2.50 | 1.58 | 0.22 | 0.27 | 0.05 | 0.08 | -0.29 | 0.22 | -0.15 | 0.00 | 0.20 | 0.07 | 0.03 | 0.04 | 0.24 | 0.06 | -0.14 | -0.13 | 0.52 | -0.11 | 0.03 | 0.01 | -0.27 | -0.26 | -0.28 | -0.29 | -0.05 |
| hd147135 | H | 6832 | 3.71 | 3.45 | -0.18 | -0.15 | -0.06 | -0.04 | 0.13 | 0.06 | 0.21 | -0.01 | 0.17 | -0.15 | -0.29 | -0.22 | 0.05 | -0.12 | -0.30 | -0.28 | | 0.42 | 0.44 | 0.10 | 0.18 | 0.40 | 0.31 | 0.29 | 0.42 |
| hd169689 | H | 4989 | 2.05 | 1.94 | 0.46 | 0.27 | 0.60 | 0.32 | 0.33 | 0.24 | 0.07 | 0.23 | 0.26 | 0.27 | -0.05 | 0.14 | 0.09 | 0.10 | -0.10 | 0.01 | 0.48 | 0.17 | 0.14 | 0.31 | 0.33 | 0.28 | 0.39 | 0.19 | 0.32 |
| hd176354 | H | 5165 | 3.82 | 0.50 | 0.64 | 0.43 | 0.54 | 0.47 | 0.93 | 0.36 | 0.47 | 0.46 | 0.66 | 0.51 | 0.56 | 0.41 | 0.41 | 0.55 | 0.74 | 0.58 | 0.41 | 0.53 | 0.32 | 0.35 | 0.68 | 0.77 | 0.66 | 0.64 | 0.99 |
| hd181433 | H | 4800 | 2.50 | 1.48 | 1.15 | 0.51 | 0.51 | 0.15 | 0.34 | 0.46 | -0.10 | 0.05 | 0.37 | 0.26 | 0.21 | -0.07 | 0.02 | 0.06 | 0.21 | -0.27 | 0.32 | -0.33 | -0.15 | -0.95 | -0.51 | -0.36 | -0.20 | -0.19 | 0.17 |
| hd181517 | H | 4895 | 2.73 | 1.27 | 0.36 | 0.15 | 0.22 | 0.25 | 0.53 | 0.18 | 0.03 | 0.10 | 0.15 | 0.18 | 0.20 | 0.11 | 0.01 | 0.15 | 0.16 | 0.08 | 0.25 | 0.20 | 0.16 | 0.14 | 0.32 | 0.46 | 0.30 | 0.26 | 0.11 |
| hd182893 | H | 4864 | 2.78 | 1.15 | 0.39 | 0.23 | 0.28 | 0.31 | 0.49 | 0.23 | 0.05 | 0.14 | 0.25 | 0.25 | 0.23 | 0.16 | 0.03 | 0.20 | 0.22 | 0.27 | 0.27 | 0.19 | 0.12 | 0.21 | 0.28 | 0.60 | 0.37 | 0.21 | 0.24 |
| hd187669a | H | 4754 | 2.50 | 1.97 | 0.36 | | | -0.11 | 0.76 | -0.01 | 0.12 | 0.09 | 0.15 | 0.33 | -0.24 | -0.22 | -0.05 | -0.07 | -0.30 | -0.51 | | 0.19 | 0.32 | -0.36 | 0.69 | 0.22 | 0.80 | 0.63 | -0.03 |
| hd188114 | H | 4594 | 2.24 | 1.32 | -0.14 | -0.15 | -0.15 | -0.07 | 0.22 | -0.19 | -0.33 | -0.24 | -0.17 | -0.24 | -0.32 | -0.28 | -0.35 | -0.29 | -0.24 | -0.33 | -0.10 | -0.29 | -0.37 | -0.11 | 0.08 | 0.18 | 0.08 | -0.11 | -0.19 |
| hd190056 | H | 4275 | 1.92 | 1.37 | -0.33 | -0.13 | -0.13 | -0.05 | 0.34 | -0.32 | -0.30 | -0.24 | -0.16 | -0.48 | -0.63 | -0.50 | -0.37 | -0.41 | 0.03 | -0.36 | -0.29 | -0.61 | -0.62 | -0.55 | -0.27 | -0.28 | -0.19 | -0.19 | -0.44 |
| hd196171 | H | 4810 | 2.59 | 1.28 | 0.11 | 0.06 | 0.12 | 0.12 | 0.39 | 0.04 | -0.09 | -0.04 | 0.02 | 0.01 | -0.01 | -0.03 | -0.09 | -0.03 | -0.01 | -0.07 | 0.12 | 0.06 | -0.08 | 0.17 | 0.19 | 0.47 | 0.26 | 0.16 | 0.07 |
| hd198232 | H | 4750 | 2.27 | 1.44 | 0.23 | 0.05 | 0.12 | 0.20 | 0.44 | 0.05 | -0.14 | -0.10 | -0.03 | 0.05 | -0.03 | -0.03 | -0.12 | -0.03 | -0.13 | -0.03 | 0.12 | 0.00 | -0.09 | 0.24 | 0.31 | 0.40 | 0.19 | 0.13 | -0.02 |
| hd199951 | H | 5009 | 2.69 | 1.49 | 0.32 | 0.04 | 0.08 | 0.14 | 0.51 | 0.04 | -0.10 | -0.03 | -0.05 | 0.08 | -0.11 | -0.03 | -0.13 | -0.04 | -0.25 | -0.14 | 0.13 | 0.14 | 0.01 | 0.26 | 0.29 | 0.64 | 0.37 | 0.13 | 0.06 |
| hd200763 | H | 4632 | 2.30 | 1.36 | 0.32 | 0.17 | 0.23 | 0.26 | 0.66 | 0.15 | -0.04 | 0.08 | 0.26 | 0.18 | 0.16 | 0.08 | -0.01 | 0.10 | 0.19 | 0.04 | 0.21 | 0.11 | 0.03 | 0.21 | 0.43 | 0.61 | 0.39 | 0.21 | 0.05 |
| hd201852 | H | 4883 | 2.65 | 1.25 | 0.16 | 0.09 | 0.12 | 0.14 | 0.28 | 0.09 | -0.09 | 0.01 | 0.05 | 0.05 | 0.02 | 0.00 | 0.01 | -0.05 | 0.19 | 0.09 | -0.01 | 0.18 | 0.34 | 0.49 | 0.35 | 0.16 | 0.11 |
| hd206642 | H | 4326 | 1.59 | 1.58 | -1.16 | -0.78 | -0.99 | -0.60 | 0.04 | -0.85 | -1.05 | -0.93 | -1.07 | -1.06 | -1.37 | -1.02 | -0.98 | -1.06 | -1.31 | -0.88 | -0.82 | -0.89 | -0.96 | -0.78 | -0.71 | -0.70 | -0.57 | -0.58 | -0.70 |
| hd209449 | H | 5685 | 4.02 | 0.78 | 0.78 | 0.56 | 0.59 | 0.48 | 0.66 | 0.46 | 0.61 | 0.51 | 0.52 | 0.53 | 0.60 | 0.50 | 0.50 | 0.61 | 0.80 | 0.74 | 0.56 | 0.57 | 0.41 | 0.45 | 0.63 | 0.67 | 0.64 | 0.59 | 1.09 |
| hd211317 | H | 5780 | 4.07 | 0.86 | 0.49 | 0.39 | 0.38 | 0.31 | 0.48 | 0.30 | 0.35 | 0.31 | 0.30 | 0.34 | 0.37 | 0.29 | 0.30 | 0.37 | 0.46 | 0.47 | 0.41 | 0.45 | 0.32 | 0.22 | 0.42 | 0.37 | 0.35 | 0.25 | 1.07 |
| hip031592 | H | 4735 | 3.09 | 0.50 | 0.58 | 0.51 | 0.47 | 0.55 | 0.93 | 0.41 | 0.47 | 0.52 | 0.90 | 0.53 | 0.60 | 0.44 | 0.40 | 0.62 | 0.80 | 0.68 | 0.38 | 0.57 | 0.32 | 0.38 | 0.69 | 0.83 | 0.86 | 0.82 | 0.48 |
| hip080242b | H | 4315 | 3.40 | 0.75 | 0.05 | 0.48 | 0.15 | 1.00 | 1.71 | 0.00 | 0.38 | 0.06 | 0.13 | 0.29 | 0.24 | 0.61 | 0.40 | 0.70 | 0.61 | 1.01 | -0.06 | 0.64 | -0.03 | 0.66 | 1.07 | 1.49 | 1.13 | 1.35 | 0.62 |
| hr2959 | H | 3916 | 2.50 | 1.25 | 0.53 | 0.68 | 0.43 | 1.10 | 1.26 | 0.32 | 0.63 | 0.60 | 0.88 | 0.58 | 0.52 | 0.72 | 0.64 | 0.83 | 0.84 | 1.01 | 0.44 | 0.70 | 0.76 | 1.01 | 1.38 | 1.46 | 1.55 | 1.57 | 0.52 |
| hr3728 | H | 4888 | 2.60 | 1.34 | -0.20 | -0.21 | -0.13 | -0.14 | 0.05 | -0.25 | -0.34 | -0.30 | -0.34 | -0.34 | -0.41 | -0.37 | -0.37 | -0.36 | -0.33 | -0.31 | -0.23 | -0.33 | -0.34 | -0.20 | -0.12 | -0.09 | -0.11 | -0.21 | -0.18 |
| hr3919 | H | 4415 | 1.90 | 1.41 | 0.29 | 0.19 | 0.30 | 0.28 | 0.72 | 0.08 | -0.02 | 0.06 | 0.36 | 0.13 | 0.12 | 0.01 | 0.04 | 0.09 | 0.64 | 0.12 | 0.10 | -0.06 | -0.18 | -0.03 | 0.17 | 0.14 | 0.16 | 0.08 | -0.12 |
| hr5480 | H | 4807 | 2.09 | 1.76 | 0.41 | 0.20 | 0.18 | 0.28 | 0.66 | 0.05 | -0.07 | -0.04 | -0.04 | 0.13 | -0.11 | 0.02 | -0.09 | 0.03 | -0.14 | -0.03 | 0.08 | 0.12 | -0.03 | 0.20 | 0.27 | 0.64 | 0.26 | 0.13 | 0.31 |
| hr7150 | H | 4541 | 1.71 | 1.82 | 0.25 | 0.18 | 0.23 | 0.27 | 0.52 | 0.01 | -0.15 | -0.04 | -0.02 | 0.11 | 0.02 | 0.02 | -0.07 | 0.03 | | 0.35 | 0.13 | 0.07 | -0.08 | 0.22 | 0.37 | 0.45 | 0.27 | 0.24 | 0.05 |
| ic4651no14527 | H | 4803 | 2.65 | 1.34 | 0.42 | 0.29 | 0.34 | 0.41 | 0.56 | 0.24 | 0.16 | 0.15 | 0.27 | 0.28 | 0.30 | 0.21 | 0.14 | 0.28 | 0.33 | 0.49 | 0.50 | 0.46 | 0.32 | 0.36 | 0.50 | 0.66 | 0.51 | 0.48 | 0.48 |
| ic4651no8540 | H | 4826 | 2.68 | 1.29 | 0.48 | 0.35 | 0.30 | 0.38 | 0.60 | 0.27 | 0.20 | 0.27 | 0.45 | 0.33 | 0.38 | 0.24 | 0.18 | 0.30 | 0.41 | 0.23 | 0.32 | 0.38 | 0.20 | 0.32 | 0.57 | 0.60 | 0.51 | 0.49 | 0.31 |
| ic4651no9025 | H | 4799 | 2.65 | 1.26 | 0.41 | 0.25 | 0.25 | 0.26 | 0.43 | 0.19 | 0.10 | 0.19 | 0.33 | 0.25 | 0.24 | 0.16 | 0.09 | 0.21 | 0.36 | 0.16 | 0.25 | 0.14 | 0.14 | 0.15 | 0.26 | 0.48 | 0.35 | 0.29 | 0.16 |
| ic4651no9791 | H | 4463 | 2.21 | 1.29 | 0.33 | 0.22 | 0.31 | 0.33 | 0.74 | 0.15 | 0.05 | 0.16 | 0.43 | 0.25 | 0.27 | 0.15 | 0.09 | 0.23 | 0.57 | 0.34 | 0.22 | 0.22 | 0.01 | 0.13 | 0.36 | 0.44 | 0.41 | 0.36 | -0.01 |
| ksihya | H | 4950 | 2.65 | 1.32 | 0.45 | 0.21 | 0.23 | 0.25 | 0.58 | 0.20 | -0.04 | 0.05 | 0.12 | 0.20 | 0.15 | 0.10 | 0.01 | 0.11 | 0.04 | 0.07 | 0.17 | 0.09 | 0.02 | 0.19 | 0.29 | 0.32 | 0.13 | 0.14 | 0.17 |
| ngc2287no107 | H | 4602 | 1.57 | 1.84 | 0.22 | 0.02 | 0.08 | 0.14 | 0.59 | -0.09 | -0.31 | -0.21 | -0.18 | -0.03 | -0.11 | -0.13 | -0.22 | -0.14 | -0.17 | 0.04 | -0.01 | -0.07 | -0.25 | -0.02 | 0.19 | 0.36 | 0.11 | 0.02 | -0.12 |
| ngc2287no204 | H | 4382 | 1.58 | 1.52 | -0.04 | -0.10 | -0.18 | -0.09 | 0.37 | -0.16 | -0.38 | -0.22 | -0.10 | -0.20 | -0.30 | -0.32 | -0.36 | -0.31 | -0.05 | -0.34 | -0.05 | -0.27 | -0.40 | -0.19 | 0.02 | 0.10 | 0.01 | -0.15 | -0.22 |
| ngc2287no21 | H | 4020 | 2.50 | 1.58 | 0.44 | 0.60 | 0.42 | 0.99 | | 0.15 | 0.50 | 0.48 | 0.75 | 0.50 | 0.52 | 0.65 | 0.59 | 0.71 | 0.73 | 0.73 | 0.38 | 0.54 | 0.66 | | | 1.39 | 1.34 | 1.33 | 1.33 | 0.42 |
| ngc2287no75 | H | 4423 | 1.51 | 1.91 | 0.21 | 0.18 | 0.17 | 0.32 | 0.59 | 0.00 | -0.19 | -0.09 | -0.11 | 0.06 | -0.04 | -0.02 | -0.11 | 0.00 | 0.38 | 0.15 | 0.03 | 0.02 | -0.24 | 0.18 | 0.41 | 0.38 | 0.32 | 0.28 | -0.09 |
| ngc2287no87 | H | 4199 | 1.41 | 1.56 | 0.06 | -0.14 | 0.05 | 0.03 | 0.79 | -0.09 | -0.27 | -0.13 | 0.07 | -0.11 | -0.25 | -0.27 | -0.25 | -0.20 | 0.03 | -0.38 | 0.06 | -0.34 | -0.30 | -0.35 | -0.02 | 0.01 | 0.11 | -0.06 | -0.21 |
| ngc2287no97 | H | 4596 | 1.84 | 1.81 | 0.24 | 0.13 | 0.15 | 0.22 | 0.57 | 0.00 | -0.09 | -0.10 | -0.09 | 0.06 | -0.01 | -0.02 | -0.11 | -0.03 | -0.01 | 0.13 | 0.15 | 0.04 | -0.17 | 0.22 | 0.37 | 0.45 | 0.30 | 0.31 | 0.03 |
| ngc3532no100 | H | 4740 | 2.05 | 1.75 | 0.30 | 0.15 | 0.14 | 0.25 | 0.65 | 0.08 | -0.10 | -0.08 | -0.06 | 0.07 | -0.02 | -0.01 | -0.13 | -0.04 | -0.11 | 0.07 | 0.08 | 0.04 | -0.07 | 0.14 | 0.31 | 0.58 | 0.20 | 0.16 | 0.07 |
| ngc3532no122 | H | 4963 | 2.43 | 1.80 | 0.36 | 0.15 | 0.13 | 0.18 | 0.40 | 0.13 | -0.07 | 0.04 | 0.04 | 0.18 | -0.07 | 0.03 | -0.12 | 0.03 | -0.15 | -0.01 | 0.06 | 0.12 | 0.17 | 0.26 | 0.42 | 0.20 | 0.49 | 0.21 | 0.24 |
| ngc3532no19 | H | 4844 | 2.23 | 1.55 | 0.25 | 0.07 | 0.17 | 0.22 | 0.55 | 0.10 | -0.06 | -0.06 | -0.05 | 0.12 | -0.01 | 0.01 | -0.10 | 0.02 | -0.10 | 0.04 | 0.13 | 0.10 | -0.07 | 0.28 | 0.37 | 0.55 | 0.29 | 0.25 | 0.11 |
| ngc3532no596 | H | 4955 | 2.37 | 1.89 | 0.33 | 0.22 | 0.15 | 0.21 | 0.34 | -0.04 | 0.00 | -0.02 | -0.08 | 0.10 | -0.09 | -0.02 | -0.09 | -0.03 | -0.22 | -0.10 | 0.44 | 0.23 | 0.05 | 0.07 | 0.23 | 0.75 | 0.49 | 0.15 | 0.06 |
| ngc3532no649 | H | 4849 | 2.29 | 1.41 | 0.04 | -0.01 | 0.02 | 0.01 | -0.01 | -0.09 | -0.18 | -0.11 | -0.07 | -0.10 | -0.18 | -0.15 | -0.19 | -0.16 | -0.16 | -0.31 | 0.13 | -0.11 | -0.16 | 0.05 | 0.07 | 0.21 | 0.16 | 0.03 | -0.03 |
| ngc4349no127 | H | 4479 | 1.58 | 1.76 | 0.30 | 0.11 | 0.15 | 0.15 | 0.20 | 0.01 | -0.20 | -0.04 | 0.04 | 0.11 | 0.07 | -0.05 | -0.12 | -0.03 | 0.08 | 0.16 | 0.15 | 0.04 | 0.03 | -0.02 | 0.40 | 0.37 | 0.25 | 0.28 | -0.11 |
| ngc6705no1286 | H | 4929 | 2.32 | 2.18 | 0.98 | 0.45 | 0.57 | 0.56 | 0.99 | 0.30 | 0.26 | 0.35 | 0.45 | 0.49 | 0.22 | 0.27 | 0.32 | 0.32 | 0.08 | 0.24 | 0.28 | 0.28 | 0.40 | 0.09 | 0.58 | 0.20 | 0.59 | 0.50 |
| ngc6705no1423 | H | 4521 | 1.83 | 1.92 | 0.57 | 0.42 | 0.53 | 0.47 | 0.81 | 0.26 | 0.19 | 0.28 | 0.41 | 0.41 | 0.36 | 0.25 | 0.25 | 0.35 | 0.93 | 0.40 | 0.39 | 0.30 | 0.17 | 0.19 | 0.70 | 0.42 | 0.46 | 0.47 | 0.29 |
| ngc6705no411 | H | 4445 | 1.84 | 2.00 | 0.75 | 0.33 | 0.47 | 0.52 | 0.84 | 0.23 | 0.17 | 0.23 | 0.35 | 0.32 | 0.26 | 0.23 | 0.24 | 0.31 | 0.47 | -0.32 | 0.32 | 0.15 | 0.05 | -0.01 | 0.51 | 0.75 | 0.51 | 0.58 | 0.27 |
| ngc6705no660 | H | 4788 | 2.13 | 1.95 | 0.89 | 0.47 | 0.47 | 0.44 | 0.88 | 0.35 | 0.24 | 0.33 | 0.42 | 0.43 | 0.37 | 0.29 | 0.25 | 0.33 | 0.37 | 0.35 | 0.42 | 0.30 | 0.23 | 0.16 | 0.50 | 0.50 | 0.36 | 0.48 | 0.25 |
| ngc6705no779 | H | 4307 | 1.65 | 1.98 | 0.59 | 0.21 | 0.47 | 0.59 | 0.86 | 0.13 | 0.02 | 0.11 | 0.21 | 0.29 | 0.25 | 0.16 | 0.14 | 0.27 | 0.41 | -0.09 | 0.25 | 0.08 | 0.02 | -0.05 | 0.34 | 0.30 | 0.36 | 0.51 | 0.45 |

| Name | Type | col3 | col4 | col5 | col6 | col7 | col8 | col9 | col10 | col11 | col12 | col13 | col14 | col15 | col16 | col17 | col18 | col19 | col20 | col21 | col22 | col23 | col24 | col25 | col26 | col27 | col28 | col29 | col30 | col31 |
|---|---|---|---|---|---|---|---|---|---|---|---|---|---|---|---|---|---|---|---|---|---|---|---|---|---|---|---|---|---|---|
| pihya | H | 4563 | 2.39 | 1.20 | 0.17 | 0.14 | 0.10 | 0.21 | 0.69 | 0.04 | -0.07 | 0.02 | 0.22 | 0.06 | 0.11 | 0.01 | -0.03 | 0.07 | 0.20 | -0.05 | 0.10 | 0.08 | -0.09 | 0.12 | 0.32 | 0.49 | 0.32 | 0.20 | 0.03 |
| sand1016 | H | 4431 | 2.16 | 1.32 | 0.24 | 0.29 | 0.34 | 0.29 | 0.59 | 0.14 | 0.08 | 0.17 | 0.49 | 0.23 | 0.24 | 0.11 | 0.10 | 0.17 | 0.51 | 0.21 | 0.22 | 0.12 | 0.00 | 0.08 | 0.37 | 0.37 | 0.37 | 0.31 | 0.03 |
| sand1054 | H | 4721 | 2.67 | 1.11 | 0.22 | 0.14 | 0.28 | 0.19 | 0.04 | 0.14 | 0.02 | 0.13 | 0.31 | 0.22 | 0.19 | 0.09 | 0.03 | 0.15 | 0.26 | 0.36 | 0.31 | 0.18 | -0.04 | 0.06 | 0.48 | 0.43 | 0.41 | 0.34 | 0.02 |
| sand1074 | H | 4718 | 2.49 | 1.37 | 0.30 | 0.19 | 0.25 | 0.26 | 0.56 | 0.13 | 0.03 | 0.09 | 0.26 | 0.17 | 0.21 | 0.08 | 0.03 | 0.13 | 0.19 | 0.12 | 0.21 | 0.14 | -0.02 | 0.08 | 0.33 | 0.43 | 0.30 | 0.30 | 0.19 |
| sand1084 | H | 4745 | 2.48 | 1.37 | 0.33 | 0.19 | 0.24 | 0.25 | 0.35 | 0.10 | 0.02 | 0.09 | 0.27 | 0.15 | 0.18 | 0.07 | 0.02 | 0.14 | 0.18 | 0.05 | 0.16 | 0.06 | -0.07 | 0.06 | 0.38 | 0.32 | 0.21 | 0.17 | 0.13 |
| sand1237 | H | 5067 | 2.85 | 1.35 | 0.33 | 0.14 | 0.22 | 0.14 | 0.18 | 0.14 | 0.00 | 0.10 | 0.17 | 0.18 | 0.14 | 0.06 | 0.04 | 0.08 | -0.03 | -0.02 | 0.25 | 0.11 | 0.03 | 0.05 | 0.11 | 0.23 | 0.16 | 0.14 | 0.80 |
| sand1279 | H | 4732 | 2.48 | 1.34 | 0.36 | 0.23 | 0.22 | 0.27 | 0.48 | 0.15 | -0.01 | 0.15 | 0.35 | 0.21 | 0.21 | 0.11 | 0.08 | 0.16 | 0.31 | 0.12 | 0.17 | 0.09 | 0.00 | 0.09 | 0.40 | 0.39 | 0.30 | 0.29 | 0.18 |
| sand364 | H | 4245 | 1.83 | 1.36 | 0.25 | 0.19 | 0.23 | 0.40 | 0.74 | 0.06 | -0.01 | 0.08 | 0.35 | 0.13 | 0.09 | 0.07 | 0.07 | 0.17 | 0.15 | 0.26 | 0.09 | -0.03 | -0.14 | 0.05 | 0.23 | 0.54 | 0.32 | 0.29 | -0.18 |
| sand978 | H | 4257 | 1.82 | 1.48 | 0.29 | 0.16 | 0.28 | 0.27 | 0.69 | 0.14 | 0.08 | 0.11 | 0.37 | 0.19 | 0.16 | 0.04 | 0.05 | 0.13 | -0.16 | 0.21 | 0.28 | 0.05 | -0.01 | -0.07 | 0.39 | 0.48 | 0.41 | 0.37 | -0.18 |
| sand989 | H | 4793 | 2.84 | 1.14 | 0.38 | 0.30 | 0.34 | 0.29 | 0.60 | 0.23 | 0.21 | 0.24 | 0.42 | 0.33 | 0.31 | 0.20 | 0.16 | 0.29 | 0.44 | 0.28 | 0.29 | 0.35 | 0.22 | 0.21 | 0.45 | 0.76 | 0.55 | 0.40 | 0.44 |
| sigpup | H | 4081 | 3.53 | 1.25 | 0.23 | 0.61 | 0.57 | 0.96 | 2.14 | 0.24 | 1.04 | 0.82 | 1.12 | 0.69 | 0.53 | 0.68 | 1.00 | 0.91 | 1.09 | 1.24 | 0.69 | 1.02 | 0.81 | 0.96 | 1.52 | 1.64 | 1.78 | 1.75 | 0.48 |
| tzfor | H | 5332 | 2.96 | 0.52 | 0.42 | 0.08 | 0.29 | 0.00 | 0.14 | 0.16 | 0.10 | 0.24 | 0.48 | 0.36 | 0.10 | 0.12 | 0.17 | 0.16 | -0.28 | -0.11 | 0.46 | 0.19 | 0.30 | -0.03 | 0.57 | 0.47 | 0.68 | 0.53 | 0.70 |
| v1045sco | H | 3789 | 1.37 | 2.80 | -0.23 | -0.10 | -0.27 | 0.67 | 2.26 | -0.69 | -0.29 | -0.42 | -0.32 | -0.14 | -0.54 | -0.16 | 0.11 | -0.06 | 0.78 | -0.45 | 0.29 | -0.41 | -0.57 | -0.58 | -0.13 | 0.06 | 0.09 | 0.05 | 0.18 |
| xcae | H | 6973 | 3.68 | 5.37 | | 0.00 | | 0.08 | | -0.14 | -0.07 | 0.19 | | -0.36 | -0.39 | -0.34 | 1.44 | 0.03 | | -0.74 | | 1.59 | 2.05 | -0.28 | | | | 1.20 | | |
| hd001142 | S | 5186 | 2.90 | 1.07 | 0.34 | 0.07 | 0.27 | 0.14 | 0.52 | 0.22 | 0.14 | 0.31 | 0.42 | 0.30 | 0.17 | 0.14 | 0.17 | 0.16 | 0.06 | 0.14 | 0.38 | 0.40 | 0.16 | 0.31 | 0.43 | 0.19 | 0.33 | 0.66 | 0.23 |
| hd002410 | S | 5281 | 2.63 | 1.58 | 0.46 | 0.24 | 0.45 | 0.20 | 0.32 | 0.34 | 0.18 | 0.42 | 0.48 | 0.36 | 0.39 | 0.23 | 0.29 | 0.22 | 0.30 | -0.05 | 0.60 | 0.36 | 0.31 | 0.31 | 0.38 | 0.36 | 0.40 | 0.31 | 0.32 |
| hd002954 | S | 6308 | 3.55 | 3.00 | 0.38 | | 0.21 | 0.20 | 0.34 | 0.11 | 0.10 | 0.27 | 0.18 | 0.46 | -0.21 | 0.02 | 0.41 | 0.09 | | | 0.06 | 0.02 | 0.27 | | 0.26 | 0.96 | -0.07 | 0.54 | | 0.87 |
| hd005418 | S | 4977 | 2.84 | 1.33 | 0.15 | 0.06 | 0.16 | 0.14 | 0.55 | 0.09 | 0.04 | 0.07 | 0.02 | 0.08 | 0.02 | 0.02 | -0.02 | 0.03 | 0.06 | 0.02 | 0.27 | 0.20 | -0.15 | 0.27 | 0.31 | 0.32 | 0.34 | 0.36 | 0.18 |
| hd006037 | S | 4566 | 2.71 | 1.26 | 0.64 | 0.44 | 0.66 | 0.56 | 1.55 | 0.32 | 0.40 | 0.48 | 0.82 | 0.54 | 0.55 | 0.37 | 0.51 | 0.59 | 0.88 | 1.65 | 0.38 | 0.49 | -0.01 | 0.19 | 0.81 | 0.56 | 0.83 | 1.47 | 0.60 |
| hd012116 | S | 4462 | 2.47 | 1.23 | -0.07 | 0.06 | 0.11 | 0.22 | 0.86 | -0.07 | 0.00 | -0.07 | 0.02 | -0.03 | -0.13 | -0.09 | -0.02 | 0.05 | 0.41 | 0.65 | -0.08 | -0.05 | -0.66 | -0.09 | 0.23 | 0.03 | 0.29 | 0.79 | 0.19 |
| hd013004 | S | 4584 | 2.77 | 1.08 | 0.46 | 0.52 | 0.54 | 0.52 | 1.33 | 0.33 | 0.35 | 0.40 | 0.63 | 0.44 | 0.44 | 0.35 | 0.38 | 0.52 | 0.92 | 1.40 | 0.26 | 0.49 | -0.12 | 0.25 | 1.16 | 0.52 | 0.72 | 1.37 | 0.52 |
| hd015533 | S | 4370 | 2.49 | 1.24 | 0.44 | 0.21 | 0.57 | 0.72 | 1.89 | 0.39 | 0.39 | 0.30 | 0.69 | 0.42 | 0.41 | 0.40 | 0.45 | 0.56 | | 2.10 | | 0.29 | -0.38 | 0.17 | 0.89 | 0.49 | 0.70 | 0.63 | 0.58 |
| hd015866 | S | 5718 | 3.84 | 1.29 | 0.70 | 0.36 | 0.44 | 0.38 | 0.56 | 0.27 | 0.42 | 0.27 | 0.25 | 0.32 | 0.35 | 0.26 | 0.30 | 0.38 | 0.53 | 0.36 | 0.68 | 0.36 | 0.37 | 0.12 | 0.38 | 0.23 | 0.38 | 0.36 | 0.63 |
| hd016150 | S | 6145 | 3.21 | 3.19 | 0.28 | | -0.43 | 0.13 | 0.03 | 0.32 | -0.35 | 0.25 | 0.65 | 0.39 | -0.23 | -0.05 | 0.27 | 0.22 | | -0.39 | 1.39 | 0.23 | 1.03 | -0.26 | 1.08 | 0.59 | 0.13 | | -0.31 |
| hd016314 | S | 6533 | 3.70 | 3.58 | 0.47 | 0.06 | 0.24 | 0.22 | 0.42 | 0.12 | 0.35 | 0.26 | 0.32 | 0.26 | 0.00 | 0.08 | 0.50 | 0.18 | -0.29 | 0.37 | | 0.28 | -0.01 | -0.16 | 0.60 | 0.32 | 0.38 | 0.45 | 0.15 |
| hd017001 | S | 4863 | 2.76 | 1.32 | 0.06 | -0.01 | 0.12 | 0.15 | 0.62 | -0.01 | -0.01 | -0.04 | -0.03 | 0.03 | -0.09 | -0.06 | -0.08 | -0.01 | 0.09 | 0.32 | 0.21 | 0.11 | -0.28 | 0.25 | 0.29 | 0.15 | 0.26 | 0.26 | 0.19 |
| hd019210 | S | 4952 | 2.65 | 1.39 | 0.31 | 0.18 | 0.32 | 0.20 | 0.47 | 0.26 | 0.11 | 0.24 | 0.27 | 0.24 | 0.24 | 0.12 | 0.11 | 0.14 | 0.29 | 0.40 | 0.31 | 0.25 | 0.02 | 0.18 | 0.43 | 0.19 | 0.32 | 0.37 | 0.19 |
| hd021585 | S | 5204 | 2.84 | 1.62 | -0.65 | -0.37 | -0.36 | -0.38 | -0.29 | -0.53 | -0.43 | -0.46 | -0.59 | -0.64 | -0.91 | -0.68 | -0.50 | -0.67 | -0.78 | -0.42 | 0.13 | -0.61 | -0.57 | -0.48 | -0.45 | -0.58 | -0.38 | -0.54 | -0.25 |
| hd021760 | S | 4627 | 2.79 | 1.03 | 0.25 | 0.30 | 0.33 | 0.36 | 1.00 | 0.15 | 0.12 | 0.15 | 0.32 | 0.21 | 0.19 | 0.12 | 0.12 | 0.27 | 0.48 | 0.79 | 0.10 | 0.19 | -0.32 | 0.03 | 0.61 | 0.17 | 0.43 | 1.05 | 0.07 |
| hd025627 | S | 4613 | 2.74 | 1.14 | 0.39 | 0.29 | 0.50 | 0.47 | 1.12 | 0.25 | 0.32 | 0.33 | 0.49 | 0.36 | 0.33 | 0.24 | 0.29 | 0.41 | 0.60 | 1.46 | 0.25 | 0.40 | -0.14 | 0.19 | 0.92 | 0.50 | 0.71 | 1.10 | 0.33 |
| hd026004 | S | 4490 | 2.40 | 1.27 | 0.05 | -0.04 | 0.12 | 0.25 | 0.82 | -0.01 | -0.07 | -0.01 | 0.14 | 0.02 | -0.04 | -0.03 | -0.02 | 0.07 | 0.27 | 0.24 | 0.10 | 0.21 | -0.41 | 0.17 | 0.45 | 0.27 | 0.39 | 0.81 | 0.26 |
| hd026625 | S | 4898 | 2.85 | 1.22 | 0.31 | 0.24 | 0.29 | 0.30 | 0.57 | 0.19 | 0.09 | 0.10 | 0.14 | 0.22 | 0.15 | 0.11 | 0.06 | 0.15 | 0.21 | 0.33 | 0.27 | 0.26 | -0.17 | 0.18 | 0.60 | 0.19 | 0.35 | 0.58 | 0.21 |
| hd033844 | S | 4792 | 3.13 | 0.79 | 0.58 | 0.41 | 0.57 | 0.53 | 1.13 | 0.39 | 0.39 | 0.43 | 0.63 | 0.45 | 0.47 | 0.38 | 0.36 | 0.55 | 0.85 | 1.11 | 0.36 | 0.57 | -0.01 | 0.36 | 0.94 | 0.41 | 0.67 | 1.34 | 0.42 |
| hd058898 | S | 4387 | 2.32 | 1.27 | 0.37 | 0.22 | 0.42 | 0.44 | 1.37 | 0.17 | 0.17 | 0.22 | 0.40 | 0.31 | 0.29 | 0.17 | 0.22 | 0.33 | 0.59 | 1.61 | 0.16 | 0.23 | -0.31 | 0.03 | 0.76 | 0.24 | 0.58 | 1.20 | 0.23 |
| hd061191 | S | 4688 | 2.82 | 1.01 | 0.35 | 0.05 | 0.36 | 0.36 | 0.93 | 0.20 | 0.14 | 0.25 | 0.37 | 0.30 | 0.24 | 0.17 | 0.16 | 0.29 | 0.60 | 0.94 | 0.10 | 0.24 | -0.11 | 0.20 | 0.64 | 0.22 | 0.45 | 0.94 | 0.27 |
| hd068667 | S | 5010 | 2.76 | 1.33 | 0.25 | 0.05 | 0.18 | 0.18 | 0.62 | 0.14 | 0.02 | 0.03 | 0.02 | 0.11 | -0.01 | 0.03 | -0.03 | 0.03 | 0.01 | 0.10 | 0.34 | 0.07 | -0.16 | 0.25 | 0.28 | 0.20 | 0.23 | 0.25 | 0.24 |
| hd070522 | S | 6122 | 3.77 | 1.94 | 0.14 | 0.03 | 0.18 | 0.16 | 0.38 | 0.15 | 0.31 | 0.17 | 0.13 | 0.16 | 0.01 | 0.02 | 0.25 | 0.06 | -0.16 | 0.11 | | 0.17 | 0.27 | -0.01 | 0.06 | -0.10 | 0.10 | 0.24 | 0.19 |
| hd077232 | S | 6773 | 3.78 | 6.50 | -0.36 | -0.12 | | 0.05 | | -1.88 | 0.32 | 0.83 | 1.07 | 0.88 | -0.14 | 0.00 | 1.15 | 0.36 | | | 1.55 | 2.63 | | 0.81 | 0.68 | 0.53 | | 0.53 |
| hd083087 | S | 4777 | 3.06 | 0.82 | 0.19 | -0.01 | 0.25 | 0.26 | 0.71 | 0.13 | 0.07 | 0.18 | 0.27 | 0.20 | 0.17 | 0.12 | 0.08 | 0.23 | 0.44 | 0.94 | 0.13 | 0.24 | -0.31 | 0.18 | 0.54 | 0.22 | 0.46 | 0.75 | 0.25 |
| hd085440 | S | 5041 | 3.39 | 1.01 | -0.06 | 0.04 | 0.05 | 0.08 | 0.26 | 0.00 | -0.06 | -0.04 | -0.08 | 0.00 | -0.09 | -0.08 | -0.10 | -0.04 | 0.03 | 0.02 | 0.11 | 0.06 | -0.28 | 0.16 | 0.27 | 0.25 | 0.25 | 0.71 | 0.02 |
| hd089280 | S | 7384 | 4.18 | 5.28 | -0.25 | | 0.34 | -0.03 | -0.66 | -0.53 | 0.05 | 0.15 | 1.07 | 0.66 | 0.47 | -0.85 | 1.08 | 0.15 | | | -0.05 | 0.48 | | 0.34 | 0.60 | 1.09 | 0.85 | | |
| hd090250 | S | 4691 | 2.63 | 1.33 | 0.22 | 0.20 | 0.35 | 0.37 | 0.90 | 0.14 | 0.12 | 0.14 | 0.24 | 0.18 | 0.23 | 0.11 | 0.12 | 0.22 | 0.47 | 0.90 | 0.17 | 0.04 | -0.26 | 0.13 | 0.50 | 0.16 | 0.37 | 0.59 | 0.19 |
| hd098579 | S | 4665 | 2.89 | 1.05 | 0.38 | 0.33 | 0.46 | 0.44 | 1.01 | 0.29 | 0.33 | 0.37 | 0.59 | 0.36 | 0.40 | 0.28 | 0.32 | 0.44 | 0.87 | 1.35 | 0.20 | 0.39 | -0.14 | 0.27 | 1.03 | 0.45 | 0.80 | 1.29 | 0.44 |
| hd101321 | S | 4757 | 2.88 | 1.08 | -0.03 | -0.05 | 0.12 | 0.11 | 0.47 | -0.04 | -0.06 | -0.04 | -0.04 | -0.05 | -0.12 | -0.11 | -0.08 | -0.01 | 0.18 | 0.27 | -0.01 | -0.06 | -0.42 | -0.10 | 0.18 | 0.05 | 0.22 | 0.47 | 0.02 |
| hd104819 | S | 4618 | 3.10 | 1.20 | 0.82 | 0.62 | 0.72 | 0.78 | 1.61 | 0.51 | 0.51 | 0.56 | 0.84 | 0.64 | 0.61 | 0.48 | 0.64 | 0.69 | 0.92 | 1.93 | 0.40 | 0.68 | 0.18 | 0.18 | 1.09 | 0.69 | 0.92 | 1.96 | 0.72 |
| hd104883 | S | 6203 | 3.60 | 4.19 | | -0.75 | 0.24 | 0.31 | -0.10 | -0.13 | 0.13 | 0.10 | 0.80 | 0.70 | -0.19 | -0.12 | 0.12 | 0.31 | | | 1.88 | | -0.49 | 0.16 | 1.17 | 0.30 | 1.89 | 1.13 | | |
| hd106972 | S | 6166 | 3.71 | 2.01 | 0.14 | 0.03 | 0.14 | 0.10 | 0.22 | 0.15 | 0.32 | 0.15 | 0.05 | 0.07 | -0.05 | -0.01 | 0.01 | 0.02 | -0.07 | 0.09 | | 0.15 | 0.50 | 0.21 | 0.22 | 0.21 | 0.14 | 0.13 | 0.24 | |
| hd107415 | S | 4797 | 2.69 | 1.27 | 0.02 | -0.09 | 0.06 | 0.06 | 0.51 | -0.09 | -0.09 | -0.10 | -0.02 | -0.07 | -0.18 | -0.13 | -0.11 | -0.10 | -0.03 | 0.21 | 0.11 | -0.02 | -0.41 | 0.14 | 0.31 | 0.15 | 0.25 | 0.20 | 0.05 |
| hd107569 | S | 6101 | 3.53 | 2.91 | 0.56 | -0.76 | 0.36 | 0.30 | 0.26 | 0.15 | 0.12 | 0.29 | 0.25 | 0.19 | 0.04 | 0.07 | 0.32 | 0.17 | -0.33 | 0.72 | 0.65 | 0.06 | 0.42 | -0.11 | 0.42 | 0.30 | 0.33 | 0.73 | 1.56 |
| hd107610 | S | 4693 | 2.74 | 1.18 | 0.59 | 0.48 | 0.58 | 0.51 | 1.19 | 0.32 | 0.31 | 0.41 | 0.59 | 0.44 | 0.43 | 0.31 | 0.36 | 0.47 | 0.72 | 1.57 | 0.30 | 0.41 | -0.02 | 0.23 | 1.05 | 0.34 | 0.57 | 1.18 | 0.42 |

| ID | | | | | | | | | | | | | | | | | | | | | | | | | | | | |
|---|---|---|---|---|---|---|---|---|---|---|---|---|---|---|---|---|---|---|---|---|---|---|---|---|---|---|---|---|
| hd112357 | S | 4969 | 3.38 | 0.92 | -0.10 | -0.13 | -0.02 | 0.03 | 0.28 | -0.08 | -0.14 | -0.11 | -0.16 | -0.09 | -0.11 | -0.14 | -0.15 | -0.08 | 0.03 | 0.07 | 0.13 | -0.04 | -0.33 | -0.05 | 0.11 | 0.11 | 0.15 | 0.51 | 0.05 |
| hd116204 | S | 4472 | 2.74 | 2.93 | 0.19 | 0.38 | 0.55 | 0.69 | 1.64 | 0.06 | 0.11 | -0.07 | -0.06 | 0.34 | -0.05 | 0.03 | 0.29 | 0.18 | | | 0.18 | -0.47 | -0.52 | 0.34 | 0.88 | 0.40 | 0.38 | 0.10 |
| hd126265 | S | 5868 | 3.84 | 1.35 | 0.11 | 0.01 | 0.11 | 0.11 | 0.21 | 0.09 | 0.17 | 0.04 | -0.06 | 0.06 | -0.08 | 0.00 | 0.03 | 0.00 | -0.22 | 0.04 | 0.45 | 0.16 | 0.12 | 0.10 | -0.10 | 0.31 | 0.18 | 0.15 | 0.08 |
| hd127740 | S | 6241 | 3.73 | 3.40 | 0.14 | -0.39 | -0.02 | 0.04 | 0.67 | 0.15 | -0.17 | 0.31 | 0.43 | 0.30 | -0.09 | -0.15 | 0.66 | -0.13 | -0.65 | 0.01 | 1.58 | 0.97 | 1.00 | -0.41 | 0.69 | 0.51 | 0.38 | 0.30 | 0.75 |
| hd128853 | S | 4819 | 3.06 | 0.97 | -0.06 | -0.14 | -0.01 | 0.12 | 0.52 | -0.07 | -0.11 | -0.13 | -0.15 | -0.07 | -0.18 | -0.12 | -0.16 | -0.05 | 0.02 | 0.19 | 0.12 | -0.06 | -0.46 | 0.09 | 0.17 | 0.16 | 0.25 | 0.44 | -0.01 |
| hd133194 | S | 6312 | 3.47 | 3.64 | 0.11 | -0.12 | 0.08 | 0.16 | 0.44 | 0.11 | 0.33 | 0.09 | 0.14 | 0.12 | -0.07 | -0.04 | 0.29 | -0.02 | -0.60 | 0.07 | | | 0.14 | 0.55 | -0.23 | | 0.51 | -0.23 | 0.37 | 0.77 |
| hd138085 | S | 4799 | 2.59 | 1.35 | -0.25 | -0.24 | -0.12 | -0.09 | 0.28 | -0.28 | -0.31 | -0.35 | -0.37 | -0.32 | -0.50 | -0.40 | -0.38 | -0.34 | -0.34 | 0.02 | 0.04 | -0.35 | -0.63 | -0.23 | -0.17 | -0.17 | -0.13 | -0.10 | -0.21 |
| hd138686 | S | 6257 | 3.25 | 6.50 | | | 0.17 | | 1.31 | 1.31 | 0.84 | 0.18 | 0.74 | -0.39 | -0.20 | 1.20 | 0.36 | | | 0.96 | | -0.70 | -0.07 | | | 1.31 | | | -0.27 |
| hd148317 | S | 5758 | 3.56 | 1.43 | 0.25 | 0.20 | 0.22 | 0.20 | 0.31 | 0.18 | 0.22 | 0.09 | 0.05 | 0.16 | 0.10 | 0.10 | 0.06 | 0.12 | 0.04 | 0.04 | 0.68 | 0.19 | 0.14 | 0.18 | 0.02 | 0.17 | 0.08 | -0.06 | 0.42 |
| hd149216 | S | 4295 | 2.65 | 0.68 | -0.29 | -0.07 | -0.27 | 0.56 | 1.22 | -0.25 | -0.10 | -0.49 | -0.59 | -0.23 | -0.42 | 0.01 | -0.21 | 0.18 | 0.30 | 0.67 | -0.11 | 0.18 | -1.01 | 0.31 | 0.30 | 0.31 | 0.27 | 0.37 | 0.57 | -0.06 |
| hd157935 | S | 6768 | 3.47 | 2.53 | -0.06 | 0.00 | 0.00 | -0.05 | 0.14 | 0.19 | 0.24 | 0.22 | 0.62 | 0.51 | -0.42 | -0.21 | 0.77 | -0.11 | | -0.18 | | 0.32 | 1.56 | 0.05 | | | 0.77 | 0.25 | -0.08 |
| hd161502 | S | 4963 | 2.99 | 1.30 | -0.05 | -0.16 | -0.09 | -0.08 | 0.18 | -0.15 | -0.29 | -0.28 | -0.32 | -0.18 | -0.31 | -0.30 | -0.36 | -0.31 | -0.39 | -0.06 | -0.02 | -0.20 | -0.51 | -0.06 | -0.02 | -0.14 | -0.23 | 0.00 | -0.18 |
| hd167576 | S | 4506 | 2.56 | 1.58 | 0.57 | 0.55 | 0.62 | 0.73 | 1.86 | 0.45 | 0.39 | 0.47 | 0.72 | 0.55 | 0.48 | 0.42 | 0.55 | 0.61 | 0.57 | 2.16 | 0.33 | 0.42 | -0.04 | 0.14 | 1.29 | 0.63 | 0.76 | 1.40 | 0.55 |
| hd172052 | S | 5544 | 4.79 | 1.47 | -0.14 | -0.38 | -0.11 | 0.06 | 1.33 | -0.32 | 1.08 | -0.21 | -0.25 | -0.22 | -0.52 | -0.13 | -0.33 | -0.41 | | 0.14 | | 0.55 | -0.09 | | 0.96 | 0.64 | 0.67 | 0.64 | 0.98 |
| hd173378 | S | 4855 | 2.89 | 1.11 | -0.12 | -0.22 | -0.04 | -0.07 | 0.29 | -0.17 | -0.26 | -0.18 | -0.19 | -0.19 | -0.27 | -0.28 | -0.27 | -0.21 | -0.18 | -0.08 | 0.11 | -0.28 | -0.57 | -0.21 | -0.08 | -0.16 | -0.06 | 0.06 | -0.14 |
| hd175940 | S | 4629 | 2.76 | 1.06 | 0.32 | 0.19 | 0.37 | 0.40 | 0.86 | 0.21 | 0.17 | 0.26 | 0.45 | 0.30 | 0.31 | 0.16 | 0.20 | 0.31 | 0.66 | 0.86 | 0.37 | 0.29 | -0.25 | 0.11 | 0.54 | 0.14 | 0.55 | 0.95 | 0.30 |
| hd182901 | S | 6491 | 4.01 | 2.33 | 0.42 | 0.19 | 0.31 | 0.28 | 0.32 | 0.27 | 0.37 | 0.45 | 0.39 | 0.24 | 0.09 | 0.17 | 0.49 | 0.18 | -0.26 | | | 0.33 | 0.23 | 0.05 | 0.67 | 0.05 | 0.60 | 0.34 | 1.40 |
| hd186535 | S | 5012 | 2.74 | 1.34 | 0.26 | 0.01 | 0.20 | 0.17 | 0.55 | 0.12 | -0.02 | 0.06 | 0.02 | 0.15 | 0.09 | 0.03 | -0.02 | 0.04 | 0.00 | 0.21 | 0.38 | 0.08 | -0.14 | 0.18 | 0.27 | 0.18 | 0.27 | 0.34 | 0.25 |
| hd188993 | S | 5673 | 3.45 | 1.59 | 0.10 | 0.00 | 0.15 | 0.13 | 0.28 | 0.16 | 0.13 | 0.04 | -0.08 | 0.10 | -0.03 | 0.00 | -0.02 | 0.02 | -0.15 | 0.02 | 0.60 | 0.07 | 0.04 | 0.11 | -0.08 | 0.07 | 0.18 | -0.17 | 0.22 |
| hd189186 | S | 4899 | 3.09 | 0.91 | -0.27 | -0.18 | -0.20 | -0.12 | 0.06 | -0.25 | -0.29 | -0.30 | -0.35 | -0.32 | -0.44 | -0.37 | -0.39 | -0.30 | -0.23 | 0.01 | -0.06 | -0.34 | -0.69 | -0.17 | -0.18 | -0.17 | -0.10 | -0.24 | -0.18 |
| hd194708 | S | 6313 | 3.67 | 3.00 | | | 0.55 | -0.51 | 0.05 | 0.82 | 0.36 | 0.37 | 0.67 | 0.03 | 0.24 | 0.58 | 0.29 | -0.33 | | | 0.10 | 1.06 | | 0.51 | 0.82 | 0.57 | | | | |
| hd200925 | S | 6642 | 3.40 | 3.10 | -0.11 | -0.08 | 0.10 | 0.02 | 0.25 | 0.14 | 0.42 | 0.21 | 0.23 | 0.13 | -0.24 | -0.08 | 0.33 | -0.02 | -0.27 | 0.13 | | 0.55 | 0.89 | 0.32 | 0.52 | 0.36 | 0.26 | 0.35 | 0.47 |
| hd204642 | S | 4660 | 2.84 | 1.09 | 0.30 | 0.31 | 0.38 | 0.40 | 0.97 | 0.23 | 0.22 | 0.23 | 0.40 | 0.30 | 0.27 | 0.21 | 0.25 | 0.34 | 0.70 | 1.29 | 0.19 | 0.30 | -0.20 | 0.20 | 0.74 | 0.24 | 0.59 | 1.13 | 0.43 |
| hd205011 | S | 4770 | 2.54 | 1.43 | 0.11 | 0.13 | 0.19 | 0.33 | 0.98 | -0.01 | -0.05 | -0.03 | -0.02 | 0.06 | -0.08 | -0.02 | -0.06 | 0.06 | 0.43 | 0.79 | 0.82 | 0.61 | 0.41 | 0.85 | 1.01 | 0.70 | 0.75 | 0.60 | 0.24 |
| hd205972 | S | 4742 | 2.89 | 1.02 | 0.28 | 0.16 | 0.32 | 0.32 | 0.92 | 0.19 | 0.13 | 0.21 | 0.36 | 0.26 | 0.21 | 0.13 | 0.15 | 0.25 | 0.43 | 0.78 | 0.15 | 0.12 | -0.22 | 0.05 | 0.63 | 0.13 | 0.42 | 0.82 | 0.09 |
| hd211607 | S | 4944 | 2.98 | 1.19 | 0.36 | 0.28 | 0.30 | 0.34 | 0.70 | 0.25 | 0.19 | 0.21 | 0.29 | 0.29 | 0.25 | 0.21 | 0.16 | 0.27 | 0.32 | 0.99 | 0.35 | 0.39 | -0.03 | 0.36 | 0.57 | 0.36 | 0.47 | 0.70 | 0.39 |
| hd212334 | S | 4722 | 2.49 | 1.38 | 0.02 | 0.00 | 0.14 | 0.18 | 0.52 | 0.00 | -0.07 | -0.01 | 0.00 | -0.02 | -0.08 | -0.05 | -0.02 | 0.02 | 0.18 | 0.28 | 0.07 | -0.15 | -0.42 | -0.03 | 0.17 | 0.02 | 0.15 | 0.41 | 0.11 |
| hd213619 | S | 6884 | 3.78 | 5.80 | -1.28 | -0.56 | 0.35 | -0.17 | -0.62 | -1.07 | 0.36 | 0.53 | 1.49 | 0.23 | | 0.07 | 1.14 | -0.07 | 0.55 | | 0.76 | 0.97 | | 0.76 | 1.40 | 0.79 | 0.87 | 1.09 | | |
| hd217590 | S | 4950 | 2.84 | 1.27 | 0.32 | 0.16 | 0.30 | 0.27 | 0.60 | 0.20 | 0.11 | 0.15 | 0.16 | 0.22 | 0.18 | 0.13 | 0.06 | 0.16 | 0.24 | 0.46 | 0.30 | 0.31 | -0.19 | 0.25 | 0.47 | 0.25 | 0.38 | 0.54 | 0.24 |
| hd219409 | S | 4634 | 2.73 | 1.08 | 0.11 | 0.18 | 0.19 | 0.26 | 0.75 | 0.02 | 0.04 | 0.06 | 0.16 | 0.12 | 0.10 | 0.04 | 0.06 | 0.16 | 0.47 | 0.71 | 0.12 | 0.05 | -0.39 | 0.05 | 0.58 | 0.12 | 0.38 | 0.77 | 0.07 |
| hd222683 | S | 4936 | 2.85 | 1.23 | 0.36 | 0.11 | 0.28 | 0.31 | 0.65 | 0.21 | 0.10 | 0.16 | 0.20 | 0.23 | 0.18 | 0.14 | 0.09 | 0.17 | 0.21 | 0.38 | 0.30 | 0.28 | -0.15 | 0.21 | 0.40 | 0.19 | 0.26 | 0.43 | 0.23 |
| hd223869 | S | 4830 | 3.24 | 0.68 | 0.16 | 0.22 | 0.25 | 0.24 | 0.63 | 0.16 | 0.07 | 0.16 | 0.29 | 0.17 | 0.13 | 0.10 | 0.06 | 0.24 | 0.40 | 0.34 | -0.01 | 0.23 | -0.27 | 0.10 | 0.29 | 0.22 | 0.48 | 1.04 | 0.23 |
| hd224349 | S | 4830 | 2.57 | 1.37 | -0.07 | -0.08 | 0.03 | 0.08 | 0.45 | -0.07 | -0.11 | -0.12 | -0.15 | -0.09 | -0.17 | -0.14 | -0.16 | -0.12 | -0.11 | -0.02 | 0.09 | -0.09 | -0.38 | 0.10 | 0.16 | 0.01 | 0.14 | 0.17 | 0.05 |
| hd225292 | S | 4940 | 2.63 | 1.36 | 0.13 | -0.02 | 0.14 | 0.10 | 0.40 | -0.02 | -0.09 | -0.05 | -0.04 | -0.01 | -0.10 | -0.10 | -0.09 | -0.06 | -0.01 | 0.07 | 0.12 | -0.06 | -0.29 | 0.07 | 0.16 | -0.02 | 0.10 | 0.07 | 0.12 |
| hr0002 | S | 4648 | 2.42 | 1.39 | 0.42 | 0.32 | 0.44 | 0.53 | 1.27 | 0.25 | 0.20 | 0.18 | 0.29 | 0.28 | 0.26 | 0.23 | 0.21 | 0.40 | 0.60 | 1.14 | 0.25 | 0.15 | -0.18 | 0.24 | 0.82 | 0.23 | 0.43 | 0.97 | 0.32 |
| hr0004 | S | 5104 | 2.70 | 1.42 | 0.24 | 0.00 | 0.18 | 0.13 | 0.46 | 0.10 | 0.00 | 0.02 | 0.00 | 0.08 | -0.06 | 0.00 | -0.05 | -0.01 | -0.10 | 0.04 | 0.36 | 0.11 | -0.03 | 0.28 | 0.28 | 0.18 | 0.23 | 0.16 | 0.16 |
| hr0016 | S | 4738 | 2.65 | 1.29 | 0.04 | 0.05 | 0.16 | 0.21 | 0.70 | 0.04 | -0.02 | 0.00 | -0.04 | 0.02 | -0.02 | -0.03 | -0.03 | 0.05 | 0.20 | 0.54 | 0.18 | 0.16 | -0.31 | 0.17 | 0.39 | 0.17 | 0.36 | 0.42 | 0.23 |
| hr0019 | S | 4898 | 2.64 | 1.38 | 0.01 | -0.02 | 0.10 | 0.07 | 0.42 | -0.02 | -0.06 | -0.02 | -0.05 | -0.04 | -0.11 | -0.10 | -0.07 | -0.06 | -0.05 | -0.04 | 0.29 | 0.00 | -0.25 | 0.13 | 0.20 | 0.10 | 0.22 | 0.25 | 0.12 |
| hr0022 | S | 4774 | 2.59 | 1.32 | 0.32 | 0.17 | 0.37 | 0.33 | 0.86 | 0.21 | 0.12 | 0.16 | 0.21 | 0.22 | 0.17 | 0.12 | 0.08 | 0.18 | 0.41 | 0.56 | 0.37 | 0.21 | -0.19 | 0.21 | 0.46 | 0.18 | 0.32 | 0.53 | 0.25 |
| hr0040 | S | 5642 | 3.07 | 0.64 | 0.24 | 0.06 | 0.30 | 0.03 | -0.11 | 0.25 | 0.12 | 0.36 | 0.40 | 0.25 | 0.18 | 0.21 | 0.22 | 0.19 | -0.02 | 0.10 | 0.79 | 0.26 | 0.29 | 0.40 | 0.42 | 0.36 | 0.38 | 0.32 | 0.46 |
| hr0059 | S | 5061 | 2.87 | 1.35 | 0.32 | 0.13 | 0.33 | 0.23 | 0.65 | 0.22 | 0.12 | 0.15 | 0.15 | 0.21 | 0.18 | 0.11 | 0.09 | 0.14 | 0.16 | 0.36 | 0.40 | 0.26 | -0.08 | 0.25 | 0.41 | 0.24 | 0.33 | 0.38 | 0.24 |
| hr0069 | S | 4888 | 2.51 | 1.42 | 0.40 | 0.21 | 0.36 | 0.34 | 0.70 | 0.27 | 0.15 | 0.26 | 0.37 | 0.30 | 0.26 | 0.18 | 0.20 | 0.23 | 0.33 | 0.60 | 0.39 | 0.26 | 0.02 | 0.23 | 0.54 | 0.21 | 0.41 | 0.68 | 0.27 |
| hr0074 | S | 4479 | 1.87 | 1.71 | 0.49 | 0.26 | 0.33 | 0.46 | 1.08 | 0.20 | 0.02 | 0.12 | 0.22 | 0.29 | 0.20 | 0.12 | 0.13 | 0.23 | 0.27 | 1.51 | 0.29 | 0.22 | -0.28 | 0.04 | 0.68 | 0.19 | 0.22 | 0.64 | 0.17 |
| hr0084 | S | 4888 | 2.58 | 1.51 | 0.59 | 0.27 | 0.37 | 0.43 | 0.90 | 0.30 | 0.19 | 0.19 | 0.29 | 0.31 | 0.31 | 0.19 | 0.18 | 0.28 | 0.33 | 1.05 | 0.40 | 0.22 | -0.09 | 0.29 | 0.72 | 0.28 | 0.31 | 0.59 | 0.39 |
| hr0101 | S | 5001 | 2.71 | 1.36 | -0.18 | -0.27 | -0.09 | -0.15 | 0.07 | -0.19 | -0.23 | -0.22 | -0.26 | -0.26 | -0.37 | -0.32 | -0.27 | -0.30 | -0.24 | -0.32 | 0.21 | -0.32 | -0.48 | -0.17 | -0.12 | -0.15 | -0.06 | -0.23 | -0.10 |
| hr0131 | S | 4701 | 2.59 | 1.29 | 0.25 | 0.15 | 0.31 | 0.43 | 1.02 | 0.16 | 0.10 | 0.04 | 0.18 | 0.18 | 0.15 | 0.14 | 0.13 | 0.21 | | 0.81 | | 0.13 | -0.44 | 0.20 | 0.55 | 0.22 | 0.34 | 0.45 | 0.24 |
| hr0135 | S | 4792 | 2.53 | 1.41 | 0.18 | 0.13 | 0.26 | 0.22 | 0.61 | 0.10 | 0.03 | 0.07 | 0.11 | 0.10 | 0.11 | 0.01 | 0.03 | 0.08 | 0.28 | 0.48 | 0.20 | -0.01 | -0.22 | -0.01 | 0.31 | 0.08 | 0.21 | 0.44 | 0.13 |
| hr0141 | S | 4646 | 2.63 | 1.23 | 0.24 | 0.12 | 0.31 | 0.26 | 0.92 | 0.15 | 0.18 | 0.22 | 0.38 | 0.23 | 0.18 | 0.09 | 0.13 | 0.21 | 0.53 | 0.75 | 0.32 | 0.30 | -0.12 | 0.16 | 0.61 | 0.22 | 0.42 | 0.81 | 0.34 |
| hr0156 | S | 4608 | 2.56 | 1.47 | 0.71 | 0.56 | 0.55 | 0.65 | 1.39 | 0.40 | 0.38 | 0.44 | 0.67 | 0.54 | 0.56 | 0.40 | 0.46 | 0.57 | 0.68 | 1.37 | 0.37 | 0.45 | -0.02 | 0.20 | 1.10 | 0.48 | 0.66 | 1.10 | 0.58 |
| hr0163 | S | 4870 | 2.66 | 1.43 | -0.56 | -0.24 | -0.28 | -0.18 | 0.06 | -0.46 | -0.37 | -0.46 | -0.63 | -0.57 | -0.85 | -0.60 | -0.54 | -0.55 | -0.56 | -0.11 | -0.01 | -0.55 | -0.54 | -0.44 | -0.40 | -0.42 | -0.33 | -0.39 | -0.13 |

| ID | | col3 | col4 | col5 | col6 | col7 | col8 | col9 | col10 | col11 | col12 | col13 | col14 | col15 | col16 | col17 | col18 | col19 | col20 | col21 | col22 | col23 | col24 | col25 | col26 | col27 | col28 | col29 | col30 |
|---|---|---|---|---|---|---|---|---|---|---|---|---|---|---|---|---|---|---|---|---|---|---|---|---|---|---|---|---|---|
| hr0165 | S | 4330 | 2.18 | 1.37 | 0.66 | 0.49 | 0.70 | 0.71 | 1.66 | 0.34 | 0.34 | 0.27 | 0.68 | 0.44 | 0.49 | 0.37 | 0.42 | 0.56 | | 2.01 | | 0.18 | -0.45 | 0.18 | 0.48 | 0.45 | 0.63 | 0.53 | 0.63 |
| hr0168 | S | 4555 | 1.65 | 1.99 | 0.37 | 0.00 | 0.14 | 0.39 | 0.97 | -0.04 | -0.12 | -0.08 | -0.08 | 0.12 | -0.14 | -0.01 | -0.06 | 0.03 | 0.01 | 0.85 | 0.28 | 0.16 | -0.34 | 0.18 | 0.41 | 0.07 | 0.17 | 0.31 | 0.14 |
| hr0175 | S | 5020 | 2.58 | 1.44 | 0.15 | 0.00 | 0.06 | 0.08 | 0.40 | 0.04 | -0.08 | -0.09 | -0.15 | 0.03 | -0.17 | -0.07 | -0.14 | -0.08 | -0.12 | 0.11 | 0.27 | 0.02 | -0.18 | 0.22 | 0.22 | 0.16 | 0.19 | 0.02 | 0.06 |
| hr0188 | S | 4792 | 2.29 | 1.59 | 0.59 | 0.17 | 0.23 | 0.27 | 0.72 | 0.16 | -0.06 | 0.03 | 0.06 | 0.20 | 0.02 | 0.03 | 0.03 | 0.08 | -0.13 | 0.64 | 0.25 | 0.15 | -0.22 | 0.13 | 0.26 | -0.03 | 0.07 | 0.14 | 0.12 |
| hr0213 | S | 4871 | 2.72 | 1.33 | 0.33 | 0.12 | 0.20 | 0.28 | 0.82 | 0.20 | 0.06 | 0.09 | 0.11 | 0.21 | 0.11 | 0.08 | 0.06 | 0.13 | 0.19 | 0.47 | 0.31 | 0.22 | -0.10 | 0.24 | 0.36 | 0.20 | 0.28 | 0.34 | 0.19 |
| hr0215 | S | 4584 | 2.28 | 2.81 | | | 0.22 | 0.49 | 0.50 | -0.67 | 0.16 | -0.08 | 0.00 | 0.48 | -0.45 | 0.09 | 0.05 | -0.10 | | | 0.04 | -1.86 | | 0.71 | -0.04 | -0.02 | | | 0.10 |
| hr0216 | S | 5014 | 2.65 | 1.28 | 0.36 | 0.19 | 0.33 | 0.25 | 0.51 | 0.27 | -0.02 | 0.11 | 0.11 | 0.20 | 0.20 | 0.12 | 0.07 | 0.13 | | 0.07 | | 0.28 | -0.22 | 0.26 | 0.37 | 0.17 | 0.29 | 0.21 | 0.22 |
| hr0224 | S | 3955 | 1.48 | 1.58 | 0.28 | 0.25 | 0.41 | 0.64 | 1.99 | 0.03 | 0.14 | 0.16 | 0.44 | 0.24 | 0.17 | 0.15 | 0.29 | 0.30 | 1.04 | 2.69 | 0.60 | 0.23 | -0.27 | 0.16 | 0.39 | 0.65 | 0.72 | 1.22 | 0.02 |
| hr0228 | S | 4898 | 3.07 | 0.98 | 0.17 | 0.22 | 0.22 | 0.21 | 0.61 | 0.15 | 0.05 | 0.12 | 0.18 | 0.17 | 0.14 | 0.06 | 0.04 | 0.14 | 0.25 | 0.36 | 0.18 | 0.19 | -0.17 | 0.07 | 0.59 | 0.22 | 0.35 | 0.62 | 0.11 |
| hr0230 | S | 6671 | 3.01 | 2.20 | | | 0.10 | 0.08 | | -0.58 | 0.51 | 0.68 | 1.03 | | 0.36 | 0.99 | 1.04 | | | 1.30 | 1.97 | | 1.12 | | | 0.28 | | | | |
| hr0231 | S | 6732 | 3.03 | 6.50 | | | -0.22 | 0.67 | -0.68 | | 1.00 | 0.65 | 0.89 | | 0.68 | 1.03 | -0.80 | | | | | | 0.85 | -0.20 | | | | | | |
| hr0249 | S | 4612 | 2.51 | 1.42 | 0.42 | 0.30 | 0.41 | 0.45 | 1.25 | 0.18 | 0.23 | 0.22 | 0.37 | 0.31 | 0.31 | 0.23 | 0.27 | 0.40 | 0.50 | 0.81 | 0.33 | 0.31 | -0.10 | 0.25 | 0.78 | 0.37 | 0.49 | 0.92 | 0.48 |
| hr0255 | S | 4842 | 2.53 | 1.43 | -0.48 | -0.34 | -0.28 | -0.20 | 0.04 | -0.41 | -0.36 | -0.44 | -0.53 | -0.48 | -0.70 | -0.50 | -0.46 | -0.48 | -0.46 | -0.27 | 0.08 | -0.52 | -0.78 | -0.37 | -0.37 | -0.47 | -0.30 | -0.47 | -0.19 |
| hr0265 | S | 4829 | 2.54 | 1.29 | -0.25 | -0.20 | -0.11 | -0.07 | 0.23 | -0.18 | -0.23 | -0.26 | -0.29 | -0.27 | -0.33 | -0.30 | -0.29 | -0.29 | | -0.03 | | -0.23 | -0.76 | -0.23 | -0.16 | -0.41 | -0.24 | -0.10 | -0.12 |
| hr0279 | S | 4841 | 2.54 | 1.36 | -0.09 | -0.06 | -0.03 | 0.03 | 0.21 | -0.12 | -0.15 | -0.18 | -0.24 | -0.15 | -0.32 | -0.20 | -0.22 | -0.17 | -0.12 | 0.03 | 0.28 | -0.15 | -0.46 | 0.10 | 0.21 | 0.02 | 0.09 | -0.02 | 0.01 |
| hr0294 | S | 4814 | 2.48 | 1.34 | -0.10 | -0.21 | -0.05 | -0.02 | 0.19 | -0.12 | -0.27 | -0.22 | -0.25 | -0.19 | -0.37 | -0.27 | -0.28 | -0.24 | -0.13 | -0.24 | 0.19 | -0.27 | -0.38 | -0.10 | 0.04 | -0.06 | -0.09 | -0.18 | -0.08 |
| hr0315 | S | 4753 | 2.76 | 1.16 | 0.15 | 0.07 | 0.21 | 0.28 | 0.73 | 0.11 | 0.02 | 0.07 | 0.14 | 0.15 | 0.10 | 0.04 | 0.01 | 0.12 | 0.31 | 0.34 | 0.16 | 0.14 | -0.26 | 0.14 | 0.41 | 0.17 | 0.34 | 0.53 | 0.16 |
| hr0320 | S | 4682 | 2.52 | 1.39 | 0.40 | 0.13 | 0.30 | 0.36 | 0.94 | 0.24 | 0.12 | 0.17 | 0.28 | 0.21 | 0.19 | 0.11 | 0.15 | 0.23 | 0.34 | 0.99 | 0.26 | 0.26 | -0.19 | 0.12 | 0.41 | 0.16 | 0.29 | 0.54 | 0.14 |
| hr0325 | S | 7461 | 4.06 | 2.86 | | | 0.12 | 0.46 | 0.62 | 1.10 | 1.00 | 0.82 | 0.68 | 1.54 | 0.51 | 1.59 | 0.60 | 0.35 | | 1.46 | 1.38 | 2.46 | | 0.97 | 1.17 | 0.99 | 1.15 | 1.61 |
| hr0334 | S | 4543 | 2.39 | 1.38 | 0.45 | 0.32 | 0.39 | 0.46 | 1.19 | 0.26 | 0.17 | 0.22 | 0.37 | 0.33 | 0.27 | 0.22 | 0.24 | 0.37 | 0.47 | 1.03 | 0.29 | 0.36 | -0.15 | 0.21 | 0.72 | 0.32 | 0.48 | 0.92 | 0.34 |
| hr0352 | S | 4658 | 2.58 | 1.30 | 0.24 | 0.22 | 0.34 | 0.36 | 0.87 | 0.17 | 0.06 | 0.13 | 0.27 | 0.20 | 0.16 | 0.12 | 0.11 | 0.25 | 0.58 | 1.12 | 0.12 | 0.09 | -0.33 | 0.11 | 0.53 | 0.17 | 0.36 | 0.79 | 0.12 |
| hr0356 | S | 4884 | 2.74 | 1.39 | -0.11 | -0.03 | 0.11 | 0.09 | 0.38 | -0.05 | -0.01 | -0.01 | -0.05 | -0.08 | -0.11 | -0.12 | -0.09 | -0.07 | 0.12 | 0.23 | 0.23 | -0.16 | -0.48 | -0.04 | 0.09 | 0.00 | 0.11 | 0.24 | 0.06 |
| hr0367 | S | 4732 | 2.35 | 1.44 | 0.48 | 0.24 | 0.42 | 0.38 | 0.97 | 0.23 | 0.09 | 0.17 | 0.31 | 0.25 | 0.30 | 0.14 | 0.14 | 0.26 | 0.33 | 0.96 | 0.31 | 0.15 | -0.16 | 0.09 | 0.54 | 0.14 | 0.27 | 0.74 | 0.21 |
| hr0371 | S | 4573 | 2.49 | 1.41 | 0.53 | 0.44 | 0.56 | 0.66 | 1.37 | 0.36 | 0.24 | 0.29 | 0.45 | 0.41 | 0.44 | 0.33 | 0.35 | 0.50 | 0.62 | 1.81 | 0.23 | 0.33 | -0.23 | 0.17 | 0.57 | 0.35 | 0.46 | 1.04 | 0.47 |
| hr0390 | S | 4704 | 2.52 | 1.38 | 0.42 | 0.32 | 0.42 | 0.45 | 1.02 | 0.27 | 0.18 | 0.22 | 0.30 | 0.32 | 0.34 | 0.22 | 0.22 | 0.33 | 0.51 | 0.65 | 0.27 | 0.33 | -0.18 | 0.26 | 0.67 | 0.32 | 0.41 | 0.66 | 0.28 |
| hr0402 | S | 4660 | 2.58 | 1.38 | 0.05 | 0.18 | 0.16 | 0.26 | 0.80 | -0.01 | 0.01 | 0.03 | 0.06 | 0.05 | -0.05 | -0.02 | 0.02 | 0.09 | 0.30 | 0.43 | -0.01 | -0.11 | -0.47 | -0.03 | 0.42 | 0.09 | 0.21 | 0.57 | 0.18 |
| hr0406 | S | 4845 | 2.94 | 1.10 | -0.05 | -0.03 | 0.01 | 0.04 | 0.38 | -0.09 | -0.08 | -0.09 | -0.10 | -0.10 | -0.09 | -0.15 | -0.13 | -0.09 | 0.02 | 0.19 | -0.03 | -0.14 | -0.49 | -0.07 | 0.14 | 0.03 | 0.15 | 0.36 | -0.01 |
| hr0407 | S | 6455 | 3.75 | 3.09 | 0.24 | -0.07 | 0.26 | 0.12 | 0.15 | 0.02 | 0.32 | 0.33 | 0.39 | 0.22 | 0.04 | -0.03 | 0.45 | 0.15 | -0.39 | | 1.71 | 0.71 | 0.20 | -0.48 | | 0.93 | 0.46 | | 1.31 |
| hr0412 | S | 4316 | 1.94 | 1.44 | -0.02 | -0.04 | 0.15 | 0.14 | 0.87 | -0.21 | -0.17 | -0.08 | 0.06 | -0.04 | -0.20 | -0.19 | -0.12 | -0.08 | 0.05 | 0.86 | -0.06 | -0.17 | -0.73 | -0.39 | 0.16 | -0.05 | 0.13 | 0.55 | -0.16 |
| hr0414 | S | 4617 | 2.50 | 1.34 | 0.23 | 0.14 | 0.30 | 0.41 | 1.01 | 0.16 | 0.09 | 0.13 | 0.25 | 0.18 | 0.19 | 0.13 | 0.13 | 0.25 | 0.45 | 1.15 | 0.11 | 0.08 | -0.19 | 0.15 | 0.59 | 0.24 | 0.35 | 0.66 | 0.21 |
| hr0426 | S | 4688 | 2.68 | 1.35 | 0.48 | 0.32 | 0.49 | 0.52 | 1.14 | 0.33 | 0.24 | 0.30 | 0.45 | 0.38 | 0.39 | 0.30 | 0.30 | 0.44 | 0.70 | 1.26 | 0.31 | 0.39 | -0.07 | 0.30 | 0.84 | 0.43 | 0.50 | 0.91 | 0.33 |
| hr0430 | S | 4879 | 2.60 | 1.41 | 0.37 | 0.24 | 0.33 | 0.32 | 0.81 | 0.21 | 0.10 | 0.12 | 0.17 | 0.21 | 0.22 | 0.11 | 0.09 | 0.17 | 0.31 | 0.73 | 0.32 | 0.10 | -0.15 | 0.18 | 0.39 | 0.17 | 0.27 | 0.52 | 0.15 |
| hr0434 | S | 4126 | 1.68 | 1.48 | -0.03 | 0.04 | 0.16 | 0.12 | 1.42 | -0.20 | -0.05 | -0.02 | 0.16 | -0.10 | -0.38 | -0.26 | -0.08 | -0.08 | 0.55 | 0.70 | 0.11 | -0.09 | -0.60 | -0.34 | 0.07 | -0.06 | 0.30 | 0.77 | -0.05 |
| hr0437 | S | 4875 | 2.01 | 1.97 | 0.40 | 0.08 | 0.16 | 0.22 | 0.56 | 0.01 | -0.12 | -0.03 | -0.08 | 0.10 | -0.10 | -0.04 | -0.05 | -0.02 | -0.14 | 0.42 | 0.49 | 0.01 | -0.30 | 0.11 | 0.15 | 0.04 | 0.13 | -0.09 | 0.10 |
| hr0442 | S | 4736 | 2.45 | 1.36 | -0.15 | -0.04 | 0.01 | 0.06 | 0.46 | -0.14 | -0.22 | -0.20 | -0.20 | -0.22 | -0.28 | -0.27 | -0.25 | -0.20 | -0.05 | 0.11 | 0.08 | -0.35 | -0.71 | -0.26 | -0.14 | -0.27 | -0.12 | -0.01 | -0.14 |
| hr0454 | S | 4357 | 2.05 | 1.96 | 0.20 | 0.10 | 0.19 | 0.36 | 1.19 | -0.14 | -0.17 | -0.23 | -0.20 | 0.10 | -0.27 | -0.07 | -0.04 | 0.05 | 0.24 | 1.06 | -0.15 | 0.00 | -0.66 | -0.07 | 0.04 | 0.08 | 0.25 | 1.53 | 0.16 |
| hr0464 | S | 4310 | 1.98 | 1.65 | 0.36 | 0.32 | 0.44 | 0.50 | 1.32 | 0.05 | 0.07 | 0.09 | 0.28 | 0.26 | 0.23 | 0.13 | 0.21 | 0.31 | 0.60 | 1.18 | 0.19 | -0.01 | -0.41 | -0.11 | 0.70 | 0.22 | 0.49 | 1.03 | 0.18 |
| hr0510 | S | 4903 | 2.36 | 1.55 | 0.30 | 0.16 | 0.19 | 0.18 | 0.47 | 0.04 | -0.03 | -0.02 | -0.06 | 0.07 | -0.01 | -0.04 | -0.05 | -0.01 | 0.03 | 0.25 | 0.27 | 0.01 | -0.32 | 0.10 | 0.27 | 0.03 | 0.10 | 0.14 | 0.08 |
| hr0521 | S | 4855 | 2.68 | 1.37 | 0.00 | -0.04 | 0.06 | 0.07 | 0.48 | -0.01 | -0.05 | -0.07 | -0.09 | -0.04 | -0.17 | -0.10 | -0.13 | -0.06 | -0.04 | 0.33 | 0.43 | -0.02 | -0.39 | 0.17 | 0.28 | 0.20 | 0.26 | 0.16 | 0.10 |
| hr0527 | S | 4837 | 2.67 | 1.33 | 0.04 | -0.05 | 0.07 | 0.13 | 0.47 | -0.03 | -0.10 | -0.09 | -0.14 | -0.04 | -0.11 | -0.08 | -0.13 | -0.06 | -0.01 | 0.19 | 0.21 | -0.01 | -0.22 | 0.18 | 0.27 | 0.17 | 0.23 | 0.13 | 0.07 |
| hr0539 | S | 4579 | 2.07 | 1.60 | 0.26 | 0.01 | 0.19 | 0.27 | 0.76 | 0.04 | -0.14 | -0.08 | 0.00 | 0.08 | -0.03 | -0.02 | -0.06 | 0.03 | 0.02 | 0.79 | 0.23 | -0.02 | -0.24 | 0.17 | 0.47 | 0.12 | 0.26 | 0.49 | 0.08 |
| hr0557 | S | 4509 | 2.31 | 1.38 | -0.14 | 0.03 | 0.13 | 0.22 | 0.71 | -0.21 | -0.15 | -0.26 | -0.25 | -0.17 | -0.23 | -0.14 | -0.14 | -0.03 | 0.14 | 0.53 | -0.17 | -0.28 | -0.71 | -0.10 | 0.05 | -0.13 | 0.05 | 0.16 | -0.07 |
| hr0594 | S | 4880 | 2.58 | 1.39 | -0.59 | -0.36 | -0.47 | -0.32 | -0.27 | -0.51 | -0.43 | -0.58 | -0.70 | -0.59 | -0.79 | -0.62 | -0.61 | -0.60 | -0.61 | -0.36 | -0.01 | -0.62 | -0.98 | -0.43 | -0.32 | -0.44 | -0.37 | -0.46 | -0.39 |
| hr0616 | S | 5061 | 2.58 | 1.30 | 0.50 | 0.24 | 0.34 | 0.31 | 0.78 | 0.29 | 0.17 | 0.29 | 0.33 | 0.32 | 0.33 | 0.18 | 0.23 | 0.26 | 0.25 | 0.38 | 0.40 | 0.32 | 0.04 | 0.22 | 0.47 | 0.18 | 0.22 | 0.46 | 0.30 |
| hr0617 | S | 4512 | 2.30 | 1.33 | 0.10 | -0.01 | 0.13 | 0.19 | 0.90 | -0.06 | -0.03 | -0.01 | 0.06 | 0.05 | -0.02 | -0.06 | -0.03 | 0.02 | 0.40 | 0.61 | 0.05 | -0.07 | -0.35 | 0.04 | 0.55 | 0.05 | 0.31 | 0.49 | 0.07 |
| hr0619 | S | 4916 | 2.77 | 1.34 | -0.15 | -0.16 | -0.04 | -0.06 | 0.34 | -0.16 | -0.21 | -0.18 | -0.22 | -0.21 | -0.32 | -0.28 | -0.25 | -0.24 | -0.14 | 0.07 | 0.18 | -0.25 | -0.47 | -0.11 | -0.09 | -0.08 | 0.01 | -0.09 | -0.04 |
| hr0621 | S | 5110 | 2.58 | 1.49 | 0.32 | 0.11 | 0.24 | 0.16 | 0.40 | 0.17 | 0.00 | 0.12 | 0.13 | 0.16 | 0.18 | 0.06 | 0.02 | 0.06 | 0.10 | 0.21 | 0.30 | 0.12 | 0.01 | 0.18 | 0.21 | 0.18 | 0.19 | 0.21 | 0.15 |
| hr0661 | S | 4502 | 2.20 | 1.41 | 0.31 | 0.17 | 0.33 | 0.36 | 0.81 | 0.20 | 0.07 | 0.15 | 0.26 | 0.27 | 0.10 | 0.10 | 0.10 | 0.21 | 0.29 | 0.59 | 0.31 | 0.27 | -0.25 | 0.18 | 0.64 | 0.27 | 0.47 | 0.63 | 0.25 |
| hr0666 | S | 4884 | 2.60 | 1.35 | 0.36 | 0.25 | 0.29 | 0.25 | 0.67 | 0.14 | 0.06 | 0.04 | 0.04 | 0.13 | 0.07 | 0.04 | 0.02 | 0.10 | 0.25 | 0.43 | 0.32 | 0.12 | -0.27 | 0.06 | 0.31 | 0.05 | 0.16 | 0.24 | 0.21 |
| hr0697 | S | 4865 | 2.65 | 1.44 | -0.27 | -0.02 | -0.01 | -0.03 | 0.33 | -0.21 | -0.21 | -0.14 | -0.22 | -0.32 | -0.55 | -0.39 | -0.24 | -0.30 | -0.23 | 0.01 | 0.12 | -0.26 | -0.68 | -0.43 | -0.28 | -0.32 | -0.22 | -0.09 | -0.08 |

| ID | | | | | | | | | | | | | | | | | | | | | | | | | | | | |
|---|---|---|---|---|---|---|---|---|---|---|---|---|---|---|---|---|---|---|---|---|---|---|---|---|---|---|---|---|
| hr0699 | S | 3927 | 1.40 | 1.65 | 0.29 | 0.35 | 0.35 | 0.77 | 2.15 | 0.18 | 0.03 | 0.13 | 0.38 | 0.28 | 0.21 | 0.21 | 0.26 | 0.33 | 0.84 | | 0.63 | 0.19 | -0.17 | 0.33 | 0.65 | 0.53 | 0.79 | 1.43 | 0.10 |
| hr0712 | S | 4572 | 2.44 | 1.29 | 0.16 | 0.09 | 0.22 | 0.24 | 0.89 | 0.07 | 0.05 | 0.08 | 0.17 | 0.13 | 0.09 | 0.01 | 0.13 | 0.14 | 0.34 | 0.68 | 0.17 | 0.10 | -0.36 | 0.03 | 0.46 | 0.10 | 0.40 | 0.67 | 0.10 |
| hr0725 | S | 4780 | 2.66 | 1.25 | 0.21 | 0.12 | 0.26 | 0.20 | 0.58 | 0.13 | 0.01 | 0.12 | 0.18 | 0.16 | 0.11 | 0.03 | 0.04 | 0.10 | 0.30 | 0.34 | 0.33 | 0.04 | -0.21 | 0.05 | 0.39 | 0.18 | 0.30 | 0.66 | 0.11 |
| hr0726 | S | 4723 | 2.52 | 1.35 | 0.35 | 0.27 | 0.37 | 0.38 | 0.91 | 0.22 | 0.07 | 0.17 | 0.24 | 0.27 | 0.25 | 0.15 | 0.14 | 0.25 | 0.44 | 0.68 | 0.29 | 0.15 | -0.16 | 0.10 | 0.51 | 0.17 | 0.29 | 0.50 | 0.13 |
| hr0738 | S | 4622 | 2.61 | 1.15 | 0.16 | 0.22 | 0.23 | 0.27 | 0.84 | 0.07 | 0.00 | 0.08 | 0.17 | 0.15 | 0.10 | 0.06 | 0.03 | 0.16 | 0.44 | 0.45 | 0.06 | 0.05 | -0.30 | 0.13 | 0.64 | 0.16 | 0.39 | 0.58 | 0.05 |
| hr0739 | S | 4803 | 2.56 | 1.33 | 0.24 | 0.15 | 0.26 | 0.26 | 0.67 | 0.13 | 0.02 | 0.06 | 0.06 | 0.16 | 0.08 | 0.04 | 0.00 | 0.11 | 0.23 | 0.50 | 0.26 | 0.18 | -0.22 | 0.17 | 0.34 | 0.18 | 0.27 | 0.31 | 0.18 |
| hr0743 | S | 5084 | 2.80 | 1.38 | 0.24 | 0.12 | 0.09 | 0.14 | 0.33 | 0.14 | 0.01 | 0.04 | -0.01 | 0.09 | 0.00 | 0.00 | -0.04 | 0.00 | -0.04 | 0.04 | 0.27 | 0.10 | -0.16 | 0.24 | 0.26 | 0.16 | 0.22 | 0.18 | 0.24 |
| hr0766 | S | 4678 | 2.57 | 1.18 | -0.03 | -0.04 | 0.09 | 0.09 | 0.53 | -0.06 | -0.10 | -0.08 | -0.07 | -0.05 | -0.15 | -0.12 | -0.10 | -0.01 | 0.15 | 0.35 | 0.00 | -0.10 | -0.52 | -0.03 | 0.18 | 0.07 | 0.17 | 0.44 | -0.03 |
| hr0768 | S | 6268 | 3.52 | 3.00 | 0.29 | | | 0.21 | 0.57 | -0.06 | 0.07 | 0.38 | 0.52 | 0.40 | 0.38 | 0.10 | 0.42 | 0.14 | -0.66 | 0.14 | 2.04 | 0.31 | 1.40 | -0.13 | 0.10 | | | 0.34 | 1.61 | 1.32 |
| hr0771 | S | 4758 | 2.56 | 1.34 | 0.21 | 0.10 | 0.21 | 0.29 | 0.96 | 0.11 | 0.04 | 0.03 | 0.03 | 0.13 | 0.07 | 0.05 | 0.00 | 0.12 | 0.26 | 0.51 | 0.21 | 0.17 | -0.41 | 0.17 | 0.43 | 0.20 | 0.30 | 0.43 | 0.12 |
| hr0808 | S | 4624 | 2.63 | 1.20 | 0.38 | 0.28 | 0.41 | 0.45 | 1.13 | 0.23 | 0.24 | 0.29 | 0.44 | 0.36 | 0.37 | 0.26 | 0.24 | 0.38 | 0.53 | 1.03 | 0.33 | 0.34 | -0.13 | 0.30 | 0.87 | 0.33 | 0.57 | 0.99 | 0.36 |
| hr0824 | S | 4628 | 2.53 | 1.37 | 0.32 | 0.29 | 0.42 | 0.41 | 0.96 | 0.21 | 0.11 | 0.18 | 0.31 | 0.26 | 0.21 | 0.14 | 0.18 | 0.28 | 0.54 | 0.91 | 0.15 | 0.13 | -0.27 | 0.11 | 0.57 | 0.27 | 0.37 | 0.80 | 0.16 |
| hr0831 | S | 6504 | 3.59 | 2.85 | 0.23 | 0.07 | 0.25 | 0.19 | 0.28 | 0.25 | 0.31 | 0.21 | 0.26 | 0.21 | 0.06 | 0.04 | 0.32 | 0.09 | -0.31 | 0.01 | | 0.33 | 0.63 | -0.04 | 0.15 | 0.35 | 0.09 | 0.37 | 0.14 |
| hr0840 | S | 6752 | 3.63 | 6.50 | | | 0.22 | -0.58 | | 0.26 | 0.88 | 0.97 | 0.46 | | 0.36 | 0.86 | 0.30 | | | 1.19 | | | | | 1.03 | 0.75 | 0.91 | | |
| hr0844 | S | 4657 | 2.39 | 1.40 | 0.16 | 0.16 | 0.26 | 0.27 | 0.74 | 0.02 | 0.06 | 0.06 | 0.12 | 0.12 | 0.06 | 0.02 | 0.05 | 0.13 | 0.41 | 0.47 | 0.08 | -0.04 | -0.35 | -0.05 | 0.43 | 0.05 | 0.20 | 0.62 | 0.02 |
| hr0850 | S | 4976 | 2.85 | 1.24 | 0.22 | 0.04 | 0.19 | 0.20 | 0.55 | 0.09 | 0.02 | 0.00 | -0.01 | 0.10 | -0.06 | 0.01 | -0.04 | 0.02 | 0.01 | 0.53 | 0.30 | 0.22 | -0.16 | 0.29 | 0.36 | 0.26 | 0.24 | 0.14 | 0.18 |
| hr0856 | S | 6413 | 3.44 | 3.61 | | | -0.15 | | 0.51 | 0.47 | 0.71 | 1.45 | 1.07 | | 0.35 | 1.24 | 0.91 | | | 0.39 | 2.23 | | | | | -0.38 | 0.07 |
| hr0900 | S | 4735 | 2.44 | 1.36 | 0.33 | 0.16 | 0.33 | 0.36 | 0.78 | 0.24 | 0.08 | 0.12 | 0.17 | 0.23 | 0.21 | 0.12 | 0.07 | 0.21 | 0.41 | 0.73 | 0.30 | 0.27 | -0.20 | 0.13 | 0.37 | 0.20 | 0.32 | 0.39 | 0.16 |
| hr0907 | S | 4751 | 2.55 | 1.42 | 0.09 | 0.08 | 0.19 | 0.20 | 0.66 | 0.02 | 0.03 | 0.03 | 0.08 | 0.04 | 0.04 | -0.02 | 0.01 | 0.07 | 0.31 | 0.47 | 0.06 | -0.04 | -0.34 | -0.02 | 0.23 | 0.02 | 0.17 | 0.38 | 0.14 |
| hr0908 | S | 4702 | 2.54 | 1.28 | -0.12 | -0.14 | -0.02 | -0.07 | 0.30 | -0.19 | -0.18 | -0.15 | -0.13 | -0.20 | -0.22 | -0.30 | -0.21 | -0.21 | 0.02 | 0.29 | -0.06 | -0.29 | -0.60 | -0.34 | -0.09 | -0.23 | -0.05 | 0.21 | -0.21 |
| hr0926 | S | 4705 | 2.39 | 1.45 | -0.03 | -0.06 | 0.06 | 0.13 | 0.39 | -0.08 | -0.09 | -0.11 | -0.15 | -0.07 | -0.13 | -0.15 | -0.11 | -0.09 | 0.06 | -0.17 | 0.10 | 0.01 | -0.38 | 0.04 | 0.25 | 0.07 | 0.14 | 0.17 | -0.04 |
| hr0931 | S | 4697 | 2.50 | 1.35 | 0.43 | 0.34 | 0.43 | 0.47 | 1.05 | 0.22 | 0.18 | 0.23 | 0.34 | 0.31 | 0.32 | 0.22 | 0.25 | 0.35 | 0.68 | 0.79 | 0.27 | 0.16 | -0.19 | 0.22 | 0.72 | 0.26 | 0.45 | 0.93 | 0.27 |
| hr0941 | S | 4857 | 2.71 | 1.28 | 0.43 | 0.17 | 0.39 | 0.41 | 0.88 | 0.24 | 0.10 | 0.16 | 0.19 | 0.27 | 0.25 | 0.18 | 0.14 | 0.26 | 0.34 | 0.81 | 0.28 | 0.29 | -0.18 | 0.20 | 0.52 | 0.24 | 0.35 | 0.66 | 0.31 |
| hr0946 | S | 4348 | 2.04 | 1.41 | 0.08 | 0.09 | 0.20 | 0.25 | 1.11 | -0.08 | -0.01 | 0.03 | 0.19 | 0.09 | -0.01 | -0.05 | 0.03 | 0.09 | 0.21 | 0.75 | 0.05 | -0.03 | -0.56 | -0.10 | 0.41 | 0.02 | 0.35 | 0.72 | 0.06 |
| hr0947 | S | 4586 | 2.18 | 1.54 | 0.23 | 0.05 | 0.16 | 0.28 | 0.78 | 0.04 | -0.07 | -0.06 | 0.01 | 0.08 | -0.01 | -0.01 | -0.04 | 0.04 | 0.03 | 0.62 | 0.25 | 0.01 | -0.34 | 0.13 | 0.42 | 0.17 | 0.30 | 0.54 | 0.06 |
| hr0951 | S | 4769 | 2.53 | 1.37 | 0.36 | 0.24 | 0.33 | 0.41 | 0.93 | 0.22 | 0.04 | 0.12 | 0.18 | 0.25 | 0.24 | 0.13 | 0.09 | 0.22 | 0.39 | 0.58 | 0.45 | 0.12 | -0.22 | 0.12 | 0.52 | 0.16 | 0.23 | 0.45 | 0.22 |
| hr0956 | S | 4959 | 2.75 | 1.38 | 0.45 | 0.22 | 0.35 | 0.36 | 0.68 | 0.29 | 0.16 | 0.25 | 0.26 | 0.33 | 0.28 | 0.19 | 0.19 | 0.26 | 0.38 | 0.21 | 0.36 | 0.34 | -0.02 | 0.25 | 0.71 | 0.23 | 0.38 | 0.57 | 0.29 |
| hr0992 | S | 4621 | 2.61 | 1.12 | 0.11 | 0.12 | 0.21 | 0.22 | 0.77 | 0.05 | 0.04 | 0.04 | 0.16 | 0.07 | -0.02 | -0.02 | 0.01 | 0.11 | 0.43 | 0.83 | 0.05 | -0.05 | -0.47 | -0.04 | 0.41 | 0.16 | 0.31 | 0.64 | 0.13 |
| hr0994 | S | 4960 | 2.67 | 1.19 | 0.35 | 0.18 | 0.34 | 0.19 | 0.60 | 0.20 | 0.14 | 0.23 | 0.32 | 0.27 | 0.19 | 0.09 | 0.12 | 0.15 | 0.14 | 0.56 | 0.45 | 0.27 | 0.05 | 0.18 | 0.48 | 0.21 | 0.30 | 0.49 | 0.29 |
| hr1000 | S | 4449 | 2.49 | 1.69 | 0.43 | 0.42 | 0.48 | 0.66 | 1.58 | 0.16 | 0.20 | 0.25 | 0.44 | 0.40 | 0.37 | 0.26 | 0.37 | 0.43 | | 1.59 | 0.21 | 0.17 | -0.26 | -0.12 | 1.00 | 0.32 | 0.53 | 1.00 | 0.39 |
| hr1007 | S | 4948 | 2.76 | 1.37 | 0.36 | 0.22 | 0.27 | 0.31 | 0.89 | 0.22 | 0.07 | 0.12 | 0.12 | 0.21 | 0.18 | 0.14 | 0.06 | 0.18 | 0.25 | 0.55 | 0.38 | 0.28 | -0.10 | 0.22 | 0.41 | 0.27 | 0.33 | 0.45 | 0.26 |
| hr1015 | S | 4385 | 2.07 | 1.50 | 0.54 | 0.28 | 0.47 | 0.47 | 1.31 | 0.15 | 0.15 | 0.24 | 0.42 | 0.37 | 0.27 | 0.17 | 0.23 | 0.32 | 0.50 | 1.35 | 0.27 | 0.19 | -0.22 | 0.06 | 0.91 | 0.23 | 0.56 | 1.08 | 0.23 |
| hr1022 | S | 4486 | 2.52 | 1.20 | 0.50 | 0.46 | 0.61 | 0.58 | 1.39 | 0.35 | 0.36 | 0.45 | 0.71 | 0.48 | 0.50 | 0.35 | 0.45 | 0.58 | 0.75 | 1.44 | 0.31 | 0.42 | -0.11 | 0.25 | 0.78 | 0.44 | 0.90 | 1.43 | 0.48 |
| hr1030 | S | 4982 | 2.20 | 1.66 | 0.29 | -0.09 | 0.04 | 0.09 | 0.20 | -0.11 | -0.15 | -0.18 | -0.27 | -0.07 | -0.21 | -0.16 | -0.25 | -0.17 | -0.28 | 0.05 | 0.33 | -0.01 | -0.40 | 0.19 | 0.20 | -0.04 | -0.01 | -0.24 | 0.05 |
| hr1050 | S | 4652 | 2.53 | 1.39 | 0.47 | 0.35 | 0.51 | 0.52 | 1.15 | 0.28 | 0.24 | 0.25 | 0.37 | 0.41 | 0.36 | 0.28 | 0.29 | 0.40 | 0.59 | 0.66 | 0.38 | 0.36 | -0.12 | 0.20 | 0.91 | 0.32 | 0.50 | 1.02 | 0.25 |
| hr1052 | S | 4163 | 1.57 | 1.59 | 0.07 | -0.04 | 0.07 | 0.25 | 1.27 | -0.12 | -0.17 | -0.08 | 0.07 | 0.01 | -0.09 | -0.11 | -0.02 | 0.03 | 0.07 | 1.15 | 0.08 | 0.00 | -0.57 | 0.06 | 0.52 | 0.13 | 0.43 | 0.65 | -0.03 |
| hr1060 | S | 4851 | 2.58 | 1.36 | 0.27 | 0.17 | 0.25 | 0.25 | 0.70 | 0.17 | 0.04 | 0.14 | 0.16 | 0.20 | 0.17 | 0.08 | 0.06 | 0.13 | 0.26 | 0.31 | 0.30 | 0.09 | -0.11 | 0.17 | 0.35 | 0.19 | 0.32 | 0.39 | 0.19 |
| hr1098 | S | 4984 | 2.58 | 1.36 | -0.03 | -0.11 | -0.01 | -0.08 | 0.10 | -0.07 | -0.25 | -0.15 | -0.22 | -0.13 | -0.26 | -0.23 | -0.24 | -0.24 | -0.22 | -0.16 | 0.15 | -0.18 | -0.36 | -0.08 | -0.05 | -0.02 | -0.02 | -0.16 | -0.09 |
| hr1108 | S | 4899 | 2.69 | 1.35 | 0.23 | 0.07 | 0.20 | 0.23 | 0.67 | 0.11 | 0.06 | 0.05 | 0.04 | 0.14 | 0.08 | 0.04 | 0.00 | 0.09 | 0.18 | 0.35 | 0.24 | 0.20 | -0.19 | 0.19 | 0.33 | 0.21 | 0.30 | 0.50 | 0.25 |
| hr1110 | S | 4948 | 2.68 | 1.33 | 0.31 | 0.13 | 0.27 | 0.25 | 0.71 | 0.18 | -0.03 | 0.07 | 0.08 | 0.17 | 0.13 | 0.07 | 0.00 | 0.11 | 0.14 | 0.33 | 0.31 | 0.20 | -0.20 | 0.15 | 0.33 | 0.13 | 0.21 | 0.34 | 0.19 |
| hr1117 | S | 4783 | 2.63 | 1.36 | 0.10 | 0.25 | 0.28 | 0.20 | 0.66 | 0.09 | -0.01 | 0.17 | 0.16 | 0.01 | -0.04 | -0.08 | 0.06 | 0.06 | 0.40 | 0.45 | 0.09 | -0.11 | -0.46 | -0.11 | 0.21 | -0.07 | 0.13 | 0.41 | 0.13 |
| hr1119 | S | 4901 | 2.71 | 1.35 | 0.15 | 0.07 | 0.19 | 0.15 | 0.52 | 0.09 | 0.01 | 0.06 | 0.05 | 0.09 | 0.04 | -0.02 | 0.04 | 0.14 | 0.27 | 0.25 | 0.16 | -0.19 | 0.19 | 0.32 | 0.22 | 0.30 | 0.25 | 0.20 |
| hr1132 | S | 4801 | 2.52 | 1.42 | -0.11 | -0.19 | 0.02 | 0.07 | 0.33 | -0.16 | -0.13 | -0.15 | -0.13 | -0.10 | -0.25 | -0.19 | -0.15 | -0.11 | -0.09 | 0.30 | 0.12 | -0.18 | -0.52 | -0.04 | 0.03 | 0.03 | 0.06 | 0.30 | -0.02 |
| hr1159 | S | 4800 | 2.53 | 1.36 | 0.01 | -0.13 | 0.10 | 0.11 | 0.41 | -0.04 | -0.12 | -0.10 | -0.12 | -0.03 | -0.17 | -0.11 | -0.15 | -0.07 | -0.08 | 0.27 | 0.11 | 0.00 | -0.42 | 0.11 | 0.14 | 0.15 | 0.19 | 0.26 | 0.01 |
| hr1256 | S | 4732 | 2.54 | 1.22 | 0.22 | 0.27 | 0.47 | 0.15 | 0.42 | 0.21 | 0.13 | 0.30 | 0.55 | 0.12 | 0.07 | -0.05 | 0.13 | 0.03 | | 0.38 | | 0.00 | -0.35 | -0.08 | 0.46 | -0.03 | 0.25 | 0.28 | 0.17 |
| hr1265 | S | 4654 | 2.33 | 1.41 | 0.18 | 0.10 | 0.17 | 0.24 | 0.84 | 0.11 | -0.06 | -0.01 | 0.02 | 0.12 | 0.03 | 0.02 | -0.03 | 0.09 | 0.30 | 0.43 | 0.11 | 0.14 | -0.31 | 0.19 | 0.43 | 0.14 | 0.30 | 0.30 | 0.07 |
| hr1267 | S | 4604 | 2.42 | 1.45 | 0.45 | 0.27 | 0.42 | 0.48 | 1.21 | 0.24 | 0.20 | 0.25 | 0.38 | 0.33 | 0.28 | 0.22 | 0.26 | 0.36 | 0.53 | 1.08 | 0.25 | 0.30 | -0.11 | 0.24 | 0.80 | 0.36 | 0.49 | 0.90 | 0.34 |
| hr1283 | S | 4736 | 2.51 | 1.38 | 0.34 | 0.29 | 0.41 | 0.35 | 0.88 | 0.24 | 0.10 | 0.19 | 0.29 | 0.24 | 0.29 | 0.14 | 0.13 | 0.24 | 0.50 | 0.53 | 0.30 | 0.11 | -0.19 | 0.12 | 0.53 | 0.12 | 0.33 | 0.56 | 0.16 |
| hr1295 | S | 4732 | 2.59 | 1.34 | 0.29 | 0.16 | 0.35 | 0.38 | 0.84 | 0.19 | 0.09 | 0.17 | 0.25 | 0.22 | 0.20 | 0.12 | 0.10 | 0.21 | 0.43 | 0.61 | 0.25 | 0.29 | -0.34 | 0.19 | 0.51 | 0.24 | 0.34 | 0.68 | 0.15 |
| hr1301 | S | 4934 | 2.62 | 1.44 | 0.12 | 0.10 | 0.26 | 0.11 | 0.36 | 0.10 | 0.07 | 0.17 | 0.19 | 0.11 | 0.16 | -0.01 | 0.06 | 0.07 | 0.24 | 0.18 | 0.18 | 0.08 | -0.21 | 0.03 | 0.25 | 0.08 | 0.22 | 0.33 | 0.18 |

| ID | | T | | | | | | | | | | | | | | | | | | | | | | | | | | |
|---|---|---|---|---|---|---|---|---|---|---|---|---|---|---|---|---|---|---|---|---|---|---|---|---|---|---|---|---|
| hr1310 | S | 4627 | 2.62 | 1.34 | 0.39 | 0.37 | 0.49 | 0.50 | 1.24 | 0.23 | 0.33 | 0.31 | 0.48 | 0.38 | 0.38 | 0.29 | 0.32 | 0.47 | 0.82 | 1.41 | 0.22 | 0.37 | -0.23 | 0.27 | 0.62 | 0.35 | 0.65 | 1.18 | 0.42 |
| hr1313 | S | 4693 | 2.69 | 1.18 | 0.48 | 0.20 | 0.42 | 0.40 | 1.00 | 0.23 | 0.23 | 0.30 | 0.48 | 0.38 | 0.36 | 0.22 | 0.24 | 0.36 | 0.57 | 1.29 | 0.31 | 0.25 | -0.16 | 0.25 | 0.78 | 0.34 | 0.51 | 1.03 | 0.41 |
| hr1318 | S | 4549 | 2.36 | 1.35 | 0.56 | 0.38 | 0.53 | 0.50 | 1.31 | 0.37 | 0.26 | 0.36 | 0.52 | 0.41 | 0.36 | 0.26 | 0.30 | 0.41 | 0.60 | 1.17 | 0.35 | 0.38 | -0.15 | 0.22 | 0.88 | 0.35 | 0.64 | 0.95 | 0.33 |
| hr1319 | S | 6606 | 4.13 | 3.89 | | | | 1.17 | | 0.55 | 0.54 | 1.34 | 1.28 | 0.52 | 0.55 | 0.36 | 1.56 | 0.58 | | | 1.46 | 1.77 | | 0.58 | | 0.19 | | | |
| hr1327 | S | 5211 | 2.79 | 1.43 | 0.24 | -0.03 | 0.06 | 0.08 | 0.25 | 0.02 | -0.03 | -0.06 | -0.10 | 0.04 | -0.09 | -0.06 | -0.11 | -0.08 | -0.23 | -0.01 | 0.36 | 0.08 | -0.16 | 0.29 | 0.12 | 0.16 | 0.15 | 0.02 | 0.16 |
| hr1343 | S | 4926 | 2.75 | 1.31 | 0.34 | 0.11 | 0.27 | 0.27 | 0.86 | 0.19 | 0.05 | 0.10 | 0.09 | 0.21 | 0.14 | 0.09 | 0.05 | 0.14 | 0.14 | 0.34 | 0.35 | 0.17 | -0.15 | 0.17 | 0.42 | 0.18 | 0.22 | 0.28 | 0.24 |
| hr1346 | S | 4901 | 2.60 | 1.49 | 0.54 | 0.28 | 0.38 | 0.40 | 0.89 | 0.24 | 0.14 | 0.17 | 0.19 | 0.29 | 0.27 | 0.17 | 0.15 | 0.24 | 0.35 | 0.86 | 0.51 | 0.19 | -0.07 | 0.22 | 0.47 | 0.24 | 0.30 | 0.60 | 0.25 |
| hr1348 | S | 4457 | 2.00 | 1.39 | -0.13 | -0.21 | -0.05 | 0.04 | 0.62 | -0.23 | -0.30 | -0.21 | -0.16 | -0.21 | -0.29 | -0.32 | -0.29 | -0.23 | 0.14 | 0.33 | -0.13 | -0.28 | -0.66 | -0.33 | 0.08 | -0.16 | -0.02 | 0.15 | -0.22 |
| hr1373 | S | 4883 | 2.62 | 1.47 | 0.51 | 0.11 | 0.30 | 0.43 | 0.87 | 0.19 | 0.15 | 0.12 | 0.19 | 0.26 | 0.22 | 0.17 | 0.14 | 0.25 | 0.30 | 1.27 | 0.32 | 0.22 | -0.11 | 0.23 | 0.42 | 0.23 | 0.31 | 0.55 | 0.32 |
| hr1407 | S | 4516 | 2.45 | 1.18 | 0.18 | 0.17 | 0.28 | 0.30 | 0.96 | 0.12 | 0.06 | 0.12 | 0.26 | 0.17 | 0.14 | 0.07 | 0.09 | 0.21 | 0.44 | 0.94 | 0.06 | 0.15 | -0.35 | 0.03 | 0.52 | 0.17 | 0.41 | 0.83 | 0.17 |
| hr1409 | S | 4836 | 2.57 | 1.49 | 0.58 | 0.27 | 0.36 | 0.47 | 1.01 | 0.25 | 0.22 | 0.19 | 0.20 | 0.31 | 0.31 | 0.22 | 0.19 | 0.31 | 0.36 | 1.20 | 0.37 | 0.22 | -0.14 | 0.28 | 0.70 | 0.29 | 0.34 | 0.55 | 0.33 |
| hr1411 | S | 4948 | 2.75 | 1.40 | 0.52 | 0.22 | 0.37 | 0.38 | 0.67 | 0.22 | 0.17 | 0.19 | 0.24 | 0.29 | 0.30 | 0.18 | 0.18 | 0.24 | 0.32 | 0.82 | 0.36 | 0.23 | -0.07 | 0.26 | 0.48 | 0.24 | 0.33 | 0.62 | 0.31 |
| hr1413 | S | 4753 | 2.73 | 1.26 | 0.03 | -0.04 | 0.10 | 0.21 | 0.55 | -0.03 | 0.00 | -0.06 | -0.03 | 0.00 | -0.11 | -0.04 | -0.08 | 0.01 | 0.09 | 0.65 | 0.12 | 0.14 | -0.41 | 0.19 | 0.35 | 0.18 | 0.28 | 0.44 | 0.10 |
| hr1421 | S | 4462 | 2.44 | 1.48 | 0.49 | 0.56 | 0.54 | 0.69 | 1.48 | 0.30 | 0.37 | 0.36 | 0.59 | 0.48 | 0.43 | 0.36 | 0.42 | 0.56 | 0.86 | 1.85 | 0.29 | 0.25 | -0.23 | 0.20 | 1.03 | 0.48 | 0.66 | 1.28 | 0.58 |
| hr1425 | S | 4748 | 2.71 | 1.32 | -0.02 | 0.00 | 0.10 | 0.18 | 0.74 | -0.02 | 0.01 | -0.01 | -0.02 | 0.00 | -0.04 | -0.05 | -0.01 | 0.04 | 0.29 | 0.39 | -0.02 | -0.07 | -0.54 | 0.03 | 0.24 | 0.14 | 0.25 | 0.45 | 0.14 |
| hr1431 | S | 4814 | 2.55 | 1.38 | 0.31 | 0.13 | 0.23 | 0.27 | 0.77 | 0.15 | 0.00 | 0.05 | 0.06 | 0.16 | 0.10 | 0.06 | 0.02 | 0.11 | 0.21 | 0.42 | 0.27 | 0.10 | -0.23 | 0.12 | 0.36 | 0.13 | 0.23 | 0.31 | 0.18 |
| hr1453 | S | 4773 | 2.84 | 1.15 | -0.09 | -0.10 | 0.00 | 0.09 | 0.36 | -0.10 | -0.14 | -0.14 | -0.12 | -0.12 | -0.13 | -0.17 | -0.16 | -0.10 | -0.02 | 0.37 | 0.01 | -0.13 | -0.53 | -0.04 | 0.11 | 0.10 | 0.14 | 0.32 | -0.08 |
| hr1455 | S | 5455 | 2.99 | 0.74 | 0.20 | 0.07 | 0.22 | 0.00 | 0.05 | 0.27 | 0.06 | 0.28 | 0.36 | 0.27 | 0.20 | 0.13 | 0.26 | 0.14 | -0.14 | -0.05 | 0.76 | 0.29 | 0.21 | 0.28 | 0.38 | 0.29 | 0.32 | 0.33 | 0.36 |
| hr1514 | S | 4969 | 2.64 | 1.30 | 0.54 | 0.26 | 0.34 | 0.36 | 0.92 | 0.31 | 0.23 | 0.27 | 0.32 | 0.35 | 0.34 | 0.23 | 0.21 | 0.30 | 0.24 | 0.86 | 0.41 | 0.36 | 0.03 | 0.39 | 0.65 | 0.34 | 0.39 | 0.67 | 0.35 |
| hr1517 | S | 4382 | 2.50 | 1.31 | 0.58 | 0.41 | 0.51 | 0.75 | 1.65 | 0.33 | 0.33 | 0.31 | 0.54 | 0.45 | 0.45 | 0.38 | 0.42 | 0.60 | 0.53 | 1.81 | 0.23 | 0.30 | -0.17 | 0.16 | 1.07 | 0.41 | 0.71 | 1.35 | 0.51 |
| hr1529 | S | 4416 | 2.55 | 1.00 | 0.24 | | 0.26 | 0.46 | 1.33 | 0.15 | 0.26 | 0.24 | 0.60 | 0.15 | -0.09 | 0.13 | 0.11 | 0.31 | 0.56 | | -0.07 | 0.39 | -0.08 | | 0.46 | 0.09 | 0.63 | 1.18 | |
| hr1533 | S | 4043 | 1.24 | 1.64 | -0.13 | -0.18 | -0.08 | 0.17 | 1.19 | -0.26 | -0.27 | -0.19 | 0.00 | -0.17 | -0.38 | -0.33 | -0.23 | -0.20 | | 0.90 | 0.08 | -0.21 | -0.71 | -0.41 | -0.23 | -0.28 | 0.10 | 0.50 | -0.19 |
| hr1535 | S | 4868 | 2.64 | 1.37 | 0.21 | 0.11 | 0.23 | 0.19 | 0.52 | 0.13 | 0.02 | 0.09 | 0.09 | 0.14 | 0.09 | 0.03 | 0.00 | 0.08 | 0.20 | 0.63 | 0.27 | 0.19 | -0.20 | 0.18 | 0.36 | 0.27 | 0.28 | 0.32 | 0.18 |
| hr1549 | S | 4818 | 2.59 | 1.38 | 0.28 | 0.08 | 0.25 | 0.29 | 0.69 | 0.16 | 0.05 | 0.05 | 0.09 | 0.17 | 0.08 | 0.06 | -0.01 | 0.11 | 0.16 | 0.68 | 0.32 | 0.14 | -0.24 | 0.14 | 0.33 | 0.15 | 0.24 | 0.40 | 0.06 |
| hr1580 | S | 4466 | 2.18 | 1.37 | -0.02 | -0.09 | 0.08 | 0.17 | 0.70 | -0.15 | -0.12 | -0.09 | -0.04 | -0.05 | -0.10 | -0.14 | -0.09 | -0.02 | 0.39 | 0.39 | 0.03 | -0.16 | -0.56 | -0.05 | 0.15 | 0.04 | 0.24 | 0.56 | 0.03 |
| hr1625 | S | 4476 | 2.41 | 1.45 | 0.53 | 0.53 | 0.62 | 0.68 | 1.61 | 0.30 | 0.36 | 0.44 | 0.64 | 0.51 | 0.53 | 0.41 | 0.47 | 0.61 | 0.76 | 1.58 | 0.28 | 0.28 | -0.18 | 0.18 | 1.22 | 0.48 | 0.72 | 1.32 | 0.57 |
| hr1628 | S | 4708 | 2.46 | 1.37 | 0.36 | 0.28 | 0.36 | 0.38 | 0.82 | 0.15 | 0.13 | 0.17 | 0.23 | 0.27 | 0.23 | 0.18 | 0.17 | 0.26 | 0.54 | 0.53 | 0.33 | 0.30 | -0.21 | 0.23 | 0.67 | 0.30 | 0.39 | 0.57 | 0.22 |
| hr1654 | S | 4005 | 1.37 | 1.73 | 0.22 | 0.31 | 0.26 | 0.48 | 1.99 | -0.01 | -0.02 | 0.02 | 0.21 | 0.18 | 0.10 | 0.08 | 0.14 | 0.23 | 0.80 | | 0.39 | 0.08 | -0.31 | 0.20 | 0.47 | 0.42 | 0.59 | 1.07 | 0.01 |
| hr1681 | S | 4671 | 2.64 | 1.30 | 0.40 | 0.34 | 0.43 | 0.49 | 1.09 | 0.27 | 0.23 | 0.28 | 0.40 | 0.34 | 0.36 | 0.28 | 0.26 | 0.43 | 0.66 | 0.89 | 0.29 | 0.36 | -0.18 | 0.33 | 0.72 | 0.39 | 0.52 | 0.86 | 0.30 |
| hr1698 | S | 4533 | 2.04 | 1.73 | 0.76 | 0.27 | 0.46 | 0.61 | 1.38 | 0.26 | 0.25 | 0.29 | 0.47 | 0.46 | 0.41 | 0.25 | 0.32 | 0.40 | 0.45 | 1.94 | 0.38 | 0.38 | -0.13 | 0.15 | 0.61 | 0.37 | 0.44 | 0.98 | 0.43 |
| hr1726 | S | 4264 | 1.92 | 1.38 | -0.07 | -0.11 | 0.11 | 0.10 | 0.99 | -0.18 | -0.15 | -0.03 | 0.13 | -0.11 | -0.29 | -0.27 | -0.15 | -0.14 | 0.54 | 1.01 | -0.01 | -0.13 | -0.63 | -0.32 | 0.10 | -0.08 | 0.19 | 0.75 | -0.05 |
| hr1739 | S | 4954 | 2.67 | 1.35 | 0.33 | 0.13 | 0.18 | 0.23 | 0.62 | 0.18 | 0.03 | 0.06 | 0.06 | 0.19 | 0.11 | 0.07 | -0.02 | 0.08 | 0.13 | 0.64 | 0.26 | 0.15 | -0.24 | 0.19 | 0.27 | 0.17 | 0.25 | 0.20 | 0.23 |
| hr1784 | S | 4839 | 2.57 | 1.37 | -0.04 | -0.10 | -0.03 | 0.05 | 0.29 | -0.11 | -0.16 | -0.20 | -0.22 | -0.10 | -0.26 | -0.17 | -0.24 | -0.14 | -0.15 | 0.00 | 0.11 | -0.14 | -0.40 | 0.14 | 0.18 | 0.11 | 0.17 | 0.06 | -0.01 |
| hr1796 | S | 4624 | 2.61 | 1.38 | 0.49 | 0.35 | 0.51 | 0.59 | 1.17 | 0.24 | 0.28 | 0.33 | 0.51 | 0.37 | 0.36 | 0.30 | 0.34 | 0.48 | 0.68 | 1.79 | 0.30 | 0.29 | -0.13 | 0.23 | 0.59 | 0.45 | 0.48 | 1.02 | 0.56 |
| hr1822 | S | 6273 | 3.51 | 3.16 | 0.35 | -0.21 | 0.09 | 0.16 | 0.50 | -0.06 | 0.32 | 0.30 | 0.40 | 0.51 | -0.01 | -0.04 | 0.47 | 0.14 | -0.37 | 0.24 | 1.78 | 0.58 | 1.40 | -0.23 | 0.31 | 1.06 | 0.44 | 0.42 | 0.88 |
| hr1831 | S | 4554 | 2.46 | 1.18 | 0.30 | 0.23 | 0.37 | 0.32 | 0.97 | 0.20 | 0.14 | 0.25 | 0.45 | 0.24 | 0.18 | 0.11 | 0.14 | 0.26 | 0.49 | 0.94 | 0.17 | 0.25 | -0.24 | 0.07 | 0.58 | 0.25 | 0.52 | 0.96 | 0.20 |
| hr1889 | S | 6379 | 3.65 | 3.63 | | -0.20 | | 0.56 | -0.28 | 0.35 | -0.18 | 0.79 | 1.04 | -0.43 | | -0.02 | 1.58 | -0.22 | | | 0.04 | | -1.29 | | | | | | |
| hr1954 | S | 4549 | 2.59 | 1.35 | 0.54 | 0.58 | 0.74 | 0.63 | 1.26 | 0.47 | 0.41 | 0.51 | 0.72 | 0.53 | 0.49 | 0.39 | 0.46 | 0.56 | 0.84 | 1.30 | 0.33 | 0.36 | -0.09 | 0.28 | 1.05 | 0.45 | 0.71 | 1.23 | 0.55 |
| hr1963 | S | 4389 | 1.98 | 1.42 | -0.32 | -0.27 | -0.14 | -0.02 | 0.50 | -0.36 | -0.33 | -0.35 | -0.35 | -0.35 | -0.45 | -0.43 | -0.36 | -0.32 | 0.00 | 0.16 | -0.24 | -0.48 | -0.78 | -0.44 | -0.24 | -0.34 | -0.17 | -0.09 | -0.26 |
| hr1986 | S | 4701 | 2.68 | 1.16 | 0.27 | 0.20 | 0.32 | 0.30 | 0.84 | 0.24 | 0.09 | 0.19 | 0.29 | 0.26 | 0.22 | 0.13 | 0.10 | 0.22 | 0.50 | 0.57 | 0.21 | 0.21 | -0.18 | 0.18 | 0.64 | 0.27 | 0.35 | 0.61 | 0.20 |
| hr1987 | S | 4988 | 2.82 | 1.27 | 0.09 | -0.14 | 0.04 | 0.08 | 0.37 | -0.04 | -0.11 | -0.10 | -0.15 | -0.04 | -0.14 | -0.10 | -0.15 | -0.09 | -0.13 | 0.27 | 0.26 | 0.00 | -0.29 | 0.19 | 0.24 | 0.15 | 0.20 | 0.21 | 0.08 |
| hr1995 | S | 4876 | 2.62 | 1.36 | 0.09 | -0.03 | 0.10 | 0.15 | 0.53 | -0.01 | -0.09 | -0.09 | -0.13 | -0.02 | -0.11 | -0.09 | -0.13 | -0.07 | -0.04 | 0.24 | 0.16 | -0.02 | -0.31 | 0.19 | 0.25 | 0.17 | 0.23 | 0.15 | 0.10 |
| hr2012 | S | 4590 | 2.14 | 1.50 | 0.25 | 0.10 | 0.23 | 0.34 | 0.97 | 0.07 | -0.03 | 0.00 | 0.03 | 0.12 | 0.06 | 0.02 | -0.03 | 0.08 | 0.28 | 0.99 | 0.27 | 0.01 | -0.28 | 0.14 | 0.47 | 0.14 | 0.25 | 0.29 | 0.08 |
| hr2076 | S | 4640 | 2.59 | 1.43 | 0.50 | 0.30 | 0.44 | 0.47 | 1.09 | 0.22 | 0.20 | 0.25 | 0.37 | 0.35 | 0.30 | 0.25 | 0.30 | 0.40 | 0.56 | 1.49 | 0.31 | 0.25 | -0.24 | 0.21 | 0.91 | 0.38 | 0.49 | 0.91 | 0.28 |
| hr2077 | S | 4768 | 2.56 | 1.34 | 0.15 | 0.03 | 0.18 | 0.24 | 0.74 | 0.08 | -0.02 | 0.01 | 0.00 | 0.09 | -0.01 | 0.01 | -0.02 | 0.07 | 0.13 | 0.43 | 0.19 | 0.18 | -0.34 | 0.14 | 0.41 | 0.21 | 0.25 | 0.48 | 0.10 |
| hr2080 | S | 4540 | 2.55 | 1.47 | 0.62 | 0.50 | 0.73 | 0.70 | 1.64 | 0.34 | 0.38 | 0.46 | 0.71 | 0.51 | 0.51 | 0.40 | 0.49 | 0.61 | 0.69 | 2.01 | 0.32 | 0.49 | -0.11 | 0.20 | 1.11 | 0.57 | 0.66 | 1.39 | 0.56 |
| hr2119 | S | 4786 | 2.94 | 1.07 | -0.03 | -0.05 | 0.12 | 0.10 | 0.47 | -0.07 | -0.03 | -0.04 | 0.01 | -0.05 | -0.05 | -0.11 | -0.06 | -0.02 | 0.16 | 0.45 | -0.02 | -0.16 | -0.48 | -0.09 | 0.20 | 0.03 | 0.20 | 0.52 | 0.01 |
| hr2136 | S | 4950 | 2.72 | 1.34 | 0.16 | 0.07 | 0.15 | 0.15 | 0.60 | 0.08 | -0.03 | -0.02 | -0.03 | 0.04 | 0.01 | -0.03 | -0.08 | -0.01 | 0.05 | 0.22 | 0.19 | 0.12 | -0.36 | 0.17 | 0.30 | 0.18 | 0.21 | 0.07 | 0.11 |
| hr2152 | S | 4609 | 2.09 | 1.58 | -0.26 | -0.19 | -0.18 | -0.15 | 0.29 | -0.39 | -0.35 | -0.35 | -0.34 | -0.37 | -0.49 | -0.45 | -0.37 | -0.38 | -0.20 | 0.27 | 0.06 | -0.46 | -0.69 | -0.39 | -0.18 | -0.30 | -0.23 | -0.22 | -0.30 |
| hr2183 | S | 4635 | 2.73 | 1.40 | 0.61 | 0.53 | 0.65 | 0.60 | 1.35 | 0.42 | 0.39 | 0.44 | 0.66 | 0.54 | 0.52 | 0.37 | 0.48 | 0.53 | 0.60 | 1.15 | 0.36 | 0.36 | -0.09 | 0.10 | 0.93 | 0.45 | 0.72 | 1.18 | 0.45 |

| ID | | | | | | | | | | | | | | | | | | | | | | | | | | | | | |
|---|---|---|---|---|---|---|---|---|---|---|---|---|---|---|---|---|---|---|---|---|---|---|---|---|---|---|---|---|---|
| hr2200 | S | 4690 | 2.49 | 1.36 | 0.35 | 0.14 | 0.28 | 0.34 | 0.92 | 0.15 | 0.05 | 0.08 | 0.13 | 0.17 | 0.09 | 0.06 | 0.01 | 0.14 | 0.31 | 0.73 | 0.18 | 0.10 | -0.21 | 0.21 | 0.64 | 0.27 | 0.37 | 0.31 | 0.19 |
| hr2218 | S | 4926 | 2.79 | 1.27 | 0.09 | -0.02 | 0.06 | 0.11 | 0.37 | 0.00 | -0.06 | -0.08 | -0.12 | -0.01 | -0.14 | -0.07 | -0.14 | -0.05 | 0.01 | 0.42 | 0.46 | -0.03 | -0.29 | 0.19 | 0.21 | 0.17 | 0.20 | 0.12 | 0.08 |
| hr2219 | S | 4685 | 2.35 | 1.41 | -0.17 | -0.05 | 0.04 | 0.05 | 0.45 | -0.12 | -0.21 | -0.18 | -0.23 | -0.19 | -0.26 | -0.26 | -0.24 | -0.18 | 0.00 | 0.16 | 0.03 | -0.37 | -0.65 | -0.29 | -0.08 | -0.29 | -0.11 | -0.07 | -0.11 |
| hr2230 | S | 5072 | 2.71 | 1.49 | 0.31 | 0.06 | 0.24 | 0.22 | 0.54 | 0.17 | 0.04 | 0.07 | 0.05 | 0.13 | 0.06 | 0.04 | 0.04 | 0.08 | 0.06 | 0.17 | 0.34 | 0.14 | -0.13 | 0.13 | 0.25 | 0.08 | 0.18 | 0.33 | 0.25 |
| hr2239 | S | 4552 | 2.43 | 1.37 | 0.29 | 0.32 | 0.47 | 0.44 | 1.25 | 0.17 | 0.18 | 0.26 | 0.37 | 0.25 | 0.16 | 0.14 | 0.20 | 0.30 | 0.53 | 1.06 | 0.10 | 0.17 | -0.40 | 0.06 | 0.58 | 0.23 | 0.32 | 1.03 | 0.42 |
| hr2243 | S | 4701 | 2.96 | 0.91 | 0.30 | 0.35 | 0.34 | 0.35 | 0.98 | 0.21 | 0.24 | 0.24 | 0.40 | 0.32 | 0.27 | 0.24 | 0.24 | 0.37 | 0.62 | 0.54 | 0.28 | 0.39 | -0.10 | 0.28 | 1.02 | 0.35 | 0.72 | 1.29 | 0.17 |
| hr2259 | S | 5076 | 3.00 | 0.96 | 0.10 | 0.01 | 0.12 | 0.05 | 0.34 | 0.10 | 0.00 | 0.08 | 0.11 | 0.11 | 0.05 | -0.01 | 0.00 | 0.00 | -0.01 | 0.06 | 0.38 | 0.18 | -0.19 | 0.23 | 0.31 | 0.22 | 0.25 | 0.43 | 0.18 |
| hr2287 | S | 6986 | 3.77 | 3.43 | | 0.38 | | 0.20 | -0.64 | 0.67 | 1.55 | 0.73 | 1.34 | 1.06 | 0.84 | 0.10 | 0.92 | 0.23 | | 0.73 | | 0.93 | | -0.37 | | | 1.16 | 0.79 | | |
| hr2302 | S | 4729 | 2.44 | 1.38 | 0.11 | 0.11 | 0.22 | 0.19 | 0.73 | 0.06 | -0.05 | 0.03 | 0.04 | 0.04 | 0.03 | -0.03 | -0.03 | 0.04 | 0.19 | 0.36 | 0.10 | -0.04 | -0.35 | 0.02 | 0.26 | 0.06 | 0.18 | 0.31 | 0.10 |
| hr2333 | S | 4736 | 2.49 | 1.38 | 0.11 | -0.08 | 0.24 | 0.22 | 0.55 | 0.06 | -0.04 | -0.02 | 0.00 | 0.04 | -0.07 | -0.03 | -0.05 | 0.02 | 0.18 | 0.53 | 0.12 | 0.08 | -0.29 | 0.11 | 0.23 | 0.10 | 0.21 | 0.33 | 0.08 |
| hr2379 | S | 4345 | 2.00 | 1.55 | 0.43 | 0.31 | 0.41 | 0.48 | 1.31 | 0.19 | 0.16 | 0.17 | 0.34 | 0.29 | 0.20 | 0.15 | 0.21 | 0.28 | 0.60 | 1.20 | 0.23 | 0.06 | -0.34 | 0.07 | 0.92 | 0.30 | 0.47 | 0.84 | 0.26 |
| hr2392 | S | 4725 | 2.53 | 1.42 | 0.04 | 0.06 | 0.23 | 0.47 | 1.20 | -0.03 | -0.01 | -0.03 | -0.15 | 0.08 | -0.18 | -0.07 | -0.15 | 0.09 | 0.14 | 0.72 | 1.20 | 0.82 | 0.70 | | 1.86 | 1.41 | 1.42 | 1.57 | 0.90 |
| hr2477 | S | 4883 | 2.85 | 1.15 | 0.11 | -0.14 | 0.10 | 0.12 | 0.53 | 0.10 | -0.03 | 0.02 | 0.01 | 0.06 | -0.03 | -0.02 | -0.06 | 0.00 | 0.09 | 0.41 | 0.17 | 0.06 | -0.23 | 0.20 | 0.34 | 0.22 | 0.29 | 0.36 | 0.10 |
| hr2478 | S | 4502 | 2.02 | 1.55 | 0.20 | -0.04 | 0.13 | 0.24 | 0.68 | -0.02 | -0.06 | -0.08 | -0.01 | -0.02 | -0.11 | -0.09 | -0.05 | 0.00 | 0.18 | 0.59 | 0.09 | 0.03 | -0.50 | 0.06 | 0.35 | 0.10 | 0.23 | 0.20 | 0.10 |
| hr2552 | S | 4960 | 2.70 | 1.40 | 0.45 | 0.20 | 0.25 | 0.34 | 0.90 | 0.25 | 0.12 | 0.16 | 0.18 | 0.25 | 0.21 | 0.14 | 0.07 | 0.16 | 0.20 | 0.41 | 0.30 | 0.19 | -0.05 | 0.29 | 0.47 | 0.15 | 0.25 | 0.39 | 0.29 |
| hr2556 | S | 4994 | 2.74 | 1.33 | 0.32 | 0.14 | 0.23 | 0.21 | 0.65 | 0.21 | 0.00 | 0.06 | 0.10 | 0.17 | 0.12 | 0.06 | 0.02 | 0.08 | 0.10 | 0.43 | 0.29 | 0.16 | -0.16 | 0.16 | 0.30 | 0.09 | 0.18 | 0.27 | 0.21 |
| hr2573 | S | 4738 | 2.50 | 1.38 | 0.26 | 0.20 | 0.34 | 0.30 | 0.78 | 0.19 | 0.08 | 0.16 | 0.25 | 0.22 | 0.22 | 0.10 | 0.10 | 0.19 | 0.42 | 0.77 | 0.25 | 0.07 | -0.19 | 0.12 | 0.46 | 0.15 | 0.28 | 0.49 | 0.19 |
| hr2574 | S | 4044 | 1.49 | 1.57 | -0.03 | -0.01 | 0.09 | 0.26 | 1.40 | -0.26 | -0.09 | -0.06 | 0.16 | -0.02 | -0.24 | -0.13 | 0.01 | 0.03 | 0.62 | 0.92 | 0.13 | -0.14 | -0.58 | -0.11 | 0.10 | 0.09 | 0.41 | 1.13 | -0.08 |
| hr2642 | S | 4614 | 2.55 | 1.43 | 0.36 | 0.31 | 0.49 | 0.56 | 1.28 | 0.38 | 0.35 | 0.45 | 0.62 | 0.48 | 0.43 | 0.34 | 0.37 | 0.43 | 0.55 | 1.56 | 0.27 | 0.38 | -0.09 | 0.29 | 0.88 | 0.45 | 0.65 | 1.21 | 0.58 |
| hr2684 | S | 4740 | 2.41 | 1.42 | 0.05 | -0.01 | 0.13 | 0.14 | 0.65 | -0.03 | -0.10 | -0.08 | -0.07 | -0.03 | -0.09 | -0.10 | -0.09 | -0.04 | 0.08 | 0.37 | 0.13 | 0.02 | -0.43 | 0.01 | 0.22 | -0.01 | 0.09 | 0.17 | 0.02 |
| hr2701 | S | 4692 | 2.57 | 1.12 | -0.08 | -0.12 | -0.01 | 0.02 | 0.46 | -0.07 | -0.25 | -0.16 | -0.16 | -0.15 | -0.22 | -0.24 | -0.23 | -0.17 | -0.01 | 0.28 | 0.00 | -0.14 | -0.57 | -0.17 | -0.06 | -0.10 | 0.05 | 0.07 | -0.18 |
| hr2728 | S | 4787 | 2.53 | 1.42 | 0.04 | -0.12 | 0.12 | 0.13 | 0.35 | -0.04 | -0.07 | -0.06 | -0.08 | -0.03 | -0.10 | -0.10 | -0.09 | -0.06 | 0.08 | 0.51 | 0.30 | -0.05 | -0.40 | 0.05 | 0.23 | 0.13 | 0.20 | 0.22 | 0.14 |
| hr2764 | S | 3937 | 0.32 | 2.98 | 0.21 | 0.24 | 0.08 | 0.61 | 1.97 | 0.06 | -0.26 | -0.16 | -0.01 | 0.17 | -0.09 | -0.08 | -0.10 | 0.08 | | | 0.37 | -0.09 | -0.43 | | 0.15 | 0.24 | 0.23 | 0.62 | 0.25 |
| hr2793 | S | 4575 | 2.38 | 1.30 | -0.16 | -0.30 | -0.06 | -0.02 | 0.49 | -0.15 | -0.22 | -0.20 | -0.18 | -0.22 | -0.27 | -0.32 | -0.25 | -0.23 | 0.06 | 0.23 | -0.16 | -0.35 | -0.65 | -0.35 | -0.10 | -0.18 | -0.06 | 0.19 | -0.28 |
| hr2877 | S | 4620 | 2.47 | 1.39 | 0.42 | 0.22 | 0.38 | 0.50 | 1.02 | 0.22 | 0.14 | 0.24 | 0.36 | 0.32 | 0.27 | 0.20 | 0.21 | 0.31 | 0.42 | 0.77 | 0.25 | 0.18 | -0.19 | 0.15 | 0.65 | 0.23 | 0.34 | 0.78 | 0.30 |
| hr2896 | S | 4829 | 2.49 | 1.53 | 0.48 | 0.29 | 0.35 | 0.41 | 0.79 | 0.23 | 0.03 | 0.18 | 0.17 | 0.28 | 0.23 | 0.13 | 0.13 | 0.21 | 0.17 | 0.61 | 0.35 | 0.10 | -0.14 | 0.06 | 0.19 | 0.13 | 0.22 | 0.53 | 0.18 |
| hr2899 | S | 4600 | 2.54 | 1.38 | 0.35 | 0.25 | 0.40 | 0.47 | 1.05 | 0.16 | 0.16 | 0.19 | 0.30 | 0.28 | 0.26 | 0.21 | 0.24 | 0.38 | 0.55 | 1.32 | 0.17 | 0.30 | -0.24 | 0.21 | 0.75 | 0.33 | 0.47 | 0.89 | 0.40 |
| hr2924 | S | 4997 | 2.64 | 1.40 | 0.43 | 0.16 | 0.28 | 0.23 | 0.46 | 0.15 | 0.03 | 0.07 | 0.05 | 0.16 | 0.04 | 0.05 | 0.01 | 0.06 | 0.11 | 0.35 | 0.29 | 0.13 | -0.20 | 0.12 | 0.37 | 0.07 | 0.10 | 0.13 | 0.19 |
| hr2970 | S | 4749 | 2.50 | 1.36 | 0.18 | 0.03 | 0.22 | 0.23 | 0.78 | 0.09 | -0.05 | -0.01 | 0.01 | 0.06 | 0.03 | 0.00 | -0.05 | 0.06 | 0.15 | 0.73 | 0.12 | 0.09 | -0.43 | 0.08 | 0.29 | 0.08 | 0.19 | 0.28 | 0.07 |
| hr2975 | S | 4357 | 2.24 | 1.42 | 0.28 | 0.29 | 0.35 | 0.64 | 1.51 | 0.17 | 0.14 | 0.14 | 0.33 | 0.28 | 0.27 | 0.27 | 0.28 | 0.42 | 0.57 | 1.05 | 0.08 | 0.23 | -0.47 | 0.15 | 0.82 | 0.27 | 0.44 | 0.98 | 0.40 |
| hr2985 | S | 4954 | 2.61 | 1.45 | 0.32 | 0.01 | 0.12 | 0.20 | 0.60 | 0.11 | -0.03 | -0.01 | -0.02 | 0.09 | 0.04 | 0.00 | -0.05 | 0.02 | 0.02 | 0.64 | 0.32 | 0.09 | -0.17 | 0.17 | 0.25 | 0.16 | 0.20 | 0.15 | 0.15 |
| hr2990 | S | 4821 | 2.72 | 1.21 | 0.35 | 0.31 | 0.23 | 0.34 | 0.78 | 0.22 | 0.04 | 0.16 | 0.19 | 0.23 | 0.18 | 0.13 | 0.05 | 0.18 | 0.28 | 0.57 | 0.27 | 0.23 | -0.06 | 0.16 | 0.54 | 0.21 | 0.30 | 0.46 | 0.23 |
| hr3044 | S | 4330 | 2.21 | 1.49 | 0.52 | 0.33 | 0.58 | 0.62 | 1.73 | 0.26 | 0.32 | 0.30 | 0.56 | 0.41 | 0.38 | 0.26 | 0.42 | 0.45 | 0.71 | 0.99 | 0.28 | 0.32 | -0.27 | 0.07 | 1.10 | 0.50 | 0.75 | 1.26 | 0.42 |
| hr3054 | S | 4660 | 2.47 | 1.46 | 0.52 | 0.18 | 0.46 | 0.39 | 0.98 | 0.20 | 0.13 | 0.22 | 0.34 | 0.26 | 0.26 | 0.14 | 0.17 | 0.26 | 0.49 | 0.58 | 0.22 | 0.08 | -0.21 | 0.10 | 0.58 | 0.17 | 0.32 | 0.56 | 0.27 |
| hr3094 | S | 4772 | 2.57 | 1.38 | 0.18 | 0.09 | 0.26 | 0.24 | 0.64 | 0.09 | 0.04 | 0.08 | 0.12 | 0.11 | 0.08 | 0.03 | 0.01 | 0.09 | 0.22 | 0.34 | 0.17 | 0.05 | -0.22 | 0.16 | 0.40 | 0.17 | 0.26 | 0.39 | 0.13 |
| hr3097 | S | 4865 | 2.79 | 1.29 | 0.38 | 0.15 | 0.34 | 0.31 | 0.71 | 0.22 | 0.13 | 0.17 | 0.20 | 0.25 | 0.20 | 0.13 | 0.10 | 0.19 | 0.33 | 0.59 | 0.30 | 0.27 | -0.09 | 0.21 | 0.57 | 0.27 | 0.30 | 0.60 | 0.27 |
| hr3110 | S | 5008 | 2.84 | 1.27 | 0.37 | -0.01 | 0.25 | 0.17 | 0.54 | 0.12 | 0.07 | 0.09 | 0.10 | 0.10 | 0.05 | 0.03 | 0.01 | 0.07 | 0.16 | 0.45 | 0.49 | 0.12 | -0.19 | 0.10 | 0.27 | 0.12 | 0.15 | 0.16 | 0.19 |
| hr3122 | S | 4422 | 2.14 | 1.41 | 0.02 | 0.03 | 0.06 | 0.24 | 0.82 | -0.08 | -0.09 | -0.07 | 0.00 | -0.02 | -0.10 | -0.09 | -0.07 | 0.03 | 0.11 | 0.50 | 0.05 | -0.03 | -0.43 | 0.13 | 0.43 | 0.15 | 0.37 | 0.51 | 0.18 |
| hr3125 | S | 4710 | 2.55 | 1.38 | 0.23 | 0.16 | 0.34 | 0.30 | 0.89 | 0.16 | 0.11 | 0.18 | 0.26 | 0.22 | 0.24 | 0.10 | 0.13 | 0.19 | 0.51 | 0.67 | 0.21 | 0.09 | -0.20 | 0.15 | 0.48 | 0.27 | 0.33 | 0.59 | 0.24 |
| hr3145 | S | 4280 | 1.86 | 1.46 | -0.18 | -0.30 | -0.02 | 0.08 | 0.76 | -0.25 | -0.21 | -0.18 | -0.11 | -0.21 | -0.32 | -0.31 | -0.25 | -0.20 | 0.02 | 0.40 | -0.16 | -0.23 | -0.77 | -0.35 | 0.06 | -0.18 | 0.03 | 0.26 | -0.18 |
| hr3149 | S | 4567 | 2.37 | 1.30 | 0.27 | 0.08 | 0.28 | 0.30 | 0.85 | 0.13 | 0.04 | 0.11 | 0.21 | 0.21 | 0.18 | 0.08 | 0.07 | 0.19 | 0.42 | 0.82 | 0.23 | 0.06 | -0.29 | 0.11 | 0.51 | 0.19 | 0.35 | 0.66 | 0.18 |
| hr3150 | S | 5702 | 3.18 | 1.70 | -0.05 | -0.06 | -0.14 | -0.05 | 0.11 | -0.03 | -0.03 | -0.13 | -0.21 | -0.08 | -0.24 | -0.20 | -0.26 | -0.23 | -0.32 | -0.14 | 0.72 | 0.03 | -0.06 | 0.13 | 0.00 | 0.00 | -0.02 | -0.37 | 0.00 |
| hr3211 | S | 4917 | 2.86 | 1.26 | 0.29 | 0.07 | 0.22 | 0.27 | 0.70 | 0.17 | 0.00 | 0.04 | 0.03 | 0.14 | 0.09 | 0.05 | -0.01 | 0.10 | 0.17 | 0.52 | 0.18 | 0.19 | -0.23 | 0.17 | 0.34 | 0.15 | 0.21 | 0.36 | 0.25 |
| hr3212 | S | 4988 | 2.72 | 1.29 | 0.13 | -0.04 | 0.11 | 0.11 | 0.37 | 0.03 | -0.10 | -0.08 | -0.13 | -0.02 | -0.12 | -0.07 | -0.17 | -0.07 | -0.07 | 0.27 | 0.47 | 0.02 | -0.31 | 0.20 | 0.24 | 0.18 | 0.19 | 0.13 | 0.13 |
| hr3216 | S | 4997 | 2.73 | 1.30 | 0.13 | 0.00 | 0.10 | 0.09 | 0.40 | 0.02 | -0.08 | -0.06 | -0.09 | 0.02 | -0.09 | -0.06 | -0.14 | -0.07 | -0.06 | 0.10 | 0.46 | 0.03 | -0.22 | 0.18 | 0.31 | 0.19 | 0.18 | 0.13 | 0.14 |
| hr3222 | S | 4971 | 2.60 | 1.23 | 0.26 | -0.01 | 0.28 | 0.08 | 0.41 | 0.16 | 0.01 | 0.20 | 0.34 | 0.17 | 0.10 | 0.02 | 0.05 | 0.03 | 0.01 | 0.13 | 0.45 | 0.31 | 0.05 | 0.28 | 0.38 | 0.14 | 0.29 | 0.21 | 0.18 |
| hr3263 | S | 4930 | 2.82 | 1.29 | 0.25 | 0.16 | 0.25 | 0.25 | 0.56 | 0.15 | 0.05 | 0.08 | 0.11 | 0.16 | 0.11 | 0.08 | 0.01 | 0.10 | 0.17 | 0.39 | 0.26 | 0.22 | -0.18 | 0.21 | 0.38 | 0.20 | 0.28 | 0.34 | 0.24 |
| hr3281 | S | 4660 | 2.49 | 1.33 | 0.08 | 0.19 | 0.28 | 0.26 | 0.73 | 0.04 | -0.03 | 0.09 | 0.09 | -0.03 | -0.18 | -0.10 | 0.00 | 0.04 | 0.36 | 0.37 | 0.04 | -0.14 | -0.55 | -0.12 | 0.19 | 0.02 | 0.14 | 0.60 | 0.10 |
| hr3289 | S | 4863 | 2.63 | 1.37 | 0.08 | 0.04 | 0.11 | 0.15 | 0.49 | 0.02 | -0.03 | -0.01 | 0.03 | 0.03 | -0.02 | -0.05 | -0.07 | -0.02 | 0.08 | 0.32 | 0.24 | 0.01 | -0.39 | 0.12 | 0.26 | 0.14 | 0.23 | 0.23 | 0.12 |
| hr3303 | S | 4934 | 2.78 | 1.31 | 0.17 | -0.08 | 0.21 | 0.15 | 0.55 | 0.09 | 0.01 | 0.06 | 0.07 | 0.09 | -0.01 | 0.00 | -0.04 | 0.02 | 0.07 | 0.25 | 0.45 | 0.13 | -0.20 | 0.22 | 0.41 | 0.22 | 0.31 | 0.31 | 0.11 |

| ID | Type | V1 | V2 | V3 | V4 | V5 | V6 | V7 | V8 | V9 | V10 | V11 | V12 | V13 | V14 | V15 | V16 | V17 | V18 | V19 | V20 | V21 | V22 | V23 | V24 | V25 | V26 | V27 | V28 | V29 |
|---|---|---|---|---|---|---|---|---|---|---|---|---|---|---|---|---|---|---|---|---|---|---|---|---|---|---|---|---|---|---|
| hr3324 | S | 4490 | 2.39 | 1.26 | 0.16 | 0.14 | 0.27 | 0.30 | 0.96 | 0.08 | 0.05 | 0.11 | 0.26 | 0.14 | 0.08 | 0.04 | 0.08 | 0.17 | 0.47 | 1.13 | 0.09 | 0.03 | -0.37 | 0.04 | 0.51 | 0.19 | 0.45 | 0.75 | 0.28 |
| hr3366 | S | 4415 | 2.20 | 1.52 | 0.45 | 0.36 | 0.38 | 0.58 | 1.49 | 0.13 | 0.22 | 0.28 | 0.44 | 0.38 | 0.35 | 0.24 | 0.34 | 0.39 | 0.59 | 1.41 | 0.30 | 0.36 | -0.24 | 0.09 | 0.97 | 0.43 | 0.55 | 1.11 | 0.41 |
| hr3369 | S | 4781 | 2.43 | 1.53 | 0.31 | 0.12 | 0.21 | 0.33 | 0.95 | 0.06 | 0.03 | 0.01 | 0.03 | 0.12 | 0.03 | 0.04 | 0.00 | 0.09 | 0.10 | 0.23 | 0.23 | 0.16 | -0.29 | 0.17 | 0.40 | 0.15 | 0.21 | 0.24 | 0.21 |
| hr3376 | S | 4553 | 2.60 | 1.49 | 0.48 | 0.43 | 0.46 | 0.64 | 1.48 | 0.29 | 0.31 | 0.38 | 0.54 | 0.43 | 0.44 | 0.33 | 0.41 | 0.50 | 0.66 | 1.45 | 0.23 | 0.40 | -0.20 | 0.22 | 0.95 | 0.49 | 0.62 | 1.19 | 0.43 |
| hr3403 | S | 4438 | 2.11 | 1.45 | 0.03 | -0.07 | 0.14 | 0.23 | 0.85 | -0.09 | -0.11 | -0.06 | -0.03 | -0.03 | -0.07 | -0.11 | -0.05 | 0.01 | 0.18 | 0.49 | 0.07 | -0.06 | -0.50 | 0.06 | 0.39 | 0.11 | 0.33 | 0.61 | 0.03 |
| hr3409 | S | 4673 | 2.38 | 1.37 | 0.29 | 0.14 | 0.29 | 0.28 | 0.72 | 0.17 | 0.02 | 0.13 | 0.23 | 0.20 | 0.16 | 0.06 | 0.04 | 0.13 | 0.40 | 0.37 | 0.25 | 0.10 | -0.21 | 0.13 | 0.47 | 0.16 | 0.35 | 0.49 | 0.09 |
| hr3418 | S | 4491 | 2.01 | 1.74 | 0.47 | 0.17 | 0.33 | 0.53 | 1.24 | 0.22 | 0.09 | 0.12 | 0.21 | 0.31 | 0.18 | 0.15 | 0.15 | 0.26 | 0.25 | 1.11 | 0.35 | 0.20 | -0.26 | 0.11 | 0.66 | 0.26 | 0.32 | 0.59 | 0.23 |
| hr3423 | S | 6574 | 3.38 | 2.19 | 0.57 | 0.11 | 0.61 | 0.23 | 0.36 | 0.27 | 0.23 | 0.25 | 0.25 | 0.27 | 0.52 | 0.03 | 1.05 | 0.11 | -0.26 | | | 0.34 | 0.57 | 0.30 | 1.36 | 0.85 | 0.61 | -0.32 | 0.05 |
| hr3424 | S | 4577 | 2.65 | 1.19 | 0.22 | 0.23 | 0.25 | 0.47 | 1.15 | 0.13 | 0.08 | 0.06 | 0.14 | 0.20 | 0.17 | 0.16 | 0.10 | 0.29 | 0.52 | 0.59 | 0.10 | 0.25 | -0.46 | 0.22 | 0.54 | 0.20 | 0.38 | 0.68 | 0.18 |
| hr3433 | S | 4876 | 2.59 | 1.40 | -0.37 | -0.41 | -0.27 | -0.21 | -0.07 | -0.39 | -0.37 | -0.45 | -0.52 | -0.43 | -0.64 | -0.51 | -0.46 | -0.46 | -0.44 | -0.21 | 0.01 | -0.46 | -0.72 | -0.34 | -0.34 | -0.34 | -0.22 | -0.31 | -0.25 |
| hr3461 | S | 4636 | 2.56 | 1.34 | 0.17 | 0.19 | 0.26 | 0.39 | 1.04 | 0.06 | 0.09 | 0.11 | 0.16 | 0.14 | 0.14 | 0.08 | 0.08 | 0.22 | 0.35 | 1.08 | 0.07 | 0.03 | -0.37 | 0.12 | 0.52 | 0.12 | 0.33 | 0.62 | 0.20 |
| hr3464 | S | 4966 | 2.49 | 1.56 | 0.34 | -0.03 | 0.11 | 0.19 | 0.65 | 0.02 | -0.08 | -0.03 | -0.06 | 0.06 | -0.11 | -0.03 | -0.09 | -0.04 | -0.15 | 0.11 | 0.22 | 0.16 | -0.18 | 0.29 | 0.53 | 0.07 | 0.15 | 0.84 | 0.14 |
| hr3475 | S | 4849 | 2.15 | 1.98 | 0.45 | 0.28 | 0.38 | 0.37 | 0.73 | 0.06 | 0.01 | 0.02 | -0.08 | 0.12 | -0.07 | 0.03 | -0.02 | 0.06 | 0.04 | 0.51 | 0.02 | -0.02 | -0.18 | 0.25 | 0.23 | 0.14 | 0.14 | 0.18 | 0.17 |
| hr3484 | S | 4968 | 2.50 | 1.41 | 0.10 | -0.21 | -0.01 | 0.03 | 0.32 | 0.00 | -0.20 | -0.16 | -0.25 | -0.06 | -0.22 | -0.15 | -0.23 | -0.16 | -0.18 | 0.13 | 0.22 | -0.11 | -0.35 | 0.16 | 0.07 | 0.04 | 0.08 | -0.12 | 0.09 |
| hr3508 | S | 4761 | 2.48 | 1.33 | -0.14 | -0.14 | -0.06 | 0.00 | 0.39 | -0.19 | -0.25 | -0.27 | -0.31 | -0.23 | -0.37 | -0.29 | -0.32 | -0.24 | -0.15 | 0.25 | 0.16 | -0.21 | -0.55 | -0.11 | -0.05 | -0.12 | -0.07 | -0.12 | -0.06 |
| hr3518 | S | 4322 | 1.95 | 1.47 | 0.21 | 0.09 | 0.20 | 0.35 | 1.08 | 0.06 | 0.04 | 0.07 | 0.20 | 0.14 | 0.10 | 0.02 | 0.04 | 0.13 | 0.39 | 1.49 | 0.15 | 0.16 | -0.46 | 0.11 | 0.63 | 0.18 | 0.45 | 0.70 | 0.09 |
| hr3529 | S | 4552 | 2.46 | 1.28 | 0.26 | 0.24 | 0.35 | 0.34 | 1.03 | 0.17 | 0.10 | 0.15 | 0.32 | 0.25 | 0.24 | 0.11 | 0.12 | 0.23 | 0.33 | 0.54 | 0.17 | 0.21 | -0.27 | 0.07 | 0.51 | 0.22 | 0.38 | 0.73 | 0.16 |
| hr3531 | S | 4989 | 2.71 | 1.27 | 0.04 | -0.18 | 0.01 | 0.01 | 0.28 | -0.07 | -0.17 | -0.14 | -0.15 | -0.05 | -0.12 | -0.14 | -0.19 | -0.15 | -0.18 | 0.11 | 0.24 | -0.11 | -0.30 | 0.16 | 0.13 | 0.19 | 0.15 | 0.15 | 0.03 |
| hr3547 | S | 4795 | 2.28 | 1.52 | 0.22 | -0.04 | 0.16 | 0.21 | 0.65 | 0.02 | -0.13 | -0.09 | -0.11 | 0.02 | -0.08 | -0.07 | -0.13 | -0.05 | -0.03 | 0.61 | 0.28 | -0.08 | -0.37 | 0.21 | 0.27 | 0.14 | 0.16 | 0.12 | 0.01 |
| hr3575 | S | 4974 | 2.63 | 1.37 | 0.24 | 0.02 | 0.14 | 0.15 | 0.54 | 0.05 | -0.07 | -0.04 | -0.12 | 0.03 | -0.03 | -0.04 | -0.13 | -0.06 | -0.08 | 0.19 | 0.28 | 0.05 | -0.19 | 0.22 | 0.32 | 0.16 | 0.21 | 0.06 | 0.11 |
| hr3621 | S | 4965 | 2.76 | 1.31 | 0.17 | -0.07 | 0.09 | 0.13 | 0.57 | 0.07 | -0.07 | -0.08 | -0.13 | 0.04 | -0.09 | -0.06 | -0.14 | -0.06 | -0.10 | 0.08 | 0.23 | 0.06 | -0.30 | 0.23 | 0.34 | 0.17 | 0.12 | 0.14 | 0.12 |
| hr3640 | S | 5076 | 2.86 | 1.33 | 0.13 | 0.08 | 0.12 | 0.10 | 0.44 | 0.01 | -0.05 | -0.05 | -0.07 | 0.02 | -0.13 | -0.06 | -0.10 | -0.07 | -0.17 | -0.04 | 0.55 | 0.05 | -0.22 | 0.24 | 0.26 | 0.22 | 0.24 | 0.04 | 0.18 |
| hr3653 | S | 4825 | 2.58 | 1.32 | -0.09 | -0.01 | -0.01 | 0.01 | 0.31 | -0.12 | -0.19 | -0.16 | -0.22 | -0.16 | -0.25 | -0.23 | -0.25 | -0.20 | -0.09 | 0.05 | 0.04 | -0.26 | -0.50 | -0.09 | 0.09 | -0.13 | -0.04 | -0.06 | -0.07 |
| hr3664 | S | 5019 | 2.61 | 1.55 | -0.59 | -0.43 | -0.30 | -0.32 | -0.16 | -0.52 | -0.41 | -0.50 | -0.64 | -0.58 | -0.85 | -0.64 | -0.55 | -0.62 | -0.69 | -0.28 | -0.08 | -0.64 | -0.65 | -0.54 | -0.46 | -0.42 | -0.38 | -0.51 | -0.30 |
| hr3681 | S | 4500 | 2.38 | 1.31 | 0.31 | 0.16 | 0.31 | 0.37 | 1.15 | 0.14 | 0.15 | 0.15 | 0.35 | 0.26 | 0.20 | 0.14 | 0.14 | 0.26 | 0.42 | 0.85 | 0.21 | 0.29 | -0.28 | 0.15 | 0.70 | 0.21 | 0.54 | 0.95 | 0.28 |
| hr3687 | S | 4665 | 2.44 | 1.28 | -0.13 | -0.20 | -0.02 | -0.04 | 0.44 | -0.18 | -0.26 | -0.20 | -0.12 | -0.24 | -0.31 | -0.28 | -0.23 | -0.10 | -0.01 | 0.08 | -0.27 | -0.58 | -0.22 | 0.03 | -0.15 | -0.04 | 0.03 | -0.15 |
| hr3706 | S | 5003 | 2.49 | 1.59 | 0.25 | 0.00 | 0.12 | 0.15 | 0.47 | 0.05 | -0.02 | -0.05 | -0.09 | 0.03 | -0.10 | -0.05 | -0.09 | -0.06 | -0.17 | 0.26 | 0.27 | 0.04 | -0.21 | 0.26 | 0.25 | 0.16 | 0.21 | 0.08 | 0.21 |
| hr3707 | S | 4666 | 2.71 | 1.13 | 0.36 | 0.30 | 0.41 | 0.41 | 0.99 | 0.28 | 0.22 | 0.30 | 0.45 | 0.33 | 0.28 | 0.22 | 0.25 | 0.37 | 0.71 | 0.83 | 0.27 | 0.33 | -0.18 | 0.21 | 0.80 | 0.37 | 0.57 | 1.03 | 0.40 |
| hr3709 | S | 4965 | 2.74 | 1.34 | 0.25 | 0.14 | 0.18 | 0.21 | 0.63 | 0.15 | -0.02 | 0.01 | -0.01 | 0.11 | -0.02 | 0.03 | -0.05 | 0.04 | 0.03 | 0.50 | 0.24 | 0.14 | -0.25 | 0.25 | 0.23 | 0.09 | 0.21 | 0.16 | 0.16 |
| hr3731 | S | 4402 | 2.17 | 1.43 | 0.24 | 0.10 | 0.27 | 0.36 | 0.95 | 0.05 | 0.08 | 0.09 | 0.28 | 0.17 | 0.10 | 0.03 | 0.10 | 0.17 | 0.49 | 1.40 | 0.20 | -0.02 | -0.40 | 0.01 | 0.50 | 0.17 | 0.39 | 0.77 | 0.13 |
| hr3733 | S | 4959 | 2.79 | 1.31 | 0.09 | -0.02 | 0.12 | 0.11 | 0.40 | -0.02 | -0.07 | -0.09 | -0.13 | -0.03 | -0.08 | -0.09 | -0.13 | -0.07 | -0.05 | 0.18 | 0.33 | 0.06 | -0.18 | 0.15 | 0.21 | 0.15 | 0.18 | 0.13 | 0.10 |
| hr3748 | S | 4097 | 1.33 | 1.91 | 0.46 | 0.04 | 0.26 | 0.54 | 1.51 | -0.03 | -0.13 | -0.06 | 0.10 | 0.17 | 0.10 | 0.04 | 0.04 | 0.12 | 0.55 | 1.27 | 0.26 | 0.06 | -0.42 | 0.20 | 0.49 | 0.19 | 0.43 | 0.73 | 0.10 |
| hr3772 | S | 4426 | 2.56 | 1.19 | 0.49 | 0.48 | 0.66 | 0.62 | 1.51 | 0.43 | 0.39 | 0.35 | 0.70 | 0.45 | 0.42 | 0.35 | 0.44 | 0.54 | | 0.93 | 0.20 | 0.38 | -0.16 | 0.18 | 1.08 | 0.41 | 0.68 | 1.51 | 0.46 |
| hr3788 | S | 4419 | 2.31 | 1.23 | -0.02 | 0.04 | 0.18 | 0.24 | 0.89 | -0.01 | 0.01 | 0.04 | 0.13 | 0.05 | -0.11 | -0.03 | -0.01 | 0.07 | 0.21 | 0.63 | 0.06 | 0.12 | -0.47 | 0.09 | 0.44 | 0.21 | 0.44 | 0.79 | 0.26 |
| hr3791 | S | 4402 | 2.12 | 1.40 | 0.35 | 0.23 | 0.41 | 0.42 | 1.17 | 0.19 | 0.17 | 0.23 | 0.43 | 0.32 | 0.27 | 0.13 | 0.18 | 0.27 | 0.52 | 1.14 | 0.38 | 0.24 | -0.25 | 0.14 | 0.74 | 0.24 | 0.52 | 0.89 | 0.23 |
| hr3800 | S | 5003 | 2.81 | 1.43 | 0.29 | 0.01 | 0.19 | 0.19 | 0.50 | 0.09 | 0.01 | 0.02 | 0.00 | 0.12 | -0.05 | 0.01 | -0.06 | 0.01 | -0.16 | 0.25 | 0.20 | 0.05 | -0.20 | 0.28 | 0.25 | 0.24 | 0.15 | 0.20 | 0.16 |
| hr3801 | S | 4861 | 2.62 | 1.37 | 0.06 | -0.08 | 0.08 | 0.10 | 0.44 | -0.02 | -0.08 | -0.09 | -0.10 | -0.03 | -0.14 | -0.10 | -0.13 | -0.08 | -0.04 | 0.09 | 0.15 | -0.04 | -0.32 | 0.13 | 0.19 | 0.13 | 0.19 | 0.16 | 0.08 |
| hr3805 | S | 4490 | 2.35 | 1.28 | 0.42 | 0.29 | 0.44 | 0.42 | 1.08 | 0.21 | 0.14 | 0.21 | 0.37 | 0.29 | 0.19 | 0.14 | 0.18 | 0.29 | 0.47 | 1.01 | 0.18 | 0.17 | -0.25 | 0.05 | 0.64 | 0.21 | 0.43 | 0.69 | 0.15 |
| hr3808 | S | 4883 | 2.46 | 1.72 | 0.77 | 0.31 | 0.37 | 0.51 | 1.09 | 0.21 | 0.17 | 0.22 | 0.23 | 0.38 | 0.27 | 0.24 | 0.23 | 0.32 | 0.25 | 0.81 | 0.40 | 0.21 | -0.05 | 0.25 | 0.64 | 0.27 | 0.33 | 0.85 | 0.45 |
| hr3809 | S | 4808 | 2.55 | 1.36 | 0.01 | -0.09 | 0.09 | 0.12 | 0.50 | -0.01 | -0.09 | -0.08 | -0.13 | -0.04 | -0.13 | -0.11 | -0.11 | -0.07 | 0.02 | 0.25 | 0.11 | -0.01 | -0.36 | 0.12 | 0.22 | 0.11 | 0.19 | 0.17 | 0.04 |
| hr3827 | S | 4708 | 2.47 | 1.28 | 0.20 | 0.13 | 0.20 | 0.22 | 0.72 | 0.09 | -0.03 | 0.05 | 0.13 | 0.13 | 0.05 | 0.02 | -0.04 | 0.08 | 0.23 | 0.30 | 0.21 | 0.17 | -0.25 | 0.16 | 0.41 | 0.20 | 0.32 | 0.40 | 0.09 |
| hr3834 | S | 4188 | 1.71 | 1.47 | -0.07 | -0.10 | 0.09 | 0.16 | 1.15 | -0.24 | -0.15 | -0.11 | 0.07 | -0.12 | -0.25 | -0.26 | -0.13 | -0.11 | 0.20 | 1.47 | -0.02 | -0.15 | -0.71 | -0.35 | 0.12 | -0.24 | 0.13 | 0.54 | -0.20 |
| hr3845 | S | 4244 | 1.71 | 1.58 | 0.18 | 0.04 | 0.12 | 0.34 | 1.27 | -0.09 | -0.13 | -0.06 | 0.11 | 0.08 | -0.05 | -0.04 | -0.01 | 0.05 | 0.24 | | 0.17 | 0.06 | -0.51 | 0.16 | 0.57 | 0.20 | 0.43 | 0.57 | 0.03 |
| hr3851 | S | 4945 | 2.65 | 1.43 | 0.33 | -0.10 | 0.20 | 0.25 | 0.60 | 0.14 | -0.01 | 0.03 | 0.01 | 0.14 | 0.10 | 0.04 | -0.04 | 0.07 | 0.10 | 0.49 | 0.25 | 0.17 | -0.13 | 0.20 | 0.31 | 0.15 | 0.20 | 0.29 | 0.17 |
| hr3903 | S | 4955 | 2.42 | 1.55 | 0.30 | 0.05 | 0.17 | 0.20 | 0.57 | 0.07 | -0.08 | -0.04 | -0.05 | 0.10 | -0.02 | -0.02 | -0.05 | -0.02 | -0.16 | 0.51 | 0.29 | 0.06 | -0.22 | 0.21 | 0.22 | 0.05 | 0.11 | 0.06 | 0.14 |
| hr3905 | S | 4471 | 2.45 | 1.57 | 0.83 | 0.67 | 0.66 | 0.81 | 1.83 | 0.41 | 0.50 | 0.48 | 0.70 | 0.56 | 0.59 | 0.44 | 0.57 | 0.63 | 0.84 | 2.24 | 0.34 | 0.35 | -0.01 | 0.04 | 1.04 | 0.56 | 0.74 | 1.41 | 0.65 |
| hr3907 | S | 4988 | 2.73 | 1.47 | 0.38 | 0.15 | 0.25 | 0.24 | 0.75 | 0.16 | 0.02 | 0.06 | 0.09 | 0.17 | 0.07 | 0.05 | 0.04 | 0.07 | -0.09 | 0.43 | 0.26 | 0.09 | -0.13 | 0.17 | 0.25 | 0.12 | 0.09 | 0.26 | 0.24 |
| hr3908 | S | 4779 | 2.56 | 1.40 | 0.39 | 0.23 | 0.42 | 0.32 | 0.83 | 0.20 | 0.11 | 0.12 | 0.22 | 0.21 | 0.19 | 0.10 | 0.10 | 0.17 | 0.32 | 0.36 | 0.21 | 0.13 | -0.23 | 0.05 | 0.43 | 0.09 | 0.23 | 0.37 | 0.18 |
| hr3911 | S | 4868 | 2.71 | 1.42 | 0.58 | 0.32 | 0.53 | 0.42 | 0.96 | 0.36 | 0.27 | 0.36 | 0.46 | 0.41 | 0.44 | 0.27 | 0.30 | 0.37 | 0.70 | 0.50 | 0.39 | 0.42 | -0.01 | 0.21 | 0.76 | 0.33 | 0.45 | 0.95 | 0.30 |
| hr3929 | S | 4630 | 2.54 | 1.30 | 0.42 | 0.29 | 0.41 | 0.52 | 1.21 | 0.23 | 0.20 | 0.20 | 0.30 | 0.30 | 0.33 | 0.25 | 0.23 | 0.40 | 0.65 | 0.92 | 0.24 | 0.19 | -0.17 | 0.27 | 0.62 | 0.36 | 0.42 | 0.90 | 0.21 |

| | | | | | | | | | | | | | | | | | | | | | | | | | | | | |
|---|---|---|---|---|---|---|---|---|---|---|---|---|---|---|---|---|---|---|---|---|---|---|---|---|---|---|---|---|
| hr3942 | S | 4702 | 2.48 | 1.35 | 0.21 | 0.20 | 0.25 | 0.32 | 0.94 | 0.14 | -0.02 | 0.06 | 0.10 | 0.15 | 0.11 | 0.05 | 0.03 | 0.15 | 0.32 | 0.45 | 0.16 | 0.08 | -0.28 | 0.14 | 0.38 | 0.08 | 0.23 | 0.38 | 0.11 |
| hr3973 | S | 4958 | 2.89 | 1.31 | 0.32 | 0.13 | 0.17 | 0.24 | 0.70 | 0.20 | 0.03 | 0.09 | 0.08 | 0.19 | 0.09 | 0.08 | 0.03 | 0.12 | 0.15 | 0.77 | 0.26 | 0.25 | -0.16 | 0.25 | 0.36 | 0.19 | 0.27 | 0.33 | 0.32 |
| hr3994 | S | 4851 | 2.74 | 1.35 | 0.61 | 0.31 | 0.40 | 0.50 | 0.94 | 0.31 | 0.21 | 0.25 | 0.30 | 0.36 | 0.39 | 0.25 | 0.26 | 0.37 | 0.54 | 1.34 | 0.34 | 0.24 | -0.07 | 0.14 | 0.50 | 0.26 | 0.32 | 0.73 | 0.35 |
| hr4006 | S | 5169 | 2.69 | 1.65 | 0.30 | 0.24 | 0.16 | 0.09 | 0.18 | 0.00 | -0.05 | 0.01 | -0.09 | -0.02 | -0.19 | -0.10 | -0.08 | -0.10 | -0.27 | 0.00 | | -0.05 | -0.15 | 0.14 | 0.17 | 0.06 | 0.17 | -0.08 | 0.07 |
| hr4032 | S | 4388 | 2.17 | 1.25 | 0.23 | 0.16 | 0.31 | 0.37 | 1.14 | 0.02 | 0.05 | 0.07 | 0.32 | 0.16 | 0.04 | 0.06 | 0.09 | 0.20 | | 1.14 | 0.02 | 0.11 | -0.48 | -0.02 | 0.49 | -0.02 | 0.41 | 1.06 | 0.13 |
| hr4052 | S | 4520 | 2.11 | 1.43 | 0.04 | 0.01 | 0.06 | 0.13 | 0.56 | -0.14 | -0.13 | -0.13 | -0.09 | -0.11 | -0.19 | -0.18 | -0.11 | -0.07 | 0.26 | 0.14 | 0.01 | -0.21 | -0.60 | -0.05 | 0.20 | -0.03 | 0.11 | 0.30 | 0.03 |
| hr4057 | S | 4304 | 1.68 | 1.58 | -0.24 | -0.12 | -0.05 | 0.03 | 0.74 | -0.41 | -0.40 | -0.41 | -0.39 | -0.33 | -0.49 | -0.37 | -0.35 | -0.27 | -0.01 | 0.41 | -0.26 | -0.36 | -0.86 | -0.25 | -0.13 | -0.24 | -0.10 | -0.08 | -0.06 |
| hr4077 | S | 4652 | 2.85 | 1.09 | 0.46 | 0.37 | 0.54 | 0.46 | 1.27 | 0.37 | 0.34 | 0.41 | 0.63 | 0.42 | 0.47 | 0.32 | 0.38 | 0.50 | 0.66 | 1.54 | 0.33 | 0.47 | -0.15 | 0.24 | 1.11 | 0.49 | 0.76 | 1.26 | 0.56 |
| hr4078 | S | 4583 | 2.54 | 1.22 | -0.02 | 0.00 | 0.11 | 0.17 | 0.82 | -0.04 | -0.06 | -0.08 | -0.01 | -0.06 | -0.04 | -0.10 | -0.04 | 0.02 | 0.24 | 0.59 | -0.06 | -0.01 | -0.40 | -0.02 | 0.27 | 0.13 | 0.30 | 0.62 | -0.05 |
| hr4085 | S | 4803 | 2.86 | 1.15 | -0.12 | -0.01 | 0.11 | 0.02 | 0.37 | -0.07 | -0.08 | -0.07 | -0.03 | -0.11 | -0.12 | -0.16 | -0.11 | -0.09 | 0.06 | 0.18 | -0.06 | -0.20 | -0.39 | -0.12 | 0.14 | 0.07 | 0.16 | 0.40 | -0.02 |
| hr4097 | S | 4509 | 2.46 | 1.41 | 0.35 | 0.32 | 0.39 | 0.59 | 1.36 | 0.19 | 0.21 | 0.20 | 0.39 | 0.29 | 0.32 | 0.26 | 0.30 | 0.45 | | 1.59 | 0.17 | 0.19 | -0.36 | 0.17 | 0.86 | 0.34 | 0.49 | 1.11 | 0.43 |
| hr4100 | S | 4986 | 2.76 | 1.24 | 0.54 | 0.23 | 0.45 | 0.35 | 0.80 | 0.34 | 0.19 | 0.25 | 0.37 | 0.32 | 0.31 | 0.20 | 0.22 | 0.26 | 0.36 | 0.50 | 0.35 | 0.31 | 0.02 | 0.17 | 0.59 | 0.13 | 0.32 | 0.50 | 0.36 |
| hr4106 | S | 4635 | 2.64 | 1.39 | 0.32 | 0.28 | 0.40 | 0.51 | 1.08 | 0.19 | 0.23 | 0.28 | 0.42 | 0.30 | 0.33 | 0.25 | 0.29 | 0.41 | 0.43 | 0.76 | 0.18 | 0.21 | -0.16 | 0.27 | 0.75 | 0.35 | 0.49 | 1.06 | 0.43 |
| hr4126 | S | 4965 | 2.71 | 1.38 | 0.39 | 0.07 | 0.20 | 0.26 | 0.67 | 0.19 | 0.06 | 0.07 | 0.07 | 0.16 | 0.11 | 0.07 | 0.02 | 0.10 | 0.19 | 0.66 | 0.46 | 0.12 | -0.15 | 0.20 | 0.33 | 0.14 | 0.24 | 0.30 | 0.22 |
| hr4146 | S | 4900 | 2.55 | 1.41 | 0.03 | -0.11 | 0.00 | 0.04 | 0.14 | -0.09 | -0.19 | -0.17 | -0.21 | -0.11 | -0.23 | -0.17 | -0.22 | -0.17 | -0.19 | 0.09 | 0.36 | -0.11 | -0.43 | 0.11 | 0.19 | 0.13 | 0.11 | -0.03 | 0.06 |
| hr4171 | S | 4952 | 2.79 | 1.34 | 0.19 | 0.07 | 0.18 | 0.19 | 0.46 | 0.07 | 0.02 | 0.00 | -0.04 | 0.08 | 0.00 | 0.00 | -0.03 | 0.05 | 0.11 | 0.51 | 0.18 | 0.12 | -0.25 | 0.17 | 0.29 | 0.12 | 0.20 | 0.18 | 0.19 |
| hr4178 | S | 4425 | 2.34 | 1.43 | 0.43 | 0.47 | 0.43 | 0.63 | 1.56 | 0.23 | 0.24 | 0.31 | 0.51 | 0.37 | 0.38 | 0.32 | 0.37 | 0.52 | 0.66 | 1.15 | 0.18 | 0.34 | -0.36 | 0.20 | 1.07 | 0.37 | 0.55 | 1.21 | 0.45 |
| hr4208 | S | 4612 | 2.78 | 1.24 | 0.49 | 0.36 | 0.54 | 0.52 | 1.26 | 0.36 | 0.31 | 0.37 | 0.58 | 0.42 | 0.39 | 0.29 | 0.35 | 0.47 | 0.62 | 1.68 | 0.26 | 0.40 | -0.05 | 0.14 | 0.92 | 0.45 | 0.64 | 1.25 | 0.44 |
| hr4209 | S | 4985 | 2.77 | 1.27 | 0.16 | -0.03 | 0.15 | 0.17 | 0.56 | 0.09 | -0.05 | -0.04 | -0.05 | 0.05 | -0.03 | -0.02 | -0.09 | -0.02 | -0.09 | 0.19 | 0.22 | 0.08 | -0.19 | 0.21 | 0.27 | 0.20 | 0.22 | 0.25 | 0.19 |
| hr4232 | S | 4340 | 1.91 | 1.57 | -0.02 | 0.05 | 0.11 | 0.28 | 0.79 | -0.19 | -0.07 | -0.07 | 0.01 | -0.03 | -0.08 | -0.11 | -0.03 | 0.04 | 0.14 | 0.78 | -0.02 | -0.06 | -0.59 | -0.09 | 0.28 | 0.04 | 0.17 | 0.52 | 0.16 |
| hr4233 | S | 4675 | 2.46 | 1.39 | 0.06 | 0.09 | 0.19 | 0.24 | 0.77 | 0.03 | -0.09 | -0.01 | 0.03 | 0.03 | 0.05 | -0.02 | -0.01 | 0.08 | 0.23 | 0.52 | 0.04 | -0.10 | -0.46 | 0.03 | 0.25 | 0.08 | 0.19 | 0.54 | 0.09 |
| hr4235 | S | 4526 | 2.32 | 1.34 | 0.05 | 0.03 | 0.10 | 0.22 | 0.77 | -0.05 | -0.09 | -0.06 | 0.01 | 0.02 | -0.05 | -0.06 | -0.07 | 0.02 | 0.15 | 0.37 | 0.06 | 0.11 | -0.46 | 0.09 | 0.38 | 0.13 | 0.30 | 0.63 | 0.14 |
| hr4242 | S | 4736 | 2.83 | 1.55 | 0.22 | 0.22 | 0.35 | 0.45 | 0.85 | 0.11 | 0.25 | 0.26 | 0.32 | 0.33 | 0.31 | 0.23 | 0.28 | 0.35 | | 0.95 | 0.28 | 0.31 | -0.10 | 0.33 | 0.77 | 0.48 | 0.61 | 1.25 | 0.53 |
| hr4253 | S | 4922 | 2.67 | 1.41 | 0.39 | 0.15 | 0.33 | 0.29 | 0.86 | 0.24 | 0.03 | 0.11 | 0.10 | 0.21 | 0.17 | 0.11 | 0.05 | 0.16 | 0.09 | 0.55 | 0.29 | 0.20 | -0.09 | 0.22 | 0.39 | 0.25 | 0.31 | 0.38 | 0.24 |
| hr4255 | S | 5193 | 2.63 | 1.61 | 0.23 | -0.05 | 0.09 | 0.05 | 0.21 | -0.02 | -0.04 | -0.12 | -0.19 | -0.04 | -0.22 | -0.12 | -0.17 | -0.15 | -0.34 | -0.15 | 0.54 | 0.01 | -0.16 | 0.25 | 0.15 | 0.19 | 0.12 | 0.07 | 0.15 |
| hr4256 | S | 4673 | 2.54 | 1.27 | 0.01 | -0.04 | 0.04 | 0.15 | 0.66 | -0.05 | -0.14 | -0.14 | -0.13 | -0.05 | -0.08 | -0.11 | -0.13 | -0.03 | 0.05 | 0.40 | 0.04 | -0.07 | -0.44 | -0.01 | 0.19 | 0.06 | 0.19 | 0.41 | -0.07 |
| hr4258 | S | 4569 | 2.39 | 1.31 | 0.09 | 0.10 | 0.13 | 0.21 | 0.74 | -0.04 | -0.09 | -0.06 | 0.00 | 0.02 | -0.04 | -0.06 | -0.05 | 0.01 | 0.28 | 0.30 | 0.13 | -0.09 | -0.40 | 0.08 | 0.48 | 0.12 | 0.30 | 0.35 | 0.00 |
| hr4264 | S | 4524 | 2.52 | 1.20 | 0.12 | 0.09 | 0.23 | 0.26 | 0.69 | 0.02 | 0.03 | 0.04 | 0.15 | 0.09 | 0.06 | 0.00 | 0.04 | 0.14 | 0.40 | 0.15 | 0.09 | 0.06 | -0.48 | 0.03 | 0.56 | 0.15 | 0.37 | 0.59 | 0.24 |
| hr4283 | S | 4892 | 2.53 | 1.53 | 0.56 | 0.26 | 0.36 | 0.44 | 0.99 | 0.27 | 0.14 | 0.20 | 0.24 | 0.29 | 0.28 | 0.21 | 0.17 | 0.30 | 0.38 | 0.85 | 0.35 | 0.34 | -0.02 | 0.28 | 0.59 | 0.26 | 0.32 | 0.55 | 0.30 |
| hr4287 | S | 4645 | 2.42 | 1.37 | 0.17 | 0.14 | 0.25 | 0.29 | 0.80 | 0.08 | 0.00 | 0.05 | 0.10 | 0.14 | 0.13 | 0.04 | 0.04 | 0.15 | 0.34 | 0.86 | 0.24 | 0.05 | -0.30 | 0.12 | 0.43 | 0.12 | 0.28 | 0.66 | 0.09 |
| hr4301 | S | 4636 | 1.96 | 1.57 | 0.30 | 0.15 | 0.16 | 0.23 | 0.66 | 0.05 | -0.11 | -0.05 | 0.01 | 0.05 | -0.03 | -0.05 | -0.08 | -0.02 | -0.03 | 0.99 | 0.23 | -0.06 | -0.36 | 0.18 | 0.35 | 0.05 | 0.17 | 0.24 | 0.05 |
| hr4305 | S | 4967 | 2.67 | 1.36 | 0.40 | 0.12 | 0.28 | 0.20 | 0.51 | 0.12 | -0.01 | 0.05 | 0.06 | 0.06 | -0.02 | -0.01 | 0.01 | 0.04 | 0.17 | 0.35 | 0.14 | 0.05 | -0.24 | 0.00 | 0.23 | -0.04 | 0.11 | 0.09 | 0.14 |
| hr4335 | S | 4523 | 2.05 | 1.50 | 0.23 | 0.09 | 0.18 | 0.28 | 0.88 | 0.07 | -0.12 | -0.03 | 0.01 | 0.10 | -0.02 | -0.01 | -0.05 | 0.05 | 0.21 | 0.78 | 0.23 | 0.10 | -0.30 | 0.11 | 0.45 | 0.15 | 0.25 | 0.39 | 0.15 |
| hr4351 | S | 4556 | 2.48 | 1.37 | 0.35 | 0.32 | 0.43 | 0.48 | 1.22 | 0.26 | 0.19 | 0.23 | 0.41 | 0.30 | 0.34 | 0.23 | 0.26 | 0.40 | | 1.11 | 0.16 | 0.21 | -0.27 | 0.20 | 0.75 | 0.30 | 0.50 | 1.05 | 0.34 |
| hr4382 | S | 4510 | 1.99 | 1.55 | -0.25 | -0.25 | -0.15 | -0.04 | 0.32 | -0.35 | -0.39 | -0.41 | -0.40 | -0.39 | -0.52 | -0.44 | -0.37 | -0.36 | -0.18 | 0.24 | -0.22 | -0.47 | -0.77 | -0.33 | -0.25 | -0.37 | -0.22 | -0.25 | -0.30 |
| hr4383 | S | 4825 | 2.57 | 1.31 | 0.45 | 0.20 | 0.45 | 0.38 | 0.94 | 0.29 | 0.22 | 0.30 | 0.47 | 0.34 | 0.35 | 0.19 | 0.24 | 0.31 | 0.41 | 0.62 | 0.38 | 0.22 | 0.03 | 0.19 | 0.67 | 0.25 | 0.37 | 0.74 | 0.29 |
| hr4400 | S | 4984 | 2.65 | 1.50 | 0.20 | -0.02 | 0.09 | 0.16 | 0.55 | 0.08 | -0.01 | -0.04 | -0.09 | 0.05 | -0.08 | -0.04 | -0.10 | -0.04 | -0.05 | -0.05 | 0.22 | 0.09 | -0.26 | 0.26 | 0.30 | 0.26 | 0.25 | 0.32 | 0.21 |
| hr4407 | S | 4836 | 2.59 | 1.35 | 0.07 | 0.12 | 0.14 | 0.14 | 0.58 | 0.01 | -0.05 | -0.03 | -0.08 | -0.02 | -0.10 | -0.08 | -0.12 | -0.05 | 0.01 | 0.62 | 0.19 | 0.09 | -0.30 | 0.18 | 0.26 | 0.14 | 0.21 | 0.15 | 0.06 |
| hr4419 | S | 4638 | 2.29 | 1.31 | 0.23 | 0.03 | 0.29 | 0.17 | 0.68 | 0.11 | 0.03 | 0.15 | 0.31 | 0.18 | 0.02 | -0.03 | 0.01 | 0.05 | 0.01 | 0.27 | 0.27 | 0.21 | -0.03 | 0.24 | 0.56 | 0.19 | 0.37 | 0.57 | 0.16 |
| hr4433 | S | 4678 | 2.38 | 1.34 | 0.05 | 0.05 | 0.07 | 0.10 | 0.45 | -0.02 | -0.13 | -0.08 | -0.06 | -0.02 | -0.10 | -0.10 | -0.13 | -0.07 | 0.03 | -0.04 | 0.08 | -0.08 | -0.38 | 0.11 | 0.42 | 0.13 | 0.23 | 0.24 | -0.06 |
| hr4452 | S | 4648 | 2.36 | 1.37 | 0.06 | 0.12 | 0.25 | 0.24 | 0.73 | 0.00 | -0.08 | 0.02 | 0.02 | -0.05 | -0.15 | -0.09 | -0.02 | 0.02 | 0.28 | 0.41 | -0.04 | -0.17 | -0.58 | -0.07 | 0.12 | -0.08 | 0.10 | 0.51 | 0.11 |
| hr4459 | S | 4814 | 2.69 | 1.30 | 0.29 | 0.15 | 0.32 | 0.31 | 0.91 | 0.17 | 0.10 | 0.13 | 0.17 | 0.18 | 0.16 | 0.11 | 0.05 | 0.18 | 0.21 | 0.89 | 0.27 | 0.20 | -0.16 | 0.20 | 0.50 | 0.27 | 0.35 | 0.63 | 0.16 |
| hr4461 | S | 4802 | 2.54 | 1.34 | -0.12 | -0.18 | -0.06 | -0.01 | 0.39 | -0.17 | -0.26 | -0.26 | -0.28 | -0.22 | -0.30 | -0.28 | -0.30 | -0.24 | -0.17 | 0.00 | 0.15 | -0.30 | -0.56 | -0.17 | -0.15 | -0.16 | -0.14 | 0.05 | -0.12 |
| hr4471 | S | 4786 | 2.47 | 1.36 | -0.01 | -0.12 | 0.04 | 0.13 | 0.45 | -0.05 | -0.16 | -0.14 | -0.14 | -0.08 | -0.19 | -0.13 | -0.16 | -0.10 | -0.10 | 0.21 | 0.07 | -0.02 | -0.56 | 0.10 | 0.17 | 0.16 | 0.15 | 0.18 | 0.05 |
| hr4474 | S | 4718 | 2.42 | 1.41 | 0.00 | -0.02 | 0.14 | 0.37 | 1.04 | -0.05 | -0.10 | -0.11 | -0.21 | 0.00 | -0.26 | -0.11 | -0.16 | 0.02 | 0.55 | 0.98 | 0.94 | 0.83 | 0.45 | | 1.37 | 1.09 | 1.09 | 1.03 | 0.17 |
| hr4478 | S | 4691 | 2.60 | 1.39 | 0.47 | 0.29 | 0.37 | 0.51 | 1.18 | 0.28 | 0.25 | 0.23 | 0.36 | 0.34 | 0.37 | 0.26 | 0.27 | 0.39 | 0.36 | 0.63 | 0.29 | 0.30 | -0.19 | 0.25 | 0.71 | 0.34 | 0.42 | 0.87 | 0.34 |
| hr4480 | S | 6672 | 3.52 | 4.72 | 0.17 | | | 0.07 | 0.42 | 0.04 | 0.05 | 0.36 | 0.48 | 0.23 | 0.03 | -0.07 | 0.49 | 0.02 | -0.43 | 0.17 | 1.88 | 0.28 | 1.21 | -0.06 | | 0.33 | 0.36 | | | |
| hr4495 | S | 4863 | 2.61 | 1.36 | 0.25 | -0.08 | 0.29 | 0.22 | 0.41 | 0.19 | 0.01 | 0.05 | 0.06 | 0.14 | 0.09 | 0.04 | -0.01 | 0.06 | 0.14 | 0.37 | 0.31 | 0.06 | -0.20 | 0.12 | 0.35 | 0.06 | 0.23 | 0.27 | 0.11 |
| hr4510 | S | 4914 | 2.69 | 1.31 | 0.14 | 0.03 | 0.17 | 0.17 | 0.49 | 0.09 | -0.05 | 0.00 | 0.00 | 0.07 | -0.04 | -0.02 | -0.08 | 0.00 | -0.02 | 0.06 | 0.25 | 0.13 | -0.21 | 0.20 | 0.28 | 0.16 | 0.26 | 0.35 | 0.15 |
| hr4518 | S | 4392 | 1.95 | 1.56 | -0.25 | -0.10 | -0.08 | 0.06 | 0.65 | -0.37 | -0.32 | -0.28 | -0.26 | -0.30 | -0.41 | -0.35 | -0.28 | -0.25 | -0.09 | 0.31 | -0.32 | -0.45 | -0.78 | -0.35 | -0.08 | -0.31 | -0.15 | -0.07 | -0.12 |

| ID | Sp | col3 | col4 | col5 | col6 | col7 | col8 | col9 | col10 | col11 | col12 | col13 | col14 | col15 | col16 | col17 | col18 | col19 | col20 | col21 | col22 | col23 | col24 | col25 | col26 | col27 | col28 | col29 |
|---|---|---|---|---|---|---|---|---|---|---|---|---|---|---|---|---|---|---|---|---|---|---|---|---|---|---|---|---|
| hr4521 | S | 4411 | 2.42 | 1.69 | 0.77 | 0.70 | 0.55 | 0.84 | 2.05 | 0.36 | 0.39 | 0.42 | 0.65 | 0.58 | 0.62 | 0.44 | 0.55 | 0.60 | 0.88 | 2.46 | 0.35 | 0.40 | -0.06 | 0.01 | 0.86 | 0.61 | 0.73 | 1.47 | 0.76 |
| hr4544 | S | 4668 | 2.50 | 1.22 | 0.07 | 0.01 | 0.12 | 0.17 | 0.62 | 0.03 | -0.10 | -0.04 | 0.01 | 0.02 | -0.07 | -0.06 | -0.08 | 0.00 | 0.12 | 0.37 | 0.08 | -0.06 | -0.42 | 0.08 | 0.28 | 0.14 | 0.27 | 0.53 | 0.06 |
| hr4558 | S | 4970 | 2.70 | 1.45 | -0.31 | -0.41 | -0.29 | -0.28 | -0.26 | -0.41 | -0.34 | -0.45 | -0.51 | -0.47 | -0.62 | -0.52 | -0.45 | -0.49 | -0.47 | -0.36 | 0.09 | -0.54 | -0.63 | -0.32 | -0.34 | -0.33 | -0.25 | -0.40 | -0.27 |
| hr4566 | S | 4565 | 2.45 | 1.42 | 0.15 | 0.36 | 0.38 | 0.42 | 1.07 | 0.02 | 0.02 | 0.10 | 0.14 | 0.12 | -0.03 | 0.05 | 0.10 | 0.17 | 0.42 | 1.44 | 0.03 | 0.16 | -0.48 | -0.02 | 0.48 | 0.18 | 0.27 | 0.90 | 0.26 |
| hr4593 | S | 4670 | 2.40 | 1.36 | 0.10 | 0.12 | 0.17 | 0.25 | 0.73 | 0.02 | -0.03 | -0.05 | -0.03 | 0.00 | 0.00 | -0.05 | -0.04 | 0.04 | 0.28 | 0.62 | -0.04 | -0.16 | -0.48 | -0.01 | 0.24 | -0.03 | 0.09 | 0.38 | 0.01 |
| hr4608 | S | 4762 | 2.46 | 1.33 | -0.27 | -0.27 | -0.12 | -0.06 | 0.24 | -0.31 | -0.31 | -0.36 | -0.50 | -0.33 | -0.46 | -0.40 | -0.39 | -0.33 | -0.18 | 0.29 | 0.29 | -0.02 | -0.14 | 0.47 | 0.58 | 0.37 | 0.46 | 0.18 | -0.11 |
| hr4609 | S | 4677 | 2.50 | 1.43 | -0.12 | 0.03 | 0.14 | 0.13 | 0.48 | -0.10 | -0.14 | -0.05 | -0.04 | -0.20 | -0.30 | -0.26 | -0.15 | -0.14 | 0.10 | 0.43 | -0.07 | -0.30 | -0.62 | -0.39 | -0.04 | -0.21 | -0.10 | 0.17 | -0.04 |
| hr4610 | S | 4433 | 2.30 | 1.34 | 0.08 | 0.06 | 0.22 | 0.31 | 0.88 | -0.03 | 0.10 | 0.08 | 0.23 | 0.08 | -0.01 | -0.03 | 0.06 | 0.12 | | 0.96 | 0.08 | 0.07 | -0.48 | -0.03 | 0.41 | 0.19 | 0.45 | 0.78 | 0.35 |
| hr4626 | S | 4553 | 2.71 | 1.12 | 0.27 | 0.24 | 0.37 | 0.45 | 1.20 | 0.15 | 0.29 | 0.25 | 0.42 | 0.29 | 0.30 | 0.22 | 0.27 | 0.37 | 0.74 | 1.25 | 0.13 | 0.24 | -0.34 | 0.17 | 0.82 | 0.39 | 0.69 | 1.19 | 0.35 |
| hr4630 | S | 4285 | 1.22 | 1.98 | 0.25 | 0.02 | 0.12 | 0.45 | 1.25 | 0.00 | -0.17 | -0.09 | -0.05 | 0.14 | 0.07 | -0.01 | 0.00 | 0.09 | 0.12 | 1.47 | 0.23 | -0.17 | -0.48 | 0.02 | 0.32 | 0.01 | 0.15 | 0.56 | 0.08 |
| hr4654 | S | 4771 | 2.46 | 1.42 | 0.14 | 0.04 | 0.22 | 0.18 | 0.44 | 0.10 | -0.07 | 0.03 | 0.04 | 0.07 | -0.03 | -0.01 | -0.06 | 0.03 | 0.15 | 0.52 | 0.28 | 0.13 | -0.26 | 0.22 | 0.37 | 0.19 | 0.28 | 0.29 | 0.12 |
| hr4655 | S | 4765 | 2.77 | 1.32 | 0.63 | 0.41 | 0.45 | 0.59 | 1.22 | 0.33 | 0.25 | 0.30 | 0.39 | 0.38 | 0.46 | 0.32 | 0.32 | 0.48 | 0.70 | 1.52 | 0.34 | 0.35 | -0.04 | 0.21 | 0.90 | 0.29 | 0.40 | 0.84 | 0.40 |
| hr4667 | S | 4861 | 2.63 | 1.35 | 0.02 | -0.07 | 0.05 | 0.09 | 0.51 | -0.03 | -0.11 | -0.16 | -0.15 | -0.06 | -0.14 | -0.12 | -0.16 | -0.10 | -0.15 | 0.01 | 0.11 | -0.04 | -0.34 | 0.18 | 0.19 | 0.16 | 0.19 | 0.17 | 0.07 |
| hr4668 | S | 4463 | 2.08 | 1.39 | 0.02 | -0.07 | 0.05 | 0.13 | 0.76 | -0.10 | -0.25 | -0.16 | -0.09 | -0.11 | -0.19 | -0.22 | -0.20 | -0.11 | -0.03 | 0.75 | -0.09 | -0.30 | -0.61 | -0.24 | 0.07 | -0.19 | -0.03 | 0.20 | -0.16 |
| hr4695 | S | 4422 | 2.06 | 1.71 | -0.26 | -0.09 | -0.13 | 0.11 | 0.61 | -0.26 | -0.33 | -0.17 | -0.15 | -0.32 | -0.50 | -0.38 | -0.20 | -0.23 | -0.07 | 0.44 | -0.19 | -0.41 | -0.95 | -0.59 | -0.32 | -0.43 | -0.28 | 0.08 | -0.17 |
| hr4697 | S | 4711 | 2.27 | 1.43 | -0.02 | -0.27 | -0.06 | -0.06 | 0.29 | -0.16 | -0.30 | -0.29 | -0.32 | -0.29 | -0.39 | -0.34 | -0.33 | -0.29 | -0.20 | 0.26 | -0.10 | -0.38 | -0.68 | -0.24 | -0.18 | -0.25 | -0.19 | -0.26 | -0.24 |
| hr4699 | S | 4724 | 2.74 | 1.20 | 0.20 | 0.14 | 0.29 | 0.25 | 0.86 | 0.14 | 0.03 | 0.12 | 0.17 | 0.19 | 0.12 | 0.06 | 0.04 | 0.14 | 0.36 | 0.53 | 0.23 | 0.14 | -0.26 | 0.17 | 0.58 | 0.23 | 0.35 | 0.58 | 0.11 |
| hr4728 | S | 4938 | 2.64 | 1.40 | 0.35 | 0.19 | 0.25 | 0.27 | 0.68 | 0.23 | 0.02 | 0.07 | 0.10 | 0.17 | 0.13 | 0.09 | 0.02 | 0.11 | 0.07 | 0.34 | 0.28 | 0.23 | -0.15 | 0.18 | 0.40 | 0.12 | 0.25 | 0.30 | 0.28 |
| hr4737 | S | 4634 | 2.53 | 1.49 | 0.55 | 0.39 | 0.48 | 0.62 | 1.32 | 0.34 | 0.29 | 0.32 | 0.49 | 0.43 | 0.43 | 0.32 | 0.34 | 0.49 | 0.54 | 1.68 | 0.30 | 0.23 | -0.14 | 0.15 | 0.97 | 0.35 | 0.53 | 0.99 | 0.46 |
| hr4753 | S | 6388 | 3.56 | 6.50 | | -0.34 | | 0.37 | 1.02 | -0.05 | -0.12 | 0.68 | 0.82 | 0.49 | -0.58 | 0.20 | 1.07 | 0.45 | | | 1.60 | | | | 0.91 | 0.59 | 0.68 | | |
| hr4772 | S | 4719 | 2.50 | 1.46 | -0.11 | 0.02 | 0.06 | 0.12 | 0.46 | -0.10 | -0.10 | -0.09 | -0.11 | -0.12 | -0.21 | -0.15 | -0.10 | -0.06 | 0.07 | 0.27 | 0.01 | -0.16 | -0.55 | -0.09 | 0.05 | 0.02 | 0.07 | 0.20 | 0.01 |
| hr4777 | S | 5003 | 2.71 | 1.38 | 0.41 | 0.16 | 0.32 | 0.28 | 0.65 | 0.24 | 0.04 | 0.14 | 0.18 | 0.21 | 0.17 | 0.12 | 0.06 | 0.15 | 0.18 | 0.37 | 0.34 | 0.22 | -0.11 | 0.18 | 0.39 | 0.10 | 0.23 | 0.25 | 0.24 |
| hr4783 | S | 4784 | 2.56 | 1.40 | 0.15 | 0.05 | 0.21 | 0.20 | 0.64 | 0.04 | -0.03 | -0.02 | 0.04 | 0.07 | 0.02 | -0.01 | -0.04 | 0.04 | 0.08 | 0.27 | 0.16 | 0.03 | -0.31 | 0.12 | 0.32 | 0.13 | 0.21 | 0.46 | 0.14 |
| hr4784 | S | 4705 | 2.49 | 1.40 | 0.08 | 0.11 | 0.22 | 0.25 | 0.67 | 0.02 | 0.01 | 0.01 | 0.01 | 0.07 | 0.05 | -0.01 | -0.01 | 0.08 | 0.17 | 0.76 | 0.17 | -0.02 | -0.41 | 0.05 | 0.41 | 0.13 | 0.29 | 0.43 | 0.14 |
| hr4786 | S | 5090 | 2.41 | 1.82 | 0.38 | 0.01 | 0.20 | 0.18 | 0.58 | 0.08 | -0.03 | -0.02 | -0.09 | 0.09 | -0.06 | -0.01 | -0.05 | -0.01 | -0.13 | 0.40 | 0.44 | 0.16 | -0.16 | 0.26 | 0.20 | 0.12 | 0.11 | -0.08 | 0.16 |
| hr4812 | S | 4719 | 2.65 | 1.35 | 0.37 | 0.29 | 0.46 | 0.41 | 0.92 | 0.18 | 0.23 | 0.28 | 0.43 | 0.32 | 0.31 | 0.21 | 0.25 | 0.33 | 0.51 | 0.48 | 0.34 | 0.39 | -0.09 | 0.33 | 0.71 | 0.35 | 0.49 | 0.90 | 0.46 |
| hr4813 | S | 4404 | 2.04 | 1.71 | 0.72 | 0.53 | 0.57 | 0.77 | 1.72 | 0.21 | 0.26 | 0.29 | 0.48 | 0.44 | 0.57 | 0.34 | 0.39 | 0.52 | 0.78 | 2.23 | 0.30 | 0.17 | -0.30 | -0.05 | 0.65 | 0.39 | 0.52 | 1.11 | 0.35 |
| hr4815 | S | 4943 | 2.78 | 1.35 | 0.36 | 0.19 | 0.29 | 0.25 | 0.63 | 0.24 | 0.08 | 0.16 | 0.18 | 0.23 | 0.16 | 0.12 | 0.08 | 0.15 | 0.27 | 0.53 | 0.25 | 0.24 | -0.12 | 0.19 | 0.53 | 0.22 | 0.28 | 0.35 | 0.22 |
| hr4840 | S | 4372 | 2.10 | 1.44 | 0.34 | 0.24 | 0.46 | 0.51 | 1.47 | 0.25 | 0.10 | 0.21 | 0.43 | 0.32 | 0.26 | 0.15 | 0.17 | 0.32 | 0.46 | 1.45 | 0.32 | 0.11 | -0.29 | 0.09 | 0.78 | 0.20 | 0.43 | 0.85 | 0.25 |
| hr4851 | S | 4190 | 1.82 | 1.45 | 0.26 | 0.19 | 0.37 | 0.42 | 1.58 | 0.08 | 0.08 | 0.11 | 0.34 | 0.18 | 0.09 | 0.05 | 0.17 | 0.22 | 0.55 | 1.85 | 0.17 | 0.11 | -0.45 | -0.08 | 0.66 | 0.12 | 0.47 | 0.93 | 0.09 |
| hr4860 | S | 4814 | 2.52 | 1.39 | -0.29 | -0.21 | -0.13 | -0.12 | 0.09 | -0.29 | -0.32 | -0.36 | -0.45 | -0.36 | -0.46 | -0.41 | -0.38 | -0.36 | -0.28 | -0.04 | 0.11 | -0.37 | -0.69 | -0.29 | -0.26 | -0.31 | -0.20 | -0.29 | -0.22 |
| hr4873 | S | 4808 | 2.79 | 1.16 | 0.07 | 0.11 | 0.11 | 0.16 | 0.52 | 0.00 | -0.03 | -0.01 | 0.00 | 0.02 | -0.06 | -0.05 | -0.07 | -0.01 | 0.09 | 0.29 | 0.15 | 0.02 | -0.37 | 0.14 | 0.37 | 0.20 | 0.33 | 0.48 | 0.11 |
| hr4877 | S | 4744 | 2.46 | 1.40 | -0.03 | -0.04 | 0.09 | 0.15 | 0.64 | -0.09 | -0.09 | -0.12 | -0.11 | -0.08 | -0.13 | -0.14 | -0.13 | -0.06 | 0.07 | 0.46 | 0.04 | -0.10 | -0.50 | 0.00 | 0.17 | -0.02 | 0.14 | 0.36 | 0.03 |
| hr4883 | S | 5564 | 2.90 | 2.86 | 0.58 | | | 0.45 | | 0.25 | 0.28 | 0.57 | 0.49 | 0.62 | 0.01 | 0.16 | 0.81 | 0.61 | | 1.97 | 0.73 | -0.16 | | 1.43 | 1.38 | 0.53 | -0.63 | 1.28 | |
| hr4894 | S | 4982 | 2.62 | 1.48 | 0.61 | 0.23 | 0.37 | 0.42 | 0.86 | 0.29 | 0.24 | 0.30 | 0.37 | 0.37 | 0.43 | 0.23 | 0.26 | 0.32 | 0.47 | 0.94 | 0.46 | 0.28 | 0.00 | 0.21 | 0.54 | 0.27 | 0.28 | 0.69 | 0.40 |
| hr4896 | S | 4702 | 2.74 | 1.15 | 0.25 | 0.18 | 0.34 | 0.27 | 0.81 | 0.20 | 0.09 | 0.21 | 0.34 | 0.21 | 0.19 | 0.09 | 0.13 | 0.21 | 0.50 | 0.93 | 0.20 | 0.18 | -0.27 | 0.08 | 0.53 | 0.20 | 0.43 | 0.80 | 0.25 |
| hr4925 | S | 4580 | 2.48 | 1.36 | 0.31 | 0.27 | 0.43 | 0.42 | 1.13 | 0.16 | 0.11 | 0.18 | 0.33 | 0.24 | 0.26 | 0.16 | 0.18 | 0.31 | 0.50 | 1.04 | 0.11 | 0.10 | -0.36 | 0.05 | 0.60 | 0.21 | 0.41 | 0.91 | 0.26 |
| hr4929 | S | 4916 | 2.61 | 1.38 | 0.22 | 0.04 | 0.19 | 0.15 | 0.47 | 0.09 | -0.03 | 0.02 | 0.02 | 0.10 | 0.05 | 0.00 | -0.03 | 0.02 | 0.03 | 0.38 | 0.26 | 0.06 | -0.25 | 0.16 | 0.21 | 0.21 | 0.23 | 0.33 | 0.10 |
| hr4932 | S | 4988 | 2.64 | 1.49 | 0.45 | 0.12 | 0.23 | 0.30 | 0.75 | 0.19 | 0.06 | 0.08 | 0.07 | 0.20 | 0.13 | 0.10 | 0.06 | 0.14 | 0.18 | 0.63 | 0.31 | 0.17 | -0.15 | 0.16 | 0.30 | 0.12 | 0.15 | 0.22 | 0.26 |
| hr4953 | S | 4816 | 2.59 | 1.24 | -0.11 | -0.17 | -0.02 | -0.09 | 0.27 | -0.11 | -0.24 | -0.15 | -0.15 | -0.18 | -0.25 | -0.28 | -0.26 | -0.23 | -0.12 | 0.07 | 0.21 | -0.29 | -0.49 | -0.21 | -0.17 | -0.18 | -0.04 | -0.03 | -0.13 |
| hr4955 | S | 4585 | 2.26 | 1.39 | 0.28 | 0.27 | 0.36 | 0.36 | 0.78 | 0.17 | 0.05 | 0.13 | 0.25 | 0.23 | 0.13 | 0.10 | 0.07 | 0.20 | 0.33 | 0.70 | 0.24 | 0.17 | -0.38 | 0.20 | 0.71 | 0.23 | 0.33 | 0.56 | 0.35 |
| hr4956 | S | 4198 | 2.05 | 1.56 | 0.72 | 0.56 | 0.70 | 0.77 | 1.98 | 0.27 | 0.30 | 0.34 | 0.66 | 0.54 | 0.55 | 0.37 | 0.49 | 0.60 | | 1.63 | 0.33 | 0.37 | -0.12 | 0.09 | 0.98 | 0.82 | 0.84 | 1.44 | 0.39 |
| hr4959 | S | 4802 | 2.59 | 1.37 | 0.12 | 0.05 | 0.16 | 0.18 | 0.61 | 0.02 | -0.02 | -0.03 | 0.00 | 0.03 | -0.03 | -0.05 | -0.03 | 0.03 | 0.14 | 0.52 | 0.12 | 0.05 | -0.30 | 0.06 | 0.27 | 0.07 | 0.17 | 0.25 | 0.08 |
| hr4960 | S | 4765 | 2.57 | 1.33 | 0.27 | 0.15 | 0.30 | 0.29 | 0.54 | 0.16 | 0.03 | 0.08 | 0.07 | 0.18 | 0.10 | 0.07 | 0.02 | 0.12 | 0.28 | 0.22 | 0.26 | 0.20 | -0.23 | 0.20 | 0.57 | 0.19 | 0.31 | 0.36 | 0.23 |
| hr4964 | S | 4464 | 2.37 | 1.48 | 0.70 | 0.39 | 0.49 | 0.58 | 1.45 | 0.21 | 0.27 | 0.30 | 0.52 | 0.41 | 0.33 | 0.25 | 0.36 | 0.42 | | 1.38 | 0.31 | 0.29 | -0.15 | 0.04 | 0.74 | 0.42 | 0.57 | 1.09 | 0.38 |
| hr4984 | S | 4825 | 2.57 | 1.36 | 0.34 | 0.12 | 0.35 | 0.33 | 0.79 | 0.23 | 0.02 | 0.11 | 0.21 | 0.20 | 0.17 | 0.10 | 0.05 | 0.15 | 0.13 | 0.43 | 0.29 | 0.14 | -0.09 | 0.14 | 0.41 | 0.11 | 0.28 | 0.49 | 0.15 |
| hr4992 | S | 4398 | 2.28 | 1.32 | 0.08 | 0.12 | 0.21 | 0.26 | 1.06 | -0.01 | 0.05 | 0.08 | 0.23 | 0.08 | 0.03 | -0.01 | 0.06 | 0.12 | | 1.01 | 0.05 | 0.02 | -0.48 | -0.04 | 0.43 | 0.11 | 0.46 | 0.86 | 0.18 |
| hr5001 | S | 4741 | 3.06 | 0.71 | 0.22 | 0.31 | 0.32 | 0.32 | 1.00 | 0.17 | 0.18 | 0.23 | 0.36 | 0.21 | 0.17 | 0.13 | 0.12 | 0.28 | 0.55 | 0.74 | 0.16 | 0.23 | -0.31 | 0.13 | 0.71 | 0.15 | 0.46 | 0.98 | 0.21 |
| hr5007 | S | 4807 | 2.84 | 1.17 | -0.09 | -0.04 | 0.02 | 0.07 | 0.45 | -0.10 | -0.12 | -0.12 | -0.14 | -0.10 | -0.20 | -0.15 | -0.16 | -0.09 | -0.13 | 0.35 | 0.04 | -0.05 | -0.42 | 0.10 | 0.39 | 0.11 | 0.25 | 0.35 | 0.01 |
| hr5020 | S | 5019 | 2.60 | 1.58 | 0.36 | 0.09 | 0.16 | 0.23 | 0.59 | 0.12 | -0.03 | 0.01 | 0.00 | 0.14 | 0.03 | 0.03 | -0.01 | 0.05 | 0.04 | 0.45 | 0.39 | 0.10 | -0.20 | 0.19 | 0.25 | 0.13 | 0.18 | 0.21 | 0.25 |

| ID | | Teff | | | | | | | | | | | | | | | | | | | | | | | | | |
|---|---|---|---|---|---|---|---|---|---|---|---|---|---|---|---|---|---|---|---|---|---|---|---|---|---|---|---|
| hr5044 | S | 4820 | 2.53 | 1.40 | 0.11 | -0.08 | 0.09 | 0.17 | 0.49 | 0.02 | -0.11 | -0.09 | -0.10 | 0.02 | -0.04 | -0.05 | -0.10 | -0.02 | 0.07 | 0.23 | 0.11 | -0.02 | -0.39 | 0.12 | 0.23 | 0.09 | 0.13 | 0.16 | 0.03 |
| hr5053 | S | 4643 | 2.27 | 1.44 | 0.01 | -0.02 | 0.09 | 0.16 | 0.60 | -0.05 | -0.08 | -0.12 | -0.05 | -0.06 | -0.09 | -0.10 | -0.11 | -0.04 | 0.13 | 0.37 | 0.02 | -0.10 | -0.48 | 0.05 | 0.25 | -0.04 | 0.18 | 0.45 | 0.06 |
| hr5067 | S | 4851 | 2.49 | 1.42 | 0.10 | -0.11 | 0.15 | 0.13 | 0.36 | 0.03 | -0.11 | -0.05 | -0.04 | 0.01 | -0.07 | -0.08 | -0.10 | -0.06 | 0.01 | 0.23 | 0.19 | -0.08 | -0.33 | 0.09 | 0.21 | 0.14 | 0.16 | 0.16 | 0.05 |
| hr5068 | S | 4730 | 2.37 | 1.59 | 0.61 | 0.32 | 0.43 | 0.55 | 1.19 | 0.25 | 0.19 | 0.26 | 0.34 | 0.35 | 0.36 | 0.23 | 0.27 | 0.36 | 0.39 | 1.94 | 0.36 | 0.18 | -0.18 | 0.13 | 0.49 | 0.20 | 0.34 | 0.72 | 0.33 |
| hr5081 | S | 4658 | 2.47 | 1.30 | 0.08 | 0.00 | 0.09 | 0.18 | 0.69 | -0.02 | -0.12 | -0.06 | 0.01 | 0.02 | -0.04 | -0.08 | -0.07 | -0.01 | 0.15 | 0.41 | 0.14 | 0.03 | -0.37 | 0.02 | 0.19 | 0.08 | 0.22 | 0.45 | -0.02 |
| hr5100 | S | 4837 | 2.64 | 1.29 | -0.10 | -0.14 | -0.03 | 0.03 | 0.44 | -0.11 | -0.20 | -0.17 | -0.22 | -0.15 | -0.20 | -0.22 | -0.24 | -0.17 | -0.11 | 0.18 | 0.08 | -0.21 | -0.48 | -0.08 | 0.02 | -0.10 | 0.03 | -0.05 | -0.08 |
| hr5111 | S | 4881 | 2.64 | 1.37 | -0.09 | -0.07 | -0.01 | -0.02 | 0.31 | -0.13 | -0.19 | -0.18 | -0.22 | -0.18 | -0.25 | -0.25 | -0.25 | -0.21 | -0.16 | 0.05 | 0.24 | -0.24 | -0.53 | -0.13 | -0.08 | -0.18 | -0.01 | -0.09 | -0.09 |
| hr5126 | S | 4519 | 2.31 | 1.44 | 0.38 | 0.23 | 0.37 | 0.53 | 1.22 | 0.20 | 0.18 | 0.20 | 0.33 | 0.32 | 0.26 | 0.21 | 0.21 | 0.37 | | 1.36 | 0.27 | 0.21 | -0.15 | 0.22 | 0.80 | 0.32 | 0.48 | 0.93 | 0.38 |
| hr5143 | S | 4819 | 2.44 | 1.42 | -0.13 | -0.19 | -0.08 | -0.05 | 0.31 | -0.19 | -0.26 | -0.24 | -0.30 | -0.24 | -0.38 | -0.31 | -0.34 | -0.27 | -0.28 | 0.06 | -0.04 | -0.19 | -0.57 | -0.12 | -0.04 | -0.08 | -0.04 | -0.21 | -0.18 |
| hr5149 | S | 4836 | 2.57 | 1.41 | 0.18 | 0.05 | 0.19 | 0.20 | 0.60 | 0.11 | 0.01 | 0.03 | 0.03 | 0.10 | 0.06 | 0.00 | -0.02 | 0.05 | 0.11 | 0.40 | 0.17 | 0.05 | -0.25 | 0.16 | 0.34 | 0.15 | 0.26 | 0.25 | 0.10 |
| hr5161 | S | 5155 | 2.72 | 1.43 | 0.25 | 0.09 | 0.12 | 0.13 | 0.51 | 0.13 | -0.02 | 0.09 | 0.06 | 0.14 | 0.00 | 0.03 | 0.00 | 0.02 | -0.07 | 0.15 | 0.40 | 0.12 | -0.09 | 0.31 | 0.29 | 0.20 | 0.27 | 0.12 | 0.17 |
| hr5180 | S | 4935 | 2.67 | 1.42 | 0.47 | 0.21 | 0.28 | 0.33 | 0.94 | 0.23 | 0.08 | 0.11 | 0.16 | 0.22 | 0.19 | 0.11 | 0.07 | 0.17 | 0.04 | 0.99 | 0.36 | 0.27 | -0.16 | 0.18 | 0.35 | 0.14 | 0.16 | 0.38 | 0.21 |
| hr5186 | S | 4698 | 2.38 | 1.43 | 0.14 | -0.07 | 0.14 | 0.22 | 0.68 | -0.05 | -0.11 | -0.08 | -0.05 | 0.00 | -0.01 | -0.07 | -0.09 | -0.01 | 0.14 | 0.43 | 0.06 | 0.02 | -0.48 | -0.01 | 0.39 | -0.02 | 0.15 | 0.25 | -0.02 |
| hr5195 | S | 4683 | 2.55 | 1.21 | -0.03 | -0.06 | 0.06 | 0.09 | 0.40 | -0.09 | -0.10 | -0.10 | -0.11 | -0.06 | -0.20 | -0.12 | -0.13 | -0.05 | 0.11 | 0.52 | 0.01 | 0.02 | -0.43 | 0.10 | 0.32 | 0.13 | 0.28 | 0.34 | -0.02 |
| hr5196 | S | 4706 | 2.47 | 1.37 | 0.25 | 0.12 | 0.27 | 0.32 | 0.88 | 0.15 | 0.03 | 0.05 | 0.11 | 0.14 | 0.11 | 0.05 | 0.03 | 0.13 | 0.26 | 0.90 | 0.22 | 0.06 | -0.45 | 0.10 | 0.47 | 0.13 | 0.21 | 0.42 | 0.16 |
| hr5205 | S | 5049 | 2.62 | 1.45 | 0.19 | -0.11 | 0.07 | 0.10 | 0.53 | 0.03 | -0.06 | -0.08 | -0.11 | 0.01 | -0.12 | -0.06 | -0.11 | -0.08 | -0.32 | 0.07 | 0.31 | 0.11 | -0.25 | 0.28 | 0.31 | 0.11 | 0.17 | 0.08 | 0.18 |
| hr5213 | S | 4921 | 2.62 | 1.39 | -0.09 | -0.04 | -0.04 | -0.07 | 0.19 | -0.11 | -0.21 | -0.17 | -0.19 | -0.18 | -0.26 | -0.27 | -0.24 | -0.24 | -0.19 | 0.02 | 0.08 | -0.23 | -0.54 | -0.15 | -0.08 | -0.09 | -0.08 | -0.10 | -0.10 |
| hr5227 | S | 4464 | 2.51 | 1.06 | 0.34 | 0.31 | 0.53 | 0.45 | 1.22 | 0.30 | 0.30 | 0.39 | 0.65 | 0.34 | 0.21 | 0.17 | 0.26 | 0.35 | 0.68 | 1.50 | 0.24 | 0.32 | -0.26 | 0.10 | 0.68 | 0.20 | 0.73 | 1.26 | 0.36 |
| hr5232 | S | 4703 | 2.47 | 1.45 | 0.44 | 0.22 | 0.40 | 0.43 | 0.98 | 0.22 | 0.12 | 0.21 | 0.27 | 0.32 | 0.29 | 0.16 | 0.17 | 0.26 | 0.39 | 0.94 | 0.31 | 0.18 | -0.27 | 0.08 | 0.65 | 0.18 | 0.40 | 0.77 | 0.19 |
| hr5235 | S | 6028 | 3.73 | 1.85 | 0.93 | 0.50 | 0.63 | 0.46 | 0.62 | 0.38 | 0.48 | 0.47 | 0.45 | 0.48 | 0.50 | 0.38 | 0.48 | 0.51 | 0.67 | 0.59 | | 0.58 | 0.60 | 0.16 | 0.04 | 0.41 | 0.32 | 0.85 | 0.46 |
| hr5276 | S | 4265 | 2.33 | 1.38 | 0.59 | 0.68 | 0.66 | 0.79 | 2.02 | 0.41 | 0.45 | 0.52 | 0.77 | 0.62 | 0.55 | 0.43 | 0.56 | 0.65 | 1.08 | 2.75 | 0.41 | 0.45 | 0.01 | 0.21 | 0.96 | 0.65 | 1.13 | 1.25 | 0.54 |
| hr5277 | S | 4684 | 2.51 | 1.35 | 0.36 | 0.13 | 0.36 | 0.49 | 1.06 | 0.21 | 0.09 | 0.09 | 0.16 | 0.23 | 0.17 | 0.14 | 0.09 | 0.26 | 0.44 | 1.01 | 0.20 | 0.13 | -0.21 | 0.09 | 0.54 | 0.19 | 0.18 | 0.72 | 0.22 |
| hr5302 | S | 4770 | 2.71 | 1.38 | -0.10 | 0.01 | 0.11 | 0.16 | 0.50 | -0.03 | -0.04 | -0.02 | -0.04 | -0.04 | -0.10 | -0.09 | -0.05 | -0.01 | 0.13 | 0.33 | -0.01 | -0.15 | -0.47 | -0.01 | 0.17 | 0.15 | 0.24 | 0.43 | 0.05 |
| hr5315 | S | 4155 | 1.62 | 1.59 | -0.25 | -0.11 | -0.06 | 0.10 | 1.03 | -0.36 | -0.26 | -0.24 | -0.12 | -0.23 | -0.39 | -0.34 | -0.19 | -0.24 | 0.02 | 0.85 | -0.12 | -0.33 | -0.86 | -0.40 | 0.04 | -0.24 | 0.01 | 0.18 | -0.22 |
| hr5340 | S | 4281 | 1.74 | 1.62 | -0.35 | -0.10 | -0.04 | 0.01 | 0.83 | -0.38 | -0.38 | -0.27 | -0.25 | -0.40 | -0.60 | -0.48 | -0.29 | -0.34 | -0.17 | 0.22 | -0.30 | -0.49 | -0.90 | -0.62 | -0.35 | -0.47 | -0.27 | 0.03 | -0.10 |
| hr5344 | S | 4833 | 2.57 | 1.35 | 0.04 | -0.04 | 0.08 | 0.13 | 0.63 | -0.05 | -0.13 | -0.11 | -0.13 | -0.03 | -0.12 | -0.10 | -0.15 | -0.08 | -0.07 | 0.18 | 0.06 | 0.06 | -0.36 | 0.18 | 0.28 | 0.11 | 0.23 | 0.11 | 0.04 |
| hr5366 | S | 4730 | 2.47 | 1.40 | 0.10 | 0.02 | 0.13 | 0.20 | 0.63 | 0.03 | -0.07 | -0.08 | -0.04 | -0.01 | -0.07 | -0.06 | -0.08 | -0.01 | 0.12 | 0.29 | 0.11 | 0.03 | -0.43 | 0.11 | 0.29 | 0.13 | 0.19 | 0.24 | 0.00 |
| hr5370 | S | 4433 | 2.38 | 1.47 | 0.43 | 0.49 | 0.65 | 0.65 | 1.53 | 0.27 | 0.34 | 0.37 | 0.61 | 0.47 | 0.49 | 0.37 | 0.45 | 0.53 | 0.61 | 1.53 | 0.23 | 0.26 | -0.20 | 0.15 | 1.15 | 0.48 | 0.60 | 1.30 | 0.46 |
| hr5383 | S | 4874 | 2.76 | 1.27 | 0.34 | 0.18 | 0.32 | 0.32 | 0.75 | 0.24 | 0.04 | 0.11 | 0.14 | 0.22 | 0.15 | 0.13 | 0.05 | 0.19 | 0.22 | 0.71 | 0.26 | 0.16 | -0.21 | 0.15 | 0.41 | 0.23 | 0.26 | 0.49 | 0.20 |
| hr5394 | S | 4440 | 2.36 | 1.39 | 0.31 | 0.15 | 0.33 | 0.39 | 1.04 | 0.06 | 0.09 | 0.13 | 0.31 | 0.20 | 0.16 | 0.08 | 0.17 | 0.24 | 0.49 | 1.19 | 0.16 | 0.21 | -0.38 | 0.02 | 0.60 | 0.25 | 0.47 | 0.82 | 0.22 |
| hr5429 | S | 4281 | 1.95 | 1.55 | 0.10 | 0.19 | 0.28 | 0.45 | 1.43 | -0.01 | 0.08 | 0.08 | 0.21 | 0.17 | 0.04 | 0.07 | 0.10 | 0.16 | 0.41 | 2.73 | 0.04 | 0.10 | -0.46 | 0.03 | 0.62 | 0.15 | 0.36 | 0.77 | 0.28 |
| hr5430 | S | 4095 | 1.45 | 1.76 | 0.21 | 0.05 | 0.17 | 0.39 | 1.50 | -0.16 | -0.07 | -0.04 | 0.16 | 0.16 | 0.08 | 0.01 | 0.11 | 0.11 | 0.46 | | 0.28 | 0.09 | -0.42 | 0.13 | 0.42 | 0.31 | 0.52 | 0.72 | 0.03 |
| hr5454 | S | 4736 | 2.65 | 1.33 | 0.08 | 0.08 | 0.20 | 0.25 | 0.73 | 0.09 | 0.03 | 0.03 | 0.04 | 0.08 | 0.03 | 0.02 | -0.01 | 0.10 | 0.26 | 0.59 | 0.15 | 0.14 | -0.30 | 0.18 | 0.38 | 0.16 | 0.32 | 0.39 | 0.19 |
| hr5481 | S | 4903 | 2.49 | 1.39 | 0.12 | 0.03 | 0.13 | 0.08 | 0.42 | -0.04 | -0.09 | -0.06 | -0.13 | -0.11 | -0.27 | -0.19 | -0.14 | -0.01 | 0.08 | 0.02 | -0.12 | -0.42 | -0.13 | 0.00 | -0.11 | 0.05 | 0.09 | 0.07 | |
| hr5487 | S | 6487 | 3.91 | 4.23 | | -0.03 | | 0.09 | -0.22 | 0.18 | -0.02 | 0.15 | 0.58 | 0.51 | -0.09 | -0.30 | 0.75 | 0.00 | -0.64 | | 1.51 | 0.46 | | -0.64 | 0.61 | 0.83 | 0.26 | -0.34 | | |
| hr5502 | S | 4864 | 2.51 | 1.46 | 0.25 | 0.06 | 0.18 | 0.22 | 0.57 | 0.10 | -0.04 | -0.04 | -0.07 | 0.07 | -0.03 | -0.01 | -0.07 | 0.01 | 0.01 | 0.64 | 0.23 | 0.02 | -0.25 | 0.20 | 0.32 | 0.15 | 0.22 | 0.24 | 0.13 |
| hr5518 | S | 4641 | 2.59 | 1.39 | 0.02 | 0.19 | 0.20 | 0.31 | 0.76 | -0.03 | 0.03 | 0.03 | 0.07 | 0.02 | -0.07 | -0.01 | 0.04 | 0.11 | 0.31 | 0.71 | 0.05 | -0.06 | -0.60 | -0.01 | 0.33 | 0.06 | 0.26 | 0.81 | 0.22 |
| hr5573 | S | 4627 | 2.51 | 1.41 | 0.35 | 0.23 | 0.30 | 0.44 | 0.95 | 0.17 | 0.14 | 0.19 | 0.28 | 0.28 | 0.27 | 0.21 | 0.21 | 0.32 | 0.44 | 1.54 | 0.24 | 0.34 | -0.13 | 0.32 | 0.74 | 0.39 | 0.53 | 0.75 | 0.36 |
| hr5600 | S | 3962 | 1.48 | 1.78 | 0.50 | 0.20 | 0.33 | 0.65 | 1.85 | 0.04 | -0.05 | 0.03 | 0.30 | 0.25 | 0.10 | 0.15 | 0.22 | 0.30 | 0.96 | | 0.44 | 0.16 | -0.34 | 0.10 | 0.71 | 0.53 | 0.72 | 1.12 | 0.06 |
| hr5601 | S | 4664 | 2.40 | 1.31 | 0.01 | -0.26 | 0.04 | 0.13 | 0.65 | -0.10 | -0.16 | -0.13 | -0.12 | -0.06 | -0.16 | -0.14 | -0.15 | -0.09 | 0.04 | 0.46 | 0.04 | 0.00 | -0.48 | 0.13 | 0.34 | 0.11 | 0.22 | 0.20 | -0.06 |
| hr5602 | S | 4920 | 2.34 | 1.70 | 0.26 | -0.03 | 0.08 | 0.18 | 0.58 | 0.08 | -0.05 | -0.08 | -0.14 | 0.05 | -0.08 | -0.05 | -0.11 | -0.04 | -0.09 | 0.43 | 0.31 | 0.06 | -0.27 | 0.29 | 0.24 | 0.09 | 0.15 | 0.09 | 0.13 |
| hr5609 | S | 4800 | 2.61 | 1.34 | 0.34 | 0.14 | 0.34 | 0.37 | 0.85 | 0.25 | 0.04 | 0.11 | 0.16 | 0.22 | 0.18 | 0.12 | 0.05 | 0.17 | 0.19 | 0.71 | 0.27 | 0.16 | -0.22 | 0.16 | 0.47 | 0.20 | 0.28 | 0.56 | 0.20 |
| hr5616 | S | 4302 | 1.93 | 1.51 | -0.03 | 0.06 | 0.13 | 0.35 | 1.00 | -0.09 | -0.14 | -0.20 | 0.00 | -0.12 | -0.13 | -0.07 | -0.03 | 0.01 | | 1.24 | | -0.24 | -0.97 | -0.06 | 0.40 | -0.05 | 0.03 | -0.16 | 0.19 |
| hr5620 | S | 4720 | 3.02 | 0.80 | 0.12 | 0.26 | 0.20 | 0.25 | 0.69 | 0.09 | 0.07 | 0.17 | 0.25 | 0.16 | 0.07 | 0.10 | 0.07 | 0.22 | 0.47 | 0.90 | 0.03 | 0.19 | -0.38 | 0.09 | 0.54 | 0.29 | 0.37 | 1.07 | 0.17 |
| hr5635 | S | 4812 | 2.54 | 1.32 | -0.18 | -0.16 | -0.08 | -0.03 | 0.24 | -0.22 | -0.26 | -0.29 | -0.34 | -0.28 | -0.38 | -0.33 | -0.32 | -0.28 | -0.21 | 0.27 | -0.08 | -0.29 | -0.65 | -0.22 | -0.17 | -0.22 | -0.16 | -0.11 | -0.07 |
| hr5648 | S | 4722 | 2.49 | 1.37 | 0.24 | 0.19 | 0.29 | 0.34 | 0.79 | 0.13 | 0.04 | 0.07 | 0.12 | 0.15 | 0.16 | 0.06 | 0.04 | 0.14 | 0.30 | 0.62 | 0.20 | 0.18 | -0.27 | 0.14 | 0.43 | 0.12 | 0.28 | 0.55 | 0.18 |
| hr5673 | S | 4399 | 2.27 | 1.37 | 0.06 | 0.10 | 0.17 | 0.24 | 1.08 | -0.01 | 0.02 | 0.03 | 0.16 | 0.10 | -0.02 | -0.01 | 0.06 | 0.13 | 0.25 | 0.81 | 0.02 | 0.01 | -0.44 | 0.06 | 0.47 | 0.21 | 0.44 | 1.00 | 0.13 |
| hr5681a | S | 4821 | 4.23 | 0.51 | -0.30 | -0.09 | 0.05 | 0.32 | 0.81 | -0.24 | 0.36 | -0.02 | -0.01 | -0.06 | -0.15 | 0.11 | 0.13 | 0.21 | 0.28 | 0.57 | 0.21 | 0.38 | -0.26 | 0.37 | 0.87 | 0.81 | 0.96 | 0.97 | 0.87 |
| hr5681b | S | 5812 | 4.49 | 0.78 | -0.39 | -0.31 | -0.28 | -0.30 | -0.15 | -0.29 | -0.27 | -0.31 | -0.38 | -0.40 | -0.48 | -0.42 | -0.37 | -0.42 | -0.50 | -0.29 | | -0.29 | | -0.25 | | 0.18 | -0.12 | -0.15 | 0.06 |
| hr5707 | S | 4783 | 2.60 | 1.44 | 0.60 | 0.40 | 0.47 | 0.57 | 1.16 | 0.30 | 0.27 | 0.29 | 0.38 | 0.41 | 0.38 | 0.29 | 0.30 | 0.43 | 0.43 | 1.18 | 0.34 | 0.36 | -0.11 | 0.21 | 0.81 | 0.29 | 0.39 | 0.82 | 0.36 |

| ID | | Col3 | Col4 | Col5 | Col6 | Col7 | Col8 | Col9 | Col10 | Col11 | Col12 | Col13 | Col14 | Col15 | Col16 | Col17 | Col18 | Col19 | Col20 | Col21 | Col22 | Col23 | Col24 | Col25 | Col26 | Col27 | Col28 | Col29 |
|---|---|---|---|---|---|---|---|---|---|---|---|---|---|---|---|---|---|---|---|---|---|---|---|---|---|---|---|---|
| hr5709 | S | 4705 | 2.57 | 1.30 | -0.14 | -0.01 | 0.07 | 0.16 | 0.50 | -0.11 | -0.06 | -0.13 | -0.17 | -0.14 | -0.23 | -0.15 | -0.15 | -0.07 | 0.16 | 0.43 | -0.02 | -0.19 | -0.53 | -0.05 | 0.27 | -0.03 | 0.12 | 0.24 | 0.02 |
| hr5744 | S | 4531 | 2.40 | 1.37 | 0.39 | 0.31 | 0.45 | 0.46 | 1.14 | 0.25 | 0.17 | 0.24 | 0.42 | 0.34 | 0.29 | 0.18 | 0.22 | 0.30 | 0.40 | 1.05 | 0.25 | 0.15 | -0.18 | 0.04 | 0.73 | 0.19 | 0.46 | 1.03 | 0.28 |
| hr5769 | S | 6040 | 3.58 | 1.79 | 0.21 | 0.01 | 0.04 | 0.09 | 0.10 | 0.06 | -0.03 | -0.05 | -0.05 | -0.02 | -0.10 | -0.09 | -0.02 | -0.07 | -0.31 | 0.02 | | -0.06 | 0.12 | -0.08 | -0.02 | -0.14 | 0.01 | -0.05 | 0.31 |
| hr5785 | S | 4798 | 2.60 | 1.34 | -0.08 | -0.06 | 0.06 | 0.08 | 0.53 | -0.13 | -0.11 | -0.15 | -0.20 | -0.11 | -0.25 | -0.16 | -0.17 | -0.11 | 0.01 | 0.28 | -0.01 | -0.14 | -0.50 | 0.06 | 0.19 | 0.03 | 0.11 | 0.04 | 0.06 |
| hr5794 | S | 4135 | 1.58 | 1.61 | 0.13 | 0.09 | 0.16 | 0.37 | 1.49 | -0.10 | -0.10 | -0.03 | 0.14 | 0.14 | -0.01 | 0.00 | 0.07 | 0.11 | 0.37 | | 0.16 | 0.13 | -0.47 | 0.14 | 0.68 | 0.18 | 0.53 | 0.76 | 0.04 |
| hr5802 | S | 4889 | 2.74 | 1.23 | 0.20 | 0.01 | 0.16 | 0.23 | 0.62 | 0.07 | -0.04 | -0.02 | -0.05 | 0.08 | -0.01 | -0.01 | -0.07 | 0.04 | 0.12 | 0.32 | 0.63 | 0.44 | 0.14 | 0.50 | 0.47 | 0.42 | 0.40 | 0.52 | 0.09 |
| hr5810 | S | 4633 | 2.56 | 1.24 | -0.02 | 0.00 | 0.03 | 0.22 | 0.58 | -0.10 | -0.11 | -0.14 | -0.12 | -0.06 | -0.16 | -0.11 | -0.11 | -0.01 | 0.20 | 0.63 | 0.19 | 0.20 | -0.43 | 0.05 | 0.29 | 0.09 | 0.21 | 0.51 | -0.01 |
| hr5811 | S | 4517 | 2.32 | 1.51 | 0.59 | 0.43 | 0.49 | 0.60 | 1.38 | 0.42 | 0.27 | 0.32 | 0.52 | 0.47 | 0.42 | 0.32 | 0.39 | 0.48 | | 1.30 | 0.33 | 0.38 | -0.13 | 0.15 | 1.00 | 0.32 | 0.59 | 1.12 | 0.33 |
| hr5828 | S | 4625 | 2.47 | 1.26 | 0.13 | 0.08 | 0.19 | 0.25 | 0.70 | 0.04 | 0.01 | 0.06 | 0.11 | 0.10 | 0.06 | 0.01 | -0.02 | 0.10 | 0.33 | 0.51 | 0.19 | 0.09 | -0.32 | 0.16 | 0.48 | 0.21 | 0.35 | 0.86 | 0.19 |
| hr5835 | S | 5172 | 2.83 | 1.30 | 0.47 | 0.12 | 0.25 | 0.17 | 0.51 | 0.20 | 0.03 | 0.12 | 0.11 | 0.18 | 0.04 | 0.08 | 0.07 | 0.09 | 0.18 | 0.04 | 0.52 | 0.19 | -0.05 | 0.19 | 0.28 | 0.11 | 0.16 | 0.15 | 0.33 |
| hr5840 | S | 4982 | 2.46 | 1.41 | 0.19 | -0.04 | 0.09 | 0.11 | 0.49 | 0.03 | -0.13 | -0.15 | -0.22 | -0.04 | -0.19 | -0.10 | -0.18 | -0.11 | -0.29 | 0.26 | 0.25 | 0.01 | -0.32 | 0.20 | 0.12 | 0.16 | 0.09 | -0.10 | 0.06 |
| hr5841 | S | 4659 | 2.43 | 1.35 | 0.37 | 0.24 | 0.41 | 0.42 | 0.93 | 0.23 | 0.08 | 0.20 | 0.28 | 0.28 | 0.29 | 0.15 | 0.14 | 0.26 | 0.38 | 0.67 | 0.26 | 0.15 | -0.27 | 0.12 | 0.56 | 0.22 | 0.33 | 0.68 | 0.08 |
| hr5855 | S | 4626 | 2.64 | 1.13 | 0.20 | 0.21 | 0.24 | 0.37 | 0.91 | 0.21 | 0.07 | 0.09 | 0.17 | 0.18 | 0.14 | 0.11 | 0.06 | 0.22 | 0.27 | 0.75 | 0.20 | 0.16 | -0.31 | 0.19 | 0.62 | 0.25 | 0.42 | 0.53 | 0.15 |
| hr5888 | S | 4715 | 2.41 | 1.41 | 0.03 | 0.04 | 0.06 | 0.17 | 0.50 | -0.07 | -0.16 | -0.16 | -0.16 | -0.09 | -0.20 | -0.15 | -0.16 | -0.09 | 0.03 | -0.12 | 0.01 | -0.07 | -0.56 | 0.00 | 0.12 | 0.01 | 0.08 | 0.18 | -0.07 |
| hr5893 | S | 4873 | 3.08 | 1.06 | 0.30 | 0.33 | 0.32 | 0.33 | 0.77 | 0.21 | 0.17 | 0.21 | 0.28 | 0.26 | 0.21 | 0.19 | 0.18 | 0.27 | 0.45 | 0.73 | 0.26 | 0.20 | -0.18 | 0.25 | 0.65 | 0.40 | 0.47 | 0.83 | 0.35 |
| hr5922 | S | 4823 | 2.61 | 1.36 | 0.33 | 0.14 | 0.24 | 0.37 | 0.93 | 0.17 | -0.01 | 0.03 | 0.06 | 0.18 | 0.11 | 0.08 | 0.00 | 0.15 | 0.10 | 0.37 | 0.25 | 0.11 | -0.25 | 0.09 | 0.38 | 0.12 | 0.18 | 0.46 | 0.13 |
| hr5940 | S | 4593 | 2.60 | 1.21 | 0.45 | 0.43 | 0.61 | 0.45 | 0.97 | 0.31 | 0.34 | 0.41 | 0.60 | 0.41 | 0.27 | 0.24 | 0.34 | 0.43 | 0.55 | 1.20 | 0.35 | 0.41 | -0.09 | 0.26 | 0.98 | 0.42 | 0.87 | 1.15 | 0.70 |
| hr5947 | S | 4364 | 1.95 | 1.55 | 0.06 | 0.03 | 0.11 | 0.30 | 0.90 | -0.06 | -0.06 | -0.08 | 0.01 | -0.01 | -0.09 | -0.09 | -0.04 | 0.04 | 0.15 | 0.91 | 0.03 | 0.05 | -0.52 | 0.05 | 0.39 | 0.07 | 0.24 | 0.54 | 0.10 |
| hr5966 | S | 4798 | 2.52 | 1.39 | 0.00 | -0.06 | 0.06 | 0.15 | 0.45 | -0.07 | -0.08 | -0.12 | -0.11 | -0.06 | -0.15 | -0.12 | -0.12 | -0.05 | 0.08 | 0.32 | 0.00 | -0.15 | -0.46 | 0.01 | 0.22 | 0.04 | 0.08 | 0.16 | -0.01 |
| hr5976 | S | 4933 | 2.78 | 1.32 | 0.39 | 0.15 | 0.27 | 0.33 | 0.83 | 0.19 | 0.07 | 0.07 | 0.09 | 0.20 | 0.15 | 0.11 | 0.04 | 0.15 | 0.12 | 0.46 | 0.31 | 0.22 | -0.16 | 0.19 | 0.45 | 0.19 | 0.19 | 0.41 | 0.21 |
| hr6038 | S | 4744 | 2.51 | 1.38 | 0.27 | 0.19 | 0.34 | 0.30 | 0.84 | 0.20 | 0.08 | 0.12 | 0.18 | 0.18 | 0.16 | 0.07 | 0.07 | 0.16 | 0.44 | 0.30 | 0.28 | 0.13 | -0.23 | 0.12 | 0.41 | 0.07 | 0.34 | 0.57 | 0.17 |
| hr6057 | S | 4560 | 2.51 | 1.26 | 0.34 | 0.32 | 0.43 | 0.39 | 1.01 | 0.25 | 0.20 | 0.23 | 0.39 | 0.28 | 0.22 | 0.13 | 0.18 | 0.28 | 0.58 | 1.33 | 0.13 | 0.23 | -0.26 | 0.03 | 0.66 | 0.23 | 0.45 | 0.94 | 0.28 |
| hr6065 | S | 4620 | 2.43 | 1.42 | 0.39 | 0.34 | 0.45 | 0.42 | 1.08 | 0.23 | 0.07 | 0.16 | 0.30 | 0.27 | 0.25 | 0.16 | 0.17 | 0.28 | 0.35 | 1.09 | 0.22 | 0.08 | -0.40 | 0.05 | 0.64 | 0.22 | 0.37 | 0.67 | 0.25 |
| hr6075 | S | 4837 | 2.53 | 1.37 | 0.12 | -0.13 | 0.12 | 0.14 | 0.24 | 0.00 | -0.07 | -0.07 | -0.10 | 0.01 | -0.11 | -0.06 | -0.12 | -0.05 | -0.06 | 0.19 | 0.19 | -0.01 | -0.27 | 0.16 | 0.38 | 0.14 | 0.20 | 0.15 | 0.10 |
| hr6104 | S | 4757 | 2.48 | 1.37 | 0.16 | 0.11 | 0.15 | 0.22 | 0.82 | 0.01 | -0.08 | -0.08 | -0.05 | 0.01 | -0.04 | -0.07 | -0.09 | 0.00 | 0.16 | 0.63 | 0.32 | 0.17 | -0.14 | 0.19 | 0.28 | 0.13 | 0.21 | 0.37 | 0.08 |
| hr6121 | S | 4682 | 2.61 | 1.36 | 0.07 | -0.07 | 0.12 | 0.26 | 0.61 | -0.02 | 0.01 | -0.04 | -0.02 | 0.04 | 0.01 | -0.01 | -0.03 | 0.08 | 0.22 | 0.45 | 0.12 | -0.04 | -0.32 | 0.16 | 0.51 | 0.11 | 0.28 | 0.39 | 0.05 |
| hr6124 | S | 5030 | 2.68 | 1.42 | 0.24 | 0.04 | 0.13 | 0.15 | 0.58 | 0.09 | -0.07 | -0.06 | -0.12 | 0.03 | -0.01 | -0.05 | -0.12 | -0.05 | -0.17 | 0.17 | 0.27 | 0.04 | -0.20 | 0.20 | 0.23 | 0.17 | 0.19 | 0.29 | 0.16 |
| hr6126 | S | 4636 | 2.38 | 1.42 | 0.52 | 0.30 | 0.40 | 0.55 | 1.09 | 0.29 | 0.21 | 0.28 | 0.39 | 0.38 | 0.26 | 0.28 | 0.30 | 0.41 | 0.51 | 1.34 | 0.30 | 0.26 | -0.17 | 0.27 | 0.80 | 0.40 | 0.39 | 0.73 | 0.27 |
| hr6130 | S | 4946 | 2.41 | 1.49 | 0.39 | 0.12 | 0.24 | 0.23 | 0.42 | 0.12 | -0.08 | 0.04 | 0.04 | 0.15 | 0.12 | 0.03 | -0.01 | 0.06 | -0.02 | 0.40 | 0.29 | 0.14 | -0.22 | 0.04 | 0.17 | 0.01 | 0.10 | 0.11 | 0.04 |
| hr6132 | S | 4962 | 2.71 | 1.34 | 0.30 | 0.06 | 0.16 | 0.17 | 0.55 | 0.10 | -0.01 | -0.04 | -0.08 | 0.03 | -0.07 | -0.03 | -0.07 | -0.01 | -0.16 | 0.00 | 0.18 | 0.03 | -0.24 | 0.16 | 0.27 | 0.12 | 0.17 | 0.17 | 0.14 |
| hr6145 | S | 4696 | 2.94 | 1.06 | 0.50 | 0.30 | 0.51 | 0.50 | 1.17 | 0.30 | 0.33 | 0.41 | 0.59 | 0.48 | 0.43 | 0.32 | 0.38 | 0.51 | 0.78 | 1.50 | 0.32 | 0.42 | -0.13 | 0.25 | 1.02 | 0.46 | 0.64 | 1.28 | 0.53 |
| hr6147 | S | 5038 | 2.57 | 1.64 | 0.69 | 0.22 | 0.33 | 0.43 | 0.96 | 0.25 | 0.17 | 0.20 | 0.18 | 0.34 | 0.24 | 0.22 | 0.20 | 0.30 | 0.16 | 0.91 | 0.36 | 0.29 | -0.04 | 0.26 | 0.43 | 0.20 | 0.25 | 0.31 | 0.25 |
| hr6148 | S | 4912 | 2.36 | 1.56 | 0.24 | -0.05 | 0.04 | 0.17 | 0.60 | 0.07 | -0.10 | -0.10 | -0.15 | 0.02 | -0.08 | -0.07 | -0.16 | -0.05 | -0.13 | 0.67 | 0.23 | 0.03 | -0.29 | 0.23 | 0.23 | 0.12 | 0.12 | 0.04 | 0.09 |
| hr6150 | S | 4735 | 2.50 | 1.38 | 0.14 | 0.08 | 0.25 | 0.28 | 0.66 | 0.07 | 0.02 | 0.06 | 0.09 | 0.09 | 0.09 | 0.01 | 0.01 | 0.10 | 0.23 | 0.29 | 0.14 | -0.03 | -0.35 | 0.05 | 0.28 | 0.07 | 0.23 | 0.50 | 0.10 |
| hr6190 | S | 4718 | 2.46 | 1.40 | 0.08 | 0.06 | 0.21 | 0.25 | 0.65 | 0.05 | -0.02 | 0.02 | 0.09 | 0.04 | -0.03 | -0.03 | -0.02 | 0.06 | 0.30 | 0.26 | 0.17 | -0.08 | -0.35 | -0.05 | 0.22 | -0.06 | 0.17 | 0.50 | 0.02 |
| hr6199 | S | 4641 | 2.48 | 1.54 | 0.15 | 0.10 | 0.23 | 0.39 | 0.90 | 0.02 | 0.02 | 0.03 | 0.07 | 0.15 | -0.04 | 0.04 | 0.04 | 0.13 | 0.21 | 0.68 | 0.06 | 0.06 | -0.35 | 0.10 | 0.05 | 0.16 | 0.19 | 0.84 | 0.23 |
| hr6220 | S | 4932 | 2.76 | 1.30 | 0.00 | -0.08 | -0.01 | 0.06 | 0.51 | -0.10 | -0.17 | -0.18 | -0.21 | -0.08 | -0.22 | -0.16 | -0.23 | -0.16 | -0.28 | 0.09 | 0.36 | -0.06 | -0.28 | 0.18 | 0.20 | 0.19 | 0.13 | -0.02 | 0.07 |
| hr6259 | S | 4756 | 2.74 | 1.14 | 0.05 | -0.05 | 0.12 | 0.10 | 0.45 | -0.01 | -0.08 | 0.01 | 0.04 | 0.02 | -0.08 | -0.07 | -0.06 | 0.01 | 0.16 | 0.35 | 0.06 | 0.07 | -0.37 | 0.06 | 0.26 | 0.10 | 0.28 | 0.43 | 0.00 |
| hr6280 | S | 4640 | 2.57 | 1.29 | 0.39 | 0.31 | 0.36 | 0.43 | 1.13 | 0.23 | 0.16 | 0.21 | 0.32 | 0.32 | 0.32 | 0.23 | 0.22 | 0.35 | 0.60 | 1.13 | 0.27 | 0.19 | -0.14 | 0.26 | 0.97 | 0.35 | 0.52 | 0.84 | 0.22 |
| hr6287 | S | 4838 | 2.61 | 1.30 | 0.05 | -0.02 | 0.04 | 0.14 | 0.29 | -0.02 | -0.11 | -0.11 | -0.15 | -0.04 | -0.18 | -0.10 | -0.16 | -0.07 | 0.01 | 0.00 | 0.10 | -0.01 | -0.38 | 0.19 | 0.26 | 0.17 | 0.22 | 0.15 | 0.16 |
| hr6292 | S | 4979 | 2.67 | 1.35 | 0.03 | -0.05 | 0.08 | 0.05 | 0.28 | -0.04 | -0.16 | -0.11 | -0.12 | -0.07 | -0.22 | -0.14 | -0.17 | -0.13 | -0.19 | -0.06 | 0.17 | -0.05 | -0.34 | 0.17 | 0.19 | 0.16 | 0.19 | 0.12 | 0.11 |
| hr6305 | S | 4991 | 2.62 | 1.47 | 0.06 | -0.11 | 0.06 | 0.04 | 0.24 | -0.09 | -0.09 | -0.11 | -0.15 | -0.07 | -0.20 | -0.16 | -0.17 | -0.14 | -0.28 | -0.07 | 0.24 | -0.05 | -0.46 | 0.14 | 0.21 | 0.13 | 0.10 | -0.02 | 0.13 |
| hr6307 | S | 4634 | 2.36 | 1.40 | 0.22 | 0.15 | 0.24 | 0.33 | 0.75 | 0.12 | 0.01 | 0.03 | 0.08 | 0.11 | 0.07 | 0.04 | 0.01 | 0.15 | 0.22 | 0.74 | 0.10 | 0.13 | -0.40 | 0.07 | 0.36 | 0.11 | 0.28 | 0.45 | 0.08 |
| hr6333 | S | 4779 | 2.59 | 1.39 | 0.00 | -0.01 | 0.11 | 0.15 | 0.55 | -0.04 | -0.07 | -0.05 | -0.07 | -0.05 | -0.07 | -0.10 | -0.06 | -0.01 | 0.14 | 0.19 | 0.06 | -0.12 | -0.42 | -0.02 | 0.21 | 0.02 | 0.12 | 0.38 | 0.07 |
| hr6342 | S | 4702 | 2.61 | 1.22 | -0.05 | 0.01 | 0.05 | 0.07 | 0.50 | -0.06 | -0.06 | -0.07 | -0.06 | -0.09 | -0.15 | -0.13 | -0.13 | -0.06 | 0.12 | 0.40 | 0.00 | 0.05 | -0.50 | 0.05 | 0.20 | 0.14 | 0.22 | 0.33 | -0.05 |
| hr6359 | S | 5078 | 2.91 | 1.25 | 0.17 | 0.05 | 0.14 | 0.08 | 0.24 | 0.08 | 0.00 | 0.01 | 0.00 | 0.08 | -0.08 | -0.02 | -0.06 | -0.04 | -0.22 | 0.11 | 0.33 | 0.09 | -0.13 | 0.27 | 0.27 | 0.19 | 0.25 | 0.29 | 0.17 |
| hr6360 | S | 4886 | 2.68 | 1.29 | 0.45 | 0.12 | 0.39 | 0.50 | 1.18 | 0.27 | 0.10 | 0.16 | 0.20 | 0.28 | 0.22 | 0.22 | 0.13 | 0.29 | 0.31 | 0.99 | 0.30 | 0.35 | -0.14 | 0.29 | 0.59 | 0.29 | 0.35 | 0.61 | 0.67 |
| hr6363 | S | 4602 | 2.48 | 1.16 | 0.27 | 0.22 | 0.31 | 0.31 | 0.75 | 0.15 | 0.08 | 0.16 | 0.31 | 0.22 | 0.21 | 0.09 | 0.09 | 0.19 | 0.43 | 0.61 | 0.24 | 0.17 | -0.23 | 0.08 | 0.74 | 0.12 | 0.36 | 0.74 | 0.19 |
| hr6364 | S | 4281 | 2.05 | 1.36 | 0.37 | 0.29 | 0.47 | 0.53 | 1.44 | 0.16 | 0.20 | 0.26 | 0.48 | 0.31 | 0.16 | 0.15 | 0.25 | 0.29 | 0.56 | 1.23 | 0.20 | 0.13 | -0.37 | 0.02 | 0.72 | 0.24 | 0.67 | 1.40 | 0.19 |
| hr6365 | S | 4889 | 2.67 | 1.39 | 0.50 | 0.21 | 0.31 | 0.48 | 1.02 | 0.19 | 0.15 | 0.12 | 0.12 | 0.22 | 0.24 | 0.15 | 0.12 | 0.27 | 0.23 | 0.71 | 0.39 | 0.28 | -0.17 | 0.16 | 0.48 | 0.17 | 0.27 | 0.52 | 0.32 |

| ID | Type | col3 | col4 | col5 | col6 | col7 | col8 | col9 | col10 | col11 | col12 | col13 | col14 | col15 | col16 | col17 | col18 | col19 | col20 | col21 | col22 | col23 | col24 | col25 | col26 | col27 | col28 | col29 | col30 |
|---|---|---|---|---|---|---|---|---|---|---|---|---|---|---|---|---|---|---|---|---|---|---|---|---|---|---|---|---|---|
| hr6388 | S | 4329 | 2.03 | 1.40 | 0.23 | 0.04 | 0.32 | 0.39 | 1.23 | 0.04 | 0.03 | 0.10 | 0.31 | 0.18 | 0.11 | 0.03 | 0.10 | 0.19 | 0.44 | 1.19 | 0.10 | 0.12 | -0.46 | 0.01 | 0.51 | 0.09 | 0.46 | 0.87 | 0.15 |
| hr6390 | S | 4697 | 2.44 | 1.29 | -0.08 | -0.09 | 0.03 | 0.05 | 0.32 | -0.14 | -0.11 | -0.13 | -0.10 | -0.11 | -0.16 | -0.20 | -0.15 | -0.12 | 0.07 | 0.20 | 0.13 | -0.15 | -0.54 | -0.11 | 0.11 | -0.03 | 0.13 | 0.29 | -0.08 |
| hr6394 | S | 6068 | 3.71 | 1.68 | -0.12 | -0.07 | -0.02 | -0.02 | 0.08 | -0.01 | 0.03 | -0.11 | -0.14 | -0.09 | -0.23 | -0.16 | -0.10 | -0.18 | -0.37 | -0.08 | 0.66 | 0.03 | 0.09 | 0.09 | -0.05 | 0.13 | -0.02 | -0.63 | -0.02 |
| hr6404 | S | 4658 | 2.65 | 1.37 | 0.45 | 0.33 | 0.47 | 0.54 | 1.20 | 0.21 | 0.25 | 0.21 | 0.34 | 0.32 | 0.32 | 0.27 | 0.27 | 0.40 | 0.48 | 1.13 | 0.25 | 0.35 | -0.18 | 0.27 | 1.05 | 0.33 | 0.53 | 0.83 | 0.39 |
| hr6415 | S | 4508 | 2.34 | 1.46 | 0.43 | 0.37 | 0.56 | 0.62 | 1.38 | 0.23 | 0.24 | 0.29 | 0.46 | 0.39 | 0.39 | 0.29 | 0.30 | 0.43 | | 1.36 | 0.22 | 0.38 | -0.22 | 0.17 | 0.93 | 0.33 | 0.53 | 1.14 | 0.43 |
| hr6443 | S | 4870 | 2.63 | 1.37 | 0.18 | -0.01 | 0.18 | 0.19 | 0.55 | 0.06 | -0.03 | 0.00 | 0.00 | 0.07 | 0.01 | -0.02 | -0.06 | 0.01 | 0.11 | 0.48 | 0.25 | 0.08 | -0.25 | 0.14 | 0.25 | 0.17 | 0.19 | 0.28 | 0.08 |
| hr6444 | S | 4773 | 2.62 | 1.33 | 0.43 | 0.20 | 0.35 | 0.44 | 1.15 | 0.19 | 0.02 | 0.04 | 0.11 | 0.23 | 0.19 | 0.12 | 0.07 | 0.22 | 0.24 | 0.76 | 0.27 | 0.18 | -0.23 | 0.05 | 0.47 | 0.19 | 0.16 | 0.56 | 0.18 |
| hr6448 | S | 4650 | 2.35 | 1.44 | 0.09 | 0.09 | 0.23 | 0.29 | 0.72 | 0.02 | -0.04 | 0.05 | 0.11 | 0.07 | 0.06 | -0.02 | 0.01 | 0.09 | 0.29 | 0.61 | 0.04 | -0.07 | -0.41 | 0.00 | 0.28 | -0.03 | 0.17 | 0.49 | 0.09 |
| hr6472 | S | 4980 | 2.88 | 1.31 | 0.37 | 0.10 | 0.27 | 0.26 | 0.57 | 0.22 | 0.08 | 0.09 | 0.11 | 0.21 | 0.13 | 0.10 | 0.04 | 0.13 | 0.09 | 0.44 | 0.24 | 0.18 | -0.16 | 0.28 | 0.40 | 0.18 | 0.23 | 0.31 | 0.29 |
| hr6488 | S | 5461 | 2.88 | 0.00 | -0.05 | -0.17 | -0.10 | -0.23 | -0.06 | -0.25 | -0.19 | -0.01 | 0.09 | 0.06 | -0.15 | -0.20 | -0.02 | -0.16 | -0.77 | 0.52 | 0.41 | -0.01 | 0.14 | 0.09 | 0.17 | 0.23 | 0.18 | 0.32 | 0.29 |
| hr6524 | S | 5162 | 2.78 | 1.14 | 0.51 | 0.19 | 0.32 | 0.26 | 0.56 | 0.34 | 0.23 | 0.34 | 0.40 | 0.36 | 0.32 | 0.24 | 0.22 | 0.28 | 0.22 | 0.53 | 0.45 | 0.42 | 0.14 | 0.39 | 0.67 | 0.36 | 0.43 | 0.58 | 0.40 |
| hr6564 | S | 4191 | 2.12 | 1.45 | 0.59 | 0.48 | 0.68 | 0.82 | 2.07 | 0.34 | 0.40 | 0.39 | 0.73 | 0.57 | 0.54 | 0.40 | 0.52 | 0.63 | 0.95 | 1.48 | 0.45 | 0.45 | -0.09 | 0.15 | 0.92 | 0.80 | 0.93 | 1.32 | 0.42 |
| hr6566 | S | 4616 | 2.58 | 1.30 | 0.02 | 0.08 | 0.18 | 0.31 | 0.86 | 0.02 | -0.04 | -0.02 | -0.01 | 0.00 | -0.02 | 0.00 | -0.01 | 0.10 | 0.16 | 0.69 | 0.00 | -0.14 | -0.54 | 0.03 | 0.24 | 0.12 | 0.18 | 0.58 | 0.09 |
| hr6575a | S | 4813 | 2.67 | 1.27 | 0.31 | 0.22 | 0.29 | 0.36 | 0.79 | 0.20 | 0.02 | 0.07 | 0.08 | 0.22 | 0.12 | 0.11 | 0.04 | 0.17 | 0.23 | 0.56 | 0.35 | 0.20 | -0.20 | 0.18 | 0.45 | 0.26 | 0.33 | 0.51 | 0.19 |
| hr6575b | S | 6616 | 3.87 | 3.62 | | -0.12 | 0.73 | 1.27 | | -0.23 | 0.43 | 0.62 | 0.93 | 0.95 | 1.19 | 0.31 | 1.34 | 0.57 | | | 1.67 | 1.14 | | | | 1.55 | 0.27 | | 1.28 |
| hr6591 | S | 4705 | 2.50 | 1.27 | 0.04 | -0.02 | 0.17 | 0.08 | 0.48 | -0.04 | -0.07 | 0.05 | 0.12 | -0.01 | -0.03 | -0.12 | -0.04 | -0.02 | 0.26 | 0.43 | 0.10 | -0.10 | -0.34 | -0.09 | 0.19 | 0.04 | 0.23 | 0.39 | 0.01 |
| hr6603 | S | 4553 | 2.41 | 1.53 | 0.58 | 0.46 | 0.50 | 0.58 | 1.33 | 0.30 | 0.26 | 0.33 | 0.50 | 0.45 | 0.43 | 0.31 | 0.38 | 0.49 | 0.70 | 1.40 | 0.28 | 0.35 | -0.20 | 0.13 | 0.74 | 0.42 | 0.58 | 0.94 | 0.47 |
| hr6606 | S | 4819 | 2.51 | 1.42 | 0.04 | -0.03 | 0.16 | 0.17 | 0.46 | -0.02 | -0.08 | -0.04 | -0.04 | -0.02 | -0.02 | -0.09 | -0.09 | -0.05 | 0.00 | 0.28 | 0.12 | 0.00 | -0.39 | 0.07 | 0.17 | 0.08 | 0.14 | 0.20 | 0.06 |
| hr6607 | S | 4745 | 2.72 | 1.24 | 0.66 | 0.35 | 0.56 | 0.59 | 1.20 | 0.37 | 0.26 | 0.29 | 0.48 | 0.42 | 0.40 | 0.35 | 0.34 | 0.50 | 0.49 | 1.10 | 0.39 | 0.52 | -0.05 | 0.31 | 0.65 | 0.35 | 0.46 | 0.71 | 0.43 |
| hr6638 | S | 4944 | 2.84 | 1.26 | 0.31 | 0.15 | 0.23 | 0.26 | 0.77 | 0.20 | 0.02 | 0.07 | 0.11 | 0.18 | 0.11 | 0.08 | 0.03 | 0.13 | 0.07 | 0.60 | 0.21 | 0.23 | -0.17 | 0.20 | 0.36 | 0.15 | 0.21 | 0.47 | 0.24 |
| hr6639 | S | 4468 | 2.46 | 1.50 | 0.24 | 0.21 | 0.38 | 0.49 | 1.23 | 0.15 | 0.13 | 0.13 | 0.27 | 0.25 | 0.23 | 0.21 | 0.28 | 0.35 | 0.52 | 1.39 | 0.04 | 0.18 | -0.35 | 0.09 | 0.72 | 0.33 | 0.54 | 0.96 | 0.32 |
| hr6644 | S | 4555 | 2.48 | 1.43 | 0.34 | 0.37 | 0.37 | 0.53 | 1.17 | 0.30 | 0.19 | 0.31 | 0.44 | 0.33 | 0.22 | 0.22 | 0.30 | 0.38 | 0.44 | 1.01 | 0.13 | 0.26 | -0.32 | 0.20 | 0.77 | 0.42 | 0.49 | 1.15 | 0.39 |
| hr6654 | S | 4661 | 2.64 | 1.20 | 0.20 | 0.07 | 0.21 | 0.22 | 0.65 | 0.05 | -0.02 | 0.04 | 0.12 | 0.09 | 0.09 | 0.01 | 0.01 | 0.11 | 0.31 | 0.46 | 0.11 | -0.03 | -0.35 | 0.07 | 0.48 | 0.16 | 0.35 | 0.69 | 0.10 |
| hr6659 | S | 4577 | 2.50 | 1.27 | 0.11 | -0.02 | 0.04 | 0.35 | 1.02 | -0.12 | -0.10 | -0.20 | -0.22 | 0.01 | -0.15 | -0.02 | -0.08 | 0.06 | 0.16 | 0.59 | 0.05 | 0.13 | -0.61 | 0.17 | 0.38 | 0.06 | 0.17 | 0.53 | 0.02 |
| hr6698 | S | 4874 | 2.54 | 1.57 | 0.57 | 0.15 | 0.32 | 0.36 | 0.78 | 0.17 | 0.08 | 0.10 | 0.21 | 0.24 | 0.20 | 0.13 | 0.11 | 0.19 | 0.22 | 0.58 | 0.37 | 0.23 | -0.25 | 0.07 | 0.30 | 0.04 | 0.15 | 0.25 | 0.17 |
| hr6703 | S | 4930 | 2.67 | 1.41 | 0.34 | 0.13 | 0.19 | 0.25 | 0.74 | 0.17 | -0.01 | 0.01 | 0.02 | 0.15 | 0.05 | 0.04 | 0.01 | 0.07 | 0.04 | 0.65 | 0.21 | 0.08 | -0.13 | 0.19 | 0.27 | 0.09 | 0.16 | 0.19 | 0.14 |
| hr6711 | S | 4999 | 2.78 | 1.37 | 0.32 | 0.11 | 0.23 | 0.26 | 0.62 | 0.17 | 0.10 | 0.11 | 0.11 | 0.19 | 0.16 | 0.09 | 0.04 | 0.11 | 0.05 | 0.11 | 0.34 | 0.17 | -0.06 | 0.21 | 0.38 | 0.20 | 0.28 | 0.37 | 0.25 |
| hr6757 | S | 4943 | 2.37 | 1.71 | 0.63 | 0.17 | 0.31 | 0.23 | 0.29 | 0.12 | 0.08 | 0.14 | 0.11 | 0.21 | -0.05 | 0.07 | 0.04 | 0.07 | -0.10 | 0.35 | 0.55 | 0.11 | -0.17 | 0.32 | 0.24 | 0.12 | 0.28 | 0.34 | 0.22 |
| hr6770 | S | 4921 | 2.49 | 1.49 | 0.30 | -0.04 | 0.25 | 0.24 | 0.52 | 0.07 | -0.05 | -0.01 | 0.01 | 0.11 | 0.01 | 0.01 | -0.05 | 0.02 | -0.06 | 0.36 | 0.31 | 0.03 | -0.22 | 0.14 | 0.24 | 0.06 | 0.15 | 0.13 | 0.21 |
| hr6793 | S | 4516 | 2.62 | 1.22 | 0.49 | 0.45 | 0.52 | 0.56 | 1.34 | 0.24 | 0.31 | 0.31 | 0.56 | 0.42 | 0.45 | 0.31 | 0.38 | 0.51 | 0.75 | 1.40 | 0.22 | 0.29 | -0.23 | 0.13 | 1.06 | 0.37 | 0.67 | 0.97 | 0.51 |
| hr6799 | S | 4487 | 2.40 | 1.31 | 0.40 | 0.25 | 0.48 | 0.48 | 1.29 | 0.20 | 0.20 | 0.24 | 0.48 | 0.34 | 0.26 | 0.18 | 0.22 | 0.34 | 0.44 | 1.39 | 0.21 | 0.14 | -0.15 | 0.06 | 0.76 | 0.14 | 0.52 | 1.06 | 0.24 |
| hr6801 | S | 4687 | 2.38 | 1.44 | 0.08 | -0.04 | 0.15 | 0.18 | 0.59 | 0.04 | -0.07 | -0.05 | -0.05 | 0.02 | -0.03 | -0.06 | -0.09 | -0.03 | 0.04 | 0.45 | 0.14 | -0.02 | -0.39 | 0.11 | 0.28 | 0.14 | 0.24 | 0.28 | 0.01 |
| hr6840 | S | 4898 | 2.66 | 1.53 | -0.44 | -0.24 | -0.26 | -0.16 | 0.12 | -0.44 | -0.49 | -0.45 | -0.60 | -0.60 | -0.95 | -0.65 | -0.53 | -0.59 | -0.72 | -0.04 | 0.11 | -0.47 | -0.68 | -0.42 | -0.23 | -0.49 | -0.30 | -0.48 | -0.15 |
| hr6853 | S | 4798 | 2.49 | 1.48 | 0.25 | 0.05 | -0.01 | 0.01 | 0.32 | -0.18 | -0.26 | -0.12 | -0.21 | -0.32 | -0.56 | -0.40 | -0.24 | -0.30 | -0.10 | -0.03 | -0.05 | -0.42 | -0.69 | -0.52 | -0.32 | -0.35 | -0.28 | -0.11 | -0.12 |
| hr6859 | S | 4202 | 1.16 | 2.11 | 0.31 | 0.04 | 0.05 | 0.36 | 1.12 | -0.20 | -0.23 | -0.17 | -0.11 | 0.10 | -0.09 | -0.06 | -0.08 | -0.02 | 0.27 | | 0.18 | -0.04 | -0.52 | 0.03 | 0.27 | 0.11 | 0.16 | 0.83 | 0.12 |
| hr6865 | S | 4646 | 2.65 | 1.11 | 0.10 | 0.08 | 0.21 | 0.25 | 0.74 | 0.02 | -0.01 | 0.03 | 0.09 | 0.09 | 0.04 | 0.02 | 0.00 | 0.12 | 0.27 | 0.44 | 0.07 | 0.08 | -0.35 | 0.11 | 0.39 | 0.13 | 0.29 | 0.62 | 0.09 |
| hr6866 | S | 4982 | 2.59 | 1.36 | 0.20 | -0.03 | 0.14 | 0.12 | 0.51 | 0.05 | -0.13 | -0.09 | -0.12 | 0.01 | -0.11 | -0.06 | -0.14 | -0.08 | -0.18 | 0.27 | 0.16 | 0.03 | -0.23 | 0.19 | 0.10 | 0.13 | 0.12 | 0.10 | 0.10 |
| hr6872 | S | 4549 | 2.16 | 1.54 | 0.41 | 0.15 | 0.30 | 0.44 | 1.09 | 0.19 | 0.08 | 0.16 | 0.29 | 0.28 | 0.22 | 0.15 | 0.13 | 0.26 | 0.33 | 1.13 | 0.29 | 0.16 | -0.20 | 0.09 | 0.68 | 0.17 | 0.29 | 0.83 | 0.18 |
| hr6884 | S | 4894 | 2.67 | 1.33 | 0.14 | -0.05 | 0.15 | 0.18 | 0.38 | 0.06 | -0.05 | -0.02 | -0.05 | 0.05 | -0.02 | -0.03 | -0.07 | -0.01 | -0.02 | 0.48 | 0.25 | 0.04 | -0.23 | 0.20 | 0.24 | 0.24 | 0.24 | 0.27 | 0.12 |
| hr6895 | S | 4484 | 2.43 | 1.19 | 0.18 | 0.15 | 0.30 | 0.31 | 1.01 | 0.09 | 0.09 | 0.15 | 0.36 | 0.15 | 0.10 | 0.06 | 0.11 | 0.21 | 0.94 | 0.88 | 0.07 | 0.12 | -0.36 | -0.01 | 0.57 | 0.16 | 0.44 | 0.80 | 0.25 |
| hr6935 | S | 4833 | 2.77 | 1.18 | 0.28 | 0.09 | 0.22 | 0.33 | 0.88 | 0.11 | 0.00 | 0.02 | 0.08 | 0.13 | 0.11 | 0.07 | -0.01 | 0.14 | 0.15 | 0.34 | 0.21 | 0.17 | -0.24 | 0.17 | 0.44 | 0.13 | 0.29 | 0.43 | 0.12 |
| hr6945 | S | 4413 | 1.94 | 1.45 | -0.19 | -0.37 | -0.08 | -0.03 | 0.40 | -0.28 | -0.31 | -0.29 | -0.21 | -0.29 | -0.43 | -0.40 | -0.29 | -0.31 | 0.08 | 0.22 | -0.04 | -0.40 | -0.85 | -0.42 | -0.22 | -0.37 | -0.14 | -0.04 | -0.23 |
| hr6970 | S | 5008 | 2.55 | 1.38 | 0.20 | 0.06 | 0.18 | 0.14 | 0.36 | 0.14 | -0.18 | -0.21 | -0.03 | -0.04 | -0.27 | -0.16 | -0.21 | -0.14 | -0.65 | -0.04 | 0.37 | -0.07 | -0.30 | 0.19 | 0.12 | -0.02 | -0.11 | 0.01 | 0.20 |
| hr6973 | S | 4241 | 1.86 | 1.62 | 0.07 | 0.12 | 0.28 | 0.43 | 1.36 | 0.02 | 0.04 | 0.06 | 0.25 | 0.16 | 0.09 | 0.05 | 0.15 | 0.18 | | 1.22 | 0.03 | 0.08 | -0.51 | -0.04 | 0.57 | 0.08 | 0.43 | 0.98 | 0.10 |
| hr7010 | S | 5085 | 2.74 | 1.46 | 0.19 | 0.06 | 0.19 | 0.08 | 0.30 | 0.06 | 0.03 | 0.07 | 0.09 | 0.03 | 0.05 | -0.04 | 0.02 | 0.00 | 0.11 | 0.20 | 0.47 | 0.03 | -0.10 | 0.08 | 0.13 | 0.10 | 0.18 | 0.22 | 0.13 |
| hr7042 | S | 4943 | 2.76 | 1.36 | 0.53 | 0.24 | 0.39 | 0.42 | 0.79 | 0.30 | 0.19 | 0.25 | 0.34 | 0.33 | 0.28 | 0.22 | 0.21 | 0.29 | 0.31 | 0.62 | 0.34 | 0.24 | -0.07 | 0.26 | 0.48 | 0.26 | 0.37 | 0.67 | 0.33 |
| hr7064 | S | 4438 | 2.20 | 1.42 | 0.21 | 0.19 | 0.25 | 0.34 | 0.93 | 0.04 | 0.03 | 0.05 | 0.17 | 0.14 | 0.12 | 0.03 | 0.05 | 0.14 | 0.21 | 1.01 | 0.13 | 0.13 | -0.36 | 0.06 | 0.57 | 0.12 | 0.36 | 0.69 | 0.21 |
| hr7120 | S | 4244 | 1.97 | 1.69 | 0.09 | -0.09 | 0.28 | 0.52 | 1.48 | -0.10 | -0.02 | 0.05 | 0.21 | 0.17 | -0.09 | 0.05 | 0.16 | 0.21 | 0.28 | 1.31 | 0.77 | 0.45 | 0.27 | 0.62 | 0.67 | 0.61 | 0.80 | 1.20 | 0.43 |
| hr7125 | S | 4354 | 1.70 | 2.19 | 0.04 | -0.20 | -0.01 | -0.08 | 0.36 | -0.51 | -0.47 | -0.44 | -0.50 | -0.25 | -0.62 | -0.48 | -0.36 | -0.34 | | | -0.50 | -1.04 | -0.32 | -0.04 | -0.29 | -0.05 | 0.59 | -0.15 |
| hr7135 | S | 4666 | 2.50 | 1.32 | -0.01 | -0.12 | 0.02 | 0.08 | 0.60 | -0.13 | -0.12 | -0.12 | -0.12 | -0.08 | -0.18 | -0.14 | -0.17 | -0.09 | -0.03 | 0.40 | 0.10 | -0.07 | -0.50 | 0.05 | 0.22 | 0.13 | 0.19 | 0.45 | -0.06 |

| ID | | T | V | B | | | | | | | | | | | | | | | | | | | | | | | | |
|---|---|---|---|---|---|---|---|---|---|---|---|---|---|---|---|---|---|---|---|---|---|---|---|---|---|---|---|---|
| hr7137 | S | 5024 | 2.36 | 1.48 | 0.28 | 0.03 | 0.16 | 0.15 | 0.41 | 0.16 | -0.11 | -0.02 | -0.02 | 0.09 | -0.05 | -0.02 | -0.06 | -0.03 | -0.11 | 0.28 | 0.42 | 0.04 | -0.19 | 0.26 | 0.18 | 0.08 | 0.10 | -0.03 | 0.08 |
| hr7144 | S | 4951 | 2.53 | 1.57 | 0.31 | | 0.35 | 0.26 | 0.84 | 0.27 | -0.04 | -0.02 | 0.04 | 0.07 | 0.00 | 0.00 | 0.10 | 0.04 | -0.18 | | 0.20 | 0.09 | 0.03 | | 0.25 | 0.27 | 0.25 | 0.26 | |
| hr7146 | S | 4803 | 2.55 | 1.38 | 0.34 | 0.07 | 0.30 | 0.32 | 0.62 | 0.22 | 0.05 | 0.13 | 0.16 | 0.22 | 0.21 | 0.09 | 0.07 | 0.14 | 0.28 | 0.63 | 0.28 | 0.21 | -0.17 | 0.10 | 0.52 | 0.15 | 0.25 | 0.52 | 0.18 |
| hr7148 | S | 4710 | 2.52 | 1.41 | 0.32 | 0.19 | 0.34 | 0.38 | 0.94 | 0.25 | 0.17 | 0.23 | 0.36 | 0.28 | 0.23 | 0.15 | 0.16 | 0.26 | 0.41 | 1.08 | 0.23 | 0.12 | -0.19 | 0.10 | 0.60 | 0.21 | 0.34 | 0.79 | 0.20 |
| hr7176 | S | 4692 | 2.47 | 1.47 | 0.56 | 0.33 | 0.39 | 0.50 | 1.11 | 0.22 | 0.16 | 0.22 | 0.31 | 0.30 | 0.32 | 0.21 | 0.27 | 0.35 | 0.56 | 1.33 | 0.25 | 0.22 | -0.22 | 0.13 | 0.59 | 0.24 | 0.39 | 0.76 | 0.24 |
| hr7180 | S | 4561 | 2.10 | 1.55 | 0.22 | 0.06 | 0.21 | 0.27 | 0.84 | 0.03 | -0.12 | -0.02 | 0.04 | 0.10 | 0.00 | -0.01 | -0.05 | 0.06 | 0.11 | | 0.25 | 0.12 | -0.21 | 0.17 | 0.45 | 0.22 | 0.32 | 0.49 | 0.21 |
| hr7186 | S | 4991 | 2.56 | 1.59 | 0.42 | 0.13 | 0.23 | 0.29 | 0.73 | 0.19 | 0.07 | 0.09 | 0.08 | 0.20 | 0.12 | 0.10 | 0.03 | 0.14 | 0.03 | 0.33 | 0.33 | 0.17 | -0.12 | 0.17 | 0.33 | 0.18 | 0.23 | 0.32 | 0.27 |
| hr7187 | S | 4952 | 2.71 | 1.42 | 0.36 | 0.10 | 0.20 | 0.27 | 0.68 | 0.16 | -0.05 | 0.04 | 0.05 | 0.16 | 0.04 | 0.05 | -0.02 | 0.08 | 0.03 | 0.61 | 0.33 | 0.10 | -0.15 | 0.16 | 0.24 | 0.09 | 0.06 | 0.25 | 0.21 |
| hr7193 | S | 4601 | 2.47 | 1.27 | 0.18 | 0.07 | 0.20 | 0.24 | 0.73 | 0.08 | -0.03 | 0.03 | 0.12 | 0.11 | 0.10 | 0.03 | 0.01 | 0.11 | 0.32 | 0.43 | 0.12 | 0.12 | -0.33 | 0.13 | 0.41 | 0.20 | 0.36 | 0.55 | 0.03 |
| hr7196 | S | 4824 | 2.57 | 1.39 | -0.03 | -0.05 | 0.10 | 0.11 | 0.39 | -0.07 | -0.09 | -0.10 | -0.12 | -0.10 | -0.18 | -0.15 | -0.14 | -0.09 | -0.03 | 0.10 | 0.04 | -0.04 | -0.46 | 0.06 | 0.14 | 0.08 | 0.18 | 0.24 | 0.05 |
| hr7204 | S | 4869 | 2.63 | 1.34 | 0.29 | 0.10 | 0.27 | 0.29 | 0.69 | 0.17 | 0.04 | 0.07 | 0.09 | 0.18 | 0.11 | 0.08 | -0.01 | 0.11 | 0.12 | 0.45 | 0.29 | 0.23 | -0.20 | 0.20 | 0.43 | 0.19 | 0.34 | 0.42 | 0.16 |
| hr7217 | S | 4766 | 2.52 | 1.41 | 0.17 | 0.06 | 0.16 | 0.23 | 0.57 | 0.09 | -0.04 | -0.05 | -0.03 | 0.06 | 0.00 | -0.02 | -0.07 | 0.03 | 0.13 | 0.41 | 0.16 | -0.01 | -0.26 | 0.11 | 0.30 | 0.08 | 0.20 | 0.32 | 0.15 |
| hr7225a | S | 4496 | 2.27 | 1.35 | -0.04 | -0.13 | -0.09 | 0.13 | 0.64 | -0.18 | -0.23 | -0.29 | -0.23 | -0.17 | -0.24 | -0.16 | -0.20 | -0.10 | -0.02 | 0.72 | -0.06 | -0.16 | -0.63 | 0.02 | 0.15 | 0.04 | 0.11 | 0.29 | -0.14 |
| hr7225b | S | 4156 | 1.91 | 1.63 | 0.36 | 0.45 | 0.58 | 0.71 | 2.06 | 0.37 | 0.26 | 0.33 | 0.62 | 0.48 | 0.42 | 0.33 | 0.43 | 0.52 | 1.06 | 3.02 | 0.32 | 0.43 | -0.21 | 0.21 | 0.82 | 0.76 | 0.86 | 1.54 | 0.38 |
| hr7234 | S | 4444 | 2.17 | 1.37 | 0.09 | -0.16 | 0.11 | 0.19 | 0.78 | -0.09 | -0.10 | -0.08 | 0.03 | -0.01 | -0.04 | -0.08 | -0.07 | 0.02 | 0.22 | 0.69 | -0.02 | -0.09 | -0.55 | 0.03 | 0.23 | 0.14 | 0.31 | 0.49 | 0.06 |
| hr7295 | S | 4875 | 2.53 | 1.64 | 0.54 | 0.09 | 0.30 | 0.47 | 0.90 | 0.16 | 0.10 | 0.13 | 0.16 | 0.26 | 0.19 | 0.17 | 0.16 | 0.24 | 0.16 | 0.98 | 0.38 | 0.18 | -0.15 | 0.20 | 0.54 | 0.28 | 0.26 | 0.89 | 0.38 |
| hr7310 | S | 4776 | 2.52 | 1.37 | 0.02 | -0.08 | 0.11 | 0.13 | 0.32 | -0.07 | -0.15 | -0.13 | -0.14 | -0.04 | -0.12 | -0.11 | -0.16 | -0.08 | -0.03 | 0.48 | 0.07 | 0.02 | -0.42 | 0.07 | 0.16 | 0.02 | 0.13 | 0.26 | 0.00 |
| hr7325 | S | 4753 | 2.45 | 1.44 | 0.04 | -0.07 | 0.11 | 0.14 | 0.39 | 0.01 | -0.13 | -0.05 | -0.08 | -0.01 | -0.10 | -0.10 | -0.11 | -0.07 | 0.00 | 0.02 | 0.16 | -0.06 | -0.35 | 0.16 | 0.34 | 0.18 | 0.26 | 0.34 | 0.07 |
| hr7331 | S | 7262 | 3.43 | 3.45 | 0.21 | 0.30 | | 0.41 | 0.21 | 0.38 | 0.31 | 0.77 | 1.00 | 0.84 | -0.13 | 0.35 | 1.20 | 0.57 | | 0.47 | | 1.36 | | | 0.21 | 0.63 | 0.62 | 0.72 | |
| hr7349 | S | 4662 | 2.47 | 1.39 | 0.11 | 0.11 | 0.24 | 0.30 | 0.70 | 0.04 | 0.04 | -0.02 | 0.05 | 0.03 | 0.02 | 0.00 | -0.02 | 0.08 | 0.20 | 0.67 | 0.12 | 0.09 | -0.39 | 0.12 | 0.41 | 0.07 | 0.28 | 0.52 | 0.10 |
| hr7359 | S | 4838 | 2.63 | 1.31 | 0.38 | 0.25 | 0.37 | 0.31 | 0.81 | 0.27 | 0.09 | 0.25 | 0.30 | 0.29 | 0.31 | 0.15 | 0.14 | 0.23 | 0.36 | 0.54 | 0.30 | 0.19 | -0.10 | 0.18 | 0.54 | 0.22 | 0.39 | 0.67 | 0.18 |
| hr7376 | S | 4679 | 2.64 | 1.17 | 0.00 | 0.00 | 0.16 | 0.11 | 0.69 | -0.02 | -0.04 | 0.02 | 0.09 | 0.00 | 0.01 | -0.08 | -0.04 | 0.03 | 0.27 | 0.22 | 0.02 | -0.08 | -0.43 | -0.08 | 0.21 | 0.04 | 0.26 | 0.57 | -0.04 |
| hr7385 | S | 4763 | 2.50 | 1.36 | 0.08 | -0.02 | 0.12 | 0.18 | 0.53 | 0.01 | -0.08 | -0.09 | -0.12 | 0.01 | -0.09 | -0.07 | -0.12 | -0.04 | 0.07 | 0.52 | 0.14 | -0.06 | -0.36 | 0.14 | 0.24 | 0.13 | 0.20 | 0.08 | 0.03 |
| hr7389 | S | 6281 | 3.64 | 3.27 | 0.50 | -0.11 | 0.06 | 0.32 | 0.51 | 0.09 | -0.06 | 0.16 | 0.11 | 0.29 | 0.13 | -0.01 | 0.32 | 0.13 | -0.43 | | | 0.11 | 0.24 | -0.35 | 0.66 | | 0.16 | | 0.77 | |
| hr7407 | S | 4740 | 2.54 | 1.38 | 0.24 | 0.16 | 0.33 | 0.30 | 0.90 | 0.14 | 0.02 | 0.10 | 0.17 | 0.20 | 0.14 | 0.09 | 0.07 | 0.16 | 0.34 | 0.48 | 0.22 | 0.15 | -0.25 | 0.16 | 0.48 | 0.19 | 0.33 | 0.46 | 0.16 |
| hr7433 | S | 4631 | 2.61 | 1.27 | 0.03 | 0.05 | 0.12 | 0.32 | 0.83 | 0.00 | -0.02 | -0.08 | -0.06 | 0.03 | -0.01 | 0.01 | -0.03 | 0.10 | 0.27 | 1.18 | 0.06 | 0.04 | -0.47 | 0.16 | 0.31 | 0.15 | 0.30 | 0.44 | 0.09 |
| hr7449 | S | 4810 | 2.67 | 1.38 | 0.38 | 0.22 | 0.32 | 0.38 | 1.01 | 0.20 | 0.11 | 0.12 | 0.23 | 0.23 | 0.16 | 0.11 | 0.10 | 0.18 | 0.37 | 1.37 | 0.65 | 0.45 | 0.27 | 0.45 | 0.71 | 0.42 | 0.48 | 0.74 | 0.27 |
| hr7465 | S | 4730 | 2.41 | 1.34 | 0.01 | -0.08 | 0.06 | 0.07 | 0.29 | -0.07 | -0.14 | -0.08 | -0.08 | -0.05 | -0.16 | -0.13 | -0.17 | -0.11 | 0.01 | 0.03 | 0.09 | -0.07 | -0.37 | 0.14 | 0.22 | 0.16 | 0.29 | 0.34 | -0.03 |
| hr7478b | S | 4872 | 2.51 | 0.80 | 0.11 | -0.21 | 0.04 | 0.01 | 0.32 | -0.10 | -0.32 | -0.20 | -0.17 | -0.04 | -0.22 | -0.27 | -0.25 | -0.18 | -0.91 | 1.11 | 1.32 | 0.10 | -0.32 | -0.49 | -0.06 | 0.44 | -0.02 | 0.24 | 0.07 |
| HR7487 | S | 5052 | 2.87 | 1.33 | 0.28 | 0.09 | 0.21 | 0.23 | 0.55 | 0.14 | 0.04 | 0.07 | 0.04 | 0.15 | 0.10 | 0.06 | 0.01 | 0.08 | -0.02 | 0.19 | 0.35 | 0.21 | -0.10 | 0.28 | 0.36 | 0.21 | 0.34 | 0.53 | 0.21 |
| hr7506 | S | 5008 | 2.69 | 1.48 | 0.30 | 0.14 | 0.21 | 0.22 | 0.39 | 0.14 | 0.02 | 0.04 | 0.06 | 0.18 | 0.01 | 0.05 | -0.01 | 0.04 | -0.01 | 0.24 | 0.50 | 0.19 | -0.20 | 0.25 | 0.31 | 0.25 | 0.19 | 0.27 | 0.28 |
| hr7517 | S | 4920 | 2.55 | 1.40 | 0.25 | -0.09 | 0.16 | 0.24 | 0.69 | 0.11 | 0.00 | -0.01 | -0.04 | 0.10 | -0.01 | 0.02 | -0.08 | 0.04 | 0.04 | 0.38 | 0.26 | 0.10 | -0.22 | 0.24 | 0.26 | 0.14 | 0.20 | 0.05 | 0.16 |
| hr7526 | S | 5035 | 3.08 | 0.99 | -0.14 | -0.21 | -0.05 | -0.12 | 0.17 | -0.08 | -0.23 | -0.12 | -0.18 | -0.16 | -0.25 | -0.23 | -0.24 | -0.21 | -0.15 | -0.16 | 0.19 | -0.16 | -0.39 | -0.07 | 0.04 | 0.02 | 0.03 | 0.09 | 0.00 |
| hr7540 | S | 4959 | 2.64 | 1.40 | 0.07 | -0.10 | 0.15 | 0.05 | 0.28 | 0.03 | -0.02 | 0.05 | 0.04 | 0.05 | 0.03 | -0.07 | -0.06 | -0.04 | 0.00 | 0.07 | 0.22 | -0.02 | -0.21 | 0.11 | 0.22 | 0.19 | 0.22 | 0.33 | 0.16 |
| hr7541 | S | 4421 | 2.39 | 1.42 | 0.42 | 0.42 | 0.41 | 0.57 | 1.50 | 0.34 | 0.27 | 0.29 | 0.53 | 0.42 | 0.35 | 0.27 | 0.37 | 0.48 | 0.59 | 1.62 | 0.25 | 0.26 | -0.16 | 0.07 | 0.99 | 0.39 | 0.67 | 1.38 | 0.46 |
| hr7561 | S | 4779 | 2.47 | 0.98 | 0.06 | -0.05 | 0.12 | 0.06 | 0.37 | 0.08 | -0.07 | 0.00 | 0.13 | 0.03 | -0.09 | -0.10 | -0.11 | -0.05 | -0.05 | 0.67 | 0.19 | 0.05 | -0.26 | 0.11 | 0.26 | 0.10 | 0.23 | 0.30 | -0.01 |
| hr7576 | S | 4337 | 2.28 | 1.69 | 0.86 | 0.67 | 0.66 | 0.88 | 1.95 | 0.43 | 0.39 | 0.42 | 0.73 | 0.60 | 0.75 | 0.44 | 0.62 | 0.68 | 0.72 | 1.65 | 0.34 | 0.36 | -0.09 | -0.06 | 0.79 | 0.66 | 0.73 | 1.28 | 0.42 |
| hr7597 | S | 5306 | 3.53 | 0.94 | 0.01 | -0.04 | 0.09 | 0.09 | 0.28 | 0.06 | 0.02 | -0.04 | -0.11 | 0.03 | -0.08 | -0.04 | -0.12 | -0.02 | -0.01 | 0.01 | 0.41 | 0.14 | 0.05 | 0.12 | 0.12 | 0.17 | 0.15 | -0.05 | 0.29 |
| hr7681 | S | 4519 | 2.52 | 1.25 | 0.17 | | 0.25 | 0.50 | 1.25 | 0.20 | 0.05 | -0.05 | 0.41 | 0.13 | 0.02 | 0.15 | 0.06 | 0.32 | 0.31 | | 0.01 | 0.23 | -0.06 | | 0.55 | 0.34 | 0.38 | 0.74 | |
| hr7712 | S | 4439 | 2.21 | 1.42 | 0.12 | 0.04 | 0.09 | 0.36 | 1.03 | -0.15 | -0.06 | -0.17 | -0.11 | -0.02 | -0.05 | -0.01 | -0.07 | 0.09 | 0.17 | 0.83 | 0.12 | -0.02 | -0.50 | 0.23 | 0.45 | 0.17 | 0.33 | 0.55 | 0.04 |
| hr7713 | S | 5060 | 2.72 | 1.41 | -0.25 | -0.31 | -0.16 | -0.22 | -0.01 | -0.27 | -0.29 | -0.29 | -0.34 | -0.33 | -0.49 | -0.40 | -0.33 | -0.38 | -0.49 | -0.34 | 0.24 | -0.35 | -0.53 | -0.17 | -0.19 | -0.17 | -0.10 | -0.25 | -0.17 |
| hr7748 | S | 4782 | 2.57 | 1.33 | 0.07 | 0.05 | 0.12 | 0.16 | 0.46 | -0.03 | -0.09 | -0.08 | -0.10 | -0.02 | -0.06 | -0.07 | -0.13 | -0.04 | 0.03 | 0.32 | 0.17 | 0.02 | -0.35 | 0.15 | 0.33 | 0.10 | 0.21 | 0.29 | 0.03 |
| hr7778 | S | 5080 | 2.64 | 1.45 | 0.29 | 0.08 | 0.38 | 0.16 | 0.52 | 0.11 | 0.01 | 0.05 | 0.07 | 0.13 | 0.06 | 0.04 | 0.02 | 0.03 | -0.07 | 0.16 | 0.46 | 0.10 | -0.08 | 0.18 | 0.20 | 0.22 | 0.20 | 0.25 | 0.32 |
| hr7788 | S | 4969 | 2.61 | 1.51 | 0.22 | -0.06 | 0.02 | 0.18 | 0.48 | -0.03 | -0.06 | -0.17 | -0.30 | -0.02 | -0.21 | -0.10 | -0.22 | -0.10 | -0.23 | 0.26 | 0.32 | 0.01 | -0.40 | 0.34 | 0.23 | 0.28 | 0.19 | 0.42 | 0.18 |
| hr7794 | S | 4866 | 2.73 | 1.25 | 0.22 | 0.08 | 0.19 | 0.25 | 0.65 | 0.12 | 0.00 | 0.02 | 0.04 | 0.10 | 0.10 | 0.04 | -0.02 | 0.08 | 0.15 | 0.39 | 0.20 | 0.20 | -0.20 | 0.19 | 0.41 | 0.15 | 0.28 | 0.31 | 0.23 |
| hr7802 | S | 4827 | 2.75 | 1.25 | 0.49 | 0.33 | 0.46 | 0.45 | 0.90 | 0.31 | 0.20 | 0.25 | 0.31 | 0.35 | 0.33 | 0.26 | 0.21 | 0.36 | 0.43 | 0.80 | 0.33 | 0.37 | -0.10 | 0.31 | 0.68 | 0.37 | 0.45 | 0.81 | 0.32 |
| hr7806 | S | 4211 | 1.86 | 1.55 | 0.23 | 0.13 | 0.33 | 0.54 | 1.64 | 0.05 | 0.06 | 0.05 | 0.27 | 0.22 | 0.18 | 0.10 | 0.14 | 0.23 | 0.52 | 0.92 | 0.19 | 0.20 | -0.40 | 0.10 | 0.51 | 0.16 | 0.57 | 1.04 | 0.11 |
| hr7820 | S | 4703 | 2.74 | 1.25 | 0.06 | 0.01 | 0.17 | 0.25 | 0.68 | 0.06 | 0.03 | 0.01 | 0.02 | 0.08 | 0.04 | 0.04 | -0.01 | 0.08 | 0.21 | 0.54 | 0.16 | 0.18 | -0.30 | 0.27 | 0.67 | 0.26 | 0.42 | 0.47 | 0.17 |
| hr7824 | S | 5029 | 2.79 | 1.29 | 0.11 | -0.02 | 0.04 | 0.07 | 0.41 | -0.03 | -0.09 | -0.08 | -0.12 | -0.01 | -0.15 | -0.10 | -0.17 | -0.10 | -0.15 | 0.00 | 0.27 | 0.02 | -0.27 | 0.21 | 0.19 | 0.17 | 0.18 | 0.15 | 0.14 |
| hr7831 | S | 4592 | 2.51 | 1.48 | 0.45 | 0.36 | 0.46 | 0.56 | 1.30 | 0.27 | 0.26 | 0.31 | 0.51 | 0.39 | 0.40 | 0.29 | 0.39 | 0.46 | 0.58 | 1.83 | 0.22 | 0.28 | -0.14 | 0.17 | 1.08 | 0.41 | 0.53 | 1.16 | 0.60 |
| hr7854 | S | 4800 | 2.57 | 1.28 | 0.13 | 0.00 | 0.12 | 0.16 | 0.49 | 0.04 | -0.09 | 0.00 | 0.02 | 0.06 | 0.00 | -0.04 | -0.08 | -0.02 | 0.07 | 0.36 | 0.17 | 0.05 | -0.22 | 0.13 | 0.31 | 0.15 | 0.28 | 0.45 | 0.07 |
| hr7897 | S | 4715 | 2.57 | 1.37 | 0.32 | 0.08 | 0.25 | 0.36 | 0.85 | 0.18 | 0.04 | 0.10 | 0.16 | 0.22 | 0.17 | 0.12 | 0.09 | 0.20 | 0.41 | 0.92 | 0.23 | 0.25 | -0.24 | 0.20 | 0.54 | 0.21 | 0.36 | 0.48 | 0.17 |
| hr7904 | S | 4728 | 2.44 | 1.39 | 0.22 | 0.10 | 0.29 | 0.25 | 0.62 | 0.11 | 0.06 | 0.10 | 0.17 | 0.16 | 0.09 | 0.03 | 0.05 | 0.11 | 0.27 | 0.56 | 0.25 | 0.13 | -0.26 | 0.16 | 0.44 | 0.17 | 0.34 | 0.53 | 0.16 |
| hr7905 | S | 4760 | 2.56 | 1.32 | -0.18 | -0.12 | -0.01 | 0.04 | 0.35 | -0.17 | -0.22 | -0.20 | -0.23 | -0.21 | -0.32 | -0.27 | -0.27 | -0.20 | -0.44 | 0.05 | 0.05 | -0.28 | -0.66 | -0.24 | -0.09 | -0.23 | -0.08 | 0.02 | -0.14 |
| hr7923 | S | 4916 | 2.61 | 1.35 | -0.16 | -0.23 | -0.03 | -0.09 | 0.25 | -0.15 | -0.23 | -0.20 | -0.24 | -0.22 | -0.30 | -0.30 | -0.29 | -0.27 | -0.41 | -0.12 | 0.25 | -0.30 | -0.33 | -0.15 | -0.05 | -0.11 | -0.04 | -0.05 | -0.06 |
| hr7939 | S | 4473 | 2.01 | 1.54 | 0.18 | 0.14 | 0.21 | 0.22 | 0.81 | 0.00 | -0.05 | -0.04 | 0.03 | 0.04 | -0.01 | -0.05 | -0.04 | 0.04 | 0.17 | 0.66 | 0.01 | -0.06 | -0.57 | -0.07 | 0.43 | -0.04 | 0.17 | 0.42 | 0.00 |
| hr7942 | S | 4701 | 2.27 | 1.42 | 0.19 | -0.04 | 0.16 | 0.19 | 0.53 | 0.07 | -0.13 | -0.06 | -0.03 | 0.04 | -0.09 | -0.06 | -0.11 | -0.04 | 0.08 | 0.46 | 0.17 | 0.05 | -0.43 | 0.19 | 0.22 | 0.10 | 0.19 | 0.34 | 0.01 |
| hr7962 | S | 4623 | 2.40 | 1.36 | 0.05 | -0.05 | 0.18 | 0.09 | 0.49 | -0.07 | 0.00 | 0.03 | 0.13 | 0.00 | 0.00 | -0.12 | -0.02 | -0.01 | 0.38 | 0.24 | 0.11 | -0.11 | -0.41 | -0.12 | 0.20 | 0.01 | 0.23 | 0.49 | 0.18 |
| hr7995 | S | 5155 | 2.86 | 1.51 | 0.24 | -0.05 | 0.04 | 0.12 | 0.32 | -0.01 | -0.10 | -0.05 | -0.11 | 0.03 | -0.12 | -0.08 | -0.14 | -0.09 | -0.20 | 0.04 | 0.30 | 0.12 | -0.19 | 0.21 | 0.09 | 0.18 | 0.13 | 0.07 | 0.23 |
| hr8000 | S | 4525 | 2.34 | 1.35 | -0.24 | -0.07 | -0.01 | -0.07 | 0.37 | -0.23 | -0.20 | -0.11 | -0.11 | -0.34 | -0.56 | -0.46 | -0.29 | -0.34 | 0.07 | 0.07 | -0.04 | -0.32 | -0.60 | -0.47 | -0.12 | -0.26 | -0.09 | 0.08 | -0.02 |
| hr8017 | S | 4441 | 2.47 | 1.45 | 0.45 | 0.41 | 0.50 | 0.72 | 1.53 | 0.35 | 0.33 | 0.35 | 0.57 | 0.45 | 0.47 | 0.36 | 0.44 | 0.54 | 0.82 | 1.70 | 0.21 | 0.26 | -0.23 | 0.21 | 0.94 | 0.40 | 0.75 | 1.26 | 0.54 |
| hr8030 | S | 4966 | 2.89 | 1.20 | 0.20 | 0.05 | 0.21 | 0.20 | 0.43 | 0.12 | -0.02 | 0.05 | 0.05 | 0.12 | 0.06 | 0.03 | -0.04 | 0.07 | -0.32 | 0.22 | 0.38 | 0.20 | -0.19 | 0.23 | 0.35 | 0.19 | 0.32 | 0.35 | 0.29 |
| hr8034a | S | 6223 | 3.74 | 2.46 | 0.43 | | | 0.19 | 0.27 | 0.11 | 0.36 | 0.37 | 0.34 | 0.35 | 0.38 | -0.05 | 0.40 | -0.08 | | 0.22 | | 1.27 | | -0.19 | 0.14 | 1.19 | 0.73 | 0.60 | 0.88 |
| hr8034b | S | 6399 | 3.89 | 1.59 | 0.10 | 0.13 | 0.13 | 0.06 | -0.03 | 0.12 | 0.09 | 0.12 | 0.12 | 0.11 | -0.05 | -0.01 | 0.24 | 0.04 | -0.23 | -0.20 | | 0.17 | 0.04 | 0.04 | 0.59 | 0.30 | 0.15 | 0.15 | 0.22 |
| hr8035 | S | 4926 | 2.85 | 1.29 | 0.42 | 0.22 | 0.36 | 0.36 | 0.77 | 0.25 | 0.14 | 0.19 | 0.23 | 0.29 | 0.27 | 0.18 | 0.18 | 0.25 | 0.36 | 0.97 | 0.35 | 0.21 | -0.14 | 0.23 | 0.60 | 0.29 | 0.36 | 0.57 | 0.29 |
| hr8072 | S | 4887 | 2.82 | 1.33 | 0.54 | 0.29 | 0.35 | 0.45 | 1.00 | 0.27 | 0.20 | 0.23 | 0.32 | 0.32 | 0.31 | 0.23 | 0.21 | 0.33 | 0.41 | 1.02 | 0.41 | 0.35 | -0.03 | 0.27 | 0.63 | 0.28 | 0.39 | 0.76 | 0.41 |
| hr8082 | S | 4819 | 2.52 | 1.39 | 0.15 | 0.08 | 0.20 | 0.17 | 0.43 | 0.08 | 0.00 | 0.06 | 0.07 | 0.10 | 0.03 | 0.00 | -0.04 | 0.04 | 0.12 | 0.24 | 0.22 | 0.13 | -0.27 | 0.18 | 0.32 | 0.13 | 0.29 | 0.49 | 0.13 |
| hr8093 | S | 4938 | 2.87 | 1.24 | 0.19 | -0.03 | 0.14 | 0.24 | 0.52 | 0.15 | -0.01 | 0.04 | 0.03 | 0.12 | 0.06 | 0.05 | -0.03 | 0.07 | 0.07 | 0.38 | 0.29 | 0.21 | -0.18 | 0.24 | 0.31 | 0.22 | 0.33 | 0.37 | 0.24 |
| hr8096 | S | 4563 | 2.59 | 1.41 | 0.37 | 0.35 | 0.50 | 0.54 | 1.34 | 0.28 | 0.23 | 0.29 | 0.45 | 0.35 | 0.36 | 0.30 | 0.34 | 0.44 | 0.65 | 1.60 | 0.12 | 0.17 | -0.31 | 0.21 | 0.85 | 0.36 | 0.56 | 1.14 | 0.44 |
| hr8115 | S | 4891 | 2.45 | 1.50 | 0.34 | -0.12 | 0.24 | 0.30 | 0.67 | 0.15 | -0.05 | -0.01 | -0.07 | 0.09 | -0.05 | 0.01 | -0.07 | 0.05 | 0.11 | 0.52 | 0.63 | 0.37 | 0.11 | 0.53 | 0.54 | 0.45 | 0.38 | 0.47 | 0.18 |
| hr8165 | S | 4689 | 2.53 | 1.44 | 0.08 | 0.19 | 0.24 | 0.26 | 0.68 | 0.02 | 0.02 | 0.05 | 0.11 | -0.02 | -0.15 | -0.10 | 0.04 | 0.01 | 0.42 | 0.51 | 0.04 | -0.12 | -0.48 | -0.10 | 0.23 | 0.06 | 0.13 | 0.70 | 0.15 |
| hr8167 | S | 5011 | 2.66 | 1.49 | 0.30 | 0.14 | 0.08 | 0.14 | 0.51 | 0.02 | -0.05 | -0.02 | -0.07 | 0.11 | -0.07 | -0.03 | -0.10 | -0.03 | -0.30 | 0.32 | 0.55 | 0.15 | -0.24 | 0.27 | 0.14 | 0.08 | 0.13 | 0.16 | 0.16 |
| hr8173 | S | 4620 | 2.47 | 1.32 | 0.21 | 0.17 | 0.37 | 0.42 | 1.06 | 0.15 | 0.11 | 0.16 | 0.24 | 0.23 | 0.16 | 0.14 | 0.08 | 0.23 | 0.52 | 0.74 | 0.17 | 0.14 | -0.28 | 0.10 | 0.53 | 0.23 | 0.38 | 0.79 | 0.18 |
| hr8179 | S | 4805 | 2.48 | 1.43 | 0.07 | 0.02 | 0.15 | 0.17 | 0.51 | -0.04 | -0.06 | -0.05 | -0.03 | -0.01 | -0.07 | -0.08 | -0.09 | -0.03 | 0.03 | 0.21 | 0.15 | 0.07 | -0.37 | 0.09 | 0.26 | 0.16 | 0.20 | 0.27 | 0.06 |
| hr8185 | S | 4691 | 2.48 | 1.37 | 0.44 | 0.18 | 0.37 | 0.39 | 0.95 | 0.28 | 0.06 | 0.20 | 0.33 | 0.28 | 0.25 | 0.14 | 0.17 | 0.29 | 0.64 | 0.93 | 0.19 | 0.22 | -0.20 | 0.07 | 0.49 | 0.12 | 0.29 | 0.77 | 0.13 |
| hr8191 | S | 6271 | 3.38 | 4.59 | 0.93 | 1.05 | | 0.24 | 0.25 | -0.21 | 0.82 | 0.24 | 0.59 | 0.67 | -0.29 | 0.01 | 0.73 | 0.12 | 0.29 | | | 0.93 | -0.32 | | 1.09 | 0.62 | 0.35 | 1.04 | 1.45 |
| hr8228 | S | 4909 | 2.85 | 1.20 | 0.34 | 0.19 | 0.32 | 0.30 | 0.64 | 0.20 | 0.06 | 0.16 | 0.20 | 0.25 | 0.21 | 0.14 | 0.09 | 0.20 | 0.33 | 0.36 | 0.37 | 0.27 | -0.11 | 0.20 | 0.52 | 0.26 | 0.42 | 0.52 | 0.24 |
| hr8252 | S | 5010 | 2.89 | 1.23 | 0.09 | -0.11 | 0.06 | 0.08 | 0.42 | 0.03 | -0.13 | -0.09 | -0.13 | -0.02 | -0.09 | -0.08 | -0.16 | -0.09 | -0.12 | 0.38 | 0.40 | 0.02 | -0.26 | 0.21 | 0.13 | 0.09 | 0.17 | 0.13 | 0.20 |
| hr8255 | S | 4640 | 2.44 | 1.37 | 0.29 | 0.13 | 0.31 | 0.38 | 0.92 | 0.12 | 0.07 | 0.07 | 0.21 | 0.17 | 0.14 | 0.10 | 0.09 | 0.19 | 0.43 | 1.07 | 0.22 | 0.06 | -0.39 | 0.07 | 0.41 | 0.10 | 0.30 | 0.69 | 0.06 |
| hr8274 | S | 4779 | 2.51 | 1.36 | 0.18 | 0.06 | 0.25 | 0.22 | 0.57 | 0.07 | -0.04 | 0.06 | 0.10 | 0.12 | 0.03 | 0.01 | -0.01 | 0.06 | 0.21 | 0.25 | 0.20 | 0.16 | -0.23 | 0.11 | 0.32 | 0.11 | 0.26 | 0.49 | 0.17 |
| hr8277 | S | 4701 | 2.59 | 1.36 | 0.13 | 0.16 | 0.22 | 0.30 | 0.87 | 0.01 | 0.02 | -0.02 | 0.04 | 0.05 | 0.02 | 0.01 | -0.01 | 0.11 | 0.27 | 0.57 | 0.08 | 0.00 | -0.39 | 0.07 | 0.32 | 0.12 | 0.25 | 0.54 | 0.12 |
| hr8288 | S | 4960 | 2.48 | 1.49 | -0.06 | -0.28 | -0.15 | -0.13 | -0.03 | -0.25 | -0.32 | -0.34 | -0.42 | -0.34 | -0.55 | -0.41 | -0.41 | -0.40 | -0.42 | -0.14 | 0.20 | -0.40 | -0.59 | -0.20 | -0.24 | -0.31 | -0.23 | -0.43 | -0.16 |
| hr8317 | S | 4588 | 2.47 | 1.38 | 0.50 | 0.30 | 0.43 | 0.59 | 1.29 | 0.25 | 0.14 | 0.21 | 0.36 | 0.32 | 0.34 | 0.27 | 0.29 | 0.42 | 0.66 | 1.48 | 0.20 | 0.23 | -0.28 | 0.12 | 0.62 | 0.29 | 0.40 | 0.78 | 0.19 |
| hr8320 | S | 4892 | 2.61 | 1.41 | -0.10 | -0.13 | 0.03 | -0.02 | 0.22 | -0.15 | -0.10 | -0.13 | -0.13 | -0.15 | -0.19 | -0.21 | -0.17 | -0.16 | -0.07 | 0.02 | 0.09 | -0.15 | -0.43 | 0.04 | 0.12 | 0.04 | 0.14 | 0.12 | 0.03 |
| hr8324 | S | 4710 | 2.55 | 1.30 | 0.43 | 0.27 | 0.37 | 0.45 | 1.14 | 0.24 | 0.13 | 0.17 | 0.26 | 0.30 | 0.32 | 0.22 | 0.20 | 0.33 | 0.52 | 1.32 | 0.24 | 0.28 | -0.23 | 0.21 | 0.70 | 0.26 | 0.40 | 0.72 | 0.27 |
| hr8325 | S | 4484 | 2.43 | 1.29 | 0.39 | 0.35 | 0.50 | 0.45 | 1.14 | 0.22 | 0.24 | 0.29 | 0.48 | 0.33 | 0.24 | 0.18 | 0.21 | 0.31 | 0.53 | 0.95 | 0.36 | 0.27 | -0.16 | 0.13 | 0.75 | 0.25 | 0.58 | 0.98 | 0.41 |
| hr8360 | S | 4527 | 2.47 | 1.48 | 0.46 | 0.49 | 0.45 | 0.63 | 1.33 | 0.33 | 0.29 | 0.34 | 0.53 | 0.42 | 0.44 | 0.32 | 0.39 | 0.47 | 0.60 | 1.17 | 0.23 | 0.35 | -0.18 | 0.17 | 0.84 | 0.49 | 0.44 | 1.31 | 0.51 |
| hr8391 | S | 6397 | 3.80 | 3.50 | 0.42 | | -0.22 | 0.30 | 0.48 | 0.06 | 0.36 | 0.40 | 0.55 | 0.57 | 0.78 | 0.07 | 0.81 | 0.39 | | | 0.67 | 0.51 | 0.03 | | 0.72 | 0.66 | 2.23 | 0.99 |
| hr8394 | S | 4811 | 2.86 | 1.16 | 0.10 | 0.03 | 0.15 | 0.19 | 0.46 | 0.06 | -0.03 | 0.04 | 0.08 | 0.07 | 0.03 | -0.02 | -0.02 | 0.03 | 0.21 | 0.41 | 0.11 | 0.11 | -0.26 | 0.08 | 0.31 | 0.22 | 0.30 | 0.33 | 0.04 |
| hr8401 | S | 4944 | 2.72 | 1.38 | 0.47 | 0.32 | 0.37 | 0.29 | 0.67 | 0.19 | 0.11 | 0.16 | 0.18 | 0.21 | 0.14 | 0.10 | 0.13 | 0.17 | 0.36 | 0.46 | 0.33 | 0.07 | -0.25 | 0.05 | 0.55 | 0.08 | 0.14 | 0.46 | 0.22 |
| hr8442 | S | 5260 | 2.55 | 1.47 | 0.46 | 0.25 | 0.33 | 0.20 | 0.40 | 0.29 | 0.17 | 0.31 | 0.34 | 0.31 | 0.28 | 0.20 | 0.20 | 0.21 | 0.16 | 0.16 | 0.59 | 0.27 | 0.24 | 0.41 | 0.37 | 0.28 | 0.34 | 0.32 | 0.34 |
| hr8453 | S | 4946 | 2.88 | 1.26 | 0.50 | 0.28 | 0.44 | 0.40 | 0.83 | 0.31 | 0.21 | 0.29 | 0.35 | 0.36 | 0.38 | 0.25 | 0.22 | 0.33 | 0.42 | 0.54 | 0.37 | 0.04 | -0.03 | 0.28 | 0.65 | 0.34 | 0.44 | 0.73 | 0.35 |
| hr8454 | S | 6253 | 2.96 | 6.50 | 0.09 | -2.20 | | -0.07 | | -1.89 | 0.20 | 0.10 | 1.32 | 0.73 | | -0.11 | 0.34 | 0.37 | -0.40 | | -0.29 | | | | | | 1.33 | -0.29 |
| hr8456 | S | 4811 | 2.54 | 1.36 | 0.33 | 0.19 | 0.40 | 0.31 | 0.77 | 0.20 | 0.08 | 0.22 | 0.26 | 0.25 | 0.20 | 0.14 | 0.09 | 0.18 | 0.35 | 0.64 | 0.38 | 0.22 | -0.13 | 0.22 | 0.52 | 0.25 | 0.38 | 0.60 | 0.25 |
| hr8461 | S | 4950 | 3.32 | 0.84 | 0.23 | 0.20 | 0.28 | 0.27 | 0.66 | 0.22 | 0.18 | 0.25 | 0.27 | 0.25 | 0.21 | 0.19 | 0.17 | 0.29 | 0.48 | 0.88 | 0.19 | 0.36 | -0.08 | 0.33 | 0.72 | 0.34 | 0.63 | 0.91 | 0.40 |
| hr8476 | S | 4734 | 2.52 | 1.44 | 0.40 | 0.37 | 0.50 | 0.41 | 0.74 | 0.24 | 0.18 | 0.30 | 0.41 | 0.32 | 0.33 | 0.21 | 0.26 | 0.32 | 0.58 | 1.01 | 0.26 | 0.20 | -0.06 | 0.19 | 1.04 | 0.35 | 0.37 | 0.96 | 0.36 |
| hr8482 | S | 4503 | 2.54 | 1.21 | 0.46 | 0.52 | 0.53 | 0.58 | 1.38 | 0.38 | 0.33 | 0.34 | 0.57 | 0.40 | 0.41 | 0.31 | 0.36 | 0.50 | 0.72 | 1.54 | 0.25 | 0.38 | -0.20 | 0.17 | 1.01 | 0.36 | 0.79 | 1.25 | 0.41 |
| hr8499 | S | 4894 | 2.63 | 1.46 | 0.41 | 0.17 | 0.36 | 0.31 | 0.99 | 0.23 | 0.04 | 0.12 | 0.18 | 0.24 | 0.19 | 0.14 | 0.10 | 0.18 | 0.21 | 0.70 | 0.32 | 0.18 | -0.14 | 0.18 | 0.45 | 0.17 | 0.26 | 0.33 | 0.27 |
| hr8500 | S | 4544 | 2.35 | 1.30 | 0.25 | 0.25 | 0.30 | 0.33 | 0.98 | 0.10 | 0.11 | 0.12 | 0.27 | 0.20 | 0.18 | 0.08 | 0.08 | 0.18 | 0.42 | 0.50 | 0.23 | 0.12 | -0.25 | 0.13 | 0.39 | 0.16 | 0.37 | 0.65 | 0.26 |
| hr8516 | S | 4552 | 2.60 | 1.12 | -0.05 | 0.12 | 0.04 | 0.23 | 0.79 | -0.07 | -0.05 | -0.10 | -0.04 | -0.07 | -0.12 | -0.08 | -0.05 | 0.04 | 0.31 | 0.50 | -0.09 | -0.15 | -0.52 | -0.09 | 0.22 | -0.02 | 0.24 | 0.57 | -0.09 |

| ID | Type | col3 | col4 | col5 | col6 | col7 | col8 | col9 | col10 | col11 | col12 | col13 | col14 | col15 | col16 | col17 | col18 | col19 | col20 | col21 | col22 | col23 | col24 | col25 | col26 | col27 | col28 | col29 | col30 |
|---|---|---|---|---|---|---|---|---|---|---|---|---|---|---|---|---|---|---|---|---|---|---|---|---|---|---|---|---|---|
| hr8538 | S | 4711 | 2.52 | 1.18 | -0.13 | -0.06 | 0.00 | -0.05 | 0.36 | -0.12 | -0.24 | -0.16 | -0.15 | -0.17 | -0.26 | -0.26 | -0.26 | -0.19 | -0.04 | 0.08 | -0.02 | -0.19 | -0.55 | -0.17 | -0.07 | -0.13 | 0.00 | 0.01 | -0.11 |
| hr8568 | S | 4695 | 2.64 | 1.33 | 0.42 | 0.41 | 0.46 | 0.49 | 1.00 | 0.28 | 0.26 | 0.31 | 0.41 | 0.38 | 0.35 | 0.29 | 0.29 | 0.39 | 0.75 | 0.76 | 0.35 | 0.31 | -0.07 | 0.33 | 0.80 | 0.40 | 0.56 | 0.82 | 0.33 |
| hr8594 | S | 4942 | 2.69 | 1.48 | 0.57 | 0.25 | 0.31 | 0.41 | 0.89 | 0.28 | 0.15 | 0.21 | 0.29 | 0.33 | 0.30 | 0.21 | 0.19 | 0.27 | 0.31 | 0.59 | 0.39 | 0.28 | -0.02 | 0.26 | 0.53 | 0.25 | 0.31 | 0.50 | 0.38 |
| hr8596 | S | 4875 | 2.81 | 1.22 | 0.34 | 0.11 | 0.31 | 0.38 | 0.82 | 0.21 | 0.05 | 0.15 | 0.18 | 0.24 | 0.23 | 0.14 | 0.08 | 0.18 | 0.32 | 0.46 | 0.27 | 0.20 | -0.14 | 0.16 | 0.39 | 0.20 | 0.29 | 0.58 | 0.19 |
| hr8617 | S | 5338 | 2.81 | 1.18 | 0.28 | 0.01 | 0.16 | 0.08 | 0.11 | 0.12 | 0.05 | 0.09 | 0.07 | 0.11 | -0.01 | 0.04 | 0.00 | 0.02 | -0.16 | -0.02 | 0.67 | 0.11 | -0.07 | 0.35 | 0.28 | 0.24 | 0.30 | -0.06 | 0.25 |
| hr8632 | S | 4256 | 1.63 | 1.62 | 0.08 | -0.01 | 0.02 | 0.25 | 1.21 | -0.19 | -0.24 | -0.19 | -0.09 | -0.02 | -0.16 | -0.13 | -0.14 | -0.05 | 0.06 | 0.74 | 0.04 | -0.15 | -0.57 | 0.10 | 0.38 | 0.10 | 0.32 | 0.54 | 0.16 |
| hr8642 | S | 4584 | 2.64 | 1.51 | 0.14 | 0.17 | 0.20 | 0.46 | 1.27 | 0.06 | 0.01 | 0.09 | 0.19 | 0.24 | 0.12 | 0.13 | 0.14 | 0.24 | 0.27 | 1.40 | 0.12 | 0.15 | -0.33 | 0.16 | 0.33 | 0.28 | 0.39 | 0.86 | 0.40 |
| hr8643 | S | 4754 | 2.67 | 1.14 | -0.21 | -0.15 | -0.15 | -0.11 | 0.32 | -0.27 | -0.28 | -0.26 | -0.27 | -0.30 | -0.42 | -0.35 | -0.34 | -0.27 | -0.12 | -0.01 | 0.07 | -0.28 | -0.62 | -0.26 | 0.02 | -0.14 | -0.08 | 0.07 | -0.21 |
| hr8656 | S | 4904 | 2.69 | 1.40 | 0.56 | 0.23 | 0.29 | 0.37 | 0.70 | 0.22 | 0.08 | 0.08 | 0.12 | 0.21 | 0.13 | 0.11 | 0.07 | 0.18 | 0.25 | 0.57 | 0.27 | 0.21 | -0.20 | 0.15 | 0.31 | 0.05 | 0.12 | 0.26 | 0.24 |
| hr8660 | S | 4738 | 2.62 | 1.33 | 0.05 | -0.02 | 0.14 | 0.21 | 0.73 | 0.02 | 0.01 | 0.01 | 0.08 | 0.04 | 0.03 | -0.01 | -0.01 | 0.07 | 0.19 | 0.30 | 0.07 | 0.00 | -0.30 | 0.13 | 0.37 | 0.13 | 0.28 | 0.55 | 0.16 |
| hr8670 | S | 4829 | 2.53 | 1.40 | -0.29 | -0.09 | -0.16 | -0.14 | 0.17 | -0.29 | -0.32 | -0.33 | -0.40 | -0.34 | -0.50 | -0.40 | -0.36 | -0.36 | -0.27 | -0.11 | 0.00 | -0.42 | -0.60 | -0.35 | -0.28 | -0.24 | -0.22 | -0.20 | -0.25 |
| hr8678 | S | 4803 | 2.55 | 1.36 | 0.11 | 0.04 | 0.20 | 0.18 | 0.56 | 0.06 | -0.04 | 0.01 | 0.04 | 0.06 | 0.00 | -0.01 | -0.04 | 0.04 | 0.10 | 0.02 | 0.18 | 0.02 | -0.32 | 0.19 | 0.37 | 0.13 | 0.22 | 0.37 | 0.16 |
| hr8703 | S | 4603 | 2.44 | 2.48 | 0.46 |  | 0.77 | 0.46 | 0.41 | -0.19 | 0.10 | 0.20 | 0.38 | 0.54 | -0.34 | -0.02 | 0.03 | 0.14 |  |  | 0.07 | 0.51 | -0.65 |  | 1.36 | 1.13 | 0.77 |  | 0.71 |
| hr8712 | S | 4641 | 2.57 | 1.40 | 0.42 | 0.31 | 0.46 | 0.48 | 1.07 | 0.25 | 0.20 | 0.30 | 0.50 | 0.35 | 0.38 | 0.26 | 0.32 | 0.44 | 0.77 | 0.88 | 0.23 | 0.23 | -0.14 | 0.21 | 0.91 | 0.33 | 0.54 | 1.13 | 0.33 |
| hr8715 | S | 7728 | 3.73 | 4.14 | 1.29 |  | 0.65 | 0.58 | 0.27 | 0.18 | 1.26 | 1.46 | 2.16 | 0.81 | 0.92 | 0.38 | 1.36 | 1.03 |  |  | 0.58 | 2.38 | 0.46 |  | 0.69 | 1.16 | 1.26 | 0.41 |  |
| hr8730 | S | 4586 | 2.49 | 1.31 | 0.32 | 0.17 | 0.32 | 0.44 | 1.06 | 0.18 | 0.09 | 0.11 | 0.23 | 0.20 | 0.18 | 0.14 | 0.12 | 0.27 | 0.61 | 0.92 | 0.14 | 0.09 | -0.36 | 0.11 | 0.57 | 0.17 | 0.35 | 0.76 | 0.12 |
| hr8742 | S | 4831 | 2.48 | 1.34 | 0.55 | 0.29 | 0.32 | 0.35 | 0.84 | 0.24 | -0.01 | 0.07 | 0.18 | 0.21 | 0.17 | 0.10 | 0.04 | 0.15 | 0.20 | 0.72 | 0.29 | 0.07 | -0.26 | 0.05 | 0.35 | 0.04 | 0.11 | 0.38 | 0.08 |
| hr8780 | S | 4668 | 2.61 | 1.32 | -0.03 | 0.07 | 0.17 | 0.23 | 0.70 | -0.06 | -0.01 | -0.05 | -0.03 | 0.00 | -0.04 | -0.03 | -0.01 | 0.04 | 0.27 | 0.49 | 0.04 | -0.11 | -0.48 | 0.07 | 0.32 | 0.11 | 0.25 | 0.52 | 0.13 |
| hr8807 | S | 4984 | 2.83 | 1.31 | -0.13 | -0.09 | 0.03 | -0.02 | 0.35 | -0.20 | -0.09 | -0.10 | -0.11 | -0.13 | -0.24 | -0.19 | -0.14 | -0.12 | -0.04 | 0.00 | 0.26 | -0.12 | -0.41 | -0.03 | 0.10 | 0.04 | 0.19 | 0.23 | 0.12 |
| hr8812 | S | 4414 | 1.73 | 1.62 | 0.16 | 0.03 | 0.09 | 0.25 | 0.75 | -0.05 | -0.23 | -0.14 | -0.07 | 0.01 | -0.11 | -0.10 | -0.14 | -0.04 | 0.02 | 0.80 | 0.11 | 0.03 | -0.45 | 0.10 | 0.44 | 0.13 | 0.21 | 0.33 | 0.10 |
| hr8839 | S | 4594 | 2.23 | 1.52 | 0.72 | 0.53 | 0.66 | 0.65 | 1.51 | 0.47 | 0.35 | 0.41 | 0.57 | 0.52 | 0.44 | 0.34 | 0.41 | 0.51 | 0.60 | 1.38 | 0.34 | 0.27 | -0.14 | 0.14 | 0.78 | 0.45 | 0.66 | 1.21 | 0.54 |
| hr8841 | S | 4624 | 2.55 | 1.38 | 0.10 | 0.04 | 0.24 | 0.35 | 0.67 | 0.08 | 0.11 | 0.13 | 0.17 | 0.19 | 0.12 | 0.11 | 0.09 | 0.18 | 0.33 | 0.41 | 0.14 | 0.14 | -0.46 | 0.07 | 0.66 | 0.22 | 0.32 | 0.75 | 0.15 |
| hr8852 | S | 4810 | 2.42 | 1.40 | -0.39 | -0.18 | -0.25 | -0.18 | 0.13 | -0.37 | -0.43 | -0.38 | -0.50 | -0.54 | -0.78 | -0.58 | -0.50 | -0.52 | -0.51 | -0.26 | -0.04 | -0.56 | -0.82 | -0.51 | -0.48 | -0.55 | -0.38 | -0.35 | -0.30 |
| hr8875 | S | 4639 | 2.56 | 1.37 | 0.21 | 0.22 | 0.37 | 0.36 | 0.86 | 0.10 | 0.09 | 0.15 | 0.29 | 0.17 | 0.20 | 0.10 | 0.15 | 0.25 | 0.51 | 0.78 | 0.11 | 0.14 | -0.33 | 0.08 | 0.47 | 0.16 | 0.38 | 0.80 | 0.11 |
| hr8878 | S | 4246 | 1.58 | 1.48 | -0.52 | -0.19 | -0.28 | -0.15 | 0.42 | -0.47 | -0.51 | -0.47 | -0.52 | -0.61 | -0.89 | -0.66 | -0.52 | -0.55 | -0.29 | -0.09 | -0.32 | -0.57 | -0.87 | -0.72 | -0.46 | -0.60 | -0.39 | -0.21 | -0.40 |
| hr8892 | S | 4545 | 2.25 | 1.33 | -0.18 | -0.21 | -0.07 | -0.06 | 0.49 | -0.26 | -0.32 | -0.25 | -0.20 | -0.27 | -0.34 | -0.35 | -0.30 | -0.25 | -0.06 | 0.38 | -0.25 | -0.32 | -0.66 | -0.32 | -0.11 | -0.22 | -0.08 | 0.08 | -0.27 |
| hr8912 | S | 4828 | 2.71 | 1.30 | 0.08 | 0.01 | 0.14 | 0.17 | 0.64 | 0.05 | 0.01 | 0.01 | 0.00 | 0.05 | -0.05 | -0.02 | -0.07 | 0.04 | 0.09 | 0.06 | 0.16 | 0.08 | -0.31 | 0.19 | 0.30 | 0.25 | 0.29 | 0.44 | 0.09 |
| hr8922 | S | 4727 | 2.51 | 1.38 | 0.07 | 0.09 | 0.23 | 0.20 | 0.46 | 0.06 | 0.06 | 0.08 | 0.15 | 0.06 | 0.04 | -0.03 | 0.05 | 0.08 | 0.27 | 0.38 | 0.11 | -0.01 | -0.33 | 0.03 | 0.33 | 0.04 | 0.26 | 0.69 | 0.06 |
| hr8923 | S | 4938 | 2.81 | 1.32 | 0.36 | 0.06 | 0.21 | 0.34 | 0.68 | 0.21 | 0.10 | 0.11 | 0.09 | 0.20 | 0.15 | 0.11 | 0.05 | 0.16 | 0.19 | 0.58 | 0.33 | 0.20 | -0.14 | 0.28 | 0.39 | 0.31 | 0.28 | 0.37 | 0.21 |
| hr8930 | S | 4613 | 2.39 | 1.39 | -0.13 | 0.06 | 0.10 | 0.11 | 0.56 | -0.13 | -0.24 | -0.09 | -0.11 | -0.22 | -0.31 | -0.28 | -0.19 | -0.16 | 0.15 | 0.15 | -0.06 | -0.45 | -0.69 | -0.43 | -0.34 | -0.25 | -0.16 | 0.16 | -0.14 |
| hr8941 | S | 4962 | 2.50 | 1.48 | 0.38 | 0.14 | 0.24 | 0.24 | 0.45 | 0.21 | 0.07 | 0.12 | 0.14 | 0.20 | 0.16 | 0.09 | 0.05 | 0.11 | 0.10 | 0.30 | 0.34 | 0.14 | -0.07 | 0.23 | 0.39 | 0.19 | 0.30 | 0.12 | 0.20 |
| hr8946 | S | 4208 | 2.06 | 1.44 | 0.44 | 0.39 | 0.52 | 0.61 | 1.82 | 0.24 | 0.22 | 0.27 | 0.53 | 0.41 | 0.37 | 0.23 | 0.37 | 0.42 | 0.73 | 1.65 | 0.28 | 0.38 | -0.28 | 0.05 | 0.94 | 0.35 | 0.72 | 1.34 | 0.27 |
| hr8948 | S | 4761 | 2.56 | 1.25 | 0.30 | 0.18 | 0.34 | 0.33 | 0.74 | 0.20 | 0.05 | 0.06 | 0.18 | 0.14 | 0.10 | 0.11 | 0.07 | 0.15 |  | 0.54 |  | 0.11 | -0.44 | 0.20 | 0.48 | 0.07 | 0.25 | 0.16 | 0.15 |
| hr8958 | S | 4708 | 2.62 | 1.35 | -0.08 | 0.08 | 0.11 | 0.19 | 0.65 | -0.05 | -0.03 | -0.08 | -0.08 | -0.06 | -0.11 | -0.08 | -0.09 | 0.01 | 0.19 | 0.45 | 0.00 | -0.14 | -0.53 | 0.00 | 0.20 | 0.06 | 0.19 | 0.42 | 0.10 |
| hr9008 | S | 4617 | 2.60 | 1.46 | 0.19 | 0.20 | 0.31 | 0.48 | 1.12 | 0.13 | 0.21 | 0.14 | 0.20 | 0.23 | 0.22 | 0.21 | 0.19 | 0.36 | 0.52 | 0.76 | 0.18 | 0.36 | -0.26 | 0.40 | 0.70 | 0.45 | 0.57 | 1.02 | 0.33 |
| hr9009 | S | 4661 | 2.57 | 1.29 | 0.26 | 0.05 | 0.30 | 0.42 | 1.01 | 0.19 | 0.11 | 0.12 | 0.18 | 0.21 | 0.23 | 0.13 | 0.10 | 0.25 | 0.45 | 0.97 | 0.17 | 0.13 | -0.28 | 0.17 | 0.59 | 0.28 | 0.34 | 0.61 | 0.20 |
| hr9012 | S | 4926 | 2.64 | 1.35 | 0.24 | -0.04 | 0.22 | 0.22 | 0.73 | 0.11 | -0.04 | 0.00 | -0.03 | 0.08 | 0.02 | 0.01 | -0.07 | 0.02 | 0.04 | 0.35 | 0.21 | 0.06 | -0.16 | 0.20 | 0.28 | 0.20 | 0.19 | 0.28 | 0.16 |
| hr9067 | S | 5026 | 2.75 | 1.36 | 0.26 | 0.07 | 0.19 | 0.18 | 0.49 | 0.12 | 0.02 | 0.09 | 0.06 | 0.16 | 0.00 | 0.05 | 0.02 | 0.07 | 0.06 | 0.17 | 0.33 | 0.15 | -0.08 | 0.23 | 0.36 | 0.20 | 0.32 | 0.29 | 0.20 |
| hr9101 | S | 4663 | 2.65 | 1.33 | 0.50 | 0.45 | 0.52 | 0.55 | 1.33 | 0.25 | 0.28 | 0.29 | 0.39 | 0.38 | 0.41 | 0.30 | 0.31 | 0.44 | 0.65 | 1.09 | 0.28 | 0.31 | -0.16 | 0.31 | 1.08 | 0.36 | 0.58 | 1.00 | 0.30 |
| hr9104 | S | 4701 | 2.49 | 1.38 | 0.24 | 0.14 | 0.23 | 0.27 | 0.70 | 0.03 | 0.03 | 0.02 | 0.07 | 0.09 | 0.09 | 0.01 | 0.04 | 0.09 | 0.25 | 0.56 | 0.16 | -0.06 | -0.22 | 0.08 | 0.35 | 0.12 | 0.29 | 0.36 | 0.08 |
| hd001638 | U | 4225 | 2.50 | 1.64 | -0.63 | -0.17 | -0.33 | 0.17 | 0.87 | -0.63 | -0.20 | -0.28 | -0.26 | -0.51 | -0.64 | -0.32 | -0.22 | -0.23 | 0.01 | -0.10 | -0.32 | -0.33 | -0.47 | -0.15 | 0.11 | 0.26 | 0.27 | 0.24 | 0.19 |
| hd004388 | U | 4640 | 2.57 | 1.23 | 0.26 | 0.27 | 0.43 | 0.25 | 1.09 | 0.17 | 0.18 | 0.21 | 0.44 | 0.24 | 0.20 | 0.16 | 0.08 | 0.22 | 0.34 | 0.11 | 0.22 | 0.17 | 0.14 | 0.16 | 0.59 | 0.67 | 0.48 | 0.57 | 0.20 |
| hd011171 | U | 6746 | 4.08 | 3.91 | 1.24 |  | 1.48 | -0.10 | 0.56 | 0.15 | 0.58 | 0.64 | 0.93 | 0.38 | 0.08 | 1.72 | 0.86 |  | 1.91 | 1.57 |  |  | 2.25 | 2.72 |  | 1.80 |  |  |  |  |
| hd037763 | U | 4620 | 3.12 | 0.75 | 0.64 | 0.72 | 0.78 | 0.69 | 1.54 | 0.53 | 0.65 | 0.72 | 1.06 | 0.72 | 0.69 | 0.50 | 0.62 | 0.71 | 1.20 | 1.25 | 0.44 | 0.61 | 0.52 | 0.40 | 0.99 | 1.44 | 1.26 | 1.50 | 0.59 |
| hd037811 | U | 5023 | 2.57 | 1.42 | 0.22 | 0.07 | 0.17 | 0.14 | 0.43 | 0.10 | -0.05 | -0.03 | -0.04 | 0.05 | -0.03 | 0.00 | -0.09 | -0.03 | -0.11 | -0.09 | 0.30 | 0.04 | -0.10 | 0.24 | 0.20 | 0.17 | 0.19 | 0.23 | 0.09 |
| hd061603 | U | 4012 | 0.65 | 2.19 | 0.97 | 0.21 | 0.30 | 0.60 | 2.38 | 0.18 | -0.06 | 0.09 | 0.30 | 0.29 | 0.21 | 0.04 | 0.09 | 0.19 | 0.68 | 0.18 | 0.78 | -0.01 | 0.06 |  | 0.22 | 0.23 | 0.38 | 0.30 | -0.17 |
| hd062412 | U | 4861 | 2.61 | 1.38 | 0.27 | 0.16 | 0.21 | 0.25 | 0.50 | 0.15 | -0.03 | 0.04 | 0.12 | 0.15 | 0.11 | 0.08 | 0.01 | 0.11 | 0.08 | 0.22 | 0.17 | 0.08 | -0.05 | 0.16 | 0.29 | 0.62 | 0.27 | 0.29 | 0.12 |
| hd065354 | U | 3832 | 0.06 | 2.28 | 0.21 | 0.04 | 0.08 | 0.51 |  | -0.23 | -0.49 | -0.42 | -0.33 | -0.01 | -0.10 | -0.12 | -0.29 | -0.05 | 0.06 | -0.29 | 0.22 | -0.41 | -0.56 |  | -0.13 | -0.02 | -0.03 | 0.02 | -0.03 |
| hd065714 | U | 5070 | 2.39 | 1.75 | 0.74 | 0.44 | 0.50 | 0.41 | 0.59 | 0.37 | 0.23 | 0.33 | 0.50 | 0.39 | 0.36 | 0.28 | 0.30 | 0.29 | 0.29 | 0.17 | 0.37 | 0.33 | 0.25 | 0.26 | 0.39 | 0.67 | 0.44 | 0.35 | 0.41 |
| hd065925 | U | 6585 | 3.63 | 6.50 | 0.64 |  | 0.32 | -0.16 | -0.07 | -0.07 | 0.51 | -0.07 | 0.89 | -0.12 | -0.54 | -0.51 | 0.86 | -0.14 | -1.10 |  | 2.04 | 0.28 |  |  | -0.80 | 0.60 | 1.09 |  |  |

| ID | | Teff | | | | | | | | | | | | | | | | | | | | | | | | | | |
|---|---|---|---|---|---|---|---|---|---|---|---|---|---|---|---|---|---|---|---|---|---|---|---|---|---|---|---|---|
| hd069511 | U | 4020 | 0.75 | 2.45 | 0.20 | -0.08 | 0.16 | 0.28 | | -0.28 | -0.26 | -0.32 | -0.22 | -0.01 | -0.11 | -0.07 | -0.21 | -0.05 | | -0.24 | 0.10 | -0.28 | -0.35 | | 0.14 | 0.20 | 0.16 | 0.30 | 0.57 |
| hd069836 | U | 4788 | 2.50 | 1.43 | 0.32 | 0.31 | 0.46 | 0.21 | 0.64 | 0.20 | 0.14 | 0.22 | 0.46 | 0.22 | 0.25 | 0.13 | 0.12 | 0.17 | 0.29 | 0.02 | 0.19 | 0.04 | 0.08 | 0.01 | 0.31 | 0.20 | 0.22 | 0.22 | 0.11 |
| hd071160 | U | 4096 | 1.55 | 1.77 | 0.35 | 0.18 | 0.35 | 0.54 | 1.28 | -0.01 | 0.06 | 0.16 | 0.37 | 0.30 | 0.17 | 0.14 | 0.21 | 0.26 | 0.28 | 0.55 | 0.33 | 0.15 | -0.03 | 0.07 | 0.37 | 0.49 | 0.56 | 0.39 | -0.09 |
| hd072320 | U | 5028 | 2.77 | 1.34 | 0.24 | 0.11 | 0.34 | 0.12 | 0.24 | 0.11 | 0.00 | -0.01 | 0.03 | 0.07 | 0.00 | 0.02 | -0.06 | -0.01 | -0.11 | -0.09 | 0.21 | 0.07 | 0.03 | 0.14 | 0.25 | 0.27 | 0.19 | 0.09 | 0.12 |
| hd072324 | U | 4781 | 2.43 | 1.54 | 0.35 | 0.17 | 0.28 | 0.30 | 0.85 | 0.13 | 0.01 | 0.03 | 0.10 | 0.14 | 0.07 | 0.07 | 0.00 | 0.11 | 0.08 | 0.27 | 0.11 | 0.12 | -0.02 | 0.20 | 0.33 | 0.70 | 0.36 | 0.23 | 0.18 |
| hd073829 | U | 4962 | 2.50 | 1.50 | 0.49 | 0.35 | 0.56 | 0.13 | 0.46 | 0.33 | 0.27 | 0.43 | 0.71 | 0.37 | 0.36 | 0.21 | 0.25 | 0.23 | 0.49 | -0.17 | 0.46 | 0.23 | 0.34 | 0.16 | 0.49 | 0.44 | 0.44 | 0.40 | 0.28 |
| hd074088 | U | 3840 | 1.06 | 1.74 | -0.19 | -0.12 | -0.14 | 0.36 | 2.11 | -0.32 | -0.28 | -0.22 | -0.12 | -0.19 | -0.47 | -0.22 | -0.19 | -0.12 | 0.12 | 0.03 | 0.01 | -0.25 | -0.51 | -0.26 | -0.19 | 0.12 | 0.23 | 0.23 | -0.34 |
| hd074165 | U | 4675 | 2.50 | 1.43 | 0.47 | 0.46 | 0.56 | 0.39 | 1.29 | 0.23 | 0.24 | 0.37 | 0.70 | 0.38 | 0.42 | 0.29 | 0.30 | 0.39 | | 0.76 | 0.24 | 0.24 | 0.07 | 0.14 | 0.67 | 0.41 | 0.49 | 0.47 | 0.19 |
| hd074166 | U | 4629 | 2.50 | 1.78 | 0.82 | 0.47 | 0.72 | 0.45 | | 0.48 | 0.49 | 0.63 | 0.81 | 0.59 | 0.45 | 0.40 | 0.48 | 0.47 | | 0.56 | 0.66 | 0.58 | 0.59 | 0.60 | 0.70 | 0.90 | 0.92 | 0.89 | 0.32 |
| hd074529 | U | 4598 | 2.38 | 1.40 | 0.43 | 0.37 | 0.48 | 0.44 | 1.27 | 0.21 | 0.25 | 0.22 | 0.51 | 0.32 | 0.34 | 0.26 | 0.23 | 0.36 | 0.44 | 0.37 | 0.26 | 0.15 | 0.04 | 0.15 | 0.59 | 0.38 | 0.40 | 0.33 | 0.08 |
| hd074900 | U | 4572 | 2.51 | 1.35 | 0.32 | 0.41 | 0.51 | 0.45 | 1.18 | 0.19 | 0.30 | 0.29 | 0.56 | 0.25 | 0.28 | 0.24 | 0.26 | 0.34 | | 0.81 | 0.11 | -0.03 | -0.06 | 0.17 | 0.54 | 0.43 | 0.54 | 0.49 | 0.05 |
| hd075058 | U | 4650 | 2.79 | 1.00 | 0.29 | 0.30 | 0.40 | 0.27 | 1.05 | 0.15 | 0.17 | 0.24 | 0.51 | 0.22 | 0.24 | 0.16 | 0.12 | 0.27 | 0.43 | 0.37 | 0.15 | 0.18 | 0.02 | 0.05 | 0.47 | 0.28 | 0.47 | 0.36 | -0.14 |
| hd076128 | U | 4487 | 2.50 | 1.39 | 0.35 | 0.38 | 0.46 | 0.51 | | 0.17 | 0.21 | 0.28 | 0.51 | 0.33 | 0.36 | 0.32 | 0.29 | 0.41 | 0.49 | 0.39 | 0.16 | 0.22 | 0.12 | 0.24 | 0.69 | 0.47 | 0.57 | 0.51 | 0.03 |
| hd078002 | U | 4859 | 2.50 | 1.41 | 0.54 | 0.31 | 0.47 | 0.32 | 0.79 | 0.20 | 0.14 | 0.12 | 0.25 | 0.25 | 0.27 | 0.18 | 0.15 | 0.23 | 0.19 | 0.16 | 0.18 | 0.09 | 0.02 | 0.03 | 0.36 | 0.25 | 0.21 | 0.15 | 0.21 |
| hd078479 | U | 4479 | 2.26 | 1.64 | 0.59 | 0.49 | 0.67 | 0.64 | 1.22 | 0.28 | 0.30 | 0.34 | 0.56 | 0.44 | 0.41 | 0.34 | 0.41 | 0.46 | 1.13 | 0.93 | 0.21 | 0.16 | 0.00 | -0.03 | 0.49 | 0.85 | 0.52 | 0.61 | 0.28 |
| hd078528 | U | 4012 | 2.50 | 1.00 | 0.75 | 0.57 | 0.79 | 1.03 | | 0.67 | 1.02 | 1.00 | 1.26 | 0.94 | 0.78 | 0.79 | 0.93 | 1.03 | | 1.06 | 0.94 | 0.82 | 1.13 | 0.93 | 1.40 | 1.60 | 1.72 | 1.80 | |
| hd078959 | U | 4150 | 1.20 | 2.16 | 0.41 | -0.05 | 0.41 | 0.23 | | 0.00 | 0.01 | 0.08 | 0.18 | 0.16 | -0.05 | -0.02 | -0.05 | -0.02 | | -0.60 | 0.60 | 0.04 | 0.20 | | 0.32 | 0.32 | 0.51 | 0.38 | |
| hd080571 | U | 4659 | 2.35 | 1.41 | 0.28 | 0.24 | 0.43 | 0.25 | 0.97 | 0.10 | 0.07 | 0.08 | 0.27 | 0.15 | 0.13 | 0.07 | 0.04 | 0.13 | 0.21 | 0.02 | 0.11 | 0.01 | -0.12 | -0.03 | 0.35 | 0.18 | 0.20 | 0.17 | -0.01 |
| hd081278 | U | 4936 | 2.50 | 1.43 | 0.54 | 0.42 | 0.61 | 0.20 | 0.78 | 0.36 | 0.16 | 0.29 | 0.57 | 0.34 | 0.40 | 0.17 | 0.21 | 0.20 | 0.35 | 0.00 | 0.35 | 0.04 | 0.17 | -0.22 | 0.02 | 0.04 | 0.20 | 0.10 | 0.19 |
| hd082395 | U | 4686 | 2.58 | 1.37 | 0.14 | 0.18 | 0.17 | 0.29 | 0.82 | 0.02 | -0.01 | -0.01 | 0.13 | 0.05 | 0.07 | 0.05 | 0.00 | 0.13 | 0.23 | 0.40 | 0.04 | 0.09 | -0.15 | 0.05 | 0.34 | 0.65 | 0.32 | 0.28 | 0.08 |
| hd082403 | U | 4724 | 2.50 | 1.33 | 0.42 | 0.33 | 0.53 | 0.22 | -0.07 | 0.29 | 0.29 | 0.43 | 0.64 | 0.35 | 0.35 | 0.22 | 0.25 | 0.31 | 0.62 | 0.48 | 0.46 | 0.39 | 0.19 | 0.28 | 0.54 | 0.37 | 0.49 | 0.52 | 0.27 |
| hd082668 | U | 4105 | 1.38 | 1.70 | 0.15 | -0.07 | 0.26 | 0.30 | | 0.02 | -0.02 | 0.12 | 0.42 | 0.15 | -0.12 | -0.05 | 0.02 | 0.07 | 0.37 | 0.12 | 0.77 | -0.03 | -0.02 | 0.28 | 0.21 | 0.20 | 0.48 | 1.05 | -0.04 |
| hd083155 | U | 5166 | 2.50 | 1.62 | 0.14 | 0.08 | | -0.04 | -0.30 | 0.03 | -0.06 | 0.12 | 0.16 | 0.04 | -0.03 | -0.03 | 0.02 | -0.06 | -0.06 | -0.37 | 0.40 | -0.13 | 0.05 | 0.04 | 0.04 | 0.00 | 0.13 | 0.22 | 0.01 |
| hd083234 | U | 4185 | 2.50 | 1.31 | 0.34 | 0.52 | 0.45 | 0.78 | | 0.24 | 0.37 | 0.40 | 0.61 | 0.48 | 0.42 | 0.54 | 0.51 | 0.64 | | 0.90 | 0.24 | 0.30 | 0.34 | 0.57 | 1.11 | 1.08 | 0.93 | 0.93 | -0.04 |
| hd084598 | U | 4896 | 2.64 | 1.40 | 0.30 | 0.15 | | 0.20 | 0.51 | 0.10 | 0.02 | -0.01 | 0.05 | 0.09 | 0.00 | 0.04 | -0.08 | 0.04 | -0.06 | 0.06 | 0.10 | 0.02 | 0.03 | 0.18 | 0.33 | 0.33 | 0.23 | 0.15 | |
| hd085552 | U | 5426 | 3.19 | 2.15 | 0.17 | 0.05 | 0.20 | 0.17 | 0.46 | 0.05 | 0.12 | 0.04 | -0.19 | 0.09 | -0.21 | -0.03 | -0.13 | -0.08 | -0.51 | -0.05 | | 0.33 | 0.25 | 0.32 | 0.55 | 0.24 | 0.38 | 0.03 | 0.17 |
| hd086757 | U | 4081 | 2.50 | 2.42 | 0.25 | 0.29 | 0.36 | 0.93 | | -0.07 | 0.30 | 0.10 | 0.27 | 0.31 | 0.34 | 0.50 | 0.45 | 0.53 | | 0.40 | 0.23 | 0.31 | 0.16 | | 1.10 | 1.04 | 1.03 | 1.04 | 1.42 |
| hd089736 | U | 3830 | 0.49 | 2.58 | 0.40 | 0.00 | 0.17 | 0.66 | | -0.14 | -0.23 | -0.22 | -0.05 | 0.05 | -0.15 | -0.01 | -0.13 | 0.05 | 0.50 | -0.22 | 0.60 | -0.29 | -0.19 | | 0.16 | 0.21 | 0.10 | 0.16 | 0.26 |
| hd095849 | U | 4472 | 1.97 | 1.64 | 0.54 | 0.41 | 0.47 | 0.49 | 1.06 | 0.28 | 0.13 | 0.22 | 0.40 | 0.32 | 0.30 | 0.23 | 0.26 | 0.33 | 0.92 | 0.62 | 0.24 | 0.12 | 0.01 | 0.02 | 0.40 | 0.64 | 0.43 | 0.53 | 0.14 |
| hd099322 | U | 4857 | 2.59 | 1.32 | 0.24 | 0.13 | 0.21 | 0.19 | 0.57 | 0.14 | -0.08 | 0.02 | 0.11 | 0.09 | 0.09 | 0.04 | -0.03 | 0.05 | 0.02 | 0.04 | 0.16 | 0.09 | -0.01 | 0.18 | 0.21 | 0.40 | 0.25 | 0.15 | 0.12 |
| hd105740 | U | 4668 | 2.79 | 1.04 | -0.43 | -0.18 | -0.18 | -0.12 | 0.28 | -0.32 | -0.35 | -0.28 | -0.39 | -0.49 | -0.75 | -0.53 | -0.43 | -0.43 | -0.33 | -0.23 | -0.10 | -0.47 | -0.54 | -0.47 | -0.45 | -0.42 | -0.27 | -0.12 | -0.26 |
| hd107446 | U | 4121 | 1.64 | 1.76 | -0.07 | 0.03 | -0.10 | 0.19 | | -0.20 | -0.19 | -0.09 | 0.10 | -0.05 | -0.18 | -0.11 | -0.08 | -0.03 | 0.01 | -0.17 | 0.05 | -0.24 | -0.25 | 0.02 | 0.18 | 0.22 | 0.29 | 0.14 | -0.48 |
| hd110458 | U | 4682 | 2.55 | 1.32 | 0.44 | 0.28 | 0.38 | 0.42 | 0.38 | 0.25 | 0.12 | 0.19 | 0.42 | 0.26 | 0.34 | 0.23 | 0.20 | 0.31 | 0.40 | 0.36 | 0.22 | 0.24 | 0.03 | 0.25 | 0.41 | 0.44 | 0.36 | 0.31 | 0.20 |
| hd111464 | U | 4170 | 1.68 | 1.56 | 0.16 | 0.15 | 0.16 | 0.39 | 0.99 | -0.05 | -0.04 | -0.04 | 0.13 | 0.08 | -0.02 | 0.03 | 0.04 | 0.12 | 0.29 | 0.37 | 0.03 | -0.15 | -0.26 | -0.06 | 0.31 | 0.52 | 0.32 | 0.39 | -0.13 |
| hd111721 | U | 4892 | 2.62 | 1.22 | -1.44 | -0.92 | -1.19 | -0.88 | | -1.00 | -1.28 | -1.16 | -1.46 | -1.35 | -1.81 | -1.41 | -1.37 | -1.44 | -1.84 | -1.12 | -0.35 | -1.26 | -1.01 | -1.18 | -1.30 | -1.09 | -1.17 | -1.38 | -1.04 |
| hd113002 | U | 5274 | 3.24 | 1.57 | -0.84 | -0.64 | -0.76 | -0.52 | -0.29 | -0.66 | -0.50 | -0.64 | -0.82 | -0.79 | -1.05 | -0.76 | -0.74 | -0.84 | -1.05 | -0.67 | -0.55 | -0.62 | -0.53 | -0.42 | -0.43 | -0.37 | -0.34 | -0.36 | 0.11 |
| hd119971 | U | 4071 | 1.56 | 1.68 | -0.52 | -0.15 | -0.30 | 0.02 | 1.01 | -0.52 | -0.44 | -0.38 | -0.43 | -0.57 | -0.90 | -0.52 | -0.45 | -0.42 | -0.37 | -0.22 | -0.35 | -0.50 | -0.59 | -0.48 | -0.35 | -0.20 | -0.20 | -0.08 | -0.14 |
| hd122721 | U | 4684 | 3.05 | 0.50 | 0.32 | 0.40 | 0.39 | 0.43 | 1.22 | 0.27 | 0.42 | 0.43 | 0.65 | 0.40 | 0.44 | 0.38 | 0.22 | 0.50 | 0.61 | 0.62 | 0.27 | 0.53 | 0.27 | 0.57 | 1.08 | 0.96 | 0.91 | 1.06 | 0.36 |
| hd124186 | U | 4417 | 2.32 | 1.53 | 0.91 | 0.57 | 0.77 | 0.69 | 1.59 | 0.45 | 0.38 | 0.42 | 0.69 | 0.56 | 0.52 | 0.43 | 0.49 | 0.55 | 0.55 | 0.73 | 0.32 | 0.30 | 0.16 | 0.09 | 0.51 | 0.98 | 0.72 | 0.91 | 0.29 |
| hd128279 | U | 5162 | 3.01 | 1.59 | -2.36 | -1.83 | | -1.72 | | -1.93 | -2.22 | -2.07 | -1.76 | -2.44 | -2.73 | -2.27 | -1.58 | -2.26 | | -2.27 | | -2.55 | | -2.88 | | | | | |
| hd138688 | U | 4191 | 1.83 | 1.38 | 0.28 | 0.22 | 0.33 | 0.40 | 0.71 | 0.10 | 0.07 | 0.19 | 0.54 | 0.22 | 0.05 | 0.13 | 0.08 | 0.20 | 0.20 | 0.14 | 0.34 | 0.06 | 0.11 | 0.34 | 0.31 | 0.43 | 0.58 | 0.55 | -0.03 |
| hd145206 | U | 4029 | 1.51 | 1.47 | 0.31 | 0.32 | 0.31 | 0.63 | 1.01 | 0.16 | 0.03 | 0.10 | 0.42 | 0.21 | 0.15 | 0.21 | 0.17 | 0.32 | | 0.44 | 0.21 | 0.03 | -0.03 | 0.24 | 0.28 | 0.29 | 0.44 | 0.36 | -0.29 |
| hd146836 | U | 6285 | 3.64 | 2.38 | -0.09 | -0.12 | -0.07 | -0.02 | 0.07 | -0.09 | -0.20 | -0.08 | -0.14 | -0.19 | -0.36 | -0.17 | 0.19 | -0.15 | -0.72 | -0.55 | | 0.24 | -0.07 | -0.15 | | -0.02 | -0.11 | -0.19 | -0.17 | |
| hd148451 | U | 5015 | 2.61 | 1.41 | -0.20 | -0.16 | -0.12 | -0.21 | -0.25 | -0.26 | -0.32 | -0.24 | -0.32 | -0.50 | -0.70 | -0.53 | -0.38 | -0.48 | -0.54 | -0.40 | -0.25 | -0.46 | -0.38 | -0.43 | -0.33 | -0.37 | -0.27 | -0.27 | 0.15 |
| hd148513 | U | 4065 | 1.63 | 1.83 | 0.73 | 0.34 | 0.50 | 0.69 | 1.41 | 0.08 | 0.09 | 0.13 | 0.37 | 0.36 | 0.27 | 0.20 | 0.27 | 0.37 | 0.58 | 0.94 | 0.24 | 0.19 | 0.01 | -0.25 | 0.33 | 0.47 | 0.41 | 0.34 | -0.13 |
| hd149447 | U | 3875 | 1.25 | 1.92 | 0.25 | 0.15 | 0.32 | 0.62 | 1.41 | 0.17 | 0.03 | 0.02 | 0.30 | 0.19 | -0.08 | 0.12 | 0.14 | 0.23 | 0.63 | 0.54 | 0.56 | -0.28 | -0.06 | 0.18 | 0.16 | 0.16 | 0.36 | 0.45 | -0.01 |
| hd150798 | U | 4135 | 2.50 | 2.46 | 0.43 | 0.29 | 0.13 | 1.10 | 2.33 | -0.02 | 0.24 | 0.08 | 0.23 | 0.31 | 0.29 | 0.47 | 0.45 | 0.56 | 0.55 | 0.93 | 0.22 | 0.16 | -0.20 | | 1.05 | 0.85 | 0.96 | 1.08 | 0.89 |
| hd152786 | U | 3851 | 0.52 | 2.17 | 0.34 | 0.07 | 0.07 | 0.45 | | -0.17 | -0.32 | -0.24 | -0.08 | -0.01 | -0.04 | -0.06 | -0.12 | 0.03 | 0.15 | -0.17 | 0.29 | -0.23 | -0.29 | | 0.11 | 0.02 | 0.19 | 0.08 | -0.16 |
| hd157457 | U | 4911 | 2.34 | 1.72 | 0.15 | -0.04 | 0.26 | 0.22 | 0.47 | 0.16 | 0.02 | 0.19 | 0.26 | 0.25 | 0.12 | 0.07 | 0.07 | 0.05 | 0.01 | -0.10 | 0.38 | 0.29 | 0.10 | 0.33 | 0.24 | 0.23 | 0.30 | 0.35 | 0.08 |
| hd162391 | U | 4724 | 2.04 | 2.17 | 0.37 | 0.23 | 0.18 | 0.40 | 0.87 | -0.02 | -0.09 | -0.09 | -0.11 | 0.13 | -0.13 | 0.03 | -0.08 | 0.06 | -0.16 | 0.24 | 0.16 | 0.25 | -0.08 | 0.30 | 0.27 | 0.26 | 0.36 | 0.22 | 0.21 |

| Name | | Teff | | | | | | | | | | | | | | | | | | | | | | | | | | | |
|---|---|---|---|---|---|---|---|---|---|---|---|---|---|---|---|---|---|---|---|---|---|---|---|---|---|---|---|---|---|
| hd162587 | U | 4764 | 1.92 | 1.76 | 0.36 | 0.00 | 0.19 | 0.25 | 0.59 | 0.13 | -0.14 | -0.08 | -0.04 | 0.09 | 0.01 | 0.00 | -0.10 | -0.01 | -0.10 | 0.03 | 0.10 | -0.04 | -0.11 | 0.13 | 0.16 | 0.31 | 0.09 | 0.03 | 0.00 |
| hd163652 | U | 4968 | 2.64 | 1.41 | -0.22 | -0.26 | -0.11 | -0.18 | -0.22 | -0.23 | -0.31 | -0.28 | -0.29 | -0.32 | -0.42 | -0.35 | -0.33 | -0.35 | -0.35 | -0.35 | -0.16 | -0.29 | -0.34 | -0.16 | -0.08 | -0.04 | -0.06 | -0.15 | 0.14 |
| hd167818 | U | 3909 | 0.64 | 2.25 | -0.04 | 0.06 | -0.09 | 0.47 | 0.77 | -0.37 | -0.43 | -0.42 | -0.43 | -0.16 | -0.23 | -0.09 | -0.24 | -0.03 | -0.13 | -0.13 | -0.17 | -0.43 | -0.59 | | -0.05 | 0.09 | -0.03 | 0.00 | -0.16 |
| hd169191 | U | 4357 | 1.99 | 1.49 | 0.12 | 0.10 | 0.13 | 0.27 | 0.98 | -0.02 | -0.11 | 0.01 | 0.13 | 0.05 | 0.00 | 0.00 | -0.02 | 0.06 | 0.51 | 0.14 | 0.09 | -0.03 | -0.16 | 0.12 | 0.40 | 0.63 | 0.40 | 0.45 | 0.20 |
| hd175545 | U | 4471 | 2.80 | 0.92 | 0.35 | 0.56 | 0.54 | 0.57 | 1.41 | 0.31 | 0.38 | 0.46 | 0.79 | 0.50 | 0.43 | 0.45 | 0.44 | 0.55 | 0.87 | 0.99 | 0.24 | 0.47 | 0.28 | 0.36 | 0.83 | 1.25 | 1.11 | 1.25 | 0.47 |
| hd183275 | U | 4743 | 2.50 | 1.56 | 0.62 | 0.42 | 0.64 | 0.49 | 1.13 | 0.38 | 0.26 | 0.40 | 0.61 | 0.47 | 0.51 | 0.33 | 0.37 | 0.42 | 0.65 | 0.32 | 0.37 | 0.27 | 0.11 | 0.09 | 0.56 | 0.30 | 0.45 | 0.92 | 0.36 |
| hd196983 | U | 4593 | 2.50 | 1.39 | 0.42 | 0.28 | 0.43 | 0.45 | 1.21 | 0.23 | 0.16 | 0.25 | 0.44 | 0.33 | 0.28 | 0.27 | 0.26 | 0.35 | 0.66 | 0.44 | 0.23 | 0.15 | -0.02 | 0.16 | 0.59 | 0.30 | 0.44 | 0.98 | 0.35 |
| hd199642 | U | 3828 | 1.32 | 1.94 | 0.16 | 0.17 | 0.35 | 0.65 | 1.49 | -0.14 | 0.01 | -0.01 | 0.22 | 0.03 | -0.20 | 0.08 | 0.09 | 0.13 | 0.68 | 0.59 | 0.30 | -0.21 | -0.29 | 0.19 | 0.18 | 0.00 | 0.31 | 0.26 | 0.08 |
| hd202320 | U | 4490 | 1.76 | 1.79 | 0.07 | 0.04 | 0.09 | 0.29 | 0.26 | -0.13 | -0.21 | -0.19 | -0.16 | -0.02 | -0.06 | -0.05 | -0.15 | -0.04 | -0.01 | 0.27 | -0.03 | 0.01 | -0.28 | 0.18 | 0.24 | 0.59 | 0.27 | 0.22 | -0.11 |
| hd203638 | U | 4532 | 2.51 | 1.47 | 0.45 | 0.45 | 0.49 | 0.56 | 1.25 | 0.22 | 0.24 | 0.26 | 0.49 | 0.33 | 0.37 | 0.31 | 0.35 | 0.47 | 0.82 | 0.92 | 0.10 | 0.17 | -0.07 | 0.17 | 0.58 | 0.80 | 0.52 | 0.61 | 0.23 |
| hd207964 | U | 6538 | 3.90 | 3.23 | | | | -0.37 | 1.24 | 0.43 | 0.26 | 0.09 | 2.04 | 1.05 | 1.71 | 0.00 | 1.25 | 0.98 | | | 0.87 | 1.54 | | | | | | | |
| hd211173 | U | 4883 | 3.11 | 1.00 | -0.12 | -0.20 | 0.15 | 0.07 | | -0.11 | -0.15 | -0.12 | -0.10 | -0.14 | -0.21 | -0.15 | -0.15 | -0.09 | 0.01 | 0.07 | 0.40 | 0.43 | 0.07 | 0.57 | 0.54 | 0.48 | 0.54 | 0.30 | 0.25 |
| hd212320 | U | 4851 | 2.40 | 1.47 | 0.11 | 0.01 | -0.08 | 0.19 | 0.50 | -0.06 | -0.17 | -0.23 | -0.30 | -0.08 | -0.23 | -0.14 | -0.25 | -0.13 | -0.14 | 0.28 | 0.69 | 0.60 | 0.46 | 0.84 | 1.01 | 1.14 | 0.77 | 0.77 | 0.00 |
| hd223094 | U | 3764 | 0.23 | 1.94 | -0.19 | -0.53 | -0.04 | 0.07 | 1.81 | -0.31 | -0.56 | -0.54 | -0.36 | -0.32 | -0.45 | -0.35 | -0.44 | -0.31 | | -0.72 | 0.14 | -0.51 | -0.86 | -0.45 | -0.24 | -0.34 | -0.14 | 0.17 | |
| hic007995 | U | 6036 | 3.85 | 0.93 | 0.68 | 1.01 | 0.81 | 0.30 | -0.06 | 0.74 | 0.86 | 1.19 | 1.60 | 0.83 | 0.99 | 0.75 | 0.92 | 0.70 | 1.46 | 0.27 | 1.07 | 0.84 | 1.20 | 0.83 | 1.09 | 1.10 | 1.17 | 1.09 | 1.01 |
| hip049418 | U | 4660 | 2.38 | 1.33 | 0.43 | 0.31 | 0.50 | 0.28 | 1.01 | 0.17 | 0.08 | 0.17 | 0.46 | 0.23 | 0.28 | 0.14 | 0.11 | 0.21 | 0.29 | 0.10 | 0.16 | -0.01 | -0.04 | 0.02 | 0.24 | 0.20 | 0.20 | 0.10 | 0.03 |
| hip051077 | U | 4589 | 2.57 | 1.29 | 0.20 | 0.36 | 0.47 | 0.39 | 1.14 | 0.14 | 0.21 | 0.21 | 0.45 | 0.26 | 0.26 | 0.25 | 0.17 | 0.29 | 0.41 | 0.26 | 0.13 | 0.03 | -0.04 | 0.26 | 0.53 | 0.47 | 0.46 | 0.46 | 0.50 |
| hip053502 | U | 4786 | 2.54 | 1.35 | 0.17 | 0.11 | 0.33 | 0.15 | 0.64 | 0.07 | -0.03 | -0.01 | 0.09 | 0.04 | 0.03 | 0.01 | -0.06 | 0.01 | 0.04 | -0.06 | 0.13 | -0.03 | -0.06 | 0.13 | 0.26 | 0.23 | 0.18 | 0.16 | 0.17 |
| hip059785 | U | 4793 | 2.57 | 1.52 | -0.22 | 0.00 | 0.06 | -0.05 | 0.08 | -0.19 | -0.15 | -0.12 | -0.17 | -0.34 | -0.50 | -0.37 | -0.25 | -0.32 | -0.22 | -0.22 | -0.27 | -0.41 | -0.37 | -0.48 | -0.19 | -0.30 | -0.17 | -0.14 | -0.13 |
| hip066936 | U | 4589 | 2.51 | 1.32 | 0.37 | 0.45 | 0.56 | 0.43 | 1.33 | 0.17 | 0.23 | 0.33 | 0.69 | 0.36 | 0.45 | 0.30 | 0.32 | 0.41 | 0.62 | 0.81 | 0.23 | 0.21 | 0.07 | 0.19 | 0.53 | 0.42 | 0.47 | 0.44 | |
| hip070306 | U | 4255 | 2.35 | 1.91 | 0.01 | -0.11 | 0.18 | 0.05 | 1.41 | -0.49 | -0.09 | -0.12 | 0.12 | 0.07 | -0.12 | -0.14 | 0.10 | 0.02 | | -0.25 | 0.04 | -0.12 | -0.24 | -0.82 | 0.05 | 0.10 | 0.26 | 0.14 | |
| hip078650 | U | 4370 | 2.25 | 1.53 | -0.01 | 0.04 | 0.32 | 0.16 | 1.15 | -0.22 | -0.05 | -0.03 | 0.24 | 0.07 | 0.02 | -0.05 | 0.07 | 0.08 | | -0.03 | 0.02 | -0.15 | -0.21 | -0.23 | 0.08 | 0.11 | 0.27 | 0.08 | |
| hip080343 | U | 4758 | 2.48 | 1.64 | 0.03 | -0.21 | 0.05 | -0.15 | 0.46 | -0.36 | -0.32 | -0.27 | -0.19 | -0.16 | -0.23 | -0.32 | -0.25 | -0.23 | -0.20 | -0.20 | 0.10 | -0.06 | -0.04 | -0.36 | -0.06 | -0.04 | -0.08 | -0.20 | |
| hip082396 | U | 4522 | 2.34 | 1.30 | 0.30 | 0.33 | 0.41 | 0.32 | 0.80 | 0.14 | 0.04 | 0.18 | 0.49 | 0.20 | 0.23 | 0.15 | 0.13 | 0.24 | 0.44 | 0.41 | 0.07 | 0.01 | -0.17 | 0.04 | 0.19 | 0.24 | 0.25 | 0.22 | 0.15 |
| hip093498 | U | 4482 | 2.54 | 1.24 | 0.60 | 0.56 | 0.73 | 0.64 | 1.60 | 0.38 | 0.43 | 0.53 | 0.92 | 0.55 | 0.57 | 0.49 | 0.53 | 0.59 | | 0.51 | 0.36 | 0.21 | 0.18 | 0.30 | 0.59 | 0.63 | 0.69 | 0.63 | |
| hip103738 | U | 5009 | 2.69 | 1.60 | 0.22 | -0.03 | 0.23 | 0.07 | 0.23 | -0.04 | -0.11 | -0.11 | -0.10 | 0.01 | -0.23 | -0.08 | -0.15 | -0.11 | -0.43 | -0.31 | 0.12 | -0.02 | -0.06 | 0.18 | 0.31 | 0.37 | 0.13 | 0.07 | 0.10 |
| hip106039 | U | 5017 | 2.83 | 1.32 | 0.17 | 0.04 | | 0.06 | 0.36 | 0.00 | -0.09 | -0.09 | -0.07 | -0.01 | -0.09 | -0.05 | -0.13 | -0.08 | -0.20 | -0.18 | 0.14 | -0.03 | -0.05 | 0.14 | 0.18 | 0.26 | 0.16 | 0.08 | 0.10 |
| hip113246 | U | 4828 | 2.55 | 1.36 | -0.06 | -0.04 | 0.14 | -0.01 | 0.29 | -0.12 | -0.19 | -0.18 | -0.17 | -0.14 | -0.22 | -0.16 | -0.23 | -0.17 | -0.23 | -0.26 | -0.02 | -0.09 | -0.15 | 0.09 | 0.15 | 0.25 | 0.23 | 0.15 | 0.29 |
| hip114119 | U | 4962 | 2.64 | 1.09 | 0.20 | 0.11 | 0.33 | 0.09 | 0.35 | 0.12 | -0.02 | 0.07 | 0.20 | 0.11 | 0.05 | 0.03 | -0.03 | 0.01 | -0.07 | -0.21 | 0.20 | 0.04 | 0.06 | 0.17 | 0.26 | 0.33 | 0.23 | 0.15 | 0.76 |
| hip115102 | U | 4578 | 2.43 | 1.36 | 0.14 | 0.20 | 0.32 | 0.24 | 0.86 | -0.02 | 0.06 | 0.05 | 0.30 | 0.08 | 0.12 | 0.06 | 0.05 | 0.12 | 0.28 | 0.19 | -0.01 | -0.14 | -0.16 | 0.02 | 0.25 | 0.22 | 0.19 | 0.17 | 0.18 |
| hip116853 | U | 4886 | 2.67 | 1.33 | 0.48 | 0.30 | 0.46 | 0.34 | 0.68 | 0.21 | 0.08 | 0.14 | 0.27 | 0.25 | 0.27 | 0.19 | 0.14 | 0.24 | 0.18 | 0.15 | 0.17 | 0.13 | 0.05 | 0.19 | 0.42 | 0.27 | 0.28 | 0.22 | 0.15 |
| hr0296 | U | 4500 | 2.19 | 1.38 | -0.09 | -0.06 | 0.10 | 0.00 | 0.69 | -0.20 | -0.11 | -0.17 | -0.04 | -0.17 | -0.24 | -0.22 | -0.23 | -0.18 | -0.04 | -0.19 | -0.18 | -0.22 | -0.35 | -0.19 | 0.15 | 0.24 | 0.06 | 0.13 | -0.06 |
| hr4321 | U | 4569 | 2.47 | 1.32 | 0.52 | 0.41 | 0.55 | 0.60 | 0.79 | 0.33 | 0.20 | 0.28 | 0.55 | 0.38 | 0.44 | 0.34 | 0.35 | 0.49 | 0.72 | 0.73 | 0.28 | 0.13 | -0.15 | 0.26 | 0.50 | 0.29 | 0.45 | 0.44 | 0.24 |
| ic2391-0022 | U | 4610 | 3.03 | 0.90 | 0.24 | 0.14 | 0.14 | 0.52 | 0.92 | 0.08 | 0.11 | 0.02 | 0.15 | 0.13 | 0.14 | 0.23 | 0.11 | 0.35 | 0.42 | 0.62 | 0.03 | 0.36 | -0.13 | 0.50 | 0.67 | 0.97 | 0.64 | 0.62 | 0.33 |
| ic2391-0026 | U | 4585 | 3.22 | 0.44 | -0.22 | -0.11 | -0.07 | 0.35 | 0.86 | -0.13 | 0.13 | -0.04 | -0.03 | -0.07 | -0.07 | 0.09 | -0.07 | 0.23 | 0.36 | 0.55 | -0.14 | 0.30 | -0.02 | 0.38 | 0.71 | 1.08 | 0.92 | 0.79 | 0.27 |
| ic2391-0044 | U | 6699 | 4.16 | 5.45 | -1.80 | | | 0.34 | | 0.40 | 0.76 | 0.69 | 0.64 | 1.30 | | 1.10 | 2.20 | 0.66 | | | 1.84 | 0.92 | 3.46 | | | 1.24 | | | |
| ic4651-E12 | U | 4829 | 2.54 | 1.37 | 0.34 | 0.14 | 0.26 | 0.27 | 0.66 | 0.16 | -0.02 | 0.01 | 0.11 | 0.11 | 0.13 | 0.06 | 0.02 | 0.10 | 0.13 | 0.19 | 0.19 | 0.05 | -0.21 | 0.15 | 0.15 | 0.05 | 0.13 | 0.10 | 0.15 |
| ic4651-E60 | U | 4760 | 2.60 | 1.23 | 0.31 | 0.16 | 0.32 | 0.30 | 0.65 | 0.21 | 0.00 | 0.11 | 0.24 | 0.19 | 0.22 | 0.13 | 0.09 | 0.18 | 0.25 | 0.27 | 0.21 | 0.11 | -0.11 | 0.19 | 0.22 | 0.11 | 0.21 | 0.10 | 0.20 |
| ic4651no7646 | U | 4829 | 2.54 | 1.47 | 0.36 | 0.15 | 0.28 | 0.31 | 0.88 | 0.14 | 0.02 | 0.04 | 0.11 | 0.12 | 0.10 | 0.08 | 0.04 | 0.14 | 0.14 | 0.41 | 0.21 | 0.19 | -0.12 | 0.14 | 0.29 | 0.04 | 0.19 | 0.44 | 0.28 |
| ic4651no9122 | U | 4694 | 2.56 | 1.36 | 0.36 | 0.24 | 0.38 | 0.35 | 0.85 | 0.25 | 0.13 | 0.23 | 0.39 | 0.27 | 0.26 | 0.21 | 0.21 | 0.28 | 0.45 | 0.25 | 0.33 | 0.14 | 0.03 | 0.27 | 0.50 | 0.31 | 0.48 | 0.79 | 0.34 |
| ngc2447No41 | U | 5058 | 2.56 | 1.54 | 0.25 | 0.03 | 0.10 | 0.14 | 0.44 | 0.06 | -0.08 | 0.00 | -0.04 | 0.05 | -0.11 | -0.02 | -0.07 | -0.03 | -0.12 | 0.15 | 0.26 | 0.04 | 0.00 | 0.32 | 0.15 | 0.22 | 0.21 | 0.18 | 0.12 |
| NGC2682ESIII-35 | U | 5046 | 3.27 | 0.99 | 0.27 | 0.22 | 0.32 | 0.24 | 0.57 | 0.26 | 0.17 | 0.30 | 0.41 | 0.27 | 0.34 | 0.21 | 0.21 | 0.28 | 0.47 | 0.24 | 0.26 | 0.32 | 0.05 | 0.25 | 0.40 | 0.25 | 0.43 | 0.80 | 0.37 |
| ngc2682No164 | U | 4734 | 2.48 | 1.37 | 0.31 | 0.10 | 0.32 | 0.26 | 0.72 | 0.19 | 0.00 | 0.14 | 0.24 | 0.18 | 0.16 | 0.11 | 0.09 | 0.19 | 0.27 | 0.21 | 0.17 | -0.01 | -0.10 | 0.19 | 0.08 | 0.23 | 0.47 | 0.13 | |
| ngc2682No286 | U | 4778 | 2.53 | 1.41 | 0.34 | 0.21 | 0.36 | 0.25 | 0.80 | 0.19 | 0.05 | 0.15 | 0.26 | 0.20 | 0.22 | 0.12 | 0.12 | 0.19 | 0.26 | 0.23 | 0.18 | 0.06 | -0.06 | 0.13 | 0.26 | 0.14 | 0.27 | 0.49 | 0.17 |
| ngc3114no181 | U | 4456 | 1.61 | 1.83 | 0.31 | 0.07 | 0.19 | 0.32 | 1.13 | 0.00 | -0.19 | -0.09 | -0.05 | 0.08 | -0.10 | 0.00 | -0.09 | 0.03 | 0.16 | 0.29 | 0.22 | 0.06 | -0.29 | 0.28 | 0.26 | 0.12 | 0.16 | 0.30 | 0.15 |
| ngc3680no26 | U | 4662 | 2.52 | 1.22 | 0.18 | 0.11 | 0.25 | 0.26 | 0.90 | 0.15 | -0.01 | 0.10 | 0.20 | 0.14 | 0.07 | 0.08 | 0.01 | 0.11 | 0.22 | 0.30 | 0.22 | 0.06 | -0.20 | 0.22 | 0.29 | 0.22 | 0.44 | 0.50 | 0.09 |
| p1955 | U | 5211 | 2.50 | 3.34 | | | -0.16 | | | -0.13 | -0.22 | 1.02 | | 0.20 | -0.06 | 0.33 | | | | | 2.38 | 1.90 | | | | -0.34 | 1.63 | | |
| txpic | U | 4435 | 2.13 | 4.25 | | | 0.69 | | -0.26 | -0.06 | -0.82 | -0.49 | | -0.76 | -0.19 | 0.42 | 0.90 | | | -1.06 | | | | | | | | 1.60 | |
| XSct | U | 4763 | 2.81 | 3.19 | 0.28 | -0.80 | | 0.16 | 1.10 | -0.14 | | -0.35 | -0.28 | -0.43 | -1.10 | -0.47 | -0.55 | -0.38 | | -0.37 | | -0.84 | 0.04 | | 0.12 | 0.68 | 0.25 | 0.28 | |

| ID | | T | | | | | | | | | | | | | | | | | | | | | | | | | | | |
|---|---|---|---|---|---|---|---|---|---|---|---|---|---|---|---|---|---|---|---|---|---|---|---|---|---|---|---|---|---|
| hd001671 | E | 6323 | 2.95 | 2.71 | 0.13 | | | 0.14 | 0.38 | 0.23 | 0.10 | 0.24 | 0.75 | 0.34 | 0.04 | 0.00 | 0.39 | 0.32 | -0.60 | -0.42 | 1.46 | 0.77 | 1.85 | | 0.01 | 0.04 | 0.17 | 0.53 | |
| hd002910 | E | 4696 | 2.10 | 1.50 | 0.36 | 0.21 | 0.21 | 0.31 | 0.79 | 0.14 | -0.03 | 0.02 | 0.16 | 0.16 | 0.14 | 0.05 | 0.00 | 0.12 | 0.13 | 0.51 | -0.03 | 0.03 | -0.11 | -0.09 | 0.29 | 0.56 | 0.21 | 0.14 | 0.04 |
| hd004188 | E | 4793 | 2.20 | 1.46 | 0.30 | 0.20 | 0.26 | 0.20 | 0.58 | 0.08 | -0.06 | 0.02 | 0.12 | 0.13 | 0.09 | 0.02 | -0.02 | 0.06 | 0.03 | 0.23 | 0.09 | 0.10 | -0.01 | 0.00 | 0.29 | 0.45 | 0.22 | 0.23 | 0.04 |
| hd004502 | E | 4570 | 0.10 | 2.61 | | 0.09 | 0.06 | | 0.00 | -0.37 | -0.22 | -0.11 | -0.08 | -0.21 | -0.47 | -0.23 | -0.38 | | | -0.65 | 0.07 | | -0.16 | | -0.46 | -0.16 | -0.11 | | |
| hd005234 | E | 4422 | 1.58 | 1.62 | 0.27 | 0.06 | 0.20 | 0.33 | 0.67 | -0.03 | -0.18 | -0.03 | 0.00 | 0.14 | 0.01 | -0.04 | -0.10 | -0.01 | 0.03 | -0.41 | 0.04 | -0.06 | -0.16 | -0.01 | 0.36 | 0.67 | 0.37 | 0.65 | -0.08 |
| hd006319 | E | 4755 | 2.44 | 1.48 | 0.38 | 0.26 | 0.40 | 0.29 | 0.81 | 0.17 | 0.20 | 0.24 | 0.41 | 0.27 | 0.25 | 0.17 | 0.18 | 0.27 | 0.37 | 0.32 | 0.14 | 0.15 | 0.11 | 0.15 | 0.51 | 0.72 | 0.57 | 0.49 | 0.19 |
| hd010380 | E | 4154 | 0.90 | 1.86 | 0.23 | 0.16 | 0.13 | 0.12 | 0.90 | -0.20 | -0.38 | -0.21 | -0.16 | -0.06 | -0.27 | -0.30 | -0.22 | -0.20 | -0.02 | -0.04 | -0.07 | -0.34 | -0.39 | -0.62 | -0.06 | -0.08 | 0.02 | -0.04 | -0.41 |
| hd011559 | E | 4947 | 2.54 | 1.37 | 0.48 | 0.26 | 0.31 | 0.30 | 0.75 | 0.22 | 0.00 | 0.14 | 0.24 | 0.24 | 0.16 | 0.12 | 0.07 | 0.15 | 0.14 | 0.07 | 0.19 | 0.16 | 0.02 | 0.03 | 0.23 | 0.52 | 0.31 | 0.24 | 0.19 |
| hd013174 | E | 6710 | 3.49 | 7.75 | -1.24 | | | 0.28 | | 1.27 | 0.77 | 1.32 | 1.34 | 0.36 | -0.11 | 0.44 | 1.56 | 0.41 | | | 0.53 | -0.04 | 2.22 | | | 0.18 | 0.78 | 1.34 | 0.80 |
| hd013480 | E | 5082 | 1.72 | 2.92 | | | | -0.03 | 0.12 | 0.09 | -0.22 | 0.20 | 0.30 | 0.57 | -0.27 | -0.06 | 0.33 | 0.01 | -0.65 | | | 2.32 | -0.08 | 0.98 | -0.59 | 0.23 | 0.24 | 0.55 | 0.13 | 0.68 |
| hd013520 | E | 4010 | 0.19 | 1.96 | 0.01 | -0.25 | -0.05 | -0.04 | 1.08 | -0.32 | -0.61 | -0.42 | -0.37 | -0.16 | -0.57 | -0.43 | -0.38 | -0.36 | 0.15 | -0.07 | 0.19 | -0.37 | -0.63 | -0.81 | -0.47 | -0.41 | -0.26 | -0.08 | -0.38 |
| hd015257 | E | 7079 | 4.80 | 4.16 | 0.69 | | | 0.14 | | 0.49 | 1.20 | 1.17 | 1.20 | 0.75 | 0.07 | 0.36 | 1.22 | 1.02 | | 0.28 | | 1.64 | 0.64 | | 1.47 | | 2.08 | | | |
| hd015596 | E | 4788 | 2.38 | 1.21 | -0.42 | -0.27 | -0.20 | -0.30 | 0.09 | -0.31 | -0.55 | -0.38 | -0.47 | -0.54 | -0.83 | -0.65 | -0.53 | -0.61 | -0.65 | -0.54 | -0.36 | -0.59 | -0.50 | -0.72 | -0.54 | -0.43 | -0.36 | -0.49 | -0.07 |
| hd018885 | E | 4670 | 2.14 | 1.51 | 0.54 | 0.40 | 0.29 | 0.39 | 0.91 | 0.20 | 0.04 | 0.14 | 0.32 | 0.27 | 0.28 | 0.16 | 0.14 | 0.25 | 0.34 | 0.69 | 0.09 | 0.11 | 0.07 | 0.03 | 0.42 | 0.62 | 0.35 | 0.43 | 0.11 |
| hd022764 | E | 4205 | 1.08 | 2.52 | 0.82 | 0.04 | 0.48 | 0.49 | 1.84 | 0.15 | -0.06 | 0.13 | 0.25 | 0.36 | 0.05 | 0.04 | 0.10 | 0.14 | 0.56 | -0.27 | 0.48 | 0.03 | 0.18 | | | 0.47 | 0.46 | 0.36 | 0.56 | 0.44 |
| hd023249 | E | 4966 | 3.27 | 0.84 | 0.41 | 0.31 | 0.36 | 0.25 | 0.68 | 0.25 | 0.10 | 0.17 | 0.29 | 0.23 | 0.20 | 0.13 | 0.10 | 0.24 | 0.43 | 0.49 | 0.09 | 0.12 | -0.04 | -0.10 | 0.27 | 0.35 | 0.37 | 0.37 | 0.25 |
| hd025602 | E | 4743 | 2.16 | 1.29 | 0.02 | -0.03 | 0.03 | -0.04 | 0.24 | -0.10 | -0.31 | -0.24 | -0.18 | -0.15 | -0.22 | -0.24 | -0.23 | -0.20 | -0.19 | -0.17 | -0.23 | -0.27 | -0.46 | -0.34 | -0.12 | -0.07 | -0.07 | -0.22 | 0.13 |
| hd026659 | E | 5170 | 2.71 | 1.43 | 0.17 | 0.09 | 0.09 | 0.02 | 0.12 | 0.03 | -0.13 | -0.04 | -0.02 | 0.03 | -0.14 | -0.06 | -0.07 | -0.07 | -0.19 | -0.23 | 0.06 | 0.02 | -0.02 | 0.14 | 0.06 | 0.29 | 0.13 | -0.07 | 0.44 |
| hd029139 | E | 3903 | -0.15 | 2.03 | 0.19 | -0.29 | 0.17 | 0.29 | 1.90 | -0.52 | -0.56 | -0.51 | -0.31 | -0.18 | -0.79 | -0.37 | -0.32 | -0.29 | 0.56 | 0.01 | 0.15 | -0.45 | -0.69 | -0.77 | -0.66 | -0.46 | -0.26 | -0.39 | -1.59 |
| hd030834 | E | 4153 | 0.66 | 1.88 | 0.23 | -0.04 | -0.04 | -0.08 | 0.76 | -0.28 | -0.48 | -0.32 | -0.35 | -0.16 | -0.45 | -0.37 | -0.38 | -0.33 | 0.00 | -0.56 | -0.14 | -0.46 | -0.53 | -0.41 | -0.11 | -0.04 | -0.12 | -0.04 | -0.24 |
| hd031444 | E | 5051 | 2.58 | 1.33 | 0.19 | 0.12 | 0.11 | 0.06 | 0.38 | 0.07 | -0.14 | -0.08 | -0.02 | 0.04 | -0.11 | -0.07 | -0.12 | -0.08 | -0.33 | -0.11 | 0.18 | 0.02 | -0.07 | 0.11 | 0.10 | 0.32 | 0.19 | 0.01 | 0.34 |
| hd033419 | E | 4674 | 2.07 | 1.55 | 0.52 | 0.41 | 0.34 | 0.37 | 0.94 | 0.19 | 0.04 | 0.16 | 0.29 | 0.28 | 0.23 | 0.15 | 0.14 | 0.24 | 0.44 | 0.34 | 0.11 | 0.13 | 0.01 | 0.00 | 0.37 | 0.45 | 0.42 | 0.34 | 0.09 |
| hd033618 | E | 4558 | 2.17 | 1.36 | 0.65 | 0.41 | 0.48 | 0.49 | 1.29 | 0.39 | 0.24 | 0.42 | 0.57 | 0.47 | 0.47 | 0.29 | 0.34 | 0.50 | 0.69 | 0.71 | 0.28 | 0.28 | 0.14 | 0.22 | 0.70 | 0.99 | 0.73 | 0.76 | 0.39 |
| hd034029 | E | 5155 | 1.93 | 2.47 | | 0.43 | -0.38 | 0.04 | 0.32 | -0.52 | -0.25 | -0.27 | -0.02 | 0.30 | -0.45 | -0.49 | -0.28 | -0.13 | -1.61 | -0.45 | 1.01 | -0.04 | -0.03 | -1.92 | 0.14 | -0.06 | -0.41 | -0.33 | -0.41 |
| hd037160 | E | 4742 | 2.20 | 1.25 | -0.32 | -0.23 | -0.14 | -0.28 | 0.15 | -0.31 | -0.47 | -0.35 | -0.37 | -0.45 | -0.67 | -0.56 | -0.47 | -0.52 | -0.49 | -0.40 | -0.45 | -0.58 | -0.66 | -0.73 | -0.49 | -0.45 | -0.40 | -0.33 | 0.08 |
| hd037638 | E | 5088 | 2.62 | 1.41 | 0.26 | 0.10 | 0.16 | 0.12 | 0.40 | 0.13 | -0.07 | -0.01 | 0.03 | 0.09 | 0.00 | 0.00 | -0.05 | -0.03 | -0.15 | 0.08 | 0.30 | 0.06 | 0.03 | 0.16 | 0.26 | 0.57 | 0.36 | 0.18 | 0.15 |
| hd038309 | E | 6927 | 4.16 | 3.90 | -0.03 | 1.04 | | 0.02 | | -0.21 | 0.20 | 0.62 | 0.73 | 0.37 | 0.05 | -0.24 | 1.42 | 0.16 | -0.34 | | | 0.78 | 2.23 | -0.81 | | 0.38 | 0.99 | 0.95 | 0.50 |
| hd039070 | E | 5086 | 2.80 | 1.26 | 0.35 | 0.19 | 0.19 | 0.21 | 0.36 | 0.20 | 0.04 | 0.12 | 0.15 | 0.18 | 0.04 | 0.09 | 0.05 | 0.09 | 0.01 | 0.14 | 0.36 | 0.22 | 0.40 | 0.22 | 0.31 | 0.68 | 0.48 | 0.31 | 0.57 |
| hd039833 | E | 5751 | 4.03 | 1.04 | 0.23 | 0.36 | 0.25 | 0.21 | 0.20 | 0.27 | 0.15 | 0.22 | 0.27 | 0.29 | 0.18 | 0.21 | 0.18 | 0.22 | 0.02 | 0.17 | 0.49 | 0.32 | 0.33 | 0.25 | 0.49 | 0.51 | 0.39 | 0.29 | 0.81 |
| hd039910 | E | 4577 | 2.02 | 1.67 | 0.49 | 0.45 | 0.43 | 0.47 | 1.08 | 0.21 | 0.17 | 0.22 | 0.36 | 0.32 | 0.25 | 0.19 | 0.25 | 0.32 | 0.57 | 0.37 | 0.07 | 0.05 | -0.05 | -0.13 | 0.43 | 0.74 | 0.40 | 0.52 | 0.16 |
| hd040801 | E | 4787 | 2.38 | 1.29 | 0.01 | 0.09 | 0.09 | 0.01 | 0.34 | -0.05 | -0.22 | -0.14 | -0.04 | -0.11 | -0.19 | -0.19 | -0.16 | -0.15 | -0.11 | 0.01 | -0.23 | -0.30 | -0.38 | -0.37 | -0.10 | 0.17 | 0.08 | 0.04 | 0.22 |
| hd042341 | E | 4635 | 2.32 | 1.57 | 0.71 | 0.54 | 0.50 | 0.51 | 1.22 | 0.40 | 0.24 | 0.32 | 0.52 | 0.45 | 0.38 | 0.28 | 0.33 | 0.39 | 0.41 | 0.77 | 0.20 | 0.14 | 0.18 | -0.13 | 0.37 | 0.74 | 0.56 | 0.65 | 0.27 |
| hd046374 | E | 4656 | 2.15 | 1.51 | 0.34 | 0.32 | 0.35 | 0.31 | 0.79 | 0.11 | 0.02 | 0.08 | 0.25 | 0.17 | 0.12 | 0.05 | 0.05 | 0.14 | 0.27 | 0.14 | 0.08 | 0.09 | -0.05 | -0.07 | 0.29 | 0.61 | 0.35 | 0.18 | 0.00 |
| hd047138 | E | 6091 | 4.61 | 0.81 | 0.57 | 0.54 | 0.44 | 0.29 | 0.21 | 0.51 | 0.82 | 0.91 | 1.02 | 0.71 | 0.63 | 0.63 | 0.76 | 0.60 | 0.71 | 0.60 | 1.22 | 1.08 | 1.20 | 1.03 | 1.55 | 1.80 | 1.64 | 1.42 | 1.18 |
| hd047366 | E | 4857 | 2.63 | 1.23 | 0.22 | 0.15 | 0.19 | 0.17 | 0.51 | 0.13 | -0.04 | 0.06 | 0.16 | 0.13 | 0.06 | 0.02 | 0.01 | 0.05 | 0.07 | 0.12 | 0.13 | 0.03 | -0.06 | 0.02 | 0.30 | 0.38 | 0.38 | 0.21 | 0.06 |
| hd050522 | E | 5095 | 2.81 | 0.97 | 0.56 | 0.36 | 0.37 | 0.34 | 0.69 | 0.34 | 0.24 | 0.35 | 0.58 | 0.43 | 0.30 | 0.28 | 0.24 | 0.34 | 0.27 | 0.20 | 0.32 | 0.35 | 0.38 | 0.27 | 0.48 | 0.79 | 0.46 | 0.66 | 0.78 |
| hd051000 | E | 5102 | 2.56 | 1.44 | 0.30 | 0.05 | 0.06 | 0.11 | 0.30 | 0.06 | -0.13 | -0.03 | -0.02 | 0.08 | -0.09 | -0.02 | -0.06 | -0.03 | -0.25 | -0.11 | 0.26 | 0.09 | 0.03 | 0.22 | 0.31 | 0.50 | 0.16 | 0.07 | 0.13 |
| hd054079 | E | 4454 | 1.49 | 1.63 | -0.10 | -0.18 | -0.07 | -0.04 | 0.50 | -0.34 | -0.39 | -0.33 | -0.29 | -0.29 | -0.42 | -0.39 | -0.39 | -0.35 | -0.23 | -0.27 | -0.22 | -0.32 | -0.50 | -0.37 | -0.14 | 0.12 | 0.01 | -0.14 | -0.22 |
| hd055280 | E | 4639 | 2.31 | 1.30 | 0.43 | 0.29 | 0.39 | 0.34 | 0.83 | 0.23 | 0.03 | 0.15 | 0.29 | 0.26 | 0.25 | 0.13 | 0.07 | 0.20 | 0.34 | 0.52 | 0.15 | 0.06 | -0.04 | -0.03 | 0.44 | 0.65 | 0.50 | 0.45 | 0.18 |
| hd057727 | E | 5007 | 2.64 | 1.32 | 0.13 | 0.05 | 0.10 | 0.07 | 0.11 | 0.08 | -0.16 | -0.04 | 0.00 | 0.05 | -0.03 | -0.04 | -0.10 | -0.06 | -0.20 | -0.06 | 0.09 | -0.05 | 0.01 | 0.06 | 0.16 | 0.32 | 0.23 | 0.11 | 0.28 |
| hd058207 | E | 4750 | 2.08 | 1.47 | 0.18 | 0.16 | 0.18 | 0.15 | 0.65 | -0.01 | -0.14 | -0.08 | 0.01 | 0.01 | -0.01 | -0.07 | -0.11 | -0.04 | -0.03 | 0.13 | -0.05 | -0.10 | -0.18 | -0.07 | 0.14 | 0.27 | 0.12 | 0.02 | -0.07 |
| hd058923 | E | 7505 | 3.94 | 3.09 | | 0.47 | | 0.85 | | 0.76 | 0.81 | 0.99 | 1.37 | 1.24 | 0.71 | 0.51 | 1.49 | 0.74 | | 0.91 | | 1.90 | 1.66 | | 0.76 | 0.84 | 2.36 | 0.91 | | |
| hd060294 | E | 4580 | 2.16 | 1.37 | 0.48 | 0.30 | 0.40 | 0.34 | 0.98 | 0.22 | 0.00 | 0.14 | 0.29 | 0.27 | 0.21 | 0.10 | 0.07 | 0.20 | 0.38 | 0.52 | 0.13 | 0.03 | -0.06 | -0.14 | 0.36 | 0.39 | 0.36 | 0.35 | 0.06 |
| hd060522 | E | 3884 | 0.10 | 2.15 | 0.40 | -0.14 | 0.34 | 0.51 | 2.44 | -0.10 | -0.42 | -0.27 | -0.07 | 0.10 | -0.48 | -0.19 | -0.08 | -0.03 | 0.73 | -0.32 | 0.76 | -0.21 | -0.37 | -0.66 | -0.14 | 0.16 | 0.03 | -0.13 | -0.46 |
| hd061363 | E | 4713 | 1.92 | 1.45 | 0.00 | -0.02 | -0.03 | -0.04 | 0.27 | -0.13 | -0.38 | -0.27 | -0.23 | -0.18 | -0.27 | -0.29 | -0.32 | -0.26 | -0.26 | -0.11 | -0.17 | -0.38 | -0.42 | -0.33 | -0.17 | 0.14 | -0.05 | -0.21 | -0.27 |
| hd062141 | E | 4922 | 2.50 | 1.37 | 0.20 | 0.04 | 0.18 | 0.08 | 0.38 | 0.08 | -0.16 | -0.10 | -0.04 | 0.05 | -0.04 | -0.07 | -0.15 | -0.09 | -0.21 | -0.11 | 0.08 | 0.05 | -0.03 | 0.08 | 0.19 | 0.42 | 0.14 | 0.03 | -0.03 |
| hd062437 | E | 7600 | 3.80 | 2.59 | 0.44 | 0.46 | | 0.70 | 0.36 | 0.85 | 0.65 | 1.18 | 1.33 | 1.04 | 0.99 | 0.47 | 1.38 | 0.81 | | | 1.64 | 1.75 | 0.82 | 0.23 | 0.88 | 1.33 | 0.86 | 2.72 | 2.20 | |
| hd062721 | E | 4002 | 0.35 | 1.81 | 0.20 | -0.06 | 0.20 | 0.20 | 1.69 | -0.13 | -0.36 | -0.20 | -0.07 | -0.05 | -0.54 | -0.30 | -0.15 | -0.20 | 0.74 | -0.52 | 0.33 | -0.22 | -0.42 | -0.81 | -0.29 | 0.06 | 0.16 | 0.27 | -0.40 |
| hd064152 | E | 4928 | 2.32 | 1.46 | 0.42 | 0.25 | 0.25 | 0.20 | 0.65 | 0.18 | -0.07 | 0.01 | 0.13 | 0.17 | 0.09 | 0.05 | 0.01 | 0.06 | -0.11 | 0.04 | 0.20 | 0.03 | -0.04 | 0.01 | 0.22 | 0.34 | 0.13 | 0.17 | 0.31 |

| ID | Type | C1 | C2 | C3 | C4 | C5 | C6 | C7 | C8 | C9 | C10 | C11 | C12 | C13 | C14 | C15 | C16 | C17 | C18 | C19 | C20 | C21 | C22 | C23 | C24 | C25 | C26 | C27 | C28 | C29 |
|---|---|---|---|---|---|---|---|---|---|---|---|---|---|---|---|---|---|---|---|---|---|---|---|---|---|---|---|---|---|---|
| hd074794 | E | 4648 | 2.09 | 1.48 | 0.41 | 0.34 | 0.32 | 0.33 | 0.97 | 0.19 | 0.00 | 0.10 | 0.28 | 0.22 | 0.22 | 0.10 | 0.08 | 0.19 | 0.38 | 0.28 | 0.05 | 0.03 | -0.10 | -0.11 | 0.32 | 0.49 | 0.32 | 0.29 | 0.06 |
| hd081688 | E | 4760 | 2.07 | 1.49 | -0.17 | -0.02 | 0.01 | -0.04 | 0.29 | -0.19 | -0.34 | -0.23 | -0.22 | -0.21 | -0.30 | -0.29 | -0.30 | -0.28 | -0.25 | -0.30 | -0.32 | -0.45 | -0.47 | -0.45 | -0.26 | -0.07 | -0.15 | -0.22 | -0.24 |
| hd082885 | E | 5412 | 3.99 | 1.06 | 0.59 | 0.44 | 0.44 | 0.37 | 0.68 | 0.38 | 0.27 | 0.32 | 0.39 | 0.42 | 0.42 | 0.29 | 0.32 | 0.39 | 0.42 | 0.59 | 0.34 | 0.44 | 0.40 | 0.07 | 0.52 | 0.65 | 0.48 | 0.67 | 0.79 |
| hd089025 | E | 6977 | 3.11 | 7.21 | 0.83 | | | 0.27 | 0.10 | -0.64 | 0.29 | 0.70 | 0.62 | 0.26 | -0.17 | -0.31 | 1.52 | 0.24 | | | -0.24 | 0.11 | | | 0.06 | 0.12 | 0.51 | 1.08 | | |
| hd093875 | E | 4556 | 2.22 | 1.35 | 0.63 | 0.45 | 0.44 | 0.53 | 1.33 | 0.35 | 0.28 | 0.36 | 0.52 | 0.43 | 0.40 | 0.27 | 0.31 | 0.49 | 1.14 | 0.64 | 0.15 | 0.27 | 0.12 | 0.19 | 0.62 | 0.97 | 0.76 | 0.68 | 0.40 |
| hd094672 | E | 6446 | 3.59 | 1.86 | 0.12 | 0.03 | -0.24 | 0.03 | -0.17 | 0.20 | 0.03 | 0.10 | -0.01 | -0.01 | -0.21 | -0.10 | 0.18 | -0.04 | -0.33 | -0.33 | | -0.05 | 0.40 | 0.23 | -0.08 | 0.06 | 0.09 | 0.35 | 0.86 | |
| hd095345 | E | 4519 | 1.73 | 1.59 | 0.14 | 0.12 | 0.13 | 0.18 | 0.77 | -0.06 | -0.14 | -0.07 | -0.04 | 0.03 | -0.15 | -0.09 | -0.11 | -0.06 | -0.04 | -0.03 | -0.01 | -0.03 | -0.27 | -0.01 | 0.29 | 0.58 | 0.32 | 0.18 | 0.01 |
| hd098262 | E | 4113 | 0.54 | 1.85 | 0.41 | -0.12 | 0.04 | 0.17 | 0.97 | -0.23 | -0.43 | -0.23 | -0.25 | 0.01 | -0.27 | -0.22 | -0.24 | -0.17 | -0.10 | -0.45 | 0.09 | -0.25 | -0.38 | -0.05 | 0.08 | 0.19 | 0.06 | 0.18 | -0.05 |
| hd098366 | E | 4697 | 2.18 | 1.33 | 0.21 | 0.15 | 0.17 | 0.15 | 0.50 | 0.00 | -0.11 | -0.04 | 0.05 | 0.04 | -0.03 | -0.06 | -0.09 | -0.01 | 0.03 | 0.09 | 0.01 | -0.03 | -0.18 | -0.14 | 0.16 | 0.51 | 0.27 | 0.20 | -0.04 |
| hd107700 | E | 6115 | 3.50 | 0.50 | 0.40 | 0.11 | 0.37 | 0.08 | -0.18 | 0.31 | 0.34 | 0.54 | 0.79 | 0.49 | 0.23 | 0.29 | 0.63 | 0.37 | 0.01 | -0.30 | 0.66 | 0.52 | 0.72 | 0.52 | 1.19 | 0.71 | 0.70 | 0.57 | 0.60 |
| hd107950 | E | 5098 | 2.27 | 1.80 | 0.44 | 0.19 | 0.20 | 0.23 | 0.44 | 0.13 | -0.04 | 0.06 | 0.02 | 0.16 | -0.03 | 0.07 | -0.01 | 0.05 | -0.24 | -0.18 | 0.47 | 0.19 | 0.06 | 0.20 | 0.17 | 0.37 | 0.31 | 0.07 | -0.01 |
| hd112127 | E | 4383 | 1.79 | 1.94 | 0.63 | 0.52 | 0.39 | 0.65 | 1.57 | 0.18 | 0.18 | 0.14 | 0.31 | 0.37 | 0.37 | 0.24 | 0.37 | 0.42 | 0.39 | 0.53 | 0.10 | 0.01 | -0.09 | -0.39 | 0.44 | 0.66 | 0.42 | 0.44 | 0.48 |
| hd115604 | E | 7314 | 3.34 | 2.09 | 0.72 | 0.65 | 0.68 | 0.63 | 0.49 | 0.70 | 0.87 | 0.72 | 0.85 | 0.69 | 0.46 | 0.54 | 0.83 | 0.67 | 0.58 | 0.65 | 1.12 | 0.92 | 0.96 | 1.68 | 1.10 | 0.96 | 0.86 | 0.69 | 1.25 |
| hd116292 | E | 4840 | 2.21 | 1.45 | 0.15 | 0.04 | 0.10 | 0.14 | 0.27 | 0.00 | -0.18 | -0.14 | -0.09 | -0.01 | -0.17 | -0.08 | -0.18 | -0.07 | -0.24 | -0.16 | -0.08 | -0.14 | -0.17 | -0.01 | 0.02 | 0.07 | 0.08 | -0.15 | 0.21 |
| hd116515 | E | 4728 | 2.01 | 1.49 | 0.08 | 0.04 | 0.15 | 0.06 | 0.32 | -0.06 | -0.16 | -0.10 | -0.05 | -0.02 | -0.11 | -0.13 | -0.13 | -0.07 | -0.08 | -0.19 | -0.08 | -0.04 | -0.22 | -0.19 | 0.04 | 0.32 | 0.12 | 0.06 | -0.13 |
| hd117710 | E | 4661 | 2.29 | 1.46 | 0.63 | 0.44 | 0.55 | 0.50 | 0.99 | 0.28 | 0.12 | 0.20 | 0.37 | 0.36 | 0.30 | 0.21 | 0.22 | 0.35 | 0.36 | 0.73 | 0.12 | 0.15 | 0.11 | -0.19 | 0.52 | 0.45 | 0.39 | 0.49 | 0.22 |
| hd120084 | E | 4806 | 2.29 | 1.44 | 0.42 | 0.24 | 0.24 | 0.27 | 0.76 | 0.13 | -0.02 | 0.04 | 0.19 | 0.21 | 0.17 | 0.09 | 0.00 | 0.13 | 0.18 | 0.04 | 0.09 | 0.12 | -0.02 | -0.01 | 0.28 | 0.47 | 0.31 | 0.24 | 0.06 |
| hd130952 | E | 4744 | 2.10 | 1.52 | -0.10 | 0.03 | 0.07 | 0.02 | 0.25 | -0.18 | -0.34 | -0.17 | -0.18 | -0.22 | -0.41 | -0.33 | -0.26 | -0.27 | -0.30 | -0.24 | -0.32 | -0.44 | -0.41 | -0.53 | -0.35 | -0.08 | -0.15 | -0.22 | -0.11 |
| hd131873 | E | 4005 | 0.58 | 1.86 | 0.02 | -0.06 | 0.05 | 0.16 | 1.57 | -0.27 | -0.39 | -0.22 | -0.19 | 0.00 | -0.29 | -0.21 | -0.17 | -0.11 | 0.23 | -0.55 | 0.18 | -0.20 | -0.39 | -0.24 | -0.05 | 0.20 | 0.21 | 0.31 | -0.01 |
| hd136202 | E | 6025 | 3.69 | 1.41 | 0.18 | 0.08 | 0.05 | 0.09 | 0.11 | 0.09 | 0.06 | 0.01 | 0.03 | 0.04 | -0.03 | -0.02 | 0.05 | -0.01 | -0.16 | -0.10 | 0.20 | 0.13 | 0.36 | 0.05 | 0.11 | 0.09 | 0.08 | -0.12 | 0.43 |
| hd136514 | E | 4417 | 1.86 | 1.54 | 0.31 | 0.31 | 0.40 | 0.35 | 1.05 | 0.08 | -0.02 | 0.09 | 0.25 | 0.17 | 0.04 | 0.01 | 0.05 | 0.15 | 0.31 | 0.29 | 0.01 | -0.13 | -0.13 | -0.29 | 0.18 | 0.38 | 0.45 | 0.41 | 0.10 |
| hd139254 | E | 4676 | 2.19 | 1.39 | 0.31 | 0.20 | 0.13 | 0.19 | 0.67 | 0.05 | -0.06 | 0.00 | 0.06 | 0.16 | 0.07 | -0.01 | -0.06 | 0.04 | 0.04 | 0.43 | 0.10 | 0.09 | -0.07 | 0.01 | 0.09 | 0.46 | 0.17 | 0.32 | -0.02 |
| hd143553 | E | 4697 | 2.26 | 1.21 | -0.03 | -0.06 | 0.05 | -0.05 | 0.27 | -0.09 | -0.29 | -0.18 | -0.09 | -0.15 | -0.20 | -0.25 | -0.27 | -0.20 | -0.14 | -0.04 | -0.18 | -0.25 | -0.37 | -0.43 | -0.10 | -0.04 | 0.02 | -0.07 | -0.29 |
| hd148604 | E | 5167 | 2.91 | 1.04 | 0.13 | 0.05 | 0.16 | 0.01 | 0.19 | 0.04 | -0.08 | 0.00 | 0.03 | 0.08 | -0.12 | -0.04 | -0.09 | -0.07 | -0.28 | 0.02 | 0.24 | 0.13 | 0.09 | 0.22 | 0.20 | 0.35 | 0.38 | 0.12 | 0.40 |
| hd148856 | E | 4903 | 1.95 | 1.62 | 0.26 | 0.10 | 0.08 | 0.12 | 0.59 | 0.05 | -0.21 | -0.12 | -0.11 | 0.04 | -0.15 | -0.07 | -0.16 | -0.06 | -0.22 | -0.16 | 0.05 | 0.01 | -0.18 | 0.05 | 0.14 | 0.49 | 0.15 | 0.03 | -0.07 |
| hd148897 | E | 4167 | -0.60 | 1.68 | -1.36 | -0.73 | -1.05 | -0.75 | -0.43 | -1.10 | -1.57 | -1.31 | -1.60 | -1.16 | -1.25 | -1.33 | -1.23 | -1.69 | -1.07 | -1.05 | -1.50 | -1.48 | -1.51 | -1.47 | -1.26 | -1.39 | -1.43 | -1.39 | |
| hd149161 | E | 3958 | -0.18 | 1.87 | 0.18 | -0.25 | 0.03 | -0.02 | 1.39 | -0.31 | -0.60 | -0.34 | -0.28 | -0.07 | -0.62 | -0.42 | -0.37 | -0.30 | 0.10 | -0.53 | 0.29 | -0.46 | -0.53 | -0.92 | -0.47 | -0.16 | -0.10 | 0.05 | -0.58 |
| hd150557 | E | 6744 | 4.10 | 3.83 | | 0.46 | | 0.35 | 0.80 | 0.40 | 0.61 | 0.81 | 1.05 | 0.66 | 0.63 | 0.10 | 1.07 | 0.61 | | -0.35 | | 1.57 | 0.40 | | 0.93 | | 0.61 | 1.43 | | |
| hd153210 | E | 4559 | 1.87 | 1.60 | 0.40 | 0.27 | 0.36 | 0.40 | 1.18 | 0.12 | 0.07 | 0.14 | 0.29 | 0.27 | 0.24 | 0.16 | 0.15 | 0.26 | 0.52 | 0.33 | -0.03 | -0.01 | -0.14 | -0.21 | 0.32 | 0.66 | 0.44 | 0.41 | 0.20 |
| hd153956 | E | 4541 | 2.03 | 1.73 | 0.76 | 0.38 | 0.37 | 0.52 | 1.31 | 0.19 | 0.13 | 0.20 | 0.33 | 0.35 | 0.24 | 0.20 | 0.26 | 0.33 | 0.60 | 0.25 | 0.09 | 0.10 | -0.05 | -0.11 | 0.39 | 0.67 | 0.36 | 0.55 | 0.26 |
| hd159353 | E | 4832 | 2.32 | 1.40 | 0.21 | 0.18 | 0.15 | 0.20 | 0.66 | 0.13 | -0.11 | 0.00 | 0.07 | 0.10 | 0.07 | 0.02 | -0.06 | 0.04 | -0.05 | -0.02 | 0.08 | 0.00 | -0.05 | 0.06 | 0.25 | 0.40 | 0.30 | 0.06 | -0.01 |
| hd159876 | E | 7217 | 3.47 | 3.59 | 0.05 | -0.06 | 0.28 | 0.24 | 0.18 | 0.14 | 0.14 | 0.37 | 0.42 | 0.21 | -0.12 | 0.01 | 0.97 | 0.37 | 0.32 | 0.18 | | 0.94 | 1.01 | 0.52 | 0.63 | 0.73 | 0.75 | 0.58 | 0.78 | |
| hd160507 | E | 4921 | 2.34 | 1.46 | 0.32 | 0.15 | 0.17 | 0.21 | 0.59 | 0.12 | -0.11 | -0.03 | 0.01 | 0.14 | 0.04 | 0.01 | -0.06 | 0.07 | 0.05 | 0.55 | 0.60 | 0.38 | 0.41 | 0.50 | 0.61 | 1.10 | 0.54 | 0.35 | 0.12 |
| hd161074 | E | 4033 | 0.44 | 1.81 | 0.52 | 0.13 | 0.37 | 0.27 | 1.66 | 0.03 | -0.28 | -0.04 | 0.04 | 0.15 | -0.24 | -0.17 | -0.07 | -0.04 | 0.51 | 0.58 | 0.37 | -0.11 | -0.24 | -0.73 | -0.15 | 0.24 | 0.09 | 0.18 | -0.39 |
| hd162757 | E | 4621 | 2.01 | 1.42 | 0.16 | 0.08 | 0.16 | 0.13 | 0.62 | -0.01 | -0.15 | -0.11 | 0.01 | 0.03 | -0.12 | -0.10 | -0.16 | -0.06 | -0.04 | 0.04 | 0.05 | -0.14 | -0.16 | -0.20 | 0.16 | 0.20 | 0.16 | 0.11 | -0.19 |
| hd163588 | E | 4459 | 1.87 | 1.50 | 0.42 | 0.27 | 0.41 | 0.36 | 1.01 | 0.17 | -0.02 | 0.09 | 0.23 | 0.19 | 0.09 | 0.03 | 0.06 | 0.14 | 0.31 | 0.16 | 0.02 | -0.13 | -0.12 | -0.26 | 0.28 | 0.53 | 0.30 | 0.39 | 0.04 |
| hd164058 | E | 3925 | -0.26 | 2.01 | 0.77 | -0.06 | 0.20 | 0.21 | 1.81 | -0.10 | -0.53 | -0.30 | -0.22 | 0.09 | -0.18 | -0.22 | -0.25 | -0.07 | 0.58 | 0.46 | 0.63 | -0.36 | -0.42 | -0.67 | -0.27 | 0.05 | -0.25 | -0.12 | -0.63 |
| hd166208 | E | 5040 | 2.33 | 1.60 | 0.71 | 0.35 | 0.33 | 0.28 | 0.59 | 0.22 | 0.08 | 0.17 | 0.25 | 0.30 | 0.10 | 0.16 | 0.16 | 0.17 | 0.04 | 0.01 | 0.29 | 0.10 | 0.10 | 0.16 | 0.31 | 0.53 | 0.28 | 0.19 | 0.25 |
| hd168723 | E | 4858 | 2.47 | 1.23 | -0.02 | -0.03 | 0.01 | -0.02 | 0.18 | -0.07 | -0.27 | -0.17 | -0.13 | -0.14 | -0.19 | -0.20 | -0.21 | -0.18 | -0.21 | -0.20 | -0.14 | -0.26 | -0.30 | -0.23 | -0.03 | 0.02 | 0.07 | -0.11 | 0.19 |
| hd169268a | E | 6812 | 3.86 | 1.48 | -0.12 | -0.67 | -0.37 | -0.11 | -0.24 | -0.57 | 0.05 | 0.13 | 0.55 | 0.04 | -0.11 | -0.53 | 0.60 | -0.15 | -0.39 | -0.47 | 1.19 | 0.03 | 0.98 | -0.28 | 0.57 | 0.78 | 0.59 | 0.69 | 1.28 |
| hd172748 | E | 6832 | 3.25 | 4.08 | 0.35 | 0.04 | 0.49 | 0.23 | 0.36 | 0.24 | 0.34 | 0.29 | 0.54 | 0.12 | 0.13 | 0.04 | 0.59 | 0.44 | 0.33 | 0.26 | | 0.73 | 0.71 | 0.35 | | 0.74 | 0.53 | 0.67 | 0.55 | |
| hd175305 | E | 4981 | 2.33 | 1.40 | -1.48 | -1.07 | | -1.08 | -0.57 | -1.16 | -1.40 | -1.25 | -1.35 | -1.41 | -1.82 | -1.41 | -1.25 | -1.47 | -1.94 | -1.58 | -1.05 | -1.51 | -1.04 | -1.40 | -1.23 | -1.20 | -1.12 | -1.13 | -0.24 | |
| hd176408 | E | 4517 | 2.05 | 1.46 | 0.55 | 0.31 | 0.51 | 0.39 | 1.10 | 0.15 | 0.05 | 0.13 | 0.33 | 0.24 | 0.18 | 0.10 | 0.16 | 0.21 | 0.35 | 0.49 | 0.00 | -0.06 | -0.09 | -0.20 | 0.31 | 0.56 | 0.45 | 0.41 | 0.10 |
| hd178596 | E | 6784 | 4.01 | 3.35 | -0.05 | | 0.14 | | 0.26 | 0.18 | 0.47 | 0.58 | 0.70 | 0.35 | -0.16 | 0.55 | 0.41 | | | 1.69 | 0.37 | 1.38 | -0.21 | 0.75 | 1.24 | 0.89 | | | | |
| hd181214 | E | 6238 | 3.38 | 2.66 | 0.30 | 0.23 | 0.00 | 0.22 | 0.15 | 0.15 | 0.07 | 0.22 | 0.11 | 0.15 | -0.04 | 0.02 | 0.42 | 0.07 | -0.25 | -0.37 | | 0.09 | 0.70 | -0.23 | 0.33 | -0.11 | 0.04 | 0.60 | 0.76 | |
| hd181984 | E | 4413 | 1.69 | 1.88 | 0.76 | 0.49 | 0.54 | 0.65 | 1.57 | 0.27 | 0.16 | 0.28 | 0.43 | 0.46 | 0.37 | 0.24 | 0.35 | 0.37 | 0.51 | 0.50 | 0.25 | 0.11 | 0.04 | -0.42 | 0.35 | 0.49 | 0.69 | 0.50 | 0.27 |
| hd184406 | E | 4487 | 1.96 | 1.51 | 0.52 | 0.32 | 0.58 | 0.38 | 1.04 | 0.23 | 0.05 | 0.20 | 0.33 | 0.29 | 0.14 | 0.07 | 0.14 | 0.20 | 0.36 | 0.76 | 0.19 | 0.00 | -0.02 | -0.41 | 0.44 | 0.61 | 0.48 | 0.58 | -0.05 |
| hd185351 | E | 4941 | 2.72 | 1.31 | 0.31 | 0.17 | 0.25 | 0.16 | 0.58 | 0.12 | -0.14 | -0.03 | 0.07 | 0.12 | 0.04 | 0.00 | -0.01 | 0.02 | -0.08 | 0.22 | 0.03 | -0.04 | -0.15 | -0.12 | 0.10 | 0.15 | 0.08 | -0.01 | 0.35 |
| hd185644 | E | 4579 | 2.18 | 1.38 | 0.36 | 0.24 | 0.40 | 0.33 | 0.90 | 0.10 | 0.04 | 0.13 | 0.23 | 0.25 | 0.20 | 0.10 | 0.05 | 0.17 | 0.36 | 0.19 | 0.04 | 0.11 | -0.05 | -0.02 | 0.36 | 0.65 | 0.40 | 0.40 | 0.10 |
| hd187764 | E | 6919 | 1.20 | 5.75 | -0.25 | | 0.02 | -0.60 | -0.19 | 0.23 | -0.04 | 0.36 | 0.17 | 0.50 | -0.74 | 0.66 | 0.32 | -0.12 | -0.73 | 1.12 | | 1.22 | | | -0.94 | 0.06 | -0.28 | -0.78 | | |

| ID | Type | T | c1 | c2 | c3 | c4 | c5 | c6 | c7 | c8 | c9 | c10 | c11 | c12 | c13 | c14 | c15 | c16 | c17 | c18 | c19 | c20 | c21 | c22 | c23 | c24 | c25 | c26 | c27 | c28 |
|---|---|---|---|---|---|---|---|---|---|---|---|---|---|---|---|---|---|---|---|---|---|---|---|---|---|---|---|---|---|---|
| hd188119 | E | 4945 | 2.33 | 1.45 | -0.14 | -0.15 | -0.12 | -0.14 | 0.00 | -0.22 | -0.35 | -0.30 | -0.30 | -0.30 | -0.43 | -0.36 | -0.33 | -0.35 | -0.40 | -0.28 | -0.18 | -0.37 | -0.37 | -0.29 | -0.20 | -0.01 | -0.09 | -0.15 | -0.20 |
| hd188947 | E | 4783 | 2.35 | 1.47 | 0.39 | 0.25 | 0.24 | 0.28 | 0.69 | 0.16 | 0.02 | 0.09 | 0.20 | 0.20 | 0.15 | 0.09 | 0.08 | 0.13 | 0.11 | 0.12 | 0.17 | 0.11 | 0.03 | 0.04 | 0.39 | 0.64 | 0.40 | 0.26 | 0.18 |
| hd189319 | E | 3862 | -0.30 | 2.25 | 0.75 | -0.13 | 0.04 | 0.44 | 2.10 | 0.11 | -0.57 | -0.35 | -0.17 | 0.09 | -0.12 | -0.21 | -0.20 | -0.08 | 0.88 | -0.47 | 0.49 | -0.36 | -0.61 | -0.80 | -0.38 | 0.08 | -0.45 | -0.26 | -0.56 |
| hd192787 | E | 4888 | 2.17 | 1.40 | 0.14 | 0.03 | 0.09 | 0.06 | 0.22 | 0.02 | -0.23 | -0.16 | -0.14 | -0.02 | -0.13 | -0.13 | -0.18 | -0.13 | -0.30 | -0.03 | -0.02 | -0.15 | -0.16 | -0.06 | 0.02 | 0.31 | 0.12 | -0.09 | -0.09 |
| hd192836 | E | 4805 | 2.51 | 1.33 | 0.43 | 0.35 | 0.40 | 0.33 | 0.70 | 0.22 | 0.12 | 0.20 | 0.36 | 0.29 | 0.28 | 0.20 | 0.12 | 0.24 | 0.34 | 0.37 | 0.26 | 0.25 | 0.15 | 0.19 | 0.55 | 0.63 | 0.59 | 0.46 | 0.18 |
| hd196134 | E | 4770 | 2.27 | 1.30 | 0.12 | 0.04 | 0.11 | 0.05 | 0.28 | -0.03 | -0.20 | -0.13 | -0.02 | -0.04 | -0.13 | -0.14 | -0.16 | -0.12 | -0.13 | -0.14 | -0.11 | -0.21 | -0.32 | -0.21 | 0.03 | 0.22 | 0.13 | -0.03 | -0.10 |
| hd197989 | E | 4713 | 2.02 | 1.47 | 0.08 | 0.06 | 0.16 | 0.11 | 0.45 | -0.09 | -0.22 | -0.13 | -0.05 | -0.04 | -0.10 | -0.13 | -0.14 | -0.08 | -0.06 | -0.01 | -0.05 | -0.17 | -0.31 | -0.21 | 0.03 | 0.32 | 0.09 | 0.01 | -0.11 |
| hd198431 | E | 4652 | 2.15 | 1.38 | 0.04 | 0.03 | 0.11 | 0.05 | 0.42 | -0.08 | -0.12 | -0.09 | 0.02 | -0.04 | -0.11 | -0.15 | -0.16 | -0.11 | -0.01 | -0.01 | -0.20 | -0.17 | -0.29 | -0.34 | 0.05 | 0.27 | 0.21 | 0.19 | -0.19 |
| hd199178 | E | 5180 | 3.53 | 3.61 | 0.59 |  | 0.40 | 0.51 | 0.74 | 0.93 | 0.47 | 0.48 | 0.30 | 1.04 | 0.42 | 0.35 | 0.52 | 0.50 |  |  | 2.15 | 0.70 | 1.19 |  | 1.90 | 1.21 | 1.41 |  | 0.45 |
| hd202109 | E | 4892 | 2.08 | 1.67 | 0.43 | 0.20 | 0.20 | 0.25 | 0.70 | 0.13 | -0.13 | -0.01 | -0.03 | 0.14 | 0.05 | 0.02 | -0.06 | 0.06 | 0.04 | 0.39 | 0.48 | 0.31 | 0.27 | 0.29 | 0.42 | 0.80 | 0.41 | 0.29 | 0.04 |
| hd209747 | E | 4068 | 0.37 | 2.19 | 0.46 | 0.18 | 0.28 | 0.34 | 1.22 | 0.11 | -0.32 | -0.17 | -0.05 | 0.19 | -0.09 | -0.13 | 0.01 | 0.06 | 0.40 | 0.21 | 0.45 | -0.24 | -0.37 | -1.03 | -0.29 | 0.13 | -0.17 | 0.14 | -0.36 |
| hd209960 | E | 4112 | 0.69 | 1.92 | 0.76 | 0.26 | 0.41 | 0.38 | 1.54 | 0.17 | -0.19 | 0.01 | 0.16 | 0.34 | 0.10 | -0.04 | 0.10 | 0.14 | 0.62 | 0.31 | 0.40 | 0.11 | -0.22 | -0.86 | -0.11 | 0.42 | 0.23 | 0.41 | -0.31 |
| hd212943 | E | 4630 | 2.12 | 1.31 | -0.05 | 0.03 | 0.07 | -0.01 | 0.32 | -0.09 | -0.20 | -0.14 | -0.06 | -0.09 | -0.19 | -0.20 | -0.20 | -0.16 | 0.02 | -0.26 | -0.20 | -0.23 | -0.30 | -0.36 | -0.05 | 0.17 | 0.12 | 0.03 | -0.16 |
| hd214448 | E | 5272 | 3.33 | 0.47 | 0.55 | 0.28 | 0.41 | 0.27 | 0.54 | 0.37 | 0.39 | 0.48 | 0.63 | 0.49 | 0.33 | 0.40 | 0.30 | 0.45 | 0.38 | 0.48 | 0.48 | 0.74 | 0.54 | 0.56 | 0.89 | 1.27 | 1.11 | 1.00 | 1.01 |
| hd214470 | E | 6668 | 4.40 | 2.69 |  |  | 0.95 |  | 0.67 |  | 1.55 | 1.80 | 1.35 | 2.51 | 0.50 | 0.93 | 0.90 |  |  |  | 2.77 | 2.29 |  |  | 1.27 |  |  |  | 2.51 |
| hd216131 | E | 4934 | 2.33 | 1.39 | 0.22 | 0.11 | 0.15 | 0.11 | 0.41 | 0.12 | -0.16 | -0.08 | -0.04 | 0.05 | -0.05 | -0.04 | -0.12 | -0.07 | -0.18 | -0.14 | 0.09 | -0.03 | -0.05 | 0.04 | 0.15 | 0.44 | 0.20 | 0.03 | 0.03 |
| hd216228 | E | 4742 | 2.20 | 1.51 | 0.28 | 0.20 | 0.26 | 0.22 | 0.57 | 0.09 | -0.06 | 0.03 | 0.12 | 0.14 | 0.09 | 0.02 | -0.01 | 0.07 | 0.10 | -0.02 | 0.02 | -0.02 | -0.10 | -0.02 | 0.29 | 0.53 | 0.29 | 0.16 | 0.00 |
| hd219418 | E | 5077 | 2.36 | 1.51 | 0.09 | -0.14 | -0.05 | -0.13 | -0.09 | -0.18 | -0.38 | -0.34 | -0.40 | -0.30 | -0.43 | -0.35 | -0.35 | -0.34 | -0.45 | -0.34 | -0.15 | -0.29 | -0.22 | -0.34 | -0.30 | -0.16 | -0.26 | -0.42 | 0.20 |
| hd219916 | E | 5124 | 2.93 | 1.27 | 0.28 | 0.16 | 0.25 | 0.15 | 0.42 | 0.19 | 0.04 | 0.13 | 0.18 | 0.19 | 0.09 | 0.10 | 0.06 | 0.10 | 0.05 | 0.11 | 0.24 | 0.24 | 0.21 | 0.21 | 0.40 | 0.59 | 0.49 | 0.22 | 0.57 |
| hd220954 | E | 4684 | 2.06 | 1.49 | 0.42 | 0.41 | 0.32 | 0.34 | 0.80 | 0.16 | 0.03 | 0.11 | 0.23 | 0.29 | 0.20 | 0.14 | 0.09 | 0.21 | 0.33 | 0.32 | 0.22 | 0.05 | 0.07 | 0.03 | 0.37 | 0.63 | 0.35 | 0.37 | 0.04 |
| hd221148 | E | 4660 | 2.72 | 1.51 | 0.98 | 0.57 | 0.66 | 0.63 | 1.41 | 0.42 | 0.38 | 0.38 | 0.62 | 0.55 | 0.46 | 0.34 | 0.51 | 0.54 | 0.45 | 1.05 | 0.26 | 0.32 | 0.18 | -0.22 | 0.69 | 0.81 | 0.59 | 0.94 | 0.54 |
| aiscl | H | 7128 | 4.62 | 6.14 |  | 0.67 |  | 0.26 |  | 0.04 | 0.36 | 0.56 | 1.27 | 0.37 | 1.50 | -0.67 | 1.44 | 0.14 |  |  | 1.44 | 2.70 |  | 1.07 | 1.29 | 1.06 |  |  |  |
| anscl | H | 4968 | 3.32 | 2.44 | 0.42 | -0.10 |  | 0.21 | 0.70 | 0.20 | 0.13 | 0.16 | 0.15 | 0.20 | 0.08 | -0.02 | 0.09 | 0.10 |  | -0.14 |  | -0.05 | -0.23 | -0.11 | 0.03 | 0.15 | 0.51 | 1.20 | -0.10 |
| bppsc | H | 4050 | 0.20 | 3.70 |  |  | 0.18 |  |  | -2.32 | -1.59 | -1.64 | -0.63 | -1.32 | -0.94 | -0.78 | -0.54 |  | -0.41 |  | -1.33 | -2.41 |  | -1.89 |  | -1.27 | -0.80 |  |  |
| cd-33_2771 | H | 4002 | 0.20 | 1.91 | 0.16 | -0.36 | 0.08 | -0.11 |  | -0.42 | -0.61 | -0.51 | -0.36 | 0.00 | -0.58 | -0.53 | -0.36 | -0.34 |  | -0.22 | 0.41 | -0.68 | -0.37 | -1.23 | -0.67 | -0.12 | -0.27 | -0.50 | -0.76 |
| cd-3814203b | H | 6297 | 4.31 | 1.12 | 0.52 | 0.44 | 0.42 | 0.31 | 0.30 | 0.42 | 0.50 | 0.50 | 0.49 | 0.49 | 0.49 | 0.42 | 0.47 | 0.46 | 0.39 | 0.39 | 0.75 | 0.63 | 0.66 | 0.57 | 0.63 | 0.59 | 0.57 | 0.59 | 1.31 |
| cfscl | H | 5127 | 4.00 | 0.92 | 0.34 | 0.31 | 0.49 | 0.08 | 0.46 | 0.23 | 0.56 | 0.56 | 0.77 | 0.40 | 0.03 | 0.13 | 0.12 | 0.18 | 0.46 | 0.35 |  | 0.82 | 0.75 | 0.33 | 1.00 | 1.17 | 1.23 | 1.81 | 1.60 |
| epsret | H | 4702 | 2.86 | 0.94 | 0.60 | 0.51 | 0.52 | 0.45 | 1.00 | 0.41 | 0.28 | 0.35 | 0.76 | 0.44 | 0.43 | 0.24 | 0.25 | 0.36 | 0.59 | 0.29 | 0.34 | 0.16 | 0.16 | -0.04 | 0.29 | 0.35 | 0.43 | 0.35 | 0.20 |
| ereri | H | 4935 | 0.10 | 3.96 |  |  | -0.06 |  | 0.02 | -2.06 | -1.12 |  | 0.14 |  | -0.71 | -0.30 | 0.00 |  | -0.68 |  | -1.02 | -0.18 |  | -1.45 | -1.15 | -1.01 | -0.99 |  |  |
| gamaps | H | 4957 | 2.52 | 1.35 | 0.39 | 0.08 | 0.19 | 0.12 | 0.46 | 0.09 | -0.17 | -0.06 | 0.02 | 0.07 | -0.01 | -0.03 | -0.08 | -0.05 | -0.16 | -0.18 | 0.12 | -0.05 | -0.08 | 0.09 | 0.07 | 0.18 | 0.04 | -0.06 | 0.36 |
| hd000344 | H | 4584 | 2.00 | 1.43 | 0.26 | 0.16 | 0.25 | 0.24 | 0.53 | 0.05 | -0.14 | 0.00 | 0.17 | 0.10 | 0.12 | 0.00 | -0.04 | 0.06 | 0.16 | -0.01 | 0.07 | -0.04 | -0.19 | -0.06 | -0.06 | 0.18 | 0.07 | 0.00 | -0.03 |
| hd000483a | H | 5718 | 3.90 | 0.50 | 0.06 | -0.50 | -0.23 | -0.24 | 0.11 | -0.42 | -0.09 | -0.24 | -0.06 | -0.04 | -0.32 | -0.40 | -0.10 | -0.20 | -0.40 | -0.31 | 0.59 | -0.03 | 0.24 | -0.86 | 0.50 | 0.66 | 1.07 | 0.83 | 0.91 |
| hd000483b | H | 5757 | 3.95 | 0.50 | 0.06 | -0.05 | -0.04 | -0.12 | 0.09 | -0.30 | 0.04 | -0.20 | 0.26 | 0.14 | -0.30 | -0.24 | 0.04 | -0.14 | -0.42 | -0.34 | -0.53 | 0.14 | 0.55 | -0.59 | 0.29 | 0.59 | 0.71 | 1.28 | 0.36 |
| hd000770 | H | 4710 | 2.05 | 1.46 | 0.04 | 0.02 | 0.12 | 0.09 | 0.29 | -0.06 | -0.24 | -0.14 | -0.06 | -0.07 | -0.05 | -0.14 | -0.14 | -0.11 | -0.04 | -0.06 | -0.12 | -0.28 | -0.33 | -0.22 | -0.12 | 0.01 | -0.10 | -0.13 | -0.09 |
| hd001690 | H | 4219 | 1.10 | 1.61 | -0.06 | -0.05 | 0.03 | 0.00 | 0.40 | -0.20 | -0.28 | -0.19 | -0.06 | -0.13 | -0.28 | -0.27 | -0.24 | -0.21 | -0.27 | -0.18 | -0.12 | -0.39 | -0.53 | -0.41 | -0.22 | -0.15 | -0.10 | -0.17 | -0.42 |
| hd002114 | H | 5148 | 2.28 | 1.65 | 0.31 | 0.08 | 0.02 | 0.11 | 0.28 | 0.07 | -0.12 | -0.08 | -0.12 | 0.01 | -0.07 | -0.04 | -0.12 | -0.10 | -0.24 | -0.18 | 0.28 | 0.08 | -0.04 | 0.23 | 0.06 | 0.30 | 0.07 | -0.02 | 0.03 |
| hd002529 | H | 4621 | 2.19 | 1.27 | 0.19 | 0.09 | 0.19 | 0.13 | 0.39 | 0.09 | -0.13 | -0.02 | 0.14 | 0.07 | 0.08 | -0.03 | -0.09 | -0.01 | 0.09 | -0.05 | 0.09 | -0.02 | -0.11 | -0.01 | 0.06 | 0.30 | 0.19 | 0.06 | -0.08 |
| hd003488 | H | 4827 | 2.36 | 1.36 | 0.09 | 0.08 | 0.16 | 0.07 | 0.37 | 0.07 | -0.12 | -0.02 | 0.02 | 0.01 | -0.05 | -0.07 | -0.09 | -0.08 | -0.05 | -0.11 | 0.13 | 0.03 | -0.08 | 0.12 | 0.23 | 0.28 | 0.25 | 0.12 | 0.00 |
| hd003919 | H | 4952 | 2.71 | 1.24 | 0.37 | 0.17 | 0.18 | 0.22 | 0.40 | 0.20 | 0.08 | 0.17 | 0.30 | 0.20 | 0.19 | 0.14 | 0.11 | 0.15 | 0.14 | -0.04 | 0.31 | 0.26 | 0.21 | 0.36 | 0.42 | 0.57 | 0.43 | 0.34 | 0.31 |
| hd004211 | H | 4531 | 1.99 | 1.45 | 0.29 | 0.34 | 0.42 | 0.26 | 0.66 | 0.17 | -0.03 | 0.18 | 0.37 | 0.16 | 0.13 | 0.04 | 0.08 | 0.13 | 0.70 | 0.32 | 0.09 | -0.04 | -0.25 | -0.13 | 0.15 | 0.20 | 0.19 | 0.21 | 0.01 |
| hd004737 | H | 5031 | 2.60 | 1.37 | 0.26 | 0.08 | 0.12 | 0.05 | 0.21 | 0.07 | -0.16 | -0.07 | -0.03 | 0.04 | -0.03 | -0.06 | -0.14 | -0.10 | -0.20 | -0.16 | 0.20 | -0.07 | -0.06 | 0.15 | 0.03 | 0.22 | 0.05 | 0.01 | 0.08 |
| hd006192 | H | 4992 | 2.57 | 1.34 | 0.28 | 0.13 | 0.06 | 0.12 | 0.24 | 0.12 | -0.10 | 0.00 | 0.05 | 0.09 | 0.05 | 0.02 | -0.07 | -0.01 | -0.06 | -0.07 | 0.38 | 0.06 | 0.06 | 0.19 | 0.22 | 0.35 | 0.22 | 0.10 | 0.12 |
| hd006245 | H | 5001 | 2.55 | 1.32 | 0.23 | 0.04 | 0.13 | 0.07 | 0.26 | 0.07 | -0.16 | -0.07 | -0.04 | 0.04 | -0.06 | -0.04 | -0.14 | -0.09 | -0.20 | -0.21 | 0.17 | -0.03 | 000 | 0.15 | 0.19 | 0.23 | 0.09 | -0.02 | 0.36 |
| hd006559 | H | 4681 | 2.25 | 1.41 | 0.37 | 0.25 | 0.34 | 0.27 | 0.52 | 0.16 | -0.05 | 0.09 | 0.28 | 0.16 | 0.20 | 0.05 | 0.05 | 0.11 | 0.30 | 0.10 | 0.22 | 0.06 | -0.07 | 0.00 | 0.05 | 0.16 | 0.08 | 0.10 | 0.00 |
| hd006793 | H | 5039 | 2.50 | 1.64 | 0.35 | 0.04 | 0.08 | 0.10 | 0.27 | 0.07 | -0.18 | -0.09 | -0.04 | 0.07 | -0.13 | -0.05 | -0.13 | -0.07 | -0.21 | -0.26 | 0.10 | 0.04 | -0.08 | 0.07 | 0.02 | 0.28 | 0.09 | -0.03 | -0.03 |
| hd007082 | H | 4969 | 2.30 | 1.49 | -0.57 | -0.37 | -0.32 | -0.41 | -0.33 | -0.51 | -0.68 | -0.53 | -0.64 | -0.73 | -0.95 | -0.77 | -0.64 | -0.74 | -0.82 | -0.63 | -0.09 | -0.44 | -0.35 | -0.22 | -0.22 | -0.11 | -0.23 | -0.43 | -0.48 |
| hd007672 | H | 4937 | 2.46 | 1.61 | -0.16 | -0.27 | -0.11 | -0.26 | 0.00 | -0.19 | -0.45 | -0.34 | -0.30 | -0.26 | -0.44 | -0.43 | -0.43 | -0.47 | -0.56 | -0.52 | -0.21 | -0.47 | -0.39 | -0.26 | -0.42 | -0.31 | -0.35 | -0.49 | 0.00 |
| hd009362 | H | 4762 | 2.04 | 1.43 | -0.15 | -0.11 | -0.05 | -0.12 | -0.21 | -0.21 | -0.42 | -0.32 | -0.31 | -0.29 | -0.35 | -0.36 | -0.37 | -0.34 | -0.28 | -0.36 | -0.17 | -0.46 | -0.52 | -0.41 | -0.35 | -0.26 | -0.33 | -0.44 | -0.31 |
| hd009611 | H | 4627 | 1.67 | 2.91 | 0.57 | -0.09 | 0.90 | 0.35 | 1.04 | 0.45 | 0.00 | 0.31 | 0.36 | 0.47 | 0.32 | -0.02 | 0.09 | 0.11 |  | -0.82 |  | 0.36 | 0.37 | -0.68 | 0.03 | 0.28 | 0.30 | 0.88 |  |
| hd009742 | H | 4629 | 2.23 | 1.32 | 0.20 | 0.14 | 0.19 | 0.10 | 0.56 | 0.07 | -0.06 | 0.03 | 0.20 | 0.08 | 0.10 | -0.03 | -0.05 | 0.02 | 0.17 | -0.14 | 0.12 | -0.02 | -0.07 | 0.05 | 0.30 | 0.39 | 0.30 | 0.10 | -0.02 |

| ID | Type | V1 | V2 | V3 | V4 | V5 | V6 | V7 | V8 | V9 | V10 | V11 | V12 | V13 | V14 | V15 | V16 | V17 | V18 | V19 | V20 | V21 | V22 | V23 | V24 | V25 | V26 | V27 | V28 |
|---|---|---|---|---|---|---|---|---|---|---|---|---|---|---|---|---|---|---|---|---|---|---|---|---|---|---|---|---|---|
| hd010042 | H | 4799 | 2.12 | 1.58 | -0.14 | 0.03 | 0.04 | -0.03 | 0.00 | -0.17 | -0.30 | -0.15 | -0.20 | -0.27 | -0.44 | -0.35 | -0.26 | -0.30 | -0.26 | -0.02 | -0.26 | -0.48 | -0.43 | -0.53 | -0.40 | -0.18 | -0.21 | -0.18 | -0.18 |
| hd010615 | H | 4420 | 1.85 | 1.43 | 0.64 | 0.35 | 0.50 | 0.48 | 0.63 | 0.37 | 0.16 | 0.26 | 0.64 | 0.39 | 0.36 | 0.23 | 0.24 | 0.31 | 0.19 | 0.21 | 0.35 | 0.16 | 0.06 | 0.04 | 0.34 | 0.44 | 0.33 | 0.29 | 0.03 |
| hd011025 | H | 4961 | 2.45 | 1.66 | 0.35 | 0.17 | 0.26 | 0.18 | 0.46 | 0.06 | -0.09 | 0.05 | 0.09 | 0.14 | -0.03 | 0.01 | -0.04 | 0.02 | -0.16 | -0.05 | 0.11 | 0.08 | -0.02 | 0.05 | 0.18 | 0.24 | 0.17 | 0.16 | 0.15 |
| hd011977 | H | 4878 | 2.32 | 1.35 | -0.05 | -0.11 | -0.04 | -0.11 | 0.08 | -0.10 | -0.35 | -0.24 | -0.26 | -0.19 | -0.26 | -0.26 | -0.32 | -0.29 | -0.33 | -0.41 | -0.02 | -0.26 | -0.25 | -0.09 | -0.03 | 0.08 | 0.02 | -0.14 | -0.11 |
| hd012055 | H | 5035 | 2.45 | 1.47 | 0.21 | 0.06 | 0.14 | 0.08 | 0.36 | 0.03 | -0.18 | -0.10 | -0.10 | 0.01 | -0.14 | -0.09 | -0.14 | -0.10 | -0.37 | -0.06 | 0.07 | -0.11 | -0.18 | 0.10 | 0.05 | 0.25 | 0.11 | -0.11 | 0.36 |
| hd012270 | H | 4992 | 2.38 | 1.39 | 0.30 | 0.08 | 0.14 | 0.08 | 0.17 | 0.06 | -0.15 | -0.10 | -0.09 | 0.01 | -0.04 | -0.05 | -0.13 | -0.08 | -0.20 | -0.18 | 0.15 | 0.07 | 0.02 | 0.13 | 0.09 | 0.31 | 0.10 | -0.02 | 0.08 |
| hd012345 | H | 5326 | 4.19 | 0.30 | -0.14 | -0.13 | -0.01 | -0.14 | -0.08 | -0.11 | -0.17 | -0.12 | -0.13 | -0.12 | -0.18 | -0.16 | -0.18 | -0.16 | -0.06 | -0.18 | -0.03 | -0.15 | -0.14 | -0.22 | -0.03 | 0.04 | 0.07 | -0.01 | 0.40 |
| hd012438 | H | 4884 | 1.89 | 1.46 | -0.55 | -0.40 | -0.36 | -0.42 | -0.46 | -0.51 | -0.68 | -0.63 | -0.71 | -0.63 | -0.81 | -0.67 | -0.65 | -0.68 | -0.77 | -0.73 | -0.56 | -0.74 | -0.77 | -0.70 | -0.77 | -0.67 | -0.65 | -0.71 | -0.31 |
| hd012524 | H | 3960 | 0.29 | 1.65 | 0.18 | -0.09 | 0.24 | 0.18 | 0.80 | 0.06 | -0.32 | -0.10 | 0.05 | -0.06 | -0.36 | -0.29 | -0.18 | -0.19 | -0.05 | 0.33 | 0.46 | -0.33 | -0.45 | -0.45 | -0.36 | -0.29 | -0.07 | -0.07 | -0.77 |
| hd013263 | H | 5038 | 2.54 | 1.34 | 0.16 | 0.02 | 0.05 | 0.03 | -0.03 | 0.03 | -0.16 | -0.06 | -0.08 | 0.01 | -0.17 | -0.14 | -0.10 | -0.18 | -0.21 | 0.31 | 0.01 | -0.03 | 0.11 | 0.12 | 0.35 | 0.27 | 0.11 | 0.47 | |
| hd013423 | H | 5037 | 2.62 | 1.33 | 0.27 | 0.05 | 0.11 | 0.06 | 0.18 | 0.04 | -0.14 | -0.05 | -0.03 | 0.01 | -0.05 | -0.07 | -0.11 | -0.08 | -0.14 | -0.15 | 0.14 | -0.05 | -0.09 | 0.04 | 0.03 | 0.26 | 0.13 | -0.05 | 0.07 |
| hd013692 | H | 4726 | 1.95 | 1.44 | -0.01 | -0.01 | 0.03 | 0.01 | 0.22 | -0.10 | -0.32 | -0.24 | -0.20 | -0.16 | -0.19 | -0.21 | -0.24 | -0.21 | -0.24 | -0.32 | -0.04 | -0.30 | -0.31 | -0.10 | -0.11 | 0.09 | -0.08 | -0.17 | -0.17 |
| hd013940 | H | 4883 | 2.42 | 1.33 | 0.17 | 0.06 | 0.11 | 0.09 | 0.19 | 0.03 | -0.15 | -0.08 | -0.03 | 0.01 | 0.00 | -0.06 | -0.14 | -0.07 | -0.10 | -0.14 | 0.15 | 0.01 | -0.05 | 0.10 | 0.09 | 0.36 | 0.17 | 0.01 | 0.06 |
| hd014703 | H | 4930 | 2.66 | 1.25 | 0.33 | 0.12 | 0.25 | 0.21 | 0.35 | 0.19 | 0.00 | 0.08 | 0.14 | 0.18 | 0.18 | 0.09 | -0.03 | 0.09 | 0.11 | 0.05 | 0.27 | 0.16 | 0.07 | 0.16 | 0.26 | 0.44 | 0.26 | 0.13 | 0.11 |
| hd014832 | H | 4804 | 2.27 | 1.40 | -0.05 | -0.04 | 0.01 | -0.06 | 0.11 | -0.09 | -0.29 | -0.19 | -0.14 | -0.18 | -0.19 | -0.25 | -0.25 | -0.23 | -0.17 | -0.18 | 0.03 | -0.30 | -0.35 | -0.27 | -0.13 | -0.02 | -0.09 | -0.18 | -0.21 |
| hd016522 | H | 4833 | 2.25 | 1.42 | -0.04 | 0.03 | 0.07 | 0.01 | 0.11 | -0.08 | -0.23 | -0.16 | -0.12 | -0.14 | -0.15 | -0.18 | -0.16 | -0.15 | -0.08 | -0.18 | -0.06 | -0.24 | -0.28 | -0.20 | -0.14 | 0.06 | -0.06 | -0.07 | -0.10 |
| hd016815 | H | 4683 | 2.11 | 1.28 | -0.17 | -0.25 | -0.09 | -0.19 | 0.07 | -0.23 | -0.45 | -0.34 | -0.29 | -0.33 | -0.38 | -0.41 | -0.40 | -0.39 | -0.34 | -0.39 | -0.22 | -0.50 | -0.50 | -0.39 | -0.32 | -0.22 | -0.22 | -0.39 | -0.38 |
| hd016975 | H | 5014 | 2.55 | 1.36 | 0.33 | 0.13 | 0.19 | 0.13 | 0.17 | 0.15 | -0.08 | 0.00 | 0.03 | 0.10 | 0.06 | 0.02 | -0.07 | -0.01 | -0.04 | -0.14 | 0.20 | 0.09 | 0.03 | 0.18 | 0.25 | 0.35 | 0.19 | 0.05 | 0.11 |
| hd017374 | H | 4824 | 2.32 | 1.40 | 0.07 | 0.07 | 0.17 | 0.06 | 0.17 | 0.02 | -0.13 | -0.04 | 0.01 | 0.00 | -0.05 | -0.09 | -0.12 | -0.08 | -0.03 | -0.10 | 0.05 | -0.06 | -0.12 | 0.07 | 0.19 | 0.23 | 0.17 | 0.03 | 0.29 |
| hd017793 | H | 5042 | 2.60 | 1.37 | 0.41 | 0.16 | 0.12 | 0.15 | 0.46 | 0.13 | -0.07 | 0.03 | 0.06 | 0.13 | 0.08 | 0.03 | -0.02 | 0.03 | -0.08 | -0.03 | 0.24 | 0.11 | -0.02 | 0.13 | 0.03 | 0.30 | 0.12 | 0.09 | 0.18 |
| hd018293 | H | 4262 | 1.19 | 1.46 | 0.55 | 0.30 | 0.43 | 0.43 | 0.69 | 0.35 | 0.04 | 0.19 | 0.52 | 0.32 | 0.34 | 0.13 | 0.10 | 0.22 | 0.18 | 0.35 | 0.33 | 0.11 | 0.01 | 0.12 | 0.20 | 0.33 | 0.25 | 0.26 | -0.18 |
| hd018322 | H | 4614 | 2.19 | 1.29 | 0.15 | 0.10 | 0.17 | 0.12 | 0.40 | 0.05 | -0.15 | -0.04 | 0.12 | 0.01 | 0.03 | -0.07 | -0.10 | -0.04 | 0.03 | -0.15 | 0.07 | -0.08 | -0.16 | 0.02 | 0.13 | 0.35 | 0.21 | -0.01 | -0.07 |
| hd020037 | H | 5077 | 2.55 | 1.38 | 0.22 | 0.03 | 0.00 | 0.02 | 0.10 | 0.04 | -0.13 | -0.09 | -0.08 | 0.01 | -0.12 | -0.08 | -0.15 | -0.11 | -0.25 | -0.24 | 0.12 | 0.03 | -0.05 | 0.14 | 0.04 | 0.30 | 0.11 | 0.08 | 0.40 |
| hd020894 | H | 5060 | 2.46 | 1.49 | 0.20 | 0.02 | 0.01 | 0.07 | 0.24 | 0.04 | -0.14 | -0.09 | -0.10 | 0.01 | -0.11 | -0.06 | -0.12 | -0.10 | -0.21 | -0.18 | 0.20 | -0.01 | -0.01 | 0.15 | 0.13 | 0.35 | 0.14 | 0.03 | 0.02 |
| hd021011 | H | 4799 | 2.22 | 1.39 | 0.14 | -0.03 | 0.08 | 0.05 | 0.16 | -0.02 | -0.24 | -0.14 | -0.10 | -0.04 | -0.11 | -0.13 | -0.19 | -0.14 | -0.13 | -0.23 | 0.02 | -0.14 | -0.17 | 0.01 | -0.01 | 0.22 | 0.09 | -0.03 | -0.15 |
| hd021430 | H | 4979 | 2.45 | 1.45 | -0.12 | -0.14 | -0.10 | -0.18 | -0.05 | -0.18 | -0.33 | -0.24 | -0.24 | -0.27 | -0.34 | -0.33 | -0.30 | -0.32 | -0.31 | -0.41 | 0.04 | -0.30 | -0.26 | -0.19 | -0.17 | 0.01 | -0.05 | -0.20 | -0.15 |
| hd022231 | H | 4591 | 1.95 | 1.42 | 0.45 | 0.30 | 0.42 | 0.37 | 0.99 | 0.27 | -0.02 | 0.10 | 0.32 | 0.26 | 0.29 | 0.11 | 0.08 | 0.19 | 0.38 | 0.24 | 0.18 | 0.05 | -0.11 | -0.04 | 0.09 | 0.13 | 0.11 | 0.07 | -0.04 |
| hd022663 | H | 4708 | 2.60 | 0.92 | 0.41 | 0.20 | 0.40 | 0.30 | 0.63 | 0.30 | 0.23 | 0.35 | 0.72 | 0.36 | 0.31 | 0.24 | 0.18 | 0.32 | 0.46 | 0.21 | 0.43 | 0.33 | 0.30 | 0.45 | 0.67 | 0.76 | 0.74 | 0.64 | 0.34 |
| hd022676 | H | 4957 | 2.56 | 1.38 | 0.34 | 0.14 | 0.23 | 0.16 | 0.42 | 0.13 | -0.05 | 0.04 | 0.11 | 0.17 | 0.07 | 0.06 | -0.04 | 0.05 | -0.06 | -0.03 | 0.24 | 0.09 | 0.00 | 0.15 | 0.26 | 0.46 | 0.17 | 0.06 | 0.12 |
| hd023319 | H | 4539 | 2.16 | 1.37 | 0.66 | 0.47 | 0.63 | 0.59 | 1.04 | 0.40 | 0.22 | 0.33 | 0.72 | 0.42 | 0.44 | 0.29 | 0.36 | 0.43 | 1.12 | 0.52 | 0.32 | 0.20 | 0.04 | 0.14 | 0.32 | 0.31 | 0.33 | 0.24 | 0.20 |
| hd023719 | H | 4939 | 2.51 | 1.38 | 0.43 | 0.21 | 0.25 | 0.26 | 0.60 | 0.22 | -0.03 | 0.08 | 0.15 | 0.19 | 0.20 | 0.11 | 0.01 | 0.11 | 0.09 | 0.03 | 0.27 | 0.16 | 0.06 | 0.18 | 0.16 | 0.42 | 0.19 | 0.12 | 0.15 |
| hd023940 | H | 4776 | 2.08 | 1.50 | -0.17 | -0.07 | -0.02 | -0.10 | -0.01 | -0.21 | -0.40 | -0.25 | -0.29 | -0.28 | -0.46 | -0.37 | -0.35 | -0.36 | -0.34 | -0.13 | -0.30 | -0.53 | -0.57 | -0.48 | -0.36 | -0.16 | -0.31 | -0.37 | -0.28 |
| hd024160 | H | 4948 | 2.33 | 1.45 | 0.31 | 0.13 | 0.18 | 0.16 | 0.41 | 0.11 | -0.14 | -0.05 | -0.02 | 0.07 | 0.02 | 0.00 | -0.09 | -0.02 | -0.12 | -0.13 | 0.15 | 0.01 | -0.03 | 0.14 | 0.09 | 0.27 | 0.11 | 0.04 | 0.10 |
| hd024706 | H | 4427 | 1.90 | 1.36 | 0.50 | 0.31 | 0.43 | 0.42 | 0.83 | 0.31 | 0.11 | 0.20 | 0.58 | 0.29 | 0.29 | 0.16 | 0.20 | 0.28 | 0.15 | 0.31 | 0.29 | 0.09 | -0.02 | 0.04 | 0.29 | 0.23 | 0.27 | 0.17 | -0.03 |
| hd024744 | H | 5846 | 4.33 | 0.25 | 0.36 | 0.07 | 0.45 | 0.20 | 0.20 | 0.24 | 0.63 | 0.66 | 0.87 | 0.53 | 0.28 | 0.45 | 0.55 | 0.42 | 0.35 | 0.12 | 0.79 | 0.95 | 0.82 | 0.73 | 1.35 | 1.57 | 1.34 | 1.31 | 1.17 |
| hd026967 | H | 4614 | 2.32 | 1.22 | 0.19 | 0.14 | 0.15 | 0.16 | 0.40 | 0.08 | -0.06 | 0.03 | 0.23 | 0.07 | 0.10 | -0.01 | -0.05 | 0.04 | 0.18 | -0.08 | 0.10 | 0.01 | -0.12 | 0.02 | 0.09 | 0.34 | 0.23 | 0.11 | -0.03 |
| hd027256 | H | 5048 | 2.43 | 1.60 | 0.48 | 0.26 | 0.28 | 0.26 | 0.63 | 0.23 | 0.02 | 0.09 | 0.12 | 0.20 | 0.15 | 0.11 | 0.04 | 0.12 | 0.06 | 0.20 | 0.21 | 0.15 | 0.11 | 0.15 | 0.26 | 0.43 | 0.25 | 0.16 | 0.18 |
| hd028028 | H | 3982 | 0.23 | 1.62 | 0.09 | -0.12 | 0.03 | 0.09 | 0.92 | 0.00 | -0.41 | -0.20 | -0.08 | -0.08 | -0.29 | -0.30 | -0.29 | -0.23 | -0.08 | -0.18 | 0.23 | -0.37 | -0.41 | -0.33 | -0.31 | -0.17 | -0.10 | -0.26 | -0.68 |
| hd028093 | H | 4951 | 2.38 | 1.47 | 0.22 | 0.07 | 0.15 | 0.07 | 0.10 | 0.00 | -0.08 | -0.04 | 0.01 | -0.07 | -0.09 | -0.11 | -0.06 | -0.09 | -0.05 | -0.19 | 0.02 | -0.16 | -0.18 | -0.06 | -0.06 | 0.12 | 0.01 | -0.02 | 0.03 |
| hd029291 | H | 4897 | 2.23 | 1.52 | 0.32 | 0.12 | 0.14 | 0.13 | 0.47 | 0.10 | -0.16 | -0.07 | -0.05 | 0.05 | 0.01 | -0.02 | -0.12 | -0.03 | -0.09 | -0.06 | 0.11 | 0.10 | -0.07 | 0.09 | 0.22 | 0.23 | 0.08 | 0.03 | 0.00 |
| hd029399 | H | 4822 | 2.84 | 1.04 | 0.43 | 0.30 | 0.38 | 0.27 | 0.72 | 0.28 | 0.18 | 0.24 | 0.49 | 0.29 | 0.36 | 0.20 | 0.18 | 0.25 | 0.43 | 0.13 | 0.27 | 0.13 | 0.08 | 0.03 | 0.26 | 0.31 | 0.28 | 0.28 | 0.17 |
| hd029751 | H | 4855 | 2.06 | 1.46 | 0.14 | -0.10 | 0.01 | -0.06 | 0.11 | -0.14 | -0.36 | -0.28 | -0.30 | -0.21 | -0.26 | -0.28 | -0.32 | -0.28 | -0.30 | -0.27 | -0.10 | -0.36 | -0.38 | -0.27 | -0.28 | -0.12 | -0.23 | -0.32 | -0.23 |
| hd030185 | H | 4851 | 2.32 | 1.35 | 0.19 | 0.04 | 0.12 | 0.06 | 0.25 | 0.00 | -0.18 | -0.10 | -0.05 | -0.06 | -0.09 | -0.11 | -0.16 | -0.12 | -0.12 | -0.18 | 0.05 | -0.10 | -0.16 | 0.02 | 0.06 | 0.17 | 0.07 | -0.07 | -0.05 |
| hd032453 | H | 5032 | 2.47 | 1.38 | 0.19 | 0.03 | 0.06 | 0.01 | 0.10 | 0.02 | -0.18 | -0.11 | -0.12 | -0.04 | -0.12 | -0.11 | -0.18 | -0.14 | -0.25 | -0.20 | 0.16 | -0.07 | -0.05 | 0.10 | 0.12 | 0.23 | 0.11 | 0.02 | 0.43 |
| hd032515 | H | 4529 | 1.96 | 1.39 | 0.40 | 0.34 | 0.43 | 0.42 | 0.90 | 0.22 | 0.04 | 0.18 | 0.47 | 0.29 | 0.33 | 0.17 | 0.18 | 0.27 | 0.56 | 0.52 | 0.18 | 0.07 | -0.10 | 0.08 | 0.21 | 0.29 | 0.19 | 0.18 | 0.05 |
| hd033285 | H | 4825 | 1.70 | 1.88 | 0.30 | 0.10 | 0.15 | 0.16 | 0.50 | 0.00 | -0.23 | -0.11 | -0.13 | 0.08 | -0.21 | -0.05 | -0.18 | -0.05 | -0.28 | -0.18 | 0.20 | 0.00 | -0.11 | 0.10 | 0.22 | 0.59 | 0.26 | -0.02 | 0.02 |
| hd034172 | H | 4990 | 2.64 | 1.33 | 0.33 | 0.15 | 0.20 | 0.16 | 0.35 | 0.15 | -0.11 | 0.03 | 0.07 | 0.13 | 0.11 | 0.04 | -0.03 | 0.03 | 0.01 | -0.04 | 0.19 | 0.09 | 0.04 | 0.15 | 0.09 | 0.35 | 0.19 | 0.12 | 0.15 |
| hd034253 | H | 5490 | 4.21 | 0.54 | -0.03 | 0.00 | -0.13 | -0.07 | 0.02 | -0.02 | -0.11 | -0.04 | -0.01 | 0.01 | -0.07 | -0.06 | -0.04 | -0.06 | 0.05 | 0.15 | 0.91 | 0.80 | 0.69 | 0.55 | 0.73 | 0.79 | 0.64 | 0.42 | 0.93 |
| hd034266 | H | 4881 | 2.33 | 1.56 | 0.31 | 0.10 | 0.19 | 0.21 | 0.36 | 0.15 | -0.03 | 0.04 | 0.09 | 0.14 | 0.04 | 0.04 | -0.04 | 0.01 | -0.03 | -0.02 | 0.30 | 0.05 | 0.00 | 0.18 | 0.34 | 0.55 | 0.27 | 0.25 | 0.08 |
| hd034649 | H | 4320 | 1.24 | 1.62 | 0.18 | 0.05 | 0.10 | 0.20 | 0.56 | 0.02 | -0.27 | -0.11 | 0.08 | 0.04 | -0.09 | -0.10 | -0.18 | -0.10 | 0.47 | -0.18 | 0.14 | -0.17 | -0.26 | 0.09 | 0.23 | 0.32 | 0.15 | 0.10 | -0.22 |

| ID | Type | Col3 | Col4 | Col5 | Col6 | Col7 | Col8 | Col9 | Col10 | Col11 | Col12 | Col13 | Col14 | Col15 | Col16 | Col17 | Col18 | Col19 | Col20 | Col21 | Col22 | Col23 | Col24 | Col25 | Col26 | Col27 | Col28 | Col29 | Col30 |
|---|---|---|---|---|---|---|---|---|---|---|---|---|---|---|---|---|---|---|---|---|---|---|---|---|---|---|---|---|---|
| hd035929 | H | 6345 | 2.06 | 4.25 | -1.30 | | | -0.03 | -0.28 | 0.01 | 0.22 | 0.37 | 0.56 | 0.16 | -0.46 | 0.19 | 0.97 | -0.09 | | | 2.10 | 0.24 | | | | | -0.19 | -0.81 | 1.15 |
| hd036189 | H | 4911 | 2.06 | 1.81 | 0.34 | 0.12 | 0.14 | 0.14 | 0.39 | 0.03 | -0.11 | -0.06 | -0.06 | 0.05 | -0.14 | -0.04 | -0.15 | -0.05 | -0.13 | -0.01 | 0.16 | 0.01 | -0.03 | 0.14 | 0.29 | 0.50 | 0.28 | 0.10 | 0.08 |
| hd036597 | H | 4573 | 1.76 | 1.55 | 0.29 | 0.11 | 0.10 | 0.19 | 0.45 | 0.03 | -0.21 | -0.07 | 0.06 | 0.04 | 0.01 | -0.07 | -0.12 | -0.06 | 0.02 | -0.11 | 0.13 | -0.02 | -0.18 | 0.01 | 0.21 | 0.26 | 0.14 | 0.08 | -0.12 |
| hd036734 | H | 4244 | 1.31 | 1.54 | 0.16 | -0.02 | 0.11 | 0.09 | 0.23 | -0.01 | -0.19 | -0.06 | 0.23 | 0.03 | -0.04 | -0.14 | -0.15 | -0.08 | -0.09 | -0.09 | 0.17 | -0.20 | -0.22 | -0.16 | 0.12 | 0.27 | 0.14 | 0.06 | -0.30 |
| hd036848 | H | 4476 | 2.33 | 1.15 | 0.86 | 0.63 | 0.82 | 0.69 | 1.36 | 0.55 | 0.51 | 0.52 | 0.99 | 0.60 | 0.60 | 0.44 | 0.51 | 0.60 | 0.65 | 0.77 | 0.47 | 0.36 | 0.28 | 0.16 | 0.52 | 0.57 | 0.60 | 0.45 | 0.19 |
| hd037811 | H | 5023 | 2.48 | 1.47 | 0.26 | 0.10 | 0.17 | 0.11 | 0.41 | 0.10 | -0.09 | -0.02 | -0.01 | 0.07 | -0.03 | -0.02 | -0.08 | -0.04 | -0.14 | -0.18 | 0.21 | 0.03 | 0.05 | 0.18 | 0.16 | 0.36 | 0.20 | 0.16 | 0.13 |
| hd039425 | H | 4566 | 2.09 | 1.39 | 0.36 | 0.34 | 0.36 | 0.42 | 0.71 | 0.23 | 0.09 | 0.21 | 0.48 | 0.29 | 0.37 | 0.18 | 0.20 | 0.29 | 0.53 | 0.57 | 0.20 | 0.14 | -0.04 | 0.09 | 0.25 | 0.22 | 0.27 | 0.17 | 0.07 |
| hd039640 | H | 4850 | 2.31 | 1.38 | 0.08 | 0.04 | 0.07 | 0.06 | 0.25 | -0.01 | -0.20 | -0.12 | -0.07 | -0.06 | -0.10 | -0.12 | -0.18 | -0.13 | -0.16 | -0.22 | 0.04 | -0.10 | -0.12 | 0.08 | 0.07 | 0.27 | 0.13 | 0.01 | 0.00 |
| hd040176 | H | 4659 | 2.03 | 1.57 | 0.92 | 0.45 | 0.55 | 0.52 | 0.92 | 0.33 | 0.10 | 0.16 | 0.37 | 0.34 | 0.39 | 0.24 | 0.21 | 0.33 | 0.45 | 0.26 | 0.21 | 0.11 | -0.06 | 0.00 | 0.23 | 0.27 | 0.18 | 0.14 | 0.20 |
| hd040657 | H | 4288 | 1.15 | 1.52 | -0.34 | -0.22 | -0.18 | -0.26 | -0.01 | -0.32 | -0.55 | -0.38 | -0.36 | -0.52 | -0.53 | -0.75 | -0.64 | -0.55 | -0.59 | -0.26 | -0.63 | -0.31 | -0.82 | -0.79 | -0.71 | -0.61 | -0.51 | -0.42 | -0.46 | -0.56 |
| hd040808 | H | 4620 | 1.64 | 1.80 | 0.47 | 0.10 | 0.21 | 0.25 | 0.57 | 0.11 | -0.12 | -0.04 | 0.00 | 0.11 | 0.05 | -0.02 | -0.09 | -0.01 | 0.00 | -0.02 | 0.20 | 0.08 | -0.14 | 0.15 | 0.22 | 0.36 | 0.16 | 0.10 | -0.08 |
| hd041047 | H | 3891 | -0.17 | 1.90 | 0.21 | -0.09 | 0.14 | 0.18 | 1.20 | 0.08 | -0.45 | -0.29 | -0.07 | -0.06 | -0.34 | -0.29 | -0.33 | -0.24 | 0.11 | -0.26 | 0.73 | -0.58 | -0.52 | -0.32 | -0.25 | -0.45 | -0.11 | -0.30 | -0.52 |
| hd042719 | H | 5699 | 3.56 | 1.16 | 0.48 | 0.46 | 0.45 | 0.35 | 0.50 | 0.33 | 0.27 | 0.27 | 0.26 | 0.31 | 0.29 | 0.26 | 0.28 | 0.31 | 0.43 | 0.41 | 0.30 | 0.26 | 0.19 | 0.18 | 0.28 | 0.30 | 0.20 | 0.19 | 1.23 |
| hd043785 | H | 4876 | 2.45 | 1.38 | 0.51 | 0.22 | 0.30 | 0.28 | 0.47 | 0.23 | -0.04 | 0.05 | 0.19 | 0.17 | 0.22 | 0.09 | 0.01 | 0.11 | 0.13 | 0.05 | 0.17 | 0.05 | -0.05 | 0.04 | 0.06 | 0.15 | 0.07 | 0.00 | 0.09 |
| hd045669 | H | 3984 | 0.39 | 1.71 | 0.27 | -0.02 | 0.16 | 0.19 | 1.01 | 0.14 | -0.34 | -0.09 | 0.07 | 0.04 | -0.20 | -0.18 | -0.20 | -0.08 | 0.14 | -0.10 | 0.42 | -0.15 | -0.23 | -0.19 | -0.06 | -0.12 | 0.01 | -0.05 | -0.21 |
| hd046116 | H | 4853 | 2.32 | 1.45 | -0.18 | -0.09 | -0.01 | -0.10 | -0.03 | -0.17 | -0.30 | -0.22 | -0.21 | -0.27 | -0.34 | -0.32 | -0.30 | -0.31 | -0.26 | -0.34 | -0.19 | -0.43 | -0.44 | -0.40 | -0.30 | -0.22 | -0.25 | -0.32 | -0.26 |
| hd046415 | H | 4818 | 2.30 | 1.37 | -0.07 | -0.07 | -0.01 | -0.07 | 0.09 | -0.11 | -0.28 | -0.17 | -0.13 | -0.19 | -0.20 | -0.25 | -0.25 | -0.24 | -0.19 | -0.30 | -0.10 | -0.31 | -0.32 | -0.25 | -0.21 | -0.01 | -0.14 | -0.24 | -0.17 |
| hd046568 | H | 4808 | 2.22 | 1.39 | -0.05 | -0.08 | 0.00 | -0.07 | 0.19 | -0.11 | -0.31 | -0.23 | -0.20 | -0.19 | -0.21 | -0.27 | -0.30 | -0.27 | -0.24 | -0.31 | -0.04 | -0.25 | -0.34 | -0.19 | -0.07 | -0.02 | -0.13 | -0.23 | -0.23 |
| hd047001 | H | 4675 | 1.98 | 1.52 | -0.07 | -0.12 | -0.05 | -0.11 | 0.26 | -0.15 | -0.33 | -0.23 | -0.17 | -0.23 | -0.32 | -0.32 | -0.33 | -0.33 | -0.28 | -0.40 | -0.01 | -0.28 | -0.33 | -0.10 | -0.07 | 0.07 | 0.01 | -0.19 | -0.25 |
| hd047536 | H | 4353 | 1.32 | 1.48 | -0.45 | -0.31 | -0.25 | -0.35 | -0.14 | -0.43 | -0.60 | -0.45 | -0.45 | -0.62 | -0.81 | -0.72 | -0.60 | -0.67 | -0.52 | -0.69 | -0.40 | -0.77 | -0.80 | -0.83 | -0.65 | -0.58 | -0.47 | -0.50 | -0.63 |
| hd047910 | H | 4869 | 2.52 | 1.31 | 0.25 | 0.14 | 0.25 | 0.14 | 0.50 | 0.13 | -0.04 | 0.10 | 0.23 | 0.12 | 0.12 | 0.02 | -0.01 | 0.03 | 0.08 | -0.02 | 0.16 | 0.04 | 0.00 | 0.10 | 0.13 | 0.22 | 0.19 | 0.06 | 0.13 |
| hd049877 | H | 3911 | -0.30 | 3.82 | | | -0.52 | -0.07 | 0.38 | -0.52 | -0.83 | -0.93 | -0.72 | -0.08 | -0.65 | -0.61 | -0.33 | -0.53 | | | -1.05 | -1.89 | | -0.48 | -0.29 | -0.41 | | | |
| hd049947 | H | 4885 | 2.24 | 1.43 | -0.05 | -0.08 | 0.00 | -0.07 | 0.02 | -0.12 | -0.27 | -0.17 | -0.17 | -0.15 | -0.21 | -0.22 | -0.22 | -0.22 | -0.29 | -0.30 | 0.05 | -0.15 | -0.21 | -0.04 | 0.01 | 0.14 | 0.05 | -0.07 | -0.08 |
| hd050310 | H | 4489 | 1.74 | 1.57 | 0.35 | 0.15 | 0.24 | 0.29 | 0.71 | 0.07 | -0.13 | -0.02 | 0.21 | 0.09 | 0.13 | -0.02 | -0.03 | 0.04 | 0.21 | 0.11 | 0.08 | -0.08 | -0.21 | -0.08 | 0.14 | 0.24 | 0.08 | 0.04 | -0.11 |
| hd050890 | H | 4733 | 1.88 | 1.98 | 0.99 | 0.28 | 0.48 | 0.37 | 0.72 | 0.18 | 0.00 | 0.17 | 0.18 | 0.37 | 0.07 | 0.10 | 0.04 | 0.17 | | 0.14 | 0.75 | 0.37 | 0.24 | 0.27 | 0.40 | 0.10 | 0.25 | 0.49 | | |
| hd054038 | H | 5028 | 2.69 | 1.40 | 0.13 | 0.10 | 0.18 | 0.09 | 0.25 | 0.09 | -0.14 | -0.02 | 0.05 | 0.13 | 0.03 | 0.01 | -0.05 | -0.01 | -0.12 | -0.09 | 0.11 | 0.00 | -0.02 | 0.06 | 0.13 | 0.25 | 0.08 | -0.05 | 0.38 |
| hd054732 | H | 4902 | 2.58 | 1.23 | 0.38 | 0.25 | 0.14 | 0.19 | 0.56 | 0.06 | -0.09 | 0.04 | 0.09 | 0.13 | 0.10 | -0.02 | -0.01 | 0.02 | 0.67 | -0.38 | 0.16 | 0.15 | 0.01 | 0.11 | 0.28 | 0.36 | 0.24 | 0.18 | 0.17 |
| hd059219 | H | 4803 | 1.85 | 2.19 | 0.73 | 0.31 | 0.29 | 0.35 | 0.79 | 0.12 | -0.08 | 0.06 | 0.04 | 0.26 | 0.01 | 0.08 | -0.02 | 0.09 | -0.08 | 0.03 | 0.12 | 0.05 | 0.10 | -0.02 | 0.31 | 0.21 | 0.30 | 0.15 | 0.05 |
| hd059894 | H | 4904 | 2.38 | 1.38 | 0.01 | 0.01 | -0.01 | -0.01 | 0.04 | -0.03 | -0.20 | -0.11 | -0.08 | -0.08 | -0.13 | -0.15 | -0.20 | -0.17 | -0.15 | -0.26 | 0.08 | -0.09 | -0.12 | 0.06 | 0.13 | 0.25 | 0.16 | 0.02 | -0.03 |
| hd060574 | H | 4929 | 2.25 | 1.47 | -0.30 | -0.21 | -0.16 | -0.23 | -0.15 | -0.30 | -0.42 | -0.37 | -0.38 | -0.42 | -0.46 | -0.44 | -0.39 | -0.43 | -0.43 | -0.46 | -0.29 | -0.54 | -0.50 | -0.46 | -0.46 | -0.35 | -0.37 | -0.38 | -0.05 |
| hd062713 | H | 4624 | 2.12 | 1.40 | 0.16 | 0.19 | 0.24 | 0.23 | 0.58 | 0.06 | -0.14 | 0.01 | 0.17 | 0.11 | 0.13 | 0.02 | -0.04 | 0.05 | 0.15 | 0.10 | 0.01 | -0.13 | -0.25 | -0.05 | -0.02 | 0.20 | 0.11 | 0.10 | -0.05 |
| hd062849 | H | 4871 | 1.71 | 1.11 | -0.47 | -0.34 | -0.42 | -0.38 | -0.18 | -0.43 | -1.03 | -0.96 | -1.12 | -0.68 | -0.87 | -0.78 | -0.97 | -0.70 | -1.05 | -0.71 | -0.74 | -1.05 | -1.16 | -0.94 | -1.15 | -1.17 | -1.17 | -1.39 | -0.31 |
| hd062897 | H | 4762 | 2.15 | 1.56 | 0.20 | 0.15 | 0.25 | 0.16 | 0.55 | 0.05 | -0.06 | 0.01 | 0.08 | 0.11 | -0.03 | -0.03 | -0.06 | -0.02 | -0.01 | -0.03 | 0.10 | -0.01 | -0.10 | 0.08 | 0.31 | 0.61 | 0.21 | 0.19 | -0.02 |
| hd063295 | H | 4736 | 2.07 | 1.45 | 0.00 | 0.01 | 0.08 | 0.05 | 0.10 | -0.08 | -0.22 | -0.15 | -0.07 | -0.12 | -0.09 | -0.18 | -0.16 | -0.15 | -0.07 | -0.21 | -0.05 | -0.26 | -0.34 | -0.17 | -0.12 | 0.02 | -0.08 | -0.12 | -0.12 |
| hd064181 | H | 5018 | 2.47 | 1.36 | 0.20 | 0.04 | 0.10 | 0.06 | 0.31 | 0.06 | -0.18 | -0.08 | -0.07 | 0.02 | -0.08 | -0.07 | -0.14 | -0.10 | -0.19 | -0.20 | 0.19 | -0.03 | 0.01 | 0.14 | 0.17 | 0.33 | 0.15 | 0.09 | 0.03 |
| hd073598 | H | 5012 | 2.71 | 1.42 | 0.62 | 0.35 | 0.40 | 0.33 | 0.61 | 0.31 | 0.18 | 0.26 | 0.41 | 0.34 | 0.39 | 0.25 | 0.21 | 0.27 | 0.26 | 0.23 | 0.36 | 0.27 | 0.19 | 0.31 | 0.38 | 0.45 | 0.34 | 0.28 | 0.37 |
| hd073665 | H | 4954 | 2.48 | 1.51 | 0.70 | 0.31 | 0.37 | 0.36 | 0.55 | 0.29 | 0.12 | 0.18 | 0.29 | 0.29 | 0.31 | 0.21 | 0.16 | 0.24 | 0.21 | 0.13 | 0.28 | 0.23 | 0.12 | 0.24 | 0.27 | 0.35 | 0.23 | 0.16 | 0.25 |
| hd073710 | H | 4900 | 2.39 | 1.57 | 0.64 | 0.36 | 0.41 | 0.37 | 0.79 | 0.33 | 0.11 | 0.22 | 0.35 | 0.33 | 0.32 | 0.23 | 0.19 | 0.27 | 0.23 | 0.23 | 0.29 | 0.20 | 0.15 | 0.22 | 0.45 | 0.38 | 0.28 | 0.33 | 0.31 |
| hd074874 | H | 5384 | 3.45 | 0.50 | 0.15 | -0.06 | 0.18 | 0.04 | 0.20 | 0.12 | 0.16 | 0.21 | 0.30 | 0.19 | -0.01 | 0.12 | 0.10 | 0.10 | -0.02 | -0.12 | 0.35 | 0.32 | 0.35 | 0.60 | 0.59 | 0.80 | 0.68 | 0.55 | 0.72 |
| hd078964b | H | 5118 | 4.15 | 1.11 | 0.23 | 0.15 | 0.24 | 0.16 | 0.38 | 0.25 | 0.09 | 0.15 | 0.28 | 0.23 | 0.20 | 0.14 | 0.08 | 0.14 | 0.05 | -0.04 | 0.28 | 0.17 | 0.21 | 0.13 | 0.26 | 0.33 | 0.45 | 0.30 | 0.61 |
| hd091267 | H | 4873 | 4.07 | 0.50 | 0.24 | 0.03 | 0.15 | 0.05 | 0.26 | 0.16 | 0.08 | 0.14 | 0.37 | 0.14 | 0.11 | 0.01 | 0.02 | 0.05 | 0.25 | 0.06 | 0.21 | -0.09 | 0.02 | -0.20 | 0.21 | 0.26 | 0.73 | 0.19 | 0.30 |
| hd104760a | H | 5824 | 4.04 | 0.86 | 0.24 | 0.22 | 0.09 | 0.14 | 0.23 | 0.20 | 0.17 | 0.12 | 0.12 | 0.20 | 0.16 | 0.15 | 0.13 | 0.18 | 0.15 | 0.18 | 0.12 | 0.21 | 0.26 | 0.18 | 0.22 | 0.26 | 0.13 | 0.15 | 0.62 |
| hd108063 | H | 5859 | 3.56 | 1.48 | 0.94 | 0.72 | 0.70 | 0.57 | 0.69 | 0.55 | 0.58 | 0.51 | 0.53 | 0.58 | 0.62 | 0.51 | 0.56 | 0.61 | 0.66 | 0.60 | 0.60 | 0.50 | 0.52 | 0.37 | 0.55 | 0.45 | 0.35 | 0.34 | 1.54 |
| hd108570 | H | 4984 | 3.20 | 0.80 | 0.11 | 0.11 | 0.11 | 0.06 | 0.13 | 0.10 | -0.04 | 0.08 | 0.15 | 0.08 | 0.08 | 0.02 | -0.04 | 0.04 | 0.13 | -0.05 | 0.17 | 0.05 | 0.02 | 0.02 | 0.21 | 0.23 | 0.36 | 0.08 | 0.48 |
| hd110291 | H | 5392 | 3.87 | 0.76 | 0.08 | 0.11 | 0.08 | 0.02 | 0.08 | 0.09 | -0.08 | 0.02 | 0.04 | 0.09 | 0.03 | 0.03 | -0.03 | 0.01 | 0.09 | -0.05 | 0.21 | 0.02 | 0.04 | -0.13 | 0.09 | 0.25 | 0.07 | 0.05 | 0.59 |
| hd114747 | H | 5073 | 3.96 | 0.50 | 0.87 | 0.46 | 0.51 | 0.40 | 0.74 | 0.43 | 0.40 | 0.43 | 0.73 | 0.47 | 0.52 | 0.34 | 0.37 | 0.46 | 0.64 | 0.44 | 0.37 | 0.32 | 0.26 | 0.07 | 0.44 | 0.65 | 0.70 | 0.48 | 0.91 |
| hd115202 | H | 4742 | 2.70 | 0.91 | 0.23 | 0.16 | 0.27 | 0.17 | 0.34 | 0.15 | 0.01 | 0.11 | 0.31 | 0.12 | 0.16 | 0.02 | -0.02 | 0.09 | 0.30 | 0.26 | 0.03 | -0.07 | -0.19 | -0.06 | 0.05 | 0.19 | 0.20 | 0.02 | -0.01 |
| hd121416 | H | 4576 | 2.07 | 1.39 | 0.37 | 0.27 | 0.35 | 0.32 | 0.57 | 0.17 | -0.02 | 0.11 | 0.34 | 0.21 | 0.26 | 0.08 | 0.07 | 0.17 | 0.40 | 0.32 | 0.09 | -0.07 | -0.15 | -0.04 | 0.11 | 0.08 | 0.15 | 0.08 | -0.09 |
| hd123517 | H | 6340 | 4.45 | 1.09 | 0.33 | 0.32 | 0.31 | 0.24 | 0.16 | 0.36 | 0.56 | 0.49 | 0.46 | 0.40 | 0.39 | 0.40 | 0.40 | 0.36 | 0.35 | 0.36 | 0.57 | 0.58 | 0.47 | 0.56 | 0.61 | 0.69 | 0.69 | 0.50 | 1.78 |
| hd144589 | H | 6486 | 4.29 | 1.31 | 0.17 | 0.12 | 0.02 | 0.09 | 0.09 | 0.18 | 0.33 | 0.24 | 0.35 | 0.19 | 0.06 | 0.14 | 0.26 | 0.08 | -0.09 | 0.03 | 0.50 | 0.42 | 0.45 | 0.45 | 0.46 | 0.46 | 0.39 | 0.39 | 0.66 |

| ID | Type | col3 | col4 | col5 | col6 | col7 | col8 | col9 | col10 | col11 | col12 | col13 | col14 | col15 | col16 | col17 | col18 | col19 | col20 | col21 | col22 | col23 | col24 | col25 | col26 | col27 | col28 | col29 | col30 | col31 |
|---|---|---|---|---|---|---|---|---|---|---|---|---|---|---|---|---|---|---|---|---|---|---|---|---|---|---|---|---|---|---|
| hd147135 | H | 6832 | 3.35 | 3.37 | -0.16 | -0.11 | -0.04 | -0.03 | 0.07 | 0.07 | 0.12 | -0.05 | 0.13 | -0.17 | -0.27 | -0.23 | 0.07 | -0.10 | -0.28 | -0.27 | | 0.33 | 0.33 | 0.01 | 0.08 | 0.30 | 0.22 | 0.20 | 0.33 |
| hd169689 | H | 4989 | 2.25 | 1.91 | 0.45 | 0.28 | 0.59 | 0.37 | 0.41 | 0.23 | 0.17 | 0.28 | 0.28 | 0.30 | -0.04 | 0.18 | 0.14 | 0.15 | -0.05 | 0.14 | 0.49 | 0.29 | 0.22 | 0.49 | 0.41 | 0.46 | 0.58 | 0.38 | 0.50 |
| hd176354 | H | 5165 | 3.40 | 0.86 | 0.69 | 0.45 | 0.53 | 0.40 | 0.78 | 0.40 | 0.31 | 0.36 | 0.53 | 0.44 | 0.52 | 0.33 | 0.33 | 0.45 | 0.65 | 0.43 | 0.34 | 0.28 | 0.20 | 0.16 | 0.37 | 0.39 | 0.33 | 0.32 | 0.84 |
| hd181433 | H | 4800 | 3.57 | 0.82 | 1.02 | 0.57 | 0.58 | 0.41 | 0.80 | 0.49 | 0.32 | 0.34 | 0.73 | 0.45 | 0.43 | 0.25 | 0.28 | 0.38 | 0.58 | 0.31 | 0.45 | 0.15 | 0.22 | -0.23 | 0.13 | 0.35 | 0.55 | 0.49 | 0.54 |
| hd181517 | H | 4895 | 2.54 | 1.35 | 0.36 | 0.15 | 0.22 | 0.20 | 0.45 | 0.17 | -0.04 | 0.06 | 0.12 | 0.15 | 0.17 | 0.07 | -0.03 | 0.10 | 0.09 | -0.03 | 0.22 | 0.11 | 0.07 | 0.05 | 0.22 | 0.33 | 0.18 | 0.15 | 0.04 |
| hd182893 | H | 4864 | 2.59 | 1.26 | 0.39 | 0.22 | 0.27 | 0.26 | 0.40 | 0.21 | -0.02 | 0.08 | 0.19 | 0.21 | 0.19 | 0.11 | -0.01 | 0.13 | 0.13 | 0.15 | 0.24 | 0.09 | 0.07 | 0.12 | 0.20 | 0.44 | 0.23 | 0.09 | 0.16 |
| hd187669a | H | 4754 | 2.50 | 1.96 | 0.36 | | | -0.14 | 0.77 | 0.02 | 0.07 | 0.09 | 0.19 | 0.32 | -0.25 | -0.26 | -0.11 | -0.11 | -0.35 | -0.58 | | 0.13 | 0.29 | -0.47 | 0.56 | 0.10 | 0.67 | 0.51 | -0.16 |
| hd188114 | H | 4594 | 1.93 | 1.40 | -0.13 | -0.17 | -0.15 | -0.14 | 0.09 | -0.19 | -0.43 | -0.30 | -0.23 | -0.27 | -0.35 | -0.35 | -0.40 | -0.37 | -0.35 | -0.47 | -0.12 | -0.40 | -0.43 | -0.23 | -0.08 | -0.02 | -0.09 | -0.26 | -0.30 |
| hd190056 | H | 4275 | 1.26 | 1.47 | -0.32 | -0.22 | -0.15 | -0.23 | 0.07 | -0.36 | -0.52 | -0.36 | -0.30 | -0.54 | -0.73 | -0.66 | -0.53 | -0.59 | -0.17 | -0.63 | -0.32 | -0.82 | -0.75 | -0.85 | -0.58 | -0.60 | -0.49 | -0.50 | -0.67 |
| hd196171 | H | 4810 | 2.31 | 1.36 | 0.12 | 0.06 | 0.11 | 0.06 | 0.27 | 0.03 | -0.18 | -0.09 | -0.02 | -0.02 | -0.04 | -0.09 | -0.13 | -0.09 | -0.11 | -0.20 | 0.10 | -0.06 | -0.14 | 0.06 | 0.04 | 0.29 | 0.11 | 0.01 | -0.03 |
| hd198232 | H | 4750 | 1.89 | 1.49 | 0.24 | 0.03 | 0.12 | 0.13 | 0.29 | 0.04 | -0.25 | -0.14 | -0.08 | 0.03 | -0.06 | -0.09 | -0.18 | -0.10 | -0.23 | -0.18 | 0.11 | -0.08 | -0.16 | 0.06 | 0.08 | 0.16 | 0.00 | -0.05 | -0.15 |
| hd199951 | H | 5009 | 2.36 | 1.57 | 0.34 | 0.04 | 0.08 | 0.08 | 0.38 | 0.03 | -0.20 | -0.08 | -0.08 | 0.05 | -0.14 | -0.07 | -0.16 | -0.09 | -0.33 | -0.28 | 0.11 | 0.01 | -0.05 | 0.14 | 0.11 | 0.42 | 0.20 | -0.04 | -0.09 |
| hd200763 | H | 4632 | 2.11 | 1.42 | 0.33 | 0.15 | 0.22 | 0.21 | 0.57 | 0.13 | -0.10 | 0.03 | 0.20 | 0.15 | 0.13 | 0.03 | -0.04 | 0.05 | 0.11 | -0.06 | 0.19 | 0.03 | -0.02 | 0.10 | 0.32 | 0.49 | 0.28 | 0.09 | -0.02 |
| hd201852 | H | 4883 | 2.44 | 1.32 | 0.17 | 0.08 | 0.12 | 0.09 | 0.19 | 0.08 | -0.16 | -0.03 | 0.02 | 0.02 | 0.00 | -0.04 | -0.13 | -0.05 | -0.07 | -0.15 | 0.17 | -0.01 | -0.05 | 0.10 | 0.15 | 0.36 | 0.22 | 0.05 | 0.03 |
| hd206642 | H | 4326 | 0.67 | 1.59 | -1.13 | -0.73 | -0.99 | -0.78 | -0.29 | -0.88 | -1.28 | -1.04 | -1.17 | -1.09 | -1.41 | -1.15 | -1.14 | -1.20 | -1.49 | -1.15 | -0.83 | -1.19 | -1.11 | -1.11 | -1.09 | -1.02 | -0.94 | -0.94 | -1.01 |
| hd209449 | H | 5685 | 3.79 | 1.01 | 0.79 | 0.56 | 0.59 | 0.46 | 0.59 | 0.44 | 0.51 | 0.45 | 0.47 | 0.48 | 0.57 | 0.44 | 0.48 | 0.55 | 0.70 | 0.63 | 0.48 | 0.43 | 0.34 | 0.32 | 0.45 | 0.49 | 0.45 | 0.43 | 1.02 |
| hd211317 | H | 5780 | 3.92 | 1.00 | 0.50 | 0.39 | 0.38 | 0.30 | 0.43 | 0.30 | 0.27 | 0.28 | 0.28 | 0.31 | 0.36 | 0.27 | 0.29 | 0.34 | 0.40 | 0.40 | 0.37 | 0.38 | 0.26 | 0.15 | 0.31 | 0.27 | 0.26 | 0.17 | 1.01 |
| hip031592 | H | 4735 | 2.84 | 0.88 | 0.56 | 0.45 | 0.42 | 0.47 | 0.80 | 0.36 | 0.31 | 0.36 | 0.73 | 0.40 | 0.50 | 0.30 | 0.30 | 0.45 | 0.66 | 0.46 | 0.29 | 0.36 | 0.13 | 0.18 | 0.40 | 0.45 | 0.47 | 0.38 | 0.26 |
| hip056713 | H | 4993 | 2.73 | 1.00 | -1.32 | -0.98 | -1.40 | -0.94 | | -0.93 | -1.34 | -1.16 | -1.35 | -1.27 | -1.70 | -1.31 | -1.34 | -1.39 | -1.93 | -1.30 | -0.95 | -1.40 | -1.04 | -1.20 | -1.09 | -1.15 | -0.97 | -0.96 | -0.23 |
| hip080242b | H | 4315 | 0.70 | 1.41 | 0.21 | 0.08 | -0.01 | 0.24 | 0.49 | -0.10 | -0.57 | -0.55 | -0.55 | -0.12 | -0.28 | -0.19 | -0.40 | -0.16 | -0.25 | -0.12 | -0.25 | -0.42 | -0.81 | -0.32 | -0.33 | -0.17 | -0.44 | -0.40 | -0.51 |
| hr2959 | H | 3916 | 0.30 | 1.73 | 0.28 | 0.06 | 0.05 | 0.29 | 0.06 | 0.01 | -0.39 | -0.21 | -0.05 | -0.01 | -0.13 | -0.16 | -0.29 | -0.10 | -0.12 | -0.11 | 0.16 | -0.26 | -0.31 | 0.09 | -0.08 | 0.01 | 0.10 | 0.01 | -0.46 |
| hr3728 | H | 4888 | 2.25 | 1.43 | -0.19 | -0.20 | -0.13 | -0.20 | -0.08 | -0.26 | -0.43 | -0.36 | -0.36 | -0.37 | -0.43 | -0.41 | -0.40 | -0.41 | -0.42 | -0.46 | -0.25 | -0.46 | -0.43 | -0.34 | -0.36 | -0.28 | -0.29 | -0.35 | -0.29 |
| hr3919 | H | 4415 | 1.79 | 1.43 | 0.29 | 0.17 | 0.29 | 0.25 | 0.67 | 0.07 | -0.03 | 0.04 | 0.32 | 0.11 | 0.10 | -0.02 | 0.01 | 0.06 | 0.61 | 0.07 | 0.09 | -0.10 | -0.21 | -0.07 | 0.12 | 0.09 | 0.11 | 0.02 | -0.16 |
| hr5480 | H | 4807 | 1.76 | 1.78 | 0.43 | 0.19 | 0.18 | 0.22 | 0.53 | 0.06 | -0.16 | -0.07 | -0.07 | 0.12 | -0.12 | -0.02 | -0.13 | -0.02 | -0.19 | -0.14 | 0.08 | 0.02 | -0.09 | 0.06 | 0.12 | 0.49 | 0.11 | -0.02 | 0.16 |
| hr7150 | H | 4541 | 1.47 | 1.82 | 0.26 | 0.16 | 0.23 | 0.22 | 0.41 | 0.01 | -0.21 | -0.06 | -0.05 | 0.11 | 0.01 | -0.02 | -0.10 | -0.01 | | 0.27 | 0.13 | 0.00 | -0.12 | 0.14 | 0.27 | 0.35 | 0.17 | 0.14 | -0.03 |
| ic4651no14527 | H | 4803 | 2.35 | 1.42 | 0.44 | 0.26 | 0.33 | 0.35 | 0.43 | 0.21 | 0.06 | 0.09 | 0.21 | 0.26 | 0.23 | 0.15 | 0.08 | 0.20 | 0.22 | 0.35 | 0.48 | 0.35 | 0.25 | 0.25 | 0.35 | 0.49 | 0.35 | 0.28 | 0.33 |
| ic4651no8540 | H | 4826 | 2.61 | 1.32 | 0.48 | 0.35 | 0.30 | 0.36 | 0.57 | 0.27 | 0.17 | 0.25 | 0.43 | 0.32 | 0.37 | 0.22 | 0.17 | 0.28 | 0.38 | 0.19 | 0.31 | 0.32 | 0.18 | 0.28 | 0.53 | 0.55 | 0.46 | 0.45 | 0.29 |
| ic4651no9025 | H | 4799 | 2.51 | 1.32 | 0.41 | 0.23 | 0.25 | 0.23 | 0.36 | 0.18 | 0.05 | 0.15 | 0.26 | 0.22 | 0.21 | 0.12 | 0.07 | 0.17 | 0.31 | 0.09 | 0.23 | 0.05 | 0.10 | 0.09 | 0.17 | 0.38 | 0.26 | 0.20 | 0.11 |
| ic4651no9791 | H | 4463 | 1.79 | 1.41 | 0.32 | 0.16 | 0.28 | 0.21 | 0.54 | 0.12 | -0.08 | 0.04 | 0.29 | 0.18 | 0.19 | 0.02 | -0.02 | 0.09 | 0.43 | 0.12 | 0.19 | 0.03 | -0.11 | -0.07 | 0.19 | 0.18 | 0.16 | 0.13 | -0.16 |
| ksihya | H | 4950 | 2.46 | 1.38 | 0.46 | 0.20 | 0.23 | 0.22 | 0.50 | 0.20 | -0.10 | 0.02 | 0.09 | 0.18 | 0.13 | 0.07 | -0.01 | 0.07 | -0.02 | -0.03 | 0.15 | 0.01 | -0.02 | 0.11 | 0.18 | 0.21 | 0.04 | 0.04 | 0.10 |
| lra01_e1_0286 | H | 4950 | 2.74 | 1.18 | -0.10 | 0.00 | 0.07 | -0.04 | -0.12 | -0.05 | -0.15 | -0.08 | -0.07 | -0.07 | -0.18 | -0.13 | -0.18 | -0.15 | | -0.20 | 0.01 | -0.22 | -0.22 | -0.17 | -0.06 | 0.14 | 0.11 | 0.07 | 0.06 |
| lra01_e2_2249 | H | 5250 | 3.86 | 0.27 | 0.43 | 0.48 | 0.39 | 0.26 | 0.50 | 0.37 | 0.52 | 0.61 | 0.90 | 0.61 | 0.56 | 0.39 | 0.32 | 0.44 | | 0.44 | 0.70 | 0.69 | 0.73 | 0.15 | 0.76 | 0.90 | 1.02 | 1.11 | 0.68 |
| lra03_e2_0678 | H | 5235 | 4.89 | 0.74 | 0.58 | 0.15 | 0.47 | -0.19 | 0.31 | 0.64 | 0.57 | 0.77 | 1.08 | 0.57 | 0.34 | 0.18 | 0.34 | 0.20 | | -0.19 | 0.85 | 0.69 | 1.04 | 0.10 | 0.59 | 0.86 | 1.19 | 1.27 | 0.61 |
| lra03_e2_1326 | H | 4970 | 3.36 | 1.04 | 0.13 | 0.14 | 0.27 | 0.12 | 0.56 | 0.11 | 0.14 | 0.27 | 0.42 | 0.26 | 0.15 | 0.14 | 0.10 | 0.17 | | 0.65 | 0.15 | 0.39 | 0.31 | 0.00 | 0.53 | 0.60 | 0.68 | 0.66 | 0.19 |
| ngc2287no107 | H | 4602 | 1.54 | 1.84 | 0.22 | 0.02 | 0.08 | 0.14 | 0.58 | -0.09 | -0.32 | -0.21 | -0.18 | -0.03 | -0.11 | -0.14 | -0.22 | -0.15 | -0.18 | 0.03 | -0.01 | -0.08 | -0.25 | -0.04 | 0.17 | 0.35 | 0.10 | 0.01 | -0.13 |
| ngc2287no204 | H | 4382 | 1.44 | 1.53 | -0.04 | -0.11 | -0.18 | -0.12 | 0.31 | -0.17 | -0.42 | -0.24 | -0.12 | -0.21 | -0.31 | -0.34 | -0.39 | -0.34 | -0.08 | -0.39 | -0.06 | -0.32 | -0.43 | -0.25 | -0.04 | 0.04 | -0.05 | -0.22 | -0.27 |
| ngc2287no21 | H | 4020 | 0.20 | 2.00 | 0.26 | 0.02 | 0.07 | 0.19 | | -0.18 | -0.49 | -0.28 | -0.18 | -0.04 | -0.17 | -0.21 | -0.31 | -0.20 | -0.22 | -0.36 | 0.13 | -0.34 | -0.34 | | 0.00 | -0.06 | -0.05 | -0.14 | -0.55 |
| ngc2287no75 | H | 4423 | 1.28 | 1.90 | 0.22 | 0.16 | 0.17 | 0.27 | 0.48 | -0.01 | -0.25 | -0.11 | -0.13 | 0.05 | -0.05 | -0.06 | -0.15 | -0.04 | 0.35 | 0.09 | 0.04 | -0.04 | -0.27 | 0.10 | 0.32 | 0.29 | 0.23 | 0.19 | -0.16 |
| ngc2287no87 | H | 4199 | 1.20 | 1.59 | 0.06 | -0.17 | 0.04 | -0.03 | 0.70 | -0.11 | -0.34 | -0.17 | 0.02 | -0.13 | -0.28 | -0.32 | -0.31 | -0.26 | -0.04 | -0.46 | 0.05 | -0.41 | -0.35 | -0.46 | -0.12 | -0.10 | 0.01 | -0.16 | -0.31 |
| ngc2287no97 | H | 4596 | 1.56 | 1.82 | 0.25 | 0.11 | 0.15 | 0.17 | 0.45 | 0.00 | -0.17 | -0.13 | -0.12 | 0.04 | -0.03 | -0.06 | -0.15 | -0.08 | -0.07 | 0.03 | 0.15 | -0.04 | -0.21 | 0.11 | 0.25 | 0.33 | 0.18 | 0.18 | -0.07 |
| ngc3532no100 | H | 4740 | 1.79 | 1.76 | 0.32 | 0.14 | 0.14 | 0.20 | 0.55 | 0.10 | -0.18 | -0.11 | -0.08 | 0.06 | -0.02 | -0.05 | -0.15 | -0.08 | -0.15 | -0.02 | 0.09 | -0.04 | -0.11 | 0.04 | 0.20 | 0.43 | 0.09 | 0.04 | -0.05 |
| ngc3532no122 | H | 4963 | 2.26 | 1.83 | 0.36 | 0.15 | 0.13 | 0.17 | 0.33 | 0.12 | -0.10 | 0.02 | 0.03 | 0.17 | -0.08 | 0.01 | -0.14 | -0.01 | -0.19 | -0.08 | 0.04 | 0.07 | 0.13 | 0.17 | 0.33 | 0.11 | 0.41 | 0.13 | 0.16 |
| ngc3532no19 | H | 4844 | 1.92 | 1.58 | 0.26 | 0.07 | 0.17 | 0.17 | 0.42 | 0.09 | -0.15 | -0.10 | -0.07 | 0.10 | -0.02 | -0.04 | -0.03 | -0.16 | -0.07 | 0.13 | -0.01 | -0.12 | 0.15 | 0.22 | 0.40 | 0.12 | 0.10 | 0.01 | |
| ngc3532no596 | H | 4955 | 2.07 | 1.92 | 0.34 | 0.22 | 0.15 | 0.17 | 0.22 | -0.04 | -0.08 | -0.06 | -0.10 | 0.08 | -0.09 | -0.05 | -0.11 | -0.05 | -0.27 | -0.21 | 0.43 | 0.12 | -0.01 | -0.07 | 0.09 | 0.60 | 0.35 | 0.01 | -0.03 |
| ngc3532no649 | H | 4849 | 2.27 | 1.41 | 0.04 | -0.01 | 0.02 | 0.00 | -0.02 | -0.08 | -0.19 | -0.11 | -0.07 | -0.10 | -0.18 | -0.15 | -0.19 | -0.17 | -0.17 | -0.31 | 0.13 | -0.11 | -0.16 | 0.05 | 0.06 | 0.21 | 0.15 | 0.02 | -0.04 |
| ngc4349no127 | H | 4479 | 1.67 | 1.76 | 0.30 | 0.11 | 0.15 | 0.17 | 0.24 | 0.01 | -0.18 | -0.03 | 0.05 | 0.12 | 0.08 | -0.04 | -0.10 | -0.02 | 0.10 | 0.18 | 0.15 | 0.06 | 0.05 | 0.01 | 0.43 | 0.41 | 0.28 | 0.31 | -0.08 |
| ngc6705no1286 | H | 4929 | 2.35 | 2.18 | 0.97 | 0.45 | 0.57 | 0.56 | 1.00 | 0.30 | 0.27 | 0.35 | 0.45 | 0.49 | 0.22 | 0.28 | 0.32 | 0.32 | 0.09 | 0.25 | 0.28 | 0.29 | 0.40 | 0.10 | 0.59 | 0.21 | 0.60 | 0.51 | |
| ngc6705no1423 | H | 4521 | 1.89 | 1.92 | 0.57 | 0.42 | 0.53 | 0.49 | 0.84 | 0.26 | 0.21 | 0.29 | 0.41 | 0.41 | 0.37 | 0.26 | 0.26 | 0.36 | 0.94 | 0.42 | 0.39 | 0.32 | 0.18 | 0.21 | 0.72 | 0.45 | 0.49 | 0.49 | 0.31 |
| ngc6705no411 | H | 4445 | 1.65 | 2.01 | 0.75 | 0.31 | 0.47 | 0.48 | 0.74 | 0.22 | 0.12 | 0.20 | 0.32 | 0.31 | 0.23 | 0.18 | 0.19 | 0.26 | 0.42 | -0.39 | 0.32 | 0.09 | 0.02 | -0.09 | 0.43 | 0.66 | 0.43 | 0.49 | 0.19 |
| ngc6705no660 | H | 4788 | 2.17 | 1.95 | 0.89 | 0.47 | 0.47 | 0.45 | 0.90 | 0.35 | 0.25 | 0.33 | 0.42 | 0.43 | 0.37 | 0.29 | 0.26 | 0.34 | 0.38 | 0.37 | 0.42 | 0.31 | 0.24 | 0.18 | 0.51 | 0.52 | 0.38 | 0.49 | 0.26 |

| name | type | c1 | c2 | c3 | c4 | c5 | c6 | c7 | c8 | c9 | c10 | c11 | c12 | c13 | c14 | c15 | c16 | c17 | c18 | c19 | c20 | c21 | c22 | c23 | c24 | c25 | c26 | c27 | c28 | c29 |
|---|---|---|---|---|---|---|---|---|---|---|---|---|---|---|---|---|---|---|---|---|---|---|---|---|---|---|---|---|---|---|
| ngc6705no779 | H | 4307 | 1.26 | 2.00 | 0.59 | 0.15 | 0.45 | 0.48 | 0.66 | 0.12 | -0.10 | 0.04 | 0.13 | 0.25 | 0.20 | 0.07 | 0.05 | 0.17 | 0.31 | -0.23 | 0.24 | -0.05 | -0.06 | -0.23 | 0.17 | 0.13 | 0.19 | 0.33 | 0.27 |
| pihya | H | 4563 | 2.11 | 1.30 | 0.17 | 0.10 | 0.09 | 0.13 | 0.56 | 0.02 | -0.16 | -0.05 | 0.12 | 0.02 | 0.06 | -0.07 | -0.09 | -0.02 | 0.09 | -0.19 | 0.07 | -0.08 | -0.17 | 0.00 | 0.05 | 0.28 | 0.14 | 0.03 | -0.08 |
| sand1016 | H | 4431 | 1.82 | 1.42 | 0.23 | 0.23 | 0.32 | 0.20 | 0.42 | 0.12 | -0.02 | 0.08 | 0.38 | 0.17 | 0.17 | 0.00 | 0.00 | 0.06 | 0.39 | 0.04 | 0.18 | -0.04 | -0.11 | -0.07 | 0.17 | 0.13 | 0.16 | 0.12 | -0.11 |
| sand1054 | H | 4721 | 2.45 | 1.23 | 0.22 | 0.11 | 0.26 | 0.13 | -0.06 | 0.11 | -0.08 | 0.06 | 0.25 | 0.16 | 0.14 | 0.02 | -0.02 | 0.07 | 0.15 | 0.23 | 0.26 | 0.07 | -0.11 | -0.07 | 0.33 | 0.26 | 0.25 | 0.19 | -0.07 |
| sand1074 | H | 4718 | 2.24 | 1.44 | 0.31 | 0.17 | 0.25 | 0.21 | 0.45 | 0.10 | -0.06 | 0.04 | 0.21 | 0.14 | 0.18 | 0.02 | -0.01 | 0.07 | 0.11 | 0.00 | 0.19 | 0.04 | -0.07 | -0.03 | 0.20 | 0.25 | 0.16 | 0.16 | 0.09 |
| sand1084 | H | 4745 | 2.30 | 1.43 | 0.34 | 0.17 | 0.23 | 0.21 | 0.27 | 0.08 | -0.04 | 0.05 | 0.22 | 0.13 | 0.16 | 0.03 | -0.01 | 0.09 | 0.11 | -0.04 | 0.14 | -0.02 | -0.11 | -0.02 | 0.28 | 0.21 | 0.11 | 0.06 | 0.06 |
| sand1237 | H | 5067 | 2.88 | 1.33 | 0.33 | 0.14 | 0.22 | 0.15 | 0.19 | 0.14 | 0.01 | 0.11 | 0.18 | 0.19 | 0.15 | 0.07 | 0.04 | 0.09 | -0.01 | 0.00 | 0.26 | 0.12 | 0.04 | 0.07 | 0.13 | 0.25 | 0.16 | 0.16 | 0.81 |
| sand1279 | H | 4732 | 2.38 | 1.37 | 0.36 | 0.22 | 0.21 | 0.25 | 0.44 | 0.16 | -0.04 | 0.13 | 0.31 | 0.20 | 0.20 | 0.09 | 0.07 | 0.14 | 0.28 | 0.07 | 0.16 | 0.05 | -0.02 | 0.05 | 0.35 | 0.33 | 0.24 | 0.23 | 0.14 |
| sand364 | H | 4245 | 1.31 | 1.46 | 0.23 | 0.10 | 0.19 | 0.23 | 0.49 | 0.01 | -0.20 | -0.06 | 0.21 | 0.05 | -0.02 | -0.09 | -0.10 | -0.01 | -0.06 | 0.01 | 0.05 | -0.22 | -0.29 | -0.25 | -0.05 | 0.25 | 0.04 | -0.01 | -0.37 |
| sand978 | H | 4257 | 1.56 | 1.53 | 0.29 | 0.12 | 0.26 | 0.19 | 0.57 | 0.12 | -0.02 | 0.05 | 0.31 | 0.15 | 0.10 | -0.04 | -0.03 | 0.04 | -0.26 | 0.08 | 0.26 | -0.05 | -0.09 | -0.23 | 0.25 | 0.35 | 0.27 | 0.20 | -0.28 |
| sand989 | H | 4793 | 2.67 | 1.13 | 0.40 | 0.29 | 0.36 | 0.23 | 0.52 | 0.24 | 0.14 | 0.23 | 0.42 | 0.32 | 0.29 | 0.14 | 0.08 | 0.22 | 0.37 | 0.18 | 0.30 | 0.25 | 0.19 | 0.06 | 0.29 | 0.61 | 0.40 | 0.24 | 0.32 |
| sigpup | H | 4081 | 1.55 | 1.11 | 0.52 | 0.32 | 0.53 | 0.47 | 1.46 | 0.56 | 0.47 | 0.70 | 1.03 | 0.56 | 0.33 | 0.31 | 0.38 | 0.49 | 1.02 | 0.80 | 0.96 | 0.62 | 0.63 | 0.70 | 0.83 | 1.03 | 1.17 | 1.15 | -0.15 |
| tzfor | H | 5332 | 3.53 | 0.50 | 0.36 | 0.02 | 0.27 | 0.07 | 0.32 | 0.08 | 0.27 | 0.28 | 0.51 | 0.36 | 0.08 | 0.14 | 0.23 | 0.20 | -0.23 | 0.06 | 0.46 | 0.39 | 0.42 | 0.17 | 0.87 | 0.79 | 0.95 | 0.85 | 0.92 |
| v1045sco | H | 3789 | -0.30 | 2.69 | -0.21 | -0.39 | -0.39 | 0.16 | 1.68 | -0.74 | -0.82 | -0.75 | -0.73 | -0.30 | -0.86 | -0.64 | -0.51 | -0.51 | 0.45 | -0.86 | 0.24 | -0.91 | -0.91 | -1.38 | -0.92 | -0.64 | -0.65 | -0.63 | -0.63 |
| v4393sgr | H | 4000 | 0.31 | 3.04 |  | -0.09 | 0.05 | 0.30 |  | 0.26 | -0.18 | -0.11 | 0.30 | 0.26 | -0.30 | -0.13 | -0.08 | 0.04 |  | -0.21 | 0.77 | -0.11 | -0.04 |  | -0.08 | 0.01 | 0.15 | 0.06 | -0.25 |
| xcae | H | 6973 | 3.15 | 5.30 |  | 0.01 |  | 0.06 |  | -0.14 | -0.26 | 0.12 |  | -0.47 | -0.40 | -0.36 | 1.44 | 0.02 |  | -0.76 |  | 1.50 | 2.04 | -0.42 |  |  |  | 1.03 |  |
| hd001142 | S | 5186 | 3.12 | 0.99 | 0.34 | 0.08 | 0.28 | 0.19 | 0.61 | 0.23 | 0.26 | 0.35 | 0.46 | 0.33 | 0.16 | 0.19 | 0.22 | 0.22 | 0.14 | 0.26 | 0.39 | 0.53 | 0.21 | 0.51 | 0.64 | 0.38 | 0.54 | 0.89 | 0.42 |
| hd002410 | S | 5281 | 3.59 | 1.15 | 0.42 | 0.28 | 0.48 | 0.33 | 0.68 | 0.40 | 0.54 | 0.59 | 0.63 | 0.47 | 0.48 | 0.37 | 0.39 | 0.42 | 0.64 | 0.28 | 0.63 | 0.74 | 0.45 | 0.68 | 0.90 | 0.85 | 0.91 | 0.90 | 0.70 |
| hd002954 | S | 6308 | 3.51 | 3.01 | 0.38 |  | 0.21 | 0.20 | 0.33 | 0.11 | 0.08 | 0.27 | 0.18 | 0.46 | -0.21 | 0.01 | 0.41 | 0.08 |  |  | 0.25 | 0.96 | -0.09 | 0.52 |  |  |  |  | 0.86 |
| hd005418 | S | 4977 | 2.78 | 1.35 | 0.16 | 0.06 | 0.16 | 0.13 | 0.53 | 0.09 | 0.01 | 0.06 | 0.01 | 0.08 | 0.02 | 0.01 | -0.03 | 0.01 | 0.04 | 0.00 | 0.27 | 0.18 | -0.16 | 0.25 | 0.28 | 0.29 | 0.31 | 0.33 | 0.16 |
| hd006037 | S | 4566 | 2.27 | 1.46 | 0.63 | 0.35 | 0.61 | 0.45 | 1.33 | 0.27 | 0.20 | 0.35 | 0.64 | 0.45 | 0.40 | 0.21 | 0.36 | 0.42 | 0.69 | 1.42 | 0.35 | 0.30 | -0.10 | -0.07 | 0.55 | 0.34 | 0.55 | 1.20 | 0.37 |
| hd012116 | S | 4462 | 2.00 | 1.34 | -0.04 | -0.02 | 0.11 | 0.05 | 0.66 | -0.08 | -0.22 | -0.14 | -0.08 | -0.10 | -0.24 | -0.27 | -0.21 | -0.17 | 0.19 | 0.38 | -0.09 | -0.27 | -0.72 | -0.44 | -0.12 | -0.32 | -0.08 | 0.39 | -0.15 |
| hd013004 | S | 4584 | 2.23 | 1.37 | 0.45 | 0.43 | 0.47 | 0.36 | 1.06 | 0.23 | 0.11 | 0.23 | 0.39 | 0.32 | 0.28 | 0.15 | 0.22 | 0.29 | 0.68 | 1.08 | 0.23 | 0.15 | -0.24 | -0.09 | 0.79 | 0.30 | 0.35 | 1.02 | 0.24 |
| hd015533 | S | 4370 | 1.81 | 1.35 | 0.47 | 0.10 | 0.52 | 0.53 | 1.58 | 0.35 | 0.10 | 0.17 | 0.54 | 0.38 | 0.27 | 0.19 | 0.24 | 0.32 |  | 1.87 |  | 0.10 | -0.48 | -0.08 | 0.54 | 0.17 | 0.36 | 0.32 | 0.26 |
| hd015866 | S | 5718 | 3.57 | 1.44 | 0.73 | 0.36 | 0.43 | 0.36 | 0.47 | 0.25 | 0.34 | 0.23 | 0.22 | 0.29 | 0.38 | 0.23 | 0.29 | 0.34 | 0.45 | 0.27 | 0.67 | 0.27 | 0.34 | -0.04 | 0.25 | 0.11 | 0.25 | 0.22 | 0.54 |
| hd016150 | S | 6145 | 2.87 | 3.00 | 0.35 |  | -0.43 | 0.13 | -0.05 | 0.40 | -0.43 | 0.24 | 0.65 | 0.39 | -0.22 | -0.02 | 0.27 | 0.25 |  | -0.41 | 1.41 | 0.15 | 1.03 | -0.32 | 0.97 | 0.47 | 0.02 |  | -0.43 |
| hd016314 | S | 6533 | 3.29 | 3.59 | 0.49 | 0.06 | 0.24 | 0.22 | 0.33 | 0.12 | 0.21 | 0.21 | 0.26 | 0.23 | 0.00 | 0.07 | 0.50 | 0.18 | -0.30 | 0.35 |  | 0.18 | -0.16 | -0.30 | 0.46 | 0.17 | 0.24 | 0.30 | 0.01 |
| hd017001 | S | 4863 | 2.44 | 1.43 | 0.07 | -0.03 | 0.11 | 0.08 | 0.49 | -0.03 | -0.13 | -0.08 | -0.06 | -0.01 | -0.11 | -0.12 | -0.13 | -0.09 | -0.03 | 0.18 | 0.21 | 0.00 | -0.33 | 0.12 | 0.13 | 0.00 | 0.10 | 0.09 | 0.08 |
| hd019210 | S | 4952 | 2.81 | 1.37 | 0.30 | 0.18 | 0.32 | 0.25 | 0.54 | 0.25 | 0.21 | 0.26 | 0.28 | 0.26 | 0.25 | 0.16 | 0.17 | 0.21 | 0.36 | 0.50 | 0.32 | 0.34 | 0.06 | 0.35 | 0.59 | 0.36 | 0.48 | 0.54 | 0.31 |
| hd021585 | S | 5204 | 2.54 | 1.68 | -0.64 | -0.35 | -0.36 | -0.41 | -0.39 | -0.54 | -0.53 | -0.48 | -0.61 | -0.67 | -0.91 | -0.70 | -0.51 | -0.69 | -0.81 | -0.49 | 0.13 | -0.73 | -0.70 | -0.62 | -0.58 | -0.71 | -0.51 | -0.67 | -0.38 |
| hd021760 | S | 4627 | 2.52 | 1.26 | 0.24 | 0.25 | 0.30 | 0.29 | 0.86 | 0.09 | -0.02 | 0.05 | 0.18 | 0.14 | 0.09 | 0.01 | 0.03 | 0.13 | 0.32 | 0.62 | 0.08 | -0.04 | -0.38 | -0.18 | 0.39 | 0.07 | 0.19 | 0.78 | -0.05 |
| hd025627 | S | 4613 | 2.61 | 1.11 | 0.41 | 0.27 | 0.52 | 0.42 | 1.05 | 0.27 | 0.25 | 0.32 | 0.52 | 0.36 | 0.31 | 0.18 | 0.22 | 0.34 | 0.53 | 1.38 | 0.26 | 0.33 | -0.16 | 0.05 | 0.79 | 0.36 | 0.59 | 0.98 | 0.19 |
| hd026004 | S | 4490 | 1.83 | 1.44 | 0.05 | -0.12 | 0.10 | 0.08 | 0.55 | -0.06 | -0.26 | -0.12 | -0.03 | -0.06 | -0.15 | -0.19 | -0.16 | -0.11 | 0.06 | -0.01 | 0.08 | -0.02 | -0.51 | -0.07 | 0.14 | -0.01 | 0.10 | 0.45 | -0.01 |
| hd026625 | S | 4898 | 2.83 | 1.23 | 0.31 | 0.24 | 0.28 | 0.29 | 0.56 | 0.19 | 0.08 | 0.10 | 0.13 | 0.22 | 0.14 | 0.10 | 0.05 | 0.15 | 0.20 | 0.32 | 0.26 | 0.25 | -0.17 | 0.17 | 0.58 | 0.17 | 0.33 | 0.57 | 0.20 |
| hd033844 | S | 4792 | 2.89 | 1.02 | 0.57 | 0.35 | 0.54 | 0.43 | 1.00 | 0.34 | 0.22 | 0.33 | 0.49 | 0.37 | 0.38 | 0.23 | 0.24 | 0.38 | 0.67 | 0.90 | 0.33 | 0.36 | -0.10 | 0.07 | 0.63 | 0.18 | 0.38 | 1.02 | 0.22 |
| hd058898 | S | 4387 | 1.85 | 1.45 | 0.36 | 0.13 | 0.37 | 0.30 | 1.14 | 0.11 | -0.01 | 0.09 | 0.21 | 0.21 | 0.15 | 0.00 | 0.07 | 0.14 | 0.38 | 1.37 | 0.14 | 0.04 | -0.47 | -0.23 | 0.46 | 0.01 | 0.28 | 0.92 | 0.00 |
| hd061191 | S | 4688 | 2.57 | 1.24 | 0.34 | 0.01 | 0.32 | 0.28 | 0.81 | 0.14 | 0.02 | 0.14 | 0.24 | 0.22 | 0.15 | 0.07 | 0.07 | 0.16 | 0.45 | 0.78 | 0.07 | 0.15 | -0.18 | 0.03 | 0.44 | 0.08 | 0.24 | 0.68 | 0.14 |
| hd068667 | S | 5010 | 2.74 | 1.34 | 0.25 | 0.05 | 0.18 | 0.17 | 0.61 | 0.14 | 0.01 | 0.03 | 0.02 | 0.12 | -0.01 | 0.02 | -0.03 | 0.03 | 0.00 | 0.09 | 0.34 | 0.07 | -0.16 | 0.24 | 0.27 | 0.19 | 0.22 | 0.24 | 0.23 |
| hd070522 | S | 6122 | 3.71 | 1.97 | 0.14 | 0.04 | 0.18 | 0.16 | 0.37 | 0.14 | 0.29 | 0.16 | 0.12 | 0.15 | 0.01 | 0.01 | 0.24 | 0.06 | -0.16 | 0.10 |  | 0.15 | 0.26 | -0.04 | 0.03 | -0.12 | 0.07 | 0.21 | 0.17 |
| hd077232 | S | 6773 | 3.64 | 7.75 | -0.37 | -0.12 |  | 0.05 |  | -1.88 | 0.24 | 0.81 | 1.07 | 0.86 | -0.13 | 0.00 | 1.14 | 0.35 |  |  | 1.55 | 2.61 |  | 0.76 | 0.63 | 0.48 |  |  | 0.48 |
| hd083087 | S | 4777 | 2.69 | 1.19 | 0.18 | -0.06 | 0.21 | 0.16 | 0.54 | 0.06 | -0.10 | 0.03 | 0.11 | 0.10 | 0.05 | -0.02 | -0.01 | 0.05 | 0.19 | 0.70 | 0.10 | -0.04 | -0.39 | -0.09 | 0.25 | -0.02 | 0.21 | 0.39 | 0.07 |
| hd085440 | S | 5041 | 3.07 | 1.25 | -0.06 | 0.01 | 0.04 | 0.01 | 0.14 | -0.04 | -0.17 | -0.10 | -0.01 | -0.11 | -0.13 | -0.15 | -0.14 | -0.13 | -0.11 | 0.10 | -0.02 | -0.32 | -0.02 | 0.10 | 0.08 | 0.07 | 0.46 | -0.09 |  |
| hd089280 | S | 7384 | 3.78 | 5.18 | -0.23 |  | 0.36 | -0.04 | -0.71 | -0.52 | -0.02 | 0.11 | 1.05 | 0.65 | 0.48 | -0.85 | 1.09 | 0.17 |  |  | -0.17 | 0.35 |  | 0.23 | 0.49 | 0.99 | 0.75 |  |  |
| hd090250 | S | 4691 | 2.34 | 1.45 | 0.22 | 0.16 | 0.33 | 0.30 | 0.77 | 0.10 | 0.01 | 0.08 | 0.17 | 0.13 | 0.17 | 0.03 | 0.05 | 0.13 | 0.36 | 0.75 | 0.16 | -0.06 | -0.31 | -0.01 | 0.33 | 0.02 | 0.21 | 0.40 | 0.08 |
| hd098579 | S | 4665 | 2.53 | 1.26 | 0.37 | 0.25 | 0.43 | 0.31 | 0.83 | 0.24 | 0.13 | 0.26 | 0.43 | 0.28 | 0.26 | 0.11 | 0.16 | 0.24 | 0.67 | 1.09 | 0.18 | 0.08 | -0.23 | -0.07 | 0.69 | 0.16 | 0.45 | 0.94 | 0.16 |
| hd101321 | S | 4757 | 2.59 | 1.24 | -0.03 | -0.08 | 0.11 | 0.03 | 0.34 | -0.09 | -0.17 | -0.10 | -0.10 | -0.11 | -0.17 | -0.19 | -0.16 | -0.12 | 0.03 | 0.13 | -0.01 | -0.18 | -0.46 | -0.28 | 0.03 | -0.10 | 0.06 | 0.27 | -0.08 |
| hd104819 | S | 4618 | 2.50 | 1.55 | 0.79 | 0.50 | 0.64 | 0.62 | 1.34 | 0.43 | 0.25 | 0.36 | 0.55 | 0.49 | 0.41 | 0.25 | 0.43 | 0.43 | 0.67 | 1.64 | 0.36 | 0.37 | -0.04 | -0.21 | 0.69 | 0.35 | 0.52 | 1.59 | 0.40 |
| hd104883 | S | 6203 | 3.46 | 4.11 |  | -0.73 | 0.25 | 0.31 | -0.14 | -0.13 | 0.12 | 0.10 | 0.80 | 0.69 | -0.18 | -0.11 | 0.13 | 0.32 |  |  | 1.89 |  | -0.51 | 0.14 | 1.15 | 0.28 | 1.87 | 1.14 |  |
| hd106972 | S | 6166 | 3.58 | 2.07 | 0.14 | 0.04 | 0.14 | 0.09 | 0.19 | 0.14 | 0.29 | 0.13 | 0.03 | 0.07 | -0.05 | -0.02 | 0.01 | 0.02 | -0.08 | 0.08 |  | 0.10 | 0.49 | 0.15 | 0.17 | 0.16 | 0.09 | 0.08 | 0.19 |
| hd107415 | S | 4797 | 2.47 | 1.34 | 0.03 | -0.10 | 0.05 | 0.02 | 0.42 | -0.10 | -0.17 | -0.13 | -0.05 | -0.09 | -0.20 | -0.17 | -0.15 | -0.16 | -0.11 | 0.12 | 0.11 | -0.09 | -0.43 | 0.05 | 0.20 | 0.04 | 0.13 | 0.08 | -0.03 |

| ID | Type | col3 | col4 | col5 | col6 | col7 | col8 | col9 | col10 | col11 | col12 | col13 | col14 | col15 | col16 | col17 | col18 | col19 | col20 | col21 | col22 | col23 | col24 | col25 | col26 | col27 | col28 | col29 | col30 | col31 |
|---|---|---|---|---|---|---|---|---|---|---|---|---|---|---|---|---|---|---|---|---|---|---|---|---|---|---|---|---|---|---|
| hd107569 | S | 6101 | 3.55 | 2.90 | 0.56 | -0.76 | 0.36 | 0.30 | 0.26 | 0.15 | 0.13 | 0.30 | 0.25 | 0.19 | 0.04 | 0.07 | 0.32 | 0.17 | -0.33 | 0.72 | 0.65 | 0.07 | 0.43 | -0.10 | 0.43 | 0.31 | 0.33 | 0.73 | 1.56 |
| hd107610 | S | 4693 | 2.59 | 1.27 | 0.59 | 0.45 | 0.56 | 0.47 | 1.11 | 0.30 | 0.26 | 0.36 | 0.52 | 0.40 | 0.38 | 0.26 | 0.32 | 0.40 | 0.66 | 1.48 | 0.29 | 0.34 | -0.06 | 0.14 | 0.94 | 0.27 | 0.47 | 1.06 | 0.34 |
| hd112357 | S | 4969 | 3.08 | 1.16 | -0.09 | -0.16 | -0.03 | -0.04 | 0.16 | -0.13 | -0.26 | -0.17 | -0.19 | -0.13 | -0.15 | -0.22 | -0.19 | -0.18 | -0.13 | -0.06 | 0.12 | -0.22 | -0.38 | -0.25 | -0.04 | -0.05 | -0.03 | 0.29 | -0.06 |
| hd116204 | S | 4472 | 1.09 | 2.72 | 0.24 | 0.29 | 0.58 | 0.26 | 0.96 | 0.08 | -0.26 | -0.16 | -0.18 | 0.24 | -0.19 | -0.25 | -0.06 | -0.17 | | | -0.19 | -0.55 | -1.15 | -0.36 | 0.22 | -0.27 | -0.30 | 0.09 | | |
| hd126265 | S | 5868 | 3.64 | 1.49 | 0.12 | 0.01 | 0.11 | 0.10 | 0.15 | 0.06 | 0.09 | 0.01 | -0.08 | 0.04 | -0.08 | -0.02 | 0.01 | -0.02 | -0.25 | -0.01 | 0.45 | 0.07 | 0.09 | -0.02 | -0.18 | 0.22 | 0.09 | 0.05 | 0.00 |
| hd127740 | S | 6241 | 3.70 | 3.41 | 0.14 | -0.39 | -0.02 | 0.04 | 0.66 | 0.15 | -0.18 | 0.30 | 0.43 | 0.30 | -0.09 | -0.15 | 0.66 | -0.13 | -0.65 | 0.00 | 1.58 | 0.96 | 1.00 | -0.42 | 0.68 | 0.49 | 0.37 | 0.29 | 0.76 |
| hd128853 | S | 4819 | 2.73 | 1.05 | -0.04 | -0.17 | -0.01 | 0.00 | 0.40 | -0.06 | -0.28 | -0.18 | -0.19 | -0.10 | -0.22 | -0.22 | -0.28 | -0.18 | -0.12 | 0.00 | 0.11 | -0.23 | -0.54 | -0.19 | -0.11 | -0.12 | -0.04 | 0.14 | -0.20 |
| hd133194 | S | 6312 | 2.68 | 3.65 | 0.13 | -0.11 | 0.09 | 0.15 | 0.26 | 0.09 | 0.08 | 0.01 | 0.05 | 0.05 | -0.07 | -0.07 | 0.29 | -0.02 | -0.60 | 0.05 | | -0.10 | 0.49 | -0.49 | | 0.23 | -0.50 | 0.10 | 0.64 | |
| hd138085 | S | 4799 | 2.17 | 1.45 | -0.24 | -0.25 | -0.13 | -0.17 | 0.12 | -0.30 | -0.45 | -0.40 | -0.42 | -0.35 | -0.52 | -0.46 | -0.43 | -0.42 | -0.47 | -0.13 | 0.05 | -0.56 | -0.68 | -0.40 | -0.37 | -0.36 | -0.33 | -0.31 | -0.34 |
| hd138686 | S | 6257 | 2.43 | 7.75 | | | 0.15 | | 1.28 | 0.95 | 0.84 | 0.19 | 0.73 | -0.39 | -0.23 | 1.17 | 0.34 | | | 0.78 | | -0.99 | -0.35 | | | 0.94 | | | -0.56 |
| hd148317 | S | 5758 | 3.43 | 1.49 | 0.25 | 0.19 | 0.21 | 0.19 | 0.27 | 0.18 | 0.17 | 0.07 | 0.03 | 0.15 | 0.09 | 0.09 | 0.06 | 0.10 | 0.01 | 0.01 | 0.67 | 0.14 | 0.12 | 0.11 | -0.04 | 0.11 | 0.02 | -0.12 | 0.38 |
| hd149216 | S | 4295 | 0.36 | 1.39 | -0.33 | -0.40 | -0.37 | -0.16 | 0.17 | -0.53 | -0.98 | -0.99 | -1.15 | -0.54 | -0.90 | -0.69 | -0.85 | -0.60 | -0.73 | -0.40 | -0.21 | -0.85 | -1.55 | -0.84 | -1.00 | -0.97 | -0.94 | -0.94 | -0.99 |
| hd157935 | S | 6768 | 3.55 | 2.60 | -0.07 | -0.01 | -0.01 | -0.05 | 0.16 | 0.18 | 0.25 | 0.21 | 0.61 | 0.50 | -0.44 | -0.22 | 0.75 | -0.13 | | -0.19 | | 0.33 | 1.55 | 0.06 | | | 0.78 | 0.26 | -0.07 | |
| hd161502 | S | 4963 | 2.69 | 1.45 | -0.04 | -0.17 | -0.09 | -0.13 | 0.06 | -0.17 | -0.39 | -0.33 | -0.35 | -0.21 | -0.37 | -0.35 | -0.40 | -0.38 | -0.50 | -0.17 | -0.02 | -0.33 | -0.54 | -0.19 | -0.17 | -0.28 | -0.37 | -0.17 | -0.28 |
| hd167576 | S | 4506 | 2.10 | 1.70 | 0.57 | 0.47 | 0.58 | 0.60 | 1.65 | 0.40 | 0.21 | 0.36 | 0.58 | 0.48 | 0.37 | 0.28 | 0.41 | 0.45 | 0.40 | 1.99 | 0.32 | 0.25 | -0.13 | -0.09 | 1.04 | 0.40 | 0.51 | 1.15 | 0.33 |
| hd172052 | S | 5544 | 0.10 | 2.74 | 0.16 | -0.35 | -0.04 | -0.10 | 0.05 | -0.39 | -0.53 | -0.40 | -0.57 | -0.28 | -0.51 | -0.48 | -0.37 | -0.47 | | -0.48 | | -0.32 | -0.02 | | -0.97 | -1.21 | -1.15 | -1.22 | -0.91 | |
| hd173378 | S | 4855 | 2.76 | 1.19 | -0.12 | -0.22 | -0.04 | -0.09 | 0.24 | -0.18 | -0.32 | -0.21 | -0.22 | -0.20 | -0.29 | -0.31 | -0.29 | -0.24 | -0.24 | -0.13 | 0.10 | -0.33 | -0.59 | -0.28 | -0.15 | -0.22 | -0.13 | -0.02 | -0.19 |
| hd175940 | S | 4629 | 2.49 | 1.27 | 0.31 | 0.14 | 0.34 | 0.33 | 0.73 | 0.16 | 0.02 | 0.15 | 0.32 | 0.22 | 0.22 | 0.06 | 0.12 | 0.17 | 0.53 | 0.70 | 0.35 | 0.06 | -0.31 | -0.07 | 0.34 | -0.01 | 0.34 | 0.71 | 0.14 |
| hd182901 | S | 6491 | 4.19 | 2.24 | 0.41 | 0.19 | 0.31 | 0.29 | 0.37 | 0.28 | 0.43 | 0.48 | 0.40 | 0.26 | 0.09 | 0.18 | 0.50 | 0.19 | -0.25 | | | 0.39 | 0.27 | 0.14 | 0.74 | 0.12 | 0.68 | 0.41 | 1.44 | |
| hd186535 | S | 5012 | 2.80 | 1.32 | 0.25 | 0.02 | 0.20 | 0.18 | 0.57 | 0.12 | 0.01 | 0.07 | 0.02 | 0.15 | 0.09 | 0.04 | -0.01 | 0.05 | 0.03 | 0.23 | 0.38 | 0.10 | -0.14 | 0.20 | 0.30 | 0.21 | 0.30 | 0.38 | 0.27 |
| hd188993 | S | 5673 | 3.30 | 1.65 | 0.11 | 0.00 | 0.15 | 0.12 | 0.23 | 0.15 | 0.07 | 0.02 | -0.10 | 0.08 | -0.03 | -0.02 | -0.04 | 0.01 | -0.18 | -0.02 | 0.60 | 0.01 | 0.00 | 0.04 | -0.15 | 0.00 | 0.11 | -0.24 | 0.16 |
| hd189186 | S | 4899 | 2.78 | 1.12 | -0.26 | -0.19 | -0.20 | -0.19 | -0.05 | -0.28 | -0.40 | -0.36 | -0.39 | -0.36 | -0.47 | -0.43 | -0.43 | -0.40 | -0.37 | -0.12 | -0.07 | -0.47 | -0.74 | -0.33 | -0.33 | -0.32 | -0.27 | -0.40 | -0.28 |
| hd194708 | S | 6313 | 3.73 | 2.99 | | | 0.54 | -0.49 | 0.05 | 0.84 | 0.36 | 0.38 | 0.67 | 0.03 | 0.24 | 0.58 | 0.29 | -0.33 | | | 0.12 | 1.06 | | 0.53 | 0.85 | 0.59 | | | | |
| hd200925 | S | 6642 | 3.03 | 3.11 | -0.10 | -0.06 | 0.11 | 0.02 | 0.18 | 0.14 | 0.31 | 0.17 | 0.17 | 0.09 | -0.23 | -0.09 | 0.33 | -0.01 | -0.27 | 0.12 | | 0.44 | 0.83 | 0.21 | 0.40 | 0.24 | 0.15 | 0.23 | 0.35 | |
| hd204642 | S | 4660 | 2.68 | 1.12 | 0.31 | 0.28 | 0.39 | 0.33 | 0.89 | 0.24 | 0.12 | 0.20 | 0.37 | 0.28 | 0.23 | 0.13 | 0.16 | 0.24 | 0.60 | 1.17 | 0.19 | 0.21 | -0.24 | 0.02 | 0.57 | 0.08 | 0.42 | 0.96 | 0.26 |
| hd205011 | S | 4770 | 2.16 | 1.52 | 0.12 | 0.10 | 0.18 | 0.25 | 0.82 | -0.03 | -0.17 | -0.08 | -0.07 | 0.03 | -0.12 | -0.09 | -0.12 | -0.03 | 0.33 | 0.62 | 0.80 | 0.49 | 0.34 | 0.74 | 0.79 | 0.50 | 0.55 | 0.39 | 0.10 |
| hd205972 | S | 4742 | 2.82 | 1.21 | 0.25 | 0.14 | 0.29 | 0.31 | 0.89 | 0.14 | 0.12 | 0.14 | 0.27 | 0.21 | 0.16 | 0.11 | 0.16 | 0.21 | 0.37 | 0.74 | 0.14 | 0.09 | -0.24 | 0.04 | 0.61 | 0.17 | 0.41 | 0.76 | 0.09 |
| hd211607 | S | 4944 | 2.84 | 1.27 | 0.37 | 0.27 | 0.30 | 0.31 | 0.64 | 0.23 | 0.12 | 0.17 | 0.25 | 0.27 | 0.23 | 0.17 | 0.12 | 0.22 | 0.25 | 0.91 | 0.35 | 0.33 | -0.05 | 0.29 | 0.49 | 0.29 | 0.37 | 0.60 | 0.34 |
| hd212334 | S | 4722 | 2.19 | 1.45 | 0.02 | -0.02 | 0.14 | 0.12 | 0.39 | -0.02 | -0.17 | -0.05 | -0.03 | -0.05 | -0.11 | -0.11 | -0.09 | -0.05 | 0.08 | 0.16 | 0.07 | -0.24 | -0.46 | -0.16 | 0.02 | -0.12 | 0.01 | 0.25 | 0.00 |
| hd213619 | S | 6884 | 4.50 | 5.79 | -1.29 | -0.59 | 0.34 | -0.17 | -0.47 | -1.08 | 0.61 | 0.56 | 1.48 | 0.33 | | 0.08 | 1.13 | -0.07 | 0.55 | | | 0.87 | 1.07 | | 1.00 | 1.64 | 1.01 | 1.11 | 1.07 | |
| hd217590 | S | 4950 | 2.80 | 1.29 | 0.32 | 0.16 | 0.30 | 0.26 | 0.59 | 0.20 | 0.09 | 0.14 | 0.16 | 0.21 | 0.17 | 0.13 | 0.05 | 0.15 | 0.22 | 0.44 | 0.30 | 0.30 | -0.20 | 0.23 | 0.44 | 0.23 | 0.35 | 0.52 | 0.23 |
| hd219409 | S | 4634 | 2.42 | 1.27 | 0.10 | 0.13 | 0.17 | 0.18 | 0.60 | -0.01 | -0.05 | -0.02 | 0.05 | 0.06 | 0.03 | -0.06 | -0.03 | 0.03 | 0.31 | 0.54 | 0.10 | -0.08 | -0.45 | -0.14 | 0.38 | 0.02 | 0.15 | 0.52 | -0.06 |
| hd222683 | S | 4936 | 2.76 | 1.28 | 0.36 | 0.11 | 0.28 | 0.29 | 0.61 | 0.20 | 0.07 | 0.14 | 0.19 | 0.22 | 0.17 | 0.11 | 0.07 | 0.15 | 0.17 | 0.34 | 0.30 | 0.24 | -0.16 | 0.16 | 0.35 | 0.14 | 0.21 | 0.37 | 0.19 |
| hd223869 | S | 4830 | 2.98 | 1.08 | 0.14 | 0.17 | 0.21 | 0.16 | 0.51 | 0.09 | -0.07 | 0.02 | 0.10 | 0.07 | 0.03 | -0.02 | -0.01 | 0.07 | 0.18 | 0.18 | -0.03 | 0.04 | -0.34 | -0.15 | 0.10 | 0.12 | 0.23 | 0.66 | 0.07 |
| hd224349 | S | 4830 | 2.28 | 1.44 | -0.07 | -0.10 | 0.02 | 0.01 | 0.33 | -0.09 | -0.20 | -0.16 | -0.18 | -0.11 | -0.20 | -0.19 | -0.20 | -0.18 | -0.20 | -0.13 | 0.09 | -0.18 | -0.42 | -0.01 | 0.02 | -0.12 | 0.00 | 0.02 | -0.05 |
| hd225292 | S | 4940 | 2.56 | 1.37 | 0.13 | -0.03 | 0.14 | 0.09 | 0.37 | -0.02 | -0.10 | -0.06 | -0.05 | -0.02 | -0.10 | -0.10 | -0.11 | -0.07 | -0.02 | 0.04 | 0.12 | -0.08 | -0.30 | 0.04 | 0.13 | -0.05 | 0.06 | 0.04 | 0.10 |
| hr0002 | S | 4648 | 2.18 | 1.42 | 0.44 | 0.29 | 0.45 | 0.46 | 1.15 | 0.26 | 0.08 | 0.15 | 0.26 | 0.25 | 0.22 | 0.14 | 0.12 | 0.29 | 0.49 | 0.99 | 0.25 | 0.05 | -0.24 | 0.04 | 0.62 | 0.03 | 0.23 | 0.77 | 0.17 |
| hr0004 | S | 5104 | 2.73 | 1.41 | 0.24 | 0.00 | 0.18 | 0.14 | 0.48 | 0.10 | 0.01 | 0.02 | 0.00 | 0.09 | -0.06 | 0.00 | -0.04 | 0.00 | -0.09 | 0.05 | 0.36 | 0.12 | -0.03 | 0.29 | 0.29 | 0.20 | 0.22 | 0.18 | 0.17 |
| hr0016 | S | 4738 | 2.28 | 1.41 | 0.05 | 0.02 | 0.15 | 0.13 | 0.54 | 0.00 | -0.12 | -0.07 | -0.09 | -0.02 | -0.07 | -0.11 | -0.11 | -0.05 | 0.06 | 0.37 | 0.18 | 0.02 | -0.38 | 0.02 | 0.20 | -0.01 | 0.14 | 0.20 | 0.09 |
| hr0019 | S | 4898 | 2.48 | 1.42 | 0.02 | -0.03 | 0.10 | 0.05 | 0.35 | -0.03 | -0.11 | -0.04 | -0.07 | -0.05 | -0.12 | -0.13 | -0.09 | -0.09 | -0.09 | -0.10 | 0.29 | -0.06 | -0.27 | 0.08 | 0.12 | 0.03 | 0.14 | 0.16 | 0.06 |
| hr0022 | S | 4774 | 2.57 | 1.33 | 0.32 | 0.17 | 0.37 | 0.32 | 0.85 | 0.21 | 0.11 | 0.15 | 0.21 | 0.22 | 0.17 | 0.11 | 0.08 | 0.17 | 0.40 | 0.55 | 0.37 | 0.16 | -0.20 | 0.20 | 0.45 | 0.17 | 0.31 | 0.52 | 0.25 |
| hr0040 | S | 5642 | 3.94 | 0.20 | 0.20 | 0.01 | 0.29 | 0.09 | 0.16 | 0.19 | 0.43 | 0.45 | 0.48 | 0.32 | 0.20 | 0.25 | 0.27 | 0.25 | 0.10 | 0.35 | 0.81 | 0.57 | 0.43 | 0.65 | 0.96 | 0.89 | 0.92 | 0.94 | 0.85 |
| hr0059 | S | 5061 | 2.83 | 1.37 | 0.32 | 0.12 | 0.33 | 0.22 | 0.63 | 0.22 | 0.10 | 0.14 | 0.13 | 0.21 | 0.18 | 0.11 | 0.08 | 0.13 | 0.14 | 0.34 | 0.40 | 0.24 | -0.08 | 0.23 | 0.39 | 0.22 | 0.31 | 0.36 | 0.23 |
| hr0069 | S | 4888 | 2.69 | 1.37 | 0.39 | 0.22 | 0.37 | 0.37 | 0.78 | 0.27 | 0.21 | 0.27 | 0.40 | 0.31 | 0.28 | 0.21 | 0.23 | 0.28 | 0.38 | 0.68 | 0.39 | 0.32 | 0.04 | 0.30 | 0.64 | 0.29 | 0.50 | 0.79 | 0.33 |
| hr0074 | S | 4479 | 1.89 | 1.71 | 0.49 | 0.26 | 0.33 | 0.47 | 1.09 | 0.20 | 0.03 | 0.12 | 0.22 | 0.29 | 0.20 | 0.12 | 0.13 | 0.23 | 0.27 | 1.52 | 0.29 | 0.23 | -0.28 | 0.04 | 0.69 | 0.20 | 0.23 | 0.65 | 0.18 |
| hr0084 | S | 4888 | 2.40 | 1.55 | 0.60 | 0.25 | 0.37 | 0.40 | 0.83 | 0.29 | 0.13 | 0.17 | 0.26 | 0.29 | 0.29 | 0.16 | 0.16 | 0.24 | 0.27 | 0.97 | 0.40 | 0.17 | -0.11 | 0.22 | 0.63 | 0.20 | 0.22 | 0.50 | 0.33 |
| hr0101 | S | 5001 | 2.69 | 1.36 | -0.18 | -0.26 | -0.09 | -0.15 | 0.06 | -0.19 | -0.24 | -0.22 | -0.26 | -0.26 | -0.37 | -0.32 | -0.27 | -0.30 | -0.24 | -0.33 | 0.21 | -0.31 | -0.49 | -0.18 | -0.13 | -0.16 | -0.07 | -0.24 | -0.11 |
| hr0131 | S | 4701 | 2.18 | 1.35 | 0.29 | 0.12 | 0.31 | 0.34 | 0.85 | 0.17 | -0.05 | 0.02 | 0.14 | 0.18 | 0.11 | 0.06 | 0.06 | 0.11 | | 0.64 | | 0.04 | -0.45 | 0.09 | 0.35 | 0.04 | 0.15 | 0.27 | 0.09 | |
| hr0135 | S | 4792 | 2.48 | 1.42 | 0.18 | 0.13 | 0.26 | 0.21 | 0.59 | 0.10 | 0.01 | 0.06 | 0.10 | 0.09 | 0.10 | 0.00 | 0.02 | 0.07 | 0.27 | 0.45 | 0.20 | -0.03 | -0.23 | -0.03 | 0.29 | 0.05 | 0.18 | 0.41 | 0.11 |
| hr0141 | S | 4646 | 2.47 | 1.31 | 0.24 | 0.09 | 0.30 | 0.23 | 0.84 | 0.14 | 0.12 | 0.18 | 0.33 | 0.20 | 0.15 | 0.04 | 0.09 | 0.16 | 0.47 | 0.66 | 0.31 | 0.23 | -0.15 | 0.09 | 0.51 | 0.14 | 0.33 | 0.70 | 0.26 |

| ID | S | v1 | v2 | v3 | v4 | v5 | v6 | v7 | v8 | v9 | v10 | v11 | v12 | v13 | v14 | v15 | v16 | v17 | v18 | v19 | v20 | v21 | v22 | v23 | v24 | v25 | v26 | v27 | v28 |
|---|---|---|---|---|---|---|---|---|---|---|---|---|---|---|---|---|---|---|---|---|---|---|---|---|---|---|---|---|---|
| hr0156 | S | 4608 | 2.36 | 1.53 | 0.71 | 0.54 | 0.53 | 0.60 | 1.30 | 0.38 | 0.31 | 0.39 | 0.61 | 0.51 | 0.51 | 0.34 | 0.41 | 0.50 | 0.62 | 1.27 | 0.36 | 0.38 | -0.06 | 0.11 | 0.99 | 0.38 | 0.55 | 0.98 | 0.48 |
| hr0163 | S | 4870 | 2.05 | 1.53 | -0.55 | -0.25 | -0.28 | -0.29 | -0.17 | -0.48 | -0.56 | -0.52 | -0.68 | -0.61 | -0.86 | -0.66 | -0.60 | -0.63 | -0.68 | -0.31 | 0.00 | -0.74 | -0.75 | -0.70 | -0.68 | -0.69 | -0.60 | -0.67 | -0.41 |
| hr0165 | S | 4330 | 1.39 | 1.41 | 0.63 | 0.37 | 0.67 | 0.52 | 1.29 | 0.30 | 0.05 | 0.17 | 0.54 | 0.38 | 0.36 | 0.15 | 0.20 | 0.32 | | 1.73 | | 0.02 | -0.56 | -0.07 | 0.11 | 0.10 | 0.26 | 0.18 | 0.28 |
| hr0168 | S | 4555 | 1.49 | 1.99 | 0.38 | -0.01 | 0.14 | 0.36 | 0.90 | -0.05 | -0.16 | -0.10 | -0.10 | 0.11 | -0.15 | -0.03 | -0.08 | 0.00 | -0.02 | 0.79 | 0.28 | 0.11 | -0.36 | 0.12 | 0.34 | 0.00 | 0.10 | 0.24 | 0.07 |
| hr0175 | S | 5020 | 2.45 | 1.46 | 0.16 | -0.01 | 0.06 | 0.06 | 0.35 | 0.03 | -0.12 | -0.10 | -0.16 | 0.02 | -0.17 | -0.08 | -0.15 | -0.10 | -0.15 | 0.07 | 0.28 | -0.01 | -0.19 | 0.18 | 0.16 | 0.10 | 0.13 | -0.05 | 0.01 |
| hr0188 | S | 4792 | 2.16 | 1.61 | 0.60 | 0.17 | 0.23 | 0.24 | 0.67 | 0.16 | -0.10 | 0.02 | 0.05 | 0.19 | 0.01 | 0.01 | 0.01 | 0.05 | -0.17 | 0.59 | 0.25 | 0.11 | -0.23 | 0.08 | 0.20 | -0.09 | 0.00 | 0.07 | 0.08 |
| hr0213 | S | 4871 | 2.50 | 1.42 | 0.34 | 0.10 | 0.20 | 0.24 | 0.72 | 0.18 | -0.02 | 0.05 | 0.07 | 0.18 | 0.09 | 0.03 | 0.03 | 0.07 | 0.10 | 0.37 | 0.30 | 0.14 | -0.13 | 0.15 | 0.24 | 0.09 | 0.16 | 0.20 | 0.11 |
| hr0215 | S | 4584 | 0.01 | 2.65 | | | 0.23 | 0.07 | -0.38 | -0.76 | -0.39 | -0.20 | -0.36 | 0.49 | -0.54 | -0.18 | -0.25 | -0.45 | | | | -0.61 | -2.09 | | -0.28 | -1.04 | -0.99 | | -0.46 |
| hr0216 | S | 5014 | 2.73 | 1.27 | 0.35 | 0.19 | 0.32 | 0.28 | 0.54 | 0.26 | 0.06 | 0.11 | 0.12 | 0.20 | 0.20 | 0.15 | 0.11 | 0.17 | | 0.14 | | 0.34 | -0.22 | 0.41 | 0.50 | 0.30 | 0.41 | 0.33 | 0.31 |
| hr0224 | S | 3955 | -0.05 | 1.60 | 0.29 | -0.06 | 0.26 | 0.16 | 1.23 | -0.14 | -0.41 | -0.21 | 0.03 | 0.03 | -0.17 | -0.30 | -0.17 | -0.17 | 0.74 | 2.22 | 0.52 | -0.24 | -0.65 | -0.47 | -0.31 | -0.03 | 0.03 | 0.59 | -0.66 |
| hr0228 | S | 4898 | 3.01 | 1.04 | 0.17 | 0.22 | 0.22 | 0.19 | 0.58 | 0.13 | 0.03 | 0.09 | 0.17 | 0.15 | 0.13 | 0.04 | 0.02 | 0.11 | 0.21 | 0.32 | 0.18 | 0.16 | -0.19 | 0.03 | 0.54 | 0.19 | 0.31 | 0.56 | 0.09 |
| hr0230 | S | 6671 | 4.89 | 1.79 | | | 0.09 | 0.50 | | 0.13 | 0.52 | 0.86 | 1.02 | | 0.43 | 1.00 | 1.05 | | | 1.29 | 1.97 | | 1.83 | | 0.95 | | | | |
| hr0231 | S | 6732 | 4.89 | 7.63 | | | -0.23 | 1.06 | -0.72 | | 0.97 | 0.86 | 0.86 | | 0.70 | 1.02 | -0.80 | | | | | | 1.48 | 0.44 | | | | | |
| hr0249 | S | 4612 | 2.14 | 1.49 | 0.44 | 0.25 | 0.41 | 0.33 | 1.07 | 0.17 | 0.07 | 0.15 | 0.30 | 0.26 | 0.24 | 0.11 | 0.13 | 0.26 | 0.34 | 0.60 | 0.33 | 0.11 | -0.18 | -0.01 | 0.51 | 0.11 | 0.22 | 0.64 | 0.22 |
| hr0255 | S | 4842 | 2.08 | 1.51 | -0.47 | -0.35 | -0.28 | -0.28 | -0.13 | -0.43 | -0.50 | -0.49 | -0.57 | -0.52 | -0.71 | -0.56 | -0.51 | -0.54 | -0.56 | -0.41 | 0.09 | -0.65 | -0.84 | -0.56 | -0.57 | -0.67 | -0.50 | -0.67 | -0.35 |
| hr0265 | S | 4829 | 2.18 | 1.34 | -0.23 | -0.20 | -0.11 | -0.13 | 0.09 | -0.18 | -0.32 | -0.28 | -0.31 | -0.27 | -0.35 | -0.34 | -0.33 | -0.36 | | -0.15 | | -0.28 | -0.76 | -0.35 | -0.33 | -0.57 | -0.40 | -0.26 | -0.25 |
| hr0279 | S | 4841 | 2.46 | 1.35 | -0.08 | -0.06 | -0.03 | -0.03 | 0.20 | -0.08 | -0.24 | -0.20 | -0.26 | -0.17 | -0.32 | -0.25 | -0.29 | -0.22 | -0.16 | -0.06 | 0.27 | -0.23 | -0.50 | -0.08 | 0.05 | -0.15 | -0.06 | -0.18 | -0.09 |
| hr0294 | S | 4814 | 2.30 | 1.40 | -0.10 | -0.22 | -0.06 | -0.05 | 0.12 | -0.14 | -0.33 | -0.25 | -0.25 | -0.21 | -0.38 | -0.30 | -0.31 | -0.28 | -0.19 | -0.30 | 0.19 | -0.33 | -0.41 | -0.17 | -0.05 | -0.14 | -0.18 | -0.27 | -0.14 |
| hr0315 | S | 4753 | 2.55 | 1.27 | 0.15 | 0.04 | 0.19 | 0.23 | 0.64 | 0.09 | -0.07 | 0.02 | 0.10 | 0.11 | 0.07 | -0.02 | -0.03 | 0.05 | 0.21 | 0.24 | 0.16 | 0.05 | -0.29 | 0.04 | 0.28 | 0.06 | 0.22 | 0.38 | 0.08 |
| hr0320 | S | 4682 | 2.17 | 1.51 | 0.40 | 0.09 | 0.28 | 0.28 | 0.78 | 0.21 | -0.02 | 0.11 | 0.19 | 0.17 | 0.13 | 0.02 | 0.08 | 0.12 | 0.21 | 0.82 | 0.25 | 0.14 | -0.23 | -0.04 | 0.22 | -0.01 | 0.11 | 0.33 | 0.02 |
| hr0325 | S | 7461 | 4.97 | 2.68 | | | 0.08 | 0.57 | 0.56 | 1.19 | 1.03 | 0.79 | 0.75 | 1.51 | 0.50 | 1.56 | 0.58 | 0.32 | | 1.38 | 1.42 | 2.42 | | 1.24 | 1.46 | 1.24 | 1.42 | 1.55 | |
| hr0334 | S | 4543 | 2.10 | 1.43 | 0.46 | 0.27 | 0.39 | 0.36 | 1.05 | 0.25 | 0.03 | 0.17 | 0.30 | 0.29 | 0.21 | 0.10 | 0.11 | 0.24 | 0.34 | 0.85 | 0.29 | 0.22 | -0.21 | -0.02 | 0.49 | 0.10 | 0.28 | 0.67 | 0.12 |
| hr0352 | S | 4658 | 2.19 | 1.44 | 0.25 | 0.17 | 0.32 | 0.29 | 0.69 | 0.13 | -0.08 | 0.06 | 0.18 | 0.14 | 0.09 | 0.02 | 0.03 | 0.12 | 0.44 | 0.93 | 0.12 | -0.06 | -0.40 | -0.07 | 0.31 | -0.02 | 0.15 | 0.53 | -0.03 |
| hr0356 | S | 4884 | 2.55 | 1.45 | -0.10 | -0.04 | 0.10 | 0.05 | 0.31 | -0.06 | -0.07 | -0.04 | -0.06 | -0.10 | -0.13 | -0.15 | -0.11 | -0.12 | 0.05 | 0.15 | 0.23 | -0.21 | -0.51 | -0.13 | 0.00 | -0.09 | 0.02 | 0.14 | -0.01 |
| hr0367 | S | 4732 | 2.26 | 1.50 | 0.47 | 0.23 | 0.40 | 0.39 | 0.94 | 0.20 | 0.10 | 0.15 | 0.28 | 0.24 | 0.28 | 0.14 | 0.16 | 0.25 | 0.31 | 0.95 | 0.31 | 0.15 | -0.16 | 0.13 | 0.56 | 0.19 | 0.28 | 0.75 | 0.24 |
| hr0371 | S | 4573 | 2.06 | 1.52 | 0.53 | 0.38 | 0.53 | 0.54 | 1.17 | 0.32 | 0.12 | 0.19 | 0.33 | 0.34 | 0.36 | 0.21 | 0.24 | 0.36 | 0.46 | 1.62 | 0.23 | 0.19 | -0.30 | -0.03 | 0.35 | 0.15 | 0.24 | 0.79 | 0.27 |
| hr0390 | S | 4704 | 2.34 | 1.39 | 0.44 | 0.30 | 0.43 | 0.38 | 0.93 | 0.29 | 0.09 | 0.20 | 0.29 | 0.30 | 0.31 | 0.15 | 0.14 | 0.25 | 0.42 | 0.54 | 0.27 | 0.23 | -0.21 | 0.10 | 0.50 | 0.15 | 0.24 | 0.49 | 0.16 |
| hr0402 | S | 4660 | 2.17 | 1.50 | 0.05 | 0.14 | 0.15 | 0.19 | 0.62 | -0.05 | -0.13 | -0.03 | -0.02 | 0.00 | -0.11 | -0.11 | -0.06 | -0.04 | 0.16 | 0.26 | -0.01 | -0.25 | -0.53 | -0.23 | 0.22 | -0.10 | 0.01 | 0.33 | 0.03 |
| hr0406 | S | 4845 | 2.73 | 1.20 | -0.05 | -0.04 | 0.01 | 0.00 | 0.29 | -0.12 | -0.16 | -0.12 | -0.13 | -0.12 | -0.11 | -0.19 | -0.17 | -0.14 | -0.07 | 0.10 | -0.03 | -0.22 | -0.52 | -0.19 | 0.03 | -0.07 | 0.03 | 0.23 | -0.08 |
| hr0407 | S | 6455 | 3.43 | 3.14 | 0.26 | -0.03 | 0.26 | 0.11 | 0.07 | 0.02 | 0.22 | 0.30 | 0.36 | 0.21 | 0.04 | -0.05 | 0.45 | 0.15 | -0.40 | | 1.71 | 0.62 | 0.09 | -0.59 | | 0.81 | 0.34 | | 1.31 |
| hr0412 | S | 4316 | 1.50 | 1.55 | -0.02 | -0.11 | 0.12 | 0.01 | 0.67 | -0.22 | -0.34 | -0.18 | -0.07 | -0.10 | -0.22 | -0.32 | -0.23 | -0.22 | -0.12 | 0.65 | -0.07 | -0.34 | -0.80 | -0.61 | -0.07 | -0.25 | -0.11 | 0.29 | -0.37 |
| hr0414 | S | 4617 | 2.13 | 1.46 | 0.24 | 0.10 | 0.28 | 0.32 | 0.85 | 0.12 | -0.02 | 0.06 | 0.16 | 0.13 | 0.13 | 0.03 | 0.05 | 0.13 | 0.32 | 0.97 | 0.11 | -0.06 | -0.26 | -0.02 | 0.38 | 0.06 | 0.18 | 0.43 | 0.07 |
| hr0426 | S | 4688 | 2.25 | 1.50 | 0.49 | 0.27 | 0.47 | 0.41 | 0.94 | 0.28 | 0.10 | 0.22 | 0.32 | 0.31 | 0.31 | 0.19 | 0.20 | 0.30 | 0.55 | 1.05 | 0.30 | 0.23 | -0.14 | 0.09 | 0.59 | 0.22 | 0.31 | 0.63 | 0.16 |
| hr0430 | S | 4879 | 2.63 | 1.39 | 0.37 | 0.25 | 0.33 | 0.33 | 0.82 | 0.22 | 0.12 | 0.13 | 0.18 | 0.22 | 0.22 | 0.12 | 0.10 | 0.18 | 0.32 | 0.74 | 0.33 | 0.11 | -0.14 | 0.20 | 0.41 | 0.19 | 0.29 | 0.55 | 0.16 |
| hr0434 | S | 4126 | 0.93 | 1.59 | -0.02 | -0.08 | 0.11 | -0.10 | 1.07 | -0.27 | -0.32 | -0.18 | -0.07 | -0.20 | -0.54 | -0.48 | -0.31 | -0.31 | 0.34 | 0.36 | 0.08 | -0.34 | -0.76 | -0.65 | -0.27 | -0.39 | -0.08 | 0.38 | -0.38 |
| hr0437 | S | 4875 | 1.96 | 1.98 | 0.40 | 0.07 | 0.16 | 0.21 | 0.54 | 0.00 | -0.14 | -0.03 | -0.09 | 0.09 | -0.10 | -0.05 | -0.06 | -0.03 | -0.15 | 0.41 | 0.49 | -0.01 | -0.31 | 0.09 | 0.13 | 0.01 | 0.11 | -0.11 | 0.08 |
| hr0442 | S | 4736 | 2.23 | 1.43 | -0.14 | -0.05 | 0.01 | 0.01 | 0.37 | -0.16 | -0.30 | -0.23 | -0.24 | -0.24 | -0.30 | -0.31 | -0.28 | -0.26 | -0.12 | 0.02 | 0.08 | -0.40 | -0.73 | -0.36 | -0.24 | -0.37 | -0.23 | -0.13 | -0.21 |
| hr0454 | S | 4357 | 1.29 | 1.97 | 0.23 | 0.01 | 0.17 | 0.16 | 0.83 | -0.16 | -0.40 | -0.32 | -0.31 | 0.05 | -0.38 | -0.23 | -0.20 | -0.13 | 0.06 | 0.76 | -0.13 | -0.21 | -0.75 | -0.39 | -0.29 | -0.25 | -0.08 | 1.22 | -0.17 |
| hr0464 | S | 4310 | 1.57 | 1.69 | 0.37 | 0.25 | 0.42 | 0.39 | 1.13 | 0.02 | -0.05 | 0.02 | 0.19 | 0.22 | 0.15 | 0.03 | 0.08 | 0.19 | 0.50 | 1.00 | 0.18 | -0.12 | -0.48 | -0.31 | 0.50 | 0.04 | 0.30 | 0.83 | 0.00 |
| hr0510 | S | 4903 | 2.34 | 1.55 | 0.31 | 0.16 | 0.19 | 0.18 | 0.47 | 0.04 | -0.03 | -0.02 | -0.06 | 0.07 | -0.01 | -0.04 | -0.05 | -0.02 | 0.03 | 0.24 | 0.27 | 0.01 | -0.32 | 0.10 | 0.26 | 0.02 | 0.09 | 0.13 | 0.07 |
| hr0521 | S | 4855 | 2.46 | 1.43 | 0.01 | -0.06 | 0.06 | 0.02 | 0.39 | -0.03 | -0.13 | -0.10 | -0.12 | -0.06 | -0.19 | -0.14 | -0.17 | -0.11 | -0.12 | 0.24 | 0.43 | -0.09 | -0.42 | 0.09 | 0.17 | 0.10 | 0.15 | 0.04 | 0.02 |
| hr0527 | S | 4837 | 2.30 | 1.43 | 0.05 | -0.06 | 0.07 | 0.07 | 0.32 | -0.06 | -0.22 | -0.14 | -0.17 | -0.08 | -0.15 | -0.14 | -0.18 | -0.15 | -0.13 | 0.05 | 0.21 | -0.07 | -0.27 | 0.03 | 0.08 | -0.01 | 0.05 | -0.06 | -0.05 |
| hr0539 | S | 4579 | 1.90 | 1.62 | 0.26 | 0.00 | 0.19 | 0.24 | 0.69 | 0.03 | -0.19 | -0.10 | -0.02 | 0.07 | -0.05 | -0.05 | -0.09 | -0.01 | -0.03 | 0.72 | 0.23 | -0.07 | -0.27 | 0.09 | 0.39 | 0.04 | 0.18 | 0.40 | 0.02 |
| hr0557 | S | 4509 | 1.69 | 1.48 | -0.11 | -0.04 | 0.13 | 0.03 | 0.45 | -0.20 | -0.41 | -0.34 | -0.34 | -0.24 | -0.33 | -0.33 | -0.36 | -0.26 | -0.11 | 0.21 | -0.17 | -0.52 | -0.81 | -0.50 | -0.37 | -0.54 | -0.37 | -0.26 | -0.39 |
| hr0594 | S | 4880 | 2.09 | 1.47 | -0.58 | -0.34 | -0.47 | -0.40 | -0.44 | -0.54 | -0.57 | -0.63 | -0.74 | -0.62 | -0.80 | -0.67 | -0.65 | -0.66 | -0.70 | -0.51 | -0.01 | -0.76 | -1.04 | -0.63 | -0.54 | -0.66 | -0.59 | -0.68 | -0.53 |
| hr0616 | S | 5061 | 2.90 | 1.17 | 0.48 | 0.26 | 0.35 | 0.36 | 0.91 | 0.31 | 0.27 | 0.34 | 0.38 | 0.36 | 0.35 | 0.24 | 0.27 | 0.34 | 0.37 | 0.52 | 0.40 | 0.45 | 0.08 | 0.37 | 0.65 | 0.33 | 0.36 | 0.65 | 0.42 |
| hr0617 | S | 4512 | 2.06 | 1.39 | 0.10 | -0.04 | 0.12 | 0.13 | 0.79 | -0.06 | -0.11 | -0.05 | 0.02 | 0.01 | -0.06 | -0.12 | -0.09 | -0.05 | 0.33 | 0.50 | 0.05 | -0.15 | -0.39 | -0.06 | 0.42 | -0.06 | 0.18 | 0.35 | -0.04 |
| hr0619 | S | 4916 | 2.56 | 1.42 | -0.14 | -0.17 | -0.05 | -0.10 | 0.26 | -0.17 | -0.28 | -0.21 | -0.23 | -0.23 | -0.34 | -0.31 | -0.27 | -0.28 | -0.20 | -0.01 | 0.18 | -0.33 | -0.50 | -0.19 | -0.19 | -0.18 | -0.10 | -0.20 | -0.11 |
| hr0621 | S | 5110 | 2.96 | 1.34 | 0.31 | 0.13 | 0.25 | 0.22 | 0.55 | 0.20 | 0.14 | 0.19 | 0.17 | 0.22 | 0.21 | 0.12 | 0.06 | 0.14 | 0.23 | 0.36 | 0.30 | 0.26 | 0.04 | 0.34 | 0.41 | 0.37 | 0.40 | 0.43 | 0.29 |
| hr0661 | S | 4502 | 2.04 | 1.46 | 0.31 | 0.15 | 0.31 | 0.32 | 0.74 | 0.18 | 0.01 | 0.11 | 0.21 | 0.24 | 0.07 | 0.06 | 0.06 | 0.15 | 0.22 | 0.52 | 0.30 | 0.21 | -0.28 | 0.08 | 0.55 | 0.19 | 0.38 | 0.53 | 0.17 |

| ID | | | | | | | | | | | | | | | | | | | | | | | | | | | | |
|---|---|---|---|---|---|---|---|---|---|---|---|---|---|---|---|---|---|---|---|---|---|---|---|---|---|---|---|---|
| hr0666 | S | 4884 | 2.50 | 1.38 | 0.37 | 0.25 | 0.29 | 0.23 | 0.63 | 0.14 | 0.03 | 0.03 | 0.03 | 0.12 | 0.06 | 0.02 | 0.01 | 0.08 | 0.22 | 0.39 | 0.32 | 0.08 | -0.28 | 0.02 | 0.26 | 0.00 | 0.11 | 0.19 | 0.18 |
| hr0697 | S | 4865 | 2.47 | 1.50 | -0.26 | -0.03 | -0.02 | -0.07 | 0.26 | -0.22 | -0.27 | -0.16 | -0.24 | -0.34 | -0.56 | -0.42 | -0.26 | -0.34 | -0.29 | -0.06 | 0.12 | -0.33 | -0.69 | -0.52 | -0.37 | -0.40 | -0.30 | -0.18 | -0.14 |
| hr0699 | S | 3927 | -0.18 | 1.68 | 0.27 | 0.02 | 0.18 | 0.24 | 1.38 | -0.01 | -0.50 | -0.25 | -0.08 | 0.06 | -0.15 | -0.26 | -0.21 | -0.15 | 0.46 | | 0.54 | -0.31 | -0.72 | -0.29 | -0.10 | -0.17 | 0.07 | 0.74 | -0.61 |
| hr0712 | S | 4572 | 2.20 | 1.38 | 0.16 | 0.06 | 0.21 | 0.19 | 0.78 | 0.04 | -0.04 | 0.03 | 0.10 | 0.08 | 0.04 | -0.06 | -0.04 | 0.05 | 0.25 | 0.56 | 0.16 | 0.00 | -0.40 | -0.09 | 0.32 | -0.02 | 0.23 | 0.51 | -0.02 |
| hr0725 | S | 4780 | 2.70 | 1.22 | 0.21 | 0.13 | 0.26 | 0.21 | 0.60 | 0.14 | 0.03 | 0.14 | 0.19 | 0.17 | 0.12 | 0.04 | 0.05 | 0.12 | 0.33 | 0.36 | 0.33 | 0.06 | -0.20 | 0.08 | 0.42 | 0.20 | 0.33 | 0.70 | 0.13 |
| hr0726 | S | 4723 | 2.45 | 1.38 | 0.35 | 0.26 | 0.37 | 0.36 | 0.88 | 0.22 | 0.04 | 0.15 | 0.21 | 0.26 | 0.24 | 0.13 | 0.12 | 0.22 | 0.41 | 0.64 | 0.29 | 0.12 | -0.17 | 0.07 | 0.47 | 0.13 | 0.25 | 0.46 | 0.10 |
| hr0738 | S | 4622 | 2.39 | 1.26 | 0.16 | 0.19 | 0.21 | 0.21 | 0.74 | 0.05 | -0.04 | 0.03 | 0.10 | 0.11 | 0.04 | -0.01 | -0.03 | 0.08 | 0.35 | 0.34 | 0.05 | -0.03 | -0.34 | 0.02 | 0.50 | 0.05 | 0.28 | 0.42 | -0.03 |
| hr0739 | S | 4803 | 2.45 | 1.37 | 0.24 | 0.14 | 0.26 | 0.24 | 0.62 | 0.12 | -0.02 | 0.04 | 0.05 | 0.14 | 0.07 | 0.02 | -0.02 | 0.08 | 0.18 | 0.45 | 0.26 | 0.14 | -0.24 | 0.12 | 0.28 | 0.12 | 0.21 | 0.24 | 0.14 |
| hr0743 | S | 5084 | 2.88 | 1.35 | 0.23 | 0.12 | 0.09 | 0.15 | 0.36 | 0.15 | 0.04 | 0.05 | 0.00 | 0.10 | 0.01 | 0.01 | -0.03 | 0.01 | -0.02 | 0.07 | 0.27 | 0.13 | -0.15 | 0.27 | 0.30 | 0.19 | 0.26 | 0.22 | 0.27 |
| hr0766 | S | 4678 | 2.38 | 1.26 | -0.03 | -0.06 | 0.09 | 0.05 | 0.45 | -0.09 | -0.17 | -0.13 | -0.10 | -0.08 | -0.12 | -0.17 | -0.14 | -0.08 | 0.07 | 0.26 | 0.00 | -0.17 | -0.54 | -0.12 | 0.08 | -0.02 | 0.05 | 0.32 | -0.10 |
| hr0768 | S | 6268 | 3.80 | 2.90 | 0.28 | | | 0.22 | 0.65 | -0.05 | 0.18 | 0.41 | 0.54 | 0.44 | 0.39 | 0.11 | 0.42 | 0.15 | -0.66 | 0.17 | 2.05 | 0.43 | 1.40 | 0.00 | 0.21 | | 0.44 | 1.73 | 1.32 |
| hr0771 | S | 4758 | 2.21 | 1.45 | 0.22 | 0.07 | 0.20 | 0.21 | 0.80 | 0.08 | -0.09 | -0.02 | -0.03 | 0.08 | 0.02 | -0.03 | -0.06 | 0.01 | 0.13 | 0.36 | 0.21 | 0.05 | -0.44 | 0.02 | 0.24 | 0.04 | 0.15 | 0.23 | -0.02 |
| hr0808 | S | 4624 | 2.29 | 1.29 | 0.40 | 0.22 | 0.41 | 0.35 | 0.96 | 0.21 | 0.08 | 0.22 | 0.37 | 0.31 | 0.29 | 0.13 | 0.11 | 0.23 | 0.37 | 0.83 | 0.32 | 0.19 | -0.20 | 0.05 | 0.59 | 0.08 | 0.30 | 0.70 | 0.11 |
| hr0824 | S | 4628 | 2.26 | 1.46 | 0.33 | 0.26 | 0.41 | 0.34 | 0.83 | 0.19 | 0.02 | 0.14 | 0.25 | 0.22 | 0.16 | 0.07 | 0.11 | 0.19 | 0.45 | 0.78 | 0.15 | 0.04 | -0.31 | -0.01 | 0.42 | 0.14 | 0.23 | 0.64 | 0.06 |
| hr0831 | S | 6504 | 3.34 | 2.87 | 0.23 | 0.07 | 0.25 | 0.19 | 0.22 | 0.24 | 0.23 | 0.19 | 0.24 | 0.18 | 0.06 | 0.03 | 0.32 | 0.08 | -0.31 | 0.00 | | 0.25 | 0.59 | -0.13 | 0.06 | 0.25 | 0.00 | 0.28 | 0.05 |
| hr0840 | S | 6752 | 4.51 | 7.75 | | | | 0.22 | -0.39 | | | 0.44 | 0.88 | 0.97 | 0.50 | | 0.40 | 0.86 | 0.31 | | | 1.30 | | | | | 1.38 | 1.09 | 1.27 |
| hr0844 | S | 4657 | 2.25 | 1.43 | 0.17 | 0.15 | 0.26 | 0.24 | 0.68 | 0.02 | 0.01 | 0.05 | 0.09 | 0.11 | 0.04 | -0.01 | 0.03 | 0.09 | 0.37 | 0.41 | 0.09 | -0.09 | -0.37 | -0.11 | 0.36 | -0.01 | 0.15 | 0.54 | -0.03 |
| hr0850 | S | 4976 | 2.64 | 1.33 | 0.22 | 0.02 | 0.19 | 0.16 | 0.47 | 0.07 | -0.07 | -0.02 | -0.03 | 0.08 | -0.08 | -0.03 | -0.07 | -0.03 | 0.43 | 0.30 | -0.13 | -0.19 | 0.21 | 0.25 | 0.16 | 0.13 | 0.03 | 0.11 | |
| hr0856 | S | 6413 | 2.67 | 3.44 | | | -0.13 | | 0.55 | 0.23 | 0.74 | 1.47 | 1.09 | | 0.35 | 1.26 | 0.94 | | | 0.15 | 2.25 | | | | | -0.59 | -0.16 | | | |
| hr0900 | S | 4735 | 2.31 | 1.41 | 0.33 | 0.14 | 0.32 | 0.33 | 0.73 | 0.23 | 0.04 | 0.10 | 0.14 | 0.21 | 0.19 | 0.09 | 0.05 | 0.18 | 0.36 | 0.66 | 0.30 | 0.21 | -0.22 | 0.07 | 0.30 | 0.13 | 0.24 | 0.31 | 0.11 |
| hr0907 | S | 4751 | 2.30 | 1.49 | 0.10 | 0.06 | 0.19 | 0.15 | 0.55 | 0.00 | -0.05 | -0.01 | 0.03 | 0.02 | 0.00 | -0.07 | -0.04 | 0.00 | 0.23 | 0.37 | 0.06 | -0.12 | -0.38 | -0.14 | 0.11 | -0.10 | 0.04 | 0.24 | 0.05 |
| hr0908 | S | 4702 | 2.52 | 1.28 | -0.12 | -0.14 | -0.02 | -0.07 | 0.29 | -0.18 | -0.19 | -0.15 | -0.13 | -0.20 | -0.22 | -0.30 | -0.22 | -0.22 | 0.02 | 0.28 | -0.06 | -0.29 | -0.60 | -0.34 | -0.10 | -0.24 | -0.06 | 0.20 | -0.22 |
| hr0926 | S | 4705 | 2.13 | 1.50 | -0.03 | -0.08 | 0.06 | 0.07 | 0.29 | -0.10 | -0.17 | -0.15 | -0.18 | -0.10 | -0.16 | -0.20 | -0.18 | -0.15 | -0.02 | -0.26 | 0.10 | -0.08 | -0.42 | -0.07 | 0.13 | -0.05 | 0.02 | 0.04 | -0.12 |
| hr0931 | S | 4697 | 2.29 | 1.42 | 0.44 | 0.32 | 0.42 | 0.43 | 0.95 | 0.20 | 0.10 | 0.19 | 0.30 | 0.27 | 0.28 | 0.16 | 0.20 | 0.28 | 0.60 | 0.68 | 0.27 | 0.12 | -0.22 | 0.12 | 0.60 | 0.16 | 0.34 | 0.80 | 0.19 |
| hr0941 | S | 4857 | 2.67 | 1.23 | 0.44 | 0.17 | 0.40 | 0.39 | 0.86 | 0.28 | 0.06 | 0.17 | 0.20 | 0.28 | 0.26 | 0.16 | 0.09 | 0.23 | 0.33 | 0.76 | 0.28 | 0.25 | -0.20 | 0.11 | 0.43 | 0.14 | 0.25 | 0.59 | 0.24 |
| hr0946 | S | 4348 | 1.59 | 1.50 | 0.08 | 0.01 | 0.17 | 0.13 | 0.89 | -0.12 | -0.20 | -0.06 | 0.05 | 0.02 | -0.10 | -0.18 | -0.09 | -0.06 | 0.05 | 0.54 | 0.05 | -0.19 | -0.64 | -0.30 | 0.18 | -0.19 | 0.11 | 0.47 | -0.14 |
| hr0947 | S | 4586 | 1.88 | 1.59 | 0.24 | 0.02 | 0.15 | 0.23 | 0.65 | 0.02 | -0.17 | -0.10 | -0.05 | 0.05 | -0.05 | -0.07 | -0.09 | -0.03 | -0.06 | 0.49 | 0.25 | -0.08 | -0.33 | -0.03 | 0.27 | 0.03 | 0.16 | 0.38 | -0.05 |
| hr0951 | S | 4769 | 2.42 | 1.41 | 0.36 | 0.23 | 0.32 | 0.39 | 0.89 | 0.21 | 0.00 | 0.10 | 0.15 | 0.24 | 0.22 | 0.11 | 0.07 | 0.19 | 0.36 | 0.53 | 0.45 | 0.08 | -0.23 | 0.07 | 0.46 | 0.10 | 0.18 | 0.38 | 0.18 |
| hr0956 | S | 4959 | 2.85 | 1.34 | 0.45 | 0.23 | 0.35 | 0.37 | 0.72 | 0.30 | 0.19 | 0.27 | 0.28 | 0.34 | 0.30 | 0.22 | 0.21 | 0.28 | 0.42 | 0.26 | 0.36 | 0.37 | -0.01 | 0.30 | 0.77 | 0.28 | 0.43 | 0.63 | 0.33 |
| hr0992 | S | 4621 | 2.32 | 1.27 | 0.10 | 0.07 | 0.19 | 0.14 | 0.63 | 0.01 | -0.10 | -0.04 | 0.07 | 0.01 | -0.09 | -0.11 | -0.06 | 0.00 | 0.30 | 0.67 | 0.04 | -0.16 | -0.53 | -0.22 | 0.23 | 0.01 | 0.13 | 0.43 | -0.01 |
| hr0994 | S | 4960 | 2.80 | 1.19 | 0.34 | 0.18 | 0.33 | 0.23 | 0.66 | 0.20 | 0.21 | 0.24 | 0.32 | 0.28 | 0.19 | 0.13 | 0.17 | 0.20 | 0.19 | 0.65 | 0.46 | 0.35 | 0.08 | 0.33 | 0.63 | 0.35 | 0.44 | 0.63 | 0.39 |
| hr1000 | S | 4449 | 1.75 | 1.76 | 0.47 | 0.34 | 0.44 | 0.51 | 1.27 | 0.14 | 0.03 | 0.16 | 0.33 | 0.32 | 0.26 | 0.13 | 0.23 | 0.29 | | 1.35 | 0.22 | 0.01 | -0.36 | -0.32 | 0.76 | 0.08 | 0.28 | 0.75 | 0.15 |
| hr1007 | S | 4948 | 2.60 | 1.43 | 0.36 | 0.21 | 0.27 | 0.29 | 0.82 | 0.21 | 0.01 | 0.10 | 0.09 | 0.20 | 0.16 | 0.11 | 0.03 | 0.14 | 0.19 | 0.48 | 0.38 | 0.21 | -0.12 | 0.15 | 0.33 | 0.19 | 0.24 | 0.36 | 0.20 |
| hr1015 | S | 4385 | 1.77 | 1.56 | 0.54 | 0.23 | 0.45 | 0.39 | 1.17 | 0.12 | 0.08 | 0.18 | 0.33 | 0.33 | 0.20 | 0.08 | 0.15 | 0.22 | 0.39 | 1.21 | 0.26 | 0.04 | -0.28 | -0.11 | 0.75 | 0.09 | 0.40 | 0.91 | 0.09 |
| hr1022 | S | 4486 | 2.12 | 1.37 | 0.49 | 0.37 | 0.56 | 0.46 | 1.19 | 0.28 | 0.18 | 0.33 | 0.53 | 0.39 | 0.38 | 0.20 | 0.30 | 0.41 | 0.56 | 1.22 | 0.28 | 0.24 | -0.27 | 0.03 | 0.53 | 0.22 | 0.63 | 1.18 | 0.27 |
| hr1030 | S | 4982 | 2.11 | 1.67 | 0.29 | -0.09 | 0.04 | 0.08 | 0.17 | -0.11 | -0.17 | -0.19 | -0.28 | -0.07 | -0.21 | -0.17 | -0.26 | -0.18 | -0.30 | 0.02 | 0.33 | -0.04 | -0.41 | 0.16 | 0.15 | -0.08 | -0.05 | -0.29 | 0.01 |
| hr1050 | S | 4652 | 2.28 | 1.46 | 0.47 | 0.32 | 0.50 | 0.46 | 1.04 | 0.26 | 0.16 | 0.21 | 0.31 | 0.38 | 0.31 | 0.21 | 0.22 | 0.32 | 0.50 | 0.55 | 0.37 | 0.27 | -0.16 | 0.09 | 0.77 | 0.20 | 0.37 | 0.86 | 0.16 |
| hr1052 | S | 4163 | 0.84 | 1.64 | 0.11 | -0.17 | 0.06 | 0.00 | 0.96 | -0.16 | -0.46 | -0.22 | -0.14 | -0.08 | -0.25 | -0.36 | -0.30 | -0.24 | -0.18 | 0.80 | 0.08 | -0.31 | -0.73 | -0.39 | 0.07 | -0.31 | -0.03 | 0.20 | -0.48 |
| hr1060 | S | 4851 | 2.63 | 1.33 | 0.27 | 0.18 | 0.25 | 0.27 | 0.72 | 0.17 | 0.06 | 0.15 | 0.17 | 0.21 | 0.18 | 0.10 | 0.07 | 0.15 | 0.29 | 0.33 | 0.30 | 0.11 | -0.10 | 0.20 | 0.38 | 0.20 | 0.35 | 0.43 | 0.21 |
| hr1098 | S | 4984 | 2.51 | 1.40 | -0.03 | -0.12 | -0.01 | -0.06 | 0.05 | -0.09 | -0.21 | -0.15 | -0.21 | -0.13 | -0.26 | -0.21 | -0.20 | -0.22 | -0.24 | -0.13 | 0.16 | -0.14 | -0.33 | 0.05 | 0.04 | 0.08 | 0.09 | -0.07 | -0.03 |
| hr1108 | S | 4899 | 2.53 | 1.41 | 0.24 | 0.06 | 0.20 | 0.20 | 0.60 | 0.10 | 0.00 | 0.02 | 0.02 | 0.12 | 0.07 | 0.01 | -0.03 | 0.05 | 0.12 | 0.29 | 0.23 | 0.14 | -0.22 | 0.13 | 0.25 | 0.16 | 0.20 | 0.39 | 0.19 |
| hr1110 | S | 4948 | 2.73 | 1.31 | 0.31 | 0.14 | 0.27 | 0.26 | 0.73 | 0.18 | -0.01 | 0.08 | 0.08 | 0.17 | 0.14 | 0.08 | 0.01 | 0.12 | 0.16 | 0.35 | 0.31 | 0.22 | -0.19 | 0.17 | 0.36 | 0.15 | 0.23 | 0.37 | 0.21 |
| hr1117 | S | 4783 | 2.54 | 1.39 | 0.10 | 0.24 | 0.28 | 0.19 | 0.62 | 0.08 | -0.02 | 0.16 | 0.15 | 0.00 | -0.05 | -0.10 | 0.05 | 0.03 | 0.37 | 0.41 | 0.09 | -0.14 | -0.47 | -0.15 | 0.16 | -0.11 | 0.08 | 0.36 | 0.09 |
| hr1119 | S | 4901 | 2.63 | 1.38 | 0.15 | 0.06 | 0.18 | 0.13 | 0.49 | 0.09 | -0.02 | 0.05 | 0.04 | 0.08 | 0.03 | 0.00 | -0.03 | 0.02 | 0.11 | 0.24 | 0.25 | 0.13 | -0.20 | 0.16 | 0.28 | 0.18 | 0.26 | 0.20 | 0.16 |
| hr1132 | S | 4801 | 2.35 | 1.43 | -0.09 | -0.20 | 0.02 | 0.00 | 0.28 | -0.13 | -0.25 | -0.17 | -0.15 | -0.13 | -0.26 | -0.25 | -0.23 | -0.19 | -0.16 | 0.18 | 0.11 | -0.30 | -0.58 | -0.25 | -0.18 | -0.17 | -0.15 | 0.10 | -0.15 |
| hr1159 | S | 4800 | 2.30 | 1.41 | 0.02 | -0.14 | 0.10 | 0.06 | 0.31 | -0.05 | -0.19 | -0.13 | -0.14 | -0.05 | -0.19 | -0.15 | -0.18 | -0.12 | -0.16 | 0.18 | 0.11 | -0.08 | -0.46 | 0.03 | 0.03 | 0.04 | 0.08 | 0.14 | -0.07 |
| hr1256 | S | 4732 | 2.76 | 1.16 | 0.20 | 0.29 | 0.47 | 0.19 | 0.51 | 0.23 | 0.22 | 0.31 | 0.62 | 0.13 | 0.10 | 0.00 | 0.19 | 0.08 | | 0.47 | | 0.04 | -0.34 | 0.01 | 0.58 | 0.08 | 0.36 | 0.39 | 0.27 |
| hr1265 | S | 4654 | 2.10 | 1.47 | 0.19 | 0.07 | 0.17 | 0.19 | 0.74 | 0.09 | -0.12 | -0.05 | -0.02 | 0.09 | 0.00 | -0.03 | -0.08 | 0.03 | 0.22 | 0.33 | 0.11 | 0.06 | -0.35 | 0.09 | 0.31 | 0.03 | 0.18 | 0.17 | -0.02 |
| hr1267 | S | 4604 | 2.19 | 1.47 | 0.47 | 0.24 | 0.42 | 0.40 | 1.10 | 0.25 | 0.10 | 0.22 | 0.35 | 0.31 | 0.24 | 0.13 | 0.16 | 0.27 | 0.43 | 0.94 | 0.25 | 0.19 | -0.16 | 0.06 | 0.60 | 0.16 | 0.30 | 0.71 | 0.15 |
| hr1283 | S | 4736 | 2.45 | 1.40 | 0.34 | 0.28 | 0.41 | 0.33 | 0.86 | 0.24 | 0.08 | 0.18 | 0.27 | 0.24 | 0.28 | 0.12 | 0.12 | 0.22 | 0.48 | 0.51 | 0.30 | 0.09 | -0.20 | 0.09 | 0.50 | 0.09 | 0.29 | 0.52 | 0.14 |

| ID | | | | | | | | | | | | | | | | | | | | | | | | | | | | |
|---|---|---|---|---|---|---|---|---|---|---|---|---|---|---|---|---|---|---|---|---|---|---|---|---|---|---|---|---|
| hr1295 | S | 4732 | 2.39 | 1.42 | 0.29 | 0.14 | 0.34 | 0.34 | 0.76 | 0.17 | 0.01 | 0.13 | 0.20 | 0.18 | 0.17 | 0.07 | 0.06 | 0.15 | 0.35 | 0.51 | 0.24 | 0.21 | -0.36 | 0.10 | 0.40 | 0.14 | 0.23 | 0.54 | 0.07 |
| hr1301 | S | 4934 | 2.85 | 1.35 | 0.12 | 0.12 | 0.27 | 0.15 | 0.45 | 0.13 | 0.15 | 0.20 | 0.23 | 0.13 | 0.18 | 0.04 | 0.09 | 0.12 | 0.32 | 0.27 | 0.18 | 0.17 | -0.18 | 0.14 | 0.36 | 0.19 | 0.36 | 0.46 | 0.27 |
| hr1310 | S | 4627 | 2.18 | 1.47 | 0.41 | 0.31 | 0.47 | 0.39 | 1.03 | 0.19 | 0.17 | 0.23 | 0.35 | 0.30 | 0.29 | 0.17 | 0.21 | 0.32 | 0.67 | 1.20 | 0.21 | 0.20 | -0.29 | 0.06 | 0.39 | 0.14 | 0.38 | 0.93 | 0.21 |
| hr1313 | S | 4693 | 2.52 | 1.27 | 0.48 | 0.17 | 0.40 | 0.35 | 0.92 | 0.21 | 0.16 | 0.25 | 0.41 | 0.34 | 0.31 | 0.17 | 0.20 | 0.29 | 0.49 | 1.19 | 0.30 | 0.15 | -0.19 | 0.15 | 0.66 | 0.24 | 0.44 | 0.90 | 0.32 |
| hr1318 | S | 4549 | 2.20 | 1.39 | 0.56 | 0.36 | 0.52 | 0.45 | 1.23 | 0.36 | 0.20 | 0.32 | 0.47 | 0.39 | 0.32 | 0.21 | 0.26 | 0.35 | 0.54 | 1.10 | 0.35 | 0.33 | -0.18 | 0.15 | 0.79 | 0.27 | 0.56 | 0.85 | 0.25 |
| hr1319 | S | 6606 | 3.55 | 4.01 | | | 1.23 | | 0.57 | 0.43 | 1.32 | 1.28 | 0.38 | 0.54 | 0.33 | 1.55 | 0.57 | | | 1.39 | 1.66 | | 0.38 | | -0.01 | | | | |
| hr1327 | S | 5211 | 2.69 | 1.46 | 0.24 | -0.04 | 0.06 | 0.07 | 0.22 | 0.01 | -0.06 | -0.07 | -0.11 | 0.03 | -0.09 | -0.07 | -0.12 | -0.08 | -0.25 | -0.04 | 0.36 | 0.04 | -0.17 | 0.26 | 0.07 | 0.12 | 0.10 | -0.03 | 0.12 |
| hr1343 | S | 4926 | 2.51 | 1.41 | 0.35 | 0.10 | 0.27 | 0.23 | 0.76 | 0.17 | -0.03 | 0.06 | 0.06 | 0.18 | 0.11 | 0.05 | 0.01 | 0.08 | 0.05 | 0.23 | 0.34 | 0.08 | -0.18 | 0.06 | 0.29 | 0.06 | 0.09 | 0.14 | 0.14 |
| hr1346 | S | 4901 | 2.54 | 1.50 | 0.54 | 0.27 | 0.38 | 0.39 | 0.87 | 0.24 | 0.12 | 0.16 | 0.18 | 0.28 | 0.27 | 0.16 | 0.14 | 0.22 | 0.34 | 0.83 | 0.51 | 0.17 | -0.08 | 0.20 | 0.44 | 0.21 | 0.28 | 0.57 | 0.24 |
| hr1348 | S | 4457 | 1.86 | 1.43 | -0.13 | -0.22 | -0.05 | 0.00 | 0.56 | -0.24 | -0.35 | -0.24 | -0.19 | -0.23 | -0.31 | -0.36 | -0.32 | -0.27 | 0.09 | 0.27 | -0.14 | -0.33 | -0.72 | -0.39 | 0.01 | -0.23 | -0.10 | 0.07 | -0.29 |
| hr1373 | S | 4883 | 2.51 | 1.46 | 0.53 | 0.11 | 0.31 | 0.39 | 0.82 | 0.23 | 0.08 | 0.11 | 0.18 | 0.25 | 0.22 | 0.14 | 0.08 | 0.21 | 0.26 | 1.19 | 0.32 | 0.14 | -0.15 | 0.10 | 0.28 | 0.09 | 0.18 | 0.42 | 0.23 |
| hr1407 | S | 4516 | 2.08 | 1.36 | 0.17 | 0.11 | 0.24 | 0.19 | 0.78 | 0.06 | -0.12 | 0.02 | 0.11 | 0.09 | 0.03 | -0.06 | -0.01 | 0.06 | 0.28 | 0.73 | 0.04 | -0.01 | -0.42 | -0.19 | 0.28 | -0.02 | 0.18 | 0.55 | -0.02 |
| hr1409 | S | 4836 | 2.39 | 1.54 | 0.59 | 0.26 | 0.35 | 0.43 | 0.93 | 0.24 | 0.16 | 0.16 | 0.17 | 0.30 | 0.29 | 0.18 | 0.15 | 0.27 | 0.30 | 1.12 | 0.37 | 0.17 | -0.16 | 0.20 | 0.60 | 0.21 | 0.25 | 0.45 | 0.27 |
| hr1411 | S | 4948 | 2.63 | 1.44 | 0.53 | 0.22 | 0.37 | 0.35 | 0.62 | 0.22 | 0.14 | 0.17 | 0.22 | 0.27 | 0.28 | 0.16 | 0.16 | 0.21 | 0.27 | 0.76 | 0.36 | 0.19 | -0.08 | 0.21 | 0.42 | 0.19 | 0.27 | 0.55 | 0.27 |
| hr1413 | S | 4753 | 2.18 | 1.44 | 0.04 | -0.09 | 0.09 | 0.11 | 0.32 | -0.07 | -0.18 | -0.14 | -0.11 | -0.07 | -0.16 | -0.16 | -0.18 | -0.14 | -0.12 | 0.40 | 0.12 | -0.07 | -0.48 | -0.05 | 0.07 | -0.08 | 0.02 | 0.13 | -0.08 |
| hr1421 | S | 4462 | 1.92 | 1.60 | 0.49 | 0.47 | 0.50 | 0.54 | 1.24 | 0.29 | 0.19 | 0.25 | 0.44 | 0.41 | 0.32 | 0.20 | 0.28 | 0.38 | 0.69 | 1.62 | 0.29 | 0.07 | -0.33 | -0.04 | 0.77 | 0.22 | 0.38 | 1.01 | 0.34 |
| hr1425 | S | 4748 | 2.32 | 1.45 | -0.01 | -0.03 | 0.10 | 0.09 | 0.58 | -0.06 | -0.12 | -0.08 | -0.08 | -0.04 | -0.10 | -0.13 | -0.09 | -0.06 | 0.15 | 0.23 | -0.02 | -0.19 | -0.58 | -0.15 | 0.04 | -0.05 | 0.05 | 0.23 | -0.01 |
| hr1431 | S | 4814 | 2.39 | 1.43 | 0.32 | 0.11 | 0.22 | 0.24 | 0.70 | 0.14 | -0.04 | 0.02 | 0.04 | 0.14 | 0.08 | 0.03 | -0.01 | 0.07 | 0.15 | 0.36 | 0.27 | 0.04 | -0.26 | 0.05 | 0.28 | 0.06 | 0.17 | 0.22 | 0.12 |
| hr1453 | S | 4773 | 2.66 | 1.11 | -0.06 | -0.11 | 0.01 | 0.02 | 0.31 | -0.07 | -0.21 | -0.15 | -0.12 | -0.13 | -0.14 | -0.23 | -0.25 | -0.17 | -0.07 | 0.25 | 0.00 | -0.22 | -0.58 | -0.22 | -0.09 | -0.11 | -0.06 | 0.14 | -0.21 |
| hr1455 | S | 5455 | 3.78 | 0.50 | 0.13 | 0.02 | 0.21 | 0.08 | 0.31 | 0.20 | 0.30 | 0.34 | 0.43 | 0.31 | 0.22 | 0.16 | 0.32 | 0.22 | -0.01 | 0.17 | 0.78 | 0.50 | 0.33 | 0.45 | 0.79 | 0.67 | 0.72 | 0.78 | 0.63 |
| hr1514 | S | 4969 | 2.50 | 1.35 | 0.55 | 0.25 | 0.33 | 0.33 | 0.86 | 0.30 | 0.18 | 0.25 | 0.29 | 0.33 | 0.31 | 0.21 | 0.19 | 0.27 | 0.19 | 0.79 | 0.41 | 0.31 | 0.01 | 0.33 | 0.57 | 0.27 | 0.32 | 0.58 | 0.30 |
| hr1517 | S | 4382 | 1.79 | 1.53 | 0.57 | 0.26 | 0.43 | 0.56 | 1.30 | 0.20 | 0.04 | 0.12 | 0.28 | 0.31 | 0.26 | 0.13 | 0.20 | 0.31 | 0.21 | 1.46 | 0.19 | 0.03 | -0.33 | -0.21 | 0.68 | 0.07 | 0.31 | 0.89 | 0.17 |
| hr1529 | S | 4416 | 2.27 | 1.42 | 0.14 | | 0.22 | 0.36 | 1.15 | -0.02 | 0.07 | -0.03 | 0.37 | -0.05 | -0.24 | -0.07 | -0.03 | 0.07 | 0.35 | | -0.11 | 0.07 | -0.22 | | 0.23 | -0.10 | 0.28 | 0.73 | |
| hr1533 | S | 4043 | 0.62 | 1.68 | -0.12 | -0.28 | -0.12 | -0.01 | 0.90 | -0.32 | -0.46 | -0.31 | -0.17 | -0.24 | -0.50 | -0.49 | -0.38 | -0.38 | | 0.64 | 0.07 | -0.42 | -0.84 | -0.71 | -0.50 | -0.55 | -0.18 | 0.22 | -0.46 |
| hr1535 | S | 4868 | 2.60 | 1.38 | 0.21 | 0.11 | 0.23 | 0.19 | 0.50 | 0.13 | 0.01 | 0.08 | 0.08 | 0.13 | 0.08 | 0.03 | 0.00 | 0.07 | 0.19 | 0.61 | 0.27 | 0.18 | -0.21 | 0.17 | 0.34 | 0.25 | 0.26 | 0.30 | 0.16 |
| hr1549 | S | 4818 | 2.47 | 1.42 | 0.28 | 0.07 | 0.25 | 0.27 | 0.63 | 0.15 | 0.01 | 0.04 | 0.07 | 0.15 | 0.08 | 0.04 | -0.03 | 0.07 | 0.11 | 0.63 | 0.32 | 0.10 | -0.25 | 0.09 | 0.26 | 0.10 | 0.18 | 0.33 | 0.02 |
| hr1580 | S | 4466 | 1.93 | 1.39 | 0.01 | -0.13 | 0.09 | 0.07 | 0.60 | -0.11 | -0.25 | -0.11 | -0.05 | -0.08 | -0.15 | -0.25 | -0.22 | -0.15 | 0.26 | 0.23 | 0.02 | -0.29 | -0.61 | -0.28 | -0.09 | -0.20 | 0.00 | 0.34 | -0.20 |
| hr1625 | S | 4476 | 1.90 | 1.58 | 0.53 | 0.44 | 0.58 | 0.53 | 1.37 | 0.22 | 0.17 | 0.31 | 0.48 | 0.42 | 0.42 | 0.25 | 0.30 | 0.43 | 0.58 | 1.34 | 0.27 | 0.11 | -0.28 | -0.07 | 0.94 | 0.23 | 0.43 | 1.05 | 0.33 |
| hr1628 | S | 4708 | 2.45 | 1.31 | 0.37 | 0.29 | 0.38 | 0.35 | 0.81 | 0.18 | 0.10 | 0.19 | 0.26 | 0.28 | 0.24 | 0.16 | 0.12 | 0.23 | 0.53 | 0.50 | 0.34 | 0.27 | -0.22 | 0.14 | 0.61 | 0.22 | 0.31 | 0.51 | 0.18 |
| hr1654 | S | 4005 | 0.30 | 1.73 | 0.21 | 0.10 | 0.16 | 0.19 | 1.47 | -0.14 | -0.38 | -0.20 | -0.08 | 0.03 | -0.13 | -0.23 | -0.17 | -0.08 | 0.55 | | 0.34 | -0.24 | -0.66 | -0.22 | -0.02 | -0.05 | 0.11 | 0.61 | -0.47 |
| hr1681 | S | 4671 | 2.22 | 1.44 | 0.41 | 0.28 | 0.40 | 0.39 | 0.90 | 0.23 | 0.09 | 0.19 | 0.29 | 0.28 | 0.28 | 0.16 | 0.16 | 0.28 | 0.51 | 0.68 | 0.29 | 0.21 | -0.25 | 0.13 | 0.48 | 0.18 | 0.32 | 0.59 | 0.14 |
| hr1698 | S | 4533 | 2.07 | 1.72 | 0.76 | 0.28 | 0.46 | 0.62 | 1.39 | 0.27 | 0.26 | 0.30 | 0.47 | 0.46 | 0.41 | 0.26 | 0.34 | 0.41 | 0.46 | 1.95 | 0.39 | 0.39 | -0.12 | 0.18 | 0.62 | 0.39 | 0.46 | 1.00 | 0.44 |
| hr1726 | S | 4264 | 1.45 | 1.49 | -0.07 | -0.18 | 0.08 | -0.03 | 0.77 | -0.23 | -0.32 | -0.14 | -0.02 | -0.18 | -0.38 | -0.41 | -0.28 | -0.29 | 0.39 | 0.79 | -0.02 | -0.37 | -0.72 | -0.53 | -0.14 | -0.30 | -0.06 | 0.46 | -0.27 |
| hr1739 | S | 4954 | 2.66 | 1.35 | 0.33 | 0.13 | 0.18 | 0.23 | 0.62 | 0.18 | 0.02 | 0.06 | 0.06 | 0.19 | 0.11 | 0.07 | -0.03 | 0.08 | 0.13 | 0.64 | 0.26 | 0.14 | -0.24 | 0.19 | 0.26 | 0.17 | 0.24 | 0.20 | 0.23 |
| hr1784 | S | 4839 | 2.25 | 1.44 | -0.03 | -0.12 | -0.03 | -0.02 | 0.16 | -0.13 | -0.27 | -0.24 | -0.25 | -0.13 | -0.28 | -0.22 | -0.28 | -0.20 | -0.25 | -0.12 | 0.11 | -0.24 | -0.44 | 0.02 | 0.03 | -0.04 | 0.02 | -0.10 | -0.11 |
| hr1796 | S | 4624 | 2.23 | 1.50 | 0.48 | 0.30 | 0.49 | 0.50 | 1.00 | 0.21 | 0.16 | 0.25 | 0.40 | 0.32 | 0.28 | 0.19 | 0.25 | 0.35 | 0.55 | 1.62 | 0.29 | 0.16 | -0.20 | 0.05 | 0.39 | 0.27 | 0.35 | 0.79 | 0.37 |
| hr1822 | S | 6273 | 3.22 | 3.20 | 0.35 | -0.22 | 0.09 | 0.15 | 0.43 | -0.07 | 0.23 | 0.29 | 0.38 | 0.49 | -0.01 | -0.05 | 0.47 | 0.14 | -0.38 | 0.21 | 1.78 | 0.51 | 1.40 | -0.34 | 0.20 | 0.95 | 0.33 | 0.31 | 0.83 |
| hr1831 | S | 4554 | 2.43 | 1.20 | 0.29 | 0.23 | 0.37 | 0.31 | 0.95 | 0.20 | 0.13 | 0.23 | 0.43 | 0.23 | 0.17 | 0.10 | 0.13 | 0.25 | 0.48 | 0.92 | 0.17 | 0.23 | -0.25 | 0.05 | 0.56 | 0.24 | 0.50 | 0.93 | 0.18 |
| hr1889 | S | 6379 | 4.45 | 3.40 | | -0.32 | | 0.54 | -0.06 | 0.35 | 0.16 | 0.88 | 1.04 | -0.32 | | 0.01 | 1.60 | -0.20 | | | | 0.39 | | -0.98 | | | | | |
| hr1954 | S | 4549 | 2.27 | 1.46 | 0.54 | 0.52 | 0.72 | 0.54 | 1.11 | 0.43 | 0.28 | 0.42 | 0.61 | 0.48 | 0.42 | 0.29 | 0.35 | 0.43 | 0.72 | 1.14 | 0.31 | 0.25 | -0.15 | 0.13 | 0.85 | 0.29 | 0.49 | 1.03 | 0.39 |
| hr1963 | S | 4389 | 1.52 | 1.50 | -0.32 | -0.32 | -0.16 | -0.14 | 0.30 | -0.39 | -0.49 | -0.41 | -0.42 | -0.39 | -0.51 | -0.53 | -0.47 | -0.45 | -0.14 | -0.03 | -0.24 | -0.63 | -0.86 | -0.64 | -0.45 | -0.55 | -0.39 | -0.32 | -0.47 |
| hr1986 | S | 4701 | 2.57 | 1.24 | 0.27 | 0.19 | 0.31 | 0.27 | 0.79 | 0.22 | 0.04 | 0.16 | 0.23 | 0.23 | 0.19 | 0.09 | 0.08 | 0.17 | 0.44 | 0.50 | 0.20 | 0.11 | -0.21 | 0.12 | 0.56 | 0.21 | 0.28 | 0.52 | 0.14 |
| hr1987 | S | 4988 | 2.59 | 1.34 | 0.10 | -0.15 | 0.04 | 0.03 | 0.28 | -0.05 | -0.18 | -0.14 | -0.18 | -0.05 | -0.15 | -0.23 | -0.18 | -0.13 | -0.20 | 0.18 | 0.26 | -0.08 | -0.32 | 0.10 | 0.12 | 0.04 | 0.09 | 0.09 | 0.01 |
| hr1995 | S | 4876 | 2.34 | 1.43 | 0.10 | -0.05 | 0.10 | 0.10 | 0.41 | -0.02 | -0.18 | -0.12 | -0.16 | -0.04 | -0.14 | -0.13 | -0.17 | -0.12 | -0.12 | 0.14 | 0.16 | -0.11 | -0.34 | 0.08 | 0.11 | 0.04 | 0.09 | 0.01 | 0.01 |
| hr2012 | S | 4590 | 1.99 | 1.53 | 0.26 | 0.08 | 0.23 | 0.30 | 0.91 | 0.07 | -0.08 | -0.02 | 0.00 | 0.11 | 0.04 | -0.01 | -0.06 | 0.04 | 0.24 | 0.92 | 0.27 | -0.03 | -0.31 | 0.06 | 0.39 | 0.07 | 0.18 | 0.21 | 0.03 |
| hr2076 | S | 4640 | 2.23 | 1.50 | 0.52 | 0.25 | 0.44 | 0.35 | 0.92 | 0.22 | 0.04 | 0.20 | 0.31 | 0.30 | 0.24 | 0.13 | 0.17 | 0.26 | 0.40 | 1.29 | 0.30 | 0.09 | -0.30 | -0.04 | 0.65 | 0.12 | 0.22 | 0.63 | 0.08 |
| hr2077 | S | 4768 | 2.32 | 1.42 | 0.16 | 0.01 | 0.17 | 0.19 | 0.64 | 0.06 | -0.07 | -0.04 | -0.03 | 0.06 | -0.04 | -0.05 | -0.06 | 0.00 | 0.03 | 0.32 | 0.19 | 0.09 | -0.38 | 0.04 | 0.28 | 0.09 | 0.14 | 0.34 | 0.01 |
| hr2080 | S | 4540 | 2.10 | 1.60 | 0.62 | 0.43 | 0.69 | 0.59 | 1.43 | 0.28 | 0.21 | 0.35 | 0.57 | 0.44 | 0.41 | 0.26 | 0.36 | 0.45 | 0.52 | 1.84 | 0.31 | 0.32 | -0.20 | -0.02 | 0.87 | 0.34 | 0.38 | 1.12 | 0.34 |
| hr2119 | S | 4786 | 2.75 | 1.18 | -0.03 | -0.07 | 0.11 | 0.05 | 0.39 | -0.10 | -0.10 | -0.08 | -0.03 | -0.08 | -0.09 | -0.16 | -0.09 | -0.08 | 0.07 | 0.36 | -0.03 | -0.22 | -0.51 | -0.20 | 0.09 | -0.06 | 0.10 | 0.38 | -0.06 |
| hr2136 | S | 4950 | 2.55 | 1.39 | 0.17 | 0.06 | 0.14 | 0.12 | 0.53 | 0.07 | -0.08 | -0.04 | -0.05 | 0.02 | 0.00 | -0.06 | -0.10 | -0.04 | -0.01 | 0.16 | 0.19 | 0.06 | -0.37 | 0.11 | 0.21 | 0.10 | 0.13 | -0.01 | 0.06 |

| ID | | | | | | | | | | | | | | | | | | | | | | | | | | | | |
|---|---|---|---|---|---|---|---|---|---|---|---|---|---|---|---|---|---|---|---|---|---|---|---|---|---|---|---|---|
| hr2152 | S | 4609 | 2.01 | 1.59 | -0.25 | -0.20 | -0.18 | -0.17 | 0.26 | -0.38 | -0.37 | -0.36 | -0.35 | -0.38 | -0.50 | -0.46 | -0.38 | -0.40 | -0.22 | 0.24 | 0.06 | -0.48 | -0.70 | -0.42 | -0.22 | -0.33 | -0.26 | -0.26 | -0.33 |
| hr2183 | S | 4635 | 2.41 | 1.52 | 0.62 | 0.48 | 0.62 | 0.52 | 1.20 | 0.39 | 0.28 | 0.36 | 0.57 | 0.49 | 0.45 | 0.27 | 0.36 | 0.42 | 0.47 | 0.99 | 0.36 | 0.25 | -0.15 | -0.08 | 0.73 | 0.30 | 0.54 | 0.97 | 0.29 |
| hr2200 | S | 4690 | 2.34 | 1.42 | 0.35 | 0.13 | 0.27 | 0.31 | 0.85 | 0.13 | -0.01 | 0.04 | 0.10 | 0.14 | 0.07 | 0.03 | -0.02 | 0.09 | 0.25 | 0.66 | 0.18 | 0.05 | -0.24 | 0.14 | 0.55 | 0.19 | 0.29 | 0.21 | 0.13 |
| hr2218 | S | 4926 | 2.47 | 1.37 | 0.10 | -0.04 | 0.06 | 0.04 | 0.24 | -0.02 | -0.17 | -0.13 | -0.15 | -0.04 | -0.16 | -0.12 | -0.18 | -0.12 | -0.10 | 0.29 | 0.46 | -0.14 | -0.33 | 0.06 | 0.05 | 0.02 | 0.04 | -0.04 | -0.03 |
| hr2219 | S | 4685 | 2.13 | 1.47 | -0.16 | -0.07 | 0.04 | 0.00 | 0.36 | -0.14 | -0.28 | -0.21 | -0.26 | -0.21 | -0.28 | -0.30 | -0.27 | -0.23 | -0.08 | 0.07 | 0.03 | -0.43 | -0.69 | -0.39 | -0.18 | -0.36 | -0.22 | -0.18 | -0.19 |
| hr2230 | S | 5072 | 2.66 | 1.50 | 0.32 | 0.06 | 0.24 | 0.21 | 0.52 | 0.17 | 0.03 | 0.06 | 0.05 | 0.12 | 0.05 | 0.03 | 0.04 | 0.07 | 0.06 | 0.15 | 0.34 | 0.12 | -0.14 | 0.11 | 0.22 | 0.06 | 0.16 | 0.30 | 0.24 |
| hr2239 | S | 4552 | 2.16 | 1.47 | 0.29 | 0.27 | 0.45 | 0.36 | 1.12 | 0.13 | 0.07 | 0.19 | 0.29 | 0.20 | 0.10 | 0.06 | 0.13 | 0.20 | 0.42 | 0.92 | 0.09 | 0.07 | -0.45 | -0.08 | 0.42 | 0.10 | 0.20 | 0.86 | 0.28 |
| hr2243 | S | 4701 | 2.68 | 1.04 | 0.31 | 0.30 | 0.34 | 0.25 | 0.84 | 0.18 | 0.08 | 0.17 | 0.31 | 0.27 | 0.20 | 0.12 | 0.11 | 0.22 | 0.46 | 0.36 | 0.26 | 0.23 | -0.16 | 0.03 | 0.74 | 0.16 | 0.45 | 1.00 | 0.01 |
| hr2259 | S | 5076 | 3.13 | 0.83 | 0.10 | 0.02 | 0.13 | 0.08 | 0.40 | 0.12 | 0.05 | 0.11 | 0.14 | 0.13 | 0.05 | 0.03 | 0.02 | 0.05 | 0.08 | 0.12 | 0.39 | 0.26 | -0.17 | 0.30 | 0.40 | 0.30 | 0.34 | 0.56 | 0.25 |
| hr2287 | S | 6986 | 4.91 | 3.03 | | 0.37 | | 0.20 | -0.41 | 0.65 | 1.56 | 0.80 | 1.36 | 1.07 | 0.85 | 0.14 | 0.93 | 0.24 | | 0.79 | | 1.09 | | 0.02 | | | 1.58 | 1.20 | |
| hr2302 | S | 4729 | 2.31 | 1.42 | 0.12 | 0.10 | 0.21 | 0.16 | 0.68 | 0.05 | -0.10 | 0.01 | 0.03 | 0.03 | 0.01 | -0.06 | -0.05 | 0.01 | 0.14 | 0.31 | 0.10 | -0.09 | -0.37 | -0.04 | 0.19 | 0.00 | 0.11 | 0.24 | 0.05 |
| hr2333 | S | 4736 | 2.27 | 1.44 | 0.12 | -0.10 | 0.23 | 0.17 | 0.45 | 0.04 | -0.11 | -0.05 | -0.03 | 0.02 | -0.08 | -0.08 | -0.09 | -0.04 | 0.11 | 0.43 | 0.12 | 0.00 | -0.33 | 0.02 | 0.13 | 0.00 | 0.13 | 0.21 | 0.00 |
| hr2379 | S | 4345 | 1.68 | 1.60 | 0.43 | 0.26 | 0.39 | 0.39 | 1.16 | 0.17 | 0.05 | 0.11 | 0.26 | 0.24 | 0.13 | 0.06 | 0.11 | 0.18 | 0.50 | 1.06 | 0.22 | -0.04 | -0.40 | -0.10 | 0.75 | 0.15 | 0.31 | 0.66 | 0.11 |
| hr2392 | S | 4725 | 1.98 | 1.54 | 0.06 | 0.02 | 0.23 | 0.35 | 0.97 | -0.06 | -0.19 | -0.12 | -0.23 | 0.05 | -0.33 | -0.17 | -0.23 | -0.04 | -0.01 | 0.48 | 1.16 | 0.57 | 0.57 | | 1.53 | 1.08 | 1.07 | 1.25 | 0.61 |
| hr2477 | S | 4883 | 2.77 | 1.20 | 0.11 | -0.14 | 0.10 | 0.10 | 0.50 | 0.09 | -0.07 | 0.00 | -0.01 | 0.04 | -0.03 | -0.04 | -0.07 | -0.03 | 0.05 | 0.37 | 0.17 | 0.03 | -0.25 | 0.16 | 0.29 | 0.18 | 0.26 | 0.30 | 0.07 |
| hr2478 | S | 4502 | 1.86 | 1.58 | 0.20 | -0.05 | 0.13 | 0.20 | 0.61 | -0.03 | -0.11 | -0.11 | -0.04 | -0.04 | -0.13 | -0.13 | -0.08 | -0.05 | 0.13 | 0.52 | 0.09 | -0.03 | -0.52 | -0.03 | 0.27 | 0.03 | 0.15 | 0.12 | 0.02 |
| hr2552 | S | 4960 | 2.59 | 1.44 | 0.45 | 0.20 | 0.25 | 0.33 | 0.86 | 0.23 | 0.08 | 0.14 | 0.16 | 0.23 | 0.20 | 0.12 | 0.05 | 0.13 | 0.15 | 0.37 | 0.30 | 0.15 | -0.06 | 0.25 | 0.41 | 0.09 | 0.19 | 0.32 | 0.25 |
| hr2556 | S | 4994 | 2.81 | 1.29 | 0.32 | 0.14 | 0.23 | 0.22 | 0.68 | 0.22 | 0.03 | 0.08 | 0.10 | 0.18 | 0.13 | 0.03 | 0.10 | 0.13 | 0.46 | 0.29 | 0.19 | -0.15 | 0.19 | 0.34 | 0.12 | 0.22 | 0.32 | 0.23 | |
| hr2573 | S | 4738 | 2.47 | 1.39 | 0.26 | 0.20 | 0.34 | 0.29 | 0.76 | 0.19 | 0.07 | 0.16 | 0.24 | 0.22 | 0.21 | 0.09 | 0.09 | 0.18 | 0.41 | 0.76 | 0.25 | 0.06 | -0.20 | 0.11 | 0.45 | 0.13 | 0.27 | 0.47 | 0.17 |
| hr2574 | S | 4044 | 0.44 | 1.63 | 0.00 | -0.21 | 0.04 | -0.09 | 0.93 | -0.31 | -0.50 | -0.29 | -0.19 | -0.15 | -0.48 | -0.47 | -0.35 | -0.35 | 0.32 | 0.45 | 0.11 | -0.49 | -0.91 | -0.73 | -0.48 | -0.48 | -0.18 | 0.54 | -0.66 |
| hr2642 | S | 4614 | 2.36 | 1.49 | 0.36 | 0.28 | 0.48 | 0.51 | 1.19 | 0.36 | 0.26 | 0.40 | 0.56 | 0.46 | 0.39 | 0.28 | 0.32 | 0.36 | 0.48 | 1.48 | 0.27 | 0.31 | -0.12 | 0.21 | 0.77 | 0.36 | 0.55 | 1.10 | 0.49 |
| hr2684 | S | 4740 | 2.28 | 1.45 | 0.06 | -0.02 | 0.13 | 0.11 | 0.60 | -0.04 | -0.14 | -0.09 | -0.09 | -0.04 | -0.11 | -0.12 | -0.12 | -0.06 | 0.03 | 0.32 | 0.13 | -0.03 | -0.44 | -0.05 | 0.16 | -0.07 | 0.02 | 0.11 | -0.03 |
| hr2701 | S | 4692 | 2.41 | 1.21 | -0.08 | -0.13 | -0.02 | -0.02 | 0.39 | -0.09 | -0.31 | -0.20 | -0.19 | -0.17 | -0.25 | -0.28 | -0.28 | -0.22 | -0.08 | 0.21 | 0.00 | -0.21 | -0.60 | -0.26 | -0.14 | -0.18 | -0.04 | -0.04 | -0.24 |
| hr2728 | S | 4787 | 2.35 | 1.47 | 0.04 | -0.14 | 0.12 | 0.09 | 0.28 | -0.05 | -0.13 | -0.08 | -0.10 | -0.05 | -0.11 | -0.14 | -0.13 | -0.10 | 0.01 | 0.44 | 0.30 | -0.11 | -0.42 | -0.03 | 0.14 | 0.05 | 0.10 | 0.12 | 0.05 |
| hr2764 | S | 3937 | -0.49 | 2.82 | 0.21 | 0.14 | 0.06 | 0.41 | 1.67 | 0.04 | -0.53 | -0.28 | -0.18 | 0.11 | -0.17 | -0.26 | -0.31 | -0.09 | | | 0.33 | -0.32 | -0.60 | | -0.25 | -0.14 | -0.13 | 0.31 | -0.16 |
| hr2793 | S | 4575 | 2.16 | 1.38 | -0.16 | -0.32 | -0.06 | -0.08 | 0.39 | -0.17 | -0.28 | -0.24 | -0.23 | -0.25 | -0.31 | -0.37 | -0.30 | -0.29 | -0.02 | 0.14 | -0.16 | -0.43 | -0.68 | -0.46 | -0.21 | -0.28 | -0.18 | 0.05 | -0.36 |
| hr2877 | S | 4620 | 2.20 | 1.48 | 0.43 | 0.19 | 0.36 | 0.43 | 0.90 | 0.20 | 0.05 | 0.17 | 0.29 | 0.27 | 0.28 | 0.12 | 0.14 | 0.23 | 0.32 | 0.64 | 0.24 | 0.17 | -0.24 | 0.03 | 0.49 | 0.10 | 0.20 | 0.61 | 0.17 |
| hr2896 | S | 4829 | 2.41 | 1.55 | 0.48 | 0.29 | 0.35 | 0.39 | 0.75 | 0.22 | 0.01 | 0.17 | 0.16 | 0.27 | 0.22 | 0.12 | 0.12 | 0.19 | 0.15 | 0.57 | 0.35 | 0.08 | -0.15 | 0.03 | 0.15 | 0.09 | 0.18 | 0.48 | 0.14 |
| hr2899 | S | 4600 | 2.31 | 1.41 | 0.37 | 0.22 | 0.41 | 0.40 | 0.94 | 0.17 | 0.05 | 0.16 | 0.26 | 0.25 | 0.21 | 0.11 | 0.14 | 0.27 | 0.44 | 1.18 | 0.17 | 0.18 | -0.29 | 0.01 | 0.56 | 0.14 | 0.27 | 0.69 | 0.21 |
| hr2924 | S | 4997 | 2.68 | 1.39 | 0.43 | 0.16 | 0.28 | 0.23 | 0.48 | 0.15 | 0.04 | 0.07 | 0.05 | 0.16 | 0.04 | 0.05 | 0.02 | 0.07 | 0.12 | 0.37 | 0.29 | 0.14 | -0.20 | 0.13 | 0.39 | 0.08 | 0.12 | 0.15 | 0.21 |
| hr2970 | S | 4749 | 2.35 | 1.41 | 0.19 | 0.02 | 0.22 | 0.20 | 0.72 | 0.07 | -0.10 | -0.04 | -0.02 | 0.04 | 0.01 | -0.03 | -0.07 | 0.02 | 0.09 | 0.66 | 0.12 | 0.04 | -0.44 | 0.02 | 0.21 | 0.01 | 0.10 | 0.20 | 0.01 |
| hr2975 | S | 4357 | 1.59 | 1.53 | 0.29 | 0.16 | 0.31 | 0.42 | 1.18 | 0.12 | -0.13 | -0.01 | 0.14 | 0.18 | 0.11 | 0.03 | 0.04 | 0.16 | 0.30 | 0.70 | 0.07 | -0.03 | -0.61 | -0.23 | 0.40 | -0.09 | 0.07 | 0.55 | 0.02 |
| hr2985 | S | 4954 | 2.52 | 1.47 | 0.32 | 0.01 | 0.12 | 0.19 | 0.56 | 0.10 | -0.06 | -0.02 | -0.02 | 0.08 | 0.04 | -0.01 | -0.06 | 0.00 | 0.00 | 0.60 | 0.32 | 0.06 | -0.18 | 0.14 | 0.21 | 0.15 | 0.16 | 0.10 | 0.12 |
| hr2990 | S | 4821 | 2.65 | 1.25 | 0.35 | 0.31 | 0.23 | 0.33 | 0.75 | 0.22 | 0.02 | 0.14 | 0.17 | 0.22 | 0.16 | 0.11 | 0.04 | 0.16 | 0.25 | 0.54 | 0.27 | 0.20 | -0.07 | 0.12 | 0.49 | 0.17 | 0.26 | 0.41 | 0.20 |
| hr3044 | S | 4330 | 1.83 | 1.54 | 0.53 | 0.24 | 0.57 | 0.48 | 1.54 | 0.24 | 0.14 | 0.21 | 0.43 | 0.35 | 0.27 | 0.10 | 0.23 | 0.28 | 0.55 | 0.77 | 0.27 | 0.15 | -0.37 | -0.19 | 0.83 | 0.23 | 0.48 | 0.99 | 0.16 |
| hr3054 | S | 4660 | 2.38 | 1.49 | 0.52 | 0.17 | 0.45 | 0.37 | 0.94 | 0.19 | 0.09 | 0.20 | 0.32 | 0.25 | 0.24 | 0.12 | 0.15 | 0.23 | 0.47 | 0.54 | 0.22 | 0.05 | -0.23 | 0.06 | 0.53 | 0.13 | 0.27 | 0.50 | 0.23 |
| hr3094 | S | 4772 | 2.48 | 1.41 | 0.18 | 0.08 | 0.25 | 0.22 | 0.60 | 0.08 | 0.01 | 0.06 | 0.10 | 0.10 | 0.06 | 0.01 | 0.00 | 0.06 | 0.18 | 0.30 | 0.16 | 0.02 | -0.23 | 0.12 | 0.35 | 0.13 | 0.23 | 0.34 | 0.10 |
| hr3097 | S | 4865 | 2.67 | 1.35 | 0.39 | 0.14 | 0.33 | 0.30 | 0.66 | 0.22 | 0.08 | 0.15 | 0.18 | 0.23 | 0.18 | 0.10 | 0.07 | 0.15 | 0.28 | 0.53 | 0.30 | 0.22 | -0.10 | 0.15 | 0.50 | 0.21 | 0.21 | 0.52 | 0.23 |
| hr3110 | S | 5008 | 2.88 | 1.25 | 0.37 | -0.01 | 0.25 | 0.18 | 0.55 | 0.12 | 0.08 | 0.10 | 0.11 | 0.10 | 0.05 | 0.04 | 0.01 | 0.08 | 0.18 | 0.47 | 0.49 | 0.14 | -0.19 | 0.12 | 0.29 | 0.14 | 0.18 | 0.18 | 0.20 |
| hr3122 | S | 4422 | 1.72 | 1.49 | 0.02 | -0.03 | 0.05 | 0.13 | 0.63 | -0.11 | -0.23 | -0.14 | -0.10 | -0.07 | -0.17 | -0.20 | -0.17 | -0.10 | -0.03 | 0.31 | 0.05 | -0.17 | -0.51 | -0.05 | 0.21 | -0.04 | 0.15 | 0.29 | -0.01 |
| hr3125 | S | 4710 | 2.40 | 1.43 | 0.23 | 0.14 | 0.33 | 0.27 | 0.82 | 0.15 | 0.06 | 0.15 | 0.23 | 0.20 | 0.21 | 0.06 | 0.09 | 0.14 | 0.46 | 0.60 | 0.21 | 0.04 | -0.22 | 0.08 | 0.40 | 0.19 | 0.27 | 0.50 | 0.18 |
| hr3145 | S | 4280 | 1.48 | 1.54 | -0.18 | -0.35 | -0.04 | -0.01 | 0.59 | -0.28 | -0.33 | -0.26 | -0.21 | -0.26 | -0.32 | -0.42 | -0.34 | -0.32 | -0.12 | 0.23 | -0.16 | -0.37 | -0.84 | -0.52 | -0.13 | -0.35 | -0.16 | 0.06 | -0.36 |
| hr3149 | S | 4567 | 2.11 | 1.39 | 0.27 | 0.05 | 0.26 | 0.23 | 0.73 | 0.11 | -0.06 | 0.06 | 0.13 | 0.17 | 0.13 | 0.01 | 0.01 | 0.09 | 0.33 | 0.69 | 0.23 | -0.02 | -0.34 | -0.01 | 0.36 | 0.06 | 0.23 | 0.49 | 0.06 |
| hr3150 | S | 5702 | 2.99 | 1.75 | -0.05 | -0.05 | -0.14 | -0.06 | 0.05 | -0.04 | -0.09 | -0.15 | -0.23 | -0.09 | -0.24 | -0.21 | -0.26 | -0.24 | -0.33 | -0.17 | 0.72 | -0.04 | -0.11 | 0.05 | -0.08 | -0.08 | -0.10 | -0.45 | -0.08 |
| hr3211 | S | 4917 | 2.75 | 1.32 | 0.29 | 0.06 | 0.21 | 0.25 | 0.65 | 0.16 | -0.04 | 0.02 | 0.02 | 0.12 | 0.07 | 0.03 | -0.03 | 0.07 | 0.11 | 0.47 | 0.18 | 0.15 | -0.25 | 0.11 | 0.28 | 0.10 | 0.15 | 0.29 | 0.21 |
| hr3212 | S | 4988 | 2.57 | 1.34 | 0.14 | -0.04 | 0.11 | 0.08 | 0.31 | 0.02 | -0.16 | -0.10 | -0.14 | -0.04 | -0.13 | -0.10 | -0.19 | -0.11 | -0.12 | 0.22 | 0.47 | -0.04 | -0.33 | 0.14 | 0.17 | 0.11 | 0.13 | 0.06 | 0.08 |
| hr3216 | S | 4997 | 2.67 | 1.32 | 0.13 | 0.00 | 0.10 | 0.08 | 0.38 | 0.02 | -0.10 | -0.07 | -0.10 | 0.01 | -0.10 | -0.07 | -0.14 | -0.08 | -0.08 | 0.08 | 0.46 | 0.01 | -0.22 | 0.16 | 0.28 | 0.16 | 0.15 | 0.10 | 0.12 |
| hr3222 | S | 4971 | 2.89 | 1.06 | 0.26 | 0.01 | 0.30 | 0.12 | 0.53 | 0.22 | 0.11 | 0.26 | 0.41 | 0.22 | 0.14 | 0.09 | 0.10 | 0.12 | 0.15 | 0.26 | 0.46 | 0.44 | 0.10 | 0.40 | 0.56 | 0.30 | 0.45 | 0.41 | 0.30 |
| hr3263 | S | 4930 | 2.74 | 1.33 | 0.26 | 0.15 | 0.25 | 0.23 | 0.53 | 0.15 | 0.01 | 0.07 | 0.11 | 0.15 | 0.10 | 0.06 | -0.01 | 0.08 | 0.14 | 0.35 | 0.26 | 0.19 | -0.19 | 0.17 | 0.33 | 0.16 | 0.23 | 0.29 | 0.21 |
| hr3281 | S | 4660 | 2.27 | 1.39 | 0.09 | 0.17 | 0.27 | 0.21 | 0.64 | 0.02 | -0.11 | 0.05 | 0.06 | -0.06 | -0.21 | -0.14 | -0.04 | -0.01 | 0.30 | 0.28 | 0.04 | -0.21 | -0.57 | -0.23 | 0.08 | -0.09 | 0.03 | 0.47 | 0.02 |

| ID | | | | | | | | | | | | | | | | | | | | | | | | | | | | |
|---|---|---|---|---|---|---|---|---|---|---|---|---|---|---|---|---|---|---|---|---|---|---|---|---|---|---|---|---|
| hr3289 | S | 4863 | 2.55 | 1.40 | 0.08 | 0.03 | 0.10 | 0.14 | 0.46 | 0.01 | -0.06 | -0.02 | 0.02 | 0.02 | -0.03 | -0.07 | -0.08 | -0.04 | 0.05 | 0.29 | 0.24 | -0.02 | -0.40 | 0.08 | 0.22 | 0.11 | 0.19 | 0.19 | 0.09 |
| hr3303 | S | 4934 | 2.79 | 1.30 | 0.17 | -0.08 | 0.21 | 0.16 | 0.56 | 0.10 | 0.01 | 0.06 | 0.07 | 0.10 | -0.01 | 0.01 | -0.03 | 0.02 | 0.08 | 0.26 | 0.45 | 0.14 | -0.20 | 0.23 | 0.41 | 0.22 | 0.32 | 0.31 | 0.12 |
| hr3324 | S | 4490 | 1.99 | 1.40 | 0.17 | 0.08 | 0.24 | 0.20 | 0.77 | 0.02 | -0.12 | 0.01 | 0.13 | 0.07 | -0.01 | -0.08 | -0.03 | 0.02 | 0.32 | 0.93 | 0.08 | -0.12 | -0.45 | -0.16 | 0.28 | -0.01 | 0.22 | 0.49 | 0.08 |
| hr3366 | S | 4415 | 1.96 | 1.54 | 0.47 | 0.31 | 0.38 | 0.49 | 1.37 | 0.14 | 0.09 | 0.24 | 0.38 | 0.34 | 0.28 | 0.14 | 0.22 | 0.27 | 0.48 | 1.27 | 0.30 | 0.24 | -0.29 | -0.12 | 0.78 | 0.23 | 0.35 | 0.91 | 0.21 |
| hr3369 | S | 4781 | 2.18 | 1.58 | 0.32 | 0.11 | 0.21 | 0.28 | 0.84 | 0.07 | -0.05 | -0.02 | 0.00 | 0.10 | 0.01 | -0.01 | -0.06 | 0.04 | 0.03 | 0.13 | 0.23 | 0.09 | -0.32 | 0.08 | 0.27 | 0.03 | 0.09 | 0.12 | 0.12 |
| hr3376 | S | 4553 | 2.24 | 1.58 | 0.49 | 0.37 | 0.44 | 0.54 | 1.31 | 0.27 | 0.19 | 0.31 | 0.44 | 0.37 | 0.36 | 0.23 | 0.31 | 0.38 | 0.54 | 1.29 | 0.22 | 0.28 | -0.26 | 0.06 | 0.77 | 0.31 | 0.44 | 1.00 | 0.26 |
| hr3403 | S | 4438 | 1.90 | 1.49 | 0.03 | -0.10 | 0.13 | 0.18 | 0.75 | -0.09 | -0.18 | -0.09 | -0.07 | -0.06 | -0.10 | -0.17 | -0.10 | -0.06 | 0.11 | 0.40 | 0.07 | -0.17 | -0.53 | -0.04 | 0.28 | 0.01 | 0.23 | 0.49 | -0.07 |
| hr3409 | S | 4673 | 2.42 | 1.35 | 0.29 | 0.15 | 0.29 | 0.29 | 0.74 | 0.18 | 0.04 | 0.14 | 0.25 | 0.21 | 0.17 | 0.07 | 0.05 | 0.15 | 0.42 | 0.39 | 0.26 | 0.12 | -0.20 | 0.15 | 0.50 | 0.18 | 0.38 | 0.52 | 0.11 |
| hr3418 | S | 4491 | 1.83 | 1.75 | 0.48 | 0.16 | 0.32 | 0.49 | 1.16 | 0.22 | 0.03 | 0.10 | 0.19 | 0.29 | 0.16 | 0.11 | 0.12 | 0.21 | 0.21 | 1.04 | 0.36 | 0.16 | -0.28 | 0.04 | 0.57 | 0.18 | 0.24 | 0.51 | 0.15 |
| hr3423 | S | 6574 | 3.52 | 2.18 | 0.55 | 0.10 | 0.61 | 0.23 | 0.38 | 0.28 | 0.28 | 0.27 | 0.27 | 0.29 | 0.52 | 0.04 | 1.06 | 0.12 | -0.25 | | | 0.38 | 0.60 | 0.34 | 1.42 | 0.90 | 0.66 | -0.27 | 0.10 |
| hr3424 | S | 4577 | 2.00 | 1.49 | 0.21 | 0.12 | 0.20 | 0.31 | 0.84 | 0.03 | -0.15 | -0.11 | -0.05 | 0.06 | 0.01 | -0.04 | -0.07 | 0.03 | 0.22 | 0.27 | 0.08 | -0.03 | -0.56 | -0.11 | 0.15 | -0.14 | 0.01 | 0.23 | -0.09 |
| hr3433 | S | 4876 | 2.16 | 1.48 | -0.35 | -0.42 | -0.27 | -0.28 | -0.23 | -0.42 | -0.50 | -0.49 | -0.56 | -0.46 | -0.65 | -0.55 | -0.50 | -0.52 | -0.53 | -0.35 | 0.01 | -0.59 | -0.78 | -0.51 | -0.54 | -0.48 | -0.42 | -0.50 | -0.38 |
| hr3461 | S | 4636 | 2.20 | 1.47 | 0.17 | 0.14 | 0.24 | 0.31 | 0.88 | 0.05 | -0.05 | 0.04 | 0.07 | 0.08 | 0.07 | -0.01 | -0.01 | 0.09 | 0.21 | 0.90 | 0.07 | -0.09 | -0.43 | -0.05 | 0.32 | -0.06 | 0.13 | 0.40 | 0.06 |
| hr3464 | S | 4966 | 2.20 | 1.61 | 0.35 | -0.04 | 0.11 | 0.15 | 0.53 | 0.00 | -0.17 | -0.05 | -0.08 | 0.04 | -0.12 | -0.06 | -0.12 | -0.08 | -0.21 | 0.01 | 0.23 | 0.06 | -0.21 | 0.15 | 0.39 | -0.06 | 0.01 | 0.69 | 0.00 |
| hr3475 | S | 4849 | 1.89 | 1.99 | 0.46 | 0.27 | 0.38 | 0.33 | 0.63 | 0.06 | -0.07 | 0.00 | -0.10 | 0.11 | -0.08 | 0.01 | -0.05 | 0.02 | -0.01 | 0.43 | 0.02 | -0.09 | -0.21 | 0.14 | 0.11 | 0.02 | 0.02 | 0.06 | 0.05 |
| hr3484 | S | 4968 | 2.46 | 1.42 | 0.10 | -0.21 | -0.01 | 0.03 | 0.31 | 0.00 | -0.21 | -0.17 | -0.25 | -0.07 | -0.22 | -0.16 | -0.23 | -0.17 | -0.19 | 0.12 | 0.22 | -0.12 | -0.35 | 0.14 | 0.05 | 0.02 | 0.06 | -0.14 | 0.08 |
| hr3508 | S | 4761 | 2.14 | 1.43 | -0.14 | -0.16 | -0.06 | -0.07 | 0.25 | -0.22 | -0.37 | -0.32 | -0.35 | -0.27 | -0.40 | -0.35 | -0.37 | -0.31 | -0.26 | 0.12 | 0.16 | -0.33 | -0.60 | -0.24 | -0.22 | -0.27 | -0.24 | -0.30 | -0.22 |
| hr3518 | S | 4322 | 1.43 | 1.56 | 0.22 | 0.00 | 0.17 | 0.20 | 0.82 | 0.01 | -0.17 | -0.04 | 0.05 | 0.07 | -0.01 | -0.12 | -0.09 | -0.04 | 0.21 | 1.25 | 0.14 | -0.03 | -0.56 | -0.17 | 0.36 | -0.06 | 0.18 | 0.42 | -0.15 |
| hr3529 | S | 4552 | 2.16 | 1.40 | 0.26 | 0.20 | 0.32 | 0.27 | 0.89 | 0.13 | 0.02 | 0.08 | 0.22 | 0.19 | 0.17 | 0.02 | 0.04 | 0.11 | 0.21 | 0.40 | 0.16 | 0.09 | -0.33 | -0.08 | 0.34 | 0.07 | 0.23 | 0.53 | 0.02 |
| hr3531 | S | 4989 | 2.62 | 1.30 | 0.05 | -0.18 | 0.01 | 0.00 | 0.24 | -0.07 | -0.20 | -0.16 | -0.16 | -0.06 | -0.13 | -0.15 | -0.20 | -0.17 | -0.21 | 0.07 | 0.24 | -0.14 | -0.31 | 0.12 | 0.09 | 0.15 | 0.11 | 0.10 | 0.00 |
| hr3547 | S | 4795 | 2.20 | 1.53 | 0.22 | -0.05 | 0.16 | 0.19 | 0.62 | 0.02 | -0.15 | -0.10 | -0.12 | 0.02 | -0.09 | -0.08 | -0.14 | -0.07 | -0.05 | 0.58 | 0.28 | -0.10 | -0.38 | 0.18 | 0.23 | 0.10 | 0.12 | 0.08 | -0.02 |
| hr3575 | S | 4974 | 2.52 | 1.40 | 0.24 | 0.02 | 0.14 | 0.13 | 0.50 | 0.04 | -0.10 | -0.06 | -0.13 | 0.02 | -0.04 | -0.05 | -0.14 | -0.07 | -0.11 | 0.15 | 0.28 | 0.02 | -0.21 | 0.18 | 0.27 | 0.11 | 0.15 | 0.00 | 0.08 |
| hr3621 | S | 4965 | 2.53 | 1.39 | 0.17 | -0.08 | 0.09 | 0.09 | 0.48 | 0.06 | -0.14 | -0.11 | -0.16 | 0.02 | -0.10 | -0.10 | -0.17 | -0.11 | -0.18 | -0.01 | 0.23 | -0.02 | -0.33 | 0.14 | 0.23 | 0.06 | 0.01 | 0.02 | 0.04 |
| hr3640 | S | 5076 | 2.67 | 1.39 | 0.14 | 0.08 | 0.12 | 0.07 | 0.37 | 0.00 | -0.12 | -0.07 | -0.09 | 0.00 | -0.14 | -0.08 | -0.12 | -0.10 | -0.22 | -0.10 | 0.54 | -0.02 | -0.25 | 0.17 | 0.17 | 0.13 | 0.14 | -0.06 | 0.12 |
| hr3653 | S | 4825 | 2.45 | 1.37 | -0.08 | -0.02 | -0.01 | -0.01 | 0.26 | -0.13 | -0.24 | -0.17 | -0.23 | -0.17 | -0.26 | -0.26 | -0.28 | -0.23 | -0.13 | 0.00 | 0.04 | -0.30 | -0.52 | -0.14 | 0.02 | -0.19 | -0.10 | -0.14 | -0.11 |
| hr3664 | S | 5019 | 2.25 | 1.61 | -0.59 | -0.41 | -0.30 | -0.36 | -0.29 | -0.53 | -0.52 | -0.53 | -0.67 | -0.60 | -0.85 | -0.67 | -0.58 | -0.65 | -0.74 | -0.38 | -0.08 | -0.76 | -0.72 | -0.70 | -0.63 | -0.57 | -0.54 | -0.67 | -0.46 |
| hr3681 | S | 4500 | 2.03 | 1.45 | 0.31 | 0.11 | 0.28 | 0.28 | 0.98 | 0.10 | 0.02 | 0.07 | 0.23 | 0.19 | 0.17 | 0.03 | 0.04 | 0.14 | 0.25 | 0.67 | 0.19 | 0.04 | -0.34 | -0.02 | 0.48 | 0.03 | 0.32 | 0.69 | 0.10 |
| hr3687 | S | 4665 | 2.31 | 1.33 | -0.13 | -0.21 | -0.03 | -0.08 | 0.39 | -0.19 | -0.30 | -0.22 | -0.15 | -0.25 | -0.32 | -0.34 | -0.30 | -0.27 | -0.15 | -0.06 | 0.07 | -0.32 | -0.60 | -0.28 | -0.04 | -0.21 | -0.10 | -0.04 | -0.19 |
| hr3706 | S | 5003 | 2.38 | 1.61 | 0.26 | 0.00 | 0.12 | 0.13 | 0.43 | 0.05 | -0.05 | -0.06 | -0.10 | 0.02 | -0.11 | -0.06 | -0.10 | -0.07 | -0.20 | 0.22 | 0.27 | 0.01 | -0.22 | 0.20 | 0.20 | 0.11 | 0.15 | 0.03 | 0.17 |
| hr3707 | S | 4666 | 2.59 | 1.09 | 0.38 | 0.29 | 0.43 | 0.36 | 0.94 | 0.31 | 0.16 | 0.30 | 0.47 | 0.33 | 0.27 | 0.17 | 0.18 | 0.31 | 0.65 | 0.75 | 0.27 | 0.28 | -0.20 | 0.08 | 0.69 | 0.24 | 0.45 | 0.92 | 0.27 |
| hr3709 | S | 4965 | 2.66 | 1.37 | 0.25 | 0.14 | 0.18 | 0.19 | 0.60 | 0.14 | -0.05 | 0.00 | -0.02 | 0.10 | -0.03 | 0.01 | -0.06 | 0.02 | 0.00 | 0.46 | 0.24 | 0.09 | -0.26 | 0.21 | 0.19 | 0.05 | 0.17 | 0.12 | 0.13 |
| hr3731 | S | 4402 | 1.87 | 1.51 | 0.24 | 0.05 | 0.25 | 0.28 | 0.80 | 0.02 | -0.08 | 0.02 | 0.18 | 0.13 | 0.03 | -0.06 | 0.02 | 0.07 | 0.37 | 1.25 | 0.19 | -0.12 | -0.46 | -0.13 | 0.34 | 0.03 | 0.23 | 0.59 | -0.01 |
| hr3733 | S | 4959 | 2.52 | 1.40 | 0.10 | -0.03 | 0.11 | 0.05 | 0.29 | -0.04 | -0.17 | -0.13 | -0.16 | -0.05 | -0.12 | -0.14 | -0.17 | -0.13 | -0.14 | 0.08 | 0.33 | -0.04 | -0.22 | 0.04 | 0.07 | 0.03 | 0.05 | -0.02 | 0.02 |
| hr3748 | S | 4097 | 0.52 | 1.88 | 0.46 | -0.08 | 0.22 | 0.31 | 1.11 | -0.05 | -0.37 | -0.17 | -0.06 | 0.10 | -0.03 | -0.16 | -0.18 | -0.08 | 0.38 | 0.94 | 0.26 | -0.17 | -0.58 | -0.08 | 0.14 | -0.14 | 0.10 | 0.40 | -0.26 |
| hr3772 | S | 4426 | 1.93 | 1.43 | 0.47 | 0.34 | 0.59 | 0.43 | 1.19 | 0.29 | 0.12 | 0.17 | 0.45 | 0.34 | 0.24 | 0.12 | 0.22 | 0.28 | | 0.60 | 0.17 | 0.12 | -0.31 | -0.17 | 0.71 | 0.09 | 0.30 | 1.15 | 0.15 |
| hr3788 | S | 4419 | 1.81 | 1.39 | -0.02 | -0.04 | 0.14 | 0.12 | 0.65 | -0.08 | -0.15 | -0.07 | -0.03 | -0.02 | -0.22 | -0.19 | -0.14 | -0.10 | -0.01 | 0.39 | 0.04 | -0.07 | -0.56 | -0.14 | 0.17 | -0.04 | 0.16 | 0.47 | 0.02 |
| hr3791 | S | 4402 | 1.87 | 1.47 | 0.34 | 0.19 | 0.39 | 0.35 | 1.05 | 0.17 | 0.07 | 0.16 | 0.34 | 0.27 | 0.21 | 0.06 | 0.11 | 0.19 | 0.42 | 1.02 | 0.37 | 0.15 | -0.31 | -0.01 | 0.60 | 0.12 | 0.38 | 0.74 | 0.11 |
| hr3800 | S | 5003 | 2.56 | 1.51 | 0.30 | -0.01 | 0.19 | 0.14 | 0.40 | 0.08 | -0.07 | -0.01 | -0.03 | 0.10 | -0.07 | -0.03 | -0.09 | -0.04 | -0.24 | 0.16 | 0.20 | -0.03 | -0.23 | 0.19 | 0.13 | 0.12 | 0.02 | 0.07 | 0.07 |
| hr3801 | S | 4861 | 2.35 | 1.43 | 0.07 | -0.09 | 0.08 | 0.05 | 0.33 | -0.03 | -0.17 | -0.13 | -0.13 | -0.05 | -0.16 | -0.14 | -0.18 | -0.13 | -0.11 | -0.01 | 0.15 | -0.12 | -0.35 | 0.03 | 0.07 | 0.01 | 0.06 | 0.03 | -0.01 |
| hr3805 | S | 4490 | 2.01 | 1.42 | 0.42 | 0.23 | 0.40 | 0.33 | 0.91 | 0.14 | -0.01 | 0.11 | 0.25 | 0.23 | 0.15 | 0.02 | 0.08 | 0.16 | 0.31 | 0.83 | 0.17 | 0.04 | -0.32 | -0.13 | 0.43 | 0.04 | 0.23 | 0.46 | -0.02 |
| hr3808 | S | 4883 | 2.34 | 1.73 | 0.78 | 0.30 | 0.37 | 0.49 | 1.04 | 0.21 | 0.14 | 0.21 | 0.22 | 0.38 | 0.27 | 0.22 | 0.22 | 0.30 | 0.23 | 0.77 | 0.41 | 0.18 | -0.06 | 0.20 | 0.59 | 0.22 | 0.28 | 0.79 | 0.41 |
| hr3809 | S | 4808 | 2.36 | 1.41 | 0.01 | -0.11 | 0.09 | 0.08 | 0.42 | -0.02 | -0.14 | -0.11 | -0.16 | -0.06 | -0.15 | -0.13 | -0.14 | -0.11 | -0.05 | 0.18 | 0.11 | -0.07 | -0.39 | 0.05 | 0.13 | 0.02 | 0.09 | 0.07 | -0.03 |
| hr3827 | S | 4708 | 2.40 | 1.31 | 0.20 | 0.12 | 0.20 | 0.20 | 0.69 | 0.09 | -0.04 | 0.03 | 0.12 | 0.12 | 0.04 | -0.01 | -0.05 | 0.05 | 0.20 | 0.27 | 0.21 | 0.14 | -0.26 | 0.13 | 0.37 | 0.17 | 0.28 | 0.36 | 0.06 |
| hr3834 | S | 4188 | 1.16 | 1.57 | -0.06 | -0.19 | 0.05 | 0.03 | 0.90 | -0.25 | -0.38 | -0.23 | -0.08 | -0.19 | -0.37 | -0.42 | -0.31 | -0.28 | 0.01 | 1.23 | -0.04 | -0.35 | -0.82 | -0.59 | -0.16 | -0.43 | -0.15 | 0.23 | -0.45 |
| hr3845 | S | 4244 | 1.28 | 1.63 | 0.18 | -0.03 | 0.10 | 0.21 | 1.06 | -0.12 | -0.24 | -0.13 | -0.02 | 0.02 | -0.14 | -0.16 | -0.13 | -0.08 | 0.10 | | 0.16 | -0.09 | -0.59 | -0.04 | 0.36 | 0.00 | 0.22 | 0.36 | -0.17 |
| hr3851 | S | 4945 | 2.57 | 1.45 | 0.33 | -0.10 | 0.20 | 0.24 | 0.57 | 0.13 | -0.04 | 0.02 | 0.00 | 0.13 | 0.10 | 0.03 | -0.05 | 0.05 | 0.08 | 0.46 | 0.25 | 0.15 | -0.15 | 0.17 | 0.27 | 0.11 | 0.16 | 0.25 | 0.15 |
| hr3903 | S | 4955 | 2.35 | 1.57 | 0.30 | 0.04 | 0.17 | 0.19 | 0.54 | 0.07 | -0.10 | -0.05 | -0.06 | 0.09 | -0.02 | -0.03 | -0.06 | -0.03 | -0.19 | 0.48 | 0.29 | 0.04 | -0.22 | 0.17 | 0.18 | 0.02 | 0.09 | 0.02 | 0.11 |
| hr3905 | S | 4471 | 1.79 | 1.68 | 0.83 | 0.58 | 0.61 | 0.67 | 1.53 | 0.39 | 0.29 | 0.36 | 0.53 | 0.47 | 0.44 | 0.25 | 0.39 | 0.41 | 0.66 | 2.02 | 0.34 | 0.14 | -0.12 | -0.27 | 0.70 | 0.25 | 0.36 | 1.09 | 0.34 |
| hr3907 | S | 4988 | 2.58 | 1.52 | 0.39 | 0.14 | 0.24 | 0.22 | 0.69 | 0.16 | -0.03 | 0.04 | 0.08 | 0.15 | 0.02 | 0.02 | 0.02 | 0.04 | -0.14 | 0.37 | 0.26 | 0.10 | -0.14 | 0.10 | 0.17 | 0.05 | 0.01 | 0.17 | 0.19 |
| hr3908 | S | 4779 | 2.46 | 1.43 | 0.40 | 0.22 | 0.42 | 0.30 | 0.79 | 0.20 | 0.08 | 0.10 | 0.21 | 0.19 | 0.18 | 0.08 | 0.09 | 0.15 | 0.29 | 0.32 | 0.21 | 0.09 | -0.24 | 0.00 | 0.38 | 0.04 | 0.18 | 0.32 | 0.15 |

| ID | | | | | | | | | | | | | | | | | | | | | | | | | | | | |
|---|---|---|---|---|---|---|---|---|---|---|---|---|---|---|---|---|---|---|---|---|---|---|---|---|---|---|---|---|
| hr3911 | S | 4868 | 2.71 | 1.42 | 0.58 | 0.32 | 0.53 | 0.42 | 0.96 | 0.36 | 0.27 | 0.36 | 0.46 | 0.41 | 0.44 | 0.27 | 0.30 | 0.37 | 0.70 | 0.50 | 0.39 | 0.42 | -0.01 | 0.21 | 0.76 | 0.33 | 0.45 | 0.95 | 0.30 |
| hr3929 | S | 4630 | 2.16 | 1.39 | 0.44 | 0.23 | 0.41 | 0.40 | 1.03 | 0.21 | 0.03 | 0.13 | 0.21 | 0.25 | 0.26 | 0.12 | 0.18 | 0.24 | 0.48 | 0.70 | 0.24 | 0.06 | -0.26 | -0.01 | 0.33 | 0.09 | 0.14 | 0.59 | 0.07 |
| hr3942 | S | 4702 | 2.29 | 1.41 | 0.22 | 0.19 | 0.24 | 0.27 | 0.86 | 0.12 | -0.09 | 0.03 | 0.07 | 0.12 | 0.08 | 0.01 | 0.00 | 0.09 | 0.25 | 0.37 | 0.16 | 0.01 | -0.31 | 0.06 | 0.28 | 0.01 | 0.18 | 0.27 | 0.04 |
| hr3973 | S | 4958 | 2.58 | 1.44 | 0.33 | 0.11 | 0.16 | 0.18 | 0.57 | 0.18 | -0.06 | 0.03 | 0.04 | 0.15 | 0.06 | 0.02 | -0.01 | 0.05 | 0.03 | 0.63 | 0.26 | 0.14 | -0.20 | 0.12 | 0.20 | 0.04 | 0.11 | 0.15 | 0.20 |
| hr3994 | S | 4851 | 2.63 | 1.39 | 0.62 | 0.30 | 0.40 | 0.48 | 0.89 | 0.30 | 0.17 | 0.23 | 0.28 | 0.35 | 0.37 | 0.23 | 0.24 | 0.33 | 0.50 | 1.29 | 0.34 | 0.20 | -0.09 | 0.09 | 0.44 | 0.21 | 0.26 | 0.66 | 0.30 |
| hr4006 | S | 5169 | 2.68 | 1.65 | 0.30 | 0.24 | 0.16 | 0.09 | 0.18 | 0.00 | -0.05 | 0.01 | -0.09 | -0.02 | -0.19 | -0.10 | -0.09 | -0.10 | -0.27 | -0.01 | | -0.05 | -0.15 | 0.14 | 0.17 | 0.05 | 0.17 | -0.09 | 0.07 |
| hr4032 | S | 4388 | 1.74 | 1.39 | 0.23 | 0.09 | 0.27 | 0.24 | 0.92 | -0.04 | -0.12 | -0.03 | 0.11 | 0.08 | -0.06 | -0.08 | -0.03 | 0.04 | | 0.91 | 0.01 | -0.16 | -0.57 | -0.24 | 0.23 | -0.22 | 0.16 | 0.76 | -0.08 |
| hr4052 | S | 4520 | 1.97 | 1.46 | 0.04 | -0.01 | 0.06 | 0.09 | 0.50 | -0.14 | -0.18 | -0.16 | -0.11 | -0.12 | -0.22 | -0.21 | -0.14 | -0.11 | 0.22 | 0.08 | 0.01 | -0.25 | -0.62 | -0.11 | 0.13 | -0.10 | 0.04 | 0.23 | -0.03 |
| hr4057 | S | 4304 | 1.08 | 1.61 | -0.22 | -0.18 | -0.06 | -0.12 | 0.47 | -0.43 | -0.55 | -0.49 | -0.49 | -0.38 | -0.56 | -0.50 | -0.48 | -0.42 | -0.15 | 0.17 | -0.26 | -0.55 | -0.96 | -0.50 | -0.40 | -0.50 | -0.36 | -0.35 | -0.33 |
| hr4077 | S | 4652 | 2.54 | 1.32 | 0.45 | 0.30 | 0.50 | 0.38 | 1.11 | 0.30 | 0.19 | 0.29 | 0.47 | 0.34 | 0.35 | 0.19 | 0.28 | 0.34 | 0.50 | 1.35 | 0.30 | 0.20 | -0.22 | 0.01 | 0.87 | 0.32 | 0.51 | 1.02 | 0.38 |
| hr4078 | S | 4583 | 2.11 | 1.36 | -0.02 | -0.05 | 0.09 | 0.07 | 0.63 | -0.09 | -0.18 | -0.15 | -0.11 | -0.12 | -0.10 | -0.21 | -0.14 | -0.12 | 0.07 | 0.39 | -0.06 | -0.18 | -0.48 | -0.23 | 0.05 | -0.08 | 0.06 | 0.35 | -0.19 |
| hr4085 | S | 4803 | 2.68 | 1.23 | -0.12 | -0.02 | 0.11 | -0.03 | 0.29 | -0.09 | -0.14 | -0.10 | -0.06 | -0.14 | -0.13 | -0.20 | -0.15 | -0.14 | -0.02 | 0.10 | -0.06 | -0.24 | -0.41 | -0.22 | 0.05 | -0.01 | 0.07 | 0.29 | -0.08 |
| hr4097 | S | 4509 | 1.95 | 1.51 | 0.38 | 0.23 | 0.37 | 0.43 | 1.12 | 0.19 | -0.02 | 0.10 | 0.26 | 0.22 | 0.19 | 0.08 | 0.11 | 0.25 | | 1.32 | 0.17 | -0.02 | -0.45 | -0.15 | 0.51 | 0.01 | 0.15 | 0.74 | 0.10 |
| hr4100 | S | 4986 | 2.86 | 1.19 | 0.54 | 0.24 | 0.45 | 0.37 | 0.85 | 0.35 | 0.22 | 0.26 | 0.41 | 0.34 | 0.33 | 0.22 | 0.23 | 0.29 | 0.41 | 0.55 | 0.35 | 0.35 | 0.03 | 0.23 | 0.65 | 0.18 | 0.38 | 0.57 | 0.40 |
| hr4106 | S | 4635 | 2.38 | 1.44 | 0.34 | 0.24 | 0.40 | 0.41 | 0.95 | 0.20 | 0.11 | 0.23 | 0.36 | 0.27 | 0.27 | 0.15 | 0.17 | 0.28 | 0.31 | 0.59 | 0.18 | 0.09 | -0.22 | 0.06 | 0.53 | 0.14 | 0.28 | 0.84 | 0.22 |
| hr4126 | S | 4965 | 2.60 | 1.42 | 0.40 | 0.07 | 0.20 | 0.24 | 0.62 | 0.18 | 0.02 | 0.05 | 0.06 | 0.15 | 0.10 | 0.05 | 0.01 | 0.08 | 0.15 | 0.61 | 0.46 | 0.09 | -0.17 | 0.15 | 0.27 | 0.09 | 0.18 | 0.24 | 0.18 |
| hr4146 | S | 4900 | 2.39 | 1.44 | 0.04 | -0.12 | 0.00 | 0.01 | 0.08 | -0.10 | -0.24 | -0.18 | -0.23 | -0.12 | -0.24 | -0.20 | -0.24 | -0.20 | -0.23 | 0.04 | 0.37 | -0.15 | -0.45 | 0.06 | 0.12 | 0.06 | 0.04 | -0.10 | 0.00 |
| hr4171 | S | 4952 | 2.54 | 1.43 | 0.20 | 0.06 | 0.18 | 0.14 | 0.36 | 0.05 | -0.06 | -0.03 | -0.06 | 0.05 | -0.02 | -0.04 | -0.07 | -0.01 | 0.02 | 0.41 | 0.18 | 0.03 | -0.29 | 0.06 | 0.17 | 0.00 | 0.07 | 0.05 | 0.10 |
| hr4178 | S | 4425 | 1.77 | 1.54 | 0.43 | 0.38 | 0.39 | 0.47 | 1.29 | 0.17 | 0.04 | 0.19 | 0.34 | 0.28 | 0.26 | 0.15 | 0.21 | 0.32 | 0.47 | 0.88 | 0.18 | 0.14 | -0.46 | -0.04 | 0.79 | 0.10 | 0.26 | 0.92 | 0.19 |
| hr4208 | S | 4612 | 2.51 | 1.39 | 0.48 | 0.31 | 0.51 | 0.45 | 1.13 | 0.32 | 0.19 | 0.28 | 0.47 | 0.36 | 0.32 | 0.19 | 0.27 | 0.35 | 0.48 | 1.54 | 0.25 | 0.28 | -0.11 | -0.04 | 0.73 | 0.31 | 0.47 | 1.07 | 0.30 |
| hr4209 | S | 4985 | 2.56 | 1.35 | 0.17 | -0.05 | 0.14 | 0.13 | 0.47 | 0.07 | -0.13 | -0.07 | -0.08 | 0.03 | -0.06 | -0.06 | -0.12 | -0.06 | -0.18 | 0.11 | 0.22 | 0.00 | -0.22 | 0.12 | 0.16 | 0.14 | 0.11 | 0.14 | 0.12 |
| hr4232 | S | 4340 | 1.50 | 1.60 | 0.03 | -0.02 | 0.12 | 0.16 | 0.63 | -0.18 | -0.25 | -0.13 | -0.06 | -0.08 | -0.16 | -0.26 | -0.20 | -0.14 | -0.01 | 0.55 | -0.02 | -0.25 | -0.66 | -0.38 | -0.03 | -0.27 | -0.14 | 0.22 | -0.15 |
| hr4233 | S | 4675 | 2.15 | 1.46 | 0.07 | 0.06 | 0.19 | 0.17 | 0.63 | 0.01 | -0.19 | -0.06 | -0.01 | 0.01 | 0.01 | -0.08 | -0.07 | 0.01 | 0.13 | 0.39 | 0.04 | -0.20 | -0.51 | -0.10 | 0.10 | -0.07 | 0.03 | 0.37 | -0.03 |
| hr4235 | S | 4526 | 1.90 | 1.44 | 0.06 | -0.02 | 0.08 | 0.11 | 0.57 | -0.07 | -0.23 | -0.13 | -0.07 | -0.04 | -0.11 | -0.16 | -0.16 | -0.10 | 0.00 | 0.19 | 0.05 | -0.04 | -0.52 | -0.08 | 0.16 | -0.07 | 0.09 | 0.39 | -0.06 |
| hr4242 | S | 4736 | 2.51 | 1.63 | 0.24 | 0.19 | 0.34 | 0.37 | 0.71 | 0.14 | 0.14 | 0.22 | 0.26 | 0.29 | 0.26 | 0.16 | 0.21 | 0.26 | | 0.81 | 0.29 | 0.21 | -0.14 | 0.21 | 0.60 | 0.33 | 0.45 | 1.07 | 0.38 |
| hr4253 | S | 4922 | 2.51 | 1.47 | 0.40 | 0.14 | 0.33 | 0.26 | 0.79 | 0.23 | -0.03 | 0.08 | 0.08 | 0.19 | 0.15 | 0.08 | 0.02 | 0.12 | 0.01 | 0.48 | 0.29 | 0.14 | -0.12 | 0.16 | 0.30 | 0.18 | 0.23 | 0.29 | 0.18 |
| hr4255 | S | 5193 | 2.51 | 1.63 | 0.23 | -0.05 | 0.09 | 0.04 | 0.17 | -0.03 | -0.08 | -0.13 | -0.20 | -0.04 | -0.22 | -0.13 | -0.17 | -0.16 | -0.36 | -0.18 | 0.54 | -0.02 | -0.18 | 0.21 | 0.09 | 0.13 | 0.06 | 0.01 | 0.11 |
| hr4256 | S | 4673 | 2.21 | 1.38 | 0.01 | -0.07 | 0.04 | 0.08 | 0.51 | -0.08 | -0.23 | -0.19 | -0.18 | -0.09 | -0.13 | -0.19 | -0.20 | -0.13 | -0.09 | 0.25 | 0.04 | -0.20 | -0.50 | -0.17 | 0.02 | -0.10 | 0.02 | 0.22 | -0.18 |
| hr4258 | S | 4569 | 2.13 | 1.39 | 0.09 | 0.06 | 0.12 | 0.14 | 0.62 | -0.05 | -0.18 | -0.11 | -0.06 | -0.02 | -0.07 | -0.13 | -0.13 | -0.06 | 0.18 | 0.18 | 0.12 | -0.17 | -0.44 | -0.04 | 0.34 | -0.01 | 0.16 | 0.20 | -0.10 |
| hr4264 | S | 4524 | 2.13 | 1.36 | 0.12 | 0.03 | 0.20 | 0.14 | 0.50 | 0.00 | -0.08 | -0.05 | 0.02 | 0.02 | -0.03 | -0.12 | -0.07 | -0.01 | 0.24 | -0.03 | 0.08 | -0.09 | -0.54 | -0.18 | 0.33 | -0.04 | 0.14 | 0.33 | 0.04 |
| hr4283 | S | 4892 | 2.42 | 1.55 | 0.57 | 0.25 | 0.35 | 0.42 | 0.94 | 0.27 | 0.12 | 0.19 | 0.23 | 0.28 | 0.27 | 0.19 | 0.15 | 0.27 | 0.35 | 0.81 | 0.35 | 0.30 | -0.03 | 0.25 | 0.54 | 0.21 | 0.26 | 0.49 | 0.26 |
| hr4287 | S | 4645 | 2.23 | 1.43 | 0.18 | 0.12 | 0.24 | 0.24 | 0.71 | 0.07 | -0.06 | 0.01 | 0.06 | 0.12 | 0.10 | 0.00 | 0.00 | 0.09 | 0.27 | 0.77 | 0.24 | -0.01 | -0.33 | 0.04 | 0.33 | 0.03 | 0.16 | 0.54 | 0.02 |
| hr4301 | S | 4636 | 1.94 | 1.57 | 0.30 | 0.15 | 0.16 | 0.22 | 0.65 | 0.05 | -0.11 | -0.05 | 0.00 | 0.05 | -0.03 | -0.06 | -0.08 | -0.03 | -0.03 | 0.99 | 0.23 | -0.06 | -0.36 | 0.17 | 0.35 | 0.04 | 0.16 | 0.23 | 0.04 |
| hr4305 | S | 4967 | 2.70 | 1.35 | 0.39 | 0.12 | 0.28 | 0.20 | 0.52 | 0.12 | 0.00 | 0.06 | 0.06 | 0.07 | -0.02 | -0.01 | 0.02 | 0.05 | 0.18 | 0.36 | 0.14 | 0.06 | -0.23 | 0.01 | 0.24 | -0.03 | 0.13 | 0.11 | 0.15 |
| hr4335 | S | 4523 | 1.89 | 1.53 | 0.23 | 0.07 | 0.18 | 0.24 | 0.81 | 0.07 | -0.17 | -0.06 | -0.02 | 0.08 | -0.04 | -0.04 | -0.08 | 0.01 | 0.16 | 0.70 | 0.23 | 0.04 | -0.33 | 0.03 | 0.37 | 0.07 | 0.17 | 0.30 | 0.07 |
| hr4351 | S | 4556 | 2.16 | 1.43 | 0.37 | 0.27 | 0.43 | 0.38 | 1.06 | 0.25 | 0.04 | 0.17 | 0.33 | 0.25 | 0.27 | 0.11 | 0.14 | 0.26 | | 0.92 | 0.16 | 0.06 | -0.33 | -0.03 | 0.50 | 0.06 | 0.25 | 0.79 | 0.10 |
| hr4382 | S | 4510 | 1.72 | 1.58 | -0.24 | -0.27 | -0.16 | -0.10 | 0.21 | -0.37 | -0.48 | -0.44 | -0.44 | -0.41 | -0.55 | -0.49 | -0.43 | -0.42 | -0.26 | 0.13 | -0.22 | -0.55 | -0.82 | -0.43 | -0.38 | -0.49 | -0.34 | -0.37 | -0.40 |
| hr4383 | S | 4825 | 2.70 | 1.25 | 0.45 | 0.22 | 0.45 | 0.40 | 1.00 | 0.29 | 0.27 | 0.33 | 0.50 | 0.36 | 0.38 | 0.23 | 0.26 | 0.34 | 0.46 | 0.69 | 0.38 | 0.27 | 0.05 | 0.26 | 0.76 | 0.32 | 0.41 | 0.83 | 0.34 |
| hr4400 | S | 4984 | 2.46 | 1.54 | 0.21 | -0.03 | 0.09 | 0.13 | 0.48 | 0.07 | -0.07 | -0.07 | -0.10 | 0.03 | -0.09 | -0.06 | -0.11 | -0.07 | -0.10 | -0.12 | 0.22 | 0.03 | -0.28 | 0.19 | 0.21 | 0.17 | 0.16 | 0.22 | 0.14 |
| hr4407 | S | 4836 | 2.38 | 1.41 | 0.08 | 0.11 | 0.14 | 0.10 | 0.49 | -0.01 | -0.12 | -0.07 | -0.10 | -0.05 | -0.13 | -0.12 | -0.15 | -0.10 | -0.06 | 0.52 | 0.19 | -0.05 | -0.33 | 0.10 | 0.16 | 0.04 | 0.10 | 0.04 | -0.02 |
| hr4419 | S | 4638 | 2.30 | 1.31 | 0.23 | 0.03 | 0.29 | 0.17 | 0.68 | 0.11 | 0.04 | 0.15 | 0.31 | 0.18 | 0.02 | -0.03 | 0.01 | 0.05 | 0.01 | 0.28 | 0.27 | 0.21 | -0.03 | 0.24 | 0.56 | 0.19 | 0.37 | 0.58 | 0.16 |
| hr4433 | S | 4678 | 2.29 | 1.36 | 0.05 | 0.04 | 0.07 | 0.09 | 0.41 | -0.03 | -0.16 | -0.10 | -0.07 | -0.03 | -0.11 | -0.12 | -0.15 | -0.09 | 0.00 | -0.07 | 0.08 | -0.11 | -0.39 | 0.08 | 0.38 | 0.09 | 0.19 | 0.19 | -0.09 |
| hr4452 | S | 4648 | 2.13 | 1.42 | 0.07 | 0.10 | 0.24 | 0.19 | 0.63 | -0.01 | -0.15 | -0.02 | -0.01 | -0.07 | -0.18 | -0.14 | -0.08 | -0.03 | 0.21 | 0.31 | -0.03 | -0.22 | -0.61 | -0.17 | 0.01 | -0.19 | 0.00 | 0.38 | 0.03 |
| hr4459 | S | 4814 | 2.48 | 1.39 | 0.29 | 0.13 | 0.31 | 0.27 | 0.82 | 0.15 | 0.03 | 0.08 | 0.12 | 0.15 | 0.13 | 0.06 | 0.01 | 0.12 | 0.11 | 0.78 | 0.26 | 0.12 | -0.19 | 0.11 | 0.38 | 0.16 | 0.23 | 0.49 | 0.09 |
| hr4461 | S | 4802 | 2.33 | 1.41 | -0.12 | -0.19 | -0.06 | -0.05 | 0.31 | -0.19 | -0.33 | -0.29 | -0.30 | -0.24 | -0.32 | -0.32 | -0.33 | -0.29 | -0.24 | -0.08 | 0.15 | -0.37 | -0.59 | -0.25 | -0.25 | -0.25 | -0.24 | -0.06 | -0.20 |
| hr4471 | S | 4786 | 2.22 | 1.41 | 0.00 | -0.14 | 0.04 | 0.08 | 0.34 | -0.06 | -0.23 | -0.17 | -0.17 | -0.10 | -0.21 | -0.17 | -0.20 | -0.15 | -0.19 | 0.12 | 0.07 | -0.17 | -0.58 | 0.01 | 0.05 | 0.04 | 0.03 | 0.05 | -0.04 |
| hr4474 | S | 4718 | 1.93 | 1.50 | 0.02 | -0.06 | 0.14 | 0.28 | 0.82 | -0.07 | -0.23 | -0.17 | -0.26 | -0.05 | -0.37 | -0.19 | -0.24 | -0.08 | 0.46 | 0.77 | 0.91 | 0.65 | 0.34 | | 1.09 | 0.82 | 0.82 | 0.77 | 0.01 |
| hr4478 | S | 4691 | 2.24 | 1.49 | 0.49 | 0.25 | 0.36 | 0.43 | 1.02 | 0.25 | 0.12 | 0.18 | 0.29 | 0.29 | 0.31 | 0.17 | 0.19 | 0.27 | 0.23 | 0.47 | 0.28 | 0.18 | -0.23 | 0.09 | 0.51 | 0.17 | 0.24 | 0.65 | 0.21 |
| hr4480 | S | 6672 | 3.41 | 4.70 | 0.17 | | | 0.07 | 0.40 | 0.04 | 0.01 | 0.35 | 0.46 | 0.22 | 0.03 | -0.08 | 0.49 | 0.01 | -0.44 | 0.17 | 1.89 | 0.25 | 1.21 | -0.10 | | 0.29 | 0.32 | | |
| hr4495 | S | 4863 | 2.56 | 1.37 | 0.25 | -0.08 | 0.29 | 0.20 | 0.39 | 0.19 | -0.01 | 0.05 | 0.05 | 0.13 | 0.09 | 0.03 | -0.02 | 0.05 | 0.13 | 0.35 | 0.31 | 0.05 | -0.20 | 0.10 | 0.33 | 0.04 | 0.21 | 0.24 | 0.10 |

| ID | | Teff | | | | | | | | | | | | | | | | | | | | | | | | | |
|---|---|---|---|---|---|---|---|---|---|---|---|---|---|---|---|---|---|---|---|---|---|---|---|---|---|---|---|
| hr4510 | S | 4914 | 2.55 | 1.36 | 0.14 | 0.02 | 0.17 | 0.14 | 0.43 | 0.08 | -0.10 | -0.02 | -0.01 | 0.05 | -0.06 | -0.04 | -0.10 | -0.03 | -0.08 | 0.01 | 0.25 | 0.08 | -0.23 | 0.14 | 0.21 | 0.09 | 0.20 | 0.27 | 0.10 |
| hr4518 | S | 4392 | 1.54 | 1.61 | -0.24 | -0.14 | -0.09 | -0.05 | 0.47 | -0.38 | -0.45 | -0.33 | -0.33 | -0.33 | -0.47 | -0.44 | -0.37 | -0.35 | -0.21 | 0.14 | -0.31 | -0.56 | -0.85 | -0.51 | -0.27 | -0.49 | -0.33 | -0.26 | -0.30 |
| hr4521 | S | 4411 | 1.75 | 1.80 | 0.76 | 0.60 | 0.50 | 0.65 | 1.75 | 0.30 | 0.16 | 0.28 | 0.47 | 0.50 | 0.44 | 0.24 | 0.35 | 0.38 | 0.69 | 2.27 | 0.34 | 0.20 | -0.19 | -0.31 | 0.53 | 0.29 | 0.40 | 1.15 | 0.43 |
| hr4544 | S | 4668 | 2.34 | 1.29 | 0.07 | -0.01 | 0.11 | 0.13 | 0.55 | 0.01 | -0.14 | -0.07 | -0.03 | 0.00 | -0.10 | -0.11 | -0.11 | -0.06 | 0.04 | 0.30 | 0.08 | -0.12 | -0.44 | 0.00 | 0.20 | 0.07 | 0.18 | 0.41 | -0.01 |
| hr4558 | S | 4970 | 2.39 | 1.52 | -0.30 | -0.40 | -0.28 | -0.32 | -0.37 | -0.43 | -0.44 | -0.48 | -0.54 | -0.49 | -0.64 | -0.55 | -0.48 | -0.54 | -0.53 | -0.45 | 0.10 | -0.61 | -0.70 | -0.45 | -0.48 | -0.47 | -0.39 | -0.54 | -0.36 |
| hr4566 | S | 4565 | 2.00 | 1.53 | 0.17 | 0.31 | 0.35 | 0.32 | 0.86 | 0.00 | -0.11 | 0.03 | 0.05 | 0.07 | -0.09 | -0.06 | -0.02 | 0.03 | 0.27 | 1.24 | 0.03 | 0.00 | -0.54 | -0.22 | 0.24 | -0.03 | 0.01 | 0.64 | 0.05 |
| hr4593 | S | 4670 | 2.16 | 1.43 | 0.10 | 0.10 | 0.16 | 0.19 | 0.63 | -0.01 | -0.11 | -0.09 | -0.07 | -0.03 | -0.03 | -0.10 | -0.09 | -0.03 | 0.19 | 0.51 | -0.04 | -0.23 | -0.51 | -0.13 | 0.12 | -0.14 | -0.03 | 0.25 | -0.08 |
| hr4608 | S | 4762 | 2.10 | 1.42 | -0.26 | -0.29 | -0.12 | -0.13 | 0.10 | -0.34 | -0.43 | -0.41 | -0.53 | -0.36 | -0.49 | -0.46 | -0.44 | -0.41 | -0.28 | 0.15 | 0.28 | -0.15 | -0.21 | 0.35 | 0.39 | 0.19 | 0.26 | -0.02 | -0.24 |
| hr4609 | S | 4677 | 2.20 | 1.52 | -0.11 | 0.01 | 0.13 | 0.06 | 0.36 | -0.13 | -0.24 | -0.09 | -0.10 | -0.24 | -0.34 | -0.32 | -0.20 | -0.21 | 0.00 | 0.30 | -0.07 | -0.40 | -0.67 | -0.53 | -0.19 | -0.35 | -0.25 | 0.01 | -0.15 |
| hr4610 | S | 4433 | 1.87 | 1.46 | 0.08 | -0.01 | 0.19 | 0.18 | 0.67 | -0.06 | -0.07 | -0.01 | 0.09 | 0.02 | -0.10 | -0.15 | -0.06 | -0.02 | | 0.75 | 0.08 | -0.18 | -0.55 | -0.23 | 0.18 | -0.02 | 0.22 | 0.52 | 0.14 |
| hr4626 | S | 4553 | 2.16 | 1.34 | 0.28 | 0.13 | 0.33 | 0.27 | 0.92 | 0.10 | 0.05 | 0.11 | 0.23 | 0.19 | 0.14 | 0.00 | 0.06 | 0.11 | 0.48 | 0.92 | 0.11 | -0.03 | -0.45 | -0.24 | 0.39 | 0.02 | 0.23 | 0.75 | -0.01 |
| hr4630 | S | 4285 | 1.18 | 1.97 | 0.25 | 0.02 | 0.12 | 0.44 | 1.24 | 0.00 | -0.18 | -0.09 | -0.05 | 0.14 | 0.07 | -0.02 | 0.00 | 0.08 | 0.12 | 1.46 | 0.23 | -0.17 | -0.49 | 0.01 | 0.30 | -0.01 | 0.14 | 0.55 | 0.07 |
| hr4654 | S | 4771 | 2.46 | 1.41 | 0.14 | 0.04 | 0.22 | 0.18 | 0.44 | 0.10 | -0.07 | 0.04 | 0.04 | 0.08 | -0.02 | -0.01 | -0.06 | 0.03 | 0.15 | 0.52 | 0.28 | 0.14 | -0.26 | 0.22 | 0.38 | 0.19 | 0.28 | 0.29 | 0.12 |
| hr4655 | S | 4765 | 2.37 | 1.47 | 0.63 | 0.36 | 0.43 | 0.49 | 1.04 | 0.30 | 0.11 | 0.21 | 0.29 | 0.33 | 0.38 | 0.21 | 0.24 | 0.35 | 0.56 | 1.33 | 0.34 | 0.12 | -0.11 | -0.01 | 0.66 | 0.10 | 0.22 | 0.59 | 0.24 |
| hr4667 | S | 4861 | 2.34 | 1.42 | 0.03 | -0.09 | 0.05 | 0.04 | 0.39 | -0.05 | -0.21 | -0.19 | -0.18 | -0.09 | -0.16 | -0.17 | -0.20 | -0.16 | -0.26 | -0.10 | 0.11 | -0.13 | -0.38 | 0.07 | 0.05 | 0.03 | 0.05 | 0.02 | -0.03 |
| hr4668 | S | 4463 | 1.93 | 1.44 | 0.02 | -0.09 | 0.04 | 0.11 | 0.69 | -0.11 | -0.31 | -0.19 | -0.13 | -0.13 | -0.22 | -0.26 | -0.24 | -0.16 | -0.09 | 0.67 | -0.09 | -0.35 | -0.64 | -0.31 | -0.01 | -0.26 | -0.11 | 0.11 | -0.23 |
| hr4695 | S | 4422 | 1.70 | 1.75 | -0.25 | -0.13 | -0.14 | 0.02 | 0.46 | -0.26 | -0.44 | -0.22 | -0.21 | -0.35 | -0.55 | -0.45 | -0.28 | -0.32 | -0.16 | 0.29 | -0.19 | -0.52 | -0.99 | -0.75 | -0.48 | -0.59 | -0.44 | -0.09 | -0.33 |
| hr4697 | S | 4711 | 2.18 | 1.45 | -0.02 | -0.27 | -0.07 | -0.08 | 0.25 | -0.17 | -0.33 | -0.30 | -0.32 | -0.29 | -0.40 | -0.36 | -0.34 | -0.31 | -0.22 | 0.23 | -0.10 | -0.41 | -0.69 | -0.28 | -0.23 | -0.29 | -0.23 | -0.30 | -0.26 |
| hr4699 | S | 4724 | 2.59 | 1.29 | 0.20 | 0.12 | 0.28 | 0.21 | 0.79 | 0.12 | 0.01 | 0.08 | 0.12 | 0.16 | 0.09 | 0.02 | 0.01 | 0.08 | 0.29 | 0.45 | 0.22 | 0.07 | -0.29 | 0.09 | 0.48 | 0.15 | 0.26 | 0.47 | 0.05 |
| hr4728 | S | 4938 | 2.61 | 1.41 | 0.35 | 0.19 | 0.25 | 0.27 | 0.66 | 0.22 | 0.01 | 0.07 | 0.10 | 0.16 | 0.12 | 0.08 | 0.02 | 0.11 | 0.06 | 0.33 | 0.28 | 0.21 | -0.15 | 0.17 | 0.38 | 0.10 | 0.23 | 0.29 | 0.26 |
| hr4737 | S | 4634 | 2.21 | 1.56 | 0.56 | 0.35 | 0.46 | 0.55 | 1.17 | 0.32 | 0.18 | 0.27 | 0.42 | 0.39 | 0.38 | 0.24 | 0.27 | 0.40 | 0.44 | 1.55 | 0.30 | 0.13 | -0.18 | 0.02 | 0.80 | 0.20 | 0.37 | 0.81 | 0.31 |
| hr4753 | S | 6388 | 4.43 | 7.75 | | -0.38 | | 0.37 | 1.22 | -0.05 | 0.21 | 0.69 | 0.81 | 0.47 | -0.58 | 0.22 | 1.07 | 0.43 | | | 1.59 | | | | 1.24 | 0.91 | 1.01 | | |
| hr4772 | S | 4719 | 2.23 | 1.51 | -0.10 | 0.00 | 0.05 | 0.07 | 0.35 | -0.11 | -0.18 | -0.12 | -0.14 | -0.14 | -0.24 | -0.20 | -0.15 | -0.13 | -0.01 | 0.17 | 0.01 | -0.24 | -0.59 | -0.20 | -0.08 | -0.10 | -0.06 | 0.07 | -0.09 |
| hr4777 | S | 5003 | 2.75 | 1.37 | 0.41 | 0.16 | 0.32 | 0.28 | 0.67 | 0.24 | 0.05 | 0.14 | 0.18 | 0.21 | 0.18 | 0.12 | 0.07 | 0.15 | 0.19 | 0.38 | 0.34 | 0.24 | -0.10 | 0.20 | 0.41 | 0.12 | 0.25 | 0.27 | 0.25 |
| hr4783 | S | 4784 | 2.34 | 1.46 | 0.16 | 0.03 | 0.20 | 0.15 | 0.55 | 0.03 | -0.10 | -0.05 | 0.01 | 0.04 | -0.01 | -0.06 | -0.07 | -0.01 | 0.00 | 0.18 | 0.16 | -0.05 | -0.34 | 0.03 | 0.21 | 0.03 | 0.10 | 0.34 | 0.07 |
| hr4784 | S | 4705 | 2.26 | 1.46 | 0.09 | 0.09 | 0.21 | 0.19 | 0.57 | 0.01 | -0.07 | -0.02 | -0.01 | 0.04 | 0.02 | -0.06 | -0.06 | 0.03 | 0.08 | 0.66 | 0.17 | -0.10 | -0.45 | -0.04 | 0.30 | 0.02 | 0.17 | 0.30 | 0.06 |
| hr4786 | S | 5090 | 2.26 | 1.84 | 0.39 | 0.01 | 0.20 | 0.16 | 0.53 | 0.07 | -0.08 | -0.03 | -0.10 | 0.08 | -0.07 | -0.02 | -0.06 | -0.03 | -0.15 | 0.35 | 0.45 | 0.11 | -0.17 | 0.20 | 0.14 | 0.05 | 0.04 | -0.15 | 0.09 |
| hr4812 | S | 4719 | 2.47 | 1.43 | 0.37 | 0.26 | 0.45 | 0.36 | 0.83 | 0.17 | 0.16 | 0.24 | 0.37 | 0.29 | 0.27 | 0.16 | 0.20 | 0.26 | 0.44 | 0.39 | 0.33 | 0.32 | -0.12 | 0.23 | 0.60 | 0.26 | 0.39 | 0.77 | 0.37 |
| hr4813 | S | 4404 | 1.58 | 1.74 | 0.72 | 0.47 | 0.55 | 0.65 | 1.51 | 0.20 | 0.12 | 0.22 | 0.37 | 0.40 | 0.49 | 0.22 | 0.30 | 0.39 | 0.67 | 2.09 | 0.31 | 0.05 | -0.38 | -0.27 | 0.44 | 0.19 | 0.31 | 0.90 | 0.15 |
| hr4815 | S | 4943 | 2.71 | 1.42 | 0.35 | 0.18 | 0.28 | 0.26 | 0.60 | 0.22 | 0.09 | 0.14 | 0.16 | 0.22 | 0.15 | 0.12 | 0.10 | 0.16 | 0.25 | 0.54 | 0.26 | 0.25 | -0.11 | 0.26 | 0.57 | 0.28 | 0.33 | 0.38 | 0.25 |
| hr4840 | S | 4372 | 1.82 | 1.50 | 0.34 | 0.19 | 0.44 | 0.43 | 1.34 | 0.24 | 0.00 | 0.15 | 0.34 | 0.28 | 0.19 | 0.07 | 0.09 | 0.23 | 0.36 | 1.32 | 0.30 | 0.02 | -0.34 | -0.07 | 0.63 | 0.06 | 0.28 | 0.69 | 0.12 |
| hr4851 | S | 4190 | 1.09 | 1.55 | 0.24 | 0.06 | 0.30 | 0.19 | 1.22 | 0.01 | -0.19 | -0.05 | 0.14 | 0.07 | -0.08 | -0.17 | -0.04 | 0.00 | 0.31 | 1.53 | 0.14 | -0.14 | -0.62 | -0.41 | 0.29 | -0.20 | 0.10 | 0.57 | -0.24 |
| hr4860 | S | 4814 | 2.22 | 1.46 | -0.28 | -0.22 | -0.14 | -0.19 | -0.02 | -0.31 | -0.42 | -0.39 | -0.48 | -0.39 | -0.52 | -0.45 | -0.42 | -0.41 | -0.36 | -0.14 | 0.11 | -0.47 | -0.73 | -0.41 | -0.39 | -0.45 | -0.34 | -0.44 | -0.32 |
| hr4873 | S | 4808 | 2.56 | 1.27 | 0.07 | 0.09 | 0.10 | 0.10 | 0.43 | -0.03 | -0.12 | -0.06 | -0.03 | -0.02 | -0.10 | -0.10 | -0.12 | -0.08 | -0.02 | 0.18 | 0.15 | -0.06 | -0.40 | 0.03 | 0.25 | 0.08 | 0.18 | 0.32 | 0.02 |
| hr4877 | S | 4744 | 2.15 | 1.47 | -0.03 | -0.06 | 0.08 | 0.08 | 0.51 | -0.12 | -0.18 | -0.16 | -0.16 | -0.11 | -0.17 | -0.19 | -0.18 | -0.14 | -0.04 | 0.33 | 0.05 | -0.20 | -0.54 | -0.13 | 0.02 | -0.17 | -0.01 | 0.20 | -0.08 |
| hr4883 | S | 5564 | 3.04 | 2.85 | 0.56 | | | 0.46 | | 0.25 | 0.34 | 0.59 | 0.49 | 0.63 | 0.01 | 0.17 | 0.82 | 0.62 | | | 1.97 | 0.74 | -0.16 | | 1.50 | 1.44 | 0.59 | -0.57 | 1.28 |
| hr4894 | S | 4982 | 2.67 | 1.46 | 0.61 | 0.24 | 0.37 | 0.43 | 0.88 | 0.30 | 0.25 | 0.31 | 0.37 | 0.38 | 0.44 | 0.24 | 0.28 | 0.33 | 0.49 | 0.97 | 0.46 | 0.30 | 0.01 | 0.24 | 0.57 | 0.29 | 0.31 | 0.72 | 0.42 |
| hr4896 | S | 4702 | 2.67 | 1.20 | 0.25 | 0.17 | 0.34 | 0.25 | 0.78 | 0.19 | 0.05 | 0.19 | 0.31 | 0.20 | 0.24 | 0.06 | 0.11 | 0.18 | 0.47 | 0.89 | 0.20 | 0.15 | -0.28 | 0.04 | 0.48 | 0.16 | 0.38 | 0.75 | 0.21 |
| hr4925 | S | 4580 | 2.18 | 1.46 | 0.31 | 0.23 | 0.41 | 0.35 | 0.99 | 0.13 | -0.01 | 0.11 | 0.25 | 0.20 | 0.20 | 0.08 | 0.10 | 0.21 | 0.41 | 0.89 | 0.11 | -0.02 | -0.42 | -0.10 | 0.43 | 0.07 | 0.24 | 0.73 | 0.11 |
| hr4929 | S | 4916 | 2.62 | 1.37 | 0.22 | 0.04 | 0.19 | 0.16 | 0.48 | 0.09 | -0.02 | 0.03 | 0.03 | 0.10 | 0.05 | 0.00 | -0.03 | 0.03 | 0.04 | 0.39 | 0.26 | 0.06 | -0.24 | 0.17 | 0.22 | 0.21 | 0.23 | 0.34 | 0.10 |
| hr4932 | S | 4988 | 2.52 | 1.52 | 0.46 | 0.12 | 0.23 | 0.28 | 0.70 | 0.18 | 0.02 | 0.06 | 0.06 | 0.19 | 0.12 | 0.09 | 0.05 | 0.11 | 0.15 | 0.58 | 0.31 | 0.13 | -0.17 | 0.11 | 0.25 | 0.06 | 0.09 | 0.16 | 0.22 |
| hr4953 | S | 4816 | 2.72 | 1.17 | -0.11 | -0.16 | -0.02 | -0.06 | 0.32 | -0.10 | -0.19 | -0.12 | -0.12 | -0.15 | -0.23 | -0.26 | -0.24 | -0.19 | -0.06 | 0.12 | 0.22 | -0.24 | -0.47 | -0.15 | -0.11 | -0.12 | 0.03 | 0.05 | -0.08 |
| hr4955 | S | 4585 | 2.20 | 1.41 | 0.28 | 0.26 | 0.36 | 0.34 | 0.76 | 0.14 | 0.07 | 0.12 | 0.23 | 0.22 | 0.12 | 0.08 | 0.06 | 0.18 | 0.31 | 0.67 | 0.24 | 0.14 | -0.38 | 0.16 | 0.68 | 0.20 | 0.30 | 0.52 | 0.32 |
| hr4956 | S | 4198 | 1.15 | 1.68 | 0.69 | 0.36 | 0.61 | 0.50 | 1.53 | 0.19 | -0.05 | 0.13 | 0.38 | 0.37 | 0.33 | 0.08 | 0.20 | 0.28 | | 1.20 | 0.29 | 0.09 | -0.35 | -0.42 | 0.51 | 0.36 | 0.28 | 0.99 | -0.03 |
| hr4959 | S | 4802 | 2.34 | 1.44 | 0.12 | 0.03 | 0.16 | 0.13 | 0.51 | 0.01 | -0.10 | -0.06 | -0.05 | 0.01 | -0.06 | -0.09 | -0.08 | -0.03 | 0.06 | 0.41 | 0.12 | -0.05 | -0.34 | -0.04 | 0.15 | -0.04 | 0.05 | 0.11 | -0.01 |
| hr4960 | S | 4765 | 2.44 | 1.37 | 0.28 | 0.15 | 0.29 | 0.26 | 0.49 | 0.16 | -0.02 | 0.06 | 0.05 | 0.17 | 0.08 | 0.04 | -0.01 | 0.08 | 0.24 | 0.16 | 0.26 | 0.15 | -0.25 | 0.15 | 0.50 | 0.13 | 0.22 | 0.29 | 0.18 |
| hr4964 | S | 4464 | 2.06 | 1.52 | 0.71 | 0.33 | 0.48 | 0.48 | 1.30 | 0.22 | 0.12 | 0.24 | 0.44 | 0.36 | 0.25 | 0.12 | 0.21 | 0.28 | | 1.20 | 0.31 | 0.16 | -0.21 | -0.19 | 0.50 | 0.19 | 0.34 | 0.85 | 0.15 |
| hr4984 | S | 4825 | 2.58 | 1.35 | 0.34 | 0.12 | 0.35 | 0.33 | 0.79 | 0.23 | 0.02 | 0.11 | 0.22 | 0.20 | 0.17 | 0.10 | 0.05 | 0.16 | 0.14 | 0.44 | 0.29 | 0.15 | -0.08 | 0.15 | 0.42 | 0.12 | 0.29 | 0.50 | 0.15 |
| hr4992 | S | 4398 | 1.77 | 1.46 | 0.08 | 0.04 | 0.17 | 0.13 | 0.82 | -0.08 | -0.13 | -0.03 | 0.06 | 0.01 | -0.08 | -0.17 | -0.09 | -0.05 | | 0.76 | 0.04 | -0.17 | -0.58 | -0.28 | 0.16 | -0.08 | 0.18 | 0.56 | -0.06 |
| hr5001 | S | 4741 | 2.84 | 1.04 | 0.20 | 0.28 | 0.28 | 0.26 | 0.89 | 0.08 | 0.05 | 0.11 | 0.19 | 0.12 | 0.07 | 0.02 | 0.05 | 0.13 | 0.38 | 0.58 | 0.13 | -0.02 | -0.37 | -0.08 | 0.47 | 0.13 | 0.21 | 0.72 | 0.07 |

| ID | | T | | | | | | | | | | | | | | | | | | | | | | | | | |
|---|---|---|---|---|---|---|---|---|---|---|---|---|---|---|---|---|---|---|---|---|---|---|---|---|---|---|---|
| hr5007 | S | 4807 | 2.55 | 1.28 | -0.09 | -0.06 | 0.01 | 0.01 | 0.33 | -0.12 | -0.21 | -0.16 | -0.18 | -0.13 | -0.23 | -0.21 | -0.21 | -0.16 | -0.25 | 0.23 | 0.04 | -0.22 | -0.46 | -0.03 | 0.24 | -0.03 | 0.09 | 0.19 | -0.09 |
| hr5020 | S | 5019 | 2.49 | 1.60 | 0.37 | 0.08 | 0.16 | 0.22 | 0.55 | 0.11 | -0.07 | 0.00 | -0.01 | 0.13 | 0.02 | 0.02 | -0.02 | 0.04 | 0.01 | 0.41 | 0.39 | 0.05 | -0.21 | 0.15 | 0.19 | 0.08 | 0.13 | 0.16 | 0.21 |
| hr5044 | S | 4820 | 2.26 | 1.47 | 0.11 | -0.10 | 0.08 | 0.12 | 0.37 | 0.01 | -0.20 | -0.12 | -0.14 | 0.00 | -0.09 | -0.10 | -0.15 | -0.08 | -0.02 | 0.12 | 0.11 | -0.11 | -0.42 | 0.02 | 0.10 | -0.03 | 0.00 | 0.02 | -0.06 |
| hr5053 | S | 4643 | 1.99 | 1.48 | 0.02 | -0.04 | 0.09 | 0.10 | 0.48 | -0.05 | -0.16 | -0.15 | -0.09 | -0.08 | -0.12 | -0.15 | -0.16 | -0.10 | 0.05 | 0.26 | 0.02 | -0.18 | -0.51 | -0.06 | 0.12 | -0.16 | 0.05 | 0.30 | -0.04 |
| hr5067 | S | 4851 | 2.45 | 1.42 | 0.10 | -0.11 | 0.15 | 0.13 | 0.34 | 0.03 | -0.12 | -0.05 | -0.04 | 0.01 | -0.07 | -0.08 | -0.10 | -0.06 | 0.01 | 0.21 | 0.19 | -0.09 | -0.33 | 0.08 | 0.20 | 0.12 | 0.14 | 0.14 | 0.04 |
| hr5068 | S | 4730 | 2.31 | 1.60 | 0.61 | 0.32 | 0.42 | 0.53 | 1.17 | 0.25 | 0.17 | 0.25 | 0.33 | 0.35 | 0.36 | 0.22 | 0.26 | 0.34 | 0.38 | 1.93 | 0.36 | 0.17 | -0.19 | 0.11 | 0.46 | 0.18 | 0.31 | 0.68 | 0.31 |
| hr5081 | S | 4658 | 2.35 | 1.34 | 0.08 | -0.01 | 0.09 | 0.15 | 0.64 | -0.03 | -0.16 | -0.09 | -0.01 | 0.00 | -0.06 | -0.11 | -0.11 | -0.05 | 0.10 | 0.36 | 0.14 | -0.01 | -0.38 | -0.03 | 0.13 | 0.02 | 0.15 | 0.38 | -0.06 |
| hr5100 | S | 4837 | 2.39 | 1.40 | -0.09 | -0.15 | -0.03 | -0.03 | 0.34 | -0.13 | -0.30 | -0.21 | -0.25 | -0.18 | -0.23 | -0.27 | -0.27 | -0.23 | -0.21 | 0.07 | 0.08 | -0.30 | -0.53 | -0.20 | -0.10 | -0.22 | -0.10 | -0.19 | -0.17 |
| hr5111 | S | 4881 | 2.54 | 1.40 | -0.08 | -0.07 | -0.02 | -0.04 | 0.27 | -0.14 | -0.22 | -0.19 | -0.23 | -0.19 | -0.25 | -0.26 | -0.27 | -0.23 | -0.19 | 0.01 | 0.24 | -0.27 | -0.54 | -0.17 | -0.13 | -0.22 | -0.05 | -0.14 | -0.13 |
| hr5126 | S | 4519 | 2.00 | 1.47 | 0.41 | 0.19 | 0.38 | 0.43 | 1.07 | 0.20 | 0.04 | 0.15 | 0.26 | 0.28 | 0.20 | 0.10 | 0.08 | 0.24 | | 1.19 | 0.27 | 0.08 | -0.22 | 0.01 | 0.56 | 0.09 | 0.25 | 0.69 | 0.15 |
| hr5143 | S | 4819 | 2.32 | 1.46 | -0.13 | -0.19 | -0.08 | -0.07 | 0.26 | -0.21 | -0.30 | -0.26 | -0.31 | -0.25 | -0.39 | -0.33 | -0.35 | -0.30 | -0.32 | 0.02 | -0.04 | -0.24 | -0.58 | -0.17 | -0.10 | -0.14 | -0.10 | -0.27 | -0.22 |
| hr5149 | S | 4836 | 2.52 | 1.42 | 0.19 | 0.04 | 0.19 | 0.19 | 0.57 | 0.11 | -0.01 | 0.03 | 0.03 | 0.09 | 0.06 | 0.00 | -0.03 | 0.04 | 0.09 | 0.38 | 0.17 | 0.04 | -0.26 | 0.15 | 0.32 | 0.13 | 0.23 | 0.23 | 0.08 |
| hr5161 | S | 5155 | 2.85 | 1.38 | 0.25 | 0.10 | 0.12 | 0.14 | 0.56 | 0.14 | 0.04 | 0.11 | 0.08 | 0.15 | 0.01 | 0.05 | 0.01 | 0.04 | -0.03 | 0.20 | 0.40 | 0.19 | -0.07 | 0.36 | 0.36 | 0.27 | 0.33 | 0.18 | 0.24 |
| hr5180 | S | 4935 | 2.56 | 1.46 | 0.47 | 0.21 | 0.27 | 0.31 | 0.89 | 0.21 | 0.04 | 0.09 | 0.14 | 0.20 | 0.18 | 0.09 | 0.05 | 0.14 | 0.00 | 0.94 | 0.36 | 0.22 | -0.17 | 0.13 | 0.29 | 0.09 | 0.11 | 0.32 | 0.17 |
| hr5186 | S | 4698 | 2.27 | 1.46 | 0.14 | -0.08 | 0.14 | 0.20 | 0.63 | -0.06 | -0.15 | -0.10 | -0.07 | -0.01 | -0.02 | -0.10 | -0.11 | -0.04 | 0.10 | 0.38 | 0.06 | -0.02 | -0.49 | -0.06 | 0.34 | -0.07 | 0.10 | 0.19 | -0.06 |
| hr5195 | S | 4683 | 2.21 | 1.33 | -0.03 | -0.09 | 0.05 | 0.00 | 0.25 | -0.12 | -0.22 | -0.16 | -0.17 | -0.10 | -0.25 | -0.20 | -0.20 | -0.16 | -0.03 | 0.36 | 0.01 | -0.11 | -0.49 | -0.06 | 0.14 | -0.03 | 0.09 | 0.13 | -0.14 |
| hr5196 | S | 4706 | 2.31 | 1.43 | 0.25 | 0.10 | 0.27 | 0.29 | 0.81 | 0.13 | -0.03 | 0.02 | 0.08 | 0.11 | 0.09 | 0.01 | 0.00 | 0.08 | 0.19 | 0.82 | 0.21 | 0.00 | -0.46 | 0.03 | 0.38 | 0.05 | 0.14 | 0.32 | 0.09 |
| hr5205 | S | 5049 | 2.38 | 1.50 | 0.20 | -0.12 | 0.07 | 0.07 | 0.44 | 0.02 | -0.13 | -0.10 | -0.13 | -0.01 | -0.13 | -0.09 | -0.13 | -0.11 | -0.39 | -0.01 | 0.31 | 0.03 | -0.27 | 0.20 | 0.19 | -0.01 | 0.05 | -0.04 | 0.09 |
| hr5213 | S | 4921 | 2.57 | 1.41 | -0.09 | -0.04 | -0.04 | -0.08 | 0.17 | -0.12 | -0.23 | -0.18 | -0.20 | -0.18 | -0.27 | -0.27 | -0.24 | -0.25 | -0.21 | 0.00 | 0.08 | -0.24 | -0.55 | -0.17 | -0.11 | -0.12 | -0.10 | -0.13 | -0.12 |
| hr5227 | S | 4464 | 2.16 | 1.31 | 0.32 | 0.22 | 0.47 | 0.35 | 1.04 | 0.20 | 0.08 | 0.23 | 0.45 | 0.23 | 0.09 | 0.01 | 0.11 | 0.17 | 0.47 | 1.28 | 0.19 | 0.13 | -0.37 | -0.14 | 0.41 | 0.01 | 0.44 | 0.99 | 0.15 |
| hr5232 | S | 4703 | 2.33 | 1.53 | 0.44 | 0.21 | 0.38 | 0.42 | 0.92 | 0.19 | 0.15 | 0.17 | 0.21 | 0.29 | 0.28 | 0.15 | 0.19 | 0.26 | 0.36 | 0.91 | 0.30 | 0.12 | -0.29 | 0.10 | 0.64 | 0.20 | 0.40 | 0.74 | 0.19 |
| hr5235 | S | 6028 | 3.74 | 1.83 | 0.94 | 0.49 | 0.63 | 0.46 | 0.62 | 0.37 | 0.54 | 0.49 | 0.47 | 0.49 | 0.51 | 0.39 | 0.48 | 0.52 | 0.67 | 0.61 | | 0.63 | 0.62 | 0.26 | 0.13 | 0.49 | 0.40 | 0.93 | 0.54 |
| hr5276 | S | 4265 | 1.36 | 1.61 | 0.58 | 0.45 | 0.54 | 0.53 | 1.54 | 0.26 | 0.03 | 0.24 | 0.42 | 0.40 | 0.24 | 0.09 | 0.24 | 0.28 | 0.77 | 2.48 | 0.34 | 0.12 | -0.27 | -0.26 | 0.41 | 0.17 | 0.57 | 0.77 | 0.06 |
| hr5277 | S | 4684 | 2.21 | 1.46 | 0.37 | 0.10 | 0.34 | 0.42 | 0.93 | 0.17 | -0.03 | 0.02 | 0.09 | 0.18 | 0.11 | 0.06 | 0.03 | 0.17 | 0.33 | 0.86 | 0.20 | 0.02 | -0.26 | -0.06 | 0.37 | 0.04 | 0.02 | 0.53 | 0.10 |
| hr5302 | S | 4770 | 2.29 | 1.49 | -0.09 | -0.02 | 0.10 | 0.06 | 0.33 | -0.06 | -0.17 | -0.08 | -0.08 | -0.09 | -0.15 | -0.17 | -0.13 | -0.10 | -0.02 | 0.16 | -0.01 | -0.28 | -0.51 | -0.19 | -0.03 | -0.11 | 0.03 | 0.20 | -0.09 |
| hr5315 | S | 4155 | 0.99 | 1.65 | -0.25 | -0.20 | -0.09 | -0.08 | 0.74 | -0.40 | -0.47 | -0.35 | -0.28 | -0.31 | -0.50 | -0.51 | -0.39 | -0.41 | -0.17 | 0.58 | -0.13 | -0.59 | -0.98 | -0.64 | -0.25 | -0.51 | -0.28 | -0.11 | -0.50 |
| hr5340 | S | 4281 | 1.20 | 1.65 | -0.34 | -0.16 | -0.05 | -0.12 | 0.59 | -0.37 | -0.50 | -0.33 | -0.35 | -0.45 | -0.67 | -0.60 | -0.44 | -0.48 | -0.30 | 0.01 | -0.31 | -0.66 | -0.98 | -0.83 | -0.59 | -0.70 | -0.49 | -0.21 | -0.34 |
| hr5344 | S | 4833 | 2.38 | 1.40 | 0.04 | -0.06 | 0.07 | 0.09 | 0.56 | -0.05 | -0.19 | -0.13 | -0.16 | -0.05 | -0.14 | -0.13 | -0.18 | -0.12 | -0.13 | 0.10 | 0.06 | -0.01 | -0.39 | 0.11 | 0.19 | 0.02 | 0.14 | 0.02 | -0.03 |
| hr5366 | S | 4730 | 2.19 | 1.47 | 0.11 | 0.00 | 0.13 | 0.13 | 0.51 | 0.01 | -0.16 | -0.12 | -0.09 | -0.03 | -0.12 | -0.12 | -0.13 | -0.08 | 0.03 | 0.18 | 0.11 | -0.07 | -0.48 | 0.00 | 0.15 | 0.00 | 0.05 | 0.09 | -0.09 |
| hr5370 | S | 4433 | 1.90 | 1.60 | 0.42 | 0.39 | 0.61 | 0.52 | 1.29 | 0.18 | 0.15 | 0.25 | 0.45 | 0.39 | 0.37 | 0.21 | 0.30 | 0.36 | 0.42 | 1.30 | 0.22 | 0.09 | -0.29 | -0.10 | 0.90 | 0.25 | 0.38 | 1.04 | 0.23 |
| hr5383 | S | 4874 | 2.60 | 1.34 | 0.35 | 0.16 | 0.31 | 0.29 | 0.68 | 0.23 | -0.02 | 0.08 | 0.10 | 0.19 | 0.14 | 0.09 | 0.02 | 0.15 | 0.15 | 0.63 | 0.26 | 0.10 | -0.23 | 0.08 | 0.32 | 0.15 | 0.17 | 0.38 | 0.14 |
| hr5394 | S | 4440 | 1.99 | 1.51 | 0.31 | 0.09 | 0.30 | 0.28 | 0.86 | 0.04 | -0.06 | 0.04 | 0.18 | 0.15 | 0.08 | -0.03 | 0.06 | 0.10 | 0.34 | 1.00 | 0.15 | 0.07 | -0.45 | -0.16 | 0.39 | 0.07 | 0.27 | 0.58 | 0.05 |
| hr5429 | S | 4281 | 1.31 | 1.61 | 0.13 | 0.10 | 0.24 | 0.24 | 1.12 | -0.04 | -0.13 | -0.02 | 0.05 | 0.10 | -0.08 | -0.10 | -0.07 | -0.02 | 0.23 | 2.63 | 0.04 | -0.11 | -0.58 | -0.29 | 0.31 | -0.14 | 0.03 | 0.46 | -0.01 |
| hr5430 | S | 4095 | 0.65 | 1.75 | 0.22 | -0.08 | 0.13 | 0.15 | 1.11 | -0.20 | -0.32 | -0.17 | -0.01 | 0.09 | -0.06 | -0.20 | -0.11 | -0.11 | 0.27 | | 0.27 | -0.13 | -0.58 | -0.17 | 0.07 | -0.03 | 0.18 | 0.39 | -0.32 |
| hr5454 | S | 4736 | 2.31 | 1.44 | 0.09 | 0.05 | 0.19 | 0.17 | 0.58 | 0.06 | -0.09 | -0.02 | -0.02 | 0.04 | -0.02 | -0.05 | -0.07 | 0.00 | 0.14 | 0.44 | 0.15 | 0.01 | -0.35 | 0.04 | 0.20 | 0.00 | 0.14 | 0.20 | 0.06 |
| hr5481 | S | 4903 | 2.42 | 1.40 | 0.12 | 0.02 | 0.13 | 0.06 | 0.39 | -0.04 | -0.11 | -0.07 | -0.14 | -0.12 | -0.27 | -0.20 | -0.15 | -0.15 | -0.03 | 0.06 | 0.03 | -0.14 | -0.43 | -0.15 | -0.03 | -0.15 | 0.02 | 0.06 | 0.05 |
| hr5487 | S | 6487 | 3.56 | 4.24 | | -0.02 | | | 0.08 | -0.31 | 0.15 | -0.12 | 0.13 | 0.56 | 0.48 | -0.09 | -0.31 | 0.75 | -0.01 | -0.64 | | 1.52 | 0.39 | | -0.75 | 0.49 | 0.71 | 0.15 | -0.46 | |
| hr5502 | S | 4864 | 2.35 | 1.49 | 0.25 | 0.05 | 0.18 | 0.19 | 0.51 | 0.09 | -0.09 | -0.06 | -0.09 | 0.07 | -0.04 | -0.03 | -0.09 | -0.02 | -0.03 | 0.57 | 0.23 | -0.02 | -0.26 | 0.15 | 0.24 | 0.08 | 0.14 | 0.16 | 0.07 |
| hr5518 | S | 4641 | 2.10 | 1.52 | 0.03 | 0.14 | 0.19 | 0.20 | 0.54 | -0.07 | -0.12 | -0.04 | -0.03 | -0.04 | -0.14 | -0.13 | -0.08 | -0.03 | 0.15 | 0.49 | 0.05 | -0.23 | -0.66 | -0.23 | 0.09 | -0.17 | -0.01 | 0.51 | 0.04 |
| hr5573 | S | 4627 | 2.23 | 1.45 | 0.37 | 0.19 | 0.30 | 0.34 | 0.82 | 0.19 | 0.01 | 0.14 | 0.23 | 0.25 | 0.22 | 0.11 | 0.10 | 0.20 | 0.31 | 1.38 | 0.24 | 0.20 | -0.18 | 0.11 | 0.52 | 0.17 | 0.31 | 0.53 | 0.15 |
| hr5600 | S | 3962 | 0.16 | 1.81 | 0.49 | -0.08 | 0.20 | 0.25 | 1.19 | -0.13 | -0.49 | -0.28 | -0.10 | 0.05 | -0.20 | -0.25 | -0.20 | -0.10 | 0.66 | | 0.36 | -0.25 | -0.63 | -0.48 | 0.10 | -0.06 | 0.07 | 0.54 | -0.53 |
| hr5601 | S | 4664 | 2.22 | 1.35 | 0.02 | -0.28 | 0.04 | 0.10 | 0.57 | -0.11 | -0.22 | -0.15 | -0.15 | -0.07 | -0.18 | -0.17 | -0.19 | -0.13 | -0.02 | 0.38 | 0.05 | -0.14 | -0.51 | 0.06 | 0.25 | 0.02 | 0.13 | 0.11 | -0.12 |
| hr5602 | S | 4920 | 2.05 | 1.73 | 0.28 | -0.04 | 0.09 | 0.14 | 0.46 | 0.08 | -0.14 | -0.11 | -0.17 | 0.03 | -0.14 | -0.08 | -0.14 | -0.07 | -0.14 | 0.33 | 0.31 | -0.02 | -0.30 | 0.16 | 0.11 | -0.04 | 0.02 | -0.04 | 0.03 |
| hr5609 | S | 4800 | 2.50 | 1.39 | 0.34 | 0.13 | 0.33 | 0.34 | 0.80 | 0.24 | 0.00 | 0.08 | 0.14 | 0.21 | 0.16 | 0.09 | 0.03 | 0.14 | 0.14 | 0.66 | 0.26 | 0.12 | -0.24 | 0.11 | 0.41 | 0.14 | 0.22 | 0.48 | 0.16 |
| hr5616 | S | 4302 | 1.23 | 1.49 | 0.03 | -0.02 | 0.11 | 0.17 | 0.67 | -0.04 | -0.36 | -0.25 | -0.10 | -0.14 | -0.23 | -0.23 | -0.20 | -0.17 | | 0.97 | | -0.37 | -1.03 | -0.30 | 0.10 | -0.35 | -0.27 | -0.46 | -0.12 |
| hr5620 | S | 4720 | 2.57 | 1.24 | 0.09 | 0.19 | 0.15 | 0.13 | 0.48 | 0.01 | -0.14 | 0.00 | 0.06 | 0.04 | -0.07 | -0.08 | -0.05 | -0.01 | 0.19 | 0.61 | 0.00 | -0.17 | -0.47 | -0.28 | 0.19 | 0.01 | 0.09 | 0.59 | -0.03 |
| hr5635 | S | 4812 | 2.21 | 1.42 | -0.17 | -0.17 | -0.08 | -0.10 | 0.11 | -0.24 | -0.37 | -0.34 | -0.37 | -0.32 | -0.40 | -0.38 | -0.36 | -0.35 | -0.31 | 0.14 | -0.08 | -0.40 | -0.69 | -0.36 | -0.33 | -0.37 | -0.33 | -0.27 | -0.19 |
| hr5648 | S | 4722 | 2.33 | 1.42 | 0.25 | 0.18 | 0.28 | 0.31 | 0.72 | 0.13 | -0.02 | 0.05 | 0.09 | 0.13 | 0.14 | 0.03 | 0.01 | 0.10 | 0.24 | 0.55 | 0.19 | 0.12 | -0.29 | 0.07 | 0.34 | 0.05 | 0.20 | 0.45 | 0.12 |
| hr5673 | S | 4399 | 1.73 | 1.50 | 0.06 | 0.02 | 0.13 | 0.10 | 0.82 | -0.07 | -0.18 | -0.07 | 0.01 | 0.02 | -0.13 | -0.17 | -0.09 | -0.05 | 0.04 | 0.56 | 0.01 | -0.18 | -0.55 | -0.17 | 0.18 | -0.04 | 0.15 | 0.66 | -0.12 |
| hr5681a | S | 4821 | 2.18 | 1.30 | -0.10 | -0.16 | 0.07 | -0.08 | 0.04 | -0.15 | -0.26 | -0.22 | -0.30 | -0.23 | -0.35 | -0.30 | -0.27 | -0.25 | -0.17 | -0.18 | 0.19 | -0.27 | -0.59 | -0.06 | -0.15 | -0.17 | -0.09 | -0.16 | -0.11 |

| ID | | T | | | | | | | | | | | | | | | | | | | | | | | | | | |
|---|---|---|---|---|---|---|---|---|---|---|---|---|---|---|---|---|---|---|---|---|---|---|---|---|---|---|---|---|
| hr5681b | S | 5812 | 4.47 | 0.82 | -0.39 | -0.31 | -0.28 | -0.30 | -0.15 | -0.29 | -0.28 | -0.31 | -0.38 | -0.40 | -0.48 | -0.42 | -0.37 | -0.42 | -0.51 | -0.29 | | -0.30 | | -0.26 | | 0.17 | -0.13 | -0.16 | 0.06 |
| hr5707 | S | 4783 | 2.36 | 1.50 | 0.61 | 0.37 | 0.46 | 0.53 | 1.05 | 0.29 | 0.20 | 0.26 | 0.33 | 0.37 | 0.34 | 0.24 | 0.26 | 0.35 | 0.35 | 1.07 | 0.35 | 0.28 | -0.15 | 0.10 | 0.68 | 0.18 | 0.27 | 0.68 | 0.27 |
| hr5709 | S | 4705 | 2.23 | 1.36 | -0.12 | -0.04 | 0.08 | 0.04 | 0.38 | -0.10 | -0.23 | -0.17 | -0.24 | -0.18 | -0.27 | -0.25 | -0.27 | -0.21 | 0.02 | 0.24 | -0.03 | -0.37 | -0.61 | -0.33 | -0.01 | -0.32 | -0.18 | -0.05 | -0.20 |
| hr5744 | S | 4531 | 2.13 | 1.47 | 0.38 | 0.27 | 0.43 | 0.39 | 1.02 | 0.22 | 0.06 | 0.16 | 0.33 | 0.30 | 0.23 | 0.09 | 0.15 | 0.20 | 0.28 | 0.91 | 0.23 | 0.02 | -0.23 | -0.09 | 0.56 | 0.04 | 0.31 | 0.84 | 0.15 |
| hr5769 | S | 6040 | 3.68 | 1.74 | 0.21 | 0.01 | 0.04 | 0.09 | 0.13 | 0.06 | 0.01 | -0.04 | -0.04 | -0.01 | -0.10 | -0.08 | -0.02 | -0.07 | -0.31 | 0.04 | | -0.03 | 0.14 | -0.02 | 0.02 | -0.10 | 0.06 | -0.01 | 0.34 |
| hr5785 | S | 4798 | 2.33 | 1.38 | -0.06 | -0.08 | 0.06 | 0.00 | 0.43 | -0.11 | -0.26 | -0.19 | -0.23 | -0.14 | -0.27 | -0.24 | -0.27 | -0.21 | -0.09 | 0.12 | -0.02 | -0.28 | -0.57 | -0.19 | -0.06 | -0.22 | -0.15 | -0.22 | -0.12 |
| hr5794 | S | 4135 | 0.83 | 1.65 | 0.12 | -0.04 | 0.11 | 0.15 | 1.12 | -0.15 | -0.35 | -0.17 | -0.04 | 0.04 | -0.16 | -0.21 | -0.14 | -0.11 | 0.15 | | 0.15 | -0.11 | -0.63 | -0.18 | 0.33 | -0.14 | 0.18 | 0.42 | -0.29 |
| hr5802 | S | 4889 | 2.50 | 1.33 | 0.21 | -0.01 | 0.16 | 0.18 | 0.52 | 0.05 | -0.12 | -0.06 | -0.08 | 0.05 | -0.04 | -0.06 | -0.10 | -0.03 | 0.01 | 0.22 | 0.61 | 0.35 | 0.09 | 0.37 | 0.33 | 0.30 | 0.27 | 0.36 | 0.00 |
| hr5810 | S | 4633 | 2.13 | 1.39 | -0.02 | -0.05 | 0.01 | 0.12 | 0.39 | -0.16 | -0.26 | -0.22 | -0.18 | -0.12 | -0.16 | -0.21 | -0.20 | -0.14 | 0.02 | 0.43 | 0.18 | 0.03 | -0.50 | -0.16 | 0.07 | -0.11 | -0.03 | 0.25 | -0.17 |
| hr5811 | S | 4517 | 1.96 | 1.57 | 0.60 | 0.38 | 0.48 | 0.52 | 1.21 | 0.39 | 0.15 | 0.26 | 0.43 | 0.42 | 0.36 | 0.23 | 0.29 | 0.38 | | 1.14 | 0.33 | 0.27 | -0.20 | 0.01 | 0.81 | 0.16 | 0.41 | 0.93 | 0.17 |
| hr5828 | S | 4625 | 2.30 | 1.33 | 0.13 | 0.06 | 0.18 | 0.21 | 0.62 | 0.02 | -0.05 | 0.02 | 0.07 | 0.08 | 0.02 | -0.03 | -0.05 | 0.05 | 0.26 | 0.42 | 0.18 | 0.03 | -0.35 | 0.09 | 0.38 | 0.15 | 0.27 | 0.72 | 0.11 |
| hr5835 | S | 5172 | 2.97 | 1.23 | 0.46 | 0.13 | 0.25 | 0.19 | 0.57 | 0.22 | 0.08 | 0.14 | 0.14 | 0.21 | 0.05 | 0.10 | 0.08 | 0.12 | 0.23 | 0.09 | 0.52 | 0.25 | -0.03 | 0.25 | 0.35 | 0.18 | 0.24 | 0.22 | 0.38 |
| hr5840 | S | 4982 | 2.32 | 1.44 | 0.20 | -0.05 | 0.08 | 0.09 | 0.44 | 0.02 | -0.17 | -0.16 | -0.23 | -0.05 | -0.19 | -0.11 | -0.19 | -0.13 | -0.33 | 0.21 | 0.25 | -0.03 | -0.34 | 0.15 | 0.05 | 0.10 | 0.03 | -0.17 | 0.01 |
| hr5841 | S | 4659 | 2.22 | 1.46 | 0.36 | 0.22 | 0.38 | 0.40 | 0.84 | 0.20 | 0.05 | 0.15 | 0.21 | 0.24 | 0.24 | 0.12 | 0.13 | 0.23 | 0.35 | 0.61 | 0.26 | 0.11 | -0.30 | 0.11 | 0.51 | 0.19 | 0.32 | 0.60 | 0.06 |
| hr5855 | S | 4626 | 2.22 | 1.35 | 0.20 | 0.14 | 0.21 | 0.26 | 0.71 | 0.12 | -0.12 | -0.03 | 0.04 | 0.10 | 0.04 | -0.02 | -0.05 | 0.06 | 0.06 | 0.53 | 0.18 | -0.07 | -0.39 | -0.04 | 0.35 | 0.03 | 0.16 | 0.24 | -0.02 |
| hr5888 | S | 4715 | 2.02 | 1.48 | 0.04 | 0.01 | 0.06 | 0.09 | 0.33 | -0.10 | -0.28 | -0.21 | -0.22 | -0.12 | -0.22 | -0.21 | -0.23 | -0.18 | -0.08 | -0.25 | 0.02 | -0.20 | -0.61 | -0.16 | -0.07 | -0.16 | -0.12 | -0.01 | -0.19 |
| hr5893 | S | 4873 | 2.92 | 1.08 | 0.31 | 0.32 | 0.33 | 0.26 | 0.70 | 0.22 | 0.08 | 0.19 | 0.28 | 0.26 | 0.19 | 0.13 | 0.09 | 0.19 | 0.36 | 0.62 | 0.26 | 0.11 | -0.22 | 0.08 | 0.49 | 0.23 | 0.30 | 0.67 | 0.23 |
| hr5922 | S | 4823 | 2.43 | 1.43 | 0.34 | 0.12 | 0.23 | 0.33 | 0.85 | 0.15 | -0.07 | 0.00 | 0.03 | 0.15 | 0.09 | 0.04 | -0.03 | 0.10 | 0.02 | 0.29 | 0.24 | 0.04 | -0.27 | 0.01 | 0.28 | 0.04 | 0.08 | 0.35 | 0.06 |
| hr5940 | S | 4593 | 2.40 | 1.32 | 0.44 | 0.39 | 0.58 | 0.39 | 0.87 | 0.28 | 0.24 | 0.34 | 0.51 | 0.36 | 0.21 | 0.17 | 0.28 | 0.34 | 0.44 | 1.08 | 0.34 | 0.31 | -0.14 | 0.14 | 0.83 | 0.31 | 0.72 | 1.01 | 0.59 |
| hr5947 | S | 4364 | 1.57 | 1.59 | 0.07 | -0.03 | 0.10 | 0.19 | 0.72 | -0.07 | -0.18 | -0.14 | -0.08 | -0.06 | -0.16 | -0.18 | -0.13 | -0.08 | 0.04 | 0.75 | 0.03 | -0.17 | -0.59 | -0.13 | 0.21 | -0.10 | 0.06 | 0.36 | -0.07 |
| hr5966 | S | 4798 | 2.23 | 1.46 | 0.01 | -0.07 | 0.06 | 0.10 | 0.33 | -0.08 | -0.17 | -0.16 | -0.15 | -0.09 | -0.17 | -0.17 | -0.16 | -0.12 | -0.01 | 0.20 | 0.00 | -0.24 | -0.50 | -0.11 | 0.08 | -0.10 | -0.06 | 0.01 | -0.11 |
| hr5976 | S | 4933 | 2.60 | 1.41 | 0.40 | 0.13 | 0.26 | 0.29 | 0.75 | 0.17 | 0.00 | 0.04 | 0.06 | 0.17 | 0.13 | 0.07 | 0.00 | 0.10 | 0.04 | 0.37 | 0.31 | 0.11 | -0.19 | 0.10 | 0.35 | 0.10 | 0.11 | 0.30 | 0.14 |
| hr6038 | S | 4744 | 2.47 | 1.39 | 0.27 | 0.19 | 0.33 | 0.29 | 0.82 | 0.20 | 0.06 | 0.11 | 0.17 | 0.17 | 0.16 | 0.07 | 0.06 | 0.15 | 0.43 | 0.28 | 0.28 | 0.12 | -0.24 | 0.11 | 0.39 | 0.05 | 0.32 | 0.55 | 0.16 |
| hr6057 | S | 4560 | 2.27 | 1.37 | 0.34 | 0.28 | 0.40 | 0.32 | 0.90 | 0.22 | 0.12 | 0.17 | 0.31 | 0.23 | 0.17 | 0.06 | 0.12 | 0.18 | 0.47 | 1.20 | 0.12 | 0.13 | -0.31 | -0.10 | 0.51 | 0.11 | 0.30 | 0.78 | 0.16 |
| hr6065 | S | 4620 | 2.28 | 1.47 | 0.40 | 0.32 | 0.44 | 0.38 | 1.01 | 0.24 | 0.02 | 0.14 | 0.26 | 0.25 | 0.22 | 0.12 | 0.14 | 0.24 | 0.30 | 1.02 | 0.22 | 0.03 | -0.42 | -0.02 | 0.55 | 0.14 | 0.29 | 0.58 | 0.18 |
| hr6075 | S | 4837 | 2.42 | 1.40 | 0.12 | -0.14 | 0.12 | 0.12 | 0.20 | 0.00 | -0.11 | -0.09 | -0.11 | 0.00 | -0.12 | -0.08 | -0.14 | -0.07 | -0.09 | 0.15 | 0.19 | -0.04 | -0.29 | 0.11 | 0.32 | 0.08 | 0.14 | 0.10 | 0.06 |
| hr6104 | S | 4757 | 2.32 | 1.42 | 0.17 | 0.09 | 0.15 | 0.20 | 0.75 | 0.00 | -0.13 | -0.10 | -0.08 | -0.01 | -0.06 | -0.10 | -0.12 | -0.05 | 0.10 | 0.56 | 0.32 | 0.12 | -0.16 | 0.13 | 0.19 | 0.06 | 0.13 | 0.27 | 0.02 |
| hr6121 | S | 4682 | 2.16 | 1.49 | 0.08 | -0.10 | 0.11 | 0.18 | 0.42 | -0.06 | -0.14 | -0.12 | -0.10 | -0.01 | -0.04 | -0.11 | -0.14 | -0.05 | 0.05 | 0.26 | 0.12 | -0.19 | -0.39 | -0.03 | 0.28 | -0.10 | 0.05 | 0.14 | -0.11 |
| hr6124 | S | 5030 | 2.57 | 1.44 | 0.25 | 0.03 | 0.13 | 0.13 | 0.54 | 0.08 | -0.11 | -0.07 | -0.13 | 0.02 | -0.01 | -0.06 | -0.13 | -0.06 | -0.20 | 0.14 | 0.27 | 0.00 | -0.22 | 0.17 | 0.17 | 0.12 | 0.14 | 0.23 | 0.12 |
| hr6126 | S | 4636 | 2.13 | 1.48 | 0.52 | 0.27 | 0.39 | 0.48 | 0.97 | 0.27 | 0.13 | 0.23 | 0.34 | 0.35 | 0.22 | 0.21 | 0.24 | 0.34 | 0.42 | 1.23 | 0.30 | 0.17 | -0.20 | 0.17 | 0.67 | 0.27 | 0.27 | 0.59 | 0.18 |
| hr6130 | S | 4946 | 2.63 | 1.43 | 0.38 | 0.13 | 0.25 | 0.27 | 0.50 | 0.13 | 0.00 | 0.07 | 0.06 | 0.17 | 0.14 | 0.06 | 0.02 | 0.10 | 0.04 | 0.49 | 0.29 | 0.22 | -0.19 | 0.13 | 0.28 | 0.12 | 0.21 | 0.23 | 0.11 |
| hr6132 | S | 4962 | 2.50 | 1.41 | 0.31 | 0.05 | 0.16 | 0.13 | 0.46 | 0.08 | -0.08 | -0.07 | -0.09 | 0.01 | -0.09 | -0.07 | -0.10 | -0.05 | -0.24 | -0.07 | 0.18 | -0.04 | -0.27 | 0.07 | 0.16 | 0.02 | 0.07 | 0.06 | 0.07 |
| hr6145 | S | 4696 | 2.54 | 1.35 | 0.49 | 0.24 | 0.46 | 0.38 | 0.97 | 0.20 | 0.15 | 0.27 | 0.39 | 0.37 | 0.30 | 0.17 | 0.26 | 0.31 | 0.57 | 1.25 | 0.29 | 0.13 | -0.22 | -0.04 | 0.71 | 0.24 | 0.40 | 0.96 | 0.30 |
| hr6147 | S | 5038 | 2.29 | 1.68 | 0.71 | 0.21 | 0.33 | 0.39 | 0.85 | 0.24 | 0.09 | 0.18 | 0.16 | 0.32 | 0.23 | 0.18 | 0.17 | 0.26 | 0.10 | 0.81 | 0.37 | 0.20 | -0.07 | 0.16 | 0.30 | 0.07 | 0.12 | 0.17 | 0.12 |
| hr6148 | S | 4912 | 2.15 | 1.60 | 0.25 | -0.06 | 0.04 | 0.14 | 0.51 | 0.06 | -0.16 | -0.12 | -0.17 | 0.00 | -0.09 | -0.09 | -0.17 | -0.09 | -0.18 | 0.59 | 0.23 | -0.04 | -0.31 | 0.13 | 0.13 | 0.02 | 0.05 | -0.06 | 0.01 |
| hr6150 | S | 4735 | 2.38 | 1.42 | 0.14 | 0.07 | 0.24 | 0.25 | 0.61 | 0.06 | -0.02 | 0.04 | 0.08 | 0.08 | 0.08 | -0.02 | -0.01 | 0.07 | 0.19 | 0.24 | 0.14 | -0.07 | -0.37 | 0.00 | 0.22 | 0.01 | 0.17 | 0.42 | 0.06 |
| hr6190 | S | 4718 | 2.36 | 1.43 | 0.08 | 0.05 | 0.21 | 0.22 | 0.61 | 0.04 | -0.06 | 0.00 | 0.06 | 0.03 | -0.04 | -0.05 | -0.04 | 0.04 | 0.26 | 0.22 | 0.17 | -0.11 | -0.37 | -0.10 | 0.17 | -0.11 | 0.12 | 0.44 | -0.02 |
| hr6199 | S | 4641 | 1.82 | 1.66 | 0.18 | 0.06 | 0.22 | 0.25 | 0.61 | -0.03 | -0.19 | -0.06 | -0.05 | 0.11 | -0.12 | -0.09 | -0.11 | -0.04 | 0.01 | 0.40 | 0.07 | -0.15 | -0.43 | -0.16 | -0.26 | -0.14 | -0.12 | 0.46 | -0.08 |
| hr6220 | S | 4932 | 2.41 | 1.40 | 0.01 | -0.09 | -0.01 | -0.01 | 0.37 | -0.12 | -0.28 | -0.22 | -0.25 | -0.11 | -0.24 | -0.21 | -0.27 | -0.23 | -0.41 | -0.04 | 0.36 | -0.19 | -0.34 | 0.05 | 0.03 | 0.03 | -0.04 | -0.19 | -0.05 |
| hr6259 | S | 4756 | 2.62 | 1.21 | 0.05 | -0.06 | 0.11 | 0.08 | 0.40 | -0.03 | -0.11 | -0.02 | 0.01 | 0.00 | -0.10 | -0.11 | -0.10 | -0.03 | 0.10 | 0.29 | 0.05 | -0.05 | -0.39 | 0.00 | 0.19 | 0.04 | 0.21 | 0.35 | -0.04 |
| hr6280 | S | 4640 | 2.30 | 1.34 | 0.41 | 0.27 | 0.36 | 0.34 | 0.99 | 0.23 | 0.04 | 0.16 | 0.27 | 0.28 | 0.27 | 0.13 | 0.11 | 0.22 | 0.48 | 0.96 | 0.27 | 0.07 | -0.20 | 0.05 | 0.75 | 0.13 | 0.29 | 0.60 | 0.06 |
| hr6287 | S | 4838 | 2.34 | 1.38 | 0.06 | -0.03 | 0.04 | 0.09 | 0.18 | -0.05 | -0.19 | -0.14 | -0.17 | -0.07 | -0.20 | -0.15 | -0.20 | -0.13 | -0.08 | -0.10 | 0.10 | -0.10 | -0.43 | 0.09 | 0.12 | 0.04 | 0.10 | 0.01 | 0.06 |
| hr6292 | S | 4979 | 2.64 | 1.36 | 0.03 | -0.06 | 0.08 | 0.04 | 0.27 | -0.04 | -0.17 | -0.11 | -0.13 | -0.08 | -0.22 | -0.14 | -0.17 | -0.13 | -0.20 | -0.07 | 0.17 | -0.06 | -0.34 | 0.15 | 0.18 | 0.15 | 0.17 | 0.10 | 0.10 |
| hr6305 | S | 4991 | 2.41 | 1.51 | 0.07 | -0.12 | 0.06 | 0.01 | 0.16 | -0.10 | -0.15 | -0.13 | -0.17 | -0.08 | -0.21 | -0.18 | -0.19 | -0.17 | -0.34 | -0.14 | 0.24 | -0.14 | -0.48 | 0.07 | 0.11 | 0.03 | 0.00 | -0.11 | 0.06 |
| hr6307 | S | 4634 | 2.00 | 1.49 | 0.23 | 0.12 | 0.23 | 0.24 | 0.59 | 0.10 | -0.11 | -0.03 | 0.01 | 0.07 | 0.01 | -0.04 | -0.06 | 0.05 | 0.09 | 0.58 | 0.10 | 0.00 | -0.46 | -0.09 | 0.18 | -0.06 | 0.09 | 0.25 | -0.05 |
| hr6333 | S | 4779 | 2.32 | 1.45 | 0.01 | -0.03 | 0.11 | 0.09 | 0.44 | -0.06 | -0.15 | -0.09 | -0.11 | -0.08 | -0.10 | -0.15 | -0.11 | -0.08 | 0.06 | 0.09 | 0.06 | -0.20 | -0.46 | -0.13 | 0.08 | -0.11 | -0.01 | 0.24 | -0.02 |
| hr6342 | S | 4702 | 2.37 | 1.31 | -0.05 | -0.01 | 0.04 | 0.01 | 0.40 | -0.07 | -0.14 | -0.11 | -0.10 | -0.12 | -0.19 | -0.19 | -0.17 | -0.13 | 0.02 | 0.29 | 0.00 | -0.05 | -0.53 | -0.06 | 0.08 | 0.02 | 0.09 | 0.19 | -0.13 |
| hr6359 | S | 5078 | 2.87 | 1.27 | 0.17 | 0.05 | 0.14 | 0.08 | 0.23 | 0.07 | -0.01 | 0.01 | -0.01 | 0.08 | -0.08 | -0.03 | -0.07 | -0.05 | -0.24 | 0.09 | 0.33 | 0.08 | -0.14 | 0.25 | 0.25 | 0.17 | 0.23 | 0.26 | 0.15 |
| hr6360 | S | 4886 | 2.49 | 1.36 | 0.46 | 0.10 | 0.38 | 0.46 | 1.10 | 0.26 | 0.04 | 0.13 | 0.17 | 0.26 | 0.20 | 0.18 | 0.08 | 0.23 | 0.23 | 0.90 | 0.30 | 0.27 | -0.16 | 0.20 | 0.49 | 0.19 | 0.27 | 0.49 | 0.59 |
| hr6363 | S | 4602 | 2.35 | 1.23 | 0.27 | 0.20 | 0.30 | 0.28 | 0.68 | 0.13 | 0.03 | 0.12 | 0.26 | 0.19 | 0.18 | 0.04 | 0.05 | 0.14 | 0.38 | 0.54 | 0.24 | 0.11 | -0.26 | 0.01 | 0.65 | 0.05 | 0.28 | 0.64 | 0.12 |

| ID | | Teff | | | | | | | | | | | | | | | | | | | | | | | | | |
|---|---|---|---|---|---|---|---|---|---|---|---|---|---|---|---|---|---|---|---|---|---|---|---|---|---|---|---|
| hr6364 | S | 4281 | 1.38 | 1.50 | 0.36 | 0.17 | 0.41 | 0.34 | 1.11 | 0.09 | -0.07 | 0.09 | 0.28 | 0.20 | 0.00 | -0.06 | 0.05 | 0.06 | 0.32 | 0.91 | 0.17 | -0.09 | -0.52 | -0.29 | 0.37 | -0.13 | 0.30 | 1.03 | -0.12 |
| hr6365 | S | 4889 | 2.24 | 1.53 | 0.52 | 0.18 | 0.30 | 0.39 | 0.84 | 0.15 | 0.02 | 0.06 | 0.06 | 0.18 | 0.19 | 0.07 | 0.05 | 0.16 | 0.07 | 0.51 | 0.39 | 0.03 | -0.23 | -0.04 | 0.25 | -0.03 | 0.05 | 0.27 | 0.16 |
| hr6388 | S | 4329 | 1.74 | 1.48 | 0.23 | -0.01 | 0.29 | 0.30 | 1.08 | 0.01 | -0.08 | 0.01 | 0.22 | 0.13 | 0.04 | -0.06 | 0.02 | 0.08 | 0.32 | 1.04 | 0.09 | 0.00 | -0.52 | -0.13 | 0.35 | -0.04 | 0.29 | 0.70 | 0.01 |
| hr6390 | S | 4697 | 2.38 | 1.30 | -0.07 | -0.09 | 0.03 | 0.04 | 0.30 | -0.14 | -0.13 | -0.13 | -0.11 | -0.11 | -0.17 | -0.21 | -0.17 | -0.15 | 0.05 | 0.17 | 0.13 | -0.17 | -0.55 | -0.13 | 0.08 | -0.06 | 0.10 | 0.26 | -0.10 |
| hr6394 | S | 6068 | 3.37 | 1.76 | -0.11 | -0.05 | -0.02 | -0.03 | -0.03 | -0.03 | -0.07 | -0.14 | -0.16 | -0.12 | -0.27 | -0.18 | -0.09 | -0.18 | -0.38 | -0.11 | 0.67 | -0.05 | 0.04 | 0.00 | -0.14 | 0.03 | -0.12 | -0.72 | -0.12 |
| hr6404 | S | 4658 | 2.25 | 1.45 | 0.48 | 0.28 | 0.47 | 0.42 | 1.01 | 0.21 | 0.09 | 0.15 | 0.26 | 0.27 | 0.26 | 0.14 | 0.13 | 0.24 | 0.31 | 0.91 | 0.24 | 0.18 | -0.26 | 0.00 | 0.76 | 0.06 | 0.22 | 0.53 | 0.18 |
| hr6415 | S | 4508 | 1.93 | 1.55 | 0.42 | 0.31 | 0.53 | 0.51 | 1.19 | 0.19 | 0.09 | 0.21 | 0.35 | 0.31 | 0.30 | 0.17 | 0.20 | 0.29 | | 1.16 | 0.21 | 0.23 | -0.28 | -0.02 | 0.71 | 0.14 | 0.32 | 0.91 | 0.24 |
| hr6443 | S | 4870 | 2.55 | 1.40 | 0.18 | -0.02 | 0.18 | 0.18 | 0.51 | 0.06 | -0.06 | -0.02 | -0.02 | 0.06 | 0.00 | -0.03 | -0.07 | -0.01 | 0.08 | 0.44 | 0.25 | 0.09 | -0.27 | 0.11 | 0.21 | 0.14 | 0.15 | 0.23 | 0.05 |
| hr6444 | S | 4773 | 2.26 | 1.47 | 0.44 | 0.17 | 0.33 | 0.36 | 0.99 | 0.14 | -0.10 | -0.01 | 0.04 | 0.18 | 0.13 | 0.03 | 0.00 | 0.11 | 0.09 | 0.58 | 0.27 | -0.03 | -0.29 | -0.14 | 0.26 | 0.01 | -0.03 | 0.34 | 0.04 |
| hr6448 | S | 4650 | 2.19 | 1.48 | 0.10 | 0.07 | 0.23 | 0.26 | 0.65 | 0.00 | -0.09 | 0.02 | 0.07 | 0.06 | 0.04 | -0.05 | -0.02 | 0.05 | 0.25 | 0.54 | 0.04 | -0.12 | -0.43 | -0.07 | 0.20 | -0.10 | 0.09 | 0.40 | 0.03 |
| hr6472 | S | 4980 | 2.66 | 1.40 | 0.38 | 0.09 | 0.27 | 0.23 | 0.48 | 0.21 | -0.01 | 0.06 | 0.07 | 0.19 | 0.11 | 0.06 | 0.02 | 0.07 | 0.01 | 0.34 | 0.24 | 0.11 | -0.17 | 0.19 | 0.29 | 0.07 | 0.12 | 0.19 | 0.21 |
| hr6488 | S | 5461 | 3.18 | 0.50 | -0.10 | -0.20 | -0.11 | -0.22 | 0.03 | -0.31 | -0.14 | -0.02 | 0.07 | 0.05 | -0.17 | -0.24 | -0.05 | -0.20 | -0.79 | 0.49 | 0.39 | 0.01 | 0.14 | -0.01 | 0.23 | 0.28 | 0.21 | 0.33 | 0.31 |
| hr6524 | S | 5162 | 3.11 | 0.92 | 0.50 | 0.22 | 0.33 | 0.31 | 0.70 | 0.37 | 0.38 | 0.42 | 0.47 | 0.43 | 0.37 | 0.31 | 0.26 | 0.39 | 0.39 | 0.70 | 0.46 | 0.62 | 0.19 | 0.55 | 0.91 | 0.55 | 0.64 | 0.85 | 0.56 |
| hr6564 | S | 4191 | 1.16 | 1.61 | 0.58 | 0.26 | 0.58 | 0.50 | 1.60 | 0.22 | 0.02 | 0.13 | 0.41 | 0.39 | 0.28 | 0.07 | 0.22 | 0.28 | 0.64 | 1.01 | 0.37 | 0.13 | -0.36 | -0.31 | 0.41 | 0.30 | 0.42 | 0.83 | -0.03 |
| hr6566 | S | 4616 | 2.03 | 1.47 | 0.02 | 0.02 | 0.15 | 0.19 | 0.61 | -0.04 | -0.20 | -0.12 | -0.12 | -0.06 | -0.10 | -0.14 | -0.12 | -0.07 | -0.04 | 0.43 | 0.00 | -0.31 | -0.62 | -0.23 | -0.03 | -0.15 | -0.08 | 0.26 | -0.11 |
| hr6575a | S | 4813 | 2.43 | 1.38 | 0.31 | 0.20 | 0.28 | 0.30 | 0.68 | 0.16 | -0.06 | 0.03 | 0.03 | 0.18 | 0.08 | 0.06 | 0.00 | 0.10 | 0.12 | 0.44 | 0.35 | 0.10 | -0.23 | 0.06 | 0.31 | 0.14 | 0.19 | 0.35 | 0.10 |
| hr6575b | S | 6616 | 4.36 | 3.41 | | -0.18 | 0.73 | 1.24 | | -0.21 | 0.61 | 0.70 | 0.93 | 1.02 | 1.22 | 0.33 | 1.35 | 0.59 | | | 1.67 | 1.29 | | | | 1.75 | 0.45 | | 1.27 |
| hr6591 | S | 4705 | 2.55 | 1.25 | 0.04 | -0.02 | 0.17 | 0.10 | 0.50 | -0.03 | -0.03 | 0.06 | 0.14 | -0.01 | -0.02 | -0.01 | -0.03 | -0.01 | 0.28 | 0.46 | 0.10 | -0.08 | -0.34 | -0.06 | 0.22 | 0.07 | 0.26 | 0.42 | 0.03 |
| hr6603 | S | 4553 | 2.20 | 1.57 | 0.58 | 0.43 | 0.49 | 0.54 | 1.23 | 0.29 | 0.19 | 0.30 | 0.45 | 0.42 | 0.39 | 0.26 | 0.33 | 0.42 | 0.64 | 1.31 | 0.28 | 0.28 | -0.24 | 0.04 | 0.63 | 0.33 | 0.47 | 0.84 | 0.37 |
| hr6606 | S | 4819 | 2.48 | 1.42 | 0.04 | -0.03 | 0.16 | 0.17 | 0.45 | -0.02 | -0.08 | -0.04 | -0.05 | -0.02 | -0.02 | -0.09 | -0.09 | -0.05 | 0.00 | 0.27 | 0.12 | -0.01 | -0.39 | 0.06 | 0.15 | 0.07 | 0.13 | 0.18 | 0.05 |
| hr6607 | S | 4745 | 2.38 | 1.40 | 0.65 | 0.30 | 0.53 | 0.51 | 1.04 | 0.32 | 0.12 | 0.20 | 0.37 | 0.36 | 0.33 | 0.25 | 0.25 | 0.37 | 0.33 | 0.91 | 0.38 | 0.37 | -0.12 | 0.12 | 0.45 | 0.18 | 0.27 | 0.47 | 0.29 |
| hr6638 | S | 4944 | 2.64 | 1.36 | 0.32 | 0.13 | 0.22 | 0.22 | 0.68 | 0.18 | -0.06 | 0.04 | 0.07 | 0.15 | 0.08 | 0.04 | -0.01 | 0.07 | -0.03 | 0.50 | 0.20 | 0.15 | -0.20 | 0.10 | 0.25 | 0.05 | 0.10 | 0.35 | 0.16 |
| hr6639 | S | 4468 | 1.79 | 1.61 | 0.27 | 0.11 | 0.35 | 0.28 | 0.91 | 0.11 | -0.12 | 0.01 | 0.11 | 0.15 | 0.09 | -0.01 | 0.04 | 0.11 | 0.26 | 1.05 | 0.04 | -0.22 | -0.50 | -0.30 | 0.30 | -0.07 | 0.13 | 0.52 | -0.08 |
| hr6644 | S | 4555 | 2.14 | 1.49 | 0.36 | 0.32 | 0.36 | 0.41 | 1.00 | 0.28 | 0.04 | 0.25 | 0.36 | 0.28 | 0.14 | 0.09 | 0.17 | 0.23 | 0.29 | 0.81 | 0.13 | 0.09 | -0.38 | -0.04 | 0.51 | 0.17 | 0.24 | 0.89 | 0.14 |
| hr6654 | S | 4661 | 2.39 | 1.32 | 0.20 | 0.05 | 0.19 | 0.15 | 0.54 | 0.02 | -0.10 | -0.02 | 0.05 | 0.05 | 0.04 | -0.06 | -0.04 | 0.02 | 0.20 | 0.34 | 0.10 | -0.12 | -0.39 | -0.07 | 0.33 | 0.03 | 0.20 | 0.51 | 0.00 |
| hr6659 | S | 4577 | 1.59 | 1.48 | 0.12 | -0.11 | 0.02 | 0.13 | 0.61 | -0.18 | -0.41 | -0.34 | -0.37 | -0.09 | -0.27 | -0.22 | -0.28 | -0.18 | -0.18 | 0.19 | 0.05 | -0.19 | -0.75 | -0.21 | -0.08 | -0.36 | -0.27 | 0.03 | -0.31 |
| hr6698 | S | 4874 | 2.49 | 1.58 | 0.57 | 0.15 | 0.32 | 0.35 | 0.76 | 0.17 | 0.07 | 0.10 | 0.21 | 0.23 | 0.20 | 0.12 | 0.10 | 0.18 | 0.21 | 0.56 | 0.37 | 0.21 | -0.26 | 0.05 | 0.27 | 0.02 | 0.13 | 0.23 | 0.16 |
| hr6703 | S | 4930 | 2.45 | 1.47 | 0.35 | 0.12 | 0.19 | 0.21 | 0.65 | 0.16 | -0.09 | -0.02 | -0.01 | 0.13 | 0.04 | 0.00 | -0.02 | 0.02 | -0.03 | 0.55 | 0.21 | 0.00 | -0.16 | 0.11 | 0.16 | -0.01 | 0.06 | 0.08 | 0.07 |
| hr6711 | S | 4999 | 2.73 | 1.39 | 0.33 | 0.10 | 0.23 | 0.26 | 0.60 | 0.17 | 0.08 | 0.11 | 0.11 | 0.19 | 0.15 | 0.08 | 0.04 | 0.10 | 0.04 | 0.09 | 0.34 | 0.15 | -0.06 | 0.19 | 0.35 | 0.18 | 0.26 | 0.34 | 0.23 |
| hr6757 | S | 4943 | 2.35 | 1.71 | 0.63 | 0.17 | 0.32 | 0.23 | 0.28 | 0.12 | 0.08 | 0.14 | 0.11 | 0.21 | -0.05 | 0.07 | 0.04 | 0.07 | -0.10 | 0.34 | 0.55 | 0.11 | -0.17 | 0.32 | 0.23 | 0.12 | 0.27 | 0.33 | 0.21 |
| hr6770 | S | 4921 | 2.45 | 1.50 | 0.30 | -0.04 | 0.25 | 0.23 | 0.51 | 0.07 | -0.06 | -0.01 | 0.00 | 0.10 | 0.01 | 0.00 | -0.06 | 0.01 | -0.07 | 0.35 | 0.31 | 0.02 | -0.22 | 0.12 | 0.22 | 0.04 | 0.13 | 0.11 | 0.20 |
| hr6793 | S | 4516 | 2.13 | 1.45 | 0.46 | 0.36 | 0.46 | 0.41 | 1.10 | 0.19 | 0.08 | 0.15 | 0.34 | 0.30 | 0.30 | 0.13 | 0.23 | 0.31 | 0.52 | 1.12 | 0.20 | 0.06 | -0.35 | -0.18 | 0.71 | 0.18 | 0.32 | 0.65 | 0.25 |
| hr6799 | S | 4487 | 2.04 | 1.46 | 0.39 | 0.18 | 0.44 | 0.37 | 1.12 | 0.15 | 0.05 | 0.13 | 0.35 | 0.26 | 0.17 | 0.06 | 0.12 | 0.20 | 0.26 | 1.20 | 0.19 | 0.00 | -0.23 | -0.14 | 0.53 | -0.04 | 0.29 | 0.80 | 0.07 |
| hr6801 | S | 4687 | 2.11 | 1.50 | 0.08 | -0.06 | 0.14 | 0.12 | 0.47 | 0.02 | -0.15 | -0.09 | -0.09 | -0.01 | -0.06 | -0.11 | -0.13 | -0.09 | -0.05 | 0.34 | 0.14 | -0.14 | -0.43 | 0.01 | 0.15 | 0.01 | 0.11 | 0.14 | -0.08 |
| hr6840 | S | 4898 | 1.97 | 1.63 | -0.42 | -0.25 | -0.25 | -0.28 | -0.13 | -0.47 | -0.70 | -0.50 | -0.66 | -0.66 | -0.97 | -0.71 | -0.60 | -0.67 | -0.84 | -0.25 | 0.12 | -0.68 | -0.81 | -0.70 | -0.54 | -0.79 | -0.60 | -0.79 | -0.47 |
| hr6853 | S | 4798 | 2.37 | 1.51 | -0.24 | 0.05 | -0.02 | -0.02 | 0.28 | -0.19 | -0.30 | -0.14 | -0.22 | -0.33 | -0.57 | -0.42 | -0.26 | -0.33 | -0.13 | -0.08 | -0.05 | -0.46 | -0.71 | -0.57 | -0.37 | -0.40 | -0.34 | -0.17 | -0.16 |
| hr6859 | S | 4202 | 0.78 | 2.09 | 0.31 | 0.00 | 0.04 | 0.23 | 0.94 | -0.21 | -0.33 | -0.22 | -0.17 | 0.07 | -0.13 | -0.14 | -0.17 | -0.10 | 0.20 | | 0.19 | -0.14 | -0.59 | -0.10 | 0.10 | -0.05 | 0.01 | 0.69 | -0.05 |
| hr6865 | S | 4646 | 2.38 | 1.25 | 0.10 | 0.04 | 0.20 | 0.18 | 0.62 | 0.00 | -0.12 | -0.04 | 0.01 | 0.04 | -0.02 | -0.06 | -0.07 | 0.02 | 0.14 | 0.30 | 0.06 | -0.04 | -0.40 | -0.04 | 0.23 | 0.03 | 0.14 | 0.42 | -0.02 |
| hr6866 | S | 4982 | 2.48 | 1.38 | 0.21 | -0.03 | 0.14 | 0.11 | 0.46 | 0.05 | -0.17 | -0.10 | -0.13 | 0.00 | -0.11 | -0.07 | -0.16 | -0.10 | -0.20 | 0.23 | 0.16 | 0.00 | -0.24 | 0.15 | 0.05 | 0.08 | 0.06 | 0.05 | 0.06 |
| hr6872 | S | 4549 | 2.06 | 1.56 | 0.41 | 0.14 | 0.30 | 0.42 | 1.04 | 0.21 | 0.04 | 0.14 | 0.27 | 0.27 | 0.20 | 0.12 | 0.11 | 0.23 | 0.30 | 1.09 | 0.29 | 0.13 | -0.21 | 0.04 | 0.63 | 0.12 | 0.24 | 0.77 | 0.13 |
| hr6884 | S | 4894 | 2.50 | 1.38 | 0.15 | -0.06 | 0.15 | 0.15 | 0.31 | 0.05 | -0.10 | -0.05 | -0.06 | 0.03 | -0.03 | -0.06 | -0.09 | -0.05 | -0.08 | 0.40 | 0.25 | -0.02 | -0.26 | 0.13 | 0.16 | 0.16 | 0.16 | 0.18 | 0.07 |
| hr6895 | S | 4484 | 2.02 | 1.37 | 0.17 | 0.08 | 0.26 | 0.20 | 0.82 | 0.02 | -0.09 | 0.04 | 0.18 | 0.07 | 0.00 | -0.08 | -0.01 | 0.06 | 0.81 | 0.67 | 0.06 | -0.06 | -0.45 | -0.25 | 0.32 | 0.01 | 0.16 | 0.51 | 0.05 |
| hr6935 | S | 4833 | 2.44 | 1.33 | 0.29 | 0.06 | 0.21 | 0.26 | 0.74 | 0.07 | -0.12 | -0.04 | 0.02 | 0.08 | 0.06 | -0.01 | -0.07 | 0.04 | -0.01 | 0.19 | 0.20 | 0.04 | -0.29 | 0.01 | 0.25 | -0.03 | 0.11 | 0.23 | 0.00 |
| hr6945 | S | 4413 | 1.77 | 1.48 | -0.19 | -0.39 | -0.08 | -0.07 | 0.33 | -0.29 | -0.36 | -0.31 | -0.24 | -0.31 | -0.46 | -0.43 | -0.33 | -0.36 | 0.03 | 0.15 | -0.04 | -0.46 | -0.88 | -0.49 | -0.30 | -0.44 | -0.22 | -0.12 | -0.31 |
| hr6970 | S | 5008 | 2.35 | 1.42 | 0.20 | 0.06 | 0.18 | 0.11 | 0.28 | 0.13 | -0.24 | -0.22 | -0.05 | -0.04 | -0.26 | -0.18 | -0.22 | -0.18 | -0.68 | -0.12 | 0.38 | -0.13 | -0.32 | 0.12 | 0.02 | -0.11 | -0.20 | -0.09 | 0.13 |
| hr6973 | S | 4241 | 1.34 | 1.66 | 0.08 | 0.04 | 0.25 | 0.27 | 1.11 | -0.02 | -0.14 | -0.02 | 0.13 | 0.09 | -0.01 | -0.09 | 0.00 | 0.03 | | 0.99 | 0.03 | -0.08 | -0.60 | -0.28 | 0.32 | -0.14 | 0.19 | 0.74 | -0.13 |
| hr7010 | S | 5085 | 2.93 | 1.39 | 0.19 | 0.07 | 0.19 | 0.12 | 0.37 | 0.07 | 0.09 | 0.10 | 0.11 | 0.05 | 0.06 | -0.01 | 0.04 | 0.04 | 0.17 | 0.27 | 0.47 | 0.14 | -0.06 | 0.16 | 0.23 | 0.19 | 0.28 | 0.32 | 0.20 |
| hr7042 | S | 4943 | 2.70 | 1.39 | 0.53 | 0.24 | 0.39 | 0.41 | 0.76 | 0.29 | 0.17 | 0.24 | 0.32 | 0.31 | 0.27 | 0.20 | 0.20 | 0.27 | 0.28 | 0.59 | 0.34 | 0.21 | -0.08 | 0.22 | 0.45 | 0.23 | 0.33 | 0.63 | 0.31 |
| hr7064 | S | 4438 | 1.88 | 1.51 | 0.21 | 0.15 | 0.23 | 0.24 | 0.78 | -0.01 | -0.06 | -0.02 | 0.08 | 0.09 | 0.05 | -0.07 | -0.03 | 0.02 | 0.08 | 0.85 | 0.12 | 0.01 | -0.43 | -0.09 | 0.39 | -0.03 | 0.19 | 0.50 | 0.05 |
| hr7120 | S | 4244 | 1.11 | 1.74 | 0.12 | -0.21 | 0.24 | 0.27 | 1.07 | -0.13 | -0.28 | -0.09 | 0.01 | 0.06 | -0.26 | -0.18 | -0.09 | -0.02 | 0.04 | 0.95 | 0.75 | 0.21 | 0.05 | 0.35 | 0.28 | 0.23 | 0.40 | 0.81 | 0.05 |

| ID | | | | | | | | | | | | | | | | | | | | | | | | | | | | |
|---|---|---|---|---|---|---|---|---|---|---|---|---|---|---|---|---|---|---|---|---|---|---|---|---|---|---|---|---|
| hr7125 | S | 4354 | 1.39 | 2.18 | 0.06 | -0.23 | -0.02 | -0.16 | 0.23 | -0.52 | -0.55 | -0.46 | -0.53 | -0.28 | -0.65 | -0.54 | -0.42 | -0.39 | | -0.59 | -1.07 | -0.44 | -0.17 | -0.42 | -0.18 | 0.46 | -0.28 |
| hr7135 | S | 4666 | 2.24 | 1.40 | 0.00 | -0.14 | 0.01 | 0.02 | 0.48 | -0.15 | -0.23 | -0.17 | -0.17 | -0.11 | -0.19 | -0.19 | -0.22 | -0.16 | -0.14 | 0.29 | 0.10 | -0.17 | -0.53 | -0.07 | 0.09 | 0.00 | 0.06 | 0.30 | -0.15 |
| hr7137 | S | 5024 | 2.54 | 1.44 | 0.28 | 0.04 | 0.17 | 0.18 | 0.48 | 0.16 | -0.05 | 0.00 | 0.01 | 0.11 | -0.04 | 0.00 | -0.05 | -0.01 | -0.07 | 0.34 | 0.42 | 0.10 | -0.17 | 0.33 | 0.27 | 0.16 | 0.19 | 0.06 | 0.14 |
| hr7144 | S | 4951 | 2.40 | 1.61 | 0.30 | | 0.35 | 0.25 | 0.78 | 0.24 | -0.07 | -0.04 | 0.00 | 0.05 | -0.01 | -0.02 | -0.11 | 0.01 | -0.23 | | 0.20 | 0.03 | 0.00 | | 0.19 | 0.20 | 0.18 | 0.19 | |
| hr7146 | S | 4803 | 2.58 | 1.37 | 0.34 | 0.08 | 0.30 | 0.32 | 0.63 | 0.22 | 0.06 | 0.14 | 0.17 | 0.22 | 0.21 | 0.10 | 0.08 | 0.15 | 0.29 | 0.64 | 0.28 | 0.22 | -0.17 | 0.11 | 0.53 | 0.16 | 0.27 | 0.54 | 0.19 |
| hr7148 | S | 4710 | 2.35 | 1.51 | 0.31 | 0.17 | 0.31 | 0.36 | 0.86 | 0.20 | 0.14 | 0.17 | 0.28 | 0.24 | 0.19 | 0.13 | 0.16 | 0.23 | 0.39 | 1.03 | 0.23 | 0.09 | -0.21 | 0.11 | 0.58 | 0.21 | 0.33 | 0.74 | 0.19 |
| hr7176 | S | 4692 | 2.31 | 1.51 | 0.57 | 0.31 | 0.38 | 0.47 | 1.04 | 0.21 | 0.11 | 0.19 | 0.28 | 0.28 | 0.29 | 0.17 | 0.23 | 0.30 | 0.51 | 1.26 | 0.25 | 0.16 | -0.25 | 0.05 | 0.51 | 0.17 | 0.31 | 0.66 | 0.19 |
| hr7180 | S | 4561 | 2.11 | 1.54 | 0.22 | 0.06 | 0.21 | 0.27 | 0.85 | 0.04 | -0.11 | -0.02 | 0.05 | 0.11 | 0.00 | -0.01 | -0.04 | 0.07 | 0.11 | | 0.25 | 0.12 | -0.21 | 0.18 | 0.46 | 0.23 | 0.33 | 0.50 | 0.21 |
| hr7186 | S | 4991 | 2.46 | 1.61 | 0.43 | 0.13 | 0.23 | 0.27 | 0.69 | 0.18 | 0.04 | 0.08 | 0.07 | 0.20 | 0.12 | 0.08 | 0.02 | 0.12 | 0.01 | 0.29 | 0.33 | 0.14 | -0.12 | 0.14 | 0.28 | 0.14 | 0.18 | 0.27 | 0.23 |
| hr7187 | S | 4952 | 2.61 | 1.46 | 0.37 | 0.10 | 0.20 | 0.25 | 0.63 | 0.16 | -0.09 | 0.02 | 0.04 | 0.15 | 0.04 | 0.03 | -0.03 | 0.06 | -0.01 | 0.57 | 0.33 | 0.06 | -0.17 | 0.12 | 0.18 | 0.05 | 0.01 | 0.19 | 0.17 |
| hr7193 | S | 4601 | 2.15 | 1.38 | 0.18 | 0.03 | 0.19 | 0.16 | 0.59 | 0.05 | -0.15 | -0.03 | 0.04 | 0.06 | 0.04 | -0.05 | -0.09 | 0.00 | 0.20 | 0.28 | 0.11 | 0.00 | -0.39 | -0.02 | 0.23 | 0.05 | 0.19 | 0.35 | -0.09 |
| hr7196 | S | 4824 | 2.37 | 1.44 | -0.03 | -0.06 | 0.10 | 0.06 | 0.31 | -0.08 | -0.16 | -0.13 | -0.14 | -0.11 | -0.19 | -0.18 | -0.17 | -0.14 | -0.10 | 0.02 | 0.04 | -0.17 | -0.49 | -0.02 | 0.05 | -0.01 | 0.08 | 0.14 | -0.03 |
| hr7204 | S | 4869 | 2.56 | 1.37 | 0.29 | 0.10 | 0.27 | 0.27 | 0.66 | 0.16 | 0.02 | 0.05 | 0.07 | 0.17 | 0.10 | 0.06 | -0.02 | 0.09 | 0.09 | 0.42 | 0.29 | 0.20 | -0.22 | 0.16 | 0.39 | 0.15 | 0.30 | 0.37 | 0.13 |
| hr7217 | S | 4766 | 2.31 | 1.46 | 0.18 | 0.05 | 0.16 | 0.19 | 0.48 | 0.07 | -0.11 | -0.07 | -0.06 | 0.04 | -0.03 | -0.06 | -0.10 | -0.02 | 0.06 | 0.32 | 0.16 | -0.08 | -0.30 | 0.03 | 0.20 | -0.01 | 0.08 | 0.21 | 0.08 |
| hr7225a | S | 4496 | 1.80 | 1.42 | 0.00 | -0.19 | -0.09 | -0.04 | 0.45 | -0.16 | -0.45 | -0.35 | -0.28 | -0.22 | -0.34 | -0.32 | -0.38 | -0.29 | -0.22 | 0.46 | -0.07 | -0.38 | -0.72 | -0.30 | -0.19 | -0.31 | -0.24 | -0.06 | -0.40 |
| hr7225b | S | 4156 | 0.97 | 1.69 | 0.37 | 0.24 | 0.51 | 0.40 | 1.61 | 0.28 | -0.13 | 0.12 | 0.35 | 0.32 | 0.14 | 0.00 | 0.09 | 0.17 | 0.79 | 2.78 | 0.28 | 0.07 | -0.46 | -0.33 | 0.29 | 0.23 | 0.33 | 1.04 | -0.13 |
| hr7234 | S | 4444 | 1.79 | 1.46 | 0.09 | -0.21 | 0.09 | 0.08 | 0.60 | -0.12 | -0.24 | -0.15 | -0.07 | -0.07 | -0.12 | -0.18 | -0.17 | -0.11 | 0.10 | 0.51 | -0.02 | -0.22 | -0.62 | -0.14 | 0.04 | -0.04 | 0.11 | 0.27 | -0.12 |
| hr7295 | S | 4875 | 2.46 | 1.63 | 0.56 | 0.09 | 0.30 | 0.44 | 0.86 | 0.17 | 0.04 | 0.12 | 0.16 | 0.26 | 0.19 | 0.14 | 0.12 | 0.20 | 0.13 | 0.91 | 0.37 | 0.12 | -0.17 | 0.08 | 0.42 | 0.16 | 0.14 | 0.79 | 0.31 |
| hr7310 | S | 4776 | 2.30 | 1.42 | 0.03 | -0.09 | 0.11 | 0.09 | 0.23 | -0.08 | -0.22 | -0.16 | -0.16 | -0.07 | -0.14 | -0.15 | -0.19 | -0.13 | -0.10 | 0.39 | 0.07 | -0.06 | -0.45 | -0.01 | 0.06 | -0.09 | 0.03 | 0.14 | -0.07 |
| hr7325 | S | 4753 | 2.32 | 1.47 | 0.05 | -0.08 | 0.11 | 0.11 | 0.34 | 0.00 | -0.15 | -0.07 | -0.08 | -0.03 | -0.11 | -0.13 | -0.14 | -0.10 | -0.04 | -0.03 | 0.16 | -0.10 | -0.37 | 0.11 | 0.28 | 0.12 | 0.20 | 0.27 | 0.03 |
| hr7331 | S | 7262 | 4.30 | 3.35 | 0.17 | 0.16 | | 0.38 | 0.31 | 0.35 | 0.61 | 0.85 | 1.01 | 0.88 | -0.15 | 0.35 | 1.18 | 0.55 | | 0.46 | | 1.53 | | 0.43 | 0.89 | 0.89 | 0.95 | | |
| hr7349 | S | 4662 | 2.10 | 1.49 | 0.12 | 0.07 | 0.23 | 0.22 | 0.53 | 0.01 | -0.09 | -0.08 | -0.02 | -0.02 | -0.03 | -0.08 | -0.09 | -0.02 | 0.07 | 0.50 | 0.12 | -0.04 | -0.44 | -0.04 | 0.21 | -0.11 | 0.07 | 0.31 | -0.03 |
| hr7359 | S | 4838 | 2.57 | 1.39 | 0.37 | 0.24 | 0.35 | 0.32 | 0.78 | 0.25 | 0.10 | 0.22 | 0.26 | 0.26 | 0.28 | 0.15 | 0.17 | 0.23 | 0.34 | 0.55 | 0.30 | 0.20 | -0.10 | 0.24 | 0.57 | 0.27 | 0.43 | 0.68 | 0.21 |
| hr7376 | S | 4679 | 2.50 | 1.24 | -0.01 | -0.02 | 0.15 | 0.08 | 0.62 | -0.03 | -0.08 | -0.01 | 0.06 | -0.03 | -0.02 | -0.11 | -0.08 | -0.03 | 0.20 | 0.16 | 0.02 | -0.14 | -0.46 | -0.17 | 0.13 | -0.03 | 0.17 | 0.47 | -0.09 |
| hr7385 | S | 4763 | 2.31 | 1.41 | 0.08 | -0.03 | 0.12 | 0.13 | 0.45 | 0.00 | -0.14 | -0.12 | -0.13 | -0.01 | -0.13 | -0.10 | -0.15 | -0.09 | 0.01 | 0.44 | 0.14 | -0.12 | -0.38 | 0.07 | 0.14 | 0.04 | 0.10 | -0.01 | -0.03 |
| hr7389 | S | 6281 | 3.61 | 3.27 | 0.51 | -0.10 | 0.06 | 0.32 | 0.50 | 0.09 | -0.08 | 0.16 | 0.11 | 0.29 | 0.13 | -0.01 | 0.32 | 0.13 | -0.43 | | 0.10 | 0.24 | -0.36 | 0.65 | | 0.15 | | 0.76 | |
| hr7407 | S | 4740 | 2.40 | 1.43 | 0.24 | 0.15 | 0.32 | 0.27 | 0.84 | 0.12 | -0.03 | 0.08 | 0.14 | 0.18 | 0.11 | 0.06 | 0.04 | 0.13 | 0.29 | 0.41 | 0.21 | 0.10 | -0.27 | 0.10 | 0.40 | 0.13 | 0.23 | 0.38 | 0.11 |
| hr7433 | S | 4631 | 2.00 | 1.47 | 0.03 | -0.02 | 0.10 | 0.17 | 0.56 | -0.07 | -0.23 | -0.18 | -0.18 | -0.05 | -0.11 | -0.14 | -0.17 | -0.08 | 0.02 | 0.88 | 0.05 | -0.21 | -0.56 | -0.11 | 0.00 | -0.14 | -0.03 | 0.09 | -0.14 |
| hr7449 | S | 4810 | 2.53 | 1.43 | 0.39 | 0.21 | 0.31 | 0.35 | 0.95 | 0.19 | 0.06 | 0.10 | 0.20 | 0.22 | 0.14 | 0.08 | 0.08 | 0.14 | 0.32 | 1.30 | 0.64 | 0.40 | 0.24 | 0.37 | 0.63 | 0.34 | 0.43 | 0.66 | 0.22 |
| hr7465 | S | 4730 | 2.37 | 1.34 | 0.01 | -0.08 | 0.06 | 0.06 | 0.27 | -0.07 | -0.15 | -0.09 | -0.09 | -0.05 | -0.17 | -0.14 | -0.18 | -0.11 | 0.01 | 0.02 | 0.09 | 0.00 | -0.37 | 0.13 | 0.20 | 0.14 | 0.27 | 0.32 | -0.04 |
| hr7478b | S | 4872 | 2.33 | 0.95 | 0.10 | -0.22 | 0.03 | -0.03 | 0.24 | -0.14 | -0.38 | -0.25 | -0.20 | -0.07 | -0.20 | -0.32 | -0.27 | -0.24 | -0.98 | 0.99 | 1.23 | -0.01 | -0.35 | -0.64 | -0.17 | 0.29 | -0.15 | 0.08 | -0.01 |
| hr7487 | S | 5052 | 2.75 | 1.37 | 0.29 | 0.09 | 0.21 | 0.21 | 0.51 | 0.14 | -0.01 | 0.06 | 0.03 | 0.14 | 0.09 | 0.04 | -0.01 | 0.05 | -0.05 | 0.15 | 0.35 | 0.16 | -0.11 | 0.24 | 0.30 | 0.15 | 0.28 | 0.46 | 0.16 |
| hr7506 | S | 5008 | 2.66 | 1.48 | 0.31 | 0.14 | 0.21 | 0.21 | 0.38 | 0.14 | 0.02 | 0.04 | 0.06 | 0.17 | 0.01 | 0.05 | -0.01 | 0.04 | -0.01 | 0.23 | 0.51 | 0.19 | -0.20 | 0.24 | 0.30 | 0.24 | 0.18 | 0.26 | 0.27 |
| hr7517 | S | 4920 | 2.39 | 1.45 | 0.26 | -0.10 | 0.16 | 0.21 | 0.62 | 0.11 | -0.06 | -0.04 | -0.06 | 0.08 | -0.07 | -0.01 | -0.09 | 0.01 | -0.02 | 0.31 | 0.26 | 0.02 | -0.24 | 0.18 | 0.18 | 0.05 | 0.12 | -0.04 | 0.10 |
| hr7526 | S | 5035 | 3.04 | 1.13 | -0.14 | -0.21 | -0.05 | -0.10 | 0.14 | -0.11 | -0.19 | -0.14 | -0.17 | -0.17 | -0.26 | -0.22 | -0.21 | -0.21 | -0.20 | -0.13 | 0.20 | -0.13 | -0.36 | 0.04 | 0.13 | 0.12 | 0.12 | 0.16 | 0.06 |
| hr7540 | S | 4959 | 2.82 | 1.34 | 0.06 | -0.09 | 0.15 | 0.08 | 0.35 | 0.04 | 0.04 | 0.07 | 0.07 | 0.07 | 0.05 | -0.04 | -0.04 | 0.01 | 0.06 | 0.14 | 0.22 | 0.11 | -0.18 | 0.18 | 0.31 | 0.27 | 0.32 | 0.43 | 0.22 |
| hr7541 | S | 4421 | 2.06 | 1.47 | 0.43 | 0.34 | 0.41 | 0.44 | 1.34 | 0.32 | 0.11 | 0.22 | 0.41 | 0.37 | 0.26 | 0.13 | 0.23 | 0.32 | 0.44 | 1.44 | 0.25 | 0.11 | -0.23 | -0.17 | 0.74 | 0.15 | 0.42 | 1.13 | 0.22 |
| hr7561 | S | 4779 | 2.68 | 0.83 | 0.07 | -0.03 | 0.13 | 0.11 | 0.46 | 0.10 | 0.02 | 0.06 | 0.20 | 0.08 | -0.05 | -0.04 | -0.06 | 0.04 | 0.08 | 0.79 | 0.21 | 0.16 | -0.22 | 0.22 | 0.41 | 0.22 | 0.38 | 0.47 | 0.08 |
| hr7576 | S | 4337 | 1.48 | 1.79 | 0.82 | 0.53 | 0.59 | 0.66 | 1.59 | 0.30 | 0.08 | 0.25 | 0.50 | 0.48 | 0.60 | 0.20 | 0.35 | 0.41 | 0.48 | 1.29 | 0.33 | 0.12 | -0.24 | -0.43 | 0.41 | 0.28 | 0.34 | 0.88 | 0.13 |
| hr7597 | S | 5306 | 3.38 | 1.07 | 0.01 | -0.05 | 0.08 | 0.06 | 0.23 | 0.03 | -0.04 | -0.08 | -0.13 | 0.00 | -0.10 | -0.07 | -0.14 | -0.06 | -0.09 | -0.07 | 0.40 | 0.06 | 0.03 | 0.03 | 0.04 | 0.09 | 0.06 | -0.13 | 0.23 |
| hr7681 | S | 4519 | 1.73 | 1.58 | 0.12 | | 0.23 | 0.32 | 0.84 | 0.04 | -0.22 | -0.31 | 0.02 | -0.07 | -0.10 | -0.15 | -0.22 | -0.04 | -0.18 | | -0.01 | -0.24 | -0.33 | | 0.04 | -0.15 | -0.15 | 0.10 | |
| hr7712 | S | 4439 | 1.48 | 1.53 | 0.13 | -0.05 | 0.06 | 0.19 | 0.68 | -0.18 | -0.31 | -0.27 | -0.10 | -0.16 | -0.18 | -0.26 | -0.12 | -0.06 | 0.50 | 0.12 | -0.24 | -0.57 | -0.13 | 0.09 | -0.16 | -0.03 | 0.16 | -0.23 | |
| hr7713 | S | 5060 | 2.71 | 1.42 | -0.25 | -0.31 | -0.16 | -0.23 | -0.02 | -0.27 | -0.29 | -0.30 | -0.34 | -0.33 | -0.49 | -0.40 | -0.33 | -0.39 | -0.49 | -0.34 | 0.24 | -0.36 | -0.53 | -0.18 | -0.19 | -0.17 | -0.11 | -0.25 | -0.17 |
| hr7748 | S | 4782 | 2.34 | 1.40 | 0.07 | 0.04 | 0.12 | 0.11 | 0.36 | -0.05 | -0.16 | -0.11 | -0.13 | -0.04 | -0.08 | -0.12 | -0.17 | -0.10 | -0.06 | 0.23 | 0.17 | -0.09 | -0.39 | 0.05 | 0.21 | -0.01 | 0.09 | 0.16 | -0.05 |
| hr7778 | S | 5080 | 2.75 | 1.41 | 0.29 | 0.08 | 0.38 | 0.17 | 0.56 | 0.12 | 0.05 | 0.07 | 0.08 | 0.15 | 0.07 | 0.05 | -0.01 | 0.05 | -0.03 | 0.20 | 0.46 | 0.15 | -0.06 | 0.23 | 0.25 | 0.23 | 0.26 | 0.32 | 0.33 |
| hr7788 | S | 4969 | 1.96 | 1.60 | 0.25 | -0.08 | 0.03 | 0.08 | 0.23 | -0.05 | -0.25 | -0.23 | -0.35 | -0.06 | -0.23 | -0.17 | -0.28 | -0.19 | -0.36 | 0.04 | 0.33 | -0.18 | -0.47 | 0.11 | -0.07 | -0.02 | -0.11 | 0.11 | -0.12 |
| hr7794 | S | 4866 | 2.53 | 1.34 | 0.22 | 0.07 | 0.18 | 0.20 | 0.56 | 0.10 | -0.08 | -0.01 | 0.01 | 0.07 | 0.05 | -0.01 | -0.06 | 0.03 | 0.06 | 0.30 | 0.20 | 0.02 | -0.23 | 0.10 | 0.30 | 0.05 | 0.19 | 0.20 | 0.15 |
| hr7802 | S | 4827 | 2.56 | 1.35 | 0.49 | 0.30 | 0.45 | 0.41 | 0.82 | 0.26 | 0.12 | 0.20 | 0.26 | 0.31 | 0.29 | 0.20 | 0.17 | 0.30 | 0.33 | 0.70 | 0.33 | 0.21 | -0.13 | 0.20 | 0.56 | 0.27 | 0.34 | 0.67 | 0.24 |
| hr7806 | S | 4211 | 1.28 | 1.62 | 0.23 | 0.02 | 0.29 | 0.35 | 1.36 | -0.01 | -0.15 | -0.06 | 0.10 | 0.14 | 0.05 | -0.08 | -0.05 | 0.04 | 0.34 | 0.65 | 0.17 | -0.01 | -0.52 | -0.19 | 0.23 | -0.09 | 0.28 | 0.75 | -0.15 |
| hr7820 | S | 4703 | 2.31 | 1.42 | 0.07 | -0.04 | 0.15 | 0.15 | 0.49 | 0.02 | -0.13 | -0.07 | -0.07 | 0.02 | -0.02 | -0.07 | -0.12 | -0.06 | 0.03 | 0.34 | 0.16 | 0.02 | -0.36 | 0.09 | 0.42 | 0.05 | 0.20 | 0.21 | 0.02 |

| ID | | T | | | | | | | | | | | | | | | | | | | | | | | | | | |
|---|---|---|---|---|---|---|---|---|---|---|---|---|---|---|---|---|---|---|---|---|---|---|---|---|---|---|---|---|
| hr7824 | S | 5029 | 2.66 | 1.33 | 0.12 | -0.03 | 0.04 | 0.05 | 0.36 | -0.03 | -0.14 | -0.10 | -0.13 | -0.02 | -0.16 | -0.11 | -0.19 | -0.12 | -0.19 | -0.05 | 0.27 | -0.02 | -0.29 | 0.17 | 0.12 | 0.10 | 0.12 | 0.08 | 0.10 |
| hr7831 | S | 4592 | 2.24 | 1.55 | 0.45 | 0.32 | 0.44 | 0.49 | 1.17 | 0.25 | 0.16 | 0.26 | 0.44 | 0.35 | 0.34 | 0.21 | 0.32 | 0.37 | 0.48 | 1.72 | 0.22 | 0.19 | -0.20 | 0.05 | 0.94 | 0.28 | 0.35 | 1.02 | 0.47 |
| hr7854 | S | 4800 | 2.59 | 1.26 | 0.13 | 0.01 | 0.12 | 0.16 | 0.50 | 0.05 | -0.08 | 0.00 | 0.04 | 0.06 | 0.01 | -0.03 | -0.07 | 0.00 | 0.09 | 0.37 | 0.17 | 0.07 | -0.22 | 0.14 | 0.33 | 0.16 | 0.29 | 0.47 | 0.08 |
| hr7897 | S | 4715 | 2.43 | 1.42 | 0.32 | 0.07 | 0.24 | 0.33 | 0.79 | 0.17 | -0.01 | 0.07 | 0.13 | 0.20 | 0.14 | 0.09 | 0.06 | 0.16 | 0.36 | 0.86 | 0.23 | 0.20 | -0.26 | 0.14 | 0.47 | 0.15 | 0.28 | 0.39 | 0.12 |
| hr7904 | S | 4728 | 2.40 | 1.40 | 0.22 | 0.10 | 0.29 | 0.24 | 0.60 | 0.11 | 0.05 | 0.09 | 0.17 | 0.15 | 0.09 | 0.03 | 0.04 | 0.10 | 0.26 | 0.54 | 0.25 | 0.11 | -0.27 | 0.14 | 0.42 | 0.15 | 0.32 | 0.51 | 0.14 |
| hr7905 | S | 4760 | 2.34 | 1.40 | -0.18 | -0.14 | -0.02 | -0.01 | 0.26 | -0.18 | -0.29 | -0.23 | -0.26 | -0.23 | -0.34 | -0.31 | -0.30 | -0.26 | -0.51 | -0.04 | 0.05 | -0.36 | -0.69 | -0.34 | -0.19 | -0.33 | -0.21 | -0.10 | -0.22 |
| hr7923 | S | 4916 | 2.60 | 1.35 | -0.16 | -0.23 | -0.03 | -0.09 | 0.24 | -0.15 | -0.23 | -0.20 | -0.24 | -0.22 | -0.30 | -0.30 | -0.29 | -0.28 | -0.41 | -0.12 | 0.25 | -0.30 | -0.33 | -0.15 | -0.06 | -0.11 | -0.05 | -0.06 | -0.07 |
| hr7939 | S | 4473 | 1.86 | 1.56 | 0.19 | 0.13 | 0.21 | 0.18 | 0.74 | -0.01 | -0.10 | -0.06 | -0.01 | 0.02 | -0.03 | -0.09 | -0.07 | -0.01 | 0.13 | 0.59 | 0.01 | -0.11 | -0.60 | -0.12 | 0.35 | -0.11 | 0.10 | 0.34 | -0.07 |
| hr7942 | S | 4701 | 2.24 | 1.43 | 0.19 | -0.04 | 0.16 | 0.19 | 0.52 | 0.06 | -0.14 | -0.07 | -0.03 | 0.04 | -0.09 | -0.07 | -0.12 | -0.05 | 0.07 | 0.44 | 0.17 | 0.04 | -0.43 | 0.17 | 0.21 | 0.08 | 0.18 | 0.32 | -0.01 |
| hr7962 | S | 4623 | 2.41 | 1.35 | 0.06 | -0.05 | 0.18 | 0.09 | 0.49 | -0.06 | 0.00 | 0.04 | 0.14 | 0.00 | 0.01 | -0.12 | -0.01 | 0.00 | 0.39 | 0.25 | 0.11 | -0.10 | -0.41 | -0.11 | 0.21 | 0.02 | 0.24 | 0.50 | 0.18 |
| hr7995 | S | 5155 | 2.72 | 1.55 | 0.24 | -0.05 | 0.04 | 0.10 | 0.27 | -0.01 | -0.14 | -0.07 | -0.13 | 0.02 | -0.13 | -0.09 | -0.15 | -0.11 | -0.24 | -0.01 | 0.30 | 0.07 | -0.21 | 0.16 | 0.03 | 0.12 | 0.06 | 0.00 | 0.17 |
| hr8000 | S | 4525 | 2.14 | 1.41 | -0.24 | -0.09 | -0.02 | -0.12 | 0.28 | -0.23 | -0.27 | -0.14 | -0.16 | -0.37 | -0.59 | -0.51 | -0.33 | -0.40 | 0.01 | -0.01 | -0.05 | -0.40 | -0.63 | -0.57 | -0.22 | -0.35 | -0.19 | -0.03 | -0.11 |
| hr8017 | S | 4441 | 1.90 | 1.59 | 0.44 | 0.32 | 0.45 | 0.55 | 1.26 | 0.29 | 0.13 | 0.21 | 0.38 | 0.37 | 0.33 | 0.18 | 0.27 | 0.33 | 0.63 | 1.43 | 0.20 | 0.07 | -0.33 | -0.05 | 0.65 | 0.13 | 0.43 | 0.96 | 0.27 |
| hr8030 | S | 4966 | 2.81 | 1.24 | 0.20 | 0.05 | 0.21 | 0.19 | 0.39 | 0.11 | -0.05 | 0.03 | 0.04 | 0.10 | 0.05 | 0.02 | -0.05 | 0.04 | -0.35 | 0.18 | 0.38 | 0.16 | -0.21 | 0.19 | 0.31 | 0.15 | 0.28 | 0.30 | 0.26 |
| hr8034a | S | 6223 | 3.87 | 2.41 | 0.42 | | | 0.19 | 0.31 | 0.11 | 0.40 | 0.38 | 0.36 | 0.38 | 0.38 | -0.05 | 0.40 | -0.08 | | 0.23 | | 1.30 | | -0.12 | 0.19 | 1.25 | 0.78 | 0.65 | 0.93 |
| hr8034b | S | 6399 | 4.29 | 1.44 | 0.07 | 0.10 | 0.12 | 0.07 | 0.08 | 0.12 | 0.22 | 0.17 | 0.17 | 0.13 | -0.05 | 0.02 | 0.25 | 0.05 | -0.21 | -0.16 | | 0.30 | 0.20 | 0.19 | 0.74 | 0.45 | 0.30 | 0.31 | 0.37 |
| hr8035 | S | 4926 | 2.80 | 1.23 | 0.44 | 0.23 | 0.37 | 0.33 | 0.75 | 0.28 | 0.09 | 0.20 | 0.26 | 0.29 | 0.29 | 0.16 | 0.15 | 0.23 | 0.36 | 0.92 | 0.35 | 0.16 | -0.17 | 0.13 | 0.51 | 0.19 | 0.23 | 0.49 | 0.21 |
| hr8072 | S | 4887 | 2.55 | 1.43 | 0.55 | 0.27 | 0.35 | 0.40 | 0.88 | 0.25 | 0.11 | 0.18 | 0.26 | 0.28 | 0.27 | 0.17 | 0.16 | 0.25 | 0.31 | 0.88 | 0.41 | 0.22 | -0.07 | 0.14 | 0.48 | 0.15 | 0.25 | 0.59 | 0.31 |
| hr8082 | S | 4819 | 2.57 | 1.37 | 0.15 | 0.09 | 0.20 | 0.18 | 0.45 | 0.08 | 0.02 | 0.07 | 0.08 | 0.11 | 0.04 | 0.01 | -0.03 | 0.06 | 0.15 | 0.26 | 0.23 | 0.15 | -0.26 | 0.20 | 0.35 | 0.16 | 0.32 | 0.53 | 0.15 |
| hr8093 | S | 4938 | 2.70 | 1.32 | 0.20 | -0.04 | 0.14 | 0.21 | 0.45 | 0.13 | -0.07 | 0.02 | 0.00 | 0.09 | 0.04 | 0.01 | -0.05 | 0.02 | 0.00 | 0.30 | 0.28 | 0.14 | -0.21 | 0.16 | 0.22 | 0.13 | 0.24 | 0.26 | 0.18 |
| hr8096 | S | 4563 | 2.10 | 1.55 | 0.39 | 0.28 | 0.46 | 0.42 | 1.10 | 0.23 | 0.08 | 0.18 | 0.32 | 0.28 | 0.25 | 0.16 | 0.22 | 0.27 | 0.47 | 1.37 | 0.12 | -0.01 | -0.39 | -0.02 | 0.58 | 0.13 | 0.29 | 0.85 | 0.20 |
| hr8115 | S | 4891 | 2.26 | 1.54 | 0.35 | -0.12 | 0.24 | 0.27 | 0.59 | 0.14 | -0.11 | -0.03 | -0.08 | 0.08 | -0.06 | -0.02 | -0.09 | 0.02 | 0.06 | 0.44 | 0.63 | 0.28 | 0.08 | 0.45 | 0.44 | 0.36 | 0.29 | 0.36 | 0.12 |
| hr8165 | S | 4689 | 2.33 | 1.49 | 0.08 | 0.17 | 0.24 | 0.21 | 0.59 | 0.01 | -0.05 | 0.03 | 0.07 | -0.04 | -0.17 | -0.14 | -0.01 | -0.05 | 0.36 | 0.42 | 0.04 | -0.19 | -0.50 | -0.19 | 0.14 | -0.03 | 0.03 | 0.59 | 0.08 |
| hr8167 | S | 5011 | 2.48 | 1.53 | 0.31 | 0.13 | 0.08 | 0.13 | 0.44 | 0.01 | -0.10 | -0.05 | -0.08 | 0.10 | -0.08 | -0.06 | -0.12 | -0.06 | -0.35 | 0.25 | 0.54 | 0.08 | -0.26 | 0.21 | 0.06 | 0.00 | 0.04 | 0.07 | 0.08 |
| hr8173 | S | 4620 | 2.15 | 1.44 | 0.22 | 0.13 | 0.35 | 0.34 | 0.91 | 0.14 | -0.01 | 0.10 | 0.15 | 0.18 | 0.09 | 0.05 | 0.00 | 0.12 | 0.40 | 0.58 | 0.16 | 0.01 | -0.35 | -0.06 | 0.35 | 0.07 | 0.20 | 0.56 | 0.06 |
| hr8179 | S | 4805 | 2.32 | 1.46 | 0.08 | 0.02 | 0.15 | 0.13 | 0.44 | -0.04 | -0.10 | -0.07 | -0.05 | -0.02 | -0.09 | -0.11 | -0.11 | -0.07 | -0.01 | 0.15 | 0.16 | 0.02 | -0.39 | 0.03 | 0.18 | 0.08 | 0.12 | 0.19 | 0.01 |
| hr8185 | S | 4691 | 2.39 | 1.40 | 0.45 | 0.17 | 0.36 | 0.37 | 0.91 | 0.27 | 0.03 | 0.18 | 0.31 | 0.26 | 0.24 | 0.12 | 0.15 | 0.26 | 0.61 | 0.89 | 0.19 | 0.18 | -0.22 | 0.03 | 0.44 | 0.07 | 0.25 | 0.71 | 0.10 |
| hr8191 | S | 6271 | 3.15 | 4.60 | 0.93 | 1.05 | | 0.24 | 0.19 | -0.20 | 0.80 | 0.24 | 0.57 | 0.66 | -0.29 | 0.00 | 0.73 | 0.12 | 0.28 | | | 0.88 | -0.40 | | 1.00 | 0.54 | 0.27 | 0.95 | 1.45 |
| hr8228 | S | 4909 | 2.78 | 1.25 | 0.34 | 0.19 | 0.31 | 0.28 | 0.61 | 0.19 | 0.03 | 0.14 | 0.19 | 0.24 | 0.20 | 0.12 | 0.07 | 0.17 | 0.29 | 0.32 | 0.37 | 0.24 | -0.12 | 0.16 | 0.48 | 0.22 | 0.37 | 0.47 | 0.21 |
| hr8252 | S | 5010 | 2.66 | 1.32 | 0.09 | -0.12 | 0.06 | 0.04 | 0.33 | 0.01 | -0.21 | -0.12 | -0.15 | -0.04 | -0.11 | -0.12 | -0.19 | -0.14 | -0.20 | 0.28 | 0.39 | -0.06 | -0.30 | 0.11 | 0.01 | -0.01 | 0.06 | 0.01 | 0.12 |
| hr8255 | S | 4640 | 2.04 | 1.49 | 0.30 | 0.10 | 0.30 | 0.29 | 0.74 | 0.09 | -0.07 | -0.01 | 0.12 | 0.12 | 0.07 | 0.00 | -0.01 | 0.08 | 0.29 | 0.88 | 0.22 | -0.08 | -0.45 | -0.12 | 0.20 | -0.05 | 0.09 | 0.44 | -0.09 |
| hr8274 | S | 4779 | 2.43 | 1.38 | 0.18 | 0.06 | 0.25 | 0.20 | 0.53 | 0.08 | -0.06 | 0.05 | 0.09 | 0.11 | 0.02 | 0.00 | -0.03 | 0.04 | 0.18 | 0.22 | 0.20 | 0.13 | -0.24 | 0.08 | 0.28 | 0.08 | 0.22 | 0.44 | 0.14 |
| hr8277 | S | 4701 | 2.21 | 1.46 | 0.14 | 0.13 | 0.21 | 0.22 | 0.71 | -0.01 | -0.11 | -0.08 | -0.03 | 0.02 | 0.03 | -0.07 | -0.08 | 0.00 | 0.15 | 0.40 | 0.08 | -0.13 | -0.44 | -0.09 | 0.13 | -0.05 | 0.06 | 0.32 | -0.01 |
| hr8288 | S | 4960 | 2.33 | 1.52 | -0.05 | -0.29 | -0.15 | -0.16 | -0.09 | -0.25 | -0.37 | -0.36 | -0.43 | -0.35 | -0.56 | -0.42 | -0.42 | -0.42 | -0.45 | -0.19 | 0.20 | -0.45 | -0.62 | -0.26 | -0.31 | -0.38 | -0.30 | -0.50 | -0.21 |
| hr8317 | S | 4588 | 2.10 | 1.49 | 0.50 | 0.24 | 0.41 | 0.51 | 1.11 | 0.22 | 0.03 | 0.14 | 0.25 | 0.27 | 0.26 | 0.16 | 0.18 | 0.29 | 0.53 | 1.30 | 0.19 | 0.09 | -0.34 | -0.07 | 0.43 | 0.11 | 0.20 | 0.56 | 0.05 |
| hr8320 | S | 4892 | 2.64 | 1.36 | -0.09 | -0.13 | 0.03 | -0.05 | 0.25 | -0.12 | -0.15 | -0.13 | -0.14 | -0.16 | -0.19 | -0.23 | -0.21 | -0.19 | -0.08 | -0.03 | 0.08 | -0.20 | -0.46 | -0.10 | 0.01 | -0.07 | 0.03 | 0.02 | -0.05 |
| hr8324 | S | 4710 | 2.36 | 1.31 | 0.45 | 0.26 | 0.38 | 0.39 | 1.05 | 0.25 | 0.03 | 0.15 | 0.25 | 0.29 | 0.30 | 0.15 | 0.12 | 0.25 | 0.44 | 1.20 | 0.24 | 0.18 | -0.27 | 0.04 | 0.53 | 0.09 | 0.23 | 0.55 | 0.14 |
| hr8325 | S | 4484 | 2.03 | 1.44 | 0.39 | 0.29 | 0.46 | 0.34 | 0.95 | 0.17 | 0.07 | 0.18 | 0.33 | 0.26 | 0.14 | 0.05 | 0.10 | 0.17 | 0.36 | 0.74 | 0.33 | 0.12 | -0.25 | -0.06 | 0.50 | 0.08 | 0.34 | 0.71 | 0.21 |
| hr8360 | S | 4527 | 1.86 | 1.60 | 0.47 | 0.41 | 0.41 | 0.47 | 1.04 | 0.31 | 0.13 | 0.23 | 0.36 | 0.34 | 0.32 | 0.16 | 0.22 | 0.28 | 0.39 | 0.89 | 0.23 | 0.06 | -0.26 | -0.10 | 0.53 | 0.20 | 0.18 | 1.00 | 0.22 |
| hr8391 | S | 6397 | 3.81 | 3.50 | 0.42 | | -0.22 | 0.30 | 0.48 | 0.06 | 0.36 | 0.40 | 0.55 | 0.57 | 0.78 | 0.07 | 0.81 | 0.39 | | | 0.67 | 0.52 | 0.03 | | 0.72 | 0.66 | 2.23 | 1.00 | |
| hr8394 | S | 4811 | 2.73 | 1.23 | 0.11 | 0.02 | 0.15 | 0.16 | 0.41 | 0.06 | -0.08 | 0.01 | 0.05 | 0.05 | 0.00 | -0.05 | -0.06 | -0.01 | 0.15 | 0.35 | 0.11 | 0.06 | -0.29 | 0.01 | 0.24 | 0.15 | 0.22 | 0.24 | -0.01 |
| hr8401 | S | 4944 | 2.76 | 1.36 | 0.47 | 0.32 | 0.37 | 0.30 | 0.68 | 0.21 | 0.13 | 0.17 | 0.19 | 0.21 | 0.15 | 0.11 | 0.14 | 0.18 | 0.38 | 0.48 | 0.33 | 0.09 | -0.25 | 0.07 | 0.58 | 0.10 | 0.16 | 0.49 | 0.23 |
| hr8442 | S | 5260 | 3.21 | 1.23 | 0.44 | 0.28 | 0.34 | 0.28 | 0.65 | 0.33 | 0.38 | 0.40 | 0.43 | 0.38 | 0.32 | 0.30 | 0.27 | 0.33 | 0.36 | 0.39 | 0.60 | 0.51 | 0.34 | 0.64 | 0.72 | 0.60 | 0.66 | 0.69 | 0.59 |
| hr8453 | S | 4946 | 2.88 | 1.26 | 0.50 | 0.28 | 0.44 | 0.40 | 0.83 | 0.31 | 0.21 | 0.29 | 0.35 | 0.36 | 0.38 | 0.25 | 0.22 | 0.33 | 0.42 | 0.54 | 0.37 | 0.40 | -0.03 | 0.28 | 0.65 | 0.34 | 0.44 | 0.73 | 0.35 |
| hr8454 | S | 6253 | 3.56 | 7.75 | 0.03 | -2.21 | | -0.07 | | -1.90 | 0.40 | 0.09 | 1.34 | 0.72 | | -0.12 | 0.34 | 0.35 | -0.41 | | | -0.07 | | | | 1.50 | -0.07 | | |
| hr8456 | S | 4811 | 2.48 | 1.43 | 0.32 | 0.18 | 0.38 | 0.32 | 0.74 | 0.18 | 0.12 | 0.20 | 0.23 | 0.23 | 0.19 | 0.14 | 0.11 | 0.18 | 0.34 | 0.64 | 0.38 | 0.22 | -0.13 | 0.28 | 0.56 | 0.31 | 0.42 | 0.62 | 0.28 |
| hr8461 | S | 4950 | 3.18 | 0.89 | 0.24 | 0.18 | 0.29 | 0.21 | 0.59 | 0.22 | 0.09 | 0.22 | 0.27 | 0.23 | 0.18 | 0.12 | 0.10 | 0.21 | 0.38 | 0.76 | 0.18 | 0.26 | -0.12 | 0.15 | 0.55 | 0.18 | 0.42 | 0.74 | 0.29 |
| hr8476 | S | 4734 | 2.53 | 1.43 | 0.40 | 0.38 | 0.50 | 0.41 | 0.75 | 0.25 | 0.19 | 0.30 | 0.42 | 0.33 | 0.33 | 0.22 | 0.26 | 0.33 | 0.58 | 1.02 | 0.26 | 0.20 | -0.06 | 0.20 | 1.05 | 0.35 | 0.37 | 0.97 | 0.36 |
| hr8482 | S | 4503 | 2.09 | 1.39 | 0.44 | 0.44 | 0.48 | 0.44 | 1.16 | 0.33 | 0.11 | 0.21 | 0.40 | 0.31 | 0.29 | 0.16 | 0.22 | 0.32 | 0.52 | 1.30 | 0.23 | 0.20 | -0.29 | -0.08 | 0.72 | 0.19 | 0.50 | 0.96 | 0.18 |
| hr8499 | S | 4894 | 2.52 | 1.50 | 0.41 | 0.16 | 0.36 | 0.29 | 0.94 | 0.22 | 0.02 | 0.10 | 0.14 | 0.23 | 0.18 | 0.12 | 0.09 | 0.16 | 0.17 | 0.64 | 0.31 | 0.14 | -0.15 | 0.13 | 0.39 | 0.11 | 0.20 | 0.27 | 0.24 |

| ID | Type | C1 | C2 | C3 | C4 | C5 | C6 | C7 | C8 | C9 | C10 | C11 | C12 | C13 | C14 | C15 | C16 | C17 | C18 | C19 | C20 | C21 | C22 | C23 | C24 | C25 | C26 | C27 | C28 |
|---|---|---|---|---|---|---|---|---|---|---|---|---|---|---|---|---|---|---|---|---|---|---|---|---|---|---|---|---|---|
| hr8500 | S | 4544 | 2.11 | 1.39 | 0.25 | 0.22 | 0.28 | 0.26 | 0.86 | 0.08 | 0.01 | 0.06 | 0.18 | 0.16 | 0.13 | 0.01 | 0.02 | 0.10 | 0.33 | 0.38 | 0.22 | 0.04 | -0.29 | 0.02 | 0.26 | 0.05 | 0.24 | 0.50 | 0.15 |
| hr8516 | S | 4552 | 2.13 | 1.33 | -0.06 | 0.06 | 0.02 | 0.10 | 0.57 | -0.13 | -0.21 | -0.20 | -0.17 | -0.14 | -0.23 | -0.22 | -0.18 | -0.14 | 0.09 | 0.27 | -0.10 | -0.33 | -0.60 | -0.37 | -0.03 | -0.27 | -0.05 | 0.25 | -0.26 |
| hr8538 | S | 4711 | 2.40 | 1.23 | -0.13 | -0.07 | 0.00 | -0.07 | 0.32 | -0.13 | -0.28 | -0.18 | -0.18 | -0.19 | -0.28 | -0.28 | -0.29 | -0.23 | -0.08 | 0.03 | -0.02 | -0.28 | -0.57 | -0.22 | -0.13 | -0.19 | -0.05 | -0.06 | -0.15 |
| hr8568 | S | 4695 | 2.37 | 1.42 | 0.43 | 0.38 | 0.45 | 0.43 | 0.88 | 0.26 | 0.16 | 0.26 | 0.34 | 0.34 | 0.36 | 0.22 | 0.23 | 0.30 | 0.66 | 0.63 | 0.34 | 0.22 | -0.11 | 0.21 | 0.65 | 0.27 | 0.41 | 0.65 | 0.22 |
| hr8594 | S | 4942 | 2.56 | 1.52 | 0.58 | 0.24 | 0.31 | 0.39 | 0.83 | 0.27 | 0.11 | 0.19 | 0.27 | 0.32 | 0.28 | 0.18 | 0.17 | 0.24 | 0.27 | 0.54 | 0.39 | 0.24 | -0.04 | 0.20 | 0.46 | 0.19 | 0.25 | 0.43 | 0.33 |
| hr8596 | S | 4875 | 2.67 | 1.36 | 0.33 | 0.10 | 0.29 | 0.36 | 0.76 | 0.16 | 0.04 | 0.10 | 0.12 | 0.22 | 0.19 | 0.12 | 0.10 | 0.15 | 0.24 | 0.42 | 0.27 | 0.17 | -0.15 | 0.16 | 0.38 | 0.21 | 0.31 | 0.54 | 0.19 |
| hr8617 | S | 5338 | 3.11 | 1.04 | 0.27 | 0.00 | 0.16 | 0.12 | 0.22 | 0.14 | 0.16 | 0.14 | 0.10 | 0.15 | 0.00 | 0.08 | 0.01 | 0.07 | -0.07 | 0.08 | 0.69 | 0.23 | -0.02 | 0.45 | 0.45 | 0.39 | 0.44 | 0.09 | 0.36 |
| hr8632 | S | 4256 | 1.23 | 1.65 | 0.11 | -0.08 | 0.03 | 0.08 | 1.04 | -0.20 | -0.42 | -0.25 | -0.18 | -0.07 | -0.25 | -0.28 | -0.32 | -0.23 | -0.09 | 0.53 | 0.04 | -0.33 | -0.66 | -0.20 | 0.09 | -0.20 | 0.02 | 0.24 | -0.14 |
| hr8642 | S | 4584 | 1.98 | 1.65 | 0.17 | 0.11 | 0.18 | 0.29 | 0.97 | 0.01 | -0.19 | -0.01 | 0.06 | 0.17 | 0.01 | -0.03 | -0.01 | 0.07 | 0.06 | 1.12 | 0.12 | -0.05 | -0.42 | -0.09 | 0.02 | -0.02 | 0.07 | 0.52 | 0.10 |
| hr8643 | S | 4754 | 2.46 | 1.23 | -0.20 | -0.16 | -0.16 | -0.16 | 0.23 | -0.29 | -0.36 | -0.29 | -0.31 | -0.32 | -0.44 | -0.39 | -0.38 | -0.33 | -0.20 | -0.10 | 0.06 | -0.37 | -0.66 | -0.36 | -0.09 | -0.24 | -0.19 | -0.05 | -0.28 |
| hr8656 | S | 4904 | 2.47 | 1.48 | 0.57 | 0.21 | 0.28 | 0.32 | 0.61 | 0.20 | 0.00 | 0.04 | 0.09 | 0.18 | 0.10 | 0.07 | 0.04 | 0.12 | 0.16 | 0.48 | 0.27 | 0.12 | -0.24 | 0.05 | 0.20 | -0.05 | 0.01 | 0.14 | 0.16 |
| hr8660 | S | 4738 | 2.28 | 1.43 | 0.06 | -0.05 | 0.14 | 0.15 | 0.58 | -0.01 | -0.11 | -0.04 | 0.01 | 0.00 | -0.02 | -0.08 | -0.08 | -0.02 | 0.07 | 0.15 | 0.08 | -0.12 | -0.35 | -0.01 | 0.20 | -0.03 | 0.11 | 0.35 | 0.03 |
| hr8670 | S | 4829 | 2.29 | 1.46 | -0.28 | -0.10 | -0.16 | -0.19 | 0.08 | -0.31 | -0.39 | -0.36 | -0.43 | -0.36 | -0.51 | -0.43 | -0.39 | -0.40 | -0.34 | -0.19 | 0.00 | -0.49 | -0.63 | -0.46 | -0.39 | -0.35 | -0.33 | -0.32 | -0.33 |
| hr8678 | S | 4803 | 2.43 | 1.40 | 0.11 | 0.03 | 0.20 | 0.15 | 0.51 | 0.05 | -0.08 | -0.01 | 0.01 | 0.04 | -0.01 | -0.04 | -0.06 | 0.00 | 0.06 | -0.02 | 0.18 | -0.02 | -0.34 | 0.14 | 0.30 | 0.07 | 0.16 | 0.30 | 0.11 |
| hr8703 | S | 4603 | 2.21 | 2.48 | 0.46 |  | 0.77 | 0.40 | 0.31 | -0.18 | 0.04 | 0.19 | 0.29 | 0.52 | -0.36 | -0.06 | -0.03 | 0.08 |  |  | 0.08 | 0.45 | -0.65 |  | 1.27 | 1.04 | 0.68 |  | 0.65 |
| hr8712 | S | 4641 | 2.29 | 1.48 | 0.43 | 0.28 | 0.44 | 0.41 | 0.95 | 0.23 | 0.10 | 0.25 | 0.42 | 0.31 | 0.32 | 0.18 | 0.25 | 0.34 | 0.68 | 0.75 | 0.23 | 0.13 | -0.19 | 0.08 | 0.76 | 0.20 | 0.39 | 0.97 | 0.20 |
| hr8715 | S | 7728 | 4.29 | 4.04 | 1.28 |  | 0.62 | 0.56 | 0.32 | 0.15 | 1.37 | 1.46 | 2.16 | 0.84 | 0.91 | 0.39 | 1.36 | 0.98 |  |  | 0.77 | 2.37 | 0.56 |  | 0.86 | 1.30 | 1.43 |  | 0.57 |
| hr8730 | S | 4586 | 2.02 | 1.47 | 0.33 | 0.11 | 0.29 | 0.33 | 0.85 | 0.13 | -0.05 | 0.01 | 0.10 | 0.13 | 0.08 | 0.01 | 0.01 | 0.11 | 0.45 | 0.69 | 0.14 | -0.08 | -0.45 | -0.11 | 0.31 | -0.06 | 0.07 | 0.45 | -0.06 |
| hr8742 | S | 4831 | 2.56 | 1.30 | 0.55 | 0.29 | 0.33 | 0.36 | 0.88 | 0.25 | 0.02 | 0.08 | 0.22 | 0.22 | 0.19 | 0.12 | 0.06 | 0.17 | 0.24 | 0.76 | 0.30 | 0.10 | -0.24 | 0.10 | 0.39 | 0.08 | 0.15 | 0.43 | 0.11 |
| hr8780 | S | 4668 | 2.13 | 1.45 | -0.01 | 0.03 | 0.16 | 0.12 | 0.49 | -0.08 | -0.15 | -0.12 | -0.11 | -0.05 | -0.11 | -0.14 | -0.11 | -0.08 | 0.10 | 0.28 | 0.04 | -0.26 | -0.54 | -0.14 | 0.08 | -0.12 | 0.01 | 0.24 | -0.05 |
| hr8807 | S | 4984 | 2.63 | 1.37 | -0.13 | -0.10 | 0.03 | -0.07 | 0.27 | -0.21 | -0.15 | -0.12 | -0.13 | -0.14 | -0.25 | -0.22 | -0.17 | -0.17 | -0.10 | -0.07 | 0.26 | -0.18 | -0.44 | -0.12 | 0.01 | -0.05 | 0.15 | 0.12 | 0.05 |
| hr8812 | S | 4414 | 1.57 | 1.63 | 0.16 | 0.01 | 0.08 | 0.22 | 0.68 | -0.05 | -0.26 | -0.15 | -0.10 | -0.01 | -0.13 | -0.14 | -0.18 | -0.08 | -0.02 | 0.73 | 0.12 | -0.02 | -0.47 | 0.03 | 0.37 | 0.05 | 0.13 | 0.26 | 0.03 |
| hr8839 | S | 4594 | 2.15 | 1.53 | 0.72 | 0.52 | 0.65 | 0.65 | 1.48 | 0.48 | 0.32 | 0.39 | 0.56 | 0.51 | 0.43 | 0.33 | 0.39 | 0.48 | 0.58 | 1.35 | 0.34 | 0.25 | -0.15 | 0.11 | 0.75 | 0.41 | 0.62 | 1.18 | 0.51 |
| hr8841 | S | 4624 | 2.28 | 1.48 | 0.10 | 0.02 | 0.22 | 0.28 | 0.54 | 0.07 | 0.01 | 0.08 | 0.10 | 0.14 | 0.06 | 0.03 | 0.02 | 0.09 | 0.23 | 0.29 | 0.14 | -0.06 | -0.49 | -0.06 | 0.51 | 0.09 | 0.22 | 0.58 | 0.05 |
| hr8852 | S | 4810 | 2.15 | 1.45 | -0.38 | -0.19 | -0.25 | -0.23 | 0.02 | -0.39 | -0.50 | -0.41 | -0.53 | -0.56 | -0.80 | -0.61 | -0.53 | -0.57 | -0.57 | -0.34 | -0.03 | -0.64 | -0.86 | -0.63 | -0.60 | -0.67 | -0.50 | -0.47 | -0.38 |
| hr8875 | S | 4639 | 2.28 | 1.48 | 0.22 | 0.18 | 0.35 | 0.30 | 0.73 | 0.08 | -0.02 | 0.09 | 0.20 | 0.13 | 0.15 | 0.02 | 0.08 | 0.16 | 0.42 | 0.64 | 0.11 | 0.02 | -0.37 | -0.06 | 0.31 | 0.02 | 0.22 | 0.60 | 0.01 |
| hr8878 | S | 4246 | 1.28 | 1.49 | -0.49 | -0.23 | -0.27 | -0.28 | 0.28 | -0.46 | -0.67 | -0.51 | -0.57 | -0.63 | -0.95 | -0.78 | -0.66 | -0.68 | -0.42 | -0.27 | -0.31 | -0.73 | -0.92 | -0.96 | -0.71 | -0.85 | -0.64 | -0.45 | -0.65 |
| hr8892 | S | 4545 | 2.09 | 1.38 | -0.18 | -0.23 | -0.08 | -0.10 | 0.42 | -0.28 | -0.37 | -0.28 | -0.24 | -0.29 | -0.36 | -0.39 | -0.33 | -0.30 | -0.12 | 0.31 | -0.25 | -0.41 | -0.69 | -0.39 | -0.19 | -0.30 | -0.16 | -0.01 | -0.33 |
| hr8912 | S | 4828 | 2.46 | 1.39 | 0.09 | -0.01 | 0.14 | 0.11 | 0.54 | 0.03 | -0.08 | -0.03 | -0.04 | 0.02 | -0.06 | -0.07 | -0.11 | -0.03 | -0.01 | -0.04 | 0.16 | -0.01 | -0.34 | 0.09 | 0.17 | 0.13 | 0.16 | 0.28 | 0.00 |
| hr8922 | S | 4727 | 2.48 | 1.38 | 0.07 | 0.09 | 0.23 | 0.19 | 0.44 | 0.06 | 0.05 | 0.08 | 0.15 | 0.06 | 0.04 | -0.03 | 0.04 | 0.08 | 0.27 | 0.37 | 0.11 | -0.02 | -0.33 | 0.02 | 0.32 | 0.02 | 0.25 | 0.68 | 0.05 |
| hr8923 | S | 4938 | 2.62 | 1.40 | 0.37 | 0.05 | 0.21 | 0.30 | 0.60 | 0.20 | 0.03 | 0.08 | 0.05 | 0.17 | 0.12 | 0.08 | 0.03 | 0.11 | 0.12 | 0.50 | 0.32 | 0.13 | -0.16 | 0.20 | 0.29 | 0.22 | 0.18 | 0.25 | 0.14 |
| hr8930 | S | 4613 | 2.10 | 1.48 | -0.13 | 0.04 | 0.09 | 0.04 | 0.44 | -0.16 | -0.34 | -0.14 | -0.17 | -0.26 | -0.40 | -0.34 | -0.25 | -0.24 | 0.06 | 0.03 | -0.07 | -0.54 | -0.74 | -0.58 | -0.48 | -0.39 | -0.30 | -0.01 | -0.24 |
| hr8941 | S | 4962 | 2.66 | 1.43 | 0.38 | 0.15 | 0.25 | 0.26 | 0.51 | 0.22 | 0.13 | 0.14 | 0.16 | 0.22 | 0.17 | 0.12 | 0.07 | 0.14 | 0.15 | 0.37 | 0.34 | 0.19 | -0.05 | 0.29 | 0.48 | 0.26 | 0.38 | 0.20 | 0.26 |
| hr8946 | S | 4208 | 1.21 | 1.60 | 0.43 | 0.21 | 0.44 | 0.34 | 1.40 | 0.14 | -0.12 | 0.07 | 0.26 | 0.25 | 0.16 | -0.05 | 0.07 | 0.12 | 0.44 | 1.25 | 0.23 | 0.06 | -0.49 | -0.44 | 0.48 | 0.03 | 0.26 | 0.89 | -0.13 |
| hr8948 | S | 4761 | 2.51 | 1.26 | 0.30 | 0.18 | 0.34 | 0.32 | 0.72 | 0.21 | 0.03 | 0.06 | 0.17 | 0.14 | 0.10 | 0.10 | 0.06 | 0.14 |  | 0.52 |  | 0.10 | -0.44 | 0.18 | 0.45 | 0.04 | 0.23 | 0.14 | 0.13 |
| hr8958 | S | 4708 | 2.16 | 1.47 | -0.07 | 0.05 | 0.10 | 0.08 | 0.46 | -0.08 | -0.18 | -0.14 | -0.17 | -0.11 | -0.17 | -0.18 | -0.17 | -0.11 | 0.03 | 0.26 | 0.00 | -0.29 | -0.60 | -0.21 | -0.02 | -0.15 | -0.04 | 0.17 | -0.07 |
| hr9008 | S | 4617 | 2.08 | 1.56 | 0.22 | 0.13 | 0.31 | 0.34 | 0.87 | 0.12 | 0.00 | 0.05 | 0.10 | 0.17 | 0.12 | 0.06 | 0.02 | 0.16 | 0.32 | 0.49 | 0.18 | 0.18 | -0.34 | 0.05 | 0.36 | 0.11 | 0.23 | 0.66 | 0.08 |
| hr9009 | S | 4661 | 2.20 | 1.44 | 0.27 | 0.01 | 0.28 | 0.32 | 0.84 | 0.14 | -0.04 | 0.04 | 0.09 | 0.15 | 0.16 | 0.03 | 0.00 | 0.12 | 0.30 | 0.78 | 0.17 | -0.02 | -0.35 | -0.01 | 0.37 | 0.09 | 0.13 | 0.36 | 0.06 |
| hr9012 | S | 4926 | 2.44 | 1.41 | 0.25 | -0.05 | 0.22 | 0.20 | 0.65 | 0.10 | -0.11 | -0.03 | -0.05 | 0.07 | 0.00 | -0.03 | -0.09 | -0.02 | -0.03 | 0.27 | 0.21 | 0.00 | -0.18 | 0.12 | 0.18 | 0.10 | 0.08 | 0.17 | 0.10 |
| hr9067 | S | 5026 | 2.82 | 1.33 | 0.26 | 0.07 | 0.19 | 0.19 | 0.51 | 0.13 | 0.05 | 0.10 | 0.07 | 0.17 | 0.06 | 0.06 | 0.03 | 0.09 | 0.08 | 0.20 | 0.33 | 0.17 | -0.07 | 0.25 | 0.39 | 0.23 | 0.35 | 0.33 | 0.22 |
| hr9101 | S | 4663 | 2.12 | 1.49 | 0.52 | 0.40 | 0.49 | 0.42 | 1.08 | 0.22 | 0.14 | 0.17 | 0.24 | 0.29 | 0.31 | 0.16 | 0.19 | 0.27 | 0.47 | 0.83 | 0.27 | 0.13 | -0.23 | 0.06 | 0.78 | 0.10 | 0.26 | 0.67 | 0.09 |
| hr9104 | S | 4701 | 2.32 | 1.44 | 0.24 | 0.12 | 0.22 | 0.23 | 0.63 | 0.02 | -0.03 | -0.01 | 0.04 | 0.07 | 0.06 | -0.03 | 0.01 | 0.04 | 0.18 | 0.48 | 0.16 | -0.12 | -0.24 | 0.00 | 0.26 | 0.04 | 0.20 | 0.26 | 0.02 |
| hd001638 | U | 4225 | 0.69 | 1.83 | -0.61 | -0.39 | -0.38 | -0.31 | 0.07 | -0.66 | -0.78 | -0.59 | -0.63 | -0.67 | -0.92 | -0.76 | -0.67 | -0.71 | -0.50 | -0.79 | -0.38 | -0.91 | -0.84 | -1.01 | -0.68 | -0.62 | -0.57 | -0.57 | -0.61 |
| hd004388 | U | 4640 | 2.26 | 1.39 | 0.24 | 0.22 | 0.42 | 0.16 | 0.94 | 0.14 | 0.04 | 0.10 | 0.31 | 0.18 | 0.10 | 0.06 | 0.00 | 0.09 | 0.19 | -0.10 | 0.17 | 0.03 | 0.05 | -0.02 | 0.38 | 0.43 | 0.27 | 0.29 | 0.06 |
| hd011171 | U | 6746 | 4.51 | 3.76 | 1.25 |  |  | 1.44 | 0.00 | 0.57 | 0.32 | 0.66 | 0.69 | 1.03 | 0.40 | 0.09 | 1.74 | 0.87 |  |  | 1.91 | 1.63 |  |  | 2.46 | 2.90 |  | 1.79 |  |
| hd037763 | U | 4620 | 2.66 | 1.07 | 0.67 | 0.66 | 0.72 | 0.58 | 1.34 | 0.52 | 0.42 | 0.54 | 0.87 | 0.56 | 0.56 | 0.34 | 0.46 | 0.53 | 1.14 | 1.07 | 0.37 | 0.35 | 0.26 | 0.18 | 0.61 | 1.08 | 0.72 | 0.92 | 0.32 |
| hd037811 | U | 5023 | 2.54 | 1.43 | 0.22 | 0.07 | 0.17 | 0.13 | 0.42 | 0.10 | -0.06 | -0.03 | -0.04 | 0.04 | -0.03 | -0.01 | -0.09 | -0.04 | -0.12 | -0.10 | 0.29 | 0.03 | -0.10 | 0.23 | 0.19 | 0.16 | 0.17 | 0.21 | 0.08 |
| hd061603 | U | 4012 | 0.39 | 2.19 | 0.97 | 0.17 | 0.28 | 0.53 | 2.26 | 0.17 | -0.14 | 0.03 | 0.24 | 0.27 | 0.17 | -0.02 | 0.02 | 0.12 | 0.62 | 0.09 | 0.77 | -0.09 | 0.00 |  | 0.11 | 0.12 | 0.28 | 0.19 | -0.29 |
| hd062412 | U | 4861 | 2.31 | 1.47 | 0.27 | 0.16 | 0.21 | 0.19 | 0.38 | 0.13 | -0.15 | -0.02 | 0.07 | 0.12 | 0.07 | 0.02 | -0.04 | 0.03 | -0.02 | 0.08 | 0.15 | -0.03 | -0.10 | 0.04 | 0.13 | 0.44 | 0.10 | 0.12 | 0.01 |
| hd065354 | U | 3832 | -0.99 | 2.36 | 0.24 | -0.03 | -0.02 | 0.33 |  | -0.29 | -0.89 | -0.71 | -0.65 | -0.06 | -0.18 | -0.29 | -0.53 | -0.23 | -0.04 | -0.21 | 0.19 | -0.79 | -0.93 |  | -0.62 | -0.52 | -0.52 | -0.43 | -0.57 |

| ID | | T | | | | | | | | | | | | | | | | | | | | | | | | | | | |
|---|---|---|---|---|---|---|---|---|---|---|---|---|---|---|---|---|---|---|---|---|---|---|---|---|---|---|---|---|---|
| hd065714 | U | 5070 | 2.65 | 1.70 | 0.73 | 0.45 | 0.50 | 0.44 | 0.70 | 0.37 | 0.31 | 0.37 | 0.52 | 0.41 | 0.37 | 0.32 | 0.32 | 0.33 | 0.35 | 0.27 | 0.38 | 0.42 | 0.29 | 0.39 | 0.52 | 0.92 | 0.57 | 0.48 | 0.51 |
| hd065925 | U | 6585 | 3.30 | 7.01 | 0.63 | | 0.32 | -0.19 | -0.15 | -0.11 | 0.36 | -0.12 | 0.89 | -0.16 | -0.55 | -0.53 | 0.85 | -0.15 | -1.10 | | 2.04 | 0.28 | | | -0.91 | 0.46 | 0.97 | | |
| hd069511 | U | 4020 | -0.02 | 2.40 | 0.23 | -0.16 | 0.16 | 0.08 | | -0.28 | -0.45 | -0.43 | -0.36 | -0.06 | -0.19 | -0.22 | -0.38 | -0.20 | | -0.43 | 0.10 | -0.49 | -0.50 | | -0.15 | -0.08 | -0.14 | 0.03 | 0.25 |
| hd069836 | U | 4788 | 2.45 | 1.49 | 0.30 | 0.30 | 0.46 | 0.23 | 0.62 | 0.18 | 0.15 | 0.20 | 0.41 | 0.21 | 0.23 | 0.14 | 0.15 | 0.18 | 0.29 | 0.02 | 0.17 | 0.06 | 0.09 | 0.06 | 0.35 | 0.24 | 0.28 | 0.27 | 0.14 |
| hd071160 | U | 4096 | 0.91 | 1.82 | 0.34 | 0.05 | 0.29 | 0.35 | 0.96 | -0.05 | -0.15 | 0.02 | 0.20 | 0.20 | 0.04 | -0.05 | 0.01 | 0.06 | 0.07 | 0.30 | 0.31 | -0.05 | -0.17 | -0.23 | 0.08 | 0.11 | 0.26 | 0.09 | -0.32 |
| hd072320 | U | 5028 | 2.64 | 1.38 | 0.24 | 0.12 | 0.34 | 0.11 | 0.19 | 0.11 | -0.04 | -0.03 | 0.02 | 0.06 | -0.01 | 0.00 | -0.08 | -0.03 | -0.15 | -0.15 | 0.20 | 0.02 | 0.01 | 0.08 | 0.18 | 0.19 | 0.12 | 0.03 | 0.09 |
| hd072324 | U | 4781 | 2.03 | 1.61 | 0.37 | 0.15 | 0.28 | 0.23 | 0.68 | 0.12 | -0.10 | -0.03 | 0.05 | 0.11 | 0.00 | 0.00 | -0.07 | 0.03 | -0.03 | 0.11 | 0.09 | -0.01 | -0.08 | 0.03 | 0.13 | 0.46 | 0.16 | 0.08 | 0.04 |
| hd073829 | U | 4962 | 3.06 | 1.27 | 0.49 | 0.41 | 0.56 | 0.25 | 0.71 | 0.35 | 0.49 | 0.57 | 0.86 | 0.46 | 0.45 | 0.34 | 0.35 | 0.38 | 0.68 | 0.15 | 0.52 | 0.49 | 0.50 | 0.44 | 0.82 | 0.94 | 0.78 | 0.73 | 0.43 |
| hd074088 | U | 3840 | -0.80 | 1.79 | -0.19 | -0.35 | -0.29 | -0.12 | 1.44 | -0.49 | -0.92 | -0.57 | -0.61 | -0.34 | -0.79 | -0.60 | -0.64 | -0.51 | -0.22 | -0.37 | -0.09 | -0.78 | -0.98 | -1.08 | -0.90 | -0.51 | -0.51 | -0.44 | -1.13 |
| hd074165 | U | 4675 | 2.26 | 1.51 | 0.47 | 0.43 | 0.56 | 0.33 | 1.17 | 0.21 | 0.20 | 0.30 | 0.61 | 0.34 | 0.37 | 0.22 | 0.24 | 0.31 | | 0.66 | 0.22 | 0.13 | 0.01 | 0.01 | 0.52 | 0.28 | 0.34 | 0.33 | 0.13 |
| hd074166 | U | 4629 | 2.58 | 1.76 | 0.82 | 0.48 | 0.72 | 0.47 | | 0.48 | 0.52 | 0.65 | 0.83 | 0.60 | 0.47 | 0.42 | 0.50 | 0.49 | | 0.59 | 0.67 | 0.61 | 0.61 | 0.63 | 0.74 | 0.94 | 0.96 | 0.94 | 0.34 |
| hd074529 | U | 4598 | 2.03 | 1.50 | 0.42 | 0.32 | 0.47 | 0.35 | 1.11 | 0.18 | 0.11 | 0.13 | 0.37 | 0.26 | 0.27 | 0.17 | 0.15 | 0.23 | 0.31 | 0.19 | 0.22 | 0.02 | -0.05 | -0.04 | 0.39 | 0.18 | 0.22 | 0.13 | -0.01 |
| hd074900 | U | 4572 | 2.04 | 1.46 | 0.33 | 0.34 | 0.50 | 0.29 | 0.95 | 0.17 | 0.09 | 0.17 | 0.44 | 0.18 | 0.18 | 0.08 | 0.09 | 0.15 | | 0.58 | 0.08 | -0.26 | -0.19 | -0.16 | 0.20 | 0.09 | 0.19 | 0.11 | -0.11 |
| hd075058 | U | 4650 | 2.43 | 1.26 | 0.27 | 0.23 | 0.38 | 0.18 | 0.86 | 0.10 | 0.00 | 0.08 | 0.31 | 0.13 | 0.14 | 0.02 | 0.02 | 0.10 | 0.26 | 0.11 | 0.07 | -0.04 | -0.13 | -0.21 | 0.19 | 0.01 | 0.16 | 0.07 | -0.21 |
| hd076128 | U | 4487 | 1.78 | 1.60 | 0.33 | 0.26 | 0.44 | 0.29 | | 0.11 | -0.04 | 0.07 | 0.29 | 0.20 | 0.21 | 0.10 | 0.09 | 0.16 | 0.22 | 0.00 | 0.10 | -0.09 | -0.16 | -0.14 | 0.27 | 0.07 | 0.16 | 0.06 | -0.15 |
| hd078002 | U | 4859 | 2.40 | 1.44 | 0.54 | 0.30 | 0.47 | 0.31 | 0.74 | 0.19 | 0.11 | 0.10 | 0.23 | 0.24 | 0.26 | 0.16 | 0.13 | 0.21 | 0.15 | 0.11 | 0.18 | 0.05 | 0.00 | -0.02 | 0.30 | 0.19 | 0.16 | 0.10 | 0.18 |
| hd078479 | U | 4479 | 1.85 | 1.70 | 0.59 | 0.43 | 0.65 | 0.53 | 1.03 | 0.27 | 0.15 | 0.25 | 0.49 | 0.39 | 0.34 | 0.23 | 0.31 | 0.35 | 1.06 | 0.79 | 0.19 | 0.04 | -0.09 | -0.22 | 0.29 | 0.60 | 0.31 | 0.40 | 0.14 |
| hd078528 | U | 4012 | 1.01 | 1.66 | 0.58 | 0.14 | 0.48 | 0.40 | | 0.46 | 0.10 | 0.28 | 0.49 | 0.29 | 0.20 | 0.05 | 0.15 | 0.26 | | 0.22 | 0.55 | -0.01 | 0.10 | 0.06 | 0.26 | 0.35 | 0.43 | 0.34 | |
| hd078959 | U | 4150 | 0.99 | 2.16 | 0.41 | -0.08 | 0.41 | 0.17 | | 0.00 | -0.04 | 0.05 | 0.13 | 0.14 | -0.09 | -0.07 | -0.10 | -0.08 | | -0.67 | 0.59 | -0.02 | 0.15 | | 0.23 | 0.23 | 0.42 | 0.29 | |
| hd080571 | U | 4659 | 2.23 | 1.44 | 0.28 | 0.23 | 0.43 | 0.22 | 0.92 | 0.10 | 0.02 | 0.06 | 0.26 | 0.14 | 0.12 | 0.04 | 0.02 | 0.10 | 0.17 | -0.04 | 0.11 | -0.04 | -0.14 | -0.08 | 0.29 | 0.12 | 0.14 | 0.11 | -0.04 |
| hd081278 | U | 4936 | 3.20 | 1.00 | 0.55 | 0.51 | 0.63 | 0.36 | 1.12 | 0.43 | 0.46 | 0.53 | 0.87 | 0.50 | 0.55 | 0.38 | 0.37 | 0.46 | 0.67 | 0.47 | 0.46 | 0.44 | 0.42 | 0.26 | 0.68 | 0.50 | 0.72 | 0.58 | 0.39 |
| hd082395 | U | 4686 | 1.98 | 1.51 | 0.23 | 0.13 | 0.16 | 0.16 | 0.56 | -0.01 | -0.20 | -0.12 | 0.02 | 0.00 | 0.00 | -0.08 | -0.11 | -0.03 | 0.04 | 0.15 | -0.01 | -0.11 | -0.26 | -0.21 | 0.04 | 0.31 | -0.02 | -0.04 | -0.14 |
| hd082403 | U | 4724 | 2.85 | 1.11 | 0.43 | 0.39 | 0.55 | 0.31 | 0.09 | 0.36 | 0.44 | 0.56 | 0.84 | 0.44 | 0.44 | 0.35 | 0.36 | 0.46 | 0.78 | 0.70 | 0.49 | 0.58 | 0.28 | 0.47 | 0.78 | 0.52 | 0.74 | 0.80 | 0.46 |
| hd082668 | U | 4105 | 0.88 | 1.72 | 0.15 | -0.16 | 0.22 | 0.15 | | 0.02 | -0.19 | 0.02 | 0.29 | 0.08 | -0.22 | -0.18 | -0.12 | -0.07 | 0.23 | -0.09 | 0.75 | -0.19 | -0.14 | 0.09 | -0.02 | -0.01 | 0.25 | 0.83 | -0.26 |
| hd083155 | U | 5166 | 3.21 | 1.36 | 0.12 | 0.11 | | 0.06 | -0.05 | 0.10 | 0.14 | 0.23 | 0.25 | 0.13 | 0.03 | 0.06 | 0.08 | 0.07 | 0.15 | -0.06 | 0.41 | 0.17 | 0.14 | 0.33 | 0.41 | 0.34 | 0.46 | 0.61 | 0.34 |
| hd083234 | U | 4185 | 1.24 | 1.74 | 0.25 | 0.21 | 0.33 | 0.30 | | 0.06 | -0.22 | -0.09 | 0.09 | 0.09 | -0.04 | -0.01 | -0.03 | 0.07 | | 0.19 | 0.06 | -0.26 | -0.25 | -0.37 | 0.23 | 0.05 | 0.03 | -0.07 | -0.48 |
| hd084598 | U | 4896 | 2.32 | 1.49 | 0.31 | 0.15 | | 0.15 | 0.37 | 0.08 | -0.10 | -0.06 | 0.01 | 0.06 | -0.03 | -0.01 | -0.13 | -0.03 | -0.17 | -0.10 | 0.07 | -0.11 | -0.04 | 0.04 | 0.16 | 0.15 | 0.08 | -0.02 | |
| hd085552 | U | 5426 | 2.29 | 2.28 | 0.20 | 0.05 | 0.20 | 0.09 | 0.15 | 0.03 | -0.15 | -0.10 | -0.29 | -0.05 | -0.26 | -0.10 | -0.15 | -0.13 | -0.59 | -0.30 | | -0.03 | 0.04 | -0.11 | 0.15 | -0.17 | -0.04 | -0.37 | -0.22 |
| hd086757 | U | 4081 | 0.32 | 2.46 | 0.22 | -0.10 | 0.27 | 0.21 | | -0.11 | -0.35 | -0.32 | -0.29 | 0.01 | -0.10 | -0.14 | -0.26 | -0.14 | | -0.40 | 0.18 | -0.43 | -0.37 | | 0.09 | -0.03 | -0.04 | 0.02 | 0.39 |
| hd089736 | U | 3830 | -0.77 | 2.52 | 0.39 | -0.17 | 0.04 | 0.32 | | -0.20 | -0.65 | -0.48 | -0.34 | -0.04 | -0.31 | -0.27 | -0.43 | -0.21 | 0.28 | -0.44 | 0.56 | -0.66 | -0.56 | | -0.32 | -0.25 | -0.38 | -0.25 | -0.28 |
| hd095849 | U | 4472 | 1.77 | 1.66 | 0.48 | 0.39 | 0.46 | 0.44 | 0.97 | 0.27 | 0.07 | 0.18 | 0.36 | 0.30 | 0.27 | 0.18 | 0.20 | 0.27 | 0.88 | 0.54 | 0.24 | 0.06 | -0.04 | -0.08 | 0.31 | 0.55 | 0.33 | 0.43 | 0.07 |
| hd099322 | U | 4857 | 2.47 | 1.36 | 0.24 | 0.13 | 0.21 | 0.17 | 0.52 | 0.13 | -0.12 | 0.00 | 0.09 | 0.07 | 0.08 | 0.02 | -0.05 | 0.02 | -0.02 | -0.02 | 0.16 | 0.04 | -0.04 | 0.13 | 0.14 | 0.33 | 0.18 | 0.09 | 0.07 |
| hd103295 | U | 4912 | 2.29 | 1.16 | -0.91 | -0.65 | -0.61 | -0.62 | -0.25 | -0.68 | -0.95 | -0.84 | -1.06 | -1.03 | -1.35 | -1.06 | -1.00 | -1.02 | -1.26 | -0.88 | -0.17 | -1.04 | -0.82 | -1.01 | -1.13 | -1.04 | -0.92 | -1.12 | -0.86 |
| hd105740 | U | 4668 | 2.26 | 1.25 | -0.42 | -0.22 | -0.19 | -0.25 | 0.07 | -0.35 | -0.53 | -0.38 | -0.47 | -0.56 | -0.79 | -0.64 | -0.52 | -0.57 | -0.54 | -0.47 | -0.10 | -0.67 | -0.61 | -0.75 | -0.70 | -0.66 | -0.55 | -0.39 | -0.45 |
| hd107446 | U | 4121 | 0.81 | 1.82 | -0.04 | -0.12 | -0.12 | -0.10 | | -0.23 | -0.49 | -0.27 | -0.14 | -0.15 | -0.38 | -0.39 | -0.37 | -0.32 | -0.28 | -0.53 | 0.02 | -0.57 | -0.47 | -0.48 | -0.31 | -0.27 | -0.19 | -0.34 | -0.85 |
| hd110458 | U | 4682 | 2.22 | 1.42 | 0.45 | 0.24 | 0.37 | 0.35 | 0.23 | 0.23 | 0.01 | 0.11 | 0.34 | 0.21 | 0.29 | 0.15 | 0.13 | 0.21 | 0.28 | 0.19 | 0.19 | 0.09 | -0.05 | 0.10 | 0.23 | 0.25 | 0.17 | 0.13 | 0.08 |
| hd111464 | U | 4170 | 0.98 | 1.63 | 0.15 | 0.03 | 0.11 | 0.21 | 0.64 | -0.10 | -0.29 | -0.19 | -0.04 | -0.01 | -0.18 | -0.17 | -0.16 | -0.08 | 0.06 | 0.08 | 0.01 | -0.36 | -0.46 | -0.42 | -0.01 | 0.19 | -0.04 | 0.05 | -0.38 |
| hd111721 | U | 4892 | 2.24 | 1.33 | -1.43 | -0.90 | -1.18 | -0.93 | | -1.00 | -1.44 | -1.20 | -1.46 | -1.41 | -1.81 | -1.43 | -1.38 | -1.47 | -1.87 | -1.24 | -0.34 | -1.43 | -1.17 | -1.35 | -1.46 | -1.25 | -1.33 | -1.54 | -1.20 |
| hd113002 | U | 5274 | 2.36 | 1.78 | -0.82 | -0.56 | -0.75 | -0.59 | -0.59 | -0.68 | -0.81 | -0.76 | -0.89 | -0.86 | -1.06 | -0.82 | -0.76 | -0.88 | -1.10 | -0.91 | -0.55 | -1.01 | -0.83 | -0.77 | -0.82 | -0.76 | -0.72 | -0.74 | -0.21 |
| hd119971 | U | 4071 | 0.20 | 1.68 | -0.53 | -0.32 | -0.37 | -0.32 | 0.41 | -0.59 | -0.87 | -0.62 | -0.74 | -0.68 | -1.07 | -0.82 | -0.77 | -0.75 | -0.69 | -0.64 | -0.40 | -0.90 | -0.91 | -0.98 | -0.92 | -0.74 | -0.73 | -0.62 | -0.73 |
| hd122721 | U | 4684 | 2.48 | 1.19 | 0.25 | 0.22 | 0.35 | 0.20 | 0.89 | 0.11 | 0.02 | 0.07 | 0.26 | 0.13 | 0.15 | 0.04 | -0.01 | 0.11 | 0.16 | 0.02 | 0.09 | -0.04 | -0.05 | 0.02 | 0.33 | 0.18 | 0.18 | 0.13 | -0.01 |
| hd124186 | U | 4417 | 1.91 | 1.63 | 0.92 | 0.50 | 0.73 | 0.58 | 1.39 | 0.40 | 0.22 | 0.31 | 0.57 | 0.48 | 0.43 | 0.30 | 0.36 | 0.40 | 0.39 | 0.54 | 0.29 | 0.16 | 0.05 | -0.12 | 0.27 | 0.74 | 0.49 | 0.66 | 0.14 |
| hd128279 | U | 5162 | 2.64 | 1.69 | -2.28 | -1.81 | | -1.74 | | -1.93 | -2.35 | -2.12 | -1.76 | -2.47 | -2.73 | -2.28 | -1.59 | -2.28 | | -2.35 | | -2.69 | | -3.02 | | | | | |
| hd138688 | U | 4191 | 1.28 | 1.51 | 0.25 | 0.11 | 0.26 | 0.23 | 0.44 | 0.06 | -0.14 | 0.02 | 0.35 | 0.12 | -0.10 | -0.06 | -0.10 | 0.00 | -0.04 | -0.14 | 0.29 | -0.16 | -0.09 | 0.05 | 0.05 | 0.11 | 0.25 | 0.13 | -0.24 |
| hd145206 | U | 4029 | 0.59 | 1.54 | 0.28 | 0.10 | 0.21 | 0.31 | 0.57 | 0.05 | -0.35 | -0.15 | 0.11 | 0.05 | -0.09 | -0.12 | -0.17 | -0.02 | | 0.02 | 0.15 | -0.35 | -0.33 | -0.30 | -0.24 | -0.23 | -0.09 | -0.17 | -0.66 |
| hd146836 | U | 6285 | 3.61 | 2.39 | -0.09 | -0.11 | -0.07 | -0.02 | 0.07 | -0.09 | -0.21 | -0.09 | -0.14 | -0.19 | -0.36 | -0.18 | 0.19 | -0.15 | -0.72 | -0.55 | | 0.23 | -0.09 | -0.16 | | -0.03 | -0.12 | -0.20 | -0.18 |
| hd148451 | U | 5015 | 2.45 | 1.45 | -0.16 | -0.16 | -0.12 | -0.22 | -0.30 | -0.27 | -0.36 | -0.26 | -0.33 | -0.51 | -0.71 | -0.55 | -0.39 | -0.50 | -0.57 | -0.46 | -0.26 | -0.52 | -0.41 | -0.50 | -0.41 | -0.44 | -0.35 | -0.34 | 0.10 |
| hd148513 | U | 4065 | 0.75 | 1.91 | 0.72 | 0.16 | 0.41 | 0.40 | 0.96 | -0.02 | -0.24 | -0.06 | 0.12 | 0.20 | 0.09 | -0.07 | 0.00 | 0.08 | 0.30 | 0.63 | 0.19 | -0.10 | -0.22 | -0.72 | -0.07 | 0.05 | -0.01 | -0.07 | -0.44 |
| hd149447 | U | 3875 | -0.25 | 1.94 | 0.20 | -0.14 | 0.14 | 0.18 | 0.85 | -0.02 | -0.50 | -0.34 | -0.08 | -0.02 | -0.38 | -0.31 | -0.32 | -0.22 | 0.23 | 0.02 | 0.47 | -0.74 | -0.46 | -0.47 | -0.49 | -0.46 | -0.26 | -0.22 | -0.67 |
| hd150798 | U | 4135 | 0.54 | 2.42 | 0.44 | -0.09 | 0.03 | 0.47 | 1.41 | -0.10 | -0.37 | -0.27 | -0.25 | 0.07 | -0.08 | -0.10 | -0.18 | -0.07 | 0.00 | 0.15 | 0.20 | -0.34 | -0.55 | | 0.11 | -0.07 | 0.05 | 0.18 | -0.04 |

| Star | | Teff | | | | | | | | | | | | | | | | | | | | | | | | | | |
|---|---|---|---|---|---|---|---|---|---|---|---|---|---|---|---|---|---|---|---|---|---|---|---|---|---|---|---|---|
| hd152786 | U | 3851 | -0.36 | 2.12 | 0.31 | -0.07 | -0.02 | 0.21 | | -0.23 | -0.59 | -0.43 | -0.30 | -0.09 | -0.16 | -0.26 | -0.36 | -0.18 | -0.03 | -0.36 | 0.26 | -0.48 | -0.51 | | -0.23 | -0.31 | -0.14 | -0.24 | -0.55 |
| hd157457 | U | 4911 | 2.58 | 1.67 | 0.14 | -0.02 | 0.26 | 0.26 | 0.56 | 0.16 | 0.08 | 0.22 | 0.29 | 0.27 | 0.14 | 0.10 | 0.10 | 0.10 | 0.08 | 0.00 | 0.38 | 0.37 | 0.12 | 0.44 | 0.36 | 0.34 | 0.42 | 0.47 | 0.19 |
| hd162391 | U | 4724 | 1.45 | 2.16 | 0.42 | 0.22 | 0.19 | 0.30 | 0.63 | 0.00 | -0.24 | -0.13 | -0.17 | 0.11 | -0.14 | -0.03 | -0.15 | -0.01 | -0.24 | 0.05 | 0.18 | 0.08 | -0.17 | 0.09 | 0.03 | 0.01 | 0.11 | -0.03 | -0.06 |
| hd162587 | U | 4764 | 1.83 | 1.76 | 0.36 | 0.00 | 0.19 | 0.23 | 0.55 | 0.13 | -0.16 | -0.09 | -0.05 | 0.09 | -0.03 | -0.01 | -0.11 | -0.02 | -0.11 | 0.00 | 0.10 | -0.07 | -0.12 | 0.10 | 0.12 | 0.27 | 0.05 | -0.01 | -0.03 |
| hd163652 | U | 4968 | 2.39 | 1.47 | -0.21 | -0.24 | -0.11 | -0.22 | -0.31 | -0.23 | -0.39 | -0.32 | -0.31 | -0.34 | -0.44 | -0.38 | -0.34 | -0.39 | -0.40 | -0.45 | -0.17 | -0.38 | -0.39 | -0.25 | -0.28 | -0.17 | -0.19 | -0.27 | 0.07 |
| hd167818 | U | 3909 | -0.90 | 2.28 | -0.02 | -0.10 | -0.21 | 0.13 | 0.12 | -0.49 | -1.05 | -0.79 | -0.87 | -0.25 | -0.44 | -0.39 | -0.62 | -0.32 | -0.37 | -0.25 | -0.28 | -0.94 | -1.04 | | -0.74 | -0.55 | -0.71 | -0.64 | -0.93 |
| hd169191 | U | 4357 | 1.44 | 1.56 | 0.13 | 0.02 | 0.10 | 0.13 | 0.72 | -0.07 | -0.28 | -0.10 | 0.00 | -0.01 | -0.14 | -0.14 | -0.18 | -0.09 | 0.35 | -0.10 | 0.06 | -0.19 | -0.27 | -0.15 | 0.14 | 0.36 | 0.12 | 0.17 | -0.05 |
| hd175545 | U | 4471 | 2.20 | 1.36 | 0.28 | 0.38 | 0.43 | 0.35 | 1.08 | 0.19 | 0.03 | 0.13 | 0.41 | 0.26 | 0.18 | 0.11 | 0.11 | 0.19 | 0.53 | 0.56 | 0.09 | 0.06 | -0.09 | -0.17 | 0.26 | 0.60 | 0.37 | 0.44 | 0.06 |
| hd183275 | U | 4743 | 2.63 | 1.52 | 0.62 | 0.43 | 0.64 | 0.52 | 1.19 | 0.39 | 0.30 | 0.43 | 0.64 | 0.48 | 0.53 | 0.36 | 0.39 | 0.46 | 0.69 | 0.38 | 0.37 | 0.31 | 0.13 | 0.15 | 0.63 | 0.37 | 0.52 | 1.00 | 0.42 |
| hd196983 | U | 4593 | 2.24 | 1.46 | 0.43 | 0.25 | 0.42 | 0.38 | 1.08 | 0.21 | 0.07 | 0.20 | 0.37 | 0.29 | 0.29 | 0.19 | 0.19 | 0.27 | 0.58 | 0.31 | 0.22 | 0.07 | -0.06 | 0.04 | 0.46 | 0.17 | 0.33 | 0.82 | 0.23 |
| hd199642 | U | 3828 | -0.38 | 1.95 | 0.14 | -0.17 | 0.14 | 0.11 | 0.88 | -0.26 | -0.60 | -0.42 | -0.22 | -0.18 | -0.57 | -0.40 | -0.43 | -0.35 | 0.23 | 0.04 | 0.20 | -0.74 | -0.80 | -0.52 | -0.54 | -0.69 | -0.43 | -0.43 | -0.67 |
| hd202320 | U | 4490 | 1.29 | 1.79 | 0.09 | 0.02 | 0.09 | 0.19 | 0.05 | -0.13 | -0.33 | -0.24 | -0.21 | -0.04 | -0.14 | -0.13 | -0.23 | -0.12 | -0.11 | 0.13 | -0.03 | -0.13 | -0.35 | 0.01 | 0.04 | 0.40 | 0.05 | 0.03 | -0.27 |
| hd203638 | U | 4532 | 1.90 | 1.62 | 0.37 | 0.36 | 0.45 | 0.40 | 0.96 | 0.18 | 0.03 | 0.10 | 0.32 | 0.25 | 0.25 | 0.14 | 0.18 | 0.27 | 0.61 | 0.68 | 0.05 | -0.02 | -0.22 | -0.12 | 0.27 | 0.35 | 0.19 | 0.26 | 0.01 |
| hd207964 | U | 6538 | 2.82 | 3.32 | | | | -0.36 | 1.02 | 0.45 | -0.12 | 0.10 | 2.04 | 1.05 | 1.72 | -0.03 | 1.25 | 0.99 | | | 0.88 | 1.54 | | | | | | | |
| hd211173 | U | 4883 | 2.66 | 1.15 | -0.10 | -0.24 | 0.14 | -0.05 | | -0.10 | -0.32 | -0.20 | -0.16 | -0.22 | -0.25 | -0.27 | -0.29 | -0.25 | -0.17 | -0.19 | 0.38 | 0.15 | 0.01 | 0.27 | 0.19 | 0.13 | 0.18 | -0.07 | -0.09 |
| hd212320 | U | 4851 | 1.76 | 1.54 | 0.13 | 0.02 | -0.07 | 0.08 | 0.24 | -0.08 | -0.36 | -0.31 | -0.36 | -0.12 | -0.26 | -0.22 | -0.32 | -0.22 | -0.27 | 0.05 | 0.69 | 0.40 | 0.33 | 0.66 | 0.70 | 0.83 | 0.45 | 0.45 | -0.22 |
| hd223094 | U | 3764 | -0.95 | 2.00 | -0.26 | -0.68 | -0.05 | -0.15 | 1.52 | -0.45 | -0.88 | -0.78 | -0.67 | -0.41 | -0.71 | -0.53 | -0.65 | -0.51 | | -0.74 | 0.09 | -0.83 | -1.19 | -1.02 | -0.66 | -0.78 | -0.60 | -0.27 | |
| hic007995 | U | 6036 | 4.89 | 0.50 | 0.55 | 0.75 | 0.79 | 0.33 | 0.24 | 0.51 | 1.17 | 1.24 | 1.60 | 0.83 | 0.86 | 0.72 | 0.97 | 0.67 | 1.34 | 0.47 | 1.05 | 1.15 | 1.42 | 0.95 | 1.65 | 1.68 | 1.79 | 1.72 | 1.27 |
| hip049418 | U | 4660 | 2.21 | 1.39 | 0.44 | 0.29 | 0.49 | 0.23 | 0.93 | 0.16 | 0.02 | 0.12 | 0.40 | 0.19 | 0.25 | 0.09 | 0.07 | 0.15 | 0.22 | 0.00 | 0.14 | -0.08 | -0.08 | -0.07 | 0.15 | 0.10 | 0.10 | 0.01 | -0.04 |
| hip051077 | U | 4589 | 2.09 | 1.43 | 0.21 | 0.28 | 0.47 | 0.21 | 0.90 | 0.11 | -0.01 | 0.08 | 0.34 | 0.17 | 0.15 | 0.08 | 0.01 | 0.09 | 0.16 | -0.04 | 0.09 | -0.22 | -0.18 | -0.08 | 0.18 | 0.11 | 0.12 | 0.10 | 0.19 |
| hip053502 | U | 4786 | 2.29 | 1.43 | 0.18 | 0.12 | 0.33 | 0.10 | 0.53 | 0.04 | -0.10 | -0.07 | 0.04 | 0.01 | -0.01 | -0.05 | -0.10 | -0.05 | -0.06 | -0.19 | 0.11 | -0.14 | -0.12 | 0.01 | 0.12 | 0.09 | 0.06 | 0.02 | 0.06 |
| hip059785 | U | 4793 | 2.24 | 1.60 | -0.21 | -0.02 | 0.06 | -0.11 | -0.04 | -0.20 | -0.25 | -0.18 | -0.21 | -0.37 | -0.57 | -0.42 | -0.29 | -0.38 | -0.33 | -0.37 | -0.28 | -0.54 | -0.43 | -0.64 | -0.35 | -0.46 | -0.33 | -0.29 | -0.20 |
| hip066936 | U | 4589 | 2.11 | 1.48 | 0.36 | 0.38 | 0.55 | 0.33 | 1.13 | 0.12 | 0.07 | 0.20 | 0.51 | 0.28 | 0.32 | 0.17 | 0.21 | 0.25 | 0.48 | 0.63 | 0.18 | 0.02 | -0.03 | -0.05 | 0.28 | 0.18 | 0.24 | 0.19 | |
| hip070306 | U | 4255 | 1.63 | 2.00 | 0.05 | -0.25 | 0.16 | -0.22 | 1.09 | -0.47 | -0.38 | -0.27 | -0.03 | -0.04 | -0.29 | -0.40 | -0.18 | -0.26 | | -0.61 | 0.01 | -0.41 | -0.43 | -1.27 | -0.41 | -0.35 | -0.20 | -0.31 | |
| hip078650 | U | 4370 | 1.75 | 1.65 | -0.02 | -0.05 | 0.30 | 0.03 | 0.91 | -0.26 | -0.21 | -0.16 | 0.11 | -0.01 | -0.09 | -0.20 | -0.08 | -0.08 | | -0.27 | -0.02 | -0.33 | -0.34 | -0.48 | -0.18 | -0.15 | 0.01 | -0.17 | |
| hip080343 | U | 4758 | 2.11 | 1.72 | 0.05 | -0.20 | 0.05 | -0.24 | 0.31 | -0.37 | -0.43 | -0.33 | -0.22 | -0.20 | -0.26 | -0.37 | -0.30 | -0.30 | -0.31 | -0.36 | 0.09 | -0.18 | -0.12 | -0.52 | -0.24 | -0.23 | -0.24 | -0.37 | |
| hip082396 | U | 4522 | 1.95 | 1.44 | 0.30 | 0.27 | 0.40 | 0.21 | 0.61 | 0.09 | -0.10 | 0.06 | 0.33 | 0.12 | 0.15 | 0.03 | 0.03 | 0.10 | 0.31 | 0.20 | 0.03 | -0.17 | -0.28 | -0.17 | -0.11 | 0.01 | 0.06 | -0.01 | -0.03 |
| hip093498 | U | 4482 | 2.10 | 1.50 | 0.56 | 0.47 | 0.68 | 0.49 | 1.35 | 0.30 | 0.19 | 0.32 | 0.72 | 0.39 | 0.37 | 0.29 | 0.35 | 0.37 | | 0.21 | 0.27 | -0.02 | -0.03 | -0.01 | 0.18 | 0.29 | 0.35 | 0.25 | |
| hip103738 | U | 5009 | 2.38 | 1.67 | 0.23 | -0.03 | 0.23 | 0.02 | 0.11 | -0.04 | -0.21 | -0.14 | -0.13 | -0.01 | -0.25 | -0.12 | -0.18 | -0.15 | -0.51 | -0.45 | 0.10 | -0.14 | -0.12 | 0.01 | 0.16 | 0.21 | 0.00 | -0.09 | -0.04 |
| hip106039 | U | 5017 | 2.55 | 1.40 | 0.18 | 0.04 | | 0.02 | 0.25 | -0.01 | -0.19 | -0.14 | -0.09 | -0.04 | -0.11 | -0.09 | -0.16 | -0.13 | -0.29 | -0.31 | 0.12 | -0.14 | -0.11 | 0.03 | 0.04 | 0.10 | 0.02 | -0.06 | -0.03 |
| hip113246 | U | 4828 | 2.18 | 1.44 | -0.05 | -0.03 | 0.14 | -0.08 | 0.14 | -0.14 | -0.32 | -0.24 | -0.21 | -0.17 | -0.26 | -0.22 | -0.27 | -0.25 | -0.36 | -0.42 | -0.05 | -0.24 | -0.22 | -0.06 | -0.03 | 0.05 | 0.04 | -0.04 | 0.13 |
| hip114119 | U | 4962 | 2.77 | 1.01 | 0.20 | 0.12 | 0.33 | 0.11 | 0.41 | 0.13 | 0.02 | 0.11 | 0.26 | 0.14 | 0.08 | 0.07 | -0.01 | 0.06 | 0.00 | -0.12 | 0.22 | 0.11 | 0.09 | 0.24 | 0.35 | 0.46 | 0.31 | 0.24 | 0.81 |
| hip115102 | U | 4578 | 1.93 | 1.49 | 0.15 | 0.14 | 0.32 | 0.11 | 0.63 | -0.04 | -0.12 | -0.06 | 0.16 | 0.01 | 0.07 | -0.07 | -0.06 | -0.02 | 0.11 | -0.06 | -0.04 | -0.33 | -0.27 | -0.22 | -0.01 | -0.05 | -0.05 | -0.09 | -0.05 |
| hip116853 | U | 4886 | 2.47 | 1.40 | 0.48 | 0.30 | 0.46 | 0.30 | 0.59 | 0.20 | 0.03 | 0.09 | 0.23 | 0.22 | 0.24 | 0.15 | 0.11 | 0.18 | 0.10 | 0.04 | 0.16 | 0.04 | 0.00 | 0.08 | 0.30 | 0.16 | 0.17 | 0.11 | 0.06 |
| hr0296 | U | 4500 | 1.78 | 1.50 | -0.09 | -0.12 | 0.09 | -0.08 | 0.52 | -0.23 | -0.26 | -0.26 | -0.15 | -0.23 | -0.31 | -0.32 | -0.32 | -0.31 | -0.19 | -0.40 | -0.22 | -0.38 | -0.44 | -0.37 | -0.08 | -0.01 | -0.13 | -0.07 | -0.25 |
| hr4321 | U | 4569 | 2.06 | 1.44 | 0.52 | 0.34 | 0.52 | 0.49 | 0.59 | 0.30 | 0.05 | 0.19 | 0.42 | 0.30 | 0.36 | 0.22 | 0.24 | 0.34 | 0.59 | 0.51 | 0.27 | -0.03 | -0.22 | 0.07 | 0.29 | 0.08 | 0.23 | 0.23 | 0.09 |
| ic2391-0022 | U | 4610 | 1.56 | 1.46 | 0.25 | 0.01 | 0.08 | 0.13 | 0.24 | -0.08 | -0.44 | -0.36 | -0.30 | -0.11 | -0.20 | -0.22 | -0.26 | -0.17 | -0.27 | -0.21 | -0.12 | -0.40 | -0.53 | -0.29 | -0.37 | -0.25 | -0.36 | -0.41 | -0.31 |
| ic2391-0026 | U | 4585 | 1.51 | 1.44 | -0.23 | -0.26 | -0.14 | -0.14 | 0.12 | -0.31 | -0.64 | -0.54 | -0.53 | -0.41 | -0.46 | -0.46 | -0.51 | -0.43 | -0.48 | -0.46 | -0.37 | -0.68 | -0.66 | -0.56 | -0.51 | -0.45 | -0.52 | -0.61 | -0.52 |
| ic2391-0044 | U | 6699 | 4.48 | 5.52 | -1.81 | | | 0.34 | | 0.39 | 0.76 | 0.69 | 0.68 | 1.32 | | 1.11 | 2.20 | 0.65 | | | 1.83 | 1.04 | 3.45 | | | 1.36 | | | |
| ic4651-E12 | U | 4829 | 2.32 | 1.43 | 0.36 | 0.13 | 0.25 | 0.22 | 0.57 | 0.14 | -0.09 | -0.02 | 0.07 | 0.08 | 0.11 | 0.02 | -0.02 | 0.05 | 0.06 | 0.09 | 0.19 | -0.04 | -0.24 | 0.06 | 0.05 | -0.06 | 0.02 | 0.00 | 0.08 |
| ic4651-E60 | U | 4760 | 2.44 | 1.30 | 0.31 | 0.14 | 0.32 | 0.26 | 0.58 | 0.19 | -0.06 | 0.07 | 0.20 | 0.16 | 0.19 | 0.09 | 0.05 | 0.13 | 0.18 | 0.19 | 0.21 | 0.04 | -0.14 | 0.12 | 0.13 | 0.04 | 0.13 | 0.02 | 0.14 |
| ic4651no7646 | U | 4829 | 2.31 | 1.53 | 0.37 | 0.14 | 0.28 | 0.27 | 0.79 | 0.13 | -0.06 | 0.01 | 0.08 | 0.09 | 0.08 | 0.04 | 0.01 | 0.08 | 0.06 | 0.30 | 0.21 | 0.11 | -0.13 | 0.05 | 0.18 | -0.07 | 0.07 | 0.31 | 0.20 |
| ic4651no9122 | U | 4694 | 2.40 | 1.42 | 0.36 | 0.22 | 0.37 | 0.31 | 0.77 | 0.24 | 0.09 | 0.19 | 0.35 | 0.24 | 0.23 | 0.16 | 0.17 | 0.22 | 0.39 | 0.16 | 0.32 | 0.09 | 0.01 | 0.19 | 0.41 | 0.23 | 0.39 | 0.69 | 0.26 |
| ngc2447No41 | U | 5058 | 2.45 | 1.56 | 0.25 | 0.03 | 0.10 | 0.12 | 0.40 | 0.05 | -0.11 | -0.01 | -0.06 | 0.04 | -0.11 | -0.03 | -0.08 | -0.04 | -0.14 | 0.11 | 0.26 | 0.01 | -0.01 | 0.28 | 0.10 | 0.16 | 0.16 | 0.13 | 0.08 |
| NGC2682ESIII-35 | U | 5046 | 3.23 | 1.02 | 0.27 | 0.22 | 0.32 | 0.23 | 0.56 | 0.26 | 0.15 | 0.29 | 0.40 | 0.26 | 0.33 | 0.20 | 0.20 | 0.27 | 0.45 | 0.22 | 0.26 | 0.30 | 0.04 | 0.22 | 0.38 | 0.23 | 0.41 | 0.77 | 0.35 |
| ngc2682No164 | U | 4734 | 2.41 | 1.39 | 0.32 | 0.09 | 0.32 | 0.25 | 0.69 | 0.18 | -0.02 | 0.12 | 0.23 | 0.17 | 0.15 | 0.10 | 0.07 | 0.17 | 0.25 | 0.17 | 0.17 | -0.04 | -0.11 | 0.07 | 0.16 | 0.05 | 0.20 | 0.43 | 0.11 |
| ngc2682No286 | U | 4778 | 2.45 | 1.47 | 0.33 | 0.20 | 0.34 | 0.26 | 0.77 | 0.17 | 0.06 | 0.13 | 0.22 | 0.19 | 0.21 | 0.13 | 0.14 | 0.20 | 0.26 | 0.23 | 0.19 | 0.07 | -0.06 | 0.18 | 0.30 | 0.19 | 0.31 | 0.51 | 0.20 |
| ngc3114no181 | U | 4456 | 1.38 | 1.82 | 0.27 | 0.06 | 0.19 | 0.30 | 1.02 | 0.00 | -0.25 | -0.12 | -0.07 | 0.07 | -0.12 | -0.03 | -0.14 | -0.01 | 0.12 | 0.21 | 0.22 | 0.00 | -0.32 | 0.21 | 0.16 | 0.02 | 0.07 | 0.20 | 0.05 |
| ngc3680no26 | U | 4662 | 2.35 | 1.29 | 0.19 | 0.09 | 0.24 | 0.21 | 0.83 | 0.13 | -0.08 | 0.06 | 0.17 | 0.12 | 0.04 | 0.03 | -0.03 | 0.06 | 0.15 | 0.20 | 0.21 | 0.00 | -0.23 | 0.15 | 0.20 | 0.13 | 0.33 | 0.38 | 0.03 |
| p1955 | U | 5211 | 0.46 | 3.26 | | | | -0.20 | | | -0.14 | -0.23 | 0.91 | | 0.11 | -0.08 | 0.27 | | | | 2.38 | 1.80 | | | -1.18 | 0.73 | | | | |

| | | | | | | | | | | | | | | | | | | | | |
|---|---|---|---|---|---|---|---|---|---|---|---|---|---|---|---|---|---|---|---|---|
| txpic | U | 4435 | 2.13 | 3.52 | | 0.74 | | -0.20 | 0.12 | -0.73 | -0.46 | | -0.69 | -0.15 | 0.57 | 1.15 | | -1.00 | | 1.81 |
| XSct | U | 4763 | 0.20 | 3.18 | 0.48 -0.86 | -0.03 | 0.25 0.09 | -0.77 | -0.82 | -0.49 | -1.06 -0.64 -0.72 -0.52 | | -1.09 | | -0.79 -0.78 | | -1.04 -0.45 | -0.93 | -0.88 |

| Keyname | Identifier |
|---|---|
| S | Spectra Source |
| T | Effective temperature (Kelvins) |
| G | Logarithm of the surface gravity (cm/s^2) |
| Vt | Mictoturbulent Velocity (km/s) |
| Na | Logarithmic sodium abundance with respect to the solar value. |
| Mg | Logarithmic magnesium abundance with respect to the solar value. |
| Al | Logarithmic aluminum abundance with respect to the solar value. |
| Si | Logarithmic silicon abundance with respect to the solar value. |
| S | Logarithmic sulfur abundance with respect to the solar value. |
| Ca | Logarithmic calcium abundance with respect to the solar value. |
| Sc | Logarithmic scandium abundance with respect to the solar value. |
| Ti | Logarithmic titanium abundance with respect to the solar value. |
| V | Logarithmic vanadium abundance with respect to the solar value. |
| Cr | Logarithmic chromium abundance with respect to the solar value. |
| Mn | Logarithmic manganese abundance with respect to the solar value. |
| Fe | Logarithmic iron abundance with respect to the solar value. |
| Co | Logarithmic cobalt abundance with respect to the solar value. |
| Ni | Logarithmic nickel abundance with respect to the solar value. |
| Cu | Logarithmic copper abundance with respect to the solar value. |
| Zn | Logarithmic zinc abundance with respect to the solar value. |
| Sr | Logarithmic strontium abundance with respect to the solar value. |
| Y | Logarithmic yttrium abundance with respect to the solar value. |
| Zr | Logarithmic zirconium abundance with respect to the solar value. |
| Ba | Logarithmic barium abundance with respect to the solar value. |
| La | Logarithmic lanthanum abundance with respect to the solar value. |
| Ce | Logarithmic cerium abundance with respect to the solar value. |
| Nd | Logarithmic neodymium abundance with respect to the solar value. |
| Sm | Logarithmic samarium abundance with respect to the solar value. |
| Eu | Logarithmic europium abundance with respect to the solar value. |

Table 5
Gradients with respect to [Fe/H] for Ionization Balance Abundances

| Element | Slope | Int | e_Slope | e_Int | Σ | Mean | σ | Slope | Int | e_Slope | e_Int | Σ | Mean | σ | N |
|---|---|---|---|---|---|---|---|---|---|---|---|---|---|---|---|
| C | 0.617 | -0.249 | 0.020 | 0.004 | 0.126 | -0.278 | 0.177 | -0.382 | -0.249 | 0.020 | 0.004 | 0.126 | -0.231 | 0.148 | 964 |
| O | 0.574 | -0.056 | 0.021 | 0.004 | 0.128 | -0.084 | 0.173 | -0.426 | -0.056 | 0.021 | 0.004 | 0.128 | -0.036 | 0.154 | 963 |
| C/O | 0.040 | 0.364 | 0.014 | 0.003 | 0.090 | 0.363 | 0.090 | | | | | | | | 963 |
| Na | 1.082 | 0.280 | 0.019 | 0.004 | 0.121 | 0.230 | 0.249 | 0.083 | 0.280 | 0.019 | 0.004 | 0.121 | 0.276 | 0.122 | 1002 |
| Mg | 0.782 | 0.124 | 0.014 | 0.003 | 0.092 | 0.088 | 0.182 | -0.218 | 0.125 | 0.014 | 0.003 | 0.092 | 0.135 | 0.102 | 997 |
| Al | 0.833 | 0.249 | 0.014 | 0.003 | 0.089 | 0.211 | 0.190 | -0.167 | 0.250 | 0.014 | 0.003 | 0.089 | 0.257 | 0.095 | 1000 |
| Si | 0.817 | 0.234 | 0.015 | 0.003 | 0.092 | 0.197 | 0.189 | -0.183 | 0.235 | 0.015 | 0.003 | 0.093 | 0.243 | 0.100 | 1004 |
| Ca | 0.886 | 0.096 | 0.009 | 0.002 | 0.054 | 0.055 | 0.187 | -0.114 | 0.096 | 0.009 | 0.002 | 0.055 | 0.101 | 0.059 | 1004 |
| Sc | 0.940 | -0.047 | 0.011 | 0.002 | 0.072 | -0.090 | 0.202 | -0.059 | -0.047 | 0.011 | 0.002 | 0.072 | -0.044 | 0.073 | 1004 |
| Ti | 0.926 | 0.037 | 0.011 | 0.002 | 0.068 | -0.006 | 0.199 | -0.073 | 0.037 | 0.011 | 0.002 | 0.069 | 0.040 | 0.070 | 1004 |
| V | 1.142 | 0.114 | 0.018 | 0.004 | 0.116 | 0.061 | 0.257 | 0.143 | 0.114 | 0.018 | 0.004 | 0.116 | 0.107 | 0.119 | 1004 |
| Cr | 1.030 | 0.127 | 0.010 | 0.002 | 0.066 | 0.080 | 0.218 | 0.031 | 0.127 | 0.010 | 0.002 | 0.066 | 0.126 | 0.067 | 1004 |
| Mn | 1.255 | 0.050 | 0.010 | 0.002 | 0.065 | -0.008 | 0.261 | 0.256 | 0.050 | 0.010 | 0.002 | 0.065 | 0.038 | 0.083 | 1004 |
| Co | 0.953 | -0.001 | 0.010 | 0.002 | 0.064 | -0.045 | 0.202 | -0.047 | -0.002 | 0.010 | 0.002 | 0.064 | 0.001 | 0.064 | 1004 |
| Ni | 1.069 | 0.065 | 0.008 | 0.002 | 0.053 | 0.016 | 0.222 | 0.070 | 0.065 | 0.008 | 0.002 | 0.053 | 0.062 | 0.055 | 1004 |
| Cu | 1.086 | 0.154 | 0.035 | 0.007 | 0.220 | 0.103 | 0.310 | 0.087 | 0.154 | 0.035 | 0.007 | 0.220 | 0.150 | 0.221 | 967 |
| Zn | 1.390 | 0.408 | 0.065 | 0.014 | 0.413 | 0.347 | 0.498 | 0.391 | 0.408 | 0.065 | 0.014 | 0.413 | 0.391 | 0.420 | 985 |
| Sr | 0.565 | 0.216 | 0.030 | 0.006 | 0.190 | 0.190 | 0.221 | -0.435 | 0.216 | 0.030 | 0.006 | 0.190 | 0.236 | 0.209 | 992 |
| Y | 1.089 | 0.024 | 0.018 | 0.004 | 0.112 | -0.026 | 0.246 | 0.090 | 0.024 | 0.018 | 0.004 | 0.112 | 0.020 | 0.114 | 1004 |
| Zr | 0.950 | -0.228 | 0.029 | 0.006 | 0.182 | -0.272 | 0.264 | -0.049 | -0.228 | 0.029 | 0.006 | 0.182 | -0.226 | 0.182 | 1004 |
| Ba | 0.919 | 0.000 | 0.031 | 0.006 | 0.193 | -0.041 | 0.268 | -0.080 | 0.000 | 0.031 | 0.006 | 0.193 | 0.004 | 0.194 | 988 |
| La | 1.370 | 0.286 | 0.027 | 0.006 | 0.170 | 0.223 | 0.324 | 0.372 | 0.286 | 0.027 | 0.006 | 0.170 | 0.269 | 0.185 | 1004 |
| Ce | 1.061 | 0.154 | 0.030 | 0.006 | 0.188 | 0.106 | 0.284 | 0.062 | 0.154 | 0.030 | 0.006 | 0.188 | 0.152 | 0.188 | 1004 |
| Nd | 1.030 | 0.201 | 0.020 | 0.004 | 0.129 | 0.154 | 0.244 | 0.031 | 0.202 | 0.020 | 0.004 | 0.129 | 0.200 | 0.129 | 1004 |
| Sm | 1.311 | 0.335 | 0.040 | 0.008 | 0.253 | 0.275 | 0.365 | 0.312 | 0.335 | 0.040 | 0.008 | 0.253 | 0.321 | 0.260 | 1003 |
| Eu | 0.955 | 0.084 | 0.025 | 0.005 | 0.159 | 0.039 | 0.249 | -0.045 | 0.084 | 0.025 | 0.005 | 0.159 | 0.086 | 0.159 | 992 |

Gradients of the form:  [Element/H] = Slope * [Fe/H] + Int
[Element/Fe] = Slope * [Fe/H] + Int

e_Slope = Standard error of the slope
e_Int = Standard error of the intercept
Σ = Standard deviation of the fit
Mean = Mean of the ratios
σ = Standard deviation of the mean
N = Number of stars

# Figures

Figure 1: Top Panel – the mass-determined gravities versus photometric temperature. The gravities with a few exception are as expected for giants. Middle Panel – The difference in the total iron abundance as determined from Fe I and Fe II versus effective temperature. The solid red line is a LOWESS smoothing of the differences for stars with [Fe/H] < -0.3. See §3.4.1 for a discussion. Bottom Panel – The difference in the mass-derived and ionization balance gravities as a function of temperature. The implication is that if one accepts the validity of the ionization balance that there is a disconnect between atmospheric model structure and mass estimates derived from isochrones.

Figure 2: LOWESS smoothed differences in Ti I – Ti II, V I – V II, Cr I – Cr II, and Fe I – Fe II as a function of effective temperature. Outliers have been eliminated in each smoothing. The top panel shows the difference for the mass-derived gravities and the bottom for the ionization balance gravities. The mass-derived gravity differences for Ti, V, and Cr are similar in form to the differences for Fe indicating a common failure is leading to the difference.

Figure 3: The HR diagram for the program stars. The stars to the right of the vertical line at log ($T_{eff}$) = 3.78 and above the diagonal line form the "clump" sample and are indicated by a "1" in last column of Table 3 while the remainder are designated by a "2".

Figure 4: The histogram of iron abundances derived from the "clump" sample — see Section 4.1 for a formal definition. The Gaussian fit to the data yields a centroid abundance [Fe/H] = 0.089 that is about 0.04 larger than the simple mean. The centroid values for Fe I and Fe II for the mass-derived gravity abundances are higher than iron centroid for the ionization derived gravities with Fe II deviating the most.

Figure 5: Average ionization balance abundances with respect to iron as a function of element (Z). The two most deviant points are sulfur and zinc. The latter is based on a small number of unreliable lines. The cause for the sulfur "overabundance" is shown in Figure 6. The error bars are one standard deviation about the mean value.

Figure 6: Sulfur abundances as a function of effective temperature. A clear-cut temperature effect is seen. The effect stems from the high potential sulfur lines being strongly blended with increasing contributions from other species as the temperature decreases.

Figure 7: [Si/Fe], [Mn/Fe], and [Co/Fe] as a function of [Fe/H]. Silicon and manganese show definite trends versus [Fe/H], but [Co/Fe] does not. The solid black line is a linear fit to the data while the red line is a LOESS smoothing. All three elements show a dip at about [Fe/H] = 0. The error bars are the mean standard deviation of each species taken across the sample. See text for discussion.

Figure 8:   [Ni/Fe] as a function of [Fe/H]. The data has been separated as a function of distance from the Sun, but there is no discernible trend. The error bars are the mean standard deviation of each species taken across the sample. See text for discussion.

Figure 9:   [Zr/Fe], [Ba/Fe], and [Nd/Fe] as a function of [Fe/H]. The solid black line is a linear fit to the data while the red line is a LOESS smoothing. There are no believable trends in the data other than increased scatter relative to that seen in the α and Fe-peak elements. The error bars are the mean standard deviation of each species taken across the sample. See text for discussion.

Figure 10:   Dwarf lithium abundances from Lambert & Reddy (2004) and giant lithium abundances from this work are shown versus mass. The abundances for both groups assume LTE. The vertical arrows show the result of standard evolution in this mass range. See text for more discussion.

Figure 11:   A bubble plot of lithium abundances in the clump region of HR diagram. Symbol size reflects the lithium abundance in that larger circles are higher abundances. The Li rich stars found in the hotter regions may represent pre first dredge-up stars. The super-Li stars span the extent of the red-giant clump.

Figure 12:   [C/Fe], [O/Fe], and C/O as a function of [Fe/H]. The trends seen here are well-known and result from the Type Ia supernovae producing relatively more iron as a function of time. The solar C/O ratio is indicated by the red cross in the bottom panel. These stars show the overall effects of standard galactic and stellar chemical evolution. The error bar in [Fe/H] is the mean standard deviation (±0.07) across the sample and the carbon and oxygen abundance uncertain is estimated as ±0.1 from the synthesis and parameter uncertainties.

Figure 13:   Binned [C/Fe], [O/Fe], [Fe/H], and CO ratios as a function of temperature and mass. The stars have been selected to have $-0.1 < [Fe/H] < 0.1$ and $1.0 < \log(L/L_\odot) < 2.5$. The increase in the [C/Fe] ratio towards cooler temperatures and lower mass indicates that the depth of mixing is a function of mass, an effect predicted by stellar evolution models. The error bar is one standard deviation around the mean value shown.

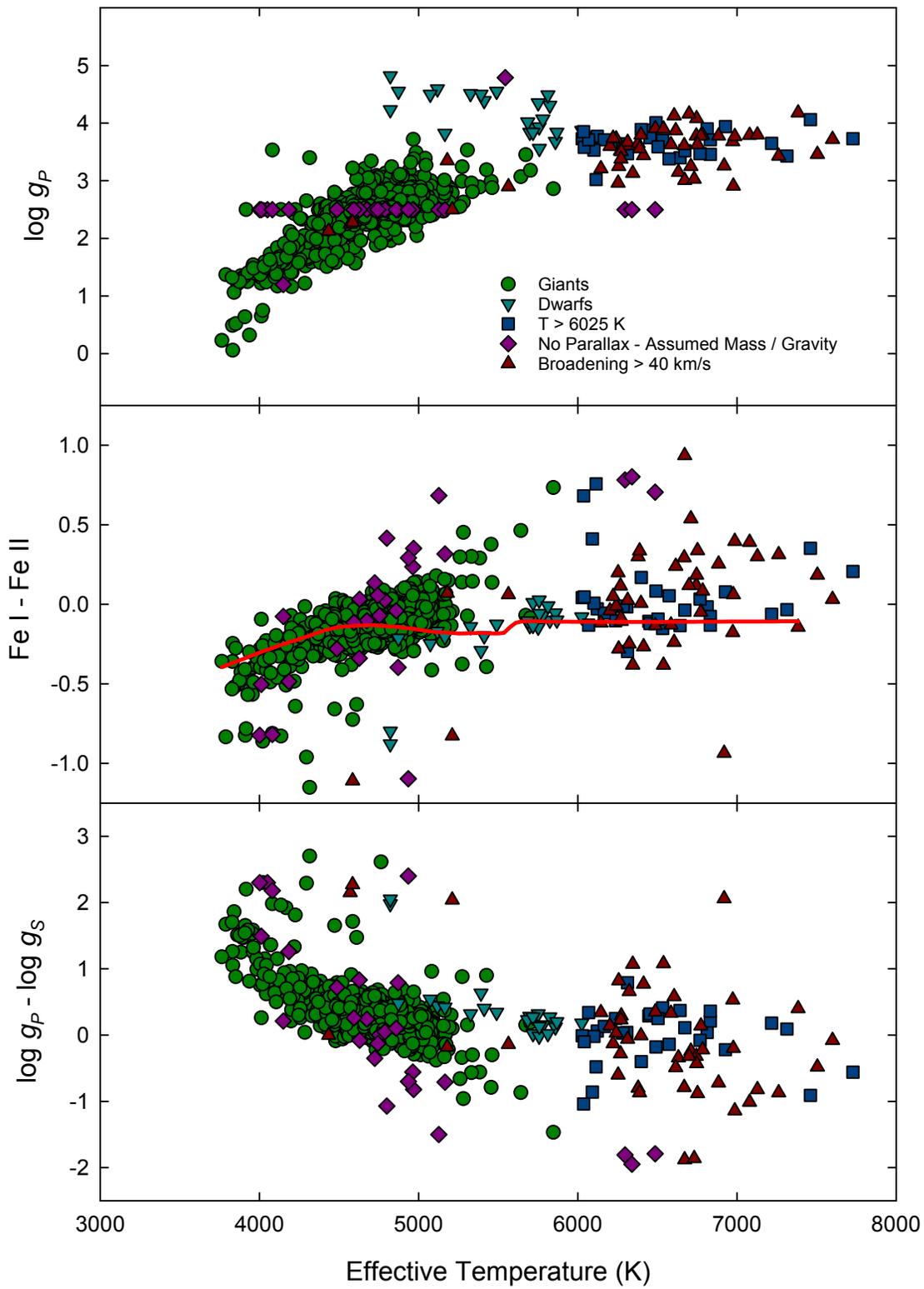

Figure 1

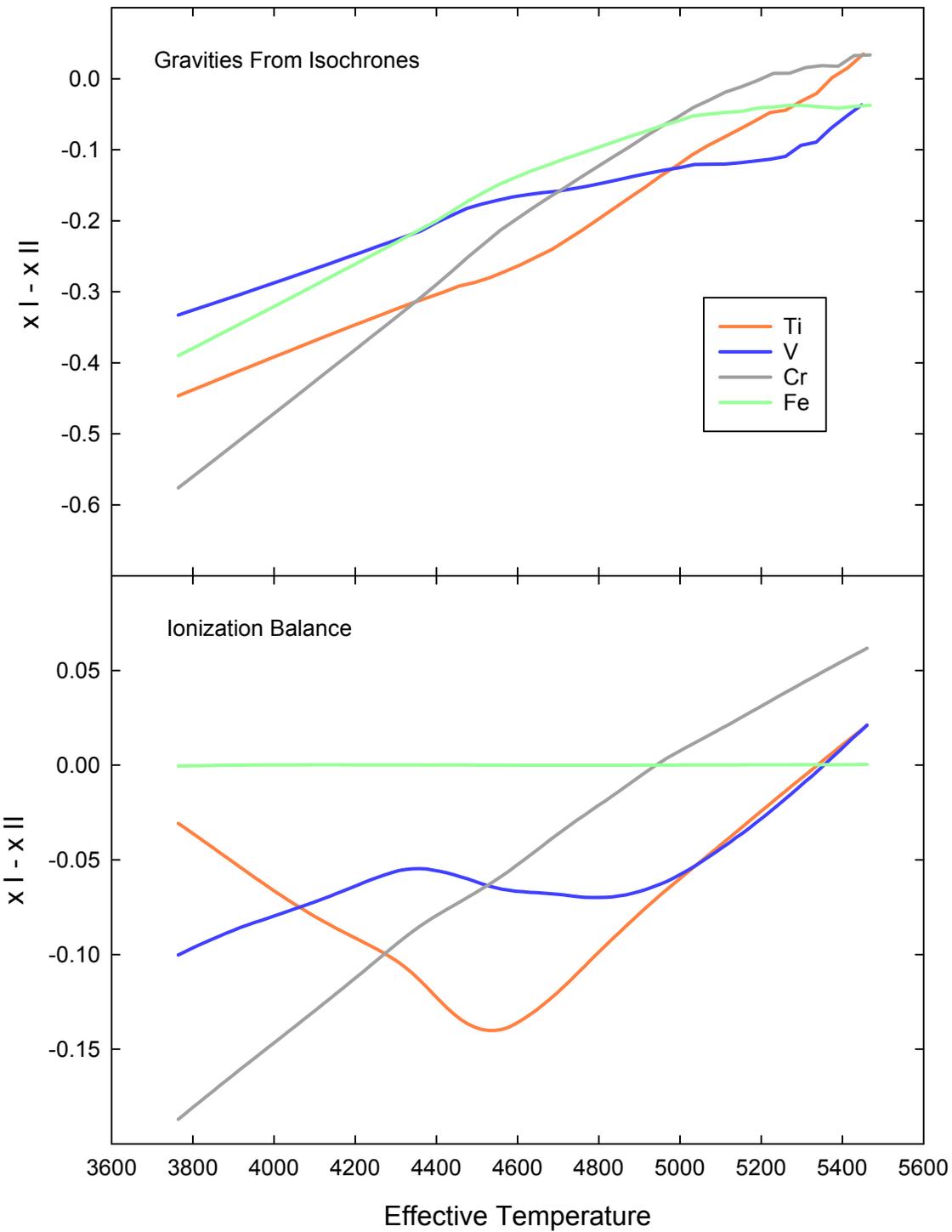

Figure 2

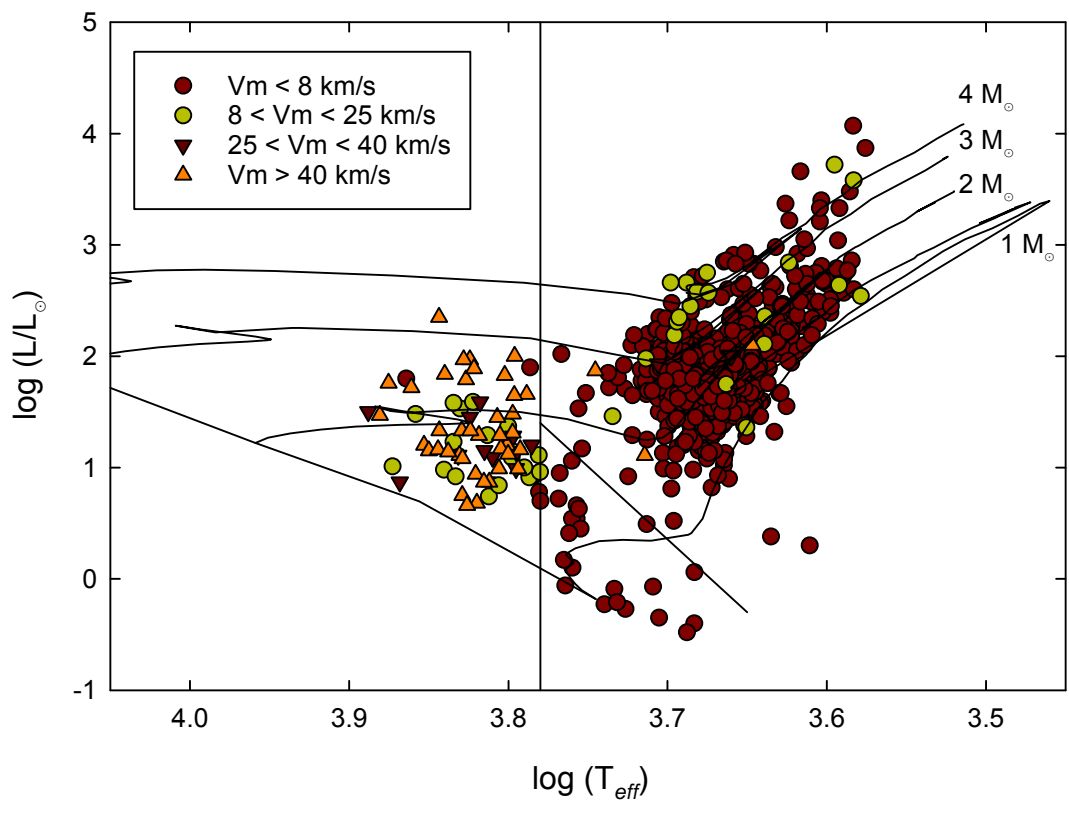

Figure 3

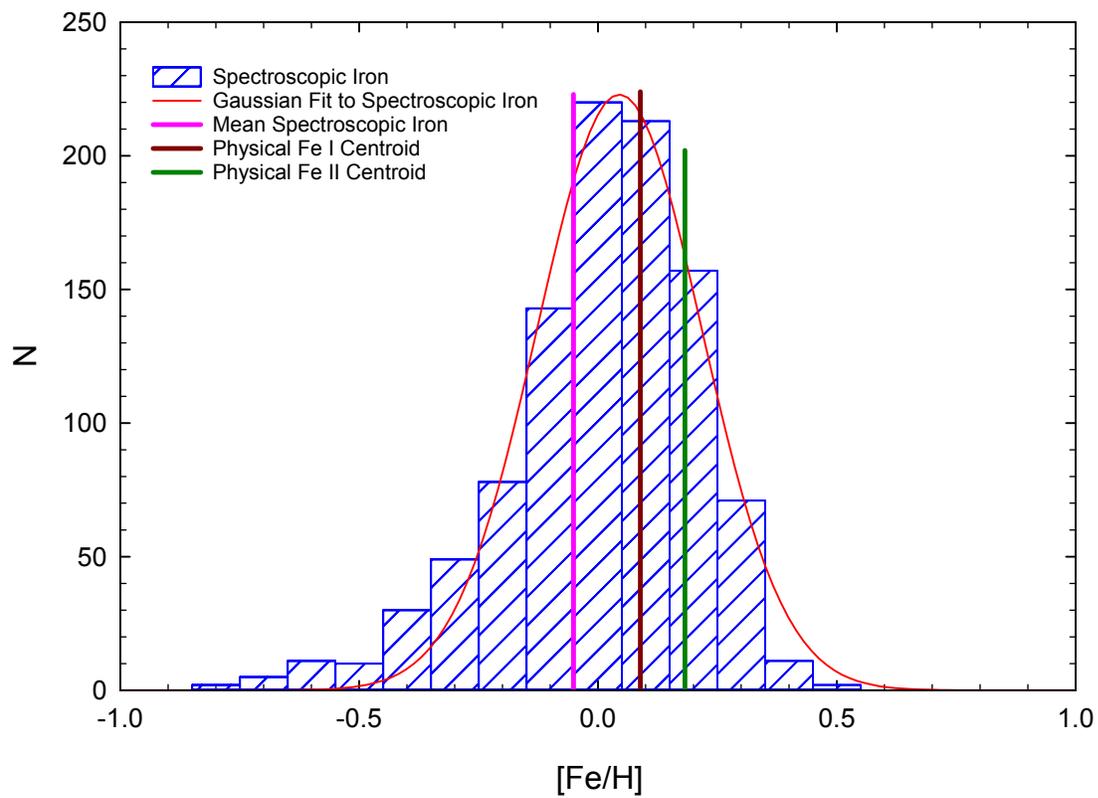

Figure 4

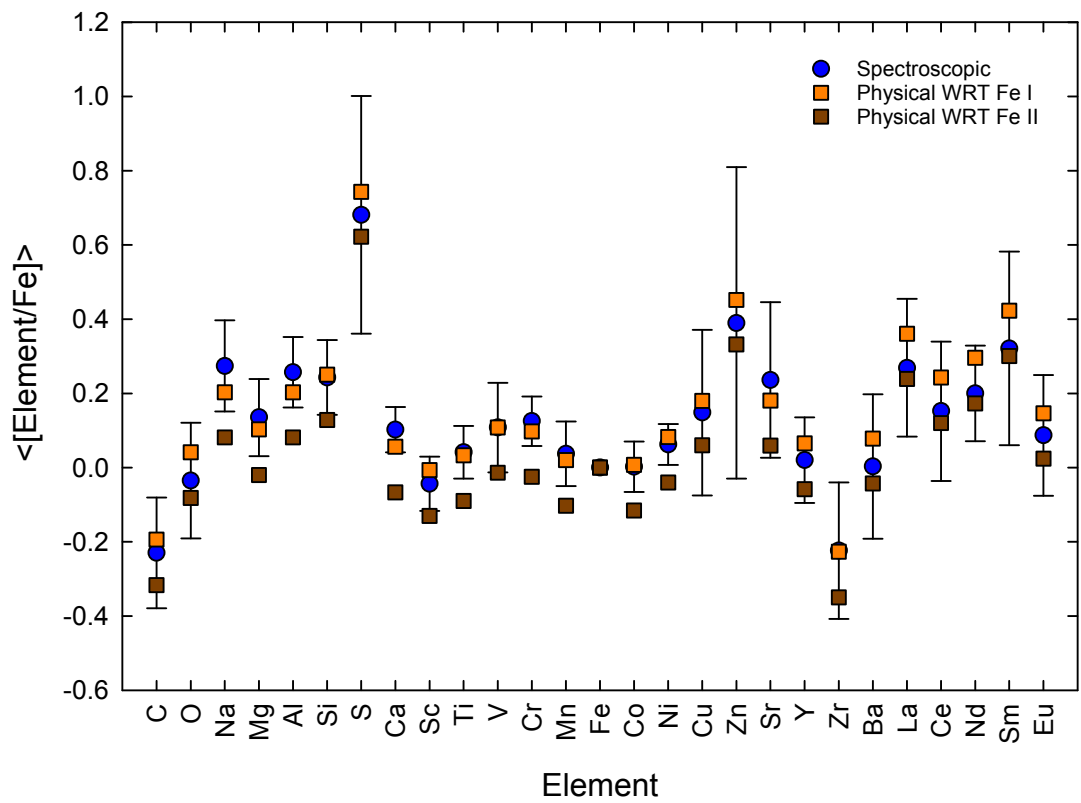

Figure 5

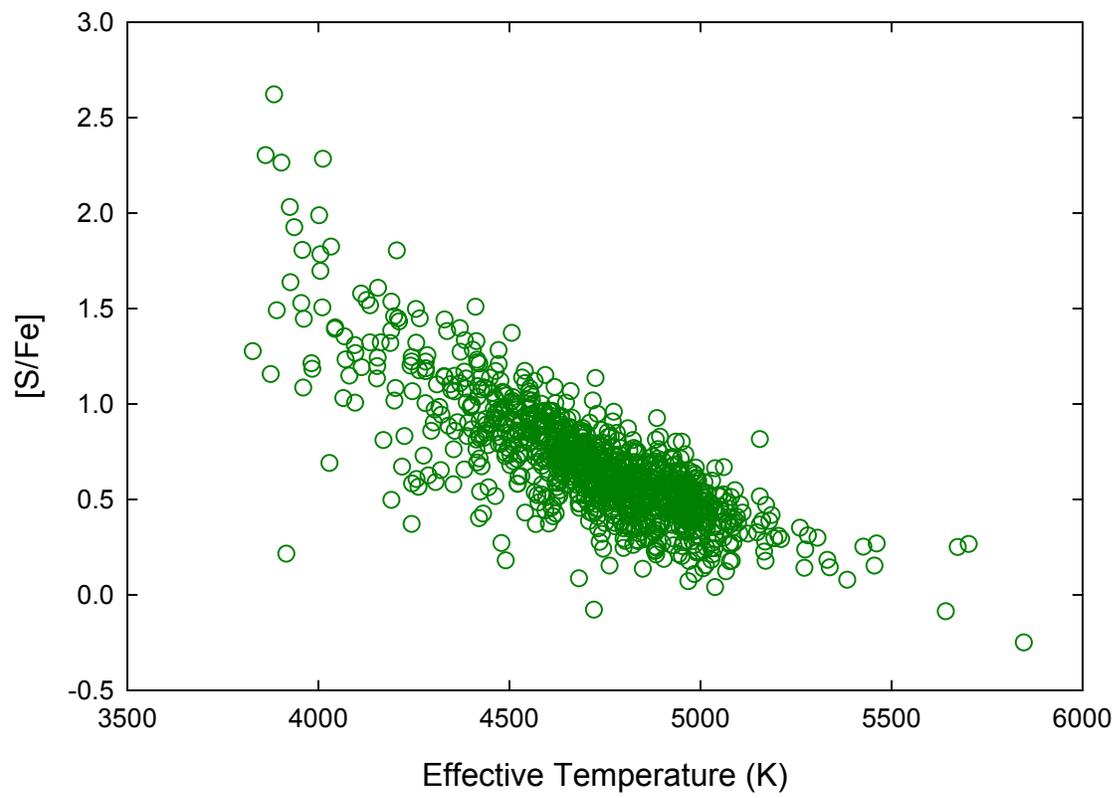

Figure 6

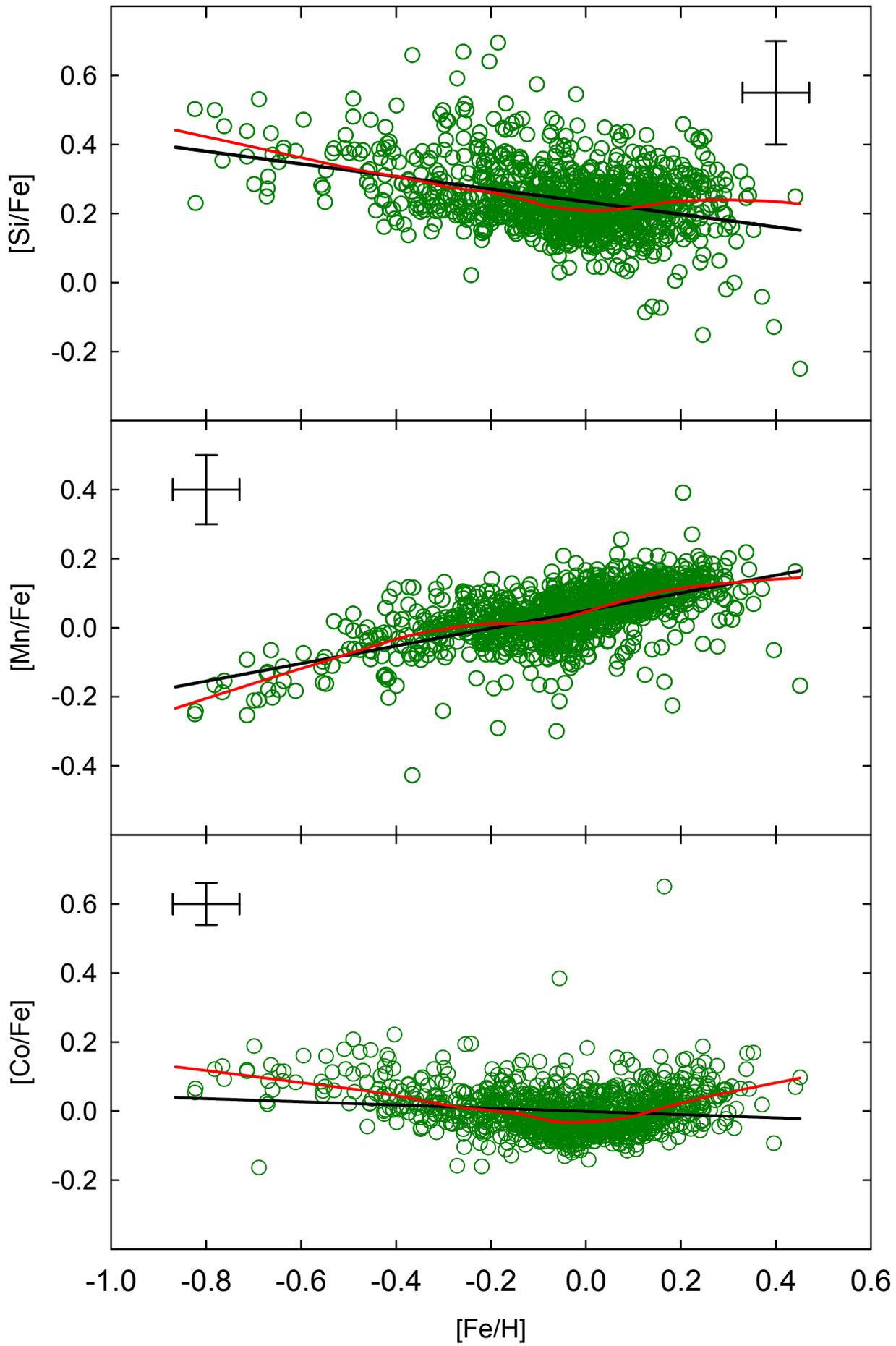

Figure 7

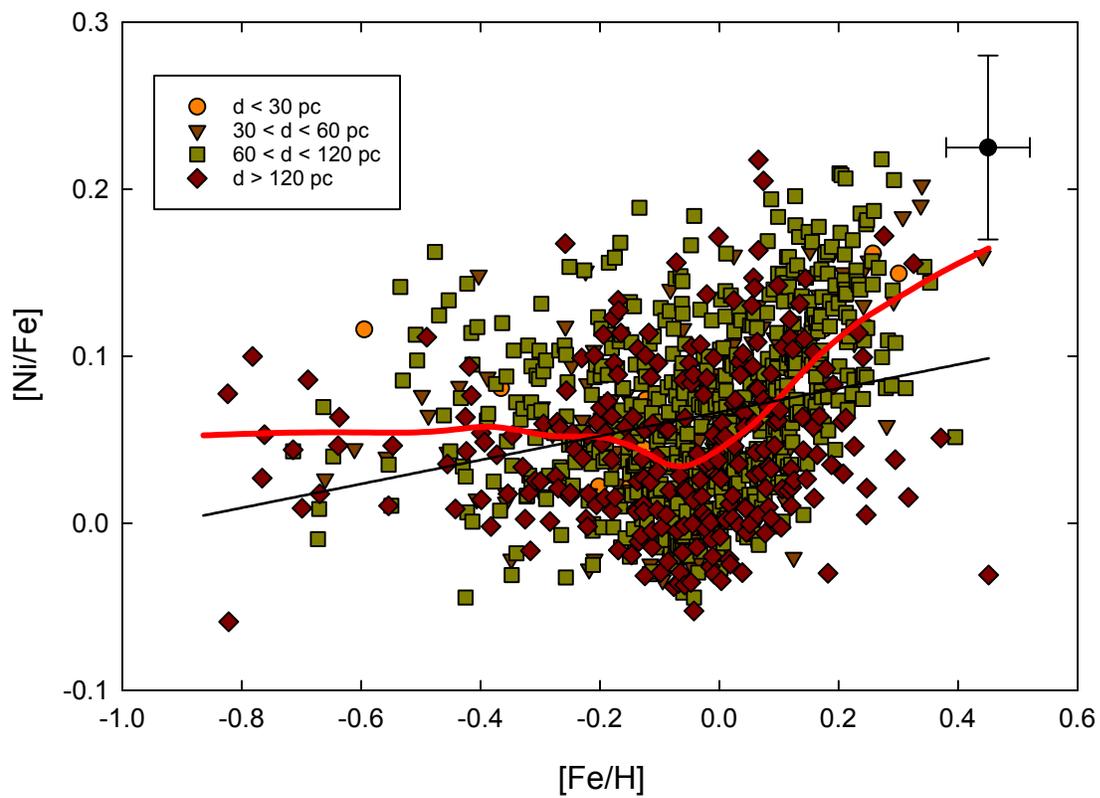

Figure 8

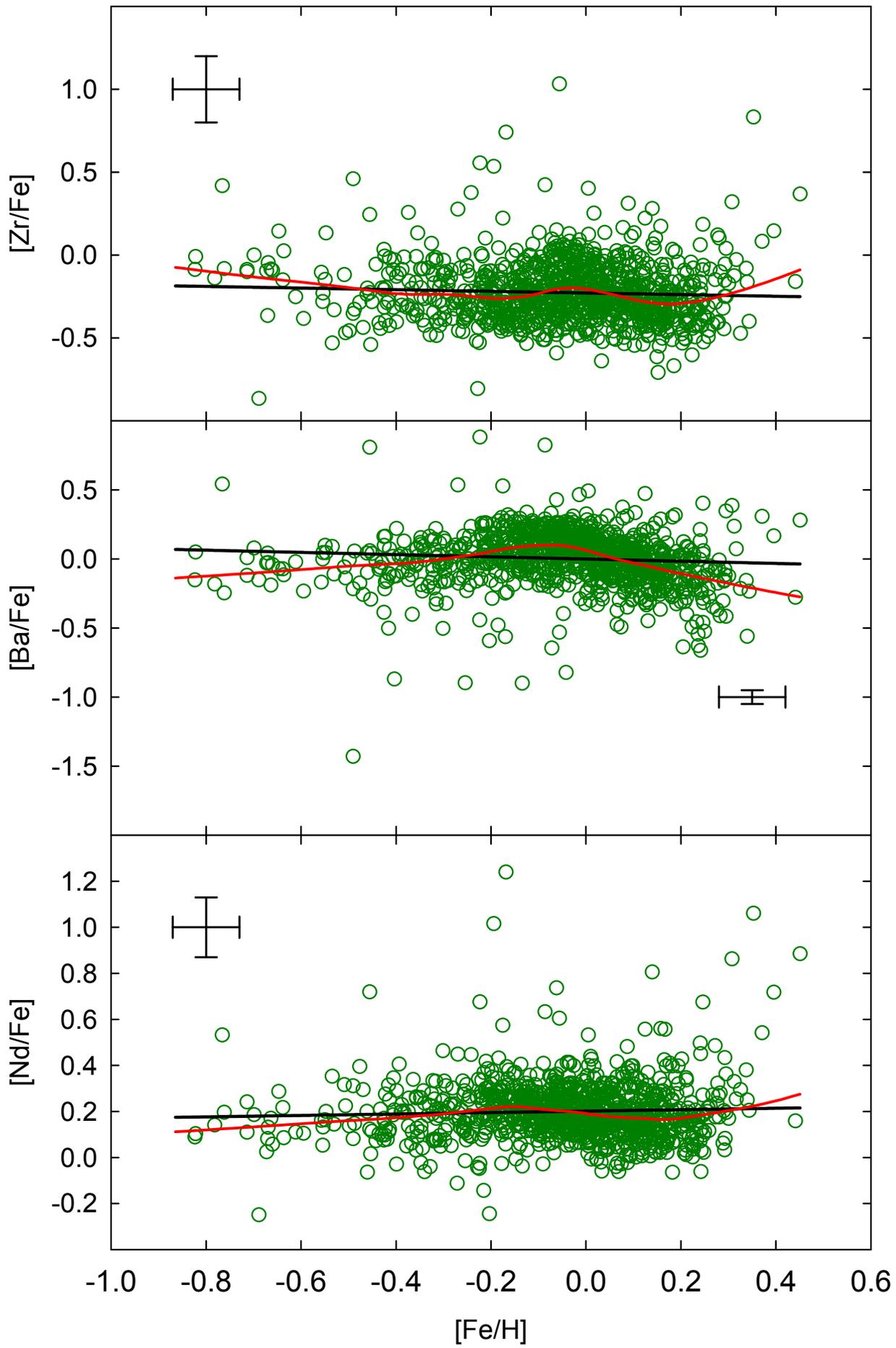

Figure 9

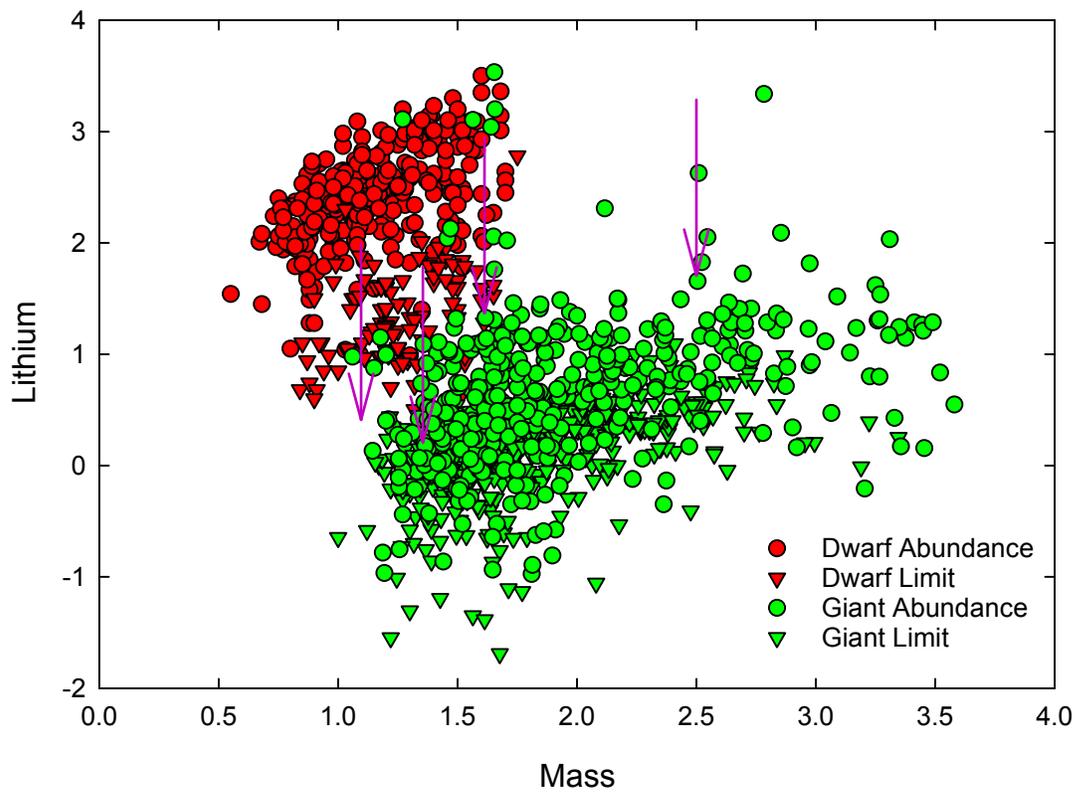

Figure 10

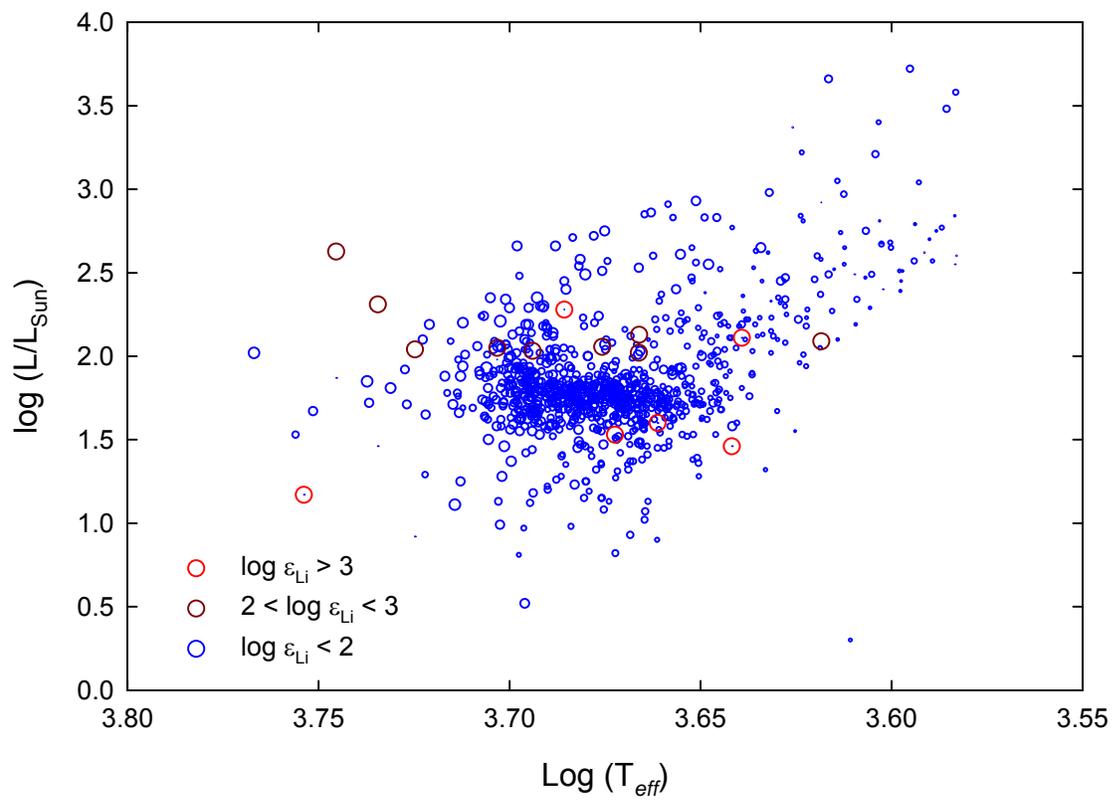

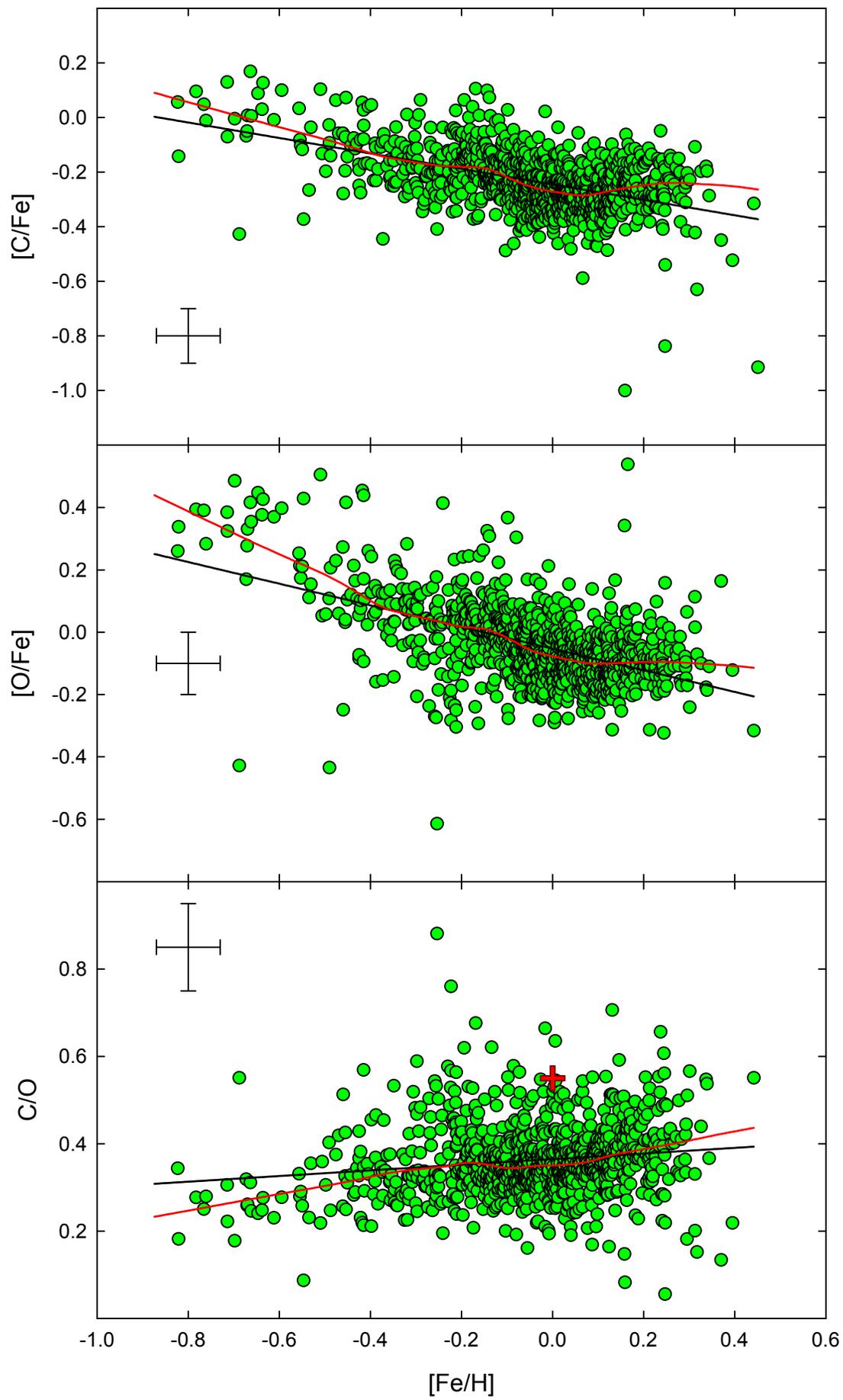

Figure 12

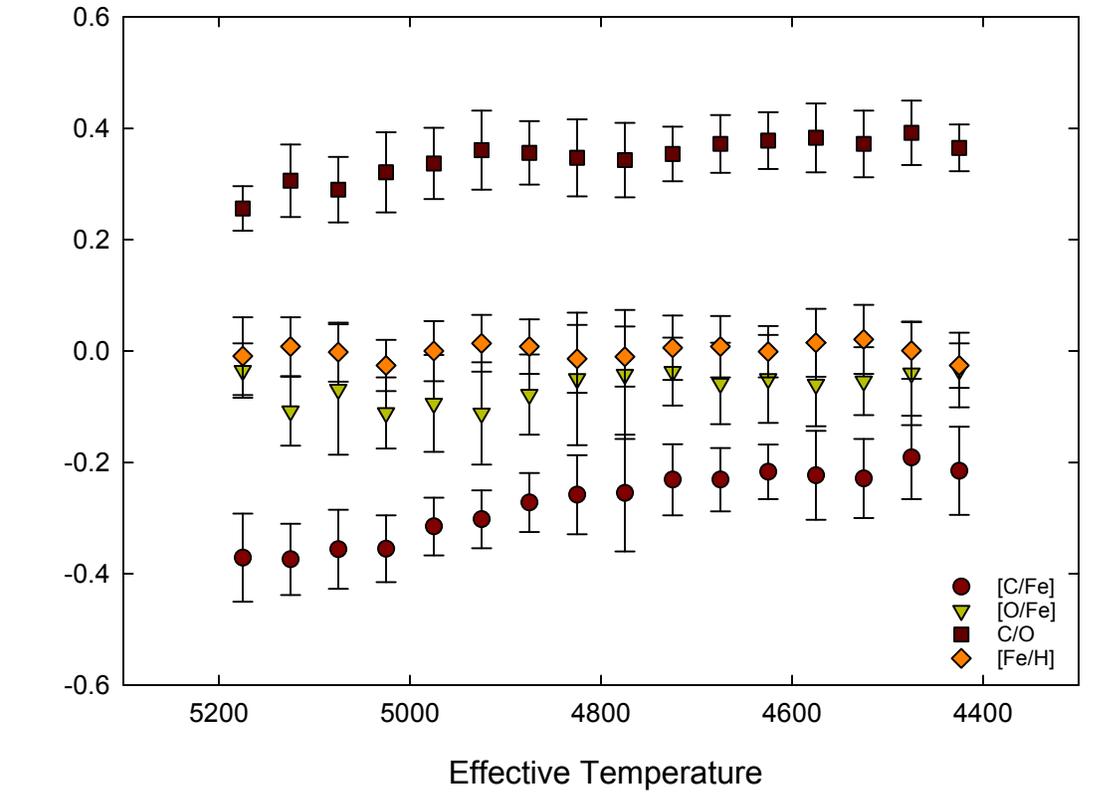
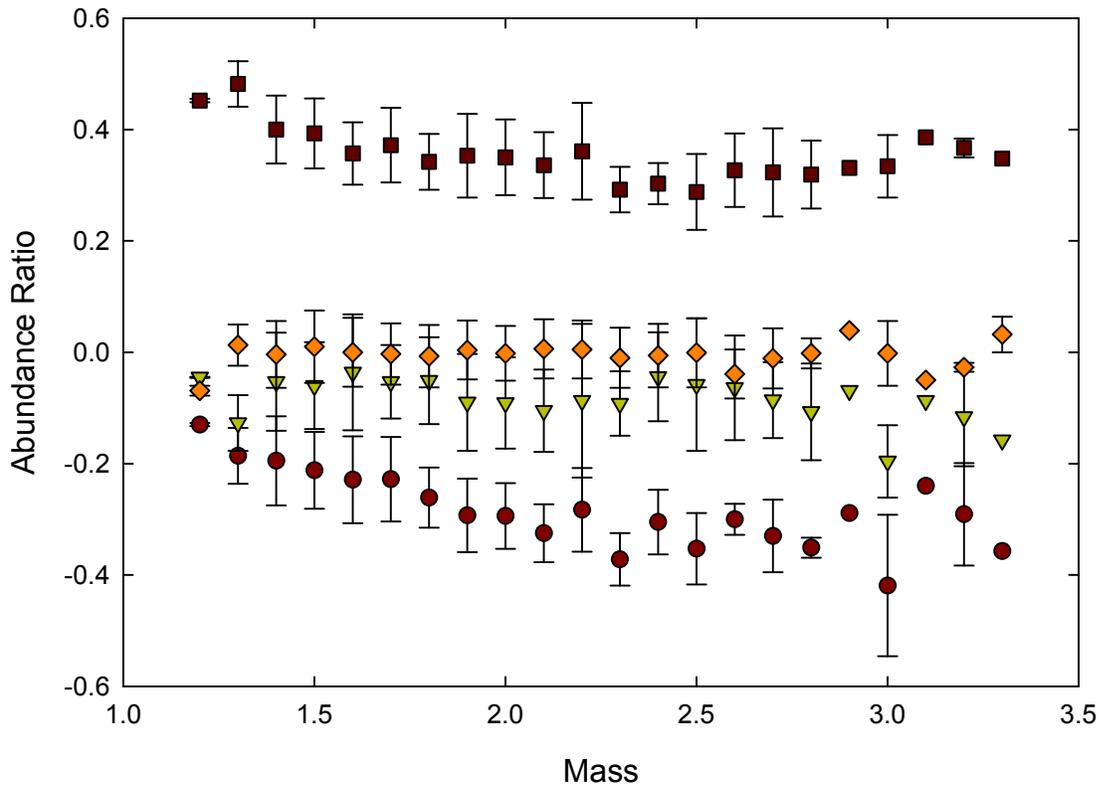

Figure 13